MINISTÉRIO DA CIÊNCIA E TECNOLOGIA

OBSERVATÓRIO NACIONAL

PROGRAMA DE PÓS-GRADUAÇÃO EM ASTRONOMIA

# UM CICLO DE MEDIDAS DO SEMIDIÂMETRO SOLAR COM ASTROLÁBIO

SÉRGIO CALDERARI BOSCARDIN

ORIENTADOR: DR. ALEXANDRE HUMBERTO ANDREI

Tese apresentada como requisito para a obtenção
do grau de Doutor em Astronomia

Rio de Janeiro, 1 de março de 2011

i

Aos meus alunos dos cursos Invest e OPJ, valentes guerreiros que lutam com coragem contra um sistema de ensino extremamente injusto. A eles o sucesso!

**Resumo**

O astrolábio do Observatório Nacional efetuou uma série de medidas do semidiâmetro solar que se estendem de 1998 a 2009 e contém 21640 observações do Sol. Com estes dados é possível determinar como o raio do Sol variou durante este tempo bem como extrair resultados sobre a mudança da elipticidade do Sol.

Consideramos como o raio do Sol tem sido medido ao longo dos últimos quatro séculos e com mais detalhe ao longo do século XX.

Mostramos os desvios no valor observado oriundos de fatores observacionais e de fatores instrumentais e como estes desvios podem ser tratados. Analisamos as relações entre as variações do semidiâmetro solar e indicadores da atividade solar. Calculamos a probabilidade de manchas solares afetarem as medidas.

À série de dados do Observatório Nacional - ON somamos todas as outras séries de dados de astrolábios cujos valores nos foram confiados. Com este conjunto foi possível estabelecer como o semidiâmetro solar se comportou nos três últimos ciclos do Sol indicando uma forte correlação com a atividade de longo termo através do número de manchas solares. A partir do comportamento do Sol e de outras indicações foi possível concluir por um mínimo de atividade solar no futuro próximo.

**Abstract**

The solar astrolabe at Observatorio Nacional has been doing a series of solar semidiameter measurements, extending from 1998 up to 2009, to a total of 21,640 observations of the Sun. Using the data series it was established the solar radius variation along that period, as well as determining results about the change of the solar ellipticity.

To the work in perspective, the measurements of the solar radius in the past four centuries are reviewed, followed by a more detailed account of the contemporaneous measurements.

The deviations of the observed values from the true values caused by observational and instrumental effects are studied and the strategies used to derive the corresponding corrections are shown. The relationships linking the variation of the semidiameter and the relevant indexes of the solar activity are analyzed.

Finally, the time series from Observatório Nacional was enchained to all the other similar astrolabe series, from which the data was graciously confided to this work. Using such much longer combined data set it was established the long term behavior of the solar semidiameter along the last three solar cycles. The outcome is a strong correlation to the long period features of the solar activity described by the annual mean of sunspots count. Putting together the long term semidiameter variation and other solar evidences we point out the near approach of a deep minimum of sunspots.

**Índice**

## Introdução

O Sol é o astro mais importante do céu. Assim, desde os primórdios da humanidade ele foi objeto de adoração e de investigação. Um conhecimento mais efetivo do Sol vem sendo feito desde que Galileu voltou seu telescópio para aquele astro em 1611 para observar uma das primeiras formas conhecidas de atividade solar: as manchas solares. Dois séculos se passaram até que Schwabe, em 1837, formulou o conceito de um ciclo regular de onze anos de atividades identificado pela observação de manchas em sua fotosfera. Mais recentemente foram também observadas variações do raio solar bem como uma modulação em seu período de rotação. Posteriormente, foram também identificados períodos de atividade solar mais longos (Ribes, 1990).

Observações do Sol mais atuais já identificaram variações também na forma do Sol. Ainda não se sabe com exatidão como estas variações ocorrem embora haja sugestões de que as mudanças da forma do Sol também ocorrem em sintonia com o ciclo de atividades. Se a identificação das variações do raio solar tem suas dificuldades, mais difícil ainda é estabelecer como o raio solar varia em relação às suas latitudes.

Várias foram as técnicas utilizadas para observar o Sol. A partir dos anos 70 a técnica mais utilizada tem sido o uso do astrolábio. O ciclo solar tem em média onze anos e a utilização de astrolábios para a observação do raio solar iniciou há 36 anos, perfazendo portanto a visão de três ciclos completos de atividade do Sol.

Com o objetivo de resolver as questões de como o raio de Sol varia, de qual é a forma do Sol e de como esta forma varia temporalmente, o ON também seguiu os pioneiros e em 1997 modificou seu astrolábio para adaptá-lo à observação e estudo do raio solar. O astrolábio é um instrumento que faz cada uma de suas medidas na mesma distância zenital eliminando problemas de refração diferencial. No ON o uso de um prisma de ângulo variável permite que se façam diversas medidas a cada sessão. A utilização de uma câmera CCD evita distorções de medida impostas pelo observador humano. A metodologia e os cálculos necessários para se obter o raio do Sol são efetuados por uma cadeia de algoritmos similares aos usados nos outros astrolábios o que permite a comparação entre seus dados. Motivados pelo trabalho em comum e na intenção de comparar dados de sítios em latitudes completamente diferentes os pesquisadores que utilizam astrolábios formaram o grupo de trabalho R2S3 da Divisão I da IAU.

Desde 1997 milhares de medidas do raio solar foram feitas no Rio de Janeiro. A partir de 1998, 1777 sessões de observação foram realizadas. Um trabalho árduo e de excelência efetuado pela incansável pesquisadora Dra. Jucira Penna, o qual deve ser elogiado. Sua pesquisa gerou diversos trabalhos publicados ao longo destes anos e possibilitou também a nossa análise, motivo pelo qual queremos aqui agradecer.

O presente trabalho põe à disposição de toda comunidade científica toda a quantidade de dados gerados nestes doze anos de pesquisa do raio do Sol. São 21640 dados de medidas independentes acompanhados de seus erros observacionais e da heliolatitude observada. Os valores posteriores a 2003 jamais haviam sido publicados.

Pudemos considerar a análise em conjunto de todas as séries de dados de astrolábios que nos foram graciosamente confiadas, o que permitiu o estudo de três ciclos completos da atividade solar. Este conjunto significa a mais longa série de estudo do raio solar feita pelo mesmo tipo de instrumento e utilizando a mesma técnica. Queremos agradecer por isso às equipes de observadores de São Paulo, de Calern na França, de Antalya na Turquia e de San Fernando na Espanha. A análise gerada a partir destes dados, aliada a outras evidências, permitiu perceber mudanças na elipticidade do Sol, correlações muito fortes da variação do raio solar com seu ciclo de atividades e a identificação de um mínimo de atividades do Sol que se aproxima.

Este trabalho está dividido em doze capítulos. No primeiro traçamos a visão histórica da observação do raio solar que é completada no segundo capítulo por uma síntese das observações mais recentes. No terceiro capítulo apresentamos as características e resultados da série de medidas efetuadas pelo astrolábio do Observatório Nacional. O quarto e o quinto capítulos analisam respectivamente erros instrumentais e de natureza observacional e a metodologia utilizada para corrigir estes desvios. No Capítulo 6 estudamos as possibilidades de manchas solares perturbarem as observações do raio solar. Os três capítulos seguintes analisam o comportamento do raio solar na comparação com alguns índices da atividade solar: o número de manchas, a irradiância solar e o seu campo magnético integrado. O Capítulo 10 analisa o comportamento temporal da granulação solar como um possível modelo para a variação do raio solar, compreendendo-o como a mudança da profundidade óptica advinda da variabilidade física dos grãos da fotosfera solar. O Capítulo 11 estuda a variação da elipticidade do Sol. São revistas algumas pesquisas sobre a forma do Sol. O último capítulo compara as diversas séries de astrolábios e analisa o seu comportamento conjunto, o que permite visualizar o comportamento do raio do Sol ao longo dos três últimos ciclos de atividade.

No âmbito deste trabalho vamos nos referir ao raio do Sol como semidiâmetro. Para isto, cabe uma explicação. Historicamente, os observadores se referiam ao raio do Sol pois imaginavam-no perfeitamente circular. Mais recentemente, percebeu-se que sua figura não era exatamente circular e, portanto, o diâmetro dependia da latitude medida. Passaram a medir-lhe os diâmetros e, por comparação com valores anteriormente determinados, dividiam-no ao meio para comparar com o raio e assim passou-se a atribuir valores ao que hoje designamos por semidiâmetro do Sol.

# 1 – Perspectiva histórica

**1.1 - Valores antigos** - Perde-se na história o momento em que, pela primeira vez, se considerou medir o diâmetro do Sol. Há pelo menos vinte e dois séculos, Aristarco de Samos, astrônomo grego (310–230 a.C.) escreveu uma obra conhecida como: "Sobre os tamanhos e distâncias entre o Sol e a Lua". Nesse tratado Aristarco realizou cálculos geométricos sobre as dimensões e distâncias daqueles dois astros. O diâmetro do Sol calculado por Aristarco foi de 30 minutos de grau.

Poucos tempo depois, Arquimedes (287-212 a.C.) calculou que o diâmetro solar tinha um valor entre 27 e 33 minutos de grau. Mais tarde, Ptolomeu (87-151) achou o valor de 31 minutos e 20 segundos de grau. Ptolomeu acompanhou a medida por um ano buscando verificar variações sem contudo percebê-las.

Muito depois, em 1591, Tycho Brahe (1546-1601) realizou onze medidas, segundo Johannes Kepler (1571-1630), obtendo um valor mínimo do raio solar de 30 minutos e 30 segundos de grau. Jean Picard (1620-1682) foi o primeiro astrônomo a observar sistematicamente o diâmetro solar com medidas quase diárias. Picard fez suas observações a partir de Paris e teve seu trabalho continuado por seu discípulo Philippe de la Hire (1640–1718). Embora os seus resultados (Wittmann,1977 e Toulmonde, 1999) indiquem um raio solar bem maior na época do Mínimo de Maunder, o fato é que não existe comprovação científica publicada disponível para tal efeito. No Capítulo 7 esta hipótese é estudada no contexto de nossas observações, e no Capítulo 12 no contexto da série combinada de observações com o astrolábio. O período conhecido como Mínimo de Maunder se situa entre 1645 e 1715, quando as manchas solares se tornaram extraordinariamente raras. Seu nome foi dado em homenagem ao astrônomo Edward W. Maunder (1851–1928) que estudou mudanças de latitudes das manchas solares (Eddy, 1976).

**1.2 – A síntese de Wittmann –** Com o surgimento do telescópio, a partir do século XVII o número de observadores do diâmetro solar multiplicou-se. Em 1977 Axel Wittmann publicou o artigo "The Diameter of the Sun" (Wittmann, 1977), onde ele cita valores de diâmetro solar obtidos por diversos pesquisadores. Sua intenção era mostrar que técnicas diferentes podem sugerir valores diferentes para o raio do Sol. Ele próprio obteve dois valores diferentes para o raio do Sol quando o observou com técnicas diferentes. Pesquisou diversos observadores começando com Auzout e Picard em 1667 e Mouton em 1670, escolheu quatro observadores no século XVIII, sete no século XIX e oito no início do século XX. E acrescentou as duas medidas feitas por ele em anos diferentes. A Tabela 1.1 traz os dados que Wittmann apresentou em 1977. Em sua primeira coluna há o nome dos observadores e a data de seus trabalhos. A segunda coluna informa o semidiâmetro solar resultante das observações e o erro associado. E a terceira coluna contém algumas notas sobre o trabalho de cada pesquisador (Wittmann,1977).

Tabela 1.1 – Dados do trabalho de Wittmann em 1977 (Wittmann, 1977).

| Referência | R (") | Notas |
|---|---|---|
| Auzout and Picard (1667) | 965.35 ± 1.70 | Primeira medida com micrômetro |
| Mouton (1670) | 960.61 ± 0.97 | A partir dos transitos de Lyon 1659-1661 |
| Bradley (1750) | 959.97 ± 0.20 | Observações de Greenwich 1750 |
| Short (1761) | 959.91 ± 0.11 | Telescope de 60 cm ($f$ = 3.7 m) com micrOmetro acromático |
| Lalande (1771) | 961.35 ± 0.31 | Heliômetro ($f$ = 3 m), irradiação estimatda 1".25 por Lalande |
| Maskelyne (1778) | 959.80 ± 0.05 | Heliômetro, medidas durante o eclipse de 24 June 1778 |
| Airy (1861) | 961.59 ± 0.20 | A partir 5181 trânsitos de Greenwich 1836-1860 |
| Secchi (1872) | 961.51 ± 0.03 | A partir de 187 trânsitos do Collegio Romano 1871-1872 |
| Rosa (1873) | 961.34 ± 0.50 | A partir de 13464 observações (Greenwich 1750-1870, etc.) |
| Fuhg (1875) | 961.50 ± 0.32 | A partir de 6827 trânsitos de Greenwich 1836-1870 |
| Thackeray (1885) | 961.25 ± 0.17 | A partir de 6148 trânsitos de Greenwich 1861-1883 |
| Auwers (1886) | 961.18 ± 0.13 | A partir de 6403 trânsitos de Greenwich 1851-1883, R = 961.29 ± 0.12 incl. outros transit. |
| Auwers (1894) | 959.63 ± 0.05 | A partir de 2849 medidas com heliômetro feitas por 29 observadores 1873 - 1886 |
| Ambronn (1905) | 959.97 ± 0.02 | A partir de 1968 medidas com heliômetro 1890 - 1902* |
| Armellini (1926) | 961.18 ± 0.06 | A partir de trânsitos de Campidoglio 1876-1900 |
| Armellini (1928) | 961.48 ± 0.11 | A partir de trânsitos de Campidoglio 1901-1911 |
| Armellini (1939) | 961.49 ± 0.30 | A partir de trânsitos de Campidoglio 1924-1937 |
| Armelini Conti (1941) | 961.65 ± 0.15 | A partir 495 de trânsitos de Monte Mario 1938-1940 |
| Gialanella (1942) | 961.38 ± 0.08 | A partir de 31200 trânsitos de Campidoglio 1874-1937 |
| Gething (1955) | 961.01 ± 0.24 | A partir de 5180 trânsitos de Greenwich 1918-1936 |
| Giannuzzi (1955) | 961.00 ± 0.20 | A partir de 10155 trânsitos de Greenwich 1851-1937 |
| Wittman (1977) | 960.00 ± 0.05 | A partir de 246 derivações fotoelétricas, Locarno 1972-1975 |
| Wittman (1977) | 960.96 ± 0.18 | A partir de 79 trânsitos de Locarno 1974-1975 |

* Com pequenas revisões (Meyermann, 1939)

Na primeira coluna o nome dos observadores e o ano em que publicaram suas observações. Na segunda coluna o semidiâmetro solar obtido e o erro associado. Na terceira colunas algumas notas sobre a observação.

A Figura 1.1 apresenta um gráfico com os dados da tabela de Wittmann. Neste gráfico apenas não aparece a medida calculada por Ausout e Picard em 1667 cujo valor R=(965,35±1,70) segundos de grau está bem acima dos demais. O gráfico torna evidente a diferença de valores encontrados de acordo com o método utilizado. O próprio Wittmann obteve dois valores diferentes ao utilizar dois métodos: R=(960,00±0,05) segundos de grau a partir de 246 medidas fotoelétricas e R=(960,96±0,18) segundos de grau a partir de observações visuais de trânsito feitas em Locarno. Por outro lado é igualmente evidente que na região definida por cada método os valores são concordantes dentro das barras de erro (ver Figura 1.5).

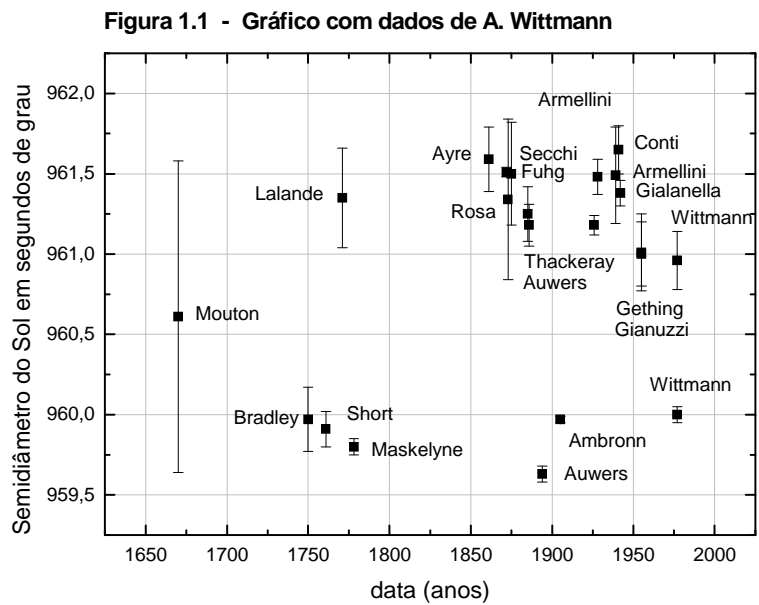

Figura 1.1 - Semidiâmetro do Sol nas observações relatadas por Wittmann (Wittmann, 1977).

**1.3 - A síntese de Toulmonde -** Em 1997 M. Toulmonde publicou um artigo intitulado "The diameter of the Sun over the past three centuries" (Toulmonde, 1997), onde cita diversas medidas feitas por vários observadores. O objetivo de Toulmonde era mostrar que, embora diferentes, estas medidas se revelavam muito semelhantes se correções por ele desenvolvidas, fossem aplicadas. Entre os observadores escolhidos por Toulmonde está o nome de Gabriel Mouton (1618–1694) que, observando de Lion, obteve as primeiras medidas do diâmetro solar considerando a excentricidade da órbita terrestre. Aparece também o nome Jean Picard e mais outros cinco observadores do século XVII. Há seis observadores no século XVIII, oito no século XIX e nove observadores no século XX (Toulmonde,1997).

A Tabela 1.2 mostra os dados que Toulmonde apresentou no seu trabalho. Na segunda e na terceira colunas aparecem o nome do observador e a data de suas observações, depois o método utilizado. A seguir o número de observações feitas, o raio solar calculado e o erro associado a cada observação individual. Toulmonde sugere diversas correções devidas à refração e à turbulência atmosférica, à uma equação pessoal e à difração da objetiva. Para ele esta última correção é muito importante para instrumentos de pequena objetiva que eram empregados nos séculos XVII e XVIII. Nos instrumentos usados a partir do século XIX, com diâmetros maiores, essa correção não lhe pareceu importante.

A fim de poder comparar os valores Toulmonde corrigiu os originais. Retirou 0,3 segundos de grau para corrigir a refração de todos os dados até o final do século XVIII e corrigiu todos os valores do efeito de difração segundo a equação 1.1, onde [Rc] é o raio corrigido, [R1] é o raio observado, [Ref] é a correção de refração e [β] a correção de difração. Esta última é da ordem de [1,22·λ/D], que corresponde a [16"/D] no filtro vermelho ou [14"/D] no filtro verde, sendo [D] o diâmetro da objetiva em centímetros. Quando [D] era desconhecido Toulmonde usou a equação 1.2 que ele considerou válida até 1870.

$$Rc = R1 + Ref - \beta \qquad (1.1)$$
$$D \approx 2,5 \cdot \exp(\text{ano}-1680)/82 \qquad (1.2)$$

Os valores de [β] são apresentados na nona coluna da tabela e na última aparece o raio corrigido. Toulmonde usou os seguintes códigos para os métodos utilizados: [PP] para tempo de trânsito projetado; [MI] para micrômetro; [DP] para tempo de trânsito observado; [HE] para heliômetro; [ME] para tempo trânsito pelo círculo meridiano e [AS] para astrolábio solar.

A Figura 1.2 contém os valores da tabela de Toulmonde anteriores à correção feita pelo autor e, por isso, mostra uma tendência de aumento dos valores de semidiâmetro solar quando se vai ao passado, particularmente nos séculos XVII e XVIII (Toulmonde, 1997). Este crescimento, contudo não aparece no levantamento de Wittmann o que ressalta o objetivo bastante diferente de cada autor.

Tabela 1.2 – Dados do trabalho de Toulmonde em 1997 (Toulmonde, 1997).

| nº | autor | data | método | N | RI(") | δR(") | D (cm) | β(") | Rc (") |
|---|---|---|---|---|---|---|---|---|---|
| 1 | Mouton | 1660 | PP | 86 | 959,4 | 3,3 | 1-2 | +8 | 951,7 |
| 2 | Auzout | 1666 | MI | 2 | 965,2 | 2 | | +5 | 960,5 |
| 3 | Picard | 1670 | MI | 304 | 964,6 | 2 | 2-3 | +5 | 959,9 |
| 4 | Richer | 1672 | DP | 26 | 961,9 | 5,2 | 2-3 | +5 | 957,2 |
| 5 | Picard | 1674 | DP | 154 | 962,9 | 3,5 | 2-3 | +5 | 958,2 |
| 6 | La Hire | 1683 | MI | 14 | 963,2 | 2,8 | 2-3 | +5 | 958,5 |
| 7 | La Hire | 1684 | DP | 304 | 965,4 | 3,8 | 2-3 | +5 | 960,7 |
| 8 | La Hire | 1701 | DP | 6980 | 963,6 | 3,8 | 2-3 | +5 | 958,9 |
| 9 | Louville | 1724 | DP | 10 | 962,4 | 2 | | 3,3 | 959,4 |
| 10 | Bouguer | 1753 | HE | 14 | 957,3 | 2 | 4/jan | +1 | 956,6 |
| 11 | Mayer | 1759 | DP | 105 | 960,4 | 1,2 | 5 | 3,2 | 957,5 |
| 12 | Lalande | 1760 | HE | 12 | 961,1 | 1,5 | 4/jan | +1 | 950,4 |
| 13 | Lalande | 1764 | HE | 12 | 961,4 | 1,5 | 4/jan | +1 | 960,7 |
| 14 | Bessel | 1824 | ME | 1698 | 960,9 | 0,5 | | 1,2 | 959,7 |
| 15 | Airy | 1837 | ME | 92 | 962,25 | 1,4 | | +1 | 961,3 |
| 16 | Goujon | 1842 | ME | 1575 | 962,20 | 0,6 | 15 | 0,9 | 961,3 |
| 17 | Smith-M | 1877 | ME | 1363 | 961,45 | 0,7 | | +1 | 960,5 |
| 18 | Auwers | 1880 | HE | 2840 | 959,63 | 0,5 | | -0,3 | 959,9 |
| 19 | Gething | 1895 | ME | 10302 | 961,04 | 0,44 | | +1 | 960,0 |
| 20 | Schur | 1896 | HE | 760 | 960,07 | 0,55 | 15 | -0,3 | 960,4 |
| 21 | Ambronn | 1897 | HE | 920 | 959,9 | 0,55 | 15 | -0,3 | 960,2 |
| 22 | Cimino | 1907 | ME | 27249 | 961,34 | 0,54 | | +1 | 960,3 |
| 23 | Smith-M | 1946 | ME | 3468 | 961,34 | 0,20 | 15 | 0,9 | 960,4 |
| 24 | Wittmann | 1974 | DP | 246 | 960,00 | 0,8 | 45 | 0,3 | 959,7 |
| 25 | Wittmann | 1978 | PP | 2159 | 960,29 | 1,8 | 45 | 0,3 | 960,0 |
| 26 | Ribes | 1981 | ME | 349 | 961,2 | 0,5 | 19 | 0,7 | 960,5 |
| 27 | Leister | 1984 | AS | 804 | 959,4 | 0,8 | 5 | -0,3 | 959,7 |
| 28 | Journet | 1986 | AS | 1176 | 959,03 | 0,4 | 5 | -0,3 | 959,3 |
| 29 | Laclare | 1987 | AS | 8000 | 959,4 | 0,3 | 5 | -0,3 | 959,7 |
| 30 | Noël | 1991 | AS | 189 | 960,8 | 0,6 | 5 | -0,3 | 961,1 |

Na segunda coluna o nome do observador, na terceira o ano de publicação dos dados, na quarta o método utilizado: [PP] para tempo de trânsito projetado; [MI] para micrômetro; [DP] para tempo de trânsito observado; [HE] para heliômetro; [ME] para tempo trânsito pelo círculo meridiano e [AS] para astrolábio solar. Na quinta coluna o número de observações, na sexta o raio solar definido por estas observações e na sétima coluna o erro associado ao raio. Na oitava coluna o diâmetro da abertura do telescópio. Na nona coluna a correção proposta por Toulmonde para a difração e na última coluna o valor do raio do Sol corrigido por ele.

A Figura 1.3 contém os dados da tabela de Toulmonde após as correções propostas por ele. O gráfico mostra os valores de semidiâmetro solar estáveis dentro de uma faixa entre 958,00 e 961,00 segundos de grau. Com sua correção Toulmonde defende a não variabilidade secular do semidiâmetro solar.

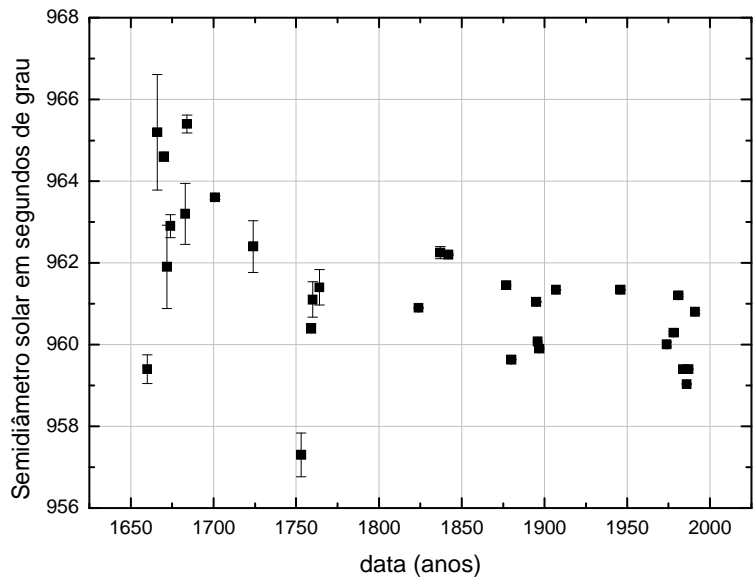

Figura 1.2 – Semidiâmetro do Sol das observações relatadas por Toulmonde (Toulmonde, 1997).

**1.4 – Outras considerações** – Embora os objetivos de Wittmann e Toulmonde fossem bem diferentes, os seus trabalhos contribuíram bastante para proporcionar uma visão ampla de como o diâmetro do Sol foi observado ao longo dos quatro últimos séculos. As duas tabelas apresentadas anteriormente mostram que o semidiâmetro solar foi observado de inúmeras maneiras com a utilização de técnicas diferentes. Sejam quais forem as técnicas utilizadas dois fatores devem ser destacados. Primeiramente o fato de que o Sol é um corpo gasoso e apresenta, por isso, diferentes diâmetros dependendo do comprimento de onda observado. Mesmo considerando apenas a faixa ótica, aparelhos observando em bandas diferentes podem apresentar pequenas diferenças nos raios medidos. Por exemplo, Wittmann em 1973 reportou a observação do Sol em dois comprimentos de onda diferentes. Para 5011,5 Å ele obteve R=(960,24±0,16) segundos de grau e para 6562,8 Å , no Hα, ele obteve R=(966,9±0,4) segundos de grau (Wittmann, 1973).

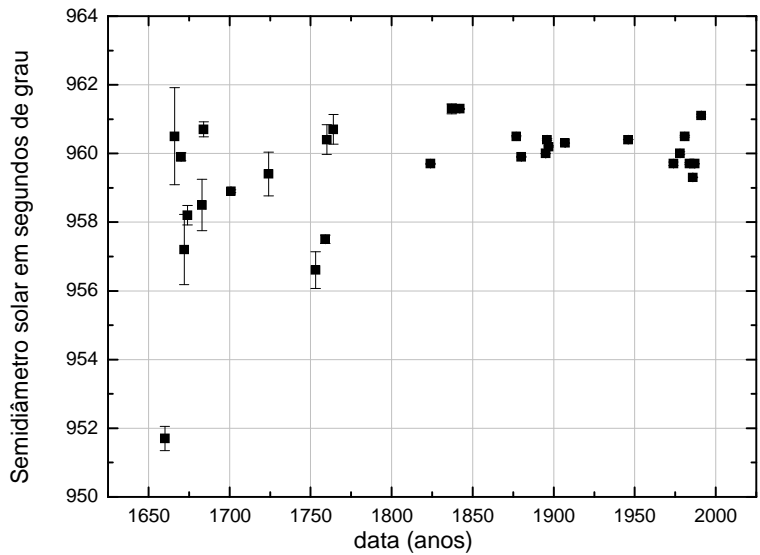

Figura 1.3 – Semidiâmetro do Sol corrigido por Toulmonde (Toulmonde, 1997).

O segundo fato trata da localização do bordo do Sol. Ao se observar o disco solar, com a intenção de medir seu diâmetro o bordo do Sol deve ser determinado. Ocorre que, qualquer que seja o método utilizado, não há uma passagem discreta da intensidade de luz observada no disco solar para aquela observada no céu. Há sim, uma passagem gradual que desenha uma curva começando com a intensidade plena do interior do Sol e decresce até à intensidade luminosa do céu. Esta curva se aproxima exponencialmente dos dois limites e é continuamente variável entre os dois pontos. Esta variação é lenta perto dos dois limites e se torna maior nos pontos intermediários. O que é adotado em todas as técnicas é admitir como bordo do Sol o ponto de inflexão desta curva, ou seja, o ponto onde a segunda derivada da curva de luz é nula. Wittmann fornece esta curva, mostrada na Figura 1.4, que foi obtida pela deriva do Sol em frente ao sensor e onde ele determinou o limbo solar no ponto de inflexão da curva (Wittmann, 1977).

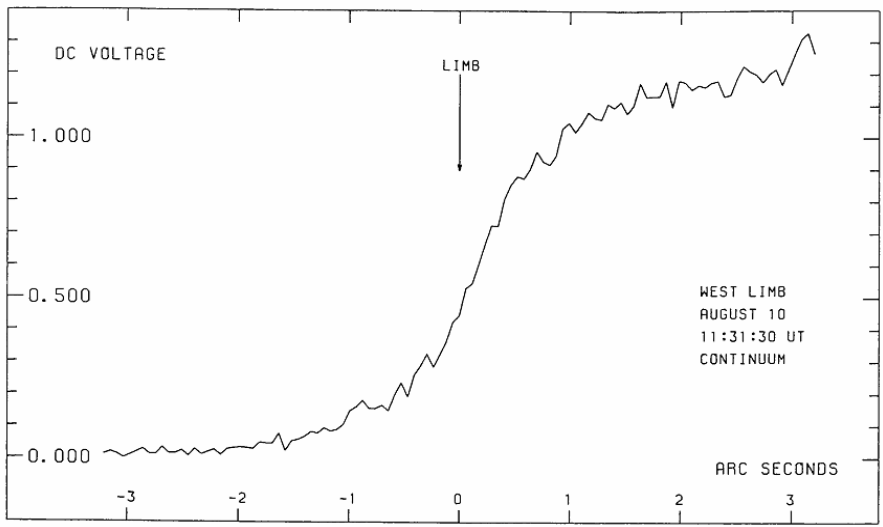

Figura 1.4 – Determinação do limbo do Sol (Wittmann, 1977).

Além destes fatos cada técnica observacional tem diferentes calibrações e sempre ocorrem desvios entre instrumentos semelhantes e mais ainda, quando se consideram técnicas diferentes.

**1.5 – As diferentes técnicas para se observar o semidiâmetro do Sol** - São as seguintes as principais técnicas que foram utilizadas para a medição do semidiâmetro do Sol:

I. Uso de micrômetro para medir o tamanho da imagem do Sol.
II. O trânsito da imagem do Sol pela objetiva do instrumento.
III. Trânsitos de Mercúrio sobre a imagem do Sol
III. Uso do Heliômetro.
IV. Medindo o tempo de totalidade de um eclipse solar.
V. Uso do Astrolábio.

O uso do micrômetro é uma das técnicas mais antigas e foi largamente utilizada durante o século XVIII. O trânsito do Sol pela objetiva do telescópio foi bastante utilizado desde os primeiros telescópios até o século XX. Não devemos esquecer do trabalho de Dicke e Goldemberg de 1967 sobre observações do Sol com o objetivo de explorar um eventual achatamento da figura do Sol a fim de explicar a precessão de mercúrio (Dicke e Goldemberg, 1967). O Heliômetro surgiu no século XVIII como um instrumento dedicado a medir pequenos ângulos ao juntar duas imagens do mesmo objeto com uma pequena diferença angular entre elas.

Os primeiros heliômetros foram usados em 1743, na Inglaterra, por Servington Savary e a seguir por Pierre Bouguer em 1748, na França. Setenta e cinco anos depois, Fraunhofer melhorou consideravelmente a performance do instrumento. A esta época, porém, devido ao seu sucesso, o heliômetro, mesmo com este nome, passou a ser usado para observar outros astros que não o Sol. Foi assim que Bessel em 1838 fez a primeira medida conhecida de paralaxe de uma estrela, a 61 Cygni (Oliveira Filho, K. S., Saraiva, M.F.O).

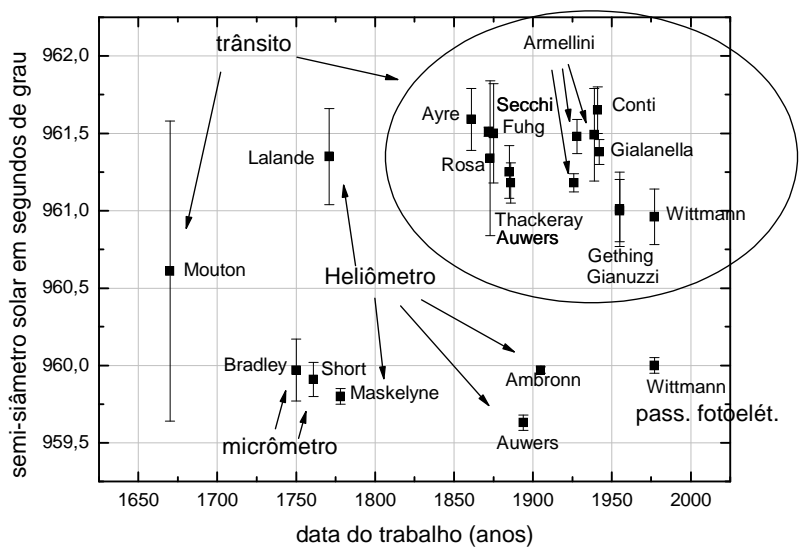

Figura 1.5 – Semidiâmetro do Sol observado por técnicas diferentes (Wittmann, 1997).

Embora fizessem medidas do diâmetro solar, as primeiras observações do Sol com astrolábio visavam a determinação de correções a elementos da órbita terrestre (Penna, 1995). As outras técnicas aqui citadas têm sido utilizadas a partir do meio do século XX e vamos comentá-las no Capítulo 2 que trata das medidas do semidiâmetro solar mais recentes.

É oportuno mostrar a Figura 1.5 que repete o gráfico do trabalho de Wittmann de 1977 destacando agora as diversas técnicas utilizadas.

## 2 – Medidas modernas

**2.1 - Medindo o tempo de totalidade de um eclipse solar -** Em 1870 Newcomb achou muito boa a idéia de medir o diâmetro do Sol a partir do tempo de totalidade de um eclipse solar. Mas achava a sua execução praticamente impossível por causa dos efeitos do limbo rugoso da Lua nos momentos de contato. O conhecimento das correções do limbo data dos anos 50, do século passado (Watts e Adams, 1950). Em 1994 Fiala, Dunham e Sofia apresentaram a medida do diâmetro do Sol a partir do tempo de eclipses solares. Eles examinaram exaustivamente a literatura astronômica a cerca de dez eclipses solares a partir de 1715. A Tabela 2.1 lista os eclipses considerados, mostra a diferença do raio solar calculado para um valor padrão de R=959,63 segundos de grau, e as correções introduzidas pelo limbo da Lua (Fiala, Dunham e Sofia).

Tabela 2.1 - Resultados de observações de eclipses.

| | | | | Correções | |
|---|---|---|---|---|---|
| | Data | Tipo do Eclipse | Num. Obs. | Raio Solar | Ecliptica Lunar |
| | | | | | Long. Lat. |
| | | | | (Todos em segundos de grau) | |
| 1715 | Mai. 3 | Total | 3 | +0,48 ± 0,2 | |
| 1925 | Jan. 24 | Total | 8 | +0,51 ± 0,08 | |
| 1976 | Out. 23 | Total | 43 | +0,04 ± 0,07 | +0,65 ± 0,10   −0,45 ± 0,09 |
| 1979 | Fev. 26 | Total | 47 | −0,11 ± 0,05 | +0,52 ± 0,09   +0,25 ± 0,05 |
| 1980 | Fev. 16 | Total | 232 | −0,03 ± 0,03 | +0,32 ± 0,03   +0,07 ± 0,04 |
| 1981 | Fev. 4 | Anular | 153 | −0,02 ± 0,03 | +0,02 ± 0,05   −0,02 ± 0,04 |
| 1983 | Jun. 11 | Total | 201 | −0,09 ± 0,02 | +0,65 ± 0,03   −0,19 ± 0,02 |
| 1984 | Mai. 30 | Anular | 51 | −0,23 ± 0,04 | +1,01 ± 0,05   −0,47 ± 0,06 |
| 1984 | Mai. 30 | Anular | 51 | −0,09 ± 0,04 | −0,12 ± 0,06   −0,03 ± 0,06 |
| 1987 | Set.23 | Anular | 123 | −0,11 ± 0,03 | +0,57 ± 0,09   +0,02 ± 0,02 |

Na coluna Raio Solar estão os desvios para o raio padrão de 959,63 segundos de grau.
Nas duas últimas as correções para a órbita da Lua.

Mencionamos por completeza que o eclipse solar de 11 de julho de 2010 foi observado em condições de totalidade na ilha de Páscoa pela equipe do Observatório Nacional utilizando o heliômetro solar. Foram feitas 13920 imagens com uma resolução temporal de décimo milésimo de segundo de tempo. Houve sucesso na observação dos instantes do segundo, terceiro e quarto contactos. O primeiro contacto será interpolado das medidas imediatas. Simultaneamente medidas heliométricas foram tomadas nas direções equatorial e polar do Sol e o diâmetro será também calculado para os dados tomados nos dias anteriores.

**2.2 - Astrolábio –** Os primeiros astrolábios solares surgiram a partir do último quartel do século XX. Tal como o heliômetro ele também utiliza duas imagens do mesmo objeto. Neste caso, porém, uma imagem é tomada diretamente e a outra após uma reflexão em um espelho horizontal. Os astrolábios observam a passagem do Sol por um almicantarado, isto é, por um círculo menor paralelo ao horizonte. Como observa a passagem do astro por um círculo de distância zenital constante, este instrumento tem a vantagem de suas medidas não serem afetadas pelos efeitos causados pela refração diferencial. Os primeiros astrolábios observavam apenas um almicantarado e podiam, assim, fazer apenas duas medidas a cada dia: uma antes da passagem meridiana e outra após. O uso de prismas refletores tornou possível observar em outras distâncias zenitais diferentes permitindo mais medidas por dia

Cada medida do astrolábio consiste em registrar os dois instantes no qual as duas imagens do Sol, a direta e a refletida vêm a se tangenciar. Nestes instantes o Sol estará tangenciando o almicantarado para o qual o astrolábio está apontado. O valor do diâmetro solar é obtido pela diferença entre estes dois instantes. Quando os bordos das duas imagens se tocam ao se aproximarem é o instante em que o primeiro bordo do Sol cruza o almicantarado. As imagens então se interpenetram e quando seus bordos opostos se tocam ao se separarem é o instante em que o outro bordo solar cruza o almicantarado. Como se conhece a cada momento o ângulo da derivação do Sol, que vem a ser o ângulo que a trajetória aparente do Sol faz com o almicantarado, pode-se calcular, a partir destes instantes, o valor do diâmetro do Sol. A Figura 2.1 mostra um esquema com o aspecto dos bordos das duas imagens e a trajetória do Sol ao passar por um círculo de altura constante.

Da mesma forma que o heliômetro, o astrolábio inicialmente foi utilizado para se melhorar o conhecimento da órbita da Terra. A acurácia dos resultados obtidos levou os pesquisadores a utilizá-lo para a determinação do diâmetro do Sol. Vários astrolábios têm sido utilizados para a determinação do diâmetro do Sol em vários sítios com latitudes bem diferentes. Foram instalados astrolábios em Santiago do Chile, em São Paulo, em Calern na França, no Rio de Janeiro, em Antalya na Turquia, em San Fernando na Espanha e em Tamanraset na Argélia.

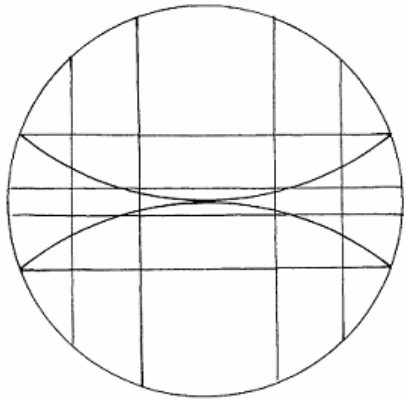 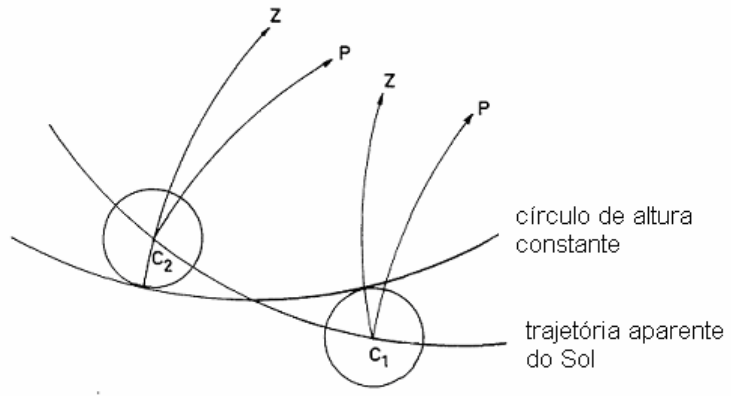

Aspecto dos bordos das imagens no campo do astrolábio. - Uma imagem direta, outra refletida.

Trajetória aparente do Sol quando da passagem pelo círculo de altura constante.

Figura 2.1 – Os bordos das duas imagens se tocando e a trajetória do Sol ao longo de um círculo de altura constante (Laclare, 1983).

Exceto o de Santiago, os demais compartilham métodos de observação detecção e análise. São Paulo em sua última fase desenvolveu uma metodologia independente de detecção e análise. Os grupos de Calern, Rio de Janeiro, Antalya, San Fernando e Tamanraset (este ainda sem série de resultados) formaram o *Reseau de Suivi au Sol du Rayon Solaire*, agregado à Comissão 8 da Divisão I da IAU.

Os dados dos astrolábios de Rio de janeiro, São Paulo, de Calern, de Antalya e de San Fernando nos foram gentilmente cedidos e são analisados no Capítulo 12. Os dados do astrolábio do Rio de Janeiro sediado no Observatório Nacional são analisados isoladamente no Capítulo 3.

Todos estes astrolábios têm sua atividade científica derivada da experiência pioneira de Benevides em Besançon (Penna, 1995) no inicio dos anos 70 e Leister em São Paulo no IAG/USP em 1973 (Leister, 1979). Em 1983 Laclare em Calern desenvolveu a metodologia de derivação do diâmetro solar.

**2.3 - Astrolábio em Santiago –** O grupo de Santiago do Departamento de Astronomia da Universidade do Chile trabalhou sempre isoladamente e seus resultados diferem dos demais em amplitude. Por esta razão seus resultados não são analisados em conjunto com os demais no Capítulo 12, porém vale o registro de seus trabalhos neste item introduzindo as observações solares com astrolábio. Em 1997 F. Noël apresentou medidas efetuadas em Santiago com seu astrolábio solar. Ele mostrou que seus resultados apresentam variações temporais similares aos valores de diâmetro solar obtido pelo magnetógrafo do telescópio solar do Observatório de Mount Wilson. A Tabela 2.2 mostra as médias anuais dos valores medidos em duas distâncias zenitais diferentes para o período entre 1990 e 1995 (Noël,1997).

Tabela 2.2 – Dados do astrolábio de Santiago (Noël,1997).

| ANO | z = 30º | n | z = 60º | n |
|---|---|---|---|---|
| 1990 | 961",071 ± 0",077 | 18 | 961",049 ± 0",100 | 44 |
| 1991 | 960,857 ± 0,080 | 29 | 960,683 ± 0,091 | 23 |
| 1992 | 960,722 ± 0,110 | 33 | 960,557 ± 0,061 | 47 |
| 1993 | 960,424 ± 0,035 | 50 | 960,527 ± 0,039 | 96 |
| 1994 | 960,230 ± 0,045 | 45 | 960,249 ± 0,044 | 79 |
| 1995 | 959,976 ± 0,044 | 46 | 960,146 ± 0,041 | 69 |
| Média | 960",446 ± 0",034 | 221 | 960,470" ± 0",027 | 358 |

Valores médios anuais do semidiâmetro solar tomados em duas distâncias zenitais diferentes,
a 30º e a 60º com os respectivos erros e o número de observações feitas para obtê-los.

Os resultados individuais das medidas de semidiâmetro solar obtidas em Santiago e reduzidas a uma unidade astronômica são mostrados no gráfico da Figura 2.2. São apresentados os resultados separados dos dois grupos de dados, a 30 e a 60 graus de distância zenital (Noël,1997).

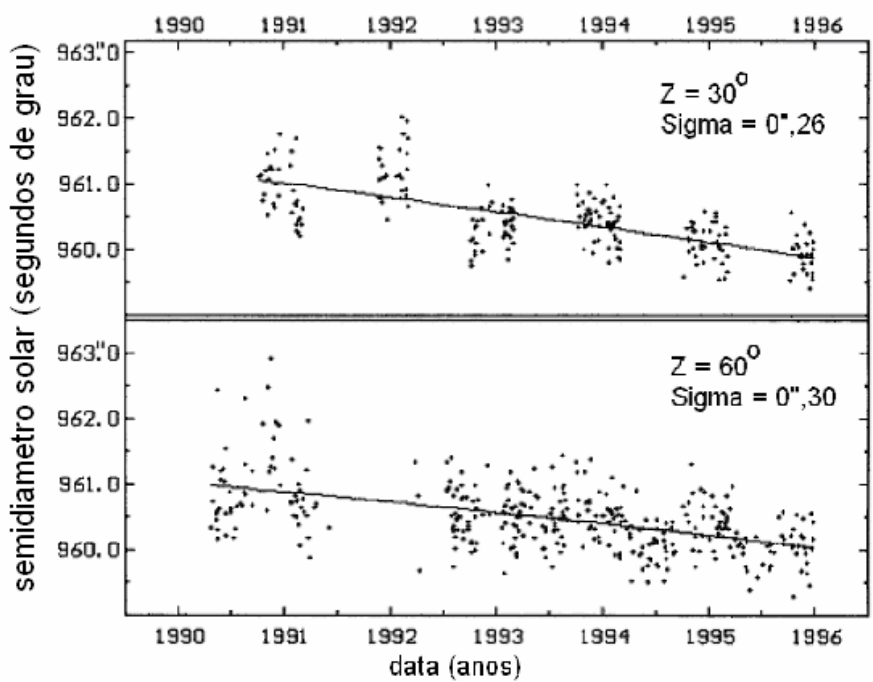

Figura 2.2 – Observações do semidiâmetro solar feitas por F. Noël em Santiago em duas distâncias zenitais diferentes. Ajustes lineares feitos a cada grupo. Sigma é o valor do desvio padrão de cada grupo (Noel, 1997).

As médias de cada par de valores anuais tomados em duas distâncias zenitais são mostradas na Figura 2.3 onde são comparadas com a curva do número de manchas solares publicada pelo *National Geophysical Data Center* em Boulder, Colorado/EUA. E resultados das medidas de semidiâmetro solar na linha espectral do ferro neutro em 525 nm efetuados com o magnetômetro do telescópio solar do observatório de Mount Wilson também foram colocados na mesma figura (Noël, 1997).

F. Noël concluiu por uma variação no semidiâmetro aparente do Sol. Praticamente a mesma variação foi observada em ambas as distâncias zenitais. Há uma variação semelhante nas medidas do magnetômetro. E o número de manchas solar também varia de forma semelhante.

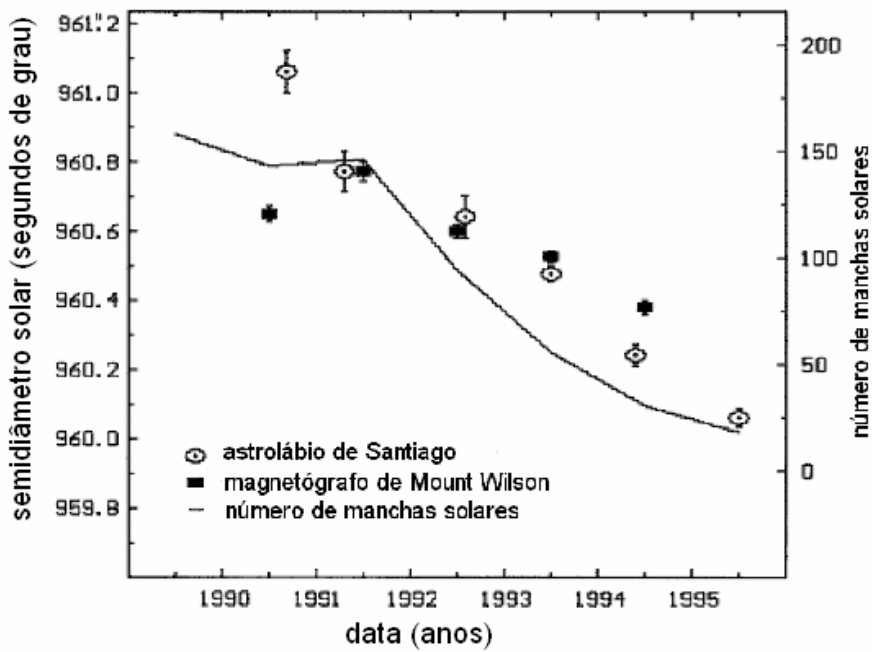

Figura 2.3 - Gráfico de Noel comparando seus dados com a atividade solar, a linha espectral do Fe neutro e o número de manchas solares (Noël, 1997).

**2.4 - Medidas da derivação solar -** Em 2000 Wittmann e Bianda publicaram os resultados de 7583 observações do diâmetro solar feitas em Izaña-Tenerife e 2470 feitas em Locarno-Suíça. As observações de Locarno foram feitas em 550 nm e as de Izaña em 486 e 583 nm. A Figura 2.4 mostra os resultados por eles obtidos. (Wittmann e Bianda, 2000).

Estas medidas foram comparadas com as de outros pesquisadores e embora haja variações significantes do diâmetro solar durante o ciclo solar de onze anos não há variações maiores que 0,05 segundos de grau. A Tabela 2.3 mostra seus dados que são comparados aos dados de Noël e de Laclare obtidos por meio de astrolábios (Wittmann e Bianda, 2000).

Mesmo abstraindo os resultados de Santiago que sempre mostram amplitudes exageradas das variações do diâmetro, os desvios padrão destas médias são de 3 a 5 vezes maiores que o erros padrão. De modo que se impõe a conclusão de que estas medidas estão revelando uma variação do semidiâmetro solar.

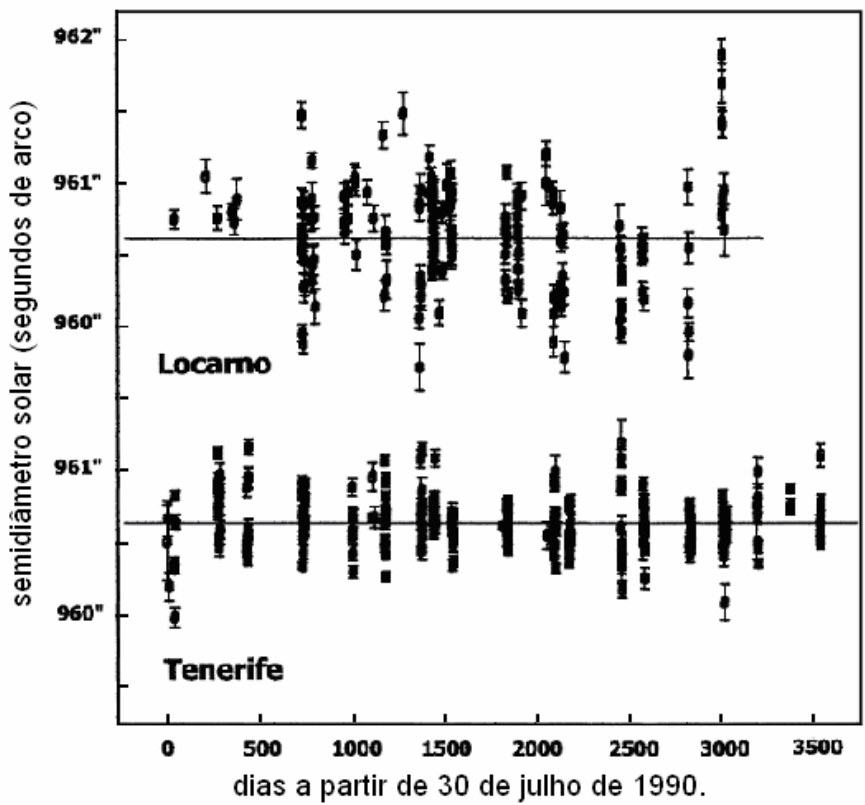

Figura 2.4 – Semidiâmetro solar observado em Locarno e em Tenerife e os erros associados (Wittmann e Bianda, 2000).

**2.5 - Trânsito meridiano -** Brown e Christensen-Dalsgaard usaram o Monitor de Diâmetro Solar do *High Altitude Observatory* no *National Center for Atmospheric Reasearch* no Colorado/EUA. Mediram o trânsito meridiano solar diariamente durante seis anos de 1981 a 1987, durante o declínio do 21º ciclo solar. Combinando as medidas obtidas com modelos da função de escurecimento do limbo solar eles obtiveram um valor para o raio equatorial solar de: R=(959,680±0,009) segundos de grau. Os valores obtidos são mostrados no gráfico da Figura 2.5 (Brown e Christensen-Dalsgaard, 1982)

Brown e Christensen-Dalsgaard concluiram que o diâmetro permaneceu essencialmente constante durante este período que cobre, em quase sua totalidade, a parte de menor atividade do ciclo solar. No entanto o último trecho do gráfico, de 1987 em diante, já na subida do ciclo 22, se afigura compatível com uma tendência de crescimento do diâmetro observado.

Tabela 2.3 – Semidiâmetro solar observado por Wittmann e Bianda e compardos com dados de outros autores. (Wittmann e Bianda, 2000).

| Ano | N | Wittmann | N | Noel | N | Bianda | N | Laclare |
|---|---|---|---|---|---|---|---|---|
| 1990 | 363 | 960,48 ± 0,04 | 124 | 961,06 ± 0,07 | 48 | 960,74 ± 0,13 | 353 | 959,38 ± 0,02 |
| 1991 | 1002 | 960,71 ± 0,02 | 104 | 960,78 ± 0,06 | 92 | 960,83 ± 0,09 | 266 | 959,44 ± 0,02 |
| 1992 | 570 | 960,67 ± 0,02 | 160 | 960,63 ± 0,06 | 322 | 960,57 ± 0,04 | 293 | 959,40 ± 0,02 |
| 1993 | 802 | 960,63 ± 0,02 | 292 | 960,49 ± 0,03 | 266 | 960,74 ± 0,05 | 347 | 959,39 ± 0,01 |
| 1994 | 1176 | 960,68 ± 0,02 | 248 | 960,24 ± 0,03 | 655 | 960,66 ± 0,03 | 267 | 959,47 ± 0,02 |
| 1995 | 481 | 960,66 ± 0,02 | 230 | 960,08 ± 0,03 | 320 | 960,57 ± 0,04 | 273 | 959,48 ± 0,02 |
| 1996 | 879 | 960,59 ± 0,02 | 246 | 960,85 ± 0,03 | 265 | 960,43 ± 0,05 | 313 | 959,47 ± 0,02 |
| 1997 | 643 | 960,62 ± 0,03 | 240 | 960,00 ± 0,03 | 276 | 960,39 ± 0,05 | 392 | 959,52 ± 0,02 |
| 1998 | 1012 | 960,60 ± 0,02 | 316 | 960,27 ± 0,03 | 226 | 961,01 ± 0,08 | 357 | 959,52 ± 0,01 |
| 1999 | 359 | 960,68 ± 0,04 | 400 | 960,47 ± 0,03 | ---- | ------------------- | ---- | ------------------- |
| 2000 | 296 | 960,67 ± 0,01 | 258 | 960,41 ± 0,03 | ---- | ------------------- | ---- | ------------------- |
| **Média:** | 7583 | 960,63 ± 0,02 | 2618 | 960,39 ± 0,01 | 2470 | 960,66 ± 0,04 | 2861 | 959,45 ± 0,02 |
| **Desvio Padrão** | | 0,064 | | 0,329 | | 0,196 | | 0,053 |

Número de observações feitas a cada ano, o valor médio do Semidiâmetro solar e o erro associado. Observações feitas por Wittmann e Bianda e comparadas com observações feitas pro Noel e Laclare.

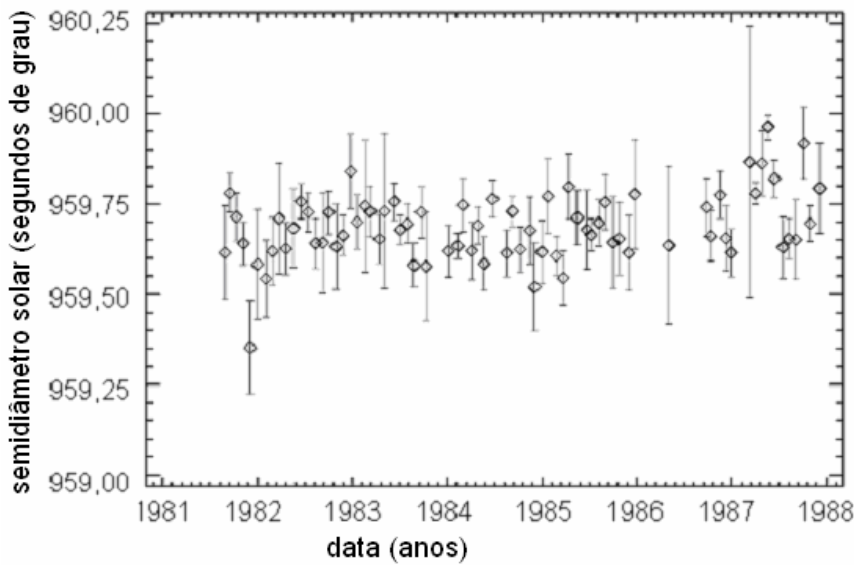

Figura 2.5 – Valores do semidiâmetro solar e erro associado obtidos pelo trânsito meridiano solar reportados (Brown e Christensen-Dalsgaard, 1982).

**2.6 - O diâmetro do Sol acima da atmosfera** – Em 2004 Kuhn, Bush, Emílio e Scherrer apresentaram valores para o semidiâmetro solar a partir de dados do *Michelson Doppler Imager* - MDI a bordo do *Solar and Heliospheric Observatory* - SOHO. Segundo os autores, é possível determinar a sensibilidade do instrumento à temperatura. E as características óticas do MDI são suficientes para se obter a acurácia desejada. O MDI indica possíveis variações do raio durante o ciclo em torno de 7 milisegundos de grau. Não há evidência de variação secular e nenhum nível de variação acima de 15 milisegundos de grau por ano. O semidiâmetro solar tem o valor de: R=(959,6 ±0,5) milisegundos de grau (Kuhn, Bush, Emilio e Scherrer, 2004).

Kuhn e colegas em 2004 trazem um trabalho exaustivo de correções sistemáticas com modelo de correções de temperatura, da distância focal e da opacidade da janela de observação, dentre os termos principais. Fica clara a vantagem de observações fora da atmosfera, mas ao mesmo tempo fica igualmente clara a vantagem de métodos instrumentais dedicados à medida da variação do diâmetro solar, mesmo que no solo.

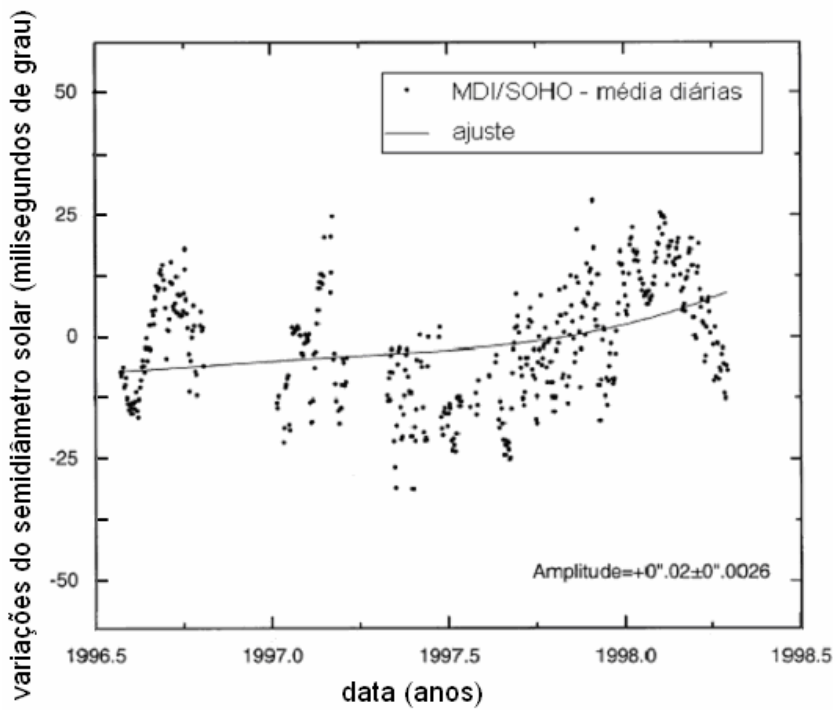

Figura 2.6 – Semidiâmetro solar medido a partir dos dados do MDI/SOHO
(Emilio, Kuhn, Bush, Scherrer, 2000).

Os autores concluem pela semelhança entre os resultados de variação do diâmetro solar dados pelo MDI/SOHO e pelo Monitor de Diâmetro Solar (Item 2.5). Aqui também concluímos por esta semelhança conforme a Figura 2.6 (Kuhn, Bush, Emilio e Scherrer, 2000). Da mesma maneira que na Figura 2.5, vê-se uma pequena variação do diâmetro solar até 1987 quando então ele começa a aumentar em consonância com a subida do ciclo 22.

# 3. A série de dados do Observatório Nacional

**3.1 - Apresentação** - A observação do Sol no Observatório Nacional começou em 1978 com a utilização de um astrolábio Danjon. Inicialmente esta observação era orientada para a determinação dos parâmetros orbitais da Terra e do sistema de referência astrométrico (Penna *et al.* 1996, 1997). Uma modificação do instrumento, que foi equipado com um prisma refletor de ângulo variável e com uma câmera CCD, permitiu que, a partir de 1997, se iniciasse um programa de acompanhamento do diâmetro do Sol em cooperação com o *Observatoire de la Cote d'Azur* – OCA (Andrei *et al.* 1996).

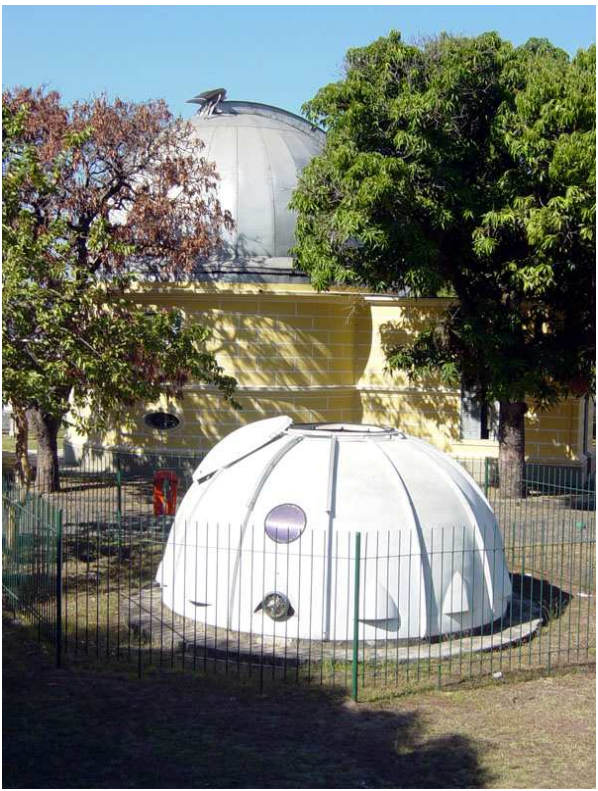

Figura 3.1 – O domo onde está instalado o astrolábio do Observatório Nacional. Sua geometria foi calculada para minimizar a refração de sala.

O astrolábio solar do Observatório Nacional (λ=+43° 13' 22,00'', Φ=-22° 53' 42,15'') consiste de um telescópio refrator na frente do qual é colocado um prisma refletor e uma bacia de mercúrio. O prisma tem duas faces refletoras que são simétricas em relação ao plano horizontal. Os raios incidentes são separados em dois feixes: Um devido à reflexão na face superior do prisma e o outro obtido por reflexão na face inferior do prisma após uma primeira reflexão numa superfície de mercúrio. Nestas condições obtêm-se duas imagens do objeto observado. Quando o objeto se move sua variação azimutal provoca deslocamentos das imagens para o mesmo lado e a variação de sua distância zenital provoca deslocamentos opostos das imagens. Assim, o instrumento detecta a coincidência de duas imagens de um objeto pontual, quando este ponto cruza uma linha de distância zenital determinada pelo ângulo do prisma. Fazendo-se variar o ângulo com o plano horizontal, simultaneamente, das duas faces do prisma, obtêm-se imagens de objetos em diferentes distâncias zenitais. O instrumento instalado no ON permite a observação de objetos entre 25º e 57º de distância zenital. O Astrolábio fornece a uma câmara *Charge-Coupled Device* – CCD as duas imagens.

A cada observação de semidiâmetro solar escolhe-se determinada distância zenital, por onde deve passar o Sol, ou seja o almicantarado de trânsito. À medida que o Sol se aproxima do almicantarado escolhido, as duas imagens se aproximam e, quando o primeiro bordo do Sol cruza o almicantarado, as duas imagens se tocam. Quando o segundo bordo do Sol cruza esta linha, as imagens se separam após os bordos das duas imagens se tocarem. Quando os bordos do Sol estão para cruzar o almicantarado, são feitas 46 imagens. O instante em que cada imagem é obtida é fornecido pelo relógio atômico da Divisão Serviço da Hora - DSHO do ON. A análise destas imagens fornece o instante em que os dois bordos, o direto e o refletido se tocaram ou se separaram.

Conhecendo-se a marcha do Sol a cada dia do ano e a cada posição que ocupa na esfera celeste, pode-se calcular, a partir do tempo que o Sol levou para cruzar totalmente o almicantarado, o seu tamanho angular. Este tamanho é então reduzido para a distância média do Sol à Terra, isto é, para uma Unidade Astronômica - UA, obtendo-se a medida angular do seu diâmetro.

As imagens obtidas são dirigidas ao CCD que tem 512 linhas e 512 colunas de *pixels*. Por conta do entrelaçamento das imagens, tomam-se apenas 256 linhas. Um *pixel* corresponde a 0,56 segundos de grau e desta forma apenas uma parte do bordo solar é tomada na imagem. Em cada imagem são identificados 256 pontos do bordo direto e 256 pontos do bordo refletido do Sol, um para cada uma das linhas da imagem. As linhas apresentam uma curva de intensidade de luz. Esta curva tem a propriedade de apresentar um extremo de sua derivada ao longo do bordo solar. A segunda derivada da função de intensidade de luz é igual a zero no bordo solar. Assim, o bordo solar é o ponto em que a curva de luz tem seu ponto de inflexão.

O bordo solar é delineado pelo conjunto destes 256 pontos. Ocorre, porém, que estes pontos formam uma curva ruidosa em função da ação da atmosfera. É preciso passar por eles uma curva mais definida que indique o bordo solar. Ajusta-se a estes pontos uma parábola e não um arco de círculo, isto porque a parábola minimiza defeitos óticos, responde melhor à forma retangular dos *pixels*, e considera o movimento do disco solar durante os 20 ms de integração da imagem no CCD (Reis Neto, 2002). Ajusta-se uma parábola para o bordo direto e uma parábola para o bordo refletido. Para cada imagem obtida existem então duas parábolas. Tomam-se as posições dos vértices destas parábolas e o instante de tempo em que as imagens foram obtidas. Estas posições em função do tempo definem duas séries de pontos por onde se ajustam duas retas. Como uma das parábolas avança e a outra recua, estas retas têm inclinações opostas. O ponto de contacto destas retas define o instante de tempo em que os bordos direto e refletido do Sol se tocam. Este é o instante em que o bordo solar cruzou o almicantarado. Da mesma forma se encontra o instante em que o segundo bordo solar cruza o almicantarado. Um conjunto de algoritmos foi desenvolvido para calcular, a partir dos instantes de passagem dos dois bordos pelo mesmo almicantarado, o diâmetro vertical observado do Sol. Após a aplicação das correções necessárias o diâmetro é reduzido para 1 UA (Sinceac, 1998).

Aos bordos do Sol ajusta-se, não apenas uma, mas, na verdade, três parábolas com critérios diferentes de desvio padrão para remoção de pontos observados. Se o número de pontos utilizados for inferior a 50, então nenhuma parábola é ajustada. Em 1997 foram definidos três níveis de critérios: 1,7, 2,0 e 2,5. Se as condições forem boas, três soluções são obtidas, caso contrário apenas as soluções para 1,7 e 2,0 alcançam resultado ou, mais raramente, apenas a primeira (Reis Neto, 2002).

As imagens obtidas são analisadas pelos programas, os quais calculam para cada observação os três valores do semidiâmetro solar e os respectivos erros de medida. Eles obtêm também a distância zenital e o azimute solar, a latitude solar na direção da qual o semidiâmetro é medido, o fator de Fried, os instantes de passagem dos bordos solares, os erros de medida destes instantes, a largura em *pixels* do bordo direto e do bordo refletido, o ajuste da parábola direta e o da refletida, o desvio padrão dos pontos da parábola direta e, da refletida, e a decalagem. Para cada seção de observações são lidos os instantes inicial e final, e as temperaturas do ar e do mercúrio, a pressão atmosférica e a umidade do ar nestes instantes. O fator de Fried descreve a qualidade do *seeing* da atmosfera, ele é definido como o comprimento de onda observado, dividido pela largura a meia altura de uma imagem pontual espalhada pela ação da atmosfera. Este fator é calculado a partir dos dados de observação (Lakhal et al., 1999). A decalagem é o desvio do centro do CCD ao ponto de contacto das duas parábolas, ou o desvio dos centros das imagens em relação ao almucantara.

Na entrada do caminho óptico há um filtro neutro deixando passar apenas $10^{-5}$ da intensidade da luz incidente. Diante da câmera do CCD, há dois filtros de luz que definem a banda de comprimentos de onda que pode ser detectada. Para estes filtros o intervalo onde 50% da luz é transmitida vai de 523,0 nm até 691,0 nm. Sendo o máximo em 563,5 nm com um índice de transmissão de 75% (Jilinski et al., 1999).

Os dados finais processados são apresentados em um arquivo para cada sessão de trabalho. A cada dia, quando possível, são realizadas duas sessões: uma antes da passagem meridiana do Sol e outra após sua passagem. Os arquivos dispõem ao analista três conjuntos de dados para cada observação com os todos os itens calculados e já citados.

**3.2 – Distribuição das medidas** – A latitude e o clima no Rio de Janeiro favorecem a observação do Sol em mais de 200 dias por ano. Em 1998 o astrolábio ganhou os filtros amarelo e azul na objetiva junto ao CCD limitando o comprimento de onda observado e a largura da banda passante. Além disso, a partir deste ano passaram-se a fazer regularmente observações nos dois lados do meridiano local. Por estes motivos os dados de 1997 foram rejeitados. Assim a série analisada de dados de observação do semidiâmetro solar do ON inicia em março de 1998. Elas foram interrompidas para manutenção do equipamento nos meses de outubro e novembro de 2001. Pelo mesmo motivo voltaram a ser interrompidas em dezembro de 2003 e novamente, por um grande intervalo, entre outubro de 2004 e maio de 2006. Em 2009 há dados registrados apenas em janeiro, agosto, outubro e novembro. Ao longo de todos estes anos um total de 21640 observações foram aproveitadas para análise. O gráfico da Figura 3.2 mostra a distribuição destas observações ao longo do tempo.

Os valores de semidiâmetro solar observados pelo astrolábio são ruidosos uma vez que a observação é feita durante o dia quando a atmosfera está sendo constantemente bombardeada pela radiação solar e pela radiação emitida pelo solo. Isto provoca fortes correntes convectivas que acabam derivando em uma agitação constante da atmosfera. A Figura 3.3 mostra todas as observações do semidiâmetro solar obtidas no ON entre 1998 e 2009.

Os valores de semidiâmetro assim obtidos oscilam em torno de sua tendência, , caracterizada como a média em torno de determinada época, e embora esta se modifique ao longo dos tempos podemos observar que a distribuição dos valores não apresenta desvio maior para um dos lados, mas assemelha-se bastante com uma distribuição normal o que pode ser visto na Figura 3.4.

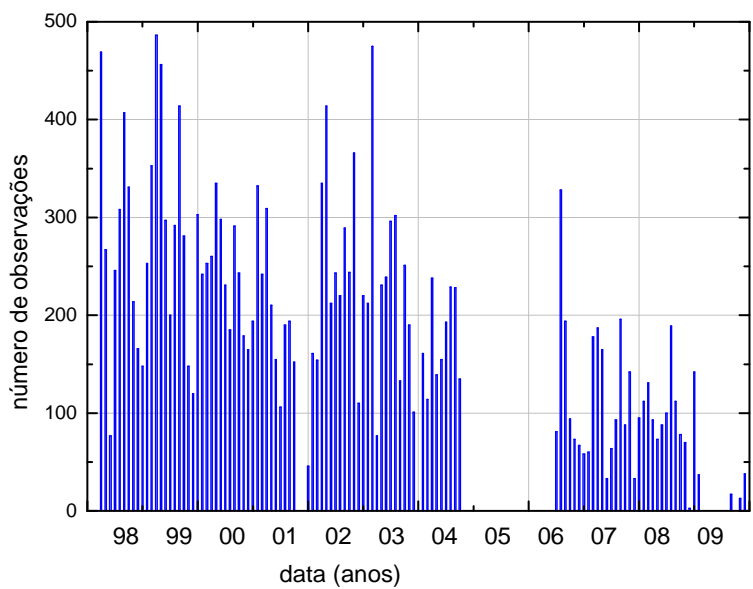

Figura 3.2 - Número de observações mensais no ON de 1998 a 2009.

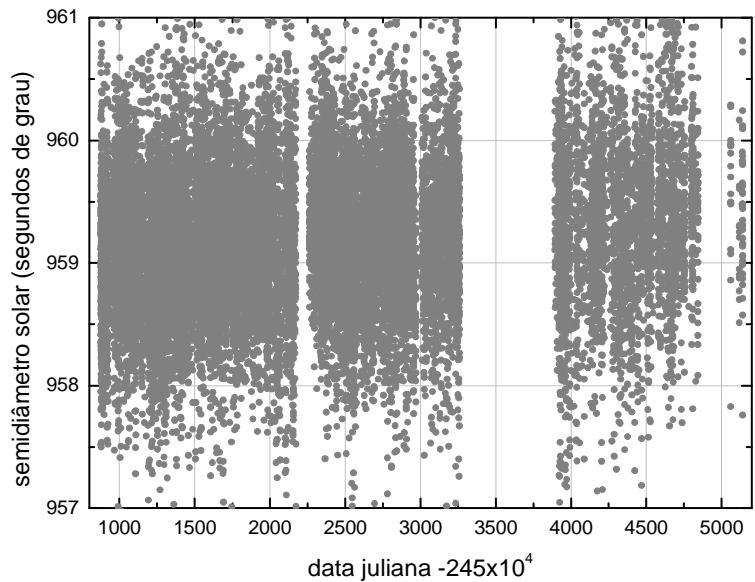

Figura 3.3 – Todas as observações do semidiâmetro do Sol obtidas no ON entre 1998 e 2008. Os espaços claros correspondem aos intervalos onde não houve observações.

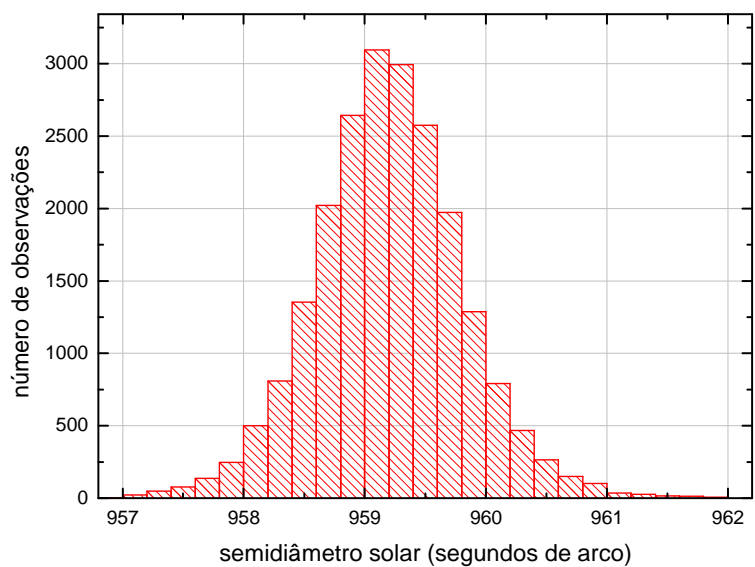

Figura 3.4 - Distribuição do semidiâmetro do Sol nas observações do ON.

As observações do semidiâmetro solar no ON são feitas em sessões distintas que ocorrem antes da passagem meridiana do Sol e após a passagem meridiana. A Figura 3.5 mostra os horários em que ocorreram as observações durante o ano de 2002. Há distintamente dois grupos de dados, um de dados observados a Leste e outro de dados observados a Oeste do meridiano local. Isto ocorre porque não é possível observar com o astrolábio uma distância zenital inferior a 25º, além disso observações muito próxima ao meridiano são longas o suficiente para não permitir o registro em tempo. Normalmente a cada dia de observação há duas sessões com aproximadamente o mesmo número de observações de modo que há um equilíbrio do número de observações efetuadas a Leste e a Oeste do meridiano local. A Figura 3.6 mostra o número de observações mensais a cada um dos lados.

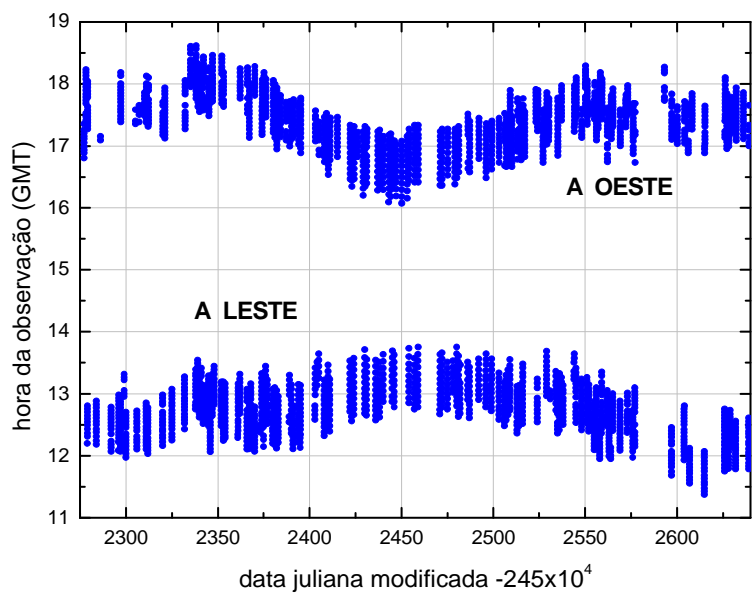

Figura 3.5 – Os horários de observação do semidiâmetro solar no ON (GMT) ao longo do ano de 2002. São duas séries: os valores observados a Leste e os valores observados a Oeste.

Os valores observados a Leste e a Oeste também são bem equilibrados e a Figura 3.7 mostra este equilíbrio ilustrando a diferença entre as médias mensais observadas. Entretanto os dados observados a Leste são bem mais ruidosos que aqueles observados a Oeste. Isto ocorre porque os primeiros são obtidos em momentos em que a atmosfera tem maiores gradientes de temperatura causando uma maior turbulência em suas camadas. Os dados observados após a passagem meridiana do Sol são obtidos em momentos em que a atmosfera está mais quente, porém mais estável. A partir de 2006 as diferenças se ampliam por efeito estatístico já que o número de observações diminuiu muito. O gráfico da Figura 3.8 mostra os desvios padrão dos valores mensais do semidiâmetro solar a Leste e a Oeste do meridiano.

Como já dissemos, o Sol pode ser observado pelo astrolábio do ON entre as distâncias zenitais de 25º e 57º. A distribuição de todas as observações ao longo das distâncias zenitais pode ser vista na Figura 3.9. Esta distribuição não é uniforme ao longo do ano trópico respondendo fortemente à sazonalidade. Este efeito pode ser visto na Figura 3.10 que mostra a distribuição das observações do semidiâmetro solar ao longo de um ano, no caso é o ano de 2002 usado como referência por ter uma grande coleção de dados sem interrupção ao longo de quase todo o tempo.

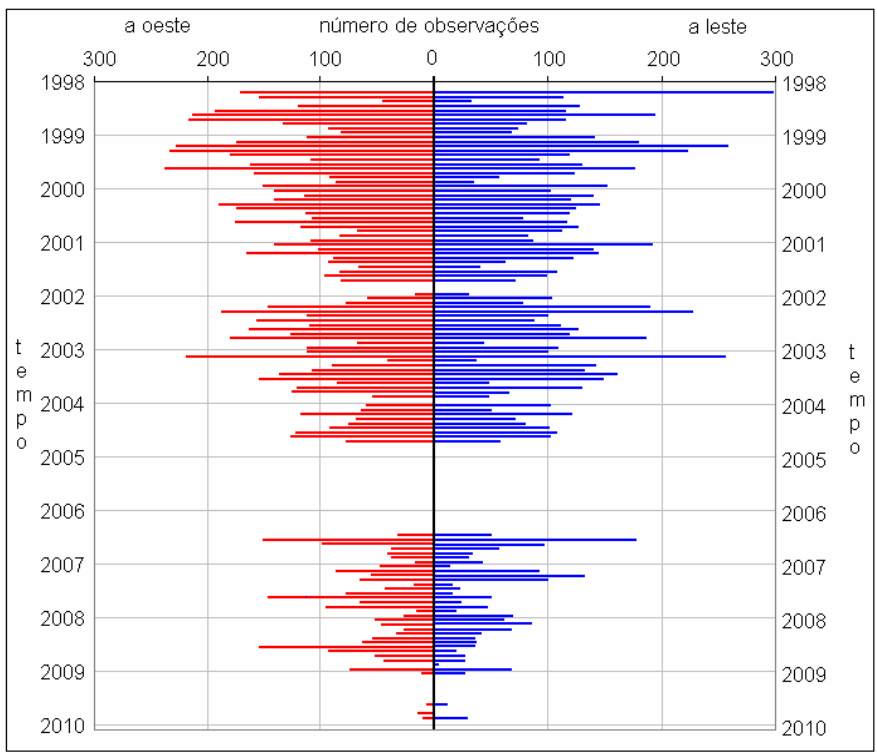

Figura 3.6 – Número de observações mensais do semidiâmetro solar no ON a Leste e a Oeste do meridiano local.

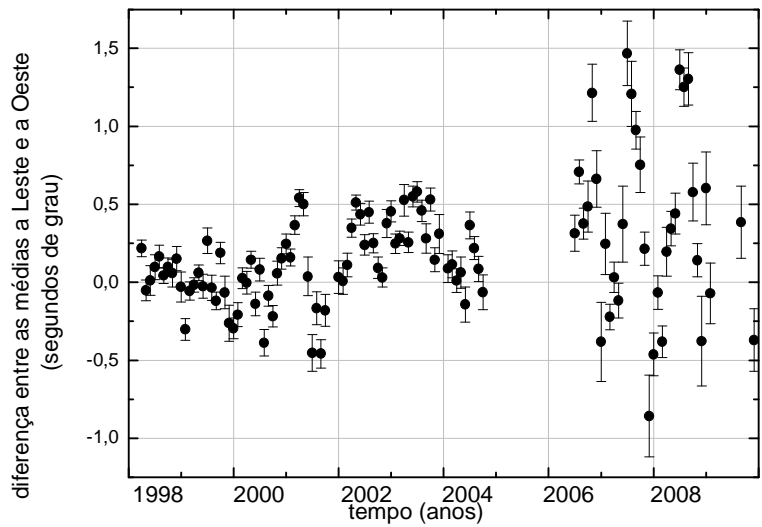

Figura 3.7 – Diferenças entre as médias mensais das observações do semidiâmetro solar no ON obtidas a Leste e a Oeste do meridiano local.

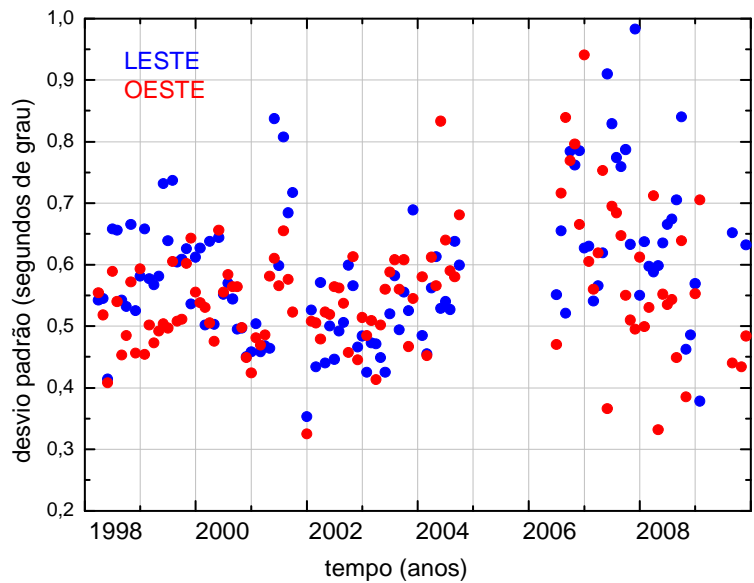

Figura 3.8 – Desvios padrão dos valores mensais do semidiâmetro solar observados no ON a Leste e a Oeste da passagem meridiano do Sol.

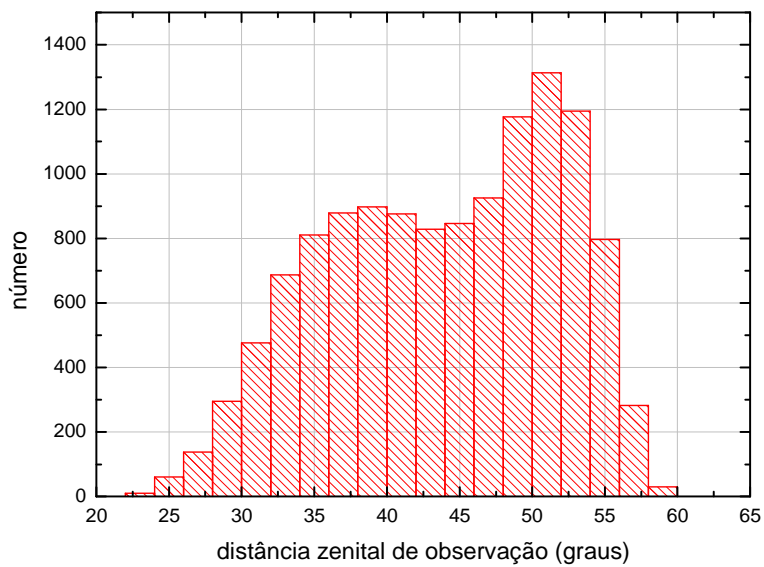

Figura 3.9 – Distribuição do número de observações do semidiâmetro solar ao longo das distâncias zenitais do Sol.

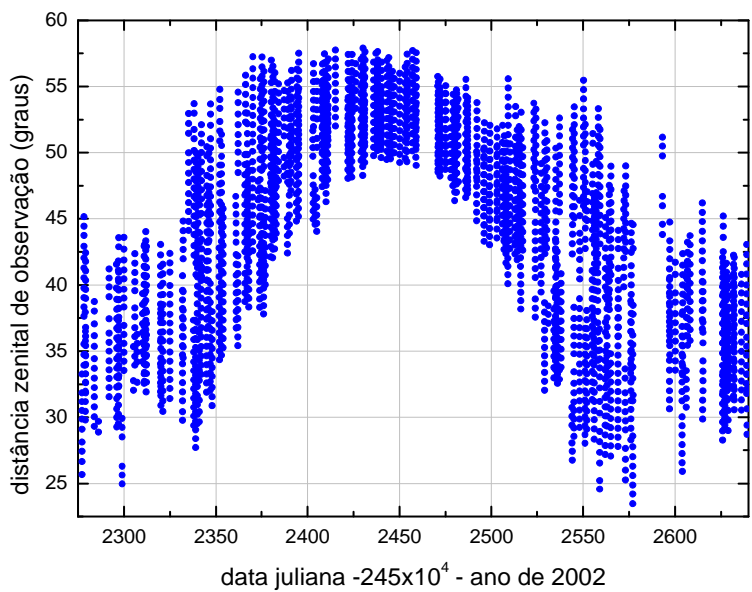

Figura 3.10 – Distribuição temporal das distâncias zenitais nas observações do semidiâmetro solar ao longo do ano de 2002 tomado como referência.

Pode-se ver que nos dias mais próximos do solstício de inverno o Sol é observado sempre em uma faixa estreita das mais altas distâncias zenitais, ocorrendo o inverso nos dias mais próximos do solstício de verão. Ou seja, há uma faixa maior e com valores menores.

Da mesma forma que há uma grande variação da distância zenital observada, há também uma variação muito grande dos azimutes observados. Não há observações apenas numa pequena faixa de 30º em torno do ponto cardeal Norte e numa faixa um pouco maior, de 135º em torno do ponto cardeal Sul. A distribuição do número de observações ao longo dos azimutes pode ser vista na Figura 3.11. Esta distribuição responde também fortemente à sazonalidade motivo pelo qual mostramos o gráfico da Figura 3.12 onde pode-se ver a distribuição temporal dos azimutes observados no ano de 2002 tomado como referência.

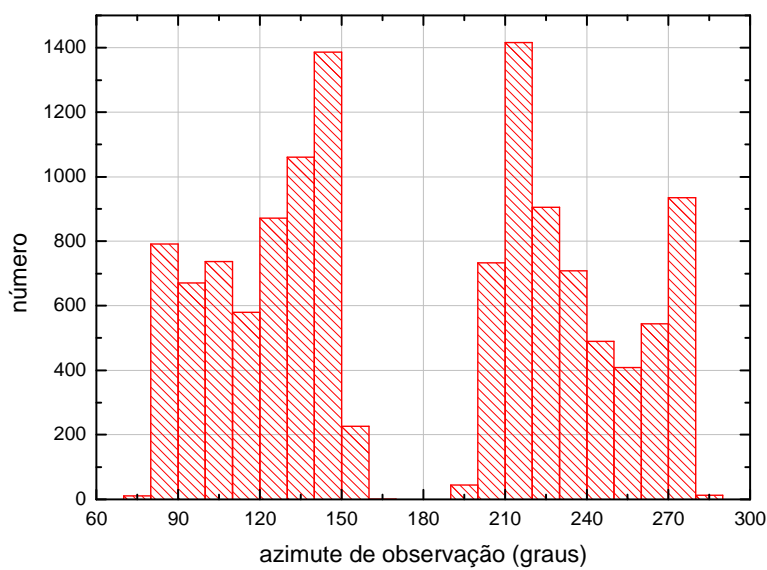

Figura 3.11 – Distribuição do número de observações do semidiâmetro solar ao longo dos azimutes. O ponto Sul está no azimute zero.

O astrolábio usa o sistema que tem como início a direção Sul, assim, o ponto cardeal Oeste está no azimute 90º, o ponto cardeal Norte no azimute 180º e o ponto cardeal Leste no azimute 270º. Pode-se ver que o Sol é observado pelo astrolábio do ON mais perto do Norte nos dias mais próximos do solstício de inverno e mais perto do Sul nos dias mais próximos do solstício de verão.

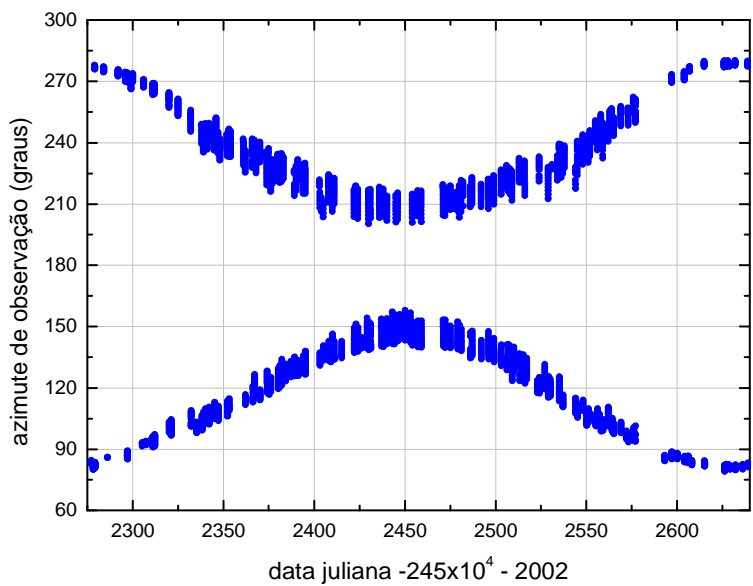

Figura 3.12 – Distribuição temporal dos azimutes de observação do semidiâmetro solar ao longo do ano de 2002 tomado como referência.

Cada observação do semidiâmetro solar tem um tempo de duração que é o tempo que o Sol leva para cruzar um almicantarado. Este tempo depende da estação e da hora do dia, de tal modo que próximo ao solstício de inverno as observações são mais longas e próximo ao solstício de verão elas são mais curtas. Este efeito pode ser visto na Figura 3.13 que mostra a duração das observações ao longo do ano de 2002 tomado como referência.

A distribuição do número de observações do semidiâmetro ao longo das durações das observações pode ser vista no gráfico da Figura 3.14. Pode-se ver que há um grande número de observações curtas e este número é sempre decrescente para observações mais longas. Isto ocorre porque quando as observações são curtas, um maior número delas pode ser feito dentro da faixa de distâncias zenitais possíveis. Ao contrário, quando as observações são longas o seu número fica bem limitado por esta faixa de possibilidades.

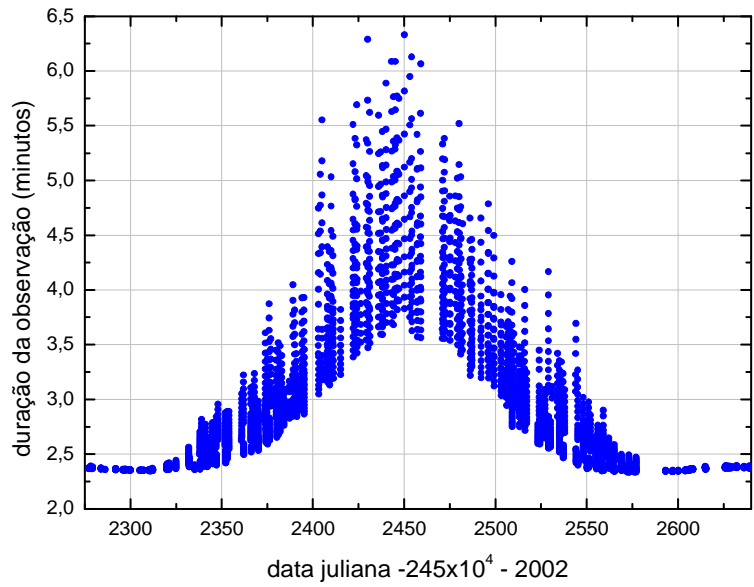

Figura 3.13 – A duração das observações ao longo do ano de 2002 tomado como referência.

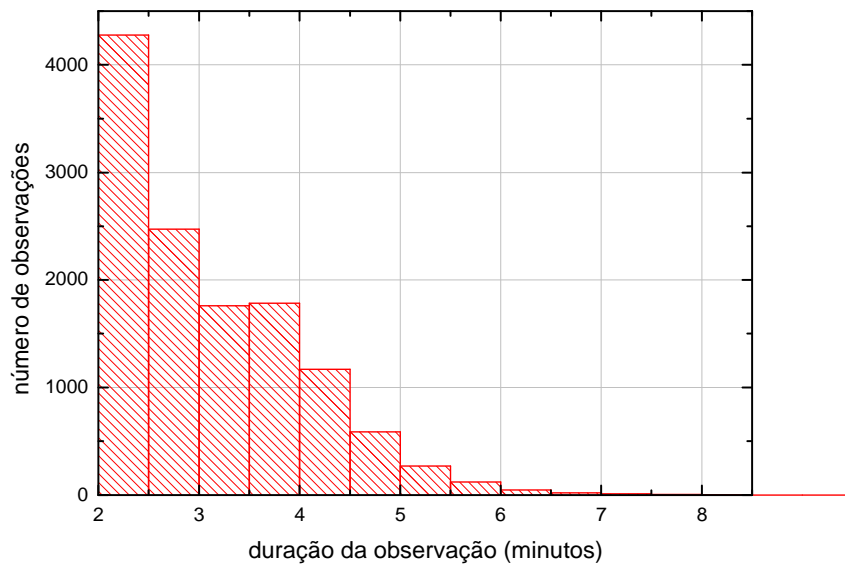

Figura 3.14 – Distribuição do número de observações do semidiâmetro solar ao longo da duração das observações.

É interessante também observar como o semidiâmetro solar medido se distribui ao longo das durações. Pode-se ver um grande espalhamento razoavelmente simétrico desta distribuição na Figura 3.15.

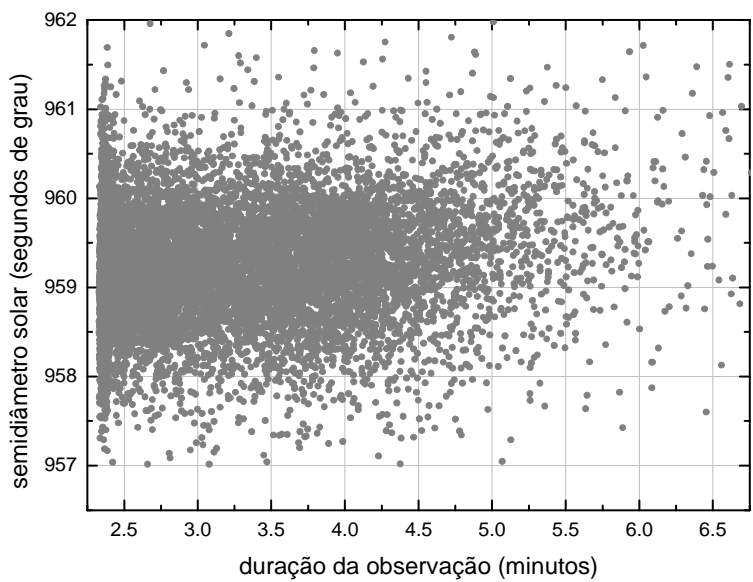

Figura 3.15 – Distribuição das observações do semidiâmetro solar ao longo das durações das observações.

**3.3 – Distribuição dos valores do semidiâmetro** – O estudo detalhado da série de dados do semidiâmetro solar do ON é objeto dos capítulos que se seguem. Nos capítulos 4, 5 e 6 serão analisados os possíveis erros da série e nos demais se procede a análise da série tendo em vista alguns indicadores de atividade do Sol. No entanto, nesta apresentação da série é oportuno colocar os valores médios e abordar sua significância.

A precisão média de uma observação isolada é de 0,606 segundos de grau. Enquanto que as análises precedentes do semidiâmetro solar, bem como as trazidas pelas observações do ON indicam que não se deve obter ao longo do ciclo solar variações maiores que poucos centésimos de segundos de grau. Deste modo adotamos considerar valores médios agrupando pelo menos 100 observações e/ou o intervalo temporal de um mês. A Tabela 3.1 lista os valores médios mensais do semidiâmetro solar observado no ON. Estes valores podem ser vistos de forma gráfica na Figura 3.16

A Tabela 3.1 e o gráfico da Figura 3.16 mostram uma tendência ascendente. No decorrer desta tese, em virtude do espaço de tempo que os dados analisados englobam, a principal ênfase está na correlação do ciclo de atividades do Sol de 11 anos. Ciclos bem mais longos, no entanto, têm sido apontados por outros autores. Abstraindo os principais ciclos estelares, os dois ciclos mais marcados da atividade solar são o ciclo principal de atividades, usualmente chamado como ciclo de manchas, de cerca de 11 anos e do ciclo de reversão do campo magnético, com cerca de 22 anos. Mesmo estes dois ciclos parecem ser modulados por envoltórios muito mais longos, de maneira que o ciclo de manchas parece ter variado num ritmo centenário entre 10 e 15 anos (Friis-Christensen e Lassen,1991). Também desta ordem, entre 60 e 90 anos (ou mesmo 150 anos) autores reconhecem o ciclo de Gleissberg (Ma, 2009). Os pesquisadores (Moussas *et al.* 2005) listam diversos outros ciclos de longa peridiocidade, 55 anos, 200 anos, e 400-500 anos (Nordemann *et al*, 2005), cujas assinaturas são reconciliáveis com mudanças climáticas. Há ciclos ainda mais longos, ao redor de 1000 anos e 2000 anos, cujas evidências seriam reconhecidas nos depósitos de carbono14 e berílio10 (Ma, 2007). Evidências geológicas apontam para ciclos ainda mais longos, como o de 100.000 anos (Rigozo *et al.* 2001). O periodograma da primeira análise de um conjunto maior dos dados da variação do diâmetro solar no Observatório Nacional, entre 1998 e 2000, conquanto no limitado intervalo apontou termos reconciliáveis com a oscilação quase-bianual (Reis Neto, 2002). Deste modo, a existência de termos seculares modulando a atividade solar e a variação do diâmetro solar encontra suporte em diversas investigações precedentes e multidiciplinares. Esta tendência ascendente encontrada na série de dados do ON pode ser a assinatura de um destes ciclos.

A maneira mais indicada de se ver como os valores do semidiâmetro solar evoluem ao longo dos tempos é compor a média móvel dos valores. A Figura 3.17 mostra a média corrida dos valores a cada 500 pontos. Foram usados 500 pontos de modo que mesmo nos trechos de maior número de observações, o intervalo considerado para média não fosse nunca inferior a um mês. Neste gráfico, a reta horizontal em torno da data Juliana menos 245x10$^4$ de 3500 corresponde ao intervalo de tempo em que não se fizeram observações entre outubro de 2004 e maio de 2006.

Tabela 3.1 – Valores médios mensais do semidiâmetro solar observados no ON.

| ano | mês | N | sd | σ | ano | mês | N | sd | σ | ano | mês | N | sd | σ |
|---|---|---|---|---|---|---|---|---|---|---|---|---|---|---|
| 1998 | mar. | 469 | 958,990 | 0,556 | 2001 | abr. | 210 | 959,275 | 0,520 | | ago. | 228 | 959,290 | 0,594 |
| | abr. | 267 | 959,014 | 0,530 | | mai. | 155 | 959,049 | 0,705 | | set. | 135 | 959,193 | 0,653 |
| | mai. | 77 | 959,149 | 0,411 | | jun. | 106 | 958,930 | 0,573 | 2006 | jun. | 81 | 959,479 | 0,552 |
| | jun. | 246 | 959,135 | 0,627 | | jul. | 190 | 959,317 | 0,732 | | jul. | 328 | 959,393 | 0,762 |
| | jul. | 308 | 959,193 | 0,591 | | ago. | 194 | 959,204 | 0,650 | | ago. | 194 | 959,206 | 0,715 |
| | ago. | 407 | 959,127 | 0,498 | | set. | 152 | 959,074 | 0,609 | | set. | 94 | 959,173 | 0,794 |
| | set. | 331 | 959,134 | 0,504 | 2002 | dez. | 46 | 959,532 | 0,355 | | out. | 73 | 959,286 | 0,929 |
| | out. | 214 | 959,108 | 0,610 | | jan. | 161 | 959,587 | 0,535 | | nov. | 67 | 959,249 | 0,760 |
| | nov. | 166 | 959,138 | 0,493 | | fev. | 154 | 959,433 | 0,496 | 2007 | dez. | 58 | 959,230 | 0,754 |
| 1999 | dez. | 148 | 959,068 | 0,588 | | mar. | 335 | 959,422 | 0,565 | | jan. | 60 | 959,301 | 0,619 |
| | jan. | 253 | 959,029 | 0,596 | | abr. | 414 | 959,335 | 0,536 | | fev. | 178 | 959,403 | 0,561 |
| | fev. | 353 | 958,925 | 0,542 | | mai. | 212 | 959,293 | 0,505 | | mar. | 187 | 959,363 | 0,582 |
| | mar. | 486 | 959,020 | 0,525 | | jun. | 243 | 959,166 | 0,513 | | abr. | 165 | 959,385 | 0,677 |
| | abr. | 456 | 958,913 | 0,538 | | jul. | 220 | 959,302 | 0,560 | | mai. | 33 | 959,626 | 0,710 |
| | mai. | 297 | 959,058 | 0,605 | | ago. | 289 | 959,256 | 0,540 | | jun. | 64 | 959,788 | 1,017 |
| | jun. | 200 | 959,098 | 0,582 | | set. | 244 | 959,213 | 0,549 | | jul. | 93 | 959,307 | 0,834 |
| | jul. | 292 | 959,204 | 0,667 | | out. | 366 | 959,336 | 0,621 | | ago. | 196 | 959,391 | 0,798 |
| | ago. | 414 | 959,066 | 0,554 | | nov. | 110 | 959,139 | 0,455 | | set. | 88 | 959,285 | 0,703 |
| | set. | 281 | 959,198 | 0,564 | 2003 | dez. | 220 | 959,308 | 0,518 | | out. | 142 | 959,051 | 0,563 |
| | out. | 148 | 959,050 | 0,612 | | jan. | 212 | 959,228 | 0,460 | | nov. | 33 | 959,531 | 0,917 |
| | nov. | 120 | 959,209 | 0,626 | | fev. | 475 | 959,267 | 0,497 | 2008 | dez. | 95 | 959,259 | 0,604 |
| 2000 | dez. | 303 | 959,197 | 0,603 | | mar. | 77 | 959,251 | 0,435 | | jan. | 112 | 959,458 | 0,579 |
| | jan. | 242 | 959,284 | 0,586 | | abr. | 231 | 959,179 | 0,475 | | fev. | 131 | 959,313 | 0,603 |
| | fev. | 253 | 959,139 | 0,515 | | mai. | 239 | 959,130 | 0,498 | | mar. | 93 | 959,501 | 0,632 |
| | mar. | 260 | 959,174 | 0,570 | | jun. | 296 | 959,097 | 0,547 | | abr. | 73 | 959,443 | 0,527 |
| | abr. | 335 | 959,211 | 0,493 | | jul. | 302 | 959,067 | 0,591 | | mai. | 88 | 959,511 | 0,625 |
| | mai. | 298 | 959,110 | 0,655 | | ago. | 133 | 959,096 | 0,572 | | jun. | 100 | 959,600 | 0,881 |
| | jun. | 231 | 959,309 | 0,555 | | set. | 251 | 959,073 | 0,582 | | jul. | 189 | 959,283 | 0,749 |
| | jul. | 185 | 959,220 | 0,609 | | out. | 190 | 959,073 | 0,480 | | ago. | 112 | 959,513 | 0,701 |
| | ago. | 291 | 959,158 | 0,558 | | nov. | 101 | 958,954 | 0,636 | | set. | 78 | 959,667 | 0,766 |
| | set. | 243 | 959,198 | 0,540 | 2004 | jan. | 161 | 959,204 | 0,512 | | out. | 70 | 959,315 | 0,422 |
| | out. | 179 | 959,222 | 0,498 | | fev. | 114 | 959,269 | 0,448 | | nov. | 3 | 959,287 | 0,486 |
| | nov. | 165 | 959,148 | 0,456 | | mar. | 238 | 959,242 | 0,595 | 2009 | dez. | 142 | 959,623 | 0,562 |
| 2001 | dez. | 194 | 959,026 | 0,456 | | abr. | 139 | 959,214 | 0,588 | | jan. | 37 | 959,384 | 0,548 |
| | jan. | 332 | 959,266 | 0,504 | | mai. | 155 | 959,319 | 0,693 | | ago. | 17 | 959,572 | 0,586 |
| | fev. | 242 | 959,221 | 0,470 | | jun. | 193 | 959,217 | 0,613 | | out. | 13 | 959,167 | 0,434 |
| | mar. | 309 | 959,159 | 0,472 | | jul. | 229 | 959,346 | 0,563 | | nov. | 38 | 959,465 | 0,621 |

A coluna [N] reporta o número de observações do mês. A coluna [sd] traz o valor médio mensal do semidiâmetro solar. A coluna [σ] informa o desvio padrão dos valores do mês.

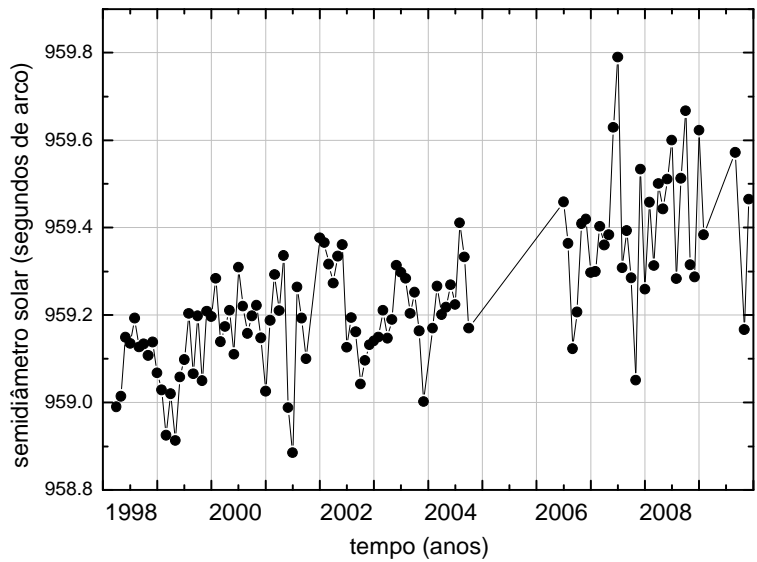

Figura 3.16 - Médias mensais do semidiâmetro solar medido no ON.

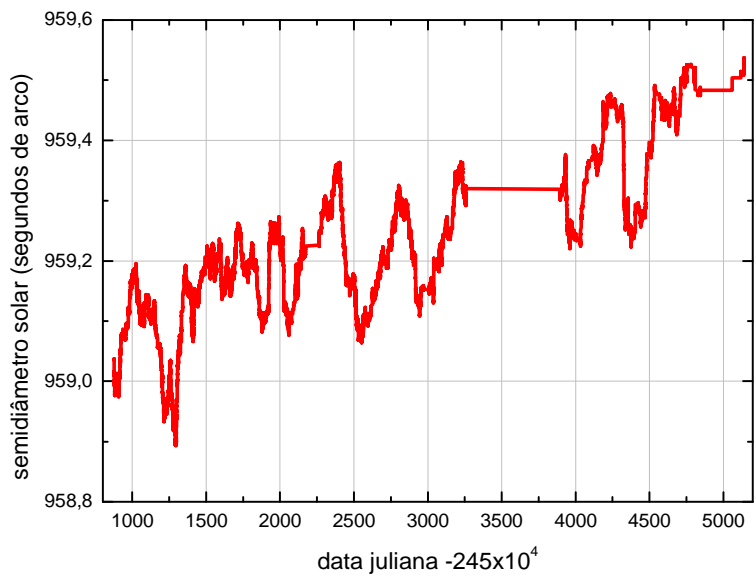

Figura 3.17 - Média corrida de 500 pontos do semidiâmetro solar observado no ON.

A média corrida não considera intervalos de tempo definidos tomando sempre a média de um número fixo de valores escolhido pelo analista, no nosso gráfico são 500 pontos. Este tipo de média não representa igualmente os diversos períodos diferentemente densos de pontos, levando a influência de pontos mais distantes para compor a média onde há uma densidade menor de pontos. Por outro lado as médias de intervalos de tempo têm o defeito oposto da média corrida de não considerarem o reduzido número de pontos dentro do intervalo escolhido. Assim, quando o número de pontos é muito baixo as médias de uma série muito ruidosa podem desviar por demais da correta tendência local. O gráfico apresentando as médias mensais mostra alguns valores de forma bastante ruidosa. O ruído grande é consequência do reduzido número de valores naqueles meses. Para ver a evolução do semidiâmetro solar sem esta oscilação provocada pelo efeito da atmosfera fazemos médias anuais. Mas, desta forma, perdemos o acompanhamento temporal mais fino. Para contornar este outro efeito podemos fazer médias anuais corridas, uma a cada mês obtendo o gráfico da Figura 3.18. Para evitar os pontos ainda muito ruidosos representamos apenas os pontos que têm mais de 500 observações. O desvio padrão dos valores observados é de 0,606 segundos de grau de modo que coleções com menos de 500 pontos têm um erro médio superior a 0,027 segundos de grau indicando pontos com ruído superior ao que desejamos.

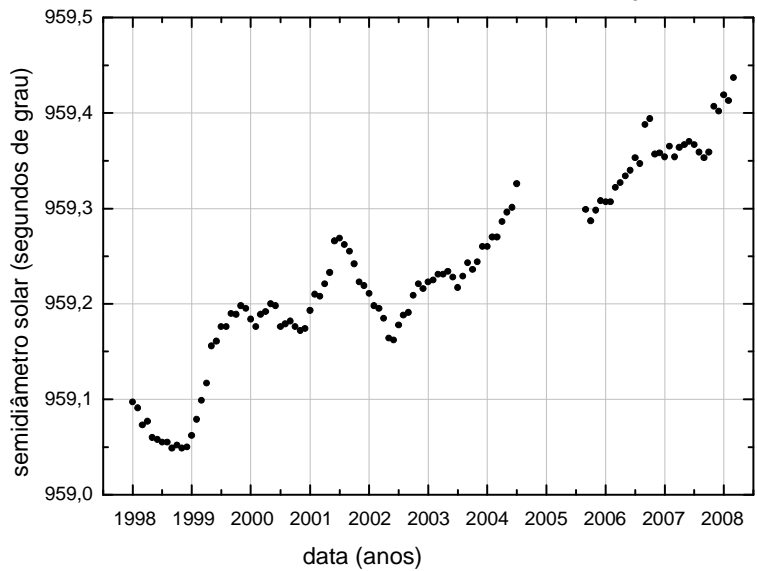

Figura 3.18 - Médias anuais corridas a cada mês do semidiâmetro solar. Apenas os pontos com mais de 500 observações.

A Figura 3.18 mostra o semidiâmetro solar variando muito rapidamente durante o máximo de atividades do Sol. Veremos depois que ele acompanha os picos de atividade do máximo do ciclo 23 que ocorreram em 2002 e 2003. Entretanto após o máximo, quando o ciclo 23 decresce e depois no mínimo entre os ciclos 23 e 24, o semidiâmetro mostra sempre uma tendência de crescimento. Há uma descontinuidade de dados e os valores em torno desta descontinuidade podem estar fornecendo valores mais erráticos em função da diminuição de dados. Se retirarmos da série os três ou quatro valores mais próximos, antes e depois deste hiato, veremos que a série apresenta boa continuidade.

A verossimilhança e as consequências dessa possível longa peridiocidade são detalhadas no Capítulo 12.

## 4. Efeitos instrumentais.

Muito embora a boa qualidade metrológica, os valores obtidos de semidiâmetro solar podem conter uma variedade de efeitos decorrentes de inúmeros fatores. Alguns erros são decorrentes da forma como se observa, outros são introduzidos pelo próprio instrumento, outros ainda pela atmosfera, outros pelas condições meteorológicas além de outros fatores. A nosso favor temos uma grande quantidade de observações que se distribuem normalmente em torno de uma tendência (ver Figura 3.4). A grande quantidade de valores observados pode ser útil também para se detectar um comportamento nos resultados que tenha sofrido a influência de determinado fator. Assim, uma inspeção adequada pode apontar um tipo de influência que tenha alterado de alguma maneira os valores medidos.

Apontamos a seguir os efeitos de natureza instrumental que introduzem erros nos cálculos do semidiâmetro solar observado pelo astrolábio do Observatório Nacional, os quais foram detectados e suas influências foram retiradas.

**4.1 - As molas do prisma -** O astrolábio do Observatório Nacional foi adaptado para observar o Sol e pode observá-lo em qualquer distância zenital escolhida entre 25º e 57º. Evidentemente as observações próximas ao horizonte não são de interesse, enquanto que as zenitais são proibidas pela configuração prismática. De qualquer forma não temos muitas observações perto do horizonte (H<30º). Os efeitos sistemáticos, se houvessem, apareceriam no ajuste de parâmetros das correções sistemáticas. Para que o astrolábio possa observar em diversas posições os espelhos do prisma objetivo devem ser deslocados de um certo ângulo, correspondente ao dobro da distância zenital até serem fixados em uma posição onde se possa observar um almicantarado desejado. Para se manter o ângulo escolhido, um conjunto de molas é utilizado. Algumas molas atuam no sentido de abrir o prisma e outras no sentido de fechá-lo. Se durante a observação, que dura poucos minutos, uma das molas sofrer alguma alteração, haverá um erro de leitura que é tanto maior quanto maior for o tempo de observação. Entretanto este erro é diferente para os lados Leste e Oeste, isto é, para as observações feitas antes do meridiano e depois do meridiano local. Isto ocorre porque no primeiro caso o Sol está diminuindo sua distância zenital e no outro caso o

Sol está aumentando a distância zenital, enquanto que a mola alterada atua num só sentido, ou de fechar ou de abrir o ângulo entre os prismas, fazendo com que a distância zenital observada, em caso de deriva, seja respectivamente aumentada em um dos lados e diminuída no outro. Assim, para um dos lados haverá um acréscimo de tempo para o Sol cruzar a linha de almicantarado desejado, enquanto que para o outro haverá um decréscimo. Esta diferença é de alguma forma descrita por uma função do tempo de observação.

O tempo de observação depende basicamente da estação do ano, mas um pouco também das horas do dia. As observações são mais curtas quanto mais próximas estão do solstício de verão, quando duram pouco mais de dois minutos e são mais longas quanto mais próximas do solstício de inverno quando podem durar até cerca de seis minutos. São também mais longas quanto mais próximas da passagem meridiana e mais curtas quanto mais afastadas estão desta passagem. Isto ocorre porque o diâmetro medido é sempre o diâmetro vertical do Sol enquanto o astro cruza um círculo de distância zenital fixa. Próximo ao solstício de verão o Sol traça grandes arcos diários que se aproximam do zênite fazendo uma trajetória em muito ascendente ou descendente, enquanto que próximo ao solstício de inverno o Sol traça arcos menores distantes do zênite fazendo trajetórias pouco ascendentes ou descendentes. E quando mais distante da passagem meridiana o Sol faz uma trajetória mais ascendente ou descendente do que quando está próximo daquela passagem. A Figura 4.1 exemplifica a duração das observações para o ano de 2003. Na figura, em algumas datas, podem ser vistas todas as observações ali efetuadas, às vezes até separadas em duas sessões, a sessão feita a Leste e a sessão feita a Oeste do meridiano local.

Se há deriva durante a observação da passagem do Sol pela altura zenital desejada as faces dos prismas tendem a se fechar, diminuindo o ângulo entre elas, fazendo com que a distância zenital observada, que deveria ser fixa, diminua ao longo da observação. Assim os valores de semidiâmetro do Sol a Leste são ligeiramente aumentados e, os valores a Oeste, diminuídos. E tais desvios são proporcionais aos tempos de observação uma vez que a ação das molas é continua durante o tempo. Embora tal desvio seja muito pequeno, ainda assim, pode ser detectado nas observações do semidiâmetro do Sol. E há evidências de que tal fato ocorreu no astrolábio do Observatório Nacional, sem dependência do processo observacional, não durante todo o tempo, mas durante alguns intervalos de tempo. As manutenções e a manipulação constante do instrumento acabam determinando os períodos em que isto ocorre.

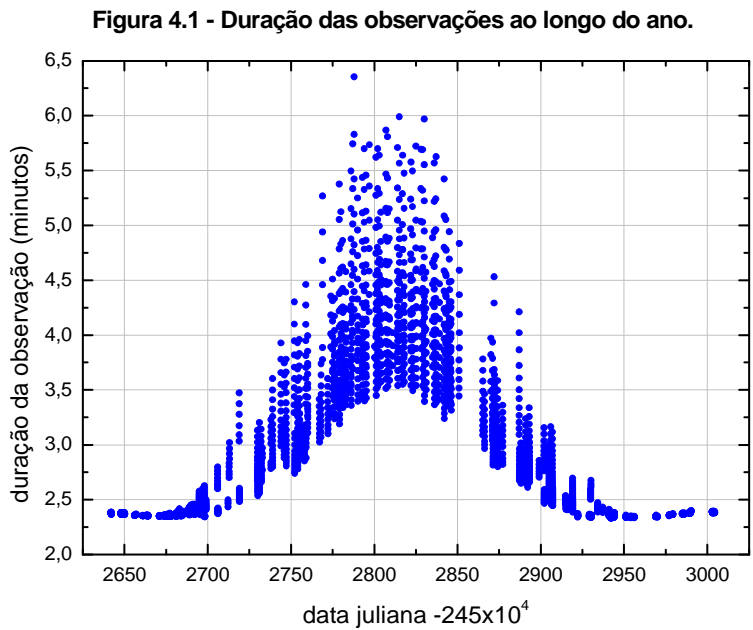

Figura 4.1 - Duração das observações em minutos para o ano de 2003.

A comparação entre observações a Leste e a Oeste feitas durante estes intervalos permite encontrar uma correlação linear entre o valor do semidiâmetro observado e a duração da respectiva medida. São encontrados dois valores similares e opostos, uma tendência proporcional à duração da observação que provoca o aumento dos valores medidos a Leste, e que tende a diminuir os valores lidos a Oeste. Os coeficientes são encontrados utilizando-se a técnica de mínimos quadrados e as correções podem ser aplicadas, recuperando os dados de cada período. Estes coeficientes que relacionam linearmente as variações do semidiâmetro solar e a duração das observações podem ser expressos pela equação 4.1. onde [$\Delta$SD] é a variação do semidiâmetro, [K] é o coeficiente calculado e [$\Delta$t] a duração da observação. Quando são encontrados coeficientes semelhantes e de sinais opostos a cada um dos lados, há a certeza de uma deriva das molas do prisma.

$$\Delta SD = K \cdot \Delta t \qquad (4.1)$$

Entre abril e outubro de 2000 o coeficiente encontrado foi de $(8,3\pm0,6).10^{-4}$ segundos de grau por segundo de tempo. Aplicada esta correção o desvio padrão dos valores de semidiâmetro solar caiu de 0,596 segundos de grau para 0,572 segundos de grau. (Reis Neto, 2002) Entretanto só foi possível encontrar estes erros porque foram denunciados pelo observador que sempre muito atento ao seu trabalho percebeu o afastamento das séries lidas a Leste e a Oeste.

Isto voltou a ocorrer em 2001 e um procedimento diferente foi adotado. Uma vez aplicada a correção o desvio padrão dos valores de semidiâmetro solar caiu de 0,610 segundos de grau para 0,594 segundos de grau.(Boscardin, 2004)

Em 2002 e 2003 aparece um descolamento entre os dados observados a Leste e aqueles observados a Oeste. Entretanto eles não puderam ser modelizados pela equação 4.1. Assim, foram corrigidos por um procedimento estatístico que considerou separadamente os valores observados a Leste e a Oeste. Cada lado de cada um dos anos foi dividido em pequenos intervalos de tempo muito aproximadamente de 30 dias. E cada um destes intervalos foi corrigido em função dos desvios que apresentavam para a tendência anual, em relação a uma distribuição normal. As tendências eram iguais a -1,0922 milisegundos de grau por dia a Leste e -0,7678 milisegundos de grau por dia a Oeste em 2002 e iguais a +0,1274 milisegundos de grau por dia a Leste e +0,1994 milisegundos de grau por dia a Oeste em 2003. A correção aos valores em cada período foi feita de modo a não alterar o valor trazido pela média ponderada de cada período. (Boscardin, 2005)

Nos demais anos não se observou a ocorrência de erros causados pelo efeito da mola do prisma.

**4.2 - A falta de horizontalidade -** Um defeito de nivelamento do aparelho observador pode ter influência na medida do semidiâmetro solar em função do azimute de observação do Sol. Um pequeno desvio da vertical do eixo em torno do qual o aparelho gira no seu apontamento azimutal pode causar erros de observação que dependerão do ângulo de apontamento. O efeito é de segunda ordem ligado à variação de focalização durante a observação e depende do azimute e de sua variação.

Os ângulos de azimute são medidos a partir da direção Sul e no sentido Sul, Oeste, Norte, Leste, retornando ao Sul. Assim, azimute zero aponta para o Sul, o azimute 90º aponta para Oeste, o azimute 180º aponta para Norte e o azimute 270º aponta para Leste. Ao procurar alguma influência deste ângulo na medida do semidiâmetro do Sol, não podemos considerar, os valores de ângulo, que não têm nenhum significado, mas sim, o desvio deste ângulo para uma direção, por exemplo, ao Norte. Ou utilizando o seno ou coseno destes ângulos. Se pudermos medir alguma correlação entre o seno do azimute observado e o semidiâmetro medido isto pode ser indicativo da ocorrência deste defeito que pode ser então corrigido.

Os dados do semidiâmetro solar do ON do ano de 2001 foram corrigidos deste efeito. Foram calculados polinômios de segundo grau ajustados aos desvios causados em função da diferença angular entre o azimute observado e a direção Norte. Os coeficientes de primeiro grau achados foram de $(6,117\pm1,606)\times10^{-2}$ segundos de grau por grau para os valores a Leste e de $(-4,896\pm1,445)\times10^{-2}$ segundos de grau por grau para os valores a Oeste. Os coeficientes para o segundo graus foram de $(-1,544\pm0,394)\times10^{-5}$ segundos de grau por grau ao quadrado para os valores a Leste e de $(1,230\pm0,354)\times10^{-5}$ segundos de grau por grau ao quadrado para os valores a Oeste. O desvio padrão dos valores corrigidos caiu de 0,591 segundos de grau para 0,580 segundos de grau. Nos demais anos de observação não foram encontrados erros causados por este efeito (Boscardin, 2005).

**4.3 - Comprimento de onda** - As discrepâncias entre as medidas do semidiâmetro solar são em muito devidas a diferenças nas características dos instrumentos, na faixa espectral de observação. Quando se corrige as medidas dos efeitos introduzidos pelas propriedades de cada instrumento reduzem-se as diferenças entre os raios medidos a um nível de incerteza relativo a cada instrumento. A Tabela 4.1 mostra as diferenças de características entre diversos instrumentos que mediram o diâmetro solar nos últimos anos (Djafer, Thuillier e Sofia, 2008).

Por esta tabela pode-se ver que alguns observam no contínuo, em diferentes comprimentos de onda, e outros no centro das linhas de Fraunhofer. Há instrumentos que usam uma banda estreita e outros uma larga faixa. A dependência das medidas com o comprimento de onda não foi rigorosamente verificada uma vez que os únicos instrumentos que mediram o diâmetro solar em diferentes comprimentos de onda foram o Coudé de Locarno na Suíça e o de Izaña em Tenerife.

Tabela 4.1 – O semidiâmetro em função de características dos instrumentos.

| Local e instrumento | λ (nm) | Δλ (nm) | Período | D (cm) | $r$ (arcsec pixel$^{-1}$) | $R$ (arc sec) | P | Referência |
|---|---|---|---|---|---|---|---|---|
| Mount Wilson | 525.02 | 0,014 | 1970-2003 | 30,48 | 9,6 x 13,1, 12,9 x 20,1[a] | 959,486 ± 0,005 | B | Ulrich & Bertello 1995, Lefebvre et al. 2006 |
| Calern: | | | | | | | | |
|   11 prismas[b] | 540 | 200 | 1989-1995 | 10 | 0,60 | 959,590 ± 0,010 | C | Laclare et al. 1999 |
|   11 prismas | 538 | 200 | 1996 | 10 | 0,74 | 959,360 ± 0,030 | C | Sinceac et al. 1998 |
|   Prismas[d] | 850 | 160 | 1996 | 10 | 0,74 | 959,385 ± 0,035 | C | Sinceac et al. 1998 |
|   11 prismas | 538 | 200 | 1996-1997 | 10 | 0,74 | 959,630 ± 0,080 | C | Chollet & Sinceac 1999 |
|   DORaySol | 548 | 60 | 2001 | 10 | 0,50 | 959,509 ± 0,014 | C | Andrei et al. 2004 |
| Rio de Janeiro: | | | | | | | | |
|   Prismas[e] | 563.5 | 168 | 1997-1998 | 10 | 0,50 | 959,200 ± 0,020 | C | Jilinski et al. 1999 |
| | | | 2001 | 10 | 0,50 | 959,190 ± 0,013 | C | Andrei et al. 2004 |
| Antalya: | | | | | | | | |
|   2 prismas | 550 | 180 | 1999-2000 | 10 | 0,78 | 959,030 ± 0,070 | C | Gölbaşi et al. 2001 |
| | | | 2001-2003 | 10 | 0,78 | 959,290 ± 0,010 | C | Kiliç et al. 2005 |
| Boulder(SDM) | 800 | 10 | 1981-1986 | 10 | 1 | 959,680 ± 0,018 | E | Brown&Christensen-Dalsgaard 1998 |
| Locarno/Izaña | 475.8 | 2.1 | 1997 | 45 | 0,179 | 959,73 ± 0,050 | A | Wittmann 1997 |
| | 486 | 2.1 | 1997 | 45 | 0,179 | 959,81 ± 0,030 | A | Wittmann & Bianda 2000 |
| | 583 | 2.1 | 2000 | 45 | 0,179 | 960,09 ± 0,040 | A | Wittmann & Bianda 2000 |
| Kitt Peak | 329.8-660 | 2 | 1981 | 152 | 0,925 | 959,620 ± 0,030 | C | Neckel 1995 |
| SDS | 620 | 80 | 1992-1996 | 13 | 0,128 | 959,561 ± 0,111 | E | Egidi et al. 2006 |
| | | | | | | 959,658 ± 0,091 | C, E | Djafer et al. 2008 |
| MDI | 676,78 | 0,0094 | 1996-2006 | 15 | 2 | 959,283 ± 0,150 | C | Kuhn et al. 2004 |

λ = comprimento de onda, Δλ = banda passante, D = diâmetro do telescópio, r = resolução espacial, R = raio do Sol, P = método:
(A) o limbo solar é definido pelo ponto de maior derivada; (B) = o limbo solar é definido pelo ponto com 25% da intensidade do centro do Sol;
(C) = o limbo solar é definido como o baricentro dos pontos em torno do máximo da primeira derivada; (E) = o limbo solar é determinado por uma trabformada de Fourier
a - O instrumento usa duas resoluções, os dados são combinados para usar a melhor resolução.
b - Medida analógica feita por CCD.
c - A abertura real do telescópio é de 5 cm x 8 cm.
d - Prisma variável motorizado.
e - Prisma variável não motorizado.

No caso do ON, com um só instrumento, tal correção não se aplica. Mesmo quando séries de diferentes astrolábios são combinadas no Capítulo 12, há uma correção de patamar que assume este tipo de efeito, caso se verificasse.

**4.4 - Efeito da simetria das imagens** – Tomamos os dados do astrolábio do ON entre 1998 a 2009 e calculamos o desvio padrão dos diâmetros solares sobre médias corridas anuais compostas mês a mês. Pudemos então perceber que de 1998 a 2005 os desvios padrão oscilam em torno de 0,576 segundos de grau. De 2006 em diante, entretanto os desvios padrão tornam-se maiores passando a oscilar em torno de 0,725 segundo de grau. Isto mostra que os dados a partir de 2006 sofreram uma perda de qualidade. A Figura 4.4 mostra estes valores.

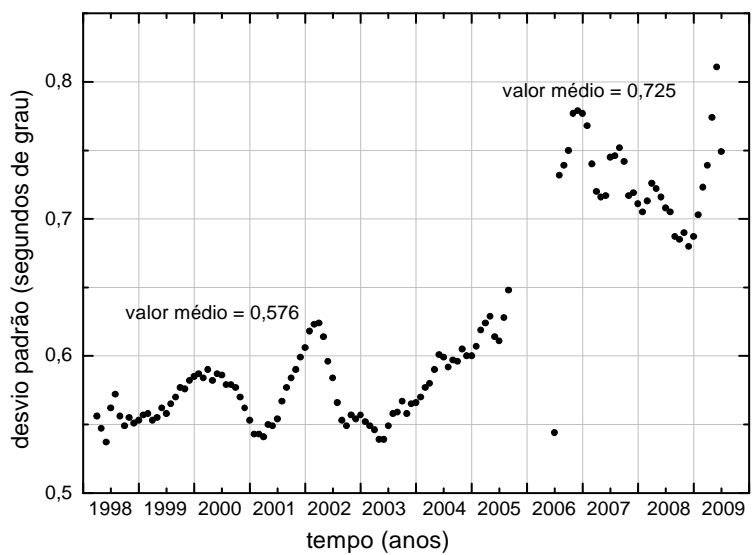

Figura 4.4 - Evolução dos desvios padrão dos valores do semidiâmetro solar. Os valores são médias anuais corridas a cada mês.

Entretanto em 2006 o astrolábio passou por uma manutenção e um novo programa de aquisição de dados foi desenvolvido. O astrolábio recebeu uma nova placa digitalizadora que permitiu maior rapidez ao descarregar os dados e, em consequência, a possibilidade de se observar mais imagens no mesmo intervalo de tempo. Próximo dos pontos de contato da imagem direta com a imagem refletida do Sol, obtém-se agora mais de 80 imagens, onde antes se obtinha apenas 46. Diante de tais melhorias era esperado um melhor desempenho do instrumento, mas, pelo contrário, e de maneira surpreendente, os desvios padrão aumentaram.

Esta descontinuidade dos valores de desvio padrão, a partir de 2005, está muito provavelmente associada aos novos equipamentos e métodos adquiridos para o astrolábio.

Retomando o fundamento das medidas, para se obter um valor do semidiâmetro solar, são observados os dois momentos em que cada um dos bordos do Sol, o bordo superior e o bordo inferior, cruza o mesmo paralelo de mesma altura zenital. Digamos que em cada um destes momentos 80 imagens são tomadas. Cada uma delas apresenta duas imagens do Sol, uma é a imagem direta e a outra é a refletida. A imagem direta é dirigida ao detetor enquanto que a imagem refletida só o faz após os raios do Sol refletirem em uma bacia de mercúrio. Quando se observa o Sol com o astrolábio sua imagens direta e refletida se deslocam igualmente em consequência de deslocamentos do astro ao longo dos azimutes. Mas elas se deslocam inversamente quando ele se desloca em distância zenital. Quando o primeiro bordo do Sol cruza determinado paralelo de distância zenital constante estas imagens se tocam. Depois as imagens se interpenetram. Quando o segundo bordo cruza o mesmo paralelo estas imagens se separam. Devemos registrar o momento exato em que estas imagens se tocam e depois o momento exato em que se separam. Este intervalo de tempo é o principal valor para o cálculo do semidiâmetro do Sol.

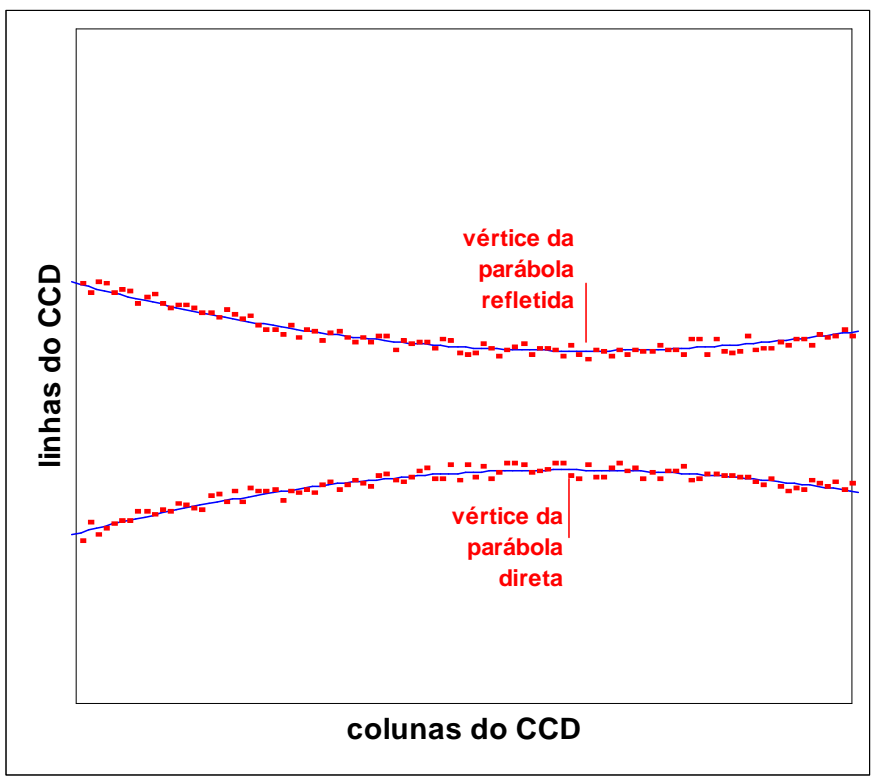

Figura 4.5 – As imagens dos bordos diretos e refletido do Sol definidos por um ponto a cada linha do CCD e as parábolas ajustadas a estes pontos. O ponto mais avançado é o vértice da parábola.

Próximo de cada um destes momentos são obtidas imagens dos dois bordos, o bordo direto e o bordo refletido. Estas imagens são a consequência do deslocamento do Sol durante a tomada das imagens. Os bordos então se aproximam, se tocam e se afastam. Duas parábolas são ajustadas aos pontos que definem os bordos do Sol. A Figura 4.5 mostra de forma esquemática as duas parábolas um instante antes de se tocarem e indica seus vértices que são os pontos extremos em distância zenital de cada uma das curvas. A câmera CCD está alinhada com a vertical de modo que suas linhas são paralelas ao eixo de variação das distâncias zenitais.

Quando estas parábolas se interpenetram é difícil separar os pontos que pertencem a ao bordo direto ou ao bordo refletido do Sol. Por isso, o exato momento de toque das parábolas é perdido por conta de imagens confusas. Mas podemos calcular este instante construindo a curva do avanço dos vértices das parábolas em função do tempo decorrido. O esquema da Figura 4.6 ilustra isto. Pode-se ver que os pontos do momento do toque não são disponíveis, mas o instante central pode ser muito bem avaliado pelo ajuste de retas aos demais pontos.

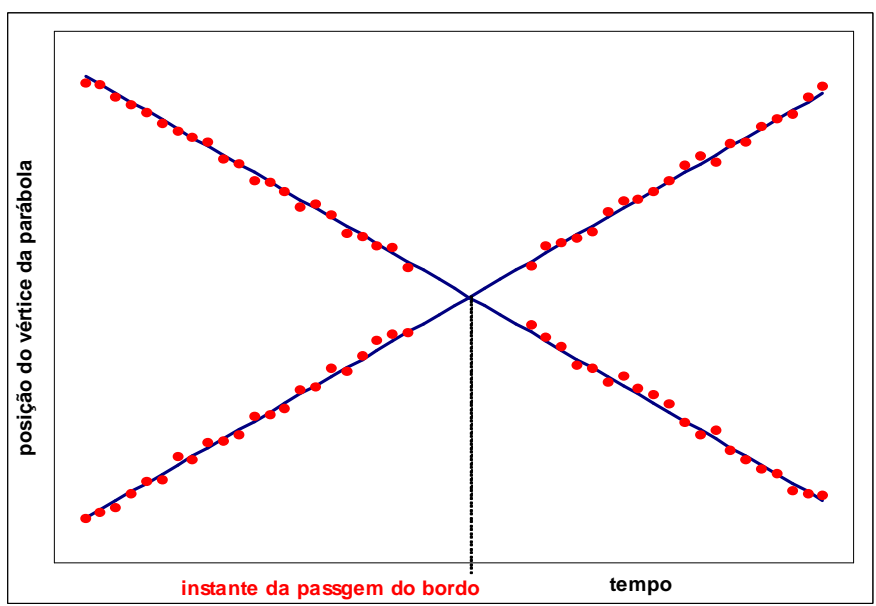

Figura 4.6 – Evolução temporal dos vértices das parábolas ajustadas aos bordos direto e refletido do Sol.

O ponto de toque destas retas tem um erro associado que decorre do erro de determinação dos vértices das parábolas. No esquema teórico da Figura 4.7 mostramos como se forma o erro de definição do instante de toque das retas. Neste esquema construímos duas retas a partir de apenas dois pontos para cada uma. Entretanto no caso real há 46 pontos para cada reta. A definição de cada um destes pontos tem um erro associado.

As retas da evolução temporal dos vértices das parábolas estão sendo construídas a partir de 46 pontos embora, atualmente tenhamos disponíveis mais de 80 pontos. A solução indicada para dar continuidade ao programa de análise e não introduzir desnecessariamente um viés sobre os resultados foi continuar a definir as retas com apenas 46 pontos.

A indeterminação na definição do momento de toque é a diferença entre dois momentos, um onde se supõem não haver erros na definição do vértice das parábolas e outro onde estes pontos, que definem as retas, têm um erro associado, conforme mostra a Figura 4.7.

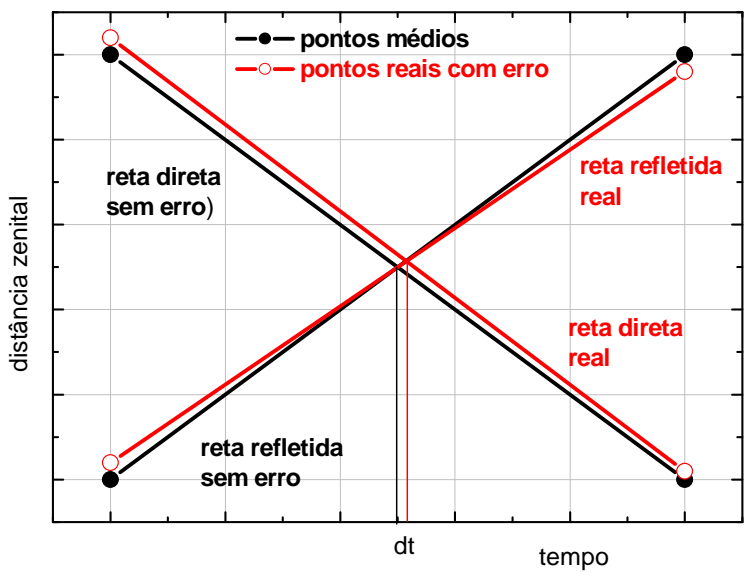

Figura 4.7 – Duas retas passando pelos pontos médios e duas retas reais definidas por pontos com erros. O esquema mostra como ocorre o erro na definição do tempo de toque das duas imagens do Sol.

Podemos simular esta indeterminação para diversas situações. Conhecemos bem o erro de definição dos pontos que traçam as retas a partir das observações. Tipicamente o desvio padrão dos pontos que definem as parábolas é de 1,037 segundos de grau para a parábola direta e de 1,093 segundos de grau para a parábola refletida.

Sorteando aleatoriamente dois valores para cada ponto e usando a transformação de Box-Muller, podemos definir cada ponto que desvia aleatoriamente da média segundo uma distribuição normal. A transformação de Box-Muller permite obter um número aleatório com distribuição normal a partir de dois números aleatórios [u] e [v], sorteados de uma distribuição uniforme entre 0 e 1. O resultado da posição do ponto em torno da média e com desvio padrão igual à unidade é igual a [x] definido pela equação 4.2

$$x = [(-2{,}0 \cdot \log(u)) \cdot \cos(2 \cdot \pi \cdot v)]^{1/2} \qquad (4.2)$$

Sorteamos então 92 números para a reta direta e 92 números para a reta refletida. Obtemos então os valores de desvio que são multiplicados pelos desvios padrão das parábolas correspondentes definindo a posição de 46 pontos que vão traçar cada uma das duas retas. Uma regressão linear para cada coleção de pontos define as duas retas conforme o esquema a seguir.

Sejam retas cujas equações são dadas por 4.3, onde [a] e [b] são definidos pelas equações 4.4 e 4.5. Na realidade há duas retas, cujos parâmetros são [$a_d$] e [$b_d$] para a reta direta e [$a_r$] e [$b_r$] para a reta refletida. O momento de encontro destas retas é dado pela equação 4.6. A Figura 4.8 mostra os pontos sorteados e as retas definidas, cada uma a partir de 46 pontos e o cálculo do momento de toque das duas retas de acordo com a equação 4.6.

$$y = a \cdot x + b \qquad (4.3)$$
$$a = (46 \cdot \Sigma xy - \Sigma x \cdot \Sigma y) / (46 \cdot \Sigma x^2 - (\Sigma x)^2) \qquad (4.4)$$
$$b = (\Sigma y - a \cdot \Sigma x) / 46 \qquad (4.5)$$
$$t = (b_r - b_d) / (a_d - a_r) \qquad (4.6)$$

Por um processo de Monte Carlo, este procedimento foi repetido dez mil vezes e os momentos em que as retas se cruzam tiveram um valor médio e uma flutuação que foi medida pelo desvio padrão dos valores. Este desvio padrão é o valor que o tempo medido desvia do valor médio. Admite-se que este desvio é o erro médio cometido na leitura do tempo. Verificamos, entretanto que as curvas obtidas tinham algumas irregularidades. Um estudo destas condições mostrou que cem mil repetições tornam estas irregularidades imperceptíveis.

A coleção de 46 pontos que definem as retas foi continuamente deslocada da situação de simetria para cada um dos lados gerando incertezas progressivamente crescentes no instante de toque das retas à medida que a situação de simetria era comprometida. Como em algumas situações não é possível aproveitar as imagens centrais dos dois bordos, que se confundem, analisamos também as configurações em que são eliminados dois, quatro ou seis pontos centrais, os quais são substituídos por outros pontos. A Figura 4.9 reproduz os principais resultados. Nela podemos ver como o erro do cálculo do ponto de toque aumenta consideravelmente com a falta de simetria das imagens escolhidas. O erro diminui quando se retiram algumas imagens centrais.

Ao se dispor então de 80 imagens no mesmo espaço de tempo em que se dispunha antes de 46, o erro que o operador comete para encontrar o exato momento de toque das imagens aumenta consideravelmente. Este erro não aumenta com o tempo, mas com o número de imagens deslocadas do centro. O número de imagens deslocadas do centro é determinante no erro de definição do momento de toque.

Todos estes cálculos foram feitos considerando-se o Sol observado em um azimute afastado de 35 graus a partir da direção Norte. A variação temporal da distância zenital do Sol, entretanto depende do azimute do Sol e da latitude de observação de acordo com a equação 4.7 onde [Z] é a distância zenital, [A] o azimute de observação e [L] a latitude local. Nesta equação [dZ/dt] é dado em segundos de grau por segundos de tempo.

$$(dZ / dt) = 15 \cdot \cos(L) \cdot \text{sen}(A) \qquad (4.7)$$

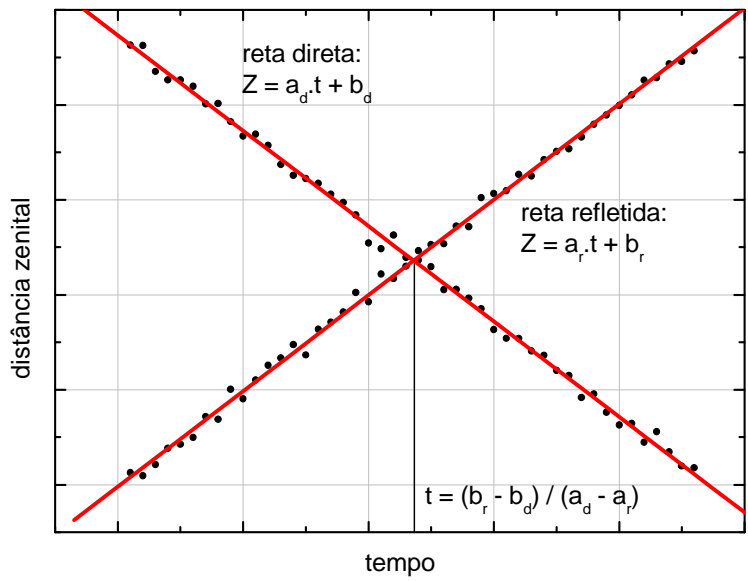

Figura 4.8 – A evolução temporal dos vértices das parábolas ajustadas aos bordos das duas imagens do Sol definem duas retas. O momento em que estas retas se tocam é o momento de toque das duas imagens.

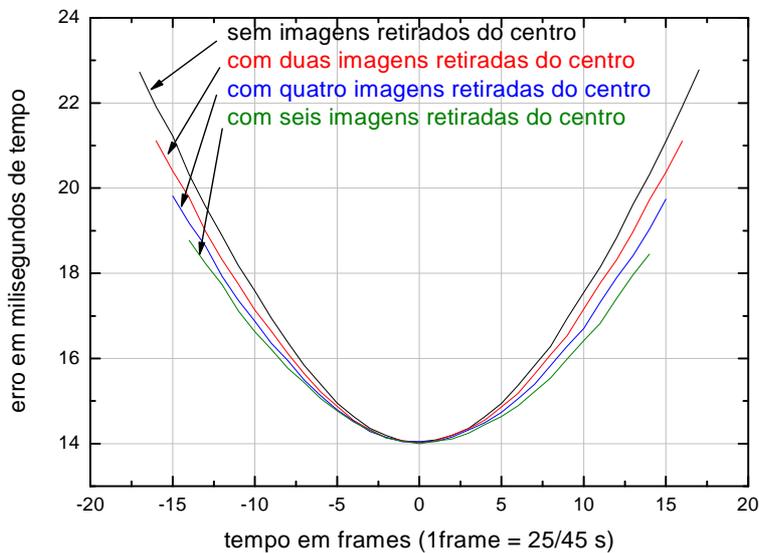

Figura 4.9 – O erro do momento de toque das imagens do Sol em função da falta de simetria dos pontos que definem as retas direta e refletida.

O valor de azimute do Sol observado nos dados do ON mais recorrente é entorno de A=±35º. Tomamos então este valor para todos os cálculos anteriores. O valor do sinal indica apenas, quando [dZ/dt] é positivo, que o Sol está subindo em direção à passagem meridiana ou quando ele é negativo que o Sol está descendo após a passagem meridiana. Na Figura 4.10 apresentamos o gráfico com a distribuição azimutal das observações do Sol no ON. O zero indica o ponto cardeal Norte, e os outros números indicam quanto a posição do azimute se afasta do Norte em graus. Pode-se ver que poucos pontos se aproximam menos do que 35 graus do ponto cardeal Norte. O número de observações entre -35 e 35 graus é pequeno, apenas 19%, e entre -30 e 30 graus é de apenas 8,3%.

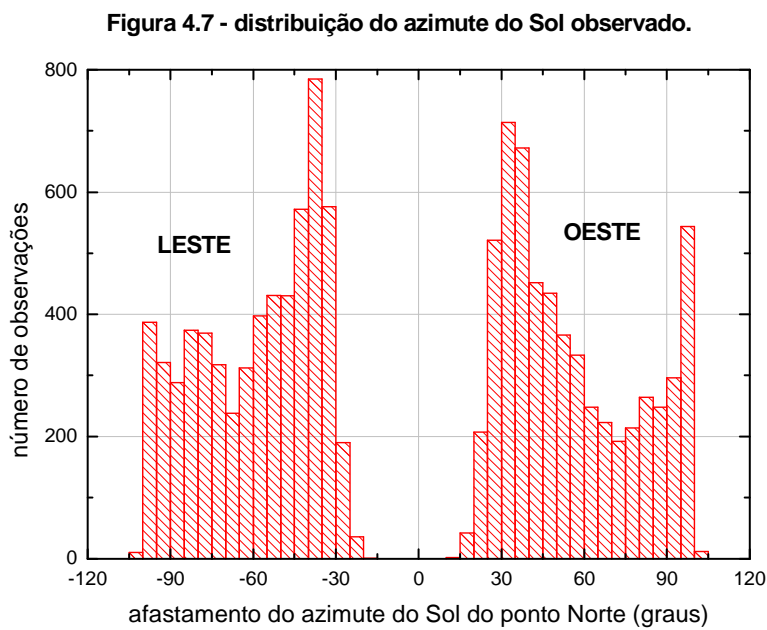

**Figura 4.7 - distribuição do azimute do Sol observado.**

Figura 4.10 - Distribuição do azimute do Sol observado.

A Figura 4.11 mostra que o erro na definição do ponto de toque cresce muito para azimutes próximos do ponto cardeal Norte. O erro cresce porque a variação temporal zenital depende do azimute. Felizmente as observações do Sol no ON não ocorrem naqueles azimutes e poucos pontos estão sujeitos a erros maiores que os calculados nas simulações.

Com base neste trabalho estamos propondo uma mudança no cálculo do semidiâmetro solar a partir das imagens que se têm disponíveis. Este cálculo é feito por meio de um programa que utiliza estas imagens. Estamos propondo algumas modificações neste algoritmo no sentido de escolher imagens mais simétricas em relação ao ponto de toque de modo a se obter erros menores no cálculo do momento de toque das parábolas que definem os bordos do Sol. As modificações necessárias foram propostas e oportunamente serão implementadas.

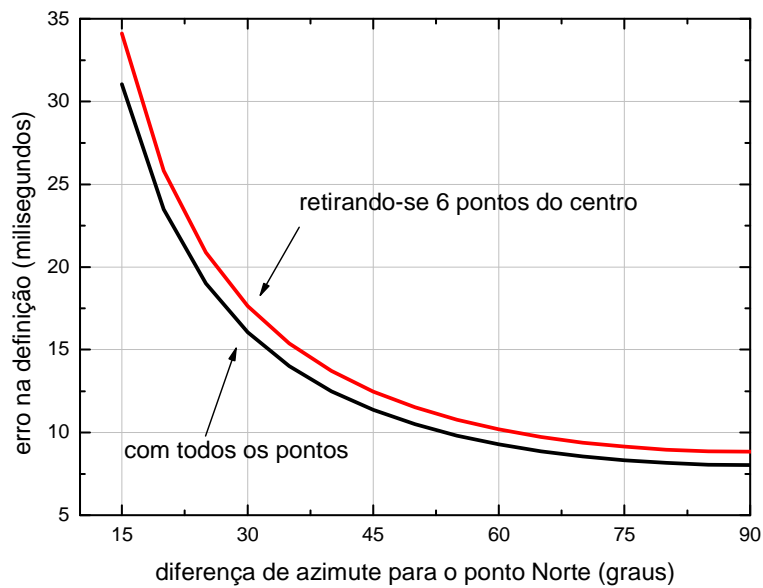

Figura 4.11 - Erro na determinação do ponto de toque em função do azimute do Sol observado.

# 5. Análise e correção de efeitos sistemáticos observacionais

Diversos fatores externos e independentes das observações podem afetar os seus resultados. Entre estes podemos citar aqueles comuns à maioria das observações astronômicas como a temperatura ambiente, a pressão atmosférica durante a observação e a turbulência da atmosfera. Mas há fatores mais específicos às observações astrométricas do diâmetro solar como a distância zenital do Sol observado, a heliolatitude observada, a duração da observação, o lado da observação, se a Leste ou a Oeste do meridiano local, uma equação pessoal para os instrumentos mais antigos, a sazonalidade e vários outros.

No caso do astrolábio do Observatório Nacional, além destes fatores, nós investigamos também a possibilidade de outros fatores interferirem na observação. São eles: o fator de Fried, o desvio padrão das parábolas ajustadas ao limbo do Sol, o desvio padrão das retas ajustadas aos vértices destas parábolas e a largura do bordo solar observado. O fator de Fried descreve a qualidade do *seeing* da atmosfera. É definido como a razão do comprimento de onda observado, dividido pela largura a meia altura de uma imagem pontual espalhada pela ação da atmosfera. Ele pode ser calculado a partir dos dados da observação (Lakhal *et al.*, 1999).

**5.1 - Observações a Leste e a Oeste do meridiano local** - Observações feitas no mesmo dia e na mesma distância zenital, porém tomadas em hemisférios diferentes, isto é, a Leste e a Oeste, poderiam apresentar medidas diferentes porque seriam afetadas por condições instrumentais e atmosféricas bem diversas. A Tabela 5.1 apresenta este efeito detectado no astrolábio do observatório de Calern ao longo do intervalo de tempo de 1975 a 1994 (Laclare, Delmas e Coin, 1996).

Contrariamente à análise dos autores, a Tabela 5.1 não traz diferenças fundamentais entre os dados obtidos a Leste e a Oeste, além da flutuação estatística esperada pelo relativamente pequeno número de observações. As diferenças reportadas são reconciliáveis às respostas a fatores como a temperatura ou a turbulência atmosférica.

Tabela 5.1 – Semidiâmetro solar médio segundo o lado observado no astrolábio de Calern. Valores em segundos de grau (Laclare, Delmas e Coin, 1996).

| Ano | Média Anual Total | N | Média Anual a Leste | N | Média Anual a Oeste | N |
|---|---|---|---|---|---|---|
| 1975 | 959,88±0,07 | 62 | | | | |
| 1976 | 959,50±0,06 | 52 | | | | |
| 1978 | 959,31±0,04 | 118 | 959,21±0,05 | 66 | 959,44±0,06 | 52 |
| 1979 | 959,16±0,03 | 113 | 959,32±0,04 | 60 | 959,19±0,05 | 53 |
| 1980 | 959,38±0,03 | 168 | 959,35±0,04 | 95 | 959,42±0,05 | 73 |
| 1981 | 959,30±0,03 | 174 | 959,31±0,03 | 104 | 959,27±0,04 | 70 |
| 1982 | 959,30±0,03 | 238 | 959,29±0,03 | 145 | 959,30±0,04 | 93 |
| 1983 | 959,53±0,02 | 265 | 959,56±0,02 | 166 | 959,47±0,03 | 99 |
| 1984 | 959,38±0,02 | 364 | 959,38±0,02 | 219 | 959,37±0,03 | 145 |
| 1985 | 959,51±0,01 | 464 | 959,54±0,02 | 279 | 959,47±0,02 | 185 |
| 1986 | 959,57±0,02 | 370 | 959,60±0,02 | 225 | 959,52±0,03 | 145 |
| 1987 | 959,50±0,02 | 341 | 959,49±0,02 | 201 | 959,52±0,03 | 140 |
| 1988 | 959,42±0,02 | 414 | 959,42±0,02 | 258 | 959,42±0,03 | 156 |
| 1989 | 959,33±0,01 | 484 | 959,34±0,02 | 303 | 959,30±0,02 | 181 |
| 1990 | 959,38±0,02 | 353 | 959,37±0,02 | 254 | 959,41±0,04 | 99 |
| 1991 | 959,44±0,02 | 266 | 959,43±0,02 | 187 | 959,45±0,03 | 79 |
| 1992 | 959,40±0,02 | 293 | 959,40±0,02 | 220 | 959,39±0,03 | 73 |
| 1993 | 959,39±0,01 | 347 | 959,39±0,02 | 262 | 959,39±0,03 | 85 |
| 1994 | 959,47±0,02 | 267 | 959,44±0,02 | 192 | 959,56±0,04 | 75 |
| Média | 959,42±0,01 | 5153 | 959,42±0,01 | 3236 | 959,41±0,01 | 1803 |

**5.2 - Influência da distância zenital** - O método de alturas iguais aplicado aos bordos opostos do Sol garante que a refração diferencial não afeta as medidas de diâmetros verticais observados pelo astrolábio. A refração diferencial não afeta os diâmetros verticais, porque estes são observados à mesma altura. Ou seja, a refração atmosférica se reduz à refração diferencial, e esta não afeta as observações. É o fundamento da medida. Entretanto medidas feitas em distâncias zenitais diferentes são afetadas pela refração atmosférica crescente com o aumento da distância zenital. A Figura 5.1 mostra os desvios do valor médio do semidiâmetro como função da distância zenital no astrolábio de Calern (Laclare, Delmas e Coin, 1996).
.
Na série do Rio de Janeiro obtida com utilização do CCD estes desvios se traduzem em índices irrelevantes ou excessivamente ruidosos, ou seja, em significado estatistico. Assim sendo e uma vez que efeitos de refração em princípio não afetam as observações diferenciais à mesma distância zenital, a interpretação da Figura 5.1 se remete mais facilmente à própria análise dos mesmos autores a respeito de diferenças de respostas do olho humano e do CCD à condições de transparência atmosférica, conforme aqui examinaremos no Item 5.4.

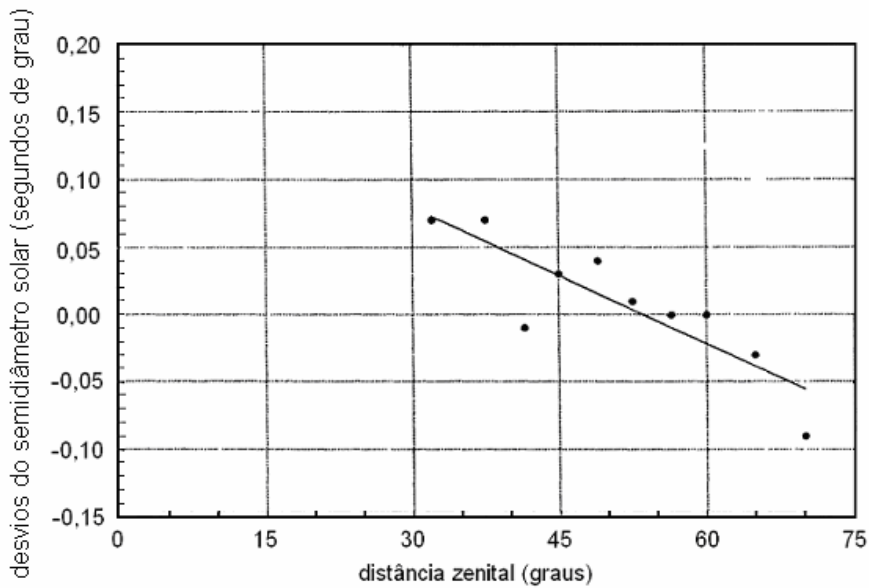

Figura 5.1 – Influência da distância zenital na medida do semidiâmetro solar (Laclare, Delmas e Coin, 1996.

**5.3 – O parâmetro de Fried –** O parâmetro que descreve a qualidade de *seeing* da atmosfera no astrolábio é o parâmetro de Fried denotado geralmente por [$r_0$] (Fried, 1966). Ele é definido como o comprimento de onda observado, dividido pela largura a meia altura de uma imagem pontual espalhada pela ação da atmosfera. No Astrolábio [$r_0$] pode ser calculado a partir dos dados. De fato, as trajetórias das imagens do Sol apresentam flutuações, as quais podem ser interpretadas como flutuações do ângulo de entrada. Sendo [$\sigma_s$] a média das flutuações sobre toda a pupila do Astrolábio cujo diâmetro é [d], o parâmetro de Fried é obtido da expressão 5.1 onde [$\lambda$] é o comprimento de onda observado e [$\sigma_s$] é expresso em segundos de grau. (Borgnino et al., 1982; Ricort, Borgnino, e Aime, 1982; Brandt, Mauter e Smartt, 1987).

$$r_0 = 8{,}25 \cdot 10^5 \cdot d^{-1/5} \cdot \lambda^{6/5} \cdot (\sigma_s^2)^{-3/5} = K_1 \cdot (\sigma_s^2)^{-3/5} \qquad (5.1)$$

Os diâmetros solares obtidos por métodos numéricos permitem calcular o parâmetro de Fried. O diâmetro medido com o astrolábio depende criticamente de uma boa determinação da trajetória das interseções nos dois trânsitos dos bordos solares pelo almicantarado. Por conta de flutuações observadas nestas trajetórias, o ponto de interseção não é bem definido. É o erro no instante desta determinação que causa o erro na medida do diâmetro solar. Usando-se técnicas de mínimos quadrados é possível determinar este erro. A Figura 5.2 mostra o erro de medida do semidiâmetro a partir de observações com câmera CCD onde somente as imagens diretas foram consideradas. (Irbah, Laclare, Borgnino e Merlin, 1994). Note-se que o erro causado pelo parâmetro de Fried depende fundamentalmente de cada instrumento e do sítio onde ele se encontra.

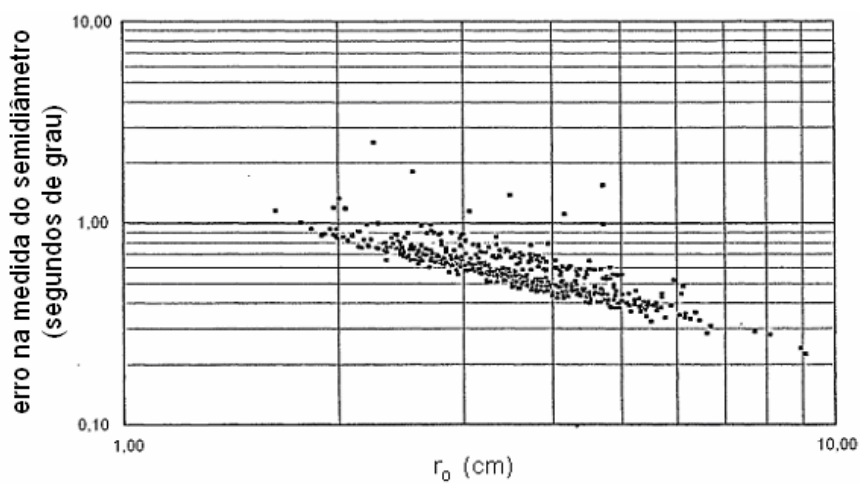

Figura 5.2 – Efeito causado pelo parâmetro de Fried (Irbah, Laclare, Borgnino e Merlin, 1994).

**5.4 – Turbulência atmosférica** - A turbulência da atmosfera modifica a imagem da forma do limbo solar. Usando o modelo solar HM98 (Hestroffer e Magnan, 1998), e o instrumento *Définition et Observation du Rayon Solaire* – DORaySol em Calern simulando o limbo solar médio e o modelo de Kolmogorov (Kolmogorov, 1941) para representar a atmosfera podemos obter a *Point Spread Function* - PSF da atmosfera e seus efeitos na determinação do ponto de inflexão da curva do limbo solar para vários valores do parâmetro de Fried. A Figura 5.3 mostra os efeito da atmosfera na PSF. A Figura 5.4 mostra os efeitos da atmosfera na forma do limbo solar (Djafer, Thuillier e Sofia, 2008).

Este efeito da turbulência atmosférica existe e nós o encontramos como pode ser visto no próximo item, mas ele é causado pelo espalhamento de luz sob dois meios diferentemente iluminados: a superfície do Sol e o fundo mais escuro do céu.

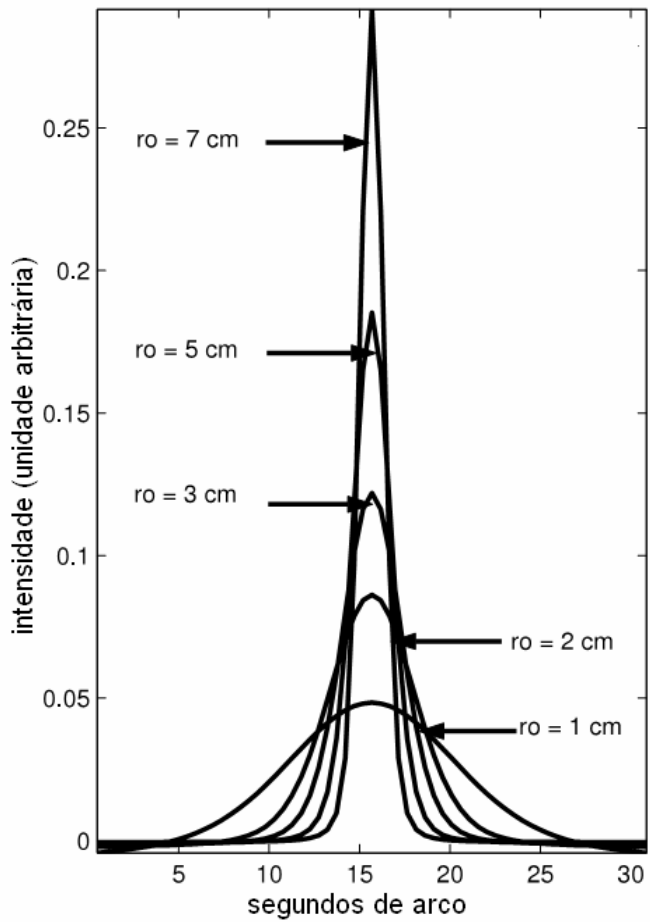

Figura 5.3 – PSF para vários valores do parâmetro de Fried (Djafer, Thuillier e Sofia, 2008).

**5.5 – Equação pessoal para as observações visuais** - As medidas mais antigas obtidas com os astrolábios foram feitas de forma visual. Ou seja, a interseção das duas imagens do bordo solar, a direta e a refletida, era estimada pelo observador. Assim, estas medidas necessitavam de ajustes diferentes de acordo com o observador que as tomou. Para se evitar efeitos do observador um sistema numérico de aquisição usando uma câmera CCD tem sido instalado. Para as observações com CCD o método numérico determina a localização dos bordos e analisa uma sequência de imagens para determinar o instante em que eles se tocam. No caso do Observatório Nacional, todas as medidas foram feitas com o auxílio da câmera CCD. Entretanto outros observatórios chegaram a fazer medidas visuais. A Figura 5.5 mostra que no observatório de Calern este efeito não é importante. Tal fato é uma combinação da experiência do observador e da resposta fisiológica de sua visão. Após a instalação de uma câmera CCD, algumas medidas foram ainda feitas de forma visual para comparação (Irbah, Laclare, Borgnino e Merlin, 1994).

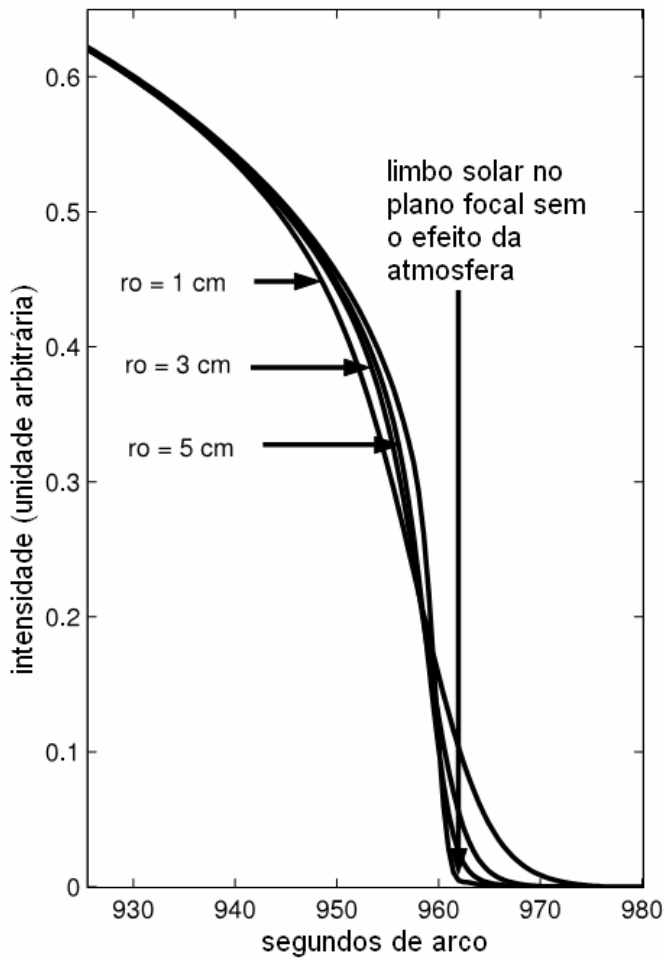

Figura 5.4 – O limbo solar para diversos valores do parâmetro de Fried (Djafer, Thuillier e Sofia, 2008).

A interpretação diferente do observador é causada pelo efeito da distância zenital nas observações visuais. Verifica-se uma diferença entre os dois métodos. Este efeito é devido à sensibilidade do olho do observador ao contraste de objetos. Por causa da absorção da atmosfera terrestre, há uma diminuição de contraste com o incremento da distância zenital. Este efeito pode ser visto na Figura 5.5 (Irbah, Laclare, Borgnino e Merlin, 1994). Tal como na série do Rio de Janeiro, a Figura 5.6 revela que as observações com CCD em Calern não mostram um efeito linearmente associado à variação de distância zenital.

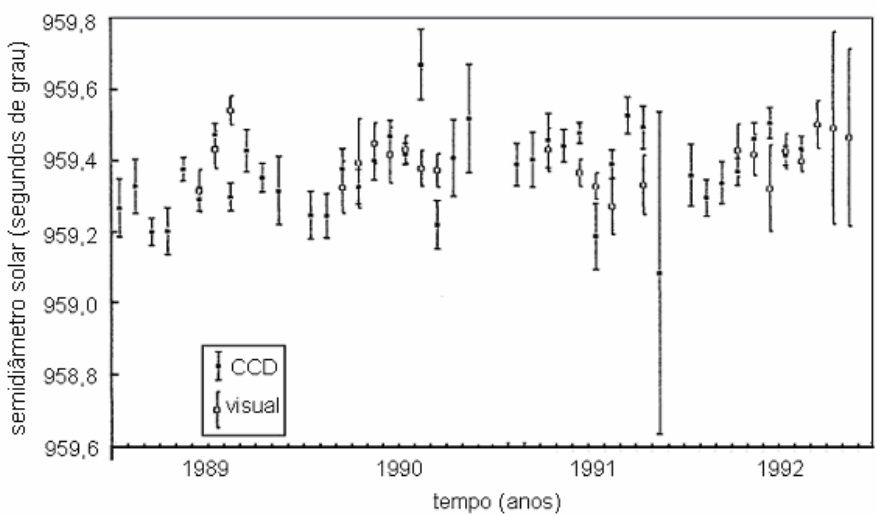

Figura 5.5 – O efeito do observador nas medidas do semidiâmetro solar (Irbah, Laclare, Borgnino e Merlin, 1994).

**5.6 - Observatório Nacional -** No Observatório Nacional nós corrigimos o efeito de diversos parâmetros sobre a série de dados do semidiâmetro solar. A modelagem dos efeitos dos parâmetros ligados a causas atmosféricas e climáticas é certamente muito complexa e possivelmente não linear. Não sendo do escopo deste trabalho definir tais funções, se é que é possível, optamos por procurar por relações lineares. Assim sendo, os coeficiente lineares impostos por estes efeitos devem ser diferentes dependendo da época em que sejam analisados, isto porque estamos observando um trecho de uma função complexa que é ajustado a uma relação linear. Por este motivo optamos por corrigir as séries anualmente.

As séries de dados observados a Leste e a Oeste mostram sofrer influências diferentes de alguns parâmetros, pois estão também em regimes diferentes daquelas funções. Os dados a Leste são muito mais ruidosos que os dados de Oeste, porque os primeiros são dados de observações feitas num momento em que a atmosfera está sofrendo a atuação de grandes gradientes de temperatura enquanto que os outros são dados de observação feitas num momento de maior estabilidade da atmosfera. Por isto optamos também por dividir a série em duas partes correspondentes aos dados observados em cada um dos lados do meridiano local.

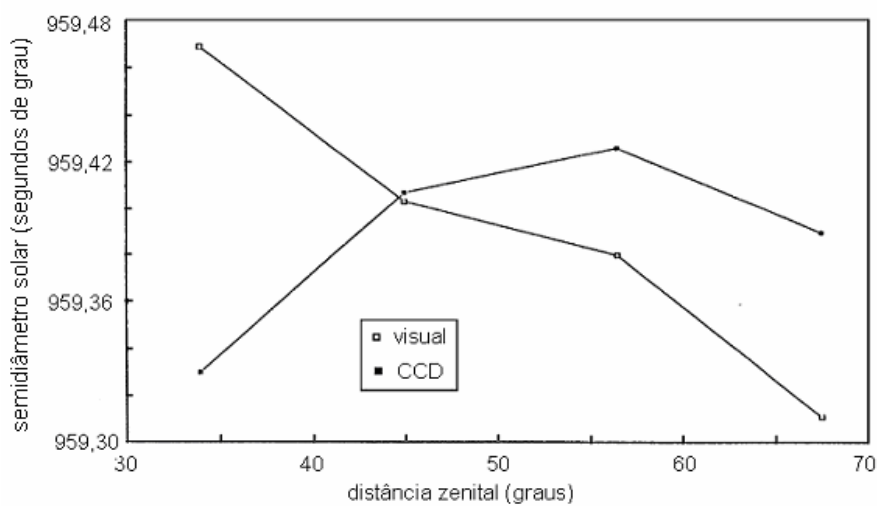

Figura 5.6 – Efeitos diferentes da distância zenital sobre as observações visual e com CCD (Irbah, Laclare, Borgnino e Merlin, 1994).

Para levantar a influência de diversos parâmetros sobre as medidas do semidiâmetro solar utilizamos o método de mínimos quadrados. Calculamos os índices de ajuste linear entre os parâmetros e o semidiâmetro solar. Para estes cálculos todos os parâmetros são normalizados, isto é, diminuídos de suas médias e divididos por seus desvios padrão. Ao selecionar os valores relevantes impusemos dois critérios. Que o coeficiente do parâmetro seja superior a um centésimo de segundo de grau por desvio padrão e que o coeficiente do parâmetro seja estatisticamente significativo, isto é, acima de três desvios padrão. Além disso, os índices só foram aceitos se, uma vez implementada a correção, a série resultante tenha o desvio padrão diminuído em relação à série não tratada.

A série corrigida tem todos os dados de semidiâmetro alterados pela subtração de um valor que é o produto do índice de ajuste pelo valor do parâmetro normalizado correspondente a cada observação do semidiâmetro. A correção é implementada com a utilização da equação 5.2 onde [∆SD] é a correção do semidiâmetro solar, [k] é o índice do ajuste linear do parâmetro e [µ] é o parâmetro normalizado (diminuído da média e dividido pelo desvio padrão).

$$\Delta SD = - k \cdot \mu \qquad (5.2)$$

Para os dados de 1998, 1999 e 2000 foram relevantes os índices relativos à temperatura durante a observação, à variação da temperatura durante a observação, ao fator de Fried e ao desvio padrão das parábolas ajustadas aos limbos da imagem do Sol. Os índices encontrados foram respectivamente de (0,050±0,006); (-0,023±0,006); (+0,032±0,006) e (-0,022±0,006) em segundos de grau por desvio padrão do parâmetro. Foram observados índices comuns aos dois lados de observação (Reis Neto, 2002).

Para os dados de 2001 foram relevantes os índices relativos à temperatura durante a observação, ao fator de Fried e ao desvio padrão das parábolas ajustadas aos limbos da imagem do Sol. As correções impostas ao semidiâmetro observado pelos parâmetros relativos foram: +0,00237; -0,07083; e -0,00213 em segundos de grau por desvio padrão do parâmetro. A aplicação destes parâmetros significou para a série uma diminuição do desvio padrão de 0,594 para 0,591 segundos de grau (Boscardin, 2004)

Os dados de 2002 foram corrigidos de modo diferente em cada um dos lados de observação. Os parâmetros considerados relevantes foram: a temperatura durante a observação, a variação da temperatura durante a observação, o fator de Fried e o desvio padrão do ajuste aos bordos das imagens do Sol. As correções impostas a Leste foram respectivamente: -0,0050; -0,0416; -0,0456 e -0,1125 todas em segundos de grau por desvio padrão do parâmetro. As correções impostas a Oeste foram respectivamente: -0,0141; -0,0083; -0,0275 e +0,0015 também em segundos de grau por desvio padrão do parâmetro (Boscardin, 2005).

Os dados de 2003 foram também corrigidos diferentemente a cada um dos lados de observação. Os parâmetros considerados relevantes foram: a temperatura durante a observação, a variação da temperatura durante a observação, o fator de Fried e o desvio padrão do ajuste aos bordos das imagens do Sol. As correções impostas a Leste foram respectivamente: -0,0763; -0,0307; +0,0113 e -0,1398 todas em segundos de grau por desvio padrão do parâmetro. As correções impostas a Oeste foram: +0,0630; +0,0117; -0,0091 e -0,0493 também em segundos de grau por desvio padrão do parâmetro (Boscardin, 2005).

A partir de 2004 a quantidade de dados observados foi menor e menos homogênea que nos anos anteriores. Assim foi impossível encontrar o ajuste linear para os efeitos dos parâmetros estudados. De qualquer maneira, estes ajustes sempre se mostram pequenos e sua aplicação, embora efetiva resulta em uma contribuição limitada diante dos erros observacionais e diante das tendências sistemáticas de origem solar de longo período.

A partir de 2004 o desvio padrão dos dados observados aumentou muito, algum tipo de erro deve ter sido introduzido na série. Conseguimos descobrir a causa de parte deste erro conforme relatamos no Capítulo 4. Estamos ainda a procura de outras causas. A falta de prática com a nova metodologia, a maior dificuldade de observação com a nova metodologia, o número muito reduzido de observações utilizando a nova metodologia (impedindo ganhar traquejo) explicam a pior qualidade das observações sob a nova metodologia. A aplicação dos parâmetros nesta parte de nossa série precisa de ser mais bem compreendida para ser efetivada. As parcelas a Leste e a Oeste da série apresentam-se muito afastadas e a aplicação da correção dos parâmetros aumenta ainda mais este afastamento.

A Tabela 5.2 apresenta um sumário dos índices lineares de correção dos efeitos de cada um dos parâmetros que foram usados para corrigir os dados observados do semidiâmetro solar do ON entre 1998 e 2003. Os valores apresentados são relativos às correções propostas e têm sinal oposto ao do índice de correlação encontrado. A correção segundo a temperatura nos diz de variações (intrínsecas e extrínsecas) do plano focal. A correção segundo a diferença de temperatura nos diz sobre modificações da distância zenital instrumental entre um bordo e outro.

Pode-se verificar que os parâmetros têm uma influência bem variável a cada época porque se referem a soluções locais de uma função mais complexa que engloba condições meteorológicas, estado da atmosfera e atuação destas variáveis sobre o todo e as partes do instrumento. Os índices são valores sempre pequenos; apenas dois casos ultrapassaram um pouco o décimo de segundo de grau por desvio padrão do parâmetro.

Tabela 5.2 – Índices de correção aplicados ao semidiâmetro observado relativos aos parâmetros.

| Ano e lado | 1998 a 2000 | 2001 | 2002-Leste | 2002-Oeste | 2003-Leste | 2003-Oeste |
|---|---|---|---|---|---|---|
| Temperatura | -0,050 | -0,0024 | -0,0050 | -0,0141 | -0,0763 | 0,0630 |
| Diferença de temperatura | 0,023 | 0 | -0,0416 | -0,0083 | -0,0307 | 0,0117 |
| Fator de Fried | -0,032 | 0,0708 | -0,0456 | -0,0275 | 0,0113 | -0,0091 |
| Desvio padrão das parábolas | 0,022 | 0,0021 | -0,1125 | 0,0015 | -0,1398 | -0,0493 |

Quando juntamos toda a série de 1998 a 2009 tínhamos apenas uma parte corrigida pelos parâmetros – de 1998 a 2003. Verificamos então a possibilidade de implementação de uma correção genérica, isto é, o uso de um único índice de correção para toda a série.

Verificamos que, tal como nos últimos anos da série, seria mais interessante fazer correções diferentes para cada um dos lados da série, isto é, os dados observados a Leste ou a Oeste do meridiano local. Os parâmetros que se mostraram relevantes para corrigir a série total foram: o fator de Fried, e a largura do bordo solar observado, ambos para os dados a Leste e a Oeste e a diferença de temperatura durante a observação, apenas para os dados observados a Leste. A largua do bordo basicamente reflete a disperção no ajuste dos limbos a parábolas (Sinceac, 1998). Os valores sugeridos para a correção são respectivamente: +0,058842; +0,121723 e +0,126271 para Leste e +0,054211 e +0,046032 para Oeste.

Os efeitos da correção dos parâmetros foram colocados no gráfico da Figura 5.7 que mostra a mudança do desvio padrão de cada um dos anos série de 2001 a 2009. O desvio padrão é sempre decrescente, ou tem um aumento insignificante, excetuado o ano de 2001. Note-se que em 2001 o efeito da variação do ângulo do prisma refletor (Item 4.1) é mais forte que no resto da série, de modo que sua modelização pela duração da observação não é satisfatória na análise global.

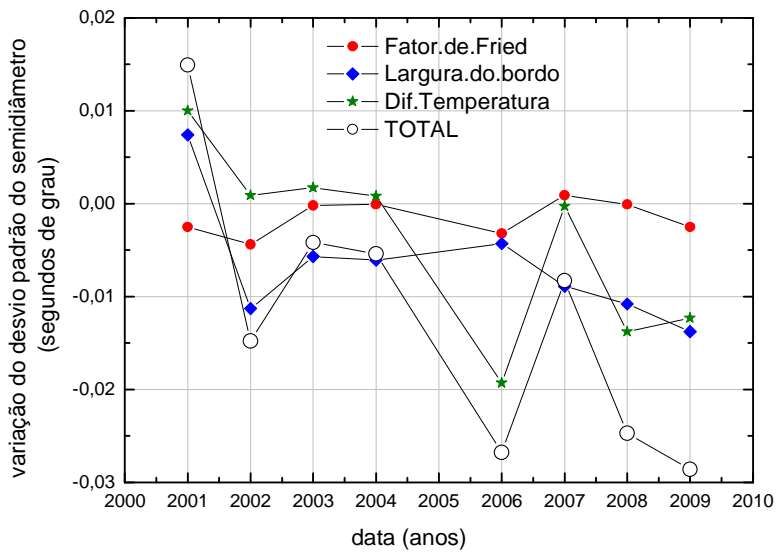

Figura 5.7 – Efeito da correção dos parâmetros no desvio padrão da série. Neste gráfico pode-se ver a atuação da correção devida a cada um dos parâmetros ano a ano da série

Outra maneira de se observar a atuação da correção dos parâmetros é através do gráfico da Figura 5.8. que mostra como o valor médio do semidiâmetro solar mudou ao longo dos anos com a introdução destas correções. O efeito de correção dos parâmetros é um efeito bastante pequeno, mas ao aplicarmos os mesmos índices para toda a série percebemos que as correções são maiores em alguns anos. Em nosso caso isto ocorre particularmente para o ano de 2001 e para os últimos anos. Pode-se ver que a principal causa para este fato é consequência da aplicação da correção da largura do bordo solar.

Podemos comparar as séries do semidiâmetro solar com e sem a correção dos parâmetros no gráfico da Figura 5.9. Os valores do semidiâmetro estão representados por médias corridas de doze meses. As maiores correções são da ordem de um décimo de segundo de grau, mas foram maiores em 2001 e tendem a aumentar o valor do semidiâmetro solar naquele ano e a diminuí-lo a partir de 2003.

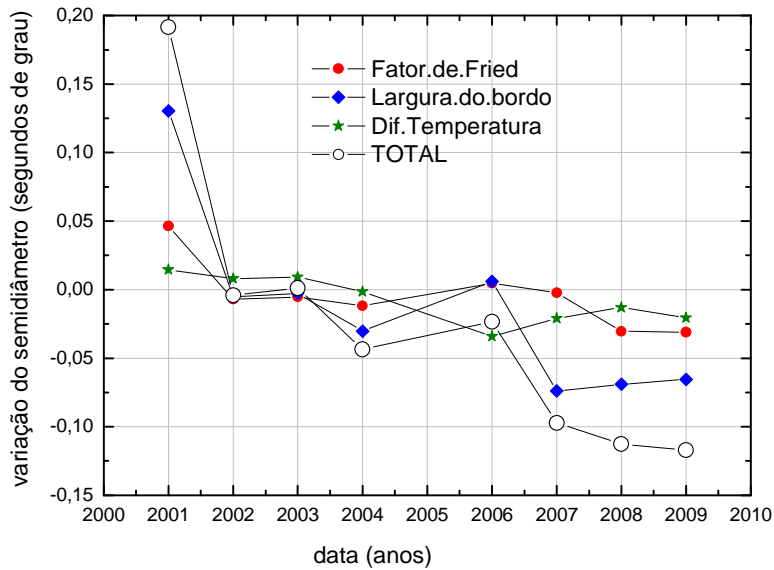

Figura 5.8 – Mudança das médias anuais do semidiâmetro em conseqüência da correção dos parâmetros.

A diferença entre a série corrigida e a série não corrigida é sempre menor que 0,1 segundo de grau, embora existam três trechos de médias caracteristicamente diferentes. De 1998 a 2000 onde não foi aplicada a correção global; de 2001 a 2003 onde foi aplicada e onde há grande densidade de medidas e de 2004 em diante onde a densidade e a homogeneidade de medidas é sensivelmente pior. No entanto o que importa caracterizar é que nos três trechos o desvio padrão relativo a diferença das séries é sempre muito pequeno, valendo 0,024 segundos de grau de 1998 a 2000; 0,062 segundos de grau de 2001 a 2003 e 0,018 segundos de grau de 2004 em diante. Ou seja, olhando a série como um todo se percebe que o uso de correções globais não modifica fundamentalmente a forma como a série se apresenta. Com exceção do ano de 2001 a série acompanha a atividade solar até 2003 e permanece crescente a partir de então, embora com valores um pouco menores. Percebemos que a correção dos parâmetros é importante quando feita localmente, isto é, separadamente para períodos anuais. Mas deixa de fazer sentido quando olhando a série como um todo. Por estes motivos descartamos de implementar estas correções de forma global.

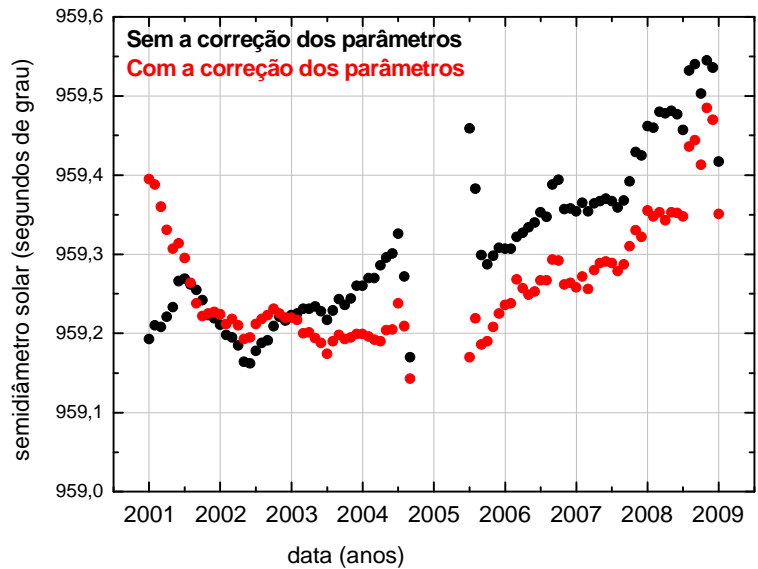

Figura 5.9 – Efeito da correção dos parâmetros no semidiâmetro solar.

Por outro lado, apenas até 2003 as correções dos parâmetros foram efetivamente implementadas, após este ano ainda não concluímos nada a favor de sua implementação. Necessitamos de mais pesquisas com os dados para implementar estas correções. Sabemos que estas correções são interessantes e nos levam a uma série isenta de pequenos erros introduzidos por diversos fatores externos, mas vale a pena ressaltar que estes erros são pequenos. Por estes motivos decidimos analisar o comportamento de nossa série sem impor este tratamento, escolhendo a série não corrigida destes efeitos.

# 6. Estudo da influência da presença de manchas no Sol

Geometricamente, o perfil de luminosidade definido pela temperatura nas manchas solares é deslocado por várias centenas de km para baixo em relação ao perfil da fotosfera uma vez que a opacidade é reduzida em manchas solares, devido à baixa densidade. Assim um interior mais profundo pode ser visto onde estão as manchas (Koskinen, H.; Vainio, R., 2009). Então, quando uma mancha solar passa pelo limbo solar, um raio ligeiramente menor é observado naquele ponto.

Qual é a probabilidade de uma mancha solar afetar a medida do semidiâmetro? Para responder a esta questão devemos nos perguntar primeiramente qual é a probabilidade de se ter uma mancha solar no limbo do Sol no instante da observação e no setor do bordo que se está observando.

A Tabela 6.1 traz a média mensal da área solar coberta por manchas em milionésimos da área visível de janeiro de 1997 a agosto de 2010. Os dados foram obtidos na página eletrônica de dados sobre manchas solares do *Royal Greenwich Observatory* – USAF/NOAA (http://solarscience.msfc.nasa.gov/greenwch.shtml). Este período cobre o período de aquisição de dados do astrolábio do ON.

Quantas manchas há no Sol? O índice diário chamado de Número de Manchas Solares, também chamado de Número de Wolf, é computado com o uso da equação 6.1 desenvolvida por Rudolph Wolf em 1848, onde [N] é o Número de Wolf, [g] é o número de grupos de manchas no disco solar, [s] é o número total de manchas individuais em todos os grupos, e [k] é uma variável de ponderação que leva em consideração as condições de observação e o tipo de telescópio. Os dados são combinados de vários observatórios, cada um com o próprio [k] para chegar-se a valor diário.

$$N = k \cdot (10g+s), \qquad (6.1)$$

Tabela 6.1 – Área média mensal do Sol coberta por manchas em milionésimos da área do Sol.

| ANO | Jan | Fev | Mar | Abr | Mai | Jun | Jul | Ago | Set | Out | Nov | Dez |
|---|---|---|---|---|---|---|---|---|---|---|---|---|
| **1997** | 10,8 | 38,0 | 51,0 | 86,8 | 116,5 | 60,7 | 38,8 | 149,5 | 726,1 | 131,4 | 632,3 | 481,0 |
| **1998** | 257,9 | 308,0 | 810,2 | 520,8 | 673,8 | 679,0 | 690,1 | 1358,9 | 1331,9 | 469,2 | 1006,6 | 1050,5 |
| **1999** | 634,5 | 937,0 | 655,7 | 460,1 | 1146,6 | 1690,3 | 1533,7 | 1397,3 | 609,9 | 1509,7 | 1871,8 | 1497,5 |
| **2000** | 927,2 | 1460,5 | 2221,5 | 1862,9 | 2182,2 | 2093,5 | 2343,9 | 1082,5 | 1872,3 | 948,8 | 1351,0 | 1024,3 |
| **2001** | 934,8 | 669,5 | 1671,0 | 1892,8 | 1109,2 | 1785,9 | 819,2 | 1619,0 | 3040,4 | 2128,9 | 2424,1 | 2354,3 |
| **2002** | 1781,4 | 1637,0 | 1367,6 | 2053,3 | 2182,2 | 1346,8 | 2196,6 | 2527,2 | 2093,5 | 1799,5 | 1843,8 | 1115,9 |
| **2003** | 987,7 | 474,0 | 1183,2 | 1107,9 | 950,6 | 1280,1 | 1177,4 | 948,8 | 593,6 | 2261,2 | 1594,1 | 632,3 |
| **2004** | 619,9 | 700,0 | 439,6 | 627,3 | 593,6 | 1239,5 | 985,4 | 566,1 | 600,4 | 796,1 | 274,6 | 632,3 |
| **2005** | 900,5 | 534,5 | 419,5 | 395,7 | 795,3 | 701,9 | 700,0 | 539,2 | 598,3 | 52,4 | 398,1 | 475,5 |
| **2006** | 150,8 | 10,5 | 78,6 | 490,9 | 188,8 | 196,9 | 214,6 | 438,5 | 193,2 | 79,5 | 528,7 | 370,3 |
| **2007** | 377,5 | 200,5 | 41,5 | 132,5 | 209,5 | 220,3 | 193,3 | 64,1 | 15,4 | 4,5 | 5,1 | 135,9 |
| **2008** | 14,5 | 8,7 | 144,5 | 28,9 | 11,1 | 9,3 | 1,8 | 0,0 | 3,3 | 15,8 | 31,7 | 4,1 |
| **2009** | 5,9 | 3,5 | 1,8 | 1,9 | 7,7 | 17,7 | 51,9 | 0,5 | 31,3 | 82,2 | 14,5 | 99,8 |
| **2010** | 223,5 | 182,0 | 222,0 | 71,4 | 55,3 | 89,6 | 187,9 | 233,9 | | | | |

Assim é fundamental conhecer o número de grupos, além do Número de Wolf, para se saber o número efetivo de manchas no Sol. As médias mensais do Número de Manchas Solares – Número de Wolf - estão na Tabela 2.2. O número de grupos de manchas é apresentado diariamente na página eletrônica *Space Weather* em uma imagem do Sol. Mostramos como exemplo a Figura 6.1 com a imagem do dia 11 de agosto de 2010. As páginas do *Space Weather* a partir de janeiro de 2001 são disponíveis. Acessando as páginas dos dias 1, 11 e 21 de todos os meses de 2001 a 2010, computamos o número de grupos. A Tabela 6.3 apresenta estes valores.

A partir dos números da Tabela 6.3 calculamos a média de grupos de manchas a cada mês considerando os valores dos dias 11 e 21 com peso 2 e os valores do dia 1 e do dia 1 do mês seguinte com peso 1. Como temos o número de grupos apenas a partir de 2001, admitimos então para os valores de 1997 a 2001 uma relação linear entre o Número de Wolf e o número de grupos, construída a partir dos valores de 2001 em diante. As médias mensais dos números de grupos de manchas de 1997 a 2010 são apresentadas na Tabela 6.4.

Tabela 6.2 – Média mensal do Número de Manchas Solares.

| ANO | Jan | Fev | Mar | Abr | Mai | Jun | Jul | Ago | Set | Out | Nov | Dez |
|---|---|---|---|---|---|---|---|---|---|---|---|---|
| **1997** | 5,7 | 7,6 | 8,7 | 15,5 | 18,5 | 12,7 | 10,4 | 24,4 | 51,3 | 22,8 | 39,0 | 41,2 |
| **1998** | 31,9 | 40,3 | 54,8 | 53,4 | 56,3 | 70,7 | 66,6 | 92,2 | 92,9 | 55,5 | 74,0 | 81,9 |
| **1999** | 62,0 | 66,3 | 68,8 | 63,7 | 106,4 | 137,7 | 113,5 | 93,7 | 71,5 | 116,7 | 133,2 | 84,6 |
| **2000** | 34,7 | 112,9 | 138,5 | 125,5 | 121,6 | 124,9 | 170,1 | 130,5 | 109,7 | 99,4 | 106,8 | 104,4 |
| **2001** | 95,6 | 80,6 | 113,5 | 107,7 | 96,6 | 134,0 | 81,8 | 106,4 | 150,7 | 125,5 | 106,5 | 132,2 |
| **2002** | 114,1 | 107,4 | 98,4 | 120,7 | 120,8 | 88,3 | 99,6 | 116,4 | 109,6 | 97,5 | 95,5 | 80,8 |
| **2003** | 79,7 | 46,0 | 61,1 | 60,0 | 54,6 | 77,4 | 83,3 | 72,7 | 48,7 | 65,5 | 67,3 | 46,5 |
| **2004** | 37,3 | 45,8 | 49,1 | 39,3 | 41,5 | 43,2 | 51,1 | 40,9 | 27,7 | 48,0 | 43,5 | 17,9 |
| **2005** | 31,3 | 29,2 | 24,5 | 24,2 | 42,7 | 39,3 | 40,1 | 36,4 | 21,9 | 8,7 | 18,0 | 41,1 |
| **2006** | 15,3 | 4,9 | 10,6 | 30,2 | 22,3 | 13,9 | 12,2 | 12,9 | 14,4 | 10,5 | 21,4 | 13,6 |
| **2007** | 16,8 | 10,7 | 4,5 | 3,4 | 11,7 | 12,1 | 9,7 | 6,0 | 2,4 | 0,9 | 1,7 | 10,1 |
| **2008** | 3,3 | 2,1 | 9,3 | 2,9 | 3,2 | 3,4 | 0,8 | 0,5 | 1,1 | 2,9 | 4,1 | 0,8 |
| **2009** | 1,3 | 1,4 | 0,7 | 0,8 | 2,9 | 2,9 | 3,2 | 0,0 | 4,3 | 4,8 | 4,1 | 10,8 |
| **2010** | 13,2 | 18,8 | 15,4 | 7,9 | 8,8 | 13,5 | 16,1 | 19,6 | 25,2 | | | |

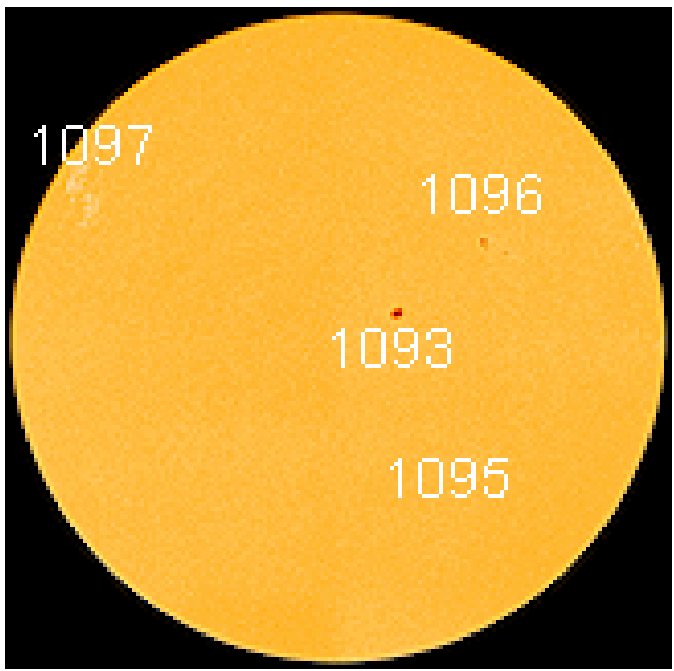

Figura 6.1 – Imagem com os grupos de manchas do Sol marcados no dia 11 de agosto de 2010 (*Space Weather*).

Com o número de grupos de manchas e com o Número de Manchas Solares segundo a equação de Wolf podemos calcular o número médio mensal de manchas individuais no Sol. Estes valores são apresentados na Tabela 6.5.

Tabela 6.3 – Grupos de Manchas Solares nos dias 1, 11 e 21 dos meses de janeiro de 2001 a dezembro de 2010.

| ANO | dia | Jan | Fev | Mar | Abr | Mai | Jun | Jul | Ago | Set | Out | Nov | Dez |
|---|---|---|---|---|---|---|---|---|---|---|---|---|---|
| **2001** | 01 | 4 | 6 | 4 | 9 | 5 | 1 | 6 | 3 | 4 | 9 | 4 | 6 |
|  | 11 | 7 | 7 | 4 | 6 | 5 | 7 | 4 | 7 | 6 | 6 | 7 | 6 |
|  | 21 | 3 | 0 | 6 | 4 | 5 | 7 | 6 | 5 | 10 | 9 | 5 | 5 |
| **2002** | 01 | 8 | 8 | 5 | 6 | 5 | 7 | 5 | 7 | 7 | 3 | 8 | 4 |
|  | 11 | 8 | 5 | 3 | 9 | 6 | 5 | 4 | 5 | 6 | 8 | 5 | 3 |
|  | 21 | 9 | 4 | 7 | 5 | 7 | 3 | 4 | 8 | 7 | 6 | 4 | 7 |
| **2003** | 01 | 3 | 4 | 3 | 6 | 6 | 3 | 4 | 3 | 4 | 3 | 5 | 7 |
|  | 11 | 7 | 8 | 5 | 2 | 2 | 4 | 2 | 4 | 2 | 2 | 3 | 3 |
|  | 21 | 5 | 2 | 2 | 5 | 4 | 4 | 4 | 2 | 3 | 2 | 3 | 3 |
| **2004** | 01 | 2 | 3 | 3 | 3 | 2 | 3 | 2 | 2 | 1 | 2 | 5 | 3 |
|  | 11 | 2 | 3 | 2 | 1 | 2 | 2 | 5 | 4 | 3 | 0 | 2 | 1 |
|  | 21 | 4 | 3 | 4 | 4 | 5 | 3 | 4 | 4 | 2 | 6 | 2 | 1 |
| **2005** | 01 | 1 | 2 | 0 | 1 | 2 | 4 | 3 | 5 | 2 | 1 | 2 | 3 |
|  | 11 | 3 | 4 | 3 | 1 | 3 | 3 | 3 | 2 | 2 | 1 | 0 | 4 |
|  | 21 | 2 | 2 | 2 | 2 | 1 | 2 | 0 | 3 | 1 | 1 | 3 | 3 |
| **2006** | 01 | 3 | 0 | 1 | 2 | 3 | 0 | 2 | 2 | 2 | 3 | 2 | 2 |
|  | 11 | 1 | 0 | 0 | 3 | 2 | 2 | 1 | 1 | 2 | 0 | 1 | 1 |
|  | 21 | 1 | 0 | 1 | 1 | 2 | 1 | 0 | 1 | 1 | 1 | 1 | 0 |
| **2007** | 01 | 2 | 2 | 1 | 1 | 2 | 1 | 2 | 0 | 1 | 1 | 0 | 1 |
|  | 11 | 3 | 0 | 1 | 0 | 1 | 1 | 1 | 1 | 0 | 0 | 0 | 1 |
|  | 21 | 1 | 2 | 0 | 0 | 1 | 0 | 0 | 0 | 0 | 0 | 0 | 0 |
| **2008** | 01 | 0 | 1 | 0 | 2 | 0 | 0 | 0 | 0 | 0 | 0 | 1 | 0 |
|  | 11 | 0 | 0 | 1 | 0 | 0 | 1 | 0 | 0 | 0 | 1 | 1 | 1 |
|  | 21 | 0 | 0 | 0 | 1 | 2 | 1 | 0 | 0 | 0 | 0 | 0 | 0 |
| **2009** | 01 | 0 | 0 | 0 | 0 | 1 | 1 | 0 | 0 | 1 | 1 | 0 | 0 |
|  | 11 | 1 | 0 | 0 | 0 | 0 | 0 | 1 | 0 | 0 | 0 | 1 | 1 |
|  | 21 | 0 | 0 | 0 | 0 | 0 | 0 | 0 | 0 | 0 | 1 | 1 | 3 |
| **2010** | 01 | 1 | 1 | 1 | 2 | 1 | 1 | 1 | 1 | 2 | 3 | 3 | 1 |
|  | 11 | 1 | 2 | 1 | 1 | 0 | 2 | 1 | 4 | 1 | 1 | 3 | 2 |
|  | 21 | 1 | 1 | 1 | 0 | 0 | 1 | 2 | 0 | 2 | 3 | 2 | 0 |

A partir da área média mensal do Sol coberta de manchas da Tabela 6.1 e do número médio mensal de manchas individuais do Sol calculamos a área média mensal de cada mancha e, supondo manchas circulares e iguais, calculamos o raio médio mensal destas manchas. A probabilidade de uma mancha de raio [r] estar em uma borda do Sol é a razão entre seu diâmetro e o arco da latitude onde ela está. A probabilidade de estar em uma ou na outra borda é o dobro. Para [N] manchas esta probabilidade é [N] vezes maior. A Tabela 6.6 apresenta o raio médio mensal das manchas individuais já multiplicado pelo número [N] de manchas.

Tabela 6.4 – Média mensal calculada de grupos de manchas solares.

| ANO | Jan | Fev | Mar | Abr | Mai | Jun | Jul | Ago | Set | Out | Nov | Dez |
|---|---|---|---|---|---|---|---|---|---|---|---|---|
| **1997** | 0,3 | 0,5 | 0,5 | 0,9 | 1,1 | 0,8 | 0,6 | 1,5 | 3,1 | 1,4 | 2,4 | 2,5 |
| **1998** | 1,9 | 2,5 | 3,3 | 3,3 | 3,4 | 4,3 | 4,1 | 5,6 | 5,7 | 3,4 | 4,5 | 5,0 |
| **1999** | 3,8 | 4,0 | 4,2 | 3,9 | 6,5 | 8,4 | 6,9 | 5,7 | 4,4 | 7,1 | 8,1 | 5,2 |
| **2000** | 2,1 | 6,9 | 8,4 | 7,6 | 7,4 | 7,6 | 10,4 | 8,0 | 6,7 | 6,1 | 6,5 | 6,4 |
| **2001** | 5,0 | 4,0 | 5,5 | 5,7 | 4,3 | 5,8 | 4,8 | 5,2 | 7,5 | 7,2 | 5,7 | 6,0 |
| **2002** | 8,3 | 5,2 | 5,2 | 6,5 | 6,3 | 4,7 | 4,7 | 6,7 | 6,0 | 6,5 | 5,0 | 4,5 |
| **2003** | 5,2 | 4,5 | 3,8 | 4,3 | 3,5 | 3,8 | 3,2 | 3,2 | 2,8 | 2,7 | 4,0 | 3,5 |
| **2004** | 2,8 | 3,0 | 3,0 | 2,5 | 3,2 | 2,5 | 3,7 | 3,2 | 2,2 | 3,2 | 2,7 | 1,3 |
| **2005** | 2,2 | 2,3 | 1,8 | 1,5 | 2,3 | 2,8 | 2,3 | 2,8 | 1,5 | 1,2 | 1,8 | 3,3 |
| **2006** | 1,2 | 0,2 | 0,8 | 2,2 | 1,8 | 1,3 | 1,0 | 1,3 | 1,8 | 1,2 | 1,3 | 1,0 |
| **2007** | 2,0 | 1,2 | 0,7 | 0,5 | 1,2 | 0,8 | 0,7 | 0,5 | 0,3 | 0,2 | 0,2 | 0,5 |
| **2008** | 0,2 | 0,2 | 0,7 | 0,7 | 0,7 | 0,7 | 0,0 | 0,0 | 0,0 | 0,5 | 0,5 | 0,3 |
| **2009** | 0,3 | 0,0 | 0,0 | 0,2 | 0,3 | 0,2 | 0,3 | 0,2 | 0,3 | 0,5 | 0,7 | 1,5 |
| **2010** | 1,0 | 1,3 | 1,2 | 0,8 | 0,3 | 1,3 | 1,3 | 1,8 | 1,8 | 2,3 | 2,3 | 1,0 |

A média dos raios médios mensais multiplicados pelo número de manchas é igual a 127,5 milésimos do raio solar. Para nossos cálculos vamos considerar que o Sol tem constantemente, durante todo o tempo em que se observou no astrolábio, uma mancha circular com 127,5 milésimos do raio solar. Esta mancha pode estar em qualquer hemisfério solar, mas para efeito de cálculo vamos considerar que ela está sempre apenas em um dos hemisférios, ou norte ou sul. Ao observar o Sol, os dois hemisférios interferem na medida. Colocando todas as manchas apenas ao norte, podemos fazer os cálculos considerando apenas este hemisfério o que simplifica os cálculos sem modificar o resultado, porque as manchas estarão todas neste hemisfério, neste modelo de equivalência.

Tabela 6.5 – Média mensal calculada do número de manchas individuais do Sol.

| ANO | Jan | Fev | Mar | Abr | Mai | Jun | Jul | Ago | Set | Out | Nov | Dez |
|---|---|---|---|---|---|---|---|---|---|---|---|---|
| **1997** | 2 | 3 | 3 | 6 | 7 | 5 | 4 | 10 | 20 | 9 | 15 | 16 |
| **1998** | 12 | 16 | 21 | 21 | 22 | 28 | 26 | 36 | 36 | 22 | 29 | 32 |
| **1999** | 24 | 26 | 27 | 25 | 42 | 54 | 44 | 37 | 28 | 46 | 52 | 33 |
| **2000** | 14 | 44 | 54 | 49 | 47 | 49 | 66 | 51 | 43 | 39 | 42 | 41 |
| **2001** | 56 | 31 | 74 | 59 | 55 | 86 | 38 | 61 | 91 | 61 | 60 | 86 |
| **2002** | 44 | 64 | 57 | 64 | 69 | 50 | 65 | 61 | 55 | 46 | 52 | 41 |
| **2003** | 35 | 6 | 33 | 27 | 25 | 46 | 57 | 48 | 25 | 47 | 39 | 15 |
| **2004** | 14 | 21 | 24 | 18 | 15 | 22 | 18 | 11 | 9 | 25 | 22 | 6 |
| **2005** | 13 | 6 | 8 | 13 | 26 | 16 | 25 | 11 | 9 | 0 | 5 | 13 |
| **2006** | 4 | 5 | 6 | 14 | 4 | 4 | 6 | 3 | 1 | 2 | 11 | 7 |
| **2007** | 0 | 1 | 1 | 2 | 2 | 7 | 3 | 3 | 1 | 1 | 2 | 5 |
| **2008** | 3 | 0 | 6 | 1 | 1 | 1 | 1 | 1 | 1 | 1 | 1 | 1 |
| **2009** | 1 | 1 | 1 | 1 | 1 | 1 | 1 | 1 | 3 | 1 | 1 | 1 |
| **2010** | 5 | 7 | 7 | 1 | 7 | 2 | 4 | 5 | | | | |

Tabela 6.6 – Raio médio mensal das manchas individuais do Sol multiplicado pelo número de manchas. Valores em milésimos do raio do Sol.

| ANO | Jan | Fev | Mar | Abr | Mai | Jun | Jul | Ago | Set | Out | Nov | Dez |
|---|---|---|---|---|---|---|---|---|---|---|---|---|
| **1997** | 5 | 11 | 13 | 23 | 29 | 17 | 13 | 38 | 121 | 34 | 98 | 88 |
| **1998** | 57 | 70 | 132 | 104 | 122 | 137 | 134 | 221 | 220 | 101 | 171 | 183 |
| **1999** | 124 | 156 | 133 | 107 | 218 | 302 | 261 | 226 | 131 | 262 | 312 | 222 |
| **2000** | 112 | 254 | 347 | 302 | 322 | 320 | 395 | 235 | 283 | 192 | 237 | 204 |
| **2001** | 228 | 143 | 350 | 335 | 247 | 391 | 178 | 315 | 525 | 359 | 381 | 449 |
| **2002** | 280 | 324 | 279 | 363 | 388 | 259 | 377 | 394 | 338 | 287 | 310 | 213 |
| **2003** | 185 | 53 | 197 | 172 | 153 | 242 | 258 | 213 | 123 | 327 | 249 | 97 |
| **2004** | 103 | 114 | 130 | 88 | 96 | 113 | 148 | 104 | 73 | 122 | 132 | 41 |
| **2005** | 108 | 56 | 57 | 70 | 144 | 106 | 133 | 78 | 72 | 4 | 43 | 78 |
| **2006** | 23 | 7 | 21 | 82 | 27 | 28 | 34 | 36 | 14 | 13 | 78 | 51 |
| **2007** | 7 | 12 | 6 | 15 | 19 | 40 | 24 | 13 | 3 | 2 | 3 | 26 |
| **2008** | 7 | 2 | 29 | 5 | 3 | 3 | 1 | 0 | 2 | 4 | 6 | 2 |
| **2009** | 2 | 2 | 1 | 1 | 3 | 5 | 7 | 1 | 9 | 9 | 4 | 10 |
| **2010** | 33 | 36 | 40 | 9 | 20 | 13 | 29 | 33 | | | | |

A probabilidade de uma mancha estar em um dos bordos do Sol depende da latitude porque a cada latitude há um paralelo com um arco de tamanho diferente. Consideramos esta probabilidade como duas vezes o diâmetro da mancha dividido pela metade deste arco. Duas vezes porque há dois bordos, e a metade do arco porque consideramos apenas o hemisfério visível do Sol. Então esta probabilidade é data pela equação 6.2, onde [r] é o raio da mancha, [R] é o raio do Sol e [α] é a latitude considerada.

$$P = 4 \cdot r / (\pi \cdot R \cdot \cos(\alpha)) \qquad (6.2)$$

Vamos dividir o Sol em setores de 5 graus e fazer este cálculo para cada setor. Considerando sempre a latitude média de cada setor, as probabilidades de uma mancha com 127,5 milésimos do raio do Sol estar em um dos bordos pode ser vista na Tabela 6.7.

Tabela 6.7 – Probabilidade de uma mancha de 127,5 milésimos do raio solar estar na borda do Sol em função de sua heliolatitude.

| setor(graus) | probabilidade |
|---|---|
| 0 -5 | 0,1625 |
| 5-10 | 0,1637 |
| 10-15 | 0,1663 |
| 15-20 | 0,1702 |
| 20-25 | 0,1757 |
| 25-30 | 0,1830 |
| 30-35 | 0,1925 |
| 35-40 | 0,2046 |
| 40-45 | 0,2202 |

Até aqui não levamos em conta a distribuição diferente de manchas ao longo dos setores. Esta distribuição é conhecida. As manchas muito raramente ocorrem em latitudes superiores a 45 graus. Usando esta distribuição e as probabilidades já calculadas podemos calcular a probabilidade de haver uma mancha ocorrendo na borda do Sol em cada um dos setores de 5 graus em que a dividimos. A Tabela 6.8 apresenta estes valores.

Entre 1998 e 2008 foram feitas no ON 21640 observações para se medir o raio solar. Conhecemos a distribuição destas observações ao longo das latitudes do Sol. A Tabela 6.9 mostra esta distribuição. Pode-se ver que um grande número de observações do semidiâmetro solar é feita acima da latitude de 45 graus, setores onde não ocorrem manchas.

Tabela 6.8 – Probabilidade de uma mancha estar na borda do Sol e em cada um dos setores de 5 graus.

| setor (graus) | distribuição das manchas (%) | probabilidade x 10³ |
|---|---|---|
| 0 -5 | 5,9 | 0,0096 |
| 5-10 | 18,2 | 0,0298 |
| 10-15 | 25,83 | 0,0430 |
| 15-20 | 23 | 0,0391 |
| 20-25 | 14,73 | 0,0259 |
| 25-30 | 7,83 | 0,0143 |
| 30-35 | 3,27 | 0,0063 |
| 35-40 | 0,93 | 0,0019 |
| 40-45 | 0,2 | 0,0004 |

Tabela 6.9 – Distribuição das observações do semidiâmetro solar ao longo dos setores entre 1998 e 2008.

| setor (graus) | número de observações | setor (graus) | número de observações |
|---|---|---|---|
| 0 -5 | 2273 | 45-50 | 1024 |
| 5-10 | 2520 | 50-55 | 1172 |
| 10-15 | 1243 | 55-60 | 1288 |
| 15-20 | 1057 | 60-65 | 1580 |
| 20-25 | 1062 | 65-70 | 2041 |
| 25-30 | 1066 | 70-75 | 1588 |
| 30-35 | 941 | 75-80 | 696 |
| 35-40 | 854 | 80-85 | 220 |
| 40-45 | 969 | 85-90 | 46 |

Cada medida do raio solar compreende a observação de cerca de 18 graus do limbo solar. Assim, uma mancha pode afetar 3,6 setores de 5 graus. A partir da distribuição das observações e considerando os setores afetados podemos calcular o número de observações do Sol que são afetadas em cada setor. Estes números podem ser vistos na Tabela 6.10

A partir do número de observações afetadas a cada setor e das probabilidades de se ter uma mancha no bordo do Sol, neste setor, calculamos o número mais provável, para cada um dos setores, de observações afetadas pela passagem de uma mancha pelo limbo solar. A soma destes números para os nove setores é de 893 observações o que significa 4,13% do total de observações efetuadas.

Tabela 6.10 – Observações do semidiâmetro solar afetadas em cada setor.

| setor (graus) | observações que o afetam |
|---|---|
| 0 -5 | 8195 |
| 5-10 | 7035 |
| 10-15 | 5821 |
| 15-20 | 4438 |
| 20-25 | 3840 |
| 25-30 | 3642 |
| 30-35 | 3470 |
| 35-40 | 3391 |
| 40-45 | 3481 |

Calculamos também o maior número de observações, em cada setor, que podem ser afetadas pela presença de uma mancha com uma probabilidade de 50% de ocorrer este ou qualquer número menor. Sua soma para os nove setores é de 891 observações o que significa 4,12% do total de observações efetuadas. Calculamos ainda o maior número de observações, em cada setor, que podem ser afetadas pela presença de uma mancha com uma probabilidade de 95% de ocorrer este ou qualquer número menor. A soma destes números em todos os setores é de 1031 observações o que significa 4,76% do total observado. A Tabela 6.11 mostra estes resultados.

Podemos, então afirmar com uma certeza de 50% que menos de 4,12% das observações feitas no astrolábio são afetadas pela presença de uma mancha solar no bordo do Sol. Com uma certeza de 95% este número é menor que 4,76% das observações. De acordo com a Tabela 6.11 cerca de 70% das possibilidades disto ocorrer se dá nos setores entre 5 e 20 graus de latitude solar.

Tabela 6.11 – Número provável de observações do semidiâmetro que foram afetadas pela passagem de uma mancha no bordo do Sol.

| setor (graus) | N | N - 50% | N - 95% |
| --- | --- | --- | --- |
| 0 -5 | 79 | 78 | 94 |
| 5-10 | 210 | 209 | 233 |
| 10-15 | 250 | 250 | 278 |
| 15-20 | 174 | 174 | 200 |
| 20-25 | 99 | 99 | 116 |
| 25-30 | 52 | 52 | 65 |
| 30-35 | 22 | 22 | 30 |
| 35-40 | 6 | 6 | 11 |
| 40-45 | 2 | 1 | 4 |
| soma | 893 | 891 | 1031 |

N = número mais provável.

N - 50% = número com certeza de 50% de ocorrer ele ou um menor

N - 95% = número com certeza de 98% de ocorrer ele ou um menor

O programa que calcula a medida do semidiâmetro descarta os pontos da parábola ajustada ao bordo do Sol cujo desvio da média ultrapassa duas vezes e meia o desvio padrão. Os pontos do bordo onde se encontra uma mancha são muitas vezes descartados por este programa. Algumas vezes a própria medida é descartada. Estas medidas tomadas pelo programa de cálculo diminuem ainda mais a possível interferência de uma mancha solar no cálculo do semidiâmetro.

Ainda assim, uma probabilidade média de 5% pode significar um número muito maior quando a atividade solar está em seu máximo com o consequente aumento do número de manchas. Cabe, portanto calcular qual o valor da influência que uma mancha pode exercer sobre a medida do semidiâmetro solar.

Considerando que uma mancha solar pode desviar a fotosfera de algumas centenas de quilômetros, isto significa que ela pode alterar a borda do Sol em poucos milésimos de seu raio o que significa um arco em torno de um segundo de grau. Elaboramos um estudo preliminar utilizando ajustes lineares a uma borda solar circular e com um deslocamento de um segundo de grau da parte da borda relativa ao tamanho de uma mancha média.

Obtivemos então que a influência de uma mancha na medida do semidiâmetro não ultrapassa dois milésimos de segundo de grau.

Desta forma, mesmo que a probabilidade de haver uma mancha no bordo solar no momento de uma medida, na época de maior atividade solar deva ser considerada, ainda assim, a influência desta mancha não irá modificar fundamentalmente os valores das medidas.

## 7. Variações do semidiâmetro relacionadas ao ciclo solar - Manchas

**7.1 – Visão geral** - O mais amplamente usado indicador do ciclo solar é o índice conhecido como número de manchas. A média mensal destes números dá uma excelente indicação da fase em que se encontra o ciclo solar. A Figura 7.1 mostra o valor médio mensal do número de manchas de janeiro de 1972 até setembro de 2010. Nela aparecem os três últimos ciclos de atividade do Sol, os ciclos 21, 22 e 23, o final do ciclo anterior a eles, o de número 20 e o começo do ciclo atual, o ciclo 24 (*Marshall Space Flight Center* da NASA).

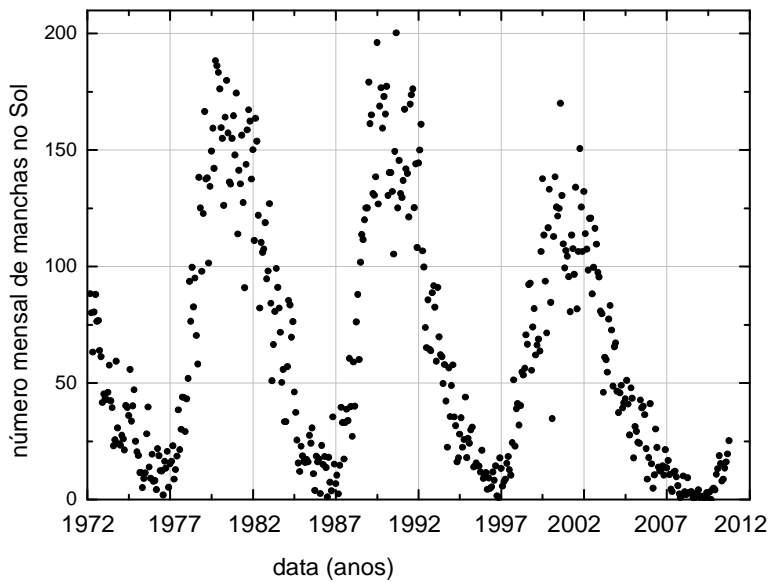

Figura 7.1 - Número de manchas no Sol - médias mensais.

O semidiâmetro do Sol vem sendo observado pelo astrolábio do Observatório Nacional desde 1998 e, portanto durante o último ciclo de atividade solar. Temos um acervo de dados de doze anos, ultrapassando um pouco o ciclo solar de onze anos. Mas, considerando que o atual ciclo de atividades do Sol parece estar se atrasando, podemos dizer que temos um ciclo completo de observações.

O comportamento do semidiâmetro em relação ao último ciclo solar, medido por sua relação com o número de manchas pode ser basicamente dividido em duas fases. A fase inicial em que o semidiâmetro se mostra acompanhando o ciclo e que perdura de 1998 até 2003 e a fase em que o semidiâmetro não mais acompanha o ciclo permanecendo em ascensão enquanto o número de manchas decresce.

A Figura 7.2 mostra este comportamento. Ela foi composta a partir de médias corridas de doze meses tanto dos valores de semidiâmetro como dos valores de número de manchas. A escala do eixo vertical está em segundos de grau e os números de mancha foram colocados em uma escala arbitrária para comparação. No eixo horizontal aparecem os anos, havendo um ponto para cada mês. Há alguns meses entre 2004 e 2006 durante os quais o Sol não foi observado gerando um hiato na série. Pode-se observar neste gráfico que o semidiâmetro acompanha o número de manchas, imitando muito bem os seus máximos até 2003.

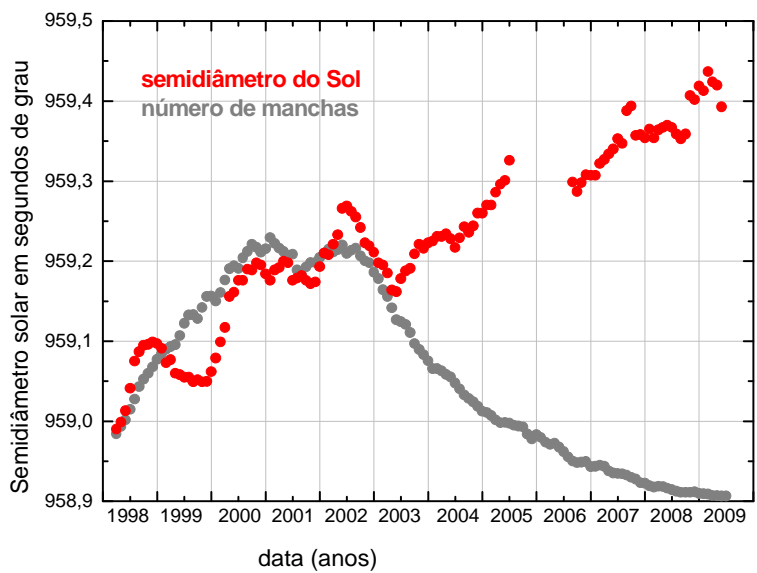

Figura 7.2 - Médias corridas de doze meses do semidiâmetro do Sol e do número de manchas.

O intervalo de tempo onde a variação do semidiâmetro e o número de manchas estão correlacionados pode ser visto em maior detalhe no gráfico da Figura 7.3. Nele os pontos representam uma média corrida dos valores considerados. Para o semidiâmetro solar consideramos a coleção completa de observações entre 1998 e 2003, com mais de 16 mil pontos. Para o número de manchas consideramos uma série com médias diárias nesse mesmo período. O raio do Sol que em 1998 parece estar adiantado em relação ao número de manchas, passa a caminhar atrás deste número copiando suas subidas e descidas cerca de 200 dias após. O pico de atividade de 2000 não foi acompanhado, mas todos os outros foram. A figura ilustra a dificuldade que se tem de certificar as relações entre os indicadores da atividade solar. Por este motivo, a série do semidiâmetro solar é comparada com outros índices da atividade solar nos capítulos seguintes, bem como a coleção de séries é a analisada em conjunto no Capítulo 12.

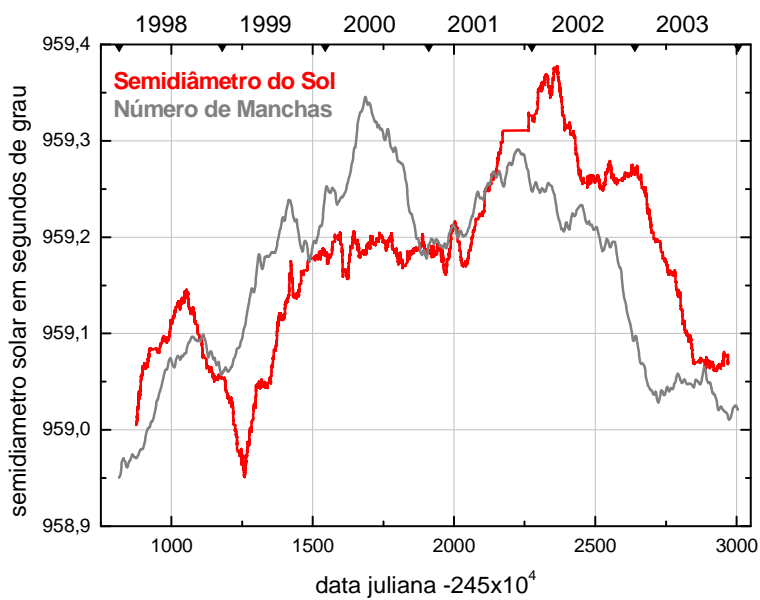

Figura 7.3 - Semidiâmetro do Sol e número de manchas – médias corridas.

**7.2 – Correlações** - Este acompanhamento do número de manchas pelo raio solar pode ser compreendido quando se fazem correlações entre as duas curvas. O menor intervalo de tempo considerado é de um mês – conforme indicado no Capítulo 3, enquanto que o maior é de um ano considerando que a série analisada contém apenas 6 anos. Assim, tomamos a duração da série de dados do semidiâmetro do início de 1998 até o final de 2003 e o dividimos inicialmente por seis, o que resulta em intervalos de um ano. Fizemos então médias corridas de um ano, a cada dez dias, tanto para o semidiâmetro como para o número de manchas. Assim fazendo obtivemos duas séries com 180 valores das quais achamos a correlação. A seguir dividimos a duração total em sete, o que resulta em intervalos pouco menores que um ano. Fizemos médias corridas destes intervalos, a cada dez dias obtendo duas séries com um pouco mais de 180 valores dos quais achamos a correlação. Depois, dividimos a duração total por oito e fizemos o mesmo, depois por nove, por dez, e continuando assim até dividir a duração total por 72, o que resulta em intervalos de um mês. Agindo assim obtivemos 67 valores de correlação que comparam as duas curvas tomadas por suas médias, desde séries anuais e diminuindo gradativamente até séries mensais. Para obter as correlações utilizamos o método de Pearson. Os valores encontrados são mostrados no gráfico da Figura 7.4. O significdo estatístico é dado pelas barras de erro que mostram o complemento da significância da correlação.

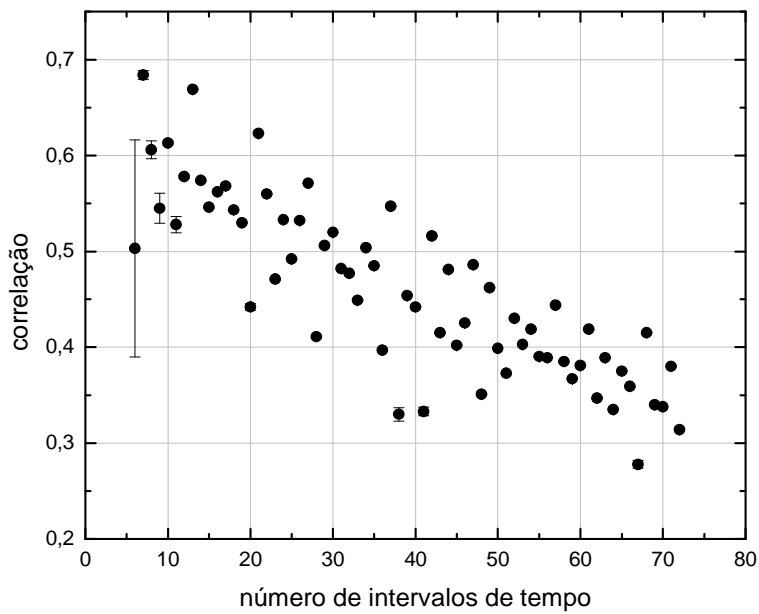

Figura 7.4 - Correlações entre o semidiâmetro do Sol e o número de manchas.

Este gráfico mostra que quando olhamos as curvas por seus aspectos mais gerais, ou seja, quando consideramos médias anuais (seis intervalos de tempo), obtemos correlações mais altas entre as duas séries. Mas quando passamos a olhar para os detalhes, ou seja, na medida em que fazemos médias com intervalos cada vez menores, obtemos correlações mais baixas. Apesar desta tendência, há uma dispersão de pontos. O reconhecimento desta dispersão mostra o acerto ao se escolher pequenos passos no cálculo das correlações. Porque os erros observacionais ora reforçam, ora esmaecem as correlações obtidas em torno de determinados períodos.

Outro número bastante significativo do ciclo solar é o fluxo rádio no comprimento de onda de 10,7 cm. Este fluxo é muito conectado com a formação de manchas solares. Pode-se preferir este indicador porque ele não depende de uma contagem subjetiva de manchas do Sol. Efetuamos então o mesmo procedimento para obter correlações da série do semidiâmetro solar com uma série de médias diárias dos valores do fluxo rádio em 10,7 cm. Obtivemos os resultados mostrados no gráfico da Figura 7.5.

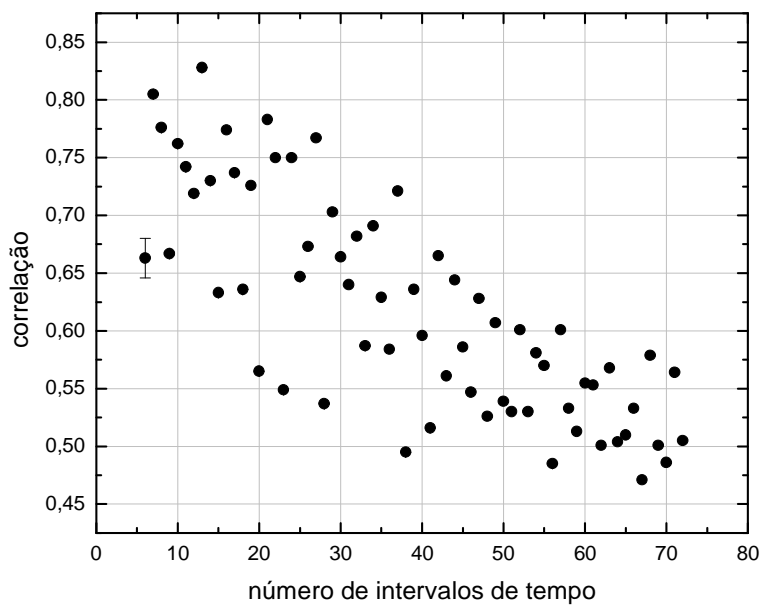

Figura 7.5 - Correlações entre o semidiâmetro do Sol e o fluxo rádio em 10,7 cm.

Este gráfico é bastante semelhante ao que mostra correlações com o número de manchas, mas com valores 36% maiores em média. As maiores correlações ultrapassam 0,80 e as menores ficam acima de 0,45. Encontramos também uma dispersão de pontos.

**7.3 – Considerando defasagens** - Estas correlações entre o semidiâmetro solar e o número de manchas ou o fluxo rádio em 10,7 cm podem ser indicativas de interações mais fortes ou mais fracas entre fenômenos físicos ligados ao ciclo de atividades do Sol e fenômenos físicos que definem o tamanho do diâmetro do Sol. Estes fenômenos podem estar de alguma forma relacionados instantaneamente, mas podem também estar relacionados com uma defasagem no tempo. Por este motivo, investigamos as possíveis mudanças nas correlações ao se considerar defasagens entre as duas curvas. Para isso, repetimos todos os procedimentos relatados para a obtenção das correlações, mas considerando, agora, defasagens entre as duas séries. Consideremos defasagens desde dois anos antes, até dois anos depois, uma a cada dez dias, o que significa cálculos para 147 defasagens diferentes. E para cada uma destas defasagens calculamos as 67 correlações conforme descrito anteriormente. O trecho de dados do semidiâmetro solar aqui estudado têm seis anos e quando o deslocamos de dois anos mantemos ainda dois terços dos dados em comum observando correlações defasadas de pelo menos 20% do ciclo de atividades do Sol.

Os resultados podem ser vistos no gráfico da Figura 7.6. Pode-se observar que as maiores correlações ocorrem quando há uma defasagem negativa de cem dias entre as duas curvas. Neste gráfico, as defasagens negativas significam que a série do semidiâmetro solar está atrasada em relação à série de número de manchas. Isto é, um atraso de cem dias significa que as duas séries têm comportamentos semelhantes, mas o semidiâmetro solar se comporta como que respondendo com um atraso de cem dias.

O mesmo procedimento foi feito em relação ao fluxo rádio em 10,7 cm e os resultados podem ser vistos no gráfico da Figura 7.7 que é bastante semelhante ao anterior como já se esperava. O mesmo atraso de cem dias é observado para as maiores correlações.

Nas Figura 7.6 e 7.7 o envoltório das correlações mais distantes de zero mostra-se bastante suave, correspondendo aos intervalos da ordem de um ano. Ao contrário, as correlações mais próximas de zero começam a mostrar uma flutuação estatística, ainda que sem borrar a tendência dominante, o que revela que a adoção do limite mínimo mensal é adequada.

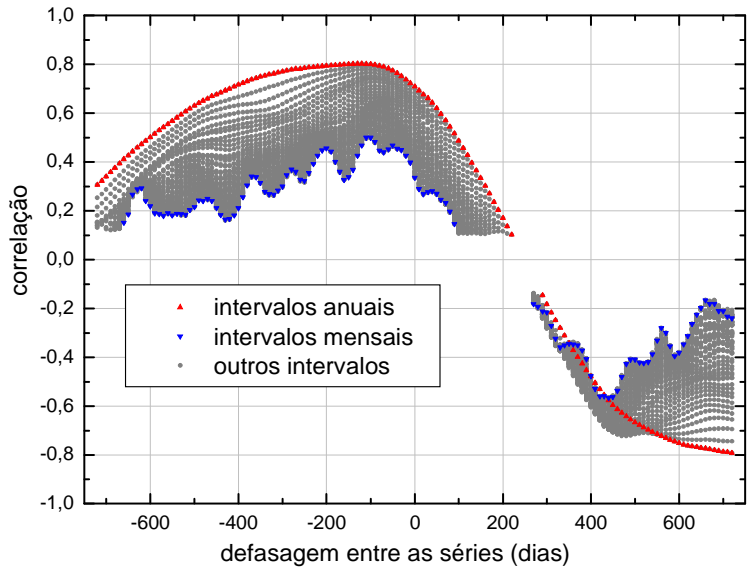

Figura 7.6 - Correlações com defasagens entre o semidiâmetro solar e o número de manchas.

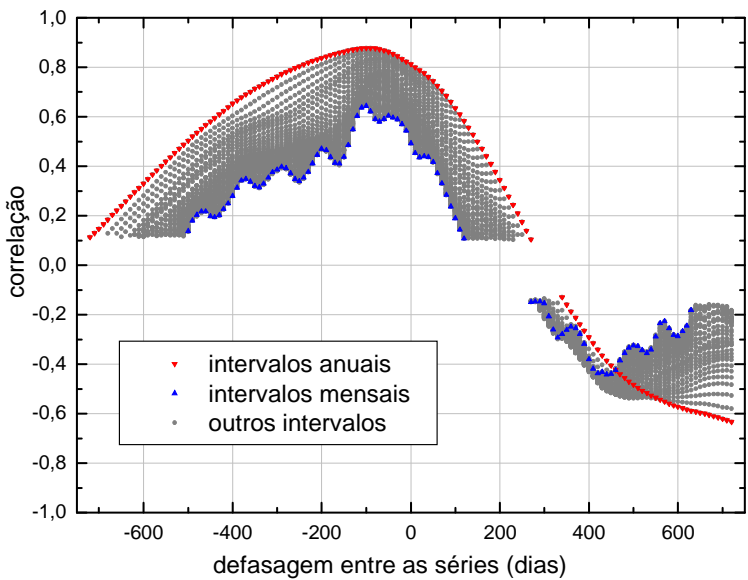

Figura 7.7 - Correlações com defasagem entre o semidiâmetro solar e o fluxo rádio em 10,7 cm.

**7.4 – O Semidiâmetro se afasta do ciclo** - A partir do meio de 2003 o semidiâmetro solar deixou de acompanhar o ciclo de atividades Enquanto o número de manchas decresceu até se anular, o semidiâmetro solar permaneceu crescendo continuamente.

O atual mínimo de atividade solar vem se mostrando mais longo e mais profundo que os anteriores. Se tomarmos os pontos do gráfico da Figura 7.1 veremos que no mínimo entre os ciclos 20 e 21 há 24 meses com número de manchas inferior a 20. No mínimo entre os ciclos 21 e 22 há 26 meses em tal condição e no mínimo entre os ciclos 22 e 23 há 30 meses. Entretanto no mínimo entre os ciclos 23 e 24 há 56 meses na mesma condição. Os mínimos anteriores apresentaram sempre menos de 500 dias sem manchas enquanto que o atual mínimo já apresentou mais de 800 dias sem manchas. Enquanto isso o semidiâmetro solar continuou aumentando de valor. Este é um comportamento não esperado, pois nós estávamos aguardando um comportamento cíclico do semidiâmetro que acompanhasse as atividades do Sol.

Durante todos os anos deste mínimo solar que se estende desde o final de 2004 até o final de 2010 o semidiâmetro do Sol vem aumentando (veja Figura 3.18). A este fato vamos acrescentar alguns outros que concordam com a observação de aumento do diâmetro solar.

A interpretação dominante sobre dados históricos indica que durante o intervalo de tempo conhecido como mínimo de Maunder o semidiâmetro do Sol era maior (Ribes, e Barthalot, 1987). Mínimo de Maunder é o nome dado ao intervalo entre 1645 e 1715, quando as manchas solares se tornaram extraordinariamente raras. A Figura 7.8 mostra os ciclos de atividade solar dos nos últimos 400 anos destacando o Mínimo de Maunder, bem como outro período de pouca atividade solar no início do século XIX conhecido como Mínimo de Dalton e o Máximo Moderno.

Komitov e Kaftan utilizando modelos de investigação de séries temporais mostraram que é altamente provável a ocorrência de um mínimo solar nas próximas décadas similar ao mínimo de Dalton. A Figura 7.9 mostra a série dos dois últimos milênios e o modelo (em preto) extrapolado até o final do século XXI e mostrando o mínimo solar (Komitov e Kaftan, 2004).

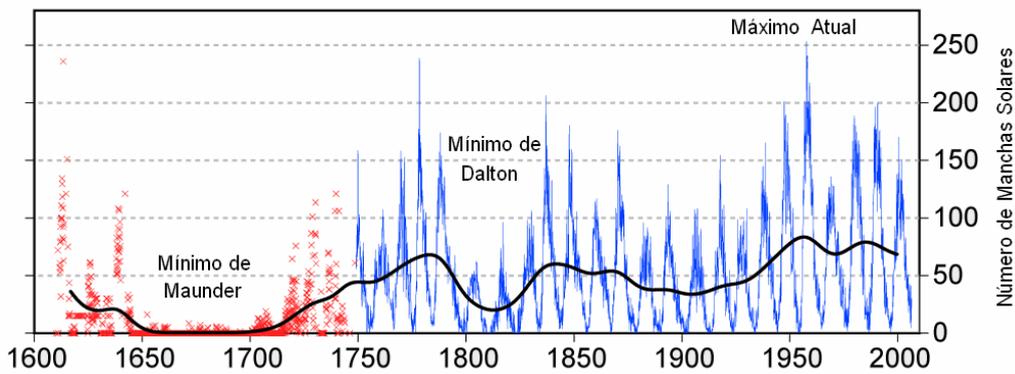

Figura 7.8 – 400 anos de observações de manchas do Sol - (Mínimo de Maunder – Wikipédia - http://pt.wikipedia.org/wiki/Mínimo_de_Maunder).

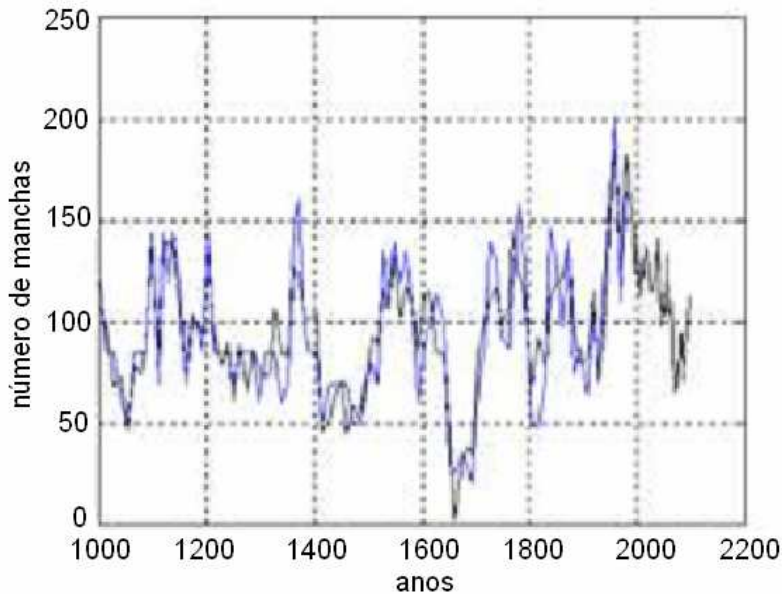

Figura 7.9 – Os dois últimos milênios e o modelo extrapolando a atividade solar até o final do século (Komitov e Kaftan, 2004).

Foram observadas mudanças na temperatura, no campo magnético e no brilho das linhas de FeI no interior de 1000 manchas solares de 1990 a 2005 (Penn e Livingston, 2008). Nas séries observadas as Umbras das manchas se tornam mais quentes a uma taxa de 45 K/ano e os campos magnéticos decrescem a 77 Gauss/ano. Esta variação ocorre independentemente do ciclo de atividades do Sol. Ajustes lineares destas séries extrapolados sugerem que poucas manchas serão visíveis depois de 2015. A Figura 7.10 mostra a variação de brilho da linha de 1565,3 nm do OH. O decréscimo na linha é constante e independente do ciclo de manchas. A Figura 7.11 mostra a variação do campo magnético no interior da mancha e a extrapolação da tendência. Nunca é observado escurecimento da fotosfera para um campo magnético abaixo de 1500 Gauss (Penn e Livingston, 2008).

A Figura 7.12 mostra o contraste de brilho entre as manchas e o brilho do Sol quieto, indicando que por volta de 2014 as manchas terão o mesmo brilho do campo solar.

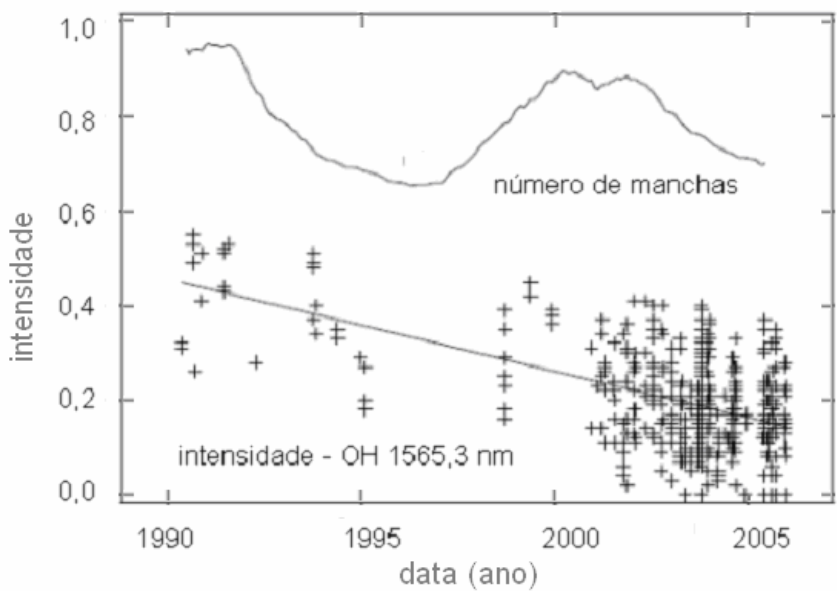

Figura 7.10 – Variação de brilho da linha de 1565,3 nm do OH (Penn e Livingston, 2008).

Estes fatos aliados ao crescimento não esperado do raio solar e a indicação dos dados de Picard (Ribes, e Barthalot, 1987) que apontam para um raio maior durante o Mínimo de Maunder, se somam ao fato de que a atividade do atual ciclo solar será certamente muito baixa conforme o gráfico da Figura 7.13 que mostra os valores previstos muito baixos e os valores ainda menores que vêm ocorrendo.

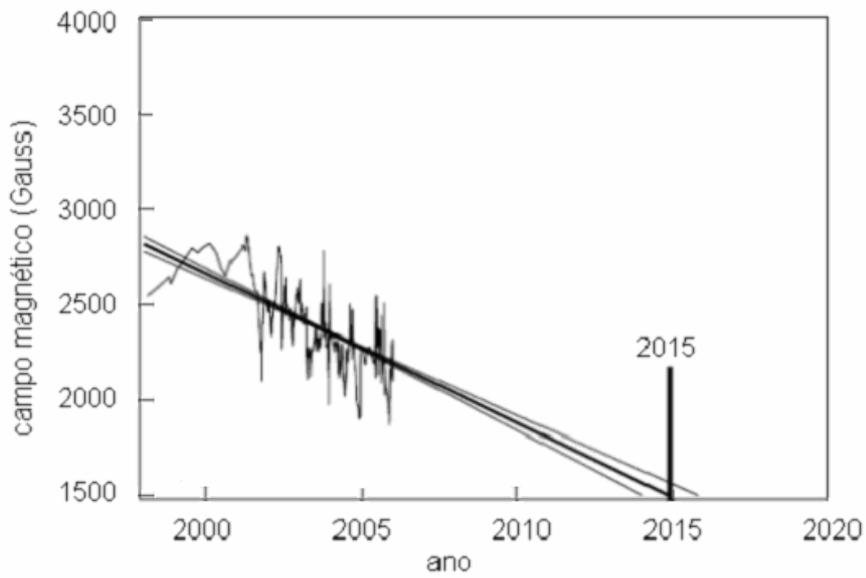

Figura 7.11 – Variação do campo magnético no interior das manchas solares (Penn e Livingston, 2008).

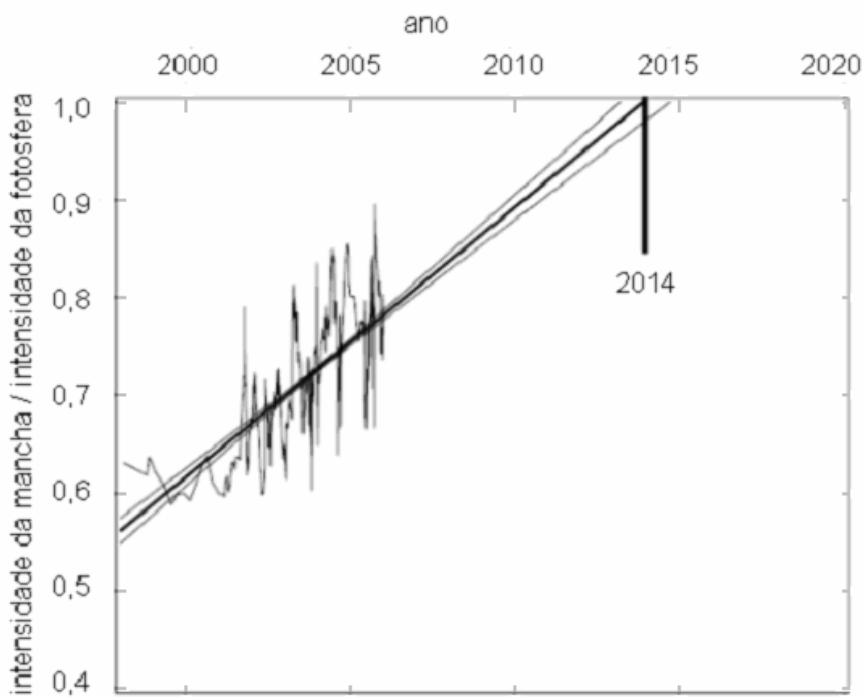

Figura 7.12 – Contraste de brilho entre manchas e o brilho do Sol quieto (Livingston e Penn, 2008).

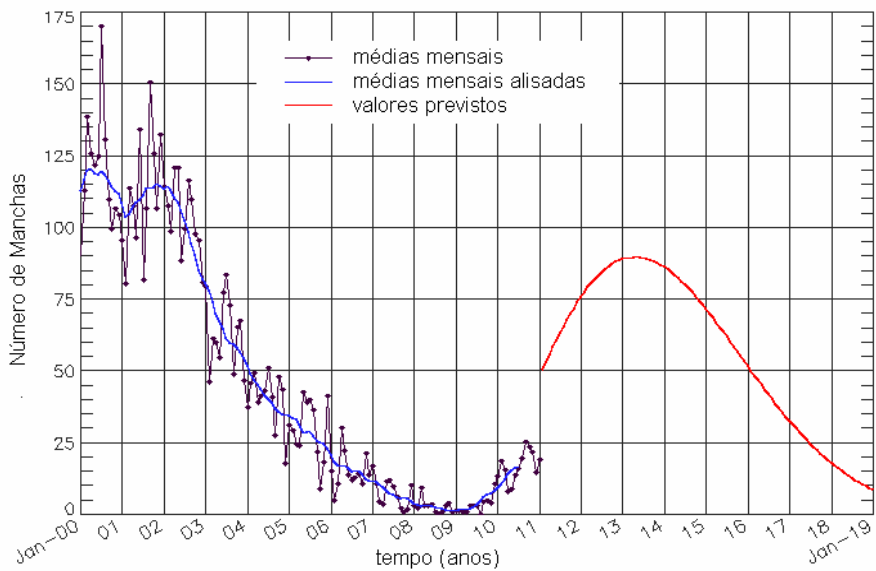

Figura 7.13– Número de Manchas do Sol – valores ocorridos e previsão (Previsão do Número de Manchas - http://www.solarcycle24.com/).

Estes fatos apontam no sentido de que estamos próximos de um mínimo solar tal como o Mínimo de Dalton ou o Mínimo de Maunder no qual o diâmetro solar foi encontrado maior tal como agora se apresenta conforme indicam nossas observações. É possível que a atividade solar diminua em função do acúmulo de parte da energia gerada de maneira constante no interior em forma gravitacional. Uma variação de 1,2 W/m$^2$ na constante solar de 1367 W/m$^2$ é observada durante o ciclo solar. Esta variação pode acumular durante meio ciclo solar um valor de 6,1·10$^{31}$ J. Este valor é cinco ordens de magnitude superior à luminosidade solar. A energia gravitacional associada a uma variação de raio solar de 8,9 km nas camadas acima de r = 0,96·R pode dar conta desta energia. De maneira análoga as mudanças no raio solar durante os setenta anos do mínimo de Maunder pode ser avaliada. Neste caso, a energia acumulada seria da ordem de 7,7·10$^{32}$ J. Este valor pode ser justificado pela energia gravitacional acumulada em variações do raio de 64 km a partir da camada r = 0,93·R (Callebaut, Makarov e Tlatov, 2002).

Um aumento de 0,1 segundo de grau no semidiâmetro do Sol pode corresponder a um aumento efetivo de 70 km. Desde 2003 nós observamos em nossa série um aumento de 0,2 segundos de grau no semidiâmetro do Sol. Pelas medidas independentes expostas acima e pelo aumento do semidiâmetro solar observado no ON, o qual se sobrepõem ao ciclo de 11 anos de atividade solar, propomos que o Sol está próximo de passar por um mínimo de atividades do qual ainda não sabemos a profundidade. Talvez se apresente semelhante ao mínimo de Dalton, ou ainda mais profundo como o mínimo de Maunder.

## 8. Variações do semidiâmetro relacionadas ao ciclo solar – Irradiância

Cada um dos diversos índices da atividade do Sol apresenta um padrão de variabilidade peculiar dentro do ciclo solar. Da mesma forma, para estudar a variabilidade do semidiâmetro dentro do ciclo solar é necessário compará-lo separadamente contra cada um daqueles índices de atividade. Em particular, a irradiância, objeto do presente capítulo, requer não só a contagem de manchas, mas também a distribuição de fáculas para explicar sua variabilidade dentro do ciclo solar (Pap et al, 2001). A irradiância total solar tem sido medida a partir do final dos anos 70 por diversos instrumentos a bordo de satélites orbitando a Terra. A Figura 8.1 a seguir mostra os dados coletados por diversos instrumentos (*Active Cavity Radiometer Irradiance Monitor* - ACRIM - http://www.acrim.com/)

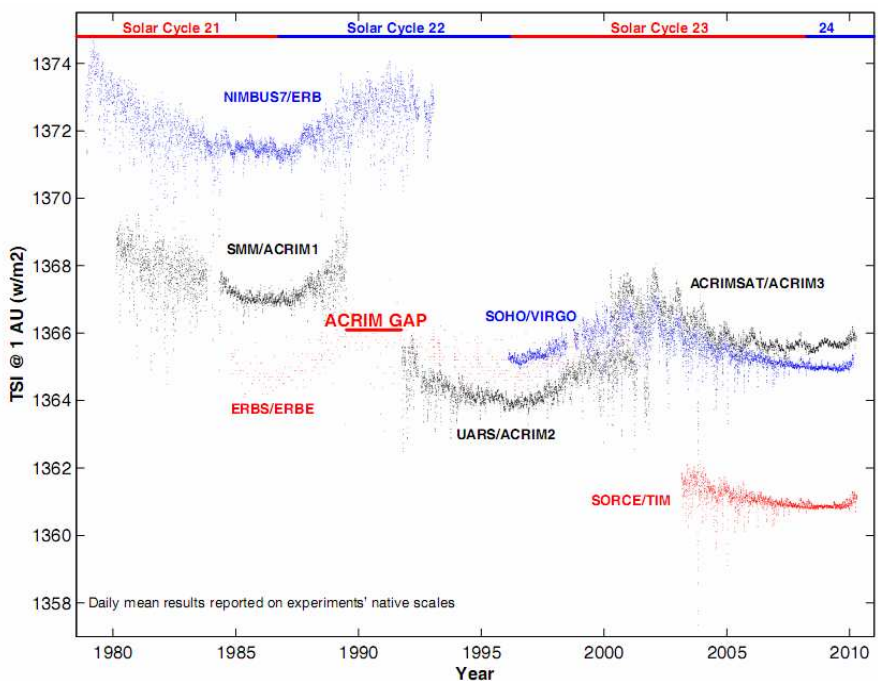

Figura 8.1 – Irradiância total do Sol de 1978 até o presente (*Active Cavity Radiometer Irradiance Monitor* – ACRIM - http://www.acrim.com/)

A partir dos valores coletados pelos diversos instrumentos, as séries foram compostas. Os dados que nós utilizamos, foram desenvolvidos pelo *National Geophysical Data Center* – NGDC e obtidos de sua página eletrônica (http://www.ngdc.noaa.gov/). Eles compõem uma série com valores médios diários. A Figura 8.2 mostra esta série de dados no intervalo de tempo de 1998 a 2003.

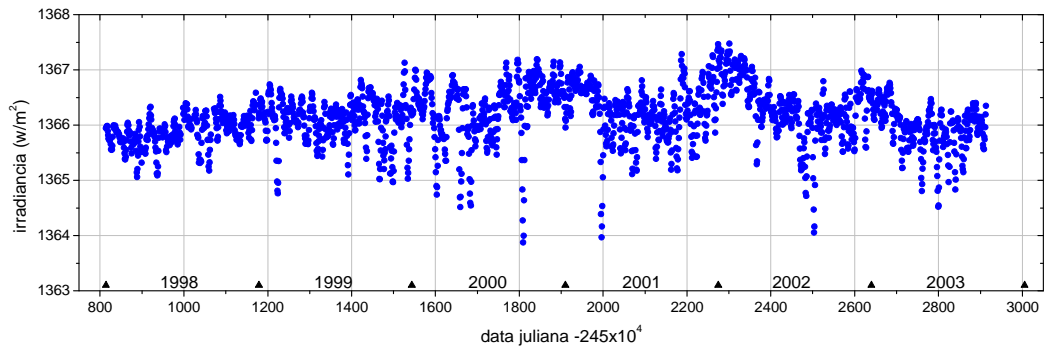

Figura 8.2 - Irradiância total do Sol - *National Geophysical Data Center* – NGDC.

**8.1 - Comparação das duas séries** - Podemos ver a série do semidiâmetro solar do ON e a série da irradiância total do Sol na Figura 8.3. Os pontos representam uma média corrida dos valores considerados. Para o semidiâmetro solar, da mesma forma que no Capítulo 7, consideramos o trecho da série de observações entre 1998 e 2003 em que há sincronismo entre as variações do semidiâmetro e o ciclo de 11 anos. Para a irradiância total do Sol consideramos a série com médias diárias no mesmo intervalo de tempo. A escala está ajustada para o semidiâmetro solar em segundos de grau e adaptada arbitrariamente para se comparar com a irradiância solar. A figura revela tendências gerais semelhantes e uma boa concordância nos picos de máximos, mas também aparecem trechos discordantes. O estudo das correlações entre as duas séries é a maneira de qualificar as relações que se apresentem.

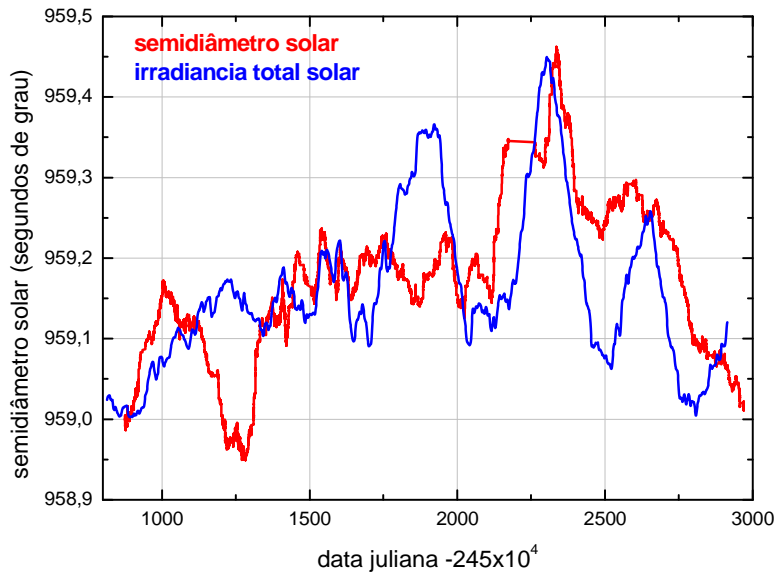

Figura 8.3 - Semidiâmetro solar e irradiância total.

**8.2 - Correlações entre as duas séries** – Da mesma maneira que fizemos na comparação entre o semidiâmetro solar e as manchas solares no Capítulo 7, (ver Item 7.2) calculamos correlações entre as duas curvas considerando intervalos de tempo desde um mês até intervalos de um ano. Assim fazendo obtivemos 67 valores de correlação entre as duas séries que comparam as duas curvas tomadas por suas médias. Os valores encontrados são mostrados na Figura 8.4. As barras de erro mostram o complemento da significância da correlação.

Este gráfico indica que embora haja uma dispersão dos valores, há uma tendência bem marcada de pontos a exemplo do que se verifica com o fluxo rádio a 10,7 cm e ao contrário do que ocorre com o número de manchas, o qual é um estimador subjetivo. O gráfico mostra que quando olhamos as curvas por seus aspectos mais gerais, ou seja, quando consideramos médias de intervalos maiores como os anuais, obtemos correlações mais altas entre as duas séries. Mas quando passamos a olhar para os detalhes, ou seja, na medida em que fazemos médias com intervalos menores, como os mensais, obtemos correlações mais baixas. Tal como no Capítulo 7 salientamos que o reconhecimento desta dispersão indica que não haveria sentido estatístico em tomar passos menores que os escolhidos.

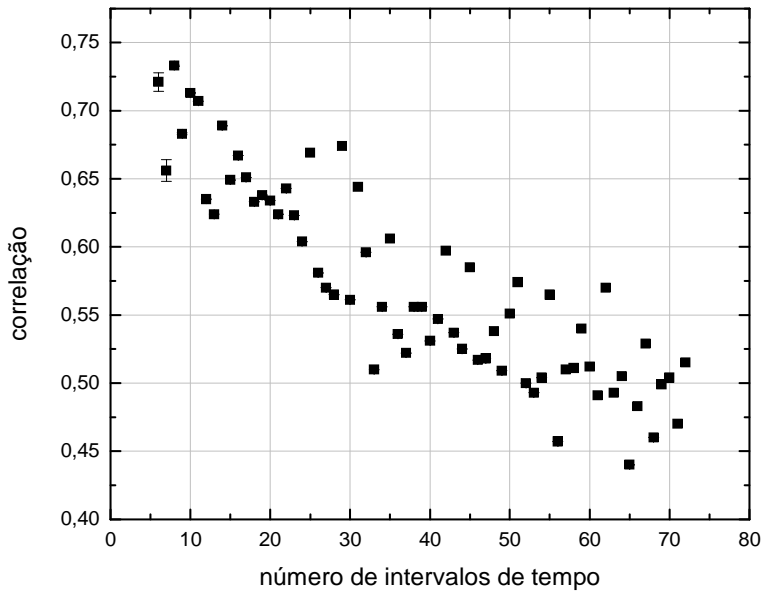

Figura 8.4 - Correlações entre o semidiâmetro do Sol e a irradiância total do Sol.

**8.3 - Correlações com as séries defasadas -** O semidiâmetro do Sol e sua irradiância fundamentalmente dependem da massa, da metalicidade e dos mecanismos de transporte radiativo da estrela. No entanto eles são modulados por fatores específicos, por exemplo, o semidiâmetro depende fortemente da rotação diferencial da estrela e a irradiância depende da atividade cromosférica. Desta maneira não há porque postular uma relação linear, nem em fase, entre semidiâmetro e irradiância embora dependam fundamentalmente das mesmas grandezas. Por este motivo investigamos as possíveis mudanças nas correlações ao se considerar defasagens entre as duas séries.

Repetimos todos os procedimentos para a obtenção das correlações, mas agora considerando defasagens entre as duas séries. Consideremos defasagens desde dois anos antes, até dois anos depois, uma a cada dez dias, o que significa cálculos para 147 defasagens diferentes. Para cada uma destas defasagens calculamos as 67 correlações conforme descrito no Capítulo 7. Como as séries estudadas têm seis anos, quando a deslocamos de dois anos mantemos ainda dois terços dos dados em comum e observamos correlações defasadas de pelo menos 20% do ciclo de atividades do Sol. Os resultados podem ser vistos na Figura 8.5.

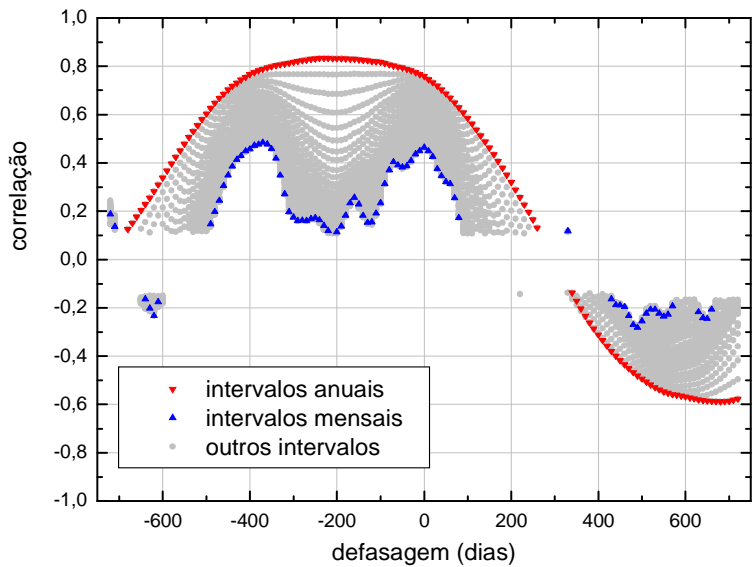

Figura 8.5 - Correlações com defasagem entre o semidiâmetro solar e a irradiância total do Sol.

Este gráfico ilustra que, excetuadas as duas maiores correlações, todas as demais ocorrem em dois pontos distintos. Um deles está onde não há defasagem entre as duas curvas e outro fica onde há uma defasagem negativa de 400 dias entre as duas curvas. Mesmo as duas famílias de maior correlação apresentam como que máximos degenerados nestes mesmos pontos. Aqui as defasagens negativas significam que a série de semidiâmetro solar está atrasada em relação à série de irradiância. Isto é, uma correlação alta com um atraso de 400 dias significa que as duas séries têm semelhanças, mas o semidiâmetro solar se comporta como que respondendo com um atraso de 400 dias. Assim, o raio do Sol reage de dois modos diferentes à irradiância solar. Um modo de reação instantânea e um modo com atraso de 400 dias.

**8.4 – Correlações quando se retiram os picos de atividade solar** – A atividade solar do ciclo 23 apresenta dois picos bem definidos. Um deles fica entre o meio de junho de 2000 e pouco depois do início de janeiro de 2001 e o outro fica entre o início de novembro de 2001 e o meio de abril de 2002. Estas datas correspondem às datas Julianas $-245 \times 10^4$ de 1709 a 1919 e de 2213 a 2378. Retiramos das duas séries de valores as medidas do semidiâmetro e da irradiância relativas à estes intervalos, e tornarmos a fazer as correlações entre as duas séries, exatamente como fizemos antes. Verificamos então os máximos das correlações se deslocam significativamente da condição de simultaneidade para a condição com atraso de 400 dias. Na série completa, os valores das maiores correlações ocorrem em torno dos dois pontos, zero dias de defasagem e -400 dias de defasagem como mostra a Figura 8.6 a seguir. Entretanto, para a série restrita, sem os picos de atividade, os valores das maiores correlações se concentram em torno de -400 dias conforme mostra a Figura 8.7.

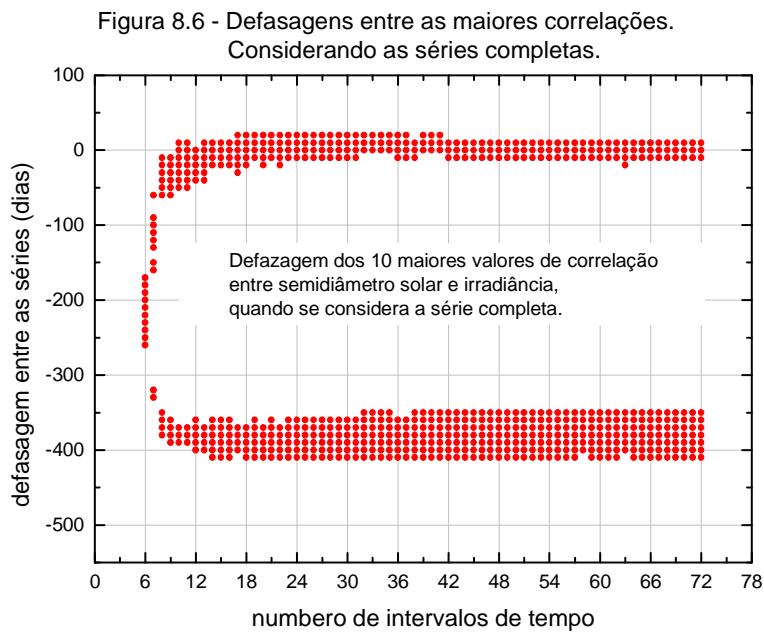

Figura 8.6 - Defasagens entre as maiores correlações considerando as séries completas.

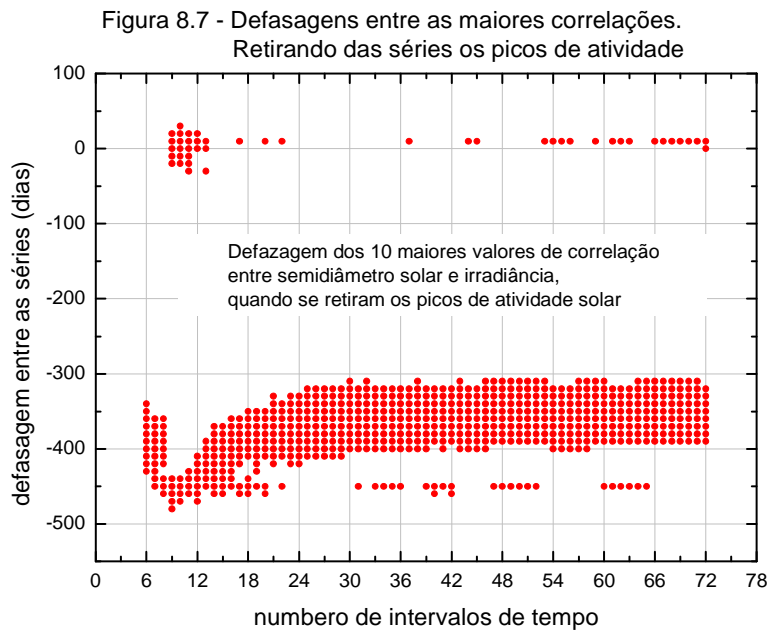

Figura 8.7 - Defasagens entre as maiores correlações. Retirando das séries os picos de atividade.

Este comportamento nos diz que há uma resposta diferente do semidiâmetro solar para os diferentes regimes de atividade do Sol. Nas atividades moduladas pelo ciclo a variação do semidiâmetro segue a variação da irradiância, e o faz com cerca de 400 dias de diferença. Nas atividades de pico a variação do semidiâmetro é imediata. Esta diferença pode explicar as divergências encontradas por outros pesquisadores nas correlações entre os dois índices. Enquanto alguns pesquisadores encontraram correlações em fase, outros encontram correlações fora de fase. Quando se têm observações bem distribuídas e frequentes é possível observar todos os detalhes, incluindo aí as variações que ocorrem durante os picos de atividades. Entretanto quando se observa por períodos não abrangentes, durante limitadas datas do ano, observam-se apenas os efeitos gerais das variações.

**8.5 - Comparando os máximos das duas séries** – Encontramos correlações entre a série do semidiâmetro solar e sua irradiância moduladas dentro do ciclo solar. Encontramos também indicações de comportamento diferente segundo se incluam ou não períodos de pico. Devemos então analisar especificamente as correlações entre as duas séries quando elas respondem aos surtos de máxima atividade. Queremos então as possíveis relações entre as datas em que ocorrem os máximos valores da série de semidiâmetro solar e as datas em que ocorrem os máximos valores da série de irradiância. Não estamos procurando pelos maiores valores de cada série, mas pelas épocas onde se deflagraram os eventos mais fortes. Em outras palavras, estamos querendo saber as datas onde ocorreram os máximos locais de cada série e não as datas onde ocorrem os maiores valores, os quais já foram abordados no item anterior.

Procuramos por máximos distantes o suficiente para serem entendidos como decorrentes de manifestações de dois eventos solares diferentes. Para tal analisamos médias diárias das duas séries e admitimos como máximos os pontos que não encontram nenhum outro valor maior no intervalo de 100 dias antes ou depois. A série do semidiâmetro apresenta-se bastante ruidosa e não é possível caracterizar corretamente médias a menos de um mês, assim, ao tomar máximos distantes pelo menos 100 dias estamos investigando um espaço pelo menos três vezes maior que o intervalo mínimo considerado. Dois picos muito próximos podem ser, na verdade, a manifestação de um mesmo evento, dois máximos a 100 dias um do outro são o resultado de eventos diferentes.

A Tabela 8.1 lista os 9 máximos encontrados na curva de semidiâmetro solar e a Tabela 8.2 lista 14 máximos encontrados na irradiância total solar. A Figura 8.8 traz a série de irradiância solar mostrando onde estão os máximos encontrados.

Tabela 8.1 – Máximos locais de semidiâmetro solar.

| N | data Juliana -245*10$^4$ (dia) | semidiâmetro do Sol (segundos de arco) |
|---|---|---|
| 1 | 1031,1138 | 959,504 |
| 2 | 1157,2000 | 959,757 |
| 3 | 1353,0667 | 959,704 |
| 4 | 1652,0167 | 959,647 |
| 5 | 1820,0600 | 959,616 |
| 6 | 2093,0473 | 959,757 |
| 7 | 2348,0488 | 960,225 |
| 8 | 2478,0552 | 960,069 |
| 9 | 2844,2144 | 959,442 |

Consideramos a possibilidade de haver alguma defasagem entre as datas em que ocorrem os máximos de semidiâmetro solar e as datas em que ocorrem os máximos de irradiância total solar. Para isto, defasamos as datas de ocorrência dos máximos do semidiâmetro desde -400 dias antes até 400 dias depois, dia a dia, obtendo assim 801 configurações diferentes entre as duas séries de datas. Como as séries têm seis anos de intervalo, os dois anos de defasagem permitem ainda comparar um terço do período total. Para cada uma das 801 configurações calculamos as diferenças em dias entre as datas dos 9 máximos de semidiâmetro solar e o máximo de irradiância solar mais próximo de cada um deles. Calculamos a média destas 9 diferenças em cada uma das configurações.

Tabela 8.2 - Máximos locais de irradiância total solar.

| N | data Juliana -245*10$^4$ (dia) | irradiância total solar (Watt/m$^2$) |
|---|---|---|
| 1 | 920,5 | 1.366,330 |
| 2 | 1004,5 | 1.366,397 |
| 3 | 1086,5 | 1.366,512 |
| 4 | 1204,5 | 1.366,735 |
| 5 | 1422,5 | 1.366,816 |
| 6 | 1526,5 | 1.367,127 |
| 7 | 1842,5 | 1.367,191 |
| 8 | 2092,5 | 1.366,809 |
| 9 | 2187,5 | 1.367,283 |
| 10 | 2301,5 | 1367,4732 |
| 11 | 2525,5 | 1366,7941 |
| 12 | 2616,5 | 1.366,982 |
| 13 | 2782,5 | 1.366,469 |
| 14 | 2889,5 | 1.366,399 |

A Figura 8.9 mostra as médias das distâncias mínimas entre os picos de semidiâmetro e irradiância para cada configuração, ou seja, para cada defasagem. Os dois valores mínimos correspondem à defasagem de -251 dias (o semidiâmetro ocorrendo antes) com um valor mínimo médio de 33,57 dias e à defasagem de 48 dias com um valor mínimo médio de 37,98 dias. Qualitativamente estes dois pontos se associam de forma natural aos dois pontos de máximo de correlação entre as séries completas de semidiâmetro solar e irradiância obtidos no Item 8.2. Ou seja, a análise destes episódios de surto é coerente tanto com a análise para a série completa como para a análise da série em que se retiraram os trechos com maior atividade.

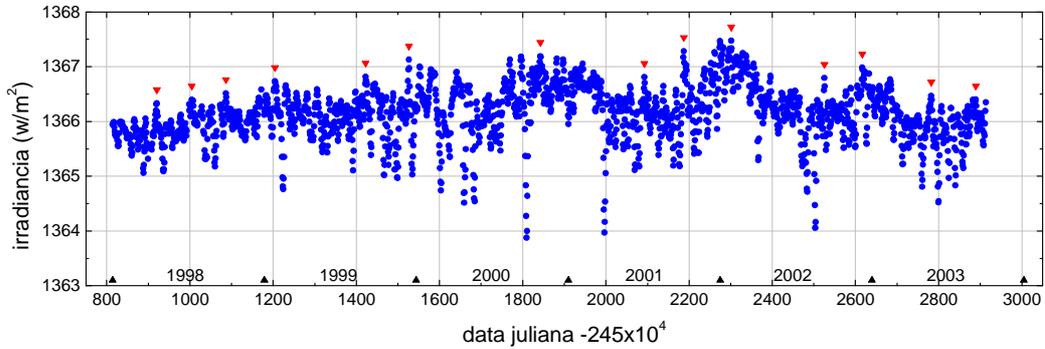

Figura 8.8 – A irradiância solar e as ocorrências máximas locais.

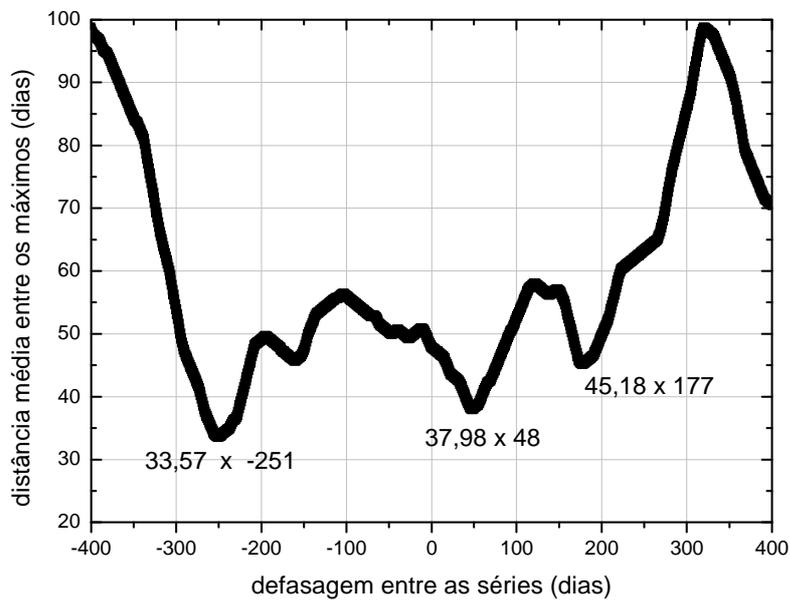

Figura 8.9 - Médias das distâncias entre os máximos das duas séries em função da defasagem entre elas.

Na Figura 8.9 aparece um terceiro mínimo. Para verificar sua importância estatística sorteamos aleatoriamente as datas de ocorrência de dez mil séries de nove máximos de semidiâmetro solar dentro do intervalo de tempo da série original. E repetimos os mesmo cálculos anteriores para cada uma delas. Obtivemos um valor médio de (62,71±19,24) dias conforme é mostrado na Figura 8.10. Desta forma o terceiro mínimo não tem significância estatística situando-se ao interno do limite de um desvio padrão, ao contrário dos dois mínimos anteriormente destacados que são estatisticamente significativos.

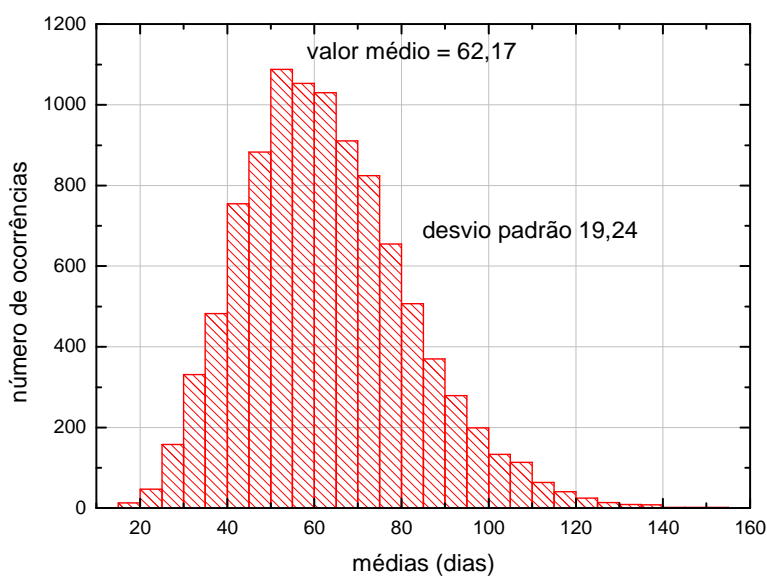

Figura 8.10 - Distribuição das médias das séries sorteadas.

Nos itens 8.3 e 8.4 encontramos duas defasagens com altas correlações entre as curvas de semidiâmetro e irradiância do Sol em torno do zero e de -400 dias. Neste item encontramos defasagens entre as datas de eventos máximos das duas séries em torno de 50 e -250 dias. A defasagem em torno de 50 dias é muito próxima do zero, mas a defasagem de -250 dias fica bem distante do valor de -400 dias. Cabe, entretanto esclarecer que, por falta de uma série mais completa com valores de irradiância, a curva da Figura 8.9 apresenta um efeito de bordas, qual seja: Quando se deslocam as curvas, os pontos extremos não encontram mais os pontos da outra curva indo buscar por pontos cada vez mais distantes o que aumenta os valores nas bordas. É possível que, retirado este efeito, o máximo encontrado em -250 dias pudesse ser deslocado para -400 dias.

# 9. Variações do semidiâmetro relacionadas ao ciclo solar – Campo Magnético

Diversas manifestações da fotosfera solar são o resultado do efeito dínamo do Sol. A principal destas manifestações é o campo magnético solar que é o principal formador das manchas solares e dos flares. Portanto, deve-se investigar também se o semidiâmetro responde também ao efeito dínamo. Para isto comparamos a série de semidiâmetro solar com a de seu campo magnético integrado. Os valores deste foram obtidos do *National Geophysical Data Center* – NGDC. Utilizamos uma série de campo magnético solar integrado com 1719 valores médios diários de 1998 a 2003. Há nesta uma média de 286 valores por ano, bem distribuídos ao longo de todo o intervalo.

**9.1 – Comparação entre as duas séries -** Uma comparação visual entre as duas séries foi feita na Figura 9.1 que mostra médias corridas das duas séries. São 330 pontos corridos para o semidiâmetro e 30 pontos corridos para o campo magnético. Desta forma, para as duas séries temos uma média corrida que abrange cerca de um mês de dados o que é necessário para se obter médias sem grande ruído. As séries são apresentadas por meio de seus desvios em relação à média e em unidades do desvio padrão de cada série.

**9.2 - Correlações entre as duas séries –** O nível de concordância entre as duas curvas pode ser compreendido quando se fazem correlações entre elas. Da mesma forma que fizemos para o número de manchas solares no Capítulo 7 e para a irradiância solar no Capítulo 8, tomamos o intervalo de tempo de dados do semidiâmetro que vai do início de 1998 até o final de 2003 e o dividimos, desde intervalos perfazendo um ano até intervalos de um mês (ver Item 7.2). Assim obtivemos 67 valores de correlação que comparam as duas curvas tomadas por suas médias, desde médias anuais, passando por intervalos de tempo cada vez menores, até às médias mensais. A Figura 9.2 mostra os valores encontrados. As barras de erro mostram o complemento da significância da correlação.

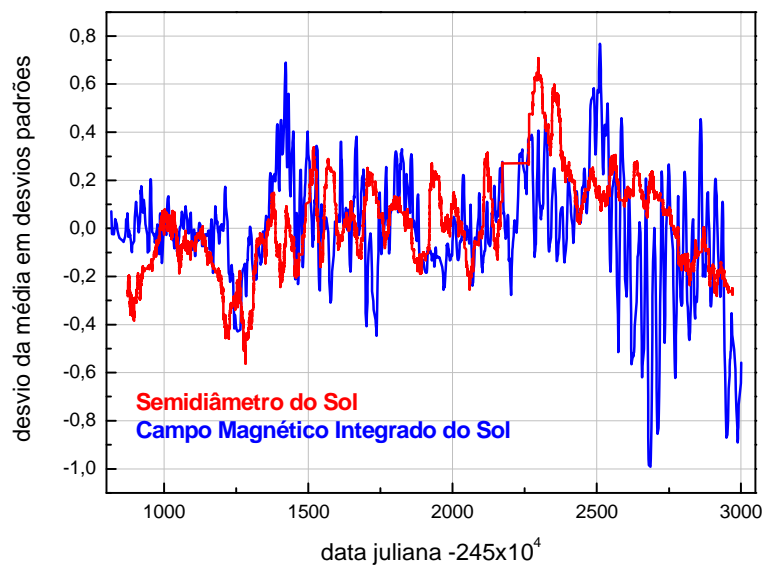

Figura 9.1 – Média corrida de 300 pontos normalizados do semidiâmetro solar e de 30 pontos, também normalizados, do campo magnético integrado.

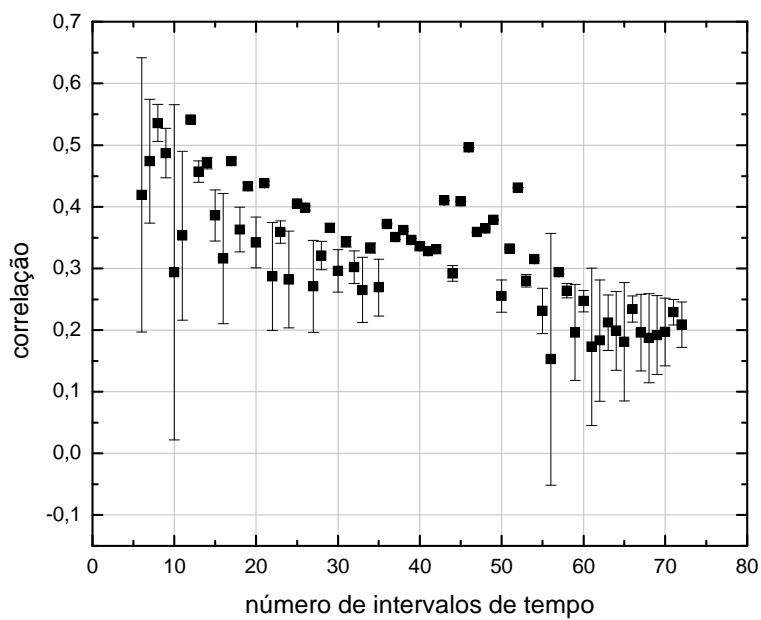

Figura 9.2 - Correlações entre o semidiâmetro do Sol e o campo magnético integrado.

As correlações entre as duas curvas não se mostram muito altas e a significância estatística é baixa. Ainda assim quando olhamos as curvas por seus aspectos mais gerais, ou seja, quando consideramos as médias anuais que são aquelas correspondentes aos números de mais baixos intervalos, obtemos correlações mais altas entre as duas séries. Quando passamos a olhar para os detalhes, ou seja, na medida em que fazemos médias para intervalos cada vez menores, obtemos correlações mais baixas. Apesar desta tendência geral, há uma dispersão de pontos.

**9.3 - Correlações com as séries defasadas -** Investigamos também as possíveis mudanças nas correlações quando se consideram defasagens entre as duas curvas. Para isso, repetimos todos os procedimentos descritos no Item 9.2, mas agora considerando defasagens entre as duas séries. Consideramos defasagens desde dois anos antes, até dois anos depois, uma a cada dez dias, o que significa 147 defasagens diferentes. Para cada uma destas defasagens calculamos as 67 correlações conforme descrito anteriormente. Os resultados podem ser vistos na Figura 9.3. Apenas os valores com significância estatística acima de 95% foram considerados, por isso, a faixa em torno de zero apresenta-se sem valores.

Nesta figura podemos ver que não há uma forte correlação entre as variações do semidiâmetro solar e as variações do campo magnético integrado solar, mas, de qualquer maneira, quando consideramos defasagens entre as duas séries há uma grande variação desta correlação. As maiores correlações entre as duas séries ocorrem quando há uma defasagem em torno de 100 dias entre elas.

Conforme precedentemente discutido, não se esperaria a mesma fase para as correlações do semidiâmetro com o campo magnético e com a irradiância, e efetivamente a defasagem aqui encontrada tem valor diferente daquela. Porém, é importante notar que ambas estão no mesmo sentido, ou seja, precedendo a variação do semidiâmetro.

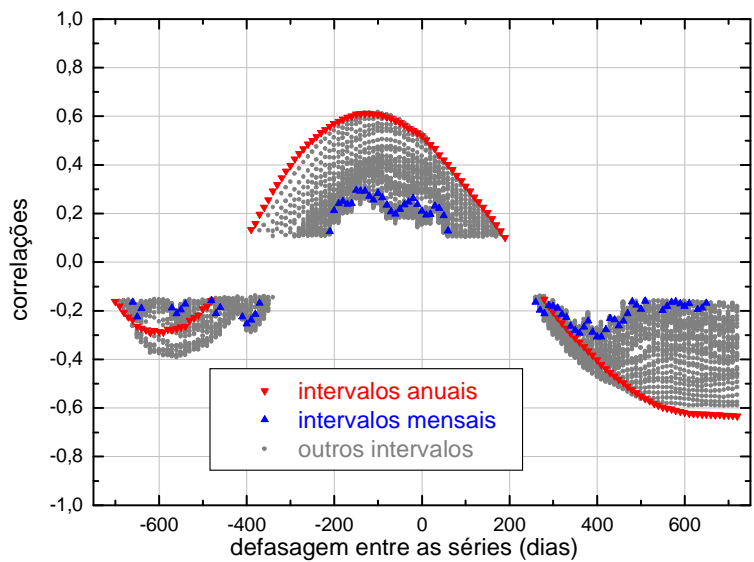

Figura 9.3 - Correlações com defasagem entre o semidiâmetro solar e o campo magnético integrado.

**9.4 - Comparando os máximos das duas séries –** Uma investigação anterior apontou uma conexão unindo picos de variações do semidiâmetro solar com picos de variações do campo magnético, erupções solares e picos de variações do campo geomagnético (Reis Neto, 2009). Isto motivou investigar também uma possível relação entre as datas de ocorrência dos máximos do semidiâmetro solar e as datas de ocorrência dos máximos do campo magnético integrado.

Inicialmente construímos uma série de semidiâmetro solar com médias diárias do início de 1998 até final de 2008. São médias ponderadas dos valores observados de 15 dias antes até quinze dias depois de cada data. Quando não foi possível obter esta média este intervalo foi estendido. A máxima extensão que se fez necessária foi de 105 dias. É relativamente fácil de se obter boas médias mensais da série de semidiâmetro, entretanto, médias diárias são muito ruidosas porque há poucos dados diários. Além de ser ineficaz, pois as perturbações do campo magnético perduram por vários dias. Ao tomarmos 15 dias para frente e 15 dias para traz, podemos obter médias com um ruído semelhante ao ruído das médias mensais.

Estas médias foram ponderadas considerando mais fortemente os valores mais próximos da data considerada e diminuindo a ponderação quando as datas se afastam. Mesmo com este intervalo, as médias de muitas datas não podem ser calculadas por falta de dados. Nestes casos estendemos o intervalo até um máximo de 105 porque assim fazendo pudemos completar quase toda a extensão de tempo considerado. Ainda assim o trecho do final de 2004 até meados de 2006 ficou sem médias.

Procuramos por correlações entre as duas séries quando elas respondem aos surtos de máxima atividade. Da mesma forma que fizemos no Capítulo 8, em relação à irradiância solar, analisamos as possíveis relações entre as datas em que ocorrem os máximos valores da série de semidiâmetro solar e as datas em que ocorrem os máximos valores da série de seu campo magnético integrado. Não estamos procurando pelos maiores valores de cada série, mas pelas épocas onde se deflagraram eventos mais fortes. Ou seja, estamos querendo saber as datas onde ocorreram os máximos locais de cada série e não as datas onde ocorrem os maiores valores. Para isso procuramos por máximos distantes o suficiente para serem entendidos como decorrentes de manifestações de dois eventos solares diferentes. Para tal analisamos médias diárias das duas séries e admitimos como máximos os pontos que não encontram nenhum outro valor maior no intervalo de 90 dias antes ou depois. Como já dissemos diante da variação característica do semidiâmetro, tanto real, como observado, só é possível se obter boas médias de períodos de pelo menos um mês, assim, ao tomar máximos distantes pelo menos 90 dias estamos investigando um espaço pelo menos três vezes maior que este intervalo. Dois picos muito próximos podem ser, na verdade, a manifestação de um mesmo evento, dois máximos a 90 dias um do outro são o resultado de eventos diferentes.

Assim fazendo encontramos 16 valores máximos na série de semidiâmetro solar. A localização temporal destes valores pode ser vista na Figura 9.4, e os valores encontrados com as datas correspondentes podem ser vistos na Tabela 9.1.

A série do campo magnético integrado do Sol com valores médios diários do início de 1996 até final de 2009 apresenta valores positivos e valores negativos. Ela pode ser vista na Figura 9.5. Entretanto, os valores acima ou abaixo de zero são apenas uma questão de polaridade do campo. Ao avaliarmos os máximos locais desta série estamos interessados apenas no módulo de seus valores e não em sua polaridade. Assim, da mesma forma que fizemos com a série de semidiâmetro solar, calculamos os valores máximos desta série, mas considerando apenas os valores absolutos da série e definindo como máximos locais os valores que não têm nenhum outro maior em um intervalo de 90 dias antes até 90 dias após a data em que ele ocorre. Assim fazendo encontramos 23 valores de máximos do campo magnético integrado. A Figura 9.6 mostra a série, agora apenas em seus valores absolutos e os máximos encontrados. E os valores dos máximos locais do campo magnético são apresentados na Tabela 9.2. As datas estão designadas pela data Juliana menos $245 \times 10^4$ e os valores estão em micro Tesla.

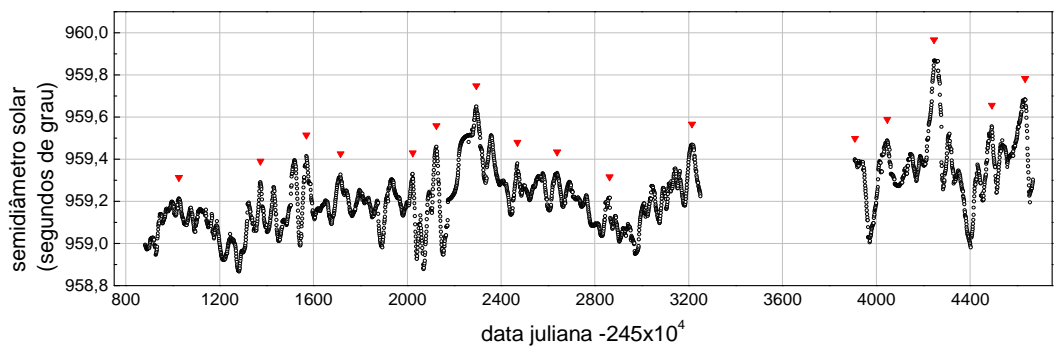

Figura 9.4 - A série do semidiâmetro solar e os máximos locais encontrados.

Queremos saber que relação há entre as datas em que ocorrem estes máximos da série de campo magnético e os máximos da série de semidiâmetro solar. Alguns máximos do semidiâmetro ocorrem bem próximos de algum máximo do campo magnético e outros ocorrem mais afastados. Para medir esta proximidade tomamos cada um dos 16 máximos de semidiâmetro e calculamos a distância em dias em que se encontra o máximo do campo magnético mais próximo dele. São 16 distâncias em dias. Achamos então a média destas distâncias.

Tabela 9.1 – Os máximos locais do semidiâmetro solar – datas em data Juliana menos $245 \times 10^4$ e os valores em segundos de grau.

| N | data | valor | N | data | valor |
|---|------|-------|---|------|-------|
| 1 | 1031,114 | 959,504 | 9 | 2844,214 | 959,442 |
| 2 | 1157,2 | 959,757 | 10 | 3166,064 | 959,884 |
| 3 | 1353,067 | 959,704 | 11 | 3934,075 | 960,12 |
| 4 | 1652,017 | 959,647 | 12 | 4035,041 | 960,164 |
| 5 | 1820,06 | 959,616 | 13 | 4315,06 | 960,776 |
| 6 | 2093,047 | 959,757 | 14 | 4428,154 | 959,937 |
| 7 | 2348,049 | 960,225 | 15 | 4636,119 | 960,038 |
| 8 | 2478,055 | 960,069 | | | |

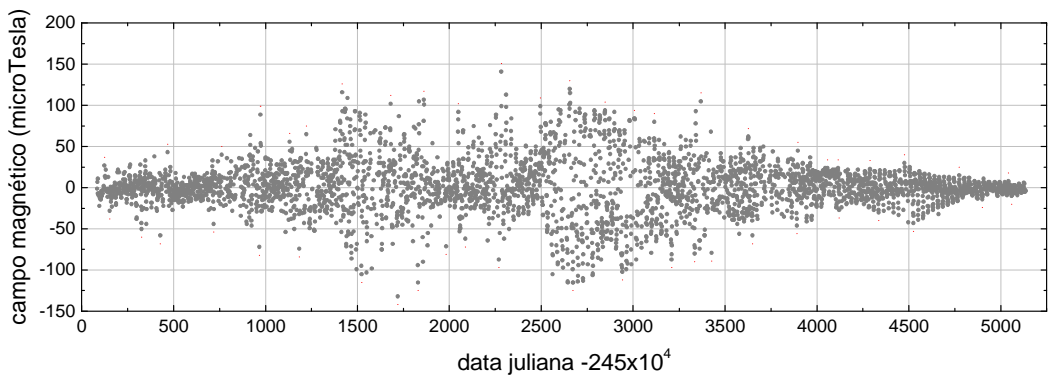

Figura 9.5 - A série do campo magnético integrado solar.

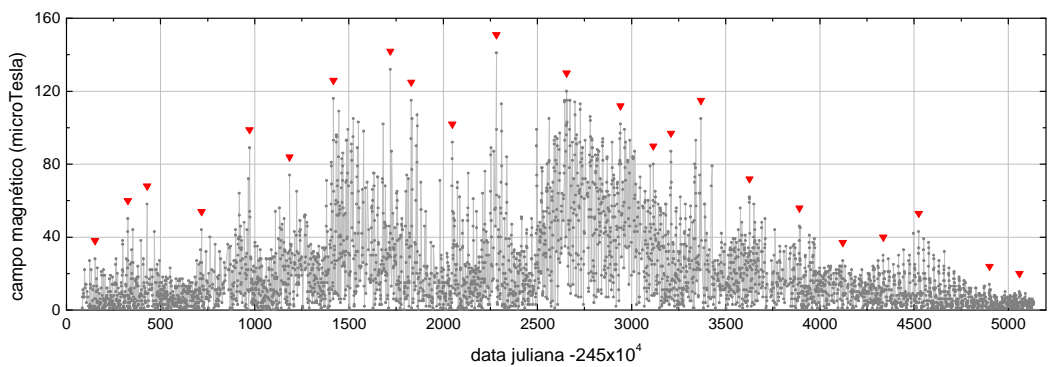

Figura 9.6 - Valores absolutos do campo magnético integrado do Sol e os máximos locais.

Tabela 9.2 – Máximos absolutos locais do campo magnético integrado. Datas em data Juliana menos 245x10$^4$ e os valores em microTeslas.

| N | data | valor | N | data | valor |
|---|------|-------|----|------|-------|
| 1 | 153  | 28    | 13 | 2942 | 102   |
| 2 | 327  | 50    | 14 | 3116 | 80    |
| 3 | 429  | 58    | 15 | 3210 | 87    |
| 4 | 718  | 44    | 16 | 3369 | 105   |
| 5 | 973  | 89    | 17 | 3627 | 62    |
| 6 | 1185 | 74    | 18 | 3892 | 46    |
| 7 | 1418 | 116   | 19 | 4122 | 27    |
| 8 | 1720 | 132   | 20 | 4337 | 30    |
| 9 | 1831 | 115   | 21 | 4525 | 43    |
| 10| 2050 | 92    | 22 | 4901 | 14    |
| 11| 2283 | 141   | 23 | 5060 | 10    |
| 12| 2656 | 120   |    |      |       |

As datas de eventos máximos da série de valores do semidiâmetro solar e do campo magnético solar podem apresentar correlações, e estas correlações podem aumentar quando se considera alguma defasagem entre as duas curvas. Por isso deslocamos a série de máximos do semidiâmetro solar desde 400 dias antes até 400 dias depois da data original, dia a dia de defasagem, e tornamos a medir a proximidade das datas dos eventos máximos do semidiâmetro solar com as datas dos eventos máximos do campo magnético, calculando a média desta proximidade a cada defasagem. Ao deslocar as duas séries alguns dados deixam de ser analisados. Ao contrário da irradiância, neste caso foi possível obter uma série bem maior de dados do campo magnético. O deslocamento de 400 dias permite que uma grande parte dos dados continue dentro do intervalo de pesquisa. A variação das médias da proximidade das datas ao longo das defasagens pode ser vista na Figura 9.7.

Nesta figura pode-se ver que há um valor mínimo, bem destacado dos demais valores, de 51,06 dias que ocorre quando as duas séries estão defasadas de 300 dias, estando a curva de semidiâmetro adiantada. Esta diferença, bem menor que as diferenças que ocorrem para outras defasagens pode estar indicando uma defasagem especialmente importante entre as curvas. Ocorrem ainda outros dois mínimos na curva da Figura 9.7. Eles têm os valores de 58,25 e 58,75 dias e se referem a defasagens de 20 dias e 265 dias respectivamente.

Avaliamos se estas diferenças em dias são estatisticamente importantes. Para tal construímos dez mil coleções de 16 datas aleatoriamente sorteadas dentro do intervalo de tempo da série de semidiâmetro do Sol. Obtendo então para cada uma delas as diferenças temporais entre estas datas sorteadas em que devem ocorrem os máximos do semidiâmetro solar e as datas mais próximas onde ocorrem os máximos do campo magnético integrado do Sol. Calculando depois o valor médio destas diferenças para cada defasagem. O gráfico da Figura 9.8 mostra a distribuição destes valores médios. O valor médio da distribuição é de (60,9±10,4) dias.

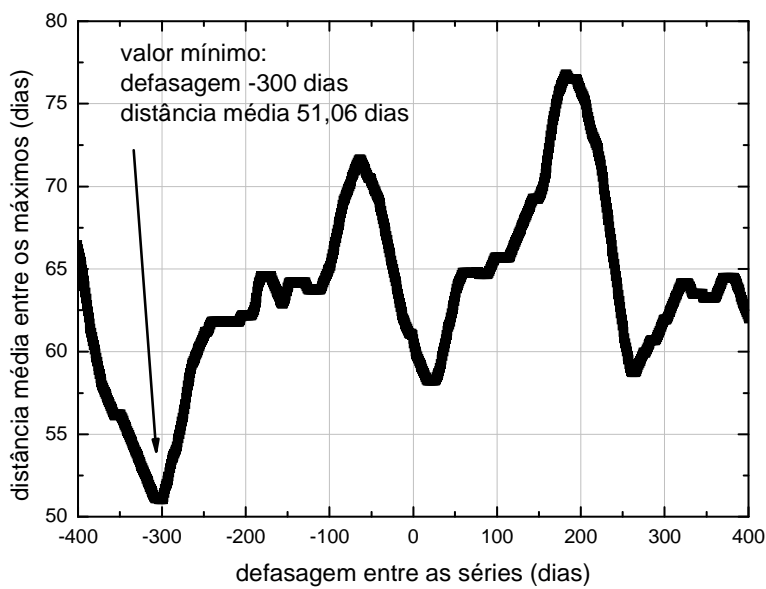

Figura 9.7 – Evolução da distância média entre os máximos locais do semidiâmetro solar e do campo magnético integrado.

O valor de 51,06 dias da Figura 9.7 ocorre apenas a 0,95 desvios padrão do valor médio da distribuição. Os outros dois pontos de mínimo ocorrem bem mais próximos do valor médio da distribuição mostrada na Figura 9.8. Desta forma nenhum deles têm significância estatística, situando-se a menos de um desvio padrão do valor médio.

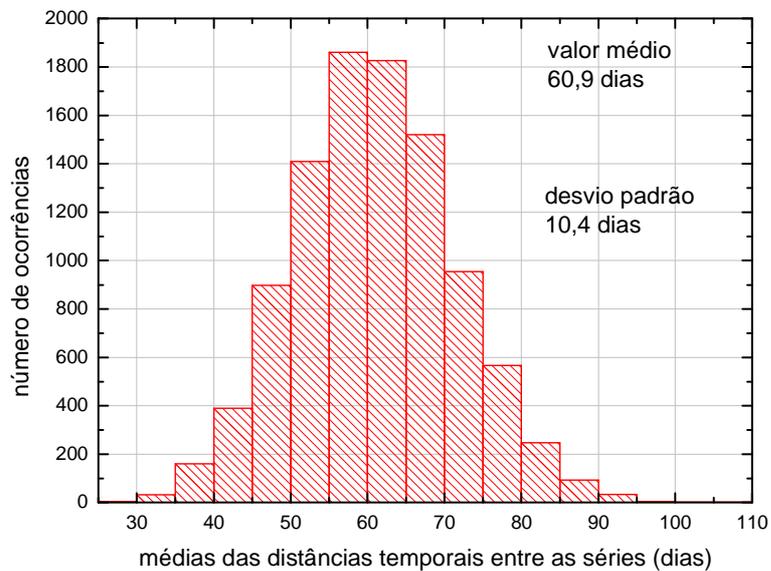

Figura 9.8 – Distribuição das médias das distâncias entre os máximos de dez mil séries sorteadas de semidiâmetro solar e os máximos do campo magnético.

Concluímos então que não há qualquer relação significativa entre os picos de ocorrência do semidiâmetro solar e os picos de ocorrência de seu campo magnético.

## 10. Variações do semidiâmetro relacionadas com a profundidade óptica.

Vimos apontando variações no semidiâmetro solar em acordo com outros observadores da ordem de 0,1 segundo de grau por ciclo solar. Variações desta ordem podem demandar uma quantidade muito grande de energia. Diversos modelos vêm sendo propostos para explicar estas variações e a energia nelas envolvida.

Considerando apenas variações de energia gravitacional, uma variação efetiva do raio solar resulta em um acúmulo de energia em conformidade com a equação 10.1 onde [ΔE] é a energia acumulada, [ΔR] é a variação do raio solar, [G] a constante universal de gravitação e [M] a massa do Sol. Assim, uma variação do raio solar da ordem de 0,1% exige uma variação de energia 75 vezes maior que a energia total irradiada (Boscardin *et al*., 2007).

$$\Delta E = \Delta R \cdot (3GM^2/5R^2) \qquad (10.1)$$

Considerando apenas a superfície irradiante, e admitindo que o fluxo por área unitária seja sempre mantido a variação do raio resulta em uma variação do fluxo irradiado de acordo com a equação 10.2 onde [I] é irradiância e [R] o raio do Sol. Isto explica metade da variação da irradiância ao longo do ciclo solar. Admitindo que a variação do raio esteja restrita apenas à zona convectiva do Sol a variação observada de 0,1% precisa de uma energia gravitacional inexpressiva, cerca de 0,01% durante o ciclo solar.

$$\Delta I = \Delta R \cdot (2I/R) \qquad (10.2)$$

Uma variação na zona convectiva pode produzir uma mudança de raio solar definido pelo equilíbrio hidrostático, sem intervenção da energia gravitacional e com variações mínimas da irradiância. A variação de irradiância é proporcional à variação de densidade e esta é descrita pela equação 10.3 onde [ρ] é a densidade e [h] a altura da camada convectiva e [$h_0$] a escala de altura.

$$d\rho = dh_0 \cdot \rho \cdot \exp[(h/h_0)^2] \qquad (10.3)$$

Há ainda algumas outras hipóteses para explicar as variações do raio solar que estão na literatura. Variações localizadas dependentes da heliolatitude em especial no bojo equatorial, em torno da "*Royal zone*", e a 60º de heliolatitude que seguem a evolução da estrutura do campo magnético ao longo do ciclo (Rozelot, 1998). Mudanças na rotação diferencial no interior da zona convectiva causadas por forças de cisalhamento e atividade magnética que levam a variações de empuxo (Brummell, Cline e Cattaneo, 2002). Auto-refração na atmosfera solar, cujo efeito no limbo é de 13 segundos de grau e varia ao longo do ciclo (Xu, 2002).

Revendo dados sismológicos do Sol de que os resultados de oscilações do modo-f que estimam mudanças do raio solar de somente 1 km/ano não excluem esforços para medir variações do raio solar na fotosfera por observações no limbo uma vez que estas são muito maiores que aquelas (Sofia *et al*, 2006). Assim, sem invocar grandes variações gravitacionais ou de irradiância, e considerando apenas a física confinada na zona convectiva, ou ainda, na fotosfera, a quantidade massiva de observações mostrando variações do raio pode ser explorada como as evidências observacionais que são.

Um outro modelo pode ser utilizado para explicar as variações do semidiâmetro solar ao longo do ciclo de atividades, já que o semidiâmetro pode ser definido como o ponto em que a profundidade óptica atinge a unidade. A profundidade óptica pode ser expressa pela equação 10.4 onde [$\tau$] é a profundidade óptica, [$\kappa$] é a opacidade, [$\rho$] a densidade do meio, [K] a constante de Boltzmann, [T] é a temperatura, [m] a massa molecular e [g] a aceleração da gravidade. Assim as variações na profundidade óptica estão associadas a variações das temperaturas e densidades nas diversas camadas da fotosfera. Na medida em que a profundidade óptica varia, o semidiâmetro observado também vai variar. A fotosfera é formada pela estrutura de grãos, onde as correntes de convecção emergem na fotosfera. Desta forma as mudanças de densidade e temperatura da fotosfera, relacionadas às variações de profundidade óptica e, portanto ao semidiâmetro observado, necessariamente refletem-se na estrutura de granulação.

$$\tau = \kappa \rho \cdot (KT/mg) \qquad (10.4)$$

Embora a granulação forme a real face da fotosfera solar, não existem registros de observações de longo termo. Dificuldades de ordem observacional e computacional para definir e seguir esta face altamente variável tornaram difícil a realização destes registros (Roudier e Reardon, 1998), mesmo sendo muito úteis para os físicos solares.

Entretanto, em anos recentes, um grande e coerente corpo de imagens em luz branca tornou-se disponível. A resolução destas imagens é da ordem de um segundo de grau, possibilitando assim, a percepção da estrutura de granulação e os grãos individuais.

Recuperamos o disco solar completo em imagem de luz branca do *Big Bear Solar Observatory* – BBSO, na Califórnia/USA, operacionalizado pelo *New Jersey Institute of Technology*.

Somente imagens no padrão *Flexible Image Transport System* - FITS foram recuperadas, podendo ser manipuladas utilizando-se o pacote de rotinas do *Image Reduction and Analysis Faclity* - IRAF. Estas imagens cobrem o início, o pico e o pós-pico do ciclo solar 23. Elas estão desigualmente, porém densamente distribuídas dos anos de 2000 a 2006. Dois tipos de imagens foram obtidos e foram tratados separadamente para controle estatístico. Imagens do disco completo, corrigidas do *dark* e do *flat field* (1261 imagens) aqui chamadas de Fl e imagens do disco completo, subtraídas do escurecimento do limbo (1341 imagens) aqui chamadas de Fr.

Todas as imagens são de 1364×1035 *pixels*. O mesmo telescópio, a mesma câmera e a mesma orientação foram mantidos. A exposição varia entre 50 e 80 ms, limitando correspondentemente o setor circular percebido da parte mais escurecida do limbo. A Figura 10.1 e a Figura 10.2 mostram respectivamente uma imagem do tipo Fl e uma imagem do tipo Fr.

Para se evitar a contaminação do escurecimento do limbo no tratamento das imagens, somente a parte central do disco solar foi recuperada para tratamento. Assim, o centro do disco solar foi achado, e em torno dele somente um setor de raio relativo de 0,35 foi considerado. Com isto a variação de intensidade foi mantida abaixo de 2% mesmo para as imagem Fl.

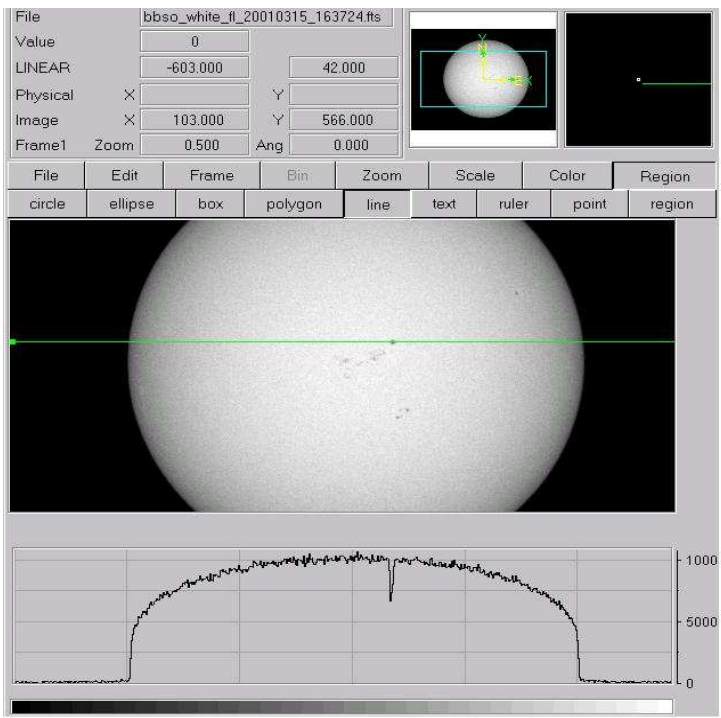

Figura 10.1 – Imagem do tipo FI. Subtraídas do *dark* e do *falt field*.

O setor considerado foi dividido em 10×10 subsetores contíguos. Cada subsetor contém 30×30 *pixels*. Três estimadores foram usados para uma avaliação do estado de granulação. Toda a análise estatística foi aplicada de maneira independente para cada subsetor, e aqueles com valores característicos se afastando da média mais do que três desvios padrão foram removidos da análise final. Para cada estimador a média dos subsetores foi recalculada após a remoção daqueles desviantes. Esta estratégia serve para descartar a presença de manchas solares na descrição da granulação solar.

Os três estimadores utilizados estão descritos na Tabela 10.1:

O modelo então supõe um campo de grãos formado por centros brilhantes e regiões escuras de contorno. Por meio de uma estatística de grandes números o balanço entre as duas estruturas ao longo dos ciclos solares pode ser acessado.

Tabela 10.1 – Os estimadores utilizados para avaliar a estrutura granular do Sol

| estimador estatístico | nome | avaliação de: |
|---|---|---|
| desvio padrão da contagem | [S] | tamanho médio dos grãos |
| diferença entre a maior e a menor contagem | [Q] | brilho dos grãos |
| grau do melhor ajuste polinomial pelas linhas e colunas | [N] | número de grãos |

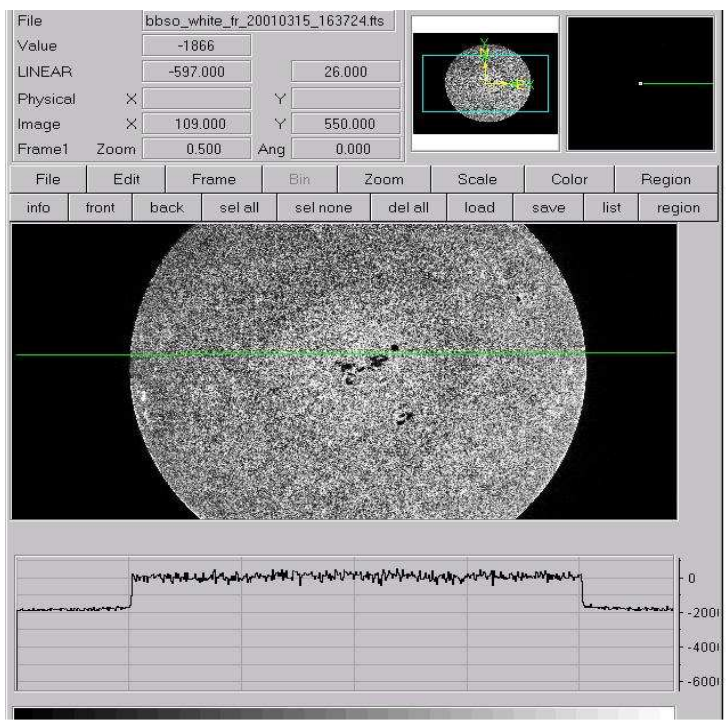

Figura 10.2 – Imagem do tipo Fr. Subtraídas do escurecimento do limbo.

Após todas as imagens [Fl] e [Fr] terem sido tratadas, e as três estatísticas obtidas para cada uma, um filtro final foi aplicado removendo, para cada ano, as imagens para as quais o valor médio das três estatísticas afastou-se mais que três desvios padrão da média anual. O número de imagens utilizadas foi de 1104 para [Fl] e 1245 para [Fr]. O número de sub-regiões usadas, isto é, sem suspeita de conter mancha ou fáculas, por imagem é muito similar, sendo 94,1% (σ = 2,6) para [Fl] e 94,6% (σ = 2,4) para [Fr].

A Figura 10.3 mostra a autocorrelação para as estatísticas [S], [Q], e [N], para imagens tipo [Fr] e a Figura 10.4 as mostra para as imagens [Fl].

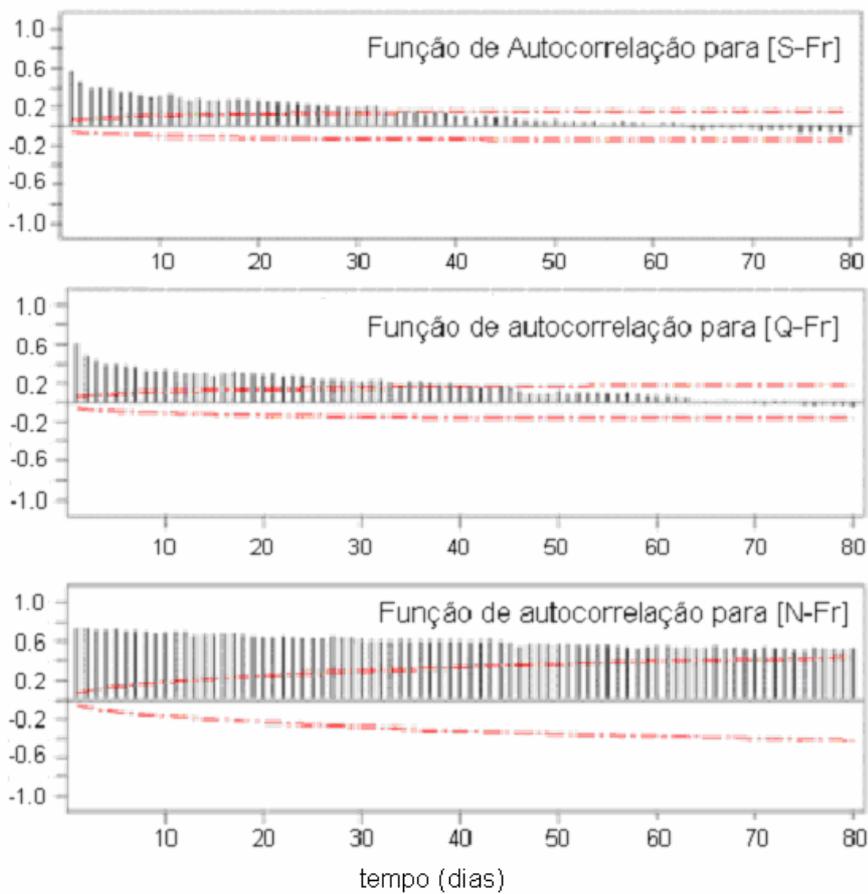

Figura 10.3 – Autocorrelação para as estatísticas das imagens Fr. A linha pontilhada é a significância estatística.

Nestes gráficos, a significância se anula em cerca de um mês. A exceção é a estatística [N] para as imagens [Fr], para a qual ela se estende além de um quarto de ano. Este comportamento é melhor entendido adiante quando da análise estatística.

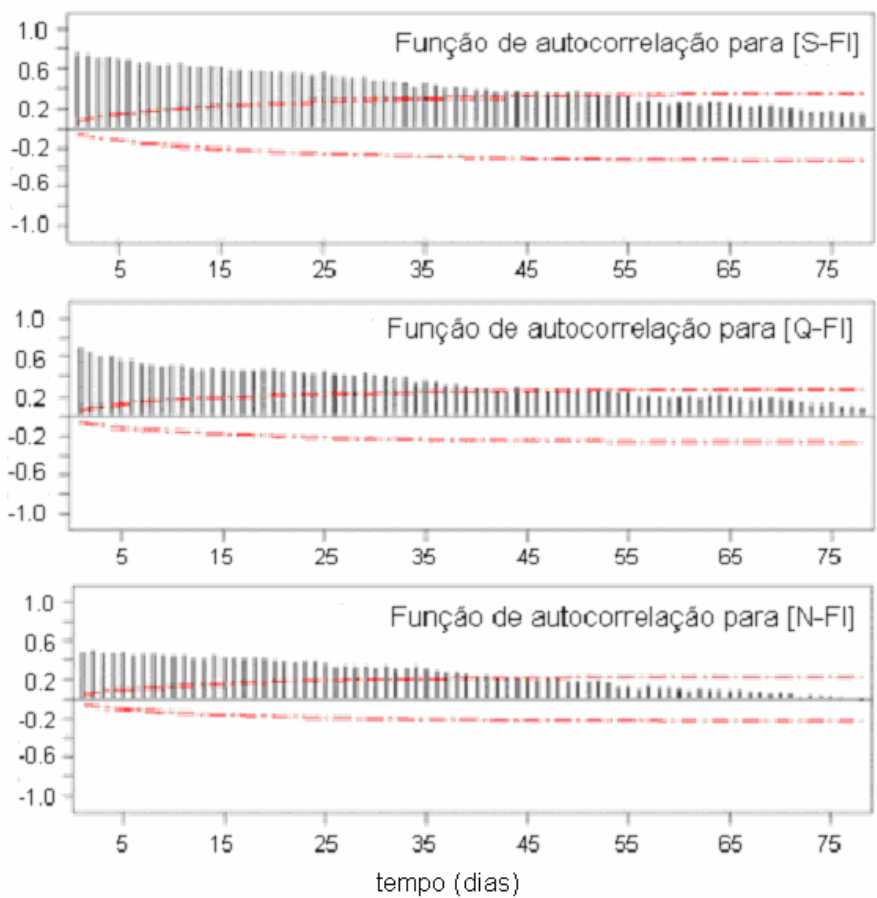

Figura 10.4 – Autocorrelação para as estatísticas das imagens Fl. A linha pontilhada é a significância estatística.

De acordo com as distribuições de autocorrelação, o teste de Run indicado pela linha tracejada vermelha mostra a presença estatisticamente significante de sinal para todas as estatísticas.

A Tabela 10.1 traz os valores médios para todas as estatísticas. A Figura 10.5 e a Figura 10.6 mostram a série temporal das estatísticas [S] e [Q] para as imagens [Fr]. Ajustou-se um polinômio de segundo grau no tempo para se investigar a existência de dependência com o ciclo solar. Note-se que este polinômio não é afetado pelas altas frequências, e a correlação de anti-fase com o ciclo solar é bastante clara. Existe uma componente anual que se deve ao fato de que os setores contêm um número sempre fixo de *pixels* enquanto que naturalmente

o tamanho angular do Sol varia ao longo do ano. Decidiu-se não remover tal componente anual desta análise, para não alterar nem minimamente os dados.

Tabela 10.1 – Valores médios anuais de todas as estatísticas.

| DATA | S-Fl | S-Fr | Q-Fl | Q-Fr | N-Fl | N-Fr |
|------|------|------|------|------|------|------|
| 2000 | 213,91899 | 154,13960 | 1089,64478 | 841,85789 | 27,00834 | 26,07540 |
| 2001 | 230,86435 | 177,77069 | 1292,35366 | 1023,89747 | 26,48127 | 26,29442 |
| 2002 | 250,62011 | 181,83760 | 1341,53689 | 985,02315 | 26,50968 | 26,32571 |
| 2003 | 232,69260 | 179,14878 | 1273,00672 | 1018,63465 | 26,47440 | 26,27116 |
| 2004 | 254,94752 | 181,06524 | 1432,81939 | 1008,43911 | 26,53336 | 26,31613 |
| 2005 | 251,04223 | 182,94556 | 1377,84262 | 1028,60612 | 26,51558 | 26,30704 |
| 2006 | 276,15892 | 226,90948 | 1616,59209 | 1247,24658 | 26,49771 | 26,19031 |

A Figura 10.7 e a Figura 10.8 mostram as séries temporais das estatísticas [S] e [Q], agora para as imagens [Fl]. Aqui a correlação de anti-fase com o ciclo solar é de novo evidente.

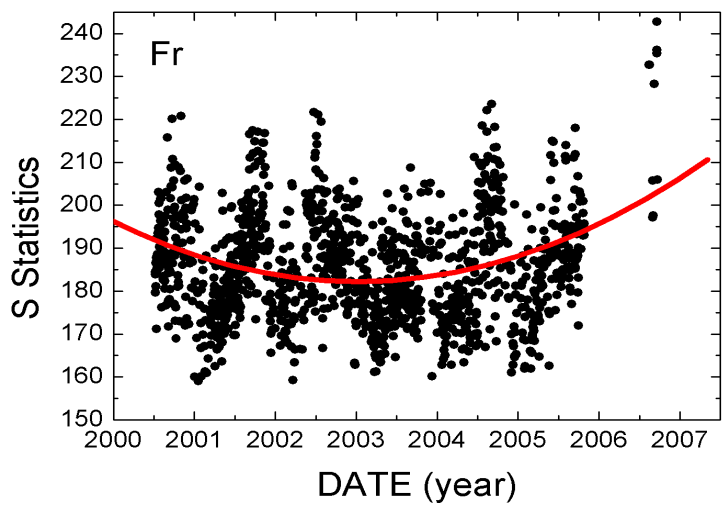

Figura 10.5 – Estatística S para as imagens Fr.

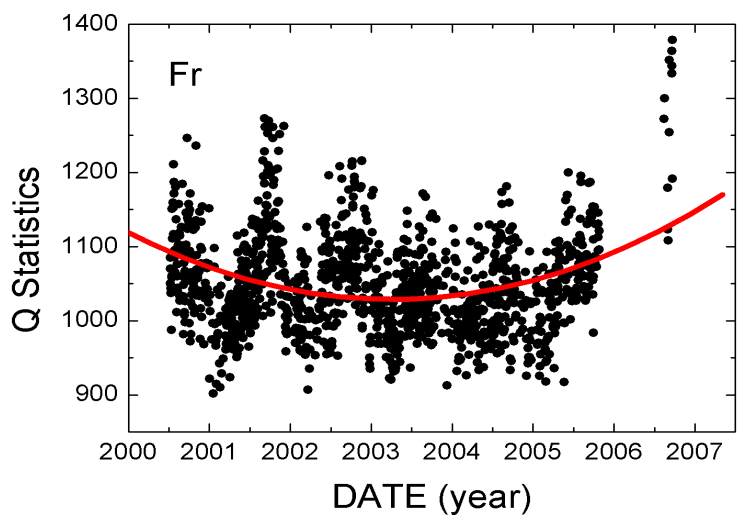

Figura 10.6 – Estatística Q para as imagens Fr.

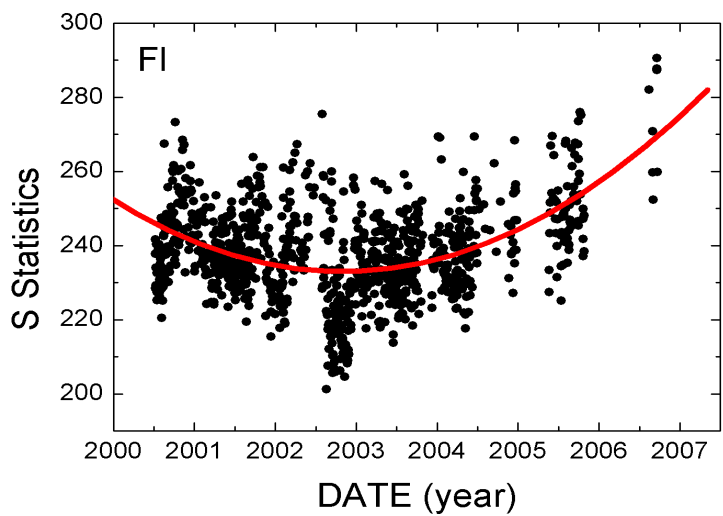

Figura 10.7 – Estatística S para as imagens Fl.

Também a posição do mínimo da curvas coincide com o máximo do ciclo solar. Para [S] com as imagens [Fr], min=2003,15; para [Q] com as imagens [Fr], min=2003,29; para [S] com as imagens [FI], min=2002,83 e para [Q] com as imagens [FI], min=2002,89. Estes valores estão enumerados em data Juliana -245x10$^4$.

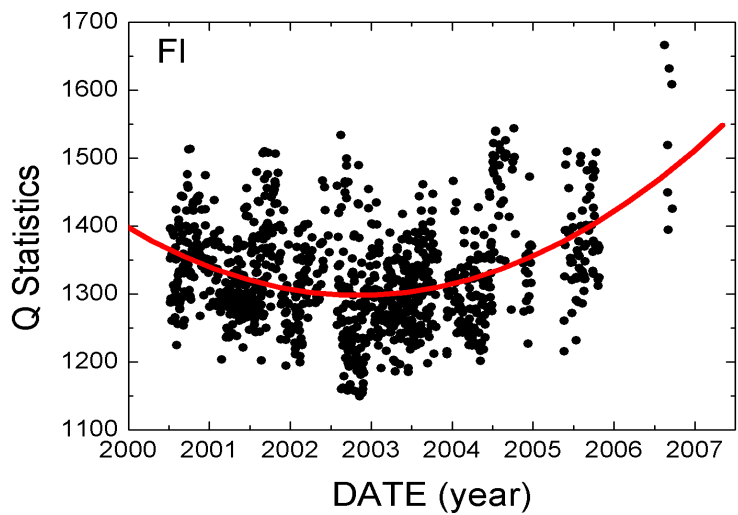

Figura 10.8 – Estatística Q para as imagens FI.

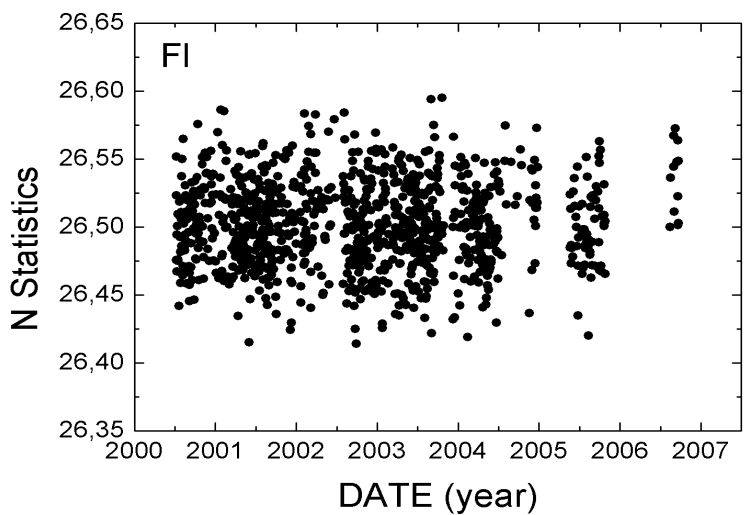

Figura 10.9 – Estatística N para as imagens FI.

Por outro lado, a Figura 10.9 e a Figura 10.10 mostram que não há variação significante para as estatísticas [N].

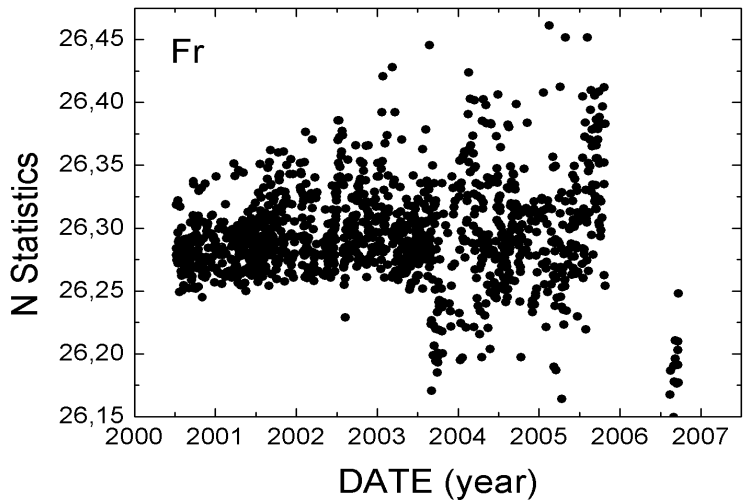

Figura 10.10 – Estatística N para as imagens Fr.

De acordo com as estatísticas, não há variação significante no número de grãos. O brilho dos grãos, ou estritamente o contraste, é mínimo no máximo de atividade solar, havendo ademais uma excelente concordância com a época obtida para a medida do máximo do diâmetro solar. Se seu número não muda e seu brilho é menor, então, o mínimo da estatística [S] mostra haver grãos de tamanhos menores no máximo solar, em excelente acordo com o máximo medido para o diâmetro.

## 11. Variações do semidiâmetro relacionadas à heliolatitude

De maneira oposta a que ocorreu com o diâmetro do Sol, a sua forma não foi objeto de estudos senão apenas nas últimas décadas. Historicamente, porque os gregos acreditavam nas idéias de Aristóteles segundo as quais o Sol era um objeto perfeito e, portanto, absolutamente circular. Esta foi uma verdade incontestável por muitos séculos porque os pensadores da Igreja Romana adotaram várias das idéias de Aristóteles, entre elas esta acerca do Sol. Além disto, as diferenças do raio solar ao longo de suas latitudes são muito pequenas e exigem instrumentos e técnicas observacionais que só se desenvolveram nos últimos cinquenta anos. Colaborou ainda para isto o fato de que a aproximação do Sol a uma esfera ser suficiente para atender os mais importantes aspectos da física solar, bem como às necessidades da astrofísica. Por estes motivos a literatura sobre o assunto é muito esparsa e bastante recente.

**11.1 - Calculando a forma do Sol –** Podemos definir o achatamento [f] do Sol pela equação 11.1 onde [$R_{eq}$] e [$R_{pol}$] são respectivamente os raios equatorial e polar do Sol. E oblacidade conforme a equação 11.2, onde [$R_o$] é o raio médio solar (Reis Neto, 2002).

$$f = (R_{eq} - R_{pol})/ R_{eq} \qquad (11.1)$$
$$\varepsilon = (R_{eq} - R_{pol}) / R_0, \qquad (11.2)$$

A Tabela 11.1 mostra as medidas de semidiâmetro solar feitas no Observatório de Calern ao longo de seis heliolatitudes diferentes em setembro de 2001. A partir dos dados desta tabela foi calculado f = (9,42 ± 3,02)·10$^{-6}$. Admitindo a figura do Sol representada pela equação 11.3 onde [$\psi$] é a colatitude, [$R_o$] o raio da esfera de mesmo volume que o elipsóide e [$P_2$] e [$P_4$] são os polinômios de Legendre de graus 2 e 4, calculou-se o termo de quadrupolo $C_2$ = -(1,1 ± 0,5)·10$^{-5}$ e o termo de hexadecapolo $C_4$ = +4,43·10$^{-6}$. Estes dados foram comparados com outras fontes conforme está na Tabela 11.2 (Rozelot, Lefebvre e Desnoux, 2003).

$$R(\psi) = R_o[1 + c_2P_2(\psi) + c_4P_4(\psi)], \qquad (11.3)$$

Tabela 11.1 – Semidiâmetro solar medido em heliolatitudes diferentes.

| Ângulo de posição (graus) | Semidiâmetro (segundos de arco) | Erro | Número de imagens |
|---|---|---|---|
| 0,0 | 959,434 | $2,8 \times 10^{-3}$ | 1980 |
| 7,2 | 959,442 | $3,2 \times 10^{-3}$ | 1320 |
| 21,7 | 959,444 | $4,1 \times 10^{-3}$ | 1056 |
| 50,0 | 959,424 | $3,3 \times 10^{-3}$ | 748 |
| 70,0 | 959,424 | $3,0 \times 10^{-3}$ | 924 |
| 90,0 | 959,425 | $2,9 \times 10^{-3}$ | 1188 |

Tabela 11.2 – Os dados de Rozelot, Lefebvre e Desnoux comparados com outros experimentos.

| experimento | $C_2$ | $C_4$ |
|---|---|---|
| MDI-SOHO | $-(5,27 \pm 0,38).10^{-6}$ | $+(1,3 \pm 0,51).10^{-6}$ |
| Pic-du-Midi | $-(1,1 \pm 0,5).10^{-5}$ | $+3,4.10^{-6}$ |
| Armstrong e Kuhn | $-(5,87).10^{-6}$ | $+0,616.10^{-6}$ |

Um modelo teórico pode explicar as distorções observadas na figura do limbo solar. Este modelo inclui um núcleo esférico com rotação uniforme encapsulado por uma tacoclina fina e prolata e uma superfície oblata, ambas girando com rotações diferentes. A Figura 11.1 mostra este esquema. A tacoclina é a região, logo abaixo da camada convectiva onde têm inicio as diferenças rotacionais do Sol em relação às latitudes. É nesta região que o campo magnético solar é formado (Rozelot, Lefebvre e Desnoux, 2003).

**11.2 A elipticidade do Sol variável ao longo da latitude e da profundidade -** A elipticidade do Sol é variável ao longo da latitude bem como da profundidade. Ela pode ser inferida com o uso de um modelo dinâmico do Sol, combinado com um modelo recente de rotação obtido de dados de heliossismologia e incluindo efeitos da rotação diferencial com a profundidade. Os valores de elipticidade podem ser calculados integrando equações diferenciais que governam fluidos e, equilíbrio hidrostático e as equações de Poisson para o potencial gravitacional. A Figura 11.2 mostra os resultados. Em torno de 70% do raio do Sol é possível se observar a passagem da zona de transição entre a rotação uniforme do núcleo do Sol e a rotação diferencial das camadas mais próximas da superfície (Godier e Rozelot, 2000).

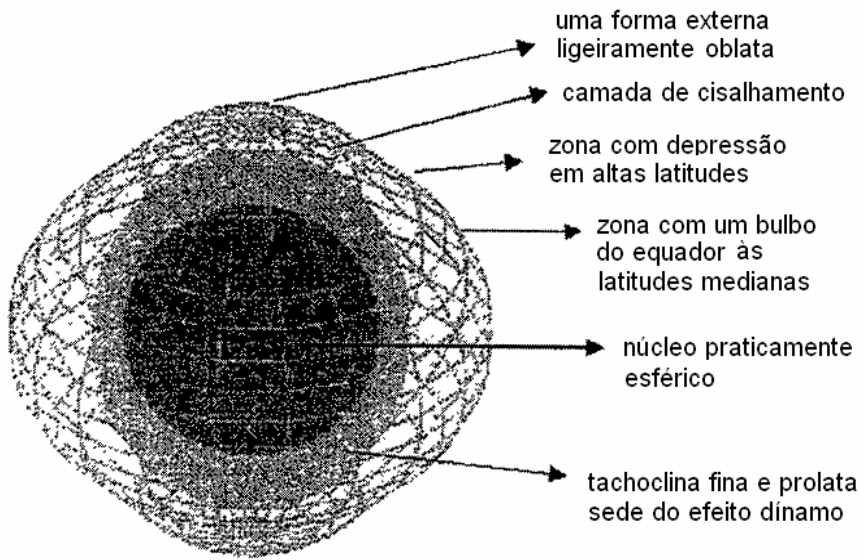

Figura 11.1 – Modelo de camadas do Sol (Rozelot, Lefebvre e Desnoux, 2003)

**11.3 - Calculando os momentos de multipolo –** Na teoria o problema da figura do Sol pode ser resolvido reduzindo-se a encontrar os equipotenciais do corpo que determinem a forma de sua superfície. A partir da esfera perfeita qualquer deformação é quantificada pelos números conhecidos como momentos de multipolo - $J_n$. Estes números são indicativos das medidas das distorções que afetam a forma do corpo: $J_1$ indica uma rotação assimétrica, $J_2$ aponta para um corpo em forma de elipsóide, $J_3$ produz uma forma de pera, e assim por diante. Quando $J_2$ é negativo, tem-se uma figura oblata: alongada na direção equatorial, quando é positivo tem-se uma figura prolata: com o diâmetro polar maior que o equatorial. Os números com índices impares são tomados como nulos quando se consideram simetrias nas latitudes e em relação ao eixo de rotação.

A Tabela 11.3 traz os resultados de diversos pesquisadores para a diferença [Δr] entre o raio equatorial e o raio polar do Sol, o achatamento [f] e o momento de quadrupolo do Sol [$J_2$] (Reis Neto, 2002).

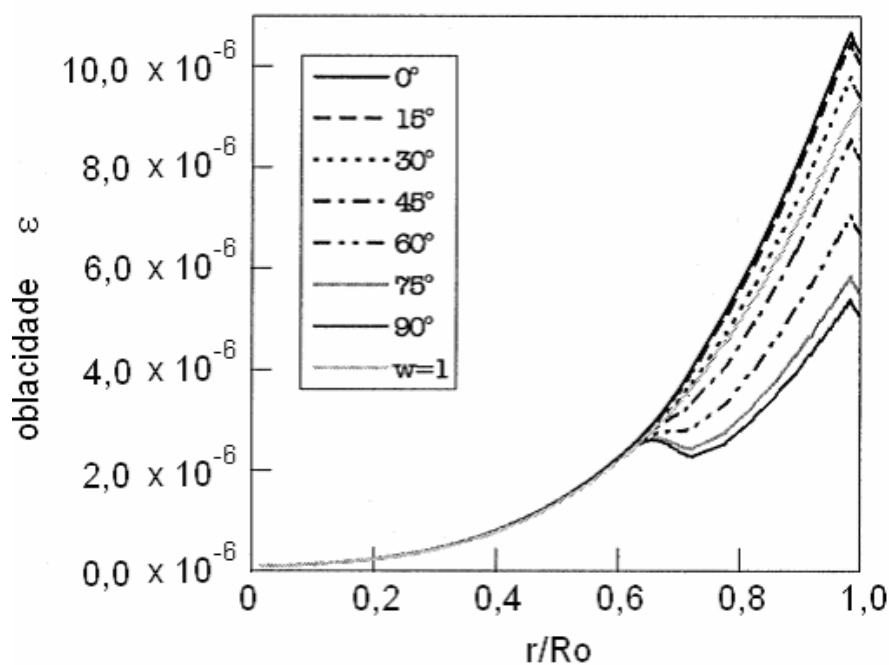

Figura 11.2 – A oblacidade varíavel do Sol ao longo da latitude e da profundidade (Godier e Rozelot, 2000).

Tabela 11.3 – A diferença [Δr] entre o raio equatorial e o raio polar do Sol, o achatamento [f] e o momento de quadrupolo do Sol [$J_2$] segundo vários pesquisadores.

| Pesquisador | Data | Δr ($10^{-3}$)" | f ($10^{-6}$) | $|J_2|$ ($10^{-6}$) |
|---|---|---|---|---|
| Reis Neto (2002) | 1998-2000 | 13 ± 4 | 13,55 ± 3,76 | 3,61 ± 2,90 |
| Rozelot & Rösh (1997) | 1996 | 13,1 ± 4,1 | - | 3,64 ± 2,84 |
| Rozelot (1996) | 1993/94 | 11,5 ± 3,4 | - | - |
| Rösh *et al.* (1996) | 1993/94 | - | 12,0 ± 3,5 | 2,57 ± 2,36 |
| Sofia *et al.* (1994) | 1994 | 8,21 ± 0,84 | 8,63 ± 0,88 | - |
| Dicke *et al.* (1987) | 1985 | 14,6 ± 2,2 | - | - |
| Beardsley (1987) | 1983 | - | 13,2 ± 1,5 | 3,4 ± 1,3 |
| Dicke *et al.* (1986) | 1984 | 5,6 ± 1,3 | 5,8 | - |
| Dicke (1981) | 1966 | | 42,3 ± 3,0 | 22,8 ± 2,0 |
| Hill & Stebbins (1975) | 1975 | 18,4 ± 12,5 | 9,6 ± 6,5 | 1,0 ± 4,3 |
| Dicke & Goldenberg (1974) | 1966 | 43,3 ± 3,3 | 45,1 ± 3,4 | 24,7 ± 2,3 |

[Δr] é dado em milésimos de segundos de grau, [f] e [$J_2$] em milionésimos.

Utilizando observações astrométricas do asteróide Ícaro de 1949 a 1987 foi possível obter o limite superior do momento de quadrupolo do Sol. O resultado formal encontrado foi de: $J_2 = -(0,6\pm5,8)\cdot10^{-6}$. Este resultado, que corresponde a uma variação secular do periélio do asteróide de $\Delta\pi = (-0,07\pm0,60)$ segundos de grau, não fornece nenhuma indicação significativa para o momento de quadrupolo do Sol, uma vez que o erro da medida é muito grande. Normalmente, os resultados de determinação astronômica de órbitas, derivados de dados astrométricos concordam melhor que o dobro do erro médio. Assim, pode-se concluir que $J_2$ muito provavelmente é menor que $2\cdot10^{-5}$ (Landgraf, 1992).

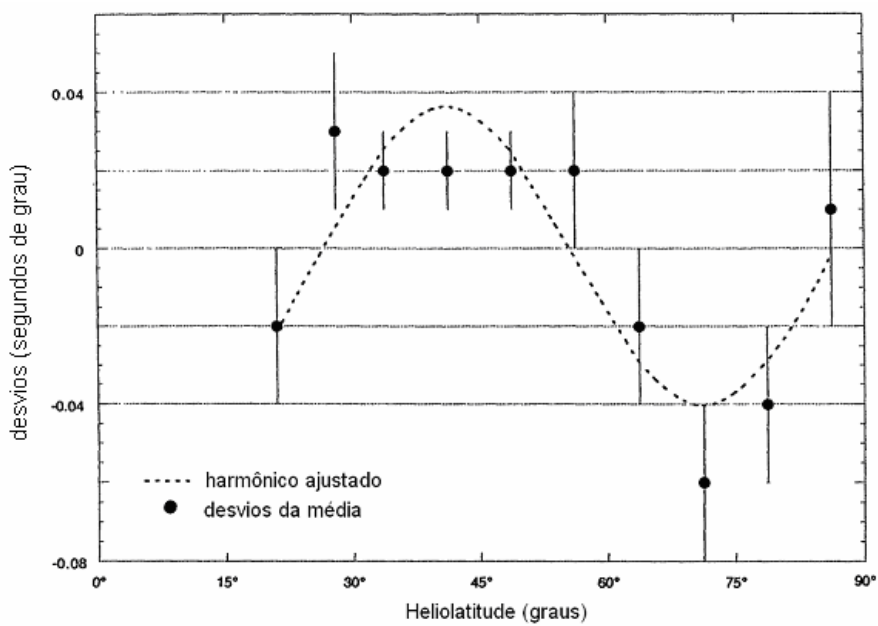

Figura 11.3 – Variações do semidiâmetro solar em função da heliolatitude no intervalo de 1978 a 1994 com mais de 5000 observações visuais (Laclare, Delmas e Coin,1996). Estes dados não mostram depleção em torno da zona *Royal* o que pode ser atribuído à amostragem limitada em apenas dez pontos.

**11.4 - Efeitos observados no astrolábio do observatório de Calern –** A latitude heliográfica de um diâmetro vertical observado com o astrolábio muda durante o ano e também com a distância zenital (ver Capítulo 3). Para estudar uma possível variação do semidiâmetro com a heliolatitude os dados foram normalizados para o zenite e agrupados em dez classes de heliolatitude porque o astrolábio de Calern utilizava dez prismas de reflexão. O resultado é mostrado na Figura 11.3 onde o raio solar varia com a heliolatitude.

Os maiores raios são encontrados em torno da Zona *Royal* e os menores em torno de 75º. A Zona *Royal*, em torno da latitude solar de 40º, é um local onde muitas das regiões ativas são encontradas. A amplitude da variação encontrada é de 0,08 segundos de grau (Laclare, Delmas e Coin,1966).

**11.5 - Medidas feitas em São Paulo –** A figura do Sol foi também traçada a partir de dados obtidos no Observatório Abraão de Moraes. Ela pode ser vista no gráfico da Figura 11.4 (Emilio e Leister, 2005).

Estes dados foram obtidos a partir de 1232 medidas independentes do semidiâmetro solar compreendidas entre 1972 e 1999. O ajuste polinomial mostra uma depleção em torno da zona *Royal*.

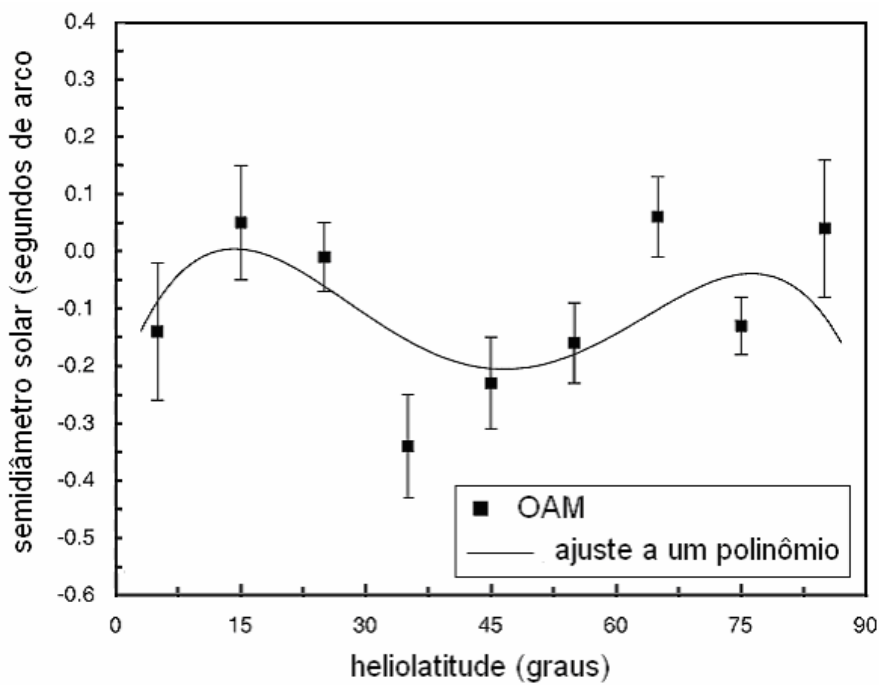

Figura 11.4 – A figura do Sol observada em São Paulo (Emilio e Leister, 2005).

**11.6 - Medidas feitas a partir de balão -** Em 1992 o *Solar Disk Sextant* – SDS a bordo de um balão fez medidas de diâmetros do Sol da altitude de 36 km no dia 30 de setembro de 1992. Foram aproveitadas 135000 medidas do diâmetro solar obtidas em diferentes heliolatitudes. Os resultados são mostrados na Figura 11.5 onde o equador é representado pelos ângulos 0 e 180 graus e os pólos pelos ângulos 90 e 270 graus. Embora o SDS tenha tomado as medidas em passos de 15, 30 e 90 graus os resultados foram agrupados em apenas 15 pontos. Este pequeno número de pontos e a significativa barra de erro a eles associada sugerem, como efetivamente é o caso, que não teria sido possível a análise detalhada do Sol de modo que o ajuste mostrado na figura corresponde a uma função seno($2\theta$).

Os valores medidos pelo SDS foram usados para calcular $\varepsilon=(8,63\pm0,88).10^{-6}$ e $J_2=(0,3\pm0,6).10^{-6}$ (Sofia, Heaps e Twigg, 1992).

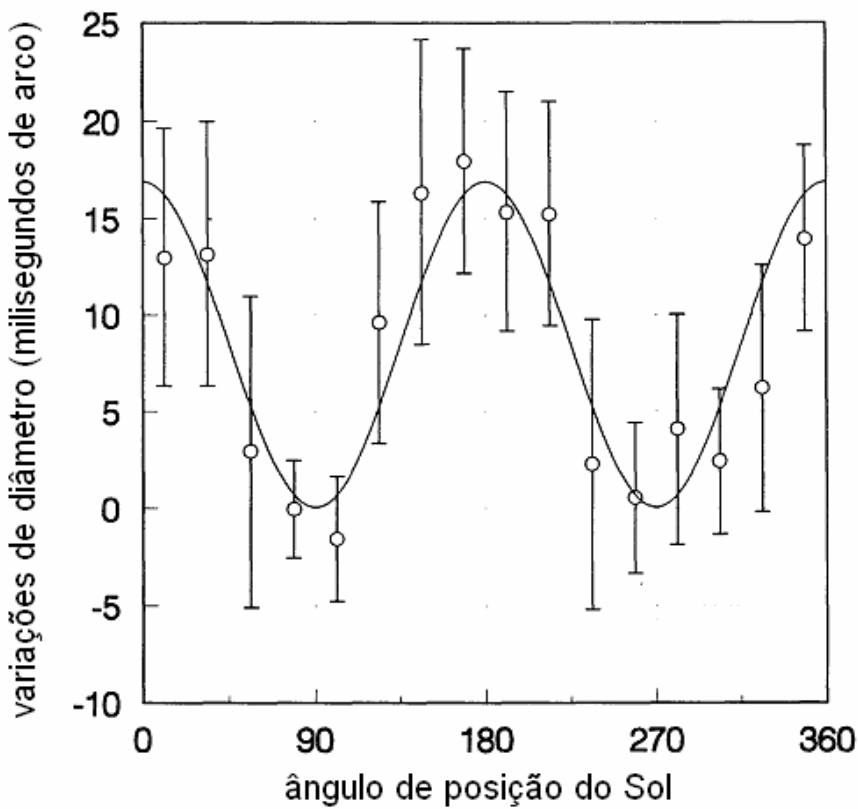

Figura 11.5 – Diferenças nas medidas do semidiâmetro solar em função da heliolatitude. (Sofia, Heaps e Twigg, 1992).

**11.7 – A figura do Sol com dados do ON –** A partir de 9112 observações independentes do semidiâmetro solar feitas no ON entre março de 1998 e dezembro de 2000 foi possível traçar a figura do Sol. Os pontos foram dispostos ao longo das heliolatitudes e agrupados em uma média corrida de 600 pontos e referidos a um quadrante. A Figura 11.6 mostra a forma do Sol estendendo o resultado do quadrante para todo o contorno (Reis Neto, 2002).

Uma depleção aparece próxima ao centro da Zona *Royal* em 36,5º com o valor de raio correspondente a 959,098 segundos de grau. Ladeando esta depleção há dois máximos, um na heliolatitude de 26,5º com o valor de raio correspondente a 959,128 segundos de grau e outro na heliolatitude de 44,5º com o valor de raio de 959,147 segundos de grau. A diferença entre o raio equatorial menos o raio polar é de (13±4) milisegundos de grau.

A variação temporal da elipticidade do Sol é apresentada no Item 11.9 utilizando os dados completos do ON tomados entre 1998 e 2009.

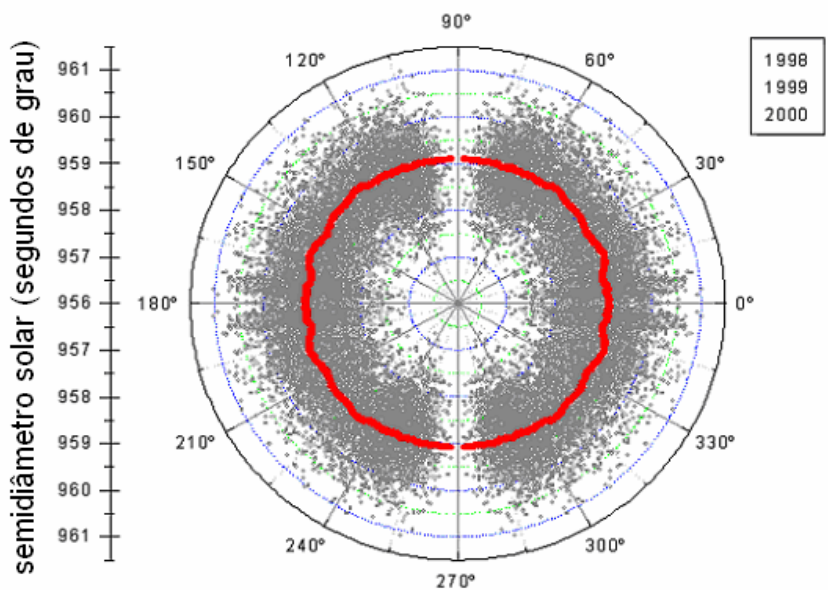

Figura 11.6 – A forma do Sol segundo dados do ON de 1998 a 2000 (Reis Neto, 2002).

**11.8 - Elipticidade solar variável –** Uma indicação pioneira de que o valor do achatamento do Sol vem se modificando ao longo do ciclo solar foi apresentada por (Dicke, Kuhn e Libbrecht, 1986). Eles compilam resultados indicando que em 1966, 1983 e 1984 a diferença entre o raio equatorial menos o raio polar era respectivamente de (41,9±3,3)·10⁻³, (19,2±1,4)·10⁻³, e (5,6±1,3)·10⁻³ segundos de grau. A sua própria medida referente a 1985 foi de (14,6±2,2)·10⁻³ segundos de grau.

Mais recentemente dados obtidos com o MDI a bordo na nave SOHO para duas épocas diferentes, 1997 e 2001 mostram valores diferentes para a elipticidade do Sol. A Figura 11.7 mostra estas mudanças do formato do Sol. (Emilio, Bush, Kuhn e Scherrer, 2007).

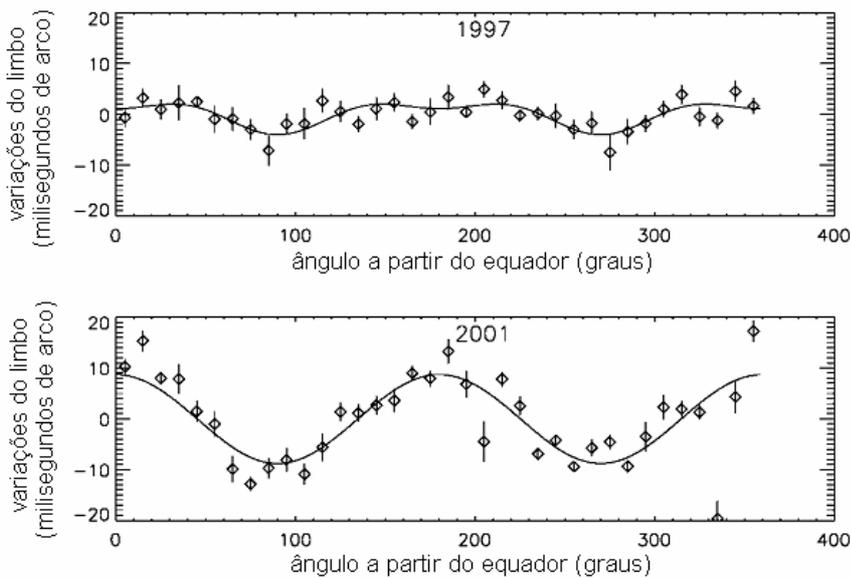

Figura 11.7 – O formato do Sol modifica-se com os anos (Emilio, Bush, Kuhn e Scherrer, 2007).

**11.9 - Dados do Observatório Nacional –** A longa série de dados do ON permite examinar ano a ano a variação da elipticidade. Inicialmente cabe colocar que ao observarmos o semidiâmetro do Sol com o astrolábio temos acesso apenas ao diâmetro vertical que o astro nos apresenta no momento da observação. Este diâmetro varia a diferentes heliolatitudes que lentamente se modificam com o correr do ano trópico. Mais lentamente também elas se modificam ao longo das horas de um dia, mas sempre dentro de uma faixa determinada pela sazonalidade. Na Figura 11.8 ilustramos como as heliolatitudes se apresentaram ao longo do ano de 2002. Elas estão separadas em duas curvas, uma para observações feitas a Leste e outra para observações feitas a Oeste do meridiano local.

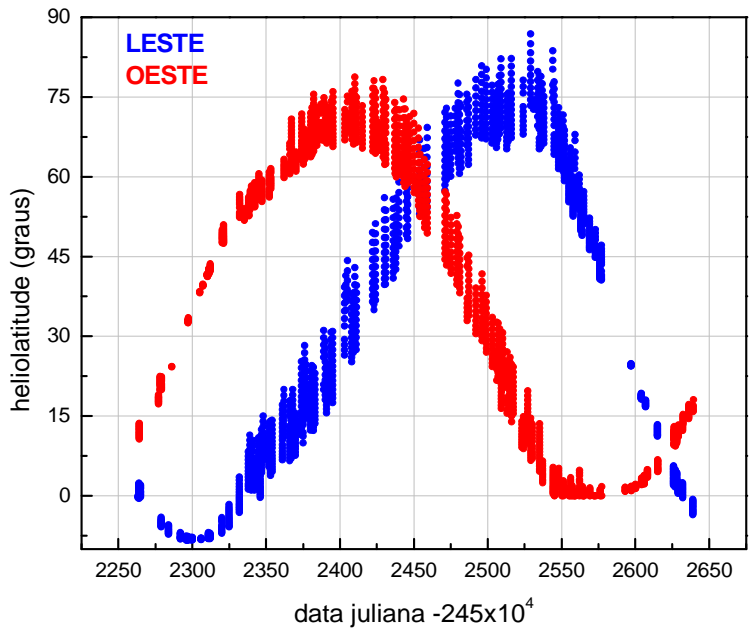

Figura 11.8 - Distribuições das heliolatitudes observadas ao longo de um ano.

Por causa destes efeitos não podemos observar igualmente todas as heliolatitudes. O gráfico da Figura 11.9 mostra a distribuição das observações no Observatório Nacional no período de 1998 a 2009. Há muitos pontos observados próximos ao equador do Sol e também muitos pontos observados próximos da heliolatitude de 70º. Há uma quantidade menor entre 10º e 60º e muito poucos pontos observados nas altas latitudes a partir de 80º.

A coleção extensa de dados do Observatório Nacional permite-nos estudar o comportamento do semidiâmetro do Sol ao longo de suas latitudes, quando se agrupa o resultado de vários anos para preencher as heliolatitudes de menor observação. Ao invés disto, neste item se tira partido desta longa série de observações para analisar ano a ano a variação da elipticidade solar.

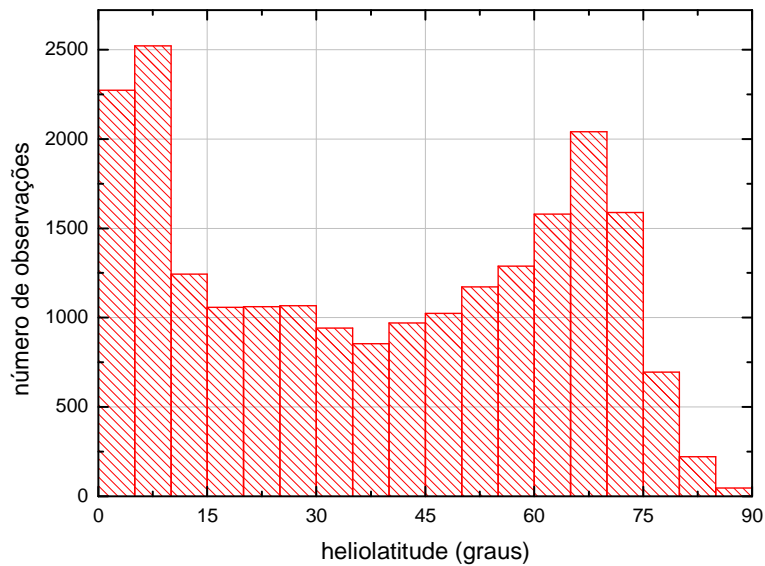

Figura 11.9 - Distribuição da heliolatitudes observadas de 1998 a 2009.

Nosso objetivo principal é estudar a varibilidade anual da elipticidade do Sol, e não exatamente determinar um valor para esta elipticidade. Para tal seria preciso um modelo que levasse em conta a elipticidade em si, os momentos de multipolo e a complexa forma solar, exigindo assim uma completeza de amostragem de heliolatitudes que não existe em trechos anuais de nossa série. Desta maneira, para a determinação da variabilidade anual da elipticidade se requer um modelo mais simples que aqui foi adotado como um ajuste linear ao longo das heliolatitudes. A Figura 11.10 apresenta as retas ajustadas aos semidiâmetros solar em função das heliolatitudes quando dividimos os dados em séries anuais.

Se tomarmos os coeficientes angulares das retas da Figura 11.10, e os colocarmos em função do tempo obtemos a interessante Figura 11.11.

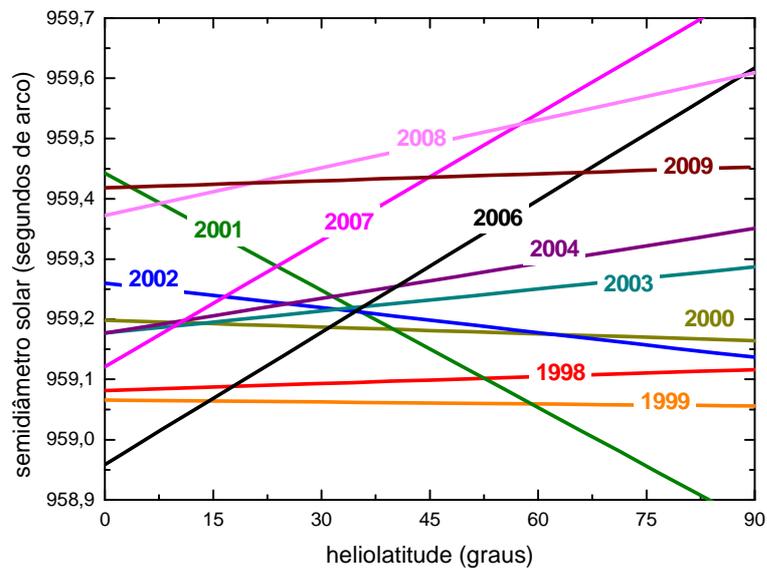

Figura 11.10 - Ajustamentos lineares ao semidiâmetro do Sol em função da heliolatitude.

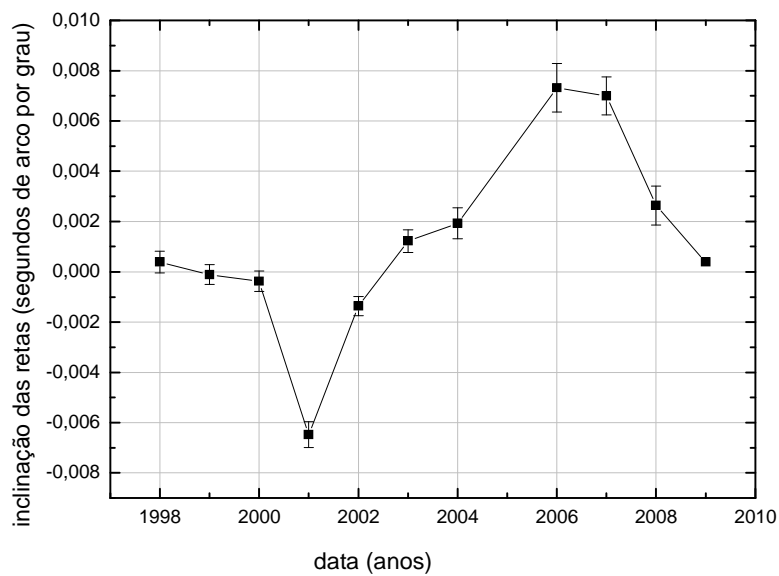

Figura 11.11 - Inclinação das retas dos ajustamentos lineares ao semidiâmetro do Sol em função da heliolatitude.

Impressiona na Figura 11.11 o fato de que há uma distribuição muito semelhante a uma senoide que tem um período de onze anos, tal como o período médio do ciclo de atividades do Sol. Nele aparecem inclinações positivas e negativas de retas ajustadas aos semidiâmetros em função das latitudes, indicando variações da elipticidade em um sentido ou no outro. Entretanto a teoria física de corpos gasosos em rotação estabelece que há sempre um diâmetro maior na direção equatorial com a consequente diminuição do diâmetro polar.

É oportuno mostrar que estas curvas ocorrem também com os dados obtidos em S.Paulo gentilmente cedidos para este trabalho por Leister e Emílio. A Figura 11.12 apresenta os ajustes de retas aos dados do semidiâmetro do Sol observados em São Paulo em função da heliolatitude e separados em séries anuais.

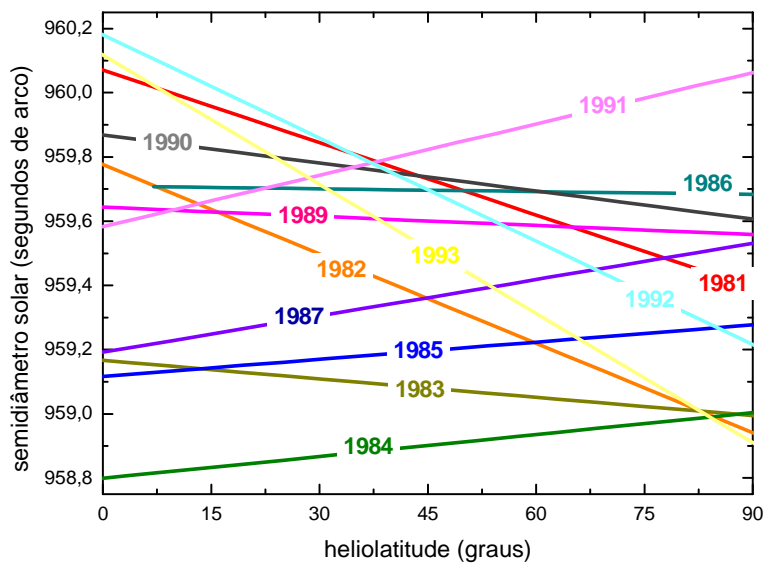

Figura 11.12 - Ajustes lineares ao semidiâmetro solar em função da heliolatitude. Valores observados em São Paulo.

A Figura 11.13 mostra o coeficiente angular das retas ajustadas aos dados anuais do semidiâmetro solar em função da heliolatitude de acordo com os valores observados em São Paulo. Compara com as do Observatório Nacional e com os valores obtidos no Cerga, dados também gentilmente cedidos pela equipe francesa. Mostra também a evolução do número de manchas do Sol numa escala arbitrária.

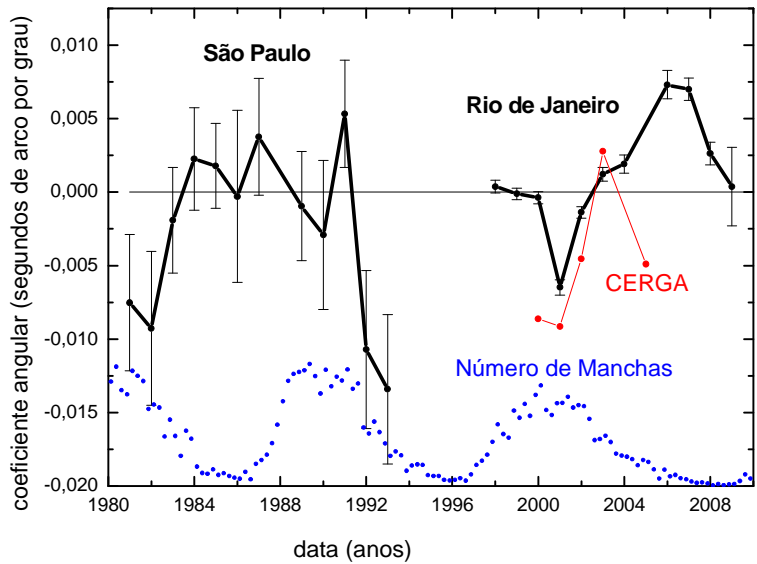

Figura 11.13 - Comparação entre os coeficientes angulares ajustados a dados de outras fontes.

Também na série de São Paulo e na série de Calern medida com CCD os coeficientes positivos continuam aparecendo, embora bem menos evidentes que na série do Rio de Janeiro. Como esperado, tais coeficientes positivos tendem a diminuir bastante quando ajustamos a figura do Sol a elipses. Ajustamos as séries anuais de semidiâmetro a elipses utilizando a técnica de mínimos quadrados e retirando fora os pontos que estavam além de dois desvios padrão. Por meio de um procedimento iterativo tais pontos foram sendo retirados até se obter uma convergência. A Figura 11.14 compara a diferença entre raio polar e raio equatorial nos dois ajustes, quais sejam, o ajuste linear e o ajuste elíptico.

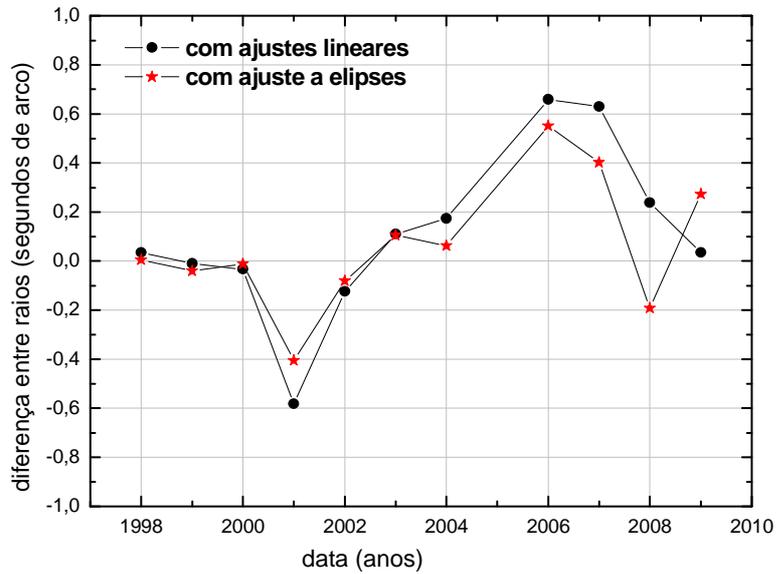

Figura 11.14 - Diferença entre raio polar e raio equatorial com ajustes lineares e com ajuste a elipses.

Na série do ON há uma diminuição média de 40% do módulo da diferença entre os raios polar e equatorial quando se utiliza ajustes a elipses em vez de ajustes lineares. O ajuste às elipses é, em presença de um grande número de dados bem distribuídos, mais adequado uma vez que a figura do Sol deve se ajustar melhor a uma elipse do que a uma variação linear do raio entre o equador e o polo.

A análise da variação anual do ajuste linear indica que a figura do Sol é mutável ao longo do ciclo de atividades do Sol apresentando em alguns anos diferenças bem marcantes entre o raio equatorial e o raio polar. Em outros anos, entretanto, estes dois raios tornam-se bastante semelhantes.

## 12. Análise de varias séries de semidiâmetro solar.

Todas as equipes participantes do grupo R2S3 da Divisão I da IAU, bem como a equipe pioneira do IAG-USP, observadores solares que utilizam o astrolábio (e por semelhança o DORAYSOL de Calern), confiaram seus dados à nossa análise. Necessário se faz agradecer a todos. Os mais antigos são os dados de observações em São Paulo analisadas por Nelson Leister e Marcelo Emilio. Seus dados trazem também informações sobre a heliolatitude observada, o que foi de grande valor para nossa análise (Capítulo 11). Estes dados se estendem de 1974 a 1994 com uma média anual de 95 observações. Outra série antiga e que é a mais longa de todas traz os dados de observações visuais feitas em Calern na França por Francis Laclare. Estes dados se estendem de 1975 a 2003 com um único ano, o de 1977, sem observações. Há nesta série uma média anual de 253 observações. Há uma outra série proveniente de Calern que é a série de dados colhidos com o uso de uma câmera CCD. Esta série se estende de 2000 a 2005 faltando observações apenas em 2004. Ela é a que possui a maior média anual de observações com 3070 valores por ano e contém também dados da heliolatitude observada. A série de valores observados em Antalya na Turquia se estende de 2000 a 2007 e tem uma média anual de 400 valores. E a série de valores observados em San Fernando na Espanha, se estende de 1972 a 1975 e sua média anual é de 133 observações. Para efeito de comparação, a série observada no Observatório Nacional se estende de 1998 a 2009, faltando dados apenas no ano de 2005. Ela tem uma média anual de 1820 valores observados. Os dados observados na França, na Turquia e na Espanha, por serem em locais com altas latitudes possuem observações feitas apenas durante os meses favoráveis quando o Sol se apresenta a uma distância zenital na qual é possível fazer observações. Os dados de São Paulo e do Rio de Janeiro encontram-se bem distribuídos ao longo do ano, uma vez que nestas latitudes é possível observar-se o Sol durante todo o ano. Análises mais completas destas séries podem ser encontradas em (Emilio e Leister, 2005), (Laclare, Delmas e Irbah, 1999), (Golbasi *et al.*, 2001). Os valores aqui discutidos foram a nós enviados através de comunicações pessoais de N. Leister, F. Cholet, F. Laclare, C. Delmas e J. Muños. A Figura 12.1 mostra, em uma escala logarítmica, o número de observações anuais feitas por cada uma destas séries.

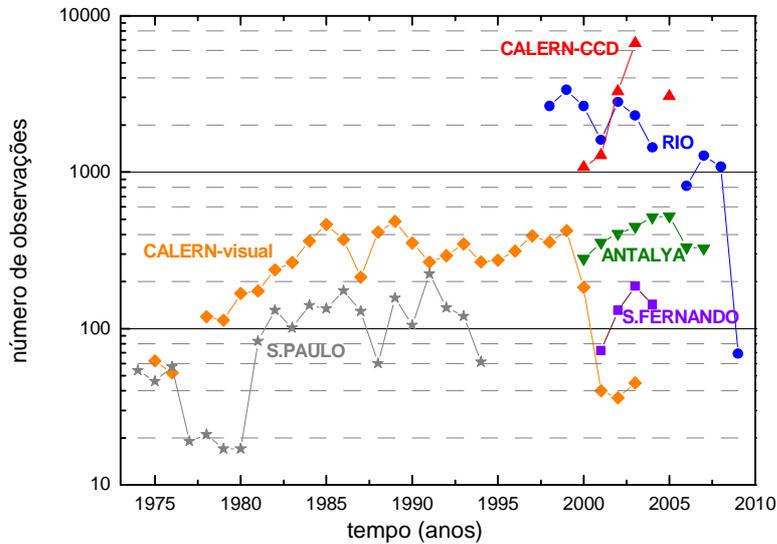

Figura 12.1 – Números de observações anuais do semidiâmetro solar em várias séries.

Para comparar e analisar a evolução de cada uma destas séries e entender o que elas têm em comum fizemos médias anuais de seus dados. A Figura 12.2 mostra as médias anuais destas séries e também, em uma outra escala adaptada para tal, a evolução do número de manchas do Sol. O erro padrão anual não é mostrado no gráfico para evitar o excesso de informação visual. Ele é mostrado na Tabela 12.2.

A primeira coisa que se destaca neste gráfico é a discordância entre os valores médios. A série de São Paulo e as séries de Calern mostram um patamar ligeiramente mais alto que a série do Rio de Janeiro. E as séries de Antalya e de San Fernando mostram valores bem mais baixos. Estes instrumentos cujos dados estamos analisando são iguais e utilizam o mesmo programa de cálculo. Do ponto de vista instrumental a exceção é o DORAYSOL que, no entanto tem a mesma abertura e distância focal dos astrolábios. Do ponto de vista de programação a exceção é São Paulo que desenvolveu uma metodologia de cálculo própria, porém repousando nos mesmos princípios que são utilizados nos demais astrolábios.

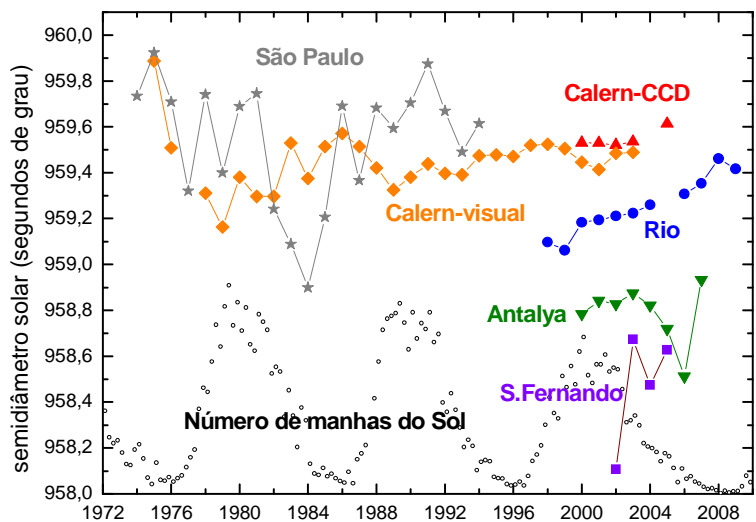

Figura 12.2 - Médias anuais de diversas séries de astrolábios.

As diferenças observadas nas séries de Antalya e San Fernando podem ser atribuídas a uma correção aplicada em distância zenital em vez de utilizar diretamente a largura do bordo. A série longa de Calern e a maior parte da série de São Paulo foram obtidas através de observações visuais o que acarreta uma diferença sistemática imposta pela diferença de resposta entre o olho humano e a câmara CCD. A série mais curta de Calern foi obtida em um sítio astronômico com um *seeing* privilegiado. A diferença no parâmetro de Fried entre o Rio de Janeiro e Calern justifica as diferenças encontradas entre as duas séries.

Há, portanto diferenças, entretanto o objetivo de todos os observadores não é determinar o real valor do diâmetro solar. Todos concordam que seu valor para a fotosfera na banda visual está em algum ponto entre 958,0 e 960,0 segundos de grau referente a uma distância astronômica. O objetivo efetivo dos observadores é conhecer a evolução temporal do diâmetro do Sol. Podemos observar esta evolução vendo como cada uma destas séries varia em relação a seu valor médio. Assim, na Figura 12.2 a curva do Rio de Janeiro apresenta uma significativa tendência linear de crescimento ao longo de toda a série no valor de 32,2 milisegundos de grau por ano. Os dados de Calern feitos com CCD apresentam uma tendência linear de crescimento de 16,1 milisegundos de grau, isto porque o último ano se destaca bem acima dos demais que tomados sem este não exibem esta tendência. Os dados de São Paulo não apresentam qualquer tendência, mas são dados referentes a duas décadas antes dos demais. Na verdade, o último ano desta série é ainda anterior ao primeiro ano da série do Rio de Janeiro. Os dados visuais de Calern apresentam dois valores iniciais mais altos e a partir daí uma série com uma leve tendência de crescimento linear no valor de 6,6 milisegundos de grau. A série de San Fernando apresenta também uma forte tendência linear de crescimento no valor de 135,9 milisegundos de grau por ano. De maneira oposta, a série de Antalya é a única a apresentar uma tendência linear decrescente de 11,8 milisegundos de grau por ano.

Tabela 12.1 – Erro padrão anual para a Figura 12.2

| ano | SP | Cal-v | ano | Rio | SP | Cal-v | Cal-C | Ant. | SF |
|---|---|---|---|---|---|---|---|---|---|
| 1974 | 0,149 | | 1992 | | 0,123 | 0,016 | | | |
| 1975 | 0,168 | 0,069 | 1993 | | 0,124 | 0,015 | | | |
| 1976 | 0,143 | 0,047 | 1994 | | 0,165 | 0,018 | | | |
| 1977 | 0,304 | | 1995 | | | 0,017 | | | |
| 1978 | 0,271 | 0,040 | 1996 | | | 0,016 | | | |
| 1979 | 0,257 | 0,033 | 1997 | | | 0,015 | | | |
| 1980 | 0,265 | 0,031 | 1998 | 0,011 | | 0,014 | | | |
| 1981 | 0,110 | 0,025 | 1999 | 0,01 | | 0,012 | | | |
| 1982 | 0,099 | 0,025 | 2000 | 0,01 | | 0,018 | 0,016 | 0,021 | |
| 1983 | 0,076 | 0,019 | 2001 | 0,014 | | 0,043 | 0,013 | 0,018 | |
| 1984 | 0,075 | 0,017 | 2002 | 0,01 | | 0,044 | 0,012 | 0,016 | 0,101 |
| 1985 | 0,069 | 0,014 | 2003 | 0,011 | | 0,038 | 0,009 | 0,018 | 0,043 |
| 1986 | 0,084 | 0,016 | 2004 | 0,015 | | | | 0,017 | 0,047 |
| 1987 | 0,089 | 0,022 | 2005 | 0,026 | | | 0,007 | 0,024 | 0,048 |
| 1988 | 0,157 | 0,016 | 2006 | | | | | 0,031 | |
| 1989 | 0,086 | 0,014 | 2007 | 0,019 | | | | 0,036 | |
| 1990 | 0,103 | 0,018 | 2008 | 0,02 | | | | | |
| 1991 | 0,080 | 0,018 | 2009 | 0,057 | | | | | |

Rio = Rio de Janeiro, SP = São Paulo, Cal-v = Calern-visual, Cal-C = Calern-CCD,
Ant. = Antalya, SF = San Fernando.

Quando estudamos isoladamente a série do Rio de Janeiro no Capítulo 3 vimos que a comparação com o ciclo de atividades do Sol apontava o acompanhamento do semidiâmetro a esta atividade no trecho do máximo de atividades do ciclo 23. Após este máximo o semidiâmetro solar se descola do ciclo de atividade e sugere acompanhar um ciclo de mais longo período. Agora, ao estudarmos as diferentes séries de semidiâmetro solar percebemos uma tendência comum de crescimento muito embora com coeficientes angulares bem diferentes. E da análise inicial das séries mostrada na da Figura 12.2 podemos concluir que as diferenças entre elas podem ser explicadas quer por fatores observacionais, quer por metodologia de cálculo. Isto sugere que as diversas curvas podem ser efetivamente combinadas formando uma única longa série de evolução temporal do semidiâmetro solar. Esta combinação também permite testar a hipótese oposta segundo a qual os resultados dos astrolábios são completamente divergentes entre si.

Partindo da hipótese de que todas as séries mostram a real variação do semidiâmetro solar e de que é irrelevante o patamar em que cada uma se situa, trouxemos todas elas para o mesmo referencial. Escolhemos como referencial fundamental o da série de Rio de Janeiro. A escolha é arbitrária, mas adequada porque ela se situa medianamente entre as demais.

Para emparelhar as diversas séries, tomamos as médias anuais de todas elas. Nos trechos comuns calculamos a diferença do valor anual entre a dada série e a do Rio de Janeiro e obtivemos a média destas diferenças. Esta média passa a corrigir toda a dada série trazendo-a para o patamar do Rio de Janeiro. Note-se que isto não altera o resultado observacional de cada uma das séries, mas assume a hipótese de que todas elas estão medindo as variações do mesmo fenômeno. Este transporte foi feito para as duas séries de Calern, a série de Antalya e a de San Fernando. A série de São Paulo não tem pontos em comum com a série do Rio de Janeiro, mas pode ser ajustada pelo largo trecho em comum que tem com a série visual de Calern e uma vez levada ao patamar desta pode depois ser trazida ao do Rio de Janeiro. Assim, todos os ajustes efetuados compreendem a soma de um valor constante para todos os pontos de cada série. Os valores destas constantes podem ser vistos na Tabela 12.2

Uma vez trazidos todos os pontos para o mesmo patamar obtivemos uma série com 77 pontos preenchendo todos os anos de 1974 a 2009, ou seja, valores preenchendo um total de 36 anos. Esta série é apresentada na Figura 12.3. Fica evidente que a primeira metade da série de 1974 a 1993 é bem mais ruidosa que a metade mais recente. Isto quantifica o grande trabalho dos pioneiros ao desenvolver o experimento e o instrumental para realizá-lo. E da mesma maneira quantifica a qualidade do trabalho das equipes que continuaram o desenvolvimento das medidas do diâmetro solar. A primeira metade da série tem um intervalo de medidas de 1,10 segundos de grau e um desvio padrão de 0,22 segundos de grau. Mesmo levando em conta a real variação do diâmetro na metade mais recente da série aqueles valores caem para a metade ou seja, um intervalo de medidas de 0,58 segundos de grau e um desvio padrão de 0,12 segundos de grau.

Tabela 12.2 – Constante de correção das séries para levá-las ao mesmo patamar do Rio de Janeiro.

| local | anos em comum nas duas séries | média local | média Rio | diferença | |
|---|---|---|---|---|---|
| Cerga-CCD | 2000 a 2003 | 959,529 | 959,203 | -0,326 | |
| Cerga-visual | 1998 a 2003 | 959,477 | 959,162 | -0,315 | |
| Antalya | 1999 a 2004 + 2006 | 958,800 | 959,206 | 0,376 | |
| San Fernando | 2002 e 2003 | 958,390 | 959,217 | 0,827 | |

| local | anos em comum nas duas séries | média local | média Cerga-vis. | diferença ao Cerga | diferença ao Rio |
|---|---|---|---|---|---|
| São Paulo | 1975,1976 + 1978 a 1994 | 959,544 | 959,430 | -0,114 | -0,429 |

Valores em segundos de grau

A busca de períodos da ordem do ciclo solar nesta série é pouco clara diante da dispersão dos valores da primeira metade. Por isso vamos procurar pelos ciclos de períodos mais longos da atividade solar, nos quais aquela dispersão pode ser naturalmente alisada. Assim sendo o método indicado é ajustar um polinômio. Ajustamos a estes pontos polinômios do terceiro ao sexto grau. Os ajustes dos pontos a curvas do terceiro e do quarto grau aparecem na Figura 12.4. Do quarto ao sexto grau os polinômios fornecem essencialmente a mesma representação, a correlação entre o ajuste ao quarto grau e o ajuste ao quinto grau é de 0,984; entre o ajuste ao quarto e ao sexto grau é de 0,846 e entre o ajuste ao quinto e ao sexto grau é de 0,895. Os ajustes polinomiais do quarto ao sexto graus são mostrados na Figura 12.5.

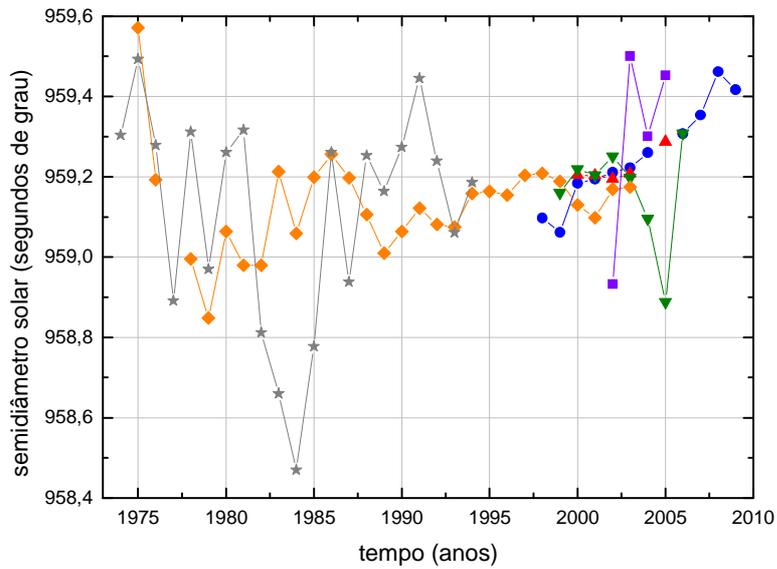

Figura 12.3 – Todas as séries de semidiâmetro solar disponíveis levadas ao patamar do Rio de Janeiro.

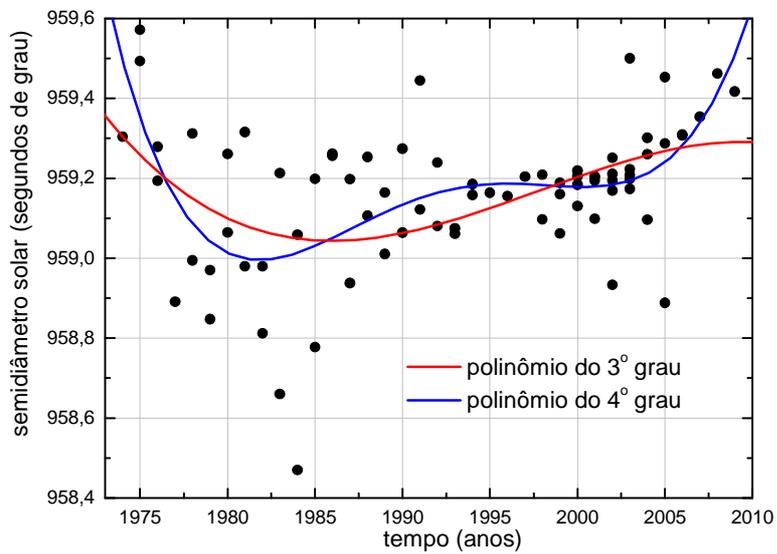

Figura 12.4 – As médias anuais de todas as séries disponíveis e polinômios do terceiro e quarto grau ajustados.

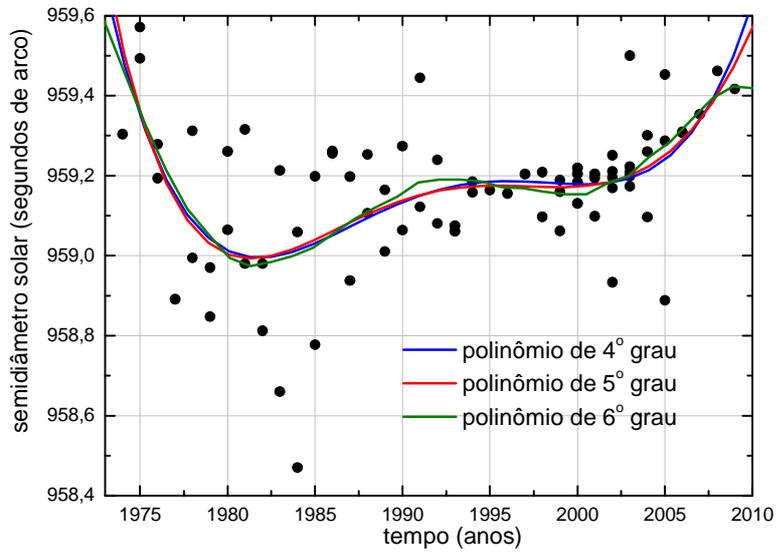

Figura 12.5 – As médias anuais de todas as séries disponíveis e polinômios do quarto, quinto e sexto graus ajustados.

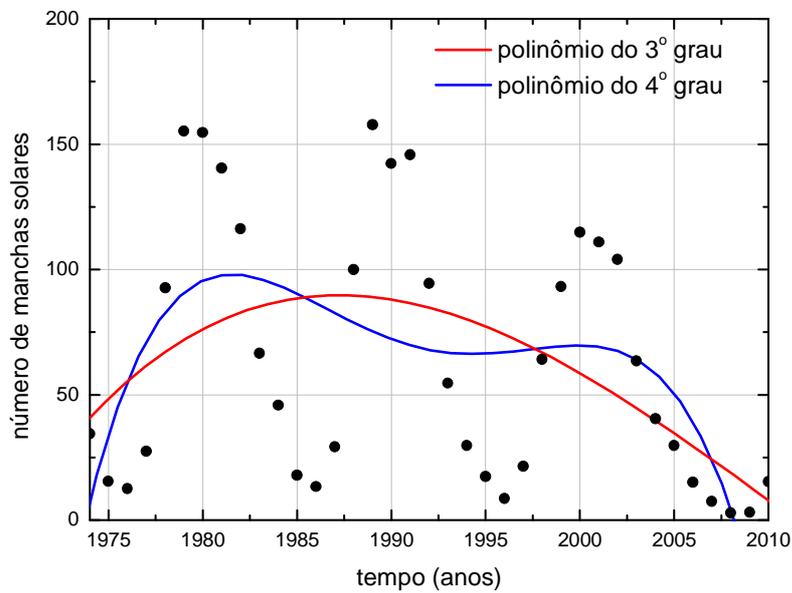

Figura 12.6 – Polinômios do terceiro e do quarto grau ajustados á série anual do número de manchas solares.

Para podermos comparar a curva composta dos semidiâmetro solares com a atividade do Sol também ajustamos polinômios do terceiro e quarto grau à serie anual de número de manchas solares. Os resultados destes ajustes aparecem na Figura 12.6. A comparação entre esta figura e a anterior mostra resultados notavelmente especulares. A correlação entre os ajustes polinomiais é de -0,986 para os dois ajustes ao quarto grau e de -0,899 para os dois ajustes ao terceiro grau.

Chamamos a atenção para o efeito de bordos que o ajuste polinomial confere o qual deve ser devidamente desconsiderado. Ao retirarmos dos ajustes ao terceiro grau um ponto a cada lado os quais extrapolam os limites da curva composta de semidiâmetro a correlação entre as duas curvas de terceiro grau é ainda maior (em módulo). O valor é de -0,985. Os ajustes aos pontos da curva composta de semidiâmetros, principalmente aquele ao quarto grau apontam no sentido do crescimento do raio na presente data. Isto reforça a idéia de que o raio do Sol pode estar seguindo o caminho de um ciclo de período mais longo que envolve os ciclos básicos de 11 anos e que caracteriza um mínimo de atividades tal como o Mínimo de Dalton, ou ainda mais profundo, como o Mínimo de Maunder.

Ao trazermos todas as séries para o mesmo patamar percebemos que o conjunto de astrolábios mediu com acerto o semidiâmetro solar e que este segue a atividade solar em fase oposta. O conjunto composto de dados dos astrolábios mostra uma relação forte com a atividade solar e isto nega a hipótese de que os astrolábios estariam apenas fazendo medidas locais diferentes ou ainda medindo algum outro fenômeno com efeito na atmosfera.

# 13 – Sumário e Perspectivas futuras

## 13.1 – Principais conclusões

O erro que um operador de astrolábio comete para encontrar o exato momento de toque das imagens no momento de passagem de um bordo solar, aumenta consideravelmente com o número de imagens obtidas por unidade de tempo. E o número de imagens deslocadas do centro é determinante no erro de definição do momento de toque.

A probabilidade de haver uma mancha no bordo solar no momento de uma medida, na época de maior atividade solar deve ser considerada, ainda assim, a influência desta mancha não modifica fundamentalmente os valores das medidas.

Entre 1998 e 2003 o semidiâmetro do Sol acompanhou a atividade solar de acordo com o número de manchas, imitando muito bem os seus máximos.

A partir de 2003 o diâmetro solar não mais acompanhou a atividade solar, mas passou a crescer constantemente. A este fato somam-se outros que indicam que estamos próximos de um mínimo solar, talvez tão profundo como o Mínimo de Dalton ou ainda mais, tal como o Mínimo de Maunder no qual o diâmetro solar foi também medido maior.

Há uma resposta diferente do semidiâmetro solar para os diferentes regimes de atividade do Sol. Nas atividades moduladas pelo ciclo a variação do semidiâmetro segue a variação da irradiância, e o faz com cerca de 400 dias de diferença. Nas atividades de pico a variação do semidiâmetro é imediata.

As maiores correlações entre o semidiâmetro solar e o campo magnético solar integrado ocorrem quando há uma defasagem em torno de 100 dias entre ambos. A mesma defasagem é observada para as maiores correlações entre o semidiâmetro e o número de manchas solares.

Não há variação significativa no número de grãos da fotosfera solar. O brilho dos grãos, ou estritamente o contraste, é mínimo no máximo de atividade solar, havendo ademais uma excelente concordância com a época obtida para a medida do máximo do diâmetro solar. Há grãos de tamanhos menores no máximo solar, também em excelente acordo com o máximo medido para o diâmetro.

A figura do Sol é mutável ao longo do ciclo de atividades do Sol apresentando em alguns anos diferenças bem marcantes entre o raio equatorial e o raio polar. Em outros anos, entretanto, estes dois raios tornam-se bastante semelhantes.

Os ajustes aos pontos da curva de semidiâmetro solar composta a partir de vários astrolábios apontam no sentido do crescimento do raio na presente data. Isto reforça a idéia de que o raio do Sol pode estar seguindo o caminho de um ciclo de período mais longo que envolve os ciclos básicos de 11 anos e que caracteriza um mínimo de atividades tal como o Mínimo de Dalton, ou ainda mais profundo, como o Mínimo de Maunder.

Ao trazermos todas as séries de astrolábios para o mesmo patamar de dados percebemos que o conjunto de astrolábios mediu com acerto o semidiâmetro solar e que este segue a atividade solar em fase oposta. O conjunto composto de dados dos astrolábios mostra uma relação forte com a atividade solar e isto nega a hipótese de que os astrolábios estariam apenas fazendo medidas locais diferentes ou ainda medindo algum outro fenômeno.

**13.2 – Perspectivas futuras** - Os astrolábios, ainda em uso provaram que fazem uma boa ciência e geraram mais de uma centena de artigos publicados. Como foi dito o seu uso já perfaz mais de três décadas. Embora sempre utilizando tecnologia inovadora algo de novo acabaria por aparecer. No caso trata-se do heliômetro solar. Este instrumento já foi utilizado para a observação do Sol no século XIX tendo obtido bons resultados considerando-se a tecnologia vigente naquela época. Mas, a partir dos anos 70 do século XX o astrolábio se impôs como o instrumento mais adequado à observação do Sol.

Recentemente a equipe do ON considerou a construção de um heliômetro já que novos materiais têm sido desenvolvidos na fabricação de vidros quase sem o efeito de dilatação e tubos de fibra carbono praticamente indeformáveis têm sido ofertados no mercado de materiais. É oportuno lembrar que há um heliômetro operando no *Pic-du-Midi* na França. Embora o astrolábio utilize medidas relativas que absorvem naturalmente os erros sistemáticos – veja-se o caso das missões astrométricas de máxima precisão – ainda assim o heliômetro tem algumas vantagens sobre o Astrolábio. Primeiramente porque não precisa esperar o Sol cruzar uma linha no céu, suas medidas são instantâneas e pode fazer quantas imagens se queira ou se tenha capacidade de guardar. Um instrumento já pronto e em fase de calibração tem feito cerca de 15 imagens por segundo. Além disso, o heliômetro é capaz de medir qualquer latitude solar que se deseje, permitindo ao observador varrer todas as heliolatitudes desejadas em um mesmo dia. Finalmente, ele pode observar o Sol em sua passagem meridiana em qualquer dia do ano, permitindo que se obtenham as melhores imagens do Sol para cada dia. Veja na Figura 13.1 o heliômetro dentro de sua cúpula.

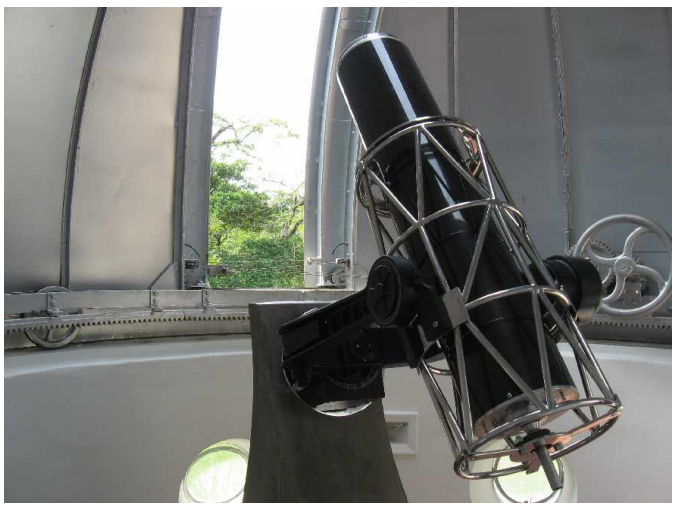

Figura 13.1 – O heliômetro do ON em sua cúpula.

O heliômetro do ON foi concebido pelo professor Victor Amorim d'Ávila e foi desenvolvido durante o decorrer da tese de doutorado de Eugênio Reis Neto (Reis Neto, 2009). O esquema da Figura 13.2 mostra a técnica adotada para se medir o diâmetro solar. Duas imagens do Sol são formadas. A distância entre os dois centros é constante e bem conhecida. Apenas uma pequena parte destas imagens é observada e corresponde ao pequeno retângulo no centro da figura. Mede-se a distância entre as bordas das duas imagens. O diâmetro do Sol é calculado pela diferença entre esta medida e a distância conhecida.

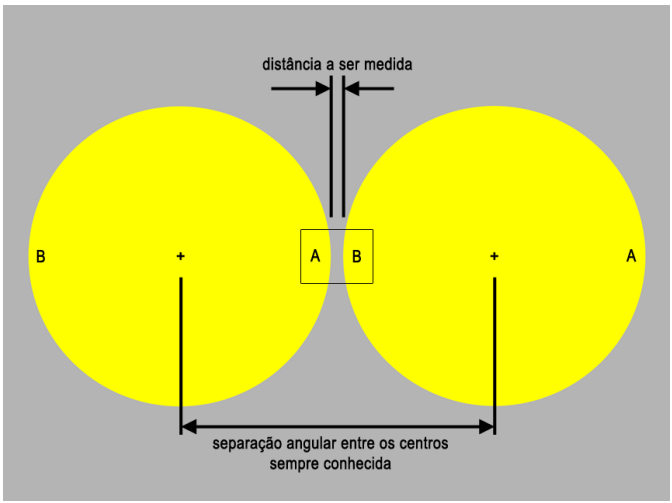

Figura 13.2 – Esquema de observação do semidiâmetro solar com o heliômetro do ON.

No dia 15 de junho de 2010 o *Centre National d'Études Spatiales* – CNES, a agência espacial da França, lançou ao espaço o satélite Picard. Ele medirá a velocidade de rotação do Sol, a intensidade de sua irradiância, a presença de manchas na fotosfera, sua forma e seus diâmetros. Além disso, o Picard vai estudar a estrutura interna do Sol. É o primeiro instrumento construído com a finalidade e a tecnologia necessária para observação do diâmetro solar e que fará isto acima da atmosfera. Suas medidas do raio solar, feitas acima da turbulenta atmosfera terrestre, serão preciosas para calibrar os instrumentos de solo que continuarão com suas observações após o encerramento da missão Picard a qual deverá se estender por cerca de três anos. A Figura 13.3 traz uma imagem do satélite Picard.

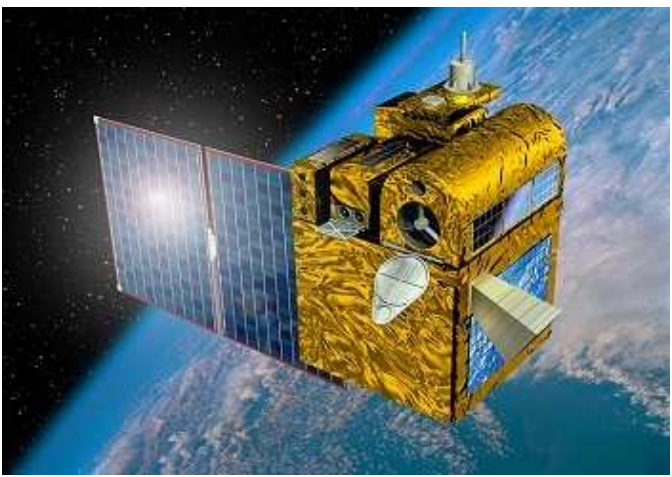

Figura 13.3 – Imagem pictórica do satélite Picard.

Também a agência espacial dos EUA no dia 11 de fevereiro de 2010 colocou em órbita da Terra o satélite *Solar Dynamics Observatory* – SDO destinado a estudar as variabilidades do Sol e seus impactos na Terra. Há a finalidade de se entender qual a estrutura do campo magnético solar e como ele é gerado. E como a energia magnética é armazenada e convertida e liberada na fotosfera e no espaço na forma de vento solar, partículas de alta energia e a irradiância. Ao mesmo tempo o SDO também fará medidas do diâmetro solar. A Figura 13.4 mostra um esquema do satélite.

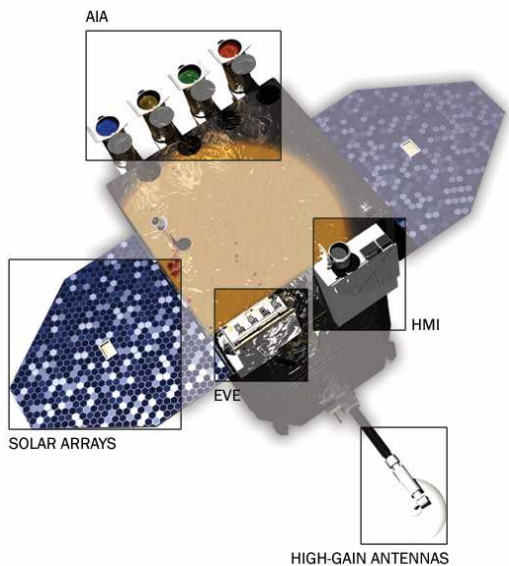

Figura 13.4 – Imagem esquemática do *Solar Dynamics Observatory.*

# 14. Bibliografia.

**Apêndice**

Com o intuito de difundir nossos dados à comunidade de Heliofísica, e em geral aos utilizadores de medidas de variações do semidiâmetro solar apresentamos neste apêndice a série completa de semidiâmetros do Sol observados no Observatório Nacional. A série compreende valores desde março de 1998 até novembro de 2008.

Eles estão apresentados separados em grupos mensais. Cada linha impressa corresponde a uma medida do semidiâmetro. A coluna [D] se refere ao dia no mês, a coluna [L] informa o lado observado, se o valor informado for [1] a observação foi feita antes da passagem meridiana do Sol se for [2] a observação foi feita após esta passagem meridiana. A coluna [SDB] é o valor em segundos de grau do semidiâmetro bruto resultante da observação. A coluna [ER] traz o erro relativo a esta observação, também em segundos de grau. A coluna [SDC] informa o valor do semidiâmetro solar, medido em segundos de grau, corrigido dos desvios instrumentais e observacionais. E a coluna [HL] mostra a heliolatitude observada em graus, e estes valores estão todos rebatidos ao primeiro quadrante. Para os anos de 2004 a 2009 não foram feitas correções dos desvios e por este motivo os valores da coluna [SDC] são iguais aos valores da coluna [SDB].

| D | L | SDB | ER | SDC | HL |
|---|---|---|---|---|---|
| 02 | 2 | 959.848 | 0.172 | 960.622 | 3.33 |
| 02 | 2 | 959.848 | 0.172 | 960.180 | 3.58 |
| 02 | 2 | 959.433 | 0.127 | 959.446 | 3.82 |
| 02 | 2 | 959.124 | 0.133 | 959.113 | 4.07 |
| 02 | 2 | 958.473 | 0.182 | 958.414 | 4.33 |
| 02 | 2 | 959.494 | 0.177 | 959.503 | 4.59 |
| 02 | 2 | 959.303 | 0.179 | 959.256 | 4.87 |
| 02 | 2 | 959.646 | 0.192 | 959.711 | 5.16 |
| 02 | 2 | 959.396 | 0.128 | 959.389 | 5.47 |
| 02 | 2 | 958.473 | 0.182 | 957.517 | 6.15 |
| 02 | 2 | 959.433 | 0.127 | 959.421 | 6.68 |
| 02 | 2 | 959.303 | 0.179 | 959.240 | 7.89 |
| 02 | 2 | 959.543 | 0.138 | 959.542 | 8.35 |
| 02 | 1 | 959.848 | 0.172 | 960.622 | 3.33 |
| 02 | 1 | 959.848 | 0.172 | 960.180 | 3.58 |
| 02 | 1 | 959.433 | 0.127 | 959.446 | 3.82 |
| 02 | 1 | 959.124 | 0.133 | 959.113 | 4.07 |
| 02 | 1 | 958.473 | 0.182 | 958.414 | 4.33 |
| 02 | 1 | 959.494 | 0.177 | 959.503 | 4.59 |
| 02 | 1 | 959.303 | 0.179 | 959.256 | 4.87 |
| 02 | 1 | 959.646 | 0.192 | 959.711 | 5.16 |
| 02 | 1 | 959.396 | 0.128 | 959.389 | 5.47 |
| 02 | 1 | 958.473 | 0.182 | 957.517 | 6.15 |
| 02 | 1 | 959.433 | 0.127 | 959.421 | 6.68 |
| 02 | 1 | 959.303 | 0.179 | 959.240 | 7.89 |
| 02 | 1 | 959.543 | 0.138 | 959.542 | 8.35 |
| 04 | 2 | 959.229 | 0.525 | 959.216 | 5.81 |
| 04 | 2 | 958.617 | 0.150 | 958.643 | 6.60 |
| 04 | 2 | 959.542 | 0.172 | 959.406 | 7.03 |
| 04 | 2 | 958.617 | 0.150 | 958.621 | 7.47 |
| 04 | 2 | 958.191 | 0.130 | 958.222 | 7.93 |
| 04 | 2 | 958.828 | 0.152 | 958.782 | 8.44 |
| 04 | 2 | 959.660 | 0.243 | 959.671 | 8.99 |
| 04 | 2 | 959.229 | 0.525 | 959.322 | 9.60 |
| 04 | 2 | 958.117 | 0.178 | 958.037 | 10.89 |
| 04 | 2 | 958.340 | 0.812 | 958.326 | 11.58 |
| 04 | 1 | 959.229 | 0.525 | 959.216 | 5.81 |
| 04 | 1 | 958.617 | 0.150 | 958.643 | 6.60 |
| 04 | 1 | 959.542 | 0.172 | 959.406 | 7.03 |
| 04 | 1 | 958.617 | 0.150 | 958.621 | 7.47 |
| 04 | 1 | 958.191 | 0.130 | 958.222 | 7.93 |
| 04 | 1 | 958.828 | 0.152 | 958.782 | 8.44 |
| 04 | 1 | 959.660 | 0.243 | 959.671 | 8.99 |
| 04 | 1 | 959.229 | 0.525 | 959.322 | 9.60 |
| 04 | 1 | 958.117 | 0.178 | 958.037 | 10.89 |
| 04 | 1 | 958.340 | 0.812 | 958.326 | 11.58 |
| 04 | 2 | 959.782 | 0.304 | 959.816 | 15.78 |
| 04 | 2 | 959.040 | 0.161 | 959.016 | 18.95 |
| 04 | 2 | 957.879 | 0.195 | 957.815 | 20.12 |
| 04 | 2 | 958.906 | 0.148 | 958.972 | 21.48 |
| 04 | 1 | 959.782 | 0.304 | 959.816 | 15.78 |
| 04 | 1 | 959.040 | 0.161 | 959.016 | 18.95 |
| 04 | 1 | 957.879 | 0.195 | 957.815 | 20.12 |
| 04 | 1 | 958.906 | 0.148 | 958.972 | 21.48 |
| 05 | 2 | 958.359 | 0.149 | 958.290 | 4.29 |
| 05 | 2 | 959.929 | 0.163 | 960.552 | 4.55 |
| 05 | 2 | 959.422 | 0.173 | 959.497 | 5.12 |
| 05 | 2 | 959.422 | 0.173 | 959.539 | 5.45 |
| 05 | 2 | 959.929 | 0.163 | 960.206 | 7.01 |
| 05 | 2 | 958.575 | 0.117 | 958.574 | 7.55 |
| 05 | 2 | 958.491 | 0.151 | 958.490 | 7.96 |
| 05 | 1 | 958.359 | 0.149 | 958.290 | 4.29 |
| 05 | 1 | 959.929 | 0.163 | 960.552 | 4.55 |
| 05 | 1 | 959.422 | 0.173 | 959.497 | 5.12 |
| 05 | 1 | 959.422 | 0.173 | 959.539 | 5.45 |
| 05 | 1 | 959.929 | 0.163 | 960.206 | 7.01 |
| 05 | 1 | 958.575 | 0.117 | 958.574 | 7.55 |
| 05 | 1 | 958.491 | 0.151 | 958.490 | 7.96 |
| 06 | 2 | 959.167 | 0.174 | 959.167 | 3.28 |
| 06 | 2 | 960.010 | 0.132 | 960.045 | 3.53 |
| 06 | 2 | 958.933 | 0.147 | 958.935 | 3.85 |

| D | L | SDB | ER | SDC | HL |
|---|---|---|---|---|---|
| 06 | 2 | 959.095 | 0.150 | 959.059 | 4.17 |
| 06 | 2 | 959.421 | 0.128 | 959.435 | 4.47 |
| 06 | 2 | 959.356 | 0.149 | 959.348 | 4.78 |
| 06 | 2 | 959.356 | 0.149 | 959.341 | 5.11 |
| 06 | 2 | 959.010 | 0.134 | 959.025 | 5.45 |
| 06 | 2 | 959.000 | 0.180 | 958.998 | 5.80 |
| 06 | 2 | 959.872 | 0.174 | 959.860 | 6.17 |
| 06 | 2 | 958.180 | 0.177 | 957.679 | 6.55 |
| 06 | 2 | 959.469 | 0.129 | 959.474 | 7.08 |
| 06 | 2 | 958.715 | 0.156 | 958.709 | 7.51 |
| 06 | 2 | 958.180 | 0.177 | 958.179 | 7.98 |
| 06 | 2 | 959.260 | 0.189 | 959.294 | 8.47 |
| 06 | 2 | 958.715 | 0.156 | 958.700 | 9.00 |
| 06 | 2 | 959.356 | 0.149 | 959.381 | 9.51 |
| 06 | 2 | 959.421 | 0.128 | 959.439 | 10.10 |
| 06 | 2 | 959.166 | 0.191 | 959.164 | 10.69 |
| 06 | 2 | 959.872 | 0.174 | 959.934 | 11.39 |
| 06 | 1 | 959.167 | 0.174 | 959.167 | 3.28 |
| 06 | 1 | 960.010 | 0.132 | 960.045 | 3.53 |
| 06 | 1 | 958.933 | 0.147 | 958.935 | 3.85 |
| 06 | 1 | 959.095 | 0.150 | 959.059 | 4.17 |
| 06 | 1 | 959.421 | 0.128 | 959.435 | 4.47 |
| 06 | 1 | 959.356 | 0.149 | 959.348 | 4.78 |
| 06 | 1 | 959.356 | 0.149 | 959.341 | 5.11 |
| 06 | 1 | 959.010 | 0.134 | 959.025 | 5.45 |
| 06 | 1 | 959.000 | 0.180 | 958.998 | 5.80 |
| 06 | 1 | 959.872 | 0.174 | 959.860 | 6.17 |
| 06 | 1 | 958.180 | 0.177 | 957.679 | 6.55 |
| 06 | 1 | 959.469 | 0.129 | 959.474 | 7.08 |
| 06 | 1 | 958.715 | 0.156 | 958.709 | 7.51 |
| 06 | 1 | 958.180 | 0.177 | 958.179 | 7.98 |
| 06 | 1 | 959.260 | 0.189 | 959.294 | 8.47 |
| 06 | 1 | 958.715 | 0.156 | 958.700 | 9.00 |
| 06 | 1 | 959.356 | 0.149 | 959.381 | 9.51 |
| 06 | 1 | 959.421 | 0.128 | 959.439 | 10.10 |
| 06 | 1 | 959.166 | 0.191 | 959.164 | 10.69 |
| 06 | 1 | 959.872 | 0.174 | 959.934 | 11.39 |
| 06 | 2 | 958.933 | 0.147 | 958.879 | 12.53 |
| 06 | 2 | 958.333 | 0.143 | 958.367 | 13.32 |
| 06 | 2 | 958.814 | 0.125 | 958.870 | 14.45 |
| 06 | 2 | 958.715 | 0.156 | 958.752 | 15.25 |
| 06 | 2 | 959.412 | 0.136 | 959.399 | 16.10 |
| 06 | 2 | 958.802 | 0.134 | 958.799 | 17.15 |
| 06 | 1 | 958.933 | 0.147 | 958.879 | 12.53 |
| 06 | 1 | 958.333 | 0.143 | 958.367 | 13.32 |
| 06 | 1 | 958.814 | 0.125 | 958.870 | 14.45 |
| 06 | 1 | 958.715 | 0.156 | 958.752 | 15.25 |
| 06 | 1 | 959.412 | 0.136 | 959.399 | 16.10 |
| 06 | 1 | 958.802 | 0.134 | 958.799 | 17.15 |
| 09 | 2 | 959.758 | 0.136 | 960.404 | 4.72 |
| 09 | 2 | 959.430 | 0.131 | 959.438 | 5.05 |
| 09 | 2 | 959.317 | 0.141 | 959.279 | 5.41 |
| 09 | 2 | 959.165 | 0.135 | 959.232 | 5.76 |
| 09 | 2 | 958.750 | 0.144 | 958.705 | 6.16 |
| 09 | 2 | 959.317 | 0.141 | 959.273 | 6.53 |
| 09 | 2 | 959.758 | 0.136 | 959.859 | 6.94 |
| 09 | 2 | 959.430 | 0.131 | 959.418 | 7.39 |
| 09 | 2 | 958.483 | 0.124 | 958.483 | 7.82 |
| 09 | 2 | 959.165 | 0.135 | 959.234 | 8.29 |
| 09 | 2 | 958.925 | 0.178 | 958.948 | 8.90 |
| 09 | 2 | 958.882 | 0.144 | 958.866 | 9.43 |
| 09 | 2 | 959.377 | 0.160 | 959.391 | 9.99 |
| 09 | 2 | 959.430 | 0.131 | 959.406 | 10.57 |
| 09 | 2 | 958.183 | 0.162 | 958.251 | 11.24 |
| 09 | 2 | 958.654 | 0.158 | 958.692 | 11.87 |
| 09 | 2 | 958.111 | 0.157 | 958.023 | 12.53 |
| 09 | 1 | 959.758 | 0.136 | 960.404 | 4.72 |
| 09 | 1 | 959.430 | 0.131 | 959.438 | 5.05 |
| 09 | 1 | 959.317 | 0.141 | 959.279 | 5.41 |
| 09 | 1 | 959.165 | 0.135 | 959.232 | 5.76 |
| 09 | 1 | 958.750 | 0.144 | 958.705 | 6.16 |

|  | 1998 - MARCO |  |  |  |  |  | 1998 - MARCO |  |  |  |
|---|---|---|---|---|---|---|---|---|---|---|
| D | L | SDB | ER | SDC | HL | D | L | SDB | ER | SDC | HL |
| 09 | 1 | 959.317 | 0.141 | 959.273 | 6.53 | 11 | 2 | 959.952 | 0.159 | 959.986 | 5.44 |
| 09 | 1 | 959.758 | 0.136 | 959.859 | 6.94 | 11 | 2 | 959.160 | 0.134 | 959.176 | 5.78 |
| 09 | 1 | 959.430 | 0.131 | 959.418 | 7.39 | 11 | 2 | 959.592 | 0.154 | 959.686 | 6.14 |
| 09 | 1 | 958.483 | 0.124 | 958.483 | 7.82 | 11 | 2 | 960.034 | 0.177 | 960.044 | 6.56 |
| 09 | 1 | 959.165 | 0.135 | 959.234 | 8.29 | 11 | 2 | 960.034 | 0.177 | 960.169 | 6.95 |
| 09 | 1 | 958.925 | 0.178 | 958.948 | 8.90 | 11 | 2 | 959.160 | 0.134 | 959.193 | 7.39 |
| 09 | 1 | 958.882 | 0.144 | 958.866 | 9.43 | 11 | 2 | 959.158 | 0.145 | 959.130 | 7.93 |
| 09 | 1 | 959.377 | 0.160 | 959.391 | 9.99 | 11 | 2 | 959.592 | 0.154 | 959.599 | 8.40 |
| 09 | 1 | 959.430 | 0.131 | 959.406 | 10.57 | 11 | 2 | 958.446 | 0.137 | 958.495 | 8.90 |
| 09 | 1 | 958.183 | 0.162 | 958.251 | 11.24 | 11 | 2 | 959.160 | 0.134 | 959.224 | 9.44 |
| 09 | 1 | 958.654 | 0.158 | 958.692 | 11.87 | 11 | 2 | 959.160 | 0.134 | 959.220 | 10.02 |
| 09 | 1 | 958.111 | 0.157 | 958.023 | 12.53 | 11 | 2 | 958.853 | 0.130 | 958.860 | 10.58 |
| 09 | 2 | 958.925 | 0.178 | 958.941 | 13.26 | 11 | 2 | 958.872 | 0.119 | 958.884 | 11.17 |
| 09 | 2 | 958.654 | 0.158 | 958.621 | 14.03 | 11 | 2 | 959.544 | 0.099 | 959.507 | 11.86 |
| 09 | 2 | 959.127 | 0.161 | 959.143 | 15.21 | 11 | 2 | 958.987 | 0.173 | 958.961 | 12.51 |
| 09 | 2 | 959.430 | 0.131 | 959.409 | 16.10 | 11 | 2 | 958.872 | 0.119 | 958.912 | 13.52 |
| 09 | 2 | 959.533 | 0.129 | 959.629 | 17.02 | 11 | 1 | 960.034 | 0.177 | 960.949 | 3.31 |
| 09 | 2 | 959.533 | 0.129 | 959.538 | 18.07 | 11 | 1 | 958.668 | 0.165 | 958.645 | 3.57 |
| 09 | 2 | 959.758 | 0.136 | 959.772 | 19.15 | 11 | 1 | 958.853 | 0.130 | 958.854 | 3.84 |
| 09 | 2 | 958.183 | 0.162 | 958.239 | 20.33 | 11 | 1 | 959.592 | 0.154 | 959.692 | 4.13 |
| 09 | 1 | 958.925 | 0.178 | 958.941 | 13.26 | 11 | 1 | 958.872 | 0.119 | 958.900 | 4.43 |
| 09 | 1 | 958.654 | 0.158 | 958.621 | 14.03 | 11 | 1 | 959.305 | 0.128 | 959.336 | 4.72 |
| 09 | 1 | 959.127 | 0.161 | 959.143 | 15.21 | 11 | 1 | 958.872 | 0.119 | 958.887 | 5.09 |
| 09 | 1 | 959.430 | 0.131 | 959.409 | 16.10 | 11 | 1 | 959.952 | 0.159 | 959.986 | 5.44 |
| 09 | 1 | 959.533 | 0.129 | 959.629 | 17.02 | 11 | 1 | 959.160 | 0.134 | 959.176 | 5.78 |
| 09 | 1 | 959.533 | 0.129 | 959.538 | 18.07 | 11 | 1 | 959.592 | 0.154 | 959.686 | 6.14 |
| 09 | 1 | 959.758 | 0.136 | 959.772 | 19.15 | 11 | 1 | 960.034 | 0.177 | 960.044 | 6.56 |
| 09 | 1 | 958.183 | 0.162 | 958.239 | 20.33 | 11 | 1 | 960.034 | 0.177 | 960.169 | 6.95 |
| 10 | 2 | 958.958 | 0.164 | 958.872 | 6.71 | 11 | 1 | 959.160 | 0.134 | 959.193 | 7.39 |
| 10 | 2 | 959.695 | 0.191 | 959.682 | 7.09 | 11 | 1 | 959.158 | 0.145 | 959.130 | 7.93 |
| 10 | 2 | 960.124 | 0.174 | 960.276 | 7.48 | 11 | 1 | 959.592 | 0.154 | 959.599 | 8.40 |
| 10 | 2 | 960.124 | 0.174 | 960.720 | 7.88 | 11 | 1 | 958.446 | 0.137 | 958.495 | 8.90 |
| 10 | 2 | 959.048 | 0.381 | 959.051 | 8.30 | 11 | 1 | 959.160 | 0.134 | 959.224 | 9.44 |
| 10 | 2 | 959.559 | 0.137 | 959.544 | 8.74 | 11 | 1 | 959.160 | 0.134 | 959.220 | 10.02 |
| 10 | 2 | 959.695 | 0.191 | 959.679 | 9.22 | 11 | 1 | 958.853 | 0.130 | 958.860 | 10.58 |
| 10 | 2 | 958.575 | 0.152 | 958.579 | 9.71 | 11 | 1 | 958.872 | 0.119 | 958.884 | 11.17 |
| 10 | 2 | 959.527 | 0.140 | 959.513 | 10.21 | 11 | 1 | 959.544 | 0.099 | 959.507 | 11.86 |
| 10 | 2 | 959.048 | 0.381 | 959.158 | 10.79 | 11 | 1 | 958.987 | 0.173 | 958.961 | 12.51 |
| 10 | 2 | 958.958 | 0.164 | 958.864 | 11.35 | 11 | 1 | 958.872 | 0.119 | 958.912 | 13.52 |
| 10 | 2 | 959.695 | 0.191 | 959.745 | 11.97 | 11 | 2 | 957.999 | 0.193 | 958.035 | 14.30 |
| 10 | 2 | 959.527 | 0.140 | 959.502 | 12.69 | 11 | 2 | 959.158 | 0.145 | 959.157 | 15.07 |
| 10 | 2 | 959.048 | 0.381 | 959.137 | 13.34 | 11 | 2 | 958.800 | 0.181 | 958.820 | 15.95 |
| 10 | 1 | 958.958 | 0.164 | 958.872 | 6.71 | 11 | 2 | 959.158 | 0.145 | 959.152 | 16.84 |
| 10 | 1 | 959.695 | 0.191 | 959.682 | 7.09 | 11 | 2 | 959.592 | 0.154 | 959.664 | 17.76 |
| 10 | 1 | 960.124 | 0.174 | 960.276 | 7.48 | 11 | 2 | 959.158 | 0.145 | 959.150 | 18.75 |
| 10 | 1 | 960.034 | 0.174 | 960.720 | 7.88 | 11 | 2 | 960.034 | 0.177 | 960.030 | 19.87 |
| 10 | 1 | 959.048 | 0.381 | 959.051 | 8.30 | 11 | 1 | 957.999 | 0.193 | 958.035 | 14.30 |
| 10 | 1 | 959.559 | 0.137 | 959.544 | 8.74 | 11 | 1 | 959.158 | 0.145 | 959.157 | 15.07 |
| 10 | 1 | 959.695 | 0.191 | 959.679 | 9.22 | 11 | 1 | 958.800 | 0.181 | 958.820 | 15.95 |
| 10 | 1 | 958.575 | 0.152 | 958.579 | 9.71 | 11 | 1 | 959.158 | 0.145 | 959.152 | 16.84 |
| 10 | 1 | 959.527 | 0.140 | 959.513 | 10.21 | 11 | 1 | 959.592 | 0.154 | 959.664 | 17.76 |
| 10 | 1 | 959.048 | 0.381 | 959.158 | 10.79 | 11 | 1 | 959.158 | 0.145 | 959.150 | 18.75 |
| 10 | 1 | 958.958 | 0.164 | 958.864 | 11.35 | 11 | 1 | 960.034 | 0.177 | 960.030 | 19.87 |
| 10 | 1 | 959.695 | 0.191 | 959.745 | 11.97 | 13 | 2 | 959.702 | 0.162 | 959.699 | 2.54 |
| 10 | 1 | 959.527 | 0.140 | 959.502 | 12.69 | 13 | 2 | 959.061 | 0.173 | 959.036 | 2.76 |
| 10 | 1 | 959.048 | 0.381 | 959.137 | 13.34 | 13 | 2 | 958.811 | 0.151 | 958.805 | 2.98 |
| 10 | 2 | 958.958 | 0.164 | 958.865 | 14.01 | 13 | 2 | 960.171 | 0.909 | 960.308 | 3.23 |
| 10 | 2 | 959.280 | 0.216 | 959.295 | 14.73 | 13 | 2 | 959.061 | 0.173 | 959.045 | 3.47 |
| 10 | 2 | 958.636 | 0.211 | 958.679 | 15.55 | 13 | 2 | 959.488 | 0.196 | 959.527 | 3.72 |
| 10 | 2 | 959.695 | 0.191 | 959.861 | 17.75 | 13 | 2 | 959.654 | 0.191 | 959.615 | 3.97 |
| 10 | 1 | 958.958 | 0.164 | 958.865 | 14.01 | 13 | 2 | 958.811 | 0.151 | 958.771 | 4.25 |
| 10 | 1 | 959.280 | 0.216 | 959.295 | 14.73 | 13 | 2 | 959.201 | 0.203 | 959.186 | 4.61 |
| 10 | 1 | 958.636 | 0.211 | 958.679 | 15.55 | 13 | 2 | 959.061 | 0.173 | 959.009 | 5.27 |
| 10 | 1 | 959.695 | 0.191 | 959.861 | 17.75 | 13 | 2 | 958.843 | 0.195 | 958.833 | 5.66 |
| 11 | 2 | 960.034 | 0.177 | 960.949 | 3.31 | 13 | 2 | 958.843 | 0.195 | 958.846 | 6.01 |
| 11 | 2 | 958.668 | 0.165 | 958.645 | 3.57 | 13 | 2 | 959.654 | 0.191 | 959.624 | 6.49 |
| 11 | 2 | 958.853 | 0.130 | 958.854 | 3.84 | 13 | 2 | 958.589 | 0.203 | 958.609 | 7.82 |
| 11 | 2 | 959.592 | 0.154 | 959.692 | 4.13 | 13 | 2 | 958.292 | 0.169 | 958.332 | 8.41 |
| 11 | 2 | 958.872 | 0.119 | 958.900 | 4.43 | 13 | 2 | 958.292 | 0.169 | 958.236 | 8.91 |
| 11 | 2 | 959.305 | 0.128 | 959.336 | 4.72 | 13 | 2 | 958.509 | 0.230 | 958.502 | 9.44 |
| 11 | 2 | 958.872 | 0.119 | 958.887 | 5.09 | 13 | 2 | 958.913 | 0.184 | 958.899 | 9.96 |

```
         1998 - MARCO                              1998 - MARCO
  D  L    SDB    ER      SDC    HL         D  L    SDB    ER      SDC    HL
 13  2  958.648  0.172  958.626  10.50     18  1  958.846  0.174  958.872   8.59
 13  2  958.661  0.168  958.710  11.10     18  1  959.382  0.147  959.416   9.04
 13  2  959.061  0.173  959.040  11.91     18  1  958.605  0.161  958.440  10.10
 13  1  959.702  0.162  959.699   2.54     18  1  958.720  0.132  958.691  15.39
 13  1  959.061  0.173  959.036   2.76     18  1  959.382  0.147  959.356  16.25
 13  1  958.811  0.151  958.805   2.98     18  2  959.549  0.155  959.557  17.04
 13  1  960.171  0.909  960.308   3.23     18  2  958.605  0.161  958.509  18.14
 13  1  959.061  0.173  959.045   3.47     18  2  959.015  0.130  958.934  19.94
 13  1  959.488  0.196  959.527   3.72     18  2  959.382  0.147  959.238  21.01
 13  1  959.654  0.191  959.615   3.97     18  2  959.549  0.155  959.587  22.10
 13  1  958.811  0.151  958.771   4.25     18  2  959.850  0.650  959.857  24.10
 13  1  959.201  0.203  959.186   4.61     18  1  959.549  0.155  959.557  17.04
 13  1  959.061  0.173  959.009   5.27     18  1  958.605  0.161  958.509  18.14
 13  1  958.843  0.195  958.833   5.66     18  1  959.015  0.130  958.934  19.94
 13  1  958.843  0.195  958.846   6.01     18  1  959.382  0.147  959.238  21.01
 13  1  959.654  0.191  959.624   6.49     18  1  959.549  0.155  959.587  22.10
 13  1  958.589  0.203  958.609   7.82     18  1  959.850  0.650  959.857  24.10
 13  1  958.292  0.169  958.332   8.41     20  2  958.685  0.132  958.665   3.20
 13  1  958.292  0.169  958.236   8.91     20  2  958.926  0.143  958.919   3.50
 13  1  958.509  0.230  958.502   9.44     20  2  959.065  0.122  959.087   3.82
 13  1  958.913  0.184  958.899   9.96     20  2  959.368  0.144  959.362   4.12
 13  1  958.648  0.172  958.626  10.50     20  2  958.989  0.136  958.958   4.42
 13  1  958.661  0.168  958.710  11.10     20  2  958.415  0.137  957.927   4.75
 13  1  959.061  0.173  959.040  11.91     20  2  958.540  0.160  958.553   5.10
 13  2  958.292  0.169  958.232  15.32     20  2  958.540  0.160  958.555   5.44
 13  1  958.292  0.169  958.232  15.32     20  2  959.223  0.161  959.259   5.79
 17  2  959.359  0.227  959.394   6.10     20  2  959.223  0.161  959.199   6.15
 17  2  959.132  0.154  959.131   6.59     20  2  958.644  0.149  958.596   6.52
 17  2  958.809  0.182  958.780   7.01     20  2  959.149  0.108  959.137   6.92
 17  2  959.359  0.227  959.348   7.42     20  2  958.415  0.137  958.254   7.32
 17  2  957.750  0.168  957.987   7.90     20  2  959.223  0.161  959.214   7.76
 17  2  957.750  0.168  957.813   8.35     20  2  959.301  0.130  959.300   8.29
 17  2  958.609  0.189  958.603   8.83     20  2  959.460  0.100  959.475   8.77
 17  2  958.809  0.182  958.820   9.33     20  2  959.223  0.161  959.207   9.30
 17  2  959.132  0.154  959.213   9.87     20  2  958.926  0.143  958.889   9.96
 17  2  958.901  0.174  958.955  10.40     20  2  959.573  0.151  959.662  10.52
 17  2  959.359  0.227  959.498  10.95     20  2  959.460  0.100  959.481  11.12
 17  2  958.513  0.172  958.534  11.67     20  2  959.573  0.151  960.078  11.72
 17  2  958.702  0.157  958.697  12.30     20  2  958.738  0.136  958.740  12.37
 17  2  958.234  0.142  958.318  13.02     20  2  959.124  0.146  959.123  13.05
 17  2  958.809  0.182  958.843  13.82     20  2  958.989  0.136  958.977  13.76
 17  2  958.470  0.225  958.372  14.77     20  2  959.051  0.142  959.033  14.59
 17  2  958.901  0.174  958.920  15.61     20  2  958.455  0.115  958.476  15.36
 17  1  959.359  0.227  959.394   6.10     20  2  958.415  0.137  958.321  16.16
 17  1  959.132  0.154  959.131   6.59     20  2  958.813  0.112  958.853  17.04
 17  1  958.809  0.182  958.780   7.01     20  1  958.685  0.132  958.665   3.20
 17  1  959.359  0.227  959.348   7.42     20  1  958.926  0.143  958.919   3.50
 17  1  957.750  0.168  957.987   7.90     20  1  959.065  0.122  959.087   3.82
 17  1  957.750  0.168  957.813   8.35     20  1  959.368  0.144  959.362   4.12
 17  1  958.609  0.189  958.603   8.83     20  1  958.989  0.136  958.958   4.42
 17  1  958.809  0.182  958.820   9.33     20  1  958.415  0.137  957.927   4.75
 17  1  959.132  0.154  959.213   9.87     20  1  958.540  0.160  958.553   5.10
 17  1  958.901  0.174  958.955  10.40     20  1  958.540  0.160  958.555   5.44
 17  1  959.359  0.227  959.498  10.95     20  1  959.223  0.161  959.259   5.79
 17  1  958.513  0.172  958.534  11.67     20  1  959.223  0.161  959.199   6.15
 17  1  958.702  0.157  958.697  12.30     20  1  958.644  0.149  958.596   6.52
 17  1  958.234  0.142  958.318  13.02     20  1  959.149  0.108  959.137   6.92
 17  1  958.809  0.182  958.843  13.82     20  1  958.415  0.137  958.254   7.32
 17  1  958.470  0.225  958.372  14.77     20  1  959.223  0.161  959.214   7.76
 17  1  958.901  0.174  958.920  15.61     20  1  959.301  0.130  959.300   8.29
 17  2  959.132  0.154  959.179  16.59     20  1  959.460  0.100  959.475   8.77
 17  2  959.873  0.166  959.803  17.57     20  1  959.223  0.161  959.207   9.30
 17  2  958.901  0.174  958.903  18.59     20  1  958.926  0.143  958.889   9.96
 17  1  959.132  0.154  959.179  16.59     20  1  959.573  0.151  959.662  10.52
 17  1  959.873  0.166  959.803  17.57     20  1  959.460  0.100  959.481  11.12
 17  1  958.901  0.174  958.903  18.59     20  1  959.573  0.151  960.078  11.72
 18  2  958.846  0.174  958.872   8.59     20  1  958.738  0.136  958.740  12.37
 18  2  959.382  0.147  959.416   9.04     20  1  959.124  0.146  959.123  13.05
 18  2  958.605  0.161  958.440  10.10     20  1  958.989  0.136  958.977  13.76
 18  2  958.720  0.132  958.691  15.39     20  1  959.051  0.142  959.033  14.59
 18  2  959.382  0.147  959.356  16.25     20  1  958.455  0.115  958.476  15.36
```

| 1998 - MARCO | | | | | | 1998 - MARCO | | | | |
|---|---|---|---|---|---|---|---|---|---|---|
| D | L | SDB | ER | SDC | HL | D | L | SDB | ER | SDC | HL |
| 20 | 1 | 958.415 | 0.137 | 958.321 | 16.16 | 25 | 2 | 959.143 | 0.119 | 960.456 | 8.63 |
| 20 | 1 | 958.813 | 0.112 | 958.853 | 17.04 | 25 | 2 | 958.409 | 0.149 | 958.452 | 11.69 |
| 20 | 2 | 959.460 | 0.100 | 959.471 | 17.91 | 25 | 2 | 958.024 | 0.143 | 957.763 | 12.31 |
| 20 | 2 | 959.060 | 0.140 | 959.057 | 18.93 | 25 | 2 | 959.009 | 0.181 | 958.947 | 12.98 |
| 20 | 2 | 958.415 | 0.137 | 958.013 | 19.92 | 25 | 2 | 958.864 | 0.177 | 958.895 | 13.65 |
| 20 | 2 | 959.301 | 0.130 | 959.269 | 21.34 | 25 | 2 | 958.824 | 0.151 | 958.795 | 16.71 |
| 20 | 2 | 959.065 | 0.122 | 959.094 | 22.53 | 25 | 2 | 958.676 | 0.110 | 958.694 | 17.54 |
| 20 | 2 | 959.390 | 0.121 | 959.397 | 23.78 | 25 | 2 | 958.529 | 0.149 | 958.560 | 18.43 |
| 20 | 1 | 959.460 | 0.100 | 959.471 | 17.91 | 25 | 1 | 959.143 | 0.119 | 959.188 | 7.71 |
| 20 | 1 | 959.060 | 0.140 | 959.057 | 18.93 | 25 | 1 | 959.143 | 0.119 | 960.456 | 8.63 |
| 20 | 1 | 958.415 | 0.137 | 958.013 | 19.92 | 25 | 1 | 958.409 | 0.149 | 958.452 | 11.69 |
| 20 | 1 | 959.301 | 0.130 | 959.269 | 21.34 | 25 | 1 | 958.024 | 0.143 | 957.763 | 12.31 |
| 20 | 1 | 959.065 | 0.122 | 959.094 | 22.53 | 25 | 1 | 959.009 | 0.181 | 958.947 | 12.98 |
| 20 | 1 | 959.390 | 0.121 | 959.397 | 23.78 | 25 | 1 | 958.864 | 0.177 | 958.895 | 13.65 |
| 24 | 2 | 958.925 | 0.130 | 958.999 | 3.97 | 25 | 1 | 958.824 | 0.151 | 958.795 | 16.71 |
| 24 | 2 | 959.667 | 0.100 | 959.651 | 4.27 | 25 | 1 | 958.676 | 0.110 | 958.694 | 17.54 |
| 24 | 2 | 959.362 | 0.154 | 959.363 | 4.60 | 25 | 1 | 958.529 | 0.149 | 958.560 | 18.43 |
| 24 | 2 | 958.925 | 0.130 | 958.961 | 4.93 | 25 | 2 | 958.409 | 0.149 | 958.445 | 19.39 |
| 24 | 2 | 959.667 | 0.100 | 959.662 | 5.27 | 25 | 2 | 958.375 | 0.165 | 958.381 | 20.39 |
| 24 | 2 | 959.305 | 0.133 | 959.274 | 5.62 | 25 | 2 | 958.355 | 0.149 | 958.348 | 21.45 |
| 24 | 2 | 959.362 | 0.154 | 959.342 | 5.97 | 25 | 2 | 959.143 | 0.119 | 959.249 | 22.57 |
| 24 | 2 | 959.165 | 0.117 | 959.173 | 6.35 | 25 | 2 | 958.409 | 0.149 | 958.411 | 23.77 |
| 24 | 2 | 958.925 | 0.130 | 958.933 | 6.75 | 25 | 2 | 958.676 | 0.110 | 958.685 | 25.07 |
| 24 | 2 | 959.502 | 0.159 | 959.484 | 7.14 | 25 | 2 | 959.143 | 0.119 | 959.099 | 26.43 |
| 24 | 2 | 959.194 | 0.144 | 959.205 | 7.57 | 25 | 2 | 958.676 | 0.110 | 958.662 | 27.86 |
| 24 | 2 | 959.223 | 0.143 | 959.243 | 8.01 | 25 | 2 | 958.024 | 0.143 | 958.129 | 29.45 |
| 24 | 2 | 959.223 | 0.143 | 959.254 | 8.45 | 25 | 1 | 958.409 | 0.149 | 958.445 | 19.39 |
| 24 | 2 | 959.088 | 0.121 | 959.070 | 8.94 | 25 | 1 | 958.375 | 0.165 | 958.381 | 20.39 |
| 24 | 2 | 958.855 | 0.120 | 958.864 | 9.43 | 25 | 1 | 958.355 | 0.149 | 958.348 | 21.45 |
| 24 | 2 | 959.223 | 0.143 | 959.249 | 10.11 | 25 | 1 | 959.143 | 0.119 | 959.249 | 22.57 |
| 24 | 2 | 959.362 | 0.154 | 959.381 | 10.66 | 25 | 1 | 958.409 | 0.149 | 958.411 | 23.77 |
| 24 | 2 | 958.855 | 0.120 | 958.841 | 11.22 | 25 | 1 | 958.676 | 0.110 | 958.685 | 25.07 |
| 24 | 2 | 958.388 | 0.144 | 958.360 | 11.79 | 25 | 1 | 959.143 | 0.119 | 959.099 | 26.43 |
| 24 | 2 | 959.193 | 0.108 | 959.193 | 12.42 | 25 | 1 | 958.676 | 0.110 | 958.662 | 27.86 |
| 24 | 2 | 958.556 | 0.141 | 958.589 | 14.64 | 25 | 1 | 958.024 | 0.143 | 958.129 | 29.45 |
| 24 | 2 | 958.855 | 0.120 | 958.854 | 15.38 | 31 | 2 | 958.440 | 0.154 | 958.409 | 5.75 |
| 24 | 2 | 958.822 | 0.139 | 958.837 | 16.55 | 31 | 2 | 958.819 | 0.138 | 958.823 | 6.12 |
| 24 | 2 | 958.556 | 0.141 | 958.520 | 17.44 | 31 | 2 | 958.819 | 0.138 | 958.824 | 6.56 |
| 24 | 2 | 958.792 | 0.144 | 958.802 | 18.36 | 31 | 2 | 958.733 | 0.115 | 958.702 | 6.96 |
| 24 | 1 | 958.925 | 0.130 | 958.999 | 3.97 | 31 | 2 | 959.399 | 0.112 | 959.443 | 7.39 |
| 24 | 1 | 959.667 | 0.100 | 959.651 | 4.27 | 31 | 2 | 959.182 | 0.125 | 959.210 | 7.81 |
| 24 | 1 | 959.362 | 0.154 | 959.363 | 4.60 | 31 | 2 | 959.298 | 0.137 | 959.302 | 8.24 |
| 24 | 1 | 958.925 | 0.130 | 958.961 | 4.93 | 31 | 2 | 959.013 | 0.120 | 959.039 | 8.69 |
| 24 | 1 | 959.667 | 0.100 | 959.662 | 5.27 | 31 | 2 | 958.638 | 0.144 | 958.681 | 9.20 |
| 24 | 1 | 959.305 | 0.133 | 959.274 | 5.62 | 31 | 2 | 959.013 | 0.120 | 959.035 | 9.80 |
| 24 | 1 | 959.362 | 0.154 | 959.342 | 5.97 | 31 | 2 | 958.440 | 0.154 | 958.465 | 10.35 |
| 24 | 1 | 959.165 | 0.117 | 959.173 | 6.35 | 31 | 2 | 959.625 | 0.139 | 959.633 | 11.01 |
| 24 | 1 | 958.925 | 0.130 | 958.933 | 6.75 | 31 | 2 | 959.321 | 0.101 | 959.327 | 11.58 |
| 24 | 1 | 959.502 | 0.159 | 959.484 | 7.14 | 31 | 2 | 958.568 | 0.143 | 958.596 | 12.17 |
| 24 | 1 | 959.194 | 0.144 | 959.205 | 7.57 | 31 | 2 | 959.559 | 0.151 | 959.505 | 12.77 |
| 24 | 1 | 959.223 | 0.143 | 959.243 | 8.01 | 31 | 2 | 959.182 | 0.125 | 959.192 | 13.45 |
| 24 | 1 | 959.223 | 0.143 | 959.254 | 8.45 | 31 | 2 | 959.321 | 0.101 | 959.336 | 14.11 |
| 24 | 1 | 959.088 | 0.121 | 959.070 | 8.94 | 31 | 2 | 958.733 | 0.115 | 958.697 | 14.88 |
| 24 | 1 | 958.855 | 0.120 | 958.864 | 9.43 | 31 | 2 | 959.298 | 0.137 | 959.264 | 15.62 |
| 24 | 1 | 959.223 | 0.143 | 959.249 | 10.11 | 31 | 2 | 959.013 | 0.120 | 959.056 | 16.55 |
| 24 | 1 | 959.362 | 0.154 | 959.381 | 10.66 | 31 | 2 | 958.733 | 0.115 | 958.695 | 17.36 |
| 24 | 1 | 958.855 | 0.120 | 958.841 | 11.22 | 31 | 2 | 959.625 | 0.139 | 959.804 | 18.20 |
| 24 | 1 | 958.388 | 0.144 | 958.360 | 11.79 | 31 | 2 | 958.921 | 0.192 | 958.939 | 19.07 |
| 24 | 1 | 959.193 | 0.108 | 959.193 | 12.42 | 31 | 2 | 958.921 | 0.192 | 958.928 | 19.99 |
| 24 | 1 | 958.556 | 0.141 | 958.589 | 14.64 | 31 | 2 | 959.107 | 0.118 | 959.086 | 21.03 |
| 24 | 1 | 958.855 | 0.120 | 958.854 | 15.38 | 31 | 1 | 958.440 | 0.154 | 958.409 | 5.75 |
| 24 | 1 | 958.822 | 0.139 | 958.837 | 16.55 | 31 | 1 | 958.819 | 0.138 | 958.823 | 6.12 |
| 24 | 1 | 958.556 | 0.141 | 958.520 | 17.44 | 31 | 1 | 958.819 | 0.138 | 958.824 | 6.56 |
| 24 | 1 | 958.792 | 0.144 | 958.802 | 18.36 | 31 | 1 | 958.733 | 0.115 | 958.702 | 6.96 |
| 24 | 2 | 959.462 | 0.118 | 959.445 | 19.30 | 31 | 1 | 959.399 | 0.112 | 959.443 | 7.39 |
| 24 | 2 | 959.319 | 0.130 | 959.332 | 20.32 | 31 | 1 | 959.182 | 0.125 | 959.210 | 7.81 |
| 24 | 2 | 958.925 | 0.130 | 958.921 | 21.41 | 31 | 1 | 959.298 | 0.137 | 959.302 | 8.24 |
| 24 | 1 | 959.462 | 0.118 | 959.445 | 19.30 | 31 | 1 | 959.013 | 0.120 | 959.039 | 8.69 |
| 24 | 1 | 959.319 | 0.130 | 959.332 | 20.32 | 31 | 1 | 958.638 | 0.144 | 958.681 | 9.20 |
| 24 | 1 | 958.925 | 0.130 | 958.921 | 21.41 | 31 | 1 | 959.013 | 0.120 | 959.035 | 9.80 |
| 25 | 2 | 959.143 | 0.119 | 959.188 | 7.71 | 31 | 1 | 958.440 | 0.154 | 958.465 | 10.35 |

```
       1998 - MARCO                              1998 - ABRIL
 D  L   SDB    ER     SDC    HL          D  L   SDB    ER     SDC    HL
31  1 959.625 0.139 959.633 11.01       12  1 958.882 0.123 958.810 16.07
31  1 959.321 0.101 959.327 11.58       12  1 959.521 0.122 959.584 16.78
31  1 958.568 0.143 958.596 12.17       12  1 958.515 0.147 958.503 17.51
31  1 959.559 0.151 959.505 12.77       12  1 959.007 0.121 958.986 18.29
31  1 959.182 0.125 959.192 13.45       12  1 959.489 0.152 959.420 19.09
31  1 959.321 0.101 959.336 14.11       12  1 959.489 0.152 959.455 19.98
31  1 958.733 0.115 958.697 14.88       12  1 958.614 0.165 958.658 20.86
31  1 959.298 0.137 959.264 15.62       12  1 959.132 0.158 959.171 21.77
31  1 959.013 0.120 959.056 16.55       12  1 958.598 0.149 958.574 22.72
31  1 958.733 0.115 958.695 17.36       12  1 959.489 0.152 959.467 23.72
31  1 959.625 0.139 959.804 18.20       12  1 959.321 0.165 959.279 24.78
31  1 958.921 0.192 958.939 19.07       12  2 958.300 0.188 958.358 25.94
31  1 958.921 0.192 958.928 19.99       12  2 959.345 0.144 959.363 27.22
31  1 959.107 0.118 959.086 21.03       12  2 958.723 0.130 958.736 28.52
31  2 959.003 0.133 958.997 22.05       12  2 958.104 0.151 958.167 29.89
31  2 958.440 0.154 958.465 23.12       12  1 958.300 0.188 958.358 25.94
31  2 959.625 0.139 959.622 24.28       12  1 959.345 0.144 959.363 27.22
31  2 959.128 0.133 959.140 25.51       12  1 958.723 0.130 958.736 28.52
31  2 958.440 0.154 958.425 26.83       12  1 958.104 0.151 958.167 29.89
31  2 958.839 0.100 958.856 28.20       13  2 959.560 0.172 960.557 18.31
31  2 958.440 0.154 958.455 29.69       13  2 959.560 0.172 959.625 21.68
31  1 959.003 0.133 958.997 22.05       13  1 959.560 0.172 960.557 18.31
31  1 958.440 0.154 958.465 23.12       13  1 959.560 0.172 959.625 21.68
31  1 959.625 0.139 959.622 24.28       13  2 958.931 0.201 958.907 32.56
31  1 959.128 0.133 959.140 25.51       13  2 958.118 0.151 958.117 34.18
31  1 958.440 0.154 958.425 26.83       13  2 958.931 0.201 959.209 37.88
31  1 958.839 0.100 958.856 28.20       13  1 958.931 0.201 958.907 32.56
31  1 958.440 0.154 958.455 29.69       13  1 958.118 0.151 958.117 34.18
                                        13  1 958.931 0.201 959.209 37.88
                                        14  2 958.578 0.163 958.557 10.39
       1998 - ABRIL                     14  2 958.535 0.158 958.496 11.37
 D  L   SDB    ER     SDC    HL         14  2 959.177 0.139 959.190 12.44
07  2 958.264 0.193 958.226 10.08       14  2 958.279 0.148 958.322 14.66
07  2 959.226 0.169 959.249 10.56       14  2 959.410 0.167 959.433 15.25
07  2 958.807 0.150 958.824 11.04       14  2 959.064 0.175 959.071 15.98
07  2 959.475 0.210 959.497 12.07       14  2 959.064 0.175 959.058 16.63
07  2 959.046 0.166 959.044 12.59       14  2 958.279 0.148 958.047 17.32
07  2 957.961 0.165 957.987 13.73       14  2 959.928 0.196 960.024 18.06
07  2 958.686 0.167 958.693 14.34       14  2 959.928 0.196 960.065 18.78
07  2 958.288 0.144 958.296 17.72       14  2 958.769 0.188 958.781 20.40
07  2 958.997 0.186 958.995 18.46       14  2 959.504 0.159 959.476 21.25
07  2 958.997 0.186 958.969 19.67       14  1 958.578 0.163 958.557 10.39
07  2 958.512 0.218 958.557 20.51       14  1 958.535 0.158 958.496 11.37
07  2 958.737 0.174 958.767 21.43       14  1 959.177 0.139 959.190 12.44
07  2 959.046 0.166 959.085 23.34       14  1 958.279 0.148 958.322 14.66
07  1 958.264 0.193 958.226 10.08       14  1 959.410 0.167 959.433 15.25
07  1 959.226 0.169 959.249 10.56       14  1 959.064 0.175 959.071 15.98
07  1 958.807 0.150 958.824 11.04       14  1 959.064 0.175 959.058 16.63
07  1 959.475 0.210 959.497 12.07       14  1 958.279 0.148 958.047 17.32
07  1 959.046 0.166 959.044 12.59       14  1 959.928 0.196 960.024 18.06
07  1 957.961 0.165 957.987 13.73       14  1 959.928 0.196 960.065 18.78
07  1 958.686 0.167 958.693 14.34       14  1 958.769 0.188 958.781 20.40
07  1 958.288 0.144 958.296 17.72       14  1 959.504 0.159 959.476 21.25
07  1 958.997 0.186 958.995 18.46       15  2 959.077 0.153 959.058 14.42
07  1 958.997 0.186 958.969 19.67       15  2 958.644 0.174 958.594 15.08
07  1 958.512 0.218 958.557 20.51       15  2 959.121 0.324 959.110 15.78
07  1 958.737 0.174 958.767 21.43       15  2 959.121 0.324 959.195 16.49
07  1 959.046 0.166 959.085 23.34       15  2 959.359 0.193 959.293 19.70
12  2 959.744 0.165 959.769 15.38       15  2 959.003 0.171 959.002 21.83
12  2 958.882 0.123 958.810 16.07       15  2 958.754 0.172 958.784 24.93
12  2 959.521 0.122 959.584 16.78       15  2 958.754 0.172 958.756 26.02
12  2 958.515 0.147 958.503 17.51       15  1 959.077 0.153 959.058 14.42
12  2 959.007 0.121 958.986 18.29       15  1 958.644 0.174 958.594 15.08
12  2 959.489 0.152 959.420 19.09       15  1 959.121 0.324 959.110 15.78
12  2 959.489 0.152 959.455 19.98       15  1 959.121 0.324 959.195 16.49
12  2 958.614 0.165 958.658 20.86       15  1 959.359 0.193 959.293 19.70
12  2 959.132 0.158 959.171 21.77       15  1 959.003 0.171 959.002 21.83
12  2 958.598 0.149 958.574 22.72       15  1 958.754 0.172 958.784 24.93
12  2 959.489 0.152 959.467 23.72       15  1 958.754 0.172 958.756 26.02
12  2 959.321 0.165 959.279 24.78       15  2 959.077 0.153 959.072 28.66
12  1 959.744 0.165 959.769 15.38       15  1 959.077 0.153 959.072 28.66
```

|      | 1998 - ABRIL |         |       |         |       |
|------|--------------|---------|-------|---------|-------|
| D    | L            | SDB     | ER    | SDC     | HL    |
| 16   | 2            | 959.161 | 0.214 | 959.212 | 13.60 |
| 16   | 2            | 959.000 | 0.167 | 958.977 | 14.83 |
| 16   | 2            | 959.446 | 0.172 | 959.500 | 16.05 |
| 16   | 2            | 959.284 | 0.182 | 959.299 | 16.72 |
| 16   | 2            | 958.972 | 0.146 | 959.909 | 17.38 |
| 16   | 2            | 959.679 | 0.178 | 959.646 | 18.07 |
| 16   | 2            | 958.401 | 0.139 | 958.504 | 22.89 |
| 16   | 2            | 958.281 | 0.186 | 958.066 | 26.55 |
| 16   | 1            | 959.161 | 0.214 | 959.212 | 13.60 |
| 16   | 1            | 959.000 | 0.167 | 958.977 | 14.83 |
| 16   | 1            | 959.446 | 0.172 | 959.500 | 16.05 |
| 16   | 1            | 959.284 | 0.182 | 959.299 | 16.72 |
| 16   | 1            | 959.972 | 0.146 | 959.909 | 17.38 |
| 16   | 1            | 959.679 | 0.178 | 959.646 | 18.07 |
| 16   | 1            | 958.401 | 0.139 | 958.504 | 22.89 |
| 16   | 1            | 958.281 | 0.186 | 958.066 | 26.55 |
| 16   | 2            | 958.705 | 0.153 | 958.650 | 27.67 |
| 16   | 2            | 959.446 | 0.172 | 959.382 | 28.84 |
| 16   | 2            | 959.583 | 0.179 | 959.530 | 30.09 |
| 16   | 1            | 958.705 | 0.153 | 958.650 | 27.67 |
| 16   | 1            | 959.446 | 0.172 | 959.382 | 28.84 |
| 16   | 1            | 959.583 | 0.179 | 959.530 | 30.09 |
| 17   | 2            | 959.000 | 0.103 | 959.021 | 13.02 |
| 17   | 2            | 959.234 | 0.113 | 959.238 | 13.59 |
| 17   | 2            | 958.927 | 0.106 | 958.925 | 14.18 |
| 17   | 2            | 959.533 | 0.119 | 959.517 | 14.82 |
| 17   | 2            | 959.572 | 0.155 | 959.580 | 15.46 |
| 17   | 2            | 959.365 | 0.169 | 959.341 | 16.10 |
| 17   | 2            | 959.130 | 0.143 | 959.128 | 16.76 |
| 17   | 2            | 958.696 | 0.168 | 958.655 | 17.49 |
| 17   | 2            | 959.394 | 0.102 | 959.417 | 18.42 |
| 17   | 2            | 958.594 | 0.133 | 958.625 | 20.90 |
| 17   | 2            | 958.840 | 0.116 | 958.816 | 21.82 |
| 17   | 2            | 958.118 | 0.151 | 958.297 | 24.92 |
| 17   | 2            | 958.957 | 0.142 | 958.962 | 26.00 |
| 17   | 1            | 959.000 | 0.103 | 959.021 | 13.02 |
| 17   | 1            | 959.234 | 0.113 | 959.238 | 13.59 |
| 17   | 1            | 958.927 | 0.106 | 958.925 | 14.18 |
| 17   | 1            | 959.533 | 0.119 | 959.517 | 14.82 |
| 17   | 1            | 959.572 | 0.155 | 959.580 | 15.46 |
| 17   | 1            | 959.365 | 0.169 | 959.341 | 16.10 |
| 17   | 1            | 959.130 | 0.143 | 959.128 | 16.76 |
| 17   | 1            | 958.696 | 0.168 | 958.655 | 17.49 |
| 17   | 1            | 959.394 | 0.102 | 959.417 | 18.42 |
| 17   | 1            | 958.594 | 0.133 | 958.625 | 20.90 |
| 17   | 1            | 958.840 | 0.116 | 958.816 | 21.82 |
| 17   | 1            | 958.118 | 0.151 | 958.297 | 24.92 |
| 17   | 1            | 958.957 | 0.142 | 958.962 | 26.00 |
| 17   | 2            | 958.957 | 0.142 | 958.954 | 28.47 |
| 17   | 1            | 958.957 | 0.142 | 958.954 | 28.47 |
| 20   | 2            | 958.707 | 0.153 | 958.725 | 13.51 |
| 20   | 2            | 959.817 | 0.133 | 959.854 | 14.09 |
| 20   | 2            | 959.016 | 0.153 | 959.048 | 14.73 |
| 20   | 2            | 959.195 | 0.141 | 959.199 | 16.79 |
| 20   | 2            | 959.403 | 0.189 | 959.481 | 17.47 |
| 20   | 2            | 959.131 | 0.205 | 959.093 | 18.20 |
| 20   | 2            | 959.131 | 0.205 | 959.088 | 19.82 |
| 20   | 2            | 959.233 | 0.164 | 959.224 | 21.52 |
| 20   | 2            | 959.131 | 0.205 | 959.113 | 22.50 |
| 20   | 2            | 957.862 | 0.197 | 957.870 | 23.48 |
| 20   | 2            | 958.995 | 0.152 | 959.004 | 24.55 |
| 20   | 2            | 958.393 | 0.258 | 958.418 | 25.98 |
| 20   | 2            | 957.862 | 0.197 | 957.873 | 27.12 |
| 20   | 2            | 958.607 | 0.228 | 958.563 | 28.34 |
| 20   | 1            | 958.707 | 0.153 | 958.725 | 13.51 |
| 20   | 1            | 959.817 | 0.133 | 959.854 | 14.09 |
| 20   | 1            | 959.016 | 0.153 | 959.048 | 14.73 |
| 20   | 1            | 959.195 | 0.141 | 959.199 | 16.79 |
| 20   | 1            | 959.403 | 0.189 | 959.481 | 17.47 |
| 20   | 1            | 959.131 | 0.205 | 959.093 | 18.20 |
| 20   | 1            | 959.131 | 0.205 | 959.088 | 19.82 |

|      | 1998 - ABRIL |         |       |         |       |
|------|--------------|---------|-------|---------|-------|
| D    | L            | SDB     | ER    | SDC     | HL    |
| 20   | 1            | 959.233 | 0.164 | 959.224 | 21.52 |
| 20   | 1            | 959.131 | 0.205 | 959.113 | 22.50 |
| 20   | 1            | 957.862 | 0.197 | 957.870 | 23.48 |
| 20   | 1            | 958.995 | 0.152 | 959.004 | 24.55 |
| 20   | 1            | 958.393 | 0.258 | 958.418 | 25.98 |
| 20   | 1            | 957.862 | 0.197 | 957.873 | 27.12 |
| 20   | 1            | 958.607 | 0.228 | 958.563 | 28.34 |
| 20   | 2            | 959.817 | 0.133 | 959.716 | 29.61 |
| 20   | 2            | 958.106 | 0.229 | 958.208 | 32.35 |
| 20   | 2            | 960.135 | 0.214 | 960.310 | 34.13 |
| 20   | 2            | 957.696 | 0.248 | 957.684 | 35.88 |
| 20   | 2            | 957.940 | 0.210 | 958.021 | 37.77 |
| 20   | 2            | 958.958 | 0.153 | 958.887 | 39.88 |
| 20   | 2            | 957.696 | 0.248 | 957.520 | 42.30 |
| 20   | 1            | 959.817 | 0.133 | 959.716 | 29.61 |
| 20   | 1            | 958.106 | 0.229 | 958.208 | 32.35 |
| 20   | 1            | 960.135 | 0.214 | 960.310 | 34.13 |
| 20   | 1            | 957.696 | 0.248 | 957.684 | 35.88 |
| 20   | 1            | 957.940 | 0.210 | 958.021 | 37.77 |
| 20   | 1            | 958.958 | 0.153 | 958.887 | 39.88 |
| 20   | 1            | 957.696 | 0.248 | 957.520 | 42.30 |
| 22   | 2            | 959.136 | 0.147 | 959.119 | 15.07 |
| 22   | 2            | 959.226 | 0.120 | 959.204 | 15.73 |
| 22   | 2            | 958.736 | 0.174 | 958.745 | 16.37 |
| 22   | 2            | 959.668 | 0.147 | 959.776 | 17.05 |
| 22   | 2            | 959.668 | 0.147 | 959.563 | 17.76 |
| 22   | 2            | 958.722 | 0.149 | 958.686 | 18.64 |
| 22   | 2            | 959.253 | 0.149 | 959.257 | 20.92 |
| 22   | 2            | 959.384 | 0.142 | 959.395 | 23.51 |
| 22   | 2            | 959.032 | 0.125 | 959.040 | 25.51 |
| 22   | 2            | 958.765 | 0.134 | 958.831 | 28.91 |
| 22   | 1            | 959.136 | 0.147 | 959.119 | 15.07 |
| 22   | 1            | 959.226 | 0.120 | 959.204 | 15.73 |
| 22   | 1            | 958.736 | 0.174 | 958.745 | 16.37 |
| 22   | 1            | 959.668 | 0.147 | 959.776 | 17.05 |
| 22   | 1            | 959.668 | 0.147 | 959.563 | 17.76 |
| 22   | 1            | 958.722 | 0.149 | 958.686 | 18.64 |
| 22   | 1            | 959.253 | 0.149 | 959.257 | 20.92 |
| 22   | 1            | 959.384 | 0.142 | 959.395 | 23.51 |
| 22   | 1            | 959.032 | 0.125 | 959.040 | 25.51 |
| 22   | 1            | 958.765 | 0.134 | 958.831 | 28.91 |
| 22   | 2            | 958.578 | 0.166 | 958.608 | 33.00 |
| 22   | 2            | 958.736 | 0.174 | 958.742 | 34.50 |
| 22   | 1            | 958.578 | 0.166 | 958.608 | 33.00 |
| 22   | 1            | 958.736 | 0.174 | 958.742 | 34.50 |

|      | 1998 - MAIO |         |       |         |       |
|------|-------------|---------|-------|---------|-------|
| D    | L           | SDB     | ER    | SDC     | HL    |
| 11   | 2           | 959.208 | 0.237 | 959.241 | 27.83 |
| 11   | 2           | 959.208 | 0.237 | 959.102 | 31.90 |
| 11   | 2           | 959.467 | 0.197 | 959.344 | 33.04 |
| 11   | 2           | 959.208 | 0.237 | 959.252 | 35.71 |
| 11   | 1           | 959.208 | 0.237 | 959.241 | 27.83 |
| 11   | 1           | 959.208 | 0.237 | 959.102 | 31.90 |
| 11   | 1           | 959.467 | 0.197 | 959.344 | 33.04 |
| 11   | 1           | 959.208 | 0.237 | 959.252 | 35.71 |
| 11   | 2           | 958.805 | 0.179 | 958.937 | 41.49 |
| 11   | 2           | 959.208 | 0.237 | 959.106 | 43.79 |
| 11   | 2           | 958.805 | 0.179 | 958.815 | 48.26 |
| 11   | 1           | 958.805 | 0.179 | 958.937 | 41.49 |
| 11   | 1           | 959.208 | 0.237 | 959.106 | 43.79 |
| 11   | 1           | 958.805 | 0.179 | 958.815 | 48.26 |
| 20   | 2           | 959.800 | 0.158 | 959.821 | 32.97 |
| 20   | 2           | 958.592 | 0.223 | 958.882 | 34.09 |
| 20   | 2           | 959.293 | 0.178 | 959.414 | 35.57 |
| 20   | 2           | 957.886 | 0.200 | 957.946 | 36.79 |
| 20   | 2           | 959.800 | 0.158 | 959.901 | 39.56 |
| 20   | 1           | 959.800 | 0.158 | 959.821 | 32.97 |
| 20   | 1           | 958.592 | 0.223 | 958.882 | 34.09 |
| 20   | 1           | 959.293 | 0.178 | 959.414 | 35.57 |

|  | 1998 - MAIO |  |  |  |  |  | 1998 - JUNHO |  |  |  |
|---|---|---|---|---|---|---|---|---|---|---|
| D | L | SDB | ER | SDC | HL | D | L | SDB | ER | SDC | HL |
| 20 | 1 | 957.886 | 0.200 | 957.946 | 36.79 | 07 | 1 | 958.358 | 0.213 | 958.478 | 43.36 |
| 20 | 1 | 959.800 | 0.158 | 959.901 | 39.56 | 07 | 1 | 959.986 | 0.219 | 960.195 | 44.68 |
| 20 | 2 | 958.592 | 0.223 | 958.672 | 43.43 | 07 | 1 | 959.916 | 0.212 | 959.763 | 45.98 |
| 20 | 2 | 958.592 | 0.223 | 958.672 | 47.30 | 07 | 2 | 959.076 | 0.227 | 959.177 | 47.45 |
| 20 | 1 | 958.592 | 0.223 | 958.672 | 43.43 | 07 | 2 | 958.900 | 0.237 | 958.820 | 48.90 |
| 20 | 1 | 958.592 | 0.223 | 958.671 | 47.30 | 07 | 2 | 958.341 | 0.191 | 958.233 | 50.49 |
| 22 | 2 | 959.185 | 0.138 | 959.230 | 33.52 | 07 | 2 | 958.955 | 0.209 | 958.974 | 52.15 |
| 22 | 2 | 959.300 | 0.164 | 959.245 | 34.58 | 07 | 2 | 959.986 | 0.219 | 960.008 | 53.92 |
| 22 | 2 | 959.185 | 0.138 | 959.214 | 35.63 | 07 | 2 | 959.503 | 0.272 | 959.671 | 55.87 |
| 22 | 2 | 959.135 | 0.162 | 959.140 | 36.80 | 07 | 1 | 959.076 | 0.227 | 959.177 | 47.45 |
| 22 | 2 | 958.979 | 0.165 | 958.978 | 37.98 | 07 | 1 | 958.900 | 0.237 | 958.820 | 48.90 |
| 22 | 2 | 958.845 | 0.181 | 958.834 | 39.25 | 07 | 1 | 958.341 | 0.191 | 958.233 | 50.49 |
| 22 | 2 | 958.756 | 0.156 | 958.781 | 40.58 | 07 | 1 | 958.955 | 0.209 | 958.974 | 52.15 |
| 22 | 1 | 959.185 | 0.138 | 959.230 | 33.52 | 07 | 1 | 959.986 | 0.219 | 960.008 | 53.92 |
| 22 | 1 | 959.300 | 0.164 | 959.245 | 34.58 | 07 | 1 | 959.503 | 0.272 | 959.671 | 55.87 |
| 22 | 1 | 959.185 | 0.138 | 959.214 | 35.63 | 08 | 2 | 959.627 | 0.194 | 959.580 | 41.75 |
| 22 | 1 | 959.135 | 0.162 | 959.140 | 36.80 | 08 | 2 | 959.127 | 0.222 | 959.076 | 42.83 |
| 22 | 1 | 958.979 | 0.165 | 958.978 | 37.98 | 08 | 2 | 960.331 | 0.240 | 960.333 | 43.98 |
| 22 | 1 | 958.845 | 0.181 | 958.834 | 39.25 | 08 | 2 | 959.860 | 0.202 | 959.922 | 45.25 |
| 22 | 1 | 958.756 | 0.156 | 958.781 | 40.58 | 08 | 2 | 959.860 | 0.202 | 959.992 | 46.52 |
| 22 | 2 | 959.393 | 0.195 | 959.439 | 42.12 | 08 | 1 | 959.627 | 0.194 | 959.580 | 41.75 |
| 22 | 2 | 958.847 | 0.151 | 958.898 | 43.69 | 08 | 1 | 959.127 | 0.222 | 959.076 | 42.83 |
| 22 | 2 | 958.746 | 0.208 | 958.403 | 45.34 | 08 | 1 | 960.331 | 0.240 | 960.333 | 43.98 |
| 22 | 2 | 959.519 | 0.180 | 959.593 | 47.16 | 08 | 1 | 959.860 | 0.202 | 959.922 | 45.25 |
| 22 | 2 | 959.519 | 0.180 | 959.640 | 49.08 | 08 | 1 | 959.860 | 0.202 | 959.992 | 46.52 |
| 22 | 1 | 959.393 | 0.195 | 959.439 | 42.12 | 08 | 2 | 959.127 | 0.222 | 959.150 | 47.93 |
| 22 | 1 | 958.847 | 0.151 | 958.898 | 43.69 | 08 | 2 | 958.850 | 0.185 | 958.818 | 49.40 |
| 22 | 1 | 958.746 | 0.208 | 958.403 | 45.34 | 08 | 2 | 959.857 | 0.180 | 959.797 | 51.11 |
| 22 | 1 | 959.519 | 0.180 | 959.593 | 47.16 | 08 | 2 | 958.673 | 0.196 | 958.519 | 52.89 |
| 22 | 1 | 959.519 | 0.180 | 959.640 | 49.08 | 08 | 2 | 958.673 | 0.196 | 958.568 | 54.81 |
| 26 | 2 | 959.325 | 0.159 | 959.469 | 33.54 | 08 | 2 | 960.331 | 0.240 | 960.343 | 56.87 |
| 26 | 2 | 959.325 | 0.159 | 959.548 | 34.49 | 08 | 1 | 959.127 | 0.222 | 959.150 | 47.93 |
| 26 | 2 | 959.325 | 0.159 | 959.372 | 35.48 | 08 | 1 | 958.850 | 0.185 | 958.818 | 49.40 |
| 26 | 2 | 959.325 | 0.159 | 959.748 | 36.56 | 08 | 1 | 959.857 | 0.180 | 959.797 | 51.11 |
| 26 | 2 | 958.991 | 0.178 | 958.959 | 37.65 | 08 | 1 | 958.673 | 0.196 | 958.519 | 52.89 |
| 26 | 2 | 959.325 | 0.159 | 959.416 | 38.92 | 08 | 1 | 958.673 | 0.196 | 958.568 | 54.81 |
| 26 | 1 | 959.325 | 0.159 | 959.469 | 33.54 | 08 | 1 | 960.331 | 0.240 | 960.343 | 56.87 |
| 26 | 1 | 959.325 | 0.159 | 959.548 | 34.49 | 09 | 2 | 960.120 | 0.112 | 959.870 | 45.28 |
| 26 | 1 | 959.325 | 0.159 | 959.372 | 35.48 | 09 | 2 | 959.027 | 0.504 | 958.906 | 46.57 |
| 26 | 1 | 959.325 | 0.159 | 959.748 | 36.56 | 09 | 1 | 960.120 | 0.112 | 959.870 | 45.28 |
| 26 | 1 | 958.991 | 0.178 | 958.959 | 37.65 | 09 | 1 | 959.027 | 0.504 | 958.906 | 46.57 |
| 26 | 1 | 959.325 | 0.159 | 959.416 | 38.92 | 09 | 2 | 959.027 | 0.504 | 958.623 | 47.93 |
|  |  |  |  |  |  | 09 | 2 | 959.027 | 0.504 | 958.257 | 49.48 |
|  |  |  |  |  |  | 09 | 1 | 959.027 | 0.504 | 958.623 | 47.93 |
|  | 1998 - JUNHO |  |  |  |  | 09 | 1 | 959.027 | 0.504 | 958.257 | 49.48 |
| D | L | SDB | ER | SDC | HL | 10 | 2 | 959.237 | 0.193 | 959.262 | 46.01 |
| 02 | 2 | 959.090 | 0.448 | 959.583 | 43.24 | 10 | 2 | 958.562 | 0.275 | 958.611 | 47.42 |
| 02 | 2 | 960.412 | 0.392 | 960.254 | 44.47 | 10 | 1 | 959.237 | 0.193 | 959.262 | 46.01 |
| 02 | 1 | 959.090 | 0.448 | 959.583 | 43.24 | 10 | 1 | 958.562 | 0.275 | 958.611 | 47.42 |
| 02 | 1 | 960.412 | 0.392 | 960.254 | 44.47 | 10 | 2 | 958.562 | 0.275 | 958.654 | 48.78 |
| 02 | 2 | 958.546 | 0.527 | 958.557 | 47.13 | 10 | 2 | 958.562 | 0.275 | 958.656 | 50.23 |
| 02 | 2 | 958.057 | 0.379 | 958.058 | 48.64 | 10 | 2 | 959.527 | 0.214 | 959.472 | 51.73 |
| 02 | 2 | 958.452 | 0.254 | 958.295 | 50.21 | 10 | 2 | 959.237 | 0.193 | 959.359 | 55.33 |
| 02 | 2 | 959.090 | 0.448 | 959.203 | 51.93 | 10 | 2 | 958.323 | 0.221 | 958.412 | 57.32 |
| 02 | 2 | 959.090 | 0.448 | 959.439 | 53.79 | 10 | 2 | 959.527 | 0.214 | 959.539 | 59.50 |
| 02 | 2 | 957.269 | 0.304 | 957.471 | 55.89 | 10 | 2 | 958.323 | 0.221 | 958.258 | 62.09 |
| 02 | 2 | 958.452 | 0.254 | 958.338 | 58.37 | 10 | 1 | 958.562 | 0.275 | 958.654 | 48.78 |
| 02 | 2 | 958.057 | 0.379 | 957.973 | 61.43 | 10 | 1 | 958.562 | 0.275 | 958.656 | 50.23 |
| 02 | 1 | 958.546 | 0.527 | 958.557 | 47.13 | 10 | 1 | 959.527 | 0.214 | 959.472 | 51.73 |
| 02 | 1 | 958.057 | 0.379 | 958.058 | 48.64 | 10 | 1 | 959.237 | 0.193 | 959.359 | 55.33 |
| 02 | 1 | 958.452 | 0.254 | 958.295 | 50.21 | 10 | 1 | 958.323 | 0.221 | 958.412 | 57.32 |
| 02 | 1 | 959.090 | 0.448 | 959.203 | 51.93 | 10 | 1 | 959.527 | 0.214 | 959.539 | 59.50 |
| 02 | 1 | 959.090 | 0.448 | 959.439 | 53.79 | 10 | 1 | 958.323 | 0.221 | 958.258 | 62.09 |
| 02 | 1 | 957.269 | 0.304 | 957.471 | 55.89 | 15 | 2 | 959.695 | 0.306 | 959.703 | 44.07 |
| 02 | 1 | 958.452 | 0.254 | 958.338 | 58.37 | 15 | 2 | 959.525 | 0.359 | 959.475 | 45.31 |
| 02 | 1 | 958.057 | 0.379 | 957.973 | 61.43 | 15 | 2 | 959.014 | 0.339 | 958.900 | 46.79 |
| 07 | 2 | 959.437 | 0.212 | 959.444 | 41.12 | 15 | 2 | 959.071 | 0.257 | 959.085 | 48.00 |
| 07 | 2 | 958.358 | 0.213 | 958.478 | 43.36 | 15 | 2 | 960.123 | 0.323 | 960.000 | 49.35 |
| 07 | 2 | 959.986 | 0.219 | 960.195 | 44.68 | 15 | 1 | 959.695 | 0.306 | 959.703 | 44.07 |
| 07 | 2 | 959.916 | 0.212 | 959.763 | 45.98 | 15 | 1 | 959.525 | 0.359 | 959.475 | 45.31 |
| 07 | 1 | 959.437 | 0.212 | 959.444 | 41.12 | 15 | 1 | 959.014 | 0.339 | 958.900 | 46.79 |

|  | 1998 - JUNHO | | | | | | 1998 - JUNHO | | | |
|---|---|---|---|---|---|---|---|---|---|---|
| D | L | SDB | ER | SDC | HL | D | L | SDB | ER | SDC | HL |
| 15 | 1 | 959.071 | 0.257 | 959.085 | 48.00 | 24 | 2 | 958.576 | 0.446 | 958.619 | 62.42 |
| 15 | 1 | 960.123 | 0.323 | 960.000 | 49.35 | 24 | 2 | 959.225 | 0.304 | 959.289 | 64.63 |
| 15 | 2 | 958.659 | 0.310 | 958.603 | 50.71 | 24 | 1 | 959.947 | 0.484 | 960.341 | 53.96 |
| 15 | 2 | 959.014 | 0.339 | 959.012 | 52.38 | 24 | 1 | 958.788 | 0.361 | 958.824 | 55.43 |
| 15 | 2 | 959.014 | 0.339 | 959.006 | 54.03 | 24 | 1 | 958.318 | 0.217 | 958.269 | 56.89 |
| 15 | 2 | 960.123 | 0.323 | 960.473 | 57.52 | 24 | 1 | 958.413 | 0.312 | 958.378 | 58.51 |
| 15 | 2 | 959.102 | 0.247 | 959.097 | 59.60 | 24 | 1 | 959.225 | 0.304 | 959.365 | 60.45 |
| 15 | 1 | 958.659 | 0.310 | 958.603 | 50.71 | 24 | 1 | 958.576 | 0.446 | 958.619 | 62.42 |
| 15 | 1 | 959.014 | 0.339 | 959.012 | 52.38 | 24 | 1 | 959.225 | 0.304 | 959.289 | 64.63 |
| 15 | 1 | 959.014 | 0.339 | 959.006 | 54.03 | 26 | 2 | 959.372 | 0.314 | 959.283 | 49.45 |
| 15 | 1 | 960.123 | 0.323 | 960.473 | 57.52 | 26 | 2 | 960.506 | 0.306 | 960.498 | 50.55 |
| 15 | 1 | 959.102 | 0.247 | 959.097 | 59.60 | 26 | 1 | 959.372 | 0.314 | 959.283 | 49.45 |
| 19 | 2 | 959.273 | 0.146 | 958.740 | 44.41 | 26 | 1 | 960.506 | 0.306 | 960.498 | 50.55 |
| 19 | 2 | 959.544 | 0.149 | 959.537 | 45.43 | 26 | 2 | 957.898 | 0.525 | 957.012 | 56.35 |
| 19 | 2 | 959.709 | 0.141 | 959.751 | 46.59 | 26 | 2 | 957.898 | 0.525 | 957.836 | 59.60 |
| 19 | 2 | 959.544 | 0.149 | 959.557 | 47.75 | 26 | 2 | 958.194 | 0.210 | 958.160 | 63.36 |
| 19 | 2 | 959.273 | 0.146 | 959.257 | 48.90 | 26 | 2 | 959.372 | 0.314 | 959.283 | 65.50 |
| 19 | 2 | 959.709 | 0.141 | 959.711 | 50.13 | 26 | 2 | 959.452 | 0.235 | 959.427 | 69.38 |
| 19 | 2 | 959.303 | 0.187 | 959.309 | 51.39 | 26 | 1 | 957.898 | 0.525 | 957.012 | 56.35 |
| 19 | 1 | 959.273 | 0.146 | 958.740 | 44.41 | 26 | 1 | 957.898 | 0.525 | 957.836 | 59.60 |
| 19 | 1 | 959.544 | 0.149 | 959.537 | 45.43 | 26 | 1 | 958.194 | 0.210 | 958.160 | 63.36 |
| 19 | 1 | 959.709 | 0.141 | 959.751 | 46.59 | 26 | 1 | 959.372 | 0.314 | 959.283 | 65.50 |
| 19 | 1 | 959.544 | 0.149 | 959.557 | 47.75 | 26 | 1 | 959.452 | 0.235 | 959.427 | 69.38 |
| 19 | 1 | 959.273 | 0.146 | 959.257 | 48.90 | 28 | 2 | 959.766 | 0.308 | 960.995 | 51.38 |
| 19 | 1 | 959.709 | 0.141 | 959.711 | 50.13 | 28 | 2 | 959.766 | 0.308 | 959.992 | 52.65 |
| 19 | 1 | 959.303 | 0.187 | 959.309 | 51.39 | 28 | 2 | 959.289 | 0.463 | 959.339 | 53.95 |
| 19 | 2 | 959.947 | 0.124 | 959.967 | 52.77 | 28 | 1 | 959.766 | 0.308 | 960.995 | 51.38 |
| 19 | 2 | 959.831 | 0.110 | 959.870 | 54.19 | 28 | 1 | 959.766 | 0.308 | 959.992 | 52.65 |
| 19 | 2 | 959.709 | 0.141 | 959.715 | 55.72 | 28 | 1 | 959.289 | 0.463 | 959.339 | 53.95 |
| 19 | 2 | 959.324 | 0.129 | 959.326 | 57.36 | 28 | 2 | 959.270 | 0.181 | 959.273 | 55.24 |
| 19 | 2 | 959.324 | 0.129 | 959.386 | 59.12 | 28 | 2 | 959.535 | 0.203 | 959.434 | 56.65 |
| 19 | 1 | 959.947 | 0.124 | 959.967 | 52.77 | 28 | 2 | 958.898 | 0.195 | 957.681 | 58.75 |
| 19 | 1 | 959.831 | 0.110 | 959.870 | 54.19 | 28 | 2 | 959.146 | 0.206 | 959.124 | 60.46 |
| 19 | 1 | 959.709 | 0.141 | 959.715 | 55.72 | 28 | 2 | 959.766 | 0.308 | 959.664 | 62.15 |
| 19 | 1 | 959.324 | 0.129 | 959.326 | 57.36 | 28 | 2 | 959.289 | 0.463 | 959.390 | 64.60 |
| 19 | 1 | 959.324 | 0.129 | 959.386 | 59.12 | 28 | 2 | 958.898 | 0.195 | 958.526 | 66.71 |
| 23 | 2 | 959.122 | 0.805 | 959.493 | 50.01 | 28 | 1 | 959.270 | 0.181 | 959.273 | 55.24 |
| 23 | 2 | 960.053 | 0.710 | 959.845 | 51.13 | 28 | 1 | 959.535 | 0.203 | 959.434 | 56.65 |
| 23 | 2 | 958.892 | 0.293 | 958.865 | 52.28 | 28 | 1 | 958.898 | 0.195 | 957.681 | 58.75 |
| 23 | 1 | 959.122 | 0.805 | 959.493 | 50.01 | 28 | 1 | 959.146 | 0.206 | 959.124 | 60.46 |
| 23 | 1 | 960.053 | 0.710 | 959.845 | 51.13 | 28 | 1 | 959.766 | 0.308 | 959.664 | 62.15 |
| 23 | 1 | 958.892 | 0.293 | 958.865 | 52.28 | 28 | 1 | 959.289 | 0.463 | 959.390 | 64.60 |
| 23 | 2 | 959.122 | 0.805 | 959.101 | 53.52 | 28 | 1 | 958.898 | 0.195 | 958.526 | 66.71 |
| 23 | 2 | 958.943 | 0.520 | 958.948 | 54.82 | 29 | 2 | 959.366 | 0.291 | 959.476 | 51.78 |
| 23 | 2 | 958.021 | 0.631 | 957.973 | 56.22 | 29 | 2 | 959.290 | 0.166 | 959.281 | 53.00 |
| 23 | 2 | 958.593 | 0.288 | 958.570 | 57.83 | 29 | 2 | 958.683 | 0.195 | 958.668 | 54.22 |
| 23 | 2 | 958.386 | 0.384 | 958.294 | 59.44 | 29 | 2 | 959.290 | 0.166 | 959.292 | 55.51 |
| 23 | 2 | 959.122 | 0.805 | 959.125 | 61.22 | 29 | 1 | 959.366 | 0.291 | 959.476 | 51.78 |
| 23 | 2 | 958.892 | 0.293 | 958.858 | 63.15 | 29 | 1 | 959.290 | 0.166 | 959.281 | 53.00 |
| 23 | 2 | 958.943 | 0.520 | 958.958 | 65.28 | 29 | 1 | 958.683 | 0.195 | 958.668 | 54.22 |
| 23 | 1 | 959.122 | 0.805 | 959.101 | 53.52 | 29 | 1 | 959.290 | 0.166 | 959.292 | 55.51 |
| 23 | 1 | 958.943 | 0.520 | 958.948 | 54.82 | 29 | 2 | 959.310 | 0.206 | 959.331 | 56.91 |
| 23 | 1 | 958.021 | 0.631 | 957.973 | 56.22 | 29 | 2 | 959.366 | 0.291 | 959.613 | 58.44 |
| 23 | 1 | 958.593 | 0.288 | 958.570 | 57.83 | 29 | 2 | 958.817 | 0.166 | 958.920 | 59.99 |
| 23 | 1 | 958.386 | 0.384 | 958.294 | 59.44 | 29 | 2 | 959.366 | 0.291 | 959.393 | 61.68 |
| 23 | 1 | 959.122 | 0.805 | 959.125 | 61.22 | 29 | 2 | 959.093 | 0.184 | 959.033 | 63.46 |
| 23 | 1 | 958.892 | 0.293 | 958.858 | 63.15 | 29 | 2 | 959.366 | 0.291 | 959.535 | 65.47 |
| 23 | 1 | 958.943 | 0.520 | 958.958 | 65.28 | 29 | 2 | 959.093 | 0.184 | 958.975 | 67.83 |
| 24 | 2 | 959.947 | 0.484 | 959.714 | 47.65 | 29 | 1 | 959.310 | 0.206 | 959.331 | 56.91 |
| 24 | 2 | 959.225 | 0.304 | 959.388 | 48.77 | 29 | 1 | 959.366 | 0.291 | 959.613 | 58.44 |
| 24 | 2 | 958.788 | 0.361 | 958.744 | 49.87 | 29 | 1 | 958.817 | 0.166 | 958.920 | 59.99 |
| 24 | 2 | 959.225 | 0.304 | 959.498 | 51.05 | 29 | 1 | 959.366 | 0.291 | 959.393 | 61.68 |
| 24 | 1 | 959.947 | 0.484 | 959.714 | 47.65 | 29 | 1 | 959.093 | 0.184 | 959.033 | 63.46 |
| 24 | 1 | 959.225 | 0.304 | 959.388 | 48.77 | 29 | 1 | 959.366 | 0.291 | 959.535 | 65.47 |
| 24 | 1 | 958.788 | 0.361 | 958.744 | 49.87 | 29 | 1 | 959.093 | 0.184 | 958.975 | 67.83 |
| 24 | 1 | 959.225 | 0.304 | 959.498 | 51.05 | 30 | 2 | 958.999 | 0.175 | 959.057 | 50.96 |
| 24 | 2 | 959.947 | 0.484 | 960.341 | 53.96 | 30 | 2 | 959.971 | 0.204 | 960.141 | 52.10 |
| 24 | 2 | 958.788 | 0.361 | 958.824 | 55.43 | 30 | 2 | 959.804 | 0.188 | 959.777 | 53.24 |
| 24 | 2 | 958.318 | 0.217 | 958.269 | 56.89 | 30 | 2 | 959.551 | 0.184 | 959.646 | 54.41 |
| 24 | 2 | 958.413 | 0.312 | 958.378 | 58.51 | 30 | 2 | 958.999 | 0.175 | 959.043 | 55.64 |
| 24 | 2 | 959.225 | 0.304 | 959.365 | 60.45 | 30 | 1 | 958.999 | 0.175 | 959.057 | 50.96 |

```
       1998 - JUNHO                              1998 - JULHO
 D  L   SDB    ER     SDC    HL        D  L   SDB    ER     SDC    HL
 30 1  959.971 0.204 960.141 52.10     03 2  958.995 0.973 958.900 54.48
 30 1  959.804 0.188 959.777 53.24     03 2  958.789 0.162 958.786 55.64
 30 1  959.551 0.184 959.646 54.41     03 1  959.277 0.278 959.694 51.15
 30 1  958.999 0.175 959.043 55.64     03 1  958.995 0.973 958.936 52.23
 30 2  959.449 0.255 959.397 57.02     03 1  958.365 0.163 958.291 53.37
 30 2  959.115 0.158 959.082 58.46     03 1  958.995 0.973 958.900 54.48
 30 2  959.115 0.158 959.206 59.94     03 1  958.789 0.162 958.786 55.64
 30 2  959.115 0.158 959.149 61.64     03 2  958.609 0.184 958.639 58.19
 30 2  959.551 0.184 959.531 63.41     03 1  958.609 0.184 958.639 58.19
 30 2  959.804 0.188 959.779 67.67     06 2  959.944 0.193 960.306 53.21
 30 1  959.449 0.255 959.397 57.02     06 2  957.950 0.250 957.969 54.31
 30 1  959.115 0.158 959.082 58.46     06 2  958.642 0.148 958.590 55.63
 30 1  959.115 0.158 959.206 59.94     06 2  958.090 0.235 958.141 56.89
 30 1  959.115 0.158 959.149 61.64     06 1  959.944 0.193 960.306 53.21
 30 1  959.551 0.184 959.531 63.41     06 1  957.950 0.250 957.969 54.31
 30 1  959.804 0.188 959.779 67.67     06 1  958.642 0.148 958.590 55.63
                                       06 1  958.090 0.235 958.141 56.89
                                       06 2  959.547 0.279 959.605 58.15
       1998 - JULHO                    06 2  958.745 0.240 958.743 59.51
 D  L   SDB    ER     SDC    HL        06 2  959.259 0.193 959.105 60.95
 01 2  959.989 0.250 960.050 50.31     06 2  958.923 0.165 959.058 62.48
 01 2  959.253 0.266 959.211 51.44     06 2  959.261 0.136 959.319 64.11
 01 2  959.253 0.266 959.391 52.55     06 2  958.339 0.228 958.316 65.83
 01 2  959.253 0.266 959.295 53.72     06 2  959.944 0.193 960.231 67.90
 01 2  959.791 0.235 959.742 55.11     06 2  958.923 0.165 958.934 70.04
 01 1  959.989 0.250 960.050 50.31     06 1  959.547 0.279 959.605 58.15
 01 1  959.253 0.266 959.211 51.44     06 1  958.745 0.240 958.743 59.51
 01 1  959.253 0.266 959.391 52.55     06 1  959.259 0.193 959.105 60.95
 01 1  959.253 0.266 959.295 53.72     06 1  958.923 0.165 959.058 62.48
 01 1  959.791 0.235 959.742 55.11     06 1  959.261 0.136 959.319 64.11
 01 2  959.015 0.280 958.961 56.41     06 1  958.339 0.228 958.316 65.83
 01 2  959.989 0.250 960.243 57.75     06 1  959.944 0.193 960.231 67.90
 01 2  959.115 0.263 959.084 60.77     06 1  958.923 0.165 958.934 70.04
 01 2  958.767 0.309 958.796 62.50     08 2  959.957 0.375 960.021 52.70
 01 2  958.482 0.282 957.712 64.32     08 2  959.957 0.375 959.995 53.71
 01 2  959.807 0.210 959.807 66.49     08 2  958.925 0.163 959.005 54.76
 01 2  958.482 0.282 958.211 68.96     08 2  959.193 0.153 959.089 55.90
 01 1  959.015 0.280 958.961 56.41     08 2  959.567 0.150 959.627 57.04
 01 1  959.989 0.250 960.243 57.75     08 2  960.391 0.141 960.446 58.31
 01 1  959.115 0.263 959.084 60.77     08 1  959.957 0.375 960.021 52.70
 01 1  958.767 0.309 958.796 62.50     08 1  959.957 0.375 959.995 53.71
 01 1  958.482 0.282 957.712 64.32     08 1  958.925 0.163 959.005 54.76
 01 1  959.807 0.210 959.807 66.49     08 1  959.193 0.153 959.089 55.90
 01 1  958.482 0.282 958.211 68.96     08 1  959.567 0.150 959.627 57.04
 02 2  960.270 0.157 960.422 50.87     08 1  960.391 0.141 960.446 58.31
 02 2  960.270 0.157 960.926 51.93     08 2  959.737 0.227 959.716 59.59
 02 2  959.513 0.146 959.528 53.08     08 2  958.721 0.273 958.720 62.33
 02 2  959.286 0.199 959.363 54.20     08 2  958.925 0.163 958.960 63.92
 02 2  960.270 0.157 960.147 55.37     08 2  959.737 0.227 959.731 65.57
 02 1  960.270 0.157 960.422 50.87     08 2  959.560 0.145 959.427 67.32
 02 1  960.270 0.157 960.926 51.93     08 1  959.737 0.227 959.716 59.59
 02 1  959.513 0.146 959.528 53.08     08 1  958.721 0.273 958.720 62.33
 02 1  959.286 0.199 959.363 54.20     08 1  958.925 0.163 958.960 63.92
 02 1  960.270 0.157 960.147 55.37     08 1  959.737 0.227 959.731 65.57
 02 2  960.021 0.183 959.968 56.70     08 1  959.560 0.145 959.427 67.32
 02 2  959.286 0.199 959.240 58.01     14 2  959.831 0.270 959.953 56.03
 02 2  959.286 0.199 959.246 59.41     14 2  958.837 0.231 958.761 57.00
 02 2  958.445 0.201 958.533 60.93     14 2  959.213 0.314 959.188 58.03
 02 2  958.975 0.154 958.918 62.51     14 2  959.828 0.374 959.719 59.06
 02 2  958.975 0.154 958.931 64.23     14 2  960.198 0.562 960.136 60.15
 02 2  959.483 0.164 959.395 66.11     14 1  959.831 0.270 959.953 56.03
 02 1  960.021 0.183 959.968 56.70     14 1  958.837 0.231 958.761 57.00
 02 1  959.286 0.199 959.240 58.01     14 1  959.213 0.314 959.188 58.03
 02 1  959.286 0.199 959.246 59.41     14 1  959.828 0.374 959.719 59.06
 02 1  958.445 0.201 958.533 60.93     14 1  960.198 0.562 960.136 60.15
 02 1  958.975 0.154 958.918 62.51     14 2  959.147 0.177 959.160 61.30
 02 1  958.975 0.154 958.931 64.23     14 2  959.828 0.374 959.694 62.50
 02 1  959.483 0.164 959.395 66.11     14 2  959.828 0.374 959.656 63.78
 03 2  959.277 0.278 959.694 51.15     14 2  958.966 0.199 958.994 65.13
 03 2  958.995 0.973 958.936 52.23     14 2  959.831 0.270 959.834 66.56
 03 2  958.365 0.163 958.291 53.37     14 2  960.198 0.562 960.165 68.14
```

```
          1998 - JULHO                                    1998 - JULHO
 D  L    SDB     ER     SDC     HL           D  L    SDB     ER     SDC     HL
14  1  959.147  0.177  959.160  61.30       27  2  959.982  0.180  959.961  70.85
14  1  959.828  0.374  959.694  62.50       27  2  959.788  0.130  959.793  72.36
14  1  959.828  0.374  959.656  63.78       27  2  959.301  0.167  959.482  74.04
14  1  958.966  0.199  958.994  65.13       27  2  959.982  0.180  959.921  75.99
14  1  959.831  0.270  959.834  66.56       27  2  959.301  0.167  959.384  77.94
14  1  960.198  0.562  960.165  68.14       27  2  959.821  0.160  959.882  80.06
15  2  959.556  0.190  959.614  56.81       27  1  959.995  0.181  960.072  69.39
15  2  958.206  0.160  957.505  58.00       27  1  959.982  0.180  959.961  70.85
15  2  959.556  0.190  960.226  59.15       27  1  959.788  0.130  959.793  72.36
15  2  958.848  0.176  958.843  60.34       27  1  959.301  0.167  959.482  74.04
15  1  959.556  0.190  959.614  56.81       27  1  959.982  0.180  959.921  75.99
15  1  958.206  0.160  957.505  58.00       27  1  959.301  0.167  959.384  77.94
15  1  959.556  0.190  960.226  59.15       27  1  959.821  0.160  959.882  80.06
15  1  958.848  0.176  958.843  60.34       28  2  959.617  0.156  959.659  62.34
15  2  959.252  0.134  959.379  61.73       28  2  960.228  0.120  960.313  63.33
15  2  958.765  0.180  958.761  63.02       28  2  960.228  0.120  960.280  64.35
15  2  958.833  0.164  958.823  64.38       28  1  959.617  0.156  959.659  62.34
15  2  959.212  0.156  959.173  65.97       28  1  960.228  0.120  960.313  63.33
15  2  958.632  0.248  958.573  67.68       28  1  960.228  0.120  960.280  64.35
15  2  958.206  0.160  958.160  69.36       28  2  958.837  0.419  958.775  65.43
15  2  958.632  0.248  958.625  71.27       28  2  959.617  0.156  959.267  73.24
15  2  958.868  0.169  958.932  73.29       28  2  958.892  0.324  959.141  78.56
15  2  958.765  0.180  958.781  75.60       28  2  957.351  0.109  957.882  80.85
15  1  959.252  0.134  959.379  61.73       28  1  958.837  0.419  958.775  65.43
15  1  958.765  0.180  958.761  63.02       28  1  959.617  0.156  959.267  73.24
15  1  958.833  0.164  958.823  64.38       28  1  958.892  0.324  959.141  78.56
15  1  959.212  0.156  959.173  65.97       28  1  957.351  0.109  957.882  80.85
15  1  958.632  0.248  958.573  67.68       31  2  959.311  0.161  958.381  71.01
15  1  958.206  0.160  958.160  69.36       31  2  959.311  0.161  959.321  72.47
15  1  958.632  0.248  958.625  71.27       31  2  959.387  0.180  959.419  73.96
15  1  958.868  0.169  958.932  73.29       31  2  959.696  0.224  959.897  79.32
15  1  958.765  0.180  958.781  75.60       31  2  959.543  0.246  959.611  81.23
17  2  960.045  0.162  960.002  55.88       31  1  959.311  0.161  958.381  71.01
17  2  960.224  0.190  960.303  56.87       31  1  959.311  0.161  959.321  72.47
17  2  959.158  0.153  959.154  57.87       31  1  959.387  0.180  959.419  73.96
17  2  959.870  0.173  959.901  58.96       31  1  959.696  0.224  959.897  79.32
17  2  960.224  0.190  960.142  60.06       31  1  959.543  0.246  959.611  81.23
17  2  959.026  0.174  959.030  61.21
17  1  960.045  0.162  960.002  55.88                1998 - AGOSTO
17  1  960.224  0.190  960.303  56.87        D  L    SDB     ER     SDC     HL
17  1  959.158  0.153  959.154  57.87       03  2  959.165  0.178  959.191  64.30
17  1  959.870  0.173  959.901  58.96       03  2  959.298  0.147  959.396  65.27
17  1  960.224  0.190  960.142  60.06       03  2  959.152  0.181  959.028  66.29
17  1  959.026  0.174  959.030  61.21       03  1  959.165  0.178  959.191  64.30
17  2  958.709  0.187  958.625  62.41       03  1  959.298  0.147  959.396  65.27
17  2  958.445  0.178  958.517  63.79       03  1  959.152  0.181  959.028  66.29
17  2  958.901  0.153  958.911  65.16       03  2  959.612  0.169  959.673  67.35
17  2  958.445  0.178  958.503  66.59       03  2  958.834  0.164  958.832  68.50
17  2  958.901  0.153  958.960  68.14       03  2  959.612  0.169  959.639  69.70
17  2  959.461  0.196  959.572  69.80       03  2  959.152  0.181  959.141  70.97
17  2  959.369  0.198  959.282  71.64       03  2  959.158  0.163  959.160  72.32
17  1  958.709  0.187  958.625  62.41       03  2  958.654  0.164  958.660  73.85
17  1  958.445  0.178  958.517  63.79       03  2  959.165  0.178  959.188  75.33
17  1  958.901  0.153  958.911  65.16       03  2  959.887  0.192  959.874  76.91
17  1  958.445  0.178  958.503  66.59       03  2  958.469  0.177  958.528  78.60
17  1  958.901  0.153  958.960  68.14       03  2  959.612  0.169  959.613  80.39
17  1  959.461  0.196  959.572  69.80       03  1  959.612  0.169  959.673  67.35
17  1  959.369  0.198  959.282  71.64       03  1  958.834  0.164  958.832  68.50
21  2  959.192  0.172  959.018  59.77       03  1  959.612  0.169  959.639  69.70
21  1  959.192  0.172  959.018  59.77       03  1  959.152  0.181  959.141  70.97
21  2  959.197  0.181  959.319  67.31       03  1  959.158  0.163  959.160  72.32
21  2  958.637  0.277  958.300  70.47       03  1  958.654  0.164  958.660  73.85
21  2  959.639  0.178  959.954  72.16       03  1  959.165  0.178  959.188  75.33
21  2  959.197  0.181  959.356  74.00       03  1  959.887  0.192  959.874  76.91
21  2  959.639  0.178  959.724  76.04       03  1  958.469  0.177  958.528  78.60
21  1  959.197  0.181  959.319  67.31       03  1  959.612  0.169  959.613  80.39
21  1  958.637  0.277  958.300  70.47       04  2  959.023  0.199  959.056  63.62
21  1  959.639  0.178  959.954  72.16       04  2  959.023  0.199  959.064  64.54
21  1  959.197  0.181  959.356  74.00       04  2  958.761  0.179  958.783  65.53
21  1  959.639  0.178  959.724  76.04       04  2  958.649  0.175  958.666  66.54
27  2  959.995  0.181  960.072  69.39
```

|    |   | 1998 - AGOSTO |       |         |       |    |   | 1998 - AGOSTO |       |         |       |
|----|---|---------|-------|---------|-------|----|---|---------|-------|---------|-------|
| D  | L | SDB     | ER    | SDC     | HL    | D  | L | SDB     | ER    | SDC     | HL    |
| 04 | 1 | 959.023 | 0.199 | 959.056 | 63.62 | 11 | 2 | 958.842 | 0.214 | 958.891 | 70.75 |
| 04 | 1 | 959.023 | 0.199 | 959.064 | 64.54 | 11 | 2 | 959.427 | 0.171 | 959.346 | 71.93 |
| 04 | 1 | 958.761 | 0.179 | 958.783 | 65.53 | 11 | 2 | 958.952 | 0.191 | 958.938 | 73.16 |
| 04 | 1 | 958.649 | 0.175 | 958.666 | 66.54 | 11 | 2 | 959.131 | 0.201 | 959.199 | 74.50 |
| 04 | 2 | 958.896 | 0.180 | 958.956 | 67.59 | 11 | 2 | 958.666 | 0.173 | 958.658 | 75.90 |
| 04 | 2 | 958.649 | 0.175 | 958.670 | 68.70 | 11 | 2 | 958.714 | 0.209 | 958.745 | 77.38 |
| 04 | 2 | 958.355 | 0.204 | 957.869 | 69.88 | 11 | 2 | 959.131 | 0.201 | 959.192 | 78.90 |
| 04 | 2 | 958.355 | 0.204 | 958.168 | 71.20 | 11 | 1 | 959.427 | 0.171 | 959.285 | 69.53 |
| 04 | 2 | 959.023 | 0.199 | 959.025 | 72.52 | 11 | 1 | 958.842 | 0.214 | 958.891 | 70.75 |
| 04 | 2 | 958.355 | 0.204 | 958.152 | 75.60 | 11 | 1 | 959.427 | 0.171 | 959.346 | 71.93 |
| 04 | 2 | 959.409 | 0.234 | 959.612 | 78.09 | 11 | 1 | 958.952 | 0.191 | 958.938 | 73.16 |
| 04 | 2 | 958.761 | 0.179 | 958.755 | 79.83 | 11 | 1 | 959.131 | 0.201 | 959.199 | 74.50 |
| 04 | 1 | 958.896 | 0.180 | 958.956 | 67.59 | 11 | 1 | 958.666 | 0.173 | 958.658 | 75.90 |
| 04 | 1 | 958.649 | 0.175 | 958.670 | 68.70 | 11 | 1 | 958.714 | 0.209 | 958.745 | 77.38 |
| 04 | 1 | 958.355 | 0.204 | 957.869 | 69.88 | 11 | 1 | 959.131 | 0.201 | 959.192 | 78.90 |
| 04 | 1 | 958.355 | 0.204 | 958.168 | 71.20 | 12 | 2 | 959.278 | 0.197 | 959.577 | 69.12 |
| 04 | 1 | 959.023 | 0.199 | 959.025 | 72.52 | 12 | 2 | 958.580 | 0.163 | 958.613 | 70.33 |
| 04 | 1 | 958.355 | 0.204 | 958.152 | 75.60 | 12 | 2 | 958.827 | 0.252 | 958.894 | 71.50 |
| 04 | 1 | 959.409 | 0.234 | 959.612 | 78.09 | 12 | 2 | 959.278 | 0.197 | 959.449 | 73.40 |
| 04 | 1 | 958.761 | 0.179 | 958.755 | 79.83 | 12 | 2 | 958.827 | 0.252 | 958.765 | 76.40 |
| 05 | 2 | 960.351 | 0.144 | 960.360 | 60.10 | 12 | 2 | 958.580 | 0.163 | 958.682 | 77.89 |
| 05 | 2 | 960.066 | 0.126 | 960.117 | 60.87 | 12 | 2 | 958.827 | 0.252 | 958.774 | 79.45 |
| 05 | 2 | 960.445 | 0.117 | 960.470 | 61.69 | 12 | 2 | 958.827 | 0.252 | 958.740 | 81.09 |
| 05 | 2 | 959.915 | 0.180 | 959.924 | 62.52 | 12 | 2 | 959.007 | 0.299 | 958.976 | 82.54 |
| 05 | 2 | 959.884 | 0.132 | 959.870 | 63.39 | 12 | 2 | 960.010 | 0.183 | 959.961 | 83.31 |
| 05 | 2 | 959.397 | 0.167 | 959.452 | 64.31 | 12 | 1 | 959.278 | 0.197 | 959.577 | 69.12 |
| 05 | 2 | 959.614 | 0.136 | 959.512 | 66.32 | 12 | 1 | 958.580 | 0.163 | 958.613 | 70.33 |
| 05 | 2 | 960.010 | 0.164 | 960.027 | 67.37 | 12 | 1 | 958.827 | 0.252 | 958.894 | 71.50 |
| 05 | 1 | 960.351 | 0.144 | 960.360 | 60.10 | 12 | 1 | 959.278 | 0.197 | 959.449 | 73.40 |
| 05 | 1 | 960.066 | 0.126 | 960.117 | 60.87 | 12 | 1 | 958.827 | 0.252 | 958.765 | 76.40 |
| 05 | 1 | 960.445 | 0.117 | 960.470 | 61.69 | 12 | 1 | 958.580 | 0.163 | 958.682 | 77.89 |
| 05 | 1 | 959.915 | 0.180 | 959.924 | 62.52 | 12 | 1 | 958.827 | 0.252 | 958.774 | 79.45 |
| 05 | 1 | 959.884 | 0.132 | 959.870 | 63.39 | 12 | 1 | 958.827 | 0.252 | 958.740 | 81.09 |
| 05 | 1 | 959.397 | 0.167 | 959.452 | 64.31 | 12 | 1 | 959.007 | 0.299 | 958.976 | 82.54 |
| 05 | 1 | 959.614 | 0.136 | 959.512 | 66.32 | 12 | 1 | 960.010 | 0.183 | 959.961 | 83.31 |
| 05 | 1 | 960.010 | 0.164 | 960.027 | 67.37 | 18 | 2 | 959.331 | 0.173 | 959.574 | 63.06 |
| 05 | 2 | 960.351 | 0.144 | 960.299 | 68.48 | 18 | 2 | 959.193 | 0.211 | 959.223 | 63.81 |
| 05 | 2 | 959.027 | 0.139 | 959.037 | 69.63 | 18 | 2 | 959.331 | 0.173 | 959.406 | 64.63 |
| 05 | 2 | 958.895 | 0.142 | 958.632 | 70.83 | 18 | 1 | 959.331 | 0.173 | 959.574 | 63.06 |
| 05 | 2 | 959.094 | 0.130 | 959.067 | 72.19 | 18 | 1 | 959.193 | 0.211 | 959.223 | 63.81 |
| 05 | 2 | 959.027 | 0.139 | 958.969 | 73.51 | 18 | 1 | 959.331 | 0.173 | 959.406 | 64.63 |
| 05 | 2 | 959.950 | 0.143 | 959.978 | 74.92 | 18 | 2 | 959.193 | 0.211 | 959.106 | 70.03 |
| 05 | 2 | 959.312 | 0.131 | 959.335 | 76.44 | 18 | 2 | 958.560 | 0.145 | 958.532 | 71.09 |
| 05 | 2 | 959.884 | 0.132 | 959.875 | 78.06 | 18 | 2 | 958.687 | 0.166 | 958.659 | 72.18 |
| 05 | 2 | 959.884 | 0.132 | 959.893 | 79.74 | 18 | 2 | 958.949 | 0.179 | 958.959 | 73.36 |
| 05 | 1 | 960.351 | 0.144 | 960.299 | 68.48 | 18 | 2 | 958.466 | 0.190 | 958.504 | 74.59 |
| 05 | 1 | 959.027 | 0.139 | 959.037 | 69.63 | 18 | 2 | 960.161 | 0.237 | 960.049 | 75.85 |
| 05 | 1 | 958.895 | 0.142 | 958.632 | 70.83 | 18 | 2 | 958.687 | 0.166 | 958.676 | 77.14 |
| 05 | 1 | 959.094 | 0.130 | 959.067 | 72.19 | 18 | 2 | 958.293 | 0.198 | 958.373 | 78.52 |
| 05 | 1 | 959.027 | 0.139 | 958.969 | 73.51 | 18 | 2 | 959.331 | 0.173 | 959.347 | 79.90 |
| 05 | 1 | 959.950 | 0.143 | 959.978 | 74.92 | 18 | 2 | 959.269 | 0.196 | 959.284 | 81.25 |
| 05 | 1 | 959.312 | 0.131 | 959.335 | 76.44 | 18 | 2 | 958.233 | 0.191 | 958.060 | 82.43 |
| 05 | 1 | 959.884 | 0.132 | 959.875 | 78.06 | 18 | 2 | 959.920 | 0.223 | 959.764 | 83.15 |
| 05 | 1 | 959.884 | 0.132 | 959.893 | 79.74 | 18 | 1 | 959.193 | 0.211 | 959.106 | 70.03 |
| 11 | 2 | 958.839 | 0.207 | 958.792 | 62.24 | 18 | 1 | 958.560 | 0.145 | 958.532 | 71.09 |
| 11 | 2 | 958.528 | 0.176 | 958.507 | 63.06 | 18 | 1 | 958.687 | 0.166 | 958.659 | 72.18 |
| 11 | 2 | 958.952 | 0.191 | 958.914 | 63.87 | 18 | 1 | 958.949 | 0.179 | 958.959 | 73.36 |
| 11 | 2 | 958.963 | 0.185 | 958.967 | 64.74 | 18 | 1 | 958.466 | 0.190 | 958.504 | 74.59 |
| 11 | 2 | 959.427 | 0.171 | 959.439 | 65.62 | 18 | 1 | 960.161 | 0.237 | 960.049 | 75.85 |
| 11 | 2 | 958.666 | 0.173 | 958.677 | 66.53 | 18 | 1 | 958.687 | 0.166 | 958.676 | 77.14 |
| 11 | 2 | 958.839 | 0.207 | 958.816 | 67.47 | 18 | 1 | 958.293 | 0.198 | 958.373 | 78.52 |
| 11 | 2 | 959.131 | 0.201 | 959.272 | 68.47 | 18 | 1 | 959.331 | 0.173 | 959.347 | 79.90 |
| 11 | 1 | 958.839 | 0.207 | 958.792 | 62.24 | 18 | 1 | 959.269 | 0.196 | 959.284 | 81.25 |
| 11 | 1 | 958.528 | 0.176 | 958.507 | 63.06 | 18 | 1 | 958.233 | 0.191 | 958.060 | 82.43 |
| 11 | 1 | 958.952 | 0.191 | 958.914 | 63.87 | 18 | 1 | 959.920 | 0.223 | 959.764 | 83.15 |
| 11 | 1 | 958.963 | 0.185 | 958.967 | 64.74 | 19 | 2 | 959.412 | 0.151 | 959.426 | 63.76 |
| 11 | 1 | 959.427 | 0.171 | 959.439 | 65.62 | 19 | 2 | 959.350 | 0.103 | 959.335 | 64.57 |
| 11 | 1 | 958.666 | 0.173 | 958.677 | 66.53 | 19 | 2 | 959.350 | 0.103 | 959.316 | 65.40 |
| 11 | 1 | 958.839 | 0.207 | 958.816 | 67.47 | 19 | 2 | 959.158 | 0.138 | 958.696 | 66.31 |
| 11 | 1 | 959.131 | 0.201 | 959.272 | 68.47 | 19 | 2 | 960.046 | 0.128 | 960.057 | 67.20 |
| 11 | 2 | 959.427 | 0.171 | 959.285 | 69.53 | 19 | 2 | 959.377 | 0.131 | 959.391 | 68.24 |

```
         1998 - AGOSTO                              1998 - AGOSTO
 D  L   SDB     ER     SDC     HL          D  L   SDB     ER     SDC     HL
19  2  959.412  0.151  959.414  69.32      21  1  958.717  0.169  958.684  67.24
19  1  959.412  0.151  959.426  63.76      21  1  959.038  0.171  959.048  69.33
19  1  959.350  0.103  959.335  64.57      21  1  959.318  0.197  959.394  70.37
19  1  959.350  0.103  959.316  65.40      21  2  958.633  0.164  958.626  71.44
19  1  959.158  0.138  958.696  66.31      21  2  959.038  0.171  958.998  72.56
19  1  960.046  0.128  960.057  67.20      21  2  959.318  0.197  959.321  73.75
19  1  959.377  0.131  959.391  68.24      21  2  958.598  0.187  958.597  74.97
19  1  959.412  0.151  959.414  69.32      21  2  958.775  0.159  958.767  76.25
19  2  960.046  0.128  960.056  70.34      21  1  958.633  0.164  958.626  71.44
19  2  959.974  0.128  959.875  71.43      21  1  959.038  0.171  958.998  72.56
19  2  959.209  0.121  959.254  72.59      21  1  959.318  0.197  959.321  73.75
19  2  959.209  0.121  959.249  73.88      21  1  958.598  0.187  958.597  74.97
19  2  959.578  0.137  959.630  75.21      21  1  958.775  0.159  958.767  76.25
19  2  960.104  0.175  960.090  76.91      24  2  958.764  0.175  958.754  62.81
19  2  960.152  0.204  960.150  78.34      24  2  959.849  0.151  959.879  63.57
19  2  960.152  0.204  960.183  79.78      24  2  958.764  0.175  958.695  64.32
19  1  960.046  0.128  960.056  70.34      24  2  958.375  0.158  958.375  65.09
19  1  959.974  0.128  959.875  71.43      24  2  958.356  0.116  958.322  65.91
19  1  959.209  0.121  959.254  72.59      24  2  958.796  0.130  958.852  66.75
19  1  959.209  0.121  959.249  73.88      24  2  958.095  0.184  958.017  68.03
19  1  959.578  0.137  959.630  75.21      24  2  958.950  0.164  958.945  68.95
19  1  960.104  0.175  960.090  76.91      24  2  959.089  0.195  959.087  69.96
19  1  960.152  0.204  960.150  78.34      24  1  958.764  0.175  958.754  62.81
19  1  960.152  0.204  960.183  79.78      24  1  959.849  0.151  959.879  63.57
20  2  959.618  0.183  959.647  63.13      24  1  958.764  0.175  958.695  64.32
20  2  959.945  0.231  960.431  63.86      24  1  958.375  0.158  958.375  65.09
20  2  959.945  0.231  960.320  64.61      24  1  958.356  0.116  958.322  65.91
20  2  959.945  0.231  960.003  65.40      24  1  958.796  0.130  958.852  66.75
20  2  959.075  0.148  959.047  67.98      24  1  958.095  0.184  958.017  68.03
20  2  958.653  0.146  958.655  68.92      24  1  958.950  0.164  958.945  68.95
20  2  959.113  0.232  959.186  69.87      24  1  959.089  0.195  959.087  69.96
20  1  959.618  0.183  959.647  63.13      24  2  958.370  0.136  958.367  71.05
20  1  959.945  0.231  960.431  63.86      24  2  959.339  0.145  959.228  72.10
20  1  959.945  0.231  960.320  64.61      24  2  959.339  0.145  959.405  73.25
20  1  959.945  0.231  960.003  65.40      24  2  958.386  0.169  958.396  74.43
20  1  959.075  0.148  959.047  67.98      24  2  959.098  0.155  959.162  75.71
20  1  958.653  0.146  958.655  68.92      24  2  958.356  0.116  958.339  76.99
20  1  959.113  0.232  959.186  69.87      24  1  958.370  0.136  958.367  71.05
20  2  958.997  0.237  958.972  70.88      24  1  959.339  0.145  959.228  72.10
20  2  959.730  0.237  959.722  71.94      24  1  959.339  0.145  959.405  73.25
20  2  959.914  0.237  959.871  73.05      24  1  958.386  0.169  958.396  74.43
20  2  959.618  0.183  959.656  74.46      24  1  959.098  0.155  959.162  75.71
20  1  958.997  0.237  958.972  70.88      24  1  958.356  0.116  958.339  76.99
20  1  959.730  0.237  959.722  71.94      25  2  958.729  0.225  958.689  61.71
20  1  959.914  0.237  959.871  73.05      25  2  959.334  0.158  959.329  62.37
20  1  959.618  0.183  959.656  74.46      25  2  959.668  0.186  959.563  63.04
21  2  958.717  0.169  958.687  57.97      25  2  959.323  0.158  959.251  63.72
21  2  958.484  0.168  958.392  58.56      25  2  958.266  0.158  958.032  64.44
21  2  958.484  0.168  958.484  59.18      25  2  959.949  0.153  959.853  65.18
21  2  959.038  0.171  958.988  59.80      25  2  958.729  0.225  958.643  66.05
21  2  958.484  0.168  957.498  60.43      25  2  958.461  0.174  958.373  66.87
21  2  958.934  0.164  958.951  61.09      25  2  959.120  0.160  959.124  67.71
21  2  959.136  0.162  959.109  61.80      25  2  959.166  0.167  959.183  68.59
21  2  958.598  0.187  958.597  62.51      25  2  959.034  0.150  959.069  69.50
21  2  958.717  0.169  958.682  63.27      25  2  958.885  0.167  958.881  70.45
21  2  959.139  0.172  959.149  64.03      25  1  958.729  0.225  958.689  61.71
21  2  959.294  0.171  959.265  64.84      25  1  959.334  0.158  959.329  62.37
21  2  958.717  0.169  958.684  67.24      25  1  959.668  0.186  959.563  63.04
21  2  959.038  0.171  959.048  69.33      25  1  959.323  0.158  959.251  63.72
21  2  959.318  0.197  959.394  70.37      25  1  958.266  0.158  958.032  64.44
21  1  958.717  0.169  958.687  57.97      25  1  959.949  0.153  959.853  65.18
21  1  958.484  0.168  958.392  58.56      25  1  958.729  0.225  958.643  66.05
21  1  958.484  0.168  958.484  59.18      25  1  958.461  0.174  958.373  66.87
21  1  959.038  0.171  958.988  59.80      25  1  959.120  0.160  959.124  67.71
21  1  958.484  0.168  957.498  60.43      25  1  959.166  0.167  959.183  68.59
21  1  958.934  0.164  958.951  61.09      25  1  959.034  0.150  959.069  69.50
21  1  959.136  0.162  959.109  61.80      25  1  958.885  0.167  958.881  70.45
21  1  958.598  0.187  958.597  62.51      25  2  958.929  0.133  958.920  71.43
21  1  958.717  0.169  958.682  63.27      25  2  959.034  0.150  958.998  72.46
21  1  959.139  0.172  959.149  64.03      25  2  959.334  0.158  959.331  73.54
21  1  959.294  0.171  959.265  64.84      25  2  958.461  0.174  958.429  74.68
```

| 1998 - AGOSTO | | | | |
|---|---|---|---|---|
| D | L | SDB | ER | SDC | HL |
| 25 | 2 | 959.166 | 0.167 | 959.180 | 75.86 |
| 25 | 2 | 959.705 | 0.176 | 959.723 | 77.08 |
| 25 | 1 | 958.929 | 0.133 | 958.920 | 71.43 |
| 25 | 1 | 959.034 | 0.150 | 958.998 | 72.46 |
| 25 | 1 | 959.334 | 0.158 | 959.331 | 73.54 |
| 25 | 1 | 958.461 | 0.174 | 958.429 | 74.68 |
| 25 | 1 | 959.166 | 0.167 | 959.180 | 75.86 |
| 25 | 1 | 959.705 | 0.176 | 959.723 | 77.08 |
| 26 | 2 | 958.816 | 0.123 | 958.777 | 62.80 |
| 26 | 2 | 959.267 | 0.110 | 959.301 | 63.54 |
| 26 | 2 | 959.751 | 0.165 | 959.647 | 64.39 |
| 26 | 2 | 959.202 | 0.135 | 959.167 | 65.16 |
| 26 | 2 | 958.555 | 0.123 | 958.454 | 65.96 |
| 26 | 2 | 959.751 | 0.165 | 959.757 | 66.94 |
| 26 | 2 | 958.555 | 0.123 | 958.392 | 67.87 |
| 26 | 2 | 958.907 | 0.150 | 958.914 | 68.79 |
| 26 | 2 | 959.751 | 0.165 | 959.773 | 69.93 |
| 26 | 2 | 959.372 | 0.130 | 959.375 | 71.01 |
| 26 | 1 | 958.816 | 0.123 | 958.777 | 62.80 |
| 26 | 1 | 959.267 | 0.110 | 959.301 | 63.54 |
| 26 | 1 | 959.751 | 0.165 | 959.647 | 64.39 |
| 26 | 1 | 959.202 | 0.135 | 959.167 | 65.16 |
| 26 | 1 | 958.555 | 0.123 | 958.454 | 65.96 |
| 26 | 1 | 959.751 | 0.165 | 959.757 | 66.94 |
| 26 | 1 | 958.555 | 0.123 | 958.392 | 67.87 |
| 26 | 1 | 958.907 | 0.150 | 958.914 | 68.79 |
| 26 | 1 | 959.751 | 0.165 | 959.773 | 69.93 |
| 26 | 1 | 959.372 | 0.130 | 959.375 | 71.01 |
| 26 | 2 | 958.979 | 0.164 | 958.977 | 72.06 |
| 26 | 2 | 958.816 | 0.123 | 958.794 | 73.16 |
| 26 | 2 | 959.499 | 0.141 | 959.599 | 74.38 |
| 26 | 2 | 959.372 | 0.130 | 959.375 | 75.63 |
| 26 | 2 | 959.202 | 0.135 | 959.177 | 76.90 |
| 26 | 2 | 959.493 | 0.121 | 959.454 | 78.26 |
| 26 | 1 | 958.979 | 0.164 | 958.977 | 72.06 |
| 26 | 1 | 958.816 | 0.123 | 958.794 | 73.16 |
| 26 | 1 | 959.499 | 0.141 | 959.599 | 74.38 |
| 26 | 1 | 959.372 | 0.130 | 959.375 | 75.63 |
| 26 | 1 | 959.202 | 0.135 | 959.177 | 76.90 |
| 26 | 1 | 959.493 | 0.121 | 959.454 | 78.26 |
| 27 | 2 | 958.983 | 0.153 | 958.963 | 61.20 |
| 27 | 2 | 959.325 | 0.125 | 959.354 | 61.81 |
| 27 | 2 | 959.096 | 0.129 | 959.139 | 62.43 |
| 27 | 2 | 959.325 | 0.125 | 959.310 | 63.07 |
| 27 | 2 | 959.096 | 0.129 | 959.139 | 63.74 |
| 27 | 2 | 959.721 | 0.128 | 959.789 | 64.43 |
| 27 | 2 | 959.325 | 0.125 | 959.365 | 65.15 |
| 27 | 2 | 958.983 | 0.153 | 958.923 | 66.04 |
| 27 | 2 | 958.998 | 0.198 | 959.026 | 66.82 |
| 27 | 2 | 959.721 | 0.128 | 960.054 | 67.63 |
| 27 | 2 | 958.998 | 0.198 | 959.019 | 68.49 |
| 27 | 2 | 959.721 | 0.128 | 960.058 | 69.37 |
| 27 | 2 | 958.998 | 0.198 | 959.008 | 70.28 |
| 27 | 1 | 958.983 | 0.153 | 958.963 | 61.20 |
| 27 | 1 | 959.325 | 0.125 | 959.354 | 61.81 |
| 27 | 1 | 959.096 | 0.129 | 959.139 | 62.43 |
| 27 | 1 | 959.325 | 0.125 | 959.310 | 63.07 |
| 27 | 1 | 959.096 | 0.129 | 959.139 | 63.74 |
| 27 | 1 | 959.721 | 0.128 | 959.789 | 64.43 |
| 27 | 1 | 959.325 | 0.125 | 959.365 | 65.15 |
| 27 | 1 | 958.983 | 0.153 | 958.923 | 66.04 |
| 27 | 1 | 958.998 | 0.198 | 959.026 | 66.82 |
| 27 | 1 | 959.721 | 0.128 | 960.054 | 67.63 |
| 27 | 1 | 958.998 | 0.198 | 959.019 | 68.49 |
| 27 | 1 | 959.721 | 0.128 | 960.058 | 69.37 |
| 27 | 1 | 958.998 | 0.198 | 959.008 | 70.28 |
| 27 | 2 | 958.786 | 0.142 | 958.632 | 71.24 |
| 27 | 2 | 959.056 | 0.129 | 959.051 | 72.31 |
| 27 | 2 | 959.721 | 0.128 | 959.847 | 73.35 |
| 27 | 2 | 959.096 | 0.129 | 959.077 | 74.46 |
| 27 | 1 | 958.786 | 0.142 | 958.632 | 71.24 |

| 1998 - AGOSTO | | | | |
|---|---|---|---|---|
| D | L | SDB | ER | SDC | HL |
| 27 | 1 | 959.056 | 0.129 | 959.051 | 72.31 |
| 27 | 1 | 959.721 | 0.128 | 959.847 | 73.35 |
| 27 | 1 | 959.096 | 0.129 | 959.077 | 74.46 |

| 1998 - SETEMBRO | | | | |
|---|---|---|---|---|
| D | L | SDB | ER | SDC | HL |
| 01 | 2 | 958.648 | 0.196 | 958.689 | 60.70 |
| 01 | 2 | 959.026 | 0.198 | 959.235 | 61.26 |
| 01 | 2 | 959.026 | 0.198 | 958.987 | 62.44 |
| 01 | 2 | 957.949 | 0.170 | 957.934 | 63.05 |
| 01 | 2 | 959.613 | 0.157 | 959.611 | 63.69 |
| 01 | 2 | 958.525 | 0.209 | 958.466 | 64.48 |
| 01 | 2 | 959.026 | 0.198 | 959.129 | 65.16 |
| 01 | 2 | 959.502 | 0.173 | 959.460 | 67.30 |
| 01 | 2 | 958.525 | 0.209 | 958.472 | 68.12 |
| 01 | 2 | 958.800 | 0.226 | 958.811 | 70.59 |
| 01 | 1 | 958.648 | 0.196 | 958.689 | 60.70 |
| 01 | 1 | 959.026 | 0.198 | 959.235 | 61.26 |
| 01 | 1 | 959.026 | 0.198 | 958.987 | 62.44 |
| 01 | 1 | 957.949 | 0.170 | 957.934 | 63.05 |
| 01 | 1 | 959.613 | 0.157 | 959.611 | 63.69 |
| 01 | 1 | 958.525 | 0.209 | 958.466 | 64.48 |
| 01 | 1 | 959.026 | 0.198 | 959.129 | 65.16 |
| 01 | 1 | 959.502 | 0.173 | 959.460 | 67.30 |
| 01 | 1 | 958.525 | 0.209 | 958.472 | 68.12 |
| 01 | 1 | 958.800 | 0.226 | 958.811 | 70.59 |
| 01 | 2 | 958.648 | 0.196 | 958.634 | 72.48 |
| 01 | 2 | 958.648 | 0.196 | 958.660 | 73.51 |
| 01 | 2 | 959.667 | 0.193 | 960.535 | 74.71 |
| 01 | 2 | 958.925 | 0.177 | 958.968 | 75.88 |
| 01 | 2 | 957.949 | 0.170 | 958.112 | 77.04 |
| 01 | 1 | 958.648 | 0.196 | 958.634 | 72.48 |
| 01 | 1 | 958.648 | 0.196 | 958.660 | 73.51 |
| 01 | 1 | 959.667 | 0.193 | 960.535 | 74.71 |
| 01 | 1 | 958.925 | 0.177 | 958.968 | 75.88 |
| 01 | 1 | 957.949 | 0.170 | 958.112 | 77.04 |
| 02 | 2 | 959.576 | 0.181 | 959.597 | 62.64 |
| 02 | 2 | 960.042 | 0.170 | 960.035 | 63.30 |
| 02 | 2 | 960.352 | 0.154 | 960.226 | 64.09 |
| 02 | 2 | 959.357 | 0.188 | 959.282 | 64.80 |
| 02 | 2 | 959.571 | 0.215 | 959.562 | 65.70 |
| 02 | 2 | 958.525 | 0.172 | 958.454 | 66.49 |
| 02 | 2 | 958.889 | 0.196 | 958.983 | 67.39 |
| 02 | 2 | 958.817 | 0.190 | 958.818 | 68.30 |
| 02 | 2 | 959.357 | 0.188 | 959.341 | 69.23 |
| 02 | 2 | 959.415 | 0.199 | 959.400 | 70.19 |
| 02 | 2 | 959.094 | 0.199 | 959.074 | 71.23 |
| 02 | 1 | 959.576 | 0.181 | 959.597 | 62.64 |
| 02 | 1 | 960.042 | 0.170 | 960.035 | 63.30 |
| 02 | 1 | 960.352 | 0.154 | 960.226 | 64.09 |
| 02 | 1 | 959.357 | 0.188 | 959.282 | 64.80 |
| 02 | 1 | 959.571 | 0.215 | 959.562 | 65.70 |
| 02 | 1 | 958.525 | 0.172 | 958.454 | 66.49 |
| 02 | 1 | 958.889 | 0.196 | 958.983 | 67.39 |
| 02 | 1 | 958.817 | 0.190 | 958.818 | 68.30 |
| 02 | 1 | 959.357 | 0.188 | 959.341 | 69.23 |
| 02 | 1 | 959.415 | 0.199 | 959.400 | 70.19 |
| 02 | 1 | 959.094 | 0.199 | 959.074 | 71.23 |
| 02 | 2 | 959.651 | 0.197 | 959.643 | 72.30 |
| 02 | 2 | 958.676 | 0.214 | 958.679 | 73.41 |
| 02 | 2 | 959.651 | 0.197 | 959.646 | 74.60 |
| 02 | 2 | 959.548 | 0.180 | 959.555 | 75.81 |
| 02 | 2 | 959.548 | 0.180 | 959.548 | 77.13 |
| 02 | 2 | 958.817 | 0.190 | 958.818 | 78.48 |
| 02 | 1 | 959.651 | 0.197 | 959.643 | 72.30 |
| 02 | 1 | 958.676 | 0.214 | 958.679 | 73.41 |
| 02 | 1 | 959.651 | 0.197 | 959.646 | 74.60 |
| 02 | 1 | 959.548 | 0.180 | 959.555 | 75.81 |
| 02 | 1 | 959.548 | 0.180 | 959.548 | 77.13 |
| 02 | 1 | 958.817 | 0.190 | 958.818 | 78.48 |

| 1998 - SETEMBRO | | | | | | 1998 - SETEMBRO | | | | |
|---|---|---|---|---|---|---|---|---|---|---|
| D | L | SDB | ER | SDC | HL | D | L | SDB | ER | SDC | HL |
| 03 | 2 | 958.966 | 0.170 | 958.972 | 63.53 | 18 | 2 | 959.577 | 0.121 | 959.551 | 62.48 |
| 03 | 2 | 958.624 | 0.168 | 958.562 | 64.18 | 18 | 2 | 959.754 | 0.105 | 959.743 | 63.10 |
| 03 | 2 | 959.263 | 0.214 | 959.300 | 64.88 | 18 | 2 | 959.066 | 0.115 | 958.944 | 63.75 |
| 03 | 2 | 959.263 | 0.214 | 959.229 | 65.59 | 18 | 2 | 959.118 | 0.208 | 959.144 | 64.46 |
| 03 | 2 | 959.526 | 0.208 | 960.172 | 66.32 | 18 | 2 | 959.118 | 0.208 | 959.194 | 65.15 |
| 03 | 2 | 959.443 | 0.160 | 959.435 | 67.08 | 18 | 2 | 959.118 | 0.208 | 959.114 | 65.89 |
| 03 | 2 | 959.263 | 0.214 | 959.262 | 67.86 | 18 | 2 | 959.118 | 0.208 | 959.163 | 66.65 |
| 03 | 2 | 959.263 | 0.214 | 959.275 | 68.67 | 18 | 2 | 959.577 | 0.121 | 959.598 | 67.46 |
| 03 | 2 | 958.624 | 0.168 | 958.532 | 69.52 | 18 | 2 | 959.577 | 0.121 | 959.537 | 68.31 |
| 03 | 2 | 958.219 | 0.174 | 958.155 | 70.42 | 18 | 2 | 959.823 | 0.164 | 960.045 | 69.20 |
| 03 | 2 | 958.928 | 0.161 | 958.925 | 71.35 | 18 | 2 | 959.823 | 0.164 | 959.922 | 70.13 |
| 03 | 1 | 958.966 | 0.170 | 958.972 | 63.53 | 18 | 1 | 959.393 | 0.119 | 959.348 | 60.77 |
| 03 | 1 | 958.624 | 0.168 | 958.562 | 64.18 | 18 | 1 | 959.066 | 0.115 | 959.018 | 61.32 |
| 03 | 1 | 959.263 | 0.214 | 959.300 | 64.88 | 18 | 1 | 959.429 | 0.102 | 959.422 | 61.90 |
| 03 | 1 | 959.263 | 0.214 | 959.229 | 65.59 | 18 | 1 | 959.577 | 0.121 | 959.551 | 62.48 |
| 03 | 1 | 959.526 | 0.208 | 960.172 | 66.32 | 18 | 1 | 959.754 | 0.105 | 959.743 | 63.10 |
| 03 | 1 | 959.443 | 0.160 | 959.435 | 67.08 | 18 | 1 | 959.066 | 0.115 | 958.944 | 63.75 |
| 03 | 1 | 959.263 | 0.214 | 959.262 | 67.86 | 18 | 1 | 959.118 | 0.208 | 959.144 | 64.46 |
| 03 | 1 | 959.263 | 0.214 | 959.275 | 68.67 | 18 | 1 | 959.118 | 0.208 | 959.194 | 65.15 |
| 03 | 1 | 958.624 | 0.168 | 958.532 | 69.52 | 18 | 1 | 959.118 | 0.208 | 959.114 | 65.89 |
| 03 | 1 | 958.219 | 0.174 | 958.155 | 70.42 | 18 | 1 | 959.118 | 0.208 | 959.163 | 66.65 |
| 03 | 1 | 958.928 | 0.161 | 958.925 | 71.35 | 18 | 1 | 959.577 | 0.121 | 959.598 | 67.46 |
| 03 | 2 | 958.219 | 0.174 | 958.152 | 72.32 | 18 | 1 | 959.577 | 0.121 | 959.537 | 68.31 |
| 03 | 2 | 959.093 | 0.154 | 959.068 | 73.32 | 18 | 1 | 959.823 | 0.164 | 960.045 | 69.20 |
| 03 | 2 | 958.966 | 0.170 | 959.004 | 74.37 | 18 | 1 | 959.823 | 0.164 | 959.922 | 70.13 |
| 03 | 2 | 958.624 | 0.168 | 958.688 | 75.48 | 18 | 2 | 959.754 | 0.105 | 959.738 | 72.18 |
| 03 | 2 | 959.526 | 0.208 | 959.756 | 76.62 | 18 | 2 | 959.118 | 0.208 | 959.143 | 73.27 |
| 03 | 2 | 958.928 | 0.161 | 958.911 | 79.03 | 18 | 2 | 959.442 | 0.227 | 959.459 | 74.39 |
| 03 | 1 | 958.219 | 0.174 | 958.152 | 72.32 | 18 | 2 | 959.066 | 0.115 | 958.984 | 75.58 |
| 03 | 1 | 959.093 | 0.154 | 959.068 | 73.32 | 18 | 2 | 958.798 | 0.112 | 958.813 | 76.82 |
| 03 | 1 | 958.966 | 0.170 | 959.004 | 74.37 | 18 | 1 | 959.754 | 0.105 | 959.738 | 72.18 |
| 03 | 1 | 958.624 | 0.168 | 958.688 | 75.48 | 18 | 1 | 959.118 | 0.208 | 959.143 | 73.27 |
| 03 | 1 | 959.526 | 0.208 | 959.756 | 76.62 | 18 | 1 | 959.442 | 0.227 | 959.459 | 74.39 |
| 03 | 1 | 958.928 | 0.161 | 958.911 | 79.03 | 18 | 1 | 959.066 | 0.115 | 958.984 | 75.58 |
| 17 | 2 | 958.879 | 0.160 | 958.865 | 61.85 | 18 | 1 | 958.798 | 0.112 | 958.813 | 76.82 |
| 17 | 2 | 959.863 | 0.181 | 959.804 | 62.40 | 23 | 2 | 958.845 | 0.172 | 958.718 | 59.82 |
| 17 | 2 | 959.581 | 0.157 | 959.628 | 62.96 | 23 | 2 | 959.795 | 0.157 | 959.900 | 60.35 |
| 17 | 2 | 959.117 | 0.222 | 959.127 | 63.55 | 23 | 2 | 959.077 | 0.180 | 959.162 | 60.87 |
| 17 | 2 | 959.190 | 0.177 | 959.167 | 64.16 | 23 | 2 | 958.123 | 0.175 | 958.418 | 61.44 |
| 17 | 2 | 957.969 | 0.188 | 957.968 | 64.80 | 23 | 1 | 958.845 | 0.172 | 958.718 | 59.82 |
| 17 | 2 | 959.581 | 0.157 | 959.706 | 66.17 | 23 | 1 | 959.795 | 0.157 | 959.900 | 60.35 |
| 17 | 2 | 958.529 | 0.204 | 958.393 | 66.92 | 23 | 1 | 959.077 | 0.180 | 959.162 | 60.87 |
| 17 | 2 | 959.002 | 0.197 | 959.007 | 67.70 | 23 | 1 | 958.123 | 0.175 | 958.418 | 61.44 |
| 17 | 2 | 958.529 | 0.204 | 958.659 | 68.50 | 24 | 2 | 959.137 | 0.138 | 959.080 | 59.80 |
| 17 | 2 | 959.117 | 0.222 | 959.148 | 69.32 | 24 | 2 | 959.464 | 0.139 | 959.455 | 60.30 |
| 17 | 2 | 958.847 | 0.155 | 958.842 | 70.20 | 24 | 2 | 958.887 | 0.147 | 958.897 | 60.83 |
| 17 | 2 | 959.892 | 0.217 | 960.002 | 71.09 | 24 | 2 | 959.137 | 0.138 | 959.122 | 61.39 |
| 17 | 1 | 958.879 | 0.160 | 958.865 | 61.85 | 24 | 2 | 958.887 | 0.147 | 958.894 | 61.95 |
| 17 | 1 | 959.863 | 0.181 | 959.804 | 62.40 | 24 | 2 | 959.536 | 0.163 | 959.518 | 62.57 |
| 17 | 1 | 959.581 | 0.157 | 959.628 | 62.96 | 24 | 2 | 959.997 | 0.175 | 959.945 | 63.20 |
| 17 | 1 | 959.117 | 0.222 | 959.127 | 63.55 | 24 | 2 | 959.544 | 0.160 | 959.557 | 63.86 |
| 17 | 1 | 959.190 | 0.177 | 959.167 | 64.16 | 24 | 2 | 958.582 | 0.160 | 958.541 | 65.38 |
| 17 | 1 | 957.969 | 0.188 | 957.968 | 64.80 | 24 | 2 | 959.997 | 0.175 | 959.957 | 66.17 |
| 17 | 1 | 959.581 | 0.157 | 959.706 | 66.17 | 24 | 2 | 959.015 | 0.163 | 959.072 | 69.37 |
| 17 | 1 | 958.529 | 0.204 | 958.393 | 66.92 | 24 | 1 | 959.137 | 0.138 | 959.080 | 59.80 |
| 17 | 1 | 959.002 | 0.197 | 959.007 | 67.70 | 24 | 1 | 959.464 | 0.139 | 959.455 | 60.30 |
| 17 | 1 | 958.529 | 0.204 | 958.659 | 68.50 | 24 | 1 | 958.887 | 0.147 | 958.897 | 60.83 |
| 17 | 1 | 959.117 | 0.222 | 959.148 | 69.32 | 24 | 1 | 959.137 | 0.138 | 959.122 | 61.39 |
| 17 | 1 | 958.847 | 0.155 | 958.842 | 70.20 | 24 | 1 | 958.887 | 0.147 | 958.894 | 61.95 |
| 17 | 1 | 959.892 | 0.217 | 960.002 | 71.09 | 24 | 1 | 959.536 | 0.163 | 959.518 | 62.57 |
| 17 | 2 | 959.400 | 0.210 | 959.320 | 72.06 | 24 | 1 | 959.997 | 0.175 | 959.945 | 63.20 |
| 17 | 2 | 959.535 | 0.151 | 959.533 | 73.09 | 24 | 1 | 959.544 | 0.160 | 959.557 | 63.86 |
| 17 | 2 | 960.241 | 0.176 | 960.451 | 74.14 | 24 | 1 | 958.582 | 0.160 | 958.541 | 65.38 |
| 17 | 2 | 959.892 | 0.217 | 959.901 | 75.23 | 24 | 1 | 959.997 | 0.175 | 959.957 | 66.17 |
| 17 | 1 | 959.400 | 0.210 | 959.320 | 72.06 | 24 | 1 | 959.015 | 0.163 | 959.072 | 69.37 |
| 17 | 1 | 959.535 | 0.151 | 959.533 | 73.09 | 24 | 2 | 958.192 | 0.266 | 958.243 | 70.59 |
| 17 | 1 | 960.241 | 0.176 | 960.451 | 74.14 | 24 | 2 | 959.193 | 0.235 | 959.219 | 71.66 |
| 17 | 1 | 959.892 | 0.217 | 959.901 | 75.23 | 24 | 2 | 958.887 | 0.147 | 958.891 | 72.75 |
| 18 | 2 | 959.393 | 0.119 | 959.348 | 60.77 | 24 | 2 | 958.813 | 0.282 | 958.709 | 75.40 |
| 18 | 2 | 959.066 | 0.115 | 959.018 | 61.32 | 24 | 2 | 959.579 | 0.150 | 959.626 | 76.71 |
| 18 | 2 | 959.429 | 0.102 | 959.422 | 61.90 | 24 | 1 | 958.192 | 0.266 | 958.243 | 70.59 |

| 1998 - SETEMBRO | | | | | | 1998 - OUTUBRO | | | | |
|---|---|---|---|---|---|---|---|---|---|---|
| D | L | SDB | ER | SDC | HL | D | L | SDB | ER | SDC | HL |
| 24 | 1 | 959.193 | 0.235 | 959.219 | 71.66 | 06 | 1 | 959.405 | 0.163 | 959.375 | 73.08 |
| 24 | 1 | 958.887 | 0.147 | 958.891 | 72.75 | 16 | 2 | 959.323 | 0.214 | 959.089 | 51.87 |
| 24 | 1 | 958.813 | 0.282 | 958.709 | 75.40 | 16 | 2 | 959.323 | 0.214 | 959.211 | 52.22 |
| 24 | 1 | 959.579 | 0.150 | 959.626 | 76.71 | 16 | 2 | 959.323 | 0.214 | 959.111 | 52.56 |
| 25 | 2 | 959.170 | 0.192 | 959.225 | 63.54 | 16 | 2 | 958.845 | 0.292 | 957.788 | 52.93 |
| 25 | 2 | 960.561 | 0.248 | 960.588 | 64.19 | 16 | 2 | 959.323 | 0.214 | 959.173 | 53.28 |
| 25 | 2 | 959.339 | 0.238 | 959.290 | 64.87 | 16 | 1 | 959.323 | 0.214 | 959.089 | 51.87 |
| 25 | 2 | 959.369 | 0.464 | 959.713 | 65.57 | 16 | 1 | 959.323 | 0.214 | 959.211 | 52.22 |
| 25 | 2 | 959.170 | 0.192 | 959.171 | 66.32 | 16 | 1 | 959.323 | 0.214 | 959.111 | 52.56 |
| 25 | 2 | 959.079 | 0.212 | 959.015 | 67.08 | 16 | 1 | 958.845 | 0.292 | 957.788 | 52.93 |
| 25 | 2 | 958.672 | 0.152 | 958.639 | 68.58 | 16 | 1 | 959.323 | 0.214 | 959.173 | 53.28 |
| 25 | 2 | 959.156 | 0.242 | 959.123 | 69.48 | 21 | 2 | 959.599 | 0.164 | 959.516 | 47.95 |
| 25 | 1 | 959.170 | 0.192 | 959.225 | 63.54 | 21 | 2 | 959.048 | 0.147 | 958.993 | 48.12 |
| 25 | 1 | 960.561 | 0.248 | 960.588 | 64.19 | 21 | 2 | 959.048 | 0.147 | 958.996 | 48.32 |
| 25 | 1 | 959.339 | 0.238 | 959.290 | 64.87 | 21 | 2 | 959.599 | 0.164 | 959.572 | 48.60 |
| 25 | 1 | 959.369 | 0.464 | 959.713 | 65.57 | 21 | 2 | 959.826 | 0.301 | 959.874 | 48.86 |
| 25 | 1 | 959.170 | 0.192 | 959.171 | 66.32 | 21 | 2 | 959.048 | 0.147 | 959.025 | 49.13 |
| 25 | 1 | 959.079 | 0.212 | 959.015 | 67.08 | 21 | 2 | 957.822 | 0.282 | 957.280 | 49.40 |
| 25 | 1 | 958.672 | 0.152 | 958.639 | 68.58 | 21 | 2 | 959.048 | 0.147 | 959.011 | 49.74 |
| 25 | 1 | 959.156 | 0.242 | 959.123 | 69.48 | 21 | 2 | 958.823 | 0.319 | 958.778 | 50.10 |
| 25 | 2 | 959.369 | 0.464 | 959.656 | 70.41 | 21 | 2 | 958.281 | 0.228 | 958.247 | 50.42 |
| 25 | 2 | 959.156 | 0.242 | 959.134 | 71.42 | 21 | 2 | 958.640 | 0.191 | 958.662 | 51.21 |
| 25 | 1 | 959.369 | 0.464 | 959.656 | 70.41 | 21 | 2 | 958.421 | 0.149 | 958.419 | 51.58 |
| 25 | 1 | 959.156 | 0.242 | 959.134 | 71.42 | 21 | 2 | 959.353 | 0.162 | 959.443 | 52.01 |
| | | | | | | 21 | 2 | 958.421 | 0.149 | 958.521 | 52.44 |
| | | | | | | 21 | 2 | 959.923 | 0.160 | 960.048 | 53.61 |
| | | 1998 - OUTUBRO | | | | 21 | 2 | 958.352 | 0.263 | 958.355 | 54.10 |
| D | L | SDB | ER | SDC | HL | 21 | 2 | 959.923 | 0.160 | 960.004 | 54.65 |
| 05 | 2 | 957.904 | 0.205 | 958.185 | 63.35 | 21 | 2 | 958.281 | 0.228 | 958.252 | 55.30 |
| 05 | 2 | 960.173 | 0.115 | 960.250 | 64.12 | 21 | 2 | 958.421 | 0.149 | 958.521 | 55.98 |
| 05 | 2 | 958.813 | 0.280 | 958.792 | 64.98 | 21 | 1 | 959.599 | 0.164 | 959.516 | 47.95 |
| 05 | 2 | 958.813 | 0.280 | 959.161 | 66.04 | 21 | 1 | 959.048 | 0.147 | 958.993 | 48.12 |
| 05 | 1 | 957.904 | 0.205 | 958.185 | 63.35 | 21 | 1 | 959.048 | 0.147 | 958.996 | 48.32 |
| 05 | 1 | 960.173 | 0.115 | 960.250 | 64.12 | 21 | 1 | 959.599 | 0.164 | 959.572 | 48.60 |
| 05 | 1 | 958.813 | 0.280 | 958.792 | 64.98 | 21 | 1 | 959.826 | 0.301 | 959.874 | 48.86 |
| 05 | 1 | 958.813 | 0.280 | 959.161 | 66.04 | 21 | 1 | 959.048 | 0.147 | 959.025 | 49.13 |
| 06 | 2 | 958.762 | 0.168 | 958.710 | 58.23 | 21 | 1 | 957.822 | 0.282 | 957.280 | 49.40 |
| 06 | 2 | 959.293 | 0.132 | 959.346 | 58.74 | 21 | 1 | 959.048 | 0.147 | 959.011 | 49.74 |
| 06 | 2 | 959.293 | 0.132 | 959.310 | 59.32 | 21 | 1 | 958.823 | 0.319 | 958.778 | 50.10 |
| 06 | 2 | 959.178 | 0.118 | 959.177 | 59.91 | 21 | 1 | 958.281 | 0.228 | 958.247 | 50.42 |
| 06 | 2 | 960.067 | 0.176 | 960.066 | 60.53 | 21 | 1 | 958.640 | 0.191 | 958.662 | 51.21 |
| 06 | 2 | 959.665 | 0.139 | 959.583 | 61.15 | 21 | 1 | 958.421 | 0.149 | 958.419 | 51.58 |
| 06 | 2 | 959.787 | 0.165 | 959.842 | 61.84 | 21 | 1 | 959.353 | 0.162 | 959.443 | 52.01 |
| 06 | 2 | 959.107 | 0.156 | 959.017 | 62.83 | 21 | 1 | 958.421 | 0.149 | 958.521 | 52.44 |
| 06 | 2 | 959.475 | 0.146 | 959.508 | 63.59 | 21 | 1 | 959.923 | 0.160 | 960.048 | 53.61 |
| 06 | 2 | 959.107 | 0.156 | 959.062 | 64.39 | 21 | 1 | 958.352 | 0.263 | 958.355 | 54.10 |
| 06 | 2 | 959.665 | 0.139 | 959.578 | 65.23 | 21 | 1 | 959.923 | 0.160 | 960.004 | 54.65 |
| 06 | 2 | 960.059 | 0.124 | 960.010 | 66.13 | 21 | 1 | 958.281 | 0.228 | 958.252 | 55.30 |
| 06 | 1 | 958.762 | 0.168 | 958.710 | 58.23 | 21 | 1 | 958.421 | 0.149 | 958.521 | 55.98 |
| 06 | 1 | 959.293 | 0.132 | 959.346 | 58.74 | 22 | 2 | 959.238 | 0.203 | 959.239 | 49.25 |
| 06 | 1 | 959.293 | 0.132 | 959.310 | 59.32 | 22 | 2 | 959.938 | 0.272 | 959.950 | 49.50 |
| 06 | 1 | 959.178 | 0.118 | 959.177 | 59.91 | 22 | 2 | 959.368 | 0.291 | 959.391 | 50.08 |
| 06 | 1 | 960.067 | 0.176 | 960.066 | 60.53 | 22 | 2 | 959.144 | 0.373 | 959.168 | 50.37 |
| 06 | 1 | 959.665 | 0.139 | 959.583 | 61.15 | 22 | 2 | 959.938 | 0.272 | 960.545 | 50.68 |
| 06 | 1 | 959.787 | 0.165 | 959.842 | 61.84 | 22 | 2 | 958.934 | 0.410 | 958.883 | 51.01 |
| 06 | 1 | 959.107 | 0.156 | 959.017 | 62.83 | 22 | 2 | 959.938 | 0.272 | 960.001 | 51.35 |
| 06 | 1 | 959.475 | 0.146 | 959.508 | 63.59 | 22 | 2 | 958.600 | 0.228 | 958.697 | 51.72 |
| 06 | 1 | 959.107 | 0.156 | 959.062 | 64.39 | 22 | 2 | 959.306 | 0.210 | 959.330 | 52.22 |
| 06 | 1 | 959.665 | 0.139 | 959.578 | 65.23 | 22 | 2 | 958.186 | 0.263 | 958.309 | 53.08 |
| 06 | 1 | 960.059 | 0.124 | 960.010 | 66.13 | 22 | 2 | 958.600 | 0.228 | 958.711 | 53.54 |
| 06 | 2 | 958.248 | 0.207 | 958.261 | 67.10 | 22 | 2 | 958.553 | 0.215 | 958.482 | 54.58 |
| 06 | 2 | 959.257 | 0.174 | 959.264 | 68.14 | 22 | 2 | 959.144 | 0.373 | 959.135 | 55.14 |
| 06 | 2 | 959.787 | 0.165 | 959.902 | 69.22 | 22 | 2 | 958.186 | 0.263 | 958.295 | 55.78 |
| 06 | 2 | 959.727 | 0.177 | 959.756 | 70.43 | 22 | 2 | 957.866 | 0.344 | 957.792 | 56.43 |
| 06 | 2 | 958.248 | 0.207 | 958.482 | 71.70 | 22 | 2 | 957.866 | 0.344 | 957.705 | 57.13 |
| 06 | 2 | 959.405 | 0.163 | 959.375 | 73.08 | 22 | 1 | 959.238 | 0.203 | 959.239 | 49.25 |
| 06 | 1 | 958.248 | 0.207 | 958.261 | 67.10 | 22 | 1 | 959.938 | 0.272 | 959.950 | 49.50 |
| 06 | 1 | 959.257 | 0.174 | 959.264 | 68.14 | 22 | 1 | 959.368 | 0.291 | 959.391 | 50.08 |
| 06 | 1 | 959.787 | 0.165 | 959.902 | 69.22 | 22 | 1 | 959.144 | 0.373 | 959.168 | 50.37 |
| 06 | 1 | 959.727 | 0.177 | 959.756 | 70.43 | 22 | 1 | 959.938 | 0.272 | 960.545 | 50.68 |
| 06 | 1 | 958.248 | 0.207 | 958.482 | 71.70 | 22 | 1 | 958.934 | 0.410 | 958.883 | 51.01 |

| | | 1998 - OUTUBRO | | | |
|----|---|---------|-------|---------|-------|
| D  | L | SDB     | ER    | SDC     | HL    |
| 22 | 1 | 959.938 | 0.272 | 960.001 | 51.35 |
| 22 | 1 | 958.600 | 0.228 | 958.697 | 51.72 |
| 22 | 1 | 959.306 | 0.210 | 959.330 | 52.22 |
| 22 | 1 | 958.186 | 0.263 | 958.309 | 53.08 |
| 22 | 1 | 958.600 | 0.228 | 958.711 | 53.54 |
| 22 | 1 | 958.553 | 0.215 | 958.482 | 54.58 |
| 22 | 1 | 959.144 | 0.373 | 959.135 | 55.14 |
| 22 | 1 | 958.186 | 0.263 | 958.295 | 55.78 |
| 22 | 1 | 957.866 | 0.344 | 957.792 | 56.43 |
| 22 | 1 | 957.866 | 0.344 | 957.705 | 57.13 |
| 23 | 2 | 958.364 | 0.188 | 958.058 | 47.15 |
| 23 | 2 | 959.668 | 0.163 | 959.712 | 47.33 |
| 23 | 2 | 960.014 | 0.139 | 960.029 | 47.50 |
| 23 | 2 | 959.373 | 0.177 | 959.358 | 47.70 |
| 23 | 2 | 959.373 | 0.177 | 959.338 | 47.93 |
| 23 | 2 | 958.575 | 0.155 | 958.592 | 48.15 |
| 23 | 2 | 960.014 | 0.139 | 960.011 | 48.39 |
| 23 | 2 | 959.049 | 0.200 | 959.052 | 48.63 |
| 23 | 2 | 959.132 | 0.224 | 959.133 | 48.88 |
| 23 | 2 | 958.526 | 0.157 | 958.548 | 49.48 |
| 23 | 2 | 959.668 | 0.163 | 959.699 | 50.42 |
| 23 | 2 | 960.014 | 0.139 | 960.080 | 50.77 |
| 23 | 2 | 959.657 | 0.156 | 959.601 | 51.20 |
| 23 | 2 | 959.419 | 0.183 | 959.447 | 52.09 |
| 23 | 2 | 959.668 | 0.163 | 959.673 | 53.11 |
| 23 | 2 | 959.540 | 0.130 | 959.558 | 53.79 |
| 23 | 2 | 959.419 | 0.183 | 959.411 | 54.47 |
| 23 | 2 | 959.049 | 0.200 | 959.056 | 55.17 |
| 23 | 2 | 960.260 | 0.191 | 960.249 | 55.82 |
| 23 | 1 | 958.364 | 0.188 | 958.058 | 47.15 |
| 23 | 1 | 959.668 | 0.163 | 959.712 | 47.33 |
| 23 | 1 | 960.014 | 0.139 | 960.029 | 47.50 |
| 23 | 1 | 959.373 | 0.177 | 959.358 | 47.70 |
| 23 | 1 | 959.373 | 0.177 | 959.338 | 47.93 |
| 23 | 1 | 958.575 | 0.155 | 958.592 | 48.15 |
| 23 | 1 | 960.014 | 0.139 | 960.011 | 48.39 |
| 23 | 1 | 959.049 | 0.200 | 959.052 | 48.63 |
| 23 | 1 | 959.132 | 0.224 | 959.133 | 48.88 |
| 23 | 1 | 958.526 | 0.157 | 958.548 | 49.48 |
| 23 | 1 | 959.668 | 0.163 | 959.699 | 50.42 |
| 23 | 1 | 960.014 | 0.139 | 960.080 | 50.77 |
| 23 | 1 | 959.657 | 0.156 | 959.601 | 51.20 |
| 23 | 1 | 959.419 | 0.183 | 959.447 | 52.09 |
| 23 | 1 | 959.668 | 0.163 | 959.673 | 53.11 |
| 23 | 1 | 959.540 | 0.130 | 959.558 | 53.79 |
| 23 | 1 | 959.419 | 0.183 | 959.411 | 54.47 |
| 23 | 1 | 959.049 | 0.200 | 959.056 | 55.17 |
| 23 | 1 | 960.260 | 0.191 | 960.249 | 55.82 |

| | | 1998 - NOVEMBRO | | | |
|----|---|---------|-------|---------|-------|
| D  | L | SDB     | ER    | SDC     | HL    |
| 06 | 2 | 958.460 | 0.146 | 958.455 | 40.17 |
| 06 | 2 | 959.163 | 0.163 | 959.166 | 40.21 |
| 06 | 2 | 959.438 | 0.221 | 959.385 | 40.27 |
| 06 | 2 | 959.484 | 0.178 | 959.493 | 40.39 |
| 06 | 2 | 958.936 | 0.175 | 958.924 | 40.46 |
| 06 | 2 | 958.936 | 0.175 | 958.849 | 40.74 |
| 06 | 2 | 959.484 | 0.178 | 959.487 | 41.10 |
| 06 | 2 | 958.517 | 0.171 | 958.511 | 42.03 |
| 06 | 2 | 959.513 | 0.129 | 959.532 | 42.25 |
| 06 | 2 | 959.513 | 0.129 | 959.562 | 42.50 |
| 06 | 2 | 959.472 | 0.157 | 959.470 | 42.77 |
| 06 | 2 | 959.251 | 0.136 | 959.263 | 43.37 |
| 06 | 2 | 959.626 | 0.156 | 959.587 | 43.70 |
| 06 | 1 | 958.460 | 0.146 | 958.455 | 40.17 |
| 06 | 1 | 959.163 | 0.163 | 959.166 | 40.21 |
| 06 | 1 | 959.438 | 0.221 | 959.385 | 40.27 |
| 06 | 1 | 959.484 | 0.178 | 959.493 | 40.39 |
| 06 | 1 | 958.936 | 0.175 | 958.924 | 40.46 |
| 06 | 1 | 958.936 | 0.175 | 958.849 | 40.74 |

| | | 1998 - NOVEMBRO | | | |
|----|---|---------|-------|---------|-------|
| D  | L | SDB     | ER    | SDC     | HL    |
| 06 | 1 | 959.484 | 0.178 | 959.487 | 41.10 |
| 06 | 1 | 958.517 | 0.171 | 958.511 | 42.03 |
| 06 | 1 | 959.513 | 0.129 | 959.532 | 42.25 |
| 06 | 1 | 959.513 | 0.129 | 959.562 | 42.50 |
| 06 | 1 | 959.472 | 0.157 | 959.470 | 42.77 |
| 06 | 1 | 959.251 | 0.136 | 959.263 | 43.37 |
| 06 | 1 | 959.626 | 0.156 | 959.587 | 43.70 |
| 16 | 2 | 960.286 | 0.197 | 960.226 | 34.62 |
| 16 | 2 | 959.541 | 0.175 | 959.560 | 34.53 |
| 16 | 2 | 959.630 | 0.214 | 959.625 | 34.50 |
| 16 | 2 | 959.630 | 0.214 | 959.634 | 34.40 |
| 16 | 2 | 959.481 | 0.167 | 959.390 | 34.40 |
| 16 | 2 | 959.771 | 0.181 | 959.819 | 34.42 |
| 16 | 2 | 958.772 | 0.219 | 958.785 | 34.44 |
| 16 | 2 | 959.175 | 0.254 | 959.214 | 34.64 |
| 16 | 2 | 958.838 | 0.163 | 958.857 | 34.83 |
| 16 | 2 | 959.023 | 0.178 | 959.017 | 35.08 |
| 16 | 2 | 958.912 | 0.266 | 958.934 | 35.22 |
| 16 | 1 | 960.286 | 0.197 | 960.226 | 34.62 |
| 16 | 1 | 959.541 | 0.175 | 959.560 | 34.53 |
| 16 | 1 | 959.630 | 0.214 | 959.625 | 34.50 |
| 16 | 1 | 959.630 | 0.214 | 959.634 | 34.40 |
| 16 | 1 | 959.481 | 0.167 | 959.390 | 34.40 |
| 16 | 1 | 959.771 | 0.181 | 959.819 | 34.42 |
| 16 | 1 | 958.772 | 0.219 | 958.785 | 34.44 |
| 16 | 1 | 959.175 | 0.254 | 959.214 | 34.64 |
| 16 | 1 | 958.838 | 0.163 | 958.857 | 34.83 |
| 16 | 1 | 959.023 | 0.178 | 959.017 | 35.08 |
| 16 | 1 | 958.912 | 0.266 | 958.934 | 35.22 |
| 17 | 2 | 960.053 | 0.164 | 960.020 | 33.99 |
| 17 | 2 | 959.709 | 0.160 | 959.660 | 33.95 |
| 17 | 2 | 959.196 | 0.260 | 959.167 | 33.79 |
| 17 | 2 | 958.679 | 0.185 | 958.673 | 33.77 |
| 17 | 2 | 958.795 | 0.144 | 958.759 | 33.75 |
| 17 | 2 | 959.709 | 0.160 | 959.764 | 33.75 |
| 17 | 2 | 958.896 | 0.132 | 958.955 | 33.76 |
| 17 | 2 | 959.709 | 0.160 | 959.669 | 33.85 |
| 17 | 2 | 958.896 | 0.132 | 958.962 | 34.22 |
| 17 | 2 | 960.053 | 0.164 | 960.061 | 34.32 |
| 17 | 1 | 960.053 | 0.164 | 960.020 | 33.99 |
| 17 | 1 | 959.709 | 0.160 | 959.660 | 33.95 |
| 17 | 1 | 959.196 | 0.260 | 959.167 | 33.79 |
| 17 | 1 | 958.679 | 0.185 | 958.673 | 33.77 |
| 17 | 1 | 958.795 | 0.144 | 958.759 | 33.75 |
| 17 | 1 | 959.709 | 0.160 | 959.764 | 33.75 |
| 17 | 1 | 958.896 | 0.132 | 958.955 | 33.76 |
| 17 | 1 | 959.709 | 0.160 | 959.669 | 33.85 |
| 17 | 1 | 958.896 | 0.132 | 958.962 | 34.22 |
| 17 | 1 | 960.053 | 0.164 | 960.061 | 34.32 |
| 18 | 2 | 959.822 | 0.121 | 959.762 | 33.24 |
| 18 | 2 | 958.384 | 0.150 | 958.424 | 33.20 |
| 18 | 2 | 958.384 | 0.150 | 958.453 | 33.14 |
| 18 | 2 | 958.652 | 0.126 | 958.700 | 33.12 |
| 18 | 2 | 959.225 | 0.127 | 959.054 | 33.11 |
| 18 | 2 | 959.535 | 0.141 | 959.553 | 33.09 |
| 18 | 2 | 959.822 | 0.121 | 959.815 | 33.10 |
| 18 | 2 | 959.391 | 0.131 | 959.414 | 33.11 |
| 18 | 2 | 959.535 | 0.141 | 959.517 | 33.13 |
| 18 | 2 | 959.535 | 0.141 | 959.540 | 33.16 |
| 18 | 2 | 959.225 | 0.127 | 959.099 | 33.20 |
| 18 | 2 | 959.839 | 0.090 | 959.864 | 33.24 |
| 18 | 2 | 959.391 | 0.131 | 959.381 | 33.38 |
| 18 | 2 | 958.522 | 0.098 | 958.460 | 33.47 |
| 18 | 2 | 958.808 | 0.129 | 958.894 | 33.56 |
| 18 | 1 | 959.822 | 0.121 | 959.762 | 33.24 |
| 18 | 1 | 958.384 | 0.150 | 958.424 | 33.20 |
| 18 | 1 | 958.384 | 0.150 | 958.453 | 33.14 |
| 18 | 1 | 958.652 | 0.126 | 958.700 | 33.12 |
| 18 | 1 | 959.225 | 0.127 | 959.054 | 33.11 |
| 18 | 1 | 959.535 | 0.141 | 959.553 | 33.09 |
| 18 | 1 | 959.822 | 0.121 | 959.815 | 33.10 |

|     | 1998 - NOVEMBRO |       |         |       |
|-----|-----------------|-------|---------|-------|
| D   | L | SDB         | ER    | SDC     | HL    |
| 18  | 1 | 959.391     | 0.131 | 959.414 | 33.11 |
| 18  | 1 | 959.535     | 0.141 | 959.517 | 33.13 |
| 18  | 1 | 959.535     | 0.141 | 959.540 | 33.16 |
| 18  | 1 | 959.225     | 0.127 | 959.099 | 33.20 |
| 18  | 1 | 959.839     | 0.090 | 959.864 | 33.24 |
| 18  | 1 | 959.391     | 0.131 | 959.381 | 33.38 |
| 18  | 1 | 958.522     | 0.098 | 958.460 | 33.47 |
| 18  | 1 | 958.808     | 0.129 | 958.894 | 33.56 |
| 24  | 2 | 959.403     | 0.180 | 959.333 | 29.66 |
| 24  | 2 | 958.280     | 0.180 | 958.318 | 29.58 |
| 24  | 2 | 959.582     | 0.180 | 959.741 | 29.51 |
| 24  | 2 | 958.178     | 0.201 | 958.188 | 29.44 |
| 24  | 2 | 958.415     | 0.164 | 958.419 | 29.28 |
| 24  | 2 | 958.415     | 0.164 | 958.479 | 29.22 |
| 24  | 2 | 959.491     | 0.209 | 959.512 | 28.84 |
| 24  | 2 | 959.582     | 0.180 | 959.726 | 28.83 |
| 24  | 1 | 959.403     | 0.180 | 959.333 | 29.66 |
| 24  | 1 | 958.280     | 0.180 | 958.318 | 29.58 |
| 24  | 1 | 959.582     | 0.180 | 959.741 | 29.51 |
| 24  | 1 | 958.178     | 0.201 | 958.188 | 29.44 |
| 24  | 1 | 958.415     | 0.164 | 958.419 | 29.28 |
| 24  | 1 | 958.415     | 0.164 | 958.479 | 29.22 |
| 24  | 1 | 959.491     | 0.209 | 959.512 | 28.84 |
| 24  | 1 | 959.582     | 0.180 | 959.726 | 28.83 |
| 26  | 2 | 959.128     | 0.246 | 959.156 | 29.23 |
| 26  | 2 | 960.246     | 0.130 | 960.321 | 28.90 |
| 26  | 2 | 959.600     | 0.109 | 959.534 | 28.70 |
| 26  | 2 | 959.269     | 0.126 | 959.228 | 28.47 |
| 26  | 2 | 959.600     | 0.109 | 959.536 | 28.39 |
| 26  | 2 | 958.635     | 0.122 | 958.655 | 28.05 |
| 26  | 2 | 958.868     | 0.158 | 958.881 | 27.90 |
| 26  | 2 | 958.722     | 0.145 | 958.774 | 27.83 |
| 26  | 2 | 959.450     | 0.108 | 959.406 | 27.77 |
| 26  | 2 | 958.374     | 0.165 | 958.486 | 27.70 |
| 26  | 2 | 960.246     | 0.130 | 960.311 | 27.53 |
| 26  | 2 | 958.181     | 0.144 | 958.100 | 27.44 |
| 26  | 1 | 959.128     | 0.246 | 959.156 | 29.23 |
| 26  | 1 | 960.246     | 0.130 | 960.321 | 28.90 |
| 26  | 1 | 959.600     | 0.109 | 959.534 | 28.70 |
| 26  | 1 | 959.269     | 0.126 | 959.228 | 28.47 |
| 26  | 1 | 959.600     | 0.109 | 959.536 | 28.39 |
| 26  | 1 | 958.635     | 0.122 | 958.655 | 28.05 |
| 26  | 1 | 958.868     | 0.158 | 958.881 | 27.90 |
| 26  | 1 | 958.722     | 0.145 | 958.774 | 27.83 |
| 26  | 1 | 959.450     | 0.108 | 959.406 | 27.77 |
| 26  | 1 | 958.374     | 0.165 | 958.486 | 27.70 |
| 26  | 1 | 960.246     | 0.130 | 960.311 | 27.53 |
| 26  | 1 | 958.181     | 0.144 | 958.100 | 27.44 |
| 30  | 2 | 959.847     | 0.364 | 959.808 | 27.04 |
| 30  | 2 | 958.613     | 0.295 | 958.850 | 26.74 |
| 30  | 2 | 958.613     | 0.295 | 958.695 | 26.30 |
| 30  | 2 | 959.344     | 0.723 | 959.450 | 25.90 |
| 30  | 1 | 959.847     | 0.364 | 959.808 | 27.04 |
| 30  | 1 | 958.613     | 0.295 | 958.850 | 26.74 |
| 30  | 1 | 958.613     | 0.295 | 958.695 | 26.30 |
| 30  | 1 | 959.344     | 0.723 | 959.450 | 25.90 |

|     | 1998 - DEZEMBRO |       |         |       |
|-----|-----------------|-------|---------|-------|
| D   | L | SDB         | ER    | SDC     | HL    |
| 01  | 2 | 959.600     | 0.265 | 959.706 | 24.95 |
| 01  | 2 | 959.600     | 0.265 | 960.628 | 24.49 |
| 01  | 2 | 958.598     | 0.483 | 958.997 | 24.39 |
| 01  | 2 | 958.028     | 0.237 | 958.022 | 23.83 |
| 01  | 1 | 959.600     | 0.265 | 959.706 | 24.95 |
| 01  | 1 | 959.600     | 0.265 | 960.628 | 24.49 |
| 01  | 1 | 958.598     | 0.483 | 958.997 | 24.39 |
| 01  | 1 | 958.028     | 0.237 | 958.022 | 23.83 |
| 02  | 2 | 959.966     | 0.129 | 959.957 | 25.84 |
| 02  | 2 | 959.038     | 0.174 | 959.030 | 25.69 |
| 02  | 2 | 959.545     | 0.181 | 959.467 | 25.41 |

|     | 1998 - DEZEMBRO |       |         |       |
|-----|-----------------|-------|---------|-------|
| D   | L | SDB         | ER    | SDC     | HL    |
| 02  | 2 | 959.070     | 0.111 | 959.141 | 25.26 |
| 02  | 2 | 958.635     | 0.144 | 958.664 | 25.12 |
| 02  | 2 | 960.077     | 0.161 | 960.072 | 24.97 |
| 02  | 2 | 959.038     | 0.174 | 958.995 | 24.53 |
| 02  | 2 | 958.837     | 0.169 | 958.820 | 23.39 |
| 02  | 2 | 959.545     | 0.181 | 959.533 | 23.18 |
| 02  | 2 | 959.769     | 0.149 | 959.829 | 22.86 |
| 02  | 1 | 959.966     | 0.129 | 959.957 | 25.84 |
| 02  | 1 | 959.038     | 0.174 | 959.030 | 25.69 |
| 02  | 1 | 959.545     | 0.181 | 959.467 | 25.41 |
| 02  | 1 | 959.070     | 0.111 | 959.141 | 25.26 |
| 02  | 1 | 958.635     | 0.144 | 958.664 | 25.12 |
| 02  | 1 | 960.077     | 0.161 | 960.072 | 24.97 |
| 02  | 1 | 959.038     | 0.174 | 958.995 | 24.53 |
| 02  | 1 | 958.837     | 0.169 | 958.820 | 23.39 |
| 02  | 1 | 959.545     | 0.181 | 959.533 | 23.18 |
| 02  | 1 | 959.769     | 0.149 | 959.829 | 22.86 |
| 03  | 2 | 959.136     | 0.206 | 959.161 | 24.55 |
| 03  | 2 | 958.639     | 0.228 | 958.612 | 24.05 |
| 03  | 2 | 959.495     | 0.183 | 959.511 | 23.90 |
| 03  | 2 | 958.400     | 0.251 | 958.443 | 23.54 |
| 03  | 2 | 958.672     | 0.217 | 958.667 | 23.42 |
| 03  | 2 | 958.243     | 0.209 | 958.306 | 23.31 |
| 03  | 2 | 958.774     | 0.208 | 958.805 | 22.97 |
| 03  | 2 | 958.639     | 0.228 | 958.603 | 22.76 |
| 03  | 1 | 959.136     | 0.206 | 959.161 | 24.55 |
| 03  | 1 | 958.639     | 0.228 | 958.612 | 24.05 |
| 03  | 1 | 959.495     | 0.183 | 959.511 | 23.90 |
| 03  | 1 | 958.400     | 0.251 | 958.443 | 23.54 |
| 03  | 1 | 958.672     | 0.217 | 958.667 | 23.42 |
| 03  | 1 | 958.243     | 0.209 | 958.306 | 23.31 |
| 03  | 1 | 958.774     | 0.208 | 958.805 | 22.97 |
| 03  | 1 | 958.639     | 0.228 | 958.603 | 22.76 |
| 04  | 2 | 958.773     | 0.169 | 958.868 | 25.11 |
| 04  | 2 | 958.107     | 0.189 | 958.063 | 24.78 |
| 04  | 2 | 959.071     | 0.139 | 959.114 | 24.32 |
| 04  | 2 | 959.161     | 0.162 | 959.174 | 24.19 |
| 04  | 2 | 958.726     | 0.156 | 958.740 | 24.05 |
| 04  | 2 | 958.107     | 0.189 | 957.987 | 23.52 |
| 04  | 2 | 959.265     | 0.184 | 959.365 | 22.62 |
| 04  | 2 | 958.186     | 0.165 | 958.285 | 22.38 |
| 04  | 2 | 958.511     | 0.189 | 958.462 | 22.27 |
| 04  | 1 | 958.773     | 0.169 | 958.868 | 25.11 |
| 04  | 1 | 958.107     | 0.189 | 958.063 | 24.78 |
| 04  | 1 | 959.071     | 0.139 | 959.114 | 24.32 |
| 04  | 1 | 959.161     | 0.162 | 959.174 | 24.19 |
| 04  | 1 | 958.726     | 0.156 | 958.740 | 24.05 |
| 04  | 1 | 958.107     | 0.189 | 957.987 | 23.52 |
| 04  | 1 | 959.265     | 0.184 | 959.365 | 22.62 |
| 04  | 1 | 958.186     | 0.165 | 958.285 | 22.38 |
| 04  | 1 | 958.511     | 0.189 | 958.462 | 22.27 |
| 08  | 2 | 959.647     | 0.149 | 959.626 | 23.05 |
| 08  | 2 | 959.041     | 0.181 | 959.050 | 22.58 |
| 08  | 2 | 959.363     | 0.164 | 959.357 | 22.28 |
| 08  | 2 | 959.063     | 0.124 | 959.067 | 22.13 |
| 08  | 2 | 959.363     | 0.164 | 959.359 | 21.84 |
| 08  | 2 | 959.182     | 0.148 | 959.234 | 21.69 |
| 08  | 2 | 958.851     | 0.135 | 958.850 | 20.97 |
| 08  | 2 | 958.851     | 0.135 | 958.788 | 20.80 |
| 08  | 2 | 958.237     | 0.197 | 958.245 | 19.65 |
| 08  | 2 | 958.336     | 0.170 | 958.361 | 19.39 |
| 08  | 1 | 959.647     | 0.149 | 959.626 | 23.05 |
| 08  | 1 | 959.041     | 0.181 | 959.050 | 22.58 |
| 08  | 1 | 959.363     | 0.164 | 959.357 | 22.28 |
| 08  | 1 | 959.063     | 0.124 | 959.067 | 22.13 |
| 08  | 1 | 959.363     | 0.164 | 959.359 | 21.84 |
| 08  | 1 | 959.182     | 0.148 | 959.234 | 21.69 |
| 08  | 1 | 958.851     | 0.135 | 958.850 | 20.97 |
| 08  | 1 | 958.851     | 0.135 | 958.788 | 20.80 |
| 08  | 1 | 958.237     | 0.197 | 958.245 | 19.65 |
| 08  | 1 | 958.336     | 0.170 | 958.361 | 19.39 |

| \  | 1998 - DEZEMBRO |  |  |  |  |  | 1999 - JANEIRO |  |  |  |
|----|----|---------|-------|---------|-------|----|----|---------|-------|---------|-------|
| D  | L  | SDB     | ER    | SDC     | HL    | D  | L  | SDB     | ER    | SDC     | HL    |
| 16 | 2  | 959.600 | 0.138 | 959.633 | 16.59 | 10 | 2  | 958.940 | 0.149 | 958.916 | 8.76  |
| 16 | 2  | 959.336 | 0.156 | 959.306 | 16.38 | 10 | 2  | 958.778 | 0.164 | 958.757 | 8.59  |
| 16 | 2  | 959.336 | 0.156 | 959.305 | 15.81 | 10 | 2  | 959.710 | 0.199 | 959.746 | 7.97  |
| 16 | 2  | 958.458 | 0.177 | 958.467 | 15.23 | 10 | 2  | 958.495 | 0.143 | 958.453 | 7.82  |
| 16 | 1  | 959.600 | 0.138 | 959.633 | 16.59 | 10 | 2  | 958.709 | 0.152 | 958.737 | 7.53  |
| 16 | 1  | 959.336 | 0.156 | 959.306 | 16.38 | 10 | 2  | 960.404 | 0.178 | 960.306 | 7.25  |
| 16 | 1  | 959.336 | 0.156 | 959.305 | 15.81 | 10 | 2  | 958.204 | 0.134 | 958.296 | 7.12  |
| 16 | 1  | 958.458 | 0.177 | 958.467 | 15.23 | 10 | 2  | 958.940 | 0.149 | 958.960 | 6.85  |
| 17 | 2  | 959.160 | 0.149 | 959.169 | 16.46 | 10 | 2  | 959.042 | 0.154 | 959.010 | 6.72  |
| 17 | 2  | 959.629 | 0.131 | 959.732 | 16.29 | 10 | 2  | 959.413 | 0.141 | 959.295 | 6.31  |
| 17 | 2  | 958.986 | 0.139 | 959.028 | 15.91 | 10 | 2  | 958.940 | 0.149 | 958.989 | 6.18  |
| 17 | 2  | 959.160 | 0.149 | 959.168 | 14.60 | 10 | 2  | 958.644 | 0.143 | 958.648 | 6.06  |
| 17 | 2  | 959.335 | 0.149 | 959.464 | 13.45 | 10 | 1  | 959.148 | 0.164 | 959.230 | 8.94  |
| 17 | 1  | 959.160 | 0.149 | 959.169 | 16.46 | 10 | 1  | 958.940 | 0.149 | 958.916 | 8.76  |
| 17 | 1  | 959.629 | 0.131 | 959.732 | 16.29 | 10 | 1  | 958.778 | 0.164 | 958.757 | 8.59  |
| 17 | 1  | 958.986 | 0.139 | 959.028 | 15.91 | 10 | 1  | 959.710 | 0.199 | 959.746 | 7.97  |
| 17 | 1  | 959.160 | 0.149 | 959.168 | 14.60 | 10 | 1  | 958.495 | 0.143 | 958.453 | 7.82  |
| 17 | 1  | 959.335 | 0.149 | 959.464 | 13.45 | 10 | 1  | 958.709 | 0.152 | 958.737 | 7.53  |
| 21 | 2  | 959.728 | 0.184 | 959.735 | 15.08 | 10 | 1  | 960.404 | 0.178 | 960.306 | 7.25  |
| 21 | 2  | 958.337 | 0.209 | 958.325 | 14.90 | 10 | 1  | 958.204 | 0.134 | 958.296 | 7.12  |
| 21 | 2  | 958.337 | 0.209 | 958.317 | 14.72 | 10 | 1  | 958.940 | 0.149 | 958.960 | 6.85  |
| 21 | 2  | 959.542 | 0.149 | 959.416 | 13.20 | 10 | 1  | 959.042 | 0.154 | 959.010 | 6.72  |
| 21 | 1  | 959.728 | 0.184 | 959.735 | 15.08 | 10 | 1  | 959.413 | 0.141 | 959.295 | 6.31  |
| 21 | 1  | 958.337 | 0.209 | 958.325 | 14.90 | 10 | 1  | 958.940 | 0.149 | 958.989 | 6.18  |
| 21 | 1  | 958.337 | 0.209 | 958.317 | 14.72 | 10 | 1  | 958.644 | 0.143 | 958.648 | 6.06  |
| 21 | 1  | 959.542 | 0.149 | 959.416 | 13.20 | 11 | 2  | 959.644 | 0.139 | 959.653 | 8.01  |
| 22 | 2  | 958.811 | 0.128 | 958.814 | 14.56 | 11 | 2  | 959.405 | 0.161 | 959.389 | 7.71  |
| 22 | 2  | 959.462 | 0.136 | 959.511 | 13.60 | 11 | 2  | 958.951 | 0.213 | 959.090 | 7.00  |
| 22 | 2  | 959.682 | 0.135 | 959.706 | 13.40 | 11 | 2  | 959.939 | 0.149 | 959.984 | 6.84  |
| 22 | 2  | 959.659 | 0.156 | 959.631 | 12.81 | 11 | 2  | 958.354 | 0.144 | 958.389 | 6.71  |
| 22 | 2  | 960.238 | 0.146 | 960.320 | 12.62 | 11 | 2  | 958.951 | 0.213 | 959.003 | 6.57  |
| 22 | 2  | 959.043 | 0.126 | 959.142 | 12.42 | 11 | 2  | 958.951 | 0.213 | 958.981 | 6.19  |
| 22 | 1  | 958.811 | 0.128 | 958.814 | 14.56 | 11 | 2  | 959.405 | 0.161 | 959.248 | 6.06  |
| 22 | 1  | 959.462 | 0.136 | 959.511 | 13.60 | 11 | 2  | 959.405 | 0.161 | 959.221 | 5.82  |
| 22 | 1  | 959.682 | 0.135 | 959.706 | 13.40 | 11 | 2  | 958.473 | 0.169 | 958.468 | 5.70  |
| 22 | 1  | 959.659 | 0.156 | 959.631 | 12.81 | 11 | 2  | 958.951 | 0.213 | 958.989 | 5.49  |
| 22 | 1  | 960.238 | 0.146 | 960.320 | 12.62 | 11 | 2  | 958.951 | 0.213 | 958.982 | 5.38  |
| 22 | 1  | 959.043 | 0.126 | 959.142 | 12.42 | 11 | 1  | 959.644 | 0.139 | 959.653 | 8.01  |
| 29 | 2  | 959.076 | 0.133 | 959.027 | 11.54 | 11 | 1  | 959.405 | 0.161 | 959.389 | 7.71  |
| 29 | 2  | 958.813 | 0.140 | 958.816 | 11.16 | 11 | 1  | 958.951 | 0.213 | 959.090 | 7.00  |
| 29 | 2  | 958.950 | 0.153 | 958.943 | 10.97 | 11 | 1  | 959.939 | 0.149 | 959.984 | 6.84  |
| 29 | 2  | 959.462 | 0.162 | 959.461 | 10.78 | 11 | 1  | 958.354 | 0.144 | 958.389 | 6.71  |
| 29 | 2  | 958.456 | 0.134 | 958.312 | 10.23 | 11 | 1  | 958.951 | 0.213 | 959.003 | 6.57  |
| 29 | 2  | 958.456 | 0.134 | 957.587 | 9.85  | 11 | 1  | 958.951 | 0.213 | 958.981 | 6.19  |
| 29 | 2  | 959.137 | 0.150 | 959.144 | 9.49  | 11 | 1  | 959.405 | 0.161 | 959.248 | 6.06  |
| 29 | 1  | 959.076 | 0.133 | 959.027 | 11.54 | 11 | 1  | 959.405 | 0.161 | 959.221 | 5.82  |
| 29 | 1  | 958.813 | 0.140 | 958.816 | 11.16 | 11 | 1  | 958.473 | 0.169 | 958.468 | 5.70  |
| 29 | 1  | 958.950 | 0.153 | 958.943 | 10.97 | 11 | 1  | 958.951 | 0.213 | 958.989 | 5.49  |
| 29 | 1  | 959.462 | 0.162 | 959.461 | 10.78 | 11 | 1  | 958.951 | 0.213 | 958.982 | 5.38  |
| 29 | 1  | 958.456 | 0.134 | 958.312 | 10.23 | 12 | 2  | 960.402 | 0.209 | 960.225 | 7.82  |
| 29 | 1  | 958.456 | 0.134 | 957.587 | 9.85  | 12 | 2  | 959.984 | 0.177 | 960.027 | 7.67  |
| 29 | 1  | 959.137 | 0.150 | 959.144 | 9.49  | 12 | 2  | 958.704 | 0.195 | 958.709 | 7.02  |
|    |    |         |       |         |       | 12 | 2  | 958.091 | 0.194 | 957.676 | 6.74  |
|    |    |         |       |         |       | 12 | 2  | 958.624 | 0.274 | 958.542 | 6.57  |
|    |    | 1999 - JANEIRO |  |         |       | 12 | 2  | 958.624 | 0.274 | 958.633 | 6.01  |
| D  | L  | SDB     | ER    | SDC     | HL    | 12 | 2  | 958.704 | 0.195 | 958.664 | 5.90  |
| 08 | 2  | 958.818 | 0.134 | 958.922 | 7.83  | 12 | 2  | 959.682 | 0.182 | 959.682 | 5.78  |
| 08 | 2  | 958.528 | 0.163 | 958.547 | 7.66  | 12 | 2  | 958.843 | 0.189 | 958.840 | 5.43  |
| 08 | 2  | 958.644 | 0.157 | 958.603 | 7.18  | 12 | 2  | 960.402 | 0.209 | 960.488 | 5.11  |
| 08 | 2  | 959.322 | 0.149 | 959.263 | 7.04  | 12 | 1  | 960.402 | 0.209 | 960.225 | 7.82  |
| 08 | 2  | 958.528 | 0.163 | 958.459 | 6.91  | 12 | 1  | 959.984 | 0.177 | 960.027 | 7.67  |
| 08 | 2  | 958.818 | 0.134 | 958.793 | 6.20  | 12 | 1  | 958.704 | 0.195 | 958.709 | 7.02  |
| 08 | 2  | 958.377 | 0.155 | 958.433 | 5.82  | 12 | 1  | 958.091 | 0.194 | 957.676 | 6.74  |
| 08 | 1  | 958.818 | 0.134 | 958.922 | 7.83  | 12 | 1  | 958.624 | 0.274 | 958.542 | 6.57  |
| 08 | 1  | 958.528 | 0.163 | 958.547 | 7.66  | 12 | 1  | 958.624 | 0.274 | 958.633 | 6.01  |
| 08 | 1  | 958.644 | 0.157 | 958.603 | 7.18  | 12 | 1  | 958.704 | 0.195 | 958.664 | 5.90  |
| 08 | 1  | 959.322 | 0.149 | 959.263 | 7.04  | 12 | 1  | 959.682 | 0.182 | 959.682 | 5.78  |
| 08 | 1  | 958.528 | 0.163 | 958.459 | 6.91  | 12 | 1  | 958.843 | 0.189 | 958.840 | 5.43  |
| 08 | 1  | 958.818 | 0.134 | 958.793 | 6.20  | 12 | 1  | 960.402 | 0.209 | 960.488 | 5.11  |
| 08 | 1  | 958.377 | 0.155 | 958.433 | 5.82  | 13 | 2  | 958.319 | 0.191 | 958.349 | 7.71  |
| 10 | 2  | 959.148 | 0.164 | 959.230 | 8.94  | 13 | 2  | 960.328 | 0.269 | 960.313 | 7.45  |

| 1999 - JANEIRO | | | | | | 1999 - JANEIRO | | | | |
|---|---|---|---|---|---|---|---|---|---|---|
| D | L | SDB | ER | SDC | HL | D | L | SDB | ER | SDC | HL |
| 13 | 2 | 957.393 | 0.192 | 957.338 | 6.95 | 22 | 2 | 958.446 | 0.143 | 958.447 | 4.96 |
| 13 | 2 | 959.687 | 0.170 | 959.879 | 6.83 | 22 | 2 | 959.282 | 0.241 | 959.302 | 4.68 |
| 13 | 2 | 960.328 | 0.269 | 960.498 | 6.72 | 22 | 2 | 959.282 | 0.241 | 959.282 | 4.59 |
| 13 | 2 | 958.424 | 0.113 | 958.504 | 6.21 | 22 | 2 | 959.054 | 0.184 | 958.980 | 4.51 |
| 13 | 2 | 958.630 | 0.235 | 958.549 | 6.10 | 22 | 2 | 958.830 | 0.161 | 958.797 | 4.43 |
| 13 | 2 | 958.630 | 0.235 | 958.666 | 5.99 | 22 | 2 | 958.052 | 0.167 | 958.159 | 4.36 |
| 13 | 2 | 957.393 | 0.192 | 957.064 | 5.79 | 22 | 2 | 959.054 | 0.184 | 959.018 | 4.30 |
| 13 | 1 | 958.319 | 0.191 | 958.349 | 7.71 | 22 | 2 | 959.069 | 0.161 | 959.106 | 4.18 |
| 13 | 1 | 960.328 | 0.269 | 960.313 | 7.45 | 22 | 2 | 959.054 | 0.184 | 959.057 | 4.12 |
| 13 | 1 | 957.393 | 0.192 | 957.338 | 6.95 | 22 | 2 | 958.523 | 0.189 | 958.599 | 4.04 |
| 13 | 1 | 959.687 | 0.170 | 959.879 | 6.83 | 22 | 2 | 958.523 | 0.189 | 958.554 | 4.00 |
| 13 | 1 | 960.328 | 0.269 | 960.498 | 6.72 | 22 | 2 | 959.407 | 0.177 | 959.876 | 3.97 |
| 13 | 1 | 958.424 | 0.113 | 958.504 | 6.21 | 22 | 2 | 958.676 | 0.150 | 958.654 | 3.94 |
| 13 | 1 | 958.630 | 0.235 | 958.549 | 6.10 | 22 | 2 | 957.754 | 0.166 | 957.390 | 3.94 |
| 13 | 1 | 958.630 | 0.235 | 958.666 | 5.99 | 22 | 2 | 958.679 | 0.234 | 958.689 | 3.95 |
| 13 | 1 | 957.393 | 0.192 | 957.064 | 5.79 | 22 | 2 | 959.069 | 0.161 | 959.105 | 3.96 |
| 18 | 2 | 959.223 | 0.150 | 959.405 | 5.10 | 22 | 1 | 959.069 | 0.161 | 959.104 | 5.07 |
| 18 | 2 | 959.223 | 0.150 | 960.504 | 4.94 | 22 | 1 | 958.446 | 0.143 | 958.447 | 4.96 |
| 18 | 2 | 957.354 | 0.229 | 957.371 | 4.87 | 22 | 1 | 959.282 | 0.241 | 959.302 | 4.68 |
| 18 | 2 | 959.223 | 0.150 | 959.811 | 4.65 | 22 | 1 | 959.282 | 0.241 | 959.282 | 4.59 |
| 18 | 2 | 958.624 | 0.211 | 958.617 | 4.47 | 22 | 1 | 959.054 | 0.184 | 958.980 | 4.51 |
| 18 | 2 | 958.183 | 0.236 | 958.235 | 4.32 | 22 | 1 | 958.830 | 0.161 | 958.797 | 4.43 |
| 18 | 2 | 958.134 | 0.202 | 958.052 | 4.28 | 22 | 1 | 958.052 | 0.167 | 958.159 | 4.36 |
| 18 | 2 | 959.223 | 0.150 | 960.334 | 4.24 | 22 | 1 | 959.054 | 0.184 | 959.018 | 4.30 |
| 18 | 1 | 959.223 | 0.150 | 959.405 | 5.10 | 22 | 1 | 959.069 | 0.161 | 959.106 | 4.18 |
| 18 | 1 | 959.223 | 0.150 | 960.504 | 4.94 | 22 | 1 | 959.054 | 0.184 | 959.057 | 4.12 |
| 18 | 1 | 957.354 | 0.229 | 957.371 | 4.87 | 22 | 1 | 958.523 | 0.189 | 958.599 | 4.04 |
| 18 | 1 | 959.223 | 0.150 | 959.811 | 4.65 | 22 | 1 | 958.523 | 0.189 | 958.554 | 4.00 |
| 18 | 1 | 958.624 | 0.211 | 958.617 | 4.47 | 22 | 1 | 959.407 | 0.177 | 959.876 | 3.97 |
| 18 | 1 | 958.183 | 0.236 | 958.235 | 4.32 | 22 | 1 | 958.676 | 0.150 | 958.654 | 3.94 |
| 18 | 1 | 958.134 | 0.202 | 958.052 | 4.28 | 22 | 1 | 957.754 | 0.166 | 957.390 | 3.94 |
| 18 | 1 | 959.223 | 0.150 | 960.334 | 4.24 | 22 | 1 | 958.679 | 0.234 | 958.689 | 3.95 |
| 19 | 2 | 959.270 | 0.177 | 959.249 | 5.69 | 22 | 1 | 959.069 | 0.161 | 959.105 | 3.96 |
| 19 | 2 | 958.950 | 0.163 | 958.956 | 5.58 | 25 | 2 | 959.716 | 0.204 | 960.646 | 4.42 |
| 19 | 2 | 959.347 | 0.161 | 959.392 | 5.36 | 25 | 2 | 959.716 | 0.204 | 959.647 | 4.32 |
| 19 | 2 | 958.057 | 0.154 | 958.258 | 5.15 | 25 | 2 | 959.133 | 0.150 | 959.155 | 4.25 |
| 19 | 2 | 958.644 | 0.166 | 958.692 | 5.05 | 25 | 2 | 959.460 | 0.155 | 959.520 | 4.18 |
| 19 | 2 | 958.057 | 0.154 | 958.098 | 4.95 | 25 | 2 | 959.000 | 0.161 | 959.029 | 4.12 |
| 19 | 2 | 958.871 | 0.166 | 958.868 | 4.86 | 25 | 2 | 958.448 | 0.162 | 958.444 | 4.06 |
| 19 | 2 | 959.209 | 0.158 | 959.219 | 4.78 | 25 | 2 | 958.272 | 0.140 | 958.240 | 4.00 |
| 19 | 2 | 958.975 | 0.145 | 959.006 | 4.69 | 25 | 2 | 958.682 | 0.146 | 958.727 | 3.94 |
| 19 | 2 | 958.639 | 0.172 | 958.513 | 4.54 | 25 | 2 | 959.460 | 0.155 | 959.568 | 3.89 |
| 19 | 2 | 959.847 | 0.145 | 959.859 | 4.47 | 25 | 2 | 959.317 | 0.181 | 959.325 | 3.82 |
| 19 | 2 | 958.057 | 0.154 | 957.796 | 4.34 | 25 | 2 | 958.888 | 0.186 | 958.867 | 3.79 |
| 19 | 2 | 958.639 | 0.172 | 958.482 | 4.20 | 25 | 2 | 958.448 | 0.162 | 958.426 | 3.77 |
| 19 | 1 | 959.270 | 0.177 | 959.249 | 5.69 | 25 | 2 | 959.133 | 0.150 | 959.188 | 3.75 |
| 19 | 1 | 958.950 | 0.163 | 958.956 | 5.58 | 25 | 2 | 958.899 | 0.184 | 958.928 | 3.74 |
| 19 | 1 | 959.347 | 0.161 | 959.392 | 5.36 | 25 | 2 | 958.272 | 0.140 | 958.337 | 3.74 |
| 19 | 1 | 958.057 | 0.154 | 958.258 | 5.15 | 25 | 2 | 958.899 | 0.184 | 958.895 | 3.74 |
| 19 | 1 | 958.644 | 0.166 | 958.692 | 5.05 | 25 | 2 | 958.272 | 0.140 | 958.312 | 3.78 |
| 19 | 1 | 958.057 | 0.154 | 958.098 | 4.95 | 25 | 2 | 958.448 | 0.162 | 958.426 | 3.81 |
| 19 | 1 | 958.871 | 0.166 | 958.868 | 4.86 | 25 | 1 | 959.716 | 0.204 | 960.646 | 4.42 |
| 19 | 1 | 959.209 | 0.158 | 959.219 | 4.78 | 25 | 1 | 959.716 | 0.204 | 959.647 | 4.32 |
| 19 | 1 | 958.975 | 0.145 | 959.006 | 4.69 | 25 | 1 | 959.133 | 0.150 | 959.155 | 4.25 |
| 19 | 1 | 958.639 | 0.172 | 958.513 | 4.54 | 25 | 1 | 959.460 | 0.155 | 959.520 | 4.18 |
| 19 | 1 | 959.847 | 0.145 | 959.859 | 4.47 | 25 | 1 | 959.000 | 0.161 | 959.029 | 4.12 |
| 19 | 1 | 958.057 | 0.154 | 957.796 | 4.34 | 25 | 1 | 958.448 | 0.162 | 958.444 | 4.06 |
| 19 | 1 | 958.639 | 0.172 | 958.482 | 4.20 | 25 | 1 | 958.272 | 0.140 | 958.240 | 4.00 |
| 21 | 2 | 959.229 | 0.118 | 959.198 | 5.09 | 25 | 1 | 958.682 | 0.146 | 958.727 | 3.94 |
| 21 | 2 | 958.609 | 0.162 | 958.624 | 4.69 | 25 | 1 | 959.460 | 0.155 | 959.568 | 3.89 |
| 21 | 2 | 958.086 | 0.142 | 958.064 | 4.31 | 25 | 1 | 959.317 | 0.181 | 959.325 | 3.82 |
| 21 | 2 | 959.899 | 0.135 | 959.907 | 4.26 | 25 | 1 | 958.888 | 0.186 | 958.867 | 3.79 |
| 21 | 2 | 958.751 | 0.119 | 958.751 | 4.13 | 25 | 1 | 958.448 | 0.162 | 958.426 | 3.77 |
| 21 | 2 | 958.609 | 0.162 | 958.584 | 4.09 | 25 | 1 | 959.133 | 0.150 | 959.188 | 3.75 |
| 21 | 1 | 959.229 | 0.118 | 959.198 | 5.09 | 25 | 1 | 958.899 | 0.184 | 958.928 | 3.74 |
| 21 | 1 | 958.609 | 0.162 | 958.624 | 4.69 | 25 | 1 | 958.272 | 0.140 | 958.337 | 3.74 |
| 21 | 1 | 958.086 | 0.142 | 958.064 | 4.31 | 25 | 1 | 958.899 | 0.184 | 958.895 | 3.74 |
| 21 | 1 | 959.899 | 0.135 | 959.907 | 4.26 | 25 | 1 | 958.272 | 0.140 | 958.312 | 3.78 |
| 21 | 1 | 958.751 | 0.119 | 958.751 | 4.13 | 25 | 1 | 958.448 | 0.162 | 958.426 | 3.81 |
| 21 | 1 | 958.609 | 0.162 | 958.584 | 4.09 | 26 | 2 | 959.080 | 0.245 | 959.149 | 3.95 |
| 22 | 2 | 959.069 | 0.161 | 959.104 | 5.07 | 26 | 2 | 958.192 | 0.207 | 958.186 | 3.90 |

|     | 1999 - JANEIRO |     |     |     |     |     | 1999 - FEVEREIRO |     |     |     |
| --- | --- | --- | --- | --- | --- | --- | --- | --- | --- | --- |
| D | L | SDB | ER | SDC | HL | D | L | SDB | ER | SDC | HL |
| 26 | 2 | 958.489 | 0.299 | 958.436 | 3.86 | 01 | 2 | 958.482 | 0.171 | 958.479 | 4.18 |
| 26 | 2 | 958.489 | 0.299 | 958.499 | 3.82 | 01 | 2 | 958.906 | 0.119 | 958.899 | 4.34 |
| 26 | 2 | 958.192 | 0.207 | 958.214 | 3.78 | 01 | 1 | 959.653 | 0.132 | 959.698 | 3.27 |
| 26 | 2 | 959.080 | 0.245 | 959.228 | 3.70 | 01 | 1 | 958.389 | 0.160 | 958.358 | 3.22 |
| 26 | 2 | 958.806 | 0.207 | 958.791 | 3.67 | 01 | 1 | 959.653 | 0.132 | 960.215 | 3.18 |
| 26 | 2 | 958.192 | 0.207 | 958.196 | 3.67 | 01 | 1 | 958.817 | 0.147 | 958.847 | 3.17 |
| 26 | 2 | 958.659 | 0.183 | 958.681 | 3.68 | 01 | 1 | 958.817 | 0.147 | 958.825 | 3.17 |
| 26 | 2 | 958.931 | 0.282 | 958.911 | 3.69 | 01 | 1 | 958.389 | 0.160 | 958.409 | 3.21 |
| 26 | 2 | 957.888 | 0.248 | 957.963 | 3.72 | 01 | 1 | 959.017 | 0.165 | 959.022 | 3.24 |
| 26 | 2 | 959.080 | 0.245 | 959.241 | 3.83 | 01 | 1 | 958.919 | 0.122 | 958.916 | 3.29 |
| 26 | 2 | 959.080 | 0.245 | 959.012 | 3.89 | 01 | 1 | 958.461 | 0.155 | 958.459 | 3.34 |
| 26 | 1 | 959.080 | 0.245 | 959.149 | 3.95 | 01 | 1 | 959.291 | 0.138 | 959.301 | 3.40 |
| 26 | 1 | 958.192 | 0.207 | 958.186 | 3.90 | 01 | 1 | 959.653 | 0.132 | 960.513 | 3.47 |
| 26 | 1 | 958.489 | 0.299 | 958.436 | 3.86 | 01 | 1 | 958.943 | 0.156 | 958.943 | 3.66 |
| 26 | 1 | 958.489 | 0.299 | 958.499 | 3.82 | 01 | 1 | 958.274 | 0.123 | 958.188 | 4.02 |
| 26 | 1 | 958.192 | 0.207 | 958.214 | 3.78 | 01 | 1 | 958.482 | 0.171 | 958.479 | 4.18 |
| 26 | 1 | 959.080 | 0.245 | 959.228 | 3.70 | 01 | 1 | 958.906 | 0.119 | 958.899 | 4.34 |
| 26 | 1 | 958.806 | 0.207 | 958.791 | 3.67 | 01 | 2 | 958.473 | 0.150 | 958.468 | 5.01 |
| 26 | 1 | 958.192 | 0.207 | 958.196 | 3.67 | 01 | 2 | 958.535 | 0.146 | 958.612 | 5.25 |
| 26 | 1 | 958.659 | 0.183 | 958.681 | 3.68 | 01 | 2 | 959.291 | 0.138 | 959.284 | 5.52 |
| 26 | 1 | 958.931 | 0.282 | 958.911 | 3.69 | 01 | 1 | 958.473 | 0.150 | 958.468 | 5.01 |
| 26 | 1 | 957.888 | 0.248 | 957.963 | 3.72 | 01 | 1 | 958.535 | 0.146 | 958.612 | 5.25 |
| 26 | 1 | 959.080 | 0.245 | 959.241 | 3.83 | 01 | 1 | 959.291 | 0.138 | 959.284 | 5.52 |
| 26 | 1 | 959.080 | 0.245 | 959.012 | 3.89 | 02 | 2 | 958.600 | 0.198 | 958.688 | 3.08 |
| 29 | 2 | 959.170 | 0.158 | 959.182 | 3.66 | 02 | 2 | 957.486 | 0.374 | 957.782 | 3.08 |
| 29 | 2 | 958.375 | 0.158 | 958.427 | 3.57 | 02 | 2 | 959.381 | 0.212 | 959.372 | 3.10 |
| 29 | 2 | 958.853 | 0.174 | 958.869 | 3.53 | 02 | 2 | 958.974 | 0.253 | 958.963 | 3.12 |
| 29 | 2 | 959.003 | 0.223 | 958.997 | 3.50 | 02 | 2 | 959.381 | 0.212 | 959.342 | 3.21 |
| 29 | 2 | 958.657 | 0.169 | 958.642 | 3.47 | 02 | 2 | 958.974 | 0.253 | 958.969 | 3.25 |
| 29 | 2 | 958.853 | 0.174 | 958.881 | 3.44 | 02 | 2 | 958.974 | 0.253 | 958.975 | 3.30 |
| 29 | 2 | 958.272 | 0.162 | 958.192 | 3.43 | 02 | 2 | 959.381 | 0.212 | 959.477 | 3.43 |
| 29 | 2 | 958.371 | 0.163 | 958.353 | 3.43 | 02 | 2 | 958.240 | 0.210 | 958.285 | 3.50 |
| 29 | 2 | 957.911 | 0.171 | 957.779 | 3.43 | 02 | 2 | 959.816 | 0.224 | 960.042 | 3.68 |
| 29 | 2 | 958.043 | 0.159 | 958.156 | 3.53 | 02 | 2 | 958.974 | 0.253 | 958.955 | 3.89 |
| 29 | 2 | 959.247 | 0.146 | 959.589 | 3.57 | 02 | 2 | 958.240 | 0.210 | 958.362 | 4.01 |
| 29 | 2 | 959.049 | 0.157 | 959.070 | 3.76 | 02 | 2 | 959.381 | 0.212 | 959.430 | 4.27 |
| 29 | 2 | 959.247 | 0.146 | 959.252 | 3.84 | 02 | 2 | 959.110 | 0.152 | 959.223 | 4.43 |
| 29 | 2 | 958.657 | 0.169 | 958.651 | 3.94 | 02 | 2 | 958.088 | 0.209 | 958.093 | 4.60 |
| 29 | 2 | 959.247 | 0.146 | 959.611 | 4.05 | 02 | 1 | 958.600 | 0.198 | 958.688 | 3.08 |
| 29 | 1 | 959.170 | 0.158 | 959.182 | 3.66 | 02 | 1 | 957.486 | 0.374 | 957.782 | 3.08 |
| 29 | 1 | 958.375 | 0.158 | 958.427 | 3.57 | 02 | 1 | 959.381 | 0.212 | 959.372 | 3.10 |
| 29 | 1 | 958.853 | 0.174 | 958.869 | 3.53 | 02 | 1 | 958.974 | 0.253 | 958.963 | 3.12 |
| 29 | 1 | 959.003 | 0.223 | 958.997 | 3.50 | 02 | 1 | 959.381 | 0.212 | 959.342 | 3.21 |
| 29 | 1 | 958.657 | 0.169 | 958.642 | 3.47 | 02 | 1 | 958.974 | 0.253 | 958.969 | 3.25 |
| 29 | 1 | 958.853 | 0.174 | 958.881 | 3.44 | 02 | 1 | 958.974 | 0.253 | 958.975 | 3.30 |
| 29 | 1 | 958.272 | 0.162 | 958.192 | 3.43 | 02 | 1 | 959.381 | 0.212 | 959.477 | 3.43 |
| 29 | 1 | 958.371 | 0.163 | 958.353 | 3.43 | 02 | 1 | 958.240 | 0.210 | 958.285 | 3.50 |
| 29 | 1 | 957.911 | 0.171 | 957.779 | 3.43 | 02 | 1 | 959.816 | 0.224 | 960.042 | 3.68 |
| 29 | 1 | 958.043 | 0.159 | 958.156 | 3.53 | 02 | 1 | 958.974 | 0.253 | 958.955 | 3.89 |
| 29 | 1 | 959.247 | 0.146 | 959.589 | 3.57 | 02 | 1 | 958.240 | 0.210 | 958.362 | 4.01 |
| 29 | 1 | 959.049 | 0.157 | 959.070 | 3.76 | 02 | 1 | 959.381 | 0.212 | 959.430 | 4.27 |
| 29 | 1 | 959.247 | 0.146 | 959.252 | 3.84 | 02 | 1 | 959.110 | 0.152 | 959.223 | 4.43 |
| 29 | 1 | 958.657 | 0.169 | 958.651 | 3.94 | 02 | 1 | 958.088 | 0.209 | 958.093 | 4.60 |
| 29 | 1 | 959.247 | 0.146 | 959.611 | 4.05 | 03 | 2 | 959.512 | 0.111 | 959.358 | 3.04 |
|  |  |  |  |  |  | 03 | 2 | 959.190 | 0.138 | 959.232 | 3.02 |
|  |  |  |  |  |  | 03 | 2 | 959.512 | 0.111 | 959.551 | 2.99 |
|     | 1999 - FEVEREIRO |     |     |     |     | 03 | 2 | 959.512 | 0.111 | 959.899 | 2.98 |
| D | L | SDB | ER | SDC | HL | 03 | 2 | 958.937 | 0.147 | 958.926 | 2.98 |
| 01 | 2 | 959.653 | 0.132 | 959.698 | 3.27 | 03 | 2 | 958.486 | 0.139 | 958.631 | 2.99 |
| 01 | 2 | 958.389 | 0.160 | 958.358 | 3.22 | 03 | 2 | 958.937 | 0.147 | 958.994 | 3.01 |
| 01 | 2 | 959.653 | 0.132 | 960.215 | 3.18 | 03 | 2 | 958.801 | 0.156 | 958.835 | 3.05 |
| 01 | 2 | 958.817 | 0.147 | 958.847 | 3.17 | 03 | 2 | 958.410 | 0.144 | 958.358 | 3.10 |
| 01 | 2 | 958.817 | 0.147 | 958.825 | 3.17 | 03 | 2 | 958.937 | 0.147 | 958.904 | 3.14 |
| 01 | 2 | 958.389 | 0.160 | 958.409 | 3.21 | 03 | 2 | 959.512 | 0.111 | 959.765 | 3.19 |
| 01 | 2 | 959.017 | 0.165 | 959.022 | 3.24 | 03 | 2 | 958.937 | 0.147 | 958.942 | 3.26 |
| 01 | 2 | 958.919 | 0.122 | 958.916 | 3.29 | 03 | 2 | 958.937 | 0.147 | 958.909 | 3.34 |
| 01 | 2 | 958.461 | 0.155 | 958.459 | 3.34 | 03 | 2 | 958.937 | 0.147 | 959.006 | 3.42 |
| 01 | 2 | 959.291 | 0.138 | 959.301 | 3.40 | 03 | 2 | 958.410 | 0.144 | 958.399 | 3.51 |
| 01 | 2 | 959.653 | 0.132 | 960.513 | 3.47 | 03 | 2 | 958.520 | 0.159 | 958.519 | 3.60 |
| 01 | 2 | 958.943 | 0.156 | 958.943 | 3.66 | 03 | 2 | 958.460 | 0.157 | 958.471 | 3.90 |
| 01 | 2 | 958.274 | 0.123 | 958.188 | 4.02 | 03 | 2 | 958.292 | 0.163 | 958.242 | 4.04 |

|    |   |    1999 - FEVEREIRO |       |         |      |    |   |    1999 - FEVEREIRO |       |         |      |
|----|---|---------|-------|---------|------|----|---|---------|-------|---------|------|
| D  | L | SDB     | ER    | SDC     | HL   | D  | L | SDB     | ER    | SDC     | HL   |
| 03 | 2 | 958.029 | 0.135 | 957.997 | 4.22 | 05 | 2 | 957.324 | 0.328 | 957.522 | 4.90 |
| 03 | 2 | 958.937 | 0.147 | 958.995 | 4.39 | 05 | 1 | 959.909 | 0.262 | 959.969 | 2.84 |
| 03 | 1 | 959.512 | 0.111 | 959.358 | 3.04 | 05 | 1 | 959.777 | 0.358 | 959.736 | 2.87 |
| 03 | 1 | 959.190 | 0.138 | 959.232 | 3.02 | 05 | 1 | 958.518 | 0.363 | 958.447 | 2.91 |
| 03 | 1 | 959.512 | 0.111 | 959.551 | 2.99 | 05 | 1 | 958.254 | 0.246 | 958.264 | 2.95 |
| 03 | 1 | 959.512 | 0.111 | 959.899 | 2.98 | 05 | 1 | 958.978 | 0.238 | 959.313 | 3.07 |
| 03 | 1 | 958.937 | 0.147 | 958.926 | 2.98 | 05 | 1 | 958.978 | 0.238 | 959.053 | 3.14 |
| 03 | 1 | 958.631 | 0.139 | 958.631 | 2.99 | 05 | 1 | 957.957 | 0.440 | 957.955 | 3.38 |
| 03 | 1 | 958.937 | 0.147 | 958.994 | 3.01 | 05 | 1 | 958.978 | 0.238 | 959.045 | 3.74 |
| 03 | 1 | 958.801 | 0.156 | 958.835 | 3.05 | 05 | 1 | 958.107 | 0.420 | 958.040 | 3.90 |
| 03 | 1 | 958.410 | 0.144 | 958.358 | 3.10 | 05 | 1 | 958.518 | 0.363 | 958.508 | 4.05 |
| 03 | 1 | 958.937 | 0.147 | 958.904 | 3.14 | 05 | 1 | 958.713 | 0.353 | 958.725 | 4.23 |
| 03 | 1 | 959.512 | 0.111 | 959.765 | 3.19 | 05 | 1 | 957.324 | 0.328 | 957.522 | 4.90 |
| 03 | 1 | 958.937 | 0.147 | 958.942 | 3.26 | 08 | 2 | 959.476 | 0.113 | 959.485 | 2.57 |
| 03 | 1 | 958.937 | 0.147 | 958.909 | 3.34 | 08 | 2 | 959.254 | 0.140 | 959.271 | 2.61 |
| 03 | 1 | 958.937 | 0.147 | 959.006 | 3.42 | 08 | 2 | 959.019 | 0.152 | 959.009 | 2.66 |
| 03 | 1 | 958.410 | 0.144 | 958.399 | 3.51 | 08 | 2 | 959.254 | 0.140 | 959.297 | 2.72 |
| 03 | 1 | 958.520 | 0.159 | 958.519 | 3.60 | 08 | 2 | 958.353 | 0.150 | 958.411 | 2.79 |
| 03 | 1 | 958.460 | 0.157 | 958.471 | 3.90 | 08 | 2 | 959.373 | 0.135 | 959.385 | 2.86 |
| 03 | 1 | 958.292 | 0.163 | 958.242 | 4.04 | 08 | 2 | 959.166 | 0.124 | 959.181 | 2.95 |
| 03 | 1 | 958.029 | 0.135 | 957.997 | 4.22 | 08 | 2 | 959.635 | 0.151 | 959.628 | 3.04 |
| 03 | 1 | 958.937 | 0.147 | 958.995 | 4.39 | 08 | 2 | 958.810 | 0.144 | 958.829 | 3.13 |
| 03 | 2 | 958.937 | 0.147 | 958.964 | 4.77 | 08 | 2 | 959.346 | 0.108 | 959.302 | 3.24 |
| 03 | 2 | 958.574 | 0.156 | 958.591 | 5.00 | 08 | 2 | 959.860 | 0.157 | 959.812 | 3.36 |
| 03 | 2 | 958.029 | 0.135 | 957.872 | 5.23 | 08 | 2 | 959.860 | 0.157 | 959.900 | 3.51 |
| 03 | 2 | 958.106 | 0.145 | 958.082 | 5.50 | 08 | 2 | 959.043 | 0.108 | 959.048 | 3.66 |
| 03 | 1 | 958.937 | 0.147 | 958.964 | 4.77 | 08 | 2 | 959.860 | 0.157 | 959.911 | 3.80 |
| 03 | 1 | 958.574 | 0.156 | 958.591 | 5.00 | 08 | 2 | 958.256 | 0.167 | 958.268 | 4.01 |
| 03 | 1 | 958.029 | 0.135 | 957.872 | 5.23 | 08 | 2 | 958.955 | 0.167 | 958.946 | 4.19 |
| 03 | 1 | 958.106 | 0.145 | 958.082 | 5.50 | 08 | 2 | 958.353 | 0.150 | 958.509 | 4.39 |
| 04 | 2 | 959.493 | 0.224 | 959.467 | 2.89 | 08 | 2 | 959.054 | 0.167 | 959.076 | 4.64 |
| 04 | 2 | 958.529 | 0.190 | 958.538 | 2.90 | 08 | 2 | 959.019 | 0.152 | 959.001 | 4.93 |
| 04 | 2 | 959.273 | 0.175 | 959.254 | 2.92 | 08 | 2 | 958.810 | 0.144 | 958.826 | 5.17 |
| 04 | 2 | 959.122 | 0.170 | 959.084 | 2.94 | 08 | 2 | 959.159 | 0.148 | 959.128 | 5.43 |
| 04 | 2 | 959.273 | 0.175 | 959.267 | 3.04 | 08 | 1 | 959.476 | 0.113 | 959.485 | 2.57 |
| 04 | 2 | 957.835 | 0.265 | 957.818 | 3.08 | 08 | 1 | 959.254 | 0.140 | 959.271 | 2.61 |
| 04 | 2 | 959.493 | 0.224 | 959.667 | 3.13 | 08 | 1 | 959.019 | 0.152 | 959.009 | 2.66 |
| 04 | 2 | 958.319 | 0.242 | 958.267 | 3.26 | 08 | 1 | 959.254 | 0.140 | 959.297 | 2.72 |
| 04 | 2 | 958.760 | 0.206 | 958.764 | 3.33 | 08 | 1 | 958.353 | 0.150 | 958.411 | 2.79 |
| 04 | 2 | 959.384 | 0.222 | 959.401 | 3.61 | 08 | 1 | 959.373 | 0.135 | 959.385 | 2.86 |
| 04 | 2 | 957.835 | 0.265 | 957.816 | 4.09 | 08 | 1 | 959.166 | 0.124 | 959.181 | 2.95 |
| 04 | 2 | 959.493 | 0.224 | 959.467 | 4.24 | 08 | 1 | 959.635 | 0.151 | 959.628 | 3.04 |
| 04 | 2 | 959.122 | 0.170 | 959.080 | 4.57 | 08 | 1 | 958.810 | 0.144 | 958.829 | 3.13 |
| 04 | 1 | 959.493 | 0.224 | 959.467 | 2.89 | 08 | 1 | 959.346 | 0.108 | 959.302 | 3.24 |
| 04 | 1 | 958.529 | 0.190 | 958.538 | 2.90 | 08 | 1 | 959.860 | 0.157 | 959.812 | 3.36 |
| 04 | 1 | 959.273 | 0.175 | 959.254 | 2.92 | 08 | 1 | 959.860 | 0.157 | 959.900 | 3.51 |
| 04 | 1 | 959.122 | 0.170 | 959.084 | 2.94 | 08 | 1 | 959.043 | 0.108 | 959.048 | 3.66 |
| 04 | 1 | 959.273 | 0.175 | 959.267 | 3.04 | 08 | 1 | 959.860 | 0.157 | 959.911 | 3.80 |
| 04 | 1 | 957.835 | 0.265 | 957.818 | 3.08 | 08 | 1 | 958.256 | 0.167 | 958.268 | 4.01 |
| 04 | 1 | 959.493 | 0.224 | 959.667 | 3.13 | 08 | 1 | 958.955 | 0.167 | 958.946 | 4.19 |
| 04 | 1 | 958.319 | 0.242 | 958.267 | 3.26 | 08 | 1 | 958.353 | 0.150 | 958.509 | 4.39 |
| 04 | 1 | 958.760 | 0.206 | 958.764 | 3.33 | 08 | 1 | 959.054 | 0.167 | 959.076 | 4.64 |
| 04 | 1 | 959.384 | 0.222 | 959.401 | 3.61 | 08 | 1 | 959.019 | 0.152 | 959.001 | 4.93 |
| 04 | 1 | 957.835 | 0.265 | 957.816 | 4.09 | 08 | 1 | 958.810 | 0.144 | 958.826 | 5.17 |
| 04 | 1 | 959.493 | 0.224 | 959.467 | 4.24 | 08 | 1 | 959.159 | 0.148 | 959.128 | 5.43 |
| 04 | 1 | 959.122 | 0.170 | 959.080 | 4.57 | 08 | 2 | 959.166 | 0.124 | 959.175 | 5.75 |
| 04 | 2 | 958.319 | 0.242 | 958.295 | 4.95 | 08 | 2 | 959.860 | 0.157 | 960.438 | 6.09 |
| 04 | 2 | 959.122 | 0.170 | 959.094 | 5.16 | 08 | 2 | 958.868 | 0.156 | 958.882 | 6.42 |
| 04 | 1 | 958.319 | 0.242 | 958.295 | 4.95 | 08 | 2 | 958.353 | 0.150 | 958.417 | 6.78 |
| 04 | 1 | 959.122 | 0.170 | 959.094 | 5.16 | 08 | 2 | 959.254 | 0.140 | 959.251 | 7.16 |
| 05 | 2 | 959.909 | 0.262 | 959.969 | 2.84 | 08 | 1 | 959.166 | 0.124 | 959.175 | 5.75 |
| 05 | 2 | 959.777 | 0.358 | 959.736 | 2.87 | 08 | 1 | 959.860 | 0.157 | 960.438 | 6.09 |
| 05 | 2 | 958.518 | 0.363 | 958.447 | 2.91 | 08 | 1 | 958.868 | 0.156 | 958.882 | 6.42 |
| 05 | 2 | 958.254 | 0.246 | 958.264 | 2.95 | 08 | 1 | 958.353 | 0.150 | 958.417 | 6.78 |
| 05 | 2 | 958.978 | 0.238 | 959.313 | 3.07 | 08 | 1 | 959.254 | 0.140 | 959.251 | 7.16 |
| 05 | 2 | 958.978 | 0.238 | 959.053 | 3.14 | 09 | 2 | 957.726 | 0.209 | 957.779 | 2.47 |
| 05 | 2 | 957.957 | 0.440 | 957.955 | 3.38 | 09 | 2 | 958.569 | 0.220 | 958.536 | 2.54 |
| 05 | 2 | 958.978 | 0.238 | 959.045 | 3.74 | 09 | 2 | 958.447 | 0.312 | 958.456 | 2.74 |
| 05 | 2 | 958.107 | 0.420 | 958.040 | 3.90 | 09 | 2 | 957.638 | 0.164 | 957.546 | 2.83 |
| 05 | 2 | 958.518 | 0.363 | 958.508 | 4.05 | 09 | 2 | 957.970 | 0.230 | 957.919 | 3.16 |
| 05 | 2 | 958.713 | 0.353 | 958.725 | 4.23 | 09 | 2 | 959.092 | 0.190 | 959.190 | 3.51 |

| 1999 - FEVEREIRO | | | | |
|---|---|---|---|---|
| D  L | SDB | ER | SDC | HL |
| 09 2 | 958.184 | 0.333 | 958.185 | 3.64 |
| 09 2 | 959.472 | 0.728 | 959.420 | 4.09 |
| 09 2 | 958.319 | 0.230 | 958.288 | 4.45 |
| 09 2 | 958.905 | 0.346 | 958.926 | 4.64 |
| 09 2 | 958.319 | 0.230 | 958.308 | 4.86 |
| 09 2 | 959.472 | 0.728 | 959.485 | 5.08 |
| 09 2 | 959.011 | 0.299 | 959.002 | 5.32 |
| 09 2 | 959.092 | 0.190 | 959.087 | 5.58 |
| 09 1 | 957.726 | 0.209 | 957.779 | 2.47 |
| 09 1 | 958.569 | 0.220 | 958.536 | 2.54 |
| 09 1 | 958.447 | 0.312 | 958.456 | 2.74 |
| 09 1 | 957.638 | 0.164 | 957.546 | 2.83 |
| 09 1 | 957.970 | 0.230 | 957.919 | 3.16 |
| 09 1 | 959.092 | 0.190 | 959.190 | 3.51 |
| 09 1 | 958.184 | 0.333 | 958.185 | 3.64 |
| 09 1 | 959.472 | 0.728 | 959.420 | 4.09 |
| 09 1 | 958.319 | 0.230 | 958.288 | 4.45 |
| 09 1 | 958.905 | 0.346 | 958.926 | 4.64 |
| 09 1 | 958.319 | 0.230 | 958.308 | 4.86 |
| 09 1 | 959.472 | 0.728 | 959.485 | 5.08 |
| 09 1 | 959.011 | 0.299 | 959.002 | 5.32 |
| 09 1 | 959.092 | 0.190 | 959.087 | 5.58 |
| 09 2 | 957.198 | 0.652 | 957.280 | 5.85 |
| 09 2 | 959.092 | 0.190 | 959.246 | 6.14 |
| 09 2 | 959.472 | 0.728 | 959.320 | 6.46 |
| 09 2 | 959.011 | 0.299 | 959.022 | 7.17 |
| 09 1 | 957.198 | 0.652 | 957.280 | 5.85 |
| 09 1 | 959.092 | 0.190 | 959.246 | 6.14 |
| 09 1 | 959.472 | 0.728 | 959.320 | 6.46 |
| 09 1 | 959.011 | 0.299 | 959.022 | 7.17 |
| 15 2 | 958.892 | 0.132 | 958.903 | 6.31 |
| 15 2 | 958.201 | 0.203 | 958.246 | 6.71 |
| 15 1 | 958.892 | 0.132 | 958.903 | 6.31 |
| 15 1 | 958.201 | 0.203 | 958.246 | 6.71 |
| 15 2 | 958.892 | 0.132 | 958.873 | 7.14 |
| 15 2 | 959.038 | 0.130 | 959.161 | 7.59 |
| 15 2 | 959.038 | 0.130 | 959.100 | 8.14 |
| 15 2 | 958.201 | 0.203 | 958.208 | 8.66 |
| 15 2 | 958.557 | 0.171 | 958.497 | 9.32 |
| 15 2 | 958.831 | 0.128 | 958.813 | 9.94 |
| 15 2 | 959.484 | 0.157 | 959.445 | 10.59 |
| 15 1 | 958.892 | 0.132 | 958.873 | 7.14 |
| 15 1 | 959.038 | 0.130 | 959.161 | 7.59 |
| 15 1 | 959.038 | 0.130 | 959.100 | 8.14 |
| 15 1 | 958.201 | 0.203 | 958.208 | 8.66 |
| 15 1 | 958.557 | 0.171 | 958.497 | 9.32 |
| 15 1 | 958.831 | 0.128 | 958.813 | 9.94 |
| 15 1 | 959.484 | 0.157 | 959.445 | 10.59 |
| 18 2 | 959.344 | 0.240 | 959.327 | 2.40 |
| 18 2 | 958.844 | 0.220 | 958.899 | 2.52 |
| 18 2 | 958.447 | 0.166 | 958.473 | 3.04 |
| 18 2 | 958.523 | 0.145 | 958.507 | 3.19 |
| 18 2 | 959.512 | 0.165 | 959.559 | 3.35 |
| 18 2 | 958.719 | 0.194 | 958.680 | 3.52 |
| 18 2 | 959.512 | 0.165 | 959.561 | 3.71 |
| 18 2 | 958.798 | 0.228 | 958.797 | 3.90 |
| 18 2 | 958.964 | 0.201 | 958.950 | 4.11 |
| 18 2 | 959.715 | 0.174 | 959.668 | 4.32 |
| 18 2 | 959.512 | 0.165 | 959.568 | 4.55 |
| 18 2 | 959.715 | 0.174 | 959.687 | 4.82 |
| 18 2 | 958.447 | 0.166 | 958.434 | 5.07 |
| 18 2 | 959.512 | 0.165 | 959.602 | 5.34 |
| 18 2 | 959.218 | 0.206 | 959.174 | 5.63 |
| 18 2 | 959.042 | 0.197 | 959.079 | 5.95 |
| 18 2 | 959.715 | 0.174 | 959.807 | 6.27 |
| 18 2 | 958.523 | 0.145 | 958.508 | 6.60 |
| 18 1 | 959.344 | 0.240 | 959.327 | 2.40 |
| 18 1 | 958.844 | 0.220 | 958.899 | 2.52 |
| 18 1 | 958.447 | 0.166 | 958.473 | 3.04 |
| 18 1 | 958.523 | 0.145 | 958.507 | 3.19 |
| 18 1 | 959.512 | 0.165 | 959.559 | 3.35 |

| 1999 - FEVEREIRO | | | | |
|---|---|---|---|---|
| D  L | SDB | ER | SDC | HL |
| 18 1 | 958.719 | 0.194 | 958.680 | 3.52 |
| 18 1 | 959.512 | 0.165 | 959.561 | 3.71 |
| 18 1 | 958.798 | 0.228 | 958.797 | 3.90 |
| 18 1 | 958.964 | 0.201 | 958.950 | 4.11 |
| 18 1 | 959.715 | 0.174 | 959.668 | 4.32 |
| 18 1 | 959.512 | 0.165 | 959.568 | 4.55 |
| 18 1 | 959.715 | 0.174 | 959.687 | 4.82 |
| 18 1 | 958.447 | 0.166 | 958.434 | 5.07 |
| 18 1 | 959.512 | 0.165 | 959.602 | 5.34 |
| 18 1 | 959.218 | 0.206 | 959.174 | 5.63 |
| 18 1 | 959.042 | 0.197 | 959.079 | 5.95 |
| 18 1 | 959.715 | 0.174 | 959.807 | 6.27 |
| 18 1 | 958.523 | 0.145 | 958.508 | 6.60 |
| 22 2 | 958.551 | 0.182 | 958.439 | 2.83 |
| 22 2 | 958.679 | 0.213 | 958.680 | 3.03 |
| 22 2 | 959.270 | 0.189 | 959.346 | 3.21 |
| 22 2 | 958.268 | 0.188 | 958.323 | 3.41 |
| 22 2 | 958.777 | 0.225 | 958.811 | 3.62 |
| 22 2 | 958.268 | 0.188 | 958.371 | 3.89 |
| 22 2 | 958.777 | 0.225 | 958.804 | 4.14 |
| 22 2 | 959.270 | 0.189 | 959.103 | 4.40 |
| 22 2 | 958.268 | 0.188 | 958.384 | 4.68 |
| 22 2 | 959.270 | 0.189 | 959.224 | 4.96 |
| 22 2 | 959.578 | 0.174 | 959.654 | 5.29 |
| 22 2 | 959.270 | 0.189 | 959.295 | 5.61 |
| 22 2 | 958.911 | 0.250 | 958.930 | 6.01 |
| 22 2 | 958.081 | 0.171 | 958.014 | 6.37 |
| 22 2 | 958.911 | 0.250 | 958.947 | 6.76 |
| 22 2 | 958.268 | 0.188 | 958.302 | 7.18 |
| 22 2 | 958.081 | 0.171 | 958.077 | 8.20 |
| 22 1 | 958.551 | 0.182 | 958.439 | 2.83 |
| 22 1 | 958.679 | 0.213 | 958.680 | 3.03 |
| 22 1 | 959.270 | 0.189 | 959.346 | 3.21 |
| 22 1 | 958.268 | 0.188 | 958.323 | 3.41 |
| 22 1 | 958.777 | 0.225 | 958.811 | 3.62 |
| 22 1 | 958.268 | 0.188 | 958.371 | 3.89 |
| 22 1 | 958.777 | 0.225 | 958.804 | 4.14 |
| 22 1 | 959.270 | 0.189 | 959.103 | 4.40 |
| 22 1 | 958.268 | 0.188 | 958.384 | 4.68 |
| 22 1 | 959.270 | 0.189 | 959.224 | 4.96 |
| 22 1 | 959.578 | 0.174 | 959.654 | 5.29 |
| 22 1 | 959.270 | 0.189 | 959.295 | 5.61 |
| 22 1 | 958.911 | 0.250 | 958.930 | 6.01 |
| 22 1 | 958.081 | 0.171 | 958.014 | 6.37 |
| 22 1 | 958.911 | 0.250 | 958.947 | 6.76 |
| 22 1 | 958.268 | 0.188 | 958.302 | 7.18 |
| 22 1 | 958.081 | 0.171 | 958.077 | 8.20 |
| 22 2 | 958.911 | 0.250 | 959.042 | 8.73 |
| 22 2 | 958.911 | 0.250 | 958.907 | 9.28 |
| 22 2 | 958.711 | 0.225 | 958.705 | 9.85 |
| 22 2 | 958.081 | 0.171 | 958.007 | 10.47 |
| 22 2 | 959.578 | 0.174 | 959.553 | 11.27 |
| 22 2 | 959.578 | 0.174 | 959.439 | 11.96 |
| 22 2 | 958.725 | 0.206 | 958.725 | 12.77 |
| 22 1 | 958.911 | 0.250 | 959.042 | 8.73 |
| 22 1 | 958.911 | 0.250 | 958.907 | 9.28 |
| 22 1 | 958.711 | 0.225 | 958.705 | 9.85 |
| 22 1 | 958.081 | 0.171 | 958.007 | 10.47 |
| 22 1 | 959.578 | 0.174 | 959.553 | 11.27 |
| 22 1 | 959.578 | 0.174 | 959.439 | 11.96 |
| 22 1 | 958.725 | 0.206 | 958.725 | 12.77 |

| 1999 - MARCO | | | | |
|---|---|---|---|---|
| D  L | SDB | ER | SDC | HL |
| 03 2 | 959.732 | 0.166 | 960.340 | 2.88 |
| 03 2 | 959.502 | 0.149 | 959.570 | 3.13 |
| 03 2 | 959.433 | 0.164 | 959.418 | 3.36 |
| 03 2 | 958.368 | 0.146 | 958.376 | 3.64 |
| 03 2 | 958.575 | 0.152 | 958.564 | 3.91 |
| 03 2 | 958.575 | 0.152 | 958.614 | 4.21 |

| 1999 - MARCO | | | | | | 1999 - MARCO | | | | |
|---|---|---|---|---|---|---|---|---|---|---|
| D | L | SDB | ER | SDC | HL | D | L | SDB | ER | SDC | HL |
| 03 | 2 | 958.256 | 0.193 | 958.268 | 4.51 | 08 | 2 | 958.698 | 0.186 | 958.710 | 18.86 |
| 03 | 2 | 959.203 | 0.134 | 959.235 | 4.82 | 08 | 2 | 959.200 | 0.124 | 959.209 | 19.92 |
| 03 | 2 | 958.878 | 0.139 | 958.920 | 5.16 | 08 | 2 | 958.240 | 0.187 | 958.238 | 21.16 |
| 03 | 2 | 958.384 | 0.163 | 958.382 | 5.51 | 08 | 2 | 959.033 | 0.194 | 958.928 | 22.41 |
| 03 | 2 | 958.667 | 0.205 | 958.650 | 5.85 | 08 | 1 | 958.806 | 0.180 | 958.893 | 14.04 |
| 03 | 2 | 959.502 | 0.149 | 959.580 | 6.24 | 08 | 1 | 959.963 | 0.177 | 959.987 | 14.88 |
| 03 | 2 | 958.384 | 0.163 | 958.387 | 6.61 | 08 | 1 | 958.204 | 0.209 | 958.190 | 15.75 |
| 03 | 2 | 959.203 | 0.134 | 959.209 | 6.99 | 08 | 1 | 958.698 | 0.186 | 958.606 | 16.76 |
| 03 | 2 | 957.825 | 0.179 | 957.878 | 7.51 | 08 | 1 | 960.170 | 0.128 | 960.361 | 17.74 |
| 03 | 2 | 959.732 | 0.166 | 959.775 | 7.94 | 08 | 1 | 958.698 | 0.186 | 958.710 | 18.86 |
| 03 | 2 | 959.696 | 0.179 | 959.649 | 8.44 | 08 | 1 | 959.200 | 0.124 | 959.209 | 19.92 |
| 03 | 2 | 957.953 | 0.267 | 958.031 | 9.00 | 08 | 1 | 958.240 | 0.187 | 958.238 | 21.16 |
| 03 | 2 | 958.713 | 0.231 | 958.717 | 9.60 | 08 | 1 | 959.033 | 0.194 | 958.928 | 22.41 |
| 03 | 1 | 959.732 | 0.166 | 960.340 | 2.88 | 10 | 2 | 959.711 | 0.171 | 959.816 | 4.46 |
| 03 | 1 | 959.502 | 0.149 | 959.570 | 3.13 | 10 | 2 | 959.056 | 0.150 | 959.060 | 4.87 |
| 03 | 1 | 959.433 | 0.164 | 959.418 | 3.36 | 10 | 2 | 958.460 | 0.171 | 958.371 | 5.25 |
| 03 | 1 | 958.368 | 0.146 | 958.376 | 3.64 | 10 | 2 | 958.668 | 0.188 | 958.689 | 5.64 |
| 03 | 1 | 958.575 | 0.152 | 958.564 | 3.91 | 10 | 2 | 959.316 | 0.169 | 959.344 | 6.10 |
| 03 | 1 | 958.575 | 0.152 | 958.614 | 4.21 | 10 | 2 | 958.882 | 0.145 | 958.877 | 6.48 |
| 03 | 1 | 958.256 | 0.193 | 958.268 | 4.51 | 10 | 2 | 959.226 | 0.184 | 959.220 | 6.89 |
| 03 | 1 | 959.203 | 0.134 | 959.235 | 4.82 | 10 | 2 | 958.492 | 0.129 | 958.487 | 7.32 |
| 03 | 1 | 958.878 | 0.139 | 958.920 | 5.16 | 10 | 2 | 959.226 | 0.184 | 959.211 | 7.77 |
| 03 | 1 | 958.384 | 0.163 | 958.382 | 5.51 | 10 | 2 | 958.668 | 0.188 | 958.640 | 8.32 |
| 03 | 1 | 958.667 | 0.205 | 958.650 | 5.85 | 10 | 2 | 958.977 | 0.165 | 958.947 | 8.97 |
| 03 | 1 | 959.502 | 0.149 | 959.580 | 6.24 | 10 | 2 | 958.668 | 0.188 | 958.648 | 9.57 |
| 03 | 1 | 958.384 | 0.163 | 958.387 | 6.61 | 10 | 2 | 958.254 | 0.176 | 958.136 | 10.48 |
| 03 | 1 | 959.203 | 0.134 | 959.209 | 6.99 | 10 | 2 | 959.711 | 0.171 | 959.757 | 11.16 |
| 03 | 1 | 957.825 | 0.179 | 957.878 | 7.51 | 10 | 2 | 959.034 | 0.149 | 959.029 | 11.78 |
| 03 | 1 | 959.732 | 0.166 | 959.775 | 7.94 | 10 | 2 | 959.151 | 0.127 | 959.170 | 12.63 |
| 03 | 1 | 959.696 | 0.179 | 959.649 | 8.44 | 10 | 1 | 959.711 | 0.171 | 959.816 | 4.46 |
| 03 | 1 | 957.953 | 0.267 | 958.031 | 9.00 | 10 | 1 | 959.056 | 0.150 | 959.060 | 4.87 |
| 03 | 1 | 958.713 | 0.231 | 958.717 | 9.60 | 10 | 1 | 958.460 | 0.171 | 958.371 | 5.25 |
| 03 | 2 | 958.713 | 0.231 | 958.729 | 11.96 | 10 | 1 | 958.668 | 0.188 | 958.689 | 5.64 |
| 03 | 2 | 958.235 | 0.214 | 958.219 | 12.99 | 10 | 1 | 959.316 | 0.169 | 959.344 | 6.10 |
| 03 | 2 | 958.713 | 0.231 | 958.696 | 13.99 | 10 | 1 | 958.882 | 0.145 | 958.877 | 6.48 |
| 03 | 2 | 958.713 | 0.231 | 958.764 | 14.88 | 10 | 1 | 959.226 | 0.184 | 959.220 | 6.89 |
| 03 | 2 | 958.713 | 0.231 | 958.776 | 15.83 | 10 | 1 | 958.492 | 0.129 | 958.487 | 7.32 |
| 03 | 2 | 958.713 | 0.231 | 958.721 | 16.81 | 10 | 1 | 959.226 | 0.184 | 959.211 | 7.77 |
| 03 | 2 | 958.316 | 0.155 | 958.289 | 17.95 | 10 | 1 | 958.668 | 0.188 | 958.640 | 8.32 |
| 03 | 1 | 958.713 | 0.231 | 958.729 | 11.96 | 10 | 1 | 958.977 | 0.165 | 958.947 | 8.97 |
| 03 | 1 | 958.235 | 0.214 | 958.219 | 12.99 | 10 | 1 | 958.668 | 0.188 | 958.648 | 9.57 |
| 03 | 1 | 958.713 | 0.231 | 958.696 | 13.99 | 10 | 1 | 958.254 | 0.176 | 958.136 | 10.48 |
| 03 | 1 | 958.713 | 0.231 | 958.764 | 14.88 | 10 | 1 | 959.711 | 0.171 | 959.757 | 11.16 |
| 03 | 1 | 958.713 | 0.231 | 958.776 | 15.83 | 10 | 1 | 959.034 | 0.149 | 959.029 | 11.78 |
| 03 | 1 | 958.713 | 0.231 | 958.721 | 16.81 | 10 | 1 | 959.151 | 0.127 | 959.170 | 12.63 |
| 03 | 1 | 958.316 | 0.155 | 958.289 | 17.95 | 10 | 2 | 959.316 | 0.169 | 959.329 | 13.54 |
| 08 | 2 | 959.895 | 0.129 | 959.925 | 5.37 | 10 | 2 | 959.528 | 0.143 | 959.510 | 14.43 |
| 08 | 2 | 960.114 | 0.129 | 960.127 | 5.74 | 10 | 2 | 959.711 | 0.171 | 959.674 | 15.28 |
| 08 | 2 | 959.200 | 0.124 | 959.241 | 6.10 | 10 | 2 | 958.882 | 0.145 | 958.865 | 16.17 |
| 08 | 2 | 959.491 | 0.210 | 959.478 | 6.51 | 10 | 2 | 958.977 | 0.165 | 958.981 | 17.15 |
| 08 | 2 | 959.535 | 0.124 | 959.562 | 6.94 | 10 | 2 | 960.730 | 0.166 | 960.765 | 18.43 |
| 08 | 2 | 959.200 | 0.124 | 959.310 | 7.39 | 10 | 2 | 959.711 | 0.171 | 959.731 | 19.70 |
| 08 | 2 | 959.094 | 0.133 | 959.105 | 7.82 | 10 | 2 | 959.711 | 0.171 | 959.822 | 20.81 |
| 08 | 2 | 959.491 | 0.210 | 959.470 | 8.33 | 10 | 1 | 959.316 | 0.169 | 959.329 | 13.54 |
| 08 | 2 | 959.033 | 0.194 | 959.050 | 8.81 | 10 | 1 | 959.528 | 0.143 | 959.510 | 14.43 |
| 08 | 2 | 960.114 | 0.129 | 960.128 | 9.33 | 10 | 1 | 959.711 | 0.171 | 959.674 | 15.28 |
| 08 | 1 | 959.895 | 0.129 | 959.925 | 5.37 | 10 | 1 | 958.882 | 0.145 | 958.865 | 16.17 |
| 08 | 1 | 960.114 | 0.129 | 960.127 | 5.74 | 10 | 1 | 958.977 | 0.165 | 958.981 | 17.15 |
| 08 | 1 | 959.200 | 0.124 | 959.241 | 6.10 | 10 | 1 | 960.730 | 0.166 | 960.765 | 18.43 |
| 08 | 1 | 959.491 | 0.210 | 959.478 | 6.51 | 10 | 1 | 959.711 | 0.171 | 959.731 | 19.70 |
| 08 | 1 | 959.535 | 0.124 | 959.562 | 6.94 | 10 | 1 | 959.711 | 0.171 | 959.822 | 20.81 |
| 08 | 1 | 959.200 | 0.124 | 959.310 | 7.39 | 12 | 2 | 958.514 | 0.247 | 958.355 | 8.55 |
| 08 | 1 | 959.094 | 0.133 | 959.105 | 7.82 | 12 | 2 | 958.683 | 0.892 | 958.961 | 9.10 |
| 08 | 1 | 959.491 | 0.210 | 959.470 | 8.33 | 12 | 2 | 958.121 | 0.178 | 958.311 | 11.18 |
| 08 | 1 | 959.033 | 0.194 | 959.050 | 8.81 | 12 | 2 | 958.121 | 0.178 | 958.257 | 12.84 |
| 08 | 1 | 960.114 | 0.129 | 960.128 | 9.33 | 12 | 1 | 958.514 | 0.247 | 958.355 | 8.55 |
| 08 | 2 | 958.806 | 0.180 | 958.893 | 14.04 | 12 | 1 | 958.683 | 0.892 | 958.961 | 9.10 |
| 08 | 2 | 959.963 | 0.177 | 959.987 | 14.88 | 12 | 1 | 958.121 | 0.178 | 958.311 | 11.18 |
| 08 | 2 | 958.204 | 0.209 | 958.190 | 15.75 | 12 | 1 | 958.121 | 0.178 | 958.257 | 12.84 |
| 08 | 2 | 958.698 | 0.186 | 958.606 | 16.76 | 16 | 2 | 958.251 | 0.189 | 958.224 | 7.34 |
| 08 | 2 | 960.170 | 0.128 | 960.361 | 17.74 | 16 | 2 | 958.251 | 0.189 | 958.223 | 7.76 |

| 1999 - MARCO | | | | | | 1999 - MARCO | | | | |
|---|---|---|---|---|---|---|---|---|---|---|
| D | L | SDB | ER | SDC | HL | D | L | SDB | ER | SDC | HL |
| 16 | 2 | 959.406 | 0.222 | 959.850 | 8.19 | 17 | 2 | 959.051 | 0.140 | 959.020 | 10.76 |
| 16 | 2 | 959.102 | 0.229 | 959.141 | 8.63 | 17 | 2 | 958.936 | 0.137 | 958.953 | 11.34 |
| 16 | 2 | 957.968 | 0.165 | 957.941 | 9.07 | 17 | 2 | 959.399 | 0.229 | 959.400 | 11.98 |
| 16 | 2 | 959.234 | 0.155 | 959.223 | 9.54 | 17 | 2 | 958.936 | 0.137 | 958.902 | 12.91 |
| 16 | 2 | 959.380 | 0.208 | 959.379 | 10.02 | 17 | 1 | 959.399 | 0.229 | 959.365 | 5.39 |
| 16 | 2 | 959.102 | 0.229 | 959.041 | 10.54 | 17 | 1 | 958.603 | 0.165 | 958.591 | 5.89 |
| 16 | 2 | 959.406 | 0.222 | 959.568 | 11.06 | 17 | 1 | 958.381 | 0.160 | 958.480 | 6.32 |
| 16 | 2 | 957.968 | 0.165 | 957.980 | 11.63 | 17 | 1 | 958.381 | 0.160 | 958.393 | 6.75 |
| 16 | 2 | 957.700 | 0.200 | 957.634 | 12.20 | 17 | 1 | 958.381 | 0.160 | 958.422 | 7.22 |
| 16 | 2 | 960.342 | 0.199 | 960.247 | 12.79 | 17 | 1 | 959.259 | 0.154 | 959.307 | 7.67 |
| 16 | 2 | 958.251 | 0.189 | 958.135 | 13.43 | 17 | 1 | 959.607 | 0.154 | 959.601 | 8.11 |
| 16 | 2 | 957.700 | 0.200 | 957.481 | 14.10 | 17 | 1 | 958.646 | 0.180 | 958.628 | 8.62 |
| 16 | 2 | 959.268 | 0.196 | 959.284 | 14.79 | 17 | 1 | 959.114 | 0.176 | 959.104 | 9.13 |
| 16 | 2 | 959.310 | 0.202 | 959.289 | 15.52 | 17 | 1 | 958.646 | 0.180 | 958.654 | 9.63 |
| 16 | 1 | 958.251 | 0.189 | 958.224 | 7.34 | 17 | 1 | 959.051 | 0.140 | 959.081 | 10.17 |
| 16 | 1 | 958.251 | 0.189 | 958.223 | 7.76 | 17 | 1 | 959.051 | 0.140 | 959.020 | 10.76 |
| 16 | 1 | 959.406 | 0.222 | 959.850 | 8.19 | 17 | 1 | 958.936 | 0.137 | 958.953 | 11.34 |
| 16 | 1 | 959.102 | 0.229 | 959.141 | 8.63 | 17 | 1 | 959.399 | 0.229 | 959.400 | 11.98 |
| 16 | 1 | 957.968 | 0.165 | 957.941 | 9.07 | 17 | 1 | 958.936 | 0.137 | 958.902 | 12.91 |
| 16 | 1 | 959.234 | 0.155 | 959.223 | 9.54 | 18 | 2 | 959.668 | 0.179 | 959.473 | 11.30 |
| 16 | 1 | 959.380 | 0.208 | 959.379 | 10.02 | 18 | 2 | 958.623 | 0.308 | 958.596 | 12.51 |
| 16 | 1 | 959.102 | 0.229 | 959.041 | 10.54 | 18 | 2 | 959.755 | 0.189 | 959.778 | 13.54 |
| 16 | 1 | 959.406 | 0.222 | 959.568 | 11.06 | 18 | 1 | 959.668 | 0.179 | 959.473 | 11.30 |
| 16 | 1 | 957.968 | 0.165 | 957.980 | 11.63 | 18 | 1 | 958.623 | 0.308 | 958.596 | 12.51 |
| 16 | 1 | 957.700 | 0.200 | 957.634 | 12.20 | 18 | 1 | 959.755 | 0.189 | 959.778 | 13.54 |
| 16 | 1 | 960.342 | 0.199 | 960.247 | 12.79 | 18 | 2 | 958.106 | 0.228 | 958.204 | 18.30 |
| 16 | 1 | 958.251 | 0.189 | 958.135 | 13.43 | 18 | 2 | 958.767 | 0.215 | 958.802 | 19.23 |
| 16 | 1 | 957.700 | 0.200 | 957.481 | 14.10 | 18 | 2 | 958.767 | 0.215 | 958.757 | 20.25 |
| 16 | 1 | 959.268 | 0.196 | 959.284 | 14.79 | 18 | 2 | 959.253 | 0.300 | 959.103 | 21.30 |
| 16 | 1 | 959.310 | 0.202 | 959.289 | 15.52 | 18 | 2 | 958.106 | 0.228 | 958.106 | 22.39 |
| 16 | 2 | 958.731 | 0.160 | 958.728 | 16.47 | 18 | 2 | 958.767 | 0.215 | 958.962 | 23.58 |
| 16 | 2 | 959.406 | 0.222 | 959.718 | 17.30 | 18 | 2 | 958.387 | 0.259 | 958.398 | 26.34 |
| 16 | 2 | 958.251 | 0.189 | 958.121 | 18.19 | 18 | 2 | 958.767 | 0.215 | 958.762 | 27.80 |
| 16 | 2 | 958.919 | 0.237 | 958.947 | 19.30 | 18 | 2 | 959.253 | 0.300 | 959.050 | 29.41 |
| 16 | 2 | 958.373 | 0.207 | 958.374 | 20.30 | 18 | 2 | 958.106 | 0.228 | 958.014 | 31.36 |
| 16 | 2 | 959.102 | 0.229 | 959.109 | 21.32 | 18 | 2 | 957.856 | 0.410 | 957.888 | 33.25 |
| 16 | 2 | 958.865 | 0.179 | 958.854 | 22.43 | 18 | 1 | 958.106 | 0.228 | 958.204 | 18.30 |
| 16 | 2 | 958.919 | 0.237 | 958.901 | 23.61 | 18 | 1 | 958.767 | 0.215 | 958.802 | 19.23 |
| 16 | 2 | 958.803 | 0.188 | 958.771 | 24.87 | 18 | 1 | 958.767 | 0.215 | 958.757 | 20.25 |
| 16 | 2 | 959.406 | 0.222 | 959.796 | 26.20 | 18 | 1 | 959.253 | 0.300 | 959.103 | 21.30 |
| 16 | 2 | 958.251 | 0.189 | 958.272 | 27.64 | 18 | 1 | 958.106 | 0.228 | 958.106 | 22.39 |
| 16 | 2 | 959.102 | 0.229 | 959.030 | 29.23 | 18 | 1 | 958.767 | 0.215 | 958.962 | 23.58 |
| 16 | 2 | 958.251 | 0.189 | 958.191 | 30.98 | 18 | 1 | 958.387 | 0.259 | 958.398 | 26.34 |
| 16 | 2 | 959.374 | 0.181 | 959.356 | 32.90 | 18 | 1 | 958.767 | 0.215 | 958.762 | 27.80 |
| 16 | 2 | 957.700 | 0.200 | 957.592 | 35.01 | 18 | 1 | 959.253 | 0.300 | 959.050 | 29.41 |
| 16 | 1 | 958.731 | 0.160 | 958.728 | 16.47 | 18 | 1 | 958.106 | 0.228 | 958.014 | 31.36 |
| 16 | 1 | 959.406 | 0.222 | 959.718 | 17.30 | 18 | 1 | 957.856 | 0.410 | 957.888 | 33.25 |
| 16 | 1 | 958.251 | 0.189 | 958.121 | 18.19 | 19 | 2 | 959.138 | 0.231 | 959.164 | 9.40 |
| 16 | 1 | 958.919 | 0.237 | 958.947 | 19.30 | 19 | 2 | 959.787 | 0.210 | 959.772 | 9.94 |
| 16 | 1 | 958.373 | 0.207 | 958.374 | 20.30 | 19 | 2 | 958.846 | 0.194 | 958.877 | 10.47 |
| 16 | 1 | 959.102 | 0.229 | 959.109 | 21.32 | 19 | 2 | 959.138 | 0.231 | 959.136 | 11.02 |
| 16 | 1 | 958.865 | 0.179 | 958.854 | 22.43 | 19 | 2 | 958.514 | 0.211 | 958.506 | 11.67 |
| 16 | 1 | 958.919 | 0.237 | 958.901 | 23.61 | 19 | 2 | 958.514 | 0.211 | 958.484 | 12.27 |
| 16 | 1 | 958.803 | 0.188 | 958.771 | 24.87 | 19 | 2 | 958.687 | 0.250 | 958.690 | 12.94 |
| 16 | 1 | 959.406 | 0.222 | 959.796 | 26.20 | 19 | 2 | 959.787 | 0.210 | 959.933 | 13.67 |
| 16 | 1 | 958.251 | 0.189 | 958.272 | 27.64 | 19 | 2 | 958.516 | 0.193 | 958.521 | 14.43 |
| 16 | 1 | 959.102 | 0.229 | 959.030 | 29.23 | 19 | 2 | 959.787 | 0.210 | 960.021 | 15.22 |
| 16 | 1 | 958.251 | 0.189 | 958.191 | 30.98 | 19 | 2 | 959.000 | 0.210 | 959.049 | 16.01 |
| 16 | 1 | 959.374 | 0.181 | 959.356 | 32.90 | 19 | 1 | 959.138 | 0.231 | 959.164 | 9.40 |
| 16 | 1 | 957.700 | 0.200 | 957.592 | 35.01 | 19 | 1 | 959.787 | 0.210 | 959.772 | 9.94 |
| 17 | 2 | 959.399 | 0.229 | 959.365 | 5.39 | 19 | 1 | 958.846 | 0.194 | 958.877 | 10.47 |
| 17 | 2 | 958.603 | 0.165 | 958.591 | 5.89 | 19 | 1 | 959.138 | 0.231 | 959.136 | 11.02 |
| 17 | 2 | 958.381 | 0.160 | 958.480 | 6.32 | 19 | 1 | 958.514 | 0.211 | 958.506 | 11.67 |
| 17 | 2 | 958.381 | 0.160 | 958.393 | 6.75 | 19 | 1 | 958.514 | 0.211 | 958.484 | 12.27 |
| 17 | 2 | 958.381 | 0.160 | 958.422 | 7.22 | 19 | 1 | 958.687 | 0.250 | 958.690 | 12.94 |
| 17 | 2 | 959.259 | 0.154 | 959.307 | 7.67 | 19 | 1 | 959.787 | 0.210 | 959.933 | 13.67 |
| 17 | 2 | 959.607 | 0.154 | 959.601 | 8.11 | 19 | 1 | 958.516 | 0.193 | 958.521 | 14.43 |
| 17 | 2 | 958.646 | 0.180 | 958.628 | 8.62 | 19 | 1 | 959.787 | 0.210 | 960.021 | 15.22 |
| 17 | 2 | 959.114 | 0.176 | 959.104 | 9.13 | 19 | 1 | 959.000 | 0.210 | 959.049 | 16.01 |
| 17 | 2 | 958.646 | 0.180 | 958.654 | 9.63 | 19 | 2 | 959.787 | 0.210 | 959.891 | 16.93 |
| 17 | 2 | 959.051 | 0.140 | 959.081 | 10.17 | 19 | 2 | 959.698 | 0.261 | 959.730 | 17.92 |

| 1999 - MARCO | | | | | | 1999 - MARCO | | | | |
|---|---|---|---|---|---|---|---|---|---|---|
| D | L | SDB | ER | SDC | HL | D | L | SDB | ER | SDC | HL |
| 19 | 2 | 959.698 | 0.261 | 959.714 | 18.85 | 23 | 2 | 959.134 | 0.155 | 959.080 | 11.78 |
| 19 | 2 | 959.138 | 0.231 | 959.159 | 20.07 | 23 | 2 | 959.637 | 0.126 | 959.599 | 13.00 |
| 19 | 2 | 958.514 | 0.211 | 958.440 | 21.25 | 23 | 2 | 959.215 | 0.151 | 959.185 | 15.57 |
| 19 | 2 | 958.781 | 0.202 | 958.804 | 22.72 | 23 | 2 | 959.637 | 0.126 | 959.621 | 16.39 |
| 19 | 2 | 959.138 | 0.231 | 959.191 | 23.97 | 23 | 2 | 959.425 | 0.180 | 959.388 | 17.29 |
| 19 | 2 | 959.515 | 0.217 | 959.517 | 25.38 | 23 | 1 | 958.901 | 0.149 | 958.883 | 8.40 |
| 19 | 2 | 958.956 | 0.190 | 958.927 | 27.04 | 23 | 1 | 959.134 | 0.155 | 959.054 | 8.91 |
| 19 | 2 | 958.335 | 0.203 | 958.385 | 29.04 | 23 | 1 | 959.215 | 0.151 | 959.207 | 9.45 |
| 19 | 2 | 959.138 | 0.231 | 959.122 | 30.93 | 23 | 1 | 958.969 | 0.157 | 958.969 | 9.98 |
| 19 | 2 | 959.138 | 0.231 | 959.177 | 33.01 | 23 | 1 | 959.450 | 0.148 | 959.450 | 11.17 |
| 19 | 2 | 959.787 | 0.210 | 960.125 | 35.30 | 23 | 1 | 959.134 | 0.155 | 959.080 | 11.78 |
| 19 | 1 | 959.787 | 0.210 | 959.891 | 16.93 | 23 | 1 | 959.637 | 0.126 | 959.599 | 13.00 |
| 19 | 1 | 959.698 | 0.261 | 959.730 | 17.92 | 23 | 1 | 959.215 | 0.151 | 959.185 | 15.57 |
| 19 | 1 | 959.698 | 0.261 | 959.714 | 18.85 | 23 | 1 | 959.637 | 0.126 | 959.621 | 16.39 |
| 19 | 1 | 959.138 | 0.231 | 959.159 | 20.07 | 23 | 1 | 959.425 | 0.180 | 959.388 | 17.29 |
| 19 | 1 | 958.514 | 0.211 | 958.440 | 21.25 | 23 | 2 | 959.637 | 0.126 | 959.721 | 18.20 |
| 19 | 1 | 958.781 | 0.202 | 958.804 | 22.72 | 23 | 2 | 958.969 | 0.157 | 959.050 | 19.52 |
| 19 | 1 | 959.138 | 0.231 | 959.191 | 23.97 | 23 | 1 | 959.637 | 0.126 | 959.721 | 18.20 |
| 19 | 1 | 959.515 | 0.217 | 959.517 | 25.38 | 23 | 1 | 958.969 | 0.157 | 959.050 | 19.52 |
| 19 | 1 | 958.956 | 0.190 | 958.927 | 27.04 | 25 | 2 | 959.476 | 0.161 | 959.472 | 10.89 |
| 19 | 1 | 958.335 | 0.203 | 958.385 | 29.04 | 25 | 2 | 958.036 | 0.200 | 958.097 | 11.45 |
| 19 | 1 | 959.138 | 0.231 | 959.122 | 30.93 | 25 | 2 | 959.149 | 0.152 | 959.135 | 11.99 |
| 19 | 1 | 959.138 | 0.231 | 959.177 | 33.01 | 25 | 2 | 959.068 | 0.188 | 959.068 | 12.56 |
| 19 | 1 | 959.787 | 0.210 | 960.125 | 35.30 | 25 | 2 | 959.807 | 0.205 | 959.942 | 13.15 |
| 22 | 2 | 958.120 | 0.239 | 958.134 | 10.33 | 25 | 2 | 958.722 | 0.133 | 958.737 | 13.76 |
| 22 | 2 | 958.824 | 0.238 | 958.834 | 10.95 | 25 | 2 | 959.091 | 0.166 | 959.090 | 14.40 |
| 22 | 2 | 959.607 | 0.172 | 959.820 | 11.52 | 25 | 2 | 959.430 | 0.150 | 959.446 | 15.08 |
| 22 | 2 | 959.087 | 0.281 | 959.141 | 12.16 | 25 | 2 | 958.917 | 0.156 | 958.955 | 15.77 |
| 22 | 2 | 959.607 | 0.172 | 959.512 | 12.81 | 25 | 2 | 958.174 | 0.159 | 958.214 | 16.50 |
| 22 | 2 | 959.087 | 0.281 | 959.158 | 13.51 | 25 | 2 | 959.345 | 0.168 | 959.350 | 17.26 |
| 22 | 2 | 959.278 | 0.180 | 959.221 | 14.19 | 25 | 2 | 959.149 | 0.152 | 959.183 | 18.07 |
| 22 | 2 | 959.087 | 0.281 | 959.112 | 14.98 | 25 | 1 | 959.476 | 0.161 | 959.472 | 10.89 |
| 22 | 2 | 959.278 | 0.180 | 959.277 | 15.84 | 25 | 1 | 958.036 | 0.200 | 958.097 | 11.45 |
| 22 | 2 | 959.087 | 0.281 | 959.058 | 16.74 | 25 | 1 | 959.149 | 0.152 | 959.135 | 11.99 |
| 22 | 2 | 959.278 | 0.180 | 959.363 | 17.75 | 25 | 1 | 959.068 | 0.188 | 959.068 | 12.56 |
| 22 | 1 | 958.120 | 0.239 | 958.134 | 10.33 | 25 | 1 | 959.807 | 0.205 | 959.942 | 13.15 |
| 22 | 1 | 958.824 | 0.238 | 958.834 | 10.95 | 25 | 1 | 958.722 | 0.133 | 958.737 | 13.76 |
| 22 | 1 | 959.607 | 0.172 | 959.820 | 11.52 | 25 | 1 | 959.091 | 0.166 | 959.090 | 14.40 |
| 22 | 1 | 959.087 | 0.281 | 959.141 | 12.16 | 25 | 1 | 959.430 | 0.150 | 959.446 | 15.08 |
| 22 | 1 | 959.607 | 0.172 | 959.512 | 12.81 | 25 | 1 | 958.917 | 0.156 | 958.955 | 15.77 |
| 22 | 1 | 959.087 | 0.281 | 959.158 | 13.51 | 25 | 1 | 958.174 | 0.159 | 958.214 | 16.50 |
| 22 | 1 | 959.278 | 0.180 | 959.221 | 14.19 | 25 | 1 | 959.345 | 0.168 | 959.350 | 17.26 |
| 22 | 1 | 959.087 | 0.281 | 959.112 | 14.98 | 25 | 1 | 959.149 | 0.152 | 959.183 | 18.07 |
| 22 | 1 | 959.278 | 0.180 | 959.277 | 15.84 | 25 | 2 | 959.807 | 0.205 | 959.647 | 18.92 |
| 22 | 1 | 959.087 | 0.281 | 959.058 | 16.74 | 25 | 2 | 959.807 | 0.205 | 959.652 | 19.82 |
| 22 | 1 | 959.278 | 0.180 | 959.363 | 17.75 | 25 | 2 | 958.722 | 0.133 | 958.752 | 20.75 |
| 22 | 2 | 958.770 | 0.187 | 958.765 | 18.86 | 25 | 2 | 958.994 | 0.185 | 958.961 | 21.76 |
| 22 | 2 | 958.824 | 0.238 | 958.878 | 20.35 | 25 | 2 | 958.714 | 0.184 | 958.700 | 22.81 |
| 22 | 2 | 958.354 | 0.199 | 958.241 | 21.47 | 25 | 2 | 959.345 | 0.168 | 959.345 | 23.92 |
| 22 | 2 | 958.661 | 0.173 | 958.612 | 22.66 | 25 | 2 | 958.603 | 0.193 | 958.630 | 25.11 |
| 22 | 2 | 959.278 | 0.180 | 959.330 | 23.99 | 25 | 2 | 958.902 | 0.197 | 958.880 | 26.38 |
| 22 | 2 | 959.278 | 0.180 | 959.240 | 25.35 | 25 | 2 | 958.174 | 0.159 | 958.265 | 27.97 |
| 22 | 2 | 958.100 | 0.289 | 957.989 | 26.96 | 25 | 2 | 959.149 | 0.152 | 959.131 | 29.44 |
| 22 | 2 | 958.120 | 0.239 | 958.202 | 30.01 | 25 | 2 | 958.902 | 0.197 | 958.876 | 31.03 |
| 22 | 2 | 958.742 | 0.187 | 958.731 | 32.08 | 25 | 2 | 959.149 | 0.152 | 959.224 | 32.81 |
| 22 | 2 | 957.860 | 0.452 | 957.777 | 34.17 | 25 | 2 | 958.174 | 0.159 | 958.173 | 34.72 |
| 22 | 1 | 958.770 | 0.187 | 958.765 | 18.86 | 25 | 2 | 957.935 | 0.167 | 957.960 | 36.84 |
| 22 | 1 | 958.824 | 0.238 | 958.878 | 20.35 | 25 | 2 | 959.484 | 0.151 | 959.490 | 39.20 |
| 22 | 1 | 958.354 | 0.199 | 958.241 | 21.47 | 25 | 1 | 959.807 | 0.205 | 959.647 | 18.92 |
| 22 | 1 | 958.661 | 0.173 | 958.612 | 22.66 | 25 | 1 | 959.807 | 0.205 | 959.652 | 19.82 |
| 22 | 1 | 959.278 | 0.180 | 959.330 | 23.99 | 25 | 1 | 958.722 | 0.133 | 958.752 | 20.75 |
| 22 | 1 | 959.278 | 0.180 | 959.240 | 25.35 | 25 | 1 | 958.994 | 0.185 | 958.961 | 21.76 |
| 22 | 1 | 958.100 | 0.289 | 957.989 | 26.96 | 25 | 1 | 958.714 | 0.184 | 958.700 | 22.81 |
| 22 | 1 | 958.120 | 0.239 | 958.202 | 30.01 | 25 | 1 | 959.345 | 0.168 | 959.345 | 23.92 |
| 22 | 1 | 958.742 | 0.187 | 958.731 | 32.08 | 25 | 1 | 958.603 | 0.193 | 958.630 | 25.11 |
| 22 | 1 | 957.860 | 0.452 | 957.777 | 34.17 | 25 | 1 | 958.902 | 0.197 | 958.880 | 26.38 |
| 23 | 2 | 958.901 | 0.149 | 958.883 | 8.40 | 25 | 1 | 958.174 | 0.159 | 958.265 | 27.97 |
| 23 | 2 | 959.134 | 0.155 | 959.054 | 8.91 | 25 | 1 | 959.149 | 0.152 | 959.131 | 29.44 |
| 23 | 2 | 959.215 | 0.151 | 959.207 | 9.45 | 25 | 1 | 958.902 | 0.197 | 958.876 | 31.03 |
| 23 | 2 | 958.969 | 0.157 | 958.969 | 9.98 | 25 | 1 | 959.149 | 0.152 | 959.224 | 32.81 |
| 23 | 2 | 959.450 | 0.148 | 959.450 | 11.17 | 25 | 1 | 958.174 | 0.159 | 958.173 | 34.72 |

| 1999 - MARCO | | | | | | | 1999 - MARCO | | | | |
|---|---|---|---|---|---|---|---|---|---|---|---|
| D | L | SDB | ER | SDC | HL | | D | L | SDB | ER | SDC | HL |
| 25 | 1 | 957.935 | 0.167 | 957.960 | 36.84 | | 29 | 2 | 959.683 | 0.141 | 959.839 | 33.61 |
| 25 | 1 | 959.484 | 0.151 | 959.490 | 39.20 | | 29 | 2 | 959.683 | 0.141 | 959.698 | 35.51 |
| 26 | 2 | 959.265 | 0.134 | 959.243 | 7.45 | | 29 | 2 | 959.563 | 0.111 | 959.548 | 37.61 |
| 26 | 2 | 958.990 | 0.114 | 959.023 | 7.94 | | 29 | 1 | 959.046 | 0.102 | 958.996 | 23.50 |
| 26 | 2 | 959.300 | 0.113 | 959.291 | 8.44 | | 29 | 1 | 959.378 | 0.134 | 959.398 | 24.63 |
| 26 | 2 | 959.300 | 0.113 | 959.286 | 8.91 | | 29 | 1 | 959.683 | 0.141 | 959.752 | 25.85 |
| 26 | 2 | 959.342 | 0.124 | 959.343 | 9.41 | | 29 | 1 | 959.456 | 0.117 | 959.498 | 27.13 |
| 26 | 2 | 958.603 | 0.106 | 958.587 | 9.98 | | 29 | 1 | 958.707 | 0.114 | 958.690 | 28.48 |
| 26 | 2 | 958.655 | 0.111 | 958.637 | 10.50 | | 29 | 1 | 958.894 | 0.107 | 958.891 | 30.30 |
| 26 | 2 | 959.342 | 0.124 | 959.344 | 11.05 | | 29 | 1 | 959.683 | 0.141 | 959.860 | 31.88 |
| 26 | 2 | 958.752 | 0.122 | 958.734 | 11.68 | | 29 | 1 | 959.683 | 0.141 | 959.839 | 33.61 |
| 26 | 2 | 959.525 | 0.116 | 959.489 | 12.29 | | 29 | 1 | 959.683 | 0.141 | 959.698 | 35.51 |
| 26 | 2 | 959.300 | 0.113 | 959.287 | 13.00 | | 29 | 1 | 959.563 | 0.111 | 959.548 | 37.61 |
| 26 | 2 | 959.525 | 0.116 | 959.482 | 13.64 | | | | | | | |
| 26 | 2 | 959.083 | 0.111 | 959.091 | 14.36 | | | | 1999 - ABRIL | | | |
| 26 | 2 | 959.525 | 0.116 | 959.518 | 15.10 | | D | L | SDB | ER | SDC | HL |
| 26 | 2 | 959.899 | 0.114 | 960.037 | 18.66 | | 01 | 2 | 958.942 | 0.188 | 958.922 | 10.34 |
| 26 | 1 | 959.265 | 0.134 | 959.243 | 7.45 | | 01 | 2 | 958.321 | 0.223 | 958.164 | 10.83 |
| 26 | 1 | 958.990 | 0.114 | 959.023 | 7.94 | | 01 | 2 | 959.211 | 0.161 | 959.203 | 11.34 |
| 26 | 1 | 959.300 | 0.113 | 959.291 | 8.44 | | 01 | 2 | 957.892 | 0.302 | 957.147 | 11.86 |
| 26 | 1 | 959.300 | 0.113 | 959.286 | 8.91 | | 01 | 2 | 958.730 | 0.220 | 958.694 | 12.39 |
| 26 | 1 | 959.342 | 0.124 | 959.343 | 9.41 | | 01 | 2 | 959.060 | 0.146 | 959.119 | 12.99 |
| 26 | 1 | 958.603 | 0.106 | 958.587 | 9.98 | | 01 | 2 | 958.803 | 0.168 | 958.774 | 13.57 |
| 26 | 1 | 958.655 | 0.111 | 958.637 | 10.50 | | 01 | 2 | 958.942 | 0.188 | 958.903 | 14.16 |
| 26 | 1 | 959.342 | 0.124 | 959.344 | 11.05 | | 01 | 2 | 958.533 | 0.229 | 958.546 | 14.79 |
| 26 | 1 | 958.752 | 0.122 | 958.734 | 11.68 | | 01 | 2 | 958.981 | 0.173 | 958.980 | 15.45 |
| 26 | 1 | 959.525 | 0.116 | 959.489 | 12.29 | | 01 | 2 | 958.408 | 0.153 | 958.414 | 16.13 |
| 26 | 1 | 959.300 | 0.113 | 959.287 | 13.00 | | 01 | 2 | 958.730 | 0.220 | 958.679 | 16.83 |
| 26 | 1 | 959.525 | 0.116 | 959.482 | 13.64 | | 01 | 2 | 959.656 | 0.137 | 959.617 | 17.55 |
| 26 | 1 | 959.083 | 0.111 | 959.091 | 14.36 | | 01 | 2 | 959.211 | 0.161 | 959.136 | 18.32 |
| 26 | 1 | 959.525 | 0.116 | 959.518 | 15.10 | | 01 | 2 | 959.901 | 0.175 | 959.966 | 19.12 |
| 26 | 1 | 959.899 | 0.114 | 960.037 | 18.66 | | 01 | 2 | 959.706 | 0.190 | 959.712 | 19.96 |
| 26 | 2 | 958.752 | 0.122 | 958.752 | 20.72 | | 01 | 2 | 959.348 | 0.156 | 959.390 | 20.86 |
| 26 | 2 | 959.525 | 0.116 | 959.462 | 21.83 | | 01 | 1 | 958.942 | 0.188 | 958.922 | 10.34 |
| 26 | 2 | 959.083 | 0.111 | 959.072 | 22.94 | | 01 | 1 | 958.321 | 0.223 | 958.164 | 10.83 |
| 26 | 2 | 958.603 | 0.106 | 958.605 | 24.17 | | 01 | 1 | 959.211 | 0.161 | 959.203 | 11.34 |
| 26 | 2 | 959.301 | 0.122 | 959.309 | 25.43 | | 01 | 1 | 957.892 | 0.302 | 957.147 | 11.86 |
| 26 | 2 | 958.806 | 0.134 | 958.857 | 26.92 | | 01 | 1 | 958.730 | 0.220 | 958.694 | 12.39 |
| 26 | 1 | 958.752 | 0.122 | 958.752 | 20.72 | | 01 | 1 | 959.060 | 0.146 | 959.119 | 12.99 |
| 26 | 1 | 959.525 | 0.116 | 959.462 | 21.83 | | 01 | 1 | 958.803 | 0.168 | 958.774 | 13.57 |
| 26 | 1 | 959.083 | 0.111 | 959.072 | 22.94 | | 01 | 1 | 958.942 | 0.188 | 958.903 | 14.16 |
| 26 | 1 | 958.603 | 0.106 | 958.605 | 24.17 | | 01 | 1 | 958.533 | 0.229 | 958.546 | 14.79 |
| 26 | 1 | 959.301 | 0.122 | 959.309 | 25.43 | | 01 | 1 | 958.981 | 0.173 | 958.980 | 15.45 |
| 26 | 1 | 958.806 | 0.134 | 958.857 | 26.92 | | 01 | 1 | 958.408 | 0.153 | 958.414 | 16.13 |
| 29 | 2 | 959.683 | 0.141 | 959.703 | 10.70 | | 01 | 1 | 958.730 | 0.220 | 958.679 | 16.83 |
| 29 | 2 | 958.489 | 0.105 | 958.442 | 11.59 | | 01 | 1 | 959.656 | 0.137 | 959.617 | 17.55 |
| 29 | 2 | 959.046 | 0.102 | 959.075 | 12.25 | | 01 | 1 | 959.211 | 0.161 | 959.136 | 18.32 |
| 29 | 2 | 959.046 | 0.102 | 959.017 | 12.87 | | 01 | 1 | 959.901 | 0.175 | 959.966 | 19.12 |
| 29 | 2 | 958.582 | 0.108 | 958.573 | 13.55 | | 01 | 1 | 959.706 | 0.190 | 959.712 | 19.96 |
| 29 | 2 | 959.046 | 0.102 | 959.112 | 14.21 | | 01 | 1 | 959.348 | 0.156 | 959.390 | 20.86 |
| 29 | 2 | 958.582 | 0.108 | 958.643 | 14.90 | | 01 | 2 | 959.241 | 0.150 | 959.236 | 21.80 |
| 29 | 2 | 959.456 | 0.117 | 959.468 | 15.63 | | 01 | 2 | 958.942 | 0.188 | 958.959 | 22.78 |
| 29 | 2 | 958.745 | 0.111 | 958.750 | 16.41 | | 01 | 2 | 959.656 | 0.137 | 959.633 | 23.81 |
| 29 | 2 | 958.873 | 0.099 | 958.876 | 17.19 | | 01 | 2 | 960.146 | 0.152 | 960.527 | 24.89 |
| 29 | 1 | 959.683 | 0.141 | 959.703 | 10.70 | | 01 | 2 | 958.603 | 0.158 | 958.605 | 26.10 |
| 29 | 1 | 958.489 | 0.105 | 958.442 | 11.59 | | 01 | 2 | 959.059 | 0.169 | 959.058 | 27.33 |
| 29 | 1 | 959.046 | 0.102 | 959.075 | 12.25 | | 01 | 2 | 959.901 | 0.175 | 959.889 | 28.64 |
| 29 | 1 | 959.046 | 0.102 | 959.017 | 12.87 | | 01 | 2 | 959.060 | 0.146 | 959.134 | 30.05 |
| 29 | 1 | 958.582 | 0.108 | 958.573 | 13.55 | | 01 | 2 | 959.574 | 0.181 | 959.513 | 31.74 |
| 29 | 1 | 959.046 | 0.102 | 959.112 | 14.21 | | 01 | 2 | 959.348 | 0.156 | 959.381 | 33.38 |
| 29 | 1 | 958.582 | 0.108 | 958.643 | 14.90 | | 01 | 2 | 959.052 | 0.201 | 959.024 | 35.21 |
| 29 | 1 | 959.456 | 0.117 | 959.468 | 15.63 | | 01 | 2 | 958.942 | 0.188 | 958.949 | 37.17 |
| 29 | 1 | 958.745 | 0.111 | 958.750 | 16.41 | | 01 | 1 | 959.241 | 0.150 | 959.236 | 21.80 |
| 29 | 1 | 958.873 | 0.099 | 958.876 | 17.19 | | 01 | 1 | 958.942 | 0.188 | 958.959 | 22.78 |
| 29 | 2 | 959.046 | 0.102 | 958.996 | 23.50 | | 01 | 1 | 959.656 | 0.137 | 959.633 | 23.81 |
| 29 | 2 | 959.378 | 0.134 | 959.398 | 24.63 | | 01 | 1 | 960.146 | 0.152 | 960.527 | 24.89 |
| 29 | 2 | 959.683 | 0.141 | 959.752 | 25.85 | | 01 | 1 | 958.603 | 0.158 | 958.605 | 26.10 |
| 29 | 2 | 959.456 | 0.117 | 959.498 | 27.13 | | 01 | 1 | 959.059 | 0.169 | 959.058 | 27.33 |
| 29 | 2 | 958.707 | 0.114 | 958.690 | 28.48 | | 01 | 1 | 959.901 | 0.175 | 959.889 | 28.64 |
| 29 | 2 | 958.894 | 0.107 | 958.891 | 30.30 | | 01 | 1 | 959.060 | 0.146 | 959.134 | 30.05 |
| 29 | 2 | 959.683 | 0.141 | 959.860 | 31.88 | | | | | | | |

| 1999 - ABRIL | | | | | | 1999 - ABRIL | | | | |
|---|---|---|---|---|---|---|---|---|---|---|
| D | L | SDB | ER | SDC | HL | D | L | SDB | ER | SDC | HL |
| 01 | 1 | 959.574 | 0.181 | 959.513 | 31.74 | 08 | 1 | 958.790 | 0.161 | 958.806 | 13.78 |
| 01 | 1 | 959.348 | 0.156 | 959.381 | 33.38 | 08 | 1 | 959.060 | 0.184 | 959.066 | 14.35 |
| 01 | 1 | 959.052 | 0.201 | 959.024 | 35.21 | 08 | 1 | 958.552 | 0.190 | 958.519 | 14.98 |
| 01 | 1 | 958.942 | 0.188 | 958.949 | 37.17 | 08 | 1 | 959.638 | 0.213 | 959.669 | 15.61 |
| 05 | 2 | 959.075 | 0.155 | 959.080 | 16.77 | 08 | 1 | 958.901 | 0.196 | 958.959 | 16.25 |
| 05 | 2 | 959.352 | 0.165 | 959.458 | 17.59 | 08 | 1 | 959.618 | 0.161 | 959.546 | 16.93 |
| 05 | 2 | 959.352 | 0.165 | 959.478 | 18.49 | 08 | 1 | 958.668 | 0.167 | 958.680 | 17.63 |
| 05 | 2 | 959.170 | 0.157 | 959.143 | 19.39 | 08 | 1 | 959.060 | 0.184 | 959.100 | 18.36 |
| 05 | 2 | 958.987 | 0.164 | 959.018 | 20.98 | 08 | 1 | 959.198 | 0.174 | 959.202 | 19.13 |
| 05 | 2 | 959.075 | 0.155 | 959.034 | 21.93 | 08 | 1 | 958.298 | 0.169 | 958.307 | 21.19 |
| 05 | 1 | 959.075 | 0.155 | 959.080 | 16.77 | 08 | 1 | 959.638 | 0.213 | 959.741 | 22.07 |
| 05 | 1 | 959.352 | 0.165 | 959.458 | 17.59 | 08 | 1 | 958.668 | 0.167 | 958.658 | 23.02 |
| 05 | 1 | 959.352 | 0.165 | 959.478 | 18.49 | 08 | 1 | 958.387 | 0.154 | 958.369 | 23.99 |
| 05 | 1 | 959.170 | 0.157 | 959.143 | 19.39 | 08 | 2 | 958.152 | 0.174 | 958.046 | 28.22 |
| 05 | 1 | 958.987 | 0.164 | 959.018 | 20.98 | 08 | 2 | 958.668 | 0.167 | 958.664 | 29.53 |
| 05 | 1 | 959.075 | 0.155 | 959.034 | 21.93 | 08 | 2 | 959.867 | 0.220 | 960.123 | 30.91 |
| 05 | 2 | 958.306 | 0.160 | 958.380 | 22.95 | 08 | 2 | 958.901 | 0.196 | 958.895 | 32.42 |
| 05 | 2 | 959.096 | 0.192 | 959.101 | 24.03 | 08 | 2 | 959.436 | 0.191 | 959.340 | 34.02 |
| 05 | 2 | 958.987 | 0.164 | 958.993 | 26.98 | 08 | 2 | 958.298 | 0.169 | 958.340 | 35.76 |
| 05 | 2 | 959.204 | 0.159 | 959.213 | 28.83 | 08 | 2 | 958.559 | 0.252 | 958.568 | 37.66 |
| 05 | 2 | 958.702 | 0.180 | 958.720 | 30.43 | 08 | 2 | 958.668 | 0.167 | 958.666 | 39.78 |
| 05 | 2 | 958.553 | 0.147 | 958.611 | 32.08 | 08 | 1 | 958.152 | 0.174 | 958.046 | 28.22 |
| 05 | 2 | 958.306 | 0.160 | 958.321 | 34.19 | 08 | 1 | 958.668 | 0.167 | 958.664 | 29.53 |
| 05 | 2 | 959.075 | 0.155 | 959.062 | 36.16 | 08 | 1 | 959.867 | 0.220 | 960.123 | 30.91 |
| 05 | 2 | 958.702 | 0.180 | 958.642 | 38.48 | 08 | 1 | 958.901 | 0.196 | 958.895 | 32.42 |
| 05 | 1 | 958.306 | 0.160 | 958.380 | 22.95 | 08 | 1 | 959.436 | 0.191 | 959.340 | 34.02 |
| 05 | 1 | 959.096 | 0.192 | 959.101 | 24.03 | 08 | 1 | 958.298 | 0.169 | 958.340 | 35.76 |
| 05 | 1 | 958.987 | 0.164 | 958.993 | 26.98 | 08 | 1 | 958.559 | 0.252 | 958.568 | 37.66 |
| 05 | 1 | 959.204 | 0.159 | 959.213 | 28.83 | 08 | 1 | 958.668 | 0.167 | 958.666 | 39.78 |
| 05 | 1 | 958.702 | 0.180 | 958.720 | 30.43 | 12 | 2 | 958.877 | 0.252 | 958.853 | 14.29 |
| 05 | 1 | 958.553 | 0.147 | 958.611 | 32.08 | 12 | 2 | 959.741 | 0.249 | 959.757 | 14.94 |
| 05 | 1 | 958.306 | 0.160 | 958.321 | 34.19 | 12 | 2 | 959.110 | 0.255 | 959.110 | 15.69 |
| 05 | 1 | 959.075 | 0.155 | 959.062 | 36.16 | 12 | 2 | 959.741 | 0.249 | 959.760 | 16.39 |
| 05 | 1 | 958.702 | 0.180 | 958.642 | 38.48 | 12 | 2 | 959.652 | 0.265 | 959.650 | 17.46 |
| 07 | 2 | 958.595 | 0.190 | 958.562 | 18.82 | 12 | 2 | 958.725 | 0.273 | 958.725 | 18.30 |
| 07 | 2 | 958.495 | 0.160 | 958.507 | 19.69 | 12 | 2 | 958.710 | 0.266 | 958.705 | 19.38 |
| 07 | 2 | 958.495 | 0.160 | 958.472 | 20.61 | 12 | 2 | 958.535 | 0.273 | 958.540 | 20.21 |
| 07 | 2 | 958.495 | 0.160 | 958.541 | 21.54 | 12 | 2 | 958.710 | 0.266 | 958.715 | 21.18 |
| 07 | 2 | 958.678 | 0.186 | 958.720 | 22.55 | 12 | 2 | 959.110 | 0.255 | 959.147 | 22.10 |
| 07 | 1 | 958.595 | 0.190 | 958.562 | 18.82 | 12 | 2 | 959.344 | 0.225 | 959.322 | 23.19 |
| 07 | 1 | 958.495 | 0.160 | 958.507 | 19.69 | 12 | 2 | 958.235 | 0.459 | 958.241 | 24.23 |
| 07 | 1 | 958.495 | 0.160 | 958.472 | 20.61 | 12 | 2 | 958.554 | 0.279 | 958.580 | 25.38 |
| 07 | 1 | 958.495 | 0.160 | 958.541 | 21.54 | 12 | 1 | 958.877 | 0.252 | 958.853 | 14.29 |
| 07 | 1 | 958.678 | 0.186 | 958.720 | 22.55 | 12 | 1 | 959.741 | 0.249 | 959.757 | 14.94 |
| 07 | 2 | 958.283 | 0.173 | 957.825 | 27.22 | 12 | 1 | 959.110 | 0.255 | 959.110 | 15.69 |
| 07 | 2 | 958.830 | 0.195 | 959.099 | 28.53 | 12 | 1 | 959.741 | 0.249 | 959.760 | 16.39 |
| 07 | 2 | 958.830 | 0.195 | 959.370 | 29.97 | 12 | 1 | 959.652 | 0.265 | 959.650 | 17.46 |
| 07 | 2 | 958.830 | 0.195 | 959.008 | 31.55 | 12 | 1 | 958.725 | 0.273 | 958.725 | 18.30 |
| 07 | 2 | 958.417 | 0.155 | 958.410 | 33.20 | 12 | 1 | 958.710 | 0.266 | 958.705 | 19.38 |
| 07 | 2 | 958.678 | 0.186 | 958.688 | 34.98 | 12 | 1 | 958.535 | 0.273 | 958.540 | 20.21 |
| 07 | 1 | 958.283 | 0.173 | 957.825 | 27.22 | 12 | 1 | 958.710 | 0.266 | 958.715 | 21.18 |
| 07 | 1 | 958.830 | 0.195 | 959.099 | 28.53 | 12 | 1 | 959.110 | 0.255 | 959.147 | 22.10 |
| 07 | 1 | 958.830 | 0.195 | 959.370 | 29.97 | 12 | 1 | 959.344 | 0.225 | 959.322 | 23.19 |
| 07 | 1 | 958.830 | 0.195 | 959.008 | 31.55 | 12 | 1 | 958.235 | 0.459 | 958.241 | 24.23 |
| 07 | 1 | 958.417 | 0.155 | 958.410 | 33.20 | 12 | 1 | 958.554 | 0.279 | 958.580 | 25.38 |
| 07 | 1 | 958.678 | 0.186 | 958.688 | 34.98 | 12 | 2 | 958.753 | 0.292 | 958.781 | 26.61 |
| 08 | 2 | 957.834 | 0.190 | 957.828 | 13.21 | 12 | 2 | 958.814 | 0.221 | 958.837 | 27.83 |
| 08 | 2 | 958.790 | 0.161 | 958.806 | 13.78 | 12 | 2 | 959.587 | 0.349 | 959.564 | 29.19 |
| 08 | 2 | 959.060 | 0.184 | 959.066 | 14.35 | 12 | 2 | 959.103 | 0.249 | 959.004 | 30.65 |
| 08 | 2 | 958.552 | 0.190 | 958.519 | 14.98 | 12 | 2 | 958.235 | 0.459 | 958.306 | 32.40 |
| 08 | 2 | 959.638 | 0.213 | 959.669 | 15.61 | 12 | 1 | 958.753 | 0.292 | 958.781 | 26.61 |
| 08 | 2 | 958.901 | 0.196 | 958.959 | 16.25 | 12 | 1 | 958.814 | 0.221 | 958.837 | 27.83 |
| 08 | 2 | 959.618 | 0.161 | 959.546 | 16.93 | 12 | 1 | 959.587 | 0.349 | 959.564 | 29.19 |
| 08 | 2 | 958.668 | 0.167 | 958.680 | 17.63 | 12 | 1 | 959.103 | 0.249 | 959.004 | 30.65 |
| 08 | 2 | 959.060 | 0.184 | 959.100 | 18.36 | 12 | 1 | 958.235 | 0.459 | 958.306 | 32.40 |
| 08 | 2 | 959.198 | 0.174 | 959.202 | 19.13 | 13 | 2 | 959.097 | 0.201 | 959.416 | 26.07 |
| 08 | 2 | 958.298 | 0.169 | 958.307 | 21.19 | 13 | 2 | 958.895 | 0.192 | 958.660 | 27.27 |
| 08 | 2 | 959.638 | 0.213 | 959.741 | 22.07 | 13 | 2 | 958.396 | 0.153 | 957.888 | 28.57 |
| 08 | 2 | 958.668 | 0.167 | 958.658 | 23.02 | 13 | 2 | 958.895 | 0.192 | 958.753 | 31.44 |
| 08 | 2 | 958.387 | 0.154 | 958.369 | 23.99 | 13 | 2 | 958.396 | 0.153 | 958.222 | 33.14 |
| 08 | 1 | 957.834 | 0.190 | 957.828 | 13.21 | 13 | 1 | 959.097 | 0.201 | 959.416 | 26.07 |

| 1999 - ABRIL | | | | | | 1999 - ABRIL | | | | |
|---|---|---|---|---|---|---|---|---|---|---|
| D | L | SDB | ER | SDC | HL | D | L | SDB | ER | SDC | HL |
| 13 | 1 | 958.895 | 0.192 | 958.660 | 27.27 | 20 | 1 | 959.648 | 0.178 | 959.765 | 17.01 |
| 13 | 1 | 958.396 | 0.153 | 957.888 | 28.57 | 20 | 1 | 958.996 | 0.237 | 958.985 | 17.73 |
| 13 | 1 | 958.895 | 0.192 | 958.753 | 31.44 | 20 | 1 | 958.996 | 0.237 | 958.991 | 18.45 |
| 13 | 1 | 958.396 | 0.153 | 958.222 | 33.14 | 20 | 1 | 958.925 | 0.192 | 958.933 | 19.24 |
| 14 | 2 | 958.693 | 0.165 | 958.700 | 12.53 | 20 | 1 | 958.763 | 0.175 | 958.746 | 20.05 |
| 14 | 2 | 959.165 | 0.340 | 959.693 | 13.15 | 20 | 1 | 958.301 | 0.250 | 958.261 | 20.90 |
| 14 | 2 | 958.834 | 0.162 | 958.837 | 15.53 | 20 | 1 | 958.601 | 0.239 | 958.605 | 21.93 |
| 14 | 2 | 959.165 | 0.340 | 959.172 | 16.25 | 20 | 1 | 958.301 | 0.250 | 958.349 | 22.87 |
| 14 | 2 | 959.165 | 0.340 | 959.258 | 16.98 | 20 | 1 | 959.244 | 0.245 | 959.377 | 23.93 |
| 14 | 2 | 959.165 | 0.340 | 959.377 | 17.70 | 20 | 1 | 958.925 | 0.192 | 958.954 | 25.09 |
| 14 | 2 | 958.693 | 0.165 | 958.686 | 18.55 | 20 | 1 | 959.550 | 0.238 | 959.430 | 27.87 |
| 14 | 2 | 958.531 | 0.176 | 958.538 | 19.39 | 20 | 2 | 959.632 | 0.185 | 959.610 | 29.20 |
| 14 | 2 | 959.069 | 0.208 | 959.006 | 20.25 | 20 | 2 | 958.996 | 0.237 | 958.971 | 30.80 |
| 14 | 2 | 958.907 | 0.164 | 958.954 | 21.15 | 20 | 2 | 959.550 | 0.238 | 959.496 | 32.41 |
| 14 | 2 | 958.445 | 0.272 | 958.227 | 22.05 | 20 | 1 | 959.632 | 0.185 | 959.610 | 29.20 |
| 14 | 2 | 958.743 | 0.179 | 958.747 | 23.04 | 20 | 1 | 958.996 | 0.237 | 958.971 | 30.80 |
| 14 | 2 | 959.069 | 0.208 | 959.049 | 24.08 | 20 | 1 | 959.550 | 0.238 | 959.496 | 32.41 |
| 14 | 2 | 958.643 | 0.158 | 958.640 | 25.17 | 23 | 2 | 959.414 | 0.164 | 959.405 | 18.27 |
| 14 | 1 | 958.693 | 0.165 | 958.700 | 12.53 | 23 | 2 | 959.824 | 0.211 | 960.603 | 19.01 |
| 14 | 1 | 959.165 | 0.340 | 959.693 | 13.15 | 23 | 2 | 959.477 | 0.218 | 959.485 | 19.82 |
| 14 | 1 | 958.834 | 0.162 | 958.837 | 15.53 | 23 | 2 | 958.799 | 0.187 | 958.758 | 20.90 |
| 14 | 1 | 959.165 | 0.340 | 959.172 | 16.25 | 23 | 2 | 959.824 | 0.211 | 959.827 | 21.75 |
| 14 | 1 | 959.165 | 0.340 | 959.258 | 16.98 | 23 | 2 | 958.891 | 0.205 | 958.847 | 22.65 |
| 14 | 1 | 959.165 | 0.340 | 959.377 | 17.70 | 23 | 2 | 959.585 | 0.256 | 959.658 | 23.69 |
| 14 | 1 | 958.693 | 0.165 | 958.686 | 18.55 | 23 | 2 | 958.550 | 0.216 | 958.597 | 24.73 |
| 14 | 1 | 958.531 | 0.176 | 958.538 | 19.39 | 23 | 2 | 958.799 | 0.187 | 958.829 | 25.94 |
| 14 | 1 | 959.069 | 0.208 | 959.006 | 20.25 | 23 | 2 | 959.197 | 0.189 | 959.219 | 27.08 |
| 14 | 1 | 958.907 | 0.164 | 958.954 | 21.15 | 23 | 2 | 959.018 | 0.228 | 959.021 | 28.29 |
| 14 | 1 | 958.445 | 0.272 | 958.227 | 22.05 | 23 | 2 | 958.799 | 0.187 | 958.821 | 29.51 |
| 14 | 1 | 958.743 | 0.179 | 958.747 | 23.04 | 23 | 1 | 959.414 | 0.164 | 959.405 | 18.27 |
| 14 | 1 | 959.069 | 0.208 | 959.049 | 24.08 | 23 | 1 | 959.824 | 0.211 | 960.603 | 19.01 |
| 14 | 1 | 958.643 | 0.158 | 958.640 | 25.17 | 23 | 1 | 959.477 | 0.218 | 959.485 | 19.82 |
| 14 | 2 | 959.069 | 0.208 | 959.058 | 26.32 | 23 | 1 | 958.799 | 0.187 | 958.758 | 20.90 |
| 14 | 2 | 958.907 | 0.164 | 958.945 | 27.67 | 23 | 1 | 959.824 | 0.211 | 959.827 | 21.75 |
| 14 | 2 | 958.724 | 0.174 | 958.714 | 29.01 | 23 | 1 | 958.891 | 0.205 | 958.847 | 22.65 |
| 14 | 2 | 958.907 | 0.164 | 958.938 | 30.32 | 23 | 1 | 959.585 | 0.256 | 959.658 | 23.69 |
| 14 | 2 | 959.069 | 0.208 | 959.081 | 31.85 | 23 | 1 | 958.550 | 0.216 | 958.597 | 24.73 |
| 14 | 2 | 958.531 | 0.176 | 958.516 | 33.46 | 23 | 1 | 958.799 | 0.187 | 958.829 | 25.94 |
| 14 | 1 | 959.069 | 0.208 | 959.058 | 26.32 | 23 | 1 | 959.197 | 0.189 | 959.219 | 27.08 |
| 14 | 1 | 958.907 | 0.164 | 958.945 | 27.67 | 23 | 1 | 959.018 | 0.228 | 959.021 | 28.29 |
| 14 | 1 | 958.724 | 0.174 | 958.714 | 29.01 | 23 | 1 | 958.799 | 0.187 | 958.821 | 29.51 |
| 14 | 1 | 958.907 | 0.164 | 958.938 | 30.32 | 23 | 2 | 959.824 | 0.211 | 959.757 | 30.81 |
| 14 | 1 | 959.069 | 0.208 | 959.081 | 31.85 | 23 | 2 | 958.357 | 0.229 | 958.070 | 32.19 |
| 14 | 1 | 958.531 | 0.176 | 958.516 | 33.46 | 23 | 2 | 958.550 | 0.216 | 958.591 | 33.62 |
| 16 | 2 | 958.376 | 0.096 | 958.565 | 19.31 | 23 | 2 | 959.824 | 0.211 | 959.718 | 35.16 |
| 16 | 2 | 958.187 | 0.199 | 958.176 | 20.07 | 23 | 2 | 959.316 | 0.165 | 959.304 | 36.82 |
| 16 | 2 | 958.376 | 0.096 | 958.520 | 20.90 | 23 | 2 | 958.891 | 0.205 | 958.879 | 38.65 |
| 16 | 2 | 959.033 | 0.264 | 959.023 | 21.70 | 23 | 1 | 959.824 | 0.211 | 959.757 | 30.81 |
| 16 | 2 | 958.209 | 0.227 | 958.200 | 23.05 | 23 | 1 | 958.357 | 0.229 | 958.070 | 32.19 |
| 16 | 1 | 958.376 | 0.096 | 958.565 | 19.31 | 23 | 1 | 958.550 | 0.216 | 958.591 | 33.62 |
| 16 | 1 | 958.187 | 0.199 | 958.176 | 20.07 | 23 | 1 | 959.824 | 0.211 | 959.718 | 35.16 |
| 16 | 1 | 958.376 | 0.096 | 958.520 | 20.90 | 23 | 1 | 959.316 | 0.165 | 959.304 | 36.82 |
| 16 | 1 | 959.033 | 0.264 | 959.023 | 21.70 | 23 | 1 | 958.891 | 0.205 | 958.879 | 38.65 |
| 16 | 1 | 958.209 | 0.227 | 958.200 | 23.05 | 26 | 2 | 960.107 | 0.239 | 960.682 | 19.95 |
| 20 | 2 | 958.996 | 0.237 | 959.002 | 15.00 | 26 | 2 | 960.107 | 0.239 | 960.328 | 20.79 |
| 20 | 2 | 958.996 | 0.237 | 958.981 | 15.64 | 26 | 2 | 959.236 | 0.232 | 959.225 | 21.60 |
| 20 | 2 | 959.648 | 0.178 | 959.683 | 16.31 | 26 | 2 | 958.926 | 0.245 | 958.944 | 22.46 |
| 20 | 2 | 959.648 | 0.178 | 959.765 | 17.01 | 26 | 2 | 959.463 | 0.301 | 959.422 | 23.32 |
| 20 | 2 | 958.996 | 0.237 | 958.985 | 17.73 | 26 | 2 | 959.754 | 0.198 | 959.756 | 24.23 |
| 20 | 2 | 958.996 | 0.237 | 958.991 | 18.45 | 26 | 2 | 958.052 | 0.198 | 958.031 | 25.19 |
| 20 | 2 | 958.925 | 0.192 | 958.933 | 19.24 | 26 | 2 | 958.915 | 0.240 | 958.873 | 26.19 |
| 20 | 2 | 958.763 | 0.175 | 958.746 | 20.05 | 26 | 2 | 958.474 | 0.299 | 958.391 | 27.29 |
| 20 | 2 | 958.301 | 0.250 | 958.261 | 20.90 | 26 | 2 | 959.483 | 0.235 | 959.521 | 28.42 |
| 20 | 2 | 958.601 | 0.239 | 958.605 | 21.93 | 26 | 2 | 958.926 | 0.245 | 959.010 | 29.58 |
| 20 | 2 | 958.301 | 0.250 | 958.349 | 22.87 | 26 | 1 | 960.107 | 0.239 | 960.682 | 19.95 |
| 20 | 2 | 959.244 | 0.245 | 959.377 | 23.93 | 26 | 1 | 960.107 | 0.239 | 960.328 | 20.79 |
| 20 | 2 | 958.925 | 0.192 | 958.954 | 25.09 | 26 | 1 | 959.236 | 0.232 | 959.225 | 21.60 |
| 20 | 2 | 959.550 | 0.238 | 959.430 | 27.87 | 26 | 1 | 958.926 | 0.245 | 958.944 | 22.46 |
| 20 | 1 | 958.996 | 0.237 | 959.002 | 15.00 | 26 | 1 | 959.463 | 0.301 | 959.422 | 23.32 |
| 20 | 1 | 958.996 | 0.237 | 958.981 | 15.64 | 26 | 1 | 959.754 | 0.198 | 959.756 | 24.23 |
| 20 | 1 | 959.648 | 0.178 | 959.683 | 16.31 | 26 | 1 | 958.052 | 0.198 | 958.031 | 25.19 |

| 1999 - ABRIL | | | | | |
|---|---|---|---|---|---|
| D | L | SDB | ER | SDC | HL |
| 26 | 1 | 958.915 | 0.240 | 958.873 | 26.19 |
| 26 | 1 | 958.474 | 0.299 | 958.391 | 27.29 |
| 26 | 1 | 959.483 | 0.235 | 959.521 | 28.42 |
| 26 | 1 | 958.926 | 0.245 | 959.010 | 29.58 |
| 26 | 2 | 959.206 | 0.265 | 959.091 | 30.81 |
| 26 | 2 | 959.483 | 0.235 | 959.610 | 32.14 |
| 26 | 2 | 958.056 | 0.238 | 958.066 | 33.55 |
| 26 | 2 | 958.926 | 0.245 | 958.947 | 35.24 |
| 26 | 2 | 959.483 | 0.235 | 959.539 | 36.90 |
| 26 | 1 | 959.206 | 0.265 | 959.091 | 30.81 |
| 26 | 1 | 959.483 | 0.235 | 959.610 | 32.14 |
| 26 | 1 | 958.056 | 0.238 | 958.066 | 33.55 |
| 26 | 1 | 958.926 | 0.245 | 958.947 | 35.24 |
| 26 | 1 | 959.483 | 0.235 | 959.539 | 36.90 |
| 28 | 2 | 959.823 | 0.320 | 959.548 | 32.60 |
| 28 | 2 | 959.067 | 0.250 | 959.156 | 33.99 |
| 28 | 2 | 960.154 | 0.259 | 960.350 | 35.48 |
| 28 | 2 | 959.067 | 0.250 | 959.216 | 37.07 |
| 28 | 2 | 959.823 | 0.320 | 959.447 | 38.83 |
| 28 | 2 | 958.560 | 0.291 | 958.660 | 40.73 |
| 28 | 1 | 959.823 | 0.320 | 959.548 | 32.60 |
| 28 | 1 | 959.067 | 0.250 | 959.156 | 33.99 |
| 28 | 1 | 960.154 | 0.259 | 960.350 | 35.48 |
| 28 | 1 | 959.067 | 0.250 | 959.216 | 37.07 |
| 28 | 1 | 959.823 | 0.320 | 959.447 | 38.83 |
| 28 | 1 | 958.560 | 0.291 | 958.660 | 40.73 |
| 29 | 2 | 959.059 | 0.320 | 959.147 | 17.59 |
| 29 | 2 | 958.021 | 0.318 | 957.874 | 18.22 |
| 29 | 2 | 958.660 | 0.334 | 958.636 | 18.85 |
| 29 | 2 | 957.665 | 0.388 | 957.818 | 19.51 |
| 29 | 2 | 958.805 | 0.856 | 958.903 | 20.19 |
| 29 | 2 | 959.835 | 0.397 | 959.786 | 20.88 |
| 29 | 2 | 958.481 | 0.384 | 958.502 | 21.60 |
| 29 | 2 | 958.021 | 0.318 | 957.971 | 22.33 |
| 29 | 2 | 958.481 | 0.384 | 958.497 | 23.20 |
| 29 | 2 | 958.084 | 0.324 | 958.119 | 23.99 |
| 29 | 2 | 958.805 | 0.856 | 958.897 | 25.84 |
| 29 | 2 | 957.403 | 0.333 | 957.354 | 26.96 |
| 29 | 2 | 958.021 | 0.318 | 958.019 | 27.97 |
| 29 | 2 | 958.660 | 0.334 | 958.640 | 28.97 |
| 29 | 2 | 958.307 | 0.262 | 958.334 | 30.04 |
| 29 | 2 | 958.778 | 0.526 | 958.770 | 31.16 |
| 29 | 1 | 959.059 | 0.320 | 959.147 | 17.59 |
| 29 | 1 | 958.021 | 0.318 | 957.874 | 18.22 |
| 29 | 1 | 958.660 | 0.334 | 958.636 | 18.85 |
| 29 | 1 | 957.665 | 0.388 | 957.818 | 19.51 |
| 29 | 1 | 958.805 | 0.856 | 958.903 | 20.19 |
| 29 | 1 | 959.835 | 0.397 | 959.786 | 20.88 |
| 29 | 1 | 958.481 | 0.384 | 958.502 | 21.60 |
| 29 | 1 | 958.021 | 0.318 | 957.971 | 22.33 |
| 29 | 1 | 958.481 | 0.384 | 958.497 | 23.20 |
| 29 | 1 | 958.084 | 0.324 | 958.119 | 23.99 |
| 29 | 1 | 958.805 | 0.856 | 958.897 | 25.84 |
| 29 | 1 | 957.403 | 0.333 | 957.354 | 26.96 |
| 29 | 1 | 958.021 | 0.318 | 958.019 | 27.97 |
| 29 | 1 | 958.660 | 0.334 | 958.640 | 28.97 |
| 29 | 1 | 958.307 | 0.262 | 958.334 | 30.04 |
| 29 | 1 | 958.778 | 0.526 | 958.770 | 31.16 |
| 29 | 2 | 960.082 | 0.266 | 960.127 | 32.38 |
| 29 | 2 | 958.805 | 0.856 | 958.845 | 33.65 |
| 29 | 2 | 959.488 | 0.316 | 959.282 | 34.98 |
| 29 | 2 | 960.082 | 0.266 | 960.091 | 36.40 |
| 29 | 2 | 959.654 | 0.468 | 959.674 | 37.92 |
| 29 | 2 | 959.488 | 0.316 | 959.512 | 39.59 |
| 29 | 2 | 959.654 | 0.468 | 959.605 | 41.39 |
| 29 | 2 | 959.488 | 0.316 | 959.528 | 43.41 |
| 29 | 2 | 959.488 | 0.316 | 959.276 | 45.64 |
| 29 | 1 | 960.082 | 0.266 | 960.127 | 32.38 |
| 29 | 1 | 958.805 | 0.856 | 958.845 | 33.65 |
| 29 | 1 | 959.488 | 0.316 | 959.282 | 34.98 |
| 29 | 1 | 960.082 | 0.266 | 960.091 | 36.40 |

| 1999 - ABRIL | | | | | |
|---|---|---|---|---|---|
| D | L | SDB | ER | SDC | HL |
| 29 | 1 | 959.654 | 0.468 | 959.674 | 37.92 |
| 29 | 1 | 959.488 | 0.316 | 959.512 | 39.59 |
| 29 | 1 | 959.654 | 0.468 | 959.605 | 41.39 |
| 29 | 1 | 959.488 | 0.316 | 959.528 | 43.41 |
| 29 | 1 | 959.488 | 0.316 | 959.276 | 45.64 |
| 30 | 2 | 958.567 | 0.212 | 958.739 | 23.25 |
| 30 | 2 | 958.342 | 0.212 | 958.385 | 24.22 |
| 30 | 2 | 959.810 | 0.222 | 959.752 | 25.19 |
| 30 | 2 | 959.013 | 0.175 | 958.914 | 26.14 |
| 30 | 2 | 957.991 | 0.159 | 957.830 | 27.21 |
| 30 | 2 | 958.567 | 0.212 | 958.766 | 28.33 |
| 30 | 2 | 959.036 | 0.254 | 959.032 | 29.55 |
| 30 | 2 | 957.991 | 0.159 | 957.970 | 30.71 |
| 30 | 2 | 958.236 | 0.172 | 958.133 | 32.32 |
| 30 | 1 | 958.567 | 0.212 | 958.739 | 23.25 |
| 30 | 1 | 958.342 | 0.212 | 958.385 | 24.22 |
| 30 | 1 | 959.810 | 0.222 | 959.752 | 25.19 |
| 30 | 1 | 959.013 | 0.175 | 958.914 | 26.14 |
| 30 | 1 | 957.991 | 0.159 | 957.830 | 27.21 |
| 30 | 1 | 958.567 | 0.212 | 958.766 | 28.33 |
| 30 | 1 | 959.036 | 0.254 | 959.032 | 29.55 |
| 30 | 1 | 957.991 | 0.159 | 957.970 | 30.71 |
| 30 | 1 | 958.236 | 0.172 | 958.133 | 32.32 |
| 30 | 2 | 959.473 | 0.359 | 959.464 | 33.70 |
| 30 | 2 | 958.264 | 0.230 | 958.281 | 35.24 |
| 30 | 2 | 958.567 | 0.212 | 958.573 | 36.78 |
| 30 | 2 | 958.264 | 0.230 | 958.298 | 38.42 |
| 30 | 2 | 957.991 | 0.159 | 957.607 | 40.36 |
| 30 | 2 | 958.489 | 0.323 | 958.526 | 42.49 |
| 30 | 2 | 959.013 | 0.175 | 958.825 | 44.75 |
| 30 | 1 | 959.473 | 0.359 | 959.464 | 33.70 |
| 30 | 1 | 958.264 | 0.230 | 958.281 | 35.24 |
| 30 | 1 | 958.567 | 0.212 | 958.573 | 36.78 |
| 30 | 1 | 958.264 | 0.230 | 958.298 | 38.42 |
| 30 | 1 | 957.991 | 0.159 | 957.607 | 40.36 |
| 30 | 1 | 958.489 | 0.323 | 958.526 | 42.49 |
| 30 | 1 | 959.013 | 0.175 | 958.825 | 44.75 |

| 1999 - MAIO | | | | | |
|---|---|---|---|---|---|
| D | L | SDB | ER | SDC | HL |
| 05 | 2 | 960.433 | 0.236 | 960.460 | 24.21 |
| 05 | 2 | 960.639 | 0.202 | 960.919 | 25.13 |
| 05 | 2 | 959.348 | 0.207 | 959.353 | 26.32 |
| 05 | 2 | 958.445 | 0.238 | 958.527 | 27.62 |
| 05 | 2 | 959.626 | 0.243 | 959.616 | 28.70 |
| 05 | 2 | 959.680 | 0.251 | 959.666 | 29.81 |
| 05 | 2 | 959.040 | 0.224 | 959.106 | 30.93 |
| 05 | 2 | 958.810 | 0.217 | 958.875 | 32.37 |
| 05 | 2 | 957.934 | 0.272 | 957.679 | 33.69 |
| 05 | 1 | 960.433 | 0.236 | 960.460 | 24.21 |
| 05 | 1 | 960.639 | 0.202 | 960.919 | 25.13 |
| 05 | 1 | 959.348 | 0.207 | 959.353 | 26.32 |
| 05 | 1 | 958.445 | 0.238 | 958.527 | 27.62 |
| 05 | 1 | 959.626 | 0.243 | 959.616 | 28.70 |
| 05 | 1 | 959.680 | 0.251 | 959.666 | 29.81 |
| 05 | 1 | 959.040 | 0.224 | 959.106 | 30.93 |
| 05 | 1 | 958.810 | 0.217 | 958.875 | 32.37 |
| 05 | 1 | 957.934 | 0.272 | 957.679 | 33.69 |
| 05 | 2 | 959.040 | 0.224 | 959.144 | 35.05 |
| 05 | 2 | 958.390 | 0.196 | 958.407 | 36.60 |
| 05 | 2 | 958.664 | 0.256 | 958.556 | 38.11 |
| 05 | 2 | 959.720 | 0.255 | 959.772 | 39.73 |
| 05 | 2 | 958.102 | 0.256 | 958.035 | 41.51 |
| 05 | 2 | 959.040 | 0.224 | 959.133 | 45.67 |
| 05 | 2 | 959.720 | 0.255 | 959.731 | 48.23 |
| 05 | 1 | 959.040 | 0.224 | 959.144 | 35.05 |
| 05 | 1 | 958.390 | 0.196 | 958.407 | 36.60 |
| 05 | 1 | 958.664 | 0.256 | 958.556 | 38.11 |
| 05 | 1 | 959.720 | 0.255 | 959.772 | 39.73 |
| 05 | 1 | 958.102 | 0.256 | 958.035 | 41.51 |

| 1999 - MAIO | | | | | | 1999 - MAIO | | | | |
|---|---|---|---|---|---|---|---|---|---|---|
| D | L | SDB | ER | SDC | HL | D | L | SDB | ER | SDC | HL |
| 05 | 1 | 959.040 | 0.224 | 959.133 | 45.67 | 12 | 1 | 958.423 | 0.229 | 958.350 | 35.36 |
| 05 | 1 | 959.720 | 0.255 | 959.731 | 48.28 | 12 | 2 | 958.747 | 0.236 | 958.769 | 36.92 |
| 10 | 2 | 959.356 | 0.193 | 959.389 | 30.89 | 12 | 2 | 959.057 | 0.221 | 959.203 | 38.36 |
| 10 | 2 | 958.921 | 0.276 | 958.928 | 31.93 | 12 | 2 | 957.849 | 0.225 | 957.787 | 39.97 |
| 10 | 2 | 959.319 | 0.160 | 959.229 | 33.02 | 12 | 2 | 958.518 | 0.523 | 958.534 | 41.57 |
| 10 | 2 | 958.434 | 0.194 | 958.462 | 34.13 | 12 | 2 | 958.432 | 0.329 | 958.434 | 43.40 |
| 10 | 2 | 959.356 | 0.193 | 959.527 | 35.31 | 12 | 1 | 958.747 | 0.236 | 958.769 | 36.92 |
| 10 | 1 | 959.356 | 0.193 | 959.389 | 30.89 | 12 | 1 | 959.057 | 0.221 | 959.203 | 38.36 |
| 10 | 1 | 958.921 | 0.276 | 958.928 | 31.93 | 12 | 1 | 957.849 | 0.225 | 957.787 | 39.97 |
| 10 | 1 | 959.319 | 0.160 | 959.229 | 33.02 | 12 | 1 | 958.518 | 0.523 | 958.534 | 41.57 |
| 10 | 1 | 958.434 | 0.194 | 958.462 | 34.13 | 12 | 1 | 958.432 | 0.329 | 958.434 | 43.40 |
| 10 | 1 | 959.356 | 0.193 | 959.527 | 35.31 | 17 | 2 | 959.576 | 0.189 | 959.578 | 31.26 |
| 10 | 2 | 957.730 | 0.214 | 957.963 | 36.55 | 17 | 2 | 959.615 | 0.173 | 959.762 | 32.38 |
| 10 | 2 | 958.594 | 0.207 | 958.677 | 37.87 | 17 | 2 | 959.615 | 0.173 | 959.665 | 33.50 |
| 10 | 2 | 958.764 | 0.216 | 958.794 | 39.28 | 17 | 2 | 959.576 | 0.189 | 959.526 | 34.64 |
| 10 | 2 | 958.594 | 0.207 | 958.678 | 40.80 | 17 | 2 | 959.263 | 0.223 | 959.265 | 35.82 |
| 10 | 2 | 959.096 | 0.269 | 959.045 | 42.43 | 17 | 2 | 959.263 | 0.223 | 959.285 | 38.48 |
| 10 | 2 | 958.288 | 0.238 | 958.193 | 44.29 | 17 | 1 | 959.576 | 0.189 | 959.578 | 31.26 |
| 10 | 2 | 958.400 | 0.213 | 958.400 | 46.28 | 17 | 1 | 959.615 | 0.173 | 959.762 | 32.38 |
| 10 | 2 | 958.594 | 0.207 | 958.615 | 48.55 | 17 | 1 | 959.615 | 0.173 | 959.665 | 33.50 |
| 10 | 2 | 959.096 | 0.269 | 959.047 | 51.10 | 17 | 1 | 959.576 | 0.189 | 959.526 | 34.64 |
| 10 | 1 | 957.730 | 0.214 | 957.963 | 36.55 | 17 | 1 | 959.263 | 0.223 | 959.265 | 35.82 |
| 10 | 1 | 958.594 | 0.207 | 958.677 | 37.87 | 17 | 1 | 959.263 | 0.223 | 959.285 | 38.48 |
| 10 | 1 | 958.764 | 0.216 | 958.794 | 39.28 | 17 | 2 | 959.259 | 0.212 | 959.167 | 39.93 |
| 10 | 1 | 958.594 | 0.207 | 958.678 | 40.80 | 17 | 2 | 958.696 | 0.170 | 958.320 | 41.44 |
| 10 | 1 | 959.096 | 0.269 | 959.045 | 42.43 | 17 | 2 | 959.466 | 0.189 | 959.481 | 43.05 |
| 10 | 1 | 958.288 | 0.238 | 958.193 | 44.29 | 17 | 2 | 959.615 | 0.173 | 959.749 | 44.97 |
| 10 | 1 | 958.400 | 0.213 | 958.400 | 46.28 | 17 | 2 | 958.874 | 0.176 | 958.899 | 46.96 |
| 10 | 1 | 958.594 | 0.207 | 958.615 | 48.55 | 17 | 2 | 959.259 | 0.212 | 959.166 | 49.12 |
| 10 | 1 | 959.096 | 0.269 | 959.047 | 51.10 | 17 | 2 | 958.958 | 0.214 | 959.019 | 51.59 |
| 11 | 2 | 958.735 | 0.219 | 958.734 | 28.94 | 17 | 1 | 959.259 | 0.212 | 959.167 | 39.93 |
| 11 | 2 | 958.570 | 0.224 | 958.548 | 30.12 | 17 | 1 | 958.696 | 0.170 | 958.320 | 41.44 |
| 11 | 2 | 958.138 | 0.231 | 958.071 | 31.36 | 17 | 1 | 959.466 | 0.189 | 959.481 | 43.05 |
| 11 | 2 | 957.925 | 0.232 | 957.841 | 32.48 | 17 | 1 | 959.615 | 0.173 | 959.749 | 44.97 |
| 11 | 2 | 957.982 | 0.317 | 957.977 | 33.61 | 17 | 1 | 958.874 | 0.176 | 958.899 | 46.96 |
| 11 | 2 | 959.204 | 0.306 | 959.442 | 34.85 | 17 | 1 | 959.259 | 0.212 | 959.166 | 49.12 |
| 11 | 2 | 958.335 | 0.289 | 958.285 | 36.49 | 17 | 1 | 958.958 | 0.214 | 959.019 | 51.59 |
| 11 | 1 | 958.735 | 0.219 | 958.734 | 28.94 | 18 | 2 | 959.120 | 0.212 | 959.233 | 30.54 |
| 11 | 1 | 958.570 | 0.224 | 958.548 | 30.12 | 18 | 2 | 960.070 | 0.208 | 960.142 | 31.54 |
| 11 | 1 | 958.138 | 0.231 | 958.071 | 31.36 | 18 | 2 | 959.781 | 0.155 | 959.901 | 32.54 |
| 11 | 1 | 957.925 | 0.232 | 957.841 | 32.48 | 18 | 2 | 959.781 | 0.155 | 959.797 | 33.61 |
| 11 | 1 | 957.982 | 0.317 | 957.977 | 33.61 | 18 | 2 | 959.514 | 0.137 | 959.329 | 34.70 |
| 11 | 1 | 959.204 | 0.306 | 959.442 | 34.85 | 18 | 2 | 959.120 | 0.212 | 959.163 | 35.87 |
| 11 | 1 | 958.335 | 0.289 | 958.285 | 36.49 | 18 | 2 | 958.052 | 0.223 | 958.168 | 37.10 |
| 11 | 2 | 958.570 | 0.224 | 958.583 | 37.91 | 18 | 2 | 959.514 | 0.137 | 959.347 | 38.39 |
| 11 | 2 | 957.589 | 0.278 | 957.743 | 39.72 | 18 | 1 | 959.120 | 0.212 | 959.233 | 30.54 |
| 11 | 2 | 957.539 | 0.273 | 957.536 | 41.38 | 18 | 1 | 960.070 | 0.208 | 960.142 | 31.54 |
| 11 | 2 | 958.360 | 0.245 | 958.449 | 43.17 | 18 | 1 | 959.781 | 0.155 | 959.901 | 32.54 |
| 11 | 2 | 957.982 | 0.317 | 958.037 | 45.07 | 18 | 1 | 959.781 | 0.155 | 959.797 | 33.61 |
| 11 | 1 | 958.570 | 0.224 | 958.583 | 37.91 | 18 | 1 | 959.514 | 0.137 | 959.329 | 34.70 |
| 11 | 1 | 957.589 | 0.278 | 957.743 | 39.72 | 18 | 1 | 959.120 | 0.212 | 959.163 | 35.87 |
| 11 | 1 | 957.539 | 0.273 | 957.536 | 41.38 | 18 | 1 | 958.052 | 0.223 | 958.168 | 37.10 |
| 11 | 1 | 958.360 | 0.245 | 958.449 | 43.17 | 18 | 1 | 959.514 | 0.137 | 959.347 | 38.39 |
| 11 | 1 | 957.982 | 0.317 | 958.037 | 45.07 | 18 | 2 | 958.645 | 0.186 | 958.659 | 39.75 |
| 12 | 2 | 959.057 | 0.221 | 959.045 | 26.82 | 18 | 2 | 958.521 | 0.187 | 958.535 | 41.63 |
| 12 | 2 | 959.443 | 0.176 | 959.842 | 27.73 | 18 | 2 | 958.052 | 0.223 | 957.941 | 43.27 |
| 12 | 2 | 959.057 | 0.221 | 959.099 | 28.69 | 18 | 2 | 958.645 | 0.186 | 958.651 | 45.10 |
| 12 | 2 | 958.609 | 0.296 | 958.635 | 29.66 | 18 | 2 | 958.645 | 0.186 | 958.660 | 47.01 |
| 12 | 2 | 958.813 | 0.199 | 958.824 | 30.72 | 18 | 1 | 958.645 | 0.186 | 958.659 | 39.75 |
| 12 | 2 | 958.747 | 0.236 | 958.731 | 31.78 | 18 | 1 | 958.521 | 0.187 | 958.535 | 41.63 |
| 12 | 2 | 959.057 | 0.221 | 958.942 | 32.89 | 18 | 1 | 958.052 | 0.223 | 957.941 | 43.27 |
| 12 | 2 | 957.849 | 0.225 | 958.038 | 34.08 | 18 | 1 | 958.645 | 0.186 | 958.651 | 45.10 |
| 12 | 2 | 958.423 | 0.229 | 958.350 | 35.36 | 18 | 1 | 958.645 | 0.186 | 958.660 | 47.01 |
| 12 | 1 | 959.057 | 0.221 | 959.045 | 26.82 | 19 | 2 | 960.665 | 0.165 | 960.649 | 31.57 |
| 12 | 1 | 959.443 | 0.176 | 959.842 | 27.73 | 19 | 2 | 959.835 | 0.214 | 959.778 | 32.62 |
| 12 | 1 | 959.057 | 0.221 | 959.099 | 28.69 | 19 | 2 | 959.183 | 0.251 | 959.184 | 33.66 |
| 12 | 1 | 958.609 | 0.296 | 958.635 | 29.66 | 19 | 2 | 958.367 | 0.292 | 957.244 | 34.79 |
| 12 | 1 | 958.813 | 0.199 | 958.824 | 30.72 | 19 | 2 | 959.650 | 0.322 | 959.597 | 36.18 |
| 12 | 1 | 958.747 | 0.236 | 958.731 | 31.78 | 19 | 2 | 958.367 | 0.292 | 958.493 | 37.48 |
| 12 | 1 | 959.057 | 0.221 | 958.942 | 32.89 | 19 | 1 | 960.665 | 0.165 | 960.649 | 31.57 |
| 12 | 1 | 957.849 | 0.225 | 958.038 | 34.08 | 19 | 1 | 959.835 | 0.214 | 959.778 | 32.62 |

| \ | \ | 1999 - MAIO | \ | \ | \ | | \ | \ | 1999 - JUNHO | \ | \ | \ |
|---|---|---|---|---|---|---|---|---|---|---|---|---|
| D | L | SDB | ER | SDC | HL | | D | L | SDB | ER | SDC | HL |
| 19 | 1 | 959.183 | 0.251 | 959.184 | 33.66 | | 01 | 2 | 959.290 | 0.352 | 959.324 | 43.47 |
| 19 | 1 | 958.367 | 0.292 | 957.244 | 34.79 | | 01 | 2 | 958.927 | 0.196 | 958.933 | 39.33 |
| 19 | 1 | 959.650 | 0.322 | 959.597 | 36.18 | | 01 | 1 | 959.290 | 0.352 | 959.291 | 40.69 |
| 19 | 1 | 958.367 | 0.292 | 958.493 | 37.48 | | 01 | 1 | 959.933 | 0.243 | 960.030 | 42.05 |
| 19 | 2 | 959.836 | 0.234 | 959.996 | 40.26 | | 01 | 1 | 959.290 | 0.352 | 959.324 | 43.47 |
| 19 | 2 | 959.835 | 0.214 | 959.827 | 43.73 | | 01 | 2 | 958.413 | 0.359 | 958.087 | 44.94 |
| 19 | 2 | 960.665 | 0.165 | 960.318 | 45.51 | | 01 | 2 | 958.915 | 0.292 | 958.908 | 47.22 |
| 19 | 2 | 959.004 | 0.254 | 959.022 | 47.63 | | 01 | 2 | 959.261 | 0.209 | 959.247 | 48.88 |
| 19 | 2 | 959.183 | 0.251 | 959.156 | 49.98 | | 01 | 2 | 959.933 | 0.243 | 959.781 | 50.62 |
| 19 | 1 | 959.836 | 0.234 | 959.996 | 40.26 | | 01 | 2 | 959.933 | 0.243 | 960.098 | 52.57 |
| 19 | 1 | 959.835 | 0.214 | 959.827 | 43.73 | | 01 | 2 | 958.850 | 0.357 | 958.785 | 54.97 |
| 19 | 1 | 960.665 | 0.165 | 960.318 | 45.51 | | 01 | 1 | 958.413 | 0.359 | 958.087 | 44.94 |
| 19 | 1 | 959.004 | 0.254 | 959.022 | 47.63 | | 01 | 1 | 958.915 | 0.292 | 958.908 | 47.22 |
| 19 | 1 | 959.183 | 0.251 | 959.156 | 49.98 | | 01 | 1 | 959.261 | 0.209 | 959.247 | 48.88 |
| 24 | 2 | 960.467 | 0.243 | 960.721 | 33.15 | | 01 | 1 | 959.933 | 0.243 | 959.781 | 50.62 |
| 24 | 2 | 960.306 | 0.250 | 960.159 | 34.23 | | 01 | 1 | 959.933 | 0.243 | 960.098 | 52.57 |
| 24 | 2 | 959.461 | 0.209 | 959.462 | 35.41 | | 01 | 1 | 958.850 | 0.357 | 958.785 | 54.97 |
| 24 | 2 | 959.633 | 0.222 | 959.603 | 36.64 | | 02 | 2 | 958.063 | 0.474 | 958.084 | 40.05 |
| 24 | 2 | 959.441 | 0.169 | 959.296 | 37.84 | | 02 | 2 | 959.788 | 0.173 | 959.824 | 43.72 |
| 24 | 2 | 959.441 | 0.169 | 959.334 | 39.16 | | 02 | 1 | 958.063 | 0.474 | 958.084 | 40.05 |
| 24 | 2 | 960.306 | 0.250 | 960.311 | 40.51 | | 02 | 1 | 959.788 | 0.173 | 959.824 | 43.72 |
| 24 | 1 | 960.467 | 0.243 | 960.721 | 33.15 | | 02 | 2 | 959.660 | 0.197 | 959.655 | 45.27 |
| 24 | 1 | 960.306 | 0.250 | 960.159 | 34.23 | | 02 | 2 | 960.403 | 0.198 | 960.847 | 46.74 |
| 24 | 1 | 959.461 | 0.209 | 959.462 | 35.41 | | 02 | 2 | 958.781 | 0.195 | 958.931 | 48.32 |
| 24 | 1 | 959.633 | 0.222 | 959.603 | 36.64 | | 02 | 2 | 959.788 | 0.173 | 959.795 | 50.01 |
| 24 | 1 | 959.441 | 0.169 | 959.296 | 37.84 | | 02 | 2 | 958.781 | 0.195 | 958.970 | 51.85 |
| 24 | 1 | 959.441 | 0.169 | 959.334 | 39.16 | | 02 | 2 | 960.403 | 0.198 | 960.348 | 54.02 |
| 24 | 1 | 960.306 | 0.250 | 960.311 | 40.51 | | 02 | 2 | 959.788 | 0.173 | 959.938 | 56.40 |
| 24 | 2 | 958.589 | 0.239 | 958.544 | 42.00 | | 02 | 2 | 958.480 | 0.259 | 958.466 | 59.34 |
| 24 | 2 | 959.897 | 0.173 | 959.959 | 43.50 | | 02 | 2 | 958.012 | 0.265 | 958.036 | 62.76 |
| 24 | 2 | 958.964 | 0.206 | 959.070 | 45.15 | | 02 | 1 | 959.660 | 0.197 | 959.655 | 45.27 |
| 24 | 2 | 958.462 | 0.265 | 958.368 | 46.89 | | 02 | 1 | 960.403 | 0.198 | 960.847 | 46.74 |
| 24 | 2 | 958.462 | 0.265 | 958.495 | 50.52 | | 02 | 1 | 958.781 | 0.195 | 958.931 | 48.32 |
| 24 | 2 | 960.467 | 0.243 | 960.521 | 53.03 | | 02 | 1 | 959.788 | 0.173 | 959.795 | 50.01 |
| 24 | 2 | 958.940 | 0.249 | 958.922 | 55.71 | | 02 | 1 | 958.781 | 0.195 | 958.970 | 51.85 |
| 24 | 2 | 959.461 | 0.209 | 959.486 | 58.99 | | 02 | 1 | 960.403 | 0.198 | 960.348 | 54.02 |
| 24 | 1 | 958.589 | 0.239 | 958.544 | 42.00 | | 02 | 1 | 959.788 | 0.173 | 959.938 | 56.40 |
| 24 | 1 | 959.897 | 0.173 | 959.959 | 43.50 | | 02 | 1 | 958.480 | 0.259 | 958.466 | 59.34 |
| 24 | 1 | 958.964 | 0.206 | 959.070 | 45.15 | | 02 | 1 | 958.012 | 0.265 | 958.036 | 62.76 |
| 24 | 1 | 958.462 | 0.265 | 958.368 | 46.89 | | 10 | 2 | 959.043 | 0.200 | 958.932 | 43.09 |
| 24 | 1 | 958.462 | 0.265 | 958.495 | 50.52 | | 10 | 2 | 958.715 | 0.189 | 958.791 | 44.24 |
| 24 | 1 | 960.467 | 0.243 | 960.521 | 53.03 | | 10 | 2 | 959.345 | 0.178 | 959.400 | 45.44 |
| 24 | 1 | 958.940 | 0.249 | 958.922 | 55.71 | | 10 | 2 | 958.715 | 0.189 | 958.792 | 46.70 |
| 24 | 1 | 959.461 | 0.209 | 959.486 | 58.99 | | 10 | 2 | 958.715 | 0.189 | 958.876 | 48.03 |
| 26 | 2 | 959.934 | 0.291 | 959.876 | 35.10 | | 10 | 1 | 959.043 | 0.200 | 958.932 | 43.09 |
| 26 | 2 | 958.398 | 0.192 | 958.408 | 36.27 | | 10 | 1 | 958.715 | 0.189 | 958.791 | 44.24 |
| 26 | 2 | 959.363 | 0.208 | 959.370 | 37.40 | | 10 | 1 | 959.345 | 0.178 | 959.400 | 45.44 |
| 26 | 2 | 960.065 | 0.224 | 960.314 | 38.61 | | 10 | 1 | 958.715 | 0.189 | 958.792 | 46.70 |
| 26 | 2 | 959.139 | 0.156 | 959.207 | 39.88 | | 10 | 1 | 958.715 | 0.189 | 958.876 | 48.03 |
| 26 | 2 | 959.934 | 0.291 | 959.793 | 41.16 | | 10 | 2 | 959.100 | 0.163 | 959.101 | 49.41 |
| 26 | 1 | 959.934 | 0.291 | 959.876 | 35.10 | | 10 | 2 | 959.111 | 0.150 | 959.121 | 50.92 |
| 26 | 1 | 958.398 | 0.192 | 958.408 | 36.27 | | 10 | 2 | 958.420 | 0.158 | 958.458 | 52.53 |
| 26 | 1 | 959.363 | 0.208 | 959.370 | 37.40 | | 10 | 2 | 958.691 | 0.217 | 958.664 | 54.23 |
| 26 | 1 | 960.065 | 0.224 | 960.314 | 38.61 | | 10 | 2 | 959.345 | 0.178 | 959.616 | 56.07 |
| 26 | 1 | 959.139 | 0.156 | 959.207 | 39.88 | | 10 | 2 | 958.420 | 0.158 | 958.427 | 58.16 |
| 26 | 1 | 959.934 | 0.291 | 959.793 | 41.16 | | 10 | 1 | 959.100 | 0.163 | 959.101 | 49.41 |
| 26 | 2 | 958.738 | 0.332 | 958.573 | 42.61 | | 10 | 1 | 959.111 | 0.150 | 959.121 | 50.92 |
| 26 | 2 | 958.121 | 0.306 | 958.090 | 44.12 | | 10 | 1 | 958.420 | 0.158 | 958.458 | 52.53 |
| 26 | 2 | 959.363 | 0.208 | 959.289 | 45.68 | | 10 | 1 | 958.691 | 0.217 | 958.664 | 54.23 |
| 26 | 2 | 958.823 | 0.206 | 958.883 | 47.43 | | 10 | 1 | 959.345 | 0.178 | 959.616 | 56.07 |
| 26 | 1 | 958.738 | 0.332 | 958.573 | 42.61 | | 10 | 1 | 958.420 | 0.158 | 958.427 | 58.16 |
| 26 | 1 | 958.121 | 0.306 | 958.090 | 44.12 | | 22 | 2 | 959.838 | 0.170 | 960.439 | 48.47 |
| 26 | 1 | 959.363 | 0.208 | 959.289 | 45.68 | | 22 | 2 | 959.685 | 0.159 | 959.744 | 49.62 |
| 26 | 1 | 958.823 | 0.206 | 958.883 | 47.43 | | 22 | 2 | 959.838 | 0.170 | 960.476 | 50.86 |
| | | | | | | | 22 | 2 | 959.838 | 0.170 | 959.849 | 52.14 |
| | | | | | | | 22 | 1 | 959.838 | 0.170 | 960.439 | 48.47 |
| | | 1999 - JUNHO | | | | | 22 | 1 | 959.685 | 0.159 | 959.744 | 49.62 |
| D | L | SDB | ER | SDC | HL | | 22 | 1 | 959.838 | 0.170 | 960.476 | 50.86 |
| 01 | 2 | 958.927 | 0.196 | 958.933 | 39.33 | | 22 | 1 | 959.838 | 0.170 | 959.849 | 52.14 |
| 01 | 2 | 959.290 | 0.352 | 959.291 | 40.69 | | 22 | 2 | 959.838 | 0.170 | 960.116 | 53.55 |
| 01 | 2 | 959.933 | 0.243 | 960.030 | 42.05 | | 22 | 2 | 958.870 | 0.195 | 958.859 | 55.00 |

```
         1999 - JUNHO                              1999 - JUNHO
 D  L    SDB     ER     SDC    HL         D  L    SDB     ER     SDC    HL
 22 2  959.576  0.139  959.600  56.49     28 1  959.235  0.178  959.291  54.83
 22 2  959.524  0.180  959.495  58.13     28 2  959.481  0.161  959.585  56.19
 22 2  958.870  0.195  959.078  59.85     28 2  959.481  0.161  959.408  57.67
 22 2  959.524  0.180  959.490  61.78     28 2  959.235  0.178  959.344  59.13
 22 2  958.870  0.195  958.865  63.86     28 2  958.991  0.160  959.066  60.76
 22 1  959.838  0.170  960.116  53.55     28 2  959.191  0.170  959.133  62.45
 22 1  958.870  0.195  958.859  55.00     28 2  959.235  0.178  959.292  64.47
 22 1  959.576  0.139  959.600  56.49     28 2  958.947  0.144  958.941  66.64
 22 1  959.524  0.180  959.495  58.13     28 1  959.481  0.161  959.585  56.19
 22 1  958.870  0.195  959.078  59.85     28 1  959.481  0.161  959.408  57.67
 22 1  959.524  0.180  959.490  61.78     28 1  959.235  0.178  959.344  59.13
 22 1  958.870  0.195  958.865  63.86     28 1  958.991  0.160  959.066  60.76
 23 2  960.567  0.162  960.656  50.53     28 1  959.191  0.170  959.133  62.45
 23 2  960.100  0.195  960.089  51.81     28 1  959.235  0.178  959.292  64.47
 23 2  960.118  0.205  960.263  53.12     28 1  958.947  0.144  958.941  66.64
 23 1  960.567  0.162  960.656  50.53     29 2  959.153  0.213  959.101  50.14
 23 1  960.100  0.195  960.089  51.81     29 2  958.708  0.154  958.557  51.22
 23 1  960.118  0.205  960.263  53.12     29 2  959.594  0.147  959.559  52.34
 23 2  959.699  0.226  959.687  54.54     29 2  959.341  0.182  959.440  53.53
 23 2  959.699  0.226  959.692  57.91     29 2  958.907  0.153  959.002  54.75
 23 2  959.057  0.161  959.060  59.72     29 1  959.153  0.213  959.101  50.14
 23 2  959.699  0.226  959.736  61.59     29 1  958.708  0.154  958.557  51.22
 23 2  959.486  0.200  959.504  63.71     29 1  959.594  0.147  959.559  52.34
 23 2  958.429  0.300  958.645  66.08     29 1  959.341  0.182  959.440  53.53
 23 1  959.699  0.226  959.687  54.54     29 1  958.907  0.153  959.002  54.75
 23 1  959.699  0.226  959.692  57.91     29 2  959.858  0.154  959.953  56.13
 23 1  959.057  0.161  959.060  59.72     29 2  958.708  0.154  958.711  57.58
 23 1  959.699  0.226  959.736  61.59     29 2  958.202  0.191  958.221  59.04
 23 1  959.486  0.200  959.504  63.71     29 2  959.764  0.161  959.721  60.60
 23 1  958.429  0.300  958.645  66.08     29 2  958.750  0.225  958.769  62.37
 24 2  959.923  0.732  959.536  53.22     29 2  959.341  0.182  959.457  64.22
 24 1  959.923  0.732  959.536  53.22     29 2  958.708  0.154  960.045  66.30
 24 2  958.502  0.265  958.471  55.96     29 1  959.858  0.154  959.953  56.13
 24 2  959.923  0.732  960.536  57.39     29 1  958.708  0.154  958.711  57.58
 24 2  958.502  0.265  958.525  58.94     29 1  958.202  0.191  958.221  59.04
 24 2  958.502  0.265  958.754  60.61     29 1  959.764  0.161  959.721  60.60
 24 2  957.919  0.283  957.977  62.47     29 1  958.750  0.225  958.769  62.37
 24 2  958.193  0.433  958.059  66.77     29 1  959.341  0.182  959.457  64.22
 24 1  958.502  0.265  958.471  55.96     29 1  959.858  0.154  960.045  66.30
 24 1  959.923  0.732  960.536  57.39
 24 1  958.502  0.265  958.525  58.94              1999 - JULHO
 24 1  958.502  0.265  958.754  60.61      D  L    SDB     ER     SDC    HL
 24 1  957.919  0.283  957.977  62.47     02 2  958.540  0.196  958.658  51.14
 24 1  958.193  0.433  958.059  66.77     02 2  959.448  0.226  959.491  52.29
 25 2  958.750  0.168  958.716  48.51     02 2  958.540  0.196  958.536  53.46
 25 2  959.394  0.184  959.437  49.58     02 2  959.190  0.250  959.233  54.68
 25 2  959.394  0.184  959.722  50.74     02 2  959.190  0.250  959.251  55.91
 25 2  958.466  0.221  958.576  51.95     02 1  958.540  0.196  958.658  51.14
 25 2  959.394  0.184  959.477  53.24     02 1  959.448  0.226  959.491  52.29
 25 1  958.750  0.168  958.716  48.51     02 1  958.540  0.196  958.536  53.46
 25 1  959.394  0.184  959.437  49.58     02 1  959.190  0.250  959.233  54.68
 25 1  959.394  0.184  959.722  50.74     02 1  959.190  0.250  959.251  55.91
 25 1  958.466  0.221  958.576  51.95     02 2  958.807  0.252  958.831  57.18
 25 1  959.394  0.184  959.477  53.24     02 2  958.540  0.196  958.656  58.55
 25 2  958.981  0.173  958.931  54.56     02 2  958.540  0.196  958.579  59.98
 25 2  958.021  0.288  958.180  55.95     02 2  957.806  0.328  957.954  61.65
 25 2  958.750  0.168  958.711  57.42     02 2  959.190  0.250  959.247  63.32
 25 2  958.750  0.168  958.696  59.00     02 2  958.891  0.290  958.896  65.15
 25 2  958.984  0.282  959.018  62.15     02 2  958.990  0.270  959.028  67.12
 25 2  959.394  0.184  959.538  64.16     02 1  958.807  0.252  958.831  57.18
 25 2  958.981  0.173  958.935  66.45     02 1  958.540  0.196  958.656  58.55
 25 1  958.981  0.173  958.931  54.56     02 1  958.540  0.196  958.579  59.98
 25 1  958.021  0.288  958.180  55.95     02 1  957.806  0.328  957.954  61.65
 25 1  958.750  0.168  958.711  57.42     02 1  959.190  0.250  959.247  63.32
 25 1  958.750  0.168  958.696  59.00     02 1  958.891  0.290  958.896  65.15
 25 1  958.984  0.282  959.018  62.15     02 1  958.990  0.270  959.028  67.12
 25 1  959.394  0.184  959.538  64.16     07 2  958.819  0.366  958.866  53.32
 25 1  958.981  0.173  958.935  66.45     07 2  958.220  0.322  958.224  56.97
 28 2  958.947  0.144  958.930  53.45     07 2  958.241  0.296  958.327  58.18
 28 2  959.235  0.178  959.291  54.83     07 1  958.819  0.366  958.866  53.32
 28 1  958.947  0.144  958.930  53.45
```

|  | 1999 - JULHO |  |  |  |
|---|---|---|---|---|
| D L | SDB | ER | SDC | HL |
| 07 1 | 958.220 | 0.322 | 958.224 | 56.97 |
| 07 1 | 958.241 | 0.296 | 958.327 | 58.18 |
| 07 2 | 957.986 | 0.403 | 957.847 | 60.83 |
| 07 2 | 959.333 | 0.260 | 959.434 | 62.24 |
| 07 2 | 958.799 | 0.252 | 958.739 | 63.84 |
| 07 2 | 958.799 | 0.252 | 958.667 | 65.69 |
| 07 2 | 958.799 | 0.252 | 958.738 | 67.50 |
| 07 2 | 957.986 | 0.403 | 957.715 | 69.78 |
| 07 1 | 957.986 | 0.403 | 957.847 | 60.83 |
| 07 1 | 959.333 | 0.260 | 959.434 | 62.24 |
| 07 1 | 958.799 | 0.252 | 958.739 | 63.84 |
| 07 1 | 958.799 | 0.252 | 958.667 | 65.69 |
| 07 1 | 958.799 | 0.252 | 958.738 | 67.50 |
| 07 1 | 957.986 | 0.403 | 957.715 | 69.78 |
| 08 2 | 959.237 | 0.190 | 959.183 | 53.68 |
| 08 2 | 958.029 | 0.380 | 958.126 | 55.75 |
| 08 2 | 958.445 | 0.291 | 958.308 | 56.89 |
| 08 1 | 959.237 | 0.190 | 959.183 | 53.68 |
| 08 1 | 958.029 | 0.380 | 958.126 | 55.75 |
| 08 1 | 958.445 | 0.291 | 958.308 | 56.89 |
| 08 2 | 958.930 | 0.252 | 958.877 | 59.45 |
| 08 2 | 957.771 | 0.281 | 957.719 | 60.74 |
| 08 2 | 959.493 | 0.290 | 959.686 | 62.05 |
| 08 2 | 959.482 | 0.498 | 959.469 | 66.34 |
| 08 2 | 958.930 | 0.252 | 958.966 | 70.09 |
| 08 2 | 958.029 | 0.380 | 958.133 | 72.27 |
| 08 2 | 959.493 | 0.290 | 959.930 | 74.80 |
| 08 1 | 958.930 | 0.252 | 958.877 | 59.45 |
| 08 1 | 957.771 | 0.281 | 957.719 | 60.74 |
| 08 1 | 959.493 | 0.290 | 959.686 | 62.05 |
| 08 1 | 959.482 | 0.498 | 959.469 | 66.34 |
| 08 1 | 958.930 | 0.252 | 958.966 | 70.09 |
| 08 1 | 958.029 | 0.380 | 958.133 | 72.27 |
| 08 1 | 959.493 | 0.290 | 959.930 | 74.80 |
| 12 2 | 959.497 | 0.267 | 959.599 | 57.42 |
| 12 2 | 960.104 | 0.221 | 960.133 | 58.61 |
| 12 2 | 960.104 | 0.221 | 960.092 | 59.81 |
| 12 1 | 959.497 | 0.267 | 959.599 | 57.42 |
| 12 1 | 960.104 | 0.221 | 960.133 | 58.61 |
| 12 1 | 960.104 | 0.221 | 960.092 | 59.81 |
| 12 2 | 959.760 | 0.346 | 959.705 | 61.10 |
| 12 2 | 960.104 | 0.221 | 960.132 | 62.40 |
| 12 2 | 959.964 | 0.200 | 959.908 | 63.80 |
| 12 2 | 959.497 | 0.267 | 959.478 | 66.88 |
| 12 2 | 959.291 | 0.223 | 959.227 | 68.70 |
| 12 2 | 959.497 | 0.267 | 959.597 | 70.66 |
| 12 2 | 959.291 | 0.223 | 959.316 | 72.79 |
| 12 1 | 959.760 | 0.346 | 959.705 | 61.10 |
| 12 1 | 960.104 | 0.221 | 960.132 | 62.40 |
| 12 1 | 959.964 | 0.200 | 959.908 | 63.80 |
| 12 1 | 959.497 | 0.267 | 959.478 | 66.88 |
| 12 1 | 959.291 | 0.223 | 959.227 | 68.70 |
| 12 1 | 959.497 | 0.267 | 959.597 | 70.66 |
| 12 1 | 959.291 | 0.223 | 959.316 | 72.79 |
| 13 2 | 959.343 | 0.305 | 959.236 | 56.95 |
| 13 2 | 959.466 | 0.197 | 959.516 | 58.08 |
| 13 2 | 959.466 | 0.197 | 959.417 | 60.45 |
| 13 1 | 959.343 | 0.305 | 959.236 | 56.95 |
| 13 1 | 959.466 | 0.197 | 959.516 | 58.08 |
| 13 1 | 959.466 | 0.197 | 959.417 | 60.45 |
| 13 2 | 958.728 | 0.484 | 958.745 | 61.72 |
| 13 2 | 958.197 | 0.334 | 958.061 | 63.04 |
| 13 2 | 958.727 | 0.253 | 958.686 | 64.44 |
| 13 2 | 959.466 | 0.197 | 960.062 | 65.97 |
| 13 2 | 958.996 | 0.429 | 958.886 | 67.58 |
| 13 1 | 958.728 | 0.484 | 958.745 | 61.72 |
| 13 1 | 958.197 | 0.334 | 958.061 | 63.04 |
| 13 1 | 958.727 | 0.253 | 958.686 | 64.44 |
| 13 1 | 959.466 | 0.197 | 960.062 | 65.97 |
| 13 1 | 958.996 | 0.429 | 958.886 | 67.58 |
| 14 2 | 960.516 | 0.229 | 960.544 | 55.99 |
|  | 1999 - JULHO |  |  |  |
| D L | SDB | ER | SDC | HL |
| 14 2 | 960.252 | 0.210 | 960.185 | 58.19 |
| 14 1 | 960.516 | 0.229 | 960.544 | 55.99 |
| 14 1 | 960.252 | 0.210 | 960.185 | 58.19 |
| 14 2 | 959.952 | 0.187 | 959.896 | 61.80 |
| 14 2 | 959.782 | 0.185 | 959.810 | 63.11 |
| 14 2 | 960.252 | 0.210 | 960.116 | 64.48 |
| 14 2 | 959.116 | 0.329 | 959.105 | 65.99 |
| 14 2 | 959.116 | 0.329 | 959.375 | 67.56 |
| 14 2 | 960.902 | 0.185 | 960.909 | 69.32 |
| 14 1 | 959.952 | 0.187 | 959.896 | 61.80 |
| 14 1 | 959.782 | 0.185 | 959.810 | 63.11 |
| 14 1 | 960.252 | 0.210 | 960.116 | 64.48 |
| 14 1 | 959.116 | 0.329 | 959.105 | 65.99 |
| 14 1 | 959.116 | 0.329 | 959.375 | 67.56 |
| 14 1 | 960.902 | 0.185 | 960.909 | 69.32 |
| 15 2 | 959.119 | 0.465 | 959.190 | 58.38 |
| 15 2 | 959.384 | 0.264 | 959.409 | 59.45 |
| 15 2 | 960.286 | 0.250 | 960.546 | 60.54 |
| 15 1 | 959.119 | 0.465 | 959.190 | 58.38 |
| 15 1 | 959.384 | 0.264 | 959.409 | 59.45 |
| 15 1 | 960.286 | 0.250 | 960.546 | 60.54 |
| 15 2 | 959.056 | 0.358 | 958.928 | 64.21 |
| 15 2 | 957.863 | 0.295 | 958.321 | 65.59 |
| 15 2 | 960.286 | 0.250 | 960.566 | 67.02 |
| 15 2 | 959.056 | 0.358 | 958.473 | 68.57 |
| 15 2 | 957.863 | 0.295 | 958.260 | 70.27 |
| 15 2 | 957.863 | 0.295 | 958.217 | 72.16 |
| 15 2 | 960.286 | 0.250 | 960.103 | 74.23 |
| 15 1 | 959.056 | 0.358 | 958.928 | 64.21 |
| 15 1 | 957.863 | 0.295 | 958.321 | 65.59 |
| 15 1 | 960.286 | 0.250 | 960.566 | 67.02 |
| 15 1 | 959.056 | 0.358 | 958.473 | 68.57 |
| 15 1 | 957.863 | 0.295 | 958.260 | 70.27 |
| 15 1 | 957.863 | 0.295 | 958.217 | 72.16 |
| 15 1 | 960.286 | 0.250 | 960.103 | 74.23 |
| 16 2 | 959.282 | 0.200 | 959.279 | 57.82 |
| 16 2 | 959.355 | 0.187 | 959.353 | 58.90 |
| 16 2 | 959.961 | 0.198 | 960.270 | 60.05 |
| 16 2 | 959.355 | 0.187 | 959.360 | 61.22 |
| 16 1 | 959.282 | 0.200 | 959.279 | 57.82 |
| 16 1 | 959.355 | 0.187 | 959.353 | 58.90 |
| 16 1 | 959.961 | 0.198 | 960.270 | 60.05 |
| 16 1 | 959.355 | 0.187 | 959.360 | 61.22 |
| 16 2 | 959.173 | 0.211 | 959.154 | 62.50 |
| 16 2 | 959.961 | 0.198 | 960.521 | 63.82 |
| 16 2 | 959.053 | 0.226 | 959.020 | 65.22 |
| 16 2 | 959.948 | 0.183 | 959.856 | 66.67 |
| 16 2 | 958.845 | 0.162 | 958.883 | 68.26 |
| 16 2 | 959.961 | 0.198 | 960.119 | 70.14 |
| 16 2 | 959.282 | 0.200 | 959.251 | 71.99 |
| 16 1 | 959.173 | 0.211 | 959.154 | 62.50 |
| 16 1 | 959.961 | 0.198 | 960.521 | 63.82 |
| 16 1 | 959.053 | 0.226 | 959.020 | 65.22 |
| 16 1 | 959.948 | 0.183 | 959.856 | 66.67 |
| 16 1 | 958.845 | 0.162 | 958.883 | 68.26 |
| 16 1 | 959.961 | 0.198 | 960.119 | 70.14 |
| 16 1 | 959.282 | 0.200 | 959.251 | 71.99 |
| 19 2 | 958.988 | 0.347 | 958.881 | 62.95 |
| 19 2 | 958.566 | 0.345 | 958.504 | 64.29 |
| 19 2 | 958.566 | 0.345 | 957.397 | 67.16 |
| 19 2 | 959.213 | 0.224 | 959.206 | 74.28 |
| 19 1 | 958.988 | 0.347 | 958.881 | 62.95 |
| 19 1 | 958.566 | 0.345 | 958.504 | 64.29 |
| 19 1 | 958.566 | 0.345 | 957.397 | 67.16 |
| 19 1 | 959.213 | 0.224 | 959.206 | 74.28 |
| 20 2 | 957.175 | 0.480 | 957.344 | 64.84 |
| 20 2 | 958.247 | 0.183 | 958.302 | 66.12 |
| 20 2 | 957.575 | 0.384 | 957.624 | 68.97 |
| 20 2 | 958.964 | 0.332 | 959.416 | 72.16 |
| 20 2 | 957.575 | 0.384 | 957.615 | 73.99 |
| 20 2 | 958.964 | 0.332 | 959.154 | 76.16 |

```
         1999 - JULHO                                    1999 - JULHO
 D  L    SDB    ER     SDC    HL            D  L    SDB    ER     SDC    HL
20  1  957.175 0.360 957.344 64.84         30  1  959.367 0.220 959.539 57.44
20  1  958.247 0.183 958.302 66.12         30  1  958.836 0.313 958.843 58.30
20  1  957.575 0.384 957.624 68.97         30  1  959.367 0.220 959.577 59.20
20  1  958.964 0.332 959.416 72.16         30  1  959.062 0.242 959.056 60.08
20  1  957.575 0.384 957.615 73.99         30  1  959.367 0.220 959.437 61.01
20  1  958.964 0.332 959.154 76.16         30  1  959.367 0.220 959.467 62.01
21  2  960.037 0.137 960.068 56.98         30  1  958.836 0.313 958.716 63.02
21  2  959.876 0.118 959.888 58.17         30  1  958.836 0.313 958.865 64.08
21  2  960.037 0.137 960.901 59.14         30  1  959.367 0.220 960.089 65.20
21  2  959.519 0.136 959.598 60.32         30  2  958.358 0.271 958.381 66.35
21  2  959.519 0.136 959.590 61.42         30  2  958.470 0.194 958.479 67.58
21  2  959.326 0.114 959.413 62.55         30  2  959.062 0.242 959.121 69.02
21  1  960.037 0.137 960.068 56.98         30  2  957.907 0.226 957.821 70.52
21  1  959.876 0.118 959.888 58.17         30  2  959.367 0.220 959.644 73.59
21  1  960.037 0.137 960.901 59.14         30  1  958.358 0.271 958.381 66.35
21  1  959.519 0.136 959.598 60.32         30  1  958.470 0.194 958.479 67.58
21  1  959.519 0.136 959.590 61.42         30  1  959.062 0.242 959.121 69.02
21  1  959.326 0.114 959.413 62.55         30  1  957.907 0.226 957.821 70.52
21  2  959.035 0.143 958.977 63.99         30  1  959.367 0.220 959.644 73.59
21  2  959.106 0.119 959.107 65.27
21  2  959.519 0.136 959.475 66.63                   1999 - AGOSTO
21  2  959.684 0.118 959.683 68.08          D  L    SDB    ER     SDC    HL
21  2  959.155 0.116 959.186 69.70         06  2  958.019 0.239 957.820 62.50
21  2  959.679 0.131 959.680 71.50         06  2  958.705 0.225 958.664 63.48
21  2  958.910 0.118 958.947 73.41         06  2  958.562 0.280 958.385 64.44
21  2  959.155 0.116 959.175 75.46         06  2  958.705 0.225 958.813 65.47
21  1  959.035 0.143 958.977 63.99         06  2  959.102 0.225 959.114 66.50
21  1  959.106 0.119 959.107 65.27         06  1  958.019 0.239 957.820 62.50
21  1  959.519 0.136 959.475 66.63         06  1  958.705 0.225 958.664 63.48
21  1  959.684 0.118 959.683 68.08         06  1  958.562 0.280 958.385 64.44
21  1  959.155 0.116 959.186 69.70         06  1  958.705 0.225 958.813 65.47
21  1  959.679 0.131 959.680 71.50         06  1  959.102 0.225 959.114 66.50
21  1  958.910 0.118 958.947 73.41         06  2  959.545 0.242 959.510 67.62
21  1  959.155 0.116 959.175 75.46         06  2  959.143 0.187 959.158 68.79
27  2  959.619 0.230 959.688 57.77         06  2  959.545 0.242 959.445 70.08
27  2  960.112 0.196 960.415 58.77         06  2  959.143 0.187 959.171 71.43
27  2  959.521 0.158 959.522 59.75         06  2  958.705 0.225 958.674 72.82
27  2  959.619 0.230 959.759 60.72         06  2  959.102 0.225 959.091 74.29
27  2  958.918 0.125 958.731 61.75         06  2  959.245 0.268 959.270 75.88
27  2  959.367 0.147 959.401 62.78         06  1  959.545 0.242 959.510 67.62
27  2  959.367 0.147 959.371 63.93         06  1  959.143 0.187 959.158 68.79
27  1  959.619 0.230 959.688 57.77         06  1  959.545 0.242 959.445 70.08
27  1  960.112 0.196 960.415 58.77         06  1  959.143 0.187 959.171 71.43
27  1  959.521 0.158 959.522 59.75         06  1  958.705 0.225 958.674 72.82
27  1  959.619 0.230 959.759 60.72         06  1  959.102 0.225 959.091 74.29
27  1  958.918 0.125 958.731 61.75         06  1  959.245 0.268 959.270 75.88
27  1  959.367 0.147 959.401 62.78         10  2  959.513 0.236 959.536 61.01
27  1  959.367 0.147 959.371 63.93         10  2  960.025 0.210 960.170 61.86
27  2  959.348 0.137 959.308 65.09         10  2  959.737 0.263 959.653 62.71
27  2  958.393 0.131 958.516 66.31         10  2  959.427 0.239 959.436 63.57
27  2  958.289 0.288 958.318 67.61         10  2  958.767 0.242 958.786 64.81
27  2  959.348 0.137 959.344 69.00         10  2  959.737 0.263 959.763 65.79
27  2  959.927 0.154 959.796 70.48         10  2  959.737 0.263 959.779 66.81
27  2  959.521 0.158 959.472 72.30         10  2  959.427 0.239 959.421 67.87
27  2  960.112 0.196 960.173 73.95         10  1  959.513 0.236 959.536 61.01
27  1  959.348 0.137 959.308 65.09         10  1  960.025 0.210 960.170 61.86
27  1  958.393 0.131 958.516 66.31         10  1  959.737 0.263 959.653 62.71
27  1  958.289 0.288 958.318 67.61         10  1  959.427 0.239 959.436 63.57
27  1  959.348 0.137 959.344 69.00         10  1  958.767 0.242 958.786 64.81
27  1  959.927 0.154 959.796 70.48         10  1  959.737 0.263 959.763 65.79
27  1  959.521 0.158 959.472 72.30         10  1  959.737 0.263 959.779 66.81
27  1  960.112 0.196 960.173 73.95         10  1  959.427 0.239 959.421 67.87
30  2  959.367 0.220 959.539 57.44         10  2  957.941 0.299 957.912 69.00
30  2  958.836 0.313 958.843 58.30         10  2  959.737 0.263 959.630 70.25
30  2  959.367 0.220 959.577 59.20         10  2  960.025 0.210 960.370 72.92
30  2  959.062 0.242 959.056 60.08         10  2  957.941 0.299 958.053 74.32
30  2  959.367 0.220 959.437 61.01         10  2  959.427 0.239 959.424 75.76
30  2  959.367 0.220 959.467 62.01         10  1  957.941 0.299 957.912 69.00
30  2  958.836 0.313 958.716 63.02         10  1  959.737 0.263 959.630 70.25
30  2  958.836 0.313 958.865 64.08         10  1  960.025 0.210 960.370 72.92
30  2  959.367 0.220 960.089 65.20
```

```
      1999 - AGOSTO                                    1999 - AGOSTO
 D  L   SDB    ER    SDC     HL              D  L   SDB    ER    SDC     HL
10  1 957.941 0.299 958.053 74.32           13  1 959.147 0.211 959.295 75.29
10  1 959.427 0.239 959.424 75.76           17  2 959.404 0.166 959.459 60.70
11  2 959.601 0.191 959.641 64.04           17  2 959.651 0.538 959.589 61.44
11  2 958.391 0.215 958.358 65.01           17  2 958.775 0.204 958.883 62.21
11  2 959.576 0.182 959.546 66.30           17  2 959.651 0.538 959.673 63.01
11  2 958.688 0.221 958.661 67.33           17  2 959.040 0.326 959.154 64.69
11  2 959.601 0.191 959.776 68.48           17  2 958.174 0.225 958.191 65.56
11  1 959.601 0.191 959.641 64.04           17  2 958.775 0.204 958.847 66.49
11  1 958.391 0.215 958.358 65.01           17  2 958.775 0.204 958.867 67.47
11  1 959.576 0.182 959.546 66.30           17  2 959.404 0.166 959.329 68.55
11  1 958.688 0.221 958.661 67.33           17  1 959.404 0.166 959.459 60.70
11  1 959.601 0.191 959.776 68.48           17  1 959.651 0.538 959.589 61.44
11  2 959.073 0.175 958.911 69.66           17  1 958.775 0.204 958.883 62.21
11  2 959.236 0.195 959.182 70.89           17  1 959.651 0.538 959.673 63.01
11  2 958.574 0.230 958.525 72.42           17  1 959.040 0.326 959.154 64.69
11  2 958.688 0.221 958.661 73.79           17  1 958.174 0.225 958.191 65.56
11  2 958.715 0.235 958.850 75.24           17  1 958.775 0.204 958.847 66.49
11  2 958.715 0.235 958.862 76.80           17  1 958.775 0.204 958.867 67.47
11  2 959.601 0.191 960.570 79.96           17  1 959.404 0.166 959.329 68.55
11  1 959.073 0.175 958.911 69.66           17  2 958.775 0.204 958.773 70.71
11  1 959.236 0.195 959.182 70.89           17  2 958.775 0.204 958.765 71.87
11  1 958.574 0.230 958.525 72.42           17  2 958.557 0.273 958.520 73.04
11  1 958.688 0.221 958.661 73.79           17  2 959.404 0.166 959.248 74.28
11  1 958.715 0.235 958.850 75.24           17  2 958.147 0.237 957.742 75.61
11  1 958.715 0.235 958.862 76.80           17  1 958.775 0.204 958.773 70.71
11  1 959.601 0.191 960.570 79.96           17  1 958.775 0.204 958.765 71.87
12  2 959.488 0.200 959.460 71.38           17  1 958.557 0.273 958.520 73.04
12  2 959.109 0.196 959.149 72.57           17  1 959.404 0.166 959.248 74.28
12  2 957.842 0.364 958.061 75.07           17  1 958.147 0.237 957.742 75.61
12  2 959.671 0.231 959.809 76.45           18  2 959.648 0.598 959.714 68.27
12  2 959.488 0.200 959.448 77.87           18  1 959.648 0.598 959.714 68.27
12  2 958.726 0.263 958.518 79.35           18  2 958.904 0.581 958.975 70.19
12  2 957.842 0.364 958.269 80.85           18  2 958.797 0.411 958.783 72.36
12  1 959.488 0.200 959.460 71.38           18  2 958.681 0.505 958.686 74.68
12  1 959.109 0.196 959.149 72.57           18  1 958.904 0.581 958.975 70.19
12  1 957.842 0.364 958.061 75.07           18  1 958.797 0.411 958.783 72.36
12  1 959.671 0.231 959.809 76.45           18  1 958.681 0.505 958.686 74.68
12  1 959.488 0.200 959.448 77.87           19  2 959.352 0.252 959.285 61.40
12  1 958.726 0.263 958.518 79.35           19  2 959.462 0.323 959.603 62.08
12  1 957.842 0.364 958.269 80.85           19  2 959.358 0.173 959.388 63.49
13  2 958.915 0.204 958.872 59.38           19  2 958.060 0.302 957.731 64.25
13  2 958.915 0.204 958.865 60.15           19  2 958.081 0.276 958.084 65.85
13  2 958.915 0.204 958.857 60.94           19  2 959.076 0.256 958.990 66.69
13  2 958.707 0.178 958.655 61.72           19  2 958.060 0.302 957.571 67.62
13  2 959.085 0.187 959.078 62.55           19  2 958.615 0.415 958.621 69.50
13  2 958.915 0.204 958.955 63.39           19  1 959.352 0.252 959.285 61.40
13  2 958.524 0.172 958.614 64.28           19  1 959.462 0.323 959.603 62.08
13  2 959.524 0.180 959.498 65.93           19  1 959.358 0.173 959.388 63.49
13  2 958.300 0.172 958.351 66.91           19  1 958.060 0.302 957.731 64.25
13  2 959.524 0.180 959.914 67.93           19  1 958.081 0.276 958.084 65.85
13  1 958.915 0.204 958.872 59.38           19  1 959.076 0.256 958.990 66.69
13  1 958.915 0.204 958.865 60.15           19  1 958.060 0.302 957.571 67.62
13  1 958.915 0.204 958.857 60.94           19  1 958.615 0.415 958.621 69.50
13  1 958.707 0.178 958.655 61.72           19  2 959.152 0.251 959.152 70.49
13  1 959.085 0.187 959.078 62.55           19  2 958.640 0.218 958.738 71.53
13  1 958.915 0.204 958.955 63.39           19  2 958.640 0.218 958.648 72.63
13  1 958.524 0.172 958.614 64.28           19  2 958.640 0.218 958.766 73.77
13  1 959.524 0.180 959.498 65.93           19  2 959.076 0.256 959.095 76.21
13  1 958.300 0.172 958.351 66.91           19  1 959.152 0.251 959.152 70.49
13  1 959.524 0.180 959.914 67.93           19  1 958.640 0.218 958.738 71.53
13  2 958.524 0.172 958.588 69.09           19  1 958.640 0.218 958.648 72.63
13  2 958.719 0.152 958.718 70.21           19  1 958.640 0.218 958.766 73.77
13  2 959.147 0.211 959.145 71.39           19  1 959.076 0.256 959.095 76.21
13  2 959.085 0.187 959.101 72.66           23  2 959.711 0.290 959.703 60.88
13  2 958.147 0.214 958.161 73.94           23  2 958.862 0.245 958.929 61.93
13  2 959.147 0.211 959.295 75.29           23  2 958.862 0.245 958.910 63.36
13  1 958.524 0.172 958.588 69.09           23  2 957.750 0.295 957.940 64.98
13  1 958.719 0.152 958.718 70.21           23  2 959.170 0.234 959.196 67.94
13  1 959.147 0.211 959.145 71.39           23  2 959.170 0.234 959.380 69.20
13  1 959.085 0.187 959.101 72.66           23  2 957.750 0.295 957.667 70.39
13  1 958.147 0.214 958.161 73.94           23  1 959.711 0.290 959.703 60.88
```

| 1999 - AGOSTO | | | | |
|---|---|---|---|---|
| D L | SDB | ER | SDC | HL |
| 23 1 | 958.862 | 0.245 | 958.929 | 61.93 |
| 23 1 | 958.862 | 0.245 | 958.910 | 63.36 |
| 23 1 | 957.750 | 0.295 | 957.940 | 64.98 |
| 23 1 | 959.170 | 0.234 | 959.196 | 67.94 |
| 23 1 | 959.170 | 0.234 | 959.380 | 69.20 |
| 23 1 | 957.750 | 0.295 | 957.667 | 70.39 |
| 23 2 | 958.359 | 0.272 | 958.493 | 72.71 |
| 23 2 | 959.711 | 0.290 | 960.739 | 76.79 |
| 23 1 | 958.359 | 0.272 | 958.493 | 72.71 |
| 23 1 | 959.711 | 0.290 | 960.739 | 76.79 |
| 24 2 | 959.488 | 0.247 | 959.834 | 61.17 |
| 24 2 | 958.339 | 0.236 | 958.329 | 61.86 |
| 24 2 | 958.677 | 0.226 | 958.643 | 62.60 |
| 24 2 | 958.395 | 0.212 | 958.379 | 63.32 |
| 24 2 | 958.229 | 0.222 | 958.229 | 64.08 |
| 24 2 | 958.929 | 0.244 | 958.915 | 64.86 |
| 24 2 | 959.182 | 0.226 | 959.186 | 65.65 |
| 24 2 | 959.258 | 0.250 | 959.267 | 66.58 |
| 24 2 | 958.395 | 0.212 | 958.380 | 67.47 |
| 24 2 | 958.445 | 0.262 | 958.440 | 68.40 |
| 24 2 | 958.698 | 0.204 | 958.695 | 69.36 |
| 24 2 | 959.352 | 0.270 | 959.385 | 70.41 |
| 24 1 | 959.488 | 0.247 | 959.834 | 61.17 |
| 24 1 | 958.339 | 0.236 | 958.329 | 61.86 |
| 24 1 | 958.677 | 0.226 | 958.643 | 62.60 |
| 24 1 | 958.395 | 0.212 | 958.379 | 63.32 |
| 24 1 | 958.229 | 0.222 | 958.229 | 64.08 |
| 24 1 | 958.929 | 0.244 | 958.915 | 64.86 |
| 24 1 | 959.182 | 0.226 | 959.186 | 65.65 |
| 24 1 | 959.258 | 0.250 | 959.267 | 66.58 |
| 24 1 | 958.395 | 0.212 | 958.380 | 67.47 |
| 24 1 | 958.445 | 0.262 | 958.440 | 68.40 |
| 24 1 | 958.698 | 0.204 | 958.695 | 69.36 |
| 24 1 | 959.352 | 0.270 | 959.385 | 70.41 |
| 24 2 | 959.488 | 0.247 | 959.514 | 71.43 |
| 24 2 | 959.182 | 0.226 | 959.179 | 73.76 |
| 24 2 | 958.445 | 0.262 | 958.485 | 74.97 |
| 24 1 | 959.488 | 0.247 | 959.514 | 71.43 |
| 24 1 | 959.182 | 0.226 | 959.179 | 73.76 |
| 24 1 | 958.445 | 0.262 | 958.485 | 74.97 |
| 25 2 | 959.682 | 0.191 | 959.767 | 63.77 |
| 25 2 | 959.014 | 0.178 | 959.233 | 64.68 |
| 25 2 | 959.682 | 0.191 | 959.658 | 65.58 |
| 25 2 | 959.526 | 0.164 | 959.448 | 66.45 |
| 25 2 | 958.810 | 0.221 | 958.843 | 67.49 |
| 25 2 | 959.682 | 0.191 | 959.983 | 68.95 |
| 25 2 | 959.014 | 0.178 | 959.035 | 70.00 |
| 25 1 | 959.682 | 0.191 | 959.767 | 63.77 |
| 25 1 | 959.014 | 0.178 | 959.233 | 64.68 |
| 25 1 | 959.682 | 0.191 | 959.658 | 65.58 |
| 25 1 | 959.526 | 0.164 | 959.448 | 66.45 |
| 25 1 | 958.810 | 0.221 | 958.843 | 67.49 |
| 25 1 | 959.682 | 0.191 | 959.983 | 68.95 |
| 25 1 | 959.014 | 0.178 | 959.035 | 70.00 |
| 25 2 | 958.416 | 0.207 | 958.374 | 71.11 |
| 25 2 | 958.718 | 0.193 | 958.708 | 72.19 |
| 25 2 | 959.014 | 0.178 | 959.133 | 73.38 |
| 25 2 | 958.416 | 0.207 | 958.425 | 74.63 |
| 25 2 | 959.014 | 0.178 | 959.066 | 75.89 |
| 25 2 | 958.718 | 0.193 | 958.698 | 77.19 |
| 25 1 | 958.416 | 0.207 | 958.374 | 71.11 |
| 25 1 | 958.718 | 0.193 | 958.708 | 72.19 |
| 25 1 | 959.014 | 0.178 | 959.133 | 73.38 |
| 25 1 | 958.416 | 0.207 | 958.425 | 74.63 |
| 25 1 | 959.014 | 0.178 | 959.066 | 75.89 |
| 25 1 | 958.718 | 0.193 | 958.698 | 77.19 |
| 26 2 | 957.973 | 0.246 | 958.034 | 62.85 |
| 26 2 | 958.849 | 0.326 | 958.846 | 63.53 |
| 26 2 | 958.937 | 0.387 | 958.897 | 64.23 |
| 26 2 | 959.845 | 0.220 | 959.714 | 65.74 |
| 26 2 | 959.845 | 0.220 | 959.827 | 66.53 |

| 1999 - AGOSTO | | | | |
|---|---|---|---|---|
| D L | SDB | ER | SDC | HL |
| 26 2 | 959.866 | 0.256 | 959.873 | 67.34 |
| 26 2 | 959.866 | 0.256 | 959.918 | 68.19 |
| 26 2 | 959.120 | 0.253 | 959.066 | 69.99 |
| 26 2 | 959.375 | 0.337 | 959.317 | 70.94 |
| 26 1 | 957.973 | 0.246 | 958.034 | 62.85 |
| 26 1 | 958.849 | 0.326 | 958.846 | 63.53 |
| 26 1 | 958.937 | 0.387 | 958.897 | 64.23 |
| 26 1 | 959.845 | 0.220 | 959.714 | 65.74 |
| 26 1 | 959.845 | 0.220 | 959.827 | 66.53 |
| 26 1 | 959.866 | 0.256 | 959.873 | 67.34 |
| 26 1 | 959.866 | 0.256 | 959.918 | 68.19 |
| 26 1 | 959.120 | 0.253 | 959.066 | 69.99 |
| 26 1 | 959.375 | 0.337 | 959.317 | 70.94 |
| 26 2 | 958.810 | 0.280 | 958.754 | 71.95 |
| 26 2 | 958.992 | 0.361 | 959.009 | 73.00 |
| 26 2 | 959.866 | 0.256 | 960.061 | 74.09 |
| 26 2 | 958.992 | 0.361 | 959.049 | 75.22 |
| 26 2 | 958.698 | 0.214 | 958.630 | 76.37 |
| 26 2 | 958.849 | 0.326 | 958.874 | 77.60 |
| 26 2 | 959.866 | 0.256 | 960.599 | 78.81 |
| 26 1 | 958.810 | 0.280 | 958.754 | 71.95 |
| 26 1 | 958.992 | 0.361 | 959.009 | 73.00 |
| 26 1 | 959.866 | 0.256 | 960.061 | 74.09 |
| 26 1 | 958.992 | 0.361 | 959.049 | 75.22 |
| 26 1 | 958.698 | 0.214 | 958.630 | 76.37 |
| 26 1 | 958.849 | 0.326 | 958.874 | 77.60 |
| 26 1 | 959.866 | 0.256 | 960.599 | 78.81 |
| 27 2 | 958.799 | 0.225 | 958.819 | 60.83 |
| 27 2 | 958.602 | 0.209 | 958.646 | 61.49 |
| 27 2 | 958.472 | 0.246 | 958.364 | 62.18 |
| 27 2 | 958.472 | 0.246 | 957.661 | 62.90 |
| 27 2 | 958.472 | 0.246 | 958.255 | 63.67 |
| 27 2 | 958.472 | 0.246 | 958.409 | 64.46 |
| 27 2 | 958.991 | 0.278 | 958.984 | 65.26 |
| 27 2 | 958.472 | 0.246 | 958.167 | 66.13 |
| 27 2 | 959.139 | 0.244 | 959.164 | 67.90 |
| 27 2 | 958.799 | 0.225 | 958.838 | 68.86 |
| 27 2 | 959.139 | 0.244 | 959.201 | 69.95 |
| 27 2 | 958.472 | 0.246 | 958.105 | 70.97 |
| 27 1 | 958.799 | 0.225 | 958.819 | 60.83 |
| 27 1 | 958.602 | 0.209 | 958.646 | 61.49 |
| 27 1 | 958.472 | 0.246 | 958.364 | 62.18 |
| 27 1 | 958.472 | 0.246 | 957.661 | 62.90 |
| 27 1 | 958.472 | 0.246 | 958.255 | 63.67 |
| 27 1 | 958.472 | 0.246 | 958.409 | 64.46 |
| 27 1 | 958.991 | 0.278 | 958.984 | 65.26 |
| 27 1 | 958.472 | 0.246 | 958.167 | 66.13 |
| 27 1 | 959.139 | 0.244 | 959.164 | 67.90 |
| 27 1 | 958.799 | 0.225 | 958.838 | 68.86 |
| 27 1 | 959.139 | 0.244 | 959.201 | 69.95 |
| 27 1 | 958.472 | 0.246 | 958.105 | 70.97 |
| 27 2 | 958.535 | 0.260 | 958.536 | 72.07 |
| 27 2 | 958.472 | 0.246 | 957.655 | 73.16 |
| 27 1 | 958.535 | 0.260 | 958.536 | 72.07 |
| 27 1 | 958.472 | 0.246 | 957.655 | 73.16 |
| 31 2 | 959.379 | 0.501 | 960.160 | 61.37 |
| 31 2 | 958.876 | 0.188 | 958.885 | 62.03 |
| 31 2 | 959.258 | 0.247 | 959.226 | 62.68 |
| 31 2 | 959.033 | 0.219 | 959.055 | 63.39 |
| 31 2 | 959.258 | 0.247 | 959.205 | 64.10 |
| 31 2 | 958.424 | 0.200 | 958.394 | 64.85 |
| 31 2 | 959.379 | 0.501 | 959.499 | 65.71 |
| 31 2 | 958.686 | 0.260 | 958.698 | 66.51 |
| 31 2 | 959.379 | 0.501 | 959.756 | 67.35 |
| 31 2 | 958.672 | 0.189 | 958.663 | 68.36 |
| 31 2 | 959.033 | 0.219 | 959.085 | 69.27 |
| 31 2 | 958.958 | 0.203 | 958.952 | 70.24 |
| 31 2 | 958.672 | 0.189 | 958.590 | 71.27 |
| 31 1 | 959.379 | 0.501 | 960.160 | 61.37 |
| 31 1 | 958.876 | 0.188 | 958.885 | 62.03 |
| 31 1 | 959.258 | 0.247 | 959.226 | 62.68 |

|    | 1999 - AGOSTO | | | | | | 1999 - SETEMBRO | | | |
|----|---|---|---|---|---|----|---|---|---|---|
| D | L | SDB | ER | SDC | HL | D | L | SDB | ER | SDC | HL |
| 31 | 1 | 959.033 | 0.219 | 959.055 | 63.39 | 02 | 1 | 958.831 | 0.158 | 958.945 | 64.74 |
| 31 | 1 | 959.258 | 0.247 | 959.205 | 64.10 | 02 | 1 | 959.149 | 0.157 | 959.207 | 65.50 |
| 31 | 1 | 958.424 | 0.200 | 958.394 | 64.85 | 02 | 1 | 958.749 | 0.142 | 958.758 | 66.23 |
| 31 | 1 | 959.379 | 0.501 | 959.499 | 65.71 | 02 | 1 | 959.372 | 0.130 | 959.325 | 66.98 |
| 31 | 1 | 958.686 | 0.260 | 958.698 | 66.51 | 02 | 1 | 959.407 | 0.240 | 959.400 | 67.77 |
| 31 | 1 | 959.379 | 0.501 | 959.756 | 67.35 | 02 | 1 | 958.749 | 0.142 | 958.771 | 68.58 |
| 31 | 1 | 958.672 | 0.189 | 958.663 | 68.36 | 02 | 1 | 959.950 | 0.146 | 960.161 | 69.43 |
| 31 | 1 | 959.033 | 0.219 | 959.085 | 69.27 | 02 | 1 | 959.149 | 0.157 | 959.230 | 70.33 |
| 31 | 1 | 958.958 | 0.203 | 958.952 | 70.24 | 02 | 1 | 958.808 | 0.146 | 958.809 | 71.25 |
| 31 | 1 | 958.672 | 0.189 | 958.590 | 71.27 | 02 | 2 | 959.837 | 0.160 | 959.742 | 72.21 |
| 31 | 2 | 959.258 | 0.247 | 959.166 | 72.29 | 02 | 2 | 959.372 | 0.130 | 959.312 | 73.22 |
| 31 | 2 | 959.379 | 0.501 | 960.007 | 73.36 | 02 | 2 | 959.569 | 0.159 | 959.630 | 74.28 |
| 31 | 2 | 959.033 | 0.219 | 959.115 | 74.55 | 02 | 2 | 958.653 | 0.187 | 958.632 | 75.38 |
| 31 | 2 | 959.258 | 0.247 | 959.244 | 75.79 | 02 | 2 | 960.537 | 0.638 | 960.835 | 76.56 |
| 31 | 2 | 959.379 | 0.501 | 959.562 | 77.04 | 02 | 1 | 959.837 | 0.160 | 959.742 | 72.21 |
| 31 | 1 | 959.258 | 0.247 | 959.166 | 72.29 | 02 | 1 | 959.372 | 0.130 | 959.312 | 73.22 |
| 31 | 1 | 959.379 | 0.501 | 960.007 | 73.36 | 02 | 1 | 959.569 | 0.159 | 959.630 | 74.28 |
| 31 | 1 | 959.033 | 0.219 | 959.115 | 74.55 | 02 | 1 | 958.653 | 0.187 | 958.632 | 75.38 |
| 31 | 1 | 959.258 | 0.247 | 959.244 | 75.79 | 02 | 1 | 960.537 | 0.638 | 960.835 | 76.56 |
| 31 | 1 | 959.379 | 0.501 | 959.562 | 77.04 | 03 | 2 | 959.241 | 0.174 | 959.164 | 61.26 |
|    |   |         |       |         |       | 03 | 2 | 959.592 | 0.168 | 959.650 | 61.89 |
|    |   |         |       |         |       | 03 | 2 | 958.795 | 0.137 | 958.878 | 62.55 |
|    | 1999 - SETEMBRO | | | | | 03 | 2 | 958.357 | 0.139 | 958.401 | 63.23 |
| D | L | SDB | ER | SDC | HL | 03 | 2 | 958.522 | 0.165 | 958.448 | 63.93 |
| 01 | 2 | 959.462 | 0.196 | 959.552 | 62.67 | 03 | 2 | 959.921 | 0.158 | 959.957 | 64.71 |
| 01 | 2 | 960.198 | 0.172 | 960.263 | 63.39 | 03 | 2 | 958.220 | 0.135 | 958.278 | 65.46 |
| 01 | 2 | 959.317 | 0.176 | 959.353 | 64.17 | 03 | 2 | 958.522 | 0.165 | 958.504 | 66.25 |
| 01 | 2 | 959.225 | 0.188 | 959.188 | 64.96 | 03 | 2 | 959.306 | 0.164 | 959.319 | 67.08 |
| 01 | 2 | 958.636 | 0.190 | 958.630 | 65.79 | 03 | 2 | 959.350 | 0.156 | 959.422 | 67.92 |
| 01 | 2 | 958.400 | 0.153 | 958.120 | 66.90 | 03 | 2 | 958.795 | 0.137 | 958.844 | 68.83 |
| 01 | 2 | 959.692 | 0.174 | 959.944 | 67.79 | 03 | 2 | 960.288 | 0.162 | 960.509 | 71.59 |
| 01 | 2 | 958.636 | 0.190 | 958.555 | 68.74 | 03 | 1 | 959.241 | 0.174 | 959.164 | 61.26 |
| 01 | 2 | 959.030 | 0.191 | 958.997 | 69.71 | 03 | 1 | 959.592 | 0.168 | 959.650 | 61.89 |
| 01 | 2 | 959.692 | 0.174 | 959.787 | 70.69 | 03 | 1 | 958.795 | 0.137 | 958.878 | 62.55 |
| 01 | 1 | 959.462 | 0.196 | 959.552 | 62.67 | 03 | 1 | 958.357 | 0.139 | 958.401 | 63.23 |
| 01 | 1 | 960.198 | 0.172 | 960.263 | 63.39 | 03 | 1 | 958.522 | 0.165 | 958.448 | 63.93 |
| 01 | 1 | 959.317 | 0.176 | 959.353 | 64.17 | 03 | 1 | 959.921 | 0.158 | 959.957 | 64.71 |
| 01 | 1 | 959.225 | 0.188 | 959.188 | 64.96 | 03 | 1 | 958.220 | 0.135 | 958.278 | 65.46 |
| 01 | 1 | 958.636 | 0.190 | 958.630 | 65.79 | 03 | 1 | 958.522 | 0.165 | 958.504 | 66.25 |
| 01 | 1 | 958.400 | 0.153 | 958.120 | 66.90 | 03 | 1 | 959.306 | 0.164 | 959.319 | 67.08 |
| 01 | 1 | 959.692 | 0.174 | 959.944 | 67.79 | 03 | 1 | 959.350 | 0.156 | 959.422 | 67.92 |
| 01 | 1 | 958.636 | 0.190 | 958.555 | 68.74 | 03 | 1 | 958.795 | 0.137 | 958.844 | 68.83 |
| 01 | 1 | 959.030 | 0.191 | 958.997 | 69.71 | 03 | 1 | 960.288 | 0.162 | 960.509 | 71.59 |
| 01 | 1 | 959.692 | 0.174 | 959.787 | 70.69 | 03 | 2 | 958.611 | 0.173 | 958.618 | 72.67 |
| 01 | 2 | 959.071 | 0.187 | 959.100 | 71.70 | 03 | 2 | 960.288 | 0.162 | 960.527 | 73.79 |
| 01 | 2 | 958.686 | 0.195 | 958.744 | 72.83 | 03 | 2 | 959.921 | 0.158 | 959.878 | 74.98 |
| 01 | 2 | 959.317 | 0.176 | 959.296 | 75.27 | 03 | 2 | 959.350 | 0.156 | 959.356 | 76.18 |
| 01 | 2 | 959.643 | 0.161 | 959.641 | 76.50 | 03 | 2 | 959.241 | 0.174 | 959.138 | 77.41 |
| 01 | 2 | 959.692 | 0.174 | 959.729 | 77.79 | 03 | 1 | 958.611 | 0.173 | 958.618 | 72.67 |
| 01 | 1 | 959.071 | 0.187 | 959.100 | 71.70 | 03 | 1 | 960.288 | 0.162 | 960.527 | 73.79 |
| 01 | 1 | 958.686 | 0.195 | 958.744 | 72.83 | 03 | 1 | 959.921 | 0.158 | 959.878 | 74.98 |
| 01 | 1 | 959.317 | 0.176 | 959.296 | 75.27 | 03 | 1 | 959.350 | 0.156 | 959.356 | 76.18 |
| 01 | 1 | 959.643 | 0.161 | 959.641 | 76.50 | 03 | 1 | 959.241 | 0.174 | 959.138 | 77.41 |
| 01 | 1 | 959.692 | 0.174 | 959.729 | 77.79 | 06 | 2 | 959.234 | 0.195 | 959.103 | 67.74 |
| 02 | 2 | 959.065 | 0.184 | 959.026 | 62.15 | 06 | 2 | 958.908 | 0.333 | 958.837 | 68.68 |
| 02 | 2 | 959.569 | 0.159 | 959.533 | 62.75 | 06 | 2 | 958.908 | 0.333 | 959.022 | 69.63 |
| 02 | 2 | 959.950 | 0.146 | 959.957 | 63.39 | 06 | 2 | 959.234 | 0.195 | 959.705 | 70.57 |
| 02 | 2 | 958.014 | 0.180 | 958.227 | 64.07 | 06 | 2 | 958.645 | 0.212 | 958.225 | 71.63 |
| 02 | 2 | 958.831 | 0.158 | 958.945 | 64.74 | 06 | 1 | 959.234 | 0.195 | 959.103 | 67.74 |
| 02 | 2 | 959.149 | 0.157 | 959.207 | 65.50 | 06 | 1 | 958.908 | 0.333 | 958.837 | 68.68 |
| 02 | 2 | 958.749 | 0.142 | 958.758 | 66.23 | 06 | 1 | 958.908 | 0.333 | 959.022 | 69.63 |
| 02 | 2 | 959.372 | 0.130 | 959.325 | 66.98 | 06 | 1 | 959.234 | 0.195 | 959.705 | 70.57 |
| 02 | 2 | 959.407 | 0.240 | 959.400 | 67.77 | 06 | 1 | 958.645 | 0.212 | 958.225 | 71.63 |
| 02 | 2 | 958.749 | 0.142 | 958.771 | 68.58 | 07 | 2 | 960.177 | 0.259 | 960.869 | 61.41 |
| 02 | 2 | 959.950 | 0.146 | 960.161 | 69.43 | 07 | 2 | 959.474 | 0.255 | 959.566 | 62.06 |
| 02 | 2 | 959.149 | 0.157 | 959.230 | 70.33 | 07 | 2 | 958.789 | 0.223 | 958.688 | 62.73 |
| 02 | 2 | 958.808 | 0.146 | 958.809 | 71.25 | 07 | 2 | 958.789 | 0.223 | 958.748 | 63.40 |
| 02 | 1 | 959.065 | 0.184 | 959.026 | 62.15 | 07 | 2 | 959.761 | 0.185 | 959.755 | 64.10 |
| 02 | 1 | 959.569 | 0.159 | 959.533 | 62.75 | 07 | 2 | 959.208 | 0.447 | 959.180 | 64.83 |
| 02 | 1 | 959.950 | 0.146 | 959.957 | 63.39 | 07 | 2 | 960.177 | 0.259 | 960.063 | 65.56 |
| 02 | 1 | 958.014 | 0.180 | 958.227 | 64.07 | 07 | 2 | 959.208 | 0.447 | 959.230 | 66.34 |

| | | 1999 - SETEMBRO | | | |
|---|---|---|---|---|---|
| D | L | SDB | ER | SDC | HL |
| 07 | 2 | 960.177 | 0.259 | 960.035 | 67.15 |
| 07 | 2 | 960.177 | 0.259 | 960.273 | 68.00 |
| 07 | 2 | 959.761 | 0.185 | 959.811 | 68.92 |
| 07 | 2 | 959.208 | 0.447 | 959.232 | 69.91 |
| 07 | 2 | 959.208 | 0.447 | 959.237 | 70.89 |
| 07 | 1 | 960.177 | 0.259 | 960.869 | 61.41 |
| 07 | 1 | 959.474 | 0.255 | 959.566 | 62.06 |
| 07 | 1 | 958.789 | 0.223 | 958.688 | 62.73 |
| 07 | 1 | 958.789 | 0.223 | 958.748 | 63.40 |
| 07 | 1 | 959.761 | 0.185 | 959.755 | 64.10 |
| 07 | 1 | 959.208 | 0.447 | 959.180 | 64.83 |
| 07 | 1 | 960.177 | 0.259 | 960.063 | 65.56 |
| 07 | 1 | 959.208 | 0.447 | 959.230 | 66.34 |
| 07 | 1 | 960.177 | 0.259 | 960.035 | 67.15 |
| 07 | 1 | 960.177 | 0.259 | 960.273 | 68.00 |
| 07 | 1 | 959.761 | 0.185 | 959.811 | 68.92 |
| 07 | 1 | 959.208 | 0.447 | 959.232 | 69.91 |
| 07 | 1 | 959.208 | 0.447 | 959.237 | 70.89 |
| 07 | 2 | 960.177 | 0.259 | 960.303 | 71.91 |
| 07 | 2 | 959.474 | 0.255 | 959.424 | 73.00 |
| 07 | 2 | 959.761 | 0.185 | 959.855 | 74.09 |
| 07 | 1 | 960.177 | 0.259 | 960.303 | 71.91 |
| 07 | 1 | 959.474 | 0.255 | 959.424 | 73.00 |
| 07 | 1 | 959.761 | 0.185 | 959.855 | 74.09 |
| 08 | 2 | 959.400 | 0.171 | 959.426 | 65.24 |
| 08 | 2 | 958.858 | 0.147 | 958.899 | 66.04 |
| 08 | 2 | 958.556 | 0.140 | 958.665 | 66.84 |
| 08 | 2 | 959.267 | 0.169 | 959.279 | 68.73 |
| 08 | 2 | 959.267 | 0.169 | 959.333 | 69.64 |
| 08 | 1 | 959.400 | 0.171 | 959.426 | 65.24 |
| 08 | 1 | 958.858 | 0.147 | 958.899 | 66.04 |
| 08 | 1 | 958.556 | 0.140 | 958.665 | 66.84 |
| 08 | 1 | 959.267 | 0.169 | 959.279 | 68.73 |
| 08 | 1 | 959.267 | 0.169 | 959.333 | 69.64 |
| 08 | 2 | 958.813 | 0.164 | 958.707 | 71.89 |
| 08 | 2 | 959.400 | 0.171 | 960.664 | 72.93 |
| 08 | 2 | 959.028 | 0.168 | 959.117 | 74.10 |
| 08 | 2 | 959.400 | 0.171 | 959.737 | 75.30 |
| 08 | 2 | 958.858 | 0.147 | 958.853 | 76.52 |
| 08 | 1 | 958.813 | 0.164 | 958.707 | 71.89 |
| 08 | 1 | 959.400 | 0.171 | 960.664 | 72.93 |
| 08 | 1 | 959.028 | 0.168 | 959.117 | 74.10 |
| 08 | 1 | 959.400 | 0.171 | 959.737 | 75.30 |
| 08 | 1 | 958.858 | 0.147 | 958.853 | 76.52 |
| 21 | 2 | 959.377 | 0.231 | 959.456 | 63.02 |
| 21 | 2 | 959.593 | 0.188 | 959.677 | 63.69 |
| 21 | 2 | 959.111 | 0.213 | 959.081 | 64.37 |
| 21 | 2 | 959.248 | 0.209 | 959.232 | 65.12 |
| 21 | 2 | 959.593 | 0.188 | 959.595 | 65.88 |
| 21 | 2 | 959.377 | 0.231 | 959.389 | 66.70 |
| 21 | 2 | 960.063 | 0.197 | 960.283 | 67.55 |
| 21 | 2 | 959.377 | 0.231 | 959.351 | 68.40 |
| 21 | 2 | 960.063 | 0.197 | 960.327 | 69.29 |
| 21 | 2 | 960.063 | 0.197 | 960.255 | 70.21 |
| 21 | 1 | 959.377 | 0.231 | 959.456 | 63.02 |
| 21 | 1 | 959.593 | 0.188 | 959.677 | 63.69 |
| 21 | 1 | 959.111 | 0.213 | 959.081 | 64.37 |
| 21 | 1 | 959.248 | 0.209 | 959.232 | 65.12 |
| 21 | 1 | 959.593 | 0.188 | 959.595 | 65.88 |
| 21 | 1 | 959.377 | 0.231 | 959.389 | 66.70 |
| 21 | 1 | 960.063 | 0.197 | 960.283 | 67.55 |
| 21 | 1 | 959.377 | 0.231 | 959.351 | 68.40 |
| 21 | 1 | 960.063 | 0.197 | 960.327 | 69.29 |
| 21 | 1 | 960.063 | 0.197 | 960.255 | 70.21 |
| 21 | 2 | 959.593 | 0.188 | 959.633 | 71.22 |
| 21 | 2 | 958.467 | 0.278 | 958.260 | 72.30 |
| 21 | 2 | 959.248 | 0.209 | 959.218 | 73.45 |
| 21 | 2 | 960.063 | 0.197 | 960.443 | 74.64 |
| 21 | 2 | 958.737 | 0.182 | 958.679 | 76.01 |
| 21 | 2 | 960.063 | 0.197 | 960.645 | 77.26 |
| 21 | 1 | 959.593 | 0.188 | 959.633 | 71.22 |

| | | 1999 - SETEMBRO | | | |
|---|---|---|---|---|---|
| D | L | SDB | ER | SDC | HL |
| 21 | 1 | 958.467 | 0.278 | 958.260 | 72.30 |
| 21 | 1 | 959.248 | 0.209 | 959.218 | 73.45 |
| 21 | 1 | 960.063 | 0.197 | 960.443 | 74.64 |
| 21 | 1 | 958.737 | 0.182 | 958.679 | 76.01 |
| 21 | 1 | 960.063 | 0.197 | 960.645 | 77.26 |
| 28 | 2 | 959.629 | 0.283 | 959.811 | 57.79 |
| 28 | 2 | 958.960 | 0.269 | 958.960 | 58.25 |
| 28 | 2 | 958.528 | 0.284 | 958.662 | 58.73 |
| 28 | 2 | 959.629 | 0.283 | 959.735 | 59.23 |
| 28 | 2 | 959.186 | 0.349 | 959.236 | 60.29 |
| 28 | 2 | 959.462 | 0.366 | 959.417 | 60.86 |
| 28 | 2 | 958.528 | 0.284 | 958.577 | 62.07 |
| 28 | 2 | 958.528 | 0.284 | 958.627 | 63.41 |
| 28 | 2 | 959.462 | 0.366 | 959.371 | 64.16 |
| 28 | 2 | 959.186 | 0.349 | 959.239 | 64.92 |
| 28 | 2 | 959.629 | 0.283 | 959.813 | 65.79 |
| 28 | 2 | 959.032 | 0.274 | 959.036 | 66.66 |
| 28 | 2 | 958.257 | 0.376 | 958.267 | 67.52 |
| 28 | 1 | 959.629 | 0.283 | 959.811 | 57.79 |
| 28 | 1 | 958.960 | 0.269 | 958.960 | 58.25 |
| 28 | 1 | 958.528 | 0.284 | 958.662 | 58.73 |
| 28 | 1 | 959.629 | 0.283 | 959.735 | 59.23 |
| 28 | 1 | 959.186 | 0.349 | 959.236 | 60.29 |
| 28 | 1 | 959.462 | 0.366 | 959.417 | 60.86 |
| 28 | 1 | 958.528 | 0.284 | 958.577 | 62.07 |
| 28 | 1 | 958.528 | 0.284 | 958.627 | 63.41 |
| 28 | 1 | 959.462 | 0.366 | 959.371 | 64.16 |
| 28 | 1 | 959.186 | 0.349 | 959.239 | 64.92 |
| 28 | 1 | 959.629 | 0.283 | 959.813 | 65.79 |
| 28 | 1 | 959.032 | 0.274 | 959.036 | 66.66 |
| 28 | 1 | 958.257 | 0.376 | 958.267 | 67.52 |
| 29 | 2 | 959.353 | 0.294 | 959.331 | 60.75 |
| 29 | 2 | 958.890 | 0.260 | 958.904 | 61.35 |
| 29 | 2 | 959.153 | 0.508 | 959.162 | 61.96 |
| 29 | 2 | 958.890 | 0.260 | 958.968 | 62.63 |
| 29 | 2 | 958.890 | 0.260 | 958.892 | 63.31 |
| 29 | 2 | 958.710 | 0.399 | 958.745 | 64.05 |
| 29 | 2 | 959.217 | 0.426 | 959.245 | 64.81 |
| 29 | 2 | 959.437 | 0.182 | 959.474 | 65.69 |
| 29 | 2 | 959.437 | 0.182 | 959.473 | 66.58 |
| 29 | 2 | 958.710 | 0.399 | 958.759 | 67.55 |
| 29 | 2 | 958.069 | 0.369 | 958.098 | 68.54 |
| 29 | 1 | 959.353 | 0.294 | 959.331 | 60.75 |
| 29 | 1 | 958.890 | 0.260 | 958.904 | 61.35 |
| 29 | 1 | 959.153 | 0.508 | 959.162 | 61.96 |
| 29 | 1 | 958.890 | 0.260 | 958.968 | 62.63 |
| 29 | 1 | 958.890 | 0.260 | 958.892 | 63.31 |
| 29 | 1 | 958.710 | 0.399 | 958.745 | 64.05 |
| 29 | 1 | 959.217 | 0.426 | 959.245 | 64.81 |
| 29 | 1 | 959.437 | 0.182 | 959.474 | 65.69 |
| 29 | 1 | 959.437 | 0.182 | 959.473 | 66.58 |
| 29 | 1 | 958.710 | 0.399 | 958.759 | 67.55 |
| 29 | 1 | 958.069 | 0.369 | 958.098 | 68.54 |
| 29 | 2 | 959.217 | 0.426 | 959.246 | 69.55 |
| 29 | 2 | 958.691 | 0.274 | 958.675 | 71.77 |
| 29 | 1 | 959.217 | 0.426 | 959.246 | 69.55 |
| 29 | 1 | 958.691 | 0.274 | 958.675 | 71.77 |

| | | 1999 - OUTUBRO | | | |
|---|---|---|---|---|---|
| D | L | SDB | ER | SDC | HL |
| 01 | 2 | 959.240 | 0.180 | 959.242 | 57.60 |
| 01 | 2 | 959.079 | 0.249 | 959.098 | 58.06 |
| 01 | 2 | 959.240 | 0.180 | 959.266 | 58.54 |
| 01 | 2 | 959.079 | 0.249 | 959.107 | 59.02 |
| 01 | 2 | 958.991 | 0.208 | 959.002 | 59.53 |
| 01 | 2 | 959.079 | 0.249 | 959.063 | 60.06 |
| 01 | 2 | 959.030 | 0.273 | 959.027 | 60.63 |
| 01 | 2 | 959.040 | 0.175 | 959.057 | 61.26 |
| 01 | 2 | 958.572 | 0.218 | 958.645 | 61.91 |
| 01 | 2 | 959.079 | 0.249 | 959.137 | 62.57 |

```
           1999  -  OUTUBRO                                1999  -  OUTUBRO
   D  L    SDB     ER     SDC    HL              D  L    SDB     ER     SDC    HL
   01 2  958.740  0.310  958.747  63.32          28 2  960.398  0.286  960.003  48.51
   01 2  958.914  0.229  958.896  64.06          28 1  958.684  0.383  958.665  46.39
   01 2  959.079  0.249  959.127  64.90          28 1  959.286  0.979  959.248  46.59
   01 2  958.740  0.310  958.723  65.73          28 1  959.521  0.210  959.788  47.05
   01 2  958.467  0.287  958.156  66.63          28 1  959.521  0.210  959.711  47.31
   01 2  959.240  0.180  959.365  67.55          28 1  958.684  0.383  958.615  47.62
   01 1  959.240  0.180  959.242  57.60          28 1  959.521  0.210  959.818  47.90
   01 1  959.079  0.249  959.098  58.06          28 1  958.340  0.326  958.172  48.19
   01 1  959.240  0.180  959.266  58.54          28 1  960.398  0.286  960.003  48.51
   01 1  959.079  0.249  959.107  59.02          29 2  958.222  0.198  958.297  44.53
   01 1  958.991  0.208  959.002  59.53          29 2  959.947  0.479  959.920  44.64
   01 1  959.079  0.249  959.063  60.06          29 2  958.901  0.258  958.925  44.89
   01 1  959.030  0.273  959.027  60.63          29 2  959.152  0.326  959.198  45.03
   01 1  959.040  0.175  959.057  61.26          29 2  958.543  0.259  958.566  45.17
   01 1  958.572  0.218  958.645  61.91          29 2  958.500  0.251  958.412  45.33
   01 1  959.079  0.249  959.137  62.57          29 2  959.947  0.479  960.124  45.49
   01 1  958.740  0.310  958.747  63.32          29 2  958.543  0.259  958.619  45.66
   01 1  958.914  0.229  958.896  64.06          29 2  958.543  0.259  958.551  46.08
   01 1  959.079  0.249  959.127  64.90          29 2  959.152  0.326  959.130  46.30
   01 1  958.740  0.310  958.723  65.73          29 2  959.591  0.259  959.576  46.52
   01 1  958.467  0.287  958.156  66.63          29 2  959.245  0.298  959.241  47.28
   01 1  959.240  0.180  959.365  67.55          29 1  958.222  0.198  958.297  44.53
   06 2  958.838  0.319  958.869  56.94          29 1  959.947  0.479  959.920  44.64
   06 2  958.836  0.387  958.607  58.86          29 1  958.901  0.258  958.925  44.89
   06 2  959.770  0.293  959.745  60.03          29 1  959.152  0.326  959.198  45.03
   06 2  959.503  0.604  959.475  64.62          29 1  958.543  0.259  958.566  45.17
   06 1  958.838  0.319  958.869  56.94          29 1  958.500  0.251  958.412  45.33
   06 1  958.836  0.387  958.607  58.86          29 1  959.947  0.479  960.124  45.49
   06 1  959.770  0.293  959.745  60.03          29 1  958.543  0.259  958.619  45.66
   06 1  959.503  0.604  959.475  64.62          29 1  958.543  0.259  958.551  46.08
   07 2  958.593  0.296  957.431  58.05          29 1  959.152  0.326  959.130  46.30
   07 2  959.271  0.271  959.220  58.55          29 1  959.591  0.259  959.576  46.52
   07 2  959.271  0.271  959.343  59.67          29 1  959.245  0.298  959.241  47.28
   07 2  959.271  0.271  960.007  60.22
   07 2  958.593  0.296  958.021  60.80                   1999  -  NOVEMBRO
   07 1  958.593  0.296  957.431  58.05          D  L    SDB     ER     SDC    HL
   07 1  959.271  0.271  959.220  58.55          03 2  958.958  0.424  958.946  41.89
   07 1  959.271  0.271  959.343  59.67          03 2  958.474  0.266  958.507  41.96
   07 1  959.271  0.271  960.007  60.22          03 2  958.557  0.206  958.522  42.12
   07 1  958.593  0.296  958.021  60.80          03 2  958.735  0.178  958.744  42.22
   13 2  959.289  0.322  959.257  51.06          03 2  957.822  0.278  957.861  42.48
   13 2  958.185  0.377  958.140  51.29          03 2  958.958  0.424  959.010  42.60
   13 2  959.008  0.408  959.064  51.80          03 2  958.958  0.424  959.054  42.88
   13 2  958.496  0.330  958.517  52.34          03 2  958.899  0.323  958.897  43.06
   13 2  958.960  0.342  958.942  52.96          03 2  958.557  0.206  958.624  43.23
   13 2  958.185  0.377  958.124  53.65          03 2  959.197  0.221  959.367  43.62
   13 2  959.331  0.344  959.480  54.43          03 2  958.329  0.286  958.383  43.84
   13 2  958.784  0.257  958.759  54.84          03 2  959.197  0.221  959.312  44.08
   13 1  959.289  0.322  959.257  51.06          03 1  958.958  0.424  958.946  41.89
   13 1  958.185  0.377  958.140  51.29          03 1  958.474  0.266  958.507  41.96
   13 1  959.008  0.408  959.064  51.80          03 1  958.557  0.206  958.522  42.12
   13 1  958.496  0.330  958.517  52.34          03 1  958.735  0.178  958.744  42.22
   13 1  958.960  0.342  958.942  52.96          03 1  957.822  0.278  957.861  42.48
   13 1  958.185  0.377  958.124  53.65          03 1  958.958  0.424  959.010  42.60
   13 1  959.331  0.344  959.480  54.43          03 1  958.958  0.424  959.054  42.88
   13 1  958.784  0.257  958.759  54.84          03 1  958.899  0.323  958.897  43.06
   14 2  958.902  0.367  958.698  57.27          03 1  958.557  0.206  958.624  43.23
   14 2  960.093  0.354  960.324  57.85          03 1  959.197  0.221  959.367  43.62
   14 2  958.902  0.367  957.331  60.49          03 1  958.329  0.286  958.383  43.84
   14 2  960.093  0.354  960.001  61.19          03 1  959.197  0.221  959.312  44.08
   14 1  958.902  0.367  958.698  57.27          08 2  959.479  0.122  959.446  39.20
   14 1  960.093  0.354  960.324  57.85          08 2  959.027  0.196  959.042  39.26
   14 1  958.902  0.367  957.331  60.49          08 2  959.479  0.122  959.481  39.34
   14 1  960.093  0.354  960.001  61.19          08 2  959.308  0.220  959.269  39.40
   28 2  958.684  0.383  958.665  46.39          08 2  959.635  0.194  959.773  39.48
   28 2  959.286  0.979  959.248  46.59          08 2  959.479  0.122  959.414  39.55
   28 2  959.521  0.210  959.788  47.05          08 2  958.515  0.203  958.503  39.64
   28 2  959.521  0.210  959.711  47.31          08 2  958.926  0.173  958.953  39.74
   28 2  958.684  0.383  958.615  47.62          08 2  958.757  0.166  958.732  40.22
   28 2  959.521  0.210  959.818  47.90          08 2  958.926  0.173  958.861  40.38
   28 2  958.340  0.326  958.172  48.19
```

|  | 1999 - NOVEMBRO | | | | |
|---|---|---|---|---|---|
| D | L | SDB | ER | SDC | HL |
| 08 | 2 | 958.399 | 0.176 | 958.433 | 40.55 |
| 08 | 1 | 959.479 | 0.122 | 959.446 | 39.20 |
| 08 | 1 | 959.027 | 0.196 | 959.042 | 39.26 |
| 08 | 1 | 959.479 | 0.122 | 959.481 | 39.34 |
| 08 | 1 | 959.308 | 0.220 | 959.269 | 39.40 |
| 08 | 1 | 959.635 | 0.194 | 959.773 | 39.48 |
| 08 | 1 | 959.479 | 0.122 | 959.414 | 39.55 |
| 08 | 1 | 958.515 | 0.203 | 958.503 | 39.64 |
| 08 | 1 | 958.926 | 0.173 | 958.953 | 39.74 |
| 08 | 1 | 958.757 | 0.166 | 958.732 | 40.22 |
| 08 | 1 | 958.926 | 0.173 | 958.861 | 40.38 |
| 08 | 1 | 958.399 | 0.176 | 958.433 | 40.55 |
| 23 | 2 | 959.437 | 0.224 | 959.466 | 30.34 |
| 23 | 2 | 959.133 | 0.253 | 957.399 | 30.27 |
| 23 | 2 | 959.437 | 0.224 | 959.551 | 30.06 |
| 23 | 2 | 959.437 | 0.224 | 959.494 | 29.98 |
| 23 | 2 | 959.385 | 0.124 | 959.368 | 29.76 |
| 23 | 2 | 960.183 | 0.262 | 959.851 | 29.75 |
| 23 | 1 | 959.437 | 0.224 | 959.466 | 30.34 |
| 23 | 1 | 959.133 | 0.253 | 957.399 | 30.27 |
| 23 | 1 | 959.437 | 0.224 | 959.551 | 30.06 |
| 23 | 1 | 959.437 | 0.224 | 959.494 | 29.98 |
| 23 | 1 | 959.385 | 0.124 | 959.368 | 29.76 |
| 23 | 1 | 960.183 | 0.262 | 959.851 | 29.75 |
| 26 | 2 | 958.943 | 0.480 | 958.941 | 28.68 |
| 26 | 2 | 959.051 | 0.228 | 959.312 | 28.59 |
| 26 | 2 | 959.051 | 0.228 | 959.054 | 28.16 |
| 26 | 2 | 959.893 | 0.514 | 959.965 | 27.83 |
| 26 | 2 | 959.051 | 0.228 | 959.196 | 27.66 |
| 26 | 1 | 958.943 | 0.480 | 958.941 | 28.68 |
| 26 | 1 | 959.051 | 0.228 | 959.312 | 28.59 |
| 26 | 1 | 959.051 | 0.228 | 959.054 | 28.16 |
| 26 | 1 | 959.893 | 0.514 | 959.965 | 27.83 |
| 26 | 1 | 959.051 | 0.228 | 959.196 | 27.66 |

|  | 1999 - DEZEMBRO | | | | |
|---|---|---|---|---|---|
| D | L | SDB | ER | SDC | HL |
| 01 | 2 | 959.781 | 0.176 | 959.761 | 27.11 |
| 01 | 2 | 959.696 | 0.191 | 959.700 | 26.97 |
| 01 | 2 | 960.079 | 0.207 | 960.080 | 26.83 |
| 01 | 2 | 958.979 | 0.173 | 959.011 | 26.66 |
| 01 | 2 | 960.079 | 0.207 | 960.095 | 26.51 |
| 01 | 2 | 958.979 | 0.173 | 958.962 | 26.37 |
| 01 | 2 | 959.069 | 0.180 | 959.204 | 26.24 |
| 01 | 2 | 959.696 | 0.191 | 959.600 | 25.79 |
| 01 | 2 | 958.836 | 0.181 | 958.353 | 25.66 |
| 01 | 2 | 959.696 | 0.191 | 959.600 | 25.53 |
| 01 | 2 | 960.284 | 0.210 | 960.332 | 25.35 |
| 01 | 2 | 959.463 | 0.249 | 959.441 | 25.23 |
| 01 | 2 | 958.836 | 0.181 | 958.400 | 25.10 |
| 01 | 2 | 960.483 | 0.256 | 960.451 | 24.99 |
| 01 | 1 | 959.781 | 0.176 | 959.761 | 27.11 |
| 01 | 1 | 959.696 | 0.191 | 959.700 | 26.97 |
| 01 | 1 | 960.079 | 0.207 | 960.080 | 26.83 |
| 01 | 1 | 958.979 | 0.173 | 959.011 | 26.66 |
| 01 | 1 | 960.079 | 0.207 | 960.095 | 26.51 |
| 01 | 1 | 958.979 | 0.173 | 958.962 | 26.37 |
| 01 | 1 | 959.069 | 0.180 | 959.204 | 26.24 |
| 01 | 1 | 959.696 | 0.191 | 959.600 | 25.79 |
| 01 | 1 | 958.836 | 0.181 | 958.353 | 25.66 |
| 01 | 1 | 959.696 | 0.191 | 959.600 | 25.53 |
| 01 | 1 | 960.284 | 0.210 | 960.332 | 25.35 |
| 01 | 1 | 959.463 | 0.249 | 959.441 | 25.23 |
| 01 | 1 | 958.836 | 0.181 | 958.400 | 25.10 |
| 01 | 1 | 960.483 | 0.256 | 960.451 | 24.99 |
| 03 | 2 | 959.679 | 0.204 | 959.748 | 24.85 |
| 03 | 2 | 958.818 | 0.312 | 958.666 | 24.70 |
| 03 | 2 | 958.818 | 0.312 | 958.592 | 24.13 |
| 03 | 2 | 959.679 | 0.204 | 959.824 | 24.01 |
| 03 | 2 | 959.679 | 0.204 | 959.570 | 23.89 |

|  | 1999 - DEZEMBRO | | | | |
|---|---|---|---|---|---|
| D | L | SDB | ER | SDC | HL |
| 03 | 2 | 959.068 | 0.258 | 959.050 | 23.77 |
| 03 | 2 | 958.818 | 0.312 | 958.258 | 23.65 |
| 03 | 2 | 959.679 | 0.204 | 959.644 | 23.54 |
| 03 | 2 | 958.946 | 0.207 | 958.960 | 23.41 |
| 03 | 2 | 958.999 | 0.269 | 959.005 | 23.12 |
| 03 | 1 | 959.679 | 0.204 | 959.748 | 24.85 |
| 03 | 1 | 958.818 | 0.312 | 958.666 | 24.70 |
| 03 | 1 | 958.818 | 0.312 | 958.592 | 24.13 |
| 03 | 1 | 959.679 | 0.204 | 959.824 | 24.01 |
| 03 | 1 | 959.679 | 0.204 | 959.570 | 23.89 |
| 03 | 1 | 959.068 | 0.258 | 959.050 | 23.77 |
| 03 | 1 | 958.818 | 0.312 | 958.258 | 23.65 |
| 03 | 1 | 959.679 | 0.204 | 959.644 | 23.54 |
| 03 | 1 | 958.946 | 0.207 | 958.960 | 23.41 |
| 03 | 1 | 958.999 | 0.269 | 959.005 | 23.12 |
| 06 | 2 | 959.827 | 0.221 | 959.845 | 24.16 |
| 06 | 2 | 959.211 | 0.146 | 959.215 | 24.00 |
| 06 | 2 | 959.238 | 0.152 | 959.252 | 23.83 |
| 06 | 2 | 960.194 | 0.200 | 960.195 | 23.66 |
| 06 | 2 | 959.238 | 0.152 | 959.286 | 23.50 |
| 06 | 2 | 959.495 | 0.157 | 959.519 | 23.35 |
| 06 | 2 | 959.211 | 0.146 | 959.009 | 23.20 |
| 06 | 2 | 958.767 | 0.233 | 958.726 | 23.05 |
| 06 | 2 | 959.211 | 0.146 | 959.187 | 22.90 |
| 06 | 2 | 959.407 | 0.160 | 959.383 | 22.74 |
| 06 | 2 | 959.820 | 0.134 | 959.749 | 22.57 |
| 06 | 2 | 958.278 | 0.165 | 958.322 | 22.43 |
| 06 | 2 | 959.211 | 0.146 | 959.118 | 22.27 |
| 06 | 2 | 958.767 | 0.233 | 958.862 | 22.13 |
| 06 | 2 | 959.634 | 0.184 | 959.721 | 21.94 |
| 06 | 2 | 959.211 | 0.146 | 959.092 | 21.80 |
| 06 | 2 | 959.407 | 0.160 | 959.369 | 21.66 |
| 06 | 2 | 959.827 | 0.221 | 959.870 | 21.52 |
| 06 | 2 | 959.407 | 0.160 | 959.401 | 21.38 |
| 06 | 2 | 959.211 | 0.146 | 959.133 | 21.23 |
| 06 | 2 | 959.211 | 0.146 | 959.221 | 20.95 |
| 06 | 2 | 959.211 | 0.146 | 959.221 | 20.95 |
| 06 | 2 | 958.634 | 0.176 | 958.548 | 20.82 |
| 06 | 2 | 959.211 | 0.146 | 959.131 | 20.69 |
| 06 | 1 | 959.827 | 0.221 | 959.845 | 24.16 |
| 06 | 1 | 959.211 | 0.146 | 959.215 | 24.00 |
| 06 | 1 | 959.238 | 0.152 | 959.252 | 23.83 |
| 06 | 1 | 960.194 | 0.200 | 960.195 | 23.66 |
| 06 | 1 | 959.238 | 0.152 | 959.286 | 23.50 |
| 06 | 1 | 959.495 | 0.157 | 959.519 | 23.35 |
| 06 | 1 | 959.211 | 0.146 | 959.009 | 23.20 |
| 06 | 1 | 958.767 | 0.233 | 958.726 | 23.05 |
| 06 | 1 | 959.211 | 0.146 | 959.187 | 22.90 |
| 06 | 1 | 959.407 | 0.160 | 959.383 | 22.74 |
| 06 | 1 | 959.820 | 0.134 | 959.749 | 22.57 |
| 06 | 1 | 958.278 | 0.165 | 958.322 | 22.43 |
| 06 | 1 | 959.211 | 0.146 | 959.118 | 22.27 |
| 06 | 1 | 958.767 | 0.233 | 958.862 | 22.13 |
| 06 | 1 | 959.634 | 0.184 | 959.721 | 21.94 |
| 06 | 1 | 959.211 | 0.146 | 959.092 | 21.80 |
| 06 | 1 | 959.407 | 0.160 | 959.369 | 21.66 |
| 06 | 1 | 959.827 | 0.221 | 959.870 | 21.52 |
| 06 | 1 | 959.407 | 0.160 | 959.401 | 21.38 |
| 06 | 1 | 959.211 | 0.146 | 959.133 | 21.23 |
| 06 | 1 | 959.211 | 0.146 | 959.221 | 20.95 |
| 06 | 1 | 959.211 | 0.146 | 959.221 | 20.95 |
| 06 | 1 | 958.634 | 0.176 | 958.548 | 20.82 |
| 06 | 1 | 959.211 | 0.146 | 959.131 | 20.69 |
| 15 | 2 | 959.553 | 0.156 | 959.579 | 17.89 |
| 15 | 2 | 959.884 | 0.116 | 960.057 | 17.69 |
| 15 | 2 | 960.873 | 0.241 | 960.942 | 17.48 |
| 15 | 2 | 959.368 | 0.126 | 959.347 | 17.30 |
| 15 | 2 | 959.377 | 0.170 | 959.419 | 17.11 |
| 15 | 2 | 959.884 | 0.116 | 959.920 | 16.92 |
| 15 | 2 | 959.377 | 0.170 | 959.450 | 16.74 |
| 15 | 2 | 959.230 | 0.115 | 959.163 | 16.50 |

| 1999 - DEZEMBRO | | | | | | 1999 - DEZEMBRO | | | | |
|---|---|---|---|---|---|---|---|---|---|---|
| D | L | SDB | ER | SDC | HL | D | L | SDB | ER | SDC | HL |
| 15 | 2 | 958.905 | 0.276 | 958.781 | 16.29 | 17 | 1 | 958.582 | 0.406 | 958.429 | 15.32 |
| 15 | 2 | 959.867 | 0.174 | 959.731 | 16.11 | 22 | 2 | 958.839 | 0.310 | 958.759 | 15.29 |
| 15 | 2 | 959.553 | 0.156 | 959.676 | 15.92 | 22 | 2 | 958.427 | 0.243 | 957.626 | 15.11 |
| 15 | 2 | 959.230 | 0.115 | 959.173 | 15.73 | 22 | 2 | 958.839 | 0.310 | 958.791 | 14.94 |
| 15 | 2 | 959.377 | 0.170 | 959.440 | 15.54 | 22 | 2 | 959.107 | 0.249 | 959.149 | 14.76 |
| 15 | 2 | 959.027 | 0.394 | 959.001 | 15.36 | 22 | 2 | 959.374 | 0.266 | 959.378 | 14.59 |
| 15 | 2 | 958.905 | 0.276 | 958.802 | 15.19 | 22 | 2 | 959.485 | 0.215 | 959.531 | 14.42 |
| 15 | 2 | 958.905 | 0.276 | 958.782 | 15.00 | 22 | 2 | 958.427 | 0.243 | 958.522 | 14.25 |
| 15 | 1 | 959.553 | 0.156 | 959.579 | 17.89 | 22 | 2 | 959.485 | 0.215 | 959.490 | 14.08 |
| 15 | 1 | 959.884 | 0.116 | 960.057 | 17.69 | 22 | 2 | 958.627 | 0.332 | 958.660 | 13.91 |
| 15 | 1 | 960.873 | 0.241 | 960.942 | 17.48 | 22 | 2 | 958.839 | 0.310 | 958.871 | 13.73 |
| 15 | 1 | 959.368 | 0.126 | 959.347 | 17.30 | 22 | 2 | 959.107 | 0.249 | 959.068 | 13.56 |
| 15 | 1 | 959.377 | 0.170 | 959.419 | 17.11 | 22 | 2 | 959.865 | 0.182 | 959.956 | 13.39 |
| 15 | 1 | 959.884 | 0.116 | 959.920 | 16.92 | 22 | 2 | 958.427 | 0.243 | 958.442 | 13.22 |
| 15 | 1 | 959.377 | 0.170 | 959.450 | 16.74 | 22 | 2 | 958.427 | 0.243 | 958.375 | 13.05 |
| 15 | 1 | 959.230 | 0.115 | 959.163 | 16.50 | 22 | 2 | 958.839 | 0.310 | 958.856 | 12.86 |
| 15 | 1 | 958.905 | 0.276 | 958.781 | 16.29 | 22 | 2 | 958.839 | 0.310 | 958.734 | 12.69 |
| 15 | 1 | 959.867 | 0.174 | 959.731 | 16.11 | 22 | 1 | 958.839 | 0.310 | 958.759 | 15.29 |
| 15 | 1 | 959.553 | 0.156 | 959.676 | 15.92 | 22 | 1 | 958.427 | 0.243 | 957.626 | 15.11 |
| 15 | 1 | 959.230 | 0.115 | 959.173 | 15.73 | 22 | 1 | 958.839 | 0.310 | 958.791 | 14.94 |
| 15 | 1 | 959.377 | 0.170 | 959.440 | 15.54 | 22 | 1 | 959.107 | 0.249 | 959.149 | 14.76 |
| 15 | 1 | 959.027 | 0.394 | 959.001 | 15.36 | 22 | 1 | 959.374 | 0.266 | 959.378 | 14.59 |
| 15 | 1 | 958.905 | 0.276 | 958.802 | 15.19 | 22 | 1 | 959.485 | 0.215 | 959.531 | 14.42 |
| 15 | 1 | 958.905 | 0.276 | 958.782 | 15.00 | 22 | 1 | 958.427 | 0.243 | 958.522 | 14.25 |
| 16 | 2 | 958.761 | 0.178 | 957.584 | 15.95 | 22 | 1 | 959.485 | 0.215 | 959.490 | 14.08 |
| 16 | 2 | 958.986 | 0.183 | 959.294 | 15.78 | 22 | 1 | 958.627 | 0.332 | 958.660 | 13.91 |
| 16 | 2 | 958.761 | 0.178 | 958.681 | 15.62 | 22 | 1 | 958.839 | 0.310 | 958.871 | 13.73 |
| 16 | 2 | 958.986 | 0.183 | 959.042 | 15.43 | 22 | 1 | 959.107 | 0.249 | 959.068 | 13.56 |
| 16 | 2 | 958.761 | 0.178 | 957.805 | 15.27 | 22 | 1 | 959.865 | 0.182 | 959.956 | 13.39 |
| 16 | 2 | 958.761 | 0.178 | 958.637 | 15.10 | 22 | 1 | 958.427 | 0.243 | 958.442 | 13.22 |
| 16 | 2 | 958.761 | 0.178 | 958.271 | 14.93 | 22 | 1 | 958.427 | 0.243 | 958.375 | 13.05 |
| 16 | 2 | 958.780 | 0.108 | 958.782 | 14.76 | 22 | 1 | 958.839 | 0.310 | 958.856 | 12.86 |
| 16 | 1 | 958.761 | 0.178 | 957.584 | 15.95 | 22 | 1 | 958.839 | 0.310 | 958.734 | 12.69 |
| 16 | 1 | 958.986 | 0.183 | 959.294 | 15.78 | 28 | 2 | 959.290 | 0.264 | 959.313 | 12.19 |
| 16 | 1 | 958.761 | 0.178 | 958.681 | 15.62 | 28 | 2 | 959.283 | 0.154 | 959.107 | 12.03 |
| 16 | 1 | 958.986 | 0.183 | 959.042 | 15.43 | 28 | 2 | 959.283 | 0.154 | 959.045 | 11.85 |
| 16 | 1 | 958.761 | 0.178 | 957.805 | 15.27 | 28 | 2 | 959.586 | 0.226 | 959.549 | 11.67 |
| 16 | 1 | 958.761 | 0.178 | 958.637 | 15.10 | 28 | 2 | 958.298 | 0.274 | 958.424 | 11.49 |
| 16 | 1 | 958.761 | 0.178 | 958.271 | 14.93 | 28 | 2 | 959.496 | 0.160 | 959.520 | 11.32 |
| 16 | 1 | 958.780 | 0.108 | 958.782 | 14.76 | 28 | 2 | 959.283 | 0.154 | 958.843 | 11.14 |
| 17 | 2 | 959.238 | 0.255 | 959.222 | 18.19 | 28 | 2 | 959.283 | 0.154 | 958.916 | 10.98 |
| 17 | 2 | 959.238 | 0.255 | 959.409 | 18.01 | 28 | 2 | 959.283 | 0.154 | 959.094 | 10.81 |
| 17 | 2 | 959.625 | 0.230 | 959.537 | 17.81 | 28 | 2 | 959.496 | 0.160 | 959.432 | 10.65 |
| 17 | 2 | 959.812 | 0.292 | 959.844 | 17.63 | 28 | 2 | 959.283 | 0.154 | 959.001 | 10.48 |
| 17 | 2 | 959.238 | 0.255 | 959.174 | 17.45 | 28 | 2 | 958.294 | 0.284 | 957.766 | 10.31 |
| 17 | 2 | 959.766 | 0.171 | 959.770 | 17.26 | 28 | 2 | 958.298 | 0.274 | 958.579 | 10.15 |
| 17 | 2 | 958.582 | 0.406 | 958.661 | 17.07 | 28 | 2 | 959.283 | 0.154 | 959.080 | 9.98 |
| 17 | 2 | 959.625 | 0.230 | 959.575 | 16.88 | 28 | 1 | 959.290 | 0.264 | 959.313 | 12.19 |
| 17 | 2 | 959.904 | 0.286 | 959.909 | 16.66 | 28 | 1 | 959.283 | 0.154 | 959.107 | 12.03 |
| 17 | 2 | 959.625 | 0.230 | 959.689 | 16.46 | 28 | 1 | 959.283 | 0.154 | 959.045 | 11.85 |
| 17 | 2 | 959.625 | 0.230 | 959.526 | 16.27 | 28 | 1 | 959.586 | 0.226 | 959.549 | 11.67 |
| 17 | 2 | 959.238 | 0.255 | 959.146 | 16.08 | 28 | 1 | 958.298 | 0.274 | 958.424 | 11.49 |
| 17 | 2 | 959.238 | 0.255 | 959.358 | 15.89 | 28 | 1 | 959.496 | 0.160 | 959.520 | 11.32 |
| 17 | 2 | 959.238 | 0.255 | 959.324 | 15.71 | 28 | 1 | 959.283 | 0.154 | 958.843 | 11.14 |
| 17 | 2 | 958.883 | 0.195 | 958.933 | 15.51 | 28 | 1 | 959.283 | 0.154 | 958.916 | 10.98 |
| 17 | 2 | 958.582 | 0.406 | 958.429 | 15.32 | 28 | 1 | 959.283 | 0.154 | 959.094 | 10.81 |
| 17 | 1 | 959.238 | 0.255 | 959.222 | 18.19 | 28 | 1 | 959.496 | 0.160 | 959.432 | 10.65 |
| 17 | 1 | 959.238 | 0.255 | 959.409 | 18.01 | 28 | 1 | 959.283 | 0.154 | 959.001 | 10.48 |
| 17 | 1 | 959.625 | 0.230 | 959.537 | 17.81 | 28 | 1 | 958.294 | 0.284 | 957.766 | 10.31 |
| 17 | 1 | 959.812 | 0.292 | 959.844 | 17.63 | 28 | 1 | 958.298 | 0.274 | 958.579 | 10.15 |
| 17 | 1 | 959.238 | 0.255 | 959.174 | 17.45 | 28 | 1 | 959.283 | 0.154 | 959.080 | 9.98 |
| 17 | 1 | 959.766 | 0.171 | 959.770 | 17.26 | 29 | 2 | 958.994 | 0.253 | 959.042 | 12.42 |
| 17 | 1 | 958.582 | 0.406 | 958.661 | 17.07 | 29 | 2 | 958.085 | 0.129 | 958.159 | 12.25 |
| 17 | 1 | 959.625 | 0.230 | 959.575 | 16.88 | 29 | 2 | 958.573 | 0.318 | 958.484 | 12.07 |
| 17 | 1 | 959.904 | 0.286 | 959.909 | 16.66 | 29 | 2 | 958.332 | 0.353 | 958.230 | 11.91 |
| 17 | 1 | 959.625 | 0.230 | 959.689 | 16.46 | 29 | 2 | 958.994 | 0.253 | 958.812 | 11.74 |
| 17 | 1 | 959.625 | 0.230 | 959.526 | 16.27 | 29 | 2 | 958.994 | 0.253 | 959.073 | 11.58 |
| 17 | 1 | 959.238 | 0.255 | 959.146 | 16.08 | 29 | 2 | 959.409 | 0.344 | 959.362 | 11.41 |
| 17 | 1 | 959.238 | 0.255 | 959.358 | 15.89 | 29 | 2 | 959.409 | 0.344 | 959.384 | 11.24 |
| 17 | 1 | 959.238 | 0.255 | 959.324 | 15.71 | 29 | 2 | 958.573 | 0.318 | 958.547 | 11.07 |
| 17 | 1 | 958.883 | 0.195 | 958.933 | 15.51 | 29 | 2 | 958.994 | 0.253 | 958.890 | 10.91 |

| 1999 - DEZEMBRO | | | | | | 2000 - JANEIRO | | | | |
|---|---|---|---|---|---|---|---|---|---|---|
| D | L | SDB | ER | SDC | HL | D | L | SDB | ER | SDC | HL |
| 29 | 2 | 958.085 | 0.129 | 958.137 | 10.58 | 05 | 1 | 959.949 | 0.153 | 959.945 | 8.39 |
| 29 | 2 | 958.573 | 0.318 | 958.663 | 10.42 | 05 | 1 | 959.949 | 0.153 | 959.640 | 8.23 |
| 29 | 2 | 959.409 | 0.344 | 959.281 | 10.25 | 05 | 1 | 960.856 | 0.154 | 960.817 | 8.07 |
| 29 | 2 | 958.573 | 0.318 | 958.574 | 10.09 | 05 | 1 | 959.949 | 0.153 | 959.615 | 7.90 |
| 29 | 2 | 958.070 | 0.224 | 957.643 | 9.93 | 05 | 1 | 960.067 | 0.132 | 960.106 | 7.75 |
| 29 | 2 | 958.573 | 0.318 | 958.486 | 9.74 | 05 | 1 | 959.949 | 0.153 | 959.887 | 7.60 |
| 29 | 2 | 958.070 | 0.224 | 957.470 | 9.57 | 05 | 1 | 958.900 | 0.154 | 958.832 | 7.45 |
| 29 | 1 | 958.994 | 0.253 | 959.042 | 12.42 | 07 | 2 | 958.458 | 0.247 | 958.421 | 8.68 |
| 29 | 1 | 958.085 | 0.129 | 958.159 | 12.25 | 07 | 2 | 958.160 | 0.693 | 958.114 | 8.53 |
| 29 | 1 | 958.573 | 0.318 | 958.484 | 12.07 | 07 | 2 | 958.287 | 0.252 | 958.296 | 8.39 |
| 29 | 1 | 958.332 | 0.353 | 958.230 | 11.91 | 07 | 2 | 958.535 | 0.197 | 958.575 | 8.25 |
| 29 | 1 | 958.994 | 0.253 | 958.812 | 11.74 | 07 | 2 | 958.535 | 0.197 | 958.579 | 8.11 |
| 29 | 1 | 958.994 | 0.253 | 959.073 | 11.58 | 07 | 2 | 958.683 | 0.241 | 958.685 | 7.96 |
| 29 | 1 | 959.409 | 0.344 | 959.362 | 11.41 | 07 | 2 | 958.799 | 0.227 | 958.835 | 7.83 |
| 29 | 1 | 959.409 | 0.344 | 959.384 | 11.24 | 07 | 2 | 958.160 | 0.693 | 957.978 | 7.69 |
| 29 | 1 | 958.573 | 0.318 | 958.547 | 11.07 | 07 | 2 | 958.897 | 0.257 | 958.901 | 7.56 |
| 29 | 1 | 958.994 | 0.253 | 958.890 | 10.91 | 07 | 2 | 958.672 | 0.226 | 958.653 | 7.42 |
| 29 | 1 | 958.085 | 0.129 | 958.137 | 10.58 | 07 | 2 | 959.058 | 0.187 | 959.047 | 7.29 |
| 29 | 1 | 958.573 | 0.318 | 958.663 | 10.42 | 07 | 2 | 958.979 | 0.182 | 958.987 | 7.16 |
| 29 | 1 | 959.409 | 0.344 | 959.281 | 10.25 | 07 | 2 | 958.458 | 0.247 | 958.436 | 7.03 |
| 29 | 1 | 958.573 | 0.318 | 958.574 | 10.09 | 07 | 2 | 958.683 | 0.241 | 958.686 | 6.89 |
| 29 | 1 | 958.070 | 0.224 | 957.643 | 9.93 | 07 | 2 | 958.279 | 0.223 | 958.252 | 6.77 |
| 29 | 1 | 958.573 | 0.318 | 958.486 | 9.74 | 07 | 2 | 958.672 | 0.226 | 958.668 | 6.64 |
| 29 | 1 | 958.070 | 0.224 | 957.470 | 9.57 | 07 | 1 | 958.458 | 0.247 | 958.421 | 8.68 |
| 30 | 2 | 959.065 | 0.206 | 959.094 | 11.92 | 07 | 1 | 958.160 | 0.693 | 958.114 | 8.53 |
| 30 | 2 | 958.269 | 0.265 | 958.353 | 11.76 | 07 | 1 | 958.287 | 0.252 | 958.296 | 8.39 |
| 30 | 2 | 958.673 | 0.240 | 958.802 | 11.60 | 07 | 1 | 958.535 | 0.197 | 958.575 | 8.25 |
| 30 | 2 | 959.016 | 0.163 | 959.018 | 11.44 | 07 | 1 | 958.535 | 0.197 | 958.579 | 8.11 |
| 30 | 2 | 957.597 | 0.184 | 957.644 | 11.27 | 07 | 1 | 958.683 | 0.241 | 958.685 | 7.96 |
| 30 | 2 | 959.255 | 0.200 | 959.185 | 11.11 | 07 | 1 | 958.799 | 0.227 | 958.835 | 7.83 |
| 30 | 2 | 959.255 | 0.200 | 959.377 | 10.95 | 07 | 1 | 958.160 | 0.693 | 957.978 | 7.69 |
| 30 | 2 | 959.255 | 0.200 | 959.318 | 10.78 | 07 | 1 | 958.897 | 0.257 | 958.901 | 7.56 |
| 30 | 2 | 958.269 | 0.265 | 958.370 | 10.62 | 07 | 1 | 958.672 | 0.226 | 958.653 | 7.42 |
| 30 | 2 | 958.269 | 0.265 | 958.077 | 10.46 | 07 | 1 | 959.058 | 0.187 | 959.047 | 7.29 |
| 30 | 2 | 958.580 | 0.228 | 958.570 | 10.30 | 07 | 1 | 958.979 | 0.182 | 958.987 | 7.16 |
| 30 | 2 | 958.673 | 0.240 | 958.841 | 9.96 | 07 | 1 | 958.458 | 0.247 | 958.436 | 7.03 |
| 30 | 2 | 957.597 | 0.184 | 957.660 | 9.80 | 07 | 1 | 958.683 | 0.241 | 958.686 | 6.89 |
| 30 | 2 | 958.673 | 0.240 | 958.691 | 9.64 | 07 | 1 | 958.279 | 0.223 | 958.252 | 6.77 |
| 30 | 2 | 958.269 | 0.265 | 958.081 | 9.49 | 07 | 1 | 958.672 | 0.226 | 958.668 | 6.64 |
| 30 | 2 | 958.269 | 0.265 | 958.138 | 9.29 | 10 | 2 | 960.412 | 0.155 | 960.394 | 7.35 |
| 30 | 2 | 958.673 | 0.240 | 958.734 | 9.14 | 10 | 2 | 960.011 | 0.158 | 960.043 | 7.20 |
| 30 | 1 | 959.065 | 0.206 | 959.094 | 11.92 | 10 | 2 | 960.124 | 0.174 | 960.166 | 7.07 |
| 30 | 1 | 958.269 | 0.265 | 958.353 | 11.76 | 10 | 2 | 959.554 | 0.173 | 959.534 | 6.93 |
| 30 | 1 | 958.673 | 0.240 | 958.802 | 11.60 | 10 | 2 | 959.228 | 0.138 | 959.248 | 6.80 |
| 30 | 1 | 959.016 | 0.163 | 959.018 | 11.44 | 10 | 2 | 959.444 | 0.157 | 959.442 | 6.67 |
| 30 | 1 | 957.597 | 0.184 | 957.644 | 11.27 | 10 | 2 | 959.343 | 0.147 | 959.357 | 6.53 |
| 30 | 1 | 959.255 | 0.200 | 959.185 | 11.11 | 10 | 2 | 959.284 | 0.177 | 959.285 | 6.40 |
| 30 | 1 | 959.255 | 0.200 | 959.377 | 10.95 | 10 | 2 | 959.649 | 0.140 | 959.630 | 6.28 |
| 30 | 1 | 959.255 | 0.200 | 959.318 | 10.78 | 10 | 2 | 958.968 | 0.136 | 958.934 | 6.15 |
| 30 | 1 | 958.269 | 0.265 | 958.370 | 10.62 | 10 | 2 | 959.458 | 0.144 | 959.454 | 6.01 |
| 30 | 1 | 958.269 | 0.265 | 958.077 | 10.46 | 10 | 2 | 959.458 | 0.144 | 959.483 | 5.88 |
| 30 | 1 | 958.580 | 0.228 | 958.570 | 10.30 | 10 | 2 | 959.210 | 0.139 | 959.216 | 5.76 |
| 30 | 1 | 958.673 | 0.240 | 958.841 | 9.96 | 10 | 2 | 959.117 | 0.154 | 959.077 | 5.64 |
| 30 | 1 | 957.597 | 0.184 | 957.660 | 9.80 | 10 | 2 | 959.284 | 0.177 | 959.292 | 5.52 |
| 30 | 1 | 958.673 | 0.240 | 958.691 | 9.64 | 10 | 2 | 958.968 | 0.136 | 958.506 | 5.41 |
| 30 | 1 | 958.269 | 0.265 | 958.081 | 9.49 | 10 | 1 | 960.412 | 0.155 | 960.394 | 7.35 |
| 30 | 1 | 958.269 | 0.265 | 958.138 | 9.29 | 10 | 1 | 960.011 | 0.158 | 960.043 | 7.20 |
| 30 | 1 | 958.673 | 0.240 | 958.734 | 9.14 | 10 | 1 | 960.124 | 0.174 | 960.166 | 7.07 |
| | | | | | | 10 | 1 | 959.554 | 0.173 | 959.534 | 6.93 |
| | | | | | | 10 | 1 | 959.228 | 0.138 | 959.248 | 6.80 |
| | 2000 - JANEIRO | | | | | 10 | 1 | 959.444 | 0.157 | 959.442 | 6.67 |
| D | L | SDB | ER | SDC | HL | 10 | 1 | 959.343 | 0.147 | 959.357 | 6.53 |
| 05 | 2 | 960.356 | 0.162 | 960.377 | 8.56 | 10 | 1 | 959.284 | 0.177 | 959.285 | 6.40 |
| 05 | 2 | 959.949 | 0.153 | 959.945 | 8.39 | 10 | 1 | 959.649 | 0.140 | 959.630 | 6.28 |
| 05 | 2 | 959.949 | 0.153 | 959.640 | 8.23 | 10 | 1 | 958.968 | 0.136 | 958.934 | 6.15 |
| 05 | 2 | 960.856 | 0.154 | 960.817 | 8.07 | 10 | 1 | 959.458 | 0.144 | 959.454 | 6.01 |
| 05 | 2 | 959.949 | 0.153 | 959.615 | 7.90 | 10 | 1 | 959.458 | 0.144 | 959.483 | 5.88 |
| 05 | 2 | 960.067 | 0.132 | 960.106 | 7.75 | 10 | 1 | 959.210 | 0.139 | 959.216 | 5.76 |
| 05 | 2 | 959.949 | 0.153 | 959.887 | 7.60 | 10 | 1 | 959.117 | 0.154 | 959.077 | 5.64 |
| 05 | 2 | 958.900 | 0.154 | 958.832 | 7.45 | 10 | 1 | 959.284 | 0.177 | 959.292 | 5.52 |
| 05 | 1 | 960.356 | 0.162 | 960.377 | 8.56 | 10 | 1 | 958.968 | 0.136 | 958.506 | 5.41 |

| 2000 - JANEIRO | | | | |
|---|---|---|---|---|
| D L | SDB | ER | SDC | HL |
| 11 2 | 959.036 | 0.152 | 959.033 | 7.64 |
| 11 2 | 959.625 | 0.165 | 959.641 | 7.51 |
| 11 2 | 959.535 | 0.288 | 959.518 | 7.38 |
| 11 2 | 958.927 | 0.221 | 958.929 | 7.25 |
| 11 2 | 959.836 | 0.218 | 959.825 | 7.12 |
| 11 2 | 959.618 | 0.519 | 959.608 | 7.00 |
| 11 2 | 959.559 | 0.197 | 959.568 | 6.88 |
| 11 2 | 958.531 | 0.214 | 958.531 | 6.76 |
| 11 2 | 958.759 | 0.205 | 958.800 | 6.64 |
| 11 2 | 959.005 | 0.148 | 958.966 | 6.52 |
| 11 2 | 959.173 | 0.224 | 959.191 | 6.40 |
| 11 2 | 959.005 | 0.148 | 959.010 | 6.28 |
| 11 2 | 958.357 | 0.344 | 958.250 | 6.17 |
| 11 2 | 958.759 | 0.205 | 958.785 | 6.05 |
| 11 2 | 959.173 | 0.224 | 959.147 | 5.94 |
| 11 2 | 958.570 | 0.167 | 958.575 | 5.83 |
| 11 2 | 958.655 | 0.170 | 958.695 | 5.73 |
| 11 2 | 960.645 | 0.361 | 960.640 | 5.62 |
| 11 1 | 959.036 | 0.152 | 959.033 | 7.64 |
| 11 1 | 959.625 | 0.165 | 959.641 | 7.51 |
| 11 1 | 959.535 | 0.288 | 959.518 | 7.38 |
| 11 1 | 958.927 | 0.221 | 958.929 | 7.25 |
| 11 1 | 959.836 | 0.218 | 959.825 | 7.12 |
| 11 1 | 959.618 | 0.519 | 959.608 | 7.00 |
| 11 1 | 959.559 | 0.197 | 959.568 | 6.88 |
| 11 1 | 958.531 | 0.214 | 958.531 | 6.76 |
| 11 1 | 958.759 | 0.205 | 958.800 | 6.64 |
| 11 1 | 959.005 | 0.148 | 958.966 | 6.52 |
| 11 1 | 959.173 | 0.224 | 959.191 | 6.40 |
| 11 1 | 959.005 | 0.148 | 959.010 | 6.28 |
| 11 1 | 958.357 | 0.344 | 958.250 | 6.17 |
| 11 1 | 958.759 | 0.205 | 958.785 | 6.05 |
| 11 1 | 959.173 | 0.224 | 959.147 | 5.94 |
| 11 1 | 958.570 | 0.167 | 958.575 | 5.83 |
| 11 1 | 958.655 | 0.170 | 958.695 | 5.73 |
| 11 1 | 960.645 | 0.361 | 960.640 | 5.62 |
| 13 2 | 959.908 | 0.212 | 959.858 | 6.68 |
| 13 2 | 958.720 | 0.237 | 958.667 | 6.56 |
| 13 2 | 958.781 | 0.128 | 958.757 | 6.45 |
| 13 2 | 957.869 | 0.376 | 957.987 | 6.33 |
| 13 2 | 958.720 | 0.237 | 958.676 | 6.22 |
| 13 2 | 959.005 | 0.335 | 959.028 | 6.11 |
| 13 2 | 958.550 | 0.310 | 958.566 | 6.01 |
| 13 2 | 958.847 | 0.193 | 958.866 | 5.90 |
| 13 2 | 958.165 | 0.176 | 958.061 | 5.80 |
| 13 2 | 959.097 | 0.187 | 959.068 | 5.70 |
| 13 2 | 958.254 | 0.191 | 958.285 | 5.60 |
| 13 2 | 959.232 | 0.203 | 959.295 | 5.50 |
| 13 2 | 959.005 | 0.335 | 958.972 | 5.40 |
| 13 2 | 959.232 | 0.203 | 959.219 | 5.31 |
| 13 2 | 958.781 | 0.128 | 958.766 | 5.22 |
| 13 2 | 958.489 | 0.251 | 958.482 | 5.13 |
| 13 2 | 959.435 | 0.150 | 959.405 | 5.04 |
| 13 1 | 959.908 | 0.212 | 959.858 | 6.68 |
| 13 1 | 958.720 | 0.237 | 958.667 | 6.56 |
| 13 1 | 958.781 | 0.128 | 958.757 | 6.45 |
| 13 1 | 957.869 | 0.376 | 957.987 | 6.33 |
| 13 1 | 958.720 | 0.237 | 958.676 | 6.22 |
| 13 1 | 959.005 | 0.335 | 959.028 | 6.11 |
| 13 1 | 958.550 | 0.310 | 958.566 | 6.01 |
| 13 1 | 958.847 | 0.193 | 958.866 | 5.90 |
| 13 1 | 958.165 | 0.176 | 958.061 | 5.80 |
| 13 1 | 959.097 | 0.187 | 959.068 | 5.70 |
| 13 1 | 958.254 | 0.191 | 958.285 | 5.60 |
| 13 1 | 959.232 | 0.203 | 959.295 | 5.50 |
| 13 1 | 959.005 | 0.335 | 958.972 | 5.40 |
| 13 1 | 959.232 | 0.203 | 959.219 | 5.31 |
| 13 1 | 958.781 | 0.128 | 958.766 | 5.22 |
| 13 1 | 958.489 | 0.251 | 958.482 | 5.13 |
| 13 1 | 959.435 | 0.150 | 959.405 | 5.04 |
| 18 2 | 959.888 | 0.143 | 959.885 | 5.94 |

| 2000 - JANEIRO | | | | |
|---|---|---|---|---|
| D L | SDB | ER | SDC | HL |
| 18 2 | 960.389 | 0.156 | 960.389 | 5.83 |
| 18 2 | 959.757 | 0.150 | 959.719 | 5.71 |
| 18 2 | 960.218 | 0.145 | 960.197 | 5.61 |
| 18 2 | 960.377 | 0.180 | 960.354 | 5.50 |
| 18 1 | 959.888 | 0.143 | 959.885 | 5.94 |
| 18 1 | 960.389 | 0.156 | 960.389 | 5.83 |
| 18 1 | 959.757 | 0.150 | 959.719 | 5.71 |
| 18 1 | 960.218 | 0.145 | 960.197 | 5.61 |
| 18 1 | 960.377 | 0.180 | 960.354 | 5.50 |
| 21 2 | 959.096 | 0.610 | 959.130 | 4.15 |
| 21 2 | 958.334 | 0.329 | 958.613 | 4.11 |
| 21 2 | 959.438 | 0.251 | 959.456 | 4.08 |
| 21 2 | 959.096 | 0.610 | 959.128 | 4.06 |
| 21 2 | 958.334 | 0.329 | 958.427 | 4.03 |
| 21 2 | 958.915 | 0.251 | 958.917 | 4.02 |
| 21 1 | 959.096 | 0.610 | 959.130 | 4.15 |
| 21 1 | 958.334 | 0.329 | 958.613 | 4.11 |
| 21 1 | 959.438 | 0.251 | 959.456 | 4.08 |
| 21 1 | 959.096 | 0.610 | 959.128 | 4.06 |
| 21 1 | 958.334 | 0.329 | 958.427 | 4.03 |
| 21 1 | 958.915 | 0.251 | 958.917 | 4.02 |
| 25 2 | 959.144 | 0.162 | 959.133 | 4.12 |
| 25 2 | 959.909 | 0.178 | 959.892 | 4.06 |
| 25 2 | 959.753 | 0.230 | 959.758 | 4.00 |
| 25 2 | 960.118 | 0.191 | 960.184 | 3.95 |
| 25 2 | 959.637 | 0.151 | 959.612 | 3.91 |
| 25 2 | 959.669 | 0.206 | 959.675 | 3.87 |
| 25 2 | 959.311 | 0.190 | 959.324 | 3.84 |
| 25 2 | 959.669 | 0.206 | 959.676 | 3.81 |
| 25 2 | 959.372 | 0.200 | 959.356 | 3.78 |
| 25 2 | 959.287 | 0.168 | 959.220 | 3.77 |
| 25 2 | 958.273 | 0.234 | 958.230 | 3.76 |
| 25 2 | 958.331 | 0.263 | 958.317 | 3.75 |
| 25 2 | 959.054 | 0.164 | 959.055 | 3.76 |
| 25 2 | 959.144 | 0.162 | 959.175 | 3.77 |
| 25 2 | 959.311 | 0.190 | 959.319 | 3.79 |
| 25 2 | 958.552 | 0.199 | 958.515 | 3.81 |
| 25 1 | 959.144 | 0.162 | 959.133 | 4.12 |
| 25 1 | 959.909 | 0.178 | 959.892 | 4.06 |
| 25 1 | 959.753 | 0.230 | 959.758 | 4.00 |
| 25 1 | 960.118 | 0.191 | 960.184 | 3.95 |
| 25 1 | 959.637 | 0.151 | 959.612 | 3.91 |
| 25 1 | 959.669 | 0.206 | 959.675 | 3.87 |
| 25 1 | 959.311 | 0.190 | 959.324 | 3.84 |
| 25 1 | 959.669 | 0.206 | 959.676 | 3.81 |
| 25 1 | 959.372 | 0.200 | 959.356 | 3.78 |
| 25 1 | 959.287 | 0.168 | 959.220 | 3.77 |
| 25 1 | 958.273 | 0.234 | 958.230 | 3.76 |
| 25 1 | 958.331 | 0.263 | 958.317 | 3.75 |
| 25 1 | 959.054 | 0.164 | 959.055 | 3.76 |
| 25 1 | 959.144 | 0.162 | 959.175 | 3.77 |
| 25 1 | 959.311 | 0.190 | 959.319 | 3.79 |
| 25 1 | 958.552 | 0.199 | 958.515 | 3.81 |

| 2000 - FEVEREIRO | | | | |
|---|---|---|---|---|
| D L | SDB | ER | SDC | HL |
| 03 2 | 959.365 | 0.234 | 959.332 | 3.12 |
| 03 2 | 958.981 | 0.218 | 958.967 | 3.16 |
| 03 2 | 960.242 | 0.160 | 960.319 | 3.26 |
| 03 2 | 958.909 | 0.176 | 958.906 | 3.33 |
| 03 2 | 959.444 | 0.207 | 959.413 | 3.48 |
| 03 2 | 958.668 | 0.203 | 958.647 | 3.57 |
| 03 2 | 959.109 | 0.244 | 959.047 | 3.77 |
| 03 2 | 958.893 | 0.313 | 958.897 | 3.89 |
| 03 2 | 958.626 | 0.260 | 958.633 | 4.02 |
| 03 1 | 959.365 | 0.234 | 959.332 | 3.12 |
| 03 1 | 958.981 | 0.218 | 958.967 | 3.16 |
| 03 1 | 960.242 | 0.160 | 960.319 | 3.26 |
| 03 1 | 958.909 | 0.176 | 958.906 | 3.33 |
| 03 1 | 959.444 | 0.207 | 959.413 | 3.48 |

| 2000 - FEVEREIRO | | | | | | 2000 - FEVEREIRO | | | | |
|---|---|---|---|---|---|---|---|---|---|---|
| D | L | SDB | ER | SDC | HL | D | L | SDB | ER | SDC | HL |
| 03 | 1 | 958.668 | 0.203 | 958.647 | 3.57 | 21 | 2 | 958.939 | 0.206 | 958.935 | 7.58 |
| 03 | 1 | 959.109 | 0.244 | 959.047 | 3.77 | 21 | 2 | 958.690 | 0.194 | 958.687 | 8.05 |
| 03 | 1 | 958.893 | 0.313 | 958.897 | 3.89 | 21 | 1 | 959.882 | 0.142 | 959.905 | 3.44 |
| 03 | 1 | 958.626 | 0.260 | 958.633 | 4.02 | 21 | 1 | 960.089 | 0.146 | 960.085 | 3.65 |
| 14 | 2 | 959.582 | 0.197 | 959.581 | 2.49 | 21 | 1 | 959.882 | 0.142 | 959.902 | 3.87 |
| 14 | 2 | 958.879 | 0.166 | 958.856 | 2.59 | 21 | 1 | 958.824 | 0.148 | 958.820 | 4.10 |
| 14 | 2 | 959.303 | 0.187 | 959.304 | 2.70 | 21 | 1 | 959.357 | 0.175 | 959.337 | 4.34 |
| 14 | 2 | 959.204 | 0.188 | 959.168 | 2.95 | 21 | 1 | 959.780 | 0.142 | 959.801 | 4.60 |
| 14 | 2 | 959.695 | 0.234 | 959.867 | 3.08 | 21 | 1 | 959.473 | 0.212 | 959.479 | 4.93 |
| 14 | 2 | 958.934 | 0.192 | 958.936 | 3.22 | 21 | 1 | 959.262 | 0.169 | 959.267 | 5.24 |
| 14 | 2 | 958.486 | 0.198 | 958.416 | 3.37 | 21 | 1 | 959.435 | 0.138 | 959.437 | 5.59 |
| 14 | 2 | 959.584 | 0.217 | 959.583 | 4.35 | 21 | 1 | 959.053 | 0.180 | 959.043 | 5.93 |
| 14 | 2 | 959.303 | 0.187 | 959.306 | 4.58 | 21 | 1 | 959.188 | 0.139 | 959.197 | 6.31 |
| 14 | 2 | 958.486 | 0.198 | 958.468 | 4.83 | 21 | 1 | 958.828 | 0.168 | 958.839 | 6.68 |
| 14 | 2 | 959.406 | 0.135 | 959.418 | 5.08 | 21 | 1 | 958.609 | 0.199 | 958.629 | 7.13 |
| 14 | 1 | 959.582 | 0.197 | 959.581 | 2.49 | 21 | 1 | 958.939 | 0.206 | 958.935 | 7.58 |
| 14 | 1 | 958.879 | 0.166 | 958.856 | 2.59 | 21 | 1 | 958.690 | 0.194 | 958.687 | 8.05 |
| 14 | 1 | 959.303 | 0.187 | 959.304 | 2.70 | 21 | 2 | 958.796 | 0.216 | 958.808 | 8.55 |
| 14 | 1 | 959.204 | 0.188 | 959.168 | 2.95 | 21 | 2 | 958.541 | 0.174 | 958.539 | 9.09 |
| 14 | 1 | 959.695 | 0.234 | 959.867 | 3.08 | 21 | 2 | 959.098 | 0.147 | 959.093 | 9.65 |
| 14 | 1 | 958.934 | 0.192 | 958.936 | 3.22 | 21 | 2 | 959.020 | 0.142 | 959.021 | 10.23 |
| 14 | 1 | 958.486 | 0.198 | 958.416 | 3.37 | 21 | 2 | 959.735 | 0.143 | 959.667 | 10.89 |
| 14 | 1 | 959.584 | 0.217 | 959.583 | 4.35 | 21 | 1 | 958.796 | 0.216 | 958.808 | 8.55 |
| 14 | 1 | 959.303 | 0.187 | 959.306 | 4.58 | 21 | 1 | 958.541 | 0.174 | 958.539 | 9.09 |
| 14 | 1 | 958.486 | 0.198 | 958.468 | 4.83 | 21 | 1 | 959.098 | 0.147 | 959.093 | 9.65 |
| 14 | 1 | 959.406 | 0.135 | 959.418 | 5.08 | 21 | 1 | 959.020 | 0.142 | 959.021 | 10.23 |
| 15 | 2 | 959.999 | 0.217 | 960.003 | 3.32 | 21 | 1 | 959.735 | 0.143 | 959.667 | 10.89 |
| 15 | 2 | 959.593 | 0.210 | 959.575 | 3.77 | 22 | 2 | 959.090 | 0.186 | 959.110 | 3.54 |
| 15 | 2 | 960.234 | 0.232 | 960.123 | 4.21 | 22 | 2 | 959.581 | 0.173 | 959.580 | 4.09 |
| 15 | 2 | 959.480 | 0.238 | 959.473 | 5.41 | 22 | 2 | 959.736 | 0.194 | 959.699 | 4.35 |
| 15 | 2 | 959.509 | 0.181 | 959.496 | 5.77 | 22 | 2 | 959.629 | 0.166 | 959.632 | 4.63 |
| 15 | 2 | 959.096 | 0.195 | 959.104 | 6.10 | 22 | 2 | 958.501 | 0.154 | 958.497 | 4.90 |
| 15 | 2 | 959.730 | 0.229 | 959.729 | 6.51 | 22 | 2 | 958.501 | 0.154 | 958.523 | 5.18 |
| 15 | 1 | 959.999 | 0.217 | 960.003 | 3.32 | 22 | 2 | 959.301 | 0.159 | 959.302 | 5.48 |
| 15 | 1 | 959.593 | 0.210 | 959.575 | 3.77 | 22 | 2 | 959.149 | 0.114 | 959.125 | 5.82 |
| 15 | 1 | 960.234 | 0.232 | 960.123 | 4.21 | 22 | 2 | 959.581 | 0.173 | 959.562 | 6.17 |
| 15 | 1 | 959.480 | 0.238 | 959.473 | 5.41 | 22 | 2 | 959.301 | 0.159 | 959.294 | 6.54 |
| 15 | 1 | 959.509 | 0.181 | 959.496 | 5.77 | 22 | 2 | 959.221 | 0.124 | 959.212 | 6.93 |
| 15 | 1 | 959.096 | 0.195 | 959.104 | 6.10 | 22 | 2 | 959.595 | 0.181 | 959.596 | 7.39 |
| 15 | 1 | 959.730 | 0.229 | 959.729 | 6.51 | 22 | 2 | 958.184 | 0.156 | 958.178 | 7.80 |
| 15 | 2 | 959.798 | 0.235 | 959.791 | 6.91 | 22 | 2 | 959.276 | 0.168 | 959.287 | 8.24 |
| 15 | 2 | 958.942 | 0.251 | 958.866 | 7.32 | 22 | 1 | 959.090 | 0.186 | 959.110 | 3.54 |
| 15 | 2 | 959.397 | 0.159 | 959.389 | 8.16 | 22 | 1 | 959.581 | 0.173 | 959.580 | 4.09 |
| 15 | 2 | 960.010 | 0.304 | 960.051 | 8.72 | 22 | 1 | 959.736 | 0.194 | 959.699 | 4.35 |
| 15 | 1 | 959.798 | 0.235 | 959.791 | 6.91 | 22 | 1 | 959.629 | 0.166 | 959.632 | 4.63 |
| 15 | 1 | 958.942 | 0.251 | 958.866 | 7.32 | 22 | 1 | 958.501 | 0.154 | 958.497 | 4.90 |
| 15 | 1 | 959.397 | 0.159 | 959.389 | 8.16 | 22 | 1 | 958.501 | 0.154 | 958.523 | 5.18 |
| 15 | 1 | 960.010 | 0.304 | 960.051 | 8.72 | 22 | 1 | 959.301 | 0.159 | 959.302 | 5.48 |
| 16 | 2 | 959.294 | 0.241 | 959.288 | 3.07 | 22 | 1 | 959.149 | 0.114 | 959.125 | 5.82 |
| 16 | 2 | 959.695 | 0.219 | 959.634 | 3.22 | 22 | 1 | 959.581 | 0.173 | 959.562 | 6.17 |
| 16 | 2 | 959.480 | 0.157 | 959.472 | 3.39 | 22 | 1 | 959.301 | 0.159 | 959.294 | 6.54 |
| 16 | 2 | 959.695 | 0.219 | 959.632 | 3.55 | 22 | 1 | 959.221 | 0.124 | 959.212 | 6.93 |
| 16 | 2 | 959.509 | 0.200 | 959.526 | 4.02 | 22 | 1 | 959.595 | 0.181 | 959.596 | 7.39 |
| 16 | 1 | 959.294 | 0.241 | 959.288 | 3.07 | 22 | 1 | 958.184 | 0.156 | 958.178 | 7.80 |
| 16 | 1 | 959.695 | 0.219 | 959.634 | 3.22 | 22 | 1 | 959.276 | 0.168 | 959.287 | 8.24 |
| 16 | 1 | 959.480 | 0.157 | 959.472 | 3.39 | 22 | 2 | 958.350 | 0.134 | 958.351 | 8.70 |
| 16 | 1 | 959.695 | 0.219 | 959.632 | 3.55 | 22 | 2 | 959.188 | 0.144 | 959.187 | 9.20 |
| 16 | 1 | 959.509 | 0.200 | 959.526 | 4.02 | 22 | 2 | 958.475 | 0.160 | 958.466 | 9.75 |
| 21 | 2 | 958.882 | 0.142 | 959.905 | 3.44 | 22 | 2 | 958.845 | 0.165 | 958.836 | 10.36 |
| 21 | 2 | 960.089 | 0.146 | 960.085 | 3.65 | 22 | 1 | 958.350 | 0.134 | 958.351 | 8.70 |
| 21 | 2 | 959.882 | 0.142 | 959.902 | 3.87 | 22 | 1 | 959.188 | 0.144 | 959.187 | 9.20 |
| 21 | 2 | 958.824 | 0.148 | 958.820 | 4.10 | 22 | 1 | 958.475 | 0.160 | 958.466 | 9.75 |
| 21 | 2 | 959.357 | 0.175 | 959.337 | 4.34 | 22 | 1 | 958.845 | 0.165 | 958.836 | 10.36 |
| 21 | 2 | 959.780 | 0.142 | 959.801 | 4.60 | 23 | 2 | 958.713 | 0.159 | 958.723 | 3.03 |
| 21 | 2 | 959.473 | 0.212 | 959.479 | 4.93 | 23 | 2 | 960.206 | 0.175 | 960.225 | 3.23 |
| 21 | 2 | 959.262 | 0.169 | 959.267 | 5.24 | 23 | 2 | 958.919 | 0.190 | 958.916 | 3.43 |
| 21 | 2 | 959.435 | 0.138 | 959.437 | 5.59 | 23 | 2 | 959.395 | 0.170 | 959.383 | 3.64 |
| 21 | 2 | 959.053 | 0.180 | 959.043 | 5.93 | 23 | 2 | 959.000 | 0.185 | 959.007 | 3.88 |
| 21 | 2 | 959.188 | 0.139 | 959.197 | 6.31 | 23 | 2 | 959.628 | 0.154 | 959.641 | 4.12 |
| 21 | 2 | 958.828 | 0.168 | 958.839 | 6.68 | 23 | 2 | 959.118 | 0.140 | 959.105 | 4.37 |
| 21 | 2 | 958.609 | 0.199 | 958.629 | 7.13 | 23 | 2 | 958.643 | 0.141 | 958.656 | 4.64 |

| 2000 - FEVEREIRO | | | | | | 2000 - FEVEREIRO | | | | |
|---|---|---|---|---|---|---|---|---|---|---|
| D | L | SDB | ER | SDC | HL | D | L | SDB | ER | SDC | HL |
| 23 | 2 | 959.118 | 0.140 | 959.132 | 4.92 | 28 | 2 | 959.189 | 0.171 | 959.231 | 8.90 |
| 23 | 2 | 958.919 | 0.190 | 958.901 | 5.23 | 28 | 2 | 958.748 | 0.163 | 958.738 | 9.42 |
| 23 | 2 | 959.000 | 0.185 | 959.008 | 5.56 | 28 | 2 | 959.189 | 0.171 | 959.197 | 9.98 |
| 23 | 2 | 959.687 | 0.177 | 959.717 | 5.91 | 28 | 1 | 959.011 | 0.137 | 959.007 | 6.69 |
| 23 | 2 | 959.243 | 0.115 | 959.272 | 6.31 | 28 | 1 | 959.296 | 0.186 | 959.268 | 7.08 |
| 23 | 2 | 959.308 | 0.203 | 959.292 | 6.74 | 28 | 1 | 960.091 | 0.177 | 960.115 | 7.47 |
| 23 | 2 | 959.018 | 0.143 | 959.016 | 7.15 | 28 | 1 | 959.946 | 0.168 | 959.909 | 7.90 |
| 23 | 2 | 958.670 | 0.131 | 958.657 | 7.59 | 28 | 1 | 958.748 | 0.163 | 958.763 | 8.36 |
| 23 | 2 | 958.794 | 0.130 | 958.806 | 8.04 | 28 | 1 | 959.189 | 0.171 | 959.231 | 8.90 |
| 23 | 2 | 958.872 | 0.165 | 958.876 | 8.52 | 28 | 1 | 958.748 | 0.163 | 958.738 | 9.42 |
| 23 | 1 | 958.713 | 0.159 | 958.723 | 3.03 | 28 | 1 | 959.189 | 0.171 | 959.197 | 9.98 |
| 23 | 1 | 960.206 | 0.175 | 960.225 | 3.23 | 28 | 2 | 959.152 | 0.140 | 959.148 | 10.55 |
| 23 | 1 | 958.919 | 0.190 | 958.916 | 3.43 | 28 | 2 | 958.353 | 0.197 | 958.362 | 11.17 |
| 23 | 1 | 959.395 | 0.170 | 959.383 | 3.64 | 28 | 2 | 958.392 | 0.155 | 958.375 | 11.82 |
| 23 | 1 | 959.000 | 0.185 | 959.007 | 3.88 | 28 | 2 | 959.296 | 0.186 | 959.341 | 12.51 |
| 23 | 1 | 959.628 | 0.154 | 959.641 | 4.12 | 28 | 2 | 959.960 | 0.203 | 960.014 | 13.24 |
| 23 | 1 | 959.118 | 0.140 | 959.105 | 4.37 | 28 | 2 | 958.597 | 0.154 | 958.592 | 14.01 |
| 23 | 1 | 958.643 | 0.141 | 958.656 | 4.64 | 28 | 2 | 958.557 | 0.224 | 958.536 | 14.85 |
| 23 | 1 | 959.118 | 0.140 | 959.132 | 4.92 | 28 | 2 | 959.429 | 0.203 | 959.478 | 15.74 |
| 23 | 1 | 958.919 | 0.190 | 958.901 | 5.23 | 28 | 1 | 959.152 | 0.140 | 959.148 | 10.55 |
| 23 | 1 | 959.000 | 0.185 | 959.008 | 5.56 | 28 | 1 | 958.353 | 0.197 | 958.362 | 11.17 |
| 23 | 1 | 959.687 | 0.177 | 959.717 | 5.91 | 28 | 1 | 958.392 | 0.155 | 958.375 | 11.82 |
| 23 | 1 | 959.243 | 0.115 | 959.272 | 6.31 | 28 | 1 | 959.296 | 0.186 | 959.341 | 12.51 |
| 23 | 1 | 959.308 | 0.203 | 959.292 | 6.74 | 28 | 1 | 959.960 | 0.203 | 960.014 | 13.24 |
| 23 | 1 | 959.018 | 0.143 | 959.016 | 7.15 | 28 | 1 | 958.597 | 0.154 | 958.592 | 14.01 |
| 23 | 1 | 958.670 | 0.131 | 958.657 | 7.59 | 28 | 1 | 958.557 | 0.224 | 958.536 | 14.85 |
| 23 | 1 | 958.794 | 0.130 | 958.806 | 8.04 | 28 | 1 | 959.429 | 0.203 | 959.478 | 15.74 |
| 23 | 1 | 958.872 | 0.165 | 958.876 | 8.52 | 29 | 2 | 958.841 | 0.191 | 958.794 | 6.57 |
| 23 | 2 | 959.469 | 0.148 | 959.476 | 9.15 | 29 | 2 | 958.486 | 0.415 | 958.472 | 6.94 |
| 23 | 2 | 958.710 | 0.145 | 958.708 | 9.70 | 29 | 2 | 959.143 | 0.192 | 959.250 | 7.74 |
| 23 | 1 | 959.469 | 0.148 | 959.476 | 9.15 | 29 | 2 | 958.297 | 0.201 | 958.324 | 8.19 |
| 23 | 1 | 958.710 | 0.145 | 958.708 | 9.70 | 29 | 2 | 958.297 | 0.201 | 958.327 | 8.64 |
| 24 | 2 | 959.094 | 0.166 | 959.095 | 3.30 | 29 | 2 | 958.225 | 0.151 | 958.256 | 9.59 |
| 24 | 2 | 959.348 | 0.253 | 959.349 | 3.49 | 29 | 1 | 958.841 | 0.191 | 958.794 | 6.57 |
| 24 | 2 | 958.581 | 0.218 | 958.574 | 3.69 | 29 | 1 | 958.486 | 0.415 | 958.472 | 6.94 |
| 24 | 2 | 958.209 | 0.242 | 958.192 | 3.90 | 29 | 1 | 959.143 | 0.192 | 959.250 | 7.74 |
| 24 | 2 | 959.660 | 0.192 | 959.794 | 4.12 | 29 | 1 | 958.297 | 0.201 | 958.324 | 8.19 |
| 24 | 2 | 959.660 | 0.192 | 959.570 | 4.35 | 29 | 1 | 958.297 | 0.201 | 958.327 | 8.64 |
| 24 | 2 | 958.923 | 0.192 | 958.910 | 4.61 | 29 | 1 | 958.225 | 0.151 | 958.256 | 9.59 |
| 24 | 2 | 959.345 | 0.203 | 959.343 | 4.87 | 29 | 2 | 958.578 | 0.146 | 958.562 | 10.75 |
| 24 | 2 | 958.762 | 0.225 | 958.762 | 5.13 | 29 | 2 | 958.185 | 0.265 | 958.188 | 11.31 |
| 24 | 2 | 958.581 | 0.218 | 958.565 | 5.71 | 29 | 2 | 959.064 | 0.214 | 959.045 | 11.91 |
| 24 | 2 | 959.250 | 0.171 | 959.233 | 6.03 | 29 | 2 | 959.143 | 0.192 | 959.229 | 12.66 |
| 24 | 2 | 959.250 | 0.171 | 959.264 | 6.35 | 29 | 2 | 960.418 | 0.244 | 960.599 | 13.38 |
| 24 | 2 | 959.471 | 0.204 | 959.512 | 6.69 | 29 | 2 | 959.462 | 0.221 | 959.322 | 14.10 |
| 24 | 2 | 958.545 | 0.204 | 958.493 | 7.07 | 29 | 1 | 958.578 | 0.146 | 958.562 | 10.75 |
| 24 | 2 | 958.998 | 0.181 | 959.012 | 7.48 | 29 | 1 | 958.185 | 0.265 | 958.188 | 11.31 |
| 24 | 2 | 958.955 | 0.216 | 958.956 | 7.87 | 29 | 1 | 959.064 | 0.214 | 959.045 | 11.91 |
| 24 | 2 | 958.292 | 0.204 | 958.284 | 8.31 | 29 | 1 | 959.143 | 0.192 | 959.229 | 12.66 |
| 24 | 1 | 959.094 | 0.166 | 959.095 | 3.30 | 29 | 1 | 960.418 | 0.244 | 960.599 | 13.38 |
| 24 | 1 | 959.348 | 0.253 | 959.349 | 3.49 | 29 | 1 | 959.462 | 0.221 | 959.322 | 14.10 |
| 24 | 1 | 958.581 | 0.218 | 958.574 | 3.69 | | | | | | |
| 24 | 1 | 958.209 | 0.242 | 958.192 | 3.90 | | | | | | |
| 24 | 1 | 959.660 | 0.192 | 959.794 | 4.12 | | | 2000 - MARCO | | | |
| 24 | 1 | 959.660 | 0.192 | 959.570 | 4.35 | D | L | SDB | ER | SDC | HL |
| 24 | 1 | 958.923 | 0.192 | 958.910 | 4.61 | 01 | 2 | 959.591 | 0.167 | 959.573 | 5.51 |
| 24 | 1 | 959.345 | 0.203 | 959.343 | 4.87 | 01 | 2 | 959.900 | 0.121 | 959.905 | 5.88 |
| 24 | 1 | 958.762 | 0.225 | 958.762 | 5.13 | 01 | 2 | 959.404 | 0.170 | 959.414 | 6.27 |
| 24 | 1 | 958.581 | 0.218 | 958.565 | 5.71 | 01 | 2 | 959.591 | 0.167 | 959.602 | 6.66 |
| 24 | 1 | 959.250 | 0.171 | 959.233 | 6.03 | 01 | 2 | 958.701 | 0.187 | 958.714 | 7.05 |
| 24 | 1 | 959.250 | 0.171 | 959.264 | 6.35 | 01 | 2 | 959.444 | 0.138 | 959.460 | 7.47 |
| 24 | 1 | 959.471 | 0.204 | 959.512 | 6.69 | 01 | 2 | 960.050 | 0.123 | 960.107 | 7.89 |
| 24 | 1 | 958.545 | 0.204 | 958.493 | 7.07 | 01 | 2 | 959.512 | 0.173 | 959.493 | 8.34 |
| 24 | 1 | 958.998 | 0.181 | 959.012 | 7.48 | 01 | 2 | 958.701 | 0.187 | 958.722 | 9.34 |
| 24 | 1 | 958.955 | 0.216 | 958.956 | 7.87 | 01 | 2 | 959.222 | 0.168 | 959.233 | 9.95 |
| 24 | 1 | 958.292 | 0.204 | 958.284 | 8.31 | 01 | 2 | 959.166 | 0.191 | 959.185 | 10.54 |
| 28 | 2 | 959.011 | 0.137 | 959.007 | 6.69 | 01 | 1 | 959.591 | 0.167 | 959.573 | 5.51 |
| 28 | 2 | 959.296 | 0.186 | 959.268 | 7.08 | 01 | 1 | 959.900 | 0.121 | 959.905 | 5.88 |
| 28 | 2 | 960.091 | 0.177 | 960.115 | 7.47 | 01 | 1 | 959.404 | 0.170 | 959.414 | 6.27 |
| 28 | 2 | 959.946 | 0.168 | 959.909 | 7.90 | 01 | 1 | 959.591 | 0.167 | 959.602 | 6.66 |
| 28 | 2 | 958.748 | 0.163 | 958.763 | 8.36 | 01 | 1 | 958.701 | 0.187 | 958.714 | 7.05 |

| 2000 - MARCO | | | | | | 2000 - MARCO | | | | |
|---|---|---|---|---|---|---|---|---|---|---|
| D | L | SDB | ER | SDC | HL | D | L | SDB | ER | SDC | HL |
| 01 | 1 | 959.444 | 0.138 | 959.460 | 7.47 | 03 | 1 | 957.639 | 0.559 | 957.620 | 4.81 |
| 01 | 1 | 960.050 | 0.123 | 960.107 | 7.89 | 03 | 1 | 958.491 | 0.215 | 958.466 | 5.10 |
| 01 | 1 | 959.512 | 0.173 | 959.493 | 8.34 | 03 | 1 | 958.851 | 0.357 | 958.844 | 5.39 |
| 01 | 1 | 958.701 | 0.187 | 958.722 | 9.34 | 03 | 1 | 960.019 | 0.187 | 959.983 | 5.69 |
| 01 | 1 | 959.222 | 0.168 | 959.233 | 9.95 | 03 | 1 | 959.145 | 0.196 | 959.170 | 6.36 |
| 01 | 1 | 959.166 | 0.191 | 959.185 | 10.54 | 03 | 1 | 959.358 | 0.206 | 959.357 | 6.71 |
| 01 | 2 | 958.599 | 0.138 | 958.621 | 11.13 | 03 | 1 | 958.491 | 0.215 | 958.454 | 7.08 |
| 01 | 2 | 959.280 | 0.134 | 959.289 | 11.77 | 03 | 1 | 958.491 | 0.215 | 958.485 | 7.47 |
| 01 | 2 | 958.663 | 0.127 | 958.657 | 12.46 | 03 | 1 | 957.260 | 0.354 | 957.263 | 7.88 |
| 01 | 2 | 959.066 | 0.130 | 959.069 | 13.17 | 03 | 1 | 959.145 | 0.196 | 959.154 | 8.31 |
| 01 | 2 | 959.835 | 0.110 | 959.782 | 13.95 | 03 | 1 | 959.724 | 0.209 | 959.728 | 8.76 |
| 01 | 2 | 958.599 | 0.138 | 958.597 | 14.89 | 03 | 1 | 959.079 | 0.241 | 959.103 | 9.23 |
| 01 | 1 | 958.599 | 0.138 | 958.621 | 11.13 | 03 | 1 | 959.047 | 0.219 | 959.029 | 9.73 |
| 01 | 1 | 959.280 | 0.134 | 959.289 | 11.77 | 03 | 1 | 959.515 | 0.230 | 959.520 | 10.25 |
| 01 | 1 | 958.663 | 0.127 | 958.657 | 12.46 | 03 | 1 | 958.706 | 0.184 | 958.706 | 10.81 |
| 01 | 1 | 959.066 | 0.130 | 959.069 | 13.17 | 09 | 2 | 959.583 | 0.200 | 959.555 | 8.91 |
| 01 | 1 | 959.835 | 0.110 | 959.782 | 13.95 | 09 | 2 | 959.199 | 0.251 | 959.199 | 9.45 |
| 01 | 1 | 958.599 | 0.138 | 958.597 | 14.89 | 09 | 2 | 959.023 | 0.210 | 959.045 | 10.05 |
| 02 | 2 | 959.458 | 0.208 | 959.475 | 4.72 | 09 | 2 | 959.241 | 0.238 | 959.248 | 10.67 |
| 02 | 2 | 958.074 | 0.184 | 958.063 | 5.06 | 09 | 2 | 959.747 | 0.209 | 959.771 | 11.36 |
| 02 | 2 | 960.461 | 0.361 | 960.417 | 5.35 | 09 | 2 | 960.398 | 0.293 | 960.450 | 12.00 |
| 02 | 2 | 959.825 | 0.192 | 959.922 | 5.67 | 09 | 2 | 959.264 | 0.341 | 959.258 | 12.68 |
| 02 | 2 | 959.254 | 0.213 | 959.264 | 5.99 | 09 | 1 | 959.583 | 0.200 | 959.555 | 8.91 |
| 02 | 2 | 959.215 | 0.195 | 959.210 | 6.34 | 09 | 1 | 959.199 | 0.251 | 959.199 | 9.45 |
| 02 | 2 | 959.449 | 0.208 | 959.443 | 6.69 | 09 | 1 | 959.023 | 0.210 | 959.045 | 10.05 |
| 02 | 2 | 959.254 | 0.213 | 959.272 | 7.07 | 09 | 1 | 959.241 | 0.238 | 959.248 | 10.67 |
| 02 | 2 | 957.735 | 0.183 | 957.820 | 7.46 | 09 | 1 | 959.747 | 0.209 | 959.771 | 11.36 |
| 02 | 2 | 958.118 | 0.204 | 958.157 | 7.87 | 09 | 1 | 960.398 | 0.293 | 960.450 | 12.00 |
| 02 | 2 | 959.569 | 0.222 | 959.545 | 8.30 | 09 | 1 | 959.264 | 0.341 | 959.258 | 12.68 |
| 02 | 2 | 958.900 | 0.389 | 958.900 | 8.75 | 09 | 2 | 958.825 | 0.630 | 958.823 | 13.44 |
| 02 | 2 | 959.101 | 0.130 | 959.108 | 9.22 | 09 | 2 | 959.488 | 0.265 | 959.485 | 14.18 |
| 02 | 2 | 959.569 | 0.222 | 959.555 | 9.70 | 09 | 2 | 959.505 | 0.293 | 959.502 | 14.94 |
| 02 | 2 | 958.769 | 0.461 | 958.753 | 10.26 | 09 | 2 | 959.894 | 0.351 | 959.914 | 15.78 |
| 02 | 2 | 959.569 | 0.222 | 959.561 | 10.80 | 09 | 2 | 959.505 | 0.293 | 959.515 | 16.66 |
| 02 | 1 | 959.458 | 0.208 | 959.475 | 4.72 | 09 | 2 | 959.488 | 0.265 | 959.486 | 17.63 |
| 02 | 1 | 958.074 | 0.184 | 958.063 | 5.06 | 09 | 2 | 958.988 | 0.328 | 959.002 | 18.66 |
| 02 | 1 | 960.461 | 0.361 | 960.417 | 5.35 | 09 | 2 | 959.894 | 0.351 | 959.913 | 19.76 |
| 02 | 1 | 959.825 | 0.192 | 959.922 | 5.67 | 09 | 1 | 958.825 | 0.630 | 958.823 | 13.44 |
| 02 | 1 | 959.254 | 0.213 | 959.264 | 5.99 | 09 | 1 | 959.488 | 0.265 | 959.485 | 14.18 |
| 02 | 1 | 959.215 | 0.195 | 959.210 | 6.34 | 09 | 1 | 959.505 | 0.293 | 959.502 | 14.94 |
| 02 | 1 | 959.449 | 0.208 | 959.443 | 6.69 | 09 | 1 | 959.894 | 0.351 | 959.914 | 15.78 |
| 02 | 1 | 959.254 | 0.213 | 959.272 | 7.07 | 09 | 1 | 959.505 | 0.293 | 959.515 | 16.66 |
| 02 | 1 | 957.735 | 0.183 | 957.820 | 7.46 | 09 | 1 | 959.488 | 0.265 | 959.486 | 17.63 |
| 02 | 1 | 958.118 | 0.204 | 958.157 | 7.87 | 09 | 1 | 958.988 | 0.328 | 959.002 | 18.66 |
| 02 | 1 | 959.569 | 0.222 | 959.545 | 8.30 | 09 | 1 | 959.894 | 0.351 | 959.913 | 19.76 |
| 02 | 1 | 958.900 | 0.389 | 958.900 | 8.75 | 14 | 2 | 957.677 | 0.294 | 957.717 | 16.84 |
| 02 | 1 | 959.101 | 0.130 | 959.108 | 9.22 | 14 | 2 | 958.948 | 0.255 | 958.939 | 24.14 |
| 02 | 1 | 959.569 | 0.222 | 959.555 | 9.70 | 14 | 2 | 957.382 | 0.705 | 957.371 | 25.69 |
| 02 | 1 | 958.769 | 0.461 | 958.753 | 10.26 | 14 | 2 | 957.868 | 0.336 | 957.904 | 27.60 |
| 02 | 1 | 959.569 | 0.222 | 959.561 | 10.80 | 14 | 1 | 957.677 | 0.294 | 957.717 | 16.84 |
| 02 | 2 | 959.101 | 0.130 | 959.067 | 11.37 | 14 | 1 | 958.948 | 0.255 | 958.939 | 24.14 |
| 02 | 2 | 958.937 | 0.180 | 958.942 | 11.96 | 14 | 1 | 957.382 | 0.705 | 957.371 | 25.69 |
| 02 | 2 | 959.253 | 0.136 | 959.240 | 12.62 | 14 | 1 | 957.868 | 0.336 | 957.904 | 27.60 |
| 02 | 1 | 959.101 | 0.130 | 959.067 | 11.37 | 17 | 2 | 958.521 | 0.151 | 958.519 | 12.09 |
| 02 | 1 | 958.937 | 0.180 | 958.942 | 11.96 | 17 | 2 | 959.951 | 0.213 | 959.910 | 12.88 |
| 02 | 1 | 959.253 | 0.136 | 959.240 | 12.62 | 17 | 2 | 959.951 | 0.213 | 960.004 | 13.80 |
| 03 | 2 | 957.639 | 0.559 | 957.620 | 4.81 | 17 | 2 | 959.839 | 0.185 | 959.763 | 14.53 |
| 03 | 2 | 958.491 | 0.215 | 958.466 | 5.10 | 17 | 2 | 959.533 | 0.176 | 959.534 | 15.31 |
| 03 | 2 | 958.851 | 0.357 | 958.844 | 5.39 | 17 | 1 | 958.521 | 0.151 | 958.519 | 12.09 |
| 03 | 2 | 960.019 | 0.187 | 959.983 | 5.69 | 17 | 1 | 959.951 | 0.213 | 959.910 | 12.88 |
| 03 | 2 | 959.145 | 0.196 | 959.170 | 6.36 | 17 | 1 | 959.951 | 0.213 | 960.004 | 13.80 |
| 03 | 2 | 959.358 | 0.206 | 959.357 | 6.71 | 17 | 1 | 959.839 | 0.185 | 959.763 | 14.53 |
| 03 | 2 | 958.491 | 0.215 | 958.454 | 7.08 | 17 | 1 | 959.533 | 0.176 | 959.534 | 15.31 |
| 03 | 2 | 958.491 | 0.215 | 958.485 | 7.47 | 17 | 2 | 958.453 | 0.249 | 958.450 | 16.27 |
| 03 | 2 | 957.260 | 0.354 | 957.263 | 7.88 | 17 | 2 | 959.035 | 0.263 | 959.064 | 17.39 |
| 03 | 2 | 959.145 | 0.196 | 959.154 | 8.31 | 17 | 2 | 958.741 | 0.218 | 958.770 | 21.17 |
| 03 | 2 | 959.724 | 0.209 | 959.728 | 8.76 | 17 | 2 | 959.533 | 0.176 | 959.562 | 22.29 |
| 03 | 2 | 959.079 | 0.241 | 959.103 | 9.23 | 17 | 2 | 959.153 | 0.164 | 959.126 | 23.54 |
| 03 | 2 | 959.047 | 0.219 | 959.029 | 9.73 | 17 | 2 | 958.851 | 0.386 | 958.820 | 24.83 |
| 03 | 2 | 959.515 | 0.230 | 959.520 | 10.25 | 17 | 2 | 957.862 | 0.217 | 957.944 | 26.28 |
| 03 | 2 | 958.706 | 0.184 | 958.706 | 10.81 | 17 | 1 | 958.453 | 0.249 | 958.450 | 16.27 |

| 2000 - MARCO | | | | | | | 2000 - MARCO | | | | |
|---|---|---|---|---|---|---|---|---|---|---|---|
| D | L | SDB | ER | SDC | HL | | D | L | SDB | ER | SDC | HL |
| 17 | 1 | 959.035 | 0.263 | 959.064 | 17.39 | | 29 | 2 | 959.444 | 0.179 | 959.429 | 28.91 |
| 17 | 1 | 958.741 | 0.218 | 958.770 | 21.17 | | 29 | 2 | 959.340 | 0.155 | 959.344 | 30.49 |
| 17 | 1 | 959.533 | 0.176 | 959.562 | 22.29 | | 29 | 2 | 959.526 | 0.192 | 959.536 | 32.15 |
| 17 | 1 | 959.153 | 0.164 | 959.126 | 23.54 | | 29 | 1 | 959.805 | 0.167 | 959.793 | 20.90 |
| 17 | 1 | 958.851 | 0.386 | 958.820 | 24.83 | | 29 | 1 | 959.053 | 0.170 | 959.075 | 21.91 |
| 17 | 1 | 957.862 | 0.217 | 957.944 | 26.28 | | 29 | 1 | 959.487 | 0.209 | 959.483 | 22.95 |
| 22 | 2 | 959.329 | 0.159 | 959.338 | 11.93 | | 29 | 1 | 959.449 | 0.169 | 959.456 | 24.14 |
| 22 | 2 | 959.471 | 0.172 | 959.471 | 15.87 | | 29 | 1 | 959.624 | 0.154 | 959.604 | 27.30 |
| 22 | 2 | 959.307 | 0.192 | 959.307 | 16.72 | | 29 | 1 | 959.444 | 0.179 | 959.429 | 28.91 |
| 22 | 2 | 958.875 | 0.177 | 958.864 | 17.59 | | 29 | 1 | 959.340 | 0.155 | 959.344 | 30.49 |
| 22 | 1 | 959.329 | 0.159 | 959.338 | 11.93 | | 29 | 1 | 959.526 | 0.192 | 959.536 | 32.15 |
| 22 | 1 | 959.471 | 0.172 | 959.471 | 15.87 | | | | | | | |
| 22 | 1 | 959.307 | 0.192 | 959.307 | 16.72 | | | | 2000 - ABRIL | | | |
| 22 | 1 | 958.875 | 0.177 | 958.864 | 17.59 | | D | L | SDB | ER | SDC | HL |
| 22 | 2 | 958.985 | 0.183 | 958.980 | 18.49 | | 05 | 2 | 959.037 | 0.165 | 959.028 | 14.80 |
| 22 | 2 | 957.854 | 0.198 | 957.890 | 19.48 | | 05 | 2 | 959.392 | 0.190 | 959.399 | 15.54 |
| 22 | 2 | 959.498 | 0.156 | 959.507 | 20.61 | | 05 | 2 | 959.549 | 0.208 | 959.550 | 16.24 |
| 22 | 2 | 960.170 | 0.187 | 960.152 | 21.77 | | 05 | 2 | 959.755 | 0.230 | 959.717 | 16.97 |
| 22 | 2 | 959.602 | 0.169 | 959.592 | 23.01 | | 05 | 2 | 959.329 | 0.178 | 959.328 | 17.75 |
| 22 | 2 | 959.810 | 0.205 | 959.809 | 24.27 | | 05 | 2 | 958.136 | 0.195 | 958.181 | 18.54 |
| 22 | 2 | 959.074 | 0.169 | 959.051 | 25.58 | | 05 | 2 | 959.618 | 0.148 | 959.611 | 19.38 |
| 22 | 1 | 958.985 | 0.183 | 958.980 | 18.49 | | 05 | 2 | 959.146 | 0.233 | 959.141 | 20.26 |
| 22 | 1 | 957.854 | 0.198 | 957.890 | 19.48 | | 05 | 2 | 959.200 | 0.114 | 959.208 | 21.30 |
| 22 | 1 | 959.498 | 0.156 | 959.507 | 20.61 | | 05 | 2 | 959.255 | 0.184 | 959.259 | 22.31 |
| 22 | 1 | 960.170 | 0.187 | 960.152 | 21.77 | | 05 | 1 | 959.037 | 0.165 | 959.028 | 14.80 |
| 22 | 1 | 959.602 | 0.169 | 959.592 | 23.01 | | 05 | 1 | 959.392 | 0.190 | 959.399 | 15.54 |
| 22 | 1 | 959.810 | 0.205 | 959.809 | 24.27 | | 05 | 1 | 959.549 | 0.208 | 959.550 | 16.24 |
| 22 | 1 | 959.074 | 0.169 | 959.051 | 25.58 | | 05 | 1 | 959.755 | 0.230 | 959.717 | 16.97 |
| 27 | 2 | 959.948 | 0.172 | 959.951 | 8.17 | | 05 | 1 | 959.329 | 0.178 | 959.328 | 17.75 |
| 27 | 2 | 959.884 | 0.169 | 959.883 | 8.71 | | 05 | 1 | 958.136 | 0.195 | 958.181 | 18.54 |
| 27 | 2 | 959.267 | 0.155 | 959.254 | 9.25 | | 05 | 1 | 959.618 | 0.148 | 959.611 | 19.38 |
| 27 | 2 | 958.895 | 0.265 | 958.886 | 9.73 | | 05 | 1 | 959.146 | 0.233 | 959.141 | 20.26 |
| 27 | 2 | 959.589 | 0.174 | 959.566 | 10.30 | | 05 | 1 | 959.200 | 0.114 | 959.208 | 21.30 |
| 27 | 2 | 959.948 | 0.172 | 959.941 | 11.17 | | 05 | 1 | 959.255 | 0.184 | 959.259 | 22.31 |
| 27 | 2 | 958.370 | 0.182 | 958.378 | 11.76 | | 05 | 2 | 959.392 | 0.190 | 959.396 | 23.35 |
| 27 | 2 | 959.287 | 0.168 | 959.318 | 12.38 | | 05 | 2 | 959.623 | 0.147 | 959.677 | 24.44 |
| 27 | 2 | 960.770 | 0.164 | 960.739 | 13.10 | | 05 | 2 | 959.197 | 0.173 | 959.197 | 25.62 |
| 27 | 2 | 958.948 | 0.203 | 958.955 | 13.78 | | 05 | 2 | 959.168 | 0.175 | 959.167 | 26.88 |
| 27 | 2 | 957.899 | 0.391 | 957.885 | 14.58 | | 05 | 2 | 959.597 | 0.124 | 959.586 | 28.23 |
| 27 | 2 | 958.923 | 0.191 | 958.921 | 15.32 | | 05 | 2 | 959.357 | 0.197 | 959.358 | 29.68 |
| 27 | 2 | 958.597 | 0.197 | 958.586 | 16.21 | | 05 | 1 | 959.392 | 0.190 | 959.396 | 23.35 |
| 27 | 2 | 958.834 | 0.161 | 958.811 | 16.96 | | 05 | 1 | 959.623 | 0.147 | 959.677 | 24.44 |
| 27 | 2 | 958.597 | 0.197 | 958.604 | 17.80 | | 05 | 1 | 959.197 | 0.173 | 959.197 | 25.62 |
| 27 | 2 | 959.155 | 0.166 | 959.161 | 18.62 | | 05 | 1 | 959.168 | 0.175 | 959.167 | 26.88 |
| 27 | 1 | 959.948 | 0.172 | 959.951 | 8.17 | | 05 | 1 | 959.597 | 0.124 | 959.586 | 28.23 |
| 27 | 1 | 959.884 | 0.169 | 959.883 | 8.71 | | 05 | 1 | 959.357 | 0.197 | 959.358 | 29.68 |
| 27 | 1 | 959.267 | 0.155 | 959.254 | 9.25 | | 10 | 2 | 959.132 | 0.150 | 959.134 | 18.89 |
| 27 | 1 | 958.895 | 0.265 | 958.886 | 9.73 | | 10 | 2 | 959.507 | 0.201 | 959.507 | 19.70 |
| 27 | 1 | 959.589 | 0.174 | 959.566 | 10.30 | | 10 | 2 | 959.334 | 0.177 | 959.342 | 20.57 |
| 27 | 1 | 959.948 | 0.172 | 959.941 | 11.17 | | 10 | 2 | 960.194 | 0.168 | 960.213 | 21.59 |
| 27 | 1 | 958.370 | 0.182 | 958.378 | 11.76 | | 10 | 2 | 959.228 | 0.160 | 959.214 | 22.60 |
| 27 | 1 | 959.287 | 0.168 | 959.318 | 12.38 | | 10 | 2 | 959.132 | 0.150 | 959.140 | 23.61 |
| 27 | 1 | 960.770 | 0.164 | 960.739 | 13.10 | | 10 | 2 | 959.132 | 0.150 | 959.137 | 24.70 |
| 27 | 1 | 958.948 | 0.203 | 958.955 | 13.78 | | 10 | 1 | 959.132 | 0.150 | 959.134 | 18.89 |
| 27 | 1 | 957.899 | 0.391 | 957.885 | 14.58 | | 10 | 1 | 959.507 | 0.201 | 959.507 | 19.70 |
| 27 | 1 | 958.923 | 0.191 | 958.921 | 15.32 | | 10 | 1 | 959.334 | 0.177 | 959.342 | 20.57 |
| 27 | 1 | 958.597 | 0.197 | 958.586 | 16.21 | | 10 | 1 | 960.194 | 0.168 | 960.213 | 21.59 |
| 27 | 1 | 958.834 | 0.161 | 958.811 | 16.96 | | 10 | 1 | 959.228 | 0.160 | 959.214 | 22.60 |
| 27 | 1 | 958.597 | 0.197 | 958.604 | 17.80 | | 10 | 1 | 959.132 | 0.150 | 959.140 | 23.61 |
| 27 | 1 | 959.155 | 0.166 | 959.161 | 18.62 | | 10 | 1 | 959.132 | 0.150 | 959.137 | 24.70 |
| 29 | 2 | 958.842 | 0.161 | 958.834 | 18.16 | | 10 | 2 | 958.853 | 0.158 | 958.884 | 25.84 |
| 29 | 2 | 960.172 | 0.158 | 960.137 | 19.01 | | 10 | 2 | 958.853 | 0.158 | 958.863 | 27.10 |
| 29 | 2 | 959.624 | 0.154 | 959.623 | 19.94 | | 10 | 2 | 959.334 | 0.177 | 959.330 | 28.38 |
| 29 | 1 | 958.842 | 0.161 | 958.834 | 18.16 | | 10 | 2 | 958.709 | 0.159 | 958.761 | 29.71 |
| 29 | 1 | 960.172 | 0.158 | 960.137 | 19.01 | | 10 | 2 | 959.132 | 0.150 | 959.179 | 31.17 |
| 29 | 1 | 959.624 | 0.154 | 959.623 | 19.94 | | 10 | 2 | 958.207 | 0.160 | 958.241 | 32.80 |
| 29 | 2 | 959.805 | 0.167 | 959.793 | 20.90 | | 10 | 2 | 959.677 | 0.204 | 959.604 | 34.47 |
| 29 | 2 | 959.053 | 0.170 | 959.075 | 21.91 | | 10 | 2 | 959.526 | 0.150 | 959.556 | 36.25 |
| 29 | 2 | 959.487 | 0.209 | 959.483 | 22.95 | | 10 | 1 | 958.853 | 0.158 | 958.884 | 25.84 |
| 29 | 2 | 959.449 | 0.169 | 959.456 | 24.14 | | 10 | 1 | 958.853 | 0.158 | 958.863 | 27.10 |
| 29 | 2 | 959.624 | 0.154 | 959.604 | 27.30 | | | | | | | |

| 2000 - ABRIL | | | | | | 2000 - ABRIL | | | | |
|---|---|---|---|---|---|---|---|---|---|---|
| D | L | SDB | ER | SDC | HL | D | L | SDB | ER | SDC | HL |
| 10 | 1 | 959.334 | 0.177 | 959.330 | 28.38 | 14 | 1 | 958.655 | 0.190 | 958.685 | 26.71 |
| 10 | 1 | 958.709 | 0.159 | 958.761 | 29.71 | 14 | 1 | 959.175 | 0.121 | 959.178 | 29.15 |
| 10 | 1 | 959.132 | 0.150 | 959.179 | 31.17 | 14 | 1 | 958.997 | 0.152 | 958.985 | 30.43 |
| 10 | 1 | 958.207 | 0.160 | 958.241 | 32.80 | 14 | 1 | 959.306 | 0.144 | 959.312 | 31.78 |
| 10 | 1 | 959.677 | 0.204 | 959.604 | 34.47 | 14 | 1 | 959.565 | 0.133 | 959.556 | 33.33 |
| 10 | 1 | 959.526 | 0.150 | 959.556 | 36.25 | 14 | 1 | 959.150 | 0.128 | 959.118 | 35.16 |
| 11 | 2 | 957.675 | 0.278 | 957.491 | 17.74 | 14 | 1 | 958.655 | 0.190 | 958.653 | 36.90 |
| 11 | 2 | 959.419 | 0.248 | 959.437 | 19.51 | 17 | 2 | 959.658 | 0.293 | 959.677 | 19.49 |
| 11 | 2 | 959.462 | 0.168 | 959.460 | 20.43 | 17 | 2 | 960.062 | 0.236 | 960.087 | 21.15 |
| 11 | 2 | 959.469 | 0.201 | 959.471 | 21.32 | 17 | 2 | 959.708 | 0.194 | 959.713 | 22.90 |
| 11 | 2 | 959.324 | 0.313 | 959.326 | 22.66 | 17 | 2 | 959.296 | 0.258 | 959.283 | 24.11 |
| 11 | 2 | 958.555 | 0.293 | 958.635 | 23.76 | 17 | 2 | 959.658 | 0.293 | 959.674 | 26.90 |
| 11 | 2 | 958.808 | 0.212 | 958.836 | 24.83 | 17 | 1 | 959.658 | 0.293 | 959.677 | 19.49 |
| 11 | 1 | 957.675 | 0.278 | 957.491 | 17.74 | 17 | 1 | 960.062 | 0.236 | 960.087 | 21.15 |
| 11 | 1 | 959.419 | 0.248 | 959.437 | 19.51 | 17 | 1 | 959.708 | 0.194 | 959.713 | 22.90 |
| 11 | 1 | 959.462 | 0.168 | 959.460 | 20.43 | 17 | 1 | 959.296 | 0.258 | 959.283 | 24.11 |
| 11 | 1 | 959.469 | 0.201 | 959.471 | 21.32 | 17 | 1 | 959.658 | 0.293 | 959.674 | 26.90 |
| 11 | 1 | 959.324 | 0.313 | 959.326 | 22.66 | 17 | 2 | 959.462 | 0.198 | 959.447 | 29.63 |
| 11 | 1 | 958.555 | 0.293 | 958.635 | 23.76 | 17 | 1 | 959.462 | 0.198 | 959.447 | 29.63 |
| 11 | 1 | 958.808 | 0.212 | 958.836 | 24.83 | 20 | 2 | 958.823 | 0.154 | 958.738 | 23.99 |
| 11 | 2 | 959.183 | 0.161 | 959.190 | 25.99 | 20 | 2 | 959.167 | 0.166 | 959.107 | 24.93 |
| 11 | 2 | 959.037 | 0.233 | 958.954 | 27.19 | 20 | 2 | 958.940 | 0.191 | 958.934 | 25.93 |
| 11 | 2 | 959.847 | 0.365 | 959.807 | 28.49 | 20 | 2 | 959.470 | 0.132 | 959.463 | 26.95 |
| 11 | 2 | 959.980 | 0.401 | 959.963 | 30.62 | 20 | 2 | 959.544 | 0.130 | 959.574 | 28.02 |
| 11 | 2 | 958.555 | 0.293 | 958.557 | 33.93 | 20 | 1 | 958.823 | 0.154 | 958.738 | 23.99 |
| 11 | 2 | 959.847 | 0.365 | 959.719 | 35.64 | 20 | 1 | 959.167 | 0.166 | 959.107 | 24.93 |
| 11 | 1 | 959.183 | 0.161 | 959.190 | 25.99 | 20 | 1 | 958.940 | 0.191 | 958.934 | 25.93 |
| 11 | 1 | 959.037 | 0.233 | 958.954 | 27.19 | 20 | 1 | 959.470 | 0.132 | 959.463 | 26.95 |
| 11 | 1 | 959.847 | 0.365 | 959.807 | 28.49 | 20 | 1 | 959.544 | 0.130 | 959.574 | 28.02 |
| 11 | 1 | 959.980 | 0.401 | 959.963 | 30.62 | 24 | 2 | 959.655 | 0.172 | 959.656 | 20.50 |
| 11 | 1 | 958.555 | 0.293 | 958.557 | 33.93 | 24 | 2 | 960.082 | 0.248 | 960.047 | 21.33 |
| 11 | 1 | 959.847 | 0.365 | 959.719 | 35.64 | 24 | 2 | 959.790 | 0.149 | 959.800 | 22.29 |
| 12 | 2 | 960.114 | 0.176 | 960.102 | 19.37 | 24 | 2 | 960.557 | 0.165 | 960.455 | 23.22 |
| 12 | 2 | 959.942 | 0.178 | 959.913 | 20.31 | 24 | 2 | 959.068 | 0.176 | 959.051 | 24.23 |
| 12 | 2 | 958.823 | 0.177 | 958.796 | 21.22 | 24 | 2 | 959.881 | 0.162 | 959.900 | 25.21 |
| 12 | 2 | 959.041 | 0.187 | 959.043 | 22.14 | 24 | 2 | 959.068 | 0.176 | 959.030 | 26.33 |
| 12 | 2 | 959.465 | 0.192 | 959.455 | 23.09 | 24 | 2 | 959.479 | 0.165 | 959.466 | 27.48 |
| 12 | 2 | 959.423 | 0.180 | 959.385 | 24.12 | 24 | 2 | 960.655 | 0.173 | 960.629 | 28.80 |
| 12 | 1 | 960.114 | 0.176 | 960.102 | 19.37 | 24 | 2 | 959.381 | 0.169 | 959.380 | 29.99 |
| 12 | 1 | 959.942 | 0.178 | 959.913 | 20.31 | 24 | 1 | 959.655 | 0.172 | 959.656 | 20.50 |
| 12 | 1 | 958.823 | 0.177 | 958.796 | 21.22 | 24 | 1 | 960.082 | 0.248 | 960.047 | 21.33 |
| 12 | 1 | 959.041 | 0.187 | 959.043 | 22.14 | 24 | 1 | 959.790 | 0.149 | 959.800 | 22.29 |
| 12 | 1 | 959.465 | 0.192 | 959.455 | 23.09 | 24 | 1 | 960.557 | 0.165 | 960.455 | 23.22 |
| 12 | 1 | 959.423 | 0.180 | 959.385 | 24.12 | 24 | 1 | 959.068 | 0.176 | 959.051 | 24.23 |
| 12 | 2 | 957.480 | 0.185 | 957.450 | 26.27 | 24 | 1 | 959.881 | 0.162 | 959.900 | 25.21 |
| 12 | 2 | 959.465 | 0.192 | 959.464 | 27.48 | 24 | 1 | 959.068 | 0.176 | 959.030 | 26.33 |
| 12 | 2 | 958.868 | 0.309 | 958.861 | 28.78 | 24 | 1 | 959.479 | 0.165 | 959.466 | 27.48 |
| 12 | 2 | 959.423 | 0.180 | 959.405 | 30.13 | 24 | 1 | 960.655 | 0.173 | 960.629 | 28.80 |
| 12 | 2 | 959.151 | 0.200 | 959.129 | 31.53 | 24 | 1 | 959.381 | 0.169 | 959.380 | 29.99 |
| 12 | 2 | 959.270 | 0.164 | 959.261 | 33.10 | 24 | 2 | 958.969 | 0.223 | 958.994 | 31.29 |
| 12 | 1 | 957.480 | 0.185 | 957.450 | 26.27 | 24 | 2 | 959.381 | 0.169 | 959.364 | 32.63 |
| 12 | 1 | 959.465 | 0.192 | 959.464 | 27.48 | 24 | 2 | 958.969 | 0.223 | 958.982 | 34.05 |
| 12 | 1 | 958.868 | 0.309 | 958.861 | 28.78 | 24 | 2 | 959.185 | 0.243 | 959.165 | 35.57 |
| 12 | 1 | 959.423 | 0.180 | 959.405 | 30.13 | 24 | 2 | 958.969 | 0.223 | 959.007 | 37.26 |
| 12 | 1 | 959.151 | 0.200 | 959.129 | 31.53 | 24 | 2 | 959.655 | 0.172 | 959.659 | 39.21 |
| 12 | 1 | 959.270 | 0.164 | 959.261 | 33.10 | 24 | 2 | 958.436 | 0.206 | 958.475 | 41.17 |
| 14 | 2 | 958.436 | 0.140 | 958.423 | 21.00 | 24 | 2 | 959.527 | 0.169 | 959.504 | 43.36 |
| 14 | 2 | 959.565 | 0.133 | 959.566 | 21.84 | 24 | 1 | 958.969 | 0.223 | 958.994 | 31.29 |
| 14 | 2 | 958.925 | 0.153 | 958.903 | 22.76 | 24 | 1 | 959.381 | 0.169 | 959.364 | 32.63 |
| 14 | 2 | 959.425 | 0.167 | 959.436 | 25.64 | 24 | 1 | 958.969 | 0.223 | 958.982 | 34.05 |
| 14 | 1 | 958.436 | 0.140 | 958.423 | 21.00 | 24 | 1 | 959.185 | 0.243 | 959.165 | 35.57 |
| 14 | 1 | 959.565 | 0.133 | 959.566 | 21.84 | 24 | 1 | 958.969 | 0.223 | 959.007 | 37.26 |
| 14 | 1 | 958.925 | 0.153 | 958.903 | 22.76 | 24 | 1 | 959.655 | 0.172 | 959.659 | 39.21 |
| 14 | 1 | 959.425 | 0.167 | 959.436 | 25.64 | 24 | 1 | 958.436 | 0.206 | 958.475 | 41.17 |
| 14 | 2 | 958.655 | 0.190 | 958.685 | 26.71 | 24 | 1 | 959.527 | 0.169 | 959.504 | 43.36 |
| 14 | 2 | 959.175 | 0.121 | 959.178 | 29.15 | 25 | 2 | 959.628 | 0.198 | 959.630 | 27.74 |
| 14 | 2 | 958.997 | 0.152 | 958.985 | 30.43 | 25 | 2 | 959.717 | 0.207 | 959.703 | 28.78 |
| 14 | 2 | 959.306 | 0.144 | 959.312 | 31.78 | 25 | 2 | 959.472 | 0.286 | 959.475 | 29.87 |
| 14 | 2 | 959.565 | 0.133 | 959.556 | 33.33 | 25 | 1 | 959.628 | 0.198 | 959.630 | 27.74 |
| 14 | 2 | 959.150 | 0.128 | 959.118 | 35.16 | 25 | 1 | 959.717 | 0.207 | 959.703 | 28.78 |
| 14 | 2 | 958.655 | 0.190 | 958.653 | 36.90 | 25 | 1 | 959.472 | 0.286 | 959.475 | 29.87 |

| 2000 - ABRIL | | | | |
|---|---|---|---|---|
| D  L | SDB | ER | SDC | HL |
| 25 2 | 958.950 | 0.196 | 958.933 | 31.01 |
| 25 2 | 959.466 | 0.226 | 959.457 | 32.22 |
| 25 2 | 960.049 | 0.273 | 960.059 | 33.52 |
| 25 2 | 959.824 | 0.260 | 959.838 | 34.92 |
| 25 2 | 959.328 | 0.194 | 959.315 | 36.39 |
| 25 2 | 959.179 | 0.320 | 959.201 | 37.98 |
| 25 2 | 960.664 | 0.241 | 960.649 | 39.71 |
| 25 2 | 959.419 | 0.277 | 959.411 | 41.59 |
| 25 2 | 959.168 | 0.245 | 959.165 | 43.70 |
| 25 1 | 958.950 | 0.196 | 958.933 | 31.01 |
| 25 1 | 959.466 | 0.226 | 959.457 | 32.22 |
| 25 1 | 960.049 | 0.273 | 960.059 | 33.52 |
| 25 1 | 959.824 | 0.260 | 959.838 | 34.92 |
| 25 1 | 959.328 | 0.194 | 959.315 | 36.39 |
| 25 1 | 959.179 | 0.320 | 959.201 | 37.98 |
| 25 1 | 960.664 | 0.241 | 960.649 | 39.71 |
| 25 1 | 959.419 | 0.277 | 959.411 | 41.59 |
| 25 1 | 959.168 | 0.245 | 959.165 | 43.70 |
| 26 2 | 959.851 | 0.209 | 959.862 | 20.94 |
| 26 2 | 959.052 | 0.210 | 959.043 | 21.75 |
| 26 2 | 958.443 | 0.175 | 958.463 | 22.68 |
| 26 2 | 959.885 | 0.177 | 959.879 | 23.59 |
| 26 2 | 959.701 | 0.162 | 959.731 | 24.51 |
| 26 2 | 958.672 | 0.193 | 958.661 | 25.49 |
| 26 2 | 959.220 | 0.201 | 959.223 | 26.51 |
| 26 2 | 958.860 | 0.198 | 958.860 | 27.59 |
| 26 2 | 959.220 | 0.201 | 959.258 | 28.79 |
| 26 2 | 959.627 | 0.252 | 959.627 | 30.41 |
| 26 1 | 959.851 | 0.209 | 959.862 | 20.94 |
| 26 1 | 959.052 | 0.210 | 959.043 | 21.75 |
| 26 1 | 958.443 | 0.175 | 958.463 | 22.68 |
| 26 1 | 959.885 | 0.177 | 959.879 | 23.59 |
| 26 1 | 959.701 | 0.162 | 959.731 | 24.51 |
| 26 1 | 958.672 | 0.193 | 958.661 | 25.49 |
| 26 1 | 959.220 | 0.201 | 959.223 | 26.51 |
| 26 1 | 958.860 | 0.198 | 958.860 | 27.59 |
| 26 1 | 959.220 | 0.201 | 959.258 | 28.79 |
| 26 1 | 959.627 | 0.252 | 959.627 | 30.41 |
| 26 2 | 958.794 | 0.187 | 958.784 | 31.72 |
| 26 2 | 959.401 | 0.224 | 959.407 | 33.05 |
| 26 2 | 960.048 | 0.175 | 960.027 | 34.45 |
| 26 2 | 959.469 | 0.202 | 959.446 | 35.96 |
| 26 2 | 958.594 | 0.187 | 958.605 | 37.55 |
| 26 2 | 959.324 | 0.212 | 959.333 | 39.31 |
| 26 2 | 958.794 | 0.187 | 958.819 | 41.22 |
| 26 2 | 958.594 | 0.187 | 958.578 | 43.38 |
| 26 1 | 958.794 | 0.187 | 958.784 | 31.72 |
| 26 1 | 959.401 | 0.224 | 959.407 | 33.05 |
| 26 1 | 960.048 | 0.175 | 960.027 | 34.45 |
| 26 1 | 959.469 | 0.202 | 959.446 | 35.96 |
| 26 1 | 958.594 | 0.187 | 958.605 | 37.55 |
| 26 1 | 959.324 | 0.212 | 959.333 | 39.31 |
| 26 1 | 958.794 | 0.187 | 958.819 | 41.22 |
| 26 1 | 958.594 | 0.187 | 958.578 | 43.38 |
| 28 2 | 959.343 | 0.186 | 959.323 | 20.48 |
| 28 2 | 960.747 | 0.136 | 960.628 | 21.26 |
| 28 2 | 959.371 | 0.149 | 959.371 | 22.06 |
| 28 2 | 959.480 | 0.186 | 959.468 | 22.93 |
| 28 2 | 959.154 | 0.223 | 959.183 | 23.79 |
| 28 2 | 959.154 | 0.223 | 959.152 | 24.73 |
| 28 2 | 959.561 | 0.189 | 959.554 | 25.67 |
| 28 2 | 958.379 | 0.175 | 958.427 | 26.66 |
| 28 2 | 958.898 | 0.168 | 958.862 | 27.70 |
| 28 2 | 959.480 | 0.186 | 959.470 | 28.81 |
| 28 2 | 958.898 | 0.168 | 958.938 | 30.17 |
| 28 2 | 960.019 | 0.260 | 959.990 | 31.40 |
| 28 1 | 959.343 | 0.186 | 959.323 | 20.48 |
| 28 1 | 960.747 | 0.136 | 960.628 | 21.26 |
| 28 1 | 959.371 | 0.149 | 959.371 | 22.06 |
| 28 1 | 959.480 | 0.186 | 959.468 | 22.93 |
| 28 1 | 959.154 | 0.223 | 959.183 | 23.79 |

| 2000 - ABRIL | | | | |
|---|---|---|---|---|
| D  L | SDB | ER | SDC | HL |
| 28 1 | 959.154 | 0.223 | 959.152 | 24.73 |
| 28 1 | 959.561 | 0.189 | 959.554 | 25.67 |
| 28 1 | 958.379 | 0.175 | 958.427 | 26.66 |
| 28 1 | 958.898 | 0.168 | 958.862 | 27.70 |
| 28 1 | 959.480 | 0.186 | 959.470 | 28.81 |
| 28 1 | 958.898 | 0.168 | 958.938 | 30.17 |
| 28 1 | 960.019 | 0.260 | 959.990 | 31.40 |
| 28 2 | 959.359 | 0.203 | 959.359 | 32.68 |
| 28 2 | 958.898 | 0.168 | 958.874 | 34.06 |
| 28 2 | 959.215 | 0.181 | 959.217 | 35.56 |
| 28 2 | 959.480 | 0.186 | 959.492 | 37.20 |
| 28 2 | 958.898 | 0.168 | 958.932 | 38.89 |
| 28 2 | 959.763 | 0.170 | 959.782 | 40.73 |
| 28 2 | 959.029 | 0.219 | 959.012 | 42.76 |
| 28 1 | 959.359 | 0.203 | 959.359 | 32.68 |
| 28 1 | 958.898 | 0.168 | 958.874 | 34.06 |
| 28 1 | 959.215 | 0.181 | 959.217 | 35.56 |
| 28 1 | 959.480 | 0.186 | 959.492 | 37.20 |
| 28 1 | 958.898 | 0.168 | 958.932 | 38.89 |
| 28 1 | 959.763 | 0.170 | 959.782 | 40.73 |
| 28 1 | 959.029 | 0.219 | 959.012 | 42.76 |

| 2000 - MAIO | | | | |
|---|---|---|---|---|
| D  L | SDB | ER | SDC | HL |
| 02 2 | 959.328 | 0.263 | 959.368 | 37.48 |
| 02 2 | 959.108 | 0.245 | 959.123 | 39.17 |
| 02 2 | 958.125 | 0.228 | 958.116 | 41.02 |
| 02 2 | 958.315 | 0.285 | 958.380 | 43.17 |
| 02 2 | 957.521 | 0.275 | 957.455 | 45.53 |
| 02 2 | 957.235 | 0.284 | 957.327 | 48.17 |
| 02 2 | 957.521 | 0.275 | 957.385 | 51.24 |
| 02 1 | 959.328 | 0.263 | 959.368 | 37.48 |
| 02 1 | 959.108 | 0.245 | 959.123 | 39.17 |
| 02 1 | 958.125 | 0.228 | 958.116 | 41.02 |
| 02 1 | 958.315 | 0.285 | 958.380 | 43.17 |
| 02 1 | 957.521 | 0.275 | 957.455 | 45.53 |
| 02 1 | 957.235 | 0.284 | 957.327 | 48.17 |
| 02 1 | 957.521 | 0.275 | 957.385 | 51.24 |
| 03 2 | 959.922 | 0.162 | 959.999 | 24.10 |
| 03 2 | 959.922 | 0.162 | 959.955 | 25.05 |
| 03 2 | 959.196 | 0.152 | 959.205 | 25.96 |
| 03 2 | 959.499 | 0.249 | 959.494 | 26.89 |
| 03 2 | 959.125 | 0.183 | 959.121 | 27.90 |
| 03 2 | 959.356 | 0.230 | 959.351 | 31.88 |
| 03 2 | 958.804 | 0.227 | 958.825 | 33.10 |
| 03 1 | 959.922 | 0.162 | 959.999 | 24.10 |
| 03 1 | 959.922 | 0.162 | 959.955 | 25.05 |
| 03 1 | 959.196 | 0.152 | 959.205 | 25.96 |
| 03 1 | 959.499 | 0.249 | 959.494 | 26.89 |
| 03 1 | 959.125 | 0.183 | 959.121 | 27.90 |
| 03 1 | 959.356 | 0.230 | 959.351 | 31.88 |
| 03 1 | 958.804 | 0.227 | 958.825 | 33.10 |
| 03 2 | 959.356 | 0.230 | 959.338 | 34.42 |
| 03 2 | 958.804 | 0.227 | 958.803 | 35.81 |
| 03 2 | 958.804 | 0.227 | 958.807 | 39.66 |
| 03 2 | 959.568 | 0.185 | 959.565 | 41.42 |
| 03 2 | 958.589 | 0.202 | 958.587 | 43.38 |
| 03 2 | 958.510 | 0.130 | 958.484 | 45.55 |
| 03 1 | 959.356 | 0.230 | 959.338 | 34.42 |
| 03 1 | 958.804 | 0.227 | 958.803 | 35.81 |
| 03 1 | 958.804 | 0.227 | 958.807 | 39.66 |
| 03 1 | 959.568 | 0.185 | 959.565 | 41.42 |
| 03 1 | 958.589 | 0.202 | 958.587 | 43.38 |
| 03 1 | 958.510 | 0.130 | 958.484 | 45.55 |
| 04 2 | 959.181 | 0.213 | 959.180 | 31.79 |
| 04 2 | 959.496 | 0.236 | 959.551 | 32.94 |
| 04 1 | 959.181 | 0.213 | 959.180 | 31.79 |
| 04 1 | 959.496 | 0.236 | 959.551 | 32.94 |
| 04 2 | 959.015 | 0.192 | 959.014 | 34.15 |
| 04 2 | 959.181 | 0.213 | 959.152 | 37.41 |

| D | L | SDB | ER | SDC | HL |
|---|---|---|---|---|---|
| 2000 - MAIO | | | | | |
| 04 | 2 | 958.569 | 0.214 | 958.540 | 38.90 |
| 04 | 2 | 959.496 | 0.236 | 959.485 | 40.52 |
| 04 | 2 | 958.409 | 0.200 | 958.411 | 42.28 |
| 04 | 2 | 959.206 | 0.205 | 959.257 | 44.16 |
| 04 | 2 | 957.709 | 0.201 | 957.732 | 48.72 |
| 04 | 1 | 959.015 | 0.192 | 959.014 | 34.15 |
| 04 | 1 | 959.181 | 0.213 | 959.152 | 37.41 |
| 04 | 1 | 958.569 | 0.214 | 958.540 | 38.90 |
| 04 | 1 | 959.496 | 0.236 | 959.485 | 40.52 |
| 04 | 1 | 958.409 | 0.200 | 958.411 | 42.28 |
| 04 | 1 | 959.206 | 0.205 | 959.257 | 44.16 |
| 04 | 1 | 957.709 | 0.201 | 957.732 | 48.72 |
| 15 | 2 | 958.643 | 0.161 | 958.656 | 27.37 |
| 15 | 2 | 960.153 | 0.170 | 960.173 | 28.23 |
| 15 | 2 | 959.083 | 0.210 | 959.078 | 29.17 |
| 15 | 2 | 959.438 | 0.182 | 959.501 | 30.15 |
| 15 | 2 | 958.202 | 0.276 | 958.260 | 31.18 |
| 15 | 2 | 959.420 | 0.198 | 959.393 | 32.21 |
| 15 | 2 | 959.016 | 0.178 | 959.011 | 33.25 |
| 15 | 2 | 958.643 | 0.161 | 958.663 | 34.40 |
| 15 | 2 | 959.778 | 0.155 | 959.780 | 35.60 |
| 15 | 2 | 959.778 | 0.155 | 959.772 | 36.79 |
| 15 | 2 | 958.888 | 0.266 | 958.889 | 38.08 |
| 15 | 1 | 958.643 | 0.161 | 958.656 | 27.37 |
| 15 | 1 | 960.153 | 0.170 | 960.173 | 28.23 |
| 15 | 1 | 959.083 | 0.210 | 959.078 | 29.17 |
| 15 | 1 | 959.438 | 0.182 | 959.501 | 30.15 |
| 15 | 1 | 958.202 | 0.276 | 958.260 | 31.18 |
| 15 | 1 | 959.420 | 0.198 | 959.393 | 32.21 |
| 15 | 1 | 959.016 | 0.178 | 959.011 | 33.25 |
| 15 | 1 | 958.643 | 0.161 | 958.663 | 34.40 |
| 15 | 1 | 959.778 | 0.155 | 959.780 | 35.60 |
| 15 | 1 | 959.778 | 0.155 | 959.772 | 36.79 |
| 15 | 1 | 958.888 | 0.266 | 958.889 | 38.08 |
| 15 | 2 | 959.183 | 0.218 | 959.177 | 39.44 |
| 15 | 2 | 959.016 | 0.178 | 959.026 | 40.96 |
| 15 | 2 | 958.643 | 0.161 | 958.657 | 42.74 |
| 15 | 2 | 958.700 | 0.178 | 958.724 | 44.43 |
| 15 | 2 | 958.918 | 0.205 | 958.911 | 46.27 |
| 15 | 1 | 959.183 | 0.218 | 959.177 | 39.44 |
| 15 | 1 | 959.016 | 0.178 | 959.026 | 40.96 |
| 15 | 1 | 958.643 | 0.161 | 958.657 | 42.74 |
| 15 | 1 | 958.700 | 0.178 | 958.724 | 44.43 |
| 15 | 1 | 958.918 | 0.205 | 958.911 | 46.27 |
| 17 | 2 | 959.168 | 0.246 | 959.157 | 34.50 |
| 17 | 2 | 959.516 | 0.141 | 959.526 | 35.73 |
| 17 | 2 | 958.871 | 0.138 | 958.849 | 36.97 |
| 17 | 2 | 958.769 | 0.128 | 958.723 | 38.22 |
| 17 | 1 | 959.168 | 0.246 | 959.157 | 34.50 |
| 17 | 1 | 959.516 | 0.141 | 959.526 | 35.73 |
| 17 | 1 | 958.871 | 0.138 | 958.849 | 36.97 |
| 17 | 1 | 958.769 | 0.128 | 958.723 | 38.22 |
| 17 | 2 | 959.093 | 0.183 | 959.114 | 39.55 |
| 17 | 2 | 959.516 | 0.141 | 959.531 | 40.98 |
| 17 | 2 | 958.945 | 0.198 | 958.957 | 44.34 |
| 17 | 2 | 959.093 | 0.183 | 959.111 | 46.17 |
| 17 | 2 | 959.382 | 0.198 | 959.386 | 49.12 |
| 17 | 2 | 959.218 | 0.168 | 959.223 | 51.57 |
| 17 | 1 | 959.093 | 0.183 | 959.114 | 39.55 |
| 17 | 1 | 959.516 | 0.141 | 959.531 | 40.98 |
| 17 | 1 | 958.945 | 0.198 | 958.957 | 44.34 |
| 17 | 1 | 959.093 | 0.183 | 959.111 | 46.17 |
| 17 | 1 | 959.382 | 0.198 | 959.386 | 49.12 |
| 17 | 1 | 959.218 | 0.168 | 959.223 | 51.57 |
| 18 | 2 | 958.841 | 0.364 | 959.014 | 42.56 |
| 18 | 2 | 958.468 | 0.406 | 958.479 | 44.15 |
| 18 | 2 | 959.263 | 0.253 | 959.255 | 45.82 |
| 18 | 2 | 960.259 | 0.232 | 960.415 | 47.65 |
| 18 | 2 | 958.841 | 0.364 | 958.912 | 49.68 |
| 18 | 1 | 958.841 | 0.364 | 959.014 | 42.56 |
| 18 | 1 | 958.468 | 0.406 | 958.479 | 44.15 |
| 2000 - MAIO | | | | | |
| 18 | 1 | 959.263 | 0.253 | 959.255 | 45.82 |
| 18 | 1 | 960.259 | 0.232 | 960.415 | 47.65 |
| 18 | 1 | 958.841 | 0.364 | 958.912 | 49.68 |
| 24 | 2 | 959.196 | 0.371 | 959.188 | 44.60 |
| 24 | 2 | 958.993 | 0.226 | 959.024 | 47.54 |
| 24 | 2 | 958.530 | 0.326 | 958.462 | 49.52 |
| 24 | 2 | 959.423 | 0.686 | 959.410 | 51.61 |
| 24 | 2 | 958.878 | 0.209 | 958.884 | 53.98 |
| 24 | 2 | 958.878 | 0.209 | 958.912 | 56.85 |
| 24 | 2 | 956.952 | 0.245 | 957.307 | 60.41 |
| 24 | 1 | 959.196 | 0.371 | 959.188 | 44.60 |
| 24 | 1 | 958.993 | 0.226 | 959.024 | 47.54 |
| 24 | 1 | 958.530 | 0.326 | 958.462 | 49.52 |
| 24 | 1 | 959.423 | 0.686 | 959.410 | 51.61 |
| 24 | 1 | 958.878 | 0.209 | 958.884 | 53.98 |
| 24 | 1 | 958.878 | 0.209 | 958.912 | 56.85 |
| 24 | 1 | 956.952 | 0.245 | 957.307 | 60.41 |
| 25 | 2 | 959.338 | 0.232 | 959.371 | 40.14 |
| 25 | 2 | 959.037 | 0.207 | 959.051 | 41.35 |
| 25 | 1 | 959.338 | 0.232 | 959.371 | 40.14 |
| 25 | 1 | 959.037 | 0.207 | 959.051 | 41.35 |
| 25 | 2 | 958.998 | 0.233 | 959.012 | 42.64 |
| 25 | 2 | 958.711 | 0.319 | 958.774 | 44.03 |
| 25 | 2 | 958.584 | 0.271 | 958.563 | 45.65 |
| 25 | 2 | 958.458 | 0.280 | 958.454 | 47.29 |
| 25 | 2 | 958.260 | 0.372 | 958.138 | 49.01 |
| 25 | 2 | 959.338 | 0.232 | 959.415 | 50.87 |
| 25 | 2 | 958.650 | 0.229 | 958.633 | 52.95 |
| 25 | 2 | 959.262 | 0.214 | 959.272 | 55.31 |
| 25 | 1 | 958.998 | 0.233 | 959.012 | 42.64 |
| 25 | 1 | 958.711 | 0.319 | 958.774 | 44.03 |
| 25 | 1 | 958.584 | 0.271 | 958.563 | 45.65 |
| 25 | 1 | 958.458 | 0.280 | 958.454 | 47.29 |
| 25 | 1 | 958.260 | 0.372 | 958.138 | 49.01 |
| 25 | 1 | 959.338 | 0.232 | 959.415 | 50.87 |
| 25 | 1 | 958.650 | 0.229 | 958.633 | 52.95 |
| 25 | 1 | 959.262 | 0.214 | 959.272 | 55.31 |
| 26 | 2 | 959.990 | 0.253 | 959.896 | 37.86 |
| 26 | 2 | 958.011 | 0.397 | 958.129 | 38.95 |
| 26 | 2 | 959.391 | 0.172 | 959.408 | 40.08 |
| 26 | 2 | 959.380 | 0.303 | 959.345 | 41.26 |
| 26 | 2 | 959.049 | 0.207 | 959.027 | 42.50 |
| 26 | 1 | 959.990 | 0.253 | 959.896 | 37.86 |
| 26 | 1 | 958.011 | 0.397 | 958.129 | 38.95 |
| 26 | 1 | 959.391 | 0.172 | 959.408 | 40.08 |
| 26 | 1 | 959.380 | 0.303 | 959.345 | 41.26 |
| 26 | 1 | 959.049 | 0.207 | 959.027 | 42.50 |
| 26 | 2 | 959.141 | 0.158 | 959.161 | 43.83 |
| 26 | 2 | 959.049 | 0.207 | 959.055 | 45.26 |
| 26 | 2 | 959.298 | 0.275 | 959.288 | 46.80 |
| 26 | 2 | 959.467 | 0.270 | 959.484 | 48.44 |
| 26 | 2 | 959.650 | 0.212 | 959.665 | 50.24 |
| 26 | 2 | 957.101 | 0.222 | 957.048 | 52.19 |
| 26 | 1 | 959.141 | 0.158 | 959.161 | 43.83 |
| 26 | 1 | 959.049 | 0.207 | 959.055 | 45.26 |
| 26 | 1 | 959.298 | 0.275 | 959.288 | 46.80 |
| 26 | 1 | 959.467 | 0.270 | 959.484 | 48.44 |
| 26 | 1 | 959.650 | 0.212 | 959.665 | 50.24 |
| 26 | 1 | 957.101 | 0.222 | 957.048 | 52.19 |
| 29 | 2 | 960.358 | 0.209 | 960.376 | 36.70 |
| 29 | 2 | 959.896 | 0.225 | 959.883 | 37.92 |
| 29 | 2 | 959.896 | 0.225 | 959.865 | 39.04 |
| 29 | 2 | 959.472 | 0.171 | 959.508 | 40.19 |
| 29 | 2 | 958.005 | 0.291 | 958.068 | 41.43 |
| 29 | 2 | 959.545 | 0.214 | 959.519 | 42.90 |
| 29 | 1 | 960.358 | 0.209 | 960.376 | 36.70 |
| 29 | 1 | 959.896 | 0.225 | 959.883 | 37.92 |
| 29 | 1 | 959.896 | 0.225 | 959.865 | 39.04 |
| 29 | 1 | 959.472 | 0.171 | 959.508 | 40.19 |
| 29 | 1 | 958.005 | 0.291 | 958.068 | 41.43 |
| 29 | 1 | 959.545 | 0.214 | 959.519 | 42.90 |

| 2000 - MAIO | | | | |
|---|---|---|---|---|
| D | L | SDB | ER | SDC | HL |
| 29 | 2 | 959.172 | 0.138 | 959.177 | 44.23 |
| 29 | 2 | 958.642 | 0.183 | 958.546 | 45.85 |
| 29 | 2 | 958.942 | 0.286 | 958.931 | 47.43 |
| 29 | 2 | 959.689 | 0.168 | 959.699 | 49.18 |
| 29 | 2 | 958.796 | 0.208 | 958.768 | 51.08 |
| 29 | 2 | 958.912 | 0.234 | 958.904 | 53.12 |
| 29 | 2 | 959.209 | 0.212 | 959.211 | 55.45 |
| 29 | 1 | 959.172 | 0.138 | 959.177 | 44.23 |
| 29 | 1 | 958.642 | 0.183 | 958.546 | 45.85 |
| 29 | 1 | 958.942 | 0.286 | 958.931 | 47.43 |
| 29 | 1 | 959.689 | 0.168 | 959.699 | 49.18 |
| 29 | 1 | 958.796 | 0.208 | 958.768 | 51.08 |
| 29 | 1 | 958.912 | 0.234 | 958.904 | 53.12 |
| 29 | 1 | 959.209 | 0.212 | 959.211 | 55.45 |
| 30 | 2 | 959.374 | 0.248 | 959.328 | 41.87 |
| 30 | 2 | 959.410 | 0.187 | 959.430 | 43.38 |
| 30 | 1 | 959.374 | 0.248 | 959.328 | 41.87 |
| 30 | 1 | 959.410 | 0.187 | 959.430 | 43.38 |
| 30 | 2 | 960.008 | 0.196 | 959.986 | 44.75 |
| 30 | 2 | 959.096 | 0.237 | 959.123 | 46.46 |
| 30 | 2 | 959.255 | 0.363 | 959.251 | 48.08 |
| 30 | 2 | 959.374 | 0.248 | 959.347 | 49.77 |
| 30 | 2 | 958.711 | 0.357 | 958.764 | 51.61 |
| 30 | 2 | 959.096 | 0.237 | 959.112 | 53.61 |
| 30 | 2 | 958.678 | 0.547 | 958.635 | 55.87 |
| 30 | 2 | 958.678 | 0.547 | 958.586 | 58.51 |
| 30 | 1 | 960.008 | 0.196 | 959.986 | 44.75 |
| 30 | 1 | 959.096 | 0.237 | 959.123 | 46.46 |
| 30 | 1 | 959.255 | 0.363 | 959.251 | 48.08 |
| 30 | 1 | 959.374 | 0.248 | 959.347 | 49.77 |
| 30 | 1 | 958.711 | 0.357 | 958.764 | 51.61 |
| 30 | 1 | 959.096 | 0.237 | 959.112 | 53.61 |
| 30 | 1 | 958.678 | 0.547 | 958.635 | 55.87 |
| 30 | 1 | 958.678 | 0.547 | 958.586 | 58.51 |
| 31 | 2 | 960.315 | 0.169 | 960.151 | 38.53 |
| 31 | 2 | 960.315 | 0.169 | 960.186 | 39.57 |
| 31 | 2 | 959.029 | 0.186 | 959.021 | 40.68 |
| 31 | 2 | 959.589 | 0.150 | 959.575 | 41.83 |
| 31 | 2 | 959.400 | 0.161 | 959.394 | 43.06 |
| 31 | 2 | 959.454 | 0.139 | 959.444 | 44.35 |
| 31 | 1 | 960.315 | 0.169 | 960.151 | 38.53 |
| 31 | 1 | 960.315 | 0.169 | 960.186 | 39.57 |
| 31 | 1 | 959.029 | 0.186 | 959.021 | 40.68 |
| 31 | 1 | 959.589 | 0.150 | 959.575 | 41.83 |
| 31 | 1 | 959.400 | 0.161 | 959.394 | 43.06 |
| 31 | 1 | 959.454 | 0.139 | 959.444 | 44.35 |
| 31 | 2 | 958.015 | 0.170 | 958.144 | 45.70 |
| 31 | 2 | 959.011 | 0.264 | 958.992 | 47.16 |
| 31 | 2 | 959.270 | 0.220 | 959.272 | 48.73 |
| 31 | 2 | 959.241 | 0.208 | 959.238 | 50.39 |
| 31 | 2 | 959.485 | 0.218 | 959.481 | 52.21 |
| 31 | 2 | 957.033 | 0.344 | 956.870 | 54.26 |
| 31 | 2 | 958.015 | 0.170 | 958.120 | 56.55 |
| 31 | 1 | 958.015 | 0.170 | 958.144 | 45.70 |
| 31 | 1 | 959.011 | 0.264 | 958.992 | 47.16 |
| 31 | 1 | 959.270 | 0.220 | 959.272 | 48.73 |
| 31 | 1 | 959.241 | 0.208 | 959.238 | 50.39 |
| 31 | 1 | 959.485 | 0.218 | 959.481 | 52.21 |
| 31 | 1 | 957.033 | 0.344 | 956.870 | 54.26 |
| 31 | 1 | 958.015 | 0.170 | 958.120 | 56.55 |

| 2000 - JUNHO | | | | |
|---|---|---|---|---|
| D | L | SDB | ER | SDC | HL |
| 05 | 2 | 959.643 | 0.267 | 959.646 | 41.78 |
| 05 | 2 | 959.671 | 0.184 | 959.665 | 42.90 |
| 05 | 2 | 959.425 | 0.156 | 959.416 | 44.12 |
| 05 | 2 | 959.223 | 0.215 | 959.191 | 45.42 |
| 05 | 1 | 959.643 | 0.267 | 959.646 | 41.78 |
| 05 | 1 | 959.671 | 0.184 | 959.665 | 42.90 |
| 05 | 1 | 959.425 | 0.156 | 959.416 | 44.12 |

| 2000 - JUNHO | | | | |
|---|---|---|---|---|
| D | L | SDB | ER | SDC | HL |
| 05 | 1 | 959.223 | 0.215 | 959.191 | 45.42 |
| 05 | 2 | 959.007 | 0.254 | 959.007 | 47.12 |
| 05 | 2 | 959.007 | 0.254 | 958.969 | 48.64 |
| 05 | 2 | 959.007 | 0.254 | 959.005 | 50.13 |
| 05 | 2 | 959.721 | 0.162 | 959.724 | 51.76 |
| 05 | 2 | 958.836 | 0.220 | 958.781 | 53.51 |
| 05 | 2 | 959.148 | 0.202 | 959.151 | 55.43 |
| 05 | 2 | 959.285 | 0.191 | 959.339 | 57.62 |
| 05 | 1 | 959.007 | 0.254 | 959.007 | 47.12 |
| 05 | 1 | 959.007 | 0.254 | 958.969 | 48.64 |
| 05 | 1 | 959.007 | 0.254 | 959.005 | 50.13 |
| 05 | 1 | 959.721 | 0.162 | 959.724 | 51.76 |
| 05 | 1 | 958.836 | 0.220 | 958.781 | 53.51 |
| 05 | 1 | 959.148 | 0.202 | 959.151 | 55.43 |
| 05 | 1 | 959.285 | 0.191 | 959.339 | 57.62 |
| 07 | 2 | 960.919 | 0.211 | 960.700 | 40.51 |
| 07 | 2 | 959.812 | 0.234 | 959.818 | 41.56 |
| 07 | 2 | 959.865 | 0.272 | 959.861 | 42.67 |
| 07 | 2 | 960.183 | 0.163 | 960.228 | 43.79 |
| 07 | 2 | 959.583 | 0.234 | 959.527 | 45.09 |
| 07 | 2 | 959.622 | 0.196 | 959.625 | 46.40 |
| 07 | 1 | 960.919 | 0.211 | 960.700 | 40.51 |
| 07 | 1 | 959.812 | 0.234 | 959.818 | 41.56 |
| 07 | 1 | 959.865 | 0.272 | 959.861 | 42.67 |
| 07 | 1 | 960.183 | 0.163 | 960.228 | 43.79 |
| 07 | 1 | 959.583 | 0.234 | 959.527 | 45.09 |
| 07 | 1 | 959.622 | 0.196 | 959.625 | 46.40 |
| 07 | 2 | 958.925 | 0.190 | 958.938 | 47.87 |
| 07 | 2 | 958.858 | 0.206 | 958.853 | 49.32 |
| 07 | 2 | 960.054 | 0.186 | 960.036 | 50.85 |
| 07 | 2 | 959.112 | 0.222 | 959.105 | 52.47 |
| 07 | 2 | 959.152 | 0.258 | 959.172 | 54.20 |
| 07 | 2 | 958.441 | 0.225 | 958.484 | 56.12 |
| 07 | 2 | 958.925 | 0.190 | 958.961 | 58.29 |
| 07 | 2 | 959.865 | 0.272 | 959.907 | 60.73 |
| 07 | 1 | 958.925 | 0.190 | 958.938 | 47.87 |
| 07 | 1 | 958.858 | 0.206 | 958.853 | 49.32 |
| 07 | 1 | 960.054 | 0.186 | 960.036 | 50.85 |
| 07 | 1 | 959.112 | 0.222 | 959.105 | 52.47 |
| 07 | 1 | 959.152 | 0.258 | 959.172 | 54.20 |
| 07 | 1 | 958.441 | 0.225 | 958.484 | 56.12 |
| 07 | 1 | 958.925 | 0.190 | 958.961 | 58.29 |
| 07 | 1 | 959.865 | 0.272 | 959.907 | 60.73 |
| 08 | 2 | 958.889 | 0.185 | 958.872 | 42.69 |
| 08 | 2 | 959.294 | 0.197 | 959.295 | 43.76 |
| 08 | 2 | 959.626 | 0.458 | 959.614 | 44.88 |
| 08 | 2 | 959.693 | 0.221 | 959.687 | 46.06 |
| 08 | 2 | 959.411 | 0.255 | 959.402 | 47.31 |
| 08 | 1 | 958.889 | 0.185 | 958.872 | 42.69 |
| 08 | 1 | 959.294 | 0.197 | 959.295 | 43.76 |
| 08 | 1 | 959.626 | 0.458 | 959.614 | 44.88 |
| 08 | 1 | 959.693 | 0.221 | 959.687 | 46.06 |
| 08 | 1 | 959.411 | 0.255 | 959.402 | 47.31 |
| 08 | 2 | 959.108 | 0.286 | 959.091 | 48.63 |
| 08 | 2 | 959.254 | 0.210 | 959.243 | 50.04 |
| 08 | 2 | 959.957 | 0.220 | 960.075 | 51.55 |
| 08 | 2 | 959.108 | 0.286 | 959.120 | 53.15 |
| 08 | 2 | 958.991 | 0.308 | 958.963 | 54.90 |
| 08 | 2 | 959.645 | 0.201 | 959.664 | 56.80 |
| 08 | 1 | 959.108 | 0.286 | 959.091 | 48.63 |
| 08 | 1 | 959.254 | 0.210 | 959.243 | 50.04 |
| 08 | 1 | 959.957 | 0.220 | 960.075 | 51.55 |
| 08 | 1 | 959.108 | 0.286 | 959.120 | 53.15 |
| 08 | 1 | 958.991 | 0.308 | 958.963 | 54.90 |
| 08 | 1 | 959.645 | 0.201 | 959.664 | 56.80 |
| 09 | 2 | 960.135 | 0.207 | 959.963 | 46.91 |
| 09 | 1 | 960.135 | 0.207 | 959.963 | 46.91 |
| 09 | 2 | 959.421 | 0.305 | 959.429 | 48.33 |
| 09 | 2 | 958.041 | 0.304 | 957.800 | 49.69 |
| 09 | 2 | 959.139 | 0.338 | 959.067 | 51.13 |
| 09 | 2 | 958.041 | 0.304 | 958.138 | 52.67 |

2000 - JUNHO

| D | L | SDB | ER | SDC | HL |
|---|---|---|---|---|---|
| 09 | 2 | 959.139 | 0.338 | 959.134 | 54.34 |
| 09 | 2 | 958.485 | 0.422 | 958.442 | 56.16 |
| 09 | 2 | 959.139 | 0.338 | 959.161 | 58.12 |
| 09 | 2 | 958.710 | 0.355 | 958.809 | 60.37 |
| 09 | 1 | 959.421 | 0.305 | 959.429 | 48.33 |
| 09 | 1 | 958.041 | 0.304 | 957.800 | 49.69 |
| 09 | 1 | 959.139 | 0.338 | 959.067 | 51.13 |
| 09 | 1 | 958.041 | 0.304 | 958.138 | 52.67 |
| 09 | 1 | 959.139 | 0.338 | 959.134 | 54.34 |
| 09 | 1 | 958.485 | 0.422 | 958.442 | 56.16 |
| 09 | 1 | 959.139 | 0.338 | 959.161 | 58.12 |
| 09 | 1 | 958.710 | 0.355 | 958.809 | 60.37 |
| 12 | 2 | 960.245 | 0.167 | 960.132 | 44.99 |
| 12 | 2 | 960.245 | 0.167 | 960.315 | 46.12 |
| 12 | 2 | 960.245 | 0.167 | 960.286 | 47.34 |
| 12 | 2 | 959.675 | 0.229 | 959.679 | 48.58 |
| 12 | 1 | 960.245 | 0.167 | 960.132 | 44.99 |
| 12 | 1 | 960.245 | 0.167 | 960.315 | 46.12 |
| 12 | 1 | 960.245 | 0.167 | 960.286 | 47.34 |
| 12 | 1 | 959.675 | 0.229 | 959.679 | 48.58 |
| 12 | 2 | 959.173 | 0.200 | 959.104 | 49.88 |
| 12 | 2 | 959.313 | 0.229 | 959.336 | 51.28 |
| 12 | 2 | 959.034 | 0.174 | 959.049 | 52.75 |
| 12 | 2 | 958.322 | 0.204 | 958.366 | 54.35 |
| 12 | 2 | 958.322 | 0.204 | 958.332 | 56.10 |
| 12 | 2 | 959.400 | 0.192 | 959.446 | 58.02 |
| 12 | 2 | 959.597 | 0.207 | 959.623 | 60.14 |
| 12 | 1 | 959.173 | 0.200 | 959.104 | 49.88 |
| 12 | 1 | 959.313 | 0.229 | 959.336 | 51.28 |
| 12 | 1 | 959.034 | 0.174 | 959.049 | 52.75 |
| 12 | 1 | 958.322 | 0.204 | 958.366 | 54.35 |
| 12 | 1 | 958.322 | 0.204 | 958.332 | 56.10 |
| 12 | 1 | 959.400 | 0.192 | 959.446 | 58.02 |
| 12 | 1 | 959.597 | 0.207 | 959.623 | 60.14 |
| 15 | 2 | 959.520 | 0.252 | 959.509 | 44.58 |
| 15 | 2 | 960.651 | 0.631 | 960.538 | 45.59 |
| 15 | 2 | 959.267 | 0.201 | 959.320 | 47.72 |
| 15 | 2 | 959.944 | 0.254 | 959.908 | 48.90 |
| 15 | 2 | 959.440 | 0.184 | 959.445 | 50.14 |
| 15 | 1 | 959.520 | 0.252 | 959.509 | 44.58 |
| 15 | 1 | 960.651 | 0.631 | 960.538 | 45.59 |
| 15 | 1 | 959.267 | 0.201 | 959.320 | 47.72 |
| 15 | 1 | 959.944 | 0.254 | 959.908 | 48.90 |
| 15 | 1 | 959.440 | 0.184 | 959.445 | 50.14 |
| 15 | 2 | 958.657 | 0.323 | 958.698 | 51.42 |
| 15 | 2 | 959.043 | 0.270 | 959.001 | 52.79 |
| 15 | 2 | 959.374 | 0.103 | 959.357 | 54.27 |
| 15 | 2 | 959.374 | 0.103 | 959.376 | 55.86 |
| 15 | 2 | 958.524 | 0.339 | 958.456 | 57.54 |
| 15 | 2 | 959.636 | 0.232 | 959.650 | 59.40 |
| 15 | 1 | 958.657 | 0.323 | 958.698 | 51.42 |
| 15 | 1 | 959.043 | 0.270 | 959.001 | 52.79 |
| 15 | 1 | 959.374 | 0.103 | 959.357 | 54.27 |
| 15 | 1 | 959.374 | 0.103 | 959.376 | 55.86 |
| 15 | 1 | 958.524 | 0.339 | 958.456 | 57.54 |
| 15 | 1 | 959.636 | 0.232 | 959.650 | 59.40 |
| 16 | 2 | 958.294 | 0.185 | 958.400 | 49.23 |
| 16 | 1 | 958.294 | 0.185 | 958.400 | 49.23 |
| 16 | 2 | 959.760 | 0.261 | 959.799 | 51.80 |
| 16 | 2 | 958.929 | 0.275 | 958.996 | 53.17 |
| 16 | 2 | 959.458 | 0.326 | 959.446 | 54.65 |
| 16 | 2 | 959.688 | 0.191 | 959.650 | 56.22 |
| 16 | 2 | 959.713 | 0.212 | 959.713 | 57.91 |
| 16 | 1 | 959.760 | 0.261 | 959.799 | 51.80 |
| 16 | 1 | 958.929 | 0.275 | 958.996 | 53.17 |
| 16 | 1 | 959.458 | 0.326 | 959.446 | 54.65 |
| 16 | 1 | 959.688 | 0.191 | 959.650 | 56.22 |
| 16 | 1 | 959.713 | 0.212 | 959.713 | 57.91 |
| 19 | 2 | 960.195 | 0.283 | 960.148 | 48.63 |
| 19 | 2 | 960.195 | 0.283 | 960.131 | 49.75 |
| 19 | 2 | 959.662 | 0.188 | 959.600 | 50.95 |

2000 - JUNHO

| D | L | SDB | ER | SDC | HL |
|---|---|---|---|---|---|
| 19 | 1 | 960.195 | 0.283 | 960.148 | 48.63 |
| 19 | 1 | 960.195 | 0.283 | 960.131 | 49.75 |
| 19 | 1 | 959.662 | 0.188 | 959.600 | 50.95 |
| 19 | 2 | 958.708 | 0.201 | 958.523 | 52.23 |
| 19 | 2 | 958.708 | 0.201 | 958.564 | 53.57 |
| 19 | 2 | 958.806 | 0.649 | 958.847 | 55.03 |
| 19 | 2 | 959.967 | 0.241 | 959.936 | 56.63 |
| 19 | 2 | 959.440 | 0.221 | 959.391 | 58.36 |
| 19 | 2 | 959.228 | 0.184 | 959.233 | 60.16 |
| 19 | 2 | 958.806 | 0.649 | 958.869 | 62.14 |
| 19 | 1 | 958.708 | 0.201 | 958.523 | 52.23 |
| 19 | 1 | 958.708 | 0.201 | 958.564 | 53.57 |
| 19 | 1 | 958.806 | 0.649 | 958.847 | 55.03 |
| 19 | 1 | 959.967 | 0.241 | 959.936 | 56.63 |
| 19 | 1 | 959.440 | 0.221 | 959.391 | 58.36 |
| 19 | 1 | 959.228 | 0.184 | 959.233 | 60.16 |
| 19 | 1 | 958.806 | 0.649 | 958.869 | 62.14 |
| 20 | 2 | 959.924 | 0.221 | 959.879 | 47.83 |
| 20 | 2 | 959.495 | 0.280 | 959.436 | 48.93 |
| 20 | 2 | 958.848 | 0.448 | 958.823 | 50.11 |
| 20 | 1 | 959.924 | 0.221 | 959.879 | 47.83 |
| 20 | 1 | 959.495 | 0.280 | 959.436 | 48.93 |
| 20 | 1 | 958.848 | 0.448 | 958.823 | 50.11 |
| 20 | 2 | 958.598 | 0.270 | 958.596 | 56.02 |
| 20 | 2 | 960.423 | 0.347 | 960.436 | 57.67 |
| 20 | 1 | 958.598 | 0.270 | 958.596 | 56.02 |
| 20 | 1 | 960.423 | 0.347 | 960.436 | 57.67 |
| 26 | 2 | 959.782 | 0.159 | 959.768 | 50.80 |
| 26 | 2 | 960.129 | 0.236 | 960.138 | 51.96 |
| 26 | 2 | 959.782 | 0.159 | 959.737 | 53.15 |
| 26 | 2 | 960.545 | 0.191 | 960.493 | 54.35 |
| 26 | 1 | 959.782 | 0.159 | 959.768 | 50.80 |
| 26 | 1 | 960.129 | 0.236 | 960.138 | 51.96 |
| 26 | 1 | 959.782 | 0.159 | 959.737 | 53.15 |
| 26 | 1 | 960.545 | 0.191 | 960.493 | 54.35 |
| 26 | 2 | 959.826 | 0.150 | 959.819 | 55.65 |
| 26 | 2 | 959.151 | 0.180 | 959.149 | 57.03 |
| 26 | 2 | 959.659 | 0.147 | 959.719 | 58.49 |
| 26 | 2 | 958.892 | 0.171 | 958.897 | 60.02 |
| 26 | 2 | 958.892 | 0.171 | 958.944 | 61.70 |
| 26 | 2 | 959.023 | 0.181 | 958.995 | 63.50 |
| 26 | 2 | 959.923 | 0.165 | 959.925 | 65.51 |
| 26 | 1 | 959.826 | 0.150 | 959.819 | 55.65 |
| 26 | 1 | 959.151 | 0.180 | 959.149 | 57.03 |
| 26 | 1 | 959.659 | 0.147 | 959.719 | 58.49 |
| 26 | 1 | 958.892 | 0.171 | 958.897 | 60.02 |
| 26 | 1 | 958.892 | 0.171 | 958.944 | 61.70 |
| 26 | 1 | 959.023 | 0.181 | 958.995 | 63.50 |
| 26 | 1 | 959.923 | 0.165 | 959.925 | 65.51 |
| 29 | 2 | 959.999 | 0.276 | 960.008 | 50.81 |
| 29 | 2 | 958.894 | 0.614 | 958.952 | 51.83 |
| 29 | 2 | 958.894 | 0.614 | 958.879 | 52.91 |
| 29 | 2 | 959.444 | 0.246 | 959.420 | 54.01 |
| 29 | 2 | 958.667 | 0.176 | 958.584 | 55.17 |
| 29 | 1 | 959.999 | 0.276 | 960.008 | 50.81 |
| 29 | 1 | 958.894 | 0.614 | 958.952 | 51.83 |
| 29 | 1 | 958.894 | 0.614 | 958.879 | 52.91 |
| 29 | 1 | 959.444 | 0.246 | 959.420 | 54.01 |
| 29 | 1 | 958.667 | 0.176 | 958.584 | 55.17 |
| 29 | 2 | 960.071 | 0.185 | 960.073 | 56.40 |
| 29 | 2 | 958.700 | 0.184 | 958.697 | 57.98 |
| 29 | 2 | 959.534 | 0.241 | 959.553 | 59.52 |
| 29 | 2 | 958.750 | 0.197 | 958.776 | 61.03 |
| 29 | 2 | 958.461 | 0.266 | 958.522 | 62.63 |
| 29 | 2 | 959.105 | 0.188 | 959.121 | 64.35 |
| 29 | 2 | 959.632 | 0.298 | 959.759 | 66.25 |
| 29 | 1 | 960.071 | 0.185 | 960.073 | 56.40 |
| 29 | 1 | 958.700 | 0.184 | 958.697 | 57.98 |
| 29 | 1 | 959.534 | 0.241 | 959.553 | 59.52 |
| 29 | 1 | 958.750 | 0.197 | 958.776 | 61.03 |
| 29 | 1 | 958.461 | 0.266 | 958.522 | 62.63 |

| 2000 - JUNHO | | | | | |
|---|---|---|---|---|---|
| D | L | SDB | ER | SDC | HL |
| 29 | 1 | 959.105 | 0.188 | 959.121 | 64.35 |
| 29 | 1 | 959.632 | 0.298 | 959.759 | 66.25 |
| 30 | 2 | 959.419 | 0.108 | 959.326 | 55.77 |
| 30 | 1 | 959.419 | 0.108 | 959.326 | 55.77 |
| 30 | 2 | 959.419 | 0.108 | 959.373 | 57.05 |
| 30 | 2 | 959.419 | 0.108 | 959.271 | 58.41 |
| 30 | 2 | 958.994 | 0.463 | 959.126 | 59.81 |
| 30 | 2 | 959.419 | 0.108 | 959.267 | 61.29 |
| 30 | 2 | 959.791 | 0.089 | 959.810 | 62.88 |
| 30 | 2 | 959.637 | 0.091 | 959.639 | 64.59 |
| 30 | 1 | 959.419 | 0.108 | 959.373 | 57.05 |
| 30 | 1 | 959.419 | 0.108 | 959.271 | 58.41 |
| 30 | 1 | 958.994 | 0.463 | 959.126 | 59.81 |
| 30 | 1 | 959.419 | 0.108 | 959.267 | 61.29 |
| 30 | 1 | 959.791 | 0.089 | 959.810 | 62.88 |
| 30 | 1 | 959.637 | 0.091 | 959.639 | 64.59 |

| 2000 - JULHO | | | | | |
|---|---|---|---|---|---|
| D | L | SDB | ER | SDC | HL |
| 07 | 2 | 959.091 | 0.181 | 959.153 | 54.73 |
| 07 | 2 | 959.030 | 0.156 | 959.053 | 55.76 |
| 07 | 2 | 958.329 | 0.150 | 958.228 | 56.84 |
| 07 | 2 | 958.702 | 0.151 | 958.683 | 57.95 |
| 07 | 1 | 959.091 | 0.181 | 959.153 | 54.73 |
| 07 | 1 | 959.030 | 0.156 | 959.053 | 55.76 |
| 07 | 1 | 958.329 | 0.150 | 958.228 | 56.84 |
| 07 | 1 | 958.702 | 0.151 | 958.683 | 57.95 |
| 07 | 2 | 959.030 | 0.156 | 959.060 | 59.13 |
| 07 | 2 | 958.610 | 0.204 | 958.644 | 60.62 |
| 07 | 2 | 958.979 | 0.148 | 958.973 | 61.97 |
| 07 | 2 | 958.329 | 0.150 | 958.263 | 63.41 |
| 07 | 2 | 957.732 | 0.147 | 957.699 | 65.50 |
| 07 | 2 | 958.610 | 0.204 | 958.596 | 67.19 |
| 07 | 1 | 959.030 | 0.156 | 959.060 | 59.13 |
| 07 | 1 | 958.610 | 0.204 | 958.644 | 60.62 |
| 07 | 1 | 958.979 | 0.148 | 958.973 | 61.97 |
| 07 | 1 | 958.329 | 0.150 | 958.263 | 63.41 |
| 07 | 1 | 957.732 | 0.147 | 957.699 | 65.50 |
| 07 | 1 | 958.610 | 0.204 | 958.596 | 67.19 |
| 10 | 2 | 960.048 | 0.153 | 959.965 | 57.21 |
| 10 | 2 | 959.484 | 0.156 | 959.511 | 58.35 |
| 10 | 2 | 960.048 | 0.153 | 960.074 | 59.64 |
| 10 | 1 | 960.048 | 0.153 | 959.965 | 57.21 |
| 10 | 1 | 959.484 | 0.156 | 959.511 | 58.35 |
| 10 | 1 | 960.048 | 0.153 | 960.074 | 59.64 |
| 10 | 2 | 958.903 | 0.136 | 958.879 | 61.02 |
| 10 | 2 | 959.154 | 0.145 | 959.135 | 62.39 |
| 10 | 2 | 958.903 | 0.136 | 958.756 | 63.91 |
| 10 | 2 | 959.016 | 0.131 | 958.991 | 65.44 |
| 10 | 2 | 959.317 | 0.132 | 959.329 | 67.07 |
| 10 | 2 | 960.048 | 0.153 | 960.028 | 68.84 |
| 10 | 2 | 959.357 | 0.120 | 959.348 | 70.76 |
| 10 | 2 | 959.154 | 0.145 | 959.140 | 72.90 |
| 10 | 1 | 958.903 | 0.136 | 958.879 | 61.02 |
| 10 | 1 | 959.154 | 0.145 | 959.135 | 62.39 |
| 10 | 1 | 958.903 | 0.136 | 958.756 | 63.91 |
| 10 | 1 | 959.016 | 0.131 | 958.991 | 65.44 |
| 10 | 1 | 959.317 | 0.132 | 959.329 | 67.07 |
| 10 | 1 | 960.048 | 0.153 | 960.028 | 68.84 |
| 10 | 1 | 959.357 | 0.120 | 959.348 | 70.76 |
| 10 | 1 | 959.154 | 0.145 | 959.140 | 72.90 |
| 11 | 2 | 959.897 | 0.135 | 959.865 | 58.17 |
| 11 | 2 | 959.508 | 0.137 | 959.522 | 59.41 |
| 11 | 1 | 959.897 | 0.135 | 959.865 | 58.17 |
| 11 | 1 | 959.508 | 0.137 | 959.522 | 59.41 |
| 11 | 2 | 959.243 | 0.151 | 959.271 | 60.65 |
| 11 | 2 | 959.605 | 0.141 | 959.595 | 62.28 |
| 11 | 2 | 958.889 | 0.122 | 958.792 | 63.78 |
| 11 | 2 | 959.648 | 0.113 | 959.683 | 65.34 |
| 11 | 2 | 958.889 | 0.122 | 958.866 | 67.34 |

| 2000 - JULHO | | | | | |
|---|---|---|---|---|---|
| D | L | SDB | ER | SDC | HL |
| 11 | 2 | 958.970 | 0.133 | 958.935 | 69.66 |
| 11 | 2 | 959.747 | 0.122 | 959.743 | 71.98 |
| 11 | 1 | 959.243 | 0.151 | 959.271 | 60.65 |
| 11 | 1 | 959.605 | 0.141 | 959.595 | 62.28 |
| 11 | 1 | 958.889 | 0.122 | 958.792 | 63.78 |
| 11 | 1 | 959.648 | 0.113 | 959.683 | 65.34 |
| 11 | 1 | 958.889 | 0.122 | 958.866 | 67.34 |
| 11 | 1 | 958.970 | 0.133 | 958.935 | 69.66 |
| 11 | 1 | 959.747 | 0.122 | 959.743 | 71.98 |
| 17 | 2 | 959.389 | 0.157 | 959.395 | 57.78 |
| 17 | 2 | 959.389 | 0.157 | 959.460 | 58.79 |
| 17 | 1 | 959.389 | 0.157 | 959.395 | 57.78 |
| 17 | 1 | 959.389 | 0.157 | 959.460 | 58.79 |
| 17 | 2 | 959.101 | 0.134 | 959.029 | 62.17 |
| 17 | 2 | 958.687 | 0.167 | 958.606 | 63.41 |
| 17 | 2 | 958.886 | 0.177 | 958.948 | 64.72 |
| 17 | 2 | 959.389 | 0.157 | 959.454 | 66.11 |
| 17 | 2 | 958.886 | 0.177 | 958.859 | 67.60 |
| 17 | 2 | 959.101 | 0.134 | 959.068 | 69.23 |
| 17 | 2 | 959.745 | 0.149 | 959.718 | 70.99 |
| 17 | 2 | 959.101 | 0.134 | 959.003 | 72.84 |
| 17 | 2 | 958.349 | 0.141 | 958.123 | 74.90 |
| 17 | 1 | 959.101 | 0.134 | 959.029 | 62.17 |
| 17 | 1 | 958.687 | 0.167 | 958.606 | 63.41 |
| 17 | 1 | 958.886 | 0.177 | 958.948 | 64.72 |
| 17 | 1 | 959.389 | 0.157 | 959.454 | 66.11 |
| 17 | 1 | 958.886 | 0.177 | 958.859 | 67.60 |
| 17 | 1 | 959.101 | 0.134 | 959.068 | 69.23 |
| 17 | 1 | 959.745 | 0.149 | 959.718 | 70.99 |
| 17 | 1 | 959.101 | 0.134 | 959.003 | 72.84 |
| 17 | 1 | 958.349 | 0.141 | 958.123 | 74.90 |
| 18 | 2 | 959.739 | 0.214 | 959.733 | 56.73 |
| 18 | 2 | 959.283 | 0.148 | 959.273 | 57.75 |
| 18 | 2 | 959.558 | 0.160 | 959.577 | 58.85 |
| 18 | 2 | 957.548 | 0.180 | 957.539 | 59.94 |
| 18 | 2 | 959.138 | 0.148 | 959.148 | 61.09 |
| 18 | 2 | 958.830 | 0.165 | 958.855 | 62.30 |
| 18 | 1 | 959.739 | 0.214 | 959.733 | 56.73 |
| 18 | 1 | 959.283 | 0.148 | 959.273 | 57.75 |
| 18 | 1 | 959.558 | 0.160 | 959.577 | 58.85 |
| 18 | 1 | 957.548 | 0.180 | 957.539 | 59.94 |
| 18 | 1 | 959.138 | 0.148 | 959.148 | 61.09 |
| 18 | 1 | 958.830 | 0.165 | 958.855 | 62.30 |
| 18 | 2 | 958.731 | 0.216 | 958.714 | 63.58 |
| 18 | 2 | 958.086 | 0.168 | 958.040 | 64.92 |
| 18 | 2 | 958.972 | 0.210 | 958.986 | 66.36 |
| 18 | 2 | 958.786 | 0.158 | 958.759 | 67.86 |
| 18 | 2 | 959.025 | 0.184 | 959.035 | 69.63 |
| 18 | 2 | 959.476 | 0.183 | 959.474 | 71.52 |
| 18 | 1 | 958.731 | 0.216 | 958.714 | 63.58 |
| 18 | 1 | 958.086 | 0.168 | 958.040 | 64.92 |
| 18 | 1 | 958.972 | 0.210 | 958.986 | 66.36 |
| 18 | 1 | 958.786 | 0.158 | 958.759 | 67.86 |
| 18 | 1 | 959.025 | 0.184 | 959.035 | 69.63 |
| 18 | 1 | 959.476 | 0.183 | 959.474 | 71.52 |
| 21 | 2 | 959.429 | 0.204 | 959.410 | 63.41 |
| 21 | 2 | 957.277 | 0.248 | 957.009 | 64.58 |
| 21 | 2 | 959.216 | 0.152 | 959.220 | 65.94 |
| 21 | 2 | 959.114 | 0.209 | 959.110 | 67.39 |
| 21 | 2 | 958.725 | 0.196 | 958.825 | 68.77 |
| 21 | 2 | 958.601 | 0.180 | 958.421 | 70.27 |
| 21 | 1 | 959.429 | 0.204 | 959.410 | 63.41 |
| 21 | 1 | 957.277 | 0.248 | 957.009 | 64.58 |
| 21 | 1 | 959.216 | 0.152 | 959.220 | 65.94 |
| 21 | 1 | 959.114 | 0.209 | 959.110 | 67.39 |
| 21 | 1 | 958.725 | 0.196 | 958.825 | 68.77 |
| 21 | 1 | 958.601 | 0.180 | 958.421 | 70.27 |
| 28 | 2 | 959.006 | 0.168 | 959.014 | 61.34 |
| 28 | 2 | 958.909 | 0.144 | 958.825 | 62.28 |
| 28 | 2 | 958.909 | 0.144 | 958.890 | 63.64 |
| 28 | 2 | 958.045 | 0.264 | 957.865 | 64.68 |

| 2000 - JULHO ||||| | 2000 - AGOSTO |||||
|---|---|---|---|---|---|---|---|---|---|---|
| D | L | SDB | ER | SDC | HL | D | L | SDB | ER | SDC | HL |
| 28 | 1 | 959.006 | 0.168 | 959.014 | 61.34 | 03 | 2 | 959.456 | 0.151 | 959.512 | 76.63 |
| 28 | 1 | 958.909 | 0.144 | 958.825 | 62.28 | 03 | 2 | 959.367 | 0.123 | 959.377 | 78.27 |
| 28 | 1 | 958.909 | 0.144 | 958.890 | 63.64 | 03 | 2 | 959.060 | 0.129 | 959.059 | 80.02 |
| 28 | 1 | 958.045 | 0.264 | 957.865 | 64.68 | 03 | 2 | 958.135 | 0.144 | 958.118 | 81.81 |
| 28 | 2 | 958.313 | 0.231 | 958.333 | 66.23 | 03 | 1 | 959.137 | 0.162 | 959.138 | 67.35 |
| 28 | 1 | 958.313 | 0.231 | 958.333 | 66.23 | 03 | 1 | 958.980 | 0.148 | 958.966 | 68.46 |
| 31 | 2 | 959.664 | 0.150 | 959.692 | 61.25 | 03 | 1 | 958.980 | 0.148 | 958.780 | 69.71 |
| 31 | 2 | 959.182 | 0.160 | 959.140 | 62.33 | 03 | 1 | 957.982 | 0.123 | 957.875 | 70.96 |
| 31 | 2 | 959.437 | 0.181 | 959.422 | 63.33 | 03 | 1 | 959.366 | 0.163 | 959.311 | 72.27 |
| 31 | 2 | 958.407 | 0.387 | 958.360 | 64.38 | 03 | 1 | 958.980 | 0.148 | 958.892 | 73.62 |
| 31 | 2 | 958.498 | 0.226 | 958.513 | 65.47 | 03 | 1 | 959.109 | 0.144 | 959.111 | 75.09 |
| 31 | 1 | 959.664 | 0.150 | 959.692 | 61.25 | 03 | 1 | 959.456 | 0.151 | 959.512 | 76.63 |
| 31 | 1 | 959.182 | 0.160 | 959.140 | 62.33 | 03 | 1 | 959.367 | 0.123 | 959.377 | 78.27 |
| 31 | 1 | 959.437 | 0.181 | 959.422 | 63.33 | 03 | 1 | 959.060 | 0.129 | 959.059 | 80.02 |
| 31 | 1 | 958.407 | 0.387 | 958.360 | 64.38 | 03 | 1 | 958.135 | 0.144 | 958.118 | 81.81 |
| 31 | 1 | 958.498 | 0.226 | 958.513 | 65.47 | 09 | 2 | 957.559 | 0.104 | 957.637 | 63.11 |
| 31 | 2 | 959.664 | 0.150 | 959.634 | 66.57 | 09 | 2 | 959.099 | 0.121 | 959.061 | 63.99 |
| 31 | 2 | 959.329 | 0.147 | 959.352 | 67.74 | 09 | 2 | 959.888 | 0.124 | 959.880 | 64.90 |
| 31 | 2 | 958.635 | 0.161 | 958.638 | 69.04 | 09 | 2 | 959.615 | 0.102 | 959.623 | 68.20 |
| 31 | 2 | 959.207 | 0.148 | 959.202 | 70.52 | 09 | 1 | 957.559 | 0.104 | 957.637 | 63.11 |
| 31 | 2 | 958.443 | 0.165 | 958.456 | 71.90 | 09 | 1 | 959.099 | 0.121 | 959.061 | 63.99 |
| 31 | 2 | 958.407 | 0.387 | 958.352 | 73.43 | 09 | 1 | 959.888 | 0.124 | 959.880 | 64.90 |
| 31 | 2 | 959.479 | 0.131 | 959.466 | 75.10 | 09 | 1 | 959.615 | 0.102 | 959.623 | 68.20 |
| 31 | 2 | 959.037 | 0.159 | 959.099 | 77.23 | 09 | 2 | 958.945 | 0.127 | 958.944 | 69.50 |
| 31 | 2 | 958.829 | 0.135 | 958.843 | 79.09 | 09 | 2 | 959.221 | 0.105 | 959.213 | 70.80 |
| 31 | 1 | 959.664 | 0.150 | 959.634 | 66.57 | 09 | 2 | 958.739 | 0.119 | 958.687 | 72.22 |
| 31 | 1 | 959.329 | 0.147 | 959.352 | 67.74 | 09 | 2 | 959.324 | 0.119 | 959.317 | 73.49 |
| 31 | 1 | 958.635 | 0.161 | 958.638 | 69.04 | 09 | 2 | 958.988 | 0.141 | 958.994 | 74.87 |
| 31 | 1 | 959.207 | 0.148 | 959.202 | 70.52 | 09 | 2 | 959.162 | 0.101 | 959.133 | 76.38 |
| 31 | 1 | 958.443 | 0.165 | 958.456 | 71.90 | 09 | 2 | 959.957 | 0.133 | 959.958 | 77.95 |
| 31 | 1 | 958.407 | 0.387 | 958.352 | 73.43 | 09 | 2 | 959.099 | 0.121 | 959.103 | 79.59 |
| 31 | 1 | 959.479 | 0.131 | 959.466 | 75.10 | 09 | 1 | 958.945 | 0.127 | 958.944 | 69.50 |
| 31 | 1 | 959.037 | 0.159 | 959.099 | 77.23 | 09 | 1 | 959.221 | 0.105 | 959.213 | 70.80 |
| 31 | 1 | 958.829 | 0.135 | 958.843 | 79.09 | 09 | 1 | 958.739 | 0.119 | 958.687 | 72.22 |
| | | | | | | 09 | 1 | 959.324 | 0.119 | 959.317 | 73.49 |
| | | | | | | 09 | 1 | 958.988 | 0.141 | 958.994 | 74.87 |
| | | 2000 - AGOSTO | | | | 09 | 1 | 959.162 | 0.101 | 959.133 | 76.38 |
| D | L | SDB | ER | SDC | HL | 09 | 1 | 959.957 | 0.133 | 959.958 | 77.95 |
| 02 | 2 | 959.977 | 0.132 | 959.988 | 62.26 | 09 | 1 | 959.099 | 0.121 | 959.103 | 79.59 |
| 02 | 2 | 959.131 | 0.126 | 959.134 | 63.23 | 14 | 2 | 960.321 | 0.186 | 960.311 | 62.55 |
| 02 | 2 | 959.281 | 0.159 | 959.276 | 64.27 | 14 | 2 | 959.853 | 0.181 | 959.875 | 63.37 |
| 02 | 2 | 959.016 | 0.134 | 959.023 | 65.55 | 14 | 2 | 959.095 | 0.197 | 959.113 | 64.21 |
| 02 | 2 | 959.526 | 0.206 | 959.547 | 66.65 | 14 | 2 | 959.290 | 0.145 | 959.250 | 65.09 |
| 02 | 1 | 959.977 | 0.132 | 959.988 | 62.26 | 14 | 2 | 959.633 | 0.158 | 959.591 | 65.99 |
| 02 | 1 | 959.131 | 0.126 | 959.134 | 63.23 | 14 | 2 | 958.519 | 0.156 | 958.501 | 66.92 |
| 02 | 1 | 959.281 | 0.159 | 959.276 | 64.27 | 14 | 2 | 959.456 | 0.174 | 959.473 | 67.93 |
| 02 | 1 | 959.016 | 0.134 | 959.023 | 65.55 | 14 | 2 | 958.600 | 0.139 | 958.570 | 68.95 |
| 02 | 1 | 959.526 | 0.206 | 959.547 | 66.65 | 14 | 1 | 960.321 | 0.186 | 960.311 | 62.55 |
| 02 | 2 | 959.062 | 0.159 | 959.052 | 67.85 | 14 | 1 | 959.853 | 0.181 | 959.875 | 63.37 |
| 02 | 2 | 959.526 | 0.206 | 959.548 | 69.08 | 14 | 1 | 959.095 | 0.197 | 959.113 | 64.21 |
| 02 | 2 | 958.443 | 0.229 | 958.359 | 70.37 | 14 | 1 | 959.290 | 0.145 | 959.250 | 65.09 |
| 02 | 2 | 959.193 | 0.201 | 959.195 | 73.59 | 14 | 1 | 959.633 | 0.158 | 959.591 | 65.99 |
| 02 | 2 | 958.443 | 0.229 | 958.473 | 75.38 | 14 | 1 | 958.519 | 0.156 | 958.501 | 66.92 |
| 02 | 2 | 958.900 | 0.150 | 958.869 | 77.05 | 14 | 1 | 959.456 | 0.174 | 959.473 | 67.93 |
| 02 | 2 | 958.145 | 0.175 | 957.824 | 78.79 | 14 | 1 | 958.600 | 0.139 | 958.570 | 68.95 |
| 02 | 2 | 958.813 | 0.137 | 958.799 | 80.88 | 14 | 2 | 958.916 | 0.190 | 958.939 | 70.29 |
| 02 | 1 | 959.062 | 0.159 | 959.052 | 67.85 | 14 | 2 | 960.107 | 0.154 | 960.104 | 71.45 |
| 02 | 1 | 959.526 | 0.206 | 959.548 | 69.08 | 14 | 2 | 958.916 | 0.190 | 958.890 | 72.65 |
| 02 | 1 | 958.443 | 0.229 | 958.359 | 70.37 | 14 | 2 | 959.148 | 0.144 | 959.162 | 73.94 |
| 02 | 1 | 959.193 | 0.201 | 959.195 | 73.59 | 14 | 2 | 959.724 | 0.155 | 959.706 | 75.30 |
| 02 | 1 | 958.443 | 0.229 | 958.473 | 75.38 | 14 | 2 | 958.797 | 0.164 | 958.761 | 76.67 |
| 02 | 1 | 958.900 | 0.150 | 958.869 | 77.05 | 14 | 2 | 958.448 | 0.147 | 958.479 | 78.11 |
| 02 | 1 | 958.145 | 0.175 | 957.824 | 78.79 | 14 | 2 | 959.073 | 0.153 | 959.082 | 79.56 |
| 02 | 1 | 958.813 | 0.137 | 958.799 | 80.88 | 14 | 1 | 958.916 | 0.190 | 958.939 | 70.29 |
| 03 | 2 | 959.137 | 0.162 | 959.138 | 67.35 | 14 | 1 | 960.107 | 0.154 | 960.104 | 71.45 |
| 03 | 2 | 958.980 | 0.148 | 958.966 | 68.46 | 14 | 1 | 958.916 | 0.190 | 958.890 | 72.65 |
| 03 | 2 | 958.980 | 0.148 | 958.780 | 69.71 | 14 | 1 | 959.148 | 0.144 | 959.162 | 73.94 |
| 03 | 2 | 957.982 | 0.123 | 957.875 | 70.96 | 14 | 1 | 959.724 | 0.155 | 959.706 | 75.30 |
| 03 | 2 | 959.366 | 0.163 | 959.311 | 72.27 | 14 | 1 | 958.797 | 0.164 | 958.761 | 76.67 |
| 03 | 2 | 958.980 | 0.148 | 958.892 | 73.62 | 14 | 1 | 958.448 | 0.147 | 958.479 | 78.11 |
| 03 | 2 | 959.109 | 0.144 | 959.111 | 75.09 | 14 | 1 | 959.073 | 0.153 | 959.082 | 79.56 |

|  | 2000 - AGOSTO | | | |
|---|---|---|---|---|
| D | L | SDB | ER | SDC | HL |
| 21 | 2 | 958.331 | 0.164 | 958.337 | 63.09 |
| 21 | 2 | 958.141 | 0.139 | 958.111 | 63.99 |
| 21 | 2 | 959.177 | 0.148 | 959.194 | 64.78 |
| 21 | 2 | 958.617 | 0.165 | 958.597 | 65.68 |
| 21 | 2 | 959.141 | 0.144 | 959.147 | 66.56 |
| 21 | 2 | 959.053 | 0.182 | 959.063 | 67.43 |
| 21 | 2 | 959.873 | 0.171 | 959.970 | 68.38 |
| 21 | 1 | 958.331 | 0.164 | 958.337 | 63.09 |
| 21 | 1 | 958.141 | 0.139 | 958.111 | 63.99 |
| 21 | 1 | 959.177 | 0.148 | 959.194 | 64.78 |
| 21 | 1 | 958.617 | 0.165 | 958.597 | 65.68 |
| 21 | 1 | 959.141 | 0.144 | 959.147 | 66.56 |
| 21 | 1 | 959.053 | 0.182 | 959.063 | 67.43 |
| 21 | 1 | 959.873 | 0.171 | 959.970 | 68.38 |
| 21 | 2 | 958.762 | 0.158 | 958.715 | 70.58 |
| 21 | 2 | 959.288 | 0.142 | 959.283 | 71.66 |
| 21 | 2 | 959.238 | 0.194 | 959.239 | 72.78 |
| 21 | 2 | 959.288 | 0.142 | 959.278 | 73.94 |
| 21 | 2 | 958.790 | 0.181 | 958.802 | 75.17 |
| 21 | 2 | 959.228 | 0.131 | 959.208 | 76.47 |
| 21 | 2 | 958.772 | 0.173 | 958.773 | 77.77 |
| 21 | 2 | 958.141 | 0.139 | 958.067 | 79.11 |
| 21 | 2 | 959.053 | 0.182 | 959.081 | 80.45 |
| 21 | 1 | 958.762 | 0.158 | 958.715 | 70.58 |
| 21 | 1 | 959.288 | 0.142 | 959.283 | 71.66 |
| 21 | 1 | 959.238 | 0.194 | 959.239 | 72.78 |
| 21 | 1 | 959.288 | 0.142 | 959.278 | 73.94 |
| 21 | 1 | 958.790 | 0.181 | 958.802 | 75.17 |
| 21 | 1 | 959.228 | 0.131 | 959.208 | 76.47 |
| 21 | 1 | 958.772 | 0.173 | 958.773 | 77.77 |
| 21 | 1 | 958.141 | 0.139 | 958.067 | 79.11 |
| 21 | 1 | 959.053 | 0.182 | 959.081 | 80.45 |
| 22 | 2 | 959.020 | 0.173 | 959.025 | 60.68 |
| 22 | 2 | 959.315 | 0.143 | 959.289 | 61.40 |
| 22 | 2 | 958.337 | 0.182 | 958.404 | 62.17 |
| 22 | 2 | 959.168 | 0.153 | 959.178 | 62.97 |
| 22 | 2 | 958.960 | 0.154 | 958.981 | 63.75 |
| 22 | 2 | 958.503 | 0.168 | 958.524 | 64.60 |
| 22 | 2 | 959.795 | 0.177 | 959.833 | 65.50 |
| 22 | 2 | 960.011 | 0.130 | 959.989 | 66.42 |
| 22 | 2 | 959.514 | 0.152 | 959.538 | 67.35 |
| 22 | 2 | 959.605 | 0.131 | 959.633 | 68.33 |
| 22 | 2 | 958.772 | 0.154 | 958.755 | 69.36 |
| 22 | 2 | 958.892 | 0.137 | 958.909 | 70.52 |
| 22 | 1 | 959.020 | 0.173 | 959.025 | 60.68 |
| 22 | 1 | 959.315 | 0.143 | 959.289 | 61.40 |
| 22 | 1 | 958.337 | 0.182 | 958.404 | 62.17 |
| 22 | 1 | 959.168 | 0.153 | 959.178 | 62.97 |
| 22 | 1 | 958.960 | 0.154 | 958.981 | 63.75 |
| 22 | 1 | 958.503 | 0.168 | 958.524 | 64.60 |
| 22 | 1 | 959.795 | 0.177 | 959.833 | 65.50 |
| 22 | 1 | 960.011 | 0.130 | 959.989 | 66.42 |
| 22 | 1 | 959.514 | 0.152 | 959.538 | 67.35 |
| 22 | 1 | 959.605 | 0.131 | 959.633 | 68.33 |
| 22 | 1 | 958.772 | 0.154 | 958.755 | 69.36 |
| 22 | 1 | 958.892 | 0.137 | 958.909 | 70.52 |
| 22 | 2 | 959.332 | 0.150 | 959.338 | 71.66 |
| 22 | 2 | 958.221 | 0.126 | 958.192 | 72.85 |
| 22 | 2 | 958.497 | 0.115 | 958.497 | 74.10 |
| 22 | 2 | 958.772 | 0.154 | 958.820 | 75.39 |
| 22 | 1 | 959.332 | 0.150 | 959.338 | 71.66 |
| 22 | 1 | 958.221 | 0.126 | 958.192 | 72.85 |
| 22 | 1 | 958.497 | 0.115 | 958.497 | 74.10 |
| 22 | 1 | 958.772 | 0.154 | 958.820 | 75.39 |
| 23 | 2 | 960.293 | 0.123 | 960.153 | 64.55 |
| 23 | 2 | 958.870 | 0.138 | 958.920 | 65.40 |
| 23 | 2 | 959.421 | 0.156 | 959.418 | 66.29 |
| 23 | 2 | 959.480 | 0.173 | 959.478 | 67.14 |
| 23 | 2 | 959.296 | 0.152 | 959.326 | 68.06 |
| 23 | 2 | 959.421 | 0.156 | 959.414 | 69.28 |
| 23 | 2 | 958.225 | 0.147 | 958.232 | 70.25 |

|  | 2000 - AGOSTO | | | |
|---|---|---|---|---|
| D | L | SDB | ER | SDC | HL |
| 23 | 1 | 960.293 | 0.123 | 960.153 | 64.55 |
| 23 | 1 | 958.870 | 0.138 | 958.920 | 65.40 |
| 23 | 1 | 959.421 | 0.156 | 959.418 | 66.29 |
| 23 | 1 | 959.480 | 0.173 | 959.478 | 67.14 |
| 23 | 1 | 959.296 | 0.152 | 959.326 | 68.06 |
| 23 | 1 | 959.421 | 0.156 | 959.414 | 69.28 |
| 23 | 1 | 958.225 | 0.147 | 958.232 | 70.25 |
| 23 | 2 | 959.057 | 0.145 | 959.086 | 71.26 |
| 23 | 2 | 959.402 | 0.142 | 959.371 | 72.37 |
| 23 | 2 | 958.573 | 0.198 | 958.614 | 73.54 |
| 23 | 2 | 958.809 | 0.161 | 958.825 | 74.70 |
| 23 | 2 | 959.015 | 0.125 | 959.006 | 75.93 |
| 23 | 2 | 959.537 | 0.163 | 959.535 | 77.21 |
| 23 | 2 | 958.661 | 0.205 | 958.659 | 78.50 |
| 23 | 1 | 959.057 | 0.145 | 959.086 | 71.26 |
| 23 | 1 | 959.402 | 0.142 | 959.371 | 72.37 |
| 23 | 1 | 958.573 | 0.198 | 958.614 | 73.54 |
| 23 | 1 | 958.809 | 0.161 | 958.825 | 74.70 |
| 23 | 1 | 959.015 | 0.125 | 959.006 | 75.93 |
| 23 | 1 | 959.537 | 0.163 | 959.535 | 77.21 |
| 23 | 1 | 958.661 | 0.205 | 958.659 | 78.50 |
| 24 | 2 | 959.250 | 0.111 | 959.237 | 76.68 |
| 24 | 2 | 960.118 | 0.160 | 960.183 | 77.91 |
| 24 | 2 | 959.250 | 0.111 | 959.203 | 79.18 |
| 24 | 2 | 959.250 | 0.111 | 959.219 | 80.43 |
| 24 | 2 | 959.514 | 0.145 | 959.515 | 81.61 |
| 24 | 1 | 959.250 | 0.111 | 959.237 | 76.68 |
| 24 | 1 | 960.118 | 0.160 | 960.183 | 77.91 |
| 24 | 1 | 959.250 | 0.111 | 959.203 | 79.18 |
| 24 | 1 | 959.250 | 0.111 | 959.219 | 80.43 |
| 24 | 1 | 959.514 | 0.145 | 959.515 | 81.61 |
| 25 | 2 | 959.342 | 0.148 | 959.329 | 68.30 |
| 25 | 2 | 958.530 | 0.222 | 958.423 | 69.21 |
| 25 | 2 | 959.545 | 0.155 | 959.556 | 70.12 |
| 25 | 1 | 959.342 | 0.148 | 959.329 | 68.30 |
| 25 | 1 | 958.530 | 0.222 | 958.423 | 69.21 |
| 25 | 1 | 959.545 | 0.155 | 959.556 | 70.12 |
| 25 | 2 | 957.937 | 0.193 | 957.758 | 71.09 |
| 25 | 2 | 959.057 | 0.160 | 959.057 | 72.10 |
| 25 | 2 | 959.057 | 0.160 | 959.046 | 73.16 |
| 25 | 2 | 958.605 | 0.258 | 958.744 | 74.26 |
| 25 | 2 | 959.968 | 0.161 | 959.930 | 75.41 |
| 25 | 2 | 959.618 | 0.189 | 959.619 | 76.60 |
| 25 | 2 | 960.204 | 0.135 | 960.204 | 77.82 |
| 25 | 2 | 959.745 | 0.135 | 959.728 | 79.08 |
| 25 | 2 | 959.753 | 0.164 | 959.755 | 80.34 |
| 25 | 2 | 959.342 | 0.148 | 959.371 | 81.51 |
| 25 | 1 | 957.937 | 0.193 | 957.758 | 71.09 |
| 25 | 1 | 959.057 | 0.160 | 959.057 | 72.10 |
| 25 | 1 | 959.057 | 0.160 | 959.046 | 73.16 |
| 25 | 1 | 958.605 | 0.258 | 958.744 | 74.26 |
| 25 | 1 | 959.968 | 0.161 | 959.930 | 75.41 |
| 25 | 1 | 959.618 | 0.189 | 959.619 | 76.60 |
| 25 | 1 | 960.204 | 0.135 | 960.204 | 77.82 |
| 25 | 1 | 959.745 | 0.135 | 959.728 | 79.08 |
| 25 | 1 | 959.753 | 0.164 | 959.755 | 80.34 |
| 25 | 1 | 959.342 | 0.148 | 959.371 | 81.51 |

|  | 2000 - SETEMBRO | | | |
|---|---|---|---|---|
| D | L | SDB | ER | SDC | HL |
| 01 | 2 | 959.313 | 0.137 | 959.300 | 69.85 |
| 01 | 2 | 958.606 | 0.125 | 958.534 | 70.78 |
| 01 | 1 | 959.313 | 0.137 | 959.300 | 69.85 |
| 01 | 1 | 958.606 | 0.125 | 958.534 | 70.78 |
| 01 | 2 | 959.663 | 0.131 | 959.660 | 71.73 |
| 01 | 2 | 958.979 | 0.149 | 958.979 | 72.72 |
| 01 | 2 | 958.413 | 0.146 | 958.469 | 73.78 |
| 01 | 2 | 959.313 | 0.137 | 959.312 | 74.87 |
| 01 | 2 | 959.313 | 0.137 | 959.332 | 76.00 |
| 01 | 2 | 959.015 | 0.132 | 959.018 | 77.16 |

| 2000 - SETEMBRO | | | | | | 2000 - SETEMBRO | | | | |
|---|---|---|---|---|---|---|---|---|---|---|
| D | L | SDB | ER | SDC | HL | D | L | SDB | ER | SDC | HL |
| 01 | 2 | 959.532 | 0.144 | 959.526 | 78.34 | 15 | 2 | 959.600 | 0.200 | 959.599 | 75.76 |
| 01 | 2 | 958.737 | 0.165 | 958.767 | 79.56 | 15 | 2 | 959.918 | 0.226 | 959.883 | 77.01 |
| 01 | 2 | 959.532 | 0.144 | 959.514 | 80.73 | 15 | 2 | 959.600 | 0.200 | 959.497 | 78.29 |
| 01 | 1 | 959.663 | 0.131 | 959.660 | 71.73 | 15 | 2 | 959.717 | 0.183 | 959.668 | 79.58 |
| 01 | 1 | 958.979 | 0.149 | 958.979 | 72.72 | 15 | 2 | 958.895 | 0.250 | 958.831 | 80.84 |
| 01 | 1 | 958.413 | 0.146 | 958.469 | 73.78 | 15 | 1 | 959.999 | 0.196 | 959.971 | 74.53 |
| 01 | 1 | 959.313 | 0.137 | 959.312 | 74.87 | 15 | 1 | 959.600 | 0.200 | 959.599 | 75.76 |
| 01 | 1 | 959.313 | 0.137 | 959.332 | 76.00 | 15 | 1 | 959.918 | 0.226 | 959.883 | 77.01 |
| 01 | 1 | 959.015 | 0.132 | 959.018 | 77.16 | 15 | 1 | 959.600 | 0.200 | 959.497 | 78.29 |
| 01 | 1 | 959.532 | 0.144 | 959.526 | 78.34 | 15 | 1 | 959.717 | 0.183 | 959.668 | 79.58 |
| 01 | 1 | 958.737 | 0.165 | 958.767 | 79.56 | 15 | 1 | 958.895 | 0.250 | 958.831 | 80.84 |
| 01 | 1 | 959.532 | 0.144 | 959.514 | 80.73 | 20 | 2 | 958.424 | 0.113 | 958.362 | 59.15 |
| 08 | 2 | 958.769 | 0.141 | 958.736 | 68.15 | 20 | 2 | 960.218 | 0.109 | 960.220 | 59.65 |
| 08 | 2 | 959.215 | 0.154 | 959.221 | 70.01 | 20 | 2 | 959.561 | 0.131 | 959.536 | 60.18 |
| 08 | 2 | 958.938 | 0.192 | 958.921 | 70.90 | 20 | 2 | 959.484 | 0.139 | 959.484 | 60.79 |
| 08 | 1 | 958.769 | 0.141 | 958.736 | 68.15 | 20 | 2 | 959.912 | 0.169 | 959.898 | 61.41 |
| 08 | 1 | 959.215 | 0.154 | 959.221 | 70.01 | 20 | 2 | 959.106 | 0.141 | 959.106 | 61.97 |
| 08 | 1 | 958.938 | 0.192 | 958.921 | 70.90 | 20 | 2 | 959.152 | 0.130 | 959.150 | 62.59 |
| 08 | 2 | 959.384 | 0.152 | 959.377 | 71.84 | 20 | 2 | 959.152 | 0.130 | 959.175 | 63.93 |
| 08 | 2 | 959.614 | 0.131 | 959.632 | 72.83 | 20 | 2 | 960.062 | 0.163 | 960.093 | 64.64 |
| 08 | 2 | 959.421 | 0.136 | 959.414 | 73.86 | 20 | 2 | 959.018 | 0.148 | 959.012 | 65.36 |
| 08 | 2 | 959.467 | 0.142 | 959.503 | 74.93 | 20 | 2 | 959.253 | 0.161 | 959.246 | 66.11 |
| 08 | 2 | 959.567 | 0.181 | 959.543 | 76.03 | 20 | 2 | 959.832 | 0.139 | 959.775 | 66.93 |
| 08 | 2 | 958.938 | 0.192 | 958.932 | 77.21 | 20 | 2 | 958.836 | 0.162 | 958.800 | 67.80 |
| 08 | 2 | 958.198 | 0.145 | 958.218 | 78.39 | 20 | 2 | 960.162 | 0.131 | 960.185 | 68.66 |
| 08 | 2 | 958.965 | 0.138 | 958.982 | 79.59 | 20 | 2 | 958.928 | 0.184 | 958.914 | 69.59 |
| 08 | 2 | 958.965 | 0.138 | 958.960 | 80.75 | 20 | 1 | 958.424 | 0.113 | 958.362 | 59.15 |
| 08 | 2 | 959.176 | 0.158 | 959.167 | 81.78 | 20 | 1 | 960.218 | 0.109 | 960.220 | 59.65 |
| 08 | 1 | 959.384 | 0.152 | 959.377 | 71.84 | 20 | 1 | 959.561 | 0.131 | 959.536 | 60.18 |
| 08 | 1 | 959.614 | 0.131 | 959.632 | 72.83 | 20 | 1 | 959.484 | 0.139 | 959.484 | 60.79 |
| 08 | 1 | 959.421 | 0.136 | 959.414 | 73.86 | 20 | 1 | 959.912 | 0.169 | 959.898 | 61.41 |
| 08 | 1 | 959.467 | 0.142 | 959.503 | 74.93 | 20 | 1 | 959.106 | 0.141 | 959.106 | 61.97 |
| 08 | 1 | 959.567 | 0.181 | 959.543 | 76.03 | 20 | 1 | 959.152 | 0.130 | 959.150 | 62.59 |
| 08 | 1 | 958.938 | 0.192 | 958.932 | 77.21 | 20 | 1 | 959.152 | 0.130 | 959.175 | 63.93 |
| 08 | 1 | 958.198 | 0.145 | 958.218 | 78.39 | 20 | 1 | 960.062 | 0.163 | 960.093 | 64.64 |
| 08 | 1 | 958.965 | 0.138 | 958.982 | 79.59 | 20 | 1 | 959.018 | 0.148 | 959.012 | 65.36 |
| 08 | 1 | 958.965 | 0.138 | 958.960 | 80.75 | 20 | 1 | 959.253 | 0.161 | 959.246 | 66.11 |
| 08 | 1 | 959.176 | 0.158 | 959.167 | 81.78 | 20 | 1 | 959.832 | 0.139 | 959.775 | 66.93 |
| 11 | 2 | 959.240 | 0.164 | 959.242 | 63.56 | 20 | 1 | 958.836 | 0.162 | 958.800 | 67.80 |
| 11 | 2 | 959.393 | 0.166 | 959.430 | 64.42 | 20 | 1 | 960.162 | 0.131 | 960.185 | 68.66 |
| 11 | 2 | 958.686 | 0.116 | 958.695 | 65.14 | 20 | 1 | 958.928 | 0.184 | 958.914 | 69.59 |
| 11 | 2 | 958.519 | 0.192 | 958.516 | 65.90 | 20 | 2 | 959.057 | 0.109 | 959.051 | 71.44 |
| 11 | 2 | 958.506 | 0.105 | 958.505 | 66.65 | 20 | 2 | 958.590 | 0.136 | 958.648 | 72.52 |
| 11 | 2 | 958.661 | 0.152 | 958.667 | 67.46 | 20 | 2 | 959.220 | 0.157 | 959.217 | 73.65 |
| 11 | 2 | 958.408 | 0.122 | 958.387 | 68.27 | 20 | 2 | 959.458 | 0.137 | 959.460 | 74.88 |
| 11 | 2 | 959.393 | 0.166 | 959.404 | 69.11 | 20 | 2 | 959.152 | 0.130 | 959.171 | 76.12 |
| 11 | 2 | 959.134 | 0.394 | 959.108 | 70.03 | 20 | 1 | 959.057 | 0.109 | 959.051 | 71.44 |
| 11 | 2 | 959.029 | 0.191 | 959.017 | 70.97 | 20 | 1 | 958.590 | 0.136 | 958.648 | 72.52 |
| 11 | 1 | 959.240 | 0.164 | 959.242 | 63.56 | 20 | 1 | 959.220 | 0.157 | 959.217 | 73.65 |
| 11 | 1 | 959.393 | 0.166 | 959.430 | 64.42 | 20 | 1 | 959.458 | 0.137 | 959.460 | 74.88 |
| 11 | 1 | 958.686 | 0.116 | 958.695 | 65.14 | 20 | 1 | 959.152 | 0.130 | 959.171 | 76.12 |
| 11 | 1 | 958.519 | 0.192 | 958.516 | 65.90 | 21 | 2 | 960.468 | 0.127 | 960.616 | 64.45 |
| 11 | 1 | 958.506 | 0.105 | 958.505 | 66.65 | 21 | 2 | 959.659 | 0.167 | 959.644 | 66.12 |
| 11 | 1 | 958.661 | 0.152 | 958.667 | 67.46 | 21 | 2 | 958.882 | 0.168 | 958.854 | 66.96 |
| 11 | 1 | 958.408 | 0.122 | 958.387 | 68.27 | 21 | 2 | 959.705 | 0.157 | 959.686 | 67.92 |
| 11 | 1 | 959.393 | 0.166 | 959.404 | 69.11 | 21 | 2 | 958.509 | 0.180 | 958.502 | 68.94 |
| 11 | 1 | 959.134 | 0.394 | 959.108 | 70.03 | 21 | 1 | 960.468 | 0.127 | 960.616 | 64.45 |
| 11 | 1 | 959.029 | 0.191 | 959.017 | 70.97 | 21 | 1 | 959.659 | 0.167 | 959.644 | 66.12 |
| 11 | 2 | 959.238 | 0.115 | 959.217 | 71.98 | 21 | 1 | 958.882 | 0.168 | 958.854 | 66.96 |
| 11 | 2 | 959.583 | 0.146 | 959.596 | 73.00 | 21 | 1 | 959.705 | 0.157 | 959.686 | 67.92 |
| 11 | 2 | 958.686 | 0.116 | 958.696 | 74.04 | 21 | 1 | 958.509 | 0.180 | 958.502 | 68.94 |
| 11 | 2 | 960.003 | 0.186 | 959.870 | 75.17 | 21 | 2 | 958.548 | 0.140 | 958.560 | 70.73 |
| 11 | 2 | 958.924 | 0.186 | 958.910 | 76.33 | 21 | 2 | 959.012 | 0.185 | 959.015 | 71.86 |
| 11 | 2 | 958.783 | 0.118 | 958.800 | 77.53 | 21 | 2 | 958.771 | 0.139 | 958.783 | 73.11 |
| 11 | 1 | 959.238 | 0.115 | 959.217 | 71.98 | 21 | 2 | 959.388 | 0.185 | 959.376 | 74.60 |
| 11 | 1 | 959.583 | 0.146 | 959.596 | 73.00 | 21 | 2 | 958.368 | 0.175 | 958.365 | 75.88 |
| 11 | 1 | 958.686 | 0.116 | 958.696 | 74.04 | 21 | 2 | 958.675 | 0.228 | 958.697 | 77.21 |
| 11 | 1 | 960.003 | 0.186 | 959.870 | 75.17 | 21 | 2 | 958.355 | 0.157 | 958.359 | 78.59 |
| 11 | 1 | 958.924 | 0.186 | 958.910 | 76.33 | 21 | 2 | 959.388 | 0.185 | 959.389 | 79.99 |
| 11 | 1 | 958.783 | 0.118 | 958.800 | 77.53 | 21 | 1 | 958.548 | 0.140 | 958.560 | 70.73 |
| 15 | 2 | 959.999 | 0.196 | 959.971 | 74.53 | 21 | 1 | 959.012 | 0.185 | 959.015 | 71.86 |

```
         2000  -  SETEMBRO                              2000  -  SETEMBRO
 D  L    SDB    ER     SDC    HL            D  L    SDB    ER     SDC    HL
 21  1  958.771 0.139 958.783 73.11         29  2  959.002 0.186 958.991 64.04
 21  1  959.388 0.185 959.376 74.60         29  2  958.331 0.164 958.346 64.82
 21  1  958.368 0.175 958.365 75.88         29  2  958.471 0.208 958.454 65.69
 21  1  958.675 0.228 958.697 77.21         29  2  959.490 0.190 959.474 66.58
 21  1  958.355 0.157 958.359 78.59         29  2  959.092 0.229 959.059 67.49
 21  1  959.388 0.185 959.389 79.99         29  2  958.632 0.183 958.623 68.46
 22  2  960.225 0.165 960.198 65.20         29  1  959.016 0.165 959.017 59.34
 22  2  959.653 0.133 959.656 65.96         29  1  958.484 0.161 958.496 59.85
 22  2  958.840 0.142 958.850 66.80         29  1  959.145 0.188 959.137 60.43
 22  2  958.973 0.145 958.969 67.65         29  1  958.854 0.199 958.845 61.05
 22  2  959.398 0.140 959.395 68.55         29  1  958.547 0.154 958.531 61.79
 22  2  959.309 0.133 959.283 69.51         29  1  957.977 0.164 958.051 62.48
 22  1  960.225 0.165 960.198 65.20         29  1  959.002 0.186 959.006 63.22
 22  1  959.653 0.133 959.656 65.96         29  1  959.002 0.186 958.991 64.04
 22  1  958.840 0.142 958.850 66.80         29  1  958.331 0.164 958.346 64.82
 22  1  958.973 0.145 958.969 67.65         29  1  958.471 0.208 958.454 65.69
 22  1  959.398 0.140 959.395 68.55         29  1  959.490 0.190 959.474 66.58
 22  1  959.309 0.133 959.283 69.51         29  1  959.092 0.229 959.059 67.49
 22  1  959.243 0.160 959.230 70.62         29  1  958.632 0.183 958.623 68.46
 22  2  959.105 0.154 959.107 71.65         29  2  959.760 0.199 959.734 69.54
 22  2  959.309 0.133 959.329 73.93         29  2  958.645 0.154 958.651 70.61
 22  2  959.541 0.140 959.503 75.12         29  2  958.944 0.200 958.955 71.76
 22  2  960.008 0.168 959.951 76.36         29  2  958.787 0.178 958.788 73.00
 22  1  959.243 0.160 959.230 70.62         29  2  959.618 0.194 959.633 74.39
 22  1  959.105 0.154 959.107 71.65         29  1  959.760 0.199 959.734 69.54
 22  1  959.309 0.133 959.329 73.93         29  1  958.645 0.154 958.651 70.61
 22  1  959.541 0.140 959.503 75.12         29  1  958.944 0.200 958.955 71.76
 22  1  960.008 0.168 959.951 76.36         29  1  958.787 0.178 958.788 73.00
 28  2  958.913 0.130 958.920 57.31         29  1  959.618 0.194 959.633 74.39
 28  2  958.252 0.154 958.241 57.81
 28  2  958.217 0.158 958.219 58.28                  2000  -  OUTUBRO
 28  2  958.776 0.157 958.800 58.77          D  L    SDB    ER     SDC    HL
 28  2  958.844 0.169 958.821 59.36         02  2  959.434 0.155 959.413 64.49
 28  2  958.252 0.154 958.262 59.91         02  2  960.383 0.206 960.428 65.39
 28  2  958.740 0.169 958.734 60.52         02  2  960.350 0.206 960.344 66.37
 28  2  958.694 0.127 958.686 61.15         02  2  959.479 0.226 959.552 67.40
 28  2  958.983 0.180 958.983 61.83         02  1  959.434 0.155 959.413 64.49
 28  2  958.933 0.161 958.949 62.51         02  1  960.383 0.206 960.428 65.39
 28  2  959.143 0.142 959.156 63.25         02  1  960.350 0.206 960.344 66.37
 28  2  959.143 0.142 959.141 63.97         02  1  959.479 0.226 959.552 67.40
 28  2  958.604 0.204 958.621 64.75         02  2  958.952 0.208 958.925 69.80
 28  2  958.314 0.145 958.313 65.61         02  2  959.179 0.196 959.141 70.97
 28  2  959.058 0.166 959.061 66.50         02  2  959.977 0.188 959.988 72.18
 28  2  958.740 0.169 958.730 67.42         02  2  959.362 0.252 959.368 73.49
 28  2  959.441 0.196 959.484 68.36         02  2  959.241 0.182 959.239 74.88
 28  1  958.913 0.130 958.920 57.31         02  2  959.714 0.165 959.758 76.30
 28  1  958.252 0.154 958.241 57.81         02  1  958.952 0.208 958.925 69.80
 28  1  958.217 0.158 958.219 58.28         02  1  959.179 0.196 959.141 70.97
 28  1  958.776 0.157 958.800 58.77         02  1  959.977 0.188 959.988 72.18
 28  1  958.844 0.169 958.821 59.36         02  1  959.362 0.252 959.368 73.49
 28  1  958.252 0.154 958.262 59.91         02  1  959.241 0.182 959.239 74.88
 28  1  958.740 0.169 958.734 60.52         02  1  959.714 0.165 959.758 76.30
 28  1  958.694 0.127 958.686 61.15         16  2  959.020 0.161 959.023 51.53
 28  1  958.983 0.180 958.983 61.83         16  2  959.390 0.167 959.398 51.84
 28  1  958.933 0.161 958.949 62.51         16  2  959.889 0.162 959.897 52.15
 28  1  959.143 0.142 959.156 63.25         16  2  959.224 0.194 959.231 52.51
 28  1  959.143 0.142 959.141 63.97         16  2  958.031 0.215 958.030 52.90
 28  1  958.604 0.204 958.621 64.75         16  2  959.010 0.186 959.012 53.28
 28  1  958.314 0.145 958.313 65.61         16  2  959.472 0.174 959.473 53.68
 28  1  959.058 0.166 959.061 66.50         16  2  959.224 0.194 959.226 54.19
 28  1  958.740 0.169 958.730 67.42         16  2  958.845 0.166 958.847 54.64
 28  1  959.441 0.196 959.484 68.36         16  2  959.224 0.194 959.235 55.11
 28  2  958.497 0.130 958.497 69.39         16  2  959.253 0.123 959.263 55.61
 28  1  958.497 0.130 958.497 69.39         16  2  959.058 0.191 959.055 56.15
 29  2  959.016 0.165 959.017 59.34         16  2  958.889 0.162 959.869 56.73
 29  2  958.484 0.161 958.496 59.85         16  2  958.031 0.215 958.075 57.34
 29  2  959.145 0.188 959.137 60.43         16  2  960.136 0.169 960.171 57.96
 29  2  958.854 0.199 958.845 61.05         16  2  959.581 0.140 959.623 58.61
 29  2  958.547 0.154 958.531 61.79         16  2  959.354 0.182 959.359 59.48
 29  2  957.977 0.164 958.051 62.48         16  2  959.472 0.174 959.503 60.29
 29  2  959.002 0.186 959.006 63.22
```

|     | 2000 - OUTUBRO |         |       |         |       |     | 2000 - OUTUBRO |         |       |         |       |
| --- | --- | --- | --- | --- | --- | --- | --- | --- | --- | --- | --- |
| D   | L   | SDB     | ER    | SDC     | HL    | D   | L   | SDB     | ER    | SDC     | HL    |
| 16  | 2   | 959.035 | 0.152 | 959.036 | 61.33 | 18  | 2   | 959.796 | 0.149 | 959.800 | 53.65 |
| 16  | 1   | 959.020 | 0.161 | 959.023 | 51.53 | 18  | 2   | 959.212 | 0.155 | 959.206 | 54.10 |
| 16  | 1   | 959.390 | 0.167 | 959.398 | 51.84 | 18  | 2   | 959.176 | 0.160 | 959.171 | 54.60 |
| 16  | 1   | 959.889 | 0.162 | 959.897 | 52.15 | 18  | 2   | 959.308 | 0.175 | 959.313 | 55.13 |
| 16  | 1   | 959.224 | 0.194 | 959.231 | 52.51 | 18  | 2   | 958.686 | 0.145 | 958.694 | 55.69 |
| 16  | 1   | 958.031 | 0.215 | 958.030 | 52.90 | 18  | 2   | 958.600 | 0.162 | 958.622 | 56.29 |
| 16  | 1   | 959.010 | 0.186 | 959.012 | 53.28 | 18  | 2   | 958.920 | 0.147 | 958.918 | 56.90 |
| 16  | 1   | 959.472 | 0.174 | 959.473 | 53.68 | 18  | 2   | 958.516 | 0.142 | 958.515 | 57.55 |
| 16  | 1   | 959.224 | 0.194 | 959.226 | 54.19 | 18  | 2   | 958.073 | 0.148 | 958.072 | 58.26 |
| 16  | 1   | 958.845 | 0.166 | 958.847 | 54.64 | 18  | 2   | 959.353 | 0.131 | 959.347 | 59.06 |
| 16  | 1   | 959.224 | 0.194 | 959.235 | 55.11 | 18  | 1   | 960.136 | 0.188 | 960.152 | 49.86 |
| 16  | 1   | 959.253 | 0.123 | 959.263 | 55.61 | 18  | 1   | 959.742 | 0.157 | 959.728 | 50.12 |
| 16  | 1   | 959.058 | 0.191 | 959.055 | 56.15 | 18  | 1   | 959.353 | 0.131 | 959.352 | 50.38 |
| 16  | 1   | 959.889 | 0.162 | 959.869 | 56.73 | 18  | 1   | 959.638 | 0.132 | 959.658 | 50.65 |
| 16  | 1   | 958.031 | 0.215 | 958.075 | 57.34 | 18  | 1   | 959.138 | 0.117 | 959.144 | 50.96 |
| 16  | 1   | 960.136 | 0.169 | 960.171 | 57.96 | 18  | 1   | 959.307 | 0.165 | 959.304 | 51.26 |
| 16  | 1   | 959.581 | 0.140 | 959.623 | 58.61 | 18  | 1   | 959.982 | 0.158 | 959.972 | 51.93 |
| 16  | 1   | 959.354 | 0.182 | 959.359 | 59.48 | 18  | 1   | 958.600 | 0.162 | 958.596 | 52.29 |
| 16  | 1   | 959.472 | 0.174 | 959.503 | 60.29 | 18  | 1   | 959.332 | 0.147 | 959.332 | 52.68 |
| 16  | 1   | 959.035 | 0.152 | 959.036 | 61.33 | 18  | 1   | 959.796 | 0.149 | 959.800 | 53.65 |
| 16  | 2   | 958.827 | 0.217 | 958.832 | 62.21 | 18  | 1   | 959.212 | 0.155 | 959.206 | 54.10 |
| 16  | 1   | 958.827 | 0.217 | 958.832 | 62.21 | 18  | 1   | 959.176 | 0.160 | 959.171 | 54.60 |
| 17  | 2   | 959.258 | 0.162 | 959.257 | 50.87 | 18  | 1   | 959.308 | 0.175 | 959.313 | 55.13 |
| 17  | 2   | 959.408 | 0.204 | 959.390 | 51.17 | 18  | 1   | 958.686 | 0.145 | 958.694 | 55.69 |
| 17  | 2   | 958.445 | 0.188 | 958.455 | 51.49 | 18  | 1   | 958.600 | 0.162 | 958.622 | 56.29 |
| 17  | 2   | 959.637 | 0.189 | 959.642 | 51.83 | 18  | 1   | 958.920 | 0.147 | 958.918 | 56.90 |
| 17  | 2   | 958.945 | 0.187 | 958.946 | 52.19 | 18  | 1   | 958.516 | 0.142 | 958.515 | 57.55 |
| 17  | 2   | 959.354 | 0.156 | 959.355 | 52.62 | 18  | 1   | 958.073 | 0.148 | 958.072 | 58.26 |
| 17  | 2   | 959.367 | 0.179 | 959.367 | 53.02 | 18  | 1   | 959.353 | 0.131 | 959.347 | 59.06 |
| 17  | 2   | 958.662 | 0.189 | 958.665 | 53.42 | 20  | 2   | 959.244 | 0.166 | 959.245 | 51.72 |
| 17  | 2   | 959.723 | 0.215 | 959.756 | 53.87 | 20  | 2   | 959.561 | 0.182 | 959.565 | 52.11 |
| 17  | 2   | 959.408 | 0.204 | 959.394 | 54.33 | 20  | 2   | 959.561 | 0.182 | 959.566 | 52.54 |
| 17  | 2   | 958.927 | 0.164 | 958.928 | 54.84 | 20  | 2   | 958.311 | 0.158 | 958.279 | 52.98 |
| 17  | 2   | 959.018 | 0.134 | 958.995 | 55.39 | 20  | 2   | 958.961 | 0.148 | 958.985 | 53.46 |
| 17  | 2   | 958.948 | 0.187 | 958.951 | 56.05 | 20  | 2   | 958.347 | 0.162 | 958.346 | 53.93 |
| 17  | 2   | 959.576 | 0.179 | 959.581 | 56.72 | 20  | 2   | 959.385 | 0.164 | 959.398 | 54.43 |
| 17  | 2   | 959.576 | 0.179 | 959.593 | 57.34 | 20  | 2   | 958.787 | 0.180 | 958.817 | 55.05 |
| 17  | 2   | 959.422 | 0.186 | 959.429 | 58.06 | 20  | 2   | 959.428 | 0.161 | 959.448 | 55.63 |
| 17  | 2   | 959.637 | 0.189 | 959.627 | 58.76 | 20  | 2   | 958.569 | 0.153 | 958.575 | 56.27 |
| 17  | 2   | 959.831 | 0.180 | 959.890 | 59.51 | 20  | 2   | 958.537 | 0.193 | 958.537 | 56.99 |
| 17  | 2   | 958.662 | 0.189 | 958.658 | 60.37 | 20  | 2   | 958.644 | 0.193 | 958.688 | 57.76 |
| 17  | 1   | 959.258 | 0.162 | 959.257 | 50.87 | 20  | 2   | 959.733 | 0.244 | 959.802 | 58.56 |
| 17  | 1   | 959.408 | 0.204 | 959.390 | 51.17 | 20  | 1   | 959.244 | 0.166 | 959.245 | 51.72 |
| 17  | 1   | 958.445 | 0.188 | 958.455 | 51.49 | 20  | 1   | 959.561 | 0.182 | 959.565 | 52.11 |
| 17  | 1   | 959.637 | 0.189 | 959.642 | 51.83 | 20  | 1   | 959.561 | 0.182 | 959.566 | 52.54 |
| 17  | 1   | 958.945 | 0.187 | 958.946 | 52.19 | 20  | 1   | 958.311 | 0.158 | 958.279 | 52.98 |
| 17  | 1   | 959.354 | 0.156 | 959.355 | 52.62 | 20  | 1   | 958.961 | 0.148 | 958.985 | 53.46 |
| 17  | 1   | 959.367 | 0.179 | 959.367 | 53.02 | 20  | 1   | 958.347 | 0.162 | 958.346 | 53.93 |
| 17  | 1   | 958.662 | 0.189 | 958.665 | 53.42 | 20  | 1   | 959.385 | 0.164 | 959.398 | 54.43 |
| 17  | 1   | 959.723 | 0.215 | 959.756 | 53.87 | 20  | 1   | 958.787 | 0.180 | 958.817 | 55.05 |
| 17  | 1   | 959.408 | 0.204 | 959.394 | 54.33 | 20  | 1   | 959.428 | 0.161 | 959.448 | 55.63 |
| 17  | 1   | 958.927 | 0.164 | 958.928 | 54.84 | 20  | 1   | 958.569 | 0.153 | 958.575 | 56.27 |
| 17  | 1   | 959.018 | 0.134 | 958.995 | 55.39 | 20  | 1   | 958.537 | 0.193 | 958.537 | 56.99 |
| 17  | 1   | 958.948 | 0.187 | 958.951 | 56.05 | 20  | 1   | 958.644 | 0.193 | 958.688 | 57.76 |
| 17  | 1   | 959.576 | 0.179 | 959.581 | 56.72 | 20  | 1   | 959.733 | 0.244 | 959.802 | 58.56 |
| 17  | 1   | 959.576 | 0.179 | 959.593 | 57.34 | 23  | 2   | 959.382 | 0.131 | 959.371 | 48.65 |
| 17  | 1   | 959.422 | 0.186 | 959.429 | 58.06 | 23  | 2   | 959.438 | 0.187 | 959.468 | 48.95 |
| 17  | 1   | 959.637 | 0.189 | 959.627 | 58.76 | 23  | 2   | 960.219 | 0.158 | 960.231 | 49.28 |
| 17  | 1   | 959.831 | 0.180 | 959.890 | 59.51 | 23  | 2   | 959.382 | 0.131 | 959.368 | 49.59 |
| 17  | 1   | 958.662 | 0.189 | 958.658 | 60.37 | 23  | 2   | 959.222 | 0.156 | 959.221 | 49.92 |
| 17  | 2   | 959.831 | 0.180 | 959.911 | 61.22 | 23  | 2   | 960.143 | 0.156 | 960.167 | 50.27 |
| 17  | 1   | 959.831 | 0.180 | 959.911 | 61.22 | 23  | 2   | 959.222 | 0.156 | 959.236 | 50.65 |
| 18  | 2   | 960.136 | 0.188 | 960.152 | 49.86 | 23  | 1   | 959.382 | 0.131 | 959.371 | 48.65 |
| 18  | 2   | 959.742 | 0.157 | 959.728 | 50.12 | 23  | 1   | 959.438 | 0.187 | 959.468 | 48.95 |
| 18  | 2   | 959.353 | 0.131 | 959.352 | 50.38 | 23  | 1   | 960.219 | 0.158 | 960.231 | 49.28 |
| 18  | 2   | 959.638 | 0.132 | 959.658 | 50.65 | 23  | 1   | 959.382 | 0.131 | 959.368 | 49.59 |
| 18  | 2   | 959.138 | 0.117 | 959.144 | 50.96 | 23  | 1   | 959.222 | 0.156 | 959.221 | 49.92 |
| 18  | 2   | 959.307 | 0.165 | 959.304 | 51.26 | 23  | 1   | 960.143 | 0.156 | 960.167 | 50.27 |
| 18  | 2   | 959.982 | 0.158 | 959.972 | 51.93 | 23  | 1   | 959.222 | 0.156 | 959.236 | 50.65 |
| 18  | 2   | 958.600 | 0.162 | 958.596 | 52.29 | 24  | 2   | 959.636 | 0.164 | 959.673 | 50.54 |
| 18  | 2   | 959.332 | 0.147 | 959.332 | 52.68 | 24  | 2   | 958.294 | 0.166 | 958.393 | 51.05 |

|     2000 - OUTUBRO     |     |     |     |     |
|----|---|---------|-------|---------|-------|
| D  | L | SDB     | ER    | SDC     | HL    |
| 24 | 2 | 958.938 | 0.184 | 958.994 | 51.54 |
| 24 | 2 | 958.938 | 0.184 | 958.956 | 52.02 |
| 24 | 2 | 959.636 | 0.164 | 959.655 | 52.61 |
| 24 | 2 | 959.079 | 0.185 | 959.013 | 53.31 |
| 24 | 2 | 959.079 | 0.185 | 959.033 | 53.91 |
| 24 | 1 | 959.636 | 0.164 | 959.673 | 50.54 |
| 24 | 1 | 958.294 | 0.166 | 958.393 | 51.05 |
| 24 | 1 | 958.938 | 0.184 | 958.994 | 51.54 |
| 24 | 1 | 958.938 | 0.184 | 958.956 | 52.02 |
| 24 | 1 | 959.636 | 0.164 | 959.655 | 52.61 |
| 24 | 1 | 959.079 | 0.185 | 959.013 | 53.31 |
| 24 | 1 | 959.079 | 0.185 | 959.033 | 53.91 |
| 26 | 2 | 959.255 | 0.166 | 959.248 | 46.48 |
| 26 | 2 | 959.449 | 0.182 | 959.487 | 46.71 |
| 26 | 2 | 959.329 | 0.180 | 959.346 | 46.93 |
| 26 | 2 | 959.449 | 0.182 | 959.480 | 47.17 |
| 26 | 2 | 959.873 | 0.210 | 959.919 | 47.43 |
| 26 | 2 | 959.277 | 0.164 | 959.275 | 47.68 |
| 26 | 2 | 958.820 | 0.227 | 958.824 | 47.98 |
| 26 | 2 | 959.449 | 0.182 | 959.452 | 48.28 |
| 26 | 2 | 959.147 | 0.187 | 959.157 | 48.67 |
| 26 | 2 | 958.908 | 0.163 | 958.912 | 49.05 |
| 26 | 2 | 957.978 | 0.207 | 958.045 | 49.48 |
| 26 | 2 | 959.379 | 0.145 | 959.355 | 49.87 |
| 26 | 2 | 959.277 | 0.164 | 959.272 | 50.29 |
| 26 | 2 | 958.908 | 0.163 | 958.909 | 50.76 |
| 26 | 2 | 958.631 | 0.182 | 958.588 | 51.25 |
| 26 | 2 | 959.005 | 0.215 | 959.043 | 51.75 |
| 26 | 1 | 959.255 | 0.166 | 959.248 | 46.48 |
| 26 | 1 | 959.449 | 0.182 | 959.487 | 46.71 |
| 26 | 1 | 959.329 | 0.180 | 959.346 | 46.93 |
| 26 | 1 | 959.449 | 0.182 | 959.480 | 47.17 |
| 26 | 1 | 959.873 | 0.210 | 959.919 | 47.43 |
| 26 | 1 | 959.277 | 0.164 | 959.275 | 47.68 |
| 26 | 1 | 958.820 | 0.227 | 958.824 | 47.98 |
| 26 | 1 | 959.449 | 0.182 | 959.452 | 48.28 |
| 26 | 1 | 959.147 | 0.187 | 959.157 | 48.67 |
| 26 | 1 | 958.908 | 0.163 | 958.912 | 49.05 |
| 26 | 1 | 957.978 | 0.207 | 958.045 | 49.48 |
| 26 | 1 | 959.379 | 0.145 | 959.355 | 49.87 |
| 26 | 1 | 959.277 | 0.164 | 959.272 | 50.29 |
| 26 | 1 | 958.908 | 0.163 | 958.909 | 50.76 |
| 26 | 1 | 958.631 | 0.182 | 958.588 | 51.25 |
| 26 | 1 | 959.005 | 0.215 | 959.043 | 51.75 |

|     2000 - NOVEMBRO    |     |     |     |     |
|----|---|---------|-------|---------|-------|
| D  | L | SDB     | ER    | SDC     | HL    |
| 03 | 2 | 959.876 | 0.194 | 959.885 | 42.46 |
| 03 | 2 | 958.442 | 0.160 | 958.453 | 42.64 |
| 03 | 2 | 959.548 | 0.151 | 959.549 | 42.82 |
| 03 | 2 | 959.084 | 0.157 | 959.086 | 43.01 |
| 03 | 2 | 959.140 | 0.144 | 959.118 | 43.22 |
| 03 | 2 | 959.531 | 0.178 | 959.521 | 43.43 |
| 03 | 2 | 958.683 | 0.162 | 958.684 | 43.69 |
| 03 | 2 | 958.475 | 0.162 | 958.485 | 43.96 |
| 03 | 1 | 959.876 | 0.194 | 959.885 | 42.46 |
| 03 | 1 | 958.442 | 0.160 | 958.453 | 42.64 |
| 03 | 1 | 959.548 | 0.151 | 959.549 | 42.82 |
| 03 | 1 | 959.084 | 0.157 | 959.086 | 43.01 |
| 03 | 1 | 959.140 | 0.144 | 959.118 | 43.22 |
| 03 | 1 | 959.531 | 0.178 | 959.521 | 43.43 |
| 03 | 1 | 958.683 | 0.162 | 958.684 | 43.69 |
| 03 | 1 | 958.475 | 0.162 | 958.485 | 43.96 |
| 09 | 2 | 958.613 | 0.185 | 958.618 | 38.23 |
| 09 | 2 | 959.178 | 0.186 | 959.181 | 38.24 |
| 09 | 2 | 959.071 | 0.118 | 959.055 | 38.26 |
| 09 | 2 | 959.220 | 0.189 | 959.220 | 38.28 |
| 09 | 2 | 959.550 | 0.209 | 959.546 | 38.31 |
| 09 | 2 | 958.712 | 0.184 | 958.707 | 38.35 |
| 09 | 2 | 958.475 | 0.184 | 958.480 | 38.39 |

|     2000 - NOVEMBRO    |     |     |     |     |
|----|---|---------|-------|---------|-------|
| D  | L | SDB     | ER    | SDC     | HL    |
| 09 | 2 | 959.233 | 0.180 | 959.240 | 38.45 |
| 09 | 2 | 958.797 | 0.144 | 958.797 | 38.51 |
| 09 | 2 | 959.593 | 0.174 | 959.600 | 38.59 |
| 09 | 2 | 958.793 | 0.183 | 958.782 | 38.68 |
| 09 | 2 | 958.699 | 0.181 | 958.704 | 38.77 |
| 09 | 2 | 958.508 | 0.173 | 958.513 | 38.88 |
| 09 | 2 | 959.115 | 0.129 | 959.097 | 38.99 |
| 09 | 2 | 958.699 | 0.181 | 958.700 | 39.13 |
| 09 | 2 | 958.561 | 0.173 | 958.568 | 39.27 |
| 09 | 2 | 959.357 | 0.190 | 959.368 | 39.42 |
| 09 | 2 | 958.745 | 0.123 | 958.741 | 39.59 |
| 09 | 1 | 958.613 | 0.185 | 958.618 | 38.23 |
| 09 | 1 | 959.178 | 0.186 | 959.181 | 38.24 |
| 09 | 1 | 959.071 | 0.118 | 959.055 | 38.26 |
| 09 | 1 | 959.220 | 0.189 | 959.220 | 38.28 |
| 09 | 1 | 959.550 | 0.209 | 959.546 | 38.31 |
| 09 | 1 | 958.712 | 0.184 | 958.707 | 38.35 |
| 09 | 1 | 958.475 | 0.184 | 958.480 | 38.39 |
| 09 | 1 | 959.233 | 0.180 | 959.240 | 38.45 |
| 09 | 1 | 958.797 | 0.144 | 958.797 | 38.51 |
| 09 | 1 | 959.593 | 0.174 | 959.600 | 38.59 |
| 09 | 1 | 958.793 | 0.183 | 958.782 | 38.68 |
| 09 | 1 | 958.699 | 0.181 | 958.704 | 38.77 |
| 09 | 1 | 958.508 | 0.173 | 958.513 | 38.88 |
| 09 | 1 | 959.115 | 0.129 | 959.097 | 38.99 |
| 09 | 1 | 958.699 | 0.181 | 958.700 | 39.13 |
| 09 | 1 | 958.561 | 0.173 | 958.568 | 39.27 |
| 09 | 1 | 959.357 | 0.190 | 959.368 | 39.42 |
| 09 | 1 | 958.745 | 0.123 | 958.741 | 39.59 |
| 10 | 2 | 960.100 | 0.135 | 960.101 | 37.69 |
| 10 | 2 | 960.254 | 0.155 | 960.208 | 37.72 |
| 10 | 2 | 959.867 | 0.206 | 959.847 | 37.74 |
| 10 | 2 | 959.224 | 0.133 | 959.228 | 37.78 |
| 10 | 2 | 959.718 | 0.140 | 959.726 | 37.82 |
| 10 | 2 | 959.603 | 0.133 | 959.598 | 37.88 |
| 10 | 2 | 959.461 | 0.117 | 959.460 | 38.02 |
| 10 | 2 | 960.075 | 0.182 | 960.074 | 38.10 |
| 10 | 2 | 958.416 | 0.171 | 958.403 | 38.19 |
| 10 | 2 | 959.670 | 0.115 | 959.669 | 38.29 |
| 10 | 2 | 959.804 | 0.209 | 959.832 | 38.40 |
| 10 | 2 | 959.410 | 0.117 | 959.409 | 38.53 |
| 10 | 2 | 959.284 | 0.119 | 959.292 | 38.67 |
| 10 | 2 | 959.320 | 0.146 | 959.305 | 38.82 |
| 10 | 2 | 959.603 | 0.133 | 959.610 | 38.98 |
| 10 | 2 | 958.929 | 0.123 | 958.937 | 39.18 |
| 10 | 1 | 960.100 | 0.135 | 960.101 | 37.69 |
| 10 | 1 | 960.254 | 0.155 | 960.208 | 37.72 |
| 10 | 1 | 959.867 | 0.206 | 959.847 | 37.74 |
| 10 | 1 | 959.224 | 0.133 | 959.228 | 37.78 |
| 10 | 1 | 959.718 | 0.140 | 959.726 | 37.82 |
| 10 | 1 | 959.603 | 0.133 | 959.598 | 37.88 |
| 10 | 1 | 959.461 | 0.117 | 959.460 | 38.02 |
| 10 | 1 | 960.075 | 0.182 | 960.074 | 38.10 |
| 10 | 1 | 958.416 | 0.171 | 958.403 | 38.19 |
| 10 | 1 | 959.670 | 0.115 | 959.669 | 38.29 |
| 10 | 1 | 959.804 | 0.209 | 959.832 | 38.40 |
| 10 | 1 | 959.410 | 0.117 | 959.409 | 38.53 |
| 10 | 1 | 959.284 | 0.119 | 959.292 | 38.67 |
| 10 | 1 | 959.320 | 0.146 | 959.305 | 38.82 |
| 10 | 1 | 959.603 | 0.133 | 959.610 | 38.98 |
| 10 | 1 | 958.929 | 0.123 | 958.937 | 39.18 |
| 17 | 2 | 959.222 | 0.178 | 959.216 | 33.69 |
| 17 | 2 | 959.651 | 0.144 | 959.635 | 33.63 |
| 17 | 2 | 959.719 | 0.178 | 959.727 | 33.50 |
| 17 | 2 | 959.424 | 0.237 | 959.411 | 33.47 |
| 17 | 2 | 958.569 | 0.141 | 958.574 | 33.45 |
| 17 | 2 | 959.590 | 0.161 | 959.615 | 33.43 |
| 17 | 2 | 959.456 | 0.164 | 959.463 | 33.42 |
| 17 | 2 | 959.374 | 0.193 | 959.377 | 33.42 |
| 17 | 2 | 959.180 | 0.170 | 959.163 | 33.47 |
| 17 | 2 | 958.648 | 0.204 | 958.647 | 33.51 |

|      2000 - NOVEMBRO      |      |      |      |      |
|----|---|---------|-------|---------|-------|
| D  | L | SDB     | ER    | SDC     | HL    |
| 17 | 2 | 959.456 | 0.164 | 959.462 | 33.55 |
| 17 | 2 | 958.454 | 0.196 | 958.462 | 33.61 |
| 17 | 2 | 959.268 | 0.189 | 959.264 | 33.69 |
| 17 | 1 | 959.222 | 0.178 | 959.216 | 33.69 |
| 17 | 1 | 959.651 | 0.144 | 959.635 | 33.63 |
| 17 | 1 | 959.719 | 0.178 | 959.727 | 33.50 |
| 17 | 1 | 959.424 | 0.237 | 959.411 | 33.47 |
| 17 | 1 | 958.569 | 0.141 | 958.574 | 33.45 |
| 17 | 1 | 959.590 | 0.161 | 959.615 | 33.43 |
| 17 | 1 | 959.456 | 0.164 | 959.463 | 33.42 |
| 17 | 1 | 959.374 | 0.193 | 959.377 | 33.42 |
| 17 | 1 | 959.180 | 0.170 | 959.163 | 33.47 |
| 17 | 1 | 958.648 | 0.204 | 958.647 | 33.51 |
| 17 | 1 | 959.456 | 0.164 | 959.462 | 33.55 |
| 17 | 1 | 958.454 | 0.196 | 958.462 | 33.61 |
| 17 | 1 | 959.268 | 0.189 | 959.264 | 33.69 |
| 21 | 2 | 958.475 | 0.162 | 958.508 | 30.88 |
| 21 | 2 | 959.071 | 0.217 | 959.075 | 30.83 |
| 21 | 2 | 960.066 | 0.154 | 960.103 | 30.75 |
| 21 | 2 | 959.897 | 0.168 | 959.943 | 30.71 |
| 21 | 2 | 959.071 | 0.217 | 959.090 | 30.69 |
| 21 | 2 | 959.454 | 0.147 | 959.458 | 30.65 |
| 21 | 2 | 959.288 | 0.178 | 959.337 | 30.66 |
| 21 | 1 | 958.475 | 0.162 | 958.508 | 30.88 |
| 21 | 1 | 959.071 | 0.217 | 959.075 | 30.83 |
| 21 | 1 | 960.066 | 0.154 | 960.103 | 30.75 |
| 21 | 1 | 959.897 | 0.168 | 959.943 | 30.71 |
| 21 | 1 | 959.071 | 0.217 | 959.090 | 30.69 |
| 21 | 1 | 959.454 | 0.147 | 959.458 | 30.65 |
| 21 | 1 | 959.288 | 0.178 | 959.337 | 30.66 |
| 28 | 2 | 959.360 | 0.171 | 959.369 | 28.51 |
| 28 | 2 | 958.824 | 0.150 | 958.815 | 28.22 |
| 28 | 2 | 959.105 | 0.168 | 959.107 | 27.93 |
| 28 | 2 | 958.824 | 0.150 | 958.819 | 27.79 |
| 28 | 2 | 958.800 | 0.138 | 958.776 | 27.63 |
| 28 | 2 | 959.579 | 0.126 | 959.545 | 27.50 |
| 28 | 2 | 958.853 | 0.148 | 958.856 | 27.37 |
| 28 | 2 | 958.832 | 0.150 | 958.834 | 27.24 |
| 28 | 2 | 958.958 | 0.177 | 958.992 | 27.12 |
| 28 | 2 | 959.446 | 0.156 | 959.443 | 27.00 |
| 28 | 2 | 959.258 | 0.109 | 959.265 | 26.88 |
| 28 | 2 | 959.861 | 0.155 | 959.922 | 26.76 |
| 28 | 2 | 959.508 | 0.181 | 959.537 | 26.54 |
| 28 | 1 | 959.360 | 0.171 | 959.369 | 28.51 |
| 28 | 1 | 958.824 | 0.150 | 958.815 | 28.22 |
| 28 | 1 | 959.105 | 0.168 | 959.107 | 27.93 |
| 28 | 1 | 958.824 | 0.150 | 958.819 | 27.79 |
| 28 | 1 | 958.800 | 0.138 | 958.776 | 27.63 |
| 28 | 1 | 959.579 | 0.126 | 959.545 | 27.50 |
| 28 | 1 | 958.853 | 0.148 | 958.856 | 27.37 |
| 28 | 1 | 958.832 | 0.150 | 958.834 | 27.24 |
| 28 | 1 | 958.958 | 0.177 | 958.992 | 27.12 |
| 28 | 1 | 959.446 | 0.156 | 959.443 | 27.00 |
| 28 | 1 | 959.258 | 0.109 | 959.265 | 26.88 |
| 28 | 1 | 959.861 | 0.155 | 959.922 | 26.76 |
| 28 | 1 | 959.508 | 0.181 | 959.537 | 26.54 |
| 29 | 2 | 959.537 | 0.112 | 959.556 | 26.69 |
| 29 | 2 | 959.170 | 0.134 | 959.164 | 26.31 |
| 29 | 2 | 959.739 | 0.129 | 959.742 | 26.07 |
| 29 | 2 | 959.071 | 0.141 | 959.063 | 25.85 |
| 29 | 2 | 959.076 | 0.160 | 959.098 | 25.74 |
| 29 | 2 | 959.261 | 0.142 | 959.261 | 25.53 |
| 29 | 2 | 959.299 | 0.160 | 959.334 | 25.13 |
| 29 | 1 | 959.537 | 0.112 | 959.556 | 26.69 |
| 29 | 1 | 959.170 | 0.134 | 959.164 | 26.31 |
| 29 | 1 | 959.739 | 0.129 | 959.742 | 26.07 |
| 29 | 1 | 959.071 | 0.141 | 959.063 | 25.85 |
| 29 | 1 | 959.076 | 0.160 | 959.098 | 25.74 |
| 29 | 1 | 959.261 | 0.142 | 959.261 | 25.53 |
| 29 | 1 | 959.299 | 0.160 | 959.334 | 25.13 |

|      2000 - DEZEMBRO      |      |      |      |      |
|----|---|---------|-------|---------|-------|
| D  | L | SDB     | ER    | SDC     | HL    |
| 05 | 2 | 960.224 | 0.127 | 960.232 | 23.71 |
| 05 | 2 | 959.550 | 0.119 | 959.548 | 23.54 |
| 05 | 2 | 959.336 | 0.136 | 959.336 | 23.38 |
| 05 | 2 | 959.317 | 0.126 | 959.309 | 23.21 |
| 05 | 2 | 959.458 | 0.091 | 959.465 | 23.04 |
| 05 | 2 | 959.395 | 0.136 | 959.386 | 22.89 |
| 05 | 2 | 959.060 | 0.146 | 959.068 | 22.74 |
| 05 | 2 | 959.359 | 0.151 | 959.357 | 22.59 |
| 05 | 2 | 959.458 | 0.091 | 959.460 | 22.44 |
| 05 | 2 | 958.710 | 0.119 | 958.857 | 22.29 |
| 05 | 2 | 959.060 | 0.146 | 959.071 | 22.14 |
| 05 | 2 | 959.060 | 0.146 | 959.069 | 22.00 |
| 05 | 2 | 958.710 | 0.119 | 958.832 | 21.84 |
| 05 | 2 | 959.415 | 0.106 | 959.417 | 21.69 |
| 05 | 2 | 959.336 | 0.136 | 959.337 | 21.55 |
| 05 | 2 | 959.038 | 0.153 | 958.964 | 21.42 |
| 05 | 2 | 959.706 | 0.148 | 959.710 | 21.28 |
| 05 | 1 | 960.224 | 0.127 | 960.232 | 23.71 |
| 05 | 1 | 959.550 | 0.119 | 959.548 | 23.54 |
| 05 | 1 | 959.336 | 0.136 | 959.336 | 23.38 |
| 05 | 1 | 959.317 | 0.126 | 959.309 | 23.21 |
| 05 | 1 | 959.458 | 0.091 | 959.465 | 23.04 |
| 05 | 1 | 959.395 | 0.136 | 959.386 | 22.89 |
| 05 | 1 | 959.060 | 0.146 | 959.068 | 22.74 |
| 05 | 1 | 959.359 | 0.151 | 959.357 | 22.59 |
| 05 | 1 | 959.458 | 0.091 | 959.460 | 22.44 |
| 05 | 1 | 958.710 | 0.119 | 958.857 | 22.29 |
| 05 | 1 | 959.060 | 0.146 | 959.071 | 22.14 |
| 05 | 1 | 959.060 | 0.146 | 959.069 | 22.00 |
| 05 | 1 | 958.710 | 0.119 | 958.832 | 21.84 |
| 05 | 1 | 959.415 | 0.106 | 959.417 | 21.69 |
| 05 | 1 | 959.336 | 0.136 | 959.337 | 21.55 |
| 05 | 1 | 959.038 | 0.153 | 958.964 | 21.42 |
| 05 | 1 | 959.706 | 0.148 | 959.710 | 21.28 |
| 07 | 2 | 959.158 | 0.215 | 959.116 | 21.76 |
| 07 | 2 | 959.042 | 0.173 | 959.069 | 21.60 |
| 07 | 2 | 958.457 | 0.159 | 958.574 | 21.44 |
| 07 | 2 | 959.912 | 0.191 | 959.922 | 21.28 |
| 07 | 2 | 958.768 | 0.147 | 958.699 | 21.12 |
| 07 | 2 | 959.430 | 0.177 | 959.403 | 20.97 |
| 07 | 2 | 959.508 | 0.190 | 959.499 | 20.81 |
| 07 | 2 | 958.457 | 0.159 | 958.454 | 20.65 |
| 07 | 2 | 958.457 | 0.159 | 958.528 | 20.50 |
| 07 | 2 | 959.013 | 0.182 | 958.980 | 20.35 |
| 07 | 2 | 958.768 | 0.147 | 958.805 | 20.20 |
| 07 | 2 | 957.955 | 0.164 | 958.168 | 20.06 |
| 07 | 2 | 959.042 | 0.173 | 959.051 | 19.91 |
| 07 | 1 | 959.158 | 0.215 | 959.116 | 21.76 |
| 07 | 1 | 959.042 | 0.173 | 959.069 | 21.60 |
| 07 | 1 | 958.457 | 0.159 | 958.574 | 21.44 |
| 07 | 1 | 959.912 | 0.191 | 959.922 | 21.28 |
| 07 | 1 | 958.768 | 0.147 | 958.699 | 21.12 |
| 07 | 1 | 959.430 | 0.177 | 959.403 | 20.97 |
| 07 | 1 | 959.508 | 0.190 | 959.499 | 20.81 |
| 07 | 1 | 958.457 | 0.159 | 958.454 | 20.65 |
| 07 | 1 | 958.457 | 0.159 | 958.528 | 20.50 |
| 07 | 1 | 959.013 | 0.182 | 958.980 | 20.35 |
| 07 | 1 | 958.768 | 0.147 | 958.805 | 20.20 |
| 07 | 1 | 957.955 | 0.164 | 958.168 | 20.06 |
| 07 | 1 | 959.042 | 0.173 | 959.051 | 19.91 |
| 08 | 2 | 958.897 | 0.150 | 958.909 | 20.49 |
| 08 | 2 | 959.210 | 0.156 | 959.281 | 20.32 |
| 08 | 2 | 959.837 | 0.172 | 959.827 | 20.17 |
| 08 | 2 | 959.034 | 0.183 | 959.019 | 20.01 |
| 08 | 2 | 958.897 | 0.150 | 958.876 | 19.85 |
| 08 | 2 | 958.897 | 0.150 | 958.834 | 19.69 |
| 08 | 2 | 959.034 | 0.183 | 958.986 | 19.53 |
| 08 | 2 | 959.837 | 0.172 | 959.828 | 19.39 |
| 08 | 2 | 958.737 | 0.179 | 958.688 | 19.23 |
| 08 | 2 | 958.319 | 0.151 | 958.150 | 19.08 |
| 08 | 2 | 960.064 | 0.187 | 960.059 | 18.93 |

| 2000 - DEZEMBRO | | | | |
|---|---|---|---|---|
| D | L | SDB | ER | SDC | HL |
| 08 | 1 | 958.897 | 0.150 | 958.909 | 20.49 |
| 08 | 1 | 959.210 | 0.156 | 959.281 | 20.32 |
| 08 | 1 | 959.837 | 0.172 | 959.827 | 20.17 |
| 08 | 1 | 959.034 | 0.183 | 959.019 | 20.01 |
| 08 | 1 | 958.897 | 0.150 | 958.876 | 19.85 |
| 08 | 1 | 958.897 | 0.150 | 958.834 | 19.69 |
| 08 | 1 | 959.034 | 0.183 | 958.986 | 19.53 |
| 08 | 1 | 959.837 | 0.172 | 959.828 | 19.39 |
| 08 | 1 | 958.737 | 0.179 | 958.688 | 19.23 |
| 08 | 1 | 958.319 | 0.151 | 958.150 | 19.08 |
| 08 | 1 | 960.064 | 0.187 | 960.059 | 18.93 |
| 14 | 2 | 958.375 | 0.147 | 958.368 | 16.76 |
| 14 | 2 | 959.002 | 0.184 | 959.003 | 16.59 |
| 14 | 2 | 958.941 | 0.182 | 958.949 | 16.40 |
| 14 | 2 | 959.831 | 0.207 | 959.892 | 16.21 |
| 14 | 2 | 958.152 | 0.144 | 958.123 | 16.03 |
| 14 | 2 | 959.143 | 0.181 | 959.134 | 15.85 |
| 14 | 2 | 959.054 | 0.164 | 959.072 | 15.68 |
| 14 | 2 | 959.557 | 0.191 | 959.553 | 15.50 |
| 14 | 2 | 959.143 | 0.181 | 959.110 | 15.33 |
| 14 | 2 | 959.440 | 0.220 | 959.366 | 15.15 |
| 14 | 1 | 958.375 | 0.147 | 958.368 | 16.76 |
| 14 | 1 | 959.002 | 0.184 | 959.003 | 16.59 |
| 14 | 1 | 958.941 | 0.182 | 958.949 | 16.40 |
| 14 | 1 | 959.831 | 0.207 | 959.892 | 16.21 |
| 14 | 1 | 958.152 | 0.144 | 958.123 | 16.03 |
| 14 | 1 | 959.143 | 0.181 | 959.134 | 15.85 |
| 14 | 1 | 959.054 | 0.164 | 959.072 | 15.68 |
| 14 | 1 | 959.557 | 0.191 | 959.553 | 15.50 |
| 14 | 1 | 959.143 | 0.181 | 959.110 | 15.33 |
| 14 | 1 | 959.440 | 0.220 | 959.366 | 15.15 |
| 22 | 2 | 959.546 | 0.164 | 959.659 | 14.11 |
| 22 | 2 | 959.865 | 0.146 | 959.764 | 13.93 |
| 22 | 2 | 958.896 | 0.133 | 958.849 | 13.74 |
| 22 | 2 | 959.332 | 0.151 | 959.264 | 13.55 |
| 22 | 2 | 959.332 | 0.151 | 959.292 | 13.36 |
| 22 | 2 | 959.146 | 0.153 | 959.194 | 13.18 |
| 22 | 2 | 958.896 | 0.133 | 958.846 | 12.99 |
| 22 | 2 | 958.767 | 0.137 | 958.749 | 12.80 |
| 22 | 2 | 958.639 | 0.133 | 958.069 | 12.62 |
| 22 | 2 | 958.794 | 0.138 | 958.834 | 12.43 |
| 22 | 2 | 958.896 | 0.133 | 958.896 | 12.23 |
| 22 | 2 | 959.987 | 0.155 | 960.052 | 12.04 |
| 22 | 1 | 959.546 | 0.164 | 959.659 | 14.11 |
| 22 | 1 | 959.865 | 0.146 | 959.764 | 13.93 |
| 22 | 1 | 958.896 | 0.133 | 958.849 | 13.74 |
| 22 | 1 | 959.332 | 0.151 | 959.264 | 13.55 |
| 22 | 1 | 959.332 | 0.151 | 959.292 | 13.36 |
| 22 | 1 | 959.146 | 0.153 | 959.194 | 13.18 |
| 22 | 1 | 958.896 | 0.133 | 958.846 | 12.99 |
| 22 | 1 | 958.767 | 0.137 | 958.749 | 12.80 |
| 22 | 1 | 958.639 | 0.133 | 958.069 | 12.62 |
| 22 | 1 | 958.794 | 0.138 | 958.834 | 12.43 |
| 22 | 1 | 958.896 | 0.133 | 958.896 | 12.23 |
| 22 | 1 | 959.987 | 0.155 | 960.052 | 12.04 |
| 27 | 2 | 959.408 | 0.140 | 959.396 | 12.89 |
| 27 | 2 | 958.779 | 0.163 | 958.671 | 12.69 |
| 27 | 2 | 959.131 | 0.161 | 959.136 | 12.51 |
| 27 | 2 | 959.899 | 0.169 | 959.895 | 12.31 |
| 27 | 2 | 959.649 | 0.147 | 959.690 | 12.12 |
| 27 | 2 | 959.310 | 0.165 | 959.355 | 11.93 |
| 27 | 2 | 959.131 | 0.161 | 959.162 | 11.75 |
| 27 | 2 | 958.421 | 0.127 | 958.455 | 11.57 |
| 27 | 2 | 959.077 | 0.154 | 959.079 | 11.37 |
| 27 | 2 | 959.294 | 0.134 | 959.294 | 11.19 |
| 27 | 2 | 959.545 | 0.135 | 959.546 | 11.01 |
| 27 | 1 | 959.408 | 0.140 | 959.396 | 12.89 |
| 27 | 1 | 958.779 | 0.163 | 958.671 | 12.69 |
| 27 | 1 | 959.131 | 0.161 | 959.136 | 12.51 |
| 27 | 1 | 959.899 | 0.169 | 959.895 | 12.31 |
| 27 | 1 | 959.649 | 0.147 | 959.690 | 12.12 |

| 2000 - DEZEMBRO | | | | |
|---|---|---|---|---|
| D | L | SDB | ER | SDC | HL |
| 27 | 1 | 959.310 | 0.165 | 959.355 | 11.93 |
| 27 | 1 | 959.131 | 0.161 | 959.162 | 11.75 |
| 27 | 1 | 958.421 | 0.127 | 958.455 | 11.57 |
| 27 | 1 | 959.077 | 0.154 | 959.079 | 11.37 |
| 27 | 1 | 959.294 | 0.134 | 959.294 | 11.19 |
| 27 | 1 | 959.545 | 0.135 | 959.546 | 11.01 |
| 28 | 2 | 959.689 | 0.202 | 959.637 | 12.53 |
| 28 | 2 | 959.402 | 0.124 | 959.427 | 12.35 |
| 28 | 2 | 959.579 | 0.122 | 959.568 | 12.16 |
| 28 | 2 | 959.296 | 0.117 | 959.265 | 11.97 |
| 28 | 2 | 959.774 | 0.157 | 959.785 | 11.78 |
| 28 | 2 | 959.583 | 0.125 | 959.622 | 11.61 |
| 28 | 2 | 958.995 | 0.160 | 958.940 | 11.42 |
| 28 | 2 | 958.794 | 0.134 | 958.791 | 11.24 |
| 28 | 2 | 959.076 | 0.116 | 959.066 | 11.06 |
| 28 | 2 | 959.296 | 0.117 | 959.239 | 10.87 |
| 28 | 2 | 959.296 | 0.117 | 959.270 | 10.69 |
| 28 | 1 | 959.689 | 0.202 | 959.637 | 12.53 |
| 28 | 1 | 959.402 | 0.124 | 959.427 | 12.35 |
| 28 | 1 | 959.579 | 0.122 | 959.568 | 12.16 |
| 28 | 1 | 959.296 | 0.117 | 959.265 | 11.97 |
| 28 | 1 | 959.774 | 0.157 | 959.785 | 11.78 |
| 28 | 1 | 959.583 | 0.125 | 959.622 | 11.61 |
| 28 | 1 | 958.995 | 0.160 | 958.940 | 11.42 |
| 28 | 1 | 958.794 | 0.134 | 958.791 | 11.24 |
| 28 | 1 | 959.076 | 0.116 | 959.066 | 11.06 |
| 28 | 1 | 959.296 | 0.117 | 959.239 | 10.87 |
| 28 | 1 | 959.296 | 0.117 | 959.270 | 10.69 |

| 2001 - JANEIRO | | | | |
|---|---|---|---|---|
| D | L | SDB | ER | SDC | HL |
| 05 | 2 | 958.357 | 0.182 | 958.650 | 20.36 |
| 05 | 2 | 958.063 | 0.170 | 958.306 | 20.50 |
| 05 | 2 | 958.517 | 0.155 | 958.718 | 20.65 |
| 05 | 2 | 959.132 | 0.153 | 959.283 | 20.81 |
| 05 | 2 | 958.727 | 0.167 | 958.988 | 20.95 |
| 05 | 2 | 958.291 | 0.161 | 958.530 | 21.11 |
| 05 | 2 | 959.210 | 0.174 | 959.490 | 21.27 |
| 05 | 2 | 959.319 | 0.179 | 959.596 | 21.43 |
| 05 | 2 | 959.681 | 0.172 | 959.881 | 21.59 |
| 05 | 2 | 958.588 | 0.160 | 958.787 | 21.75 |
| 05 | 2 | 959.113 | 0.182 | 959.395 | 21.92 |
| 05 | 2 | 958.404 | 0.175 | 958.637 | 22.09 |
| 05 | 2 | 958.512 | 0.166 | 958.742 | 22.26 |
| 05 | 2 | 959.105 | 0.191 | 959.386 | 22.42 |
| 05 | 2 | 958.832 | 0.164 | 959.060 | 22.60 |
| 05 | 2 | 958.881 | 0.183 | 959.106 | 22.76 |
| 05 | 2 | 958.581 | 0.169 | 958.779 | 22.93 |
| 05 | 2 | 958.757 | 0.182 | 959.023 | 23.09 |
| 05 | 2 | 959.453 | 0.188 | 959.655 | 23.25 |
| 05 | 2 | 958.990 | 0.222 | 959.299 | 23.43 |
| 08 | 1 | 958.755 | 0.171 | 958.815 | 5.09 |
| 08 | 1 | 957.624 | 0.153 | 957.591 | 5.25 |
| 08 | 1 | 958.595 | 0.133 | 958.457 | 5.39 |
| 08 | 1 | 959.284 | 0.143 | 959.197 | 5.53 |
| 08 | 1 | 959.119 | 0.177 | 959.153 | 5.68 |
| 08 | 1 | 958.887 | 0.160 | 958.887 | 5.82 |
| 08 | 1 | 958.514 | 0.178 | 958.533 | 5.98 |
| 08 | 1 | 959.753 | 0.155 | 959.746 | 6.11 |
| 08 | 1 | 958.750 | 0.166 | 958.723 | 6.25 |
| 08 | 1 | 959.409 | 0.172 | 959.346 | 6.38 |
| 08 | 1 | 959.136 | 0.146 | 959.032 | 6.51 |
| 08 | 1 | 958.869 | 0.140 | 958.763 | 6.64 |
| 08 | 1 | 958.827 | 0.147 | 958.777 | 6.77 |
| 08 | 1 | 959.468 | 0.174 | 959.421 | 6.90 |
| 08 | 1 | 958.731 | 0.207 | 958.741 | 7.03 |
| 08 | 1 | 959.415 | 0.192 | 959.384 | 7.15 |
| 09 | 1 | 959.588 | 0.143 | 959.581 | 4.44 |
| 09 | 1 | 959.132 | 0.147 | 959.100 | 4.60 |
| 09 | 1 | 959.665 | 0.158 | 959.511 | 4.75 |

| 2001 - JANEIRO | | | | | | 2001 - JANEIRO | | | | |
|---|---|---|---|---|---|---|---|---|---|---|
| D | L | SDB | ER | SDC | HL | D | L | SDB | ER | SDC | HL |
| 09 | 1 | 959.237 | 0.133 | 959.198 | 4.91 | 12 | 2 | 958.376 | 0.158 | 958.581 | 26.35 |
| 09 | 1 | 958.641 | 0.161 | 958.658 | 5.06 | 12 | 2 | 959.571 | 0.177 | 959.834 | 26.47 |
| 09 | 1 | 958.643 | 0.143 | 958.557 | 5.21 | 12 | 2 | 959.087 | 0.170 | 959.293 | 26.60 |
| 09 | 1 | 959.349 | 0.145 | 959.288 | 5.36 | 12 | 2 | 958.639 | 0.167 | 958.837 | 26.72 |
| 09 | 1 | 959.080 | 0.157 | 959.011 | 5.52 | 12 | 2 | 958.900 | 0.179 | 959.132 | 26.85 |
| 09 | 1 | 959.773 | 0.163 | 959.815 | 5.65 | 12 | 2 | 958.626 | 0.181 | 958.899 | 26.98 |
| 09 | 1 | 958.813 | 0.189 | 958.831 | 5.80 | 15 | 1 | 959.478 | 0.178 | 959.526 | 5.74 |
| 09 | 1 | 958.790 | 0.161 | 958.812 | 5.94 | 15 | 1 | 960.044 | 0.150 | 960.040 | 5.89 |
| 09 | 1 | 959.618 | 0.166 | 959.599 | 6.08 | 15 | 1 | 959.424 | 0.179 | 959.437 | 6.02 |
| 09 | 1 | 960.027 | 0.160 | 960.014 | 6.24 | 15 | 1 | 958.918 | 0.146 | 958.906 | 6.15 |
| 09 | 1 | 958.563 | 0.123 | 958.369 | 6.39 | 15 | 1 | 959.301 | 0.173 | 959.344 | 6.29 |
| 09 | 1 | 958.952 | 0.147 | 958.888 | 6.52 | 15 | 1 | 959.416 | 0.169 | 959.356 | 6.41 |
| 09 | 1 | 959.495 | 0.166 | 959.387 | 6.66 | 15 | 1 | 958.848 | 0.162 | 958.835 | 6.52 |
| 09 | 1 | 958.557 | 0.136 | 958.307 | 6.79 | 15 | 1 | 958.462 | 0.170 | 958.512 | 6.64 |
| 09 | 1 | 959.350 | 0.169 | 959.259 | 6.92 | 15 | 1 | 958.919 | 0.166 | 958.871 | 6.76 |
| 10 | 1 | 959.554 | 0.128 | 959.431 | 5.86 | 15 | 1 | 958.795 | 0.194 | 958.881 | 6.87 |
| 10 | 1 | 959.016 | 0.140 | 958.898 | 6.00 | 15 | 1 | 959.490 | 0.177 | 959.495 | 6.99 |
| 10 | 1 | 958.976 | 0.134 | 958.890 | 6.14 | 15 | 1 | 958.272 | 0.187 | 958.294 | 7.09 |
| 10 | 1 | 959.508 | 0.124 | 959.346 | 6.28 | 15 | 1 | 958.718 | 0.198 | 958.735 | 7.20 |
| 10 | 1 | 959.381 | 0.135 | 959.265 | 6.41 | 15 | 1 | 959.236 | 0.164 | 959.208 | 7.30 |
| 10 | 1 | 959.506 | 0.148 | 959.425 | 6.54 | 15 | 1 | 958.681 | 0.164 | 958.646 | 7.40 |
| 10 | 1 | 959.569 | 0.136 | 959.409 | 6.67 | 15 | 1 | 958.709 | 0.156 | 958.639 | 7.50 |
| 10 | 1 | 959.528 | 0.129 | 959.392 | 6.83 | 15 | 1 | 959.002 | 0.162 | 958.957 | 7.59 |
| 10 | 1 | 959.149 | 0.126 | 958.959 | 6.95 | 15 | 1 | 958.149 | 0.171 | 958.138 | 7.68 |
| 10 | 1 | 959.223 | 0.127 | 959.070 | 7.06 | 15 | 1 | 958.755 | 0.171 | 958.721 | 7.76 |
| 10 | 1 | 959.540 | 0.154 | 959.480 | 7.18 | 15 | 2 | 959.031 | 0.200 | 959.311 | 27.26 |
| 10 | 1 | 959.325 | 0.145 | 959.210 | 7.28 | 15 | 2 | 958.517 | 0.182 | 958.770 | 27.32 |
| 10 | 2 | 958.762 | 0.154 | 958.884 | 23.49 | 15 | 2 | 958.621 | 0.177 | 958.863 | 27.39 |
| 10 | 2 | 960.204 | 0.161 | 960.386 | 23.76 | 15 | 2 | 959.224 | 0.194 | 959.490 | 27.46 |
| 10 | 2 | 959.324 | 0.166 | 959.503 | 23.87 | 15 | 2 | 958.713 | 0.181 | 958.988 | 27.53 |
| 10 | 2 | 959.141 | 0.162 | 959.351 | 23.99 | 15 | 2 | 959.149 | 0.171 | 959.389 | 27.61 |
| 10 | 2 | 958.743 | 0.151 | 958.913 | 24.19 | 15 | 2 | 958.829 | 0.196 | 959.117 | 27.72 |
| 10 | 2 | 959.539 | 0.162 | 959.706 | 24.31 | 15 | 2 | 959.053 | 0.186 | 959.350 | 27.81 |
| 10 | 2 | 959.918 | 0.185 | 960.179 | 24.47 | 15 | 2 | 958.444 | 0.172 | 958.698 | 27.90 |
| 10 | 2 | 959.114 | 0.173 | 959.409 | 24.65 | 15 | 2 | 959.294 | 0.156 | 959.469 | 27.99 |
| 11 | 1 | 959.223 | 0.154 | 959.242 | 4.04 | 15 | 2 | 959.206 | 0.166 | 959.464 | 28.08 |
| 11 | 1 | 959.691 | 0.158 | 959.697 | 4.55 | 16 | 1 | 959.265 | 0.141 | 959.220 | 5.92 |
| 11 | 1 | 959.781 | 0.151 | 959.755 | 4.72 | 16 | 1 | 959.343 | 0.150 | 959.312 | 6.07 |
| 11 | 1 | 960.019 | 0.145 | 959.922 | 4.87 | 16 | 1 | 958.610 | 0.138 | 958.551 | 6.20 |
| 11 | 1 | 959.211 | 0.164 | 959.259 | 5.06 | 16 | 1 | 959.531 | 0.168 | 959.575 | 6.33 |
| 11 | 1 | 959.412 | 0.174 | 959.467 | 5.20 | 16 | 1 | 959.807 | 0.170 | 959.835 | 6.46 |
| 11 | 1 | 960.151 | 0.148 | 960.088 | 5.34 | 16 | 1 | 959.187 | 0.149 | 959.079 | 6.60 |
| 11 | 1 | 958.950 | 0.173 | 958.958 | 5.51 | 16 | 1 | 959.099 | 0.163 | 959.089 | 6.72 |
| 11 | 1 | 959.305 | 0.142 | 959.229 | 5.64 | 16 | 1 | 958.963 | 0.197 | 959.021 | 6.83 |
| 11 | 1 | 959.298 | 0.158 | 959.183 | 5.78 | 16 | 1 | 959.330 | 0.169 | 959.292 | 6.94 |
| 11 | 1 | 959.918 | 0.152 | 959.728 | 5.95 | 16 | 1 | 959.296 | 0.175 | 959.216 | 7.06 |
| 11 | 1 | 959.879 | 0.175 | 959.880 | 6.09 | 16 | 1 | 959.504 | 0.199 | 959.536 | 7.17 |
| 11 | 1 | 958.547 | 0.171 | 958.544 | 6.25 | 16 | 1 | 959.179 | 0.168 | 959.166 | 7.28 |
| 11 | 1 | 959.189 | 0.155 | 959.121 | 6.38 | 16 | 1 | 958.661 | 0.189 | 958.682 | 7.38 |
| 11 | 1 | 959.218 | 0.152 | 959.201 | 6.51 | 16 | 1 | 959.053 | 0.157 | 958.963 | 7.48 |
| 11 | 1 | 959.139 | 0.164 | 959.077 | 6.63 | 16 | 1 | 959.188 | 0.197 | 959.180 | 7.58 |
| 11 | 2 | 959.614 | 0.161 | 959.823 | 24.32 | 16 | 1 | 958.858 | 0.193 | 958.829 | 7.69 |
| 11 | 2 | 959.286 | 0.145 | 959.465 | 24.42 | 16 | 1 | 958.588 | 0.205 | 958.552 | 7.77 |
| 11 | 2 | 959.965 | 0.141 | 960.133 | 24.67 | 16 | 1 | 958.850 | 0.173 | 958.766 | 7.86 |
| 11 | 2 | 959.164 | 0.176 | 959.458 | 24.81 | 16 | 1 | 958.374 | 0.182 | 958.337 | 7.94 |
| 11 | 2 | 959.284 | 0.145 | 959.405 | 25.06 | 16 | 1 | 958.727 | 0.183 | 958.679 | 8.01 |
| 11 | 2 | 960.103 | 0.153 | 960.332 | 25.19 | 16 | 2 | 959.306 | 0.159 | 959.518 | 27.94 |
| 11 | 2 | 959.570 | 0.148 | 959.748 | 25.32 | 16 | 2 | 959.159 | 0.152 | 959.373 | 27.97 |
| 11 | 2 | 958.715 | 0.178 | 958.974 | 25.45 | 16 | 2 | 958.829 | 0.197 | 959.265 | 28.01 |
| 11 | 2 | 959.994 | 0.163 | 960.215 | 25.73 | 16 | 2 | 959.056 | 0.142 | 959.254 | 28.06 |
| 11 | 2 | 959.158 | 0.161 | 959.396 | 25.86 | 16 | 2 | 958.592 | 0.173 | 958.822 | 28.11 |
| 11 | 2 | 959.366 | 0.161 | 959.550 | 26.43 | 16 | 2 | 959.108 | 0.198 | 959.411 | 28.16 |
| 11 | 2 | 959.216 | 0.169 | 959.424 | 26.57 | 16 | 2 | 958.641 | 0.174 | 958.897 | 28.22 |
| 11 | 2 | 959.533 | 0.198 | 959.741 | 26.71 | 16 | 2 | 958.090 | 0.152 | 958.274 | 28.29 |
| 11 | 2 | 960.298 | 0.228 | 960.547 | 26.85 | 16 | 2 | 958.999 | 0.159 | 959.251 | 28.36 |
| 12 | 2 | 959.163 | 0.152 | 959.855 | 25.64 | 16 | 2 | 959.163 | 0.153 | 959.354 | 28.44 |
| 12 | 2 | 958.856 | 0.179 | 959.159 | 25.75 | 16 | 2 | 958.509 | 0.162 | 958.721 | 28.52 |
| 12 | 2 | 959.111 | 0.151 | 959.274 | 25.86 | 16 | 2 | 959.337 | 0.177 | 959.546 | 28.61 |
| 12 | 2 | 958.969 | 0.144 | 959.158 | 25.97 | 17 | 1 | 959.924 | 0.160 | 959.970 | 5.78 |
| 12 | 2 | 959.339 | 0.155 | 959.552 | 26.09 | 17 | 1 | 959.072 | 0.159 | 959.120 | 5.91 |
| 12 | 2 | 959.451 | 0.172 | 959.689 | 26.21 | 17 | 1 | 960.143 | 0.172 | 960.182 | 6.06 |

| 2001 - JANEIRO | | | | | | 2001 - JANEIRO | | | | |
|---|---|---|---|---|---|---|---|---|---|---|
| D | L | SDB | ER | SDC | HL | D | L | SDB | ER | SDC | HL |
| 17 | 1 | 959.713 | 0.179 | 959.740 | 6.19 | 19 | 2 | 959.392 | 0.167 | 959.612 | 30.88 |
| 17 | 1 | 959.404 | 0.208 | 959.542 | 6.33 | 23 | 2 | 959.538 | 0.174 | 959.768 | 33.44 |
| 17 | 1 | 958.501 | 0.209 | 958.623 | 6.46 | 23 | 2 | 959.028 | 0.187 | 959.316 | 33.46 |
| 17 | 1 | 958.936 | 0.201 | 959.035 | 6.58 | 23 | 2 | 959.345 | 0.152 | 959.542 | 33.47 |
| 17 | 1 | 960.336 | 0.193 | 960.410 | 6.70 | 23 | 2 | 959.453 | 0.144 | 959.624 | 33.50 |
| 17 | 1 | 958.005 | 0.241 | 958.113 | 6.81 | 23 | 2 | 959.001 | 0.155 | 959.162 | 33.53 |
| 17 | 1 | 959.144 | 0.169 | 959.158 | 6.93 | 23 | 2 | 958.628 | 0.155 | 958.829 | 33.56 |
| 17 | 1 | 958.473 | 0.170 | 958.484 | 7.04 | 23 | 2 | 958.990 | 0.175 | 959.221 | 33.61 |
| 17 | 1 | 958.855 | 0.157 | 958.792 | 7.14 | 23 | 2 | 959.153 | 0.179 | 959.420 | 33.66 |
| 17 | 1 | 959.761 | 0.156 | 959.691 | 7.25 | 23 | 2 | 959.097 | 0.185 | 959.395 | 33.71 |
| 17 | 1 | 959.133 | 0.170 | 959.121 | 7.34 | 23 | 2 | 959.332 | 0.172 | 959.580 | 33.76 |
| 17 | 2 | 959.849 | 0.267 | 960.196 | 28.73 | 23 | 2 | 958.753 | 0.172 | 959.007 | 33.82 |
| 17 | 2 | 958.131 | 0.173 | 958.352 | 28.75 | 23 | 2 | 958.909 | 0.167 | 959.145 | 33.89 |
| 17 | 2 | 959.061 | 0.171 | 959.352 | 28.78 | 24 | 2 | 959.162 | 0.163 | 959.399 | 34.25 |
| 17 | 2 | 959.524 | 0.154 | 959.709 | 28.82 | 24 | 2 | 959.426 | 0.184 | 959.692 | 34.22 |
| 17 | 2 | 958.883 | 0.192 | 959.158 | 28.86 | 24 | 2 | 958.786 | 0.163 | 958.980 | 34.20 |
| 17 | 2 | 959.086 | 0.175 | 959.380 | 28.91 | 24 | 2 | 959.103 | 0.169 | 959.360 | 34.18 |
| 17 | 2 | 958.404 | 0.171 | 958.614 | 28.96 | 24 | 2 | 958.940 | 0.165 | 959.155 | 34.17 |
| 17 | 2 | 958.807 | 0.147 | 959.026 | 29.03 | 24 | 2 | 958.984 | 0.155 | 959.178 | 34.17 |
| 17 | 2 | 959.095 | 0.150 | 959.281 | 29.09 | 24 | 2 | 959.030 | 0.174 | 959.244 | 34.18 |
| 17 | 2 | 959.269 | 0.166 | 959.494 | 29.16 | 24 | 2 | 959.481 | 0.175 | 959.779 | 34.19 |
| 17 | 2 | 959.566 | 0.160 | 959.824 | 29.24 | 24 | 2 | 958.932 | 0.161 | 959.100 | 34.21 |
| 17 | 2 | 958.697 | 0.193 | 959.021 | 29.31 | 24 | 2 | 958.929 | 0.177 | 959.167 | 34.24 |
| 17 | 2 | 959.041 | 0.183 | 959.286 | 29.39 | 24 | 2 | 959.094 | 0.164 | 959.292 | 34.27 |
| 17 | 2 | 958.690 | 0.172 | 958.939 | 29.48 | 24 | 2 | 959.184 | 0.184 | 959.464 | 34.30 |
| 17 | 2 | 959.389 | 0.164 | 959.625 | 29.57 | 25 | 1 | 959.370 | 0.139 | 959.270 | 7.76 |
| 17 | 2 | 958.731 | 0.178 | 958.971 | 29.67 | 25 | 1 | 959.935 | 0.146 | 959.846 | 7.83 |
| 17 | 2 | 958.832 | 0.189 | 959.072 | 29.78 | 25 | 1 | 959.494 | 0.164 | 959.409 | 7.90 |
| 18 | 1 | 959.131 | 0.197 | 959.199 | 6.03 | 25 | 1 | 958.593 | 0.172 | 958.569 | 7.96 |
| 18 | 1 | 960.124 | 0.220 | 960.179 | 6.16 | 25 | 1 | 959.312 | 0.157 | 959.243 | 8.01 |
| 18 | 1 | 959.568 | 0.179 | 959.614 | 6.30 | 25 | 1 | 958.597 | 0.249 | 958.583 | 8.06 |
| 18 | 1 | 959.961 | 0.154 | 959.925 | 6.42 | 26 | 1 | 960.509 | 0.152 | 960.475 | 7.57 |
| 18 | 1 | 959.414 | 0.143 | 959.392 | 6.54 | 26 | 1 | 960.050 | 0.126 | 959.902 | 7.65 |
| 18 | 1 | 959.568 | 0.154 | 959.577 | 6.66 | 26 | 1 | 958.958 | 0.195 | 958.934 | 7.72 |
| 18 | 1 | 959.154 | 0.143 | 958.944 | 6.77 | 26 | 1 | 958.887 | 0.184 | 958.874 | 7.80 |
| 18 | 1 | 959.646 | 0.167 | 959.566 | 6.88 | 26 | 1 | 959.260 | 0.191 | 959.229 | 7.87 |
| 18 | 1 | 959.260 | 0.193 | 959.331 | 7.00 | 26 | 1 | 960.115 | 0.147 | 959.958 | 8.09 |
| 18 | 1 | 959.770 | 0.182 | 959.807 | 7.10 | 26 | 1 | 959.898 | 0.211 | 959.943 | 8.14 |
| 18 | 1 | 959.795 | 0.169 | 958.726 | 7.19 | 26 | 1 | 959.508 | 0.171 | 959.415 | 8.17 |
| 18 | 1 | 958.844 | 0.178 | 958.850 | 7.29 | 26 | 1 | 958.732 | 0.197 | 958.634 | 8.21 |
| 18 | 1 | 959.687 | 0.191 | 959.668 | 7.39 | 29 | 1 | 960.053 | 0.154 | 959.954 | 7.94 |
| 18 | 1 | 959.611 | 0.193 | 959.638 | 7.49 | 29 | 1 | 959.707 | 0.158 | 959.611 | 8.00 |
| 18 | 1 | 959.628 | 0.193 | 959.674 | 7.58 | 29 | 1 | 958.797 | 0.156 | 958.673 | 8.05 |
| 19 | 1 | 959.737 | 0.169 | 959.655 | 6.17 | 29 | 1 | 959.671 | 0.167 | 959.594 | 8.10 |
| 19 | 1 | 960.484 | 0.197 | 960.584 | 6.29 | 29 | 1 | 959.447 | 0.192 | 959.374 | 8.14 |
| 19 | 1 | 959.525 | 0.206 | 959.595 | 6.41 | 29 | 1 | 959.203 | 0.163 | 959.047 | 8.17 |
| 19 | 1 | 960.451 | 0.238 | 960.562 | 6.53 | 29 | 1 | 958.630 | 0.161 | 958.492 | 8.20 |
| 19 | 1 | 958.961 | 0.175 | 959.019 | 6.64 | 29 | 1 | 959.386 | 0.147 | 959.239 | 8.22 |
| 19 | 1 | 959.884 | 0.223 | 959.988 | 6.76 | 29 | 1 | 959.389 | 0.179 | 959.295 | 8.24 |
| 19 | 1 | 959.059 | 0.167 | 959.050 | 6.87 | 29 | 1 | 957.966 | 0.178 | 957.883 | 8.25 |
| 19 | 1 | 959.551 | 0.164 | 959.531 | 6.97 | 29 | 1 | 959.153 | 0.165 | 958.947 | 8.26 |
| 19 | 1 | 959.274 | 0.194 | 959.349 | 7.09 | 29 | 1 | 959.603 | 0.160 | 959.459 | 8.26 |
| 19 | 1 | 959.293 | 0.157 | 959.223 | 7.19 | 29 | 1 | 959.305 | 0.185 | 959.217 | 8.25 |
| 19 | 1 | 959.218 | 0.189 | 959.204 | 7.29 | 29 | 1 | 959.344 | 0.159 | 959.147 | 8.23 |
| 19 | 1 | 959.886 | 0.187 | 959.930 | 7.38 | 29 | 2 | 959.422 | 0.255 | 959.583 | 38.42 |
| 19 | 1 | 958.764 | 0.166 | 958.733 | 7.48 | 29 | 2 | 958.802 | 0.228 | 959.133 | 38.30 |
| 19 | 1 | 959.539 | 0.191 | 959.540 | 7.56 | 29 | 2 | 958.441 | 0.236 | 958.742 | 38.18 |
| 19 | 1 | 959.120 | 0.153 | 959.047 | 7.65 | 29 | 2 | 958.978 | 0.233 | 959.305 | 38.07 |
| 19 | 2 | 960.674 | 0.210 | 960.994 | 30.35 | 29 | 2 | 959.784 | 0.319 | 960.151 | 37.99 |
| 19 | 2 | 958.860 | 0.188 | 959.142 | 30.35 | 29 | 2 | 958.110 | 0.304 | 958.397 | 37.85 |
| 19 | 2 | 960.067 | 0.244 | 960.456 | 30.36 | 29 | 2 | 959.902 | 0.350 | 960.257 | 37.79 |
| 19 | 2 | 958.736 | 0.163 | 958.972 | 30.38 | 29 | 2 | 959.409 | 0.226 | 959.557 | 37.74 |
| 19 | 2 | 957.863 | 0.197 | 958.187 | 30.40 | 31 | 1 | 959.193 | 0.146 | 959.071 | 8.02 |
| 19 | 2 | 959.683 | 0.195 | 959.967 | 30.44 | 31 | 1 | 958.771 | 0.185 | 958.753 | 8.07 |
| 19 | 2 | 958.618 | 0.177 | 958.869 | 30.47 | 31 | 1 | 959.747 | 0.196 | 959.745 | 8.10 |
| 19 | 2 | 959.199 | 0.177 | 959.477 | 30.52 | 31 | 1 | 958.789 | 0.164 | 958.720 | 8.14 |
| 19 | 2 | 959.012 | 0.176 | 959.259 | 30.57 | 31 | 1 | 959.835 | 0.178 | 959.723 | 8.16 |
| 19 | 2 | 959.377 | 0.154 | 959.584 | 30.62 | 31 | 1 | 960.326 | 0.193 | 960.281 | 8.19 |
| 19 | 2 | 959.914 | 0.189 | 960.217 | 30.68 | 31 | 1 | 958.767 | 0.177 | 958.670 | 8.20 |
| 19 | 2 | 960.111 | 0.194 | 960.403 | 30.74 | 31 | 1 | 958.630 | 0.168 | 958.585 | 8.22 |
| 19 | 2 | 959.568 | 0.188 | 959.874 | 30.81 | 31 | 1 | 959.655 | 0.172 | 959.568 | 8.22 |

| 2001 - JANEIRO | | | | |
|---|---|---|---|---|
| D | L | SDB | ER | SDC | HL |
| 31 | 1 | 959.723 | 0.178 | 959.659 | 8.22 |
| 31 | 1 | 959.947 | 0.177 | 959.879 | 8.22 |
| 31 | 1 | 959.675 | 0.180 | 959.574 | 8.21 |
| 31 | 1 | 959.374 | 0.173 | 959.260 | 8.19 |
| 31 | 1 | 959.375 | 0.210 | 959.277 | 8.16 |
| 31 | 1 | 958.999 | 0.202 | 958.912 | 8.13 |
| 31 | 1 | 959.970 | 0.224 | 959.938 | 8.09 |
| 31 | 1 | 958.799 | 0.199 | 958.716 | 8.03 |
| 31 | 1 | 958.601 | 0.176 | 958.421 | 7.97 |

| 2001 - FEVEREIRO | | | | |
|---|---|---|---|---|
| D | L | SDB | ER | SDC | HL |
| 01 | 2 | 958.678 | 0.165 | 958.883 | 41.28 |
| 01 | 2 | 959.057 | 0.166 | 959.220 | 41.03 |
| 01 | 2 | 958.569 | 0.173 | 958.839 | 40.84 |
| 01 | 2 | 958.993 | 0.185 | 959.262 | 40.67 |
| 01 | 2 | 959.077 | 0.171 | 959.296 | 40.50 |
| 01 | 2 | 959.222 | 0.173 | 959.472 | 40.36 |
| 01 | 2 | 959.407 | 0.170 | 959.639 | 40.24 |
| 01 | 2 | 959.102 | 0.177 | 959.356 | 40.12 |
| 01 | 2 | 959.416 | 0.174 | 959.639 | 40.02 |
| 02 | 1 | 959.383 | 0.152 | 959.241 | 8.18 |
| 02 | 1 | 959.221 | 0.170 | 959.122 | 8.18 |
| 02 | 1 | 959.078 | 0.160 | 958.975 | 8.17 |
| 02 | 1 | 958.734 | 0.192 | 958.571 | 8.15 |
| 02 | 1 | 959.314 | 0.162 | 959.191 | 8.12 |
| 02 | 1 | 959.912 | 0.178 | 959.778 | 8.05 |
| 02 | 1 | 958.741 | 0.174 | 958.579 | 8.00 |
| 02 | 1 | 959.190 | 0.206 | 959.057 | 7.93 |
| 02 | 1 | 960.333 | 0.155 | 960.136 | 7.83 |
| 02 | 1 | 959.431 | 0.182 | 959.300 | 7.66 |
| 02 | 1 | 959.651 | 0.173 | 959.467 | 7.55 |
| 02 | 1 | 959.409 | 0.175 | 959.248 | 7.44 |
| 02 | 1 | 959.307 | 0.180 | 959.133 | 7.32 |
| 02 | 1 | 959.856 | 0.183 | 959.641 | 7.19 |
| 09 | 1 | 960.028 | 0.156 | 959.827 | 7.99 |
| 09 | 1 | 959.762 | 0.146 | 959.550 | 7.98 |
| 09 | 1 | 960.265 | 0.144 | 960.028 | 7.97 |
| 09 | 1 | 959.534 | 0.138 | 959.332 | 7.95 |
| 09 | 1 | 959.498 | 0.142 | 959.310 | 7.92 |
| 09 | 1 | 959.685 | 0.141 | 959.521 | 7.89 |
| 09 | 1 | 959.838 | 0.142 | 959.534 | 7.85 |
| 09 | 1 | 959.668 | 0.167 | 959.501 | 7.80 |
| 09 | 1 | 959.261 | 0.178 | 959.143 | 7.75 |
| 09 | 1 | 958.839 | 0.188 | 958.690 | 7.69 |
| 09 | 1 | 959.148 | 0.186 | 958.998 | 7.63 |
| 09 | 1 | 959.157 | 0.196 | 958.862 | 7.55 |
| 09 | 1 | 959.277 | 0.160 | 959.048 | 7.38 |
| 09 | 1 | 959.289 | 0.162 | 958.990 | 7.28 |
| 09 | 1 | 959.179 | 0.155 | 958.893 | 7.13 |
| 12 | 1 | 960.024 | 0.149 | 959.754 | 6.97 |
| 12 | 1 | 959.957 | 0.145 | 959.689 | 6.84 |
| 12 | 1 | 959.144 | 0.138 | 958.846 | 6.71 |
| 12 | 1 | 959.366 | 0.148 | 959.108 | 6.56 |
| 12 | 1 | 959.306 | 0.147 | 959.015 | 6.42 |
| 12 | 1 | 960.053 | 0.160 | 959.772 | 6.26 |
| 13 | 1 | 959.579 | 0.158 | 959.383 | 7.37 |
| 13 | 1 | 958.752 | 0.189 | 959.608 | 7.28 |
| 13 | 1 | 959.847 | 0.146 | 959.629 | 7.19 |
| 13 | 1 | 959.527 | 0.164 | 959.366 | 7.09 |
| 13 | 1 | 959.352 | 0.152 | 959.100 | 6.98 |
| 13 | 1 | 959.609 | 0.163 | 959.410 | 6.86 |
| 13 | 1 | 959.532 | 0.167 | 959.302 | 6.74 |
| 13 | 1 | 959.805 | 0.152 | 959.520 | 6.60 |
| 13 | 1 | 959.166 | 0.148 | 958.897 | 6.45 |
| 13 | 1 | 959.208 | 0.165 | 958.921 | 6.28 |
| 13 | 1 | 959.877 | 0.173 | 959.646 | 6.10 |
| 13 | 1 | 959.542 | 0.152 | 959.244 | 5.91 |
| 13 | 2 | 959.353 | 0.144 | 959.485 | 48.66 |
| 13 | 2 | 958.669 | 0.140 | 958.851 | 48.40 |

| 2001 - FEVEREIRO | | | | |
|---|---|---|---|---|
| D | L | SDB | ER | SDC | HL |
| 13 | 2 | 959.400 | 0.166 | 959.618 | 48.13 |
| 13 | 2 | 958.685 | 0.128 | 958.817 | 47.91 |
| 13 | 2 | 958.379 | 0.187 | 958.682 | 47.70 |
| 13 | 2 | 958.578 | 0.180 | 958.873 | 47.50 |
| 13 | 2 | 959.060 | 0.116 | 959.047 | 47.31 |
| 13 | 2 | 958.863 | 0.164 | 958.965 | 47.15 |
| 13 | 2 | 958.645 | 0.170 | 958.866 | 46.99 |
| 13 | 2 | 959.537 | 0.203 | 959.804 | 46.85 |
| 13 | 2 | 958.976 | 0.141 | 959.105 | 46.72 |
| 13 | 2 | 959.495 | 0.181 | 959.717 | 46.59 |
| 13 | 2 | 959.479 | 0.154 | 959.678 | 46.48 |
| 13 | 2 | 958.868 | 0.206 | 959.084 | 46.37 |
| 13 | 2 | 959.419 | 0.159 | 959.626 | 46.27 |
| 13 | 2 | 958.876 | 0.138 | 959.013 | 46.18 |
| 13 | 2 | 958.703 | 0.159 | 958.874 | 46.10 |
| 13 | 2 | 958.637 | 0.165 | 958.863 | 46.02 |
| 14 | 1 | 959.948 | 0.144 | 959.706 | 7.35 |
| 14 | 1 | 959.062 | 0.142 | 958.823 | 7.27 |
| 14 | 1 | 960.015 | 0.153 | 959.766 | 7.18 |
| 14 | 1 | 959.438 | 0.159 | 959.209 | 7.08 |
| 14 | 1 | 959.626 | 0.173 | 959.445 | 6.97 |
| 14 | 1 | 959.630 | 0.187 | 959.451 | 6.76 |
| 14 | 1 | 959.408 | 0.146 | 959.134 | 6.63 |
| 14 | 1 | 958.849 | 0.154 | 958.555 | 6.49 |
| 14 | 1 | 959.321 | 0.166 | 959.060 | 6.34 |
| 14 | 1 | 959.506 | 0.187 | 959.281 | 6.18 |
| 14 | 1 | 958.477 | 0.200 | 958.249 | 5.98 |
| 14 | 1 | 958.930 | 0.228 | 958.721 | 5.78 |
| 14 | 1 | 959.620 | 0.167 | 959.307 | 5.58 |
| 14 | 1 | 959.128 | 0.174 | 958.794 | 5.36 |
| 14 | 1 | 959.099 | 0.168 | 958.804 | 5.14 |
| 14 | 1 | 960.305 | 0.160 | 959.919 | 4.90 |
| 14 | 2 | 959.533 | 0.176 | 959.815 | 48.99 |
| 14 | 2 | 958.673 | 0.169 | 958.886 | 48.75 |
| 14 | 2 | 959.034 | 0.157 | 959.251 | 48.49 |
| 14 | 2 | 959.554 | 0.153 | 959.750 | 48.29 |
| 14 | 2 | 959.208 | 0.138 | 959.377 | 48.09 |
| 14 | 2 | 959.468 | 0.157 | 959.663 | 47.89 |
| 14 | 2 | 958.861 | 0.144 | 958.994 | 47.70 |
| 14 | 2 | 959.411 | 0.131 | 959.517 | 47.55 |
| 14 | 2 | 958.971 | 0.180 | 959.237 | 47.40 |
| 14 | 2 | 959.246 | 0.158 | 959.465 | 47.27 |
| 14 | 2 | 958.746 | 0.155 | 958.954 | 47.14 |
| 14 | 2 | 960.020 | 0.141 | 960.127 | 47.02 |
| 15 | 1 | 959.418 | 0.106 | 959.005 | 7.06 |
| 15 | 1 | 959.707 | 0.123 | 959.374 | 6.95 |
| 15 | 1 | 959.611 | 0.136 | 959.334 | 6.84 |
| 15 | 1 | 960.012 | 0.139 | 959.680 | 6.73 |
| 15 | 1 | 959.715 | 0.116 | 959.358 | 6.61 |
| 15 | 1 | 959.951 | 0.119 | 959.610 | 6.48 |
| 15 | 1 | 959.580 | 0.140 | 959.256 | 6.30 |
| 15 | 1 | 959.370 | 0.128 | 959.036 | 6.14 |
| 15 | 1 | 959.765 | 0.151 | 959.472 | 5.98 |
| 15 | 1 | 959.673 | 0.141 | 959.361 | 5.78 |
| 15 | 1 | 959.502 | 0.140 | 959.223 | 5.57 |
| 15 | 1 | 958.627 | 0.143 | 958.340 | 5.37 |
| 15 | 1 | 958.373 | 0.132 | 958.024 | 4.88 |
| 15 | 2 | 958.705 | 0.162 | 958.885 | 51.88 |
| 15 | 2 | 958.680 | 0.163 | 958.853 | 51.48 |
| 15 | 2 | 958.581 | 0.151 | 958.789 | 51.12 |
| 15 | 2 | 959.343 | 0.156 | 959.589 | 50.77 |
| 15 | 2 | 959.654 | 0.155 | 959.814 | 50.44 |
| 15 | 2 | 958.655 | 0.143 | 958.858 | 50.10 |
| 15 | 2 | 958.116 | 0.165 | 958.306 | 49.80 |
| 15 | 2 | 958.399 | 0.129 | 958.412 | 49.52 |
| 15 | 2 | 958.673 | 0.135 | 958.810 | 49.28 |
| 15 | 2 | 959.231 | 0.143 | 959.377 | 49.06 |
| 15 | 2 | 959.394 | 0.133 | 959.534 | 48.85 |
| 15 | 2 | 958.745 | 0.143 | 958.743 | 48.64 |
| 16 | 1 | 959.705 | 0.152 | 959.446 | 6.70 |
| 16 | 1 | 959.655 | 0.150 | 959.412 | 6.57 |

| 2001 - FEVEREIRO | | | | | | 2001 - FEVEREIRO | | | | |
|---|---|---|---|---|---|---|---|---|---|---|
| D | L | SDB | ER | SDC | HL | D | L | SDB | ER | SDC | HL |
| 16 | 1 | 959.830 | 0.176 | 959.630 | 6.43 | 21 | 1 | 959.452 | 0.173 | 959.174 | 2.97 |
| 16 | 1 | 959.387 | 0.159 | 959.132 | 6.14 | 21 | 1 | 959.796 | 0.172 | 959.423 | 2.54 |
| 16 | 1 | 959.645 | 0.143 | 959.327 | 5.96 | 21 | 1 | 959.088 | 0.193 | 958.845 | 2.17 |
| 16 | 1 | 958.850 | 0.164 | 958.616 | 5.78 | 21 | 1 | 960.042 | 0.167 | 959.642 | 1.80 |
| 16 | 1 | 959.244 | 0.165 | 958.987 | 5.39 | 21 | 1 | 959.618 | 0.193 | 959.351 | 1.40 |
| 16 | 1 | 960.509 | 0.150 | 960.181 | 5.16 | 21 | 2 | 959.398 | 0.179 | 959.655 | 54.97 |
| 16 | 1 | 960.020 | 0.162 | 959.711 | 4.93 | 21 | 2 | 958.798 | 0.169 | 959.022 | 54.51 |
| 16 | 1 | 958.759 | 0.183 | 958.462 | 4.70 | 21 | 2 | 959.662 | 0.192 | 959.812 | 54.09 |
| 16 | 1 | 960.151 | 0.164 | 959.846 | 4.44 | 21 | 2 | 958.692 | 0.170 | 958.914 | 53.68 |
| 16 | 1 | 959.176 | 0.160 | 958.803 | 4.16 | 21 | 2 | 958.491 | 0.184 | 958.653 | 53.32 |
| 19 | 1 | 959.201 | 0.162 | 958.987 | 5.76 | 21 | 2 | 959.686 | 0.214 | 959.889 | 52.69 |
| 19 | 1 | 959.332 | 0.177 | 959.143 | 5.55 | 21 | 2 | 958.738 | 0.191 | 958.991 | 52.39 |
| 19 | 1 | 958.773 | 0.193 | 958.623 | 5.36 | 21 | 2 | 959.544 | 0.197 | 959.798 | 52.13 |
| 19 | 1 | 959.093 | 0.146 | 958.768 | 5.16 | 21 | 2 | 958.721 | 0.172 | 958.916 | 51.89 |
| 19 | 1 | 960.285 | 0.163 | 959.997 | 4.91 | 21 | 2 | 958.746 | 0.200 | 958.977 | 51.64 |
| 19 | 1 | 959.570 | 0.149 | 959.291 | 4.68 | 21 | 2 | 958.083 | 0.228 | 958.309 | 51.40 |
| 19 | 1 | 958.985 | 0.146 | 958.689 | 4.45 | 21 | 2 | 959.546 | 0.247 | 959.796 | 51.17 |
| 19 | 1 | 960.178 | 0.142 | 959.856 | 4.16 | 21 | 2 | 958.919 | 0.225 | 959.175 | 50.97 |
| 19 | 1 | 959.432 | 0.172 | 959.158 | 3.87 | 21 | 2 | 959.013 | 0.209 | 959.234 | 50.79 |
| 19 | 1 | 959.384 | 0.199 | 959.188 | 3.58 | 23 | 2 | 959.005 | 0.184 | 959.221 | 54.94 |
| 19 | 1 | 958.688 | 0.155 | 958.325 | 3.27 | 23 | 2 | 959.003 | 0.138 | 959.169 | 54.56 |
| 19 | 1 | 958.483 | 0.181 | 958.213 | 2.92 | 23 | 2 | 958.770 | 0.196 | 959.063 | 54.19 |
| 19 | 1 | 959.137 | 0.141 | 958.705 | 2.59 | 23 | 2 | 958.836 | 0.158 | 959.001 | 53.82 |
| 19 | 1 | 959.372 | 0.155 | 959.042 | 2.15 | 23 | 2 | 958.793 | 0.171 | 958.977 | 53.50 |
| 19 | 2 | 959.275 | 0.173 | 959.556 | 52.21 | 23 | 2 | 958.855 | 0.159 | 959.032 | 53.20 |
| 19 | 2 | 959.359 | 0.174 | 959.622 | 51.90 | 23 | 2 | 959.144 | 0.184 | 959.361 | 52.90 |
| 19 | 2 | 959.172 | 0.146 | 959.350 | 51.63 | 23 | 2 | 959.568 | 0.155 | 959.773 | 52.64 |
| 19 | 2 | 958.694 | 0.175 | 958.919 | 51.35 | 23 | 2 | 958.988 | 0.145 | 959.157 | 52.40 |
| 19 | 2 | 958.920 | 0.160 | 959.149 | 51.09 | 23 | 2 | 958.961 | 0.159 | 959.137 | 52.17 |
| 19 | 2 | 959.994 | 0.150 | 960.188 | 50.86 | 23 | 2 | 958.688 | 0.142 | 958.858 | 51.95 |
| 19 | 2 | 960.026 | 0.174 | 960.249 | 50.60 | 23 | 2 | 959.837 | 0.152 | 960.024 | 51.75 |
| 19 | 2 | 958.809 | 0.157 | 958.989 | 50.39 | 28 | 1 | 959.918 | 0.171 | 959.594 | 1.11 |
| 19 | 2 | 959.561 | 0.156 | 959.726 | 50.19 | 28 | 1 | 960.368 | 0.150 | 960.061 | 0.70 |
| 19 | 2 | 958.581 | 0.190 | 958.831 | 50.01 | 28 | 1 | 959.008 | 0.143 | 958.616 | 0.09 |
| 19 | 2 | 958.604 | 0.166 | 958.844 | 49.84 | 28 | 1 | 959.284 | 0.156 | 958.927 | 0.55 |
| 19 | 2 | 960.162 | 0.160 | 960.388 | 49.68 | 28 | 1 | 959.126 | 0.182 | 958.818 | 1.03 |
| 20 | 1 | 958.912 | 0.134 | 958.556 | 4.08 | 28 | 1 | 959.401 | 0.172 | 959.095 | 1.54 |
| 20 | 1 | 958.528 | 0.169 | 958.206 | 3.80 | 28 | 1 | 960.378 | 0.186 | 960.044 | 2.07 |
| 20 | 1 | 958.854 | 0.160 | 958.544 | 3.52 | 28 | 1 | 959.871 | 0.219 | 959.631 | 2.67 |
| 20 | 1 | 959.338 | 0.168 | 959.017 | 3.20 | 28 | 1 | 958.879 | 0.215 | 958.569 | 3.26 |
| 20 | 1 | 959.216 | 0.157 | 958.938 | 2.79 | 28 | 1 | 958.426 | 0.183 | 958.080 | 3.85 |
| 20 | 1 | 960.039 | 0.153 | 959.645 | 2.45 | 28 | 1 | 958.782 | 0.213 | 958.462 | 4.54 |
| 20 | 1 | 958.973 | 0.158 | 958.589 | 2.07 | | | | | | |
| 20 | 1 | 958.886 | 0.173 | 958.568 | 1.68 | | | | | | |
| 20 | 1 | 959.692 | 0.181 | 959.387 | 1.26 | | | 2001 - MARCO | | | |
| 20 | 1 | 959.640 | 0.162 | 959.279 | 0.81 | D | L | SDB | ER | SDC | HL |
| 20 | 1 | 959.332 | 0.189 | 958.952 | 0.32 | 01 | 2 | 959.371 | 0.188 | 959.615 | 55.66 |
| 20 | 1 | 960.006 | 0.149 | 959.595 | 0.20 | 01 | 2 | 959.161 | 0.176 | 959.394 | 55.30 |
| 20 | 2 | 959.119 | 0.142 | 959.258 | 56.04 | 01 | 2 | 959.062 | 0.166 | 959.273 | 54.98 |
| 20 | 2 | 960.482 | 0.146 | 960.630 | 52.38 | 01 | 2 | 959.167 | 0.196 | 959.433 | 54.66 |
| 20 | 2 | 959.152 | 0.142 | 959.288 | 52.11 | 01 | 2 | 958.965 | 0.178 | 959.179 | 54.38 |
| 20 | 2 | 959.311 | 0.164 | 959.485 | 51.84 | 01 | 2 | 959.449 | 0.166 | 959.655 | 54.11 |
| 20 | 2 | 959.459 | 0.182 | 959.677 | 51.57 | 01 | 2 | 958.777 | 0.172 | 958.954 | 53.88 |
| 20 | 2 | 959.657 | 0.175 | 959.923 | 51.34 | 01 | 2 | 959.092 | 0.169 | 959.311 | 53.61 |
| 20 | 2 | 957.716 | 0.170 | 957.950 | 51.10 | 01 | 2 | 958.823 | 0.184 | 959.052 | 53.39 |
| 20 | 2 | 960.001 | 0.186 | 960.256 | 50.89 | 01 | 2 | 959.051 | 0.182 | 959.272 | 53.18 |
| 20 | 2 | 959.440 | 0.189 | 959.707 | 50.68 | 01 | 2 | 959.064 | 0.198 | 959.265 | 52.98 |
| 20 | 2 | 958.570 | 0.180 | 958.753 | 50.48 | 02 | 1 | 960.475 | 0.192 | 960.147 | 2.31 |
| 20 | 2 | 959.421 | 0.194 | 959.642 | 50.31 | 02 | 1 | 959.030 | 0.158 | 958.612 | 1.55 |
| 20 | 2 | 959.055 | 0.144 | 959.170 | 50.10 | 02 | 1 | 959.480 | 0.204 | 959.229 | 1.19 |
| 20 | 2 | 959.177 | 0.200 | 959.414 | 49.94 | 02 | 1 | 959.327 | 0.175 | 959.013 | 0.84 |
| 21 | 1 | 960.238 | 0.199 | 960.095 | 5.60 | 02 | 1 | 958.530 | 0.222 | 958.271 | 0.44 |
| 21 | 1 | 959.101 | 0.158 | 958.818 | 5.41 | 02 | 1 | 959.455 | 0.213 | 959.092 | 0.55 |
| 21 | 1 | 959.636 | 0.135 | 959.204 | 5.19 | 02 | 1 | 959.494 | 0.217 | 959.161 | 0.98 |
| 21 | 1 | 959.061 | 0.172 | 958.817 | 4.99 | 02 | 1 | 959.106 | 0.218 | 958.658 | 2.28 |
| 21 | 1 | 959.366 | 0.159 | 959.077 | 4.78 | 02 | 1 | 959.078 | 0.221 | 958.789 | 3.54 |
| 21 | 1 | 959.720 | 0.177 | 959.470 | 4.56 | 02 | 1 | 959.190 | 0.215 | 958.854 | 4.15 |
| 21 | 1 | 959.906 | 0.199 | 959.597 | 4.34 | 02 | 1 | 959.159 | 0.215 | 958.771 | 4.95 |
| 21 | 1 | 959.651 | 0.171 | 959.384 | 4.09 | 05 | 2 | 957.996 | 0.193 | 958.292 | 59.15 |
| 21 | 1 | 958.521 | 0.153 | 958.136 | 3.83 | 05 | 2 | 958.634 | 0.176 | 958.873 | 58.65 |
| 21 | 1 | 959.451 | 0.170 | 959.159 | 3.56 | 05 | 2 | 958.883 | 0.172 | 959.063 | 58.26 |

```
        2001  -  MARCO                              2001  -  MARCO
D   L    SDB    ER     SDC    HL           D   L    SDB    ER     SDC    HL
05  2  958.580 0.161 958.720 57.85         14  1  958.909 0.152 958.585 11.89
05  2  959.488 0.170 959.714 57.48         14  1  959.420 0.151 959.087 12.83
05  2  959.897 0.195 960.079 57.10         14  1  959.133 0.149 958.835 13.77
05  2  958.515 0.152 958.680 56.55         14  1  958.998 0.162 958.686 14.79
05  2  959.213 0.162 959.346 56.26         14  1  958.753 0.190 958.524 15.85
05  2  959.436 0.186 959.659 55.94         14  1  958.625 0.206 958.439 16.97
05  2  959.701 0.183 959.924 55.65         14  2  957.617 0.171 957.847 64.96
05  2  958.790 0.172 958.956 55.36         14  2  959.397 0.167 959.576 64.34
05  2  959.059 0.162 959.225 55.06         14  2  958.646 0.174 958.849 63.71
05  2  959.204 0.163 959.404 54.82         14  2  959.231 0.179 959.444 63.14
05  2  959.600 0.231 959.866 54.58         14  2  960.099 0.253 960.464 62.59
06  1  959.775 0.170 959.452  0.27         14  2  958.457 0.191 958.717 62.08
06  1  959.601 0.172 959.258  0.14         14  2  958.890 0.187 959.093 61.59
06  1  959.665 0.161 959.311  0.57         14  2  959.415 0.173 959.581 61.11
06  1  959.131 0.176 958.857  0.99         14  2  959.200 0.193 959.454 60.64
06  1  959.382 0.194 959.100  1.41         14  2  959.188 0.155 959.314 60.21
06  1  960.238 0.160 959.846  1.86         14  2  958.146 0.193 958.385 59.77
06  1  958.805 0.163 958.450  2.35         14  2  959.384 0.193 959.610 59.01
06  1  959.181 0.188 958.858  2.85         15  1  959.398 0.166 959.042  6.74
06  1  959.777 0.165 959.369  3.40         15  1  959.179 0.161 958.851  7.33
06  1  960.370 0.184 960.010  4.00         15  1  959.198 0.166 958.889  7.97
06  1  960.362 0.171 959.994  4.70         15  1  959.111 0.169 958.823  8.65
06  1  960.080 0.191 959.741  5.36         15  1  958.574 0.175 958.280  9.33
06  1  959.289 0.180 958.855  6.09         15  1  960.062 0.160 959.752 10.07
06  1  959.501 0.178 959.176  6.77         15  1  959.196 0.141 958.797 10.84
06  2  959.021 0.144 959.174 61.91         15  1  959.082 0.150 958.642 11.63
06  2  959.305 0.209 959.603 61.31         15  1  959.833 0.157 959.470 12.71
06  2  959.049 0.253 959.389 60.80         15  1  958.907 0.185 958.621 13.72
06  2  958.883 0.199 959.105 60.31         15  1  958.916 0.150 958.591 14.74
06  2  959.350 0.199 959.623 59.81         15  1  959.918 0.188 959.720 15.80
06  2  959.406 0.160 959.556 59.36         15  1  959.132 0.172 958.778 16.89
06  2  959.591 0.183 959.846 58.92         15  1  959.284 0.194 959.075 18.06
06  2  959.312 0.209 959.613 58.52         15  2  959.124 0.158 959.260 60.11
06  2  958.692 0.214 958.985 58.08         15  2  958.407 0.200 958.654 59.72
06  2  959.074 0.219 959.340 57.70         15  2  958.898 0.189 959.054 59.27
06  2  960.436 0.206 960.726 57.35         15  2  959.007 0.179 959.216 58.81
06  2  959.488 0.174 959.704 57.03         15  2  959.545 0.236 959.806 58.47
07  2  958.857 0.195 959.165 60.07         15  2  958.229 0.215 958.411 58.11
07  2  959.203 0.208 959.408 59.59         15  2  959.414 0.181 958.644 57.72
07  2  958.845 0.175 958.996 59.18         15  2  958.013 0.188 958.258 57.41
07  2  958.918 0.139 959.008 58.80         15  2  959.077 0.184 959.287 57.12
07  2  959.158 0.182 959.423 58.21         15  2  959.441 0.174 959.631 56.80
07  2  958.279 0.205 958.564 57.86         15  2  958.046 0.208 958.265 56.53
12  2  957.975 0.166 958.140 69.65         15  2  958.788 0.209 959.007 56.27
12  2  959.829 0.168 960.035 68.79         16  1  960.027 0.177 959.690  7.16
12  2  959.027 0.158 959.209 67.93         16  1  959.593 0.189 959.320  9.45
12  2  959.215 0.196 959.443 61.29         16  1  960.991 0.191 960.710 10.24
12  2  958.243 0.189 958.488 60.84         16  1  959.584 0.185 959.260 11.26
13  2  958.312 0.217 958.544 71.39         16  1  959.342 0.218 959.064 12.17
13  2  958.264 0.167 958.385 70.39         16  1  959.679 0.240 959.260 13.12
13  2  958.992 0.184 959.173 69.45         16  1  960.273 0.273 959.849 14.06
13  2  958.499 0.183 958.714 68.57         16  2  958.431 0.196 959.307 66.21
13  2  958.311 0.188 958.529 67.70         16  2  958.431 0.154 958.600 65.49
13  2  958.748 0.181 958.898 66.94         16  2  958.933 0.161 959.122 64.87
13  2  959.005 0.156 959.173 66.20         16  2  958.578 0.185 958.829 64.25
13  2  958.212 0.184 958.442 65.50         16  2  959.204 0.228 959.501 63.71
13  2  958.364 0.186 958.613 64.84         16  2  959.030 0.189 959.179 63.20
13  2  958.219 0.170 958.334 64.22         16  2  958.761 0.195 958.929 62.65
13  2  958.598 0.194 958.857 63.63         16  2  958.532 0.209 958.759 62.16
13  2  958.506 0.193 958.772 63.08         16  2  958.901 0.205 959.160 61.72
13  2  958.860 0.210 959.122 62.56         16  2  957.801 0.319 958.109 61.23
13  2  958.865 0.183 959.031 62.07         16  2  958.313 0.223 958.630 60.76
14  1  960.183 0.150 959.783  5.45         16  2  959.183 0.189 959.451 60.32
14  1  959.339 0.134 958.937  6.02         16  2  959.639 0.183 959.884 59.90
14  1  959.653 0.163 959.337  6.63         16  2  959.345 0.253 959.708 59.53
14  1  959.522 0.171 959.210  7.25         19  2  959.668 0.163 959.683 57.31
14  1  959.901 0.131 959.456  7.87         19  2  958.681 0.171 958.848 57.01
14  1  958.977 0.142 958.546  8.54         19  2  959.477 0.191 959.544 56.50
14  1  959.049 0.156 958.707  9.30         19  2  959.266 0.227 959.387 56.19
14  1  959.509 0.155 959.174 10.17         20  1  959.967 0.154 959.419  6.00
14  1  958.944 0.152 958.641 11.02         20  1  959.709 0.153 959.335  6.58
```

| 2001 - MARCO | | | | | | 2001 - MARCO | | | | |
|---|---|---|---|---|---|---|---|---|---|---|
| D | L | SDB | ER | SDC | HL | D | L | SDB | ER | SDC | HL |
| 20 | 1 | 959.777 | 0.143 | 959.377 | 7.14 | 22 | 2 | 959.444 | 0.167 | 959.593 | 67.02 |
| 20 | 1 | 960.302 | 0.129 | 959.874 | 7.74 | 22 | 2 | 958.523 | 0.162 | 958.621 | 66.38 |
| 20 | 1 | 958.886 | 0.130 | 958.420 | 8.35 | 22 | 2 | 958.391 | 0.163 | 958.502 | 65.75 |
| 20 | 1 | 959.434 | 0.170 | 959.127 | 9.05 | 22 | 2 | 958.834 | 0.174 | 959.024 | 64.55 |
| 20 | 1 | 959.688 | 0.133 | 959.248 | 9.66 | 22 | 2 | 959.549 | 0.136 | 959.596 | 64.02 |
| 20 | 1 | 959.518 | 0.149 | 959.141 | 10.33 | 22 | 2 | 958.747 | 0.162 | 958.919 | 63.47 |
| 20 | 1 | 959.468 | 0.157 | 959.115 | 11.09 | 22 | 2 | 959.690 | 0.173 | 959.877 | 62.97 |
| 20 | 1 | 959.544 | 0.150 | 959.203 | 11.84 | 23 | 1 | 959.279 | 0.167 | 958.939 | 10.89 |
| 20 | 1 | 959.320 | 0.163 | 958.992 | 12.65 | 23 | 1 | 959.659 | 0.168 | 959.300 | 11.63 |
| 20 | 1 | 959.422 | 0.182 | 959.145 | 13.50 | 23 | 1 | 959.770 | 0.162 | 959.428 | 12.38 |
| 20 | 1 | 958.637 | 0.162 | 958.353 | 14.35 | 23 | 1 | 959.786 | 0.171 | 959.496 | 13.22 |
| 20 | 1 | 959.588 | 0.188 | 959.370 | 15.31 | 23 | 1 | 958.854 | 0.144 | 958.481 | 14.09 |
| 20 | 2 | 959.185 | 0.125 | 959.219 | 68.41 | 23 | 1 | 959.547 | 0.136 | 959.170 | 14.98 |
| 20 | 2 | 959.825 | 0.127 | 959.768 | 67.63 | 23 | 1 | 959.663 | 0.180 | 959.333 | 15.89 |
| 20 | 2 | 959.017 | 0.128 | 959.069 | 66.95 | 23 | 1 | 959.307 | 0.138 | 958.866 | 17.23 |
| 20 | 2 | 959.249 | 0.129 | 959.289 | 66.29 | 23 | 1 | 959.261 | 0.162 | 959.015 | 18.28 |
| 20 | 2 | 959.115 | 0.139 | 959.185 | 65.62 | 23 | 1 | 959.674 | 0.162 | 959.234 | 19.45 |
| 20 | 2 | 959.232 | 0.138 | 959.292 | 64.98 | 23 | 2 | 957.755 | 0.192 | 957.987 | 70.68 |
| 20 | 2 | 959.135 | 0.137 | 959.249 | 64.41 | 23 | 2 | 958.396 | 0.173 | 958.547 | 69.85 |
| 20 | 2 | 959.147 | 0.155 | 959.243 | 63.87 | 23 | 2 | 958.612 | 0.192 | 958.865 | 69.11 |
| 20 | 2 | 959.295 | 0.149 | 959.426 | 63.32 | 23 | 2 | 958.364 | 0.179 | 958.539 | 68.35 |
| 20 | 2 | 959.276 | 0.151 | 959.436 | 62.81 | 23 | 2 | 958.655 | 0.172 | 958.699 | 67.65 |
| 20 | 2 | 959.450 | 0.140 | 959.532 | 62.18 | 23 | 2 | 958.410 | 0.196 | 958.577 | 66.96 |
| 20 | 2 | 958.725 | 0.163 | 958.879 | 61.71 | 23 | 2 | 959.175 | 0.226 | 959.342 | 66.33 |
| 21 | 1 | 960.937 | 0.154 | 960.485 | 7.63 | 23 | 2 | 958.430 | 0.206 | 958.664 | 65.74 |
| 21 | 1 | 960.439 | 0.162 | 960.091 | 8.22 | 23 | 2 | 958.203 | 0.167 | 958.355 | 65.12 |
| 21 | 1 | 959.576 | 0.148 | 959.144 | 8.88 | 23 | 2 | 959.480 | 0.166 | 959.540 | 64.52 |
| 21 | 1 | 959.370 | 0.142 | 958.921 | 9.50 | 23 | 2 | 958.914 | 0.207 | 959.138 | 63.99 |
| 21 | 1 | 959.622 | 0.167 | 959.232 | 10.16 | 23 | 2 | 958.811 | 0.171 | 959.010 | 62.80 |
| 21 | 1 | 959.436 | 0.175 | 959.090 | 10.85 | 23 | 2 | 959.160 | 0.184 | 959.360 | 62.34 |
| 21 | 1 | 959.392 | 0.152 | 958.969 | 11.58 | 23 | 2 | 958.995 | 0.175 | 959.165 | 61.91 |
| 21 | 1 | 959.981 | 0.146 | 959.637 | 12.35 | 26 | 1 | 960.506 | 0.175 | 960.120 | 8.40 |
| 21 | 1 | 959.296 | 0.176 | 959.032 | 13.14 | 26 | 1 | 959.682 | 0.181 | 959.348 | 9.07 |
| 21 | 1 | 959.484 | 0.164 | 959.137 | 13.97 | 26 | 1 | 959.282 | 0.170 | 958.934 | 9.72 |
| 21 | 1 | 960.562 | 0.173 | 960.253 | 14.84 | 26 | 1 | 959.490 | 0.174 | 959.168 | 10.40 |
| 21 | 1 | 959.284 | 0.146 | 958.903 | 17.15 | 26 | 1 | 959.516 | 0.167 | 959.156 | 11.11 |
| 21 | 1 | 959.597 | 0.158 | 959.281 | 18.22 | 26 | 1 | 959.586 | 0.173 | 959.251 | 11.81 |
| 21 | 1 | 959.508 | 0.151 | 959.169 | 19.41 | 26 | 1 | 958.816 | 0.193 | 958.585 | 12.56 |
| 21 | 1 | 960.135 | 0.183 | 959.959 | 20.62 | 26 | 1 | 959.127 | 0.205 | 958.900 | 13.35 |
| 21 | 2 | 959.156 | 0.182 | 959.341 | 69.26 | 26 | 1 | 959.612 | 0.161 | 959.325 | 14.17 |
| 21 | 2 | 958.861 | 0.147 | 958.929 | 68.45 | 26 | 1 | 959.362 | 0.171 | 959.044 | 15.03 |
| 21 | 2 | 959.185 | 0.164 | 959.383 | 67.74 | 26 | 1 | 959.540 | 0.184 | 959.333 | 15.96 |
| 21 | 2 | 959.675 | 0.151 | 959.825 | 67.00 | 26 | 1 | 959.199 | 0.167 | 958.859 | 16.95 |
| 21 | 2 | 958.783 | 0.164 | 958.943 | 66.37 | 26 | 1 | 958.951 | 0.196 | 958.755 | 18.02 |
| 21 | 2 | 958.882 | 0.164 | 959.070 | 65.75 | 26 | 1 | 958.650 | 0.193 | 958.473 | 19.11 |
| 21 | 2 | 958.884 | 0.141 | 959.020 | 65.15 | 26 | 1 | 959.130 | 0.214 | 959.019 | 20.39 |
| 21 | 2 | 958.691 | 0.130 | 958.800 | 64.59 | 27 | 1 | 959.583 | 0.166 | 959.316 | 10.40 |
| 21 | 2 | 959.283 | 0.125 | 959.265 | 64.03 | 27 | 1 | 960.607 | 0.204 | 960.409 | 11.10 |
| 21 | 2 | 958.165 | 0.185 | 958.378 | 63.50 | 27 | 1 | 959.246 | 0.140 | 958.894 | 11.85 |
| 21 | 2 | 958.946 | 0.147 | 959.021 | 62.98 | 27 | 1 | 959.855 | 0.148 | 959.529 | 12.67 |
| 21 | 2 | 959.291 | 0.144 | 959.398 | 62.50 | 27 | 1 | 959.866 | 0.181 | 959.633 | 13.53 |
| 22 | 1 | 959.843 | 0.149 | 959.505 | 7.39 | 27 | 1 | 959.789 | 0.150 | 959.617 | 14.39 |
| 22 | 1 | 959.702 | 0.170 | 959.289 | 7.95 | 27 | 1 | 959.374 | 0.173 | 959.149 | 15.28 |
| 22 | 1 | 959.684 | 0.174 | 959.424 | 8.54 | 27 | 1 | 959.072 | 0.147 | 958.743 | 16.20 |
| 22 | 1 | 959.708 | 0.172 | 959.433 | 9.17 | 27 | 1 | 959.356 | 0.130 | 959.042 | 17.20 |
| 22 | 1 | 959.414 | 0.177 | 959.116 | 9.82 | 27 | 1 | 959.332 | 0.201 | 959.167 | 18.27 |
| 22 | 1 | 959.699 | 0.164 | 959.404 | 10.52 | 27 | 1 | 958.878 | 0.164 | 958.655 | 19.34 |
| 22 | 1 | 960.027 | 0.189 | 959.727 | 11.21 | 27 | 1 | 958.507 | 0.152 | 958.299 | 20.49 |
| 22 | 1 | 959.658 | 0.188 | 959.346 | 11.96 | 27 | 1 | 959.325 | 0.223 | 959.240 | 21.98 |
| 22 | 1 | 959.147 | 0.172 | 958.821 | 12.79 | 27 | 1 | 959.136 | 0.162 | 958.911 | 23.35 |
| 22 | 1 | 959.502 | 0.159 | 959.188 | 13.61 | 27 | 2 | 958.621 | 0.158 | 958.719 | 68.56 |
| 22 | 1 | 959.573 | 0.181 | 959.337 | 14.53 | 27 | 2 | 959.473 | 0.179 | 959.552 | 67.84 |
| 22 | 1 | 959.797 | 0.158 | 959.511 | 15.45 | 27 | 2 | 958.782 | 0.182 | 958.991 | 67.09 |
| 22 | 1 | 960.339 | 0.205 | 960.139 | 16.67 | 27 | 2 | 958.569 | 0.172 | 958.779 | 66.48 |
| 22 | 1 | 959.325 | 0.144 | 958.974 | 17.76 | 27 | 2 | 959.001 | 0.155 | 959.155 | 65.88 |
| 22 | 1 | 959.545 | 0.160 | 959.244 | 18.89 | 27 | 2 | 958.739 | 0.171 | 958.922 | 65.32 |
| 22 | 2 | 958.988 | 0.159 | 959.166 | 70.77 | 27 | 2 | 958.623 | 0.164 | 958.733 | 64.77 |
| 22 | 2 | 959.694 | 0.157 | 959.814 | 69.93 | 27 | 2 | 959.495 | 0.193 | 959.680 | 64.21 |
| 22 | 2 | 958.973 | 0.157 | 959.079 | 69.15 | 27 | 2 | 959.609 | 0.175 | 959.769 | 63.70 |
| 22 | 2 | 959.139 | 0.164 | 959.341 | 68.41 | 27 | 2 | 959.127 | 0.210 | 959.268 | 63.18 |
| 22 | 2 | 960.167 | 0.175 | 960.402 | 67.69 | 27 | 2 | 958.765 | 0.181 | 958.886 | 62.73 |

|  | 2001 - ABRIL |  |  |  |
|---|---|---|---|---|
| D | L | SDB | ER | SDC | HL |
| 02 | 1 | 959.761 | 0.157 | 959.452 | 10.35 |
| 02 | 1 | 960.583 | 0.150 | 960.170 | 10.97 |
| 02 | 1 | 959.704 | 0.200 | 959.473 | 11.71 |
| 02 | 1 | 960.263 | 0.191 | 959.999 | 12.42 |
| 02 | 1 | 959.865 | 0.168 | 959.541 | 13.15 |
| 02 | 1 | 959.094 | 0.179 | 958.735 | 13.92 |
| 02 | 1 | 959.489 | 0.167 | 959.249 | 14.75 |
| 02 | 1 | 959.537 | 0.161 | 959.243 | 15.57 |
| 02 | 1 | 960.349 | 0.151 | 960.041 | 16.47 |
| 02 | 1 | 959.214 | 0.185 | 959.008 | 17.39 |
| 02 | 1 | 959.445 | 0.185 | 959.292 | 18.37 |
| 02 | 1 | 959.553 | 0.152 | 959.323 | 19.46 |
| 02 | 1 | 959.887 | 0.143 | 959.595 | 20.54 |
| 02 | 1 | 959.194 | 0.166 | 959.031 | 21.76 |
| 02 | 1 | 959.641 | 0.180 | 959.527 | 23.01 |
| 02 | 2 | 959.303 | 0.132 | 959.407 | 73.35 |
| 02 | 2 | 959.616 | 0.135 | 959.648 | 72.51 |
| 02 | 2 | 958.276 | 0.168 | 958.429 | 71.68 |
| 02 | 2 | 958.440 | 0.157 | 958.579 | 70.93 |
| 02 | 2 | 959.701 | 0.151 | 959.863 | 70.22 |
| 02 | 2 | 959.170 | 0.132 | 959.266 | 69.49 |
| 02 | 2 | 958.997 | 0.149 | 959.151 | 68.81 |
| 02 | 2 | 959.327 | 0.164 | 959.337 | 68.10 |
| 02 | 2 | 958.937 | 0.168 | 959.090 | 67.49 |
| 02 | 2 | 959.485 | 0.173 | 959.633 | 66.88 |
| 02 | 2 | 959.733 | 0.189 | 959.967 | 66.23 |
| 02 | 2 | 959.738 | 0.171 | 959.889 | 65.69 |
| 03 | 1 | 960.305 | 0.181 | 960.000 | 10.95 |
| 03 | 1 | 959.380 | 0.189 | 959.052 | 11.78 |
| 03 | 1 | 959.832 | 0.162 | 959.426 | 12.51 |
| 03 | 1 | 960.014 | 0.226 | 959.742 | 15.73 |
| 03 | 1 | 959.013 | 0.186 | 958.774 | 16.65 |
| 03 | 1 | 959.298 | 0.184 | 959.036 | 17.54 |
| 03 | 1 | 958.946 | 0.195 | 958.705 | 18.52 |
| 04 | 1 | 959.849 | 0.139 | 959.397 | 12.15 |
| 04 | 1 | 959.365 | 0.164 | 958.983 | 12.85 |
| 04 | 1 | 959.798 | 0.171 | 959.452 | 13.59 |
| 04 | 1 | 959.154 | 0.202 | 958.807 | 14.38 |
| 04 | 1 | 959.312 | 0.183 | 959.020 | 15.18 |
| 04 | 1 | 959.114 | 0.175 | 958.662 | 15.98 |
| 04 | 1 | 959.301 | 0.164 | 958.895 | 16.83 |
| 04 | 1 | 959.230 | 0.132 | 958.833 | 17.71 |
| 04 | 1 | 959.483 | 0.173 | 959.286 | 18.69 |
| 04 | 1 | 959.505 | 0.147 | 959.198 | 19.67 |
| 04 | 1 | 960.157 | 0.144 | 959.842 | 20.72 |
| 04 | 1 | 959.275 | 0.162 | 958.998 | 21.84 |
| 04 | 1 | 959.246 | 0.160 | 958.956 | 23.02 |
| 04 | 1 | 959.772 | 0.190 | 959.695 | 24.30 |
| 04 | 1 | 958.753 | 0.206 | 958.736 | 25.59 |
| 05 | 2 | 958.871 | 0.160 | 958.906 | 66.41 |
| 05 | 2 | 959.256 | 0.180 | 959.490 | 64.26 |
| 05 | 2 | 958.977 | 0.182 | 959.157 | 63.33 |
| 05 | 2 | 959.137 | 0.173 | 959.298 | 62.89 |
| 05 | 2 | 959.953 | 0.189 | 960.175 | 62.36 |
| 05 | 2 | 959.387 | 0.189 | 959.546 | 61.93 |
| 06 | 2 | 957.749 | 0.203 | 957.884 | 76.59 |
| 06 | 2 | 958.183 | 0.168 | 958.243 | 74.56 |
| 06 | 2 | 958.540 | 0.150 | 958.640 | 73.70 |
| 06 | 2 | 959.150 | 0.175 | 959.285 | 72.88 |
| 06 | 2 | 958.436 | 0.227 | 958.700 | 72.09 |
| 06 | 2 | 958.525 | 0.214 | 958.761 | 71.33 |
| 06 | 2 | 957.695 | 0.193 | 957.899 | 70.61 |
| 06 | 2 | 958.910 | 0.155 | 958.954 | 69.93 |
| 06 | 2 | 959.017 | 0.154 | 959.123 | 69.00 |
| 06 | 2 | 958.365 | 0.176 | 958.545 | 68.38 |
| 06 | 2 | 958.184 | 0.180 | 958.377 | 67.77 |
| 06 | 2 | 959.185 | 0.178 | 959.329 | 67.16 |
| 06 | 2 | 960.160 | 0.223 | 960.424 | 66.60 |
| 06 | 2 | 958.437 | 0.220 | 958.575 | 66.08 |
| 11 | 1 | 959.731 | 0.138 | 959.366 | 10.99 |
| 11 | 1 | 960.161 | 0.136 | 959.794 | 11.61 |

|  | 2001 - ABRIL |  |  |  |
|---|---|---|---|---|
| D | L | SDB | ER | SDC | HL |
| 11 | 1 | 960.020 | 0.140 | 959.729 | 12.28 |
| 11 | 1 | 959.292 | 0.137 | 958.920 | 13.17 |
| 11 | 1 | 959.601 | 0.130 | 959.294 | 14.04 |
| 11 | 1 | 960.158 | 0.140 | 959.801 | 15.14 |
| 11 | 1 | 960.216 | 0.117 | 959.780 | 15.96 |
| 11 | 1 | 959.019 | 0.141 | 958.657 | 16.95 |
| 11 | 1 | 959.567 | 0.161 | 959.353 | 17.97 |
| 11 | 1 | 959.404 | 0.136 | 959.006 | 18.86 |
| 11 | 1 | 959.507 | 0.140 | 959.283 | 19.84 |
| 11 | 1 | 959.571 | 0.155 | 959.347 | 20.91 |
| 11 | 1 | 959.454 | 0.159 | 959.221 | 21.95 |
| 11 | 1 | 959.174 | 0.142 | 958.957 | 23.02 |
| 11 | 1 | 959.066 | 0.135 | 958.754 | 24.72 |
| 12 | 2 | 958.462 | 0.211 | 958.664 | 70.98 |
| 12 | 2 | 958.064 | 0.271 | 958.154 | 70.18 |
| 12 | 2 | 959.620 | 0.178 | 959.749 | 69.47 |
| 12 | 2 | 959.246 | 0.198 | 959.386 | 68.84 |
| 12 | 2 | 958.480 | 0.211 | 958.662 | 68.20 |
| 12 | 2 | 958.647 | 0.184 | 958.759 | 67.63 |
| 12 | 2 | 959.190 | 0.190 | 959.339 | 67.10 |
| 12 | 2 | 960.053 | 0.187 | 960.237 | 66.55 |
| 12 | 2 | 957.972 | 0.205 | 958.100 | 66.03 |
| 12 | 2 | 958.717 | 0.189 | 958.878 | 65.54 |
| 12 | 2 | 958.686 | 0.214 | 958.805 | 65.03 |
| 12 | 2 | 959.019 | 0.213 | 959.219 | 64.55 |
| 19 | 1 | 960.682 | 0.184 | 960.411 | 14.60 |
| 19 | 1 | 959.673 | 0.205 | 959.328 | 15.50 |
| 19 | 1 | 959.284 | 0.181 | 959.003 | 16.20 |
| 19 | 1 | 959.803 | 0.189 | 959.553 | 16.95 |
| 19 | 1 | 960.046 | 0.167 | 959.693 | 17.89 |
| 19 | 1 | 959.218 | 0.178 | 959.006 | 18.70 |
| 19 | 1 | 959.069 | 0.159 | 958.816 | 19.62 |
| 19 | 1 | 958.892 | 0.218 | 958.643 | 20.70 |
| 19 | 1 | 959.354 | 0.170 | 959.179 | 21.84 |
| 19 | 1 | 959.533 | 0.166 | 959.329 | 22.89 |
| 19 | 1 | 959.007 | 0.243 | 958.845 | 24.00 |
| 19 | 1 | 959.876 | 0.186 | 959.793 | 25.07 |
| 19 | 1 | 959.494 | 0.187 | 959.412 | 26.24 |
| 19 | 1 | 959.235 | 0.249 | 959.182 | 27.50 |
| 19 | 1 | 959.369 | 0.220 | 959.374 | 28.80 |
| 19 | 1 | 959.509 | 0.217 | 959.545 | 30.14 |
| 19 | 2 | 958.926 | 0.205 | 959.150 | 74.69 |
| 19 | 2 | 958.921 | 0.174 | 958.992 | 73.80 |
| 19 | 2 | 959.517 | 0.165 | 959.626 | 72.87 |
| 19 | 2 | 959.247 | 0.190 | 959.422 | 71.26 |
| 19 | 2 | 958.552 | 0.231 | 958.783 | 70.55 |
| 19 | 2 | 959.854 | 0.224 | 959.968 | 69.80 |
| 19 | 2 | 959.418 | 0.211 | 959.553 | 69.14 |
| 19 | 2 | 959.947 | 0.185 | 960.045 | 68.47 |
| 19 | 2 | 959.619 | 0.215 | 959.828 | 67.75 |
| 20 | 1 | 960.028 | 0.138 | 959.585 | 14.98 |
| 20 | 1 | 959.186 | 0.134 | 958.788 | 15.74 |
| 20 | 1 | 959.446 | 0.139 | 959.103 | 16.46 |
| 20 | 1 | 959.120 | 0.123 | 958.703 | 17.27 |
| 20 | 1 | 959.489 | 0.170 | 959.252 | 18.06 |
| 20 | 1 | 959.634 | 0.139 | 959.389 | 18.95 |
| 20 | 1 | 959.179 | 0.156 | 958.936 | 19.81 |
| 20 | 1 | 959.490 | 0.148 | 959.194 | 20.84 |
| 20 | 1 | 959.353 | 0.122 | 959.092 | 21.79 |
| 20 | 1 | 959.662 | 0.195 | 959.528 | 23.91 |
| 20 | 1 | 959.613 | 0.169 | 959.521 | 24.99 |
| 20 | 1 | 959.477 | 0.145 | 959.341 | 26.13 |
| 20 | 2 | 959.713 | 0.259 | 959.705 | 72.45 |
| 20 | 2 | 958.044 | 0.276 | 958.074 | 68.72 |
| 20 | 2 | 958.523 | 0.208 | 958.482 | 68.02 |
| 20 | 2 | 958.213 | 0.247 | 958.293 | 67.39 |
| 23 | 1 | 959.682 | 0.166 | 959.681 | 28.80 |
| 23 | 1 | 959.977 | 0.168 | 959.937 | 30.08 |
| 23 | 1 | 959.565 | 0.169 | 959.446 | 31.42 |
| 23 | 1 | 959.222 | 0.175 | 959.242 | 32.89 |
| 23 | 1 | 959.697 | 0.170 | 959.780 | 34.44 |

| 2001 - ABRIL |||||
|---|---|---|---|---|
| D | L | SDB | ER | SDC | HL |
| 23 | 1 | 959.091 | 0.181 | 959.299 | 36.14 |
| 23 | 2 | 959.511 | 0.257 | 959.662 | 73.72 |
| 23 | 2 | 958.471 | 0.150 | 958.516 | 71.99 |
| 23 | 2 | 959.842 | 0.152 | 959.832 | 71.22 |
| 23 | 2 | 958.879 | 0.148 | 958.892 | 70.51 |
| 23 | 2 | 959.471 | 0.169 | 959.591 | 69.78 |
| 23 | 2 | 959.261 | 0.159 | 959.330 | 69.06 |
| 23 | 2 | 958.048 | 0.180 | 958.162 | 68.40 |
| 23 | 2 | 959.296 | 0.185 | 959.418 | 67.55 |
| 23 | 2 | 958.979 | 0.198 | 959.123 | 66.91 |
| 23 | 2 | 958.402 | 0.271 | 958.576 | 66.31 |
| 24 | 1 | 959.680 | 0.204 | 959.455 | 15.53 |
| 24 | 1 | 960.386 | 0.192 | 960.156 | 17.11 |
| 24 | 1 | 960.002 | 0.195 | 959.821 | 18.74 |
| 24 | 1 | 958.746 | 0.147 | 958.410 | 19.57 |
| 24 | 1 | 959.148 | 0.212 | 959.047 | 20.42 |
| 24 | 1 | 959.850 | 0.162 | 959.640 | 21.31 |
| 24 | 1 | 959.941 | 0.149 | 959.694 | 22.23 |
| 24 | 1 | 959.255 | 0.181 | 959.106 | 23.38 |
| 24 | 1 | 959.062 | 0.177 | 958.940 | 24.36 |
| 24 | 1 | 960.963 | 0.260 | 960.980 | 26.63 |
| 24 | 1 | 958.612 | 0.184 | 958.540 | 27.82 |
| 24 | 2 | 959.340 | 0.130 | 959.310 | 73.39 |
| 24 | 2 | 959.285 | 0.124 | 959.254 | 72.59 |
| 24 | 2 | 959.949 | 0.163 | 960.069 | 71.76 |
| 24 | 2 | 959.110 | 0.142 | 959.165 | 70.98 |
| 24 | 2 | 959.319 | 0.180 | 959.409 | 70.24 |
| 24 | 2 | 959.400 | 0.143 | 959.437 | 69.54 |
| 24 | 2 | 959.195 | 0.163 | 959.248 | 68.84 |
| 24 | 2 | 959.672 | 0.173 | 959.775 | 68.19 |
| 24 | 2 | 958.978 | 0.170 | 959.072 | 67.56 |
| 24 | 2 | 959.983 | 0.180 | 960.047 | 66.93 |
| 25 | 1 | 960.089 | 0.140 | 959.803 | 16.51 |
| 25 | 1 | 959.398 | 0.186 | 959.224 | 17.23 |
| 25 | 1 | 959.284 | 0.159 | 959.096 | 24.10 |
| 25 | 1 | 960.130 | 0.184 | 960.152 | 28.70 |
| 25 | 1 | 959.969 | 0.179 | 959.922 | 29.92 |
| 25 | 1 | 959.816 | 0.178 | 959.759 | 31.25 |
| 26 | 1 | 959.872 | 0.182 | 959.681 | 20.55 |
| 26 | 1 | 959.943 | 0.185 | 959.764 | 22.24 |
| 26 | 1 | 960.332 | 0.167 | 960.160 | 23.23 |
| 26 | 1 | 960.035 | 0.177 | 959.880 | 24.87 |
| 26 | 1 | 959.321 | 0.186 | 959.147 | 26.04 |
| 26 | 1 | 960.146 | 0.198 | 960.071 | 27.20 |
| 26 | 1 | 959.480 | 0.211 | 959.512 | 28.36 |
| 26 | 1 | 959.673 | 0.206 | 959.723 | 29.63 |
| 26 | 1 | 958.949 | 0.236 | 959.092 | 30.94 |
| 26 | 1 | 959.746 | 0.209 | 959.868 | 32.32 |
| 26 | 1 | 959.975 | 0.167 | 959.981 | 33.76 |
| 27 | 1 | 959.507 | 0.199 | 959.497 | 26.74 |
| 27 | 1 | 958.988 | 0.192 | 958.978 | 27.97 |
| 27 | 1 | 958.911 | 0.165 | 958.911 | 29.26 |
| 27 | 1 | 959.219 | 0.181 | 959.272 | 30.53 |
| 27 | 1 | 960.005 | 0.199 | 960.126 | 31.89 |
| 27 | 1 | 959.052 | 0.195 | 959.236 | 33.36 |
| 27 | 1 | 957.857 | 0.197 | 958.061 | 34.89 |
| 27 | 1 | 958.682 | 0.168 | 958.894 | 36.58 |
| 27 | 2 | 958.382 | 0.202 | 958.512 | 74.86 |
| 27 | 2 | 958.476 | 0.226 | 958.706 | 73.86 |
| 27 | 2 | 959.479 | 0.210 | 959.615 | 73.01 |
| 27 | 2 | 959.548 | 0.181 | 959.653 | 71.01 |
| 27 | 2 | 958.497 | 0.177 | 958.618 | 70.27 |
| 27 | 2 | 958.929 | 0.236 | 959.091 | 69.59 |
| 27 | 2 | 959.692 | 0.226 | 959.813 | 68.90 |
| 27 | 2 | 958.148 | 0.203 | 958.265 | 68.24 |
| 27 | 2 | 958.482 | 0.198 | 959.014 | 67.61 |
| 27 | 2 | 959.114 | 0.236 | 959.272 | 66.99 |
| 27 | 2 | 958.762 | 0.271 | 958.853 | 66.43 |

| 2001 - MAIO |||||
|---|---|---|---|---|
| D | L | SDB | ER | SDC | HL |
| 02 | 1 | 959.205 | 0.247 | 959.356 | 31.37 |
| 02 | 1 | 959.285 | 0.330 | 959.473 | 32.65 |
| 02 | 1 | 959.087 | 0.320 | 959.359 | 35.54 |
| 02 | 1 | 958.297 | 0.412 | 958.762 | 38.92 |
| 03 | 1 | 958.714 | 0.347 | 958.891 | 31.20 |
| 03 | 1 | 957.708 | 0.371 | 957.965 | 34.21 |
| 03 | 1 | 958.167 | 0.311 | 958.435 | 35.61 |
| 03 | 1 | 959.814 | 0.346 | 960.121 | 37.10 |
| 03 | 1 | 959.366 | 0.328 | 959.723 | 38.72 |
| 03 | 2 | 959.450 | 0.233 | 959.594 | 75.14 |
| 03 | 2 | 958.345 | 0.212 | 958.446 | 74.18 |
| 03 | 2 | 958.625 | 0.203 | 958.720 | 73.09 |
| 03 | 2 | 959.108 | 0.200 | 959.134 | 70.44 |
| 03 | 2 | 959.254 | 0.211 | 959.347 | 69.55 |
| 03 | 2 | 958.517 | 0.238 | 958.700 | 68.85 |
| 03 | 2 | 958.651 | 0.233 | 958.731 | 68.03 |
| 04 | 1 | 958.955 | 0.255 | 958.876 | 23.50 |
| 04 | 1 | 958.714 | 0.257 | 958.879 | 24.49 |
| 04 | 1 | 958.178 | 0.253 | 958.015 | 25.52 |
| 04 | 1 | 958.558 | 0.262 | 958.413 | 26.51 |
| 04 | 1 | 958.071 | 0.291 | 958.003 | 27.58 |
| 09 | 2 | 958.864 | 0.195 | 958.947 | 75.15 |
| 09 | 2 | 958.963 | 0.214 | 959.071 | 74.15 |
| 09 | 2 | 958.873 | 0.234 | 959.018 | 73.19 |
| 09 | 2 | 958.526 | 0.209 | 958.624 | 72.26 |
| 09 | 2 | 958.786 | 0.266 | 958.932 | 71.38 |
| 09 | 2 | 958.340 | 0.220 | 958.480 | 70.54 |
| 09 | 2 | 959.069 | 0.232 | 959.142 | 69.73 |
| 09 | 2 | 959.129 | 0.207 | 959.229 | 68.88 |
| 10 | 2 | 958.342 | 0.179 | 958.370 | 78.89 |
| 10 | 2 | 959.298 | 0.230 | 959.395 | 77.71 |
| 10 | 2 | 959.154 | 0.210 | 959.230 | 76.52 |
| 10 | 2 | 959.439 | 0.226 | 959.543 | 75.37 |
| 10 | 2 | 959.510 | 0.166 | 959.447 | 74.28 |
| 10 | 2 | 959.702 | 0.207 | 959.776 | 73.23 |
| 10 | 2 | 958.972 | 0.199 | 959.105 | 72.31 |
| 10 | 2 | 959.048 | 0.208 | 959.178 | 71.39 |
| 10 | 2 | 958.906 | 0.199 | 958.934 | 70.52 |
| 10 | 2 | 960.128 | 0.231 | 960.257 | 69.67 |
| 11 | 2 | 957.525 | 0.186 | 957.519 | 82.63 |
| 11 | 2 | 957.369 | 0.202 | 957.338 | 81.26 |
| 11 | 2 | 958.794 | 0.187 | 958.902 | 79.62 |
| 11 | 2 | 957.741 | 0.175 | 957.820 | 78.39 |
| 11 | 2 | 958.582 | 0.160 | 958.630 | 77.25 |
| 11 | 2 | 958.592 | 0.153 | 958.611 | 76.09 |
| 11 | 2 | 958.038 | 0.197 | 958.170 | 75.03 |
| 11 | 2 | 958.726 | 0.152 | 958.762 | 74.08 |
| 11 | 2 | 958.936 | 0.142 | 958.908 | 73.10 |
| 11 | 2 | 958.219 | 0.175 | 958.279 | 72.21 |
| 11 | 2 | 958.492 | 0.187 | 958.493 | 71.32 |
| 11 | 2 | 958.764 | 0.272 | 958.947 | 69.96 |
| 17 | 2 | 957.552 | 0.180 | 957.534 | 78.36 |
| 17 | 2 | 958.589 | 0.208 | 958.660 | 77.14 |
| 17 | 2 | 958.837 | 0.218 | 958.988 | 75.93 |
| 17 | 2 | 959.981 | 0.226 | 959.946 | 74.82 |
| 17 | 2 | 959.332 | 0.226 | 959.398 | 73.69 |
| 17 | 2 | 959.706 | 0.214 | 959.855 | 72.63 |
| 17 | 2 | 959.269 | 0.243 | 959.430 | 71.66 |
| 17 | 2 | 959.269 | 0.227 | 959.411 | 70.58 |
| 17 | 2 | 959.557 | 0.200 | 959.585 | 69.67 |
| 17 | 2 | 960.007 | 0.226 | 960.080 | 68.79 |
| 17 | 2 | 959.896 | 0.225 | 959.951 | 67.94 |
| 18 | 1 | 958.448 | 0.133 | 958.223 | 28.60 |
| 18 | 1 | 958.358 | 0.142 | 958.248 | 29.62 |
| 18 | 1 | 958.680 | 0.118 | 958.501 | 30.67 |
| 18 | 1 | 958.482 | 0.150 | 958.442 | 31.71 |
| 18 | 1 | 958.006 | 0.152 | 957.998 | 32.75 |
| 18 | 1 | 957.796 | 0.145 | 957.773 | 33.90 |
| 18 | 1 | 957.247 | 0.152 | 957.218 | 35.04 |
| 18 | 1 | 958.435 | 0.177 | 958.596 | 37.86 |
| 18 | 1 | 958.059 | 0.161 | 958.177 | 39.46 |

| 2001 - MAIO | | | | | | 2001 - MAIO | | | | |
|---|---|---|---|---|---|---|---|---|---|---|
| D | L | SDB | ER | SDC | HL | D | L | SDB | ER | SDC | HL |
| 18 | 1 | 958.034 | 0.173 | 958.228 | 41.05 | 30 | 1 | 958.257 | 0.163 | 958.476 | 44.62 |
| 18 | 2 | 958.676 | 0.156 | 958.670 | 72.80 | 30 | 1 | 959.465 | 0.152 | 959.666 | 46.29 |
| 18 | 2 | 957.562 | 0.191 | 957.556 | 71.74 | 30 | 1 | 957.928 | 0.162 | 958.215 | 47.83 |
| 18 | 2 | 958.124 | 0.141 | 958.074 | 70.62 | 30 | 1 | 958.100 | 0.163 | 958.428 | 49.53 |
| 18 | 2 | 958.744 | 0.130 | 958.661 | 69.71 | 30 | 1 | 957.787 | 0.173 | 958.182 | 51.47 |
| 18 | 2 | 958.962 | 0.140 | 958.847 | 68.85 | 30 | 2 | 959.298 | 0.168 | 959.287 | 71.78 |
| 18 | 2 | 959.285 | 0.154 | 959.326 | 68.02 | 30 | 2 | 958.961 | 0.168 | 958.985 | 70.57 |
| 18 | 2 | 959.456 | 0.198 | 959.450 | 67.17 | 30 | 2 | 959.272 | 0.166 | 959.292 | 69.47 |
| 18 | 2 | 958.985 | 0.179 | 958.913 | 66.28 | 30 | 2 | 958.463 | 0.187 | 958.480 | 68.48 |
| 21 | 1 | 960.578 | 0.165 | 960.555 | 32.91 | 30 | 2 | 958.248 | 0.156 | 958.202 | 67.47 |
| 21 | 1 | 960.086 | 0.177 | 960.107 | 33.94 | 30 | 2 | 958.929 | 0.176 | 958.881 | 66.49 |
| 21 | 1 | 959.832 | 0.167 | 959.837 | 35.04 | 30 | 2 | 958.436 | 0.198 | 958.383 | 65.54 |
| 21 | 1 | 959.523 | 0.166 | 959.478 | 36.16 | 30 | 2 | 958.901 | 0.337 | 958.765 | 64.62 |
| 21 | 1 | 959.265 | 0.154 | 959.318 | 37.40 | | | | | | |
| 21 | 1 | 959.086 | 0.170 | 959.261 | 38.64 | | | 2001 - JUNHO | | | |
| 21 | 1 | 959.151 | 0.179 | 959.361 | 39.98 | D | L | SDB | ER | SDC | HL |
| 21 | 1 | 959.322 | 0.152 | 959.390 | 41.38 | 01 | 1 | 959.444 | 0.155 | 959.440 | 38.02 |
| 21 | 1 | 958.553 | 0.165 | 958.782 | 42.93 | 01 | 1 | 958.941 | 0.164 | 958.943 | 39.06 |
| 21 | 1 | 958.121 | 0.188 | 958.490 | 44.63 | 01 | 1 | 958.747 | 0.154 | 958.761 | 40.23 |
| 21 | 1 | 958.243 | 0.141 | 958.522 | 46.45 | 01 | 1 | 957.806 | 0.166 | 957.910 | 41.35 |
| 21 | 2 | 960.099 | 0.148 | 960.028 | 75.95 | 01 | 1 | 958.112 | 0.183 | 958.243 | 42.62 |
| 21 | 2 | 959.460 | 0.164 | 959.450 | 74.35 | 01 | 1 | 957.967 | 0.175 | 958.121 | 43.93 |
| 21 | 2 | 959.572 | 0.190 | 959.658 | 73.15 | 01 | 1 | 958.703 | 0.165 | 958.914 | 45.35 |
| 21 | 2 | 959.742 | 0.157 | 959.737 | 72.09 | 01 | 1 | 958.471 | 0.182 | 958.738 | 46.85 |
| 21 | 2 | 959.569 | 0.174 | 959.626 | 71.06 | 01 | 1 | 958.202 | 0.192 | 958.517 | 48.44 |
| 21 | 2 | 959.896 | 0.147 | 959.847 | 69.93 | 01 | 1 | 958.467 | 0.146 | 958.742 | 50.09 |
| 21 | 2 | 959.908 | 0.209 | 960.055 | 69.00 | 01 | 2 | 960.126 | 0.141 | 960.064 | 73.64 |
| 21 | 2 | 959.784 | 0.167 | 959.833 | 68.13 | 01 | 2 | 958.798 | 0.157 | 958.718 | 72.33 |
| 21 | 2 | 960.284 | 0.195 | 960.374 | 67.25 | 01 | 2 | 959.731 | 0.177 | 959.621 | 71.16 |
| 21 | 2 | 959.604 | 0.199 | 959.616 | 66.31 | 01 | 2 | 959.696 | 0.165 | 959.708 | 70.02 |
| 22 | 1 | 958.943 | 0.180 | 959.096 | 39.79 | 01 | 2 | 959.550 | 0.164 | 959.340 | 68.91 |
| 22 | 1 | 959.333 | 0.150 | 959.467 | 41.13 | 01 | 2 | 959.718 | 0.174 | 959.710 | 67.63 |
| 22 | 1 | 959.678 | 0.193 | 959.952 | 42.93 | 01 | 2 | 958.754 | 0.183 | 958.747 | 66.51 |
| 22 | 1 | 958.164 | 0.144 | 958.510 | 48.53 | 04 | 2 | 958.324 | 0.184 | 958.295 | 74.16 |
| 22 | 1 | 958.732 | 0.140 | 959.091 | 50.66 | 04 | 2 | 958.968 | 0.161 | 958.879 | 72.78 |
| 22 | 1 | 959.075 | 0.151 | 959.602 | 52.96 | 04 | 2 | 958.567 | 0.192 | 958.516 | 71.53 |
| 22 | 2 | 959.041 | 0.137 | 958.943 | 71.94 | 04 | 2 | 958.933 | 0.191 | 958.974 | 70.24 |
| 22 | 2 | 958.187 | 0.193 | 958.278 | 70.92 | 04 | 2 | 959.550 | 0.187 | 959.538 | 69.08 |
| 22 | 2 | 958.993 | 0.164 | 958.959 | 69.94 | 04 | 2 | 959.242 | 0.169 | 959.111 | 67.98 |
| 22 | 2 | 957.957 | 0.161 | 957.941 | 69.00 | 04 | 2 | 959.797 | 0.206 | 959.657 | 66.90 |
| 22 | 2 | 959.096 | 0.151 | 959.067 | 68.05 | 05 | 2 | 957.956 | 0.253 | 957.973 | 70.85 |
| 22 | 2 | 959.057 | 0.195 | 958.946 | 67.22 | 05 | 2 | 958.075 | 0.314 | 958.128 | 69.64 |
| 22 | 2 | 958.936 | 0.247 | 958.870 | 66.34 | 05 | 2 | 957.754 | 0.170 | 957.628 | 68.54 |
| 28 | 2 | 959.056 | 0.169 | 959.011 | 73.21 | 06 | 1 | 960.488 | 0.205 | 960.428 | 41.50 |
| 28 | 2 | 959.020 | 0.166 | 958.978 | 72.03 | 06 | 1 | 959.251 | 0.169 | 959.317 | 42.65 |
| 29 | 1 | 960.649 | 0.181 | 960.533 | 35.29 | 06 | 1 | 959.538 | 0.148 | 959.626 | 43.78 |
| 29 | 1 | 960.576 | 0.134 | 960.478 | 36.30 | 06 | 1 | 959.417 | 0.129 | 959.461 | 44.97 |
| 29 | 1 | 960.340 | 0.141 | 960.297 | 37.34 | 06 | 1 | 958.870 | 0.129 | 958.932 | 46.47 |
| 29 | 1 | 960.598 | 0.165 | 960.682 | 38.40 | 06 | 1 | 959.341 | 0.129 | 959.487 | 47.97 |
| 29 | 1 | 959.240 | 0.141 | 959.252 | 39.52 | 06 | 1 | 958.800 | 0.140 | 959.027 | 49.39 |
| 29 | 1 | 959.775 | 0.135 | 959.771 | 40.79 | 06 | 1 | 958.862 | 0.132 | 959.046 | 50.95 |
| 29 | 1 | 959.733 | 0.137 | 959.753 | 42.13 | 06 | 1 | 958.215 | 0.163 | 958.531 | 52.60 |
| 29 | 1 | 958.913 | 0.138 | 959.040 | 43.51 | 06 | 1 | 957.639 | 0.135 | 957.988 | 54.37 |
| 29 | 1 | 958.969 | 0.132 | 959.045 | 44.90 | 06 | 1 | 958.159 | 0.169 | 958.627 | 56.30 |
| 29 | 1 | 959.491 | 0.130 | 959.601 | 46.36 | 06 | 2 | 957.390 | 0.146 | 957.232 | 69.95 |
| 29 | 1 | 959.137 | 0.136 | 959.286 | 47.97 | 06 | 2 | 957.915 | 0.153 | 957.834 | 68.73 |
| 29 | 2 | 958.626 | 0.154 | 958.607 | 74.93 | 06 | 2 | 958.810 | 0.169 | 958.781 | 67.57 |
| 29 | 2 | 958.696 | 0.174 | 958.598 | 73.60 | 06 | 2 | 958.498 | 0.145 | 958.440 | 66.50 |
| 29 | 2 | 959.027 | 0.152 | 958.962 | 72.39 | 06 | 2 | 958.962 | 0.167 | 958.924 | 65.33 |
| 29 | 2 | 959.360 | 0.131 | 959.190 | 71.24 | 06 | 2 | 958.515 | 0.200 | 958.549 | 64.37 |
| 29 | 2 | 958.820 | 0.134 | 958.663 | 70.04 | 06 | 2 | 958.522 | 0.295 | 958.375 | 62.52 |
| 29 | 2 | 959.321 | 0.162 | 959.301 | 69.02 | 08 | 1 | 958.474 | 0.213 | 958.511 | 41.20 |
| 29 | 2 | 959.844 | 0.136 | 959.776 | 68.01 | 08 | 1 | 958.659 | 0.213 | 958.713 | 42.26 |
| 29 | 2 | 958.965 | 0.164 | 958.982 | 67.01 | 08 | 1 | 958.924 | 0.181 | 959.006 | 43.42 |
| 29 | 2 | 958.285 | 0.230 | 958.250 | 65.92 | 08 | 1 | 958.798 | 0.169 | 958.898 | 44.61 |
| 29 | 2 | 959.248 | 0.255 | 959.090 | 65.04 | 08 | 1 | 958.365 | 0.174 | 958.529 | 45.82 |
| 30 | 1 | 960.432 | 0.142 | 960.371 | 38.37 | 08 | 1 | 958.497 | 0.149 | 958.607 | 47.23 |
| 30 | 1 | 960.297 | 0.160 | 960.316 | 39.48 | 08 | 1 | 957.394 | 0.230 | 957.616 | 48.64 |
| 30 | 1 | 959.904 | 0.176 | 959.973 | 40.66 | 08 | 1 | 957.921 | 0.239 | 958.201 | 50.08 |
| 30 | 1 | 959.230 | 0.186 | 959.374 | 41.91 | 08 | 1 | 957.630 | 0.206 | 957.932 | 51.71 |
| 30 | 1 | 960.240 | 0.146 | 960.320 | 43.24 | | | | | | |

| 2001 - JUNHO | | | | | |
|---|---|---|---|---|---|
| D | L | SDB | ER | SDC | HL |
| 08 | 1 | 958.150 | 0.241 | 958.533 | 53.38 |
| 08 | 2 | 959.432 | 0.222 | 959.547 | 74.22 |
| 08 | 2 | 959.411 | 0.184 | 959.413 | 72.73 |
| 11 | 1 | 958.213 | 0.163 | 958.294 | 44.52 |
| 11 | 1 | 958.821 | 0.166 | 958.880 | 45.70 |
| 11 | 1 | 959.315 | 0.223 | 959.460 | 47.01 |
| 11 | 1 | 959.036 | 0.235 | 959.260 | 48.30 |
| 11 | 1 | 958.890 | 0.175 | 959.109 | 49.59 |
| 11 | 1 | 958.462 | 0.200 | 958.733 | 50.96 |
| 11 | 1 | 958.545 | 0.197 | 958.864 | 52.50 |
| 11 | 1 | 958.289 | 0.201 | 958.703 | 54.16 |
| 11 | 1 | 958.322 | 0.222 | 958.689 | 55.93 |
| 11 | 2 | 958.464 | 0.188 | 958.463 | 70.07 |
| 11 | 2 | 959.177 | 0.176 | 959.199 | 68.77 |
| 11 | 2 | 958.798 | 0.191 | 958.856 | 67.45 |
| 11 | 2 | 959.291 | 0.197 | 959.326 | 66.27 |
| 11 | 2 | 959.799 | 0.226 | 959.874 | 65.15 |
| 11 | 2 | 959.297 | 0.184 | 959.226 | 63.92 |
| 11 | 2 | 959.494 | 0.190 | 959.488 | 62.63 |
| 15 | 2 | 958.745 | 0.168 | 958.718 | 72.41 |
| 15 | 2 | 960.016 | 0.192 | 959.921 | 70.83 |
| 15 | 2 | 958.997 | 0.198 | 959.003 | 69.34 |
| 15 | 2 | 959.284 | 0.192 | 959.255 | 67.93 |
| 15 | 2 | 958.890 | 0.173 | 958.817 | 65.50 |
| 15 | 2 | 958.558 | 0.158 | 958.493 | 64.29 |
| 15 | 2 | 958.423 | 0.164 | 958.346 | 63.16 |
| 15 | 2 | 959.211 | 0.170 | 959.147 | 62.05 |
| 15 | 2 | 959.547 | 0.169 | 959.515 | 61.04 |
| 16 | 2 | 958.871 | 0.161 | 958.785 | 69.81 |
| 16 | 2 | 959.976 | 0.202 | 959.963 | 68.38 |
| 16 | 2 | 958.944 | 0.149 | 958.897 | 67.00 |
| 16 | 2 | 958.856 | 0.142 | 958.751 | 65.65 |
| 16 | 2 | 958.529 | 0.178 | 958.446 | 64.34 |
| 16 | 2 | 958.796 | 0.189 | 958.785 | 63.22 |
| 16 | 2 | 959.289 | 0.237 | 959.326 | 61.92 |
| 25 | 2 | 959.671 | 0.186 | 959.619 | 66.42 |
| 25 | 2 | 958.673 | 0.172 | 958.608 | 64.91 |
| 25 | 2 | 959.290 | 0.215 | 959.196 | 63.49 |
| 25 | 2 | 960.051 | 0.190 | 960.039 | 62.19 |
| 25 | 2 | 959.572 | 0.178 | 959.532 | 60.93 |
| 25 | 2 | 959.231 | 0.167 | 959.190 | 59.74 |
| 25 | 2 | 959.768 | 0.211 | 959.767 | 58.61 |
| 30 | 2 | 959.462 | 0.167 | 959.377 | 67.67 |
| 30 | 2 | 958.695 | 0.180 | 958.683 | 65.90 |
| 30 | 2 | 959.040 | 0.200 | 958.978 | 64.22 |
| 30 | 2 | 958.999 | 0.173 | 958.957 | 62.64 |
| 30 | 2 | 959.134 | 0.179 | 959.069 | 61.24 |
| 30 | 2 | 959.354 | 0.164 | 959.288 | 59.87 |
| 30 | 2 | 959.082 | 0.207 | 959.130 | 58.57 |
| 30 | 2 | 959.555 | 0.192 | 959.502 | 57.35 |
| 30 | 2 | 959.207 | 0.181 | 959.171 | 56.22 |
| 30 | 2 | 958.877 | 0.211 | 958.783 | 55.13 |

| 2001 - JULHO | | | | | |
|---|---|---|---|---|---|
| D | L | SDB | ER | SDC | HL |
| 02 | 1 | 959.995 | 0.217 | 960.146 | 54.76 |
| 02 | 1 | 959.467 | 0.178 | 959.636 | 56.28 |
| 02 | 1 | 959.447 | 0.538 | 959.476 | 58.20 |
| 02 | 1 | 959.448 | 0.233 | 959.771 | 62.81 |
| 10 | 1 | 960.465 | 0.196 | 960.602 | 60.34 |
| 10 | 1 | 959.203 | 0.173 | 959.289 | 61.61 |
| 10 | 1 | 959.247 | 0.197 | 959.465 | 62.85 |
| 10 | 1 | 959.275 | 0.168 | 959.519 | 64.21 |
| 10 | 1 | 959.880 | 0.158 | 960.165 | 65.61 |
| 10 | 1 | 959.067 | 0.151 | 959.304 | 67.12 |
| 10 | 1 | 959.061 | 0.166 | 959.447 | 68.75 |
| 10 | 1 | 959.301 | 0.158 | 959.696 | 70.51 |
| 10 | 1 | 958.607 | 0.175 | 959.158 | 72.41 |
| 10 | 2 | 957.289 | 0.170 | 957.195 | 59.05 |
| 10 | 2 | 958.449 | 0.176 | 958.336 | 57.52 |

| 2001 - JULHO | | | | | |
|---|---|---|---|---|---|
| D | L | SDB | ER | SDC | HL |
| 10 | 2 | 959.390 | 0.201 | 959.363 | 56.10 |
| 10 | 2 | 958.478 | 0.219 | 958.479 | 54.77 |
| 10 | 2 | 957.572 | 0.230 | 957.580 | 53.53 |
| 10 | 2 | 957.969 | 0.240 | 957.945 | 52.36 |
| 10 | 2 | 958.729 | 0.211 | 958.685 | 51.20 |
| 10 | 2 | 960.025 | 0.208 | 960.039 | 50.12 |
| 10 | 2 | 959.274 | 0.233 | 959.240 | 49.08 |
| 10 | 2 | 960.028 | 0.268 | 959.872 | 47.10 |
| 11 | 1 | 960.919 | 0.150 | 961.097 | 61.84 |
| 11 | 1 | 960.707 | 0.160 | 960.895 | 63.10 |
| 11 | 1 | 959.477 | 0.147 | 959.763 | 65.88 |
| 11 | 1 | 959.019 | 0.183 | 959.361 | 67.51 |
| 11 | 1 | 959.818 | 0.179 | 960.230 | 69.11 |
| 11 | 1 | 959.433 | 0.165 | 959.833 | 70.98 |
| 11 | 1 | 958.986 | 0.151 | 959.424 | 72.86 |
| 11 | 1 | 958.758 | 0.189 | 959.344 | 74.96 |
| 11 | 2 | 957.519 | 0.174 | 957.317 | 56.35 |
| 11 | 2 | 958.803 | 0.189 | 958.583 | 54.93 |
| 11 | 2 | 958.989 | 0.176 | 958.827 | 53.69 |
| 11 | 2 | 958.837 | 0.186 | 958.780 | 52.41 |
| 11 | 2 | 958.202 | 0.190 | 958.155 | 51.26 |
| 11 | 2 | 959.107 | 0.178 | 958.995 | 50.15 |
| 11 | 2 | 958.778 | 0.196 | 958.680 | 49.08 |
| 11 | 2 | 960.407 | 0.201 | 960.212 | 48.11 |
| 16 | 2 | 959.658 | 0.170 | 959.559 | 54.72 |
| 16 | 2 | 959.392 | 0.162 | 959.264 | 53.22 |
| 16 | 2 | 959.686 | 0.200 | 959.628 | 51.79 |
| 16 | 2 | 959.762 | 0.183 | 959.688 | 50.54 |
| 16 | 2 | 959.886 | 0.170 | 959.789 | 49.25 |
| 16 | 2 | 960.144 | 0.207 | 960.166 | 48.07 |
| 16 | 2 | 959.570 | 0.157 | 959.467 | 46.98 |
| 16 | 2 | 958.632 | 0.180 | 958.583 | 45.94 |
| 16 | 2 | 959.739 | 0.182 | 959.684 | 44.89 |
| 16 | 2 | 959.512 | 0.202 | 959.501 | 43.86 |
| 17 | 1 | 960.985 | 0.193 | 960.971 | 57.83 |
| 17 | 1 | 959.974 | 0.182 | 959.990 | 58.81 |
| 17 | 1 | 959.009 | 0.155 | 959.021 | 59.78 |
| 17 | 1 | 958.863 | 0.148 | 958.926 | 60.91 |
| 17 | 1 | 957.953 | 0.127 | 957.974 | 62.07 |
| 17 | 1 | 958.577 | 0.141 | 958.642 | 63.21 |
| 17 | 1 | 957.940 | 0.122 | 957.943 | 64.36 |
| 17 | 1 | 958.246 | 0.139 | 958.367 | 65.77 |
| 17 | 1 | 957.309 | 0.139 | 957.444 | 67.21 |
| 17 | 1 | 958.183 | 0.149 | 958.409 | 68.61 |
| 17 | 1 | 957.386 | 0.151 | 957.596 | 70.09 |
| 17 | 2 | 959.212 | 0.139 | 959.053 | 52.31 |
| 17 | 2 | 959.785 | 0.167 | 959.728 | 50.87 |
| 17 | 2 | 958.307 | 0.175 | 958.283 | 49.59 |
| 17 | 2 | 959.183 | 0.182 | 959.170 | 48.38 |
| 17 | 2 | 959.841 | 0.171 | 959.755 | 47.22 |
| 17 | 2 | 959.320 | 0.167 | 959.256 | 46.18 |
| 17 | 2 | 960.368 | 0.180 | 960.341 | 45.14 |
| 17 | 2 | 959.315 | 0.174 | 959.166 | 44.15 |
| 17 | 2 | 959.717 | 0.242 | 959.698 | 43.21 |
| 17 | 2 | 960.264 | 0.237 | 960.138 | 42.27 |
| 18 | 1 | 960.074 | 0.141 | 960.080 | 61.11 |
| 18 | 1 | 960.135 | 0.132 | 960.202 | 62.17 |
| 18 | 1 | 959.544 | 0.109 | 959.484 | 63.26 |
| 18 | 1 | 959.211 | 0.120 | 959.223 | 64.44 |
| 18 | 1 | 959.111 | 0.130 | 959.242 | 65.68 |
| 18 | 1 | 958.619 | 0.119 | 958.669 | 66.97 |
| 18 | 1 | 959.223 | 0.135 | 959.326 | 68.33 |
| 18 | 1 | 958.160 | 0.134 | 958.207 | 69.82 |
| 18 | 1 | 958.328 | 0.143 | 958.635 | 71.39 |
| 18 | 1 | 958.163 | 0.162 | 958.538 | 73.08 |
| 18 | 2 | 958.712 | 0.169 | 958.559 | 52.08 |
| 18 | 2 | 959.046 | 0.172 | 958.954 | 50.74 |
| 18 | 2 | 959.086 | 0.153 | 958.917 | 49.46 |
| 18 | 2 | 959.342 | 0.169 | 959.265 | 48.25 |
| 18 | 2 | 959.574 | 0.176 | 959.513 | 45.87 |
| 18 | 2 | 959.425 | 0.166 | 959.359 | 44.84 |

|     | 2001 - JULHO | | | | |     | 2001 - JULHO | | | | |
| --- | --- | --- | --- | --- | --- | --- | --- | --- | --- | --- | --- |
| D | L | SDB | ER | SDC | HL | D | L | SDB | ER | SDC | HL |
| 18 | 2 | 959.521 | 0.166 | 959.370 | 43.84 | 27 | 2 | 959.419 | 0.188 | 959.350 | 40.42 |
| 18 | 2 | 959.070 | 0.171 | 958.999 | 42.86 | 27 | 2 | 959.502 | 0.191 | 959.382 | 39.41 |
| 18 | 2 | 959.556 | 0.215 | 959.489 | 41.90 | 27 | 2 | 958.524 | 0.167 | 958.450 | 38.39 |
| 19 | 1 | 958.392 | 0.156 | 958.392 | 59.90 | 27 | 2 | 959.212 | 0.173 | 959.096 | 37.46 |
| 19 | 1 | 958.753 | 0.158 | 958.813 | 60.91 | 27 | 2 | 960.211 | 0.183 | 960.124 | 36.58 |
| 19 | 1 | 958.404 | 0.133 | 958.437 | 62.05 | 30 | 1 | 960.371 | 0.192 | 960.413 | 63.75 |
| 19 | 1 | 958.110 | 0.135 | 958.181 | 63.16 | 30 | 1 | 958.871 | 0.185 | 958.935 | 64.75 |
| 19 | 1 | 958.467 | 0.160 | 958.651 | 64.31 | 30 | 1 | 959.441 | 0.198 | 959.566 | 65.74 |
| 19 | 1 | 957.625 | 0.166 | 957.799 | 65.70 | 30 | 1 | 959.417 | 0.193 | 959.592 | 66.78 |
| 19 | 1 | 957.106 | 0.193 | 957.422 | 66.96 | 30 | 1 | 958.721 | 0.194 | 958.856 | 67.88 |
| 19 | 1 | 957.551 | 0.141 | 957.666 | 68.33 | 30 | 1 | 959.067 | 0.231 | 959.347 | 69.00 |
| 20 | 1 | 960.063 | 0.152 | 960.092 | 60.50 | 30 | 1 | 958.429 | 0.215 | 958.689 | 70.21 |
| 20 | 1 | 959.630 | 0.157 | 959.691 | 61.54 | 30 | 1 | 958.091 | 0.220 | 958.411 | 71.41 |
| 20 | 1 | 958.979 | 0.189 | 959.146 | 62.65 | 30 | 1 | 958.811 | 0.217 | 959.177 | 72.80 |
| 20 | 1 | 959.092 | 0.141 | 959.151 | 63.74 | 30 | 1 | 959.703 | 0.227 | 960.082 | 74.26 |
| 20 | 1 | 958.574 | 0.147 | 958.605 | 64.92 | 30 | 1 | 958.799 | 0.213 | 959.235 | 76.15 |
| 20 | 1 | 958.749 | 0.152 | 958.757 | 66.18 | 31 | 1 | 959.260 | 0.182 | 959.318 | 65.85 |
| 20 | 1 | 958.212 | 0.163 | 958.436 | 67.45 | 31 | 1 | 960.211 | 0.184 | 960.344 | 68.10 |
| 20 | 1 | 958.704 | 0.173 | 958.965 | 68.94 | 31 | 1 | 959.883 | 0.184 | 959.998 | 69.28 |
| 20 | 1 | 958.850 | 0.191 | 959.155 | 70.39 | 31 | 1 | 959.126 | 0.172 | 959.311 | 70.78 |
| 20 | 2 | 958.788 | 0.201 | 958.756 | 50.04 | 31 | 1 | 959.423 | 0.165 | 959.664 | 72.02 |
| 20 | 2 | 959.707 | 0.167 | 959.615 | 48.73 | 31 | 1 | 958.729 | 0.156 | 958.949 | 73.46 |
| 20 | 2 | 959.274 | 0.167 | 959.191 | 47.49 | 31 | 1 | 960.030 | 0.142 | 960.184 | 74.90 |
| 20 | 2 | 959.514 | 0.179 | 959.480 | 46.30 | 31 | 1 | 959.300 | 0.186 | 959.660 | 76.46 |
| 20 | 2 | 958.976 | 0.150 | 958.872 | 45.19 | 31 | 1 | 958.645 | 0.189 | 959.065 | 78.13 |
| 20 | 2 | 960.019 | 0.203 | 959.996 | 44.15 | 31 | 2 | 960.356 | 0.258 | 960.271 | 48.79 |
| 20 | 2 | 959.336 | 0.180 | 959.264 | 43.12 | 31 | 2 | 959.845 | 0.244 | 959.849 | 47.19 |
| 20 | 2 | 959.874 | 0.183 | 959.812 | 42.11 | 31 | 2 | 958.682 | 0.166 | 958.519 | 45.70 |
| 25 | 1 | 959.683 | 0.155 | 959.615 | 65.02 | 31 | 2 | 959.881 | 0.177 | 959.788 | 44.37 |
| 25 | 1 | 959.620 | 0.142 | 959.729 | 66.14 | 31 | 2 | 959.646 | 0.185 | 959.530 | 43.07 |
| 25 | 1 | 959.695 | 0.144 | 959.761 | 67.33 | 31 | 2 | 959.625 | 0.184 | 959.559 | 41.86 |
| 25 | 1 | 959.853 | 0.141 | 959.882 | 68.51 | 31 | 2 | 958.848 | 0.142 | 958.662 | 40.70 |
| 25 | 1 | 959.709 | 0.146 | 959.822 | 69.80 | 31 | 2 | 960.015 | 0.189 | 960.008 | 39.62 |
| 25 | 1 | 959.585 | 0.157 | 959.800 | 71.15 | | | | | | |
| 25 | 1 | 959.186 | 0.155 | 959.454 | 72.62 | | | 2001 - AGOSTO | | | |
| 25 | 1 | 959.413 | 0.168 | 959.688 | 74.26 | D | L | SDB | ER | SDC | HL |
| 26 | 1 | 960.650 | 0.160 | 960.613 | 62.06 | 01 | 1 | 960.422 | 0.133 | 960.354 | 65.11 |
| 26 | 1 | 960.012 | 0.169 | 960.081 | 63.06 | 01 | 1 | 960.707 | 0.131 | 960.685 | 66.18 |
| 26 | 1 | 959.506 | 0.174 | 959.603 | 64.16 | 01 | 1 | 959.756 | 0.151 | 959.778 | 67.20 |
| 26 | 1 | 959.427 | 0.154 | 959.525 | 65.21 | 01 | 1 | 959.507 | 0.161 | 959.581 | 68.30 |
| 26 | 1 | 958.977 | 0.174 | 959.122 | 66.31 | 01 | 1 | 959.390 | 0.163 | 959.545 | 69.45 |
| 26 | 1 | 960.031 | 0.168 | 960.186 | 67.45 | 01 | 1 | 959.385 | 0.140 | 959.498 | 70.66 |
| 26 | 1 | 958.919 | 0.169 | 959.125 | 68.74 | 01 | 1 | 958.785 | 0.159 | 958.895 | 71.91 |
| 26 | 1 | 960.092 | 0.172 | 960.263 | 70.06 | 01 | 1 | 959.435 | 0.164 | 959.650 | 73.33 |
| 26 | 1 | 958.864 | 0.186 | 959.109 | 72.04 | 01 | 1 | 958.937 | 0.168 | 959.211 | 74.73 |
| 26 | 1 | 958.299 | 0.155 | 958.512 | 73.65 | 01 | 1 | 959.299 | 0.177 | 959.596 | 76.24 |
| 26 | 2 | 960.558 | 0.201 | 960.429 | 45.73 | 01 | 1 | 959.676 | 0.165 | 959.887 | 77.92 |
| 26 | 2 | 959.480 | 0.176 | 959.355 | 44.46 | 01 | 1 | 959.097 | 0.216 | 959.444 | 79.70 |
| 26 | 2 | 959.432 | 0.154 | 959.344 | 43.28 | 01 | 2 | 959.716 | 0.170 | 959.571 | 46.89 |
| 26 | 2 | 959.824 | 0.145 | 959.582 | 42.08 | 01 | 2 | 959.724 | 0.177 | 959.535 | 45.43 |
| 26 | 2 | 959.976 | 0.147 | 959.718 | 41.05 | 01 | 2 | 959.667 | 0.155 | 959.531 | 44.06 |
| 26 | 2 | 959.628 | 0.198 | 959.539 | 40.01 | 01 | 2 | 958.580 | 0.173 | 958.465 | 42.82 |
| 26 | 2 | 959.431 | 0.169 | 959.306 | 38.80 | 01 | 2 | 959.027 | 0.166 | 958.895 | 41.66 |
| 26 | 2 | 960.683 | 0.208 | 960.644 | 37.86 | 01 | 2 | 957.659 | 0.203 | 957.604 | 40.45 |
| 26 | 2 | 959.136 | 0.190 | 959.067 | 37.00 | 01 | 2 | 958.926 | 0.182 | 958.905 | 39.38 |
| 26 | 2 | 959.002 | 0.165 | 958.832 | 36.02 | 01 | 2 | 959.619 | 0.179 | 959.520 | 38.38 |
| 27 | 1 | 960.652 | 0.145 | 960.595 | 63.11 | 01 | 2 | 959.436 | 0.175 | 959.389 | 37.41 |
| 27 | 1 | 960.998 | 0.137 | 960.917 | 64.09 | 01 | 2 | 959.566 | 0.205 | 959.515 | 36.47 |
| 27 | 1 | 960.857 | 0.133 | 960.911 | 65.14 | 01 | 2 | 959.836 | 0.147 | 959.682 | 35.54 |
| 27 | 1 | 959.702 | 0.137 | 959.777 | 66.27 | 01 | 2 | 959.320 | 0.166 | 959.220 | 34.67 |
| 27 | 1 | 959.774 | 0.126 | 959.766 | 67.46 | 01 | 2 | 959.750 | 0.161 | 959.664 | 33.79 |
| 27 | 1 | 958.914 | 0.123 | 958.950 | 68.76 | 02 | 1 | 960.392 | 0.157 | 960.408 | 64.30 |
| 27 | 1 | 959.034 | 0.138 | 959.149 | 70.09 | 02 | 1 | 960.638 | 0.172 | 960.698 | 65.27 |
| 27 | 1 | 959.098 | 0.134 | 959.263 | 71.54 | 02 | 1 | 960.615 | 0.182 | 960.692 | 66.34 |
| 27 | 1 | 958.873 | 0.150 | 959.124 | 72.97 | 02 | 1 | 958.869 | 0.152 | 958.976 | 67.42 |
| 27 | 1 | 958.805 | 0.155 | 960.083 | 74.52 | 02 | 1 | 959.025 | 0.153 | 959.129 | 68.48 |
| 27 | 2 | 959.398 | 0.150 | 959.246 | 46.35 | 02 | 1 | 958.939 | 0.162 | 959.048 | 69.69 |
| 27 | 2 | 959.319 | 0.169 | 959.189 | 45.03 | 02 | 1 | 958.213 | 0.169 | 958.389 | 70.94 |
| 27 | 2 | 959.919 | 0.156 | 959.751 | 43.82 | 02 | 1 | 959.325 | 0.183 | 959.594 | 72.23 |
| 27 | 2 | 959.584 | 0.161 | 959.443 | 42.61 | 02 | 1 | 959.578 | 0.199 | 959.832 | 73.61 |
| 27 | 2 | 959.536 | 0.159 | 959.401 | 41.50 | | | | | | |

|  | 2001 - AGOSTO |  |  |  |  |  | 2001 - AGOSTO |  |  |  |
|---|---|---|---|---|---|---|---|---|---|---|
| D | L | SDB | ER | SDC | HL | D | L | SDB | ER | SDC | HL |
| 02 | 1 | 959.015 | 0.156 | 959.267 | 75.14 | 16 | 2 | 959.885 | 0.222 | 959.818 | 25.28 |
| 02 | 1 | 958.732 | 0.186 | 959.097 | 76.70 | 16 | 2 | 959.591 | 0.175 | 959.400 | 24.52 |
| 02 | 2 | 960.055 | 0.168 | 959.927 | 42.84 | 17 | 1 | 960.195 | 0.197 | 960.234 | 65.04 |
| 02 | 2 | 960.085 | 0.193 | 959.963 | 41.56 | 17 | 1 | 959.488 | 0.165 | 959.457 | 65.88 |
| 02 | 2 | 959.551 | 0.209 | 959.538 | 40.40 | 17 | 1 | 958.721 | 0.180 | 958.748 | 66.76 |
| 02 | 2 | 959.654 | 0.195 | 959.549 | 39.22 | 17 | 1 | 958.742 | 0.170 | 958.744 | 67.73 |
| 02 | 2 | 960.337 | 0.222 | 960.364 | 37.32 | 17 | 1 | 958.114 | 0.214 | 958.210 | 69.29 |
| 02 | 2 | 959.735 | 0.191 | 959.666 | 36.31 | 17 | 1 | 958.945 | 0.188 | 959.008 | 70.29 |
| 02 | 2 | 959.493 | 0.211 | 959.453 | 35.36 | 17 | 1 | 958.685 | 0.192 | 958.868 | 71.32 |
| 02 | 2 | 959.144 | 0.224 | 959.121 | 34.44 | 17 | 1 | 959.522 | 0.178 | 959.640 | 72.44 |
| 02 | 2 | 960.393 | 0.198 | 960.350 | 33.44 | 17 | 1 | 959.037 | 0.179 | 959.158 | 73.60 |
| 02 | 2 | 959.853 | 0.215 | 959.787 | 32.60 | 17 | 1 | 958.455 | 0.160 | 958.546 | 74.88 |
| 03 | 1 | 959.423 | 0.184 | 959.452 | 64.70 | 17 | 1 | 959.251 | 0.208 | 959.520 | 76.09 |
| 03 | 1 | 959.035 | 0.176 | 959.245 | 74.83 | 17 | 1 | 958.498 | 0.179 | 958.725 | 77.40 |
| 03 | 1 | 959.148 | 0.158 | 959.357 | 76.32 | 17 | 2 | 959.910 | 0.256 | 959.982 | 31.42 |
| 03 | 1 | 959.533 | 0.185 | 959.912 | 77.87 | 17 | 2 | 958.544 | 0.214 | 958.577 | 30.31 |
| 03 | 1 | 958.700 | 0.163 | 959.088 | 79.58 | 17 | 2 | 958.817 | 0.244 | 958.845 | 29.28 |
| 03 | 2 | 959.615 | 0.205 | 959.483 | 40.43 | 17 | 2 | 959.853 | 0.156 | 959.675 | 27.25 |
| 03 | 2 | 959.137 | 0.173 | 959.044 | 39.28 | 17 | 2 | 959.085 | 0.188 | 958.935 | 26.43 |
| 03 | 2 | 959.845 | 0.165 | 959.721 | 36.58 | 17 | 2 | 960.098 | 0.182 | 960.052 | 25.62 |
| 03 | 2 | 959.538 | 0.167 | 959.384 | 35.60 | 17 | 2 | 959.951 | 0.184 | 959.811 | 24.84 |
| 03 | 2 | 958.655 | 0.198 | 958.673 | 34.69 | 17 | 2 | 959.116 | 0.216 | 959.144 | 24.09 |
| 03 | 2 | 959.013 | 0.172 | 958.920 | 33.80 | 17 | 2 | 959.107 | 0.183 | 959.004 | 23.36 |
| 03 | 2 | 959.568 | 0.199 | 959.543 | 32.92 | 17 | 2 | 959.777 | 0.178 | 959.735 | 22.64 |
| 03 | 2 | 958.893 | 0.192 | 958.750 | 32.08 | 17 | 2 | 959.087 | 0.165 | 959.010 | 21.95 |
| 03 | 2 | 959.211 | 0.205 | 959.158 | 31.26 | 17 | 2 | 960.095 | 0.178 | 960.007 | 21.27 |
| 09 | 1 | 959.816 | 0.168 | 959.806 | 64.69 | 17 | 2 | 960.002 | 0.207 | 960.003 | 20.63 |
| 09 | 1 | 959.733 | 0.157 | 959.669 | 65.52 | 20 | 1 | 960.045 | 0.196 | 960.094 | 65.75 |
| 09 | 1 | 959.382 | 0.187 | 959.472 | 66.40 | 20 | 1 | 960.215 | 0.176 | 960.168 | 66.53 |
| 09 | 1 | 959.199 | 0.159 | 959.212 | 67.35 | 20 | 1 | 959.522 | 0.177 | 959.591 | 68.67 |
| 09 | 1 | 958.254 | 0.167 | 958.340 | 68.46 | 20 | 1 | 959.055 | 0.168 | 959.066 | 69.54 |
| 09 | 1 | 958.754 | 0.191 | 958.899 | 69.53 | 20 | 1 | 959.389 | 0.175 | 959.453 | 70.54 |
| 09 | 1 | 959.339 | 0.180 | 959.469 | 70.62 | 20 | 1 | 958.177 | 0.174 | 958.252 | 71.60 |
| 09 | 1 | 959.425 | 0.171 | 959.573 | 71.74 | 20 | 1 | 957.904 | 0.192 | 958.073 | 72.64 |
| 09 | 1 | 958.882 | 0.179 | 959.024 | 73.09 | 20 | 1 | 958.450 | 0.213 | 958.658 | 73.71 |
| 09 | 1 | 959.576 | 0.161 | 959.746 | 74.32 | 20 | 1 | 958.165 | 0.206 | 958.332 | 74.88 |
| 10 | 1 | 959.342 | 0.153 | 959.321 | 70.29 | 20 | 1 | 958.258 | 0.194 | 958.533 | 76.07 |
| 10 | 1 | 959.046 | 0.194 | 959.074 | 71.74 | 20 | 1 | 958.688 | 0.161 | 958.868 | 77.33 |
| 10 | 1 | 958.905 | 0.206 | 959.084 | 72.93 | 20 | 2 | 960.930 | 0.192 | 960.922 | 21.57 |
| 10 | 1 | 959.256 | 0.176 | 959.388 | 74.22 | 20 | 2 | 960.833 | 0.207 | 960.822 | 20.27 |
| 10 | 1 | 959.001 | 0.172 | 959.206 | 75.51 | 24 | 1 | 958.577 | 0.173 | 958.535 | 65.12 |
| 10 | 1 | 958.590 | 0.209 | 958.864 | 76.93 | 24 | 1 | 958.677 | 0.155 | 958.623 | 65.89 |
| 10 | 1 | 959.073 | 0.179 | 959.366 | 78.46 | 24 | 1 | 958.738 | 0.197 | 958.766 | 66.60 |
| 10 | 2 | 959.278 | 0.165 | 959.156 | 40.15 | 24 | 1 | 958.742 | 0.162 | 958.673 | 67.43 |
| 10 | 2 | 959.188 | 0.174 | 959.043 | 38.76 | 24 | 1 | 958.717 | 0.180 | 958.741 | 68.20 |
| 10 | 2 | 959.782 | 0.150 | 959.528 | 37.51 | 24 | 1 | 958.649 | 0.165 | 958.623 | 69.21 |
| 10 | 2 | 959.755 | 0.208 | 959.710 | 36.36 | 24 | 1 | 959.260 | 0.168 | 959.261 | 70.11 |
| 10 | 2 | 959.253 | 0.167 | 959.136 | 35.09 | 24 | 1 | 958.969 | 0.181 | 958.980 | 71.06 |
| 10 | 2 | 958.995 | 0.160 | 958.777 | 34.06 | 24 | 1 | 958.433 | 0.178 | 958.533 | 72.03 |
| 10 | 2 | 959.450 | 0.208 | 959.318 | 31.84 | 24 | 1 | 958.411 | 0.185 | 958.516 | 73.20 |
| 10 | 2 | 959.899 | 0.226 | 959.820 | 30.94 | 24 | 1 | 958.759 | 0.204 | 958.875 | 74.34 |
| 10 | 2 | 960.261 | 0.237 | 960.119 | 30.07 | 28 | 2 | 958.513 | 0.191 | 958.339 | 26.48 |
| 10 | 2 | 959.788 | 0.217 | 959.700 | 29.25 | 28 | 2 | 959.631 | 0.244 | 959.623 | 25.46 |
| 16 | 1 | 959.554 | 0.218 | 959.544 | 65.95 | 28 | 2 | 958.276 | 0.218 | 958.220 | 23.62 |
| 16 | 1 | 958.879 | 0.195 | 958.905 | 66.74 | 28 | 2 | 959.194 | 0.201 | 959.065 | 22.75 |
| 16 | 1 | 958.100 | 0.151 | 958.008 | 67.58 | 28 | 2 | 958.579 | 0.221 | 958.447 | 21.68 |
| 16 | 1 | 958.079 | 0.200 | 958.138 | 68.43 | 28 | 2 | 958.950 | 0.222 | 958.858 | 20.79 |
| 16 | 1 | 958.770 | 0.171 | 958.813 | 70.59 | 28 | 2 | 959.908 | 0.253 | 959.758 | 20.01 |
| 16 | 1 | 958.068 | 0.190 | 958.199 | 71.66 | 28 | 2 | 958.333 | 0.303 | 958.285 | 18.04 |
| 16 | 1 | 958.589 | 0.156 | 958.671 | 72.81 | 28 | 2 | 959.245 | 0.234 | 959.155 | 17.42 |
| 16 | 1 | 959.263 | 0.151 | 959.248 | 73.97 | 29 | 2 | 958.830 | 0.207 | 958.804 | 27.90 |
| 16 | 1 | 958.954 | 0.151 | 959.017 | 75.12 | 29 | 2 | 958.643 | 0.167 | 958.448 | 26.75 |
| 16 | 2 | 959.520 | 0.182 | 959.482 | 34.40 | 29 | 2 | 959.758 | 0.201 | 959.737 | 25.71 |
| 16 | 2 | 959.572 | 0.173 | 959.442 | 33.22 | 29 | 2 | 958.608 | 0.170 | 958.518 | 24.71 |
| 16 | 2 | 959.125 | 0.207 | 959.115 | 32.14 | 29 | 2 | 958.984 | 0.175 | 958.903 | 23.82 |
| 16 | 2 | 960.443 | 0.182 | 960.261 | 31.09 | 29 | 2 | 958.464 | 0.174 | 958.328 | 22.96 |
| 16 | 2 | 959.688 | 0.217 | 959.519 | 29.99 | 29 | 2 | 959.265 | 0.195 | 959.234 | 22.11 |
| 16 | 2 | 959.042 | 0.183 | 958.983 | 28.87 | 29 | 2 | 959.355 | 0.216 | 959.364 | 21.30 |
| 16 | 2 | 959.123 | 0.251 | 959.174 | 27.99 | 29 | 2 | 959.101 | 0.245 | 959.002 | 14.34 |
| 16 | 2 | 958.322 | 0.206 | 958.262 | 26.94 | 29 | 2 | 959.264 | 0.266 | 959.186 | 13.73 |
| 16 | 2 | 960.128 | 0.168 | 960.025 | 26.07 | 29 | 2 | 959.411 | 0.231 | 959.313 | 13.22 |

| 2001 - AGOSTO | | | | | |
|---|---|---|---|---|---|
| D | L | SDB | ER | SDC | HL |
| 31 | 1 | 958.024 | 0.167 | 957.911 | 66.37 |
| 31 | 1 | 957.840 | 0.224 | 957.866 | 67.08 |
| 31 | 1 | 958.451 | 0.187 | 958.409 | 67.86 |
| 31 | 1 | 957.593 | 0.183 | 957.633 | 68.58 |
| 31 | 1 | 957.631 | 0.182 | 957.634 | 69.45 |
| 31 | 1 | 958.246 | 0.179 | 958.273 | 70.33 |
| 31 | 1 | 957.605 | 0.195 | 957.718 | 71.26 |
| 31 | 1 | 957.836 | 0.183 | 957.938 | 72.24 |
| 31 | 1 | 959.030 | 0.217 | 959.217 | 73.29 |
| 31 | 1 | 957.541 | 0.167 | 957.608 | 74.34 |
| 31 | 2 | 959.010 | 0.198 | 958.964 | 21.62 |
| 31 | 2 | 959.427 | 0.193 | 959.405 | 20.59 |
| 31 | 2 | 959.219 | 0.209 | 959.194 | 19.71 |
| 31 | 2 | 960.332 | 0.182 | 960.240 | 18.93 |
| 31 | 2 | 959.748 | 0.220 | 959.696 | 18.14 |
| 31 | 2 | 959.047 | 0.232 | 959.038 | 14.62 |
| 31 | 2 | 959.458 | 0.197 | 959.407 | 14.04 |
| 31 | 2 | 960.291 | 0.277 | 960.260 | 13.50 |

| 2001 - SETEMBRO | | | | | |
|---|---|---|---|---|---|
| D | L | SDB | ER | SDC | HL |
| 03 | 1 | 959.526 | 0.167 | 959.462 | 68.58 |
| 03 | 1 | 959.346 | 0.182 | 959.360 | 69.40 |
| 03 | 1 | 959.606 | 0.170 | 959.599 | 70.39 |
| 03 | 1 | 959.315 | 0.150 | 959.297 | 71.23 |
| 03 | 1 | 959.359 | 0.145 | 959.380 | 72.10 |
| 03 | 1 | 959.447 | 0.154 | 959.430 | 73.03 |
| 03 | 1 | 959.005 | 0.174 | 959.084 | 73.99 |
| 03 | 1 | 958.788 | 0.139 | 958.743 | 75.03 |
| 03 | 1 | 959.669 | 0.178 | 959.812 | 76.10 |
| 03 | 1 | 959.085 | 0.162 | 959.202 | 77.21 |
| 03 | 1 | 959.432 | 0.182 | 959.616 | 78.44 |
| 03 | 1 | 958.809 | 0.191 | 959.036 | 79.70 |
| 03 | 1 | 958.331 | 0.174 | 958.515 | 81.07 |
| 03 | 1 | 958.446 | 0.170 | 958.682 | 82.53 |
| 03 | 1 | 958.693 | 0.154 | 958.833 | 84.03 |
| 03 | 2 | 957.783 | 0.236 | 957.774 | 22.77 |
| 03 | 2 | 957.978 | 0.229 | 958.002 | 21.84 |
| 03 | 2 | 958.392 | 0.216 | 958.379 | 20.97 |
| 03 | 2 | 957.965 | 0.247 | 958.004 | 20.12 |
| 03 | 2 | 959.337 | 0.174 | 959.277 | 19.30 |
| 03 | 2 | 959.243 | 0.181 | 959.196 | 18.53 |
| 03 | 2 | 958.316 | 0.260 | 958.391 | 17.79 |
| 03 | 2 | 958.912 | 0.212 | 958.858 | 17.06 |
| 03 | 2 | 958.485 | 0.254 | 958.505 | 16.37 |
| 03 | 2 | 959.970 | 0.182 | 959.849 | 15.68 |
| 03 | 2 | 959.268 | 0.234 | 959.317 | 14.95 |
| 03 | 2 | 960.211 | 0.207 | 960.203 | 14.36 |
| 03 | 2 | 959.034 | 0.220 | 959.039 | 13.76 |
| 03 | 2 | 959.169 | 0.193 | 959.135 | 13.21 |
| 03 | 2 | 958.936 | 0.198 | 958.866 | 12.67 |
| 03 | 2 | 959.941 | 0.263 | 959.995 | 12.06 |
| 04 | 1 | 958.573 | 0.135 | 958.379 | 68.37 |
| 04 | 1 | 958.560 | 0.143 | 958.442 | 69.33 |
| 04 | 1 | 959.391 | 0.140 | 959.313 | 70.09 |
| 04 | 1 | 958.558 | 0.152 | 958.569 | 70.90 |
| 04 | 1 | 959.274 | 0.144 | 959.208 | 71.79 |
| 04 | 1 | 959.368 | 0.141 | 959.338 | 72.81 |
| 04 | 1 | 958.797 | 0.134 | 958.687 | 73.82 |
| 04 | 1 | 958.502 | 0.167 | 958.518 | 74.83 |
| 04 | 1 | 957.719 | 0.179 | 957.817 | 75.89 |
| 04 | 1 | 957.947 | 0.162 | 958.014 | 77.02 |
| 04 | 1 | 958.219 | 0.157 | 958.306 | 78.23 |
| 04 | 1 | 958.730 | 0.151 | 958.883 | 79.47 |
| 04 | 1 | 958.038 | 0.146 | 958.114 | 80.80 |
| 04 | 1 | 957.757 | 0.152 | 957.828 | 82.33 |
| 04 | 1 | 958.919 | 0.166 | 959.109 | 83.91 |
| 04 | 2 | 958.541 | 0.189 | 958.450 | 21.60 |
| 04 | 2 | 959.756 | 0.218 | 959.686 | 20.30 |
| 04 | 2 | 959.344 | 0.175 | 959.214 | 19.47 |

| 2001 - SETEMBRO | | | | | |
|---|---|---|---|---|---|
| D | L | SDB | ER | SDC | HL |
| 04 | 2 | 959.444 | 0.208 | 959.394 | 18.60 |
| 04 | 2 | 958.410 | 0.188 | 958.353 | 17.81 |
| 04 | 2 | 958.988 | 0.163 | 958.842 | 17.06 |
| 04 | 2 | 959.406 | 0.192 | 959.334 | 16.36 |
| 04 | 2 | 959.161 | 0.198 | 959.061 | 15.65 |
| 04 | 2 | 958.996 | 0.234 | 958.907 | 14.92 |
| 04 | 2 | 959.082 | 0.177 | 958.914 | 14.16 |
| 04 | 2 | 959.101 | 0.211 | 959.062 | 13.55 |
| 04 | 2 | 959.334 | 0.179 | 959.204 | 12.97 |
| 04 | 2 | 959.158 | 0.158 | 958.945 | 12.39 |
| 04 | 2 | 959.754 | 0.188 | 959.644 | 11.81 |
| 04 | 2 | 959.244 | 0.175 | 959.081 | 11.27 |
| 04 | 2 | 959.026 | 0.202 | 958.885 | 10.78 |
| 05 | 2 | 958.878 | 0.194 | 958.740 | 20.50 |
| 05 | 2 | 959.181 | 0.172 | 959.050 | 19.56 |
| 05 | 2 | 959.064 | 0.179 | 958.939 | 18.75 |
| 05 | 2 | 958.757 | 0.195 | 958.682 | 17.95 |
| 05 | 2 | 959.339 | 0.193 | 959.280 | 17.18 |
| 05 | 2 | 959.746 | 0.172 | 959.636 | 16.46 |
| 05 | 2 | 959.350 | 0.168 | 959.214 | 15.58 |
| 05 | 2 | 958.954 | 0.173 | 958.853 | 14.94 |
| 05 | 2 | 958.932 | 0.214 | 958.942 | 14.12 |
| 05 | 2 | 959.599 | 0.168 | 959.477 | 13.47 |
| 05 | 2 | 960.013 | 0.196 | 959.853 | 12.86 |
| 05 | 2 | 959.619 | 0.167 | 959.407 | 12.29 |
| 05 | 2 | 959.751 | 0.206 | 959.738 | 11.73 |
| 05 | 2 | 959.220 | 0.169 | 959.092 | 10.93 |
| 05 | 2 | 958.855 | 0.196 | 958.842 | 10.41 |
| 05 | 2 | 958.647 | 0.169 | 958.515 | 9.90 |
| 10 | 1 | 959.807 | 0.175 | 959.823 | 69.87 |
| 10 | 1 | 958.546 | 0.155 | 958.527 | 70.74 |
| 10 | 1 | 959.619 | 0.164 | 959.621 | 71.58 |
| 10 | 1 | 959.899 | 0.181 | 959.985 | 72.52 |
| 10 | 1 | 959.185 | 0.158 | 959.247 | 73.48 |
| 10 | 1 | 959.395 | 0.170 | 959.470 | 74.51 |
| 10 | 1 | 959.706 | 0.205 | 959.831 | 75.55 |
| 10 | 1 | 959.402 | 0.153 | 959.458 | 76.61 |
| 10 | 1 | 959.099 | 0.169 | 959.242 | 77.80 |
| 10 | 1 | 958.309 | 0.143 | 958.339 | 79.00 |
| 10 | 1 | 958.598 | 0.188 | 958.773 | 80.24 |
| 10 | 1 | 957.921 | 0.181 | 958.111 | 81.70 |
| 10 | 2 | 959.152 | 0.316 | 959.042 | 15.77 |
| 10 | 2 | 958.933 | 0.179 | 958.828 | 15.02 |
| 10 | 2 | 959.761 | 0.181 | 959.632 | 14.34 |
| 10 | 2 | 958.557 | 0.176 | 958.453 | 13.69 |
| 10 | 2 | 959.146 | 0.184 | 959.060 | 13.05 |
| 10 | 2 | 958.560 | 0.191 | 958.472 | 12.43 |
| 13 | 1 | 958.458 | 0.190 | 958.459 | 67.42 |
| 13 | 1 | 958.056 | 0.167 | 957.970 | 68.17 |
| 13 | 1 | 958.706 | 0.150 | 958.626 | 68.92 |
| 13 | 1 | 958.871 | 0.164 | 958.869 | 69.69 |
| 13 | 1 | 959.281 | 0.150 | 959.194 | 70.50 |
| 13 | 1 | 959.308 | 0.163 | 959.277 | 71.39 |
| 13 | 1 | 958.430 | 0.204 | 958.488 | 72.26 |
| 13 | 1 | 958.730 | 0.186 | 958.738 | 73.36 |
| 13 | 1 | 959.064 | 0.187 | 959.145 | 74.36 |
| 13 | 1 | 958.263 | 0.158 | 958.283 | 75.46 |
| 13 | 1 | 958.696 | 0.165 | 958.800 | 77.75 |
| 13 | 2 | 960.375 | 0.172 | 960.270 | 17.15 |
| 13 | 2 | 959.509 | 0.183 | 959.403 | 16.27 |
| 13 | 2 | 959.011 | 0.176 | 958.874 | 15.47 |
| 13 | 2 | 959.721 | 0.168 | 959.602 | 14.71 |
| 13 | 2 | 960.000 | 0.170 | 959.926 | 13.95 |
| 13 | 2 | 959.709 | 0.170 | 959.592 | 13.25 |
| 13 | 2 | 959.223 | 0.169 | 959.094 | 12.58 |
| 13 | 2 | 958.792 | 0.179 | 958.666 | 11.93 |
| 13 | 2 | 958.915 | 0.195 | 958.831 | 11.32 |
| 13 | 2 | 959.183 | 0.196 | 959.104 | 10.73 |
| 13 | 2 | 959.949 | 0.208 | 959.852 | 10.12 |
| 19 | 1 | 960.475 | 0.168 | 960.411 | 66.64 |
| 19 | 1 | 960.565 | 0.151 | 960.473 | 67.38 |

```
       2001  -  SETEMBRO                                    2001  -  DEZEMBRO
 D  L    SDB     ER      SDC     HL              D  L    SDB     ER      SDC     HL
19  1  960.413  0.174  960.368  68.10           21  1  959.636  0.103  959.749   1.87
19  1  959.776  0.156  959.765  68.83           21  1  960.397  0.113  960.501   1.68
19  1  960.285  0.196  960.354  69.61           21  1  959.654  0.109  959.804   1.51
19  1  960.312  0.155  960.255  70.58           21  1  959.286  0.103  959.429   1.32
19  1  959.001  0.175  958.976  71.53           21  1  958.993  0.126  959.247   0.94
19  1  958.803  0.168  958.825  72.47           21  1  959.357  0.105  959.388   0.74
19  1  959.925  0.174  959.991  73.47           21  1  958.640  0.113  958.674   0.56
19  1  959.562  0.158  959.552  74.63           21  1  959.855  0.116  959.922   0.37
19  1  958.755  0.221  958.943  75.76           21  1  959.910  0.096  959.962   0.18
19  1  959.072  0.164  959.098  76.88           21  1  959.389  0.121  959.568   0.02
19  1  960.102  0.227  960.308  78.14           21  1  959.388  0.115  959.546   0.22
19  1  958.721  0.200  958.831  79.48
19  1  959.043  0.174  959.213  82.49                     2002  -  JANEIRO
19  2  960.176  0.168  960.081  12.81            D  L    SDB     ER      SDC     HL
19  2  958.544  0.196  958.548  12.06           02  2  959.565  0.124  959.585  17.27
19  2  958.785  0.174  958.645  11.37           02  2  960.051  0.141  960.076  17.49
19  2  959.271  0.195  959.158  10.71           02  2  959.107  0.142  959.157  17.67
19  2  958.622  0.175  958.544  10.09           02  2  960.193  0.135  960.231  17.82
19  2  959.683  0.162  959.561   9.47           02  2  959.929  0.126  959.949  18.04
19  2  960.005  0.162  959.868   8.88           02  2  959.373  0.142  959.412  18.22
19  2  959.368  0.177  959.322   8.34           02  2  959.529  0.127  959.549  18.37
19  2  958.910  0.164  958.784   7.80           02  2  959.433  0.138  959.477  18.52
19  2  958.816  0.170  958.682   7.29           02  2  958.864  0.126  958.889  18.68
19  2  959.366  0.147  959.215   6.76           02  2  958.717  0.159  958.773  19.01
19  2  959.328  0.183  959.268   6.29           03  2  958.969  0.157  958.996  19.79
19  2  959.512  0.181  959.425   5.82           03  2  959.594  0.211  959.608  19.97
19  2  958.901  0.205  958.704   5.38           03  2  959.332  0.204  959.383  20.16
19  2  959.701  0.172  959.611   4.80           03  2  960.346  0.193  960.359  20.65
19  2  958.855  0.207  958.770   4.40           03  2  959.520  0.153  959.577  20.81
20  1  957.015  0.158  957.068  75.08           03  2  959.964  0.131  959.991  20.99
20  1  959.400  0.184  959.438  76.22           03  2  958.876  0.133  958.897  21.16
21  1  957.523  0.196  957.589  78.98           03  2  958.997  0.129  959.024  21.34
                                                03  2  960.096  0.124  960.142  21.52
                                                03  2  960.223  0.118  960.243  21.71
       2001  -  DEZEMBRO                        03  2  959.854  0.131  959.886  21.92
 D  L    SDB     ER      SDC     HL             03  2  959.166  0.123  959.180  22.12
19  1  959.210  0.130  959.546   0.03           03  2  958.951  0.129  958.984  22.30
19  1  959.191  0.253  959.691   0.30           03  2  959.510  0.122  959.536  22.48
20  1  959.536  0.107  959.698   2.41           04  1  959.379  0.105  959.501   4.04
20  1  959.694  0.105  959.858   2.22           04  1  959.686  0.128  959.917   4.37
20  1  959.182  0.125  959.467   2.03           04  1  959.711  0.207  960.173   4.53
20  1  959.860  0.109  960.125   1.85           04  1  959.184  0.132  959.555   4.76
20  1  959.773  0.121  960.042   1.67           04  1  959.960  0.106  959.960   4.94
20  1  959.219  0.139  959.582   1.49           04  1  960.597  0.167  960.734   5.08
20  1  959.119  0.120  959.447   1.30           04  1  959.861  0.141  960.024   5.23
20  1  959.332  0.125  959.656   1.10           04  1  959.252  0.137  959.511   5.38
20  1  959.411  0.160  959.831   0.91           04  1  958.814  0.161  959.075   5.55
20  1  959.406  0.106  959.680   0.72           04  1  959.520  0.124  959.787   5.71
20  1  959.215  0.159  959.649   0.51           04  1  959.341  0.157  959.722   5.85
20  1  959.161  0.114  959.369   0.24           04  2  959.448  0.317  959.466  19.96
20  1  959.644  0.122  959.924   0.05           04  2  958.538  0.220  958.557  20.13
20  1  958.853  0.114  959.192   0.13           04  2  958.545  0.154  958.584  20.28
20  1  959.171  0.131  959.537   0.33           04  2  958.971  0.184  959.003  20.44
20  1  959.059  0.111  959.372   0.52           04  2  959.527  0.150  959.553  20.59
20  2  959.921  0.233  959.937  10.71           04  2  958.826  0.133  958.827  20.76
20  2  959.508  0.167  959.487  10.92           04  2  959.020  0.130  959.033  20.92
20  2  959.322  0.170  959.339  11.11           04  2  959.070  0.146  959.106  21.07
20  2  959.868  0.143  959.879  11.32           04  2  958.754  0.120  958.785  21.58
20  2  959.183  0.117  959.163  11.51           04  2  959.177  0.135  959.227  21.74
20  2  958.882  0.114  958.886  11.69           04  2  959.625  0.126  959.650  21.90
20  2  959.488  0.114  959.498  11.87           04  2  958.769  0.151  958.817  22.07
20  2  958.751  0.108  958.767  12.07           04  2  958.848  0.144  958.904  22.24
20  2  959.253  0.135  959.267  12.33           04  2  958.765  0.127  958.783  22.39
20  2  959.816  0.135  959.850  12.52           09  1  958.744  0.128  959.062   5.46
20  2  959.347  0.153  959.367  12.69           09  1  959.049  0.118  959.355   5.60
20  2  959.141  0.118  959.150  12.89           09  1  959.427  0.124  959.764   5.80
20  2  959.016  0.124  959.043  13.09           09  1  959.042  0.113  959.380   5.97
20  2  959.428  0.124  959.455  13.28           09  1  960.258  0.133  960.592   6.12
20  2  959.550  0.117  959.559  13.46           09  1  959.824  0.154  960.213   6.30
20  2  959.204  0.131  959.232  13.64           09  1  958.955  0.133  959.299   6.44
21  1  959.111  0.083  959.131   2.05
```

| 2002 - JANEIRO | | | | | | 2002 - JANEIRO | | | | |
|---|---|---|---|---|---|---|---|---|---|---|
| D | L | SDB | ER | SDC | HL | D | L | SDB | ER | SDC | HL |
| 09 | 1 | 959.732 | 0.125 | 960.059 | 6.57 | 24 | 1 | 958.790 | 0.130 | 959.180 | 8.24 |
| 09 | 1 | 959.276 | 0.146 | 959.620 | 6.70 | 24 | 1 | 958.885 | 0.167 | 959.386 | 8.20 |
| 09 | 1 | 959.618 | 0.139 | 959.999 | 6.83 | 24 | 1 | 959.132 | 0.158 | 959.609 | 7.99 |
| 09 | 1 | 959.093 | 0.152 | 959.511 | 6.96 | 24 | 1 | 958.985 | 0.143 | 959.427 | 7.90 |
| 09 | 1 | 959.423 | 0.161 | 959.840 | 7.09 | 24 | 1 | 960.496 | 0.162 | 960.997 | 7.80 |
| 09 | 1 | 959.187 | 0.161 | 959.652 | 7.21 | 25 | 1 | 959.048 | 0.133 | 959.408 | 7.80 |
| 11 | 2 | 960.074 | 0.281 | 960.142 | 24.22 | 25 | 1 | 959.026 | 0.114 | 959.311 | 7.88 |
| 11 | 2 | 958.499 | 0.234 | 958.549 | 24.34 | 25 | 1 | 960.118 | 0.139 | 960.474 | 7.95 |
| 17 | 1 | 959.999 | 0.103 | 960.271 | 6.93 | 25 | 1 | 960.135 | 0.123 | 960.496 | 8.02 |
| 17 | 1 | 959.867 | 0.095 | 960.004 | 7.03 | 25 | 1 | 959.313 | 0.250 | 959.745 | 8.06 |
| 17 | 1 | 959.312 | 0.124 | 959.554 | 7.13 | 25 | 1 | 958.792 | 0.151 | 959.191 | 8.12 |
| 17 | 1 | 960.601 | 0.112 | 960.856 | 7.23 | 25 | 1 | 959.269 | 0.136 | 959.613 | 8.16 |
| 17 | 1 | 959.072 | 0.112 | 959.409 | 7.46 | 25 | 1 | 960.917 | 0.155 | 961.302 | 8.20 |
| 17 | 1 | 959.157 | 0.151 | 959.502 | 7.60 | 25 | 1 | 960.335 | 0.146 | 960.689 | 8.26 |
| 17 | 1 | 958.933 | 0.127 | 959.273 | 7.69 | 25 | 1 | 959.155 | 0.163 | 959.614 | 8.28 |
| 17 | 1 | 959.520 | 0.131 | 959.877 | 7.76 | 25 | 1 | 959.047 | 0.157 | 959.489 | 8.30 |
| 17 | 1 | 959.343 | 0.129 | 959.724 | 7.85 | 25 | 1 | 959.699 | 0.128 | 960.105 | 8.31 |
| 17 | 1 | 959.048 | 0.108 | 959.322 | 7.92 | 25 | 1 | 959.285 | 0.108 | 959.582 | 8.32 |
| 17 | 1 | 959.924 | 0.121 | 960.237 | 7.98 | 25 | 1 | 960.264 | 0.150 | 960.732 | 8.32 |
| 17 | 1 | 958.556 | 0.133 | 958.897 | 8.05 | 30 | 2 | 959.974 | 0.130 | 959.990 | 38.41 |
| 17 | 1 | 959.797 | 0.114 | 960.134 | 8.11 | 30 | 2 | 959.721 | 0.137 | 959.750 | 38.34 |
| 21 | 1 | 959.954 | 0.218 | 960.396 | 7.54 | 30 | 2 | 959.273 | 0.150 | 959.303 | 38.17 |
| 21 | 1 | 958.957 | 0.258 | 959.494 | 7.74 | 31 | 1 | 958.707 | 0.139 | 958.797 | 8.20 |
| 21 | 1 | 958.152 | 0.198 | 958.648 | 7.83 | 31 | 1 | 959.557 | 0.108 | 959.584 | 8.22 |
| 21 | 1 | 959.010 | 0.180 | 959.469 | 7.91 | 31 | 1 | 959.199 | 0.106 | 959.183 | 8.23 |
| 21 | 1 | 958.668 | 0.167 | 959.053 | 7.97 | 31 | 1 | 959.459 | 0.125 | 959.491 | 8.23 |
| 21 | 1 | 959.542 | 0.166 | 959.965 | 8.03 | 31 | 1 | 959.078 | 0.122 | 959.150 | 8.23 |
| 21 | 1 | 959.247 | 0.164 | 959.677 | 8.09 | 31 | 1 | 959.093 | 0.127 | 959.227 | 8.20 |
| 21 | 1 | 959.213 | 0.151 | 959.625 | 8.15 | 31 | 1 | 959.921 | 0.123 | 960.044 | 8.19 |
| 21 | 1 | 959.410 | 0.141 | 959.782 | 8.20 | 31 | 1 | 958.982 | 0.115 | 959.110 | 8.15 |
| 21 | 1 | 959.964 | 0.125 | 960.309 | 8.24 | 31 | 1 | 959.413 | 0.135 | 959.615 | 8.12 |
| 21 | 1 | 959.640 | 0.132 | 959.972 | 8.28 | 31 | 1 | 959.155 | 0.118 | 959.442 | 8.07 |
| 21 | 1 | 959.126 | 0.134 | 959.481 | 8.31 | 31 | 1 | 959.908 | 0.121 | 960.119 | 8.02 |
| 21 | 1 | 958.802 | 0.133 | 959.166 | 8.34 | 31 | 1 | 958.969 | 0.129 | 959.240 | 7.96 |
| 21 | 1 | 958.564 | 0.167 | 959.021 | 8.36 | | | | | | |
| 21 | 1 | 959.059 | 0.193 | 959.577 | 8.37 | | | 2002 - FEVEREIRO | | | |
| 21 | 1 | 958.581 | 0.167 | 959.033 | 8.37 | D | L | SDB | ER | SDC | HL |
| 22 | 1 | 959.676 | 0.104 | 959.910 | 7.64 | 01 | 2 | 959.714 | 0.145 | 959.718 | 39.81 |
| 22 | 1 | 959.388 | 0.128 | 959.668 | 7.72 | 01 | 2 | 958.628 | 0.117 | 958.613 | 39.64 |
| 22 | 1 | 960.051 | 0.098 | 960.238 | 7.79 | 01 | 2 | 959.477 | 0.142 | 959.493 | 39.57 |
| 22 | 1 | 958.823 | 0.127 | 959.181 | 7.87 | 01 | 2 | 959.023 | 0.145 | 959.020 | 39.52 |
| 22 | 1 | 960.866 | 0.121 | 961.199 | 7.94 | 01 | 2 | 958.852 | 0.137 | 958.862 | 39.48 |
| 22 | 1 | 959.190 | 0.143 | 959.575 | 8.00 | 04 | 2 | 960.036 | 0.127 | 960.075 | 41.74 |
| 22 | 1 | 959.329 | 0.122 | 959.688 | 8.07 | 04 | 2 | 960.877 | 0.137 | 960.916 | 41.64 |
| 22 | 1 | 959.017 | 0.145 | 959.386 | 8.12 | 04 | 2 | 958.991 | 0.121 | 959.018 | 41.55 |
| 22 | 1 | 959.445 | 0.171 | 959.831 | 8.18 | 04 | 2 | 959.479 | 0.140 | 959.505 | 41.47 |
| 22 | 1 | 959.398 | 0.147 | 959.792 | 8.22 | 04 | 2 | 959.714 | 0.131 | 959.735 | 41.41 |
| 22 | 1 | 959.067 | 0.124 | 959.446 | 8.26 | 04 | 2 | 958.891 | 0.146 | 958.948 | 41.34 |
| 22 | 1 | 958.963 | 0.144 | 959.414 | 8.29 | 04 | 2 | 958.978 | 0.107 | 958.981 | 41.29 |
| 22 | 1 | 959.416 | 0.155 | 959.867 | 8.31 | 04 | 2 | 958.621 | 0.153 | 958.621 | 41.24 |
| 22 | 1 | 958.465 | 0.123 | 958.844 | 8.33 | 05 | 1 | 959.117 | 0.117 | 959.181 | 8.08 |
| 22 | 1 | 958.664 | 0.151 | 959.056 | 8.35 | 05 | 1 | 958.323 | 0.136 | 958.429 | 8.06 |
| 22 | 1 | 958.764 | 0.140 | 959.123 | 8.36 | 05 | 1 | 959.503 | 0.129 | 959.558 | 8.02 |
| 22 | 1 | 959.139 | 0.145 | 959.556 | 8.36 | 05 | 1 | 959.204 | 0.105 | 959.184 | 7.99 |
| 22 | 1 | 959.363 | 0.156 | 959.828 | 8.35 | 05 | 1 | 960.021 | 0.115 | 960.008 | 7.94 |
| 22 | 2 | 960.085 | 0.106 | 960.071 | 32.54 | 05 | 1 | 958.973 | 0.128 | 959.038 | 7.89 |
| 22 | 2 | 959.342 | 0.127 | 959.353 | 32.57 | 05 | 1 | 958.883 | 0.133 | 958.936 | 7.83 |
| 22 | 2 | 958.985 | 0.141 | 959.008 | 32.60 | 05 | 1 | 959.563 | 0.117 | 959.673 | 7.77 |
| 22 | 2 | 959.192 | 0.128 | 959.209 | 32.64 | 05 | 1 | 959.645 | 0.142 | 959.837 | 7.65 |
| 22 | 2 | 959.243 | 0.107 | 959.222 | 32.69 | 05 | 1 | 960.087 | 0.114 | 960.218 | 7.56 |
| 22 | 2 | 958.502 | 0.117 | 958.506 | 32.73 | 05 | 1 | 959.902 | 0.154 | 960.158 | 7.47 |
| 22 | 2 | 959.702 | 0.140 | 959.731 | 32.79 | 05 | 1 | 959.465 | 0.133 | 959.731 | 7.35 |
| 22 | 2 | 959.343 | 0.124 | 959.347 | 32.85 | 05 | 1 | 959.851 | 0.156 | 960.147 | 7.24 |
| 22 | 2 | 959.510 | 0.116 | 959.515 | 32.91 | 05 | 1 | 959.715 | 0.139 | 959.983 | 7.10 |
| 22 | 2 | 959.296 | 0.138 | 959.326 | 32.97 | 05 | 1 | 959.425 | 0.119 | 959.155 | 6.95 |
| 22 | 2 | 959.156 | 0.119 | 959.155 | 33.03 | 05 | 2 | 959.467 | 0.149 | 959.516 | 42.71 |
| 22 | 2 | 959.705 | 0.092 | 959.604 | 33.10 | 05 | 2 | 960.063 | 0.137 | 960.112 | 42.58 |
| 22 | 2 | 958.977 | 0.118 | 958.958 | 33.18 | 05 | 2 | 958.799 | 0.133 | 958.829 | 42.46 |
| 22 | 2 | 959.707 | 0.114 | 959.687 | 33.40 | 05 | 2 | 958.928 | 0.131 | 958.970 | 42.35 |
| 22 | 2 | 960.672 | 0.151 | 960.690 | 33.48 | 05 | 2 | 959.592 | 0.134 | 959.602 | 42.25 |
| 24 | 1 | 959.749 | 0.124 | 960.164 | 8.27 | | | | | | |

| 2002 - FEVEREIRO | | | | | | 2002 - FEVEREIRO | | | | |
|---|---|---|---|---|---|---|---|---|---|---|
| D | L | SDB | ER | SDC | HL | D | L | SDB | ER | SDC | HL |
| 05 | 2 | 959.769 | 0.138 | 959.811 | 42.16 | 15 | 2 | 958.831 | 0.179 | 958.893 | 48.29 |
| 05 | 2 | 958.380 | 0.172 | 958.446 | 41.97 | 15 | 2 | 959.213 | 0.152 | 959.257 | 48.12 |
| 05 | 2 | 959.629 | 0.151 | 959.671 | 41.91 | 15 | 2 | 958.916 | 0.149 | 958.954 | 47.94 |
| 05 | 2 | 958.637 | 0.133 | 958.673 | 41.86 | 15 | 2 | 958.669 | 0.265 | 958.737 | 47.76 |
| 05 | 2 | 958.775 | 0.159 | 958.849 | 41.81 | 15 | 2 | 958.845 | 0.145 | 958.884 | 47.62 |
| 05 | 2 | 958.991 | 0.148 | 959.047 | 41.78 | 15 | 2 | 958.923 | 0.145 | 958.962 | 47.49 |
| 05 | 2 | 959.679 | 0.140 | 959.728 | 41.74 | 19 | 1 | 958.886 | 0.131 | 958.989 | 5.85 |
| 05 | 2 | 959.809 | 0.141 | 959.845 | 41.69 | 19 | 1 | 958.995 | 0.128 | 959.163 | 5.68 |
| 06 | 1 | 959.263 | 0.109 | 959.380 | 8.06 | 19 | 1 | 959.030 | 0.141 | 959.309 | 5.50 |
| 06 | 1 | 959.420 | 0.113 | 959.543 | 8.03 | 19 | 1 | 959.547 | 0.104 | 959.448 | 5.23 |
| 06 | 1 | 960.057 | 0.103 | 960.197 | 8.01 | 19 | 1 | 959.530 | 0.120 | 959.520 | 5.03 |
| 06 | 1 | 959.288 | 0.136 | 959.480 | 7.97 | 19 | 1 | 958.872 | 0.123 | 958.912 | 4.80 |
| 06 | 1 | 959.322 | 0.146 | 959.565 | 7.93 | 19 | 1 | 958.863 | 0.118 | 958.863 | 4.56 |
| 06 | 1 | 960.089 | 0.235 | 960.487 | 7.89 | 19 | 1 | 959.566 | 0.120 | 959.579 | 4.31 |
| 06 | 1 | 959.085 | 0.126 | 959.188 | 7.80 | 19 | 1 | 958.844 | 0.128 | 958.911 | 4.07 |
| 06 | 1 | 959.574 | 0.141 | 959.770 | 7.73 | 19 | 1 | 959.310 | 0.124 | 959.443 | 3.80 |
| 06 | 1 | 959.009 | 0.126 | 959.216 | 7.65 | 19 | 1 | 959.033 | 0.117 | 959.202 | 3.44 |
| 06 | 1 | 960.183 | 0.123 | 960.378 | 7.56 | 19 | 1 | 959.495 | 0.142 | 959.760 | 3.14 |
| 06 | 1 | 959.318 | 0.115 | 959.408 | 7.47 | 19 | 1 | 959.556 | 0.120 | 959.722 | 2.79 |
| 06 | 1 | 959.347 | 0.126 | 959.492 | 7.37 | 19 | 1 | 958.805 | 0.123 | 958.984 | 2.45 |
| 06 | 1 | 959.830 | 0.113 | 959.964 | 7.26 | 19 | 1 | 958.784 | 0.151 | 959.012 | 2.10 |
| 06 | 1 | 959.330 | 0.136 | 959.501 | 7.14 | 19 | 1 | 959.134 | 0.166 | 959.503 | 1.67 |
| 06 | 1 | 959.103 | 0.141 | 959.386 | 7.02 | 26 | 1 | 958.547 | 0.208 | 958.840 | 3.14 |
| 06 | 2 | 960.633 | 0.133 | 960.671 | 43.61 | 26 | 1 | 959.704 | 0.120 | 959.720 | 2.70 |
| 06 | 2 | 959.942 | 0.168 | 960.012 | 43.45 | 26 | 1 | 959.575 | 0.084 | 959.464 | 2.38 |
| 06 | 2 | 957.974 | 0.128 | 958.007 | 43.31 | 26 | 1 | 959.454 | 0.100 | 959.437 | 2.08 |
| 06 | 2 | 959.875 | 0.189 | 959.957 | 43.18 | 26 | 1 | 958.819 | 0.116 | 958.920 | 1.76 |
| 06 | 2 | 959.309 | 0.127 | 959.341 | 43.05 | 26 | 1 | 959.578 | 0.139 | 959.774 | 1.41 |
| 06 | 2 | 959.521 | 0.138 | 959.552 | 42.94 | 26 | 1 | 959.498 | 0.125 | 959.768 | 1.02 |
| 06 | 2 | 959.393 | 0.123 | 959.399 | 42.82 | 26 | 1 | 958.745 | 0.118 | 958.932 | 0.62 |
| 06 | 2 | 959.198 | 0.171 | 959.273 | 42.71 | 26 | 1 | 959.586 | 0.132 | 959.859 | 0.20 |
| 06 | 2 | 959.647 | 0.168 | 959.707 | 42.61 | 26 | 1 | 958.881 | 0.129 | 959.222 | 0.21 |
| 06 | 2 | 959.605 | 0.169 | 959.661 | 42.43 | 26 | 1 | 959.803 | 0.150 | 960.154 | 0.69 |
| 06 | 2 | 959.464 | 0.172 | 959.526 | 42.38 | 26 | 1 | 958.793 | 0.188 | 959.148 | 1.20 |
| 06 | 2 | 959.496 | 0.176 | 959.564 | 42.34 | 26 | 1 | 959.006 | 0.152 | 959.360 | 1.73 |
| 06 | 2 | 959.490 | 0.160 | 959.558 | 42.27 | 26 | 1 | 958.762 | 0.159 | 959.131 | 2.26 |
| 14 | 1 | 959.534 | 0.167 | 959.958 | 7.05 | 26 | 1 | 959.028 | 0.140 | 959.351 | 2.94 |
| 14 | 1 | 960.055 | 0.198 | 960.409 | 6.93 | 26 | 1 | 958.841 | 0.157 | 959.234 | 3.53 |
| 14 | 1 | 960.542 | 0.134 | 960.544 | 6.79 | 26 | 2 | 959.474 | 0.138 | 959.501 | 56.67 |
| 14 | 1 | 959.577 | 0.107 | 959.577 | 6.67 | 26 | 2 | 959.095 | 0.129 | 959.109 | 56.19 |
| 14 | 1 | 960.101 | 0.114 | 960.193 | 6.51 | 26 | 2 | 959.749 | 0.144 | 959.776 | 55.79 |
| 14 | 1 | 959.409 | 0.117 | 959.555 | 6.37 | 26 | 2 | 958.917 | 0.123 | 958.931 | 55.43 |
| 14 | 1 | 960.071 | 0.119 | 960.296 | 6.21 | 26 | 2 | 959.344 | 0.146 | 959.383 | 55.10 |
| 14 | 1 | 959.653 | 0.148 | 959.960 | 6.04 | 26 | 2 | 959.149 | 0.196 | 959.206 | 54.79 |
| 14 | 1 | 959.415 | 0.152 | 959.792 | 5.86 | 26 | 2 | 960.023 | 0.200 | 960.081 | 54.18 |
| 14 | 1 | 958.999 | 0.126 | 959.358 | 5.66 | 26 | 2 | 959.448 | 0.147 | 959.487 | 53.26 |
| 14 | 1 | 959.358 | 0.139 | 959.720 | 5.46 | 26 | 2 | 959.511 | 0.152 | 959.544 | 53.02 |
| 14 | 1 | 959.773 | 0.150 | 960.095 | 4.90 | 26 | 2 | 958.889 | 0.163 | 958.940 | 52.80 |
| 14 | 1 | 959.180 | 0.143 | 959.457 | 4.65 | 26 | 2 | 959.136 | 0.183 | 959.193 | 52.59 |
| 14 | 1 | 959.637 | 0.172 | 960.010 | 4.36 | 26 | 2 | 959.448 | 0.149 | 959.486 | 52.38 |
| 14 | 1 | 959.546 | 0.207 | 960.031 | 4.08 | | | | | | |
| 14 | 2 | 959.865 | 0.190 | 959.940 | 49.99 | | | | | | |
| 14 | 2 | 958.392 | 0.155 | 958.428 | 49.66 | | | 2002 - MARCO | | | |
| 14 | 2 | 959.473 | 0.183 | 959.535 | 49.37 | D | L | SDB | ER | SDC | HL |
| 14 | 2 | 959.600 | 0.177 | 959.668 | 49.11 | 01 | 2 | 959.347 | 0.157 | 959.366 | 53.99 |
| 14 | 2 | 959.012 | 0.151 | 959.055 | 48.83 | 01 | 2 | 958.811 | 0.146 | 958.842 | 53.76 |
| 14 | 2 | 959.080 | 0.150 | 959.111 | 48.60 | 01 | 2 | 958.982 | 0.132 | 959.007 | 53.54 |
| 14 | 2 | 958.996 | 0.192 | 959.077 | 48.38 | 01 | 2 | 958.760 | 0.148 | 958.778 | 53.34 |
| 14 | 2 | 959.317 | 0.182 | 959.366 | 48.16 | 01 | 2 | 959.317 | 0.174 | 959.329 | 53.15 |
| 14 | 2 | 959.565 | 0.124 | 959.559 | 47.75 | 01 | 2 | 958.834 | 0.132 | 958.835 | 52.92 |
| 14 | 2 | 958.524 | 0.175 | 958.543 | 47.58 | 01 | 2 | 958.791 | 0.155 | 958.817 | 52.74 |
| 15 | 2 | 958.673 | 0.184 | 958.733 | 50.97 | 01 | 2 | 959.157 | 0.165 | 959.183 | 52.55 |
| 15 | 2 | 958.872 | 0.247 | 958.913 | 50.59 | 01 | 2 | 959.644 | 0.179 | 959.657 | 52.38 |
| 15 | 2 | 958.775 | 0.155 | 958.816 | 50.24 | 01 | 2 | 958.273 | 0.135 | 958.245 | 52.07 |
| 15 | 2 | 959.250 | 0.189 | 959.323 | 49.93 | 01 | 2 | 958.917 | 0.140 | 958.912 | 51.93 |
| 15 | 2 | 958.854 | 0.172 | 958.927 | 49.64 | 01 | 2 | 958.960 | 0.154 | 958.979 | 51.79 |
| 15 | 2 | 958.404 | 0.161 | 958.459 | 49.37 | 04 | 1 | 959.201 | 0.148 | 959.271 | 1.26 |
| 15 | 2 | 959.646 | 0.149 | 959.688 | 49.11 | 04 | 1 | 958.442 | 0.105 | 958.465 | 1.78 |
| 15 | 2 | 959.125 | 0.163 | 959.174 | 48.86 | 04 | 1 | 958.921 | 0.130 | 959.037 | 2.24 |
| 15 | 2 | 959.543 | 0.155 | 959.592 | 48.66 | 04 | 1 | 959.031 | 0.149 | 959.200 | 2.95 |
| 15 | 2 | 959.168 | 0.156 | 959.223 | 48.47 | 04 | 1 | 959.357 | 0.122 | 959.515 | 3.52 |

|   |   | 2002 - MARCO |   |   |   |   |   | 2002 - MARCO |   |   |   |
|---|---|---|---|---|---|---|---|---|---|---|---|
| D | L | SDB | ER | SDC | HL | D | L | SDB | ER | SDC | HL |
| 04 | 1 | 958.948 | 0.118 | 959.065 | 4.10 | 06 | 2 | 958.999 | 0.121 | 959.093 | 56.89 |
| 04 | 1 | 959.327 | 0.119 | 959.498 | 4.69 | 06 | 2 | 959.478 | 0.129 | 959.604 | 56.59 |
| 04 | 1 | 959.267 | 0.158 | 959.493 | 5.33 | 06 | 2 | 958.905 | 0.123 | 958.998 | 56.31 |
| 04 | 1 | 959.266 | 0.164 | 959.513 | 6.03 | 06 | 2 | 959.264 | 0.125 | 959.383 | 56.02 |
| 04 | 1 | 958.531 | 0.172 | 958.812 | 6.76 | 06 | 2 | 959.049 | 0.130 | 958.166 | 55.74 |
| 04 | 1 | 959.493 | 0.152 | 959.711 | 7.48 | 06 | 2 | 959.121 | 0.161 | 959.258 | 55.48 |
| 04 | 1 | 959.030 | 0.168 | 959.364 | 8.32 | 06 | 2 | 958.910 | 0.143 | 959.047 | 55.22 |
| 04 | 2 | 960.118 | 0.122 | 960.104 | 56.13 | 06 | 2 | 959.805 | 0.126 | 959.897 | 54.99 |
| 04 | 2 | 959.216 | 0.124 | 959.214 | 55.82 | 06 | 2 | 958.227 | 0.147 | 958.318 | 54.75 |
| 04 | 2 | 959.073 | 0.151 | 959.109 | 55.48 | 06 | 2 | 959.580 | 0.141 | 959.696 | 54.52 |
| 04 | 2 | 959.075 | 0.157 | 959.111 | 55.21 | 06 | 2 | 958.622 | 0.163 | 958.744 | 54.31 |
| 04 | 2 | 958.797 | 0.136 | 958.814 | 54.96 | 07 | 1 | 959.412 | 0.122 | 959.587 | 3.61 |
| 04 | 2 | 958.985 | 0.150 | 959.007 | 54.70 | 07 | 1 | 959.268 | 0.126 | 959.506 | 4.70 |
| 04 | 2 | 959.561 | 0.154 | 959.597 | 54.46 | 07 | 1 | 958.871 | 0.144 | 959.190 | 5.35 |
| 04 | 2 | 958.911 | 0.148 | 958.915 | 54.25 | 07 | 1 | 959.056 | 0.125 | 959.320 | 6.05 |
| 04 | 2 | 958.953 | 0.147 | 958.969 | 54.04 | 07 | 1 | 959.106 | 0.151 | 959.466 | 6.81 |
| 04 | 2 | 958.662 | 0.151 | 958.645 | 53.83 | 07 | 1 | 958.719 | 0.175 | 959.135 | 8.42 |
| 04 | 2 | 959.218 | 0.190 | 959.245 | 53.64 | 07 | 1 | 958.344 | 0.215 | 958.799 | 9.49 |
| 04 | 2 | 959.008 | 0.193 | 959.040 | 53.45 | 07 | 1 | 958.198 | 0.198 | 958.576 | 10.44 |
| 04 | 2 | 959.312 | 0.198 | 959.331 | 53.23 | 07 | 2 | 960.049 | 0.138 | 960.133 | 57.85 |
| 04 | 2 | 959.751 | 0.238 | 959.764 | 53.07 | 07 | 2 | 959.281 | 0.151 | 959.383 | 57.11 |
| 04 | 2 | 958.554 | 0.213 | 958.593 | 52.79 | 07 | 2 | 958.774 | 0.178 | 958.888 | 56.77 |
| 04 | 2 | 958.647 | 0.204 | 958.723 | 52.65 | 07 | 2 | 958.585 | 0.166 | 958.699 | 56.20 |
| 05 | 1 | 959.804 | 0.103 | 959.786 | 1.81 | 07 | 2 | 959.016 | 0.161 | 959.135 | 55.91 |
| 05 | 1 | 959.351 | 0.107 | 959.346 | 2.27 | 07 | 2 | 959.693 | 0.123 | 959.782 | 55.50 |
| 05 | 1 | 958.717 | 0.117 | 958.777 | 2.77 | 07 | 2 | 959.746 | 0.121 | 959.841 | 55.03 |
| 05 | 1 | 959.273 | 0.113 | 959.362 | 3.27 | 08 | 1 | 960.259 | 0.129 | 960.261 | 1.09 |
| 05 | 1 | 958.929 | 0.125 | 959.079 | 3.86 | 08 | 1 | 959.777 | 0.143 | 959.771 | 1.50 |
| 05 | 1 | 959.762 | 0.121 | 959.869 | 4.44 | 08 | 1 | 958.620 | 0.129 | 958.703 | 1.91 |
| 05 | 1 | 959.529 | 0.109 | 959.639 | 5.05 | 08 | 1 | 959.898 | 0.132 | 959.999 | 2.34 |
| 05 | 1 | 960.126 | 0.136 | 960.325 | 5.70 | 08 | 1 | 959.794 | 0.121 | 959.882 | 2.85 |
| 05 | 1 | 958.918 | 0.181 | 959.276 | 6.40 | 08 | 1 | 960.311 | 0.120 | 960.480 | 3.35 |
| 05 | 1 | 959.707 | 0.195 | 960.010 | 7.13 | 08 | 1 | 960.467 | 0.109 | 960.621 | 3.92 |
| 05 | 1 | 959.508 | 0.122 | 959.708 | 7.91 | 08 | 1 | 959.956 | 0.109 | 960.177 | 4.48 |
| 05 | 1 | 959.640 | 0.167 | 959.935 | 8.73 | 08 | 1 | 960.393 | 0.117 | 960.650 | 5.10 |
| 05 | 1 | 959.274 | 0.157 | 959.524 | 9.56 | 08 | 1 | 960.163 | 0.134 | 960.479 | 5.72 |
| 05 | 1 | 959.216 | 0.203 | 959.612 | 10.47 | 08 | 1 | 960.480 | 0.141 | 960.784 | 6.38 |
| 05 | 1 | 958.595 | 0.191 | 958.969 | 11.45 | 08 | 1 | 959.754 | 0.150 | 960.107 | 7.06 |
| 05 | 2 | 959.683 | 0.131 | 959.771 | 57.32 | 08 | 2 | 958.872 | 0.161 | 959.024 | 58.90 |
| 05 | 2 | 959.038 | 0.119 | 959.126 | 56.89 | 08 | 2 | 958.351 | 0.176 | 958.509 | 58.50 |
| 05 | 2 | 959.005 | 0.143 | 959.112 | 56.58 | 08 | 2 | 959.146 | 0.171 | 959.291 | 58.14 |
| 05 | 2 | 958.680 | 0.126 | 958.767 | 56.27 | 08 | 2 | 958.206 | 0.175 | 958.332 | 57.79 |
| 05 | 2 | 958.990 | 0.125 | 959.081 | 55.97 | 08 | 2 | 958.429 | 0.161 | 958.568 | 57.45 |
| 05 | 2 | 959.159 | 0.139 | 959.271 | 55.68 | 08 | 2 | 960.311 | 0.160 | 960.443 | 56.58 |
| 05 | 2 | 959.346 | 0.116 | 959.393 | 55.41 | 08 | 2 | 958.944 | 0.134 | 959.064 | 56.13 |
| 05 | 2 | 958.944 | 0.119 | 959.035 | 55.16 | 08 | 2 | 958.562 | 0.132 | 958.657 | 55.86 |
| 05 | 2 | 958.396 | 0.154 | 958.507 | 54.92 | 08 | 2 | 959.266 | 0.116 | 959.374 | 55.61 |
| 05 | 2 | 958.244 | 0.127 | 958.316 | 54.70 | 08 | 2 | 959.209 | 0.130 | 959.323 | 55.37 |
| 05 | 2 | 958.950 | 0.124 | 959.036 | 54.48 | 08 | 2 | 959.075 | 0.135 | 959.208 | 55.13 |
| 05 | 2 | 959.196 | 0.139 | 959.249 | 54.27 | 08 | 2 | 959.078 | 0.123 | 959.185 | 54.91 |
| 05 | 2 | 958.764 | 0.165 | 958.867 | 54.07 | 08 | 2 | 958.816 | 0.118 | 958.935 | 54.69 |
| 05 | 2 | 958.854 | 0.171 | 958.956 | 53.86 | 08 | 2 | 958.950 | 0.132 | 959.018 | 54.49 |
| 06 | 1 | 959.511 | 0.111 | 959.466 | 1.09 | 08 | 2 | 959.300 | 0.123 | 959.413 | 54.29 |
| 06 | 1 | 958.622 | 0.105 | 958.543 | 1.53 | 11 | 1 | 958.904 | 0.162 | 959.120 | 2.16 |
| 06 | 1 | 960.116 | 0.125 | 960.099 | 2.00 | 11 | 1 | 958.153 | 0.165 | 958.461 | 2.57 |
| 06 | 1 | 959.378 | 0.091 | 959.311 | 2.45 | 11 | 1 | 958.510 | 0.225 | 958.824 | 3.09 |
| 06 | 1 | 959.664 | 0.083 | 959.593 | 2.93 | 11 | 1 | 959.407 | 0.125 | 959.394 | 3.67 |
| 06 | 1 | 959.105 | 0.117 | 959.265 | 3.43 | 11 | 1 | 959.268 | 0.127 | 959.272 | 4.19 |
| 06 | 1 | 960.072 | 0.121 | 960.172 | 3.96 | 11 | 1 | 959.251 | 0.131 | 959.319 | 4.72 |
| 06 | 1 | 959.563 | 0.103 | 959.612 | 4.57 | 11 | 1 | 959.282 | 0.114 | 959.342 | 5.26 |
| 06 | 1 | 959.168 | 0.110 | 959.301 | 5.19 | 11 | 1 | 960.178 | 0.126 | 960.259 | 5.86 |
| 06 | 1 | 959.541 | 0.132 | 959.804 | 5.84 | 11 | 1 | 959.283 | 0.142 | 959.388 | 6.46 |
| 06 | 1 | 959.302 | 0.141 | 959.491 | 6.55 | 11 | 1 | 959.069 | 0.163 | 959.252 | 7.08 |
| 06 | 1 | 960.289 | 0.144 | 960.598 | 7.27 | 11 | 1 | 959.047 | 0.153 | 959.230 | 7.78 |
| 06 | 1 | 959.609 | 0.140 | 959.853 | 8.06 | 11 | 1 | 958.724 | 0.163 | 959.012 | 8.50 |
| 06 | 1 | 958.965 | 0.136 | 959.232 | 8.89 | 11 | 1 | 958.770 | 0.145 | 959.009 | 9.27 |
| 06 | 1 | 959.061 | 0.202 | 959.456 | 9.78 | 11 | 1 | 958.449 | 0.155 | 958.660 | 10.06 |
| 06 | 2 | 959.660 | 0.120 | 959.736 | 58.29 | 11 | 1 | 959.110 | 0.157 | 959.425 | 10.90 |
| 06 | 2 | 958.433 | 0.105 | 958.509 | 57.92 | 11 | 2 | 959.870 | 0.131 | 959.993 | 60.56 |
| 06 | 2 | 958.726 | 0.102 | 958.790 | 57.55 | 11 | 2 | 959.151 | 0.153 | 959.293 | 60.10 |
| 06 | 2 | 958.488 | 0.100 | 958.558 | 57.22 | 11 | 2 | 959.884 | 0.160 | 960.032 | 59.69 |

2002 - MARCO

| D | L | SDB | ER | SDC | HL |
|---|---|---|---|---|---|
| 11 | 2 | 959.117 | 0.143 | 959.246 | 59.27 |
| 11 | 2 | 959.980 | 0.156 | 960.122 | 58.88 |
| 11 | 2 | 958.747 | 0.151 | 958.875 | 58.47 |
| 11 | 2 | 958.589 | 0.166 | 958.742 | 58.14 |
| 11 | 2 | 959.059 | 0.140 | 959.161 | 57.81 |
| 11 | 2 | 959.389 | 0.168 | 959.554 | 57.50 |
| 11 | 2 | 958.679 | 0.153 | 958.811 | 57.19 |
| 11 | 2 | 958.363 | 0.172 | 958.514 | 56.89 |
| 11 | 2 | 958.869 | 0.156 | 959.016 | 56.56 |
| 11 | 2 | 958.864 | 0.184 | 959.021 | 56.01 |
| 11 | 2 | 958.584 | 0.320 | 958.740 | 55.76 |
| 12 | 1 | 958.730 | 0.110 | 958.643 | 0.26 |
| 12 | 1 | 959.343 | 0.117 | 959.324 | 0.05 |
| 12 | 1 | 959.448 | 0.121 | 959.463 | 0.38 |
| 12 | 1 | 959.789 | 0.097 | 959.758 | 0.74 |
| 12 | 1 | 959.974 | 0.129 | 960.019 | 1.10 |
| 12 | 1 | 959.604 | 0.097 | 959.610 | 1.47 |
| 12 | 1 | 959.515 | 0.128 | 959.609 | 1.85 |
| 12 | 1 | 959.662 | 0.125 | 959.839 | 2.26 |
| 12 | 1 | 959.136 | 0.138 | 959.392 | 2.69 |
| 12 | 1 | 959.030 | 0.157 | 959.300 | 3.14 |
| 12 | 1 | 960.416 | 0.135 | 960.395 | 3.68 |
| 12 | 1 | 959.614 | 0.122 | 959.585 | 4.19 |
| 12 | 1 | 959.963 | 0.114 | 959.933 | 4.71 |
| 12 | 1 | 960.023 | 0.121 | 960.050 | 5.25 |
| 12 | 1 | 959.143 | 0.140 | 959.226 | 5.81 |
| 12 | 1 | 959.404 | 0.118 | 959.478 | 6.40 |
| 12 | 2 | 958.445 | 0.160 | 958.571 | 59.25 |
| 12 | 2 | 959.369 | 0.145 | 959.507 | 58.89 |
| 12 | 2 | 959.236 | 0.164 | 959.362 | 58.46 |
| 12 | 2 | 958.579 | 0.170 | 958.710 | 58.11 |
| 12 | 2 | 958.991 | 0.198 | 959.153 | 57.78 |
| 12 | 2 | 958.934 | 0.156 | 959.077 | 57.46 |
| 12 | 2 | 958.387 | 0.200 | 958.529 | 56.25 |
| 13 | 1 | 959.066 | 0.119 | 959.031 | 4.41 |
| 13 | 1 | 959.434 | 0.131 | 959.420 | 4.91 |
| 13 | 1 | 958.620 | 0.145 | 958.710 | 5.50 |
| 13 | 1 | 960.602 | 0.103 | 960.608 | 6.06 |
| 13 | 1 | 958.627 | 0.124 | 958.752 | 6.62 |
| 13 | 1 | 960.306 | 0.126 | 960.458 | 7.22 |
| 13 | 1 | 959.335 | 0.116 | 959.499 | 7.86 |
| 13 | 1 | 959.401 | 0.112 | 959.584 | 8.50 |
| 13 | 1 | 960.184 | 0.139 | 960.526 | 9.19 |
| 13 | 1 | 960.084 | 0.124 | 960.341 | 9.94 |
| 13 | 1 | 959.333 | 0.110 | 959.625 | 10.76 |
| 13 | 1 | 958.544 | 0.189 | 958.972 | 11.57 |
| 13 | 1 | 958.332 | 0.165 | 958.788 | 12.43 |
| 13 | 2 | 959.411 | 0.114 | 959.478 | 59.38 |
| 13 | 2 | 958.607 | 0.143 | 958.692 | 58.99 |
| 13 | 2 | 958.829 | 0.143 | 958.908 | 58.61 |
| 13 | 2 | 958.869 | 0.160 | 958.972 | 58.29 |
| 13 | 2 | 959.344 | 0.147 | 959.459 | 57.96 |
| 13 | 2 | 958.957 | 0.158 | 959.072 | 57.63 |
| 13 | 2 | 959.990 | 0.137 | 960.061 | 57.33 |
| 13 | 2 | 959.418 | 0.149 | 959.488 | 57.05 |
| 13 | 2 | 959.558 | 0.151 | 959.647 | 56.78 |
| 13 | 2 | 960.846 | 0.130 | 960.945 | 56.44 |
| 13 | 2 | 960.678 | 0.151 | 960.777 | 56.18 |
| 13 | 2 | 959.823 | 0.147 | 959.933 | 55.82 |
| 13 | 2 | 959.474 | 0.144 | 959.566 | 55.55 |
| 13 | 2 | 959.189 | 0.137 | 959.293 | 55.31 |
| 14 | 1 | 960.232 | 0.109 | 960.333 | 6.74 |
| 14 | 1 | 960.593 | 0.115 | 960.700 | 7.36 |
| 14 | 1 | 960.696 | 0.105 | 960.849 | 8.02 |
| 14 | 1 | 959.532 | 0.139 | 959.759 | 8.71 |
| 14 | 1 | 959.495 | 0.114 | 959.694 | 9.44 |
| 14 | 1 | 960.438 | 0.127 | 960.713 | 10.24 |
| 14 | 1 | 960.565 | 0.110 | 960.811 | 11.06 |
| 14 | 1 | 960.168 | 0.148 | 960.506 | 11.90 |
| 14 | 1 | 958.770 | 0.158 | 959.179 | 13.00 |
| 14 | 1 | 959.647 | 0.164 | 960.096 | 14.00 |

| D | L | SDB | ER | SDC | HL |
|---|---|---|---|---|---|
| 14 | 1 | 959.404 | 0.176 | 959.836 | 15.03 |
| 18 | 1 | 959.377 | 0.161 | 959.614 | 5.44 |
| 18 | 1 | 959.725 | 0.185 | 960.059 | 5.94 |
| 18 | 1 | 959.096 | 0.172 | 959.458 | 6.48 |
| 18 | 1 | 959.380 | 0.220 | 958.778 | 7.11 |
| 18 | 1 | 958.913 | 0.103 | 958.910 | 7.86 |
| 18 | 1 | 959.465 | 0.114 | 959.464 | 8.46 |
| 18 | 1 | 958.886 | 0.120 | 958.958 | 9.16 |
| 18 | 1 | 958.737 | 0.138 | 958.873 | 9.82 |
| 18 | 1 | 959.388 | 0.118 | 959.510 | 10.50 |
| 18 | 1 | 959.296 | 0.123 | 959.434 | 11.21 |
| 18 | 1 | 959.876 | 0.126 | 960.074 | 11.99 |
| 18 | 1 | 959.106 | 0.144 | 959.392 | 12.83 |
| 18 | 1 | 958.782 | 0.128 | 959.032 | 13.67 |
| 18 | 2 | 958.448 | 0.156 | 958.609 | 60.88 |
| 18 | 2 | 959.056 | 0.141 | 959.167 | 60.40 |
| 18 | 2 | 959.263 | 0.156 | 959.410 | 60.00 |
| 18 | 2 | 958.591 | 0.143 | 958.726 | 59.61 |
| 18 | 2 | 958.372 | 0.161 | 958.513 | 59.26 |
| 18 | 2 | 959.021 | 0.162 | 959.162 | 58.89 |
| 18 | 2 | 958.346 | 0.158 | 958.493 | 58.54 |
| 18 | 2 | 958.975 | 0.124 | 959.080 | 57.71 |
| 18 | 2 | 959.968 | 0.151 | 960.092 | 57.39 |
| 18 | 2 | 959.195 | 0.128 | 959.299 | 57.09 |
| 18 | 2 | 959.296 | 0.146 | 959.659 | 56.70 |
| 18 | 2 | 958.963 | 0.155 | 959.110 | 56.45 |
| 18 | 2 | 959.310 | 0.127 | 959.401 | 56.19 |
| 19 | 1 | 958.630 | 0.153 | 958.901 | 3.84 |
| 19 | 1 | 959.734 | 0.153 | 960.064 | 4.27 |
| 19 | 1 | 959.403 | 0.133 | 959.675 | 4.74 |
| 19 | 1 | 959.416 | 0.143 | 959.781 | 5.19 |
| 19 | 1 | 959.488 | 0.141 | 959.884 | 5.68 |
| 19 | 1 | 958.653 | 0.162 | 959.092 | 6.25 |
| 19 | 1 | 959.877 | 0.213 | 960.385 | 6.80 |
| 19 | 1 | 958.485 | 0.180 | 958.956 | 7.39 |
| 19 | 1 | 959.507 | 0.142 | 959.644 | 8.16 |
| 19 | 1 | 959.658 | 0.115 | 959.780 | 8.82 |
| 19 | 1 | 959.492 | 0.119 | 959.566 | 9.50 |
| 19 | 1 | 958.888 | 0.185 | 959.137 | 10.16 |
| 19 | 1 | 959.136 | 0.138 | 959.348 | 10.85 |
| 19 | 1 | 959.067 | 0.139 | 959.330 | 11.60 |
| 19 | 1 | 959.630 | 0.132 | 959.933 | 12.36 |
| 19 | 1 | 959.237 | 0.149 | 959.590 | 13.20 |
| 19 | 1 | 959.302 | 0.157 | 959.728 | 14.07 |
| 19 | 2 | 958.629 | 0.152 | 958.774 | 61.54 |
| 19 | 2 | 959.175 | 0.146 | 959.301 | 61.13 |
| 19 | 2 | 959.442 | 0.151 | 959.568 | 60.72 |
| 19 | 2 | 959.130 | 0.134 | 959.230 | 59.93 |
| 19 | 2 | 958.797 | 0.143 | 958.923 | 59.55 |
| 19 | 2 | 959.224 | 0.142 | 959.343 | 59.17 |
| 19 | 2 | 959.442 | 0.166 | 959.579 | 58.84 |
| 19 | 2 | 959.563 | 0.178 | 959.656 | 58.52 |
| 19 | 2 | 959.269 | 0.181 | 959.388 | 58.14 |
| 19 | 2 | 958.911 | 0.166 | 959.036 | 57.77 |
| 20 | 1 | 959.553 | 0.114 | 959.657 | 4.58 |
| 20 | 1 | 959.465 | 0.113 | 959.587 | 5.02 |
| 20 | 1 | 958.786 | 0.100 | 959.019 | 5.49 |
| 20 | 1 | 959.870 | 0.120 | 960.080 | 5.97 |
| 20 | 1 | 959.111 | 0.121 | 959.813 | 6.50 |
| 20 | 1 | 959.923 | 0.127 | 960.194 | 7.02 |
| 20 | 1 | 958.898 | 0.130 | 959.181 | 7.58 |
| 20 | 1 | 959.214 | 0.177 | 959.590 | 8.21 |
| 20 | 1 | 959.736 | 0.119 | 959.845 | 8.97 |
| 20 | 1 | 959.047 | 0.101 | 959.070 | 9.59 |
| 20 | 1 | 959.588 | 0.091 | 959.620 | 10.23 |
| 20 | 1 | 959.111 | 0.115 | 959.229 | 10.96 |
| 20 | 1 | 959.177 | 0.113 | 959.321 | 11.69 |
| 20 | 1 | 958.604 | 0.106 | 958.787 | 12.48 |
| 20 | 1 | 959.844 | 0.161 | 960.153 | 13.32 |
| 20 | 1 | 959.272 | 0.133 | 959.591 | 14.20 |
| 27 | 1 | 958.835 | 0.103 | 958.958 | 6.77 |

| 2002 - MARCO | | | | | | 2002 - ABRIL | | | | |
|---|---|---|---|---|---|---|---|---|---|---|
| D | L | SDB | ER | SDC | HL | D | L | SDB | ER | SDC | HL |
|---|---|---|---|---|---|---|---|---|---|---|---|
| 27 | 1 | 959.981 | 0.099 | 960.084 | 7.29 | 02 | 1 | 958.803 | 0.120 | 958.988 | 11.87 |
| 27 | 1 | 959.199 | 0.084 | 959.342 | 7.80 | 02 | 1 | 958.941 | 0.137 | 959.178 | 12.57 |
| 27 | 1 | 959.925 | 0.120 | 960.214 | 8.34 | 02 | 1 | 960.535 | 0.136 | 960.811 | 13.40 |
| 27 | 1 | 959.711 | 0.120 | 960.037 | 9.01 | 02 | 1 | 958.823 | 0.119 | 959.042 | 14.20 |
| 27 | 1 | 959.570 | 0.141 | 959.986 | 9.62 | 02 | 1 | 959.780 | 0.160 | 960.103 | 15.03 |
| 27 | 1 | 960.272 | 0.118 | 960.376 | 10.50 | 02 | 1 | 959.774 | 0.128 | 959.806 | 16.21 |
| 27 | 1 | 960.498 | 0.135 | 960.597 | 11.21 | 02 | 1 | 959.577 | 0.119 | 959.604 | 17.12 |
| 27 | 1 | 959.083 | 0.126 | 959.201 | 12.08 | 02 | 1 | 959.207 | 0.106 | 959.207 | 18.05 |
| 27 | 1 | 959.530 | 0.126 | 959.644 | 12.84 | 02 | 2 | 959.138 | 0.150 | 959.184 | 70.84 |
| 27 | 1 | 959.330 | 0.117 | 959.417 | 13.60 | 02 | 2 | 959.307 | 0.116 | 959.290 | 70.10 |
| 27 | 1 | 959.252 | 0.115 | 959.404 | 14.44 | 02 | 2 | 958.917 | 0.119 | 958.931 | 69.41 |
| 27 | 1 | 959.436 | 0.121 | 959.679 | 15.33 | 02 | 2 | 958.908 | 0.142 | 958.940 | 68.75 |
| 27 | 1 | 958.720 | 0.108 | 958.988 | 16.19 | 02 | 2 | 958.507 | 0.137 | 958.509 | 68.11 |
| 27 | 1 | 958.372 | 0.130 | 958.743 | 17.09 | 02 | 2 | 958.692 | 0.150 | 958.730 | 67.48 |
| 28 | 1 | 960.797 | 0.105 | 960.740 | 11.13 | 02 | 2 | 959.212 | 0.140 | 959.239 | 66.80 |
| 28 | 1 | 959.845 | 0.134 | 959.903 | 11.82 | 02 | 2 | 957.403 | 0.160 | 957.416 | 66.22 |
| 28 | 1 | 960.454 | 0.124 | 960.490 | 12.53 | 02 | 2 | 958.901 | 0.174 | 958.915 | 65.56 |
| 28 | 1 | 959.975 | 0.110 | 959.983 | 13.31 | 02 | 2 | 960.066 | 0.168 | 960.103 | 65.05 |
| 28 | 1 | 959.709 | 0.106 | 959.764 | 14.14 | 02 | 2 | 958.991 | 0.174 | 959.021 | 64.55 |
| 28 | 1 | 960.704 | 0.118 | 960.820 | 14.96 | 02 | 2 | 959.227 | 0.144 | 959.267 | 63.99 |
| 28 | 1 | 960.233 | 0.139 | 960.413 | 15.84 | 02 | 2 | 959.256 | 0.138 | 959.283 | 63.47 |
| 28 | 1 | 960.423 | 0.152 | 960.650 | 16.78 | 02 | 2 | 959.428 | 0.166 | 959.473 | 63.00 |
| 28 | 1 | 959.984 | 0.119 | 960.242 | 17.78 | 02 | 2 | 958.526 | 0.136 | 958.540 | 62.50 |
| 28 | 1 | 959.791 | 0.139 | 960.062 | 18.87 | 03 | 1 | 959.553 | 0.084 | 959.514 | 8.36 |
| 28 | 1 | 959.783 | 0.147 | 960.113 | 19.99 | 03 | 1 | 959.642 | 0.089 | 959.111 | 8.94 |
| 28 | 2 | 958.919 | 0.188 | 959.068 | 63.29 | 03 | 1 | 959.489 | 0.099 | 959.557 | 9.53 |
| 28 | 2 | 959.253 | 0.195 | 959.408 | 62.84 | 03 | 1 | 959.466 | 0.109 | 959.619 | 10.14 |
| 28 | 2 | 958.555 | 0.142 | 958.688 | 62.30 | 03 | 1 | 959.720 | 0.125 | 959.943 | 10.74 |
| 28 | 2 | 959.085 | 0.138 | 959.224 | 61.90 | 03 | 1 | 959.405 | 0.106 | 959.661 | 11.40 |
| 28 | 2 | 959.317 | 0.127 | 959.436 | 61.52 | 03 | 1 | 959.666 | 0.124 | 959.919 | 12.17 |
| 28 | 2 | 959.177 | 0.128 | 959.296 | 61.10 | 03 | 1 | 959.740 | 0.133 | 960.066 | 12.86 |
| 28 | 2 | 959.083 | 0.175 | 959.232 | 60.66 | 03 | 1 | 959.240 | 0.134 | 959.595 | 13.64 |
| 28 | 2 | 959.490 | 0.150 | 959.608 | 60.21 | 03 | 1 | 959.260 | 0.146 | 959.622 | 14.39 |
| 28 | 2 | 958.896 | 0.204 | 959.051 | 59.82 | 03 | 1 | 958.647 | 0.159 | 959.061 | 15.19 |
| | | | | | | 03 | 1 | 959.841 | 0.126 | 959.922 | 16.21 |
| | | | | | | 03 | 1 | 959.314 | 0.116 | 959.415 | 17.10 |
| | | 2002 - ABRIL | | | | 03 | 1 | 958.792 | 0.117 | 958.942 | 18.01 |
| D | L | SDB | ER | SDC | HL | 03 | 1 | 959.296 | 0.123 | 959.486 | 19.02 |
| 01 | 1 | 959.501 | 0.099 | 959.438 | 6.61 | 03 | 1 | 959.642 | 0.095 | 959.793 | 20.02 |
| 01 | 1 | 959.673 | 0.128 | 959.737 | 7.16 | 05 | 1 | 959.459 | 0.134 | 959.377 | 7.64 |
| 01 | 1 | 959.506 | 0.130 | 959.529 | 7.73 | 05 | 1 | 959.690 | 0.148 | 959.681 | 8.19 |
| 01 | 1 | 960.192 | 0.109 | 960.183 | 8.34 | 05 | 1 | 959.588 | 0.134 | 959.541 | 8.79 |
| 01 | 1 | 959.810 | 0.103 | 959.870 | 8.97 | 05 | 1 | 960.188 | 0.112 | 960.099 | 9.33 |
| 01 | 1 | 959.542 | 0.115 | 959.625 | 9.59 | 05 | 1 | 959.228 | 0.148 | 959.212 | 9.90 |
| 01 | 1 | 959.394 | 0.149 | 959.565 | 10.24 | 05 | 1 | 959.489 | 0.127 | 959.452 | 10.50 |
| 01 | 1 | 959.849 | 0.101 | 960.007 | 10.93 | 05 | 1 | 959.459 | 0.101 | 959.344 | 11.14 |
| 01 | 1 | 959.185 | 0.148 | 959.363 | 11.63 | 05 | 1 | 960.073 | 0.134 | 960.109 | 11.78 |
| 01 | 1 | 959.323 | 0.131 | 959.519 | 12.31 | 05 | 1 | 959.702 | 0.135 | 959.744 | 12.47 |
| 01 | 1 | 959.149 | 0.150 | 959.479 | 13.08 | 05 | 1 | 959.801 | 0.130 | 959.867 | 13.14 |
| 01 | 1 | 959.734 | 0.106 | 959.711 | 15.00 | 05 | 1 | 959.884 | 0.126 | 959.941 | 13.86 |
| 01 | 1 | 958.958 | 0.122 | 958.984 | 15.94 | 05 | 1 | 959.512 | 0.123 | 959.633 | 14.70 |
| 01 | 1 | 959.705 | 0.104 | 959.672 | 16.92 | 05 | 1 | 959.320 | 0.128 | 959.480 | 15.49 |
| 01 | 1 | 959.809 | 0.117 | 959.865 | 17.84 | 05 | 2 | 960.061 | 0.166 | 960.117 | 64.63 |
| 01 | 2 | 959.537 | 0.119 | 959.619 | 66.32 | 05 | 2 | 959.013 | 0.165 | 959.069 | 64.13 |
| 01 | 2 | 959.057 | 0.138 | 959.179 | 65.77 | 05 | 2 | 958.174 | 0.174 | 958.241 | 63.68 |
| 01 | 2 | 959.494 | 0.139 | 959.583 | 65.22 | 05 | 2 | 959.608 | 0.149 | 959.656 | 63.23 |
| 01 | 2 | 958.941 | 0.204 | 959.016 | 64.12 | 05 | 2 | 959.302 | 0.175 | 959.363 | 62.80 |
| 01 | 2 | 958.953 | 0.113 | 959.029 | 62.54 | 05 | 2 | 958.531 | 0.189 | 958.597 | 62.37 |
| 01 | 2 | 959.403 | 0.116 | 959.499 | 62.13 | 05 | 2 | 958.874 | 0.179 | 958.889 | 61.95 |
| 01 | 2 | 958.941 | 0.144 | 959.069 | 61.74 | 05 | 2 | 959.635 | 0.213 | 959.700 | 61.54 |
| 01 | 2 | 958.911 | 0.130 | 959.006 | 61.33 | 05 | 2 | 958.925 | 0.222 | 958.983 | 61.05 |
| 01 | 2 | 959.494 | 0.128 | 959.589 | 60.93 | 08 | 1 | 959.060 | 0.128 | 958.982 | 11.62 |
| 01 | 2 | 959.501 | 0.130 | 959.582 | 60.57 | 08 | 1 | 959.169 | 0.127 | 959.224 | 12.32 |
| 02 | 1 | 959.306 | 0.124 | 959.262 | 7.03 | 08 | 1 | 959.316 | 0.103 | 959.348 | 13.01 |
| 02 | 1 | 959.870 | 0.112 | 959.860 | 7.56 | 09 | 1 | 959.195 | 0.143 | 959.259 | 12.76 |
| 02 | 1 | 960.149 | 0.113 | 960.139 | 8.10 | 09 | 1 | 958.484 | 0.152 | 958.602 | 13.45 |
| 02 | 1 | 959.696 | 0.134 | 959.700 | 8.68 | 09 | 1 | 959.384 | 0.128 | 959.503 | 14.23 |
| 02 | 1 | 959.929 | 0.129 | 959.942 | 9.23 | 09 | 1 | 958.832 | 0.136 | 958.976 | 15.05 |
| 02 | 1 | 959.392 | 0.135 | 959.483 | 9.87 | 09 | 1 | 959.225 | 0.157 | 959.521 | 15.81 |
| 02 | 1 | 958.794 | 0.128 | 958.894 | 10.56 | 09 | 1 | 959.313 | 0.147 | 959.347 | 16.81 |
| 02 | 1 | 960.025 | 0.120 | 960.185 | 11.19 | 09 | 1 | 959.126 | 0.127 | 959.129 | 17.74 |

| 2002 - ABRIL | | | | | | 2002 - ABRIL | | | | |
|---|---|---|---|---|---|---|---|---|---|---|
| D | L | SDB | ER | SDC | HL | D | L | SDB | ER | SDC | HL |
| 09 | 1 | 959.607 | 0.138 | 959.607 | 18.65 | 12 | 1 | 960.565 | 0.156 | 960.592 | 14.63 |
| 09 | 1 | 959.365 | 0.127 | 959.320 | 19.70 | 12 | 1 | 960.482 | 0.164 | 960.559 | 15.39 |
| 09 | 1 | 960.241 | 0.144 | 960.228 | 20.70 | 12 | 1 | 960.039 | 0.135 | 960.098 | 16.19 |
| 09 | 1 | 959.490 | 0.158 | 959.558 | 21.75 | 12 | 1 | 959.867 | 0.163 | 959.953 | 17.01 |
| 09 | 1 | 959.210 | 0.130 | 959.263 | 22.87 | 12 | 1 | 959.807 | 0.149 | 959.930 | 17.86 |
| 09 | 1 | 959.969 | 0.176 | 960.113 | 24.05 | 12 | 1 | 960.464 | 0.143 | 960.593 | 18.76 |
| 09 | 1 | 959.494 | 0.137 | 959.634 | 25.27 | 12 | 1 | 960.715 | 0.156 | 960.684 | 20.03 |
| 09 | 2 | 958.475 | 0.165 | 958.499 | 71.78 | 12 | 1 | 959.377 | 0.155 | 959.340 | 21.04 |
| 09 | 2 | 958.979 | 0.172 | 959.047 | 71.04 | 12 | 1 | 960.044 | 0.142 | 960.067 | 22.20 |
| 09 | 2 | 958.483 | 0.160 | 958.531 | 70.31 | 12 | 1 | 960.275 | 0.167 | 960.456 | 23.40 |
| 09 | 2 | 957.938 | 0.172 | 957.966 | 69.59 | 12 | 1 | 959.999 | 0.162 | 960.092 | 24.67 |
| 09 | 2 | 957.781 | 0.174 | 957.796 | 68.96 | 15 | 1 | 959.302 | 0.113 | 959.166 | 12.37 |
| 09 | 2 | 958.400 | 0.175 | 958.473 | 68.25 | 15 | 1 | 959.788 | 0.139 | 959.692 | 13.03 |
| 09 | 2 | 959.658 | 0.169 | 959.673 | 67.59 | 15 | 1 | 959.316 | 0.158 | 959.214 | 13.71 |
| 09 | 2 | 958.510 | 0.185 | 958.577 | 66.83 | 15 | 1 | 959.517 | 0.123 | 959.360 | 14.39 |
| 09 | 2 | 958.770 | 0.166 | 958.795 | 66.28 | 15 | 1 | 959.987 | 0.134 | 959.831 | 15.10 |
| 09 | 2 | 959.224 | 0.160 | 959.272 | 65.74 | 15 | 1 | 959.348 | 0.121 | 959.177 | 15.83 |
| 09 | 2 | 958.808 | 0.146 | 958.832 | 65.16 | 15 | 1 | 959.398 | 0.140 | 959.289 | 16.61 |
| 09 | 2 | 958.845 | 0.156 | 958.873 | 64.69 | 15 | 1 | 959.487 | 0.116 | 959.319 | 17.39 |
| 09 | 2 | 959.253 | 0.153 | 959.293 | 64.09 | 15 | 1 | 959.201 | 0.141 | 959.141 | 18.22 |
| 09 | 2 | 959.410 | 0.159 | 959.451 | 63.64 | 15 | 1 | 959.483 | 0.119 | 959.378 | 19.17 |
| 09 | 2 | 959.538 | 0.179 | 959.579 | 63.14 | 15 | 1 | 959.546 | 0.132 | 959.526 | 20.05 |
| 10 | 1 | 958.973 | 0.203 | 959.132 | 9.98 | 15 | 1 | 958.983 | 0.174 | 959.087 | 21.00 |
| 10 | 1 | 959.676 | 0.169 | 959.888 | 11.20 | 15 | 1 | 959.456 | 0.163 | 959.530 | 22.24 |
| 10 | 1 | 960.000 | 0.117 | 959.921 | 11.91 | 15 | 1 | 958.969 | 0.154 | 958.888 | 23.54 |
| 10 | 1 | 959.173 | 0.147 | 959.144 | 13.37 | 15 | 2 | 959.198 | 0.168 | 959.245 | 71.97 |
| 10 | 1 | 959.930 | 0.170 | 959.986 | 14.06 | 15 | 2 | 959.495 | 0.156 | 959.516 | 71.20 |
| 10 | 1 | 959.659 | 0.124 | 959.725 | 14.77 | 15 | 2 | 958.673 | 0.171 | 958.687 | 70.39 |
| 10 | 1 | 959.571 | 0.112 | 959.580 | 15.56 | 15 | 2 | 958.580 | 0.198 | 958.625 | 69.71 |
| 10 | 1 | 959.461 | 0.128 | 959.538 | 16.35 | 15 | 2 | 958.916 | 0.147 | 958.901 | 68.77 |
| 10 | 1 | 959.479 | 0.171 | 959.658 | 17.19 | 15 | 2 | 958.589 | 0.183 | 958.650 | 68.14 |
| 10 | 1 | 958.233 | 0.152 | 958.389 | 18.07 | 15 | 2 | 958.889 | 0.169 | 958.894 | 67.35 |
| 10 | 1 | 959.297 | 0.199 | 959.515 | 19.01 | 15 | 2 | 958.284 | 0.187 | 958.338 | 66.78 |
| 10 | 1 | 959.235 | 0.137 | 959.457 | 19.96 | 15 | 2 | 959.503 | 0.141 | 959.531 | 66.00 |
| 10 | 1 | 959.734 | 0.219 | 960.026 | 20.97 | 15 | 2 | 958.709 | 0.145 | 958.730 | 65.43 |
| 10 | 1 | 959.471 | 0.178 | 959.482 | 22.17 | 15 | 2 | 959.389 | 0.221 | 959.434 | 64.89 |
| 10 | 2 | 959.318 | 0.160 | 959.308 | 68.66 | 15 | 2 | 959.436 | 0.183 | 959.469 | 64.22 |
| 10 | 2 | 958.404 | 0.188 | 958.425 | 68.05 | 15 | 2 | 958.666 | 0.174 | 958.710 | 63.71 |
| 10 | 2 | 957.645 | 0.220 | 957.690 | 67.45 | 16 | 1 | 958.796 | 0.198 | 958.636 | 12.41 |
| 10 | 2 | 958.856 | 0.152 | 958.886 | 66.71 | 16 | 1 | 960.306 | 0.140 | 960.103 | 13.13 |
| 10 | 2 | 959.951 | 0.154 | 959.974 | 66.18 | 16 | 1 | 959.873 | 0.123 | 959.618 | 13.82 |
| 10 | 2 | 959.074 | 0.158 | 959.091 | 65.67 | 16 | 1 | 959.790 | 0.148 | 959.593 | 14.51 |
| 10 | 2 | 958.832 | 0.158 | 958.828 | 65.18 | 16 | 1 | 959.579 | 0.126 | 959.348 | 15.23 |
| 10 | 2 | 958.807 | 0.174 | 958.854 | 64.67 | 16 | 1 | 959.507 | 0.116 | 959.224 | 15.97 |
| 10 | 2 | 959.301 | 0.157 | 959.316 | 64.18 | 16 | 1 | 959.620 | 0.139 | 959.434 | 16.73 |
| 10 | 2 | 959.048 | 0.163 | 959.100 | 63.74 | 16 | 1 | 959.246 | 0.143 | 959.083 | 17.52 |
| 10 | 2 | 959.027 | 0.172 | 959.047 | 63.28 | 16 | 1 | 960.093 | 0.159 | 959.990 | 18.35 |
| 10 | 2 | 959.566 | 0.212 | 959.624 | 62.85 | 16 | 1 | 960.341 | 0.168 | 960.342 | 19.21 |
| 10 | 2 | 960.049 | 0.212 | 960.107 | 62.39 | 16 | 1 | 959.366 | 0.183 | 959.358 | 20.15 |
| 11 | 1 | 960.079 | 0.126 | 960.001 | 16.77 | 16 | 1 | 959.381 | 0.139 | 959.290 | 21.13 |
| 11 | 1 | 959.398 | 0.155 | 959.571 | 17.91 | 16 | 1 | 959.764 | 0.160 | 959.746 | 22.27 |
| 11 | 1 | 959.034 | 0.160 | 959.161 | 18.89 | 16 | 1 | 958.922 | 0.198 | 959.007 | 23.37 |
| 11 | 1 | 959.019 | 0.172 | 959.203 | 19.91 | 16 | 1 | 959.226 | 0.169 | 959.312 | 24.50 |
| 11 | 1 | 959.522 | 0.184 | 959.718 | 20.93 | 16 | 2 | 959.259 | 0.133 | 959.254 | 73.01 |
| 11 | 1 | 960.368 | 0.123 | 960.249 | 21.99 | 16 | 2 | 958.822 | 0.161 | 958.841 | 72.24 |
| 11 | 1 | 959.122 | 0.137 | 959.061 | 23.12 | 16 | 2 | 958.631 | 0.172 | 958.630 | 71.44 |
| 11 | 1 | 959.811 | 0.161 | 959.801 | 24.30 | 16 | 2 | 959.132 | 0.158 | 959.139 | 70.66 |
| 11 | 1 | 960.107 | 0.136 | 960.099 | 25.58 | 16 | 2 | 959.337 | 0.187 | 959.291 | 69.88 |
| 11 | 1 | 959.543 | 0.192 | 959.658 | 26.89 | 16 | 2 | 959.022 | 0.174 | 959.049 | 69.11 |
| 11 | 1 | 959.901 | 0.198 | 960.042 | 28.26 | 16 | 2 | 959.378 | 0.139 | 959.386 | 68.45 |
| 11 | 2 | 959.474 | 0.186 | 959.510 | 68.86 | 16 | 2 | 958.543 | 0.161 | 958.545 | 67.82 |
| 11 | 2 | 959.995 | 0.176 | 960.057 | 68.20 | 16 | 2 | 958.501 | 0.170 | 958.507 | 67.21 |
| 11 | 2 | 959.102 | 0.203 | 959.189 | 67.63 | 16 | 2 | 958.418 | 0.167 | 958.436 | 66.63 |
| 11 | 2 | 959.211 | 0.206 | 959.304 | 67.07 | 16 | 2 | 958.860 | 0.180 | 958.860 | 65.99 |
| 11 | 2 | 958.304 | 0.186 | 958.371 | 66.48 | 16 | 2 | 960.887 | 0.212 | 960.911 | 64.82 |
| 11 | 2 | 958.680 | 0.155 | 958.710 | 65.95 | 16 | 2 | 959.418 | 0.178 | 959.448 | 64.23 |
| 11 | 2 | 958.547 | 0.175 | 958.588 | 65.42 | 17 | 1 | 959.150 | 0.138 | 958.920 | 12.27 |
| 11 | 2 | 958.456 | 0.148 | 958.484 | 64.84 | 17 | 1 | 959.882 | 0.162 | 959.669 | 12.93 |
| 11 | 2 | 959.100 | 0.174 | 959.135 | 64.33 | 17 | 1 | 959.153 | 0.132 | 958.926 | 13.61 |
| 11 | 2 | 958.506 | 0.180 | 958.558 | 63.84 | 17 | 1 | 959.114 | 0.148 | 958.924 | 14.28 |
| 11 | 2 | 959.334 | 0.171 | 959.392 | 63.32 | 17 | 1 | 959.678 | 0.150 | 959.469 | 15.01 |

| 2002 - ABRIL | | | | | | 2002 - ABRIL | | | | |
|---|---|---|---|---|---|---|---|---|---|---|
| D | L | SDB | ER | SDC | HL | D | L | SDB | ER | SDC | HL |
| 17 | 1 | 959.786 | 0.162 | 959.614 | 15.74 | 24 | 2 | 958.989 | 0.278 | 959.026 | 68.01 |
| 17 | 1 | 959.364 | 0.136 | 959.134 | 16.55 | 24 | 2 | 958.929 | 0.178 | 958.948 | 67.38 |
| 17 | 1 | 959.590 | 0.142 | 959.388 | 17.34 | 24 | 2 | 959.543 | 0.209 | 959.581 | 66.80 |
| 17 | 1 | 959.866 | 0.149 | 959.691 | 18.17 | 25 | 1 | 959.685 | 0.145 | 959.549 | 18.09 |
| 17 | 1 | 958.824 | 0.143 | 958.640 | 19.06 | 25 | 1 | 959.395 | 0.143 | 959.320 | 18.92 |
| 17 | 1 | 960.553 | 0.138 | 960.429 | 19.94 | 25 | 1 | 959.890 | 0.166 | 959.757 | 19.97 |
| 17 | 1 | 959.916 | 0.127 | 959.858 | 20.88 | 25 | 1 | 959.869 | 0.151 | 959.725 | 20.83 |
| 17 | 1 | 959.421 | 0.159 | 959.375 | 21.88 | 25 | 1 | 959.662 | 0.141 | 959.517 | 21.78 |
| 17 | 1 | 959.490 | 0.172 | 959.486 | 22.91 | 25 | 1 | 960.438 | 0.171 | 960.398 | 22.75 |
| 17 | 1 | 959.893 | 0.148 | 959.904 | 24.16 | 25 | 1 | 959.455 | 0.140 | 959.501 | 24.00 |
| 17 | 2 | 959.460 | 0.158 | 959.467 | 75.49 | 25 | 1 | 959.457 | 0.129 | 959.428 | 25.19 |
| 17 | 2 | 959.657 | 0.208 | 959.708 | 74.59 | 25 | 1 | 959.246 | 0.129 | 959.202 | 26.42 |
| 17 | 2 | 958.653 | 0.244 | 958.716 | 73.75 | 25 | 1 | 959.287 | 0.157 | 959.261 | 27.70 |
| 17 | 2 | 958.901 | 0.198 | 958.914 | 72.95 | 25 | 1 | 958.999 | 0.149 | 958.963 | 28.98 |
| 17 | 2 | 958.734 | 0.171 | 958.754 | 72.16 | 25 | 2 | 959.060 | 0.183 | 959.099 | 73.43 |
| 17 | 2 | 958.157 | 0.216 | 958.189 | 71.42 | 25 | 2 | 959.168 | 0.157 | 959.194 | 72.60 |
| 17 | 2 | 959.130 | 0.237 | 959.180 | 70.66 | 25 | 2 | 958.797 | 0.139 | 958.810 | 71.78 |
| 17 | 2 | 957.584 | 0.225 | 957.627 | 69.93 | 25 | 2 | 958.851 | 0.149 | 958.900 | 71.00 |
| 17 | 2 | 958.947 | 0.218 | 959.000 | 69.09 | 25 | 2 | 959.172 | 0.161 | 959.191 | 70.26 |
| 17 | 2 | 958.683 | 0.193 | 958.723 | 68.45 | 25 | 2 | 958.729 | 0.203 | 958.766 | 69.55 |
| 17 | 2 | 959.327 | 0.160 | 959.354 | 67.84 | 25 | 2 | 959.453 | 0.174 | 959.465 | 68.82 |
| 17 | 2 | 959.140 | 0.198 | 959.166 | 67.06 | 25 | 2 | 959.325 | 0.158 | 959.355 | 67.99 |
| 17 | 2 | 958.066 | 0.212 | 958.092 | 66.42 | 26 | 1 | 959.379 | 0.147 | 959.338 | 15.57 |
| 17 | 2 | 958.855 | 0.216 | 958.898 | 65.88 | 26 | 1 | 959.945 | 0.141 | 959.867 | 16.32 |
| 17 | 2 | 958.024 | 0.166 | 958.011 | 65.31 | 26 | 1 | 959.644 | 0.133 | 959.559 | 17.05 |
| 17 | 2 | 959.212 | 0.194 | 959.257 | 64.80 | 26 | 1 | 959.780 | 0.129 | 959.565 | 17.86 |
| 18 | 1 | 960.876 | 0.137 | 960.792 | 13.61 | 26 | 1 | 960.427 | 0.157 | 960.288 | 18.62 |
| 18 | 1 | 959.594 | 0.132 | 959.430 | 14.31 | 26 | 1 | 959.077 | 0.143 | 958.883 | 19.43 |
| 18 | 1 | 960.254 | 0.141 | 960.084 | 15.02 | 26 | 1 | 959.179 | 0.160 | 958.989 | 20.26 |
| 18 | 1 | 960.154 | 0.131 | 959.963 | 15.72 | 26 | 1 | 959.962 | 0.138 | 959.745 | 21.21 |
| 18 | 1 | 960.344 | 0.145 | 960.175 | 16.48 | 26 | 1 | 960.254 | 0.109 | 959.977 | 22.20 |
| 18 | 1 | 960.347 | 0.126 | 960.163 | 17.28 | 26 | 1 | 959.678 | 0.165 | 959.546 | 23.19 |
| 18 | 1 | 959.698 | 0.169 | 959.586 | 18.10 | 26 | 1 | 959.786 | 0.140 | 959.626 | 24.22 |
| 18 | 1 | 960.167 | 0.185 | 960.136 | 19.04 | 26 | 2 | 959.200 | 0.112 | 959.166 | 72.97 |
| 18 | 1 | 959.877 | 0.177 | 959.874 | 19.97 | 26 | 2 | 959.273 | 0.120 | 959.269 | 72.19 |
| 18 | 1 | 960.024 | 0.178 | 960.085 | 20.93 | 26 | 2 | 959.449 | 0.104 | 959.427 | 71.38 |
| 18 | 1 | 959.330 | 0.182 | 959.414 | 21.97 | 26 | 2 | 959.116 | 0.109 | 959.087 | 70.64 |
| 18 | 1 | 959.761 | 0.186 | 959.857 | 23.03 | 26 | 2 | 958.504 | 0.138 | 958.500 | 69.76 |
| 19 | 2 | 959.471 | 0.193 | 959.509 | 72.98 | 26 | 2 | 959.050 | 0.116 | 959.033 | 69.03 |
| 19 | 2 | 959.779 | 0.184 | 959.816 | 72.13 | 26 | 2 | 958.728 | 0.119 | 958.716 | 68.30 |
| 19 | 2 | 959.157 | 0.214 | 959.188 | 71.35 | 26 | 2 | 959.809 | 0.117 | 959.791 | 67.58 |
| 19 | 2 | 959.833 | 0.158 | 959.848 | 70.55 | 26 | 2 | 959.675 | 0.133 | 959.650 | 66.95 |
| 19 | 2 | 958.883 | 0.179 | 958.917 | 69.87 | 26 | 2 | 958.605 | 0.136 | 958.604 | 66.32 |
| 19 | 2 | 959.059 | 0.174 | 959.092 | 69.10 | 29 | 1 | 959.471 | 0.134 | 959.268 | 17.93 |
| 19 | 2 | 960.103 | 0.157 | 960.110 | 68.48 | 29 | 1 | 959.215 | 0.114 | 958.971 | 18.73 |
| 19 | 2 | 959.268 | 0.165 | 959.288 | 67.78 | 29 | 1 | 959.288 | 0.118 | 958.985 | 19.51 |
| 19 | 2 | 959.992 | 0.173 | 960.017 | 67.03 | 29 | 1 | 959.521 | 0.110 | 959.253 | 20.30 |
| 19 | 2 | 959.682 | 0.172 | 959.689 | 66.44 | 29 | 1 | 959.508 | 0.129 | 959.262 | 21.12 |
| 19 | 2 | 959.862 | 0.185 | 959.893 | 65.88 | 29 | 1 | 959.039 | 0.147 | 958.814 | 21.99 |
| 22 | 2 | 959.223 | 0.173 | 959.231 | 73.62 | 29 | 1 | 958.996 | 0.122 | 958.754 | 22.92 |
| 22 | 2 | 959.212 | 0.141 | 959.201 | 72.79 | 29 | 1 | 960.164 | 0.120 | 959.884 | 23.86 |
| 22 | 2 | 958.713 | 0.186 | 958.720 | 71.96 | 29 | 1 | 959.137 | 0.115 | 958.891 | 24.86 |
| 22 | 2 | 959.023 | 0.143 | 959.015 | 70.16 | 29 | 1 | 959.111 | 0.146 | 958.936 | 25.90 |
| 22 | 2 | 958.679 | 0.155 | 958.689 | 69.49 | 29 | 1 | 959.691 | 0.188 | 959.664 | 27.01 |
| 22 | 2 | 958.989 | 0.191 | 959.023 | 68.83 | 29 | 1 | 960.744 | 0.166 | 960.665 | 28.18 |
| 22 | 2 | 958.837 | 0.169 | 958.869 | 68.19 | 29 | 1 | 959.658 | 0.157 | 959.692 | 29.37 |
| 22 | 2 | 958.549 | 0.194 | 958.588 | 67.54 | 29 | 1 | 959.407 | 0.139 | 959.335 | 30.74 |
| 22 | 2 | 958.915 | 0.183 | 958.941 | 66.92 | 29 | 2 | 959.487 | 0.190 | 959.514 | 74.19 |
| 24 | 1 | 958.761 | 0.160 | 958.703 | 19.97 | 29 | 2 | 959.338 | 0.181 | 959.359 | 73.33 |
| 24 | 1 | 959.303 | 0.146 | 959.318 | 20.95 | 29 | 2 | 959.394 | 0.176 | 959.437 | 72.28 |
| 24 | 1 | 959.154 | 0.193 | 959.152 | 26.05 | 29 | 2 | 959.194 | 0.166 | 959.230 | 71.47 |
| 24 | 1 | 958.919 | 0.130 | 958.799 | 27.20 | 29 | 2 | 958.836 | 0.149 | 958.866 | 70.74 |
| 24 | 1 | 959.455 | 0.152 | 959.395 | 28.40 | 29 | 2 | 959.206 | 0.153 | 959.235 | 70.00 |
| 24 | 1 | 959.609 | 0.172 | 959.598 | 29.73 | 29 | 2 | 959.517 | 0.174 | 959.553 | 69.28 |
| 24 | 1 | 959.257 | 0.191 | 959.323 | 31.13 | 29 | 2 | 959.641 | 0.159 | 959.637 | 68.56 |
| 24 | 2 | 959.023 | 0.174 | 959.049 | 74.89 | 29 | 2 | 959.448 | 0.197 | 959.488 | 67.89 |
| 24 | 2 | 958.010 | 0.156 | 958.022 | 74.01 | 29 | 2 | 959.275 | 0.147 | 959.272 | 67.24 |
| 24 | 2 | 957.655 | 0.177 | 957.666 | 73.06 | 29 | 2 | 960.177 | 0.167 | 960.185 | 66.61 |
| 24 | 2 | 960.023 | 0.204 | 960.023 | 70.08 | 30 | 1 | 958.829 | 0.156 | 958.682 | 17.52 |
| 24 | 2 | 959.780 | 0.187 | 959.799 | 69.29 | 30 | 1 | 959.031 | 0.154 | 958.868 | 18.34 |
| 24 | 2 | 959.271 | 0.198 | 959.283 | 68.65 | 30 | 1 | 960.088 | 0.150 | 959.908 | 19.10 |

|  | 2002 - ABRIL |  |  |  |
|---|---|---|---|---|
| D | L | SDB | ER | SDC | HL |
| 30 | 1 | 959.351 | 0.150 | 959.155 | 19.95 |
| 30 | 1 | 959.435 | 0.131 | 959.191 | 20.83 |
| 30 | 1 | 959.967 | 0.122 | 959.701 | 21.67 |
| 30 | 1 | 959.295 | 0.118 | 959.030 | 22.58 |
| 30 | 1 | 960.205 | 0.116 | 959.944 | 23.48 |
| 30 | 1 | 959.574 | 0.117 | 959.282 | 24.40 |
| 30 | 1 | 959.333 | 0.114 | 959.057 | 25.38 |
| 30 | 1 | 959.278 | 0.101 | 958.951 | 26.42 |
| 30 | 1 | 959.543 | 0.136 | 959.357 | 27.48 |
| 30 | 1 | 959.395 | 0.142 | 959.235 | 28.61 |
| 30 | 1 | 959.549 | 0.132 | 959.412 | 29.76 |
| 30 | 1 | 959.563 | 0.127 | 959.468 | 30.94 |
| 30 | 2 | 958.757 | 0.118 | 958.727 | 75.99 |
| 30 | 2 | 958.619 | 0.115 | 958.545 | 74.99 |
| 30 | 2 | 959.269 | 0.132 | 959.219 | 74.05 |
| 30 | 2 | 959.801 | 0.119 | 959.798 | 72.84 |
| 30 | 2 | 959.153 | 0.111 | 959.119 | 71.97 |
| 30 | 2 | 958.483 | 0.081 | 958.373 | 71.10 |
| 30 | 2 | 958.839 | 0.096 | 958.804 | 70.32 |
| 30 | 2 | 958.927 | 0.100 | 958.873 | 69.56 |
| 30 | 2 | 959.338 | 0.103 | 959.270 | 68.87 |
| 30 | 2 | 959.362 | 0.114 | 959.325 | 68.14 |
| 30 | 2 | 959.323 | 0.119 | 959.274 | 67.49 |
| 30 | 2 | 958.747 | 0.146 | 958.710 | 66.84 |
| 30 | 2 | 959.291 | 0.136 | 959.247 | 66.25 |
| 30 | 2 | 958.812 | 0.189 | 958.774 | 65.65 |

|  | 2002 - MAIO |  |  |  |
|---|---|---|---|---|
| D | L | SDB | ER | SDC | HL |
| 08 | 1 | 958.970 | 0.121 | 958.758 | 26.42 |
| 08 | 1 | 960.102 | 0.142 | 959.955 | 27.39 |
| 08 | 1 | 959.588 | 0.162 | 959.504 | 28.35 |
| 08 | 1 | 959.315 | 0.133 | 959.172 | 29.37 |
| 08 | 1 | 958.653 | 0.194 | 958.719 | 31.98 |
| 08 | 1 | 959.662 | 0.169 | 959.780 | 33.32 |
| 08 | 1 | 958.928 | 0.151 | 958.993 | 34.75 |
| 08 | 1 | 959.535 | 0.169 | 959.573 | 36.20 |
| 08 | 1 | 958.781 | 0.145 | 958.843 | 37.64 |
| 08 | 1 | 958.903 | 0.178 | 959.012 | 39.26 |
| 08 | 2 | 958.778 | 0.143 | 958.780 | 72.58 |
| 08 | 2 | 959.580 | 0.173 | 959.632 | 71.72 |
| 08 | 2 | 959.320 | 0.137 | 959.328 | 70.83 |
| 08 | 2 | 959.377 | 0.141 | 959.378 | 69.98 |
| 08 | 2 | 959.517 | 0.129 | 959.511 | 69.21 |
| 08 | 2 | 959.590 | 0.179 | 959.639 | 68.39 |
| 08 | 2 | 959.287 | 0.161 | 959.310 | 67.70 |
| 08 | 2 | 958.849 | 0.185 | 958.859 | 66.92 |
| 09 | 1 | 958.674 | 0.236 | 958.795 | 34.90 |
| 09 | 1 | 959.392 | 0.124 | 959.375 | 36.52 |
| 09 | 1 | 959.628 | 0.133 | 959.562 | 38.03 |
| 09 | 1 | 958.877 | 0.161 | 958.874 | 39.69 |
| 09 | 1 | 959.289 | 0.132 | 959.208 | 41.40 |
| 10 | 1 | 959.402 | 0.125 | 959.441 | 37.21 |
| 10 | 1 | 959.720 | 0.098 | 959.535 | 38.82 |
| 10 | 1 | 959.636 | 0.098 | 959.401 | 40.51 |
| 10 | 1 | 960.733 | 0.108 | 960.529 | 42.29 |
| 10 | 1 | 959.762 | 0.097 | 959.564 | 44.22 |
| 10 | 2 | 958.847 | 0.128 | 958.863 | 75.14 |
| 10 | 2 | 959.966 | 0.137 | 959.987 | 74.10 |
| 10 | 2 | 959.543 | 0.125 | 959.551 | 72.00 |
| 10 | 2 | 959.664 | 0.118 | 959.665 | 71.08 |
| 10 | 2 | 959.779 | 0.137 | 959.805 | 70.18 |
| 10 | 2 | 959.339 | 0.132 | 959.364 | 69.35 |
| 10 | 2 | 959.525 | 0.140 | 959.543 | 68.46 |
| 13 | 1 | 959.161 | 0.144 | 959.100 | 25.18 |
| 13 | 1 | 959.270 | 0.137 | 959.157 | 26.02 |
| 13 | 1 | 959.699 | 0.114 | 959.488 | 26.90 |
| 13 | 1 | 959.309 | 0.096 | 958.992 | 27.84 |
| 13 | 1 | 960.152 | 0.131 | 960.003 | 28.78 |
| 13 | 1 | 959.960 | 0.108 | 959.662 | 29.75 |

|  | 2002 - MAIO |  |  |  |
|---|---|---|---|---|
| D | L | SDB | ER | SDC | HL |
| 13 | 1 | 959.232 | 0.117 | 958.981 | 30.75 |
| 13 | 1 | 959.373 | 0.145 | 959.221 | 31.82 |
| 13 | 1 | 960.160 | 0.124 | 959.935 | 32.98 |
| 13 | 1 | 959.211 | 0.116 | 958.984 | 34.16 |
| 13 | 1 | 959.267 | 0.126 | 959.077 | 35.38 |
| 13 | 1 | 960.076 | 0.144 | 960.030 | 36.66 |
| 13 | 1 | 958.424 | 0.144 | 958.308 | 38.05 |
| 13 | 2 | 959.116 | 0.152 | 959.135 | 76.10 |
| 13 | 2 | 959.213 | 0.148 | 959.238 | 75.08 |
| 13 | 2 | 959.017 | 0.151 | 959.049 | 74.00 |
| 13 | 2 | 958.438 | 0.139 | 958.451 | 73.03 |
| 13 | 2 | 959.335 | 0.143 | 959.348 | 72.04 |
| 13 | 2 | 958.778 | 0.150 | 958.783 | 71.05 |
| 13 | 2 | 959.820 | 0.129 | 959.819 | 70.17 |
| 13 | 2 | 959.395 | 0.141 | 959.412 | 69.36 |
| 13 | 2 | 958.142 | 0.151 | 958.151 | 68.49 |
| 13 | 2 | 958.239 | 0.175 | 958.235 | 67.76 |
| 13 | 2 | 959.799 | 0.243 | 959.832 | 67.01 |
| 14 | 1 | 958.848 | 0.099 | 958.616 | 26.95 |
| 14 | 1 | 959.535 | 0.119 | 959.412 | 27.84 |
| 14 | 1 | 959.387 | 0.108 | 959.190 | 28.77 |
| 14 | 1 | 960.290 | 0.122 | 960.101 | 29.73 |
| 14 | 1 | 960.184 | 0.126 | 959.982 | 30.74 |
| 14 | 1 | 960.025 | 0.124 | 959.826 | 31.82 |
| 14 | 1 | 959.343 | 0.122 | 959.167 | 32.94 |
| 14 | 1 | 959.869 | 0.121 | 959.700 | 34.06 |
| 14 | 1 | 959.820 | 0.123 | 959.626 | 35.34 |
| 14 | 1 | 959.434 | 0.139 | 959.375 | 36.60 |
| 14 | 1 | 958.583 | 0.124 | 958.460 | 37.95 |
| 14 | 2 | 958.996 | 0.151 | 959.023 | 75.83 |
| 14 | 2 | 959.465 | 0.179 | 959.473 | 74.78 |
| 14 | 2 | 959.523 | 0.171 | 959.556 | 73.77 |
| 14 | 2 | 958.587 | 0.147 | 958.613 | 72.79 |
| 14 | 2 | 959.240 | 0.152 | 959.260 | 71.83 |
| 14 | 2 | 958.873 | 0.157 | 958.886 | 70.88 |
| 14 | 2 | 958.105 | 0.135 | 958.073 | 70.02 |
| 14 | 2 | 959.571 | 0.136 | 959.582 | 69.16 |
| 14 | 2 | 959.088 | 0.173 | 959.117 | 68.35 |
| 14 | 2 | 958.913 | 0.205 | 958.910 | 67.61 |
| 14 | 2 | 959.549 | 0.175 | 959.596 | 66.88 |
| 14 | 2 | 960.130 | 0.355 | 960.051 | 66.14 |
| 15 | 1 | 959.285 | 0.104 | 959.034 | 29.10 |
| 15 | 1 | 959.542 | 0.093 | 959.256 | 30.06 |
| 15 | 1 | 960.246 | 0.099 | 959.977 | 31.07 |
| 15 | 1 | 960.095 | 0.098 | 959.792 | 32.16 |
| 15 | 1 | 960.316 | 0.122 | 960.109 | 33.27 |
| 15 | 1 | 959.968 | 0.113 | 959.785 | 34.49 |
| 15 | 1 | 959.047 | 0.126 | 958.904 | 35.72 |
| 15 | 1 | 959.810 | 0.111 | 959.695 | 36.97 |
| 15 | 1 | 959.444 | 0.103 | 959.308 | 38.31 |
| 15 | 1 | 959.096 | 0.134 | 959.001 | 39.71 |
| 15 | 1 | 959.613 | 0.107 | 959.465 | 41.24 |
| 15 | 1 | 959.224 | 0.105 | 959.161 | 42.92 |
| 15 | 2 | 959.901 | 0.147 | 959.932 | 78.73 |
| 15 | 2 | 958.863 | 0.138 | 958.882 | 77.51 |
| 15 | 2 | 958.541 | 0.164 | 958.573 | 76.38 |
| 15 | 2 | 958.803 | 0.155 | 958.835 | 75.29 |
| 15 | 2 | 959.317 | 0.151 | 959.341 | 74.24 |
| 15 | 2 | 959.326 | 0.167 | 959.370 | 73.26 |
| 15 | 2 | 959.147 | 0.120 | 959.109 | 72.29 |
| 15 | 2 | 958.378 | 0.150 | 958.408 | 71.36 |
| 15 | 2 | 958.535 | 0.142 | 958.559 | 70.47 |
| 15 | 2 | 960.004 | 0.164 | 960.027 | 69.61 |
| 15 | 2 | 958.737 | 0.176 | 958.780 | 68.78 |
| 15 | 2 | 960.048 | 0.158 | 960.063 | 67.98 |
| 15 | 2 | 959.552 | 0.171 | 959.574 | 66.79 |
| 16 | 1 | 959.844 | 0.116 | 959.627 | 27.49 |
| 16 | 1 | 959.360 | 0.082 | 959.097 | 28.48 |
| 16 | 1 | 959.908 | 0.089 | 959.574 | 29.79 |
| 16 | 1 | 960.029 | 0.095 | 959.774 | 30.95 |
| 16 | 1 | 959.504 | 0.093 | 959.197 | 32.03 |

|       | 2002 - MAIO |       |       |       |
|-------|-------------|-------|-------|-------|
| D  L  | SDB         | ER    | SDC   | HL    |
| 16 1  | 960.177     | 0.096 | 959.858 | 33.11 |
| 16 1  | 959.992     | 0.090 | 959.670 | 34.28 |
| 16 1  | 960.342     | 0.086 | 960.003 | 35.48 |
| 16 1  | 960.656     | 0.089 | 960.276 | 36.76 |
| 16 1  | 960.420     | 0.089 | 960.101 | 38.17 |
| 16 1  | 959.727     | 0.086 | 959.392 | 39.59 |
| 16 2  | 959.866     | 0.099 | 959.851 | 74.96 |
| 16 2  | 959.244     | 0.076 | 959.166 | 73.92 |
| 16 2  | 958.993     | 0.105 | 958.958 | 72.92 |
| 16 2  | 958.917     | 0.091 | 958.869 | 71.90 |
| 16 2  | 959.256     | 0.093 | 959.176 | 70.96 |
| 16 2  | 959.471     | 0.098 | 959.416 | 69.96 |
| 16 2  | 959.195     | 0.136 | 959.190 | 69.01 |
| 16 2  | 959.147     | 0.135 | 959.142 | 68.15 |
| 16 2  | 959.306     | 0.110 | 959.282 | 67.28 |
| 16 2  | 958.994     | 0.125 | 959.001 | 66.47 |
| 20 2  | 958.834     | 0.137 | 958.838 | 73.36 |
| 20 2  | 957.820     | 0.198 | 957.855 | 72.27 |
| 20 2  | 959.656     | 0.152 | 959.673 | 71.26 |
| 20 2  | 958.805     | 0.142 | 958.828 | 70.30 |
| 20 2  | 959.164     | 0.163 | 959.192 | 69.42 |
| 20 2  | 959.694     | 0.146 | 959.736 | 68.54 |
| 20 2  | 958.710     | 0.177 | 958.763 | 67.72 |
| 20 2  | 958.568     | 0.160 | 958.602 | 66.94 |
| 20 2  | 958.720     | 0.199 | 958.778 | 66.18 |
| 20 2  | 958.089     | 0.273 | 958.121 | 65.41 |
| 27 1  | 959.916     | 0.126 | 959.742 | 36.53 |
| 27 1  | 960.200     | 0.120 | 959.902 | 37.66 |
| 27 1  | 959.690     | 0.112 | 959.413 | 38.88 |
| 27 1  | 960.217     | 0.128 | 959.996 | 40.09 |
| 27 1  | 960.184     | 0.134 | 959.920 | 41.62 |
| 27 1  | 959.498     | 0.149 | 959.253 | 43.01 |
| 27 1  | 959.775     | 0.128 | 959.520 | 44.43 |
| 27 1  | 960.383     | 0.157 | 960.169 | 45.94 |
| 27 1  | 960.122     | 0.131 | 959.870 | 47.62 |
| 27 1  | 959.789     | 0.154 | 959.625 | 49.51 |
| 27 2  | 959.219     | 0.181 | 959.301 | 76.64 |
| 27 2  | 960.010     | 0.182 | 960.073 | 75.22 |
| 27 2  | 959.469     | 0.150 | 959.532 | 73.96 |
| 27 2  | 958.358     | 0.161 | 958.389 | 72.74 |
| 27 2  | 959.144     | 0.138 | 959.176 | 71.63 |
| 27 2  | 959.474     | 0.140 | 959.524 | 70.52 |
| 27 2  | 959.187     | 0.150 | 959.241 | 69.41 |
| 27 2  | 959.311     | 0.157 | 959.373 | 68.37 |
| 27 2  | 958.756     | 0.170 | 958.811 | 67.41 |
| 27 2  | 958.734     | 0.171 | 958.770 | 66.52 |
| 27 2  | 959.251     | 0.183 | 959.323 | 65.67 |
| 27 2  | 959.422     | 0.188 | 959.462 | 64.81 |
| 28 1  | 959.263     | 0.118 | 959.159 | 34.99 |
| 28 1  | 959.879     | 0.120 | 959.755 | 35.98 |
| 28 1  | 959.965     | 0.113 | 959.782 | 37.00 |
| 28 1  | 958.987     | 0.118 | 958.738 | 38.13 |
| 28 1  | 959.393     | 0.130 | 959.182 | 39.35 |
| 28 1  | 959.325     | 0.138 | 959.121 | 40.52 |
| 28 1  | 959.508     | 0.116 | 959.296 | 41.79 |
| 28 1  | 959.139     | 0.141 | 958.969 | 43.09 |
| 28 1  | 959.502     | 0.139 | 959.322 | 44.50 |
| 28 1  | 960.361     | 0.153 | 960.193 | 45.99 |
| 28 1  | 960.159     | 0.116 | 959.918 | 47.58 |
| 28 1  | 959.181     | 0.139 | 958.978 | 49.27 |
| 28 2  | 959.423     | 0.171 | 959.476 | 78.06 |
| 28 2  | 958.782     | 0.166 | 958.841 | 76.60 |
| 28 2  | 959.198     | 0.181 | 959.274 | 75.25 |
| 28 2  | 958.302     | 0.154 | 958.359 | 73.82 |
| 28 2  | 958.854     | 0.163 | 958.892 | 72.60 |
| 28 2  | 958.589     | 0.218 | 958.627 | 71.48 |
| 28 2  | 959.297     | 0.163 | 959.342 | 70.32 |
| 28 2  | 958.354     | 0.133 | 958.375 | 69.25 |
| 28 2  | 959.711     | 0.212 | 959.762 | 68.27 |
| 28 2  | 958.907     | 0.180 | 958.996 | 67.34 |
| 28 2  | 959.554     | 0.186 | 959.624 | 66.42 |

|       | 2002 - MAIO |       |       |       |
|-------|-------------|-------|-------|-------|
| D  L  | SDB         | ER    | SDC   | HL    |
| 29 1  | 959.444     | 0.141 | 959.275 | 36.79 |
| 29 1  | 959.194     | 0.124 | 959.090 | 37.95 |
| 29 1  | 959.215     | 0.126 | 959.001 | 39.26 |
| 29 1  | 959.486     | 0.147 | 959.340 | 40.42 |
| 29 1  | 958.872     | 0.135 | 958.699 | 41.69 |
| 29 1  | 959.486     | 0.140 | 959.340 | 43.10 |
| 29 1  | 960.118     | 0.145 | 960.015 | 44.53 |
| 29 1  | 958.655     | 0.127 | 958.450 | 46.04 |
| 29 1  | 959.410     | 0.142 | 959.225 | 47.62 |
| 29 1  | 960.177     | 0.146 | 959.966 | 49.32 |
| 29 1  | 958.750     | 0.128 | 958.559 | 51.15 |
| 29 2  | 960.663     | 0.171 | 960.711 | 75.98 |
| 29 2  | 959.043     | 0.158 | 959.116 | 74.69 |
| 29 2  | 959.334     | 0.133 | 959.363 | 73.34 |
| 29 2  | 959.359     | 0.137 | 959.399 | 72.15 |
| 29 2  | 959.047     | 0.136 | 959.093 | 71.01 |
| 29 2  | 958.316     | 0.158 | 958.375 | 69.91 |
| 29 2  | 959.388     | 0.130 | 959.408 | 68.88 |
| 29 2  | 959.025     | 0.140 | 959.077 | 67.92 |
| 29 2  | 959.618     | 0.142 | 959.638 | 66.98 |
| 29 2  | 958.872     | 0.160 | 958.916 | 66.10 |
| 29 2  | 959.064     | 0.166 | 959.128 | 65.25 |
| 29 2  | 959.660     | 0.177 | 959.716 | 64.43 |
| 31 2  | 959.999     | 0.138 | 960.005 | 70.17 |
| 31 2  | 959.705     | 0.146 | 959.742 | 69.07 |
| 31 2  | 959.850     | 0.166 | 959.899 | 67.97 |
| 31 2  | 958.895     | 0.150 | 958.945 | 66.95 |
| 31 2  | 958.051     | 0.162 | 958.100 | 66.00 |
| 31 2  | 958.316     | 0.163 | 958.377 | 64.94 |

|       | 2002 - JUNHO |       |       |       |
|-------|--------------|-------|-------|-------|
| D  L  | SDB          | ER    | SDC   | HL    |
| 03 2  | 958.421      | 0.179 | 958.466 | 78.27 |
| 03 2  | 959.289      | 0.126 | 959.309 | 76.59 |
| 03 2  | 959.261      | 0.168 | 959.332 | 75.10 |
| 03 2  | 958.693      | 0.142 | 958.738 | 73.68 |
| 03 2  | 958.623      | 0.187 | 958.692 | 72.36 |
| 03 2  | 959.408      | 0.150 | 959.459 | 71.11 |
| 03 2  | 959.294      | 0.139 | 959.346 | 69.95 |
| 03 2  | 958.777      | 0.142 | 958.814 | 68.84 |
| 03 2  | 959.691      | 0.180 | 959.747 | 67.75 |
| 03 2  | 959.319      | 0.157 | 959.350 | 66.73 |
| 03 2  | 959.541      | 0.150 | 959.542 | 65.59 |
| 03 2  | 959.240      | 0.164 | 959.277 | 64.64 |
| 03 2  | 959.431      | 0.253 | 959.479 | 63.74 |
| 04 1  | 960.210      | 0.135 | 960.156 | 39.75 |
| 04 1  | 959.474      | 0.096 | 959.283 | 40.83 |
| 04 1  | 959.571      | 0.110 | 959.441 | 41.96 |
| 04 1  | 959.505      | 0.152 | 959.501 | 43.12 |
| 04 1  | 959.339      | 0.110 | 959.152 | 44.38 |
| 04 1  | 959.235      | 0.150 | 959.013 | 45.70 |
| 04 1  | 959.465      | 0.118 | 959.261 | 47.06 |
| 04 1  | 959.313      | 0.135 | 959.189 | 48.55 |
| 04 1  | 959.124      | 0.130 | 958.975 | 50.15 |
| 04 1  | 959.356      | 0.120 | 959.194 | 51.88 |
| 04 1  | 959.622      | 0.116 | 959.465 | 53.78 |
| 04 1  | 959.631      | 0.132 | 959.533 | 56.06 |
| 04 2  | 959.932      | 0.122 | 959.947 | 75.78 |
| 04 2  | 959.585      | 0.155 | 959.631 | 74.22 |
| 04 2  | 958.764      | 0.144 | 958.797 | 72.84 |
| 04 2  | 959.151      | 0.148 | 959.158 | 71.56 |
| 04 2  | 959.763      | 0.131 | 959.783 | 70.36 |
| 04 2  | 959.520      | 0.119 | 959.527 | 69.21 |
| 04 2  | 959.696      | 0.140 | 959.716 | 68.09 |
| 04 2  | 958.674      | 0.129 | 958.693 | 67.05 |
| 04 2  | 959.232      | 0.124 | 959.219 | 66.01 |
| 04 2  | 959.299      | 0.151 | 959.343 | 65.00 |
| 04 2  | 958.607      | 0.156 | 958.650 | 64.09 |
| 04 2  | 959.250      | 0.162 | 959.280 | 63.22 |
| 04 2  | 959.195      | 0.189 | 959.149 | 62.33 |

| 2002 - JUNHO | | | | | | 2002 - JUNHO | | | | |
|---|---|---|---|---|---|---|---|---|---|---|
| D | L | SDB | ER | SDC | HL | D | L | SDB | ER | SDC | HL |
| 05 | 1 | 959.235 | 0.144 | 959.115 | 39.80 | 12 | 2 | 959.306 | 0.196 | 959.328 | 60.01 |
| 05 | 1 | 959.735 | 0.118 | 959.554 | 40.87 | 14 | 1 | 958.629 | 0.128 | 958.739 | 46.99 |
| 05 | 1 | 959.754 | 0.130 | 959.577 | 41.95 | 14 | 1 | 958.588 | 0.129 | 958.672 | 48.22 |
| 05 | 1 | 958.915 | 0.125 | 958.772 | 43.25 | 14 | 1 | 959.668 | 0.112 | 959.694 | 49.48 |
| 05 | 1 | 959.657 | 0.123 | 959.473 | 44.50 | 14 | 1 | 959.379 | 0.126 | 959.464 | 50.80 |
| 05 | 1 | 959.665 | 0.100 | 959.434 | 45.80 | 14 | 1 | 959.050 | 0.149 | 959.139 | 52.23 |
| 05 | 1 | 959.127 | 0.115 | 958.933 | 47.12 | 14 | 1 | 959.425 | 0.116 | 959.362 | 53.71 |
| 05 | 1 | 959.448 | 0.100 | 959.230 | 48.55 | 14 | 1 | 959.569 | 0.142 | 959.512 | 55.29 |
| 05 | 1 | 959.743 | 0.146 | 959.646 | 50.09 | 14 | 1 | 959.060 | 0.148 | 959.008 | 57.05 |
| 05 | 1 | 959.125 | 0.125 | 958.931 | 51.92 | 14 | 1 | 959.687 | 0.140 | 959.681 | 58.99 |
| 05 | 1 | 959.891 | 0.116 | 959.733 | 53.69 | 14 | 2 | 959.961 | 0.174 | 960.013 | 72.56 |
| 05 | 2 | 959.608 | 0.178 | 959.664 | 74.54 | 14 | 2 | 958.464 | 0.141 | 958.491 | 71.03 |
| 05 | 2 | 960.287 | 0.136 | 960.287 | 73.11 | 14 | 2 | 959.356 | 0.133 | 959.376 | 69.58 |
| 05 | 2 | 958.966 | 0.162 | 959.003 | 71.78 | 14 | 2 | 959.583 | 0.155 | 959.628 | 68.22 |
| 05 | 2 | 958.874 | 0.166 | 958.955 | 70.51 | 14 | 2 | 958.563 | 0.172 | 958.595 | 66.91 |
| 05 | 2 | 959.664 | 0.177 | 959.726 | 69.35 | 14 | 2 | 958.948 | 0.140 | 958.936 | 65.62 |
| 05 | 2 | 960.333 | 0.191 | 960.375 | 68.23 | 14 | 2 | 959.954 | 0.119 | 959.973 | 64.45 |
| 05 | 2 | 958.602 | 0.164 | 958.650 | 66.15 | 14 | 2 | 958.672 | 0.170 | 958.735 | 63.38 |
| 05 | 2 | 958.598 | 0.175 | 958.640 | 65.17 | 14 | 2 | 959.222 | 0.144 | 959.266 | 62.36 |
| 05 | 2 | 959.168 | 0.163 | 959.216 | 64.24 | 14 | 2 | 958.966 | 0.160 | 959.009 | 61.35 |
| 05 | 2 | 958.145 | 0.181 | 958.217 | 63.30 | 14 | 2 | 958.633 | 0.172 | 958.676 | 60.41 |
| 05 | 2 | 958.936 | 0.236 | 958.969 | 62.43 | 17 | 2 | 957.875 | 0.188 | 957.926 | 74.57 |
| 10 | 1 | 959.104 | 0.128 | 959.083 | 40.96 | 17 | 2 | 959.353 | 0.195 | 959.417 | 72.74 |
| 10 | 1 | 959.663 | 0.118 | 959.580 | 42.01 | 17 | 2 | 958.589 | 0.160 | 958.633 | 71.06 |
| 10 | 1 | 958.753 | 0.109 | 958.659 | 43.14 | 17 | 2 | 959.568 | 0.161 | 959.632 | 69.53 |
| 10 | 1 | 959.301 | 0.111 | 959.228 | 44.27 | 17 | 2 | 957.961 | 0.211 | 958.031 | 68.09 |
| 10 | 1 | 960.029 | 0.117 | 959.913 | 45.42 | 17 | 2 | 958.159 | 0.175 | 958.235 | 66.73 |
| 10 | 1 | 958.754 | 0.142 | 958.708 | 46.63 | 17 | 2 | 958.904 | 0.215 | 958.985 | 65.48 |
| 10 | 1 | 959.429 | 0.113 | 959.295 | 47.97 | 17 | 2 | 958.800 | 0.208 | 958.893 | 64.23 |
| 10 | 1 | 959.757 | 0.112 | 959.673 | 49.29 | 17 | 2 | 959.379 | 0.222 | 959.440 | 63.07 |
| 10 | 1 | 959.618 | 0.109 | 959.461 | 50.74 | 17 | 2 | 959.210 | 0.185 | 959.284 | 61.90 |
| 10 | 1 | 958.969 | 0.107 | 958.788 | 52.28 | 17 | 2 | 958.541 | 0.220 | 958.634 | 60.88 |
| 10 | 1 | 960.003 | 0.108 | 959.843 | 53.99 | 17 | 2 | 957.997 | 0.197 | 958.070 | 59.89 |
| 10 | 1 | 959.760 | 0.131 | 959.730 | 55.79 | 17 | 2 | 958.287 | 0.186 | 958.322 | 58.91 |
| 11 | 1 | 958.852 | 0.216 | 958.996 | 44.78 | 18 | 2 | 960.510 | 0.206 | 960.572 | 72.79 |
| 11 | 1 | 959.042 | 0.165 | 959.191 | 46.00 | 18 | 2 | 957.667 | 0.230 | 957.723 | 71.05 |
| 11 | 1 | 958.578 | 0.184 | 958.654 | 47.28 | 18 | 2 | 959.960 | 0.180 | 960.029 | 69.31 |
| 11 | 2 | 957.806 | 0.157 | 957.864 | 73.95 | 18 | 2 | 958.904 | 0.189 | 958.973 | 67.66 |
| 11 | 2 | 959.683 | 0.153 | 959.721 | 72.32 | 18 | 2 | 959.967 | 0.167 | 960.011 | 66.29 |
| 11 | 2 | 957.823 | 0.187 | 957.891 | 70.86 | 18 | 2 | 959.724 | 0.175 | 959.792 | 64.96 |
| 11 | 2 | 958.819 | 0.140 | 958.844 | 69.46 | 18 | 2 | 960.129 | 0.186 | 960.184 | 63.70 |
| 11 | 2 | 958.896 | 0.126 | 958.908 | 68.14 | 18 | 2 | 960.115 | 0.199 | 960.182 | 62.56 |
| 11 | 2 | 958.698 | 0.168 | 958.748 | 66.92 | 18 | 2 | 958.727 | 0.171 | 958.774 | 61.24 |
| 11 | 2 | 958.413 | 0.135 | 958.443 | 65.73 | 18 | 2 | 958.722 | 0.171 | 958.762 | 60.16 |
| 11 | 2 | 959.696 | 0.166 | 959.733 | 64.63 | 18 | 2 | 958.979 | 0.245 | 959.075 | 59.17 |
| 11 | 2 | 959.038 | 0.147 | 959.075 | 63.60 | 18 | 2 | 959.365 | 0.243 | 959.448 | 58.16 |
| 11 | 2 | 958.660 | 0.227 | 958.715 | 62.54 | 19 | 1 | 959.188 | 0.156 | 959.356 | 49.12 |
| 11 | 2 | 959.449 | 0.179 | 959.516 | 61.59 | 19 | 1 | 959.638 | 0.126 | 959.726 | 50.41 |
| 11 | 2 | 958.893 | 0.221 | 958.953 | 60.58 | 19 | 1 | 959.624 | 0.118 | 959.668 | 52.40 |
| 12 | 1 | 959.411 | 0.146 | 959.478 | 42.60 | 19 | 1 | 959.234 | 0.131 | 959.245 | 53.92 |
| 12 | 1 | 959.191 | 0.143 | 959.183 | 43.70 | 19 | 1 | 959.087 | 0.126 | 959.125 | 55.37 |
| 12 | 1 | 958.993 | 0.119 | 958.892 | 44.79 | 19 | 1 | 959.454 | 0.114 | 959.498 | 56.89 |
| 12 | 1 | 959.794 | 0.137 | 959.721 | 45.96 | 19 | 1 | 958.782 | 0.117 | 958.828 | 58.54 |
| 12 | 1 | 958.935 | 0.154 | 958.845 | 47.17 | 19 | 1 | 958.880 | 0.160 | 959.069 | 60.33 |
| 12 | 1 | 959.100 | 0.154 | 959.048 | 48.43 | 19 | 1 | 958.160 | 0.157 | 958.345 | 62.27 |
| 12 | 1 | 958.956 | 0.142 | 958.830 | 49.77 | 19 | 2 | 959.228 | 0.144 | 959.280 | 70.37 |
| 12 | 1 | 959.707 | 0.119 | 959.557 | 51.19 | 19 | 2 | 959.115 | 0.132 | 959.117 | 68.81 |
| 12 | 1 | 959.921 | 0.107 | 959.756 | 52.66 | 19 | 2 | 958.093 | 0.178 | 958.163 | 67.41 |
| 12 | 1 | 960.052 | 0.112 | 959.857 | 54.30 | 19 | 2 | 959.010 | 0.173 | 959.073 | 66.06 |
| 12 | 1 | 959.279 | 0.120 | 959.131 | 56.00 | 19 | 2 | 958.955 | 0.182 | 959.024 | 64.72 |
| 12 | 2 | 959.154 | 0.179 | 959.187 | 72.72 | 20 | 1 | 959.989 | 0.157 | 960.226 | 48.39 |
| 12 | 2 | 958.993 | 0.174 | 959.050 | 71.18 | 20 | 1 | 958.056 | 0.161 | 958.300 | 49.54 |
| 12 | 2 | 958.769 | 0.144 | 958.801 | 69.75 | 20 | 1 | 959.107 | 0.132 | 959.233 | 50.74 |
| 12 | 2 | 959.434 | 0.181 | 959.460 | 68.44 | 20 | 1 | 958.946 | 0.129 | 959.049 | 51.99 |
| 12 | 2 | 958.311 | 0.154 | 958.354 | 67.19 | 20 | 1 | 959.370 | 0.177 | 959.493 | 53.36 |
| 12 | 2 | 959.113 | 0.138 | 959.093 | 66.01 | 20 | 1 | 959.013 | 0.164 | 959.198 | 54.75 |
| 12 | 2 | 958.054 | 0.166 | 958.116 | 64.88 | 20 | 1 | 959.137 | 0.167 | 959.301 | 56.24 |
| 12 | 2 | 959.066 | 0.156 | 959.116 | 63.83 | 20 | 1 | 958.694 | 0.254 | 958.979 | 57.88 |
| 12 | 2 | 958.658 | 0.194 | 958.700 | 62.82 | 20 | 1 | 958.540 | 0.193 | 958.774 | 59.58 |
| 12 | 2 | 958.869 | 0.204 | 958.936 | 61.84 | 20 | 1 | 958.356 | 0.202 | 958.688 | 61.43 |
| 12 | 2 | 958.909 | 0.220 | 958.969 | 60.91 | 20 | 2 | 959.943 | 0.145 | 960.007 | 68.57 |

| 2002 - JUNHO | | | | | |
|---|---|---|---|---|---|
| D | L | SDB | ER | SDC | HL |
| 20 | 2 | 959.582 | 0.190 | 959.690 | 67.08 |
| 20 | 2 | 959.346 | 0.174 | 959.422 | 65.67 |
| 20 | 2 | 959.240 | 0.189 | 959.334 | 64.36 |
| 20 | 2 | 959.259 | 0.166 | 959.340 | 63.16 |
| 20 | 2 | 959.332 | 0.228 | 959.456 | 61.98 |
| 20 | 2 | 958.489 | 0.190 | 958.593 | 60.87 |
| 20 | 2 | 957.928 | 0.220 | 958.025 | 59.80 |
| 20 | 2 | 958.653 | 0.224 | 958.743 | 58.80 |
| 21 | 2 | 959.364 | 0.142 | 959.379 | 71.28 |
| 21 | 2 | 959.510 | 0.132 | 959.562 | 69.56 |
| 21 | 2 | 959.101 | 0.186 | 959.191 | 67.96 |
| 21 | 2 | 958.863 | 0.127 | 958.897 | 66.51 |
| 21 | 2 | 958.839 | 0.140 | 958.904 | 65.10 |
| 21 | 2 | 959.043 | 0.140 | 959.089 | 63.79 |
| 21 | 2 | 959.568 | 0.137 | 959.607 | 62.55 |
| 21 | 2 | 959.310 | 0.157 | 959.387 | 61.37 |
| 21 | 2 | 958.966 | 0.178 | 959.024 | 60.27 |
| 21 | 2 | 958.599 | 0.179 | 958.663 | 59.20 |
| 24 | 2 | 959.019 | 0.232 | 959.123 | 71.93 |
| 24 | 2 | 959.179 | 0.172 | 959.271 | 70.07 |
| 24 | 2 | 958.347 | 0.143 | 958.433 | 68.38 |
| 24 | 2 | 959.739 | 0.154 | 959.825 | 66.81 |
| 24 | 2 | 958.390 | 0.192 | 958.489 | 65.34 |
| 24 | 2 | 959.116 | 0.168 | 959.202 | 63.95 |
| 24 | 2 | 959.557 | 0.154 | 959.637 | 62.66 |
| 24 | 2 | 958.743 | 0.177 | 958.842 | 61.41 |
| 24 | 2 | 958.454 | 0.163 | 958.539 | 60.23 |
| 24 | 2 | 959.042 | 0.182 | 959.127 | 59.12 |
| 24 | 2 | 959.033 | 0.175 | 959.130 | 58.01 |
| 24 | 2 | 957.433 | 0.181 | 957.504 | 57.01 |
| 27 | 2 | 959.125 | 0.168 | 959.278 | 69.11 |
| 27 | 2 | 958.584 | 0.216 | 958.774 | 67.28 |
| 27 | 2 | 959.394 | 0.169 | 959.553 | 65.58 |
| 27 | 2 | 959.222 | 0.150 | 959.357 | 64.08 |
| 27 | 2 | 959.970 | 0.169 | 960.104 | 62.59 |
| 27 | 2 | 958.308 | 0.167 | 958.429 | 61.25 |
| 27 | 2 | 958.372 | 0.186 | 958.518 | 59.99 |
| 27 | 2 | 958.717 | 0.257 | 958.888 | 58.76 |
| 27 | 2 | 959.394 | 0.177 | 959.551 | 57.64 |
| 27 | 2 | 959.444 | 0.231 | 959.607 | 56.56 |
| 27 | 2 | 959.613 | 0.223 | 959.769 | 55.54 |
| 27 | 2 | 959.109 | 0.267 | 959.233 | 54.59 |
| 28 | 1 | 958.958 | 0.167 | 959.235 | 52.99 |
| 28 | 1 | 959.200 | 0.178 | 959.460 | 54.18 |
| 28 | 1 | 958.959 | 0.144 | 959.084 | 55.43 |
| 28 | 1 | 959.117 | 0.135 | 959.254 | 56.70 |
| 28 | 1 | 958.508 | 0.170 | 958.665 | 58.03 |
| 28 | 1 | 959.835 | 0.134 | 959.845 | 59.48 |
| 28 | 1 | 959.437 | 0.133 | 959.567 | 61.00 |
| 28 | 1 | 959.289 | 0.126 | 959.397 | 62.59 |
| 28 | 1 | 958.727 | 0.144 | 958.918 | 64.49 |
| 28 | 1 | 959.972 | 0.134 | 960.149 | 66.91 |
| 28 | 1 | 959.525 | 0.183 | 959.668 | 64.22 |
| 28 | 2 | 959.431 | 0.190 | 959.599 | 62.68 |
| 28 | 2 | 958.373 | 0.205 | 958.541 | 61.25 |
| 28 | 2 | 958.480 | 0.229 | 958.633 | 60.01 |
| 28 | 2 | 958.486 | 0.171 | 958.609 | 58.80 |
| 28 | 2 | 958.930 | 0.203 | 959.096 | 57.69 |
| 28 | 2 | 958.778 | 0.205 | 958.951 | 56.62 |
| 28 | 2 | 959.316 | 0.193 | 959.456 | 55.56 |
| 28 | 2 | 958.408 | 0.216 | 958.572 | 54.55 |
| 28 | 2 | 959.301 | 0.209 | 959.428 | 53.59 |
| 28 | 2 | 959.516 | 0.225 | 959.680 | 52.56 |

| 2002 - JULHO | | | | | |
|---|---|---|---|---|---|
| D | L | SDB | ER | SDC | HL |
| 01 | 1 | 959.411 | 0.147 | 959.620 | 52.89 |
| 01 | 1 | 958.991 | 0.152 | 959.180 | 53.94 |
| 01 | 1 | 958.731 | 0.148 | 958.962 | 55.12 |
| 01 | 1 | 958.857 | 0.157 | 959.009 | 56.33 |

| 2002 - JULHO | | | | | |
|---|---|---|---|---|---|
| D | L | SDB | ER | SDC | HL |
| 01 | 1 | 959.590 | 0.125 | 959.709 | 57.57 |
| 01 | 1 | 959.209 | 0.117 | 959.314 | 58.86 |
| 01 | 1 | 959.216 | 0.147 | 959.380 | 60.23 |
| 01 | 1 | 959.058 | 0.129 | 959.181 | 61.74 |
| 01 | 1 | 959.424 | 0.158 | 959.588 | 63.56 |
| 01 | 1 | 959.618 | 0.140 | 959.755 | 65.30 |
| 01 | 2 | 960.240 | 0.195 | 960.392 | 63.29 |
| 01 | 2 | 958.785 | 0.180 | 958.917 | 61.67 |
| 01 | 2 | 958.975 | 0.167 | 959.101 | 60.29 |
| 01 | 2 | 958.807 | 0.164 | 958.951 | 58.92 |
| 01 | 2 | 959.277 | 0.168 | 959.408 | 57.69 |
| 01 | 2 | 957.941 | 0.182 | 958.091 | 56.49 |
| 01 | 2 | 959.946 | 0.159 | 960.064 | 55.35 |
| 01 | 2 | 958.278 | 0.182 | 958.414 | 54.29 |
| 01 | 2 | 958.946 | 0.153 | 959.045 | 53.27 |
| 01 | 2 | 959.206 | 0.184 | 959.329 | 52.29 |
| 01 | 2 | 958.567 | 0.185 | 958.696 | 51.35 |
| 01 | 2 | 959.594 | 0.225 | 959.716 | 50.44 |
| 03 | 1 | 958.095 | 0.178 | 958.403 | 54.10 |
| 03 | 1 | 959.448 | 0.129 | 959.633 | 55.31 |
| 03 | 1 | 959.071 | 0.135 | 959.308 | 56.45 |
| 03 | 1 | 959.144 | 0.130 | 959.334 | 57.77 |
| 03 | 1 | 958.343 | 0.148 | 958.564 | 59.09 |
| 03 | 1 | 959.743 | 0.153 | 959.941 | 60.47 |
| 03 | 1 | 959.314 | 0.126 | 959.497 | 61.88 |
| 03 | 1 | 958.664 | 0.101 | 958.823 | 63.67 |
| 03 | 1 | 960.655 | 0.151 | 960.836 | 65.48 |
| 03 | 1 | 959.212 | 0.121 | 959.379 | 67.29 |
| 03 | 1 | 959.511 | 0.126 | 959.650 | 69.25 |
| 03 | 2 | 960.656 | 0.216 | 960.872 | 62.33 |
| 03 | 2 | 959.391 | 0.159 | 959.583 | 60.84 |
| 03 | 2 | 958.661 | 0.188 | 958.852 | 59.27 |
| 03 | 2 | 958.888 | 0.168 | 959.065 | 57.96 |
| 03 | 2 | 959.461 | 0.168 | 959.626 | 56.68 |
| 03 | 2 | 958.346 | 0.182 | 958.542 | 55.48 |
| 03 | 2 | 958.905 | 0.177 | 959.075 | 54.37 |
| 03 | 2 | 958.617 | 0.178 | 958.814 | 53.29 |
| 03 | 2 | 958.493 | 0.195 | 958.696 | 52.24 |
| 03 | 2 | 958.881 | 0.187 | 959.032 | 51.27 |
| 03 | 2 | 958.520 | 0.208 | 958.702 | 50.33 |
| 03 | 2 | 958.637 | 0.227 | 958.825 | 49.41 |
| 15 | 1 | 959.696 | 0.139 | 959.906 | 57.40 |
| 15 | 1 | 959.648 | 0.145 | 959.833 | 58.47 |
| 15 | 1 | 959.454 | 0.122 | 959.549 | 59.57 |
| 15 | 1 | 960.152 | 0.115 | 960.259 | 60.68 |
| 15 | 1 | 959.392 | 0.110 | 959.495 | 61.86 |
| 15 | 1 | 959.765 | 0.114 | 959.865 | 63.05 |
| 15 | 1 | 959.576 | 0.131 | 959.676 | 64.38 |
| 15 | 1 | 959.466 | 0.146 | 959.626 | 65.65 |
| 15 | 1 | 959.753 | 0.164 | 959.930 | 67.12 |
| 15 | 1 | 959.158 | 0.119 | 959.229 | 68.64 |
| 15 | 1 | 959.981 | 0.141 | 960.114 | 70.24 |
| 15 | 1 | 959.742 | 0.146 | 959.845 | 72.09 |
| 15 | 2 | 959.105 | 0.243 | 959.282 | 57.14 |
| 15 | 2 | 958.542 | 0.168 | 958.701 | 55.56 |
| 15 | 2 | 959.932 | 0.188 | 960.115 | 54.10 |
| 15 | 2 | 958.918 | 0.168 | 959.076 | 52.69 |
| 15 | 2 | 957.940 | 0.168 | 958.097 | 51.42 |
| 15 | 2 | 958.219 | 0.176 | 958.362 | 49.89 |
| 15 | 2 | 959.626 | 0.188 | 959.808 | 48.70 |
| 15 | 2 | 958.911 | 0.161 | 959.055 | 47.61 |
| 15 | 2 | 958.244 | 0.227 | 958.425 | 46.56 |
| 15 | 2 | 959.638 | 0.197 | 959.832 | 45.55 |
| 15 | 2 | 959.411 | 0.172 | 959.586 | 44.47 |
| 16 | 1 | 959.694 | 0.121 | 959.820 | 59.40 |
| 16 | 1 | 959.724 | 0.117 | 959.823 | 60.50 |
| 16 | 1 | 959.442 | 0.105 | 959.503 | 61.57 |
| 16 | 1 | 959.459 | 0.095 | 959.497 | 62.75 |
| 16 | 1 | 959.313 | 0.136 | 959.360 | 63.94 |
| 16 | 1 | 958.853 | 0.131 | 958.900 | 65.22 |
| 16 | 1 | 959.418 | 0.152 | 959.544 | 66.52 |

| 2002 - JULHO | | | | | |
|---|---|---|---|---|---|
| D | L | SDB | ER | SDC | HL |
| 16 | 1 | 959.271 | 0.107 | 959.265 | 67.93 |
| 16 | 1 | 960.090 | 0.115 | 960.057 | 69.55 |
| 16 | 1 | 960.677 | 0.119 | 960.685 | 71.17 |
| 16 | 1 | 959.158 | 0.137 | 959.232 | 72.89 |
| 16 | 2 | 958.898 | 0.216 | 959.082 | 56.63 |
| 16 | 2 | 959.775 | 0.173 | 959.934 | 55.09 |
| 16 | 2 | 959.197 | 0.201 | 959.369 | 53.63 |
| 16 | 2 | 959.130 | 0.169 | 959.283 | 52.29 |
| 16 | 2 | 958.161 | 0.145 | 958.309 | 50.98 |
| 16 | 2 | 958.768 | 0.144 | 958.915 | 49.76 |
| 16 | 2 | 958.742 | 0.117 | 958.846 | 48.56 |
| 16 | 2 | 958.492 | 0.148 | 958.645 | 47.39 |
| 16 | 2 | 958.913 | 0.142 | 959.059 | 46.34 |
| 16 | 2 | 958.466 | 0.158 | 958.606 | 45.29 |
| 16 | 2 | 959.110 | 0.161 | 959.243 | 44.34 |
| 16 | 2 | 959.415 | 0.159 | 959.553 | 43.41 |
| 19 | 1 | 958.936 | 0.147 | 959.156 | 60.12 |
| 19 | 1 | 959.340 | 0.163 | 959.640 | 61.24 |
| 19 | 1 | 959.234 | 0.160 | 959.544 | 62.30 |
| 19 | 1 | 959.342 | 0.146 | 959.465 | 63.44 |
| 19 | 1 | 959.166 | 0.160 | 959.248 | 64.63 |
| 19 | 1 | 959.358 | 0.156 | 959.472 | 65.89 |
| 19 | 1 | 959.185 | 0.137 | 959.351 | 67.13 |
| 19 | 1 | 959.813 | 0.163 | 960.128 | 68.46 |
| 19 | 1 | 959.659 | 0.187 | 959.981 | 69.89 |
| 19 | 1 | 958.493 | 0.183 | 958.889 | 71.43 |
| 19 | 2 | 959.128 | 0.242 | 959.322 | 52.37 |
| 19 | 2 | 959.269 | 0.158 | 959.446 | 50.97 |
| 19 | 2 | 959.481 | 0.172 | 959.639 | 49.66 |
| 19 | 2 | 960.033 | 0.188 | 960.228 | 48.45 |
| 19 | 2 | 959.009 | 0.184 | 959.196 | 47.20 |
| 19 | 2 | 958.479 | 0.206 | 958.640 | 45.82 |
| 19 | 2 | 959.509 | 0.135 | 959.659 | 44.76 |
| 19 | 2 | 959.175 | 0.204 | 959.375 | 43.73 |
| 19 | 2 | 958.983 | 0.163 | 959.157 | 42.78 |
| 19 | 2 | 959.657 | 0.156 | 959.792 | 41.87 |
| 22 | 1 | 959.860 | 0.128 | 960.000 | 64.83 |
| 22 | 1 | 959.440 | 0.121 | 959.514 | 65.99 |
| 22 | 1 | 960.411 | 0.130 | 960.442 | 67.24 |
| 22 | 1 | 959.993 | 0.100 | 959.992 | 68.57 |
| 22 | 1 | 959.815 | 0.126 | 959.916 | 69.93 |
| 22 | 1 | 959.976 | 0.130 | 960.126 | 71.39 |
| 22 | 1 | 960.227 | 0.144 | 960.493 | 72.90 |
| 23 | 2 | 958.668 | 0.266 | 958.847 | 52.70 |
| 23 | 2 | 958.867 | 0.166 | 959.022 | 51.11 |
| 23 | 2 | 958.827 | 0.163 | 958.982 | 49.68 |
| 23 | 2 | 959.534 | 0.176 | 959.702 | 48.35 |
| 23 | 2 | 959.417 | 0.138 | 959.547 | 47.09 |
| 23 | 2 | 959.486 | 0.157 | 959.623 | 45.89 |
| 23 | 2 | 958.792 | 0.174 | 958.935 | 44.78 |
| 23 | 2 | 958.980 | 0.186 | 959.148 | 43.66 |
| 23 | 2 | 958.904 | 0.146 | 959.028 | 42.62 |
| 23 | 2 | 958.708 | 0.154 | 958.851 | 41.60 |
| 23 | 2 | 958.901 | 0.163 | 959.011 | 40.65 |
| 23 | 2 | 959.113 | 0.156 | 959.235 | 39.75 |
| 24 | 1 | 958.671 | 0.167 | 958.710 | 62.53 |
| 24 | 1 | 959.432 | 0.133 | 959.416 | 63.57 |
| 24 | 1 | 958.896 | 0.131 | 958.900 | 64.63 |
| 24 | 1 | 959.235 | 0.126 | 959.262 | 65.70 |
| 24 | 1 | 958.672 | 0.137 | 958.748 | 66.88 |
| 24 | 1 | 959.567 | 0.157 | 959.620 | 68.10 |
| 24 | 1 | 959.269 | 0.127 | 959.220 | 69.39 |
| 24 | 1 | 958.661 | 0.165 | 958.828 | 70.80 |
| 24 | 1 | 959.107 | 0.165 | 959.247 | 72.30 |
| 24 | 1 | 959.075 | 0.192 | 959.306 | 73.89 |
| 24 | 1 | 959.359 | 0.170 | 959.497 | 75.63 |
| 24 | 1 | 958.662 | 0.162 | 958.889 | 77.58 |
| 24 | 2 | 958.094 | 0.170 | 958.197 | 50.62 |
| 24 | 2 | 959.060 | 0.184 | 959.201 | 49.17 |
| 24 | 2 | 958.801 | 0.185 | 958.930 | 47.84 |
| 24 | 2 | 958.599 | 0.157 | 958.703 | 46.61 |

| 2002 - JULHO | | | | | |
|---|---|---|---|---|---|
| D | L | SDB | ER | SDC | HL |
| 24 | 2 | 958.577 | 0.198 | 958.713 | 45.45 |
| 24 | 2 | 958.736 | 0.170 | 958.859 | 44.32 |
| 24 | 2 | 958.803 | 0.162 | 958.920 | 43.22 |
| 24 | 2 | 957.725 | 0.181 | 957.854 | 42.17 |
| 24 | 2 | 959.294 | 0.197 | 959.436 | 41.17 |
| 24 | 2 | 958.678 | 0.174 | 958.775 | 40.23 |
| 25 | 1 | 959.260 | 0.161 | 959.368 | 60.63 |
| 25 | 1 | 959.651 | 0.139 | 959.694 | 61.56 |
| 25 | 1 | 959.509 | 0.133 | 959.549 | 65.06 |
| 25 | 1 | 959.352 | 0.156 | 959.450 | 66.31 |
| 25 | 1 | 959.447 | 0.141 | 959.556 | 67.56 |
| 25 | 1 | 959.163 | 0.167 | 959.241 | 68.90 |
| 25 | 1 | 959.084 | 0.144 | 959.177 | 70.25 |
| 25 | 1 | 958.989 | 0.128 | 959.091 | 71.72 |
| 25 | 1 | 959.750 | 0.160 | 959.951 | 73.78 |
| 25 | 1 | 959.783 | 0.166 | 959.941 | 75.46 |
| 25 | 2 | 958.623 | 0.179 | 958.705 | 48.47 |
| 25 | 2 | 960.157 | 0.163 | 960.234 | 46.97 |
| 25 | 2 | 958.997 | 0.140 | 959.069 | 45.62 |
| 25 | 2 | 958.748 | 0.137 | 958.826 | 44.39 |
| 25 | 2 | 959.606 | 0.137 | 959.678 | 43.17 |
| 25 | 2 | 959.087 | 0.165 | 959.165 | 42.11 |
| 25 | 2 | 958.207 | 0.188 | 958.323 | 41.10 |
| 25 | 2 | 958.702 | 0.148 | 958.786 | 40.03 |
| 25 | 2 | 958.921 | 0.182 | 959.010 | 39.05 |
| 25 | 2 | 959.534 | 0.163 | 959.623 | 38.14 |
| 26 | 1 | 959.654 | 0.135 | 959.780 | 61.71 |
| 26 | 1 | 959.641 | 0.116 | 959.686 | 62.71 |
| 26 | 1 | 959.427 | 0.129 | 959.486 | 63.73 |
| 26 | 1 | 958.346 | 0.146 | 958.390 | 64.79 |
| 26 | 1 | 958.874 | 0.135 | 958.881 | 66.10 |
| 26 | 1 | 958.314 | 0.139 | 958.382 | 67.39 |
| 26 | 1 | 959.729 | 0.142 | 959.839 | 68.59 |
| 26 | 1 | 958.391 | 0.117 | 958.557 | 73.25 |
| 30 | 1 | 959.342 | 0.137 | 959.301 | 65.94 |
| 30 | 1 | 959.124 | 0.114 | 959.055 | 67.07 |
| 30 | 1 | 959.471 | 0.154 | 959.485 | 68.15 |
| 30 | 1 | 959.985 | 0.152 | 960.032 | 69.30 |
| 30 | 1 | 959.892 | 0.130 | 959.929 | 70.56 |
| 30 | 1 | 959.903 | 0.135 | 960.035 | 72.62 |
| 30 | 1 | 959.167 | 0.130 | 959.218 | 74.03 |
| 30 | 1 | 959.180 | 0.143 | 959.245 | 75.50 |
| 30 | 2 | 958.367 | 0.214 | 958.473 | 44.74 |
| 30 | 2 | 960.441 | 0.183 | 960.535 | 43.23 |
| 30 | 2 | 958.562 | 0.153 | 958.643 | 37.15 |
| 30 | 2 | 958.670 | 0.193 | 958.776 | 36.18 |
| 30 | 2 | 959.103 | 0.197 | 959.208 | 35.36 |
| 30 | 2 | 957.885 | 0.178 | 957.983 | 34.51 |
| 30 | 2 | 959.098 | 0.155 | 959.190 | 33.70 |
| 30 | 2 | 958.268 | 0.188 | 958.359 | 32.94 |
| 31 | 1 | 959.714 | 0.124 | 959.580 | 62.33 |
| 31 | 1 | 959.256 | 0.103 | 958.993 | 63.22 |
| 31 | 1 | 959.438 | 0.108 | 959.210 | 64.18 |
| 31 | 1 | 959.383 | 0.121 | 959.179 | 65.14 |
| 31 | 1 | 960.020 | 0.100 | 959.791 | 66.12 |
| 31 | 1 | 959.711 | 0.101 | 959.471 | 67.19 |
| 31 | 1 | 959.577 | 0.105 | 959.317 | 68.27 |
| 31 | 1 | 960.439 | 0.110 | 960.213 | 69.43 |
| 31 | 1 | 959.750 | 0.114 | 959.582 | 70.62 |
| 31 | 1 | 960.234 | 0.129 | 960.040 | 71.84 |
| 31 | 1 | 959.898 | 0.141 | 959.835 | 73.14 |
| 31 | 1 | 960.427 | 0.103 | 960.354 | 74.53 |
| 31 | 2 | 960.508 | 0.178 | 960.541 | 45.25 |
| 31 | 2 | 959.803 | 0.153 | 959.781 | 43.93 |
| 31 | 2 | 959.002 | 0.149 | 959.043 | 42.71 |
| 31 | 2 | 958.522 | 0.145 | 958.532 | 41.55 |
| 31 | 2 | 958.289 | 0.150 | 958.324 | 40.44 |
| 31 | 2 | 959.025 | 0.155 | 959.066 | 39.37 |
| 31 | 2 | 958.624 | 0.180 | 958.672 | 38.40 |
| 31 | 2 | 958.109 | 0.172 | 958.157 | 37.42 |
| 31 | 2 | 959.096 | 0.148 | 959.131 | 36.45 |

```
        2002 - JULHO                                    2002 - AGOSTO
 D  L    SDB    ER     SDC    HL              D  L    SDB    ER     SDC    HL
31  2  958.808 0.146 958.843 35.56            12  2  958.409 0.176 958.543 29.46
31  2  958.893 0.140 958.921 34.73            12  2  960.037 0.135 960.146 28.60
31  2  958.851 0.153 958.885 33.89            12  2  958.840 0.143 958.943 27.76
                                              12  2  958.380 0.146 958.496 27.03
                                              16  1  958.868 0.137 958.964 66.29
        2002 - AGOSTO                         16  1  959.200 0.121 959.273 67.15
 D  L    SDB    ER     SDC    HL              16  1  959.499 0.125 959.644 68.10
05  1  959.231 0.133 959.079 67.82            16  1  958.913 0.137 959.100 69.05
05  1  959.500 0.132 959.404 68.88            16  1  959.511 0.136 959.706 70.14
05  1  959.207 0.136 959.128 70.00            16  1  959.157 0.138 959.349 71.20
05  1  958.924 0.126 958.840 71.13            16  1  958.739 0.144 958.941 72.26
05  1  959.243 0.135 959.240 72.41            16  1  959.582 0.114 959.682 73.38
05  1  958.815 0.158 958.847 73.74            16  1  959.262 0.113 959.292 74.59
05  1  960.741 0.181 960.844 75.17            16  1  959.394 0.132 959.476 75.85
05  1  959.427 0.130 959.287 76.68            16  1  959.357 0.127 959.432 77.20
05  1  958.898 0.151 958.756 78.28            16  2  958.581 0.175 958.679 33.55
05  2  959.034 0.111 959.028 39.06            16  2  959.568 0.181 959.690 32.44
05  2  958.740 0.172 958.810 37.95            16  2  958.654 0.168 958.769 31.36
05  2  959.652 0.173 959.728 36.80            16  2  959.350 0.151 959.459 30.26
05  2  958.875 0.166 958.913 35.80            16  2  958.384 0.182 958.504 29.33
05  2  958.901 0.137 958.940 34.73            16  2  958.284 0.183 958.416 28.37
05  2  959.287 0.124 959.307 33.82            16  2  958.627 0.193 958.765 27.47
05  2  958.472 0.124 958.479 32.93            16  2  958.698 0.163 958.831 26.50
05  2  958.661 0.136 958.686 32.04            16  2  959.053 0.136 959.161 25.73
05  2  958.977 0.117 958.964 31.22            16  2  958.613 0.178 958.753 24.98
05  2  959.758 0.142 959.783 30.45            16  2  959.085 0.142 959.137 24.26
09  1  959.301 0.121 959.195 67.25            19  1  958.732 0.124 958.699 65.49
09  1  959.128 0.109 958.977 68.19            19  1  959.733 0.148 959.721 66.30
09  1  959.596 0.103 959.501 69.15            19  1  959.746 0.143 959.742 67.11
09  1  959.486 0.113 959.462 70.16            19  1  959.130 0.168 959.202 67.96
09  1  958.756 0.133 958.829 71.23            19  1  959.062 0.188 959.191 68.87
09  1  960.396 0.151 960.458 72.39            19  1  958.902 0.147 959.013 69.87
09  1  960.220 0.145 960.239 73.55            19  1  959.314 0.178 959.482 70.84
09  1  959.783 0.133 959.908 74.78            19  1  959.655 0.127 959.631 71.95
09  1  959.986 0.135 960.245 76.08            19  1  959.061 0.133 959.048 72.97
09  1  959.157 0.113 959.242 77.72            19  1  959.707 0.153 959.728 74.09
09  1  959.156 0.107 959.083 79.23            19  1  959.708 0.160 959.726 75.25
09  1  959.467 0.087 959.312 80.87            19  2  960.333 0.208 960.488 32.63
09  2  959.564 0.121 959.629 41.73            19  2  959.772 0.183 959.887 27.13
09  2  958.570 0.174 958.660 40.32            19  2  959.599 0.138 959.702 26.24
09  2  958.703 0.104 958.742 39.05            19  2  958.924 0.136 959.021 25.42
09  2  959.478 0.151 959.523 37.88            19  2  959.155 0.163 959.277 24.61
09  2  959.415 0.141 959.496 36.77            19  2  958.126 0.158 958.255 23.85
09  2  958.966 0.113 959.003 35.67            19  2  958.226 0.220 958.387 21.72
09  2  959.508 0.113 959.538 34.66            20  1  957.445 0.129 957.468 67.01
09  2  959.222 0.111 959.216 33.44            20  1  959.353 0.164 959.485 67.83
09  2  959.265 0.090 959.266 32.52            20  1  959.117 0.162 959.234 68.68
09  2  959.109 0.124 959.186 31.64            20  1  959.417 0.157 959.595 69.60
09  2  959.225 0.105 959.271 30.78            20  1  959.622 0.164 959.799 70.54
09  2  959.402 0.105 959.435 29.97            20  1  959.553 0.161 959.757 71.51
09  2  958.956 0.119 959.014 29.16            20  1  959.403 0.162 959.495 72.68
12  1  959.479 0.126 959.443 67.09            20  1  959.118 0.193 959.258 73.75
12  1  958.923 0.122 958.857 68.00            20  1  960.110 0.143 960.151 74.92
12  1  960.059 0.122 960.087 68.97            20  1  959.869 0.147 959.879 76.11
12  1  959.111 0.114 959.184 69.95            20  1  959.923 0.126 959.891 77.35
12  1  959.853 0.109 959.849 71.13            20  2  960.248 0.196 960.390 32.48
12  1  959.136 0.122 959.134 72.19            20  2  959.493 0.166 959.603 31.35
12  1  959.496 0.122 959.460 73.32            20  2  958.983 0.133 959.062 30.29
12  1  959.523 0.096 959.381 74.51            20  2  959.721 0.147 959.806 29.30
12  1  958.773 0.116 958.690 75.83            20  2  960.263 0.195 960.373 28.33
12  1  959.215 0.154 959.198 77.17            20  2  958.661 0.159 958.751 27.44
12  1  959.003 0.133 958.950 78.59            20  2  958.025 0.222 958.165 26.55
12  1  959.892 0.132 959.901 80.15            20  2  958.120 0.238 958.266 25.68
12  2  960.312 0.154 960.435 37.67            20  2  958.691 0.184 958.780 24.85
12  2  959.431 0.153 959.566 36.45            20  2  959.715 0.165 959.813 23.93
12  2  959.627 0.137 959.736 35.33            20  2  959.820 0.199 959.955 23.16
12  2  959.122 0.157 959.237 34.29            20  2  959.007 0.182 959.136 22.45
12  2  959.117 0.172 959.257 33.24            20  2  958.265 0.214 958.407 21.80
12  2  958.824 0.177 958.957 32.24            21  1  958.526 0.232 958.687 68.06
12  2  958.414 0.162 958.510 31.17            21  1  959.541 0.174 959.757 68.89
12  2  959.030 0.159 959.164 30.29            21  1  958.319 0.148 958.441 69.79
```

|  | 2002 - AGOSTO | | | | |  | 2002 - AGOSTO | | | |
|---|---|---|---|---|---|---|---|---|---|---|
| D | L | SDB | ER | SDC | HL | D | L | SDB | ER | SDC | HL |
| 21 | 1 | 959.482 | 0.149 | 959.617 | 70.72 | 26 | 2 | 958.827 | 0.191 | 958.961 | 21.01 |
| 21 | 1 | 959.238 | 0.148 | 959.353 | 71.63 | 26 | 2 | 958.770 | 0.150 | 958.900 | 18.20 |
| 21 | 1 | 958.135 | 0.191 | 958.339 | 72.61 | 26 | 2 | 959.402 | 0.147 | 959.519 | 17.55 |
| 21 | 1 | 959.341 | 0.169 | 959.509 | 73.71 | 26 | 2 | 959.087 | 0.149 | 959.216 | 16.11 |
| 21 | 1 | 959.242 | 0.183 | 959.524 | 74.90 | 26 | 2 | 958.645 | 0.148 | 958.761 | 15.52 |
| 21 | 1 | 958.838 | 0.111 | 958.748 | 76.24 | 27 | 1 | 958.721 | 0.151 | 958.797 | 66.69 |
| 21 | 1 | 959.044 | 0.153 | 958.993 | 77.55 | 27 | 1 | 959.244 | 0.162 | 959.388 | 67.42 |
| 21 | 1 | 958.986 | 0.139 | 958.944 | 78.99 | 27 | 1 | 959.383 | 0.140 | 959.497 | 68.21 |
| 21 | 2 | 960.946 | 0.209 | 961.110 | 31.40 | 27 | 1 | 959.731 | 0.144 | 959.877 | 69.02 |
| 21 | 2 | 959.287 | 0.200 | 959.444 | 30.29 | 27 | 1 | 960.005 | 0.171 | 960.072 | 69.93 |
| 21 | 2 | 960.385 | 0.186 | 960.537 | 29.25 | 27 | 1 | 958.930 | 0.148 | 958.947 | 70.81 |
| 21 | 2 | 957.871 | 0.169 | 957.997 | 28.24 | 27 | 1 | 959.247 | 0.176 | 959.315 | 72.56 |
| 21 | 2 | 957.920 | 0.164 | 958.053 | 27.29 | 27 | 1 | 959.321 | 0.148 | 959.345 | 73.54 |
| 21 | 2 | 958.181 | 0.158 | 958.314 | 26.34 | 27 | 1 | 959.409 | 0.168 | 959.496 | 74.64 |
| 21 | 2 | 959.240 | 0.161 | 959.378 | 25.50 | 27 | 1 | 959.085 | 0.228 | 959.386 | 75.73 |
| 21 | 2 | 959.556 | 0.198 | 959.712 | 24.67 | 27 | 2 | 959.944 | 0.236 | 960.130 | 24.29 |
| 21 | 2 | 958.589 | 0.168 | 958.728 | 23.75 | 27 | 2 | 959.499 | 0.203 | 959.656 | 23.34 |
| 21 | 2 | 959.436 | 0.154 | 959.575 | 23.00 | 27 | 2 | 958.619 | 0.190 | 958.776 | 22.49 |
| 21 | 2 | 959.146 | 0.134 | 959.254 | 22.27 | 27 | 2 | 958.791 | 0.161 | 958.923 | 21.69 |
| 21 | 2 | 959.102 | 0.153 | 959.229 | 21.60 | 27 | 2 | 958.496 | 0.199 | 958.665 | 20.92 |
| 22 | 1 | 959.059 | 0.128 | 959.201 | 69.22 | 27 | 2 | 958.680 | 0.188 | 958.830 | 20.15 |
| 22 | 1 | 958.108 | 0.158 | 958.364 | 70.12 | 27 | 2 | 958.680 | 0.228 | 958.856 | 18.76 |
| 22 | 1 | 959.352 | 0.148 | 959.574 | 71.10 | 27 | 2 | 958.978 | 0.185 | 959.136 | 18.13 |
| 22 | 1 | 959.628 | 0.166 | 959.669 | 72.13 | 27 | 2 | 958.769 | 0.171 | 958.902 | 17.51 |
| 22 | 1 | 959.365 | 0.127 | 959.300 | 73.12 | 27 | 2 | 959.223 | 0.176 | 959.356 | 16.93 |
| 22 | 1 | 959.869 | 0.123 | 959.788 | 74.15 | 27 | 2 | 958.015 | 0.178 | 958.148 | 16.35 |
| 22 | 1 | 958.507 | 0.140 | 958.477 | 75.31 | 28 | 2 | 959.399 | 0.199 | 959.546 | 26.06 |
| 22 | 1 | 959.555 | 0.150 | 959.553 | 76.50 | 28 | 2 | 959.082 | 0.164 | 959.230 | 25.06 |
| 22 | 1 | 959.622 | 0.168 | 959.690 | 77.80 | 28 | 2 | 958.817 | 0.206 | 959.003 | 24.13 |
| 22 | 1 | 959.975 | 0.159 | 960.103 | 79.08 | 28 | 2 | 958.449 | 0.178 | 958.597 | 23.29 |
| 22 | 1 | 959.137 | 0.149 | 959.333 | 80.49 | 28 | 2 | 959.693 | 0.198 | 959.872 | 22.46 |
| 22 | 1 | 959.589 | 0.195 | 959.773 | 82.13 | 28 | 2 | 959.687 | 0.177 | 959.816 | 21.66 |
| 22 | 2 | 959.913 | 0.159 | 960.033 | 20.95 | 28 | 2 | 958.359 | 0.170 | 958.500 | 20.92 |
| 22 | 2 | 958.914 | 0.147 | 959.020 | 20.23 | 28 | 2 | 959.063 | 0.161 | 959.204 | 20.17 |
| 22 | 2 | 958.879 | 0.131 | 958.980 | 19.53 | 28 | 2 | 958.425 | 0.223 | 958.584 | 19.47 |
| 22 | 2 | 958.995 | 0.159 | 959.116 | 18.85 | 28 | 2 | 959.706 | 0.217 | 959.870 | 18.72 |
| 22 | 2 | 958.995 | 0.150 | 959.122 | 18.26 | 28 | 2 | 959.151 | 0.173 | 959.311 | 17.89 |
| 22 | 2 | 958.851 | 0.166 | 958.958 | 17.61 | 28 | 2 | 959.823 | 0.166 | 959.977 | 17.23 |
| 22 | 2 | 959.063 | 0.179 | 959.182 | 17.03 | 28 | 2 | 959.161 | 0.182 | 959.309 | 16.62 |
| 22 | 2 | 959.051 | 0.171 | 959.176 | 16.45 | 28 | 2 | 958.691 | 0.197 | 958.845 | 16.04 |
| 23 | 2 | 959.340 | 0.172 | 959.475 | 30.65 | 28 | 2 | 957.917 | 0.188 | 958.071 | 15.46 |
| 23 | 2 | 958.725 | 0.168 | 958.848 | 29.56 | 29 | 1 | 959.369 | 0.159 | 959.519 | 67.29 |
| 23 | 2 | 959.801 | 0.173 | 959.911 | 28.52 | 29 | 1 | 958.631 | 0.142 | 958.695 | 68.11 |
| 23 | 2 | 958.753 | 0.186 | 958.875 | 27.55 | 29 | 1 | 959.314 | 0.191 | 959.551 | 68.96 |
| 23 | 2 | 958.510 | 0.195 | 958.619 | 26.61 | 29 | 1 | 959.425 | 0.173 | 959.638 | 69.78 |
| 23 | 2 | 958.856 | 0.140 | 958.946 | 25.68 | 29 | 1 | 958.800 | 0.213 | 959.080 | 70.66 |
| 23 | 2 | 958.663 | 0.151 | 958.764 | 24.81 | 29 | 1 | 959.951 | 0.145 | 959.954 | 71.62 |
| 23 | 2 | 958.329 | 0.182 | 958.474 | 23.97 | 29 | 1 | 960.194 | 0.152 | 960.218 | 72.58 |
| 23 | 2 | 959.028 | 0.188 | 959.172 | 23.18 | 29 | 1 | 958.568 | 0.131 | 958.568 | 73.58 |
| 23 | 2 | 958.712 | 0.189 | 958.826 | 22.28 | 29 | 1 | 959.063 | 0.132 | 959.076 | 74.59 |
| 23 | 2 | 959.805 | 0.187 | 959.950 | 21.58 | 29 | 1 | 959.519 | 0.131 | 959.548 | 75.67 |
| 23 | 2 | 958.321 | 0.169 | 958.441 | 20.84 | 29 | 1 | 958.722 | 0.139 | 958.777 | 76.90 |
| 23 | 2 | 958.586 | 0.192 | 958.707 | 20.19 | 29 | 1 | 959.713 | 0.192 | 959.942 | 78.14 |
| 26 | 1 | 958.332 | 0.140 | 958.428 | 65.20 | 29 | 1 | 959.523 | 0.184 | 959.750 | 79.41 |
| 26 | 1 | 958.675 | 0.139 | 958.793 | 65.88 | 29 | 1 | 959.411 | 0.160 | 959.600 | 80.77 |
| 26 | 1 | 959.975 | 0.138 | 960.099 | 66.71 | 29 | 1 | 959.362 | 0.204 | 959.656 | 82.19 |
| 26 | 1 | 958.896 | 0.153 | 959.022 | 67.44 | 29 | 2 | 959.373 | 0.167 | 959.497 | 24.67 |
| 26 | 1 | 959.712 | 0.132 | 959.801 | 68.35 | 29 | 2 | 959.159 | 0.178 | 959.283 | 23.74 |
| 26 | 1 | 957.812 | 0.259 | 957.733 | 69.36 | 29 | 2 | 958.930 | 0.206 | 959.134 | 22.88 |
| 26 | 1 | 959.319 | 0.165 | 959.488 | 70.23 | 29 | 2 | 958.584 | 0.169 | 958.707 | 22.06 |
| 26 | 1 | 959.486 | 0.153 | 959.656 | 71.19 | 29 | 2 | 958.885 | 0.160 | 958.996 | 21.26 |
| 26 | 1 | 959.152 | 0.173 | 959.295 | 72.19 | 29 | 2 | 959.752 | 0.167 | 959.862 | 20.45 |
| 26 | 1 | 960.066 | 0.150 | 960.175 | 73.22 | 29 | 2 | 958.859 | 0.163 | 958.981 | 19.71 |
| 26 | 1 | 960.205 | 0.159 | 960.385 | 74.26 | 29 | 2 | 959.101 | 0.197 | 959.222 | 18.95 |
| 26 | 1 | 959.574 | 0.182 | 959.798 | 75.38 | 29 | 2 | 959.340 | 0.154 | 959.456 | 18.12 |
| 26 | 2 | 959.339 | 0.146 | 959.450 | 27.04 | 29 | 2 | 959.108 | 0.167 | 959.230 | 17.44 |
| 26 | 2 | 958.775 | 0.126 | 958.868 | 25.32 | 29 | 2 | 959.625 | 0.183 | 959.766 | 16.79 |
| 26 | 2 | 959.022 | 0.137 | 959.108 | 24.42 | 29 | 2 | 959.169 | 0.155 | 959.272 | 16.21 |
| 26 | 2 | 958.751 | 0.130 | 958.862 | 23.50 | 29 | 2 | 958.639 | 0.157 | 958.756 | 15.58 |
| 26 | 2 | 959.116 | 0.163 | 959.233 | 22.67 | 29 | 2 | 959.596 | 0.141 | 959.707 | 14.99 |
| 26 | 2 | 959.028 | 0.165 | 959.157 | 21.78 | 29 | 2 | 959.049 | 0.158 | 959.165 | 13.92 |

| 2002 - AGOSTO | | | | |
|---|---|---|---|---|
| D | L | SDB | ER | SDC | HL |
| 30 | 2 | 959.277 | 0.143 | 959.368 | 24.22 |
| 30 | 2 | 958.693 | 0.194 | 958.834 | 23.23 |
| 30 | 2 | 959.451 | 0.128 | 959.549 | 22.32 |
| 30 | 2 | 959.040 | 0.124 | 959.133 | 21.36 |
| 30 | 2 | 958.715 | 0.100 | 958.758 | 20.58 |
| 30 | 2 | 959.441 | 0.119 | 959.527 | 19.78 |
| 30 | 2 | 958.629 | 0.120 | 958.683 | 19.02 |
| 30 | 2 | 958.881 | 0.139 | 958.991 | 18.23 |
| 30 | 2 | 960.220 | 0.123 | 960.318 | 17.40 |
| 30 | 2 | 958.992 | 0.140 | 959.096 | 16.70 |
| 30 | 2 | 959.001 | 0.115 | 959.087 | 16.08 |
| 30 | 2 | 959.061 | 0.145 | 959.191 | 15.48 |

| 2002 - SETEMBRO | | | | |
|---|---|---|---|---|
| D | L | SDB | ER | SDC | HL |
| 05 | 2 | 959.173 | 0.216 | 959.330 | 13.98 |
| 05 | 2 | 959.066 | 0.171 | 959.198 | 13.40 |
| 05 | 2 | 958.483 | 0.150 | 958.603 | 12.72 |
| 05 | 2 | 958.889 | 0.140 | 958.997 | 12.17 |
| 05 | 2 | 958.661 | 0.168 | 958.781 | 11.63 |
| 05 | 2 | 958.959 | 0.159 | 959.086 | 11.09 |
| 05 | 2 | 959.153 | 0.171 | 959.292 | 10.58 |
| 05 | 2 | 958.617 | 0.150 | 958.749 | 10.04 |
| 05 | 2 | 959.645 | 0.161 | 959.764 | 9.60 |
| 05 | 2 | 959.305 | 0.148 | 959.424 | 9.13 |
| 06 | 1 | 958.533 | 0.134 | 958.754 | 68.01 |
| 06 | 1 | 958.590 | 0.153 | 958.805 | 68.81 |
| 06 | 1 | 959.084 | 0.134 | 959.280 | 69.59 |
| 06 | 1 | 959.206 | 0.143 | 959.388 | 70.46 |
| 06 | 1 | 958.969 | 0.131 | 959.212 | 71.34 |
| 06 | 1 | 958.773 | 0.126 | 959.039 | 72.26 |
| 06 | 1 | 959.342 | 0.330 | 959.613 | 73.20 |
| 06 | 1 | 958.822 | 0.109 | 959.099 | 75.73 |
| 06 | 1 | 959.240 | 0.127 | 959.492 | 76.91 |
| 06 | 1 | 959.495 | 0.126 | 959.764 | 78.16 |
| 06 | 2 | 958.777 | 0.200 | 958.899 | 14.89 |
| 06 | 2 | 960.296 | 0.192 | 960.425 | 14.25 |
| 06 | 2 | 959.562 | 0.172 | 959.659 | 13.59 |
| 06 | 2 | 958.367 | 0.195 | 958.476 | 12.91 |
| 06 | 2 | 959.334 | 0.178 | 959.456 | 12.31 |
| 06 | 2 | 958.311 | 0.162 | 958.420 | 11.77 |
| 06 | 2 | 958.608 | 0.168 | 958.711 | 11.24 |
| 06 | 2 | 959.361 | 0.160 | 959.464 | 10.74 |
| 06 | 2 | 959.375 | 0.156 | 959.472 | 10.25 |
| 06 | 2 | 958.291 | 0.155 | 958.400 | 9.77 |
| 06 | 2 | 959.334 | 0.150 | 959.436 | 9.29 |
| 06 | 2 | 958.281 | 0.142 | 958.377 | 8.85 |
| 09 | 2 | 959.679 | 0.162 | 959.759 | 19.76 |
| 09 | 2 | 959.615 | 0.161 | 959.701 | 18.84 |
| 09 | 2 | 959.040 | 0.175 | 959.126 | 18.00 |
| 09 | 2 | 959.093 | 0.186 | 959.173 | 17.20 |
| 09 | 2 | 959.775 | 0.188 | 959.868 | 16.44 |
| 09 | 2 | 958.484 | 0.184 | 958.578 | 15.73 |
| 09 | 2 | 959.090 | 0.160 | 959.147 | 15.04 |
| 09 | 2 | 959.066 | 0.138 | 959.117 | 14.37 |
| 09 | 2 | 958.295 | 0.195 | 958.389 | 13.70 |
| 09 | 2 | 958.637 | 0.190 | 958.737 | 13.03 |
| 09 | 2 | 957.572 | 0.165 | 957.634 | 12.42 |
| 09 | 2 | 958.941 | 0.145 | 958.996 | 11.83 |
| 09 | 2 | 959.091 | 0.207 | 959.139 | 11.28 |
| 09 | 2 | 958.961 | 0.228 | 959.052 | 10.46 |
| 09 | 2 | 959.282 | 0.199 | 959.335 | 9.91 |
| 11 | 1 | 958.956 | 0.140 | 959.113 | 73.40 |
| 11 | 1 | 958.578 | 0.145 | 958.779 | 74.48 |
| 11 | 1 | 958.797 | 0.182 | 959.127 | 75.60 |
| 11 | 1 | 959.049 | 0.167 | 959.329 | 76.71 |
| 11 | 1 | 958.816 | 0.138 | 959.019 | 77.90 |
| 11 | 1 | 959.161 | 0.162 | 959.380 | 79.10 |
| 11 | 1 | 959.212 | 0.152 | 959.437 | 80.40 |
| 11 | 1 | 958.211 | 0.154 | 958.444 | 81.75 |

| 2002 - SETEMBRO | | | | |
|---|---|---|---|---|
| D | L | SDB | ER | SDC | HL |
| 11 | 1 | 959.185 | 0.149 | 959.460 | 83.28 |
| 11 | 1 | 958.928 | 0.147 | 959.177 | 84.98 |
| 11 | 1 | 959.569 | 0.169 | 959.807 | 86.88 |
| 11 | 2 | 959.630 | 0.175 | 959.722 | 14.59 |
| 11 | 2 | 959.348 | 0.114 | 959.369 | 13.85 |
| 11 | 2 | 958.839 | 0.140 | 958.892 | 13.20 |
| 11 | 2 | 959.534 | 0.141 | 959.593 | 12.56 |
| 11 | 2 | 958.065 | 0.160 | 958.130 | 11.94 |
| 11 | 2 | 958.242 | 0.166 | 958.319 | 11.37 |
| 11 | 2 | 958.354 | 0.184 | 958.449 | 10.80 |
| 11 | 2 | 958.920 | 0.159 | 958.952 | 10.23 |
| 11 | 2 | 958.890 | 0.140 | 958.934 | 9.68 |
| 11 | 2 | 959.040 | 0.180 | 959.114 | 9.12 |
| 11 | 2 | 958.983 | 0.156 | 959.064 | 8.53 |
| 11 | 2 | 958.704 | 0.160 | 958.754 | 8.07 |
| 11 | 2 | 958.736 | 0.155 | 958.804 | 7.62 |
| 11 | 2 | 959.139 | 0.144 | 959.189 | 7.16 |
| 11 | 2 | 959.207 | 0.192 | 959.282 | 6.68 |
| 12 | 2 | 959.188 | 0.118 | 959.202 | 13.40 |
| 12 | 2 | 959.453 | 0.182 | 959.525 | 12.77 |
| 12 | 2 | 958.472 | 0.141 | 958.525 | 12.11 |
| 12 | 2 | 958.919 | 0.163 | 958.997 | 11.47 |
| 12 | 2 | 959.232 | 0.153 | 959.297 | 10.89 |
| 12 | 2 | 958.924 | 0.150 | 958.980 | 10.23 |
| 12 | 2 | 959.553 | 0.157 | 959.628 | 9.39 |
| 12 | 2 | 959.700 | 0.176 | 959.782 | 8.85 |
| 12 | 2 | 959.546 | 0.153 | 959.603 | 8.32 |
| 12 | 2 | 959.924 | 0.186 | 960.006 | 7.86 |
| 12 | 2 | 959.253 | 0.204 | 959.348 | 7.41 |
| 12 | 2 | 959.028 | 0.140 | 959.080 | 6.97 |
| 13 | 1 | 959.114 | 0.145 | 959.244 | 72.64 |
| 13 | 1 | 959.124 | 0.126 | 959.197 | 73.57 |
| 13 | 1 | 959.186 | 0.132 | 959.353 | 74.56 |
| 16 | 1 | 958.878 | 0.159 | 959.137 | 69.61 |
| 16 | 1 | 958.683 | 0.157 | 958.957 | 71.28 |
| 16 | 1 | 958.971 | 0.208 | 959.255 | 72.51 |
| 16 | 1 | 959.293 | 0.136 | 959.587 | 73.48 |
| 16 | 1 | 959.354 | 0.150 | 959.676 | 74.50 |
| 16 | 1 | 958.976 | 0.169 | 959.354 | 75.58 |
| 16 | 1 | 959.984 | 0.184 | 960.382 | 76.67 |
| 16 | 1 | 959.396 | 0.202 | 959.824 | 78.20 |
| 16 | 1 | 958.815 | 0.204 | 959.213 | 79.78 |
| 17 | 1 | 959.434 | 0.133 | 959.620 | 65.24 |
| 17 | 1 | 958.396 | 0.123 | 958.580 | 65.86 |
| 17 | 1 | 957.877 | 0.148 | 958.111 | 66.49 |
| 17 | 1 | 959.273 | 0.180 | 959.522 | 67.19 |
| 17 | 1 | 958.960 | 0.133 | 959.070 | 68.96 |
| 17 | 1 | 958.096 | 0.140 | 958.285 | 69.73 |
| 17 | 1 | 958.823 | 0.141 | 958.985 | 70.55 |
| 17 | 1 | 958.929 | 0.146 | 959.168 | 71.40 |
| 17 | 1 | 959.503 | 0.142 | 959.682 | 72.30 |
| 17 | 1 | 959.172 | 0.188 | 959.450 | 73.24 |
| 17 | 1 | 959.222 | 0.145 | 959.479 | 74.25 |
| 17 | 1 | 959.499 | 0.184 | 959.852 | 75.36 |
| 17 | 1 | 959.257 | 0.156 | 959.553 | 76.47 |
| 17 | 1 | 959.733 | 0.189 | 960.061 | 77.71 |
| 17 | 1 | 959.414 | 0.149 | 959.640 | 78.97 |
| 17 | 2 | 959.180 | 0.163 | 959.251 | 13.51 |
| 17 | 2 | 958.896 | 0.164 | 958.967 | 12.72 |
| 17 | 2 | 959.044 | 0.170 | 959.115 | 11.99 |
| 17 | 2 | 959.795 | 0.162 | 959.848 | 11.34 |
| 17 | 2 | 958.957 | 0.131 | 958.972 | 10.72 |
| 17 | 2 | 959.124 | 0.143 | 959.178 | 10.09 |
| 17 | 2 | 958.487 | 0.192 | 958.553 | 9.52 |
| 17 | 2 | 958.703 | 0.179 | 958.757 | 8.87 |
| 17 | 2 | 958.886 | 0.175 | 958.971 | 8.31 |
| 17 | 2 | 957.973 | 0.169 | 958.044 | 7.67 |
| 17 | 2 | 959.146 | 0.178 | 959.236 | 7.14 |
| 17 | 2 | 959.237 | 0.141 | 959.275 | 6.49 |
| 17 | 2 | 958.316 | 0.170 | 958.379 | 6.02 |
| 17 | 2 | 959.094 | 0.158 | 959.144 | 5.58 |

```
          2002 - SETEMBRO                              2002 - SETEMBRO
 D  L    SDB     ER     SDC    HL          D  L    SDB     ER     SDC    HL
17  2  958.571  0.166  958.627   4.99      27  1  959.386  0.171  959.701  66.40
17  2  959.554  0.192  959.616   4.60      27  1  958.352  0.141  958.675  67.15
18  1  959.555  0.216  959.818  65.65      27  1  959.045  0.136  959.372  67.89
18  1  959.175  0.148  959.386  66.60      27  1  958.678  0.173  959.061  69.62
18  1  959.616  0.409  959.934  67.24      27  1  959.184  0.201  959.623  70.79
18  1  959.632  0.169  959.763  68.14      27  1  959.674  0.149  960.044  71.85
18  1  959.851  0.131  959.975  68.86      27  1  959.331  0.187  959.707  72.94
18  1  959.931  0.164  960.162  69.62      27  1  959.099  0.193  959.470  74.04
18  1  959.204  0.129  959.402  70.40      27  1  959.074  0.170  959.434  75.22
18  1  959.742  0.151  960.007  71.21      27  1  958.813  0.156  959.053  76.52
18  1  959.251  0.161  959.509  72.06      27  1  958.663  0.134  959.016  77.86
18  1  959.883  0.136  960.034  73.06      27  2  959.533  0.185  959.604   3.32
18  1  959.559  0.143  959.814  74.03      27  2  959.333  0.178  959.430   2.85
18  1  960.458  0.122  960.614  75.14      27  2  959.739  0.221  959.861   2.48
18  1  959.685  0.137  959.900  76.34      27  2  959.701  0.188  959.810   2.12
18  1  958.533  0.154  958.779  77.54      27  2  958.793  0.199  958.896   1.78
18  1  960.306  0.155  960.552  78.88      27  2  958.363  0.179  958.461   1.47
19  1  958.998  0.127  959.285  68.89      27  2  959.184  0.170  959.185   1.17
19  1  958.702  0.130  959.007  69.66      27  2  959.273  0.164  959.347   0.88
19  1  959.251  0.124  959.469  70.77      27  2  959.247  0.144  959.315   0.61
19  1  959.514  0.116  959.703  71.68      27  2  958.925  0.179  959.018   0.33
19  1  959.253  0.136  959.542  72.60      27  2  958.567  0.171  958.647   0.08
19  1  959.619  0.123  959.883  73.63      27  2  959.235  0.169  959.315   0.00
19  1  959.259  0.152  959.568  74.66      27  2  958.233  0.164  958.325   0.00
19  1  958.323  0.134  959.654  76.22      30  1  958.501  0.110  958.850  61.54
19  1  959.194  0.143  959.583  77.47      30  1  957.917  0.128  958.240  62.13
19  2  958.752  0.119  958.747   6.53      30  1  959.175  0.125  959.481  62.77
19  2  958.706  0.149  958.738   6.06      30  1  958.840  0.122  959.195  63.40
19  2  959.551  0.135  959.551   5.63      30  1  958.207  0.156  958.628  64.03
19  2  958.275  0.144  958.250   5.14      30  1  957.016  0.136  957.402  64.68
19  2  958.518  0.158  958.556   4.70      30  1  960.208  0.152  960.665  65.39
19  2  959.297  0.157  959.353   4.24      30  1  959.532  0.150  959.973  66.19
19  2  959.089  0.187  959.176   3.87      30  1  959.078  0.126  959.445  67.00
19  2  959.342  0.137  959.360   3.51      30  1  959.156  0.179  959.603  67.82
19  2  958.336  0.137  958.360   3.10      30  1  957.084  0.134  957.329  68.72
19  2  958.515  0.124  958.540   2.77      30  1  959.506  0.138  959.862  69.77
19  2  959.089  0.126  959.114   2.45      30  1  958.997  0.160  959.391  70.81
20  1  959.526  0.107  959.658  66.30      30  1  958.760  0.160  959.174  71.89
20  1  959.417  0.102  959.534  67.00      30  1  959.521  0.180  959.983  73.00
20  1  959.446  0.122  959.662  67.67      30  2  958.920  0.122  958.922   5.16
20  1  960.065  0.113  960.277  68.46      30  2  959.357  0.217  959.378   4.71
20  1  959.857  0.086  959.963  69.28      30  2  959.147  0.123  959.162   4.19
20  1  959.355  0.123  959.603  70.14      30  2  960.017  0.104  960.010   3.79
20  1  959.192  0.112  959.412  71.04      30  2  958.771  0.144  958.789   3.41
20  1  959.025  0.115  959.320  71.92      30  2  958.636  0.133  958.672   3.05
20  1  960.266  0.105  960.536  72.96      30  2  958.506  0.162  958.533   2.34
20  1  959.179  0.134  959.515  73.98      30  2  959.347  0.155  959.380   1.78
20  1  958.944  0.143  959.328  75.08
26  1  958.013  0.197  958.342  74.73
26  1  958.614  0.174  958.951  75.95              2002 - OUTUBRO
26  1  959.633  0.188  959.957  77.28      D  L    SDB     ER     SDC    HL
26  1  957.834  0.166  958.156  80.24      02  2  958.531  0.154  958.604   1.47
26  1  959.516  0.280  959.896  81.86      02  2  959.634  0.149  959.676   1.17
26  1  957.201  0.389  957.638  83.70      02  2  959.055  0.164  959.115   0.88
26  2  958.815  0.211  958.916   5.36      02  2  959.406  0.199  959.478   0.57
26  2  959.170  0.180  959.240   4.83      02  2  959.394  0.184  959.480   0.31
26  2  958.899  0.202  959.000   4.39      02  2  957.620  0.133  957.650   0.00
26  2  959.161  0.163  959.237   3.95      02  2  958.074  0.170  958.142   0.00
26  2  958.717  0.141  958.769   3.54      02  2  959.864  0.164  959.932   0.00
26  2  958.572  0.173  958.662   3.16      02  2  959.925  0.163  959.999   0.00
26  2  959.145  0.131  959.210   2.71      02  2  959.244  0.154  959.306   0.00
26  2  958.847  0.135  958.918   2.35      02  2  959.292  0.152  959.340   0.00
26  2  959.072  0.153  959.138   1.94      02  2  959.435  0.179  959.507   0.00
26  2  959.165  0.140  959.231   1.62      02  2  960.429  0.192  960.482   0.00
26  2  958.851  0.153  958.930   1.31      03  1  958.916  0.168  959.230  61.07
26  2  959.296  0.192  959.400   0.99      03  1  958.271  0.184  958.623  61.63
26  2  958.730  0.127  958.765   0.71      03  1  959.570  0.138  959.899  62.23
26  2  958.663  0.195  958.761   0.43      03  1  959.064  0.161  959.423  62.83
27  1  958.204  0.146  958.461  64.27      03  1  959.202  0.157  959.588  63.48
27  1  958.269  0.166  958.580  65.06      03  1  958.728  0.136  959.076  64.15
27  1  958.892  0.146  959.164  65.72      03  1  959.989  0.172  960.428  64.92
```

| 2002 - OUTUBRO | | | | | | 2002 - OUTUBRO | | | | |
|---|---|---|---|---|---|---|---|---|---|---|
| D | L | SDB | ER | SDC | HL | D | L | SDB | ER | SDC | HL |
| 03 | 1 | 958.820 | 0.167 | 959.293 | 65.83 | 08 | 1 | 960.099 | 0.130 | 960.529 | 59.29 |
| 03 | 1 | 959.737 | 0.154 | 960.171 | 66.70 | 08 | 1 | 959.337 | 0.144 | 959.761 | 59.89 |
| 03 | 1 | 960.017 | 0.204 | 960.487 | 67.55 | 08 | 1 | 959.242 | 0.132 | 959.611 | 60.54 |
| 03 | 1 | 959.039 | 0.168 | 959.422 | 68.45 | 08 | 1 | 959.373 | 0.155 | 959.804 | 61.22 |
| 03 | 1 | 958.635 | 0.188 | 959.076 | 69.44 | 08 | 1 | 959.665 | 0.199 | 960.110 | 61.95 |
| 03 | 1 | 958.963 | 0.198 | 959.406 | 70.51 | 08 | 1 | 959.183 | 0.162 | 959.623 | 62.74 |
| 03 | 1 | 958.630 | 0.193 | 959.028 | 71.63 | 08 | 1 | 959.122 | 0.166 | 959.600 | 63.55 |
| 03 | 2 | 959.790 | 0.186 | 959.873 | 2.78 | 08 | 1 | 959.922 | 0.223 | 960.417 | 64.42 |
| 03 | 2 | 959.095 | 0.163 | 959.159 | 2.42 | 08 | 1 | 959.352 | 0.173 | 959.807 | 65.46 |
| 03 | 2 | 959.272 | 0.153 | 959.327 | 2.01 | 08 | 2 | 960.006 | 0.180 | 960.054 | 3.41 |
| 03 | 2 | 959.118 | 0.164 | 959.185 | 1.68 | 08 | 2 | 959.468 | 0.160 | 959.497 | 3.02 |
| 03 | 2 | 959.514 | 0.192 | 959.586 | 1.37 | 08 | 2 | 959.455 | 0.140 | 959.471 | 2.68 |
| 03 | 2 | 958.443 | 0.168 | 958.516 | 1.09 | 08 | 2 | 957.941 | 0.184 | 957.976 | 2.35 |
| 03 | 2 | 959.020 | 0.166 | 959.097 | 0.79 | 08 | 2 | 959.843 | 0.186 | 959.884 | 2.02 |
| 03 | 2 | 958.859 | 0.152 | 958.898 | 0.51 | 08 | 2 | 958.704 | 0.167 | 958.733 | 1.66 |
| 03 | 2 | 958.816 | 0.151 | 958.882 | 0.17 | 08 | 2 | 958.924 | 0.144 | 958.940 | 1.36 |
| 03 | 2 | 958.920 | 0.138 | 958.960 | 0.00 | 08 | 2 | 959.430 | 0.163 | 959.471 | 1.07 |
| 03 | 2 | 958.752 | 0.132 | 958.793 | 0.00 | 08 | 2 | 958.347 | 0.166 | 958.369 | 0.78 |
| 03 | 2 | 958.895 | 0.157 | 958.968 | 0.00 | 08 | 2 | 958.708 | 0.200 | 958.743 | 0.52 |
| 03 | 2 | 958.819 | 0.165 | 958.905 | 0.00 | 08 | 2 | 958.834 | 0.161 | 958.850 | 0.25 |
| 03 | 2 | 959.290 | 0.136 | 959.344 | 0.00 | 08 | 2 | 958.128 | 0.175 | 958.175 | 0.00 |
| 04 | 1 | 959.171 | 0.143 | 959.589 | 59.26 | 09 | 2 | 960.526 | 0.213 | 960.607 | 1.91 |
| 04 | 1 | 959.356 | 0.173 | 959.789 | 59.78 | 09 | 2 | 960.295 | 0.209 | 960.363 | 1.59 |
| 04 | 1 | 959.729 | 0.174 | 960.235 | 60.32 | 09 | 2 | 959.545 | 0.200 | 959.595 | 1.27 |
| 04 | 1 | 959.266 | 0.146 | 959.570 | 60.90 | 09 | 2 | 959.489 | 0.187 | 959.558 | 0.99 |
| 04 | 1 | 960.359 | 0.123 | 960.665 | 61.49 | 09 | 2 | 959.018 | 0.184 | 959.080 | 0.68 |
| 04 | 1 | 960.194 | 0.171 | 960.585 | 62.12 | 09 | 2 | 958.201 | 0.213 | 958.269 | 0.43 |
| 04 | 1 | 959.795 | 0.148 | 960.167 | 62.78 | 09 | 2 | 960.057 | 0.203 | 960.131 | 0.19 |
| 04 | 1 | 959.395 | 0.163 | 959.815 | 63.45 | 09 | 2 | 959.013 | 0.211 | 959.075 | 0.00 |
| 04 | 1 | 958.779 | 0.158 | 959.154 | 64.17 | 09 | 2 | 959.359 | 0.211 | 959.446 | 0.00 |
| 04 | 1 | 959.260 | 0.203 | 959.709 | 65.10 | 09 | 2 | 959.482 | 0.244 | 959.568 | 0.00 |
| 04 | 1 | 959.358 | 0.160 | 959.758 | 65.94 | 10 | 1 | 959.585 | 0.128 | 960.060 | 52.60 |
| 04 | 1 | 959.361 | 0.175 | 959.790 | 66.81 | 10 | 1 | 959.564 | 0.147 | 959.720 | 52.91 |
| 04 | 1 | 958.892 | 0.178 | 959.318 | 67.75 | 10 | 1 | 959.771 | 0.130 | 959.891 | 53.25 |
| 04 | 1 | 959.554 | 0.249 | 960.050 | 68.74 | 10 | 1 | 959.183 | 0.114 | 959.305 | 53.59 |
| 04 | 2 | 959.910 | 0.160 | 959.965 | 1.49 | 10 | 1 | 959.996 | 0.136 | 960.237 | 53.94 |
| 04 | 2 | 958.663 | 0.155 | 958.718 | 1.17 | 10 | 1 | 959.626 | 0.125 | 959.875 | 54.32 |
| 04 | 2 | 959.705 | 0.185 | 959.790 | 0.88 | 10 | 1 | 959.283 | 0.152 | 959.590 | 54.78 |
| 04 | 2 | 958.794 | 0.194 | 958.885 | 0.58 | 10 | 1 | 959.089 | 0.138 | 959.447 | 55.31 |
| 07 | 1 | 958.265 | 0.150 | 958.489 | 55.24 | 10 | 1 | 959.686 | 0.149 | 960.078 | 55.76 |
| 07 | 1 | 958.830 | 0.125 | 959.034 | 55.62 | 10 | 1 | 959.534 | 0.120 | 959.853 | 56.24 |
| 07 | 1 | 958.221 | 0.126 | 958.465 | 56.03 | 10 | 1 | 959.478 | 0.112 | 959.834 | 56.73 |
| 07 | 1 | 958.924 | 0.151 | 959.239 | 56.45 | 10 | 1 | 959.658 | 0.150 | 960.075 | 57.28 |
| 07 | 1 | 959.129 | 0.138 | 959.412 | 56.88 | 10 | 2 | 960.025 | 0.185 | 960.075 | 0.12 |
| 07 | 1 | 959.035 | 0.132 | 959.369 | 57.32 | 10 | 2 | 958.702 | 0.183 | 958.752 | 0.00 |
| 07 | 1 | 959.448 | 0.169 | 959.333 | 57.82 | 10 | 2 | 959.489 | 0.189 | 959.517 | 0.00 |
| 07 | 1 | 959.251 | 0.167 | 959.697 | 58.30 | 10 | 2 | 958.778 | 0.168 | 958.840 | 0.00 |
| 07 | 1 | 958.490 | 0.183 | 958.923 | 58.86 | 10 | 2 | 958.028 | 0.182 | 958.097 | 0.00 |
| 07 | 1 | 958.505 | 0.162 | 958.923 | 59.43 | 10 | 2 | 959.835 | 0.164 | 959.859 | 0.00 |
| 07 | 1 | 959.046 | 0.138 | 959.427 | 60.00 | 10 | 2 | 958.741 | 0.139 | 958.766 | 0.00 |
| 07 | 1 | 959.027 | 0.196 | 959.520 | 60.61 | 10 | 2 | 959.650 | 0.185 | 959.719 | 0.00 |
| 07 | 1 | 959.313 | 0.160 | 959.718 | 61.36 | 10 | 2 | 959.399 | 0.160 | 959.437 | 0.00 |
| 07 | 1 | 958.572 | 0.212 | 959.066 | 62.03 | 10 | 2 | 958.628 | 0.197 | 958.660 | 0.00 |
| 07 | 1 | 958.539 | 0.198 | 959.026 | 62.76 | 10 | 2 | 958.658 | 0.184 | 958.721 | 0.00 |
| 07 | 2 | 958.118 | 0.172 | 958.169 | 1.42 | 10 | 2 | 959.409 | 0.175 | 959.465 | 0.00 |
| 07 | 2 | 959.136 | 0.206 | 959.217 | 1.11 | 11 | 1 | 958.112 | 0.156 | 958.478 | 54.83 |
| 07 | 2 | 957.957 | 0.199 | 957.994 | 0.83 | 11 | 1 | 959.169 | 0.158 | 959.549 | 55.38 |
| 07 | 2 | 959.527 | 0.197 | 959.589 | 0.56 | 11 | 1 | 957.765 | 0.181 | 958.172 | 55.87 |
| 07 | 2 | 959.895 | 0.167 | 959.931 | 0.30 | 11 | 1 | 959.614 | 0.164 | 960.017 | 56.35 |
| 07 | 2 | 958.204 | 0.198 | 958.271 | 0.07 | 11 | 1 | 959.487 | 0.135 | 959.868 | 56.86 |
| 07 | 2 | 958.330 | 0.176 | 958.372 | 0.00 | 11 | 1 | 958.686 | 0.162 | 959.125 | 57.42 |
| 07 | 2 | 957.782 | 0.203 | 957.836 | 0.00 | 11 | 1 | 958.695 | 0.272 | 959.277 | 58.12 |
| 07 | 2 | 959.346 | 0.185 | 959.412 | 0.00 | 11 | 1 | 959.610 | 0.200 | 960.083 | 59.06 |
| 07 | 2 | 957.287 | 0.232 | 957.353 | 0.00 | 11 | 1 | 958.680 | 0.164 | 959.119 | 59.95 |
| 07 | 2 | 959.410 | 0.224 | 959.458 | 0.00 | 11 | 1 | 958.197 | 0.247 | 958.676 | 61.49 |
| 08 | 1 | 959.097 | 0.135 | 959.444 | 56.29 | 11 | 1 | 959.355 | 0.205 | 959.811 | 62.46 |
| 08 | 1 | 959.175 | 0.159 | 959.620 | 56.75 | 11 | 1 | 957.838 | 0.311 | 958.345 | 63.51 |
| 08 | 1 | 958.980 | 0.141 | 959.411 | 57.22 | 11 | 1 | 958.783 | 0.352 | 959.347 | 64.65 |
| 08 | 1 | 959.830 | 0.137 | 960.248 | 57.70 | 11 | 1 | 958.733 | 0.212 | 959.297 | 66.78 |
| 08 | 1 | 958.828 | 0.115 | 959.132 | 58.21 | 11 | 1 | 959.390 | 0.227 | 959.978 | 67.96 |
| 08 | 1 | 959.303 | 0.138 | 959.662 | 58.73 | 11 | 2 | 960.205 | 0.156 | 960.246 | 0.72 |

| 2002 - OUTUBRO | | | | | | 2002 - OUTUBRO | | | | |
|---|---|---|---|---|---|---|---|---|---|---|
| D | L | SDB | ER | SDC | HL | D | L | SDB | ER | SDC | HL |
| 11 | 2 | 959.115 | 0.130 | 959.125 | 0.44 | 16 | 2 | 958.467 | 0.162 | 958.521 | 0.00 |
| 11 | 2 | 958.725 | 0.140 | 958.740 | 0.19 | 16 | 2 | 959.649 | 0.146 | 959.678 | 0.00 |
| 11 | 2 | 958.545 | 0.129 | 958.541 | 0.00 | 17 | 1 | 958.362 | 0.162 | 958.804 | 51.03 |
| 11 | 2 | 958.860 | 0.146 | 958.893 | 0.00 | 17 | 1 | 959.111 | 0.137 | 959.541 | 51.45 |
| 11 | 2 | 958.828 | 0.137 | 958.842 | 0.00 | 17 | 1 | 958.068 | 0.130 | 958.488 | 51.90 |
| 11 | 2 | 959.427 | 0.134 | 959.460 | 0.00 | 17 | 1 | 959.287 | 0.179 | 959.739 | 52.34 |
| 11 | 2 | 959.013 | 0.142 | 959.039 | 0.00 | 17 | 1 | 958.664 | 0.148 | 959.108 | 52.83 |
| 11 | 2 | 959.341 | 0.141 | 959.374 | 0.00 | 17 | 1 | 959.159 | 0.146 | 959.545 | 53.31 |
| 11 | 2 | 958.876 | 0.165 | 958.883 | 0.00 | 17 | 1 | 959.305 | 0.185 | 959.749 | 53.83 |
| 11 | 2 | 959.569 | 0.129 | 959.584 | 0.00 | 17 | 1 | 958.221 | 0.182 | 958.689 | 54.42 |
| 11 | 2 | 959.022 | 0.122 | 959.025 | 0.00 | 17 | 1 | 958.394 | 0.161 | 958.855 | 55.04 |
| 11 | 2 | 959.230 | 0.128 | 959.246 | 0.00 | 17 | 1 | 958.263 | 0.169 | 958.744 | 55.66 |
| 14 | 1 | 958.785 | 0.153 | 959.196 | 53.88 | 17 | 1 | 957.832 | 0.230 | 958.299 | 56.35 |
| 14 | 1 | 958.632 | 0.151 | 959.077 | 54.35 | 17 | 1 | 958.884 | 0.279 | 959.382 | 57.19 |
| 14 | 1 | 958.412 | 0.170 | 958.865 | 54.86 | 17 | 2 | 959.812 | 0.138 | 959.856 | 0.51 |
| 14 | 1 | 959.161 | 0.153 | 959.579 | 55.37 | 17 | 2 | 959.146 | 0.151 | 959.203 | 0.29 |
| 14 | 1 | 959.271 | 0.200 | 959.734 | 55.97 | 17 | 2 | 958.400 | 0.159 | 958.463 | 0.09 |
| 14 | 1 | 959.191 | 0.163 | 959.657 | 56.57 | 17 | 2 | 959.306 | 0.146 | 959.363 | 0.00 |
| 14 | 1 | 958.887 | 0.224 | 959.371 | 57.20 | 17 | 2 | 959.385 | 0.174 | 959.455 | 0.00 |
| 14 | 1 | 959.825 | 0.152 | 960.260 | 57.89 | 17 | 2 | 958.831 | 0.152 | 958.895 | 0.00 |
| 14 | 1 | 959.115 | 0.213 | 959.589 | 58.65 | 17 | 2 | 958.955 | 0.162 | 959.031 | 0.00 |
| 14 | 1 | 960.501 | 0.183 | 960.933 | 59.53 | 17 | 2 | 959.379 | 0.169 | 959.442 | 0.00 |
| 14 | 1 | 959.225 | 0.242 | 959.700 | 60.33 | 17 | 2 | 958.417 | 0.166 | 958.486 | 0.00 |
| 14 | 2 | 960.209 | 0.215 | 960.284 | 4.76 | 17 | 2 | 959.885 | 0.198 | 959.972 | 0.00 |
| 14 | 2 | 959.800 | 0.178 | 959.875 | 4.30 | 17 | 2 | 958.274 | 0.241 | 958.380 | 0.00 |
| 14 | 2 | 958.276 | 0.166 | 958.313 | 3.84 | 17 | 2 | 957.933 | 0.217 | 958.033 | 0.00 |
| 14 | 2 | 957.886 | 0.169 | 957.942 | 3.04 | 21 | 1 | 959.561 | 0.301 | 960.010 | 46.27 |
| 14 | 2 | 959.775 | 0.173 | 959.818 | 2.64 | 21 | 1 | 959.149 | 0.187 | 959.548 | 46.56 |
| 14 | 2 | 960.011 | 0.161 | 960.060 | 1.74 | 21 | 1 | 957.664 | 0.161 | 958.116 | 46.82 |
| 14 | 2 | 959.129 | 0.151 | 959.147 | 1.42 | 21 | 1 | 959.435 | 0.150 | 959.834 | 47.10 |
| 14 | 2 | 959.541 | 0.196 | 959.603 | 1.13 | 21 | 1 | 958.569 | 0.158 | 959.011 | 47.41 |
| 14 | 2 | 958.465 | 0.218 | 958.508 | 0.86 | 21 | 1 | 959.453 | 0.145 | 959.914 | 47.73 |
| 14 | 2 | 957.995 | 0.214 | 958.051 | 0.56 | 21 | 1 | 960.354 | 0.153 | 960.820 | 48.06 |
| 14 | 2 | 959.666 | 0.193 | 959.709 | 0.31 | 21 | 1 | 958.971 | 0.159 | 959.427 | 48.41 |
| 14 | 2 | 960.003 | 0.170 | 960.065 | 0.01 | 21 | 1 | 959.655 | 0.174 | 960.132 | 48.78 |
| 14 | 2 | 959.161 | 0.183 | 959.223 | 0.00 | 21 | 1 | 959.910 | 0.156 | 960.383 | 49.19 |
| 14 | 2 | 959.274 | 0.206 | 959.323 | 0.00 | 21 | 1 | 958.876 | 0.184 | 959.355 | 49.61 |
| 15 | 1 | 957.168 | 0.266 | 957.676 | 49.85 | 21 | 1 | 959.244 | 0.187 | 959.740 | 50.05 |
| 15 | 1 | 958.606 | 0.174 | 958.981 | 50.26 | 21 | 1 | 958.624 | 0.178 | 959.104 | 50.57 |
| 15 | 1 | 958.317 | 0.135 | 958.710 | 50.57 | 21 | 1 | 958.820 | 0.175 | 959.316 | 51.07 |
| 15 | 1 | 958.832 | 0.152 | 959.219 | 50.88 | 21 | 1 | 959.368 | 0.165 | 959.770 | 51.59 |
| 15 | 1 | 958.954 | 0.183 | 959.367 | 51.21 | 21 | 1 | 959.241 | 0.263 | 959.722 | 52.17 |
| 15 | 1 | 958.862 | 0.187 | 959.307 | 51.57 | 21 | 2 | 960.894 | 0.174 | 960.947 | 1.51 |
| 15 | 1 | 958.749 | 0.186 | 959.258 | 51.93 | 21 | 2 | 958.869 | 0.166 | 958.910 | 1.24 |
| 15 | 1 | 959.273 | 0.159 | 959.716 | 52.35 | 21 | 2 | 958.576 | 0.143 | 958.591 | 0.98 |
| 15 | 1 | 959.948 | 0.186 | 960.408 | 52.81 | 21 | 2 | 958.875 | 0.136 | 958.897 | 0.75 |
| 15 | 1 | 959.022 | 0.188 | 959.499 | 53.33 | 21 | 2 | 959.177 | 0.162 | 959.224 | 0.50 |
| 15 | 1 | 958.972 | 0.197 | 959.500 | 53.83 | 21 | 2 | 958.988 | 0.149 | 959.022 | 0.30 |
| 15 | 1 | 959.789 | 0.183 | 960.290 | 54.32 | 21 | 2 | 957.993 | 0.152 | 958.014 | 0.09 |
| 15 | 1 | 958.353 | 0.228 | 958.909 | 54.88 | 21 | 2 | 958.768 | 0.287 | 958.852 | 0.00 |
| 15 | 1 | 959.998 | 0.231 | 960.513 | 55.60 | 21 | 2 | 959.252 | 0.160 | 959.255 | 0.00 |
| 15 | 1 | 959.475 | 0.198 | 959.979 | 56.22 | 21 | 2 | 958.963 | 0.158 | 958.997 | 0.00 |
| 15 | 1 | 959.011 | 0.209 | 959.519 | 56.92 | 21 | 2 | 958.523 | 0.200 | 958.550 | 0.00 |
| 16 | 1 | 958.492 | 0.187 | 958.914 | 48.98 | 24 | 2 | 958.953 | 0.175 | 959.022 | 0.00 |
| 16 | 1 | 959.111 | 0.137 | 959.605 | 49.26 | 24 | 2 | 959.067 | 0.195 | 959.123 | 0.00 |
| 16 | 1 | 958.937 | 0.125 | 959.401 | 49.62 | 24 | 2 | 958.337 | 0.196 | 958.393 | 0.00 |
| 16 | 1 | 959.271 | 0.213 | 959.835 | 49.98 | 24 | 2 | 959.038 | 0.192 | 959.125 | 0.00 |
| 16 | 1 | 958.639 | 0.331 | 959.182 | 51.86 | 24 | 2 | 957.972 | 0.168 | 958.015 | 0.00 |
| 16 | 1 | 959.205 | 0.154 | 959.720 | 52.39 | 24 | 2 | 958.843 | 0.177 | 958.912 | 0.00 |
| 16 | 1 | 959.484 | 0.166 | 960.027 | 52.86 | 24 | 2 | 959.388 | 0.152 | 959.445 | 0.00 |
| 16 | 2 | 959.649 | 0.154 | 959.703 | 1.71 | 24 | 2 | 958.397 | 0.171 | 958.466 | 0.00 |
| 16 | 2 | 959.215 | 0.160 | 959.281 | 1.38 | 24 | 2 | 958.836 | 0.153 | 958.867 | 0.00 |
| 16 | 2 | 958.351 | 0.155 | 958.380 | 1.10 | 24 | 2 | 959.415 | 0.156 | 959.471 | 0.00 |
| 16 | 2 | 959.372 | 0.153 | 959.401 | 0.84 | 24 | 2 | 959.135 | 0.151 | 959.191 | 0.00 |
| 16 | 2 | 959.070 | 0.119 | 959.074 | 0.60 | 25 | 1 | 958.823 | 0.120 | 959.126 | 44.78 |
| 16 | 2 | 959.651 | 0.153 | 959.705 | 0.36 | 25 | 1 | 959.951 | 0.144 | 960.284 | 45.08 |
| 16 | 2 | 958.704 | 0.137 | 958.733 | 0.00 | 25 | 1 | 958.599 | 0.121 | 958.918 | 45.60 |
| 16 | 2 | 958.692 | 0.143 | 958.734 | 0.00 | 25 | 1 | 959.382 | 0.131 | 959.736 | 45.98 |
| 16 | 2 | 959.394 | 0.159 | 959.448 | 0.00 | 25 | 1 | 958.975 | 0.159 | 959.401 | 46.35 |
| 16 | 2 | 958.316 | 0.164 | 958.357 | 0.00 | 25 | 1 | 958.916 | 0.147 | 959.320 | 46.75 |
| 16 | 2 | 958.648 | 0.164 | 958.696 | 0.00 | 25 | 1 | 959.923 | 0.168 | 960.426 | 47.17 |

| 2002 - OUTUBRO | | | | |
|---|---|---|---|---|
| D  L | SDB | ER | SDC | HL |
| 25  1 | 959.019 | 0.140 | 959.481 | 48.14 |
| 25  1 | 958.544 | 0.205 | 959.088 | 48.75 |
| 25  1 | 959.358 | 0.181 | 959.878 | 49.43 |
| 25  1 | 958.783 | 0.167 | 959.272 | 50.09 |
| 25  2 | 959.692 | 0.142 | 959.730 | 0.00 |
| 25  2 | 958.478 | 0.162 | 958.542 | 0.00 |
| 25  2 | 958.339 | 0.163 | 958.409 | 0.00 |
| 25  2 | 959.236 | 0.170 | 959.312 | 0.00 |
| 25  2 | 958.693 | 0.160 | 958.750 | 0.00 |
| 25  2 | 959.121 | 0.130 | 959.140 | 0.00 |
| 25  2 | 959.345 | 0.151 | 959.401 | 0.00 |
| 25  2 | 959.272 | 0.159 | 959.328 | 0.00 |
| 25  2 | 959.257 | 0.186 | 959.312 | 0.00 |
| 25  2 | 959.218 | 0.176 | 959.297 | 0.00 |
| 28  1 | 957.857 | 0.183 | 958.247 | 40.76 |
| 28  1 | 960.086 | 0.189 | 960.530 | 41.12 |
| 28  1 | 960.610 | 0.158 | 961.003 | 41.31 |
| 28  1 | 959.628 | 0.138 | 959.972 | 41.56 |
| 28  1 | 959.892 | 0.151 | 960.307 | 41.76 |
| 28  1 | 958.792 | 0.153 | 959.134 | 42.06 |
| 28  1 | 959.719 | 0.145 | 960.041 | 42.30 |
| 28  1 | 959.745 | 0.121 | 960.052 | 42.56 |
| 28  1 | 959.202 | 0.157 | 959.625 | 42.82 |
| 28  1 | 959.301 | 0.150 | 959.697 | 43.13 |
| 28  1 | 959.003 | 0.156 | 959.440 | 43.44 |
| 28  1 | 958.686 | 0.131 | 959.053 | 43.78 |
| 28  1 | 958.976 | 0.161 | 959.415 | 44.15 |
| 28  1 | 958.735 | 0.165 | 959.231 | 44.65 |
| 28  2 | 959.597 | 0.134 | 959.631 | 1.19 |
| 28  2 | 959.046 | 0.139 | 959.093 | 0.96 |
| 28  2 | 959.709 | 0.161 | 959.781 | 0.76 |
| 28  2 | 959.119 | 0.158 | 959.185 | 0.37 |
| 28  2 | 958.737 | 0.150 | 958.797 | 0.20 |
| 28  2 | 958.532 | 0.149 | 958.566 | 0.04 |
| 28  2 | 959.166 | 0.130 | 959.188 | 0.00 |
| 28  2 | 958.783 | 0.148 | 958.836 | 0.00 |
| 28  2 | 959.123 | 0.140 | 959.157 | 0.00 |
| 28  2 | 959.807 | 0.153 | 959.871 | 0.00 |
| 28  2 | 958.772 | 0.148 | 959.830 | 0.00 |
| 28  2 | 959.162 | 0.189 | 959.238 | 0.00 |
| 28  2 | 959.673 | 0.159 | 959.724 | 0.00 |
| 28  2 | 958.716 | 0.176 | 958.761 | 0.00 |
| 28  2 | 958.238 | 0.161 | 958.302 | 0.00 |
| 29  1 | 958.541 | 0.165 | 958.926 | 40.58 |
| 29  1 | 959.839 | 0.158 | 960.260 | 40.76 |
| 29  1 | 959.732 | 0.154 | 960.146 | 40.98 |
| 29  1 | 959.055 | 0.216 | 959.507 | 41.28 |
| 29  1 | 959.366 | 0.161 | 959.803 | 41.57 |
| 29  1 | 959.750 | 0.173 | 960.291 | 42.22 |
| 29  1 | 958.934 | 0.154 | 959.397 | 43.50 |
| 29  1 | 958.849 | 0.135 | 959.260 | 43.94 |
| 29  1 | 959.023 | 0.158 | 959.470 | 44.40 |
| 29  1 | 959.086 | 0.155 | 959.543 | 44.87 |
| 29  1 | 958.706 | 0.206 | 959.210 | 45.39 |
| 29  1 | 958.827 | 0.167 | 959.245 | 45.93 |
| 29  1 | 958.447 | 0.217 | 958.915 | 46.48 |
| 29  1 | 958.490 | 0.208 | 959.008 | 47.13 |
| 29  2 | 960.328 | 0.144 | 960.360 | 1.73 |
| 29  2 | 959.176 | 0.133 | 959.193 | 0.00 |
| 29  2 | 958.559 | 0.140 | 958.594 | 0.00 |
| 29  2 | 959.691 | 0.113 | 959.661 | 0.00 |
| 29  2 | 959.244 | 0.135 | 959.258 | 0.00 |

| 2002 - NOVEMBRO | | | | |
|---|---|---|---|---|
| D  L | SDB | ER | SDC | HL |
| 14  2 | 958.078 | 0.195 | 958.154 | 0.84 |
| 14  2 | 959.275 | 0.157 | 959.388 | 0.90 |
| 14  2 | 959.457 | 0.144 | 959.537 | 1.02 |
| 14  2 | 958.991 | 0.172 | 959.082 | 1.09 |
| 14  2 | 959.223 | 0.142 | 959.309 | 1.43 |

| 2002 - NOVEMBRO | | | | |
|---|---|---|---|---|
| D  L | SDB | ER | SDC | HL |
| 14  2 | 959.249 | 0.153 | 959.348 | 1.52 |
| 14  2 | 958.382 | 0.180 | 958.491 | 1.61 |
| 18  1 | 958.929 | 0.183 | 958.829 | 24.85 |
| 18  1 | 959.169 | 0.269 | 959.091 | 24.78 |
| 18  1 | 959.293 | 0.154 | 959.133 | 24.69 |
| 18  1 | 959.831 | 0.144 | 959.644 | 24.64 |
| 18  1 | 959.801 | 0.144 | 959.676 | 24.59 |
| 18  1 | 958.470 | 0.160 | 958.356 | 24.54 |
| 18  1 | 959.132 | 0.165 | 959.023 | 24.50 |
| 18  1 | 959.138 | 0.153 | 958.999 | 24.47 |
| 18  1 | 959.934 | 0.157 | 959.813 | 24.41 |
| 18  1 | 959.793 | 0.187 | 959.657 | 24.40 |
| 18  1 | 959.772 | 0.218 | 959.642 | 24.39 |
| 18  1 | 958.623 | 0.197 | 958.559 | 24.38 |
| 18  1 | 959.873 | 0.201 | 959.746 | 24.39 |
| 18  1 | 958.757 | 0.205 | 958.628 | 24.40 |
| 18  2 | 959.200 | 0.162 | 959.366 | 0.97 |
| 18  2 | 959.108 | 0.157 | 959.274 | 1.00 |
| 18  2 | 959.468 | 0.156 | 959.633 | 1.04 |
| 18  2 | 958.738 | 0.202 | 958.914 | 1.08 |
| 18  2 | 959.726 | 0.154 | 959.886 | 1.13 |
| 18  2 | 959.030 | 0.186 | 959.201 | 1.18 |
| 18  2 | 959.024 | 0.161 | 959.185 | 1.24 |
| 18  2 | 959.377 | 0.141 | 959.535 | 1.30 |
| 18  2 | 959.132 | 0.176 | 959.292 | 1.36 |
| 18  2 | 958.890 | 0.123 | 959.022 | 1.50 |
| 18  2 | 958.385 | 0.132 | 958.505 | 1.58 |
| 18  2 | 958.478 | 0.147 | 958.592 | 1.65 |
| 18  2 | 959.805 | 0.141 | 959.914 | 1.82 |
| 21  2 | 958.778 | 0.143 | 958.909 | 1.37 |
| 21  2 | 959.109 | 0.148 | 959.226 | 1.42 |
| 21  2 | 958.880 | 0.132 | 958.981 | 1.47 |
| 21  2 | 958.872 | 0.110 | 958.954 | 1.53 |
| 21  2 | 959.160 | 0.126 | 959.230 | 1.60 |
| 21  2 | 959.203 | 0.137 | 959.276 | 1.66 |
| 21  2 | 959.028 | 0.149 | 959.108 | 1.73 |
| 21  2 | 958.926 | 0.172 | 959.030 | 1.82 |
| 21  2 | 958.545 | 0.144 | 958.641 | 1.90 |
| 21  2 | 959.123 | 0.162 | 959.252 | 2.07 |
| 21  2 | 959.261 | 0.129 | 959.374 | 2.18 |
| 25  1 | 958.923 | 0.159 | 958.729 | 19.29 |
| 25  1 | 959.358 | 0.173 | 959.138 | 19.16 |
| 25  1 | 958.810 | 0.121 | 958.575 | 19.07 |
| 25  1 | 958.920 | 0.136 | 958.733 | 18.99 |
| 25  1 | 959.982 | 0.142 | 959.732 | 18.90 |
| 25  1 | 959.242 | 0.159 | 959.077 | 18.82 |
| 25  1 | 959.697 | 0.122 | 959.549 | 18.75 |
| 25  1 | 959.114 | 0.153 | 958.915 | 18.67 |
| 25  1 | 959.446 | 0.132 | 959.240 | 18.60 |
| 25  1 | 959.237 | 0.150 | 959.002 | 18.54 |
| 25  1 | 959.211 | 0.133 | 959.018 | 18.44 |
| 25  1 | 959.203 | 0.126 | 959.038 | 18.39 |
| 25  1 | 959.822 | 0.153 | 959.638 | 18.34 |
| 25  1 | 959.204 | 0.142 | 959.024 | 18.30 |
| 25  1 | 959.827 | 0.150 | 959.685 | 18.27 |
| 25  1 | 959.424 | 0.176 | 959.239 | 18.24 |
| 25  1 | 959.842 | 0.152 | 959.672 | 18.22 |
| 25  2 | 960.223 | 0.155 | 960.387 | 2.10 |
| 25  2 | 959.337 | 0.142 | 959.489 | 2.17 |
| 25  2 | 958.577 | 0.137 | 958.738 | 2.26 |
| 25  2 | 958.665 | 0.131 | 958.823 | 2.36 |
| 25  2 | 959.916 | 0.211 | 960.096 | 2.44 |
| 25  2 | 959.112 | 0.110 | 959.243 | 2.70 |
| 25  2 | 958.613 | 0.140 | 958.766 | 2.80 |
| 25  2 | 959.022 | 0.120 | 959.177 | 2.90 |
| 25  2 | 959.133 | 0.146 | 959.305 | 3.00 |
| 25  2 | 959.373 | 0.132 | 959.549 | 3.13 |
| 25  2 | 958.239 | 0.118 | 958.424 | 3.24 |
| 25  2 | 959.308 | 0.178 | 959.508 | 3.38 |
| 27  2 | 958.699 | 0.127 | 958.844 | 2.25 |
| 27  2 | 958.647 | 0.178 | 958.790 | 2.33 |

| 2002 - NOVEMBRO | | | | | |
|---|---|---|---|---|---|
| D | L | SDB | ER | SDC | HL |
| 27 | 2 | 958.451 | 0.141 | 958.578 | 2.41 |
| 27 | 2 | 959.529 | 0.144 | 959.650 | 2.51 |
| 27 | 2 | 958.798 | 0.111 | 958.906 | 2.59 |
| 27 | 2 | 959.175 | 0.098 | 959.264 | 2.97 |
| 27 | 2 | 959.469 | 0.118 | 959.545 | 3.08 |
| 27 | 2 | 958.522 | 0.120 | 958.624 | 3.18 |
| 27 | 2 | 958.934 | 0.104 | 959.010 | 3.44 |
| 27 | 2 | 958.375 | 0.131 | 958.459 | 3.57 |
| 27 | 2 | 959.609 | 0.140 | 959.716 | 3.69 |
| 27 | 2 | 959.129 | 0.143 | 959.275 | 3.85 |
| 28 | 1 | 958.609 | 0.101 | 958.256 | 18.04 |
| 28 | 1 | 960.395 | 0.100 | 960.063 | 17.90 |
| 28 | 1 | 959.152 | 0.112 | 958.872 | 17.78 |
| 28 | 1 | 959.636 | 0.090 | 959.467 | 17.65 |
| 28 | 1 | 959.787 | 0.128 | 959.628 | 17.53 |
| 28 | 1 | 959.652 | 0.126 | 959.506 | 17.41 |
| 28 | 1 | 958.525 | 0.156 | 958.402 | 17.29 |
| 28 | 1 | 959.688 | 0.131 | 959.450 | 17.15 |
| 28 | 1 | 960.114 | 0.118 | 959.863 | 17.04 |
| 28 | 1 | 959.481 | 0.120 | 959.218 | 16.93 |
| 28 | 1 | 958.690 | 0.133 | 958.440 | 16.82 |
| 28 | 1 | 959.076 | 0.141 | 958.842 | 16.71 |
| 29 | 2 | 958.776 | 0.145 | 958.913 | 3.47 |
| 29 | 2 | 959.541 | 0.159 | 959.676 | 3.61 |
| 29 | 2 | 959.361 | 0.114 | 959.457 | 3.74 |
| 29 | 2 | 958.887 | 0.147 | 959.014 | 3.86 |
| 29 | 2 | 959.000 | 0.175 | 959.137 | 3.98 |
| 29 | 2 | 958.006 | 0.169 | 958.152 | 4.11 |
| 29 | 2 | 958.449 | 0.142 | 958.593 | 4.23 |
| 29 | 2 | 958.654 | 0.151 | 958.822 | 4.37 |
| 29 | 2 | 958.117 | 0.135 | 958.273 | 4.50 |
| 29 | 2 | 958.388 | 0.161 | 958.526 | 4.65 |
| 29 | 2 | 958.941 | 0.132 | 959.050 | 4.79 |
| 29 | 2 | 959.029 | 0.159 | 959.150 | 4.93 |

| 2002 - DEZEMBRO | | | | | |
|---|---|---|---|---|---|
| D | L | SDB | ER | SDC | HL |
| 06 | 1 | 960.238 | 0.095 | 960.217 | 13.39 |
| 06 | 1 | 960.270 | 0.082 | 960.179 | 13.22 |
| 06 | 1 | 959.446 | 0.111 | 959.335 | 13.05 |
| 06 | 1 | 959.591 | 0.117 | 959.492 | 12.88 |
| 06 | 1 | 959.012 | 0.109 | 958.912 | 12.71 |
| 06 | 1 | 958.067 | 0.132 | 958.031 | 12.53 |
| 06 | 1 | 959.002 | 0.120 | 959.007 | 12.36 |
| 06 | 1 | 958.386 | 0.132 | 958.415 | 12.17 |
| 06 | 1 | 959.322 | 0.130 | 959.344 | 12.00 |
| 06 | 1 | 959.162 | 0.121 | 959.178 | 11.83 |
| 06 | 1 | 959.431 | 0.123 | 959.404 | 11.67 |
| 06 | 1 | 959.482 | 0.119 | 959.418 | 11.51 |
| 06 | 1 | 959.081 | 0.122 | 959.028 | 11.36 |
| 06 | 1 | 959.346 | 0.113 | 959.312 | 11.21 |
| 06 | 2 | 958.533 | 0.140 | 958.788 | 4.55 |
| 06 | 2 | 960.375 | 0.120 | 960.637 | 4.69 |
| 06 | 2 | 958.938 | 0.121 | 959.192 | 4.84 |
| 06 | 2 | 958.385 | 0.126 | 958.649 | 4.99 |
| 06 | 2 | 959.007 | 0.142 | 959.276 | 5.14 |
| 06 | 2 | 959.145 | 0.136 | 959.420 | 5.30 |
| 06 | 2 | 958.926 | 0.144 | 959.223 | 5.46 |
| 06 | 2 | 958.752 | 0.118 | 958.998 | 5.76 |
| 06 | 2 | 959.104 | 0.112 | 959.301 | 5.93 |
| 06 | 2 | 958.815 | 0.123 | 959.031 | 6.10 |
| 06 | 2 | 958.874 | 0.151 | 959.095 | 6.31 |
| 06 | 2 | 959.372 | 0.147 | 959.587 | 6.47 |
| 06 | 2 | 958.702 | 0.146 | 958.928 | 6.62 |
| 06 | 2 | 958.674 | 0.120 | 958.912 | 6.78 |
| 17 | 1 | 958.834 | 0.106 | 958.684 | 5.64 |
| 17 | 1 | 959.318 | 0.102 | 959.251 | 5.44 |
| 17 | 1 | 959.269 | 0.104 | 959.218 | 5.24 |
| 17 | 1 | 959.387 | 0.104 | 959.350 | 5.06 |
| 17 | 1 | 959.842 | 0.094 | 959.846 | 4.88 |

| 2002 - DEZEMBRO | | | | | |
|---|---|---|---|---|---|
| D | L | SDB | ER | SDC | HL |
| 17 | 1 | 958.996 | 0.084 | 959.053 | 4.69 |
| 17 | 1 | 959.388 | 0.095 | 959.428 | 4.51 |
| 17 | 1 | 958.656 | 0.104 | 958.711 | 4.33 |
| 17 | 1 | 959.482 | 0.113 | 959.544 | 4.14 |
| 17 | 1 | 959.020 | 0.126 | 959.075 | 3.94 |
| 17 | 1 | 959.192 | 0.131 | 959.254 | 3.74 |
| 17 | 1 | 959.751 | 0.116 | 959.665 | 3.50 |
| 17 | 1 | 958.544 | 0.118 | 958.496 | 3.31 |
| 17 | 1 | 959.913 | 0.122 | 959.888 | 3.11 |
| 17 | 1 | 959.340 | 0.094 | 959.371 | 2.92 |
| 17 | 1 | 958.352 | 0.143 | 958.365 | 2.74 |
| 17 | 1 | 959.140 | 0.117 | 959.161 | 2.56 |
| 17 | 1 | 959.148 | 0.144 | 959.176 | 2.38 |
| 17 | 1 | 958.204 | 0.145 | 958.256 | 2.20 |
| 17 | 1 | 958.961 | 0.127 | 959.004 | 2.01 |
| 17 | 2 | 960.352 | 0.109 | 960.572 | 9.54 |
| 17 | 2 | 959.903 | 0.114 | 960.126 | 9.74 |
| 17 | 2 | 958.720 | 0.120 | 958.976 | 9.93 |
| 17 | 2 | 958.825 | 0.113 | 959.032 | 10.50 |
| 17 | 2 | 959.254 | 0.178 | 959.400 | 10.68 |
| 17 | 2 | 958.050 | 0.138 | 958.217 | 10.88 |
| 17 | 2 | 958.220 | 0.123 | 958.363 | 11.10 |
| 17 | 2 | 958.486 | 0.132 | 958.665 | 11.29 |
| 17 | 2 | 958.995 | 0.139 | 959.170 | 11.49 |
| 17 | 2 | 958.930 | 0.156 | 959.151 | 11.68 |
| 17 | 2 | 959.409 | 0.133 | 959.586 | 11.90 |
| 17 | 2 | 959.846 | 0.159 | 960.049 | 12.10 |
| 17 | 2 | 959.085 | 0.134 | 959.322 | 12.32 |
| 17 | 2 | 959.531 | 0.164 | 959.799 | 12.62 |
| 17 | 2 | 957.917 | 0.126 | 958.085 | 12.84 |
| 17 | 2 | 958.471 | 0.128 | 958.688 | 13.05 |
| 18 | 1 | 958.643 | 0.118 | 958.660 | 4.94 |
| 18 | 1 | 959.242 | 0.121 | 959.252 | 4.76 |
| 18 | 1 | 959.286 | 0.098 | 959.371 | 4.56 |
| 18 | 1 | 960.065 | 0.120 | 960.199 | 4.38 |
| 18 | 1 | 959.450 | 0.138 | 959.647 | 4.20 |
| 18 | 1 | 960.181 | 0.127 | 960.327 | 4.00 |
| 18 | 1 | 958.616 | 0.144 | 958.760 | 3.82 |
| 18 | 1 | 959.613 | 0.120 | 959.828 | 3.64 |
| 18 | 1 | 959.345 | 0.153 | 959.312 | 3.41 |
| 18 | 1 | 959.349 | 0.114 | 959.325 | 3.23 |
| 18 | 1 | 959.480 | 0.108 | 959.489 | 3.05 |
| 18 | 1 | 959.747 | 0.128 | 959.794 | 2.86 |
| 18 | 1 | 959.040 | 0.139 | 959.112 | 2.66 |
| 18 | 1 | 958.958 | 0.127 | 959.139 | 2.22 |
| 18 | 1 | 959.059 | 0.157 | 959.258 | 2.03 |
| 18 | 1 | 960.351 | 0.132 | 960.582 | 1.85 |
| 18 | 1 | 959.294 | 0.116 | 959.485 | 1.65 |
| 18 | 2 | 959.319 | 0.143 | 959.646 | 9.36 |
| 18 | 2 | 959.054 | 0.133 | 959.374 | 9.54 |
| 18 | 2 | 959.301 | 0.113 | 959.603 | 9.74 |
| 18 | 2 | 958.930 | 0.167 | 959.331 | 9.93 |
| 18 | 2 | 958.983 | 0.112 | 959.280 | 10.13 |
| 18 | 2 | 958.719 | 0.151 | 959.023 | 10.32 |
| 18 | 2 | 958.983 | 0.118 | 959.263 | 10.52 |
| 18 | 2 | 959.094 | 0.136 | 959.374 | 10.70 |
| 18 | 2 | 959.350 | 0.151 | 959.628 | 10.90 |
| 18 | 2 | 958.187 | 0.124 | 958.450 | 11.09 |
| 18 | 2 | 958.987 | 0.129 | 959.274 | 11.27 |
| 18 | 2 | 958.681 | 0.138 | 958.962 | 11.46 |
| 18 | 2 | 959.483 | 0.117 | 959.751 | 11.65 |
| 18 | 2 | 958.335 | 0.117 | 958.629 | 11.83 |
| 18 | 2 | 959.217 | 0.151 | 959.507 | 12.02 |
| 18 | 2 | 958.606 | 0.147 | 958.924 | 12.20 |
| 18 | 2 | 958.919 | 0.125 | 959.224 | 12.38 |
| 18 | 2 | 958.708 | 0.152 | 959.023 | 12.57 |
| 19 | 1 | 959.575 | 0.127 | 959.623 | 4.41 |
| 19 | 1 | 960.230 | 0.110 | 960.349 | 4.23 |
| 19 | 1 | 959.414 | 0.126 | 959.581 | 4.04 |
| 19 | 1 | 959.822 | 0.129 | 960.010 | 3.86 |
| 19 | 1 | 958.506 | 0.151 | 958.720 | 3.66 |

```
         2002 - DEZEMBRO                              2002 - DEZEMBRO
 D  L    SDB     ER      SDC     HL         D  L    SDB     ER      SDC     HL
19  1  959.537  0.158  959.587   3.44      23  2  958.901  0.152  959.174  13.96
19  1  958.995  0.135  959.051   3.24      23  2  958.618  0.155  958.891  14.16
19  1  959.907  0.130  959.968   3.06      23  2  959.574  0.170  959.876  14.35
19  1  959.844  0.157  959.925   2.85      23  2  959.075  0.166  959.357  14.55
19  1  959.296  0.114  959.485   2.64      23  2  958.420  0.168  958.738  14.73
19  1  958.056  0.136  958.230   2.42      23  2  958.871  0.166  959.181  14.92
19  1  959.742  0.147  959.947   2.21      23  2  960.063  0.181  960.384  15.11
19  1  959.143  0.141  959.369   1.95      23  2  960.458  0.162  960.788  15.32
19  2  958.078  0.166  958.396   9.97      27  2  959.064  0.240  959.349  14.52
19  2  959.768  0.139  960.079  10.16      27  2  958.590  0.176  958.878  14.72
19  2  958.339  0.146  958.638  10.35      27  2  959.200  0.131  959.454  14.92
19  2  958.677  0.149  958.970  10.54      27  2  958.634  0.176  958.903  15.09
19  2  959.289  0.146  959.584  10.72      27  2  958.799  0.133  959.041  15.28
19  2  958.790  0.148  959.069  10.90      27  2  958.641  0.155  958.892  15.47
19  2  958.663  0.151  958.926  11.08      27  2  958.162  0.147  958.406  15.67
19  2  957.844  0.138  958.094  11.26      27  2  959.066  0.152  959.302  15.86
19  2  958.332  0.143  958.579  11.45      27  2  958.833  0.139  959.068  16.05
19  2  959.245  0.125  959.458  11.65      27  2  958.303  0.147  958.515  16.26
19  2  958.390  0.147  958.613  11.83      27  2  958.481  0.121  958.691  16.45
19  2  958.993  0.142  959.210  12.03      27  2  959.211  0.131  959.441  16.63
19  2  959.092  0.145  959.330  12.21      27  2  958.407  0.174  958.649  16.82
19  2  958.582  0.146  958.811  12.40      27  2  959.198  0.167  959.451  17.00
20  1  959.569  0.094  959.662   2.64      27  2  958.638  0.133  958.883  17.19
20  1  959.686  0.098  959.846   2.46      30  1  959.321  0.141  959.341   0.69
20  1  960.073  0.097  960.259   2.27      30  1  959.159  0.159  959.005   0.89
20  1  959.251  0.102  959.445   2.06      30  1  960.570  0.128  960.530   1.07
20  1  959.704  0.105  959.840   1.87      30  1  959.595  0.122  959.584   1.27
20  1  959.333  0.102  959.460   1.69      30  1  959.623  0.126  959.636   1.48
20  1  959.537  0.107  959.662   1.50      30  1  959.001  0.124  959.040   1.65
20  1  959.438  0.115  959.583   1.32      30  1  959.507  0.142  959.580   1.85
20  1  958.885  0.117  959.092   1.12      30  1  959.768  0.129  959.861   2.06
20  1  958.431  0.110  959.657   0.89      30  1  958.525  0.114  958.691   2.22
20  1  959.820  0.106  960.010   0.70      30  1  959.460  0.166  959.357   2.48
20  1  959.471  0.123  959.659   0.48      30  1  960.112  0.132  960.058   2.65
20  2  958.956  0.126  959.307  11.17      30  1  959.970  0.124  959.994   2.82
20  2  958.297  0.121  958.638  11.35      30  1  959.371  0.143  959.440   3.01
20  2  959.175  0.117  959.506  11.53      30  1  959.542  0.120  959.642   3.18
20  2  959.230  0.100  959.542  11.72      30  1  959.270  0.136  959.386   3.35
20  2  958.125  0.123  958.473  11.90      30  1  959.087  0.099  959.268   3.54
20  2  958.420  0.121  958.778  12.10      30  2  958.966  0.146  959.323  15.88
20  2  959.018  0.115  959.372  12.32      30  2  959.714  0.182  960.078  16.05
20  2  958.097  0.134  958.480  12.53      30  2  957.810  0.129  958.161  16.36
20  2  959.071  0.128  959.465  12.73      30  2  958.526  0.116  958.870  16.72
20  2  959.032  0.145  959.453  12.96      30  2  959.389  0.147  959.736  17.01
20  2  958.728  0.162  959.150  13.15      30  2  959.504  0.115  959.836  18.11
20  2  958.237  0.195  958.683  13.33
23  1  960.014  0.133  959.907   2.43              2003 - JANEIRO
23  1  959.364  0.127  959.288   2.24       D  L    SDB     ER      SDC     HL
23  1  959.586  0.119  959.495   2.05      02  1  959.051  0.139  959.168   2.20
23  1  959.191  0.132  959.125   1.87      02  1  959.186  0.119  959.319   2.41
23  1  960.256  0.134  960.211   1.68      02  1  959.294  0.120  959.475   2.61
23  1  959.506  0.128  959.486   1.50      02  1  958.924  0.114  959.143   2.77
23  1  959.031  0.125  959.108   1.32      02  1  959.249  0.124  959.455   2.95
23  1  959.274  0.137  959.411   1.14      02  1  959.311  0.100  959.493   3.12
23  1  958.937  0.132  959.051   0.95      02  1  958.752  0.106  958.940   3.32
23  1  959.595  0.114  959.663   0.76      02  1  958.934  0.118  959.137   3.49
23  1  959.426  0.142  959.624   0.58      02  1  958.350  0.104  958.590   3.66
23  1  959.288  0.141  959.179   0.34      02  1  959.200  0.101  959.124   3.84
23  1  959.542  0.125  959.495   0.16      02  1  959.251  0.119  959.478   4.03
23  1  959.261  0.132  959.247   0.02      02  1  959.468  0.111  959.727   4.23
23  1  960.153  0.155  960.223   0.21      02  1  959.998  0.128  960.245   4.43
23  1  959.463  0.138  959.512   0.40      02  2  959.294  0.169  959.592  17.22
23  2  959.653  0.137  959.973  12.26      02  2  958.981  0.155  959.275  17.51
23  2  958.704  0.113  959.008  12.46      02  2  958.325  0.146  958.615  17.68
23  2  959.798  0.146  959.110  12.67      02  2  958.965  0.143  959.261  17.86
23  2  959.084  0.126  959.383  12.86      02  2  959.929  0.122  960.218  18.05
23  2  959.424  0.120  959.722  13.05      02  2  958.806  0.142  959.108  18.23
23  2  959.371  0.146  959.681  13.23      02  2  959.266  0.135  959.556  18.42
23  2  959.186  0.134  959.453  13.42      02  2  958.681  0.137  958.974  18.59
23  2  959.289  0.136  959.546  13.60      02  2  959.366  0.152  959.670  18.76
23  2  959.156  0.172  959.423  13.78
```

| 2003 - JANEIRO | | | | | | 2003 - JANEIRO | | | | |
|---|---|---|---|---|---|---|---|---|---|---|
| D | L | SDB | ER | SDC | HL | D | L | SDB | ER | SDC | HL |
| 02 | 2 | 958.714 | 0.163 | 959.026 | 18.95 | 10 | 1 | 959.778 | 0.119 | 959.711 | 6.14 |
| 08 | 1 | 958.640 | 0.133 | 958.661 | 3.89 | 10 | 2 | 959.913 | 0.151 | 960.164 | 23.67 |
| 08 | 1 | 958.610 | 0.122 | 958.672 | 4.05 | 10 | 2 | 959.493 | 0.164 | 959.748 | 23.80 |
| 08 | 1 | 958.382 | 0.140 | 958.470 | 4.20 | 10 | 2 | 959.649 | 0.151 | 959.891 | 23.94 |
| 08 | 1 | 959.292 | 0.156 | 959.389 | 4.35 | 10 | 2 | 959.724 | 0.149 | 959.962 | 24.07 |
| 08 | 1 | 958.709 | 0.126 | 958.833 | 4.50 | 10 | 2 | 958.158 | 0.159 | 958.359 | 24.83 |
| 08 | 1 | 960.128 | 0.114 | 960.270 | 4.65 | 10 | 2 | 959.120 | 0.143 | 959.324 | 24.99 |
| 08 | 1 | 959.352 | 0.124 | 959.498 | 4.79 | 10 | 2 | 957.970 | 0.132 | 958.195 | 25.13 |
| 08 | 1 | 959.477 | 0.126 | 959.652 | 4.94 | 10 | 2 | 958.432 | 0.145 | 958.673 | 25.29 |
| 08 | 1 | 958.685 | 0.148 | 958.656 | 5.10 | 10 | 2 | 958.360 | 0.139 | 958.600 | 25.43 |
| 08 | 1 | 959.352 | 0.122 | 959.362 | 5.24 | 10 | 2 | 959.210 | 0.159 | 959.457 | 25.60 |
| 08 | 1 | 958.805 | 0.154 | 958.848 | 5.38 | 10 | 2 | 958.382 | 0.157 | 958.631 | 25.74 |
| 08 | 1 | 958.852 | 0.143 | 958.924 | 5.57 | 10 | 2 | 958.512 | 0.139 | 958.741 | 25.90 |
| 08 | 1 | 959.631 | 0.155 | 959.731 | 5.71 | 10 | 2 | 959.167 | 0.148 | 959.392 | 26.04 |
| 08 | 1 | 958.029 | 0.135 | 958.153 | 5.84 | 16 | 2 | 960.101 | 0.223 | 960.248 | 27.71 |
| 08 | 1 | 959.042 | 0.144 | 959.185 | 6.01 | 16 | 2 | 958.678 | 0.181 | 958.811 | 27.78 |
| 08 | 1 | 959.762 | 0.166 | 959.919 | 6.14 | 16 | 2 | 959.132 | 0.172 | 959.279 | 27.85 |
| 08 | 2 | 958.898 | 0.175 | 959.117 | 22.24 | 16 | 2 | 959.039 | 0.189 | 959.188 | 27.92 |
| 08 | 2 | 959.437 | 0.152 | 959.649 | 22.37 | 16 | 2 | 959.416 | 0.159 | 959.557 | 28.00 |
| 08 | 2 | 959.265 | 0.133 | 959.460 | 22.50 | 16 | 2 | 958.321 | 0.198 | 958.469 | 28.08 |
| 08 | 2 | 959.784 | 0.157 | 959.995 | 22.63 | 16 | 2 | 959.583 | 0.143 | 959.694 | 28.17 |
| 08 | 2 | 959.229 | 0.161 | 959.438 | 22.77 | 16 | 2 | 959.234 | 0.142 | 959.343 | 28.27 |
| 08 | 2 | 959.321 | 0.119 | 959.504 | 22.91 | 16 | 2 | 959.581 | 0.163 | 959.688 | 28.35 |
| 08 | 2 | 959.003 | 0.126 | 959.192 | 23.05 | 16 | 2 | 958.870 | 0.145 | 958.967 | 28.45 |
| 08 | 2 | 959.348 | 0.125 | 959.524 | 23.19 | 16 | 2 | 958.312 | 0.144 | 958.437 | 28.54 |
| 08 | 2 | 959.205 | 0.136 | 959.384 | 23.33 | 16 | 2 | 958.325 | 0.149 | 958.473 | 28.64 |
| 08 | 2 | 958.354 | 0.153 | 958.540 | 23.50 | 17 | 1 | 958.932 | 0.109 | 959.002 | 6.74 |
| 08 | 2 | 959.529 | 0.152 | 959.715 | 23.70 | 17 | 1 | 959.375 | 0.118 | 959.378 | 6.86 |
| 08 | 2 | 958.802 | 0.155 | 958.990 | 23.84 | 17 | 1 | 958.500 | 0.130 | 958.513 | 6.97 |
| 08 | 2 | 958.080 | 0.155 | 958.261 | 23.98 | 17 | 1 | 959.622 | 0.149 | 959.569 | 7.08 |
| 08 | 2 | 959.521 | 0.174 | 959.718 | 24.13 | 17 | 1 | 958.785 | 0.144 | 958.815 | 7.19 |
| 08 | 2 | 959.382 | 0.134 | 958.577 | 24.28 | 17 | 1 | 959.181 | 0.134 | 959.141 | 7.30 |
| 09 | 1 | 959.594 | 0.110 | 959.670 | 4.30 | 17 | 1 | 959.776 | 0.133 | 959.688 | 7.40 |
| 09 | 1 | 959.902 | 0.102 | 959.874 | 4.46 | 17 | 1 | 959.674 | 0.140 | 959.623 | 7.51 |
| 09 | 1 | 959.365 | 0.124 | 959.204 | 4.62 | 17 | 1 | 959.849 | 0.115 | 959.789 | 7.60 |
| 09 | 1 | 959.530 | 0.130 | 959.411 | 4.76 | 17 | 1 | 959.310 | 0.130 | 959.292 | 7.68 |
| 09 | 1 | 960.217 | 0.128 | 960.097 | 4.91 | 17 | 1 | 959.493 | 0.130 | 959.530 | 7.78 |
| 09 | 1 | 959.932 | 0.145 | 959.917 | 5.06 | 17 | 1 | 959.378 | 0.165 | 959.468 | 7.86 |
| 09 | 1 | 959.491 | 0.128 | 959.543 | 5.25 | 21 | 1 | 958.822 | 0.106 | 958.856 | 8.10 |
| 09 | 1 | 959.247 | 0.137 | 959.259 | 5.39 | 21 | 1 | 958.978 | 0.099 | 959.025 | 8.15 |
| 09 | 1 | 959.279 | 0.109 | 959.299 | 5.53 | 21 | 1 | 958.901 | 0.104 | 958.979 | 8.20 |
| 09 | 1 | 959.316 | 0.114 | 959.333 | 5.69 | 21 | 1 | 959.687 | 0.094 | 959.748 | 8.24 |
| 09 | 1 | 959.723 | 0.130 | 959.724 | 5.82 | 21 | 1 | 959.376 | 0.104 | 959.454 | 8.28 |
| 09 | 1 | 959.666 | 0.133 | 959.670 | 5.97 | 21 | 1 | 959.178 | 0.118 | 959.266 | 8.31 |
| 09 | 1 | 959.205 | 0.135 | 959.237 | 6.11 | 21 | 1 | 959.067 | 0.113 | 959.128 | 8.33 |
| 09 | 1 | 959.187 | 0.127 | 959.234 | 6.25 | 21 | 2 | 959.426 | 0.171 | 959.604 | 31.55 |
| 09 | 2 | 960.203 | 0.124 | 960.323 | 24.00 | 21 | 2 | 958.670 | 0.158 | 958.837 | 31.55 |
| 09 | 2 | 959.074 | 0.139 | 959.188 | 24.15 | 21 | 2 | 959.156 | 0.135 | 959.316 | 31.55 |
| 09 | 2 | 958.568 | 0.164 | 959.708 | 24.30 | 21 | 2 | 959.034 | 0.136 | 959.183 | 31.57 |
| 09 | 2 | 959.557 | 0.190 | 959.722 | 24.45 | 21 | 2 | 958.922 | 0.132 | 959.058 | 31.58 |
| 09 | 2 | 959.873 | 0.166 | 960.043 | 24.60 | 21 | 2 | 958.930 | 0.133 | 959.073 | 31.61 |
| 09 | 2 | 958.946 | 0.179 | 959.194 | 24.74 | 21 | 2 | 958.946 | 0.111 | 959.090 | 31.64 |
| 09 | 2 | 958.923 | 0.157 | 959.112 | 24.89 | 21 | 2 | 959.146 | 0.136 | 959.300 | 31.68 |
| 09 | 2 | 959.501 | 0.159 | 959.648 | 25.11 | 21 | 2 | 958.887 | 0.124 | 959.034 | 31.73 |
| 09 | 2 | 959.681 | 0.172 | 959.813 | 25.26 | 21 | 2 | 958.648 | 0.130 | 958.775 | 31.77 |
| 09 | 2 | 959.703 | 0.181 | 959.843 | 25.41 | 21 | 2 | 959.129 | 0.143 | 959.248 | 31.82 |
| 09 | 2 | 958.272 | 0.178 | 958.408 | 25.56 | 23 | 2 | 959.263 | 0.107 | 959.387 | 33.21 |
| 09 | 2 | 958.772 | 0.167 | 958.939 | 25.72 | 23 | 2 | 959.219 | 0.124 | 959.302 | 33.25 |
| 09 | 2 | 958.408 | 0.171 | 958.574 | 25.88 | 23 | 2 | 958.775 | 0.120 | 958.960 | 33.30 |
| 10 | 1 | 958.818 | 0.104 | 958.771 | 4.32 | 23 | 2 | 958.775 | 0.141 | 958.828 | 33.35 |
| 10 | 1 | 959.362 | 0.098 | 959.181 | 4.48 | 23 | 2 | 959.316 | 0.111 | 959.376 | 33.41 |
| 10 | 1 | 959.196 | 0.109 | 959.042 | 4.63 | 23 | 2 | 959.645 | 0.161 | 959.711 | 33.47 |
| 10 | 1 | 959.842 | 0.117 | 959.726 | 4.77 | 23 | 2 | 959.234 | 0.113 | 959.288 | 33.54 |
| 10 | 1 | 959.359 | 0.105 | 959.279 | 4.93 | 23 | 2 | 958.360 | 0.129 | 958.442 | 33.61 |
| 10 | 1 | 959.231 | 0.099 | 959.263 | 5.09 | 23 | 2 | 958.707 | 0.138 | 958.811 | 33.70 |
| 10 | 1 | 959.043 | 0.120 | 959.064 | 5.27 | 23 | 2 | 959.086 | 0.128 | 959.197 | 33.78 |
| 10 | 1 | 958.792 | 0.103 | 958.898 | 5.43 | 23 | 2 | 958.866 | 0.114 | 958.920 | 33.87 |
| 10 | 1 | 958.940 | 0.097 | 958.932 | 5.58 | 23 | 2 | 958.631 | 0.117 | 958.742 | 33.97 |
| 10 | 1 | 959.528 | 0.102 | 959.477 | 5.72 | 23 | 2 | 959.003 | 0.103 | 959.048 | 34.07 |
| 10 | 1 | 959.584 | 0.087 | 959.552 | 5.88 | 23 | 2 | 958.379 | 0.128 | 958.454 | 34.16 |
| 10 | 1 | 959.542 | 0.098 | 959.482 | 6.01 | 24 | 1 | 959.736 | 0.117 | 959.437 | 7.71 |

| \    2003 - JANEIRO    |   |         |       |         |       |
|---|---|---|---|---|---|
| D | L | SDB | ER | SDC | HL |
| 24 | 1 | 959.863 | 0.109 | 959.607 | 7.78 |
| 24 | 1 | 959.359 | 0.116 | 959.124 | 7.85 |
| 24 | 1 | 959.285 | 0.102 | 959.093 | 7.92 |
| 24 | 1 | 959.368 | 0.091 | 959.211 | 7.98 |
| 24 | 1 | 959.641 | 0.084 | 959.524 | 8.04 |
| 24 | 1 | 959.339 | 0.089 | 959.256 | 8.09 |
| 24 | 1 | 959.762 | 0.110 | 959.605 | 8.15 |
| 24 | 1 | 959.628 | 0.113 | 959.377 | 8.19 |
| 24 | 1 | 958.818 | 0.105 | 958.603 | 8.23 |
| 24 | 1 | 958.965 | 0.085 | 958.737 | 8.26 |
| 24 | 1 | 959.847 | 0.089 | 958.658 | 8.29 |
| 24 | 2 | 960.084 | 0.157 | 960.191 | 33.82 |
| 24 | 2 | 959.129 | 0.147 | 959.228 | 33.81 |
| 24 | 2 | 959.063 | 0.111 | 959.124 | 33.86 |
| 24 | 2 | 958.470 | 0.114 | 958.544 | 33.89 |
| 24 | 2 | 958.418 | 0.106 | 958.441 | 33.92 |
| 24 | 2 | 959.010 | 0.107 | 959.021 | 33.96 |
| 24 | 2 | 958.491 | 0.148 | 958.530 | 34.01 |
| 24 | 2 | 959.383 | 0.133 | 959.417 | 34.06 |
| 24 | 2 | 959.972 | 0.120 | 960.031 | 34.11 |
| 24 | 2 | 958.688 | 0.116 | 958.743 | 34.17 |
| 24 | 2 | 958.789 | 0.136 | 958.851 | 34.23 |
| 24 | 2 | 958.837 | 0.132 | 958.902 | 34.29 |
| 31 | 1 | 958.726 | 0.132 | 958.640 | 8.15 |
| 31 | 1 | 960.170 | 0.114 | 960.051 | 8.18 |
| 31 | 1 | 959.445 | 0.155 | 959.323 | 8.20 |
| 31 | 1 | 959.685 | 0.130 | 959.594 | 8.22 |
| 31 | 1 | 959.464 | 0.132 | 959.422 | 8.23 |
| 31 | 1 | 960.081 | 0.150 | 960.065 | 8.23 |
| 31 | 1 | 959.334 | 0.149 | 959.359 | 8.23 |
| 31 | 1 | 958.529 | 0.145 | 958.487 | 8.22 |
| 31 | 1 | 959.147 | 0.129 | 959.094 | 8.20 |
| 31 | 1 | 959.359 | 0.153 | 959.287 | 8.18 |
| 31 | 1 | 959.310 | 0.120 | 959.185 | 8.15 |
| 31 | 1 | 959.262 | 0.146 | 959.156 | 8.11 |
| 31 | 1 | 959.069 | 0.130 | 959.015 | 8.06 |
| 31 | 2 | 959.392 | 0.118 | 959.427 | 38.91 |
| 31 | 2 | 959.371 | 0.134 | 959.411 | 38.84 |
| 31 | 2 | 959.182 | 0.132 | 959.225 | 38.78 |
| 31 | 2 | 958.505 | 0.130 | 958.534 | 38.73 |
| 31 | 2 | 958.665 | 0.125 | 958.672 | 38.69 |
| 31 | 2 | 958.875 | 0.107 | 958.880 | 38.65 |
| 31 | 2 | 958.544 | 0.127 | 958.553 | 38.62 |
| 31 | 2 | 958.903 | 0.145 | 958.922 | 38.60 |
| 31 | 2 | 959.381 | 0.123 | 959.402 | 38.59 |
| 31 | 2 | 958.999 | 0.145 | 959.037 | 38.58 |
| 31 | 2 | 959.845 | 0.147 | 959.895 | 38.58 |
| 31 | 2 | 958.721 | 0.122 | 958.757 | 38.59 |

| \    2003 - FEVEREIRO    |   |         |       |         |       |
|---|---|---|---|---|---|
| D | L | SDB | ER | SDC | HL |
| 03 | 1 | 959.465 | 0.107 | 959.297 | 8.17 |
| 03 | 1 | 959.509 | 0.099 | 959.388 | 8.16 |
| 03 | 1 | 959.024 | 0.105 | 958.953 | 8.15 |
| 03 | 1 | 959.837 | 0.106 | 959.851 | 8.13 |
| 03 | 1 | 959.879 | 0.121 | 959.832 | 8.10 |
| 03 | 1 | 958.895 | 0.121 | 958.993 | 8.07 |
| 03 | 1 | 959.567 | 0.120 | 959.656 | 8.03 |
| 03 | 1 | 959.615 | 0.135 | 959.640 | 7.98 |
| 03 | 1 | 959.183 | 0.117 | 959.302 | 7.93 |
| 03 | 1 | 959.163 | 0.136 | 959.259 | 7.86 |
| 03 | 1 | 958.825 | 0.111 | 958.737 | 7.77 |
| 03 | 1 | 958.958 | 0.118 | 958.934 | 7.69 |
| 03 | 1 | 959.277 | 0.127 | 959.235 | 7.59 |
| 03 | 1 | 959.314 | 0.124 | 959.287 | 7.49 |
| 03 | 1 | 958.127 | 0.100 | 958.107 | 7.37 |
| 03 | 1 | 959.417 | 0.101 | 959.479 | 7.24 |
| 03 | 1 | 959.652 | 0.127 | 959.718 | 7.10 |
| 03 | 2 | 959.049 | 0.134 | 959.132 | 40.94 |
| 03 | 2 | 958.437 | 0.125 | 958.518 | 40.85 |

| \    2003 - FEVEREIRO    |   |         |       |         |       |
|---|---|---|---|---|---|
| D | L | SDB | ER | SDC | HL |
| 03 | 2 | 959.157 | 0.122 | 959.232 | 40.78 |
| 03 | 2 | 959.927 | 0.126 | 959.998 | 40.71 |
| 03 | 2 | 958.269 | 0.143 | 958.379 | 40.65 |
| 03 | 2 | 959.256 | 0.131 | 959.397 | 40.59 |
| 03 | 2 | 959.166 | 0.111 | 959.316 | 40.54 |
| 03 | 2 | 958.712 | 0.144 | 958.830 | 40.50 |
| 03 | 2 | 958.728 | 0.140 | 958.847 | 40.47 |
| 03 | 2 | 958.501 | 0.130 | 958.622 | 40.44 |
| 03 | 2 | 958.879 | 0.142 | 958.997 | 40.42 |
| 03 | 2 | 959.418 | 0.147 | 959.558 | 40.41 |
| 03 | 2 | 959.947 | 0.139 | 960.092 | 40.40 |
| 03 | 2 | 959.043 | 0.121 | 959.196 | 40.40 |
| 04 | 1 | 959.484 | 0.104 | 959.410 | 8.11 |
| 04 | 1 | 959.125 | 0.089 | 959.082 | 8.09 |
| 04 | 1 | 959.762 | 0.103 | 959.722 | 8.05 |
| 04 | 1 | 959.356 | 0.090 | 959.398 | 8.02 |
| 04 | 1 | 959.289 | 0.086 | 959.345 | 7.97 |
| 04 | 1 | 959.173 | 0.103 | 959.262 | 7.92 |
| 04 | 1 | 959.621 | 0.107 | 959.786 | 7.86 |
| 04 | 1 | 959.163 | 0.103 | 959.253 | 7.79 |
| 04 | 1 | 959.910 | 0.115 | 960.013 | 7.71 |
| 04 | 1 | 959.796 | 0.101 | 959.672 | 7.61 |
| 04 | 1 | 958.985 | 0.102 | 958.919 | 7.52 |
| 04 | 1 | 959.013 | 0.124 | 959.006 | 7.39 |
| 04 | 1 | 960.086 | 0.133 | 960.070 | 7.27 |
| 04 | 1 | 958.817 | 0.097 | 958.831 | 7.14 |
| 04 | 1 | 959.393 | 0.129 | 959.474 | 6.83 |
| 04 | 2 | 958.971 | 0.182 | 959.108 | 41.88 |
| 04 | 2 | 959.391 | 0.152 | 959.506 | 41.76 |
| 04 | 2 | 958.662 | 0.140 | 958.759 | 41.65 |
| 04 | 2 | 959.054 | 0.131 | 959.136 | 41.54 |
| 04 | 2 | 958.326 | 0.125 | 958.391 | 41.45 |
| 04 | 2 | 958.589 | 0.154 | 958.673 | 41.37 |
| 04 | 2 | 958.748 | 0.141 | 958.857 | 41.29 |
| 04 | 2 | 959.504 | 0.153 | 959.663 | 41.21 |
| 04 | 2 | 958.651 | 0.141 | 958.808 | 41.08 |
| 04 | 2 | 958.122 | 0.154 | 958.289 | 41.05 |
| 05 | 1 | 958.385 | 0.084 | 958.424 | 8.07 |
| 05 | 1 | 959.427 | 0.098 | 959.383 | 8.04 |
| 05 | 1 | 959.270 | 0.095 | 959.348 | 8.00 |
| 05 | 1 | 960.868 | 0.105 | 960.892 | 7.96 |
| 05 | 1 | 959.211 | 0.106 | 959.205 | 7.91 |
| 05 | 1 | 958.840 | 0.091 | 958.897 | 7.85 |
| 05 | 1 | 960.172 | 0.112 | 960.199 | 7.78 |
| 05 | 1 | 959.452 | 0.102 | 959.322 | 7.68 |
| 05 | 1 | 959.719 | 0.127 | 959.548 | 7.59 |
| 05 | 1 | 958.961 | 0.110 | 958.802 | 7.49 |
| 05 | 1 | 958.360 | 0.112 | 958.257 | 7.38 |
| 05 | 1 | 958.882 | 0.135 | 958.803 | 7.26 |
| 05 | 1 | 959.048 | 0.099 | 959.026 | 7.12 |
| 05 | 1 | 959.300 | 0.120 | 959.361 | 6.98 |
| 05 | 1 | 959.831 | 0.105 | 959.885 | 6.82 |
| 05 | 1 | 959.321 | 0.122 | 959.363 | 6.65 |
| 05 | 2 | 958.710 | 0.118 | 958.790 | 42.18 |
| 05 | 2 | 959.422 | 0.155 | 959.503 | 42.07 |
| 05 | 2 | 959.082 | 0.128 | 959.153 | 41.98 |
| 05 | 2 | 959.057 | 0.115 | 959.122 | 41.90 |
| 05 | 2 | 958.404 | 0.134 | 958.481 | 41.83 |
| 05 | 2 | 959.429 | 0.106 | 959.507 | 41.77 |
| 05 | 2 | 958.589 | 0.166 | 958.726 | 41.72 |
| 05 | 2 | 958.103 | 0.141 | 958.251 | 41.67 |
| 05 | 2 | 959.034 | 0.163 | 959.195 | 41.63 |
| 05 | 2 | 959.444 | 0.157 | 959.581 | 41.60 |
| 05 | 2 | 958.482 | 0.159 | 958.642 | 41.58 |
| 05 | 2 | 959.753 | 0.123 | 959.840 | 41.55 |
| 05 | 2 | 959.597 | 0.139 | 959.597 | 41.54 |
| 05 | 2 | 958.643 | 0.129 | 958.746 | 41.54 |
| 05 | 2 | 959.056 | 0.115 | 959.146 | 41.54 |
| 06 | 1 | 959.007 | 0.101 | 958.867 | 8.09 |
| 06 | 1 | 959.580 | 0.086 | 959.385 | 8.07 |
| 06 | 1 | 959.355 | 0.093 | 959.142 | 8.04 |

| 2003 - FEVEREIRO | | | | | | 2003 - FEVEREIRO | | | | |
|---|---|---|---|---|---|---|---|---|---|---|
| D | L | SDB | ER | SDC | HL | D | L | SDB | ER | SDC | HL |
| 06 | 1 | 959.817 | 0.099 | 959.654 | 8.02 | 11 | 2 | 959.542 | 0.109 | 959.687 | 47.13 |
| 06 | 1 | 959.175 | 0.110 | 959.038 | 7.98 | 11 | 2 | 958.752 | 0.116 | 958.891 | 46.89 |
| 06 | 1 | 959.480 | 0.113 | 959.355 | 7.93 | 11 | 2 | 959.070 | 0.134 | 959.220 | 46.66 |
| 06 | 1 | 958.974 | 0.121 | 958.884 | 7.86 | 11 | 2 | 959.334 | 0.119 | 959.468 | 46.47 |
| 06 | 1 | 959.373 | 0.100 | 959.311 | 7.80 | 11 | 2 | 959.199 | 0.126 | 959.327 | 46.29 |
| 06 | 1 | 959.066 | 0.109 | 958.933 | 7.72 | 11 | 2 | 959.617 | 0.134 | 959.733 | 46.10 |
| 06 | 1 | 959.232 | 0.121 | 959.111 | 7.64 | 11 | 2 | 959.148 | 0.124 | 959.264 | 45.95 |
| 06 | 1 | 959.668 | 0.120 | 959.514 | 7.56 | 11 | 2 | 958.842 | 0.123 | 958.961 | 45.80 |
| 06 | 1 | 958.809 | 0.106 | 958.630 | 7.46 | 11 | 2 | 959.189 | 0.116 | 959.337 | 45.67 |
| 06 | 1 | 959.895 | 0.128 | 959.727 | 7.36 | 11 | 2 | 959.199 | 0.126 | 959.316 | 45.53 |
| 06 | 2 | 959.115 | 0.163 | 959.280 | 44.19 | 11 | 2 | 958.921 | 0.117 | 959.066 | 45.42 |
| 06 | 2 | 959.532 | 0.120 | 959.681 | 43.92 | 11 | 2 | 959.300 | 0.130 | 959.440 | 45.31 |
| 06 | 2 | 959.441 | 0.116 | 959.590 | 43.72 | 11 | 2 | 959.286 | 0.140 | 959.461 | 45.21 |
| 06 | 2 | 959.485 | 0.133 | 959.643 | 43.52 | 11 | 2 | 958.989 | 0.135 | 959.171 | 45.12 |
| 06 | 2 | 958.763 | 0.154 | 958.936 | 43.35 | 12 | 1 | 958.181 | 0.134 | 958.174 | 6.72 |
| 06 | 2 | 958.963 | 0.152 | 959.115 | 43.19 | 12 | 1 | 960.079 | 0.115 | 960.057 | 6.56 |
| 06 | 2 | 958.879 | 0.147 | 959.024 | 43.04 | 12 | 1 | 958.427 | 0.128 | 958.503 | 6.40 |
| 06 | 2 | 959.167 | 0.137 | 959.289 | 42.91 | 12 | 1 | 959.516 | 0.156 | 959.643 | 6.24 |
| 06 | 2 | 958.846 | 0.132 | 958.942 | 42.80 | 12 | 1 | 959.346 | 0.102 | 959.140 | 6.03 |
| 06 | 2 | 958.701 | 0.117 | 958.786 | 42.69 | 12 | 1 | 959.804 | 0.126 | 959.625 | 5.84 |
| 06 | 2 | 958.470 | 0.107 | 958.544 | 42.60 | 12 | 1 | 959.017 | 0.114 | 958.881 | 5.64 |
| 07 | 1 | 959.202 | 0.114 | 958.966 | 7.95 | 12 | 1 | 958.777 | 0.115 | 958.722 | 5.42 |
| 07 | 1 | 959.501 | 0.111 | 959.284 | 7.91 | 12 | 1 | 959.354 | 0.122 | 959.418 | 5.19 |
| 07 | 1 | 959.047 | 0.120 | 958.868 | 7.86 | 12 | 1 | 959.236 | 0.140 | 959.233 | 4.94 |
| 07 | 1 | 959.621 | 0.120 | 959.503 | 7.80 | 12 | 1 | 959.011 | 0.121 | 959.033 | 4.45 |
| 07 | 1 | 959.157 | 0.115 | 959.102 | 7.74 | 12 | 1 | 959.644 | 0.120 | 959.650 | 4.13 |
| 07 | 1 | 959.627 | 0.111 | 959.532 | 7.64 | 12 | 1 | 959.518 | 0.139 | 959.550 | 3.80 |
| 07 | 1 | 959.322 | 0.106 | 959.158 | 7.54 | 12 | 1 | 960.389 | 0.143 | 960.422 | 3.43 |
| 07 | 1 | 959.253 | 0.135 | 959.035 | 7.44 | 12 | 2 | 959.060 | 0.579 | 959.201 | 48.18 |
| 07 | 1 | 960.376 | 0.115 | 960.164 | 7.34 | 12 | 2 | 959.126 | 0.153 | 959.262 | 47.90 |
| 07 | 1 | 959.949 | 0.097 | 959.724 | 7.22 | 12 | 2 | 958.938 | 0.123 | 959.044 | 47.64 |
| 07 | 1 | 959.530 | 0.105 | 959.333 | 7.09 | 12 | 2 | 958.701 | 0.116 | 958.801 | 47.41 |
| 07 | 1 | 960.052 | 0.112 | 959.882 | 6.95 | 12 | 2 | 958.251 | 0.138 | 958.357 | 47.18 |
| 07 | 2 | 959.216 | 0.126 | 959.360 | 43.94 | 12 | 2 | 958.197 | 0.126 | 958.278 | 46.98 |
| 07 | 2 | 958.916 | 0.142 | 959.059 | 43.78 | 12 | 2 | 959.678 | 0.156 | 959.760 | 46.79 |
| 07 | 2 | 959.281 | 0.156 | 959.424 | 43.64 | 12 | 2 | 958.760 | 0.125 | 958.823 | 46.62 |
| 07 | 2 | 959.952 | 0.160 | 960.091 | 43.47 | 12 | 2 | 959.801 | 0.144 | 959.902 | 46.46 |
| 07 | 2 | 958.957 | 0.122 | 959.067 | 43.35 | 12 | 2 | 959.000 | 0.114 | 959.132 | 46.31 |
| 07 | 2 | 958.854 | 0.123 | 958.959 | 43.25 | 12 | 2 | 958.958 | 0.150 | 959.049 | 46.18 |
| 07 | 2 | 958.572 | 0.139 | 958.679 | 43.15 | 12 | 2 | 958.931 | 0.135 | 959.056 | 46.05 |
| 07 | 2 | 960.176 | 0.127 | 960.281 | 43.06 | 12 | 2 | 960.002 | 0.131 | 960.138 | 45.94 |
| 07 | 2 | 958.604 | 0.137 | 958.743 | 42.98 | 12 | 2 | 958.662 | 0.158 | 958.822 | 45.83 |
| 07 | 2 | 958.657 | 0.117 | 958.794 | 42.91 | 13 | 1 | 958.900 | 0.093 | 958.904 | 7.55 |
| 07 | 2 | 957.944 | 0.152 | 958.100 | 42.85 | 13 | 1 | 959.514 | 0.106 | 959.447 | 7.48 |
| 07 | 2 | 958.597 | 0.122 | 958.756 | 42.79 | 13 | 1 | 959.888 | 0.112 | 959.828 | 7.40 |
| 10 | 1 | 959.068 | 0.112 | 958.967 | 7.41 | 13 | 1 | 960.204 | 0.104 | 960.189 | 7.32 |
| 10 | 1 | 959.085 | 0.107 | 959.061 | 7.31 | 13 | 1 | 959.744 | 0.107 | 959.785 | 7.23 |
| 10 | 1 | 959.547 | 0.126 | 959.532 | 7.20 | 13 | 1 | 959.610 | 0.092 | 959.684 | 7.13 |
| 10 | 1 | 959.082 | 0.110 | 959.078 | 7.08 | 13 | 1 | 959.594 | 0.114 | 959.793 | 6.99 |
| 10 | 1 | 959.550 | 0.110 | 959.482 | 6.94 | 13 | 1 | 959.707 | 0.117 | 959.882 | 6.86 |
| 10 | 1 | 960.633 | 0.121 | 960.515 | 6.78 | 13 | 1 | 959.756 | 0.122 | 959.818 | 6.73 |
| 10 | 1 | 959.248 | 0.118 | 959.151 | 6.61 | 13 | 1 | 959.704 | 0.126 | 959.782 | 6.59 |
| 10 | 1 | 960.078 | 0.117 | 959.983 | 6.44 | 13 | 1 | 959.267 | 0.109 | 959.377 | 6.44 |
| 10 | 1 | 959.509 | 0.117 | 959.442 | 6.27 | 13 | 1 | 959.695 | 0.086 | 959.769 | 6.28 |
| 10 | 1 | 959.753 | 0.109 | 959.708 | 6.07 | 13 | 2 | 958.700 | 0.136 | 958.789 | 47.64 |
| 10 | 1 | 959.012 | 0.138 | 958.985 | 5.87 | 13 | 2 | 958.284 | 0.151 | 958.385 | 47.45 |
| 10 | 1 | 959.191 | 0.142 | 959.171 | 5.64 | 13 | 2 | 958.299 | 0.133 | 958.396 | 47.26 |
| 10 | 2 | 958.993 | 0.144 | 959.112 | 45.37 | 13 | 2 | 959.139 | 0.151 | 959.258 | 47.09 |
| 10 | 2 | 957.892 | 0.149 | 958.004 | 45.24 | 13 | 2 | 958.852 | 0.163 | 959.016 | 46.92 |
| 10 | 2 | 958.936 | 0.142 | 959.052 | 45.10 | 13 | 2 | 959.196 | 0.133 | 959.324 | 46.73 |
| 10 | 2 | 958.621 | 0.134 | 958.737 | 44.98 | 13 | 2 | 959.631 | 0.135 | 959.778 | 46.59 |
| 10 | 2 | 958.547 | 0.131 | 958.678 | 44.86 | 13 | 2 | 958.436 | 0.158 | 958.602 | 46.46 |
| 10 | 2 | 958.834 | 0.115 | 958.961 | 44.75 | 13 | 2 | 958.807 | 0.124 | 958.971 | 46.34 |
| 10 | 2 | 958.968 | 0.131 | 959.108 | 44.63 | 13 | 2 | 959.404 | 0.128 | 959.545 | 46.22 |
| 10 | 2 | 958.821 | 0.136 | 958.972 | 44.55 | 13 | 2 | 959.487 | 0.137 | 959.623 | 46.12 |
| 10 | 2 | 959.081 | 0.127 | 959.238 | 44.48 | 13 | 2 | 959.193 | 0.144 | 959.322 | 46.03 |
| 10 | 2 | 958.866 | 0.138 | 959.014 | 44.40 | 14 | 1 | 959.628 | 0.097 | 959.580 | 7.06 |
| 10 | 2 | 958.685 | 0.139 | 958.832 | 44.35 | 14 | 1 | 959.905 | 0.106 | 959.853 | 6.94 |
| 10 | 2 | 958.573 | 0.155 | 958.731 | 44.30 | 14 | 1 | 959.218 | 0.114 | 959.193 | 6.83 |
| 11 | 2 | 960.678 | 0.111 | 960.837 | 47.68 | 14 | 1 | 959.731 | 0.128 | 959.732 | 6.70 |
| 11 | 2 | 959.878 | 0.132 | 960.042 | 47.38 | 14 | 1 | 960.069 | 0.143 | 960.051 | 6.53 |

| 2003 - FEVEREIRO | | | | | |
|---|---|---|---|---|---|
| D | L | SDB | ER | SDC | HL |
| 14 | 1 | 959.287 | 0.142 | 959.289 | 6.37 |
| 14 | 1 | 959.280 | 0.105 | 959.243 | 6.16 |
| 14 | 1 | 959.471 | 0.148 | 959.387 | 5.99 |
| 14 | 1 | 958.846 | 0.135 | 958.763 | 5.79 |
| 14 | 1 | 960.106 | 0.138 | 960.014 | 5.60 |
| 14 | 1 | 959.488 | 0.102 | 959.396 | 5.34 |
| 14 | 1 | 959.542 | 0.126 | 959.462 | 5.10 |
| 14 | 2 | 959.225 | 0.150 | 959.378 | 48.93 |
| 14 | 2 | 958.751 | 0.162 | 958.908 | 48.68 |
| 14 | 2 | 958.665 | 0.132 | 958.814 | 48.44 |
| 14 | 2 | 959.030 | 0.129 | 959.163 | 48.22 |
| 14 | 2 | 959.246 | 0.126 | 959.391 | 47.98 |
| 14 | 2 | 958.472 | 0.123 | 958.614 | 47.78 |
| 14 | 2 | 958.584 | 0.132 | 958.725 | 47.60 |
| 17 | 1 | 959.339 | 0.133 | 959.571 | 5.70 |
| 17 | 1 | 959.427 | 0.122 | 959.650 | 5.49 |
| 17 | 1 | 958.690 | 0.114 | 958.912 | 5.30 |
| 17 | 1 | 958.774 | 0.133 | 959.001 | 5.09 |
| 17 | 1 | 959.054 | 0.131 | 959.288 | 4.84 |
| 17 | 1 | 958.908 | 0.144 | 959.155 | 4.57 |
| 17 | 1 | 958.728 | 0.123 | 958.983 | 4.31 |
| 17 | 1 | 959.107 | 0.173 | 959.335 | 4.02 |
| 17 | 1 | 959.260 | 0.151 | 959.508 | 3.71 |
| 18 | 1 | 958.207 | 0.112 | 958.081 | 6.24 |
| 18 | 1 | 959.509 | 0.106 | 959.462 | 6.10 |
| 18 | 1 | 959.460 | 0.098 | 959.427 | 5.94 |
| 18 | 1 | 959.056 | 0.110 | 959.145 | 5.78 |
| 18 | 1 | 960.008 | 0.107 | 960.013 | 5.60 |
| 18 | 1 | 959.832 | 0.150 | 959.948 | 5.41 |
| 18 | 1 | 959.432 | 0.115 | 959.520 | 5.20 |
| 18 | 1 | 959.539 | 0.138 | 959.686 | 4.98 |
| 18 | 1 | 959.418 | 0.121 | 959.558 | 4.73 |
| 18 | 1 | 958.679 | 0.128 | 958.663 | 4.44 |
| 18 | 1 | 959.435 | 0.119 | 959.414 | 4.17 |
| 18 | 1 | 959.642 | 0.120 | 959.696 | 3.91 |
| 18 | 1 | 959.845 | 0.128 | 959.932 | 3.58 |
| 18 | 1 | 959.876 | 0.118 | 960.003 | 3.26 |
| 18 | 1 | 959.823 | 0.125 | 959.963 | 2.93 |
| 18 | 1 | 959.856 | 0.131 | 960.045 | 2.58 |
| 18 | 2 | 960.889 | 0.137 | 961.009 | 51.78 |
| 18 | 2 | 959.405 | 0.145 | 959.536 | 51.42 |
| 18 | 2 | 959.138 | 0.153 | 959.265 | 51.12 |
| 18 | 2 | 959.342 | 0.132 | 959.458 | 50.85 |
| 18 | 2 | 958.454 | 0.153 | 958.576 | 50.59 |
| 18 | 2 | 958.720 | 0.146 | 958.846 | 50.33 |
| 18 | 2 | 959.128 | 0.142 | 959.261 | 50.11 |
| 18 | 2 | 958.187 | 0.156 | 958.338 | 49.90 |
| 18 | 2 | 959.061 | 0.152 | 959.226 | 49.70 |
| 18 | 2 | 958.571 | 0.149 | 958.725 | 49.50 |
| 18 | 2 | 958.831 | 0.173 | 959.012 | 49.31 |
| 18 | 2 | 958.808 | 0.179 | 959.003 | 49.13 |
| 18 | 2 | 958.671 | 0.163 | 958.867 | 48.97 |
| 18 | 2 | 958.484 | 0.159 | 958.662 | 48.82 |
| 18 | 2 | 959.077 | 0.171 | 959.261 | 48.70 |
| 19 | 1 | 959.279 | 0.113 | 959.236 | 5.66 |
| 19 | 1 | 959.163 | 0.117 | 959.224 | 5.47 |
| 19 | 1 | 959.281 | 0.128 | 959.382 | 5.23 |
| 19 | 1 | 959.753 | 0.111 | 959.821 | 5.03 |
| 19 | 1 | 959.516 | 0.121 | 959.627 | 4.81 |
| 19 | 1 | 959.424 | 0.119 | 959.518 | 4.57 |
| 19 | 1 | 958.534 | 0.147 | 958.690 | 4.32 |
| 19 | 1 | 959.306 | 0.129 | 959.186 | 3.99 |
| 19 | 1 | 958.864 | 0.118 | 958.769 | 3.72 |
| 19 | 1 | 958.789 | 0.121 | 958.713 | 3.42 |
| 19 | 1 | 959.018 | 0.129 | 958.998 | 3.12 |
| 19 | 1 | 958.834 | 0.159 | 958.896 | 2.79 |
| 19 | 1 | 958.853 | 0.130 | 958.890 | 2.44 |
| 19 | 1 | 959.281 | 0.143 | 959.336 | 2.06 |
| 19 | 1 | 958.535 | 0.124 | 958.659 | 1.68 |
| 19 | 1 | 958.545 | 0.154 | 958.670 | 1.27 |
| 19 | 2 | 960.228 | 0.174 | 960.376 | 53.30 |
| 19 | 2 | 959.390 | 0.164 | 959.535 | 52.89 |
| 19 | 2 | 959.277 | 0.133 | 959.393 | 52.54 |
| 19 | 2 | 957.874 | 0.144 | 957.996 | 52.13 |
| 19 | 2 | 959.398 | 0.142 | 959.494 | 51.79 |
| 19 | 2 | 958.815 | 0.171 | 958.904 | 51.51 |
| 19 | 2 | 958.572 | 0.135 | 958.647 | 51.24 |
| 19 | 2 | 958.845 | 0.142 | 958.936 | 50.98 |
| 20 | 1 | 959.490 | 0.129 | 959.518 | 4.79 |
| 20 | 1 | 958.523 | 0.135 | 958.555 | 4.54 |
| 20 | 1 | 959.885 | 0.125 | 959.724 | 4.25 |
| 20 | 1 | 960.795 | 0.123 | 960.647 | 3.99 |
| 20 | 1 | 960.128 | 0.156 | 960.032 | 3.71 |
| 20 | 1 | 959.591 | 0.132 | 959.548 | 3.39 |
| 20 | 1 | 959.307 | 0.159 | 959.318 | 3.01 |
| 20 | 1 | 959.948 | 0.177 | 960.009 | 2.67 |
| 20 | 1 | 959.663 | 0.157 | 959.695 | 2.28 |
| 20 | 1 | 960.319 | 0.137 | 960.361 | 1.90 |
| 20 | 1 | 959.001 | 0.149 | 959.070 | 1.50 |
| 20 | 1 | 959.750 | 0.135 | 959.832 | 1.04 |
| 20 | 2 | 959.326 | 0.136 | 959.493 | 52.80 |
| 20 | 2 | 958.455 | 0.135 | 958.620 | 52.47 |
| 20 | 2 | 958.434 | 0.176 | 958.596 | 52.17 |
| 20 | 2 | 958.530 | 0.137 | 958.686 | 51.88 |
| 20 | 2 | 958.984 | 0.139 | 959.127 | 51.61 |
| 20 | 2 | 957.992 | 0.142 | 958.152 | 51.33 |
| 20 | 2 | 958.805 | 0.138 | 958.978 | 51.01 |
| 20 | 2 | 958.529 | 0.142 | 958.711 | 50.80 |
| 20 | 2 | 958.710 | 0.141 | 958.888 | 50.59 |
| 20 | 2 | 958.590 | 0.170 | 958.782 | 50.34 |
| 20 | 2 | 958.900 | 0.150 | 959.094 | 50.15 |
| 20 | 2 | 958.757 | 0.221 | 958.967 | 49.98 |
| 21 | 1 | 958.805 | 0.152 | 958.883 | 4.69 |
| 21 | 1 | 959.370 | 0.127 | 959.440 | 4.40 |
| 21 | 1 | 959.419 | 0.113 | 959.482 | 4.14 |
| 21 | 1 | 959.199 | 0.111 | 959.271 | 3.88 |
| 21 | 1 | 959.841 | 0.112 | 959.894 | 3.61 |
| 21 | 1 | 959.443 | 0.107 | 959.500 | 3.32 |
| 21 | 1 | 959.100 | 0.157 | 959.188 | 2.95 |
| 21 | 1 | 959.248 | 0.125 | 959.369 | 2.58 |
| 21 | 1 | 959.908 | 0.132 | 960.048 | 2.23 |
| 21 | 1 | 959.273 | 0.153 | 959.416 | 1.86 |
| 21 | 1 | 959.184 | 0.154 | 959.347 | 1.46 |
| 21 | 1 | 959.561 | 0.148 | 959.731 | 1.04 |
| 21 | 2 | 959.175 | 0.133 | 959.345 | 51.79 |
| 21 | 2 | 959.054 | 0.119 | 959.213 | 51.54 |
| 21 | 2 | 959.101 | 0.126 | 959.265 | 51.29 |
| 21 | 2 | 959.204 | 0.147 | 959.380 | 51.06 |
| 21 | 2 | 959.721 | 0.158 | 959.903 | 50.83 |
| 21 | 2 | 959.528 | 0.136 | 959.701 | 50.57 |
| 21 | 2 | 959.047 | 0.146 | 959.212 | 50.40 |
| 21 | 2 | 959.593 | 0.167 | 959.754 | 50.23 |
| 21 | 2 | 958.624 | 0.123 | 958.769 | 50.08 |
| 21 | 2 | 959.359 | 0.164 | 959.534 | 49.93 |
| 21 | 2 | 959.407 | 0.133 | 959.565 | 49.79 |
| 24 | 1 | 957.036 | 0.156 | 957.119 | 2.22 |
| 24 | 1 | 958.452 | 0.172 | 958.522 | 1.82 |
| 24 | 1 | 959.202 | 0.143 | 959.090 | 1.37 |
| 24 | 1 | 958.883 | 0.146 | 958.840 | 0.97 |
| 24 | 1 | 958.815 | 0.144 | 958.711 | 0.54 |
| 24 | 1 | 959.636 | 0.109 | 959.654 | 0.08 |
| 24 | 1 | 958.960 | 0.133 | 959.039 | 0.38 |
| 24 | 1 | 958.816 | 0.108 | 958.867 | 0.92 |
| 24 | 1 | 959.548 | 0.134 | 959.610 | 1.46 |
| 24 | 1 | 958.774 | 0.138 | 958.864 | 2.02 |
| 24 | 1 | 959.071 | 0.174 | 959.188 | 2.59 |
| 24 | 1 | 959.197 | 0.172 | 959.287 | 3.22 |
| 24 | 1 | 959.880 | 0.173 | 959.976 | 3.91 |
| 24 | 2 | 960.393 | 0.133 | 960.487 | 54.68 |
| 24 | 2 | 960.061 | 0.148 | 960.172 | 54.34 |
| 24 | 2 | 959.489 | 0.143 | 959.604 | 53.97 |
| 24 | 2 | 959.174 | 0.160 | 959.287 | 53.67 |

| 2003 - FEVEREIRO | | | | | | 2003 - FEVEREIRO | | | | |
|---|---|---|---|---|---|---|---|---|---|---|
| D | L | SDB | ER | SDC | HL | D | L | SDB | ER | SDC | HL |
| 24 | 2 | 959.193 | 0.157 | 959.320 | 53.39 | 27 | 1 | 959.329 | 0.116 | 959.345 | 0.72 |
| 24 | 2 | 959.076 | 0.163 | 959.214 | 53.10 | 27 | 1 | 959.608 | 0.143 | 959.670 | 0.27 |
| 24 | 2 | 959.630 | 0.189 | 959.771 | 52.83 | 27 | 1 | 959.113 | 0.142 | 959.134 | 0.16 |
| 24 | 2 | 960.275 | 0.211 | 960.462 | 52.48 | 27 | 1 | 959.759 | 0.148 | 959.830 | 0.62 |
| 24 | 2 | 958.414 | 0.225 | 958.608 | 52.22 | 27 | 1 | 959.568 | 0.141 | 959.613 | 1.11 |
| 24 | 2 | 959.065 | 0.177 | 959.188 | 51.75 | 27 | 1 | 959.006 | 0.150 | 958.970 | 1.75 |
| 24 | 2 | 959.541 | 0.139 | 959.649 | 51.56 | 27 | 1 | 959.758 | 0.139 | 959.719 | 2.31 |
| 24 | 2 | 959.650 | 0.126 | 959.766 | 51.38 | 27 | 1 | 958.549 | 0.118 | 958.518 | 2.92 |
| 24 | 2 | 959.660 | 0.123 | 959.781 | 51.21 | 27 | 1 | 959.149 | 0.154 | 959.126 | 3.53 |
| 24 | 2 | 959.586 | 0.152 | 959.738 | 51.05 | 27 | 1 | 958.605 | 0.154 | 958.635 | 4.34 |
| 24 | 2 | 960.318 | 0.183 | 960.485 | 50.89 | 27 | 1 | 959.326 | 0.161 | 959.375 | 5.05 |
| 25 | 1 | 958.810 | 0.136 | 958.686 | 3.72 | 27 | 1 | 959.082 | 0.149 | 959.156 | 5.80 |
| 25 | 1 | 959.257 | 0.097 | 959.057 | 3.42 | 27 | 2 | 959.509 | 0.154 | 959.760 | 50.89 |
| 25 | 1 | 959.065 | 0.129 | 958.881 | 3.13 | 27 | 2 | 959.062 | 0.163 | 959.327 | 50.78 |
| 25 | 1 | 959.627 | 0.110 | 959.528 | 2.83 | 27 | 2 | 958.900 | 0.175 | 959.185 | 50.69 |
| 25 | 1 | 959.603 | 0.112 | 959.436 | 2.49 | 27 | 2 | 958.616 | 0.171 | 958.900 | 50.61 |
| 25 | 1 | 958.958 | 0.115 | 958.910 | 2.13 | 27 | 2 | 959.016 | 0.168 | 959.296 | 50.53 |
| 25 | 1 | 959.595 | 0.145 | 959.485 | 1.79 | 27 | 2 | 960.125 | 0.210 | 960.421 | 50.45 |
| 25 | 1 | 958.994 | 0.144 | 959.017 | 1.43 | 28 | 1 | 959.254 | 0.148 | 959.102 | 3.09 |
| 25 | 1 | 958.821 | 0.127 | 958.783 | 1.05 | 28 | 1 | 958.980 | 0.144 | 958.825 | 2.81 |
| 25 | 1 | 959.609 | 0.147 | 959.563 | 0.62 | 28 | 1 | 959.953 | 0.130 | 959.793 | 2.53 |
| 25 | 1 | 959.827 | 0.111 | 959.846 | 0.19 | 28 | 1 | 959.607 | 0.128 | 959.438 | 2.24 |
| 25 | 1 | 959.531 | 0.160 | 959.578 | 0.28 | 28 | 1 | 958.886 | 0.143 | 958.702 | 1.89 |
| 25 | 1 | 958.894 | 0.125 | 958.772 | 0.92 | 28 | 1 | 959.886 | 0.102 | 959.750 | 1.56 |
| 25 | 1 | 959.512 | 0.111 | 959.353 | 1.46 | 28 | 1 | 959.508 | 0.126 | 959.422 | 1.19 |
| 25 | 1 | 958.958 | 0.132 | 958.821 | 2.04 | 28 | 1 | 959.691 | 0.148 | 959.626 | 0.80 |
| 25 | 2 | 959.995 | 0.111 | 960.070 | 54.92 | 28 | 1 | 958.895 | 0.193 | 958.828 | 0.31 |
| 25 | 2 | 959.219 | 0.121 | 959.306 | 54.55 | 28 | 1 | 959.836 | 0.138 | 959.817 | 0.10 |
| 25 | 2 | 959.496 | 0.165 | 959.602 | 54.19 | 28 | 1 | 959.922 | 0.129 | 960.010 | 0.58 |
| 25 | 2 | 959.425 | 0.150 | 959.554 | 53.85 | 28 | 1 | 959.074 | 0.159 | 958.999 | 1.13 |
| 25 | 2 | 958.592 | 0.143 | 958.703 | 53.49 | | | | | | |
| 25 | 2 | 958.943 | 0.146 | 959.059 | 53.07 | | | | | | |
| 25 | 2 | 959.318 | 0.174 | 959.471 | 52.78 | | | 2003 - MARCO | | | |
| 25 | 2 | 957.973 | 0.147 | 958.131 | 52.47 | D | L | SDB | ER | SDC | HL |
| 25 | 2 | 959.001 | 0.157 | 959.122 | 52.25 | 07 | 1 | 958.553 | 0.146 | 958.470 | 5.18 |
| 25 | 2 | 959.518 | 0.144 | 959.622 | 51.96 | 07 | 1 | 959.684 | 0.155 | 959.591 | 5.85 |
| 25 | 2 | 958.548 | 0.164 | 958.677 | 51.76 | 07 | 1 | 959.884 | 0.156 | 959.829 | 6.54 |
| 25 | 2 | 959.164 | 0.157 | 959.283 | 51.56 | 07 | 1 | 959.616 | 0.135 | 959.602 | 7.44 |
| 25 | 2 | 958.980 | 0.152 | 959.118 | 51.39 | 07 | 1 | 959.449 | 0.132 | 959.476 | 8.19 |
| 25 | 2 | 959.560 | 0.150 | 959.681 | 51.23 | 07 | 1 | 958.665 | 0.141 | 958.754 | 9.21 |
| 25 | 2 | 958.970 | 0.147 | 959.118 | 51.07 | 07 | 1 | 959.281 | 0.148 | 959.427 | 10.07 |
| 25 | 2 | 959.608 | 0.149 | 959.738 | 50.88 | 07 | 1 | 958.690 | 0.138 | 958.848 | 10.97 |
| 26 | 1 | 959.177 | 0.140 | 959.016 | 2.92 | 07 | 2 | 959.298 | 0.144 | 959.549 | 55.91 |
| 26 | 1 | 959.343 | 0.103 | 959.255 | 2.56 | 07 | 2 | 958.763 | 0.134 | 959.016 | 55.64 |
| 26 | 1 | 958.791 | 0.101 | 958.705 | 2.21 | 07 | 2 | 958.954 | 0.138 | 959.218 | 55.38 |
| 26 | 1 | 959.192 | 0.135 | 959.075 | 1.86 | 07 | 2 | 959.103 | 0.121 | 959.362 | 55.12 |
| 26 | 1 | 959.758 | 0.117 | 959.664 | 1.52 | 07 | 2 | 958.672 | 0.126 | 958.947 | 54.87 |
| 26 | 1 | 959.179 | 0.130 | 959.104 | 1.13 | 07 | 2 | 958.843 | 0.131 | 959.139 | 54.64 |
| 26 | 1 | 959.606 | 0.157 | 959.602 | 0.71 | 07 | 2 | 958.876 | 0.147 | 959.211 | 54.39 |
| 26 | 1 | 959.337 | 0.154 | 959.341 | 0.22 | 07 | 2 | 958.641 | 0.151 | 958.915 | 54.09 |
| 26 | 1 | 958.727 | 0.100 | 958.743 | 0.23 | 07 | 2 | 958.776 | 0.143 | 959.024 | 53.90 |
| 26 | 1 | 959.250 | 0.135 | 959.259 | 0.80 | 07 | 2 | 958.528 | 0.157 | 958.811 | 53.70 |
| 26 | 1 | 959.441 | 0.097 | 959.365 | 1.41 | 07 | 2 | 958.692 | 0.132 | 958.977 | 53.53 |
| 26 | 1 | 958.707 | 0.187 | 958.597 | 2.05 | 13 | 2 | 958.602 | 0.151 | 958.995 | 59.45 |
| 26 | 1 | 959.226 | 0.119 | 959.121 | 2.63 | 13 | 2 | 958.832 | 0.147 | 959.159 | 59.01 |
| 26 | 1 | 958.768 | 0.116 | 958.722 | 3.32 | 13 | 2 | 958.456 | 0.204 | 958.823 | 58.43 |
| 26 | 1 | 959.098 | 0.112 | 959.090 | 4.05 | 13 | 2 | 958.380 | 0.158 | 958.747 | 58.07 |
| 26 | 1 | 959.490 | 0.123 | 959.490 | 4.81 | 13 | 2 | 958.431 | 0.171 | 958.751 | 57.01 |
| 26 | 2 | 959.202 | 0.136 | 959.288 | 53.34 | 13 | 2 | 959.174 | 0.163 | 959.513 | 56.73 |
| 26 | 2 | 958.640 | 0.126 | 958.748 | 53.06 | 14 | 1 | 959.103 | 0.124 | 959.044 | 8.80 |
| 26 | 2 | 958.877 | 0.144 | 958.975 | 52.81 | 14 | 1 | 959.930 | 0.157 | 959.917 | 9.56 |
| 26 | 2 | 958.545 | 0.140 | 958.650 | 52.60 | 14 | 1 | 958.881 | 0.141 | 958.884 | 10.34 |
| 26 | 2 | 959.153 | 0.122 | 959.209 | 52.34 | 14 | 1 | 960.398 | 0.152 | 960.229 | 11.34 |
| 26 | 2 | 959.121 | 0.122 | 959.197 | 52.15 | 14 | 1 | 959.002 | 0.124 | 958.847 | 12.20 |
| 26 | 2 | 959.099 | 0.107 | 959.183 | 51.96 | 14 | 1 | 959.438 | 0.113 | 959.337 | 13.18 |
| 26 | 2 | 960.147 | 0.147 | 960.233 | 51.75 | 14 | 1 | 959.715 | 0.138 | 959.680 | 14.17 |
| 26 | 2 | 958.852 | 0.136 | 958.938 | 51.58 | 14 | 1 | 959.688 | 0.143 | 959.677 | 15.21 |
| 26 | 2 | 959.354 | 0.129 | 959.442 | 51.39 | 14 | 1 | 959.460 | 0.141 | 959.501 | 16.32 |
| 26 | 2 | 959.201 | 0.139 | 959.287 | 51.24 | 20 | 1 | 959.331 | 0.101 | 959.150 | 17.11 |
| 26 | 2 | 959.029 | 0.159 | 959.151 | 51.12 | 20 | 1 | 959.204 | 0.109 | 958.994 | 18.21 |
| 26 | 2 | 958.982 | 0.179 | 959.143 | 50.99 | 20 | 1 | 959.039 | 0.131 | 958.882 | 19.65 |

| 2003 - MARCO | | | | |
|---|---|---|---|---|
| D | L | SDB | ER | SDC | HL |
| 20 | 1 | 959.629 | 0.125 | 959.506 | 21.32 |
| 20 | 1 | 959.817 | 0.127 | 959.734 | 22.75 |
| 20 | 1 | 958.592 | 0.157 | 958.511 | 24.18 |
| 20 | 2 | 959.114 | 0.148 | 959.289 | 62.80 |
| 20 | 2 | 959.704 | 0.115 | 959.876 | 62.31 |
| 20 | 2 | 959.053 | 0.135 | 959.232 | 61.82 |
| 20 | 2 | 959.027 | 0.141 | 959.212 | 61.37 |
| 20 | 2 | 958.838 | 0.158 | 959.042 | 60.94 |
| 20 | 2 | 958.433 | 0.139 | 958.651 | 60.52 |
| 20 | 2 | 958.316 | 0.152 | 958.594 | 59.97 |
| 20 | 2 | 958.301 | 0.138 | 958.612 | 59.59 |
| 20 | 2 | 959.279 | 0.162 | 959.570 | 59.14 |
| 20 | 2 | 958.052 | 0.169 | 958.323 | 58.80 |
| 31 | 1 | 959.958 | 0.144 | 959.900 | 8.09 |
| 31 | 1 | 959.529 | 0.154 | 959.438 | 8.70 |
| 31 | 1 | 959.021 | 0.142 | 958.973 | 9.41 |
| 31 | 1 | 959.560 | 0.119 | 959.429 | 10.05 |
| 31 | 1 | 959.252 | 0.131 | 959.086 | 10.66 |
| 31 | 1 | 959.834 | 0.136 | 959.646 | 11.33 |
| 31 | 1 | 959.415 | 0.121 | 959.214 | 12.03 |
| 31 | 1 | 960.514 | 0.132 | 960.309 | 12.73 |
| 31 | 1 | 960.029 | 0.128 | 959.847 | 13.48 |
| 31 | 1 | 959.846 | 0.100 | 959.707 | 14.41 |
| 31 | 1 | 959.231 | 0.119 | 959.079 | 15.35 |
| 31 | 1 | 959.443 | 0.118 | 959.348 | 16.37 |
| 31 | 1 | 959.101 | 0.141 | 958.996 | 17.32 |
| 31 | 1 | 958.783 | 0.137 | 958.688 | 18.41 |
| 31 | 2 | 959.540 | 0.146 | 959.772 | 68.24 |
| 31 | 2 | 958.546 | 0.157 | 958.785 | 67.52 |
| 31 | 2 | 959.188 | 0.163 | 959.432 | 66.89 |
| 31 | 2 | 958.999 | 0.155 | 959.263 | 66.31 |
| 31 | 2 | 958.829 | 0.145 | 959.083 | 65.77 |
| 31 | 2 | 959.139 | 0.132 | 959.402 | 65.23 |
| 31 | 2 | 958.398 | 0.159 | 958.706 | 64.70 |
| 31 | 2 | 959.725 | 0.177 | 960.022 | 64.17 |
| 31 | 2 | 959.203 | 0.187 | 959.515 | 63.65 |
| 31 | 2 | 959.972 | 0.192 | 960.300 | 63.16 |
| 31 | 2 | 959.070 | 0.128 | 959.320 | 62.59 |
| 31 | 2 | 958.999 | 0.149 | 959.283 | 62.17 |
| 31 | 2 | 959.010 | 0.138 | 959.342 | 61.30 |

| 2003 - ABRIL | | | | |
|---|---|---|---|---|
| D | L | SDB | ER | SDC | HL |
| 01 | 1 | 958.875 | 0.109 | 958.891 | 11.64 |
| 01 | 1 | 959.747 | 0.158 | 959.853 | 12.37 |
| 01 | 1 | 959.402 | 0.086 | 959.351 | 16.33 |
| 01 | 1 | 959.046 | 0.117 | 958.878 | 18.51 |
| 01 | 1 | 958.698 | 0.112 | 958.601 | 19.57 |
| 01 | 2 | 959.501 | 0.120 | 959.539 | 68.77 |
| 01 | 2 | 959.244 | 0.115 | 959.232 | 68.05 |
| 01 | 2 | 959.152 | 0.116 | 959.184 | 67.35 |
| 01 | 2 | 959.479 | 0.134 | 959.517 | 66.75 |
| 01 | 2 | 959.873 | 0.117 | 959.911 | 66.15 |
| 01 | 2 | 959.067 | 0.131 | 959.097 | 65.56 |
| 01 | 2 | 959.369 | 0.187 | 959.437 | 65.05 |
| 01 | 2 | 958.981 | 0.160 | 959.037 | 64.49 |
| 01 | 2 | 959.799 | 0.172 | 959.880 | 64.00 |
| 01 | 2 | 958.966 | 0.148 | 958.993 | 63.44 |
| 01 | 2 | 958.109 | 0.146 | 958.097 | 62.98 |
| 01 | 2 | 959.568 | 0.130 | 959.566 | 62.55 |
| 01 | 2 | 958.750 | 0.144 | 958.754 | 62.12 |
| 02 | 1 | 958.574 | 0.103 | 958.344 | 8.86 |
| 02 | 1 | 959.468 | 0.113 | 959.263 | 9.46 |
| 02 | 1 | 959.152 | 0.130 | 958.963 | 10.05 |
| 02 | 1 | 960.047 | 0.120 | 959.880 | 10.74 |
| 02 | 1 | 960.160 | 0.116 | 960.023 | 11.38 |
| 02 | 1 | 959.337 | 0.118 | 959.242 | 12.11 |
| 02 | 1 | 959.482 | 0.138 | 959.423 | 12.85 |
| 02 | 1 | 959.202 | 0.125 | 959.195 | 13.65 |
| 02 | 1 | 959.369 | 0.102 | 959.069 | 14.65 |

| 2003 - ABRIL | | | | |
|---|---|---|---|---|
| D | L | SDB | ER | SDC | HL |
| 02 | 1 | 959.592 | 0.119 | 959.333 | 15.48 |
| 02 | 1 | 958.918 | 0.103 | 958.623 | 16.43 |
| 02 | 1 | 958.965 | 0.113 | 958.715 | 17.35 |
| 02 | 1 | 959.575 | 0.122 | 959.308 | 18.50 |
| 02 | 2 | 959.606 | 0.128 | 959.572 | 71.89 |
| 02 | 2 | 960.723 | 0.148 | 960.697 | 71.02 |
| 02 | 2 | 959.001 | 0.148 | 958.980 | 70.28 |
| 02 | 2 | 958.668 | 0.161 | 958.669 | 69.59 |
| 02 | 2 | 958.364 | 0.191 | 958.384 | 68.93 |
| 02 | 2 | 958.993 | 0.159 | 959.031 | 68.29 |
| 02 | 2 | 958.862 | 0.165 | 958.967 | 67.60 |
| 02 | 2 | 958.346 | 0.157 | 958.377 | 66.93 |
| 02 | 2 | 958.458 | 0.117 | 958.444 | 66.29 |
| 02 | 2 | 959.030 | 0.144 | 959.052 | 65.73 |
| 02 | 2 | 959.116 | 0.139 | 959.190 | 65.19 |
| 03 | 1 | 959.200 | 0.122 | 959.168 | 10.68 |
| 03 | 1 | 959.027 | 0.107 | 959.014 | 13.38 |
| 03 | 1 | 959.112 | 0.125 | 959.166 | 14.17 |
| 03 | 1 | 959.859 | 0.142 | 959.762 | 15.22 |
| 08 | 1 | 958.268 | 0.119 | 958.027 | 11.08 |
| 08 | 1 | 959.275 | 0.097 | 959.098 | 11.71 |
| 08 | 1 | 958.406 | 0.100 | 958.254 | 12.42 |
| 08 | 1 | 958.983 | 0.090 | 958.967 | 16.11 |
| 08 | 1 | 959.631 | 0.117 | 959.613 | 16.96 |
| 08 | 1 | 959.997 | 0.102 | 959.811 | 18.01 |
| 08 | 1 | 959.050 | 0.129 | 958.814 | 18.94 |
| 08 | 1 | 959.110 | 0.108 | 959.017 | 19.98 |
| 09 | 1 | 958.671 | 0.101 | 958.478 | 11.36 |
| 09 | 1 | 958.837 | 0.100 | 958.664 | 12.03 |
| 09 | 1 | 959.558 | 0.111 | 959.430 | 12.68 |
| 09 | 1 | 959.295 | 0.113 | 959.140 | 13.42 |
| 09 | 1 | 959.092 | 0.121 | 958.951 | 14.13 |
| 09 | 1 | 959.418 | 0.115 | 959.302 | 14.91 |
| 09 | 1 | 958.971 | 0.130 | 958.936 | 15.72 |
| 09 | 1 | 959.553 | 0.116 | 959.498 | 16.60 |
| 09 | 1 | 959.768 | 0.130 | 959.720 | 17.46 |
| 09 | 1 | 959.654 | 0.142 | 959.627 | 18.35 |
| 09 | 1 | 959.039 | 0.145 | 959.073 | 19.31 |
| 09 | 1 | 960.045 | 0.103 | 959.847 | 20.39 |
| 09 | 1 | 958.263 | 0.115 | 958.016 | 21.51 |
| 09 | 1 | 959.889 | 0.115 | 959.691 | 22.65 |
| 09 | 1 | 960.038 | 0.121 | 959.889 | 23.83 |
| 09 | 1 | 959.212 | 0.123 | 959.033 | 25.20 |
| 14 | 1 | 958.813 | 0.174 | 958.701 | 15.86 |
| 14 | 1 | 959.330 | 0.139 | 959.292 | 16.63 |
| 14 | 1 | 959.238 | 0.106 | 959.241 | 17.45 |
| 14 | 1 | 959.139 | 0.130 | 959.090 | 18.27 |
| 14 | 1 | 959.077 | 0.154 | 959.034 | 19.15 |
| 14 | 1 | 959.301 | 0.135 | 959.287 | 20.10 |
| 14 | 1 | 959.134 | 0.124 | 958.951 | 21.39 |
| 14 | 1 | 959.465 | 0.140 | 959.312 | 22.45 |
| 14 | 1 | 959.241 | 0.104 | 959.084 | 23.60 |
| 14 | 1 | 959.698 | 0.139 | 959.493 | 24.76 |
| 14 | 1 | 959.063 | 0.135 | 958.895 | 26.09 |
| 14 | 1 | 959.157 | 0.120 | 959.007 | 27.46 |
| 14 | 1 | 959.131 | 0.127 | 958.964 | 28.99 |
| 16 | 1 | 960.110 | 0.120 | 959.947 | 13.51 |
| 16 | 1 | 958.855 | 0.129 | 958.655 | 14.19 |
| 16 | 1 | 959.276 | 0.134 | 959.116 | 14.99 |
| 16 | 1 | 959.365 | 0.130 | 959.183 | 15.75 |
| 16 | 1 | 959.616 | 0.116 | 959.468 | 16.54 |
| 16 | 1 | 959.348 | 0.096 | 959.186 | 17.39 |
| 16 | 1 | 958.883 | 0.131 | 958.888 | 18.26 |
| 16 | 1 | 958.952 | 0.121 | 958.923 | 19.16 |
| 16 | 1 | 959.979 | 0.111 | 959.960 | 20.06 |
| 16 | 1 | 959.666 | 0.101 | 959.578 | 24.60 |
| 16 | 1 | 958.918 | 0.103 | 958.734 | 25.97 |
| 16 | 1 | 959.068 | 0.085 | 958.879 | 27.23 |
| 17 | 1 | 959.715 | 0.105 | 959.632 | 13.87 |
| 17 | 1 | 958.948 | 0.103 | 958.845 | 14.58 |
| 17 | 1 | 959.696 | 0.128 | 959.607 | 15.34 |

| 2003 - ABRIL | | | | | | 2003 - ABRIL | | | | |
|---|---|---|---|---|---|---|---|---|---|---|
| D | L | SDB | ER | SDC | HL | D | L | SDB | ER | SDC | HL |
| 17 | 1 | 960.224 | 0.150 | 960.198 | 16.11 | 25 | 1 | 959.118 | 0.126 | 958.970 | 21.16 |
| 17 | 1 | 959.156 | 0.157 | 959.114 | 16.90 | 25 | 1 | 958.980 | 0.136 | 958.889 | 22.09 |
| 17 | 1 | 958.718 | 0.148 | 958.646 | 17.74 | 25 | 1 | 960.169 | 0.143 | 960.035 | 23.11 |
| 17 | 1 | 959.010 | 0.137 | 959.040 | 18.76 | 25 | 1 | 959.264 | 0.153 | 959.162 | 24.10 |
| 17 | 1 | 959.235 | 0.144 | 959.199 | 19.71 | 25 | 1 | 959.406 | 0.124 | 959.357 | 25.18 |
| 17 | 1 | 959.414 | 0.123 | 959.408 | 20.69 | 25 | 1 | 959.442 | 0.173 | 959.405 | 26.57 |
| 17 | 1 | 959.001 | 0.123 | 959.009 | 21.65 | 25 | 1 | 959.973 | 0.138 | 959.901 | 27.83 |
| 17 | 1 | 959.347 | 0.122 | 959.357 | 22.70 | 25 | 1 | 959.961 | 0.116 | 959.853 | 29.08 |
| 17 | 1 | 960.544 | 0.164 | 960.428 | 23.95 | 25 | 1 | 959.487 | 0.118 | 959.338 | 30.40 |
| 17 | 1 | 959.550 | 0.132 | 959.438 | 25.11 | 25 | 2 | 958.333 | 0.123 | 958.453 | 74.89 |
| 17 | 1 | 958.510 | 0.193 | 958.372 | 26.37 | 25 | 2 | 958.833 | 0.120 | 958.959 | 74.00 |
| 17 | 1 | 959.283 | 0.122 | 959.169 | 27.65 | 25 | 2 | 958.877 | 0.154 | 959.010 | 73.17 |
| 17 | 2 | 959.236 | 0.175 | 959.335 | 75.11 | 25 | 2 | 958.866 | 0.160 | 959.006 | 72.38 |
| 17 | 2 | 959.500 | 0.164 | 959.592 | 74.14 | 25 | 2 | 958.398 | 0.157 | 958.535 | 71.61 |
| 17 | 2 | 959.159 | 0.163 | 959.218 | 73.25 | 25 | 2 | 959.086 | 0.138 | 959.233 | 70.83 |
| 17 | 2 | 959.311 | 0.137 | 959.354 | 72.48 | 25 | 2 | 959.324 | 0.124 | 959.486 | 70.00 |
| 17 | 2 | 958.924 | 0.143 | 958.964 | 71.72 | 25 | 2 | 959.662 | 0.121 | 959.828 | 69.30 |
| 17 | 2 | 958.472 | 0.140 | 958.523 | 70.93 | 25 | 2 | 958.931 | 0.137 | 959.111 | 68.60 |
| 17 | 2 | 958.990 | 0.150 | 959.047 | 70.20 | 25 | 2 | 958.895 | 0.160 | 959.098 | 67.96 |
| 17 | 2 | 959.026 | 0.176 | 959.102 | 69.40 | 25 | 2 | 959.680 | 0.165 | 959.892 | 67.29 |
| 17 | 2 | 958.893 | 0.170 | 958.949 | 68.76 | 28 | 2 | 958.694 | 0.120 | 958.846 | 72.44 |
| 17 | 2 | 959.336 | 0.148 | 959.411 | 68.13 | 28 | 2 | 959.407 | 0.119 | 959.563 | 71.64 |
| 17 | 2 | 958.517 | 0.169 | 958.605 | 67.53 | 28 | 2 | 958.236 | 0.121 | 958.427 | 70.82 |
| 22 | 1 | 959.544 | 0.141 | 959.518 | 21.82 | 28 | 2 | 958.208 | 0.128 | 958.424 | 70.06 |
| 22 | 1 | 959.103 | 0.135 | 959.053 | 22.94 | 28 | 2 | 959.417 | 0.156 | 959.645 | 68.75 |
| 22 | 1 | 959.373 | 0.133 | 959.286 | 23.98 | 28 | 2 | 959.317 | 0.152 | 959.560 | 68.04 |
| 22 | 1 | 959.855 | 0.119 | 959.776 | 25.05 | 28 | 2 | 959.141 | 0.121 | 959.372 | 67.39 |
| 22 | 1 | 958.909 | 0.127 | 958.790 | 26.14 | 28 | 2 | 959.364 | 0.147 | 959.600 | 66.76 |
| 22 | 1 | 959.414 | 0.120 | 959.287 | 27.34 | 28 | 2 | 957.847 | 0.169 | 958.102 | 66.11 |
| 22 | 1 | 959.443 | 0.136 | 959.314 | 28.58 | 29 | 1 | 960.088 | 0.127 | 960.049 | 19.15 |
| 22 | 1 | 959.254 | 0.116 | 959.161 | 29.89 | 29 | 1 | 959.359 | 0.114 | 959.379 | 20.11 |
| 22 | 1 | 959.521 | 0.109 | 958.495 | 31.35 | 29 | 1 | 959.417 | 0.123 | 959.474 | 23.94 |
| 22 | 1 | 959.240 | 0.104 | 959.107 | 33.04 | 29 | 1 | 959.608 | 0.108 | 959.576 | 24.95 |
| 22 | 2 | 959.892 | 0.208 | 960.078 | 72.59 | 29 | 1 | 959.458 | 0.128 | 959.405 | 25.96 |
| 22 | 2 | 959.426 | 0.202 | 959.638 | 71.79 | 29 | 1 | 959.580 | 0.099 | 959.535 | 27.12 |
| 22 | 2 | 958.202 | 0.293 | 958.414 | 70.72 | 29 | 1 | 959.462 | 0.110 | 959.391 | 28.31 |
| 22 | 2 | 958.834 | 0.158 | 959.028 | 69.50 | 29 | 1 | 958.660 | 0.095 | 958.605 | 29.55 |
| 22 | 2 | 958.530 | 0.210 | 958.722 | 68.84 | 29 | 1 | 959.298 | 0.113 | 959.190 | 30.92 |
| 22 | 2 | 958.585 | 0.225 | 958.791 | 68.18 | 29 | 1 | 959.498 | 0.120 | 959.383 | 32.32 |
| 22 | 2 | 958.848 | 0.178 | 959.042 | 67.58 | 29 | 1 | 959.831 | 0.119 | 959.700 | 33.84 |
| 22 | 2 | 958.922 | 0.164 | 959.096 | 66.95 | 29 | 1 | 958.025 | 0.142 | 957.893 | 35.36 |
| 22 | 2 | 958.931 | 0.199 | 959.133 | 66.40 | 29 | 2 | 958.674 | 0.153 | 958.870 | 70.68 |
| 22 | 2 | 959.261 | 0.166 | 959.440 | 65.75 | 29 | 2 | 959.647 | 0.142 | 959.856 | 69.97 |
| 22 | 2 | 958.310 | 0.194 | 958.518 | 65.18 | 29 | 2 | 959.617 | 0.171 | 959.832 | 69.17 |
| 22 | 2 | 959.209 | 0.175 | 959.440 | 64.68 | 29 | 2 | 958.935 | 0.165 | 959.154 | 68.43 |
| 24 | 1 | 958.207 | 0.129 | 958.105 | 15.85 | 29 | 2 | 960.331 | 0.230 | 960.573 | 66.23 |
| 24 | 1 | 959.216 | 0.125 | 959.112 | 16.56 | 30 | 1 | 958.192 | 0.200 | 958.200 | 19.84 |
| 24 | 1 | 959.506 | 0.147 | 959.354 | 17.27 | 30 | 1 | 958.637 | 0.202 | 958.661 | 20.65 |
| 24 | 1 | 959.180 | 0.161 | 958.989 | 18.09 | 30 | 1 | 958.841 | 0.164 | 958.865 | 22.49 |
| 24 | 1 | 959.685 | 0.191 | 959.502 | 18.91 | 30 | 1 | 959.471 | 0.162 | 959.519 | 23.47 |
| 24 | 1 | 958.913 | 0.142 | 958.756 | 19.70 | 30 | 1 | 958.700 | 0.179 | 958.731 | 26.68 |
| 24 | 1 | 959.176 | 0.154 | 959.086 | 20.57 | 30 | 1 | 959.451 | 0.146 | 959.444 | 27.77 |
| 24 | 1 | 958.818 | 0.114 | 958.750 | 21.57 | 30 | 1 | 959.356 | 0.153 | 959.320 | 28.90 |
| 24 | 1 | 959.499 | 0.136 | 959.428 | 22.54 | 30 | 1 | 959.744 | 0.179 | 959.684 | 30.10 |
| 24 | 1 | 958.662 | 0.130 | 958.579 | 23.66 | 30 | 1 | 959.548 | 0.171 | 959.519 | 31.52 |
| 24 | 1 | 959.618 | 0.117 | 959.497 | 24.97 | 30 | 2 | 959.786 | 0.166 | 959.968 | 79.21 |
| 24 | 1 | 959.003 | 0.155 | 958.895 | 26.17 | 30 | 2 | 958.685 | 0.193 | 958.891 | 76.88 |
| 24 | 1 | 959.714 | 0.137 | 959.596 | 27.41 | 30 | 2 | 959.349 | 0.166 | 959.547 | 75.75 |
| 24 | 1 | 959.705 | 0.141 | 959.542 | 28.69 | 30 | 2 | 958.958 | 0.225 | 959.166 | 74.80 |
| 24 | 1 | 959.427 | 0.141 | 959.265 | 29.99 | 30 | 2 | 958.702 | 0.194 | 958.914 | 73.93 |
| 24 | 2 | 959.505 | 0.160 | 959.646 | 71.26 | 30 | 2 | 959.006 | 0.192 | 959.226 | 72.97 |
| 24 | 2 | 959.168 | 0.148 | 959.295 | 70.54 | 30 | 2 | 958.614 | 0.210 | 958.822 | 72.17 |
| 24 | 2 | 959.209 | 0.176 | 959.369 | 69.78 | | | | | | |
| 24 | 2 | 959.322 | 0.165 | 959.513 | 69.09 | | | | | | |
| 24 | 2 | 959.043 | 0.175 | 959.229 | 68.44 | 2003 - MAIO | | | | | |
| 24 | 2 | 957.921 | 0.150 | 958.111 | 67.77 | D | L | SDB | ER | SDC | HL |
| 24 | 2 | 959.404 | 0.173 | 959.602 | 67.16 | 07 | 2 | 960.105 | 0.122 | 960.171 | 73.29 |
| 24 | 2 | 958.564 | 0.135 | 958.776 | 66.50 | 07 | 2 | 959.286 | 0.148 | 959.339 | 72.42 |
| 24 | 2 | 958.965 | 0.146 | 959.182 | 65.84 | 07 | 2 | 959.558 | 0.146 | 959.627 | 71.57 |
| 24 | 2 | 958.683 | 0.192 | 958.918 | 65.29 | 07 | 2 | 958.696 | 0.143 | 958.782 | 70.72 |
| 25 | 1 | 958.779 | 0.101 | 958.621 | 20.29 | 07 | 2 | 958.848 | 0.170 | 958.939 | 69.92 |

| 2003 - MAIO | | | | | | 2003 - MAIO | | | | |
|---|---|---|---|---|---|---|---|---|---|---|
| D | L | SDB | ER | SDC | HL | D | L | SDB | ER | SDC | HL |
| 07 | 2 | 958.883 | 0.146 | 958.981 | 69.15 | 19 | 1 | 959.198 | 0.168 | 958.780 | 32.41 |
| 07 | 2 | 958.080 | 0.174 | 958.196 | 68.37 | 19 | 1 | 960.339 | 0.136 | 959.915 | 33.49 |
| 07 | 2 | 959.186 | 0.181 | 959.325 | 67.62 | 19 | 1 | 959.418 | 0.134 | 958.935 | 34.57 |
| 07 | 2 | 958.483 | 0.240 | 958.630 | 66.93 | 19 | 1 | 959.466 | 0.135 | 958.954 | 35.76 |
| 08 | 2 | 960.042 | 0.148 | 960.120 | 76.97 | 19 | 1 | 958.870 | 0.140 | 958.347 | 36.95 |
| 08 | 2 | 959.645 | 0.142 | 959.705 | 75.83 | 19 | 1 | 959.677 | 0.122 | 959.144 | 38.25 |
| 08 | 2 | 958.583 | 0.143 | 958.640 | 74.66 | 19 | 1 | 959.317 | 0.143 | 958.775 | 39.58 |
| 08 | 2 | 958.994 | 0.159 | 959.042 | 73.64 | 19 | 1 | 959.528 | 0.149 | 959.009 | 41.02 |
| 08 | 2 | 960.012 | 0.134 | 960.062 | 72.58 | 19 | 1 | 959.402 | 0.142 | 958.894 | 42.53 |
| 08 | 2 | 959.209 | 0.144 | 959.265 | 71.70 | 19 | 1 | 958.932 | 0.182 | 958.394 | 44.21 |
| 08 | 2 | 959.316 | 0.177 | 959.402 | 70.82 | 19 | 1 | 959.062 | 0.141 | 958.544 | 45.96 |
| 08 | 2 | 958.368 | 0.156 | 958.468 | 70.00 | 19 | 2 | 958.306 | 0.195 | 958.297 | 73.36 |
| 08 | 2 | 959.453 | 0.176 | 959.582 | 69.16 | 19 | 2 | 959.383 | 0.164 | 959.348 | 72.18 |
| 08 | 2 | 959.484 | 0.196 | 959.648 | 68.20 | 19 | 2 | 958.264 | 0.221 | 958.256 | 71.17 |
| 08 | 2 | 958.122 | 0.169 | 958.295 | 67.41 | 19 | 2 | 958.857 | 0.178 | 958.846 | 70.14 |
| 09 | 1 | 958.534 | 0.161 | 958.334 | 31.17 | 19 | 2 | 958.922 | 0.177 | 958.900 | 69.21 |
| 09 | 1 | 959.555 | 0.148 | 959.449 | 32.31 | 19 | 2 | 958.724 | 0.150 | 958.710 | 68.33 |
| 09 | 1 | 959.674 | 0.132 | 959.455 | 37.39 | 19 | 2 | 959.748 | 0.176 | 959.761 | 67.52 |
| 09 | 1 | 960.316 | 0.141 | 960.093 | 38.95 | 19 | 2 | 958.784 | 0.177 | 958.783 | 66.73 |
| 09 | 1 | 959.786 | 0.166 | 959.528 | 40.64 | 19 | 2 | 958.608 | 0.165 | 958.631 | 65.99 |
| 09 | 1 | 959.983 | 0.110 | 959.724 | 42.47 | 19 | 2 | 959.527 | 0.204 | 959.574 | 65.22 |
| 12 | 2 | 958.857 | 0.195 | 958.927 | 74.07 | 20 | 1 | 959.475 | 0.168 | 959.275 | 29.39 |
| 12 | 2 | 960.229 | 0.177 | 960.268 | 73.11 | 20 | 1 | 959.373 | 0.207 | 959.118 | 30.73 |
| 12 | 2 | 959.523 | 0.154 | 959.569 | 72.17 | 20 | 1 | 958.910 | 0.137 | 958.646 | 31.71 |
| 12 | 2 | 958.731 | 0.177 | 958.795 | 71.29 | 20 | 1 | 959.704 | 0.154 | 959.405 | 32.72 |
| 12 | 2 | 959.485 | 0.163 | 959.550 | 70.43 | 20 | 1 | 959.056 | 0.159 | 958.787 | 33.86 |
| 12 | 2 | 958.821 | 0.167 | 958.925 | 69.60 | 20 | 1 | 959.314 | 0.150 | 959.048 | 35.01 |
| 12 | 2 | 958.974 | 0.190 | 959.078 | 68.79 | 20 | 1 | 959.167 | 0.156 | 958.881 | 36.24 |
| 12 | 2 | 958.053 | 0.255 | 958.197 | 67.97 | 20 | 1 | 959.489 | 0.141 | 959.251 | 37.48 |
| 12 | 2 | 959.271 | 0.245 | 959.455 | 67.22 | 20 | 1 | 959.185 | 0.123 | 958.922 | 38.77 |
| 12 | 2 | 959.213 | 0.243 | 959.406 | 66.48 | 20 | 1 | 959.136 | 0.123 | 958.725 | 40.10 |
| 14 | 1 | 958.822 | 0.172 | 958.814 | 25.66 | 20 | 1 | 960.629 | 0.137 | 960.188 | 41.59 |
| 14 | 1 | 959.258 | 0.151 | 959.189 | 26.48 | 20 | 1 | 959.896 | 0.155 | 959.530 | 43.14 |
| 14 | 1 | 959.724 | 0.159 | 959.445 | 33.65 | 20 | 1 | 959.687 | 0.138 | 959.420 | 44.92 |
| 14 | 1 | 959.250 | 0.155 | 959.107 | 34.86 | 20 | 2 | 959.470 | 0.112 | 959.481 | 75.35 |
| 14 | 1 | 959.365 | 0.176 | 959.310 | 36.45 | 20 | 2 | 958.734 | 0.158 | 958.743 | 74.19 |
| 14 | 1 | 959.947 | 0.179 | 959.832 | 37.87 | 20 | 2 | 959.712 | 0.145 | 959.724 | 73.12 |
| 15 | 1 | 959.987 | 0.110 | 959.601 | 28.40 | 20 | 2 | 959.014 | 0.169 | 959.056 | 72.11 |
| 15 | 1 | 959.948 | 0.135 | 959.527 | 29.34 | 20 | 2 | 959.548 | 0.185 | 959.592 | 69.96 |
| 15 | 1 | 959.818 | 0.127 | 959.422 | 30.32 | 20 | 2 | 957.116 | 0.159 | 957.140 | 69.02 |
| 15 | 1 | 959.782 | 0.117 | 959.359 | 31.33 | 20 | 2 | 959.228 | 0.224 | 959.298 | 68.17 |
| 15 | 1 | 959.863 | 0.129 | 959.445 | 32.40 | 21 | 1 | 960.940 | 0.122 | 960.595 | 30.39 |
| 15 | 1 | 959.807 | 0.139 | 959.393 | 33.68 | 21 | 1 | 958.929 | 0.158 | 958.538 | 31.39 |
| 15 | 1 | 959.007 | 0.165 | 958.645 | 34.91 | 21 | 1 | 958.662 | 0.113 | 958.267 | 32.37 |
| 15 | 1 | 959.982 | 0.113 | 959.669 | 36.24 | 21 | 1 | 959.734 | 0.127 | 959.334 | 33.37 |
| 15 | 1 | 959.473 | 0.153 | 959.251 | 37.86 | 21 | 1 | 958.872 | 0.133 | 958.444 | 34.43 |
| 15 | 1 | 959.921 | 0.147 | 959.733 | 39.38 | 21 | 1 | 959.786 | 0.140 | 959.390 | 35.50 |
| 15 | 2 | 959.087 | 0.203 | 959.148 | 73.92 | 21 | 1 | 959.044 | 0.126 | 958.621 | 36.66 |
| 16 | 1 | 958.982 | 0.117 | 958.690 | 27.27 | 21 | 1 | 959.227 | 0.120 | 958.855 | 37.88 |
| 16 | 1 | 959.638 | 0.130 | 959.307 | 28.16 | 21 | 1 | 959.824 | 0.156 | 959.323 | 39.20 |
| 16 | 1 | 959.411 | 0.144 | 959.092 | 29.06 | 21 | 1 | 959.427 | 0.113 | 958.949 | 40.61 |
| 16 | 1 | 959.622 | 0.124 | 959.254 | 29.99 | 21 | 1 | 959.214 | 0.140 | 958.770 | 42.07 |
| 16 | 1 | 959.728 | 0.124 | 959.336 | 30.98 | 21 | 1 | 959.243 | 0.125 | 958.774 | 43.68 |
| 16 | 1 | 959.738 | 0.113 | 959.329 | 32.23 | 21 | 2 | 959.708 | 0.155 | 959.667 | 76.64 |
| 16 | 1 | 959.370 | 0.134 | 958.917 | 33.30 | 21 | 2 | 959.027 | 0.214 | 959.065 | 75.39 |
| 16 | 1 | 959.667 | 0.122 | 959.216 | 34.46 | 21 | 2 | 959.563 | 0.136 | 959.546 | 73.32 |
| 16 | 1 | 960.044 | 0.146 | 959.607 | 35.67 | 21 | 2 | 957.409 | 0.142 | 957.385 | 72.31 |
| 16 | 1 | 959.728 | 0.108 | 959.286 | 36.93 | 21 | 2 | 958.937 | 0.140 | 958.957 | 71.34 |
| 16 | 1 | 959.862 | 0.163 | 959.405 | 38.29 | 21 | 2 | 959.135 | 0.152 | 959.149 | 70.35 |
| 16 | 2 | 960.026 | 0.166 | 960.028 | 76.21 | 21 | 2 | 959.115 | 0.158 | 959.117 | 69.42 |
| 16 | 2 | 959.324 | 0.124 | 959.318 | 75.06 | 21 | 2 | 958.900 | 0.174 | 958.918 | 68.55 |
| 16 | 2 | 959.425 | 0.139 | 959.437 | 73.91 | 21 | 2 | 958.280 | 0.174 | 958.341 | 66.87 |
| 16 | 2 | 959.338 | 0.148 | 959.365 | 72.86 | 22 | 1 | 959.016 | 0.140 | 958.716 | 29.39 |
| 16 | 2 | 959.557 | 0.159 | 959.600 | 71.80 | 22 | 1 | 959.108 | 0.168 | 958.832 | 30.29 |
| 16 | 2 | 958.994 | 0.138 | 959.051 | 70.80 | 22 | 1 | 959.416 | 0.135 | 959.103 | 31.28 |
| 16 | 2 | 959.281 | 0.125 | 959.332 | 69.90 | 22 | 1 | 959.773 | 0.127 | 959.487 | 32.25 |
| 16 | 2 | 959.322 | 0.187 | 959.418 | 69.00 | 22 | 1 | 959.497 | 0.142 | 959.182 | 33.33 |
| 16 | 2 | 958.609 | 0.149 | 958.685 | 68.08 | 22 | 1 | 959.581 | 0.144 | 959.218 | 34.54 |
| 16 | 2 | 959.134 | 0.170 | 959.238 | 67.28 | 22 | 1 | 959.454 | 0.125 | 959.073 | 35.62 |
| 19 | 1 | 959.994 | 0.149 | 959.650 | 30.33 | 22 | 1 | 959.726 | 0.141 | 959.324 | 36.80 |
| 19 | 1 | 959.316 | 0.129 | 958.999 | 31.41 | 22 | 1 | 959.546 | 0.143 | 959.089 | 38.04 |

| 2003 - MAIO | | | | |
|---|---|---|---|---|
| D | L | SDB | ER | SDC | HL |
| 22 | 1 | 959.698 | 0.126 | 959.176 | 39.35 |
| 22 | 1 | 959.818 | 0.113 | 959.309 | 40.70 |
| 22 | 2 | 958.731 | 0.161 | 958.687 | 76.59 |
| 22 | 2 | 958.745 | 0.142 | 958.712 | 75.34 |
| 22 | 2 | 958.672 | 0.169 | 958.659 | 74.16 |
| 22 | 2 | 958.847 | 0.148 | 958.846 | 73.02 |
| 22 | 2 | 958.376 | 0.180 | 958.392 | 71.94 |
| 22 | 2 | 958.705 | 0.170 | 958.722 | 70.82 |
| 22 | 2 | 958.189 | 0.168 | 958.228 | 69.82 |
| 22 | 2 | 958.671 | 0.201 | 958.725 | 68.76 |
| 22 | 2 | 959.291 | 0.174 | 959.329 | 67.87 |
| 22 | 2 | 959.731 | 0.169 | 959.773 | 66.98 |
| 23 | 1 | 959.991 | 0.088 | 959.635 | 30.97 |
| 23 | 1 | 959.252 | 0.101 | 958.910 | 31.91 |
| 23 | 1 | 959.533 | 0.096 | 959.160 | 32.89 |
| 23 | 1 | 960.275 | 0.089 | 959.932 | 33.89 |
| 23 | 1 | 959.602 | 0.107 | 959.232 | 34.94 |
| 23 | 1 | 960.012 | 0.085 | 959.541 | 36.10 |
| 23 | 1 | 959.736 | 0.085 | 959.237 | 37.25 |
| 23 | 1 | 960.082 | 0.091 | 959.579 | 38.44 |
| 23 | 1 | 959.660 | 0.098 | 959.132 | 39.70 |
| 23 | 1 | 959.132 | 0.088 | 958.623 | 41.02 |
| 23 | 1 | 959.964 | 0.089 | 959.441 | 42.43 |
| 23 | 2 | 957.707 | 0.109 | 957.696 | 74.51 |
| 23 | 2 | 958.594 | 0.128 | 958.625 | 73.38 |
| 23 | 2 | 959.113 | 0.120 | 959.166 | 72.25 |
| 23 | 2 | 958.631 | 0.134 | 958.706 | 71.13 |
| 23 | 2 | 958.450 | 0.107 | 958.510 | 70.11 |
| 23 | 2 | 960.094 | 0.121 | 960.188 | 69.09 |
| 23 | 2 | 959.087 | 0.130 | 959.196 | 68.14 |
| 23 | 2 | 958.713 | 0.153 | 958.832 | 67.19 |
| 23 | 2 | 958.476 | 0.139 | 959.597 | 66.35 |
| 26 | 1 | 959.389 | 0.139 | 959.049 | 32.51 |
| 26 | 1 | 959.864 | 0.187 | 959.482 | 33.44 |
| 26 | 1 | 960.561 | 0.148 | 960.178 | 34.48 |
| 26 | 1 | 959.197 | 0.124 | 958.800 | 36.08 |
| 26 | 1 | 959.205 | 0.164 | 958.748 | 37.23 |
| 26 | 1 | 959.934 | 0.159 | 959.451 | 38.37 |
| 26 | 1 | 959.660 | 0.160 | 959.207 | 39.64 |
| 26 | 1 | 959.617 | 0.137 | 959.110 | 41.04 |
| 26 | 1 | 959.770 | 0.163 | 959.222 | 42.59 |
| 26 | 1 | 959.846 | 0.155 | 959.319 | 44.00 |
| 26 | 1 | 959.592 | 0.186 | 959.071 | 45.53 |
| 26 | 1 | 960.075 | 0.138 | 959.581 | 47.16 |
| 26 | 1 | 959.568 | 0.144 | 959.085 | 48.98 |
| 26 | 2 | 959.063 | 0.202 | 958.976 | 74.68 |
| 26 | 2 | 958.364 | 0.125 | 958.298 | 70.79 |
| 26 | 2 | 958.336 | 0.188 | 958.310 | 69.73 |
| 26 | 2 | 958.907 | 0.157 | 958.867 | 68.52 |
| 26 | 2 | 958.395 | 0.145 | 958.334 | 67.54 |
| 26 | 2 | 958.021 | 0.253 | 958.022 | 66.65 |
| 26 | 2 | 958.631 | 0.146 | 958.585 | 65.76 |
| 26 | 2 | 959.653 | 0.216 | 959.636 | 64.94 |
| 27 | 1 | 960.096 | 0.162 | 959.797 | 38.17 |
| 27 | 1 | 960.018 | 0.137 | 959.692 | 39.34 |
| 27 | 1 | 959.756 | 0.144 | 959.383 | 40.56 |
| 27 | 1 | 959.337 | 0.145 | 958.913 | 41.92 |
| 27 | 1 | 959.279 | 0.147 | 958.832 | 43.62 |
| 27 | 1 | 958.606 | 0.160 | 958.227 | 45.17 |
| 27 | 1 | 959.653 | 0.147 | 959.289 | 46.74 |
| 27 | 1 | 958.921 | 0.153 | 958.626 | 48.56 |
| 27 | 1 | 959.013 | 0.130 | 958.732 | 50.53 |
| 27 | 2 | 958.798 | 0.147 | 958.763 | 73.62 |
| 27 | 2 | 958.782 | 0.179 | 958.771 | 72.45 |
| 27 | 2 | 959.223 | 0.209 | 959.219 | 69.98 |
| 27 | 2 | 959.347 | 0.201 | 959.356 | 68.95 |
| 28 | 1 | 959.388 | 0.176 | 959.224 | 41.49 |
| 28 | 1 | 959.155 | 0.172 | 959.007 | 42.86 |
| 28 | 1 | 959.758 | 0.126 | 959.552 | 44.39 |
| 28 | 1 | 959.205 | 0.117 | 959.022 | 45.94 |
| 28 | 1 | 959.774 | 0.160 | 959.669 | 47.64 |

| 2003 - MAIO | | | | |
|---|---|---|---|---|
| D | L | SDB | ER | SDC | HL |
| 28 | 1 | 960.144 | 0.149 | 959.976 | 49.36 |
| 28 | 1 | 958.781 | 0.134 | 958.602 | 51.29 |
| 28 | 1 | 959.377 | 0.122 | 959.052 | 53.35 |
| 30 | 1 | 959.470 | 0.139 | 959.241 | 38.19 |
| 30 | 1 | 959.957 | 0.154 | 959.635 | 39.31 |
| 30 | 1 | 959.484 | 0.132 | 959.159 | 40.49 |
| 30 | 1 | 959.256 | 0.115 | 958.902 | 41.73 |
| 30 | 1 | 959.869 | 0.142 | 959.458 | 43.04 |
| 30 | 1 | 959.136 | 0.123 | 958.705 | 44.53 |
| 30 | 1 | 960.664 | 0.121 | 960.208 | 46.00 |
| 30 | 1 | 959.810 | 0.127 | 959.340 | 47.55 |
| 30 | 1 | 959.898 | 0.103 | 959.462 | 49.23 |
| 30 | 2 | 959.470 | 0.132 | 959.419 | 71.75 |
| 30 | 2 | 959.075 | 0.167 | 959.029 | 70.60 |
| 30 | 2 | 958.716 | 0.144 | 958.667 | 69.52 |
| 30 | 2 | 958.628 | 0.187 | 958.604 | 68.48 |
| 30 | 2 | 959.026 | 0.169 | 959.006 | 67.44 |
| 30 | 2 | 959.743 | 0.181 | 959.733 | 66.46 |
| 30 | 2 | 959.402 | 0.166 | 959.386 | 65.45 |
| 30 | 2 | 958.837 | 0.174 | 958.827 | 64.56 |
| 30 | 2 | 959.282 | 0.193 | 959.287 | 63.68 |

| 2003 - JUNHO | | | | |
|---|---|---|---|---|
| D | L | SDB | ER | SDC | HL |
| 02 | 1 | 959.779 | 0.139 | 959.570 | 37.14 |
| 02 | 1 | 959.050 | 0.191 | 958.873 | 38.18 |
| 02 | 1 | 959.720 | 0.149 | 959.545 | 39.24 |
| 02 | 1 | 959.804 | 0.148 | 959.527 | 40.34 |
| 02 | 1 | 960.234 | 0.173 | 959.869 | 41.46 |
| 02 | 1 | 959.808 | 0.152 | 959.445 | 42.64 |
| 02 | 1 | 959.563 | 0.155 | 959.199 | 43.90 |
| 02 | 1 | 959.773 | 0.139 | 959.363 | 45.20 |
| 02 | 1 | 959.435 | 0.125 | 959.040 | 46.60 |
| 02 | 1 | 959.928 | 0.141 | 959.563 | 48.10 |
| 02 | 1 | 960.342 | 0.171 | 959.922 | 49.68 |
| 02 | 1 | 960.663 | 0.146 | 960.184 | 51.41 |
| 02 | 2 | 959.091 | 0.182 | 959.079 | 70.38 |
| 02 | 2 | 959.010 | 0.167 | 958.998 | 69.25 |
| 02 | 2 | 958.298 | 0.163 | 958.300 | 68.19 |
| 02 | 2 | 958.123 | 0.179 | 958.135 | 67.18 |
| 02 | 2 | 958.623 | 0.145 | 958.606 | 66.23 |
| 02 | 2 | 959.593 | 0.168 | 959.592 | 65.26 |
| 02 | 2 | 958.848 | 0.204 | 958.867 | 64.38 |
| 02 | 2 | 959.055 | 0.200 | 959.079 | 63.53 |
| 02 | 2 | 959.354 | 0.241 | 959.392 | 62.71 |
| 03 | 1 | 959.893 | 0.144 | 959.642 | 40.47 |
| 03 | 1 | 959.843 | 0.146 | 959.519 | 41.58 |
| 03 | 1 | 959.340 | 0.119 | 959.039 | 42.78 |
| 03 | 1 | 959.543 | 0.143 | 959.180 | 44.00 |
| 03 | 1 | 959.892 | 0.124 | 959.472 | 45.27 |
| 03 | 1 | 959.702 | 0.139 | 959.268 | 46.61 |
| 03 | 1 | 960.155 | 0.169 | 959.703 | 48.04 |
| 03 | 1 | 960.626 | 0.157 | 960.168 | 49.58 |
| 03 | 1 | 959.835 | 0.153 | 959.494 | 51.27 |
| 03 | 1 | 960.001 | 0.137 | 959.573 | 53.09 |
| 03 | 2 | 958.162 | 0.152 | 958.147 | 71.39 |
| 03 | 2 | 958.021 | 0.160 | 957.948 | 70.24 |
| 03 | 2 | 959.442 | 0.199 | 959.406 | 69.11 |
| 03 | 2 | 958.337 | 0.153 | 958.294 | 68.04 |
| 03 | 2 | 958.116 | 0.184 | 958.133 | 62.74 |
| 03 | 2 | 957.874 | 0.303 | 957.890 | 61.92 |
| 04 | 1 | 959.923 | 0.130 | 959.652 | 37.99 |
| 04 | 1 | 958.950 | 0.160 | 958.719 | 39.03 |
| 04 | 1 | 959.394 | 0.151 | 959.114 | 40.08 |
| 04 | 1 | 959.463 | 0.141 | 959.131 | 41.15 |
| 04 | 1 | 958.713 | 0.179 | 958.385 | 42.30 |
| 04 | 1 | 958.557 | 0.140 | 958.190 | 43.51 |
| 04 | 1 | 959.077 | 0.141 | 958.691 | 44.81 |
| 04 | 1 | 959.168 | 0.184 | 958.744 | 46.13 |
| 04 | 1 | 959.296 | 0.164 | 958.841 | 47.51 |

| 2003 - JUNHO | | | | | | 2003 - JUNHO | | | | |
|---|---|---|---|---|---|---|---|---|---|---|
| D | L | SDB | ER | SDC | HL | D | L | SDB | ER | SDC | HL |
| 04 | 1 | 959.799 | 0.163 | 959.363 | 49.01 | 12 | 1 | 960.174 | 0.143 | 959.674 | 50.34 |
| 04 | 1 | 959.874 | 0.146 | 959.467 | 50.62 | 12 | 1 | 960.949 | 0.130 | 960.431 | 51.80 |
| 04 | 1 | 960.135 | 0.120 | 959.672 | 52.36 | 12 | 1 | 959.381 | 0.118 | 958.823 | 53.38 |
| 04 | 2 | 958.587 | 0.160 | 958.550 | 70.97 | 12 | 1 | 960.703 | 0.130 | 960.110 | 55.03 |
| 04 | 2 | 959.200 | 0.144 | 959.174 | 69.69 | 12 | 1 | 959.851 | 0.103 | 959.247 | 56.82 |
| 04 | 2 | 959.340 | 0.188 | 959.329 | 68.55 | 12 | 2 | 957.779 | 0.149 | 957.704 | 70.29 |
| 04 | 2 | 958.577 | 0.123 | 958.544 | 67.48 | 12 | 2 | 958.560 | 0.167 | 958.494 | 68.89 |
| 04 | 2 | 959.583 | 0.158 | 959.580 | 66.43 | 12 | 2 | 959.373 | 0.152 | 959.330 | 67.61 |
| 04 | 2 | 959.057 | 0.150 | 959.063 | 65.41 | 12 | 2 | 959.780 | 0.148 | 959.751 | 66.31 |
| 04 | 2 | 957.963 | 0.138 | 957.952 | 64.42 | 12 | 2 | 959.370 | 0.132 | 959.342 | 65.10 |
| 04 | 2 | 959.234 | 0.163 | 959.180 | 63.53 | 12 | 2 | 959.234 | 0.134 | 959.212 | 63.97 |
| 04 | 2 | 958.856 | 0.181 | 958.884 | 62.70 | 12 | 2 | 959.431 | 0.147 | 959.417 | 62.89 |
| 04 | 2 | 957.869 | 0.243 | 957.906 | 61.81 | 12 | 2 | 959.686 | 0.161 | 959.683 | 61.78 |
| 06 | 1 | 958.126 | 0.157 | 957.899 | 44.98 | 12 | 2 | 958.966 | 0.156 | 958.965 | 60.76 |
| 06 | 1 | 959.463 | 0.158 | 959.222 | 46.27 | 12 | 2 | 959.185 | 0.192 | 959.201 | 59.77 |
| 06 | 1 | 958.987 | 0.131 | 958.648 | 47.95 | 13 | 1 | 958.709 | 0.146 | 958.484 | 42.93 |
| 06 | 1 | 958.413 | 0.161 | 958.010 | 49.42 | 13 | 1 | 959.154 | 0.172 | 958.898 | 43.96 |
| 06 | 1 | 959.784 | 0.165 | 959.435 | 50.99 | 13 | 1 | 959.471 | 0.139 | 959.456 | 45.09 |
| 06 | 1 | 959.739 | 0.146 | 959.249 | 52.72 | 13 | 1 | 959.273 | 0.121 | 958.971 | 46.23 |
| 06 | 1 | 958.741 | 0.158 | 958.215 | 54.58 | 13 | 1 | 959.260 | 0.140 | 958.974 | 47.43 |
| 09 | 2 | 958.893 | 0.206 | 958.875 | 65.15 | 13 | 1 | 959.563 | 0.127 | 959.196 | 48.76 |
| 09 | 2 | 958.681 | 0.212 | 958.678 | 64.07 | 13 | 1 | 959.785 | 0.116 | 959.432 | 50.09 |
| 10 | 1 | 957.957 | 0.155 | 957.619 | 42.34 | 13 | 1 | 959.829 | 0.134 | 959.408 | 51.48 |
| 10 | 1 | 959.538 | 0.145 | 959.203 | 43.40 | 13 | 1 | 959.349 | 0.138 | 958.907 | 52.96 |
| 10 | 1 | 958.985 | 0.146 | 958.642 | 44.50 | 13 | 1 | 959.846 | 0.147 | 959.418 | 54.58 |
| 10 | 1 | 959.366 | 0.160 | 959.000 | 45.65 | 13 | 2 | 959.455 | 0.130 | 959.374 | 68.48 |
| 10 | 1 | 959.581 | 0.134 | 959.223 | 46.85 | 13 | 2 | 959.451 | 0.152 | 959.416 | 67.17 |
| 10 | 1 | 959.300 | 0.153 | 958.906 | 48.11 | 13 | 2 | 959.312 | 0.155 | 959.282 | 65.88 |
| 10 | 1 | 959.804 | 0.132 | 959.388 | 49.45 | 13 | 2 | 959.001 | 0.157 | 958.941 | 64.69 |
| 10 | 1 | 959.379 | 0.132 | 958.920 | 50.85 | 13 | 2 | 959.767 | 0.163 | 959.744 | 63.58 |
| 10 | 1 | 959.647 | 0.144 | 959.161 | 52.37 | 13 | 2 | 959.300 | 0.146 | 959.280 | 62.41 |
| 10 | 1 | 959.477 | 0.151 | 958.972 | 54.07 | 13 | 2 | 959.097 | 0.145 | 959.073 | 61.37 |
| 10 | 1 | 959.727 | 0.136 | 959.252 | 55.81 | 13 | 2 | 959.241 | 0.183 | 959.230 | 60.40 |
| 10 | 2 | 960.650 | 0.209 | 960.626 | 69.54 | 13 | 2 | 959.084 | 0.143 | 959.068 | 59.45 |
| 10 | 2 | 959.746 | 0.202 | 959.720 | 68.27 | 16 | 1 | 960.016 | 0.165 | 959.728 | 45.60 |
| 10 | 2 | 957.955 | 0.204 | 957.934 | 67.04 | 16 | 1 | 958.784 | 0.131 | 958.474 | 46.86 |
| 10 | 2 | 958.837 | 0.184 | 958.831 | 65.87 | 16 | 1 | 959.047 | 0.148 | 958.731 | 47.96 |
| 10 | 2 | 958.434 | 0.177 | 958.418 | 64.72 | 16 | 1 | 959.060 | 0.174 | 958.748 | 49.22 |
| 10 | 2 | 958.236 | 0.181 | 958.232 | 63.69 | 16 | 1 | 960.125 | 0.174 | 959.784 | 50.48 |
| 10 | 2 | 958.706 | 0.199 | 958.716 | 62.71 | 16 | 1 | 959.245 | 0.182 | 958.895 | 51.80 |
| 10 | 2 | 959.089 | 0.192 | 959.095 | 61.78 | 16 | 1 | 959.110 | 0.135 | 958.733 | 53.16 |
| 10 | 2 | 958.783 | 0.183 | 958.797 | 60.89 | 16 | 1 | 959.328 | 0.123 | 958.957 | 54.63 |
| 10 | 2 | 959.291 | 0.250 | 959.321 | 60.00 | 16 | 1 | 959.713 | 0.135 | 959.251 | 56.21 |
| 10 | 2 | 959.267 | 0.221 | 959.316 | 59.19 | 16 | 1 | 958.483 | 0.193 | 958.005 | 57.91 |
| 11 | 1 | 959.175 | 0.129 | 958.847 | 42.99 | 16 | 1 | 959.508 | 0.132 | 959.041 | 59.75 |
| 11 | 1 | 959.440 | 0.100 | 959.037 | 44.06 | 16 | 2 | 959.197 | 0.185 | 959.141 | 68.01 |
| 11 | 1 | 959.772 | 0.103 | 959.357 | 45.18 | 16 | 2 | 959.389 | 0.219 | 959.370 | 66.71 |
| 11 | 1 | 959.751 | 0.122 | 959.306 | 46.35 | 16 | 2 | 957.909 | 0.174 | 957.846 | 65.44 |
| 11 | 1 | 959.456 | 0.100 | 959.013 | 47.55 | 16 | 2 | 958.151 | 0.193 | 958.111 | 64.29 |
| 11 | 1 | 959.696 | 0.109 | 959.225 | 48.85 | 16 | 2 | 960.047 | 0.208 | 960.012 | 63.13 |
| 11 | 1 | 959.728 | 0.099 | 959.221 | 50.19 | 16 | 2 | 959.175 | 0.183 | 959.162 | 62.07 |
| 11 | 1 | 959.982 | 0.112 | 959.435 | 51.65 | 16 | 2 | 958.510 | 0.196 | 958.481 | 61.06 |
| 11 | 1 | 960.603 | 0.097 | 960.067 | 53.22 | 16 | 2 | 957.331 | 0.267 | 957.332 | 60.05 |
| 11 | 1 | 960.264 | 0.093 | 959.689 | 54.82 | 16 | 2 | 957.865 | 0.186 | 957.859 | 59.08 |
| 11 | 1 | 960.587 | 0.101 | 960.007 | 56.60 | 16 | 2 | 959.663 | 0.188 | 959.679 | 58.22 |
| 11 | 2 | 959.473 | 0.141 | 959.434 | 70.56 | 16 | 2 | 957.342 | 0.282 | 957.368 | 57.24 |
| 11 | 2 | 960.075 | 0.106 | 960.033 | 69.21 | 17 | 1 | 959.366 | 0.145 | 959.131 | 45.95 |
| 11 | 2 | 959.188 | 0.125 | 959.197 | 67.94 | 17 | 1 | 960.088 | 0.142 | 959.818 | 47.01 |
| 11 | 2 | 959.595 | 0.150 | 959.585 | 66.73 | 17 | 1 | 958.633 | 0.130 | 958.321 | 48.14 |
| 11 | 2 | 958.053 | 0.134 | 958.073 | 65.61 | 17 | 1 | 959.311 | 0.139 | 959.054 | 49.49 |
| 11 | 2 | 958.643 | 0.174 | 958.669 | 64.55 | 17 | 1 | 960.099 | 0.139 | 959.766 | 50.74 |
| 11 | 2 | 958.971 | 0.176 | 959.012 | 63.53 | 17 | 1 | 959.988 | 0.153 | 959.666 | 52.03 |
| 11 | 2 | 958.864 | 0.142 | 958.899 | 62.55 | 17 | 1 | 959.219 | 0.129 | 958.892 | 53.42 |
| 11 | 2 | 958.487 | 0.103 | 958.501 | 61.59 | 17 | 1 | 959.040 | 0.154 | 958.664 | 54.88 |
| 11 | 2 | 958.469 | 0.189 | 958.516 | 60.66 | 17 | 1 | 960.100 | 0.146 | 959.740 | 56.44 |
| 11 | 2 | 958.769 | 0.165 | 958.841 | 59.79 | 17 | 1 | 960.609 | 0.152 | 960.225 | 58.15 |
| 12 | 1 | 959.625 | 0.106 | 959.210 | 43.94 | 17 | 1 | 959.898 | 0.141 | 959.479 | 59.95 |
| 12 | 1 | 959.932 | 0.153 | 959.549 | 45.11 | 17 | 2 | 958.566 | 0.166 | 958.516 | 65.10 |
| 12 | 1 | 960.239 | 0.157 | 959.839 | 46.38 | 17 | 2 | 959.498 | 0.293 | 959.497 | 63.83 |
| 12 | 1 | 959.792 | 0.141 | 959.388 | 47.64 | 17 | 2 | 959.172 | 0.207 | 959.148 | 62.69 |
| 12 | 1 | 960.098 | 0.113 | 959.696 | 48.94 | 17 | 2 | 958.460 | 0.190 | 958.443 | 61.66 |

|  | 2003 - JUNHO | | | | |
|---|---|---|---|---|---|
| D | L | SDB | ER | SDC | HL |
| 17 | 2 | 958.493 | 0.197 | 958.475 | 60.64 |
| 17 | 2 | 959.511 | 0.179 | 959.502 | 59.69 |
| 17 | 2 | 960.169 | 0.214 | 960.158 | 58.70 |
| 17 | 2 | 959.097 | 0.189 | 959.085 | 57.79 |
| 17 | 2 | 959.047 | 0.192 | 959.060 | 56.90 |
| 18 | 1 | 959.433 | 0.118 | 959.253 | 45.39 |
| 18 | 1 | 958.491 | 0.138 | 958.304 | 46.54 |
| 18 | 1 | 959.598 | 0.150 | 959.439 | 47.63 |
| 18 | 1 | 959.426 | 0.112 | 959.299 | 48.77 |
| 18 | 1 | 960.438 | 0.095 | 960.237 | 49.94 |
| 18 | 1 | 959.224 | 0.115 | 958.999 | 51.30 |
| 18 | 1 | 959.888 | 0.116 | 959.677 | 52.64 |
| 18 | 1 | 960.203 | 0.119 | 959.915 | 54.05 |
| 18 | 1 | 958.973 | 0.118 | 958.605 | 55.54 |
| 18 | 1 | 959.610 | 0.124 | 959.303 | 57.14 |
| 18 | 2 | 958.107 | 0.138 | 958.074 | 64.29 |
| 18 | 2 | 958.493 | 0.207 | 958.505 | 63.14 |
| 18 | 2 | 958.373 | 0.178 | 958.376 | 61.01 |
| 18 | 2 | 959.326 | 0.178 | 959.342 | 59.98 |
| 18 | 2 | 959.177 | 0.152 | 959.177 | 58.99 |
| 18 | 2 | 958.822 | 0.200 | 958.841 | 58.05 |
| 18 | 2 | 958.174 | 0.222 | 958.202 | 57.16 |
| 20 | 1 | 959.035 | 0.137 | 958.877 | 46.27 |
| 20 | 1 | 959.223 | 0.137 | 959.062 | 47.32 |
| 20 | 1 | 958.741 | 0.139 | 958.564 | 48.43 |
| 20 | 1 | 959.666 | 0.138 | 959.507 | 49.60 |
| 23 | 1 | 959.285 | 0.162 | 958.999 | 48.54 |
| 23 | 1 | 959.541 | 0.134 | 959.258 | 49.71 |
| 23 | 1 | 958.782 | 0.144 | 958.494 | 50.86 |
| 23 | 1 | 958.691 | 0.169 | 958.374 | 52.00 |
| 23 | 1 | 959.371 | 0.165 | 959.005 | 53.30 |
| 23 | 1 | 959.341 | 0.136 | 958.933 | 54.58 |
| 23 | 1 | 959.618 | 0.160 | 959.215 | 55.92 |
| 23 | 1 | 959.334 | 0.159 | 958.971 | 57.45 |
| 23 | 1 | 960.147 | 0.152 | 959.686 | 59.00 |
| 23 | 1 | 959.661 | 0.163 | 959.200 | 60.67 |
| 23 | 1 | 959.971 | 0.120 | 959.503 | 62.47 |
| 23 | 2 | 959.753 | 0.183 | 959.726 | 63.46 |
| 23 | 2 | 959.017 | 0.207 | 958.990 | 62.16 |
| 23 | 2 | 959.075 | 0.175 | 959.064 | 60.99 |
| 23 | 2 | 959.361 | 0.158 | 959.308 | 59.85 |
| 23 | 2 | 958.981 | 0.205 | 958.972 | 58.77 |
| 23 | 2 | 958.815 | 0.203 | 958.787 | 57.71 |
| 23 | 2 | 959.236 | 0.181 | 959.211 | 56.73 |
| 23 | 2 | 958.031 | 0.162 | 958.019 | 55.80 |
| 23 | 2 | 958.969 | 0.191 | 958.970 | 54.90 |
| 23 | 2 | 958.794 | 0.225 | 958.804 | 54.05 |
| 24 | 1 | 958.848 | 0.137 | 958.660 | 49.83 |
| 24 | 1 | 959.561 | 0.155 | 959.399 | 51.02 |
| 24 | 1 | 959.671 | 0.146 | 959.425 | 52.20 |
| 24 | 1 | 958.555 | 0.142 | 958.302 | 53.40 |
| 24 | 1 | 959.414 | 0.157 | 959.186 | 54.70 |
| 24 | 1 | 960.004 | 0.127 | 959.713 | 56.02 |
| 24 | 1 | 959.676 | 0.118 | 959.344 | 57.50 |
| 24 | 1 | 958.986 | 0.131 | 958.639 | 59.00 |
| 24 | 1 | 959.868 | 0.145 | 959.479 | 60.61 |
| 24 | 1 | 959.517 | 0.151 | 959.147 | 62.32 |
| 24 | 1 | 959.862 | 0.159 | 959.464 | 64.21 |
| 24 | 2 | 960.267 | 0.158 | 960.224 | 65.52 |
| 24 | 2 | 959.763 | 0.148 | 959.729 | 64.16 |
| 24 | 2 | 959.417 | 0.164 | 959.384 | 62.85 |
| 24 | 2 | 959.228 | 0.126 | 959.207 | 61.64 |
| 24 | 2 | 959.138 | 0.117 | 959.113 | 60.46 |
| 24 | 2 | 959.037 | 0.164 | 959.025 | 59.35 |
| 24 | 2 | 958.669 | 0.122 | 958.658 | 58.31 |
| 24 | 2 | 959.411 | 0.166 | 959.402 | 57.26 |
| 24 | 2 | 958.900 | 0.170 | 958.904 | 56.29 |
| 24 | 2 | 959.167 | 0.144 | 959.158 | 55.35 |
| 24 | 2 | 958.479 | 0.205 | 958.497 | 54.43 |
| 26 | 1 | 958.549 | 0.156 | 958.332 | 52.38 |
| 26 | 1 | 960.051 | 0.140 | 959.679 | 53.50 |

|  | 2003 - JUNHO | | | | |
|---|---|---|---|---|---|
| D | L | SDB | ER | SDC | HL |
| 26 | 1 | 959.071 | 0.128 | 958.700 | 54.71 |
| 26 | 1 | 959.609 | 0.154 | 959.195 | 56.05 |
| 26 | 1 | 959.573 | 0.155 | 959.074 | 57.35 |
| 26 | 1 | 960.501 | 0.187 | 960.007 | 58.81 |
| 26 | 1 | 959.496 | 0.149 | 958.982 | 60.31 |
| 26 | 1 | 960.094 | 0.166 | 959.571 | 61.90 |
| 26 | 1 | 959.803 | 0.155 | 959.270 | 63.68 |
| 26 | 2 | 959.280 | 0.175 | 959.166 | 63.95 |
| 26 | 2 | 959.406 | 0.130 | 959.294 | 62.51 |
| 26 | 2 | 958.975 | 0.165 | 958.899 | 61.25 |
| 26 | 2 | 959.523 | 0.142 | 959.445 | 60.02 |
| 26 | 2 | 958.929 | 0.166 | 958.880 | 58.85 |
| 26 | 2 | 958.637 | 0.154 | 958.574 | 57.75 |
| 26 | 2 | 959.390 | 0.190 | 959.350 | 56.63 |
| 26 | 2 | 959.469 | 0.174 | 959.421 | 55.60 |
| 26 | 2 | 959.537 | 0.179 | 959.529 | 54.61 |
| 26 | 2 | 958.946 | 0.159 | 958.942 | 53.63 |
| 26 | 2 | 958.819 | 0.190 | 958.846 | 52.69 |
| 27 | 1 | 959.540 | 0.108 | 959.433 | 50.69 |
| 27 | 1 | 959.416 | 0.123 | 959.255 | 51.75 |
| 27 | 1 | 959.465 | 0.110 | 959.307 | 52.89 |
| 27 | 1 | 959.872 | 0.114 | 959.621 | 54.09 |
| 27 | 1 | 959.704 | 0.106 | 959.396 | 55.47 |
| 27 | 1 | 959.640 | 0.115 | 959.307 | 56.88 |
| 27 | 1 | 959.445 | 0.111 | 959.060 | 58.56 |
| 27 | 1 | 959.458 | 0.090 | 959.085 | 60.05 |
| 27 | 1 | 959.904 | 0.120 | 959.428 | 61.67 |
| 27 | 1 | 959.743 | 0.093 | 959.230 | 63.40 |
| 27 | 2 | 959.753 | 0.142 | 959.672 | 61.83 |
| 27 | 2 | 958.862 | 0.128 | 958.784 | 60.50 |
| 27 | 2 | 959.599 | 0.154 | 959.561 | 59.23 |
| 27 | 2 | 959.257 | 0.142 | 959.209 | 57.95 |
| 27 | 2 | 958.702 | 0.121 | 958.659 | 56.79 |
| 27 | 2 | 959.038 | 0.156 | 959.011 | 55.52 |
| 27 | 2 | 959.322 | 0.154 | 959.307 | 54.50 |
| 27 | 2 | 959.503 | 0.147 | 959.467 | 53.49 |
| 27 | 2 | 958.968 | 0.181 | 958.974 | 52.55 |

|  | 2003 - JULHO | | | | |
|---|---|---|---|---|---|
| D | L | SDB | ER | SDC | HL |
| 01 | 1 | 959.678 | 0.178 | 959.271 | 53.78 |
| 01 | 1 | 959.780 | 0.181 | 959.264 | 54.84 |
| 01 | 1 | 958.822 | 0.146 | 958.315 | 55.95 |
| 01 | 1 | 959.420 | 0.127 | 958.884 | 57.13 |
| 01 | 1 | 958.903 | 0.126 | 958.355 | 58.43 |
| 01 | 1 | 959.663 | 0.126 | 959.110 | 59.75 |
| 01 | 1 | 959.808 | 0.130 | 959.227 | 61.15 |
| 01 | 1 | 959.965 | 0.153 | 959.356 | 62.62 |
| 01 | 1 | 960.021 | 0.167 | 959.428 | 64.20 |
| 01 | 1 | 960.206 | 0.165 | 959.653 | 65.93 |
| 01 | 2 | 959.235 | 0.348 | 959.169 | 60.90 |
| 01 | 2 | 959.500 | 0.199 | 959.436 | 59.59 |
| 01 | 2 | 958.637 | 0.188 | 958.638 | 58.34 |
| 01 | 2 | 958.859 | 0.128 | 958.789 | 57.14 |
| 01 | 2 | 959.678 | 0.150 | 959.582 | 56.01 |
| 01 | 2 | 959.184 | 0.164 | 959.130 | 54.95 |
| 01 | 2 | 959.049 | 0.159 | 958.989 | 53.91 |
| 01 | 2 | 959.036 | 0.144 | 958.985 | 52.91 |
| 01 | 2 | 958.492 | 0.217 | 958.507 | 51.96 |
| 01 | 2 | 958.881 | 0.205 | 958.873 | 51.05 |
| 01 | 2 | 958.432 | 0.192 | 958.383 | 50.17 |
| 02 | 1 | 959.587 | 0.148 | 959.144 | 51.93 |
| 02 | 1 | 958.466 | 0.168 | 958.005 | 52.95 |
| 02 | 1 | 959.679 | 0.369 | 959.249 | 53.99 |
| 02 | 1 | 959.289 | 0.140 | 958.786 | 55.10 |
| 02 | 1 | 959.871 | 0.136 | 959.375 | 56.23 |
| 02 | 1 | 959.513 | 0.136 | 958.995 | 57.44 |
| 02 | 1 | 959.239 | 0.193 | 958.674 | 58.68 |
| 02 | 1 | 959.296 | 0.132 | 958.746 | 59.97 |
| 02 | 1 | 959.466 | 0.129 | 958.869 | 61.33 |

|  | 2003 - JULHO |  |  |  |  |  | 2003 - JULHO |  |  |  |
|---|---|---|---|---|---|---|---|---|---|---|
| D | L | SDB | ER | SDC | HL | D | L | SDB | ER | SDC | HL |
| 02 | 1 | 959.740 | 0.131 | 959.119 | 62.81 | 09 | 1 | 960.757 | 0.141 | 960.028 | 68.25 |
| 02 | 1 | 959.679 | 0.130 | 959.072 | 64.36 | 09 | 1 | 960.429 | 0.129 | 959.724 | 69.96 |
| 02 | 1 | 960.524 | 0.129 | 959.915 | 66.05 | 09 | 1 | 959.515 | 0.103 | 958.828 | 71.84 |
| 02 | 2 | 959.119 | 0.142 | 958.994 | 58.32 | 09 | 2 | 957.456 | 0.128 | 957.440 | 54.27 |
| 02 | 2 | 959.208 | 0.133 | 959.093 | 57.10 | 09 | 2 | 959.476 | 0.114 | 959.438 | 53.08 |
| 02 | 2 | 959.492 | 0.161 | 959.431 | 55.93 | 09 | 2 | 959.353 | 0.101 | 959.337 | 51.94 |
| 02 | 2 | 958.582 | 0.141 | 958.501 | 54.81 | 09 | 2 | 958.959 | 0.110 | 958.936 | 50.82 |
| 02 | 2 | 958.536 | 0.137 | 958.464 | 53.79 | 09 | 2 | 959.551 | 0.121 | 959.558 | 49.79 |
| 02 | 2 | 958.311 | 0.142 | 958.235 | 52.75 | 09 | 2 | 958.369 | 0.116 | 958.368 | 48.78 |
| 02 | 2 | 958.896 | 0.164 | 958.843 | 51.73 | 09 | 2 | 959.048 | 0.137 | 959.073 | 47.80 |
| 02 | 2 | 958.302 | 0.138 | 958.233 | 50.79 | 09 | 2 | 959.121 | 0.101 | 959.123 | 46.89 |
| 02 | 2 | 959.673 | 0.194 | 959.653 | 49.90 | 09 | 2 | 958.953 | 0.172 | 959.006 | 46.03 |
| 02 | 2 | 959.020 | 0.203 | 959.036 | 49.04 | 09 | 2 | 958.594 | 0.153 | 958.654 | 45.19 |
| 04 | 1 | 958.799 | 0.127 | 958.365 | 55.29 | 09 | 2 | 958.947 | 0.194 | 959.036 | 44.38 |
| 04 | 1 | 959.603 | 0.127 | 959.134 | 56.42 | 10 | 2 | 957.672 | 0.129 | 957.602 | 55.87 |
| 04 | 1 | 959.281 | 0.145 | 958.854 | 57.66 | 10 | 2 | 959.501 | 0.152 | 959.403 | 54.56 |
| 04 | 1 | 960.026 | 0.173 | 959.529 | 58.89 | 10 | 2 | 958.949 | 0.148 | 958.940 | 53.32 |
| 04 | 1 | 958.998 | 0.125 | 958.501 | 60.25 | 10 | 2 | 958.308 | 0.144 | 958.259 | 51.69 |
| 04 | 1 | 959.529 | 0.152 | 958.984 | 61.64 | 10 | 2 | 959.224 | 0.166 | 959.180 | 50.56 |
| 04 | 1 | 959.140 | 0.170 | 958.555 | 63.10 | 10 | 2 | 959.555 | 0.126 | 959.495 | 49.49 |
| 04 | 1 | 959.706 | 0.150 | 959.119 | 64.68 | 10 | 2 | 959.592 | 0.180 | 959.575 | 48.42 |
| 04 | 1 | 959.887 | 0.136 | 959.289 | 68.17 | 10 | 2 | 959.877 | 0.142 | 959.835 | 47.37 |
| 04 | 2 | 958.619 | 0.196 | 958.468 | 58.88 | 10 | 2 | 958.645 | 0.146 | 958.616 | 46.33 |
| 04 | 2 | 959.231 | 0.141 | 959.083 | 57.54 | 10 | 2 | 959.037 | 0.158 | 959.024 | 45.43 |
| 04 | 2 | 959.094 | 0.172 | 958.971 | 56.28 | 14 | 1 | 958.794 | 0.155 | 958.394 | 58.74 |
| 04 | 2 | 959.137 | 0.153 | 959.015 | 55.01 | 14 | 1 | 958.006 | 0.145 | 957.564 | 59.79 |
| 04 | 2 | 959.359 | 0.154 | 959.258 | 53.85 | 14 | 1 | 959.344 | 0.149 | 958.892 | 60.89 |
| 04 | 2 | 959.681 | 0.153 | 959.586 | 52.72 | 14 | 1 | 958.589 | 0.161 | 958.165 | 62.02 |
| 04 | 2 | 958.834 | 0.181 | 958.748 | 51.68 | 14 | 1 | 958.843 | 0.134 | 958.372 | 63.17 |
| 04 | 2 | 960.184 | 0.157 | 960.127 | 50.62 | 14 | 1 | 959.595 | 0.136 | 959.090 | 64.50 |
| 04 | 2 | 958.310 | 0.167 | 958.256 | 49.64 | 14 | 1 | 960.085 | 0.137 | 959.588 | 65.92 |
| 04 | 2 | 959.117 | 0.196 | 959.082 | 48.72 | 14 | 1 | 959.849 | 0.120 | 959.345 | 67.35 |
| 07 | 1 | 959.395 | 0.092 | 958.903 | 56.88 | 14 | 1 | 960.151 | 0.165 | 959.847 | 68.88 |
| 07 | 1 | 959.846 | 0.086 | 959.326 | 57.99 | 15 | 1 | 958.996 | 0.158 | 958.569 | 59.51 |
| 07 | 1 | 959.677 | 0.095 | 959.166 | 59.17 | 15 | 1 | 958.829 | 0.182 | 958.425 | 60.60 |
| 07 | 1 | 960.035 | 0.117 | 959.497 | 60.46 | 15 | 1 | 959.372 | 0.138 | 958.899 | 61.69 |
| 07 | 1 | 959.221 | 0.097 | 958.630 | 61.72 | 15 | 1 | 959.980 | 0.148 | 959.620 | 62.82 |
| 07 | 1 | 959.962 | 0.095 | 959.381 | 63.10 | 15 | 1 | 959.268 | 0.124 | 958.881 | 64.02 |
| 07 | 1 | 959.654 | 0.127 | 959.007 | 64.60 | 15 | 1 | 959.869 | 0.144 | 959.393 | 65.36 |
| 07 | 1 | 960.359 | 0.106 | 959.756 | 66.16 | 15 | 1 | 960.216 | 0.165 | 959.724 | 66.70 |
| 07 | 1 | 960.316 | 0.087 | 959.736 | 67.82 | 15 | 1 | 959.589 | 0.168 | 959.134 | 68.12 |
| 07 | 1 | 959.943 | 0.108 | 959.368 | 69.60 | 15 | 1 | 960.768 | 0.179 | 960.239 | 69.73 |
| 08 | 1 | 959.340 | 0.164 | 959.056 | 56.35 | 15 | 1 | 960.120 | 0.159 | 959.599 | 71.35 |
| 08 | 1 | 958.901 | 0.213 | 958.728 | 57.56 | 15 | 1 | 958.483 | 0.156 | 958.016 | 73.18 |
| 08 | 1 | 959.266 | 0.198 | 959.023 | 58.69 | 15 | 2 | 961.000 | 0.183 | 960.843 | 53.86 |
| 08 | 1 | 959.787 | 0.152 | 959.461 | 59.87 | 15 | 2 | 957.937 | 0.199 | 957.824 | 52.49 |
| 08 | 1 | 959.324 | 0.165 | 958.974 | 61.09 | 15 | 2 | 957.574 | 0.213 | 957.429 | 51.24 |
| 08 | 1 | 958.819 | 0.140 | 958.366 | 62.39 | 15 | 2 | 958.991 | 0.182 | 958.846 | 50.01 |
| 08 | 1 | 959.767 | 0.147 | 959.219 | 63.71 | 15 | 2 | 960.134 | 0.205 | 960.028 | 48.91 |
| 08 | 1 | 959.509 | 0.162 | 958.874 | 65.14 | 15 | 2 | 959.159 | 0.190 | 959.018 | 47.75 |
| 08 | 1 | 959.640 | 0.131 | 958.979 | 66.66 | 15 | 2 | 958.119 | 0.172 | 958.010 | 46.75 |
| 08 | 1 | 959.371 | 0.172 | 958.740 | 68.28 | 15 | 2 | 958.987 | 0.159 | 958.874 | 45.75 |
| 08 | 1 | 959.823 | 0.162 | 959.194 | 70.11 | 15 | 2 | 957.763 | 0.155 | 957.639 | 44.75 |
| 08 | 2 | 959.388 | 0.117 | 959.276 | 57.09 | 15 | 2 | 959.184 | 0.161 | 959.065 | 43.84 |
| 08 | 2 | 959.542 | 0.120 | 959.445 | 55.70 | 15 | 2 | 958.167 | 0.201 | 958.125 | 42.98 |
| 08 | 2 | 958.538 | 0.138 | 958.457 | 54.39 | 15 | 2 | 958.698 | 0.174 | 958.584 | 42.09 |
| 08 | 2 | 958.791 | 0.122 | 958.715 | 53.17 | 15 | 2 | 958.927 | 0.162 | 958.868 | 41.29 |
| 08 | 2 | 959.662 | 0.168 | 959.631 | 51.97 | 16 | 1 | 958.643 | 0.118 | 958.232 | 60.05 |
| 08 | 2 | 958.931 | 0.133 | 958.898 | 50.84 | 16 | 1 | 959.464 | 0.105 | 958.991 | 61.11 |
| 08 | 2 | 959.295 | 0.133 | 959.264 | 49.81 | 16 | 1 | 959.375 | 0.162 | 958.851 | 62.22 |
| 08 | 2 | 958.714 | 0.140 | 958.697 | 48.80 | 16 | 1 | 959.248 | 0.204 | 958.760 | 63.40 |
| 08 | 2 | 959.518 | 0.139 | 959.504 | 47.83 | 16 | 1 | 958.031 | 0.165 | 957.472 | 64.60 |
| 08 | 2 | 958.741 | 0.139 | 958.737 | 46.87 | 16 | 1 | 960.978 | 0.169 | 960.431 | 65.89 |
| 09 | 1 | 959.655 | 0.130 | 959.058 | 57.59 | 16 | 1 | 960.072 | 0.127 | 959.512 | 68.83 |
| 09 | 1 | 958.751 | 0.200 | 958.145 | 58.73 | 16 | 1 | 960.957 | 0.124 | 960.431 | 70.35 |
| 09 | 1 | 959.582 | 0.129 | 959.042 | 59.93 | 16 | 1 | 959.582 | 0.112 | 959.062 | 72.00 |
| 09 | 1 | 959.297 | 0.122 | 958.700 | 61.16 | 16 | 1 | 960.041 | 0.131 | 959.523 | 73.92 |
| 09 | 1 | 960.541 | 0.116 | 959.916 | 62.42 | 16 | 2 | 959.295 | 0.186 | 959.164 | 52.48 |
| 09 | 1 | 960.210 | 0.152 | 959.511 | 63.72 | 16 | 2 | 958.245 | 0.195 | 958.100 | 51.20 |
| 09 | 1 | 959.972 | 0.168 | 959.281 | 65.13 | 16 | 2 | 958.770 | 0.180 | 958.632 | 50.01 |
| 09 | 1 | 960.135 | 0.139 | 959.459 | 66.66 | 16 | 2 | 958.896 | 0.156 | 958.752 | 48.78 |

| 2003 - JULHO | | | | | |
|---|---|---|---|---|---|
| D | L | SDB | ER | SDC | HL |
| 16 | 2 | 959.978 | 0.117 | 959.845 | 47.67 |
| 16 | 2 | 958.646 | 0.155 | 958.530 | 46.61 |
| 16 | 2 | 959.113 | 0.171 | 959.007 | 45.62 |
| 16 | 2 | 958.072 | 0.155 | 957.969 | 44.64 |
| 16 | 2 | 958.057 | 0.172 | 958.008 | 43.73 |
| 16 | 2 | 958.837 | 0.167 | 958.757 | 42.82 |
| 16 | 2 | 959.425 | 0.128 | 959.327 | 41.98 |
| 16 | 2 | 958.677 | 0.179 | 958.634 | 41.15 |
| 18 | 2 | 958.757 | 0.190 | 958.643 | 51.48 |
| 18 | 2 | 959.184 | 0.148 | 959.076 | 50.18 |
| 18 | 2 | 958.904 | 0.173 | 958.867 | 48.92 |
| 18 | 2 | 958.845 | 0.179 | 958.827 | 47.73 |
| 18 | 2 | 958.689 | 0.179 | 958.675 | 46.55 |
| 18 | 2 | 959.775 | 0.167 | 959.759 | 45.48 |
| 18 | 2 | 959.964 | 0.152 | 959.955 | 44.44 |
| 18 | 2 | 959.367 | 0.141 | 959.360 | 43.46 |
| 18 | 2 | 958.830 | 0.181 | 958.815 | 42.50 |
| 18 | 2 | 959.743 | 0.180 | 959.736 | 41.60 |
| 18 | 2 | 959.107 | 0.179 | 959.111 | 40.72 |
| 21 | 1 | 959.149 | 0.110 | 958.877 | 60.98 |
| 21 | 1 | 959.728 | 0.151 | 959.377 | 61.98 |
| 21 | 1 | 959.224 | 0.133 | 958.933 | 63.00 |
| 21 | 1 | 959.895 | 0.122 | 959.565 | 64.05 |
| 21 | 1 | 959.936 | 0.109 | 959.607 | 65.16 |
| 21 | 1 | 958.476 | 0.148 | 958.135 | 66.44 |
| 21 | 1 | 960.016 | 0.114 | 959.643 | 67.68 |
| 21 | 1 | 960.250 | 0.142 | 959.857 | 69.04 |
| 21 | 1 | 960.072 | 0.145 | 959.680 | 70.43 |
| 21 | 1 | 958.821 | 0.168 | 958.429 | 71.94 |
| 21 | 1 | 960.947 | 0.160 | 960.580 | 73.62 |
| 21 | 1 | 960.131 | 0.117 | 959.743 | 75.46 |
| 21 | 2 | 959.635 | 0.183 | 959.674 | 48.43 |
| 21 | 2 | 959.839 | 0.195 | 959.907 | 47.21 |
| 21 | 2 | 959.266 | 0.175 | 959.317 | 44.90 |
| 21 | 2 | 958.225 | 0.181 | 958.299 | 43.82 |
| 21 | 2 | 959.834 | 0.156 | 959.925 | 42.83 |
| 21 | 2 | 959.130 | 0.180 | 959.223 | 41.79 |
| 21 | 2 | 959.968 | 0.163 | 960.059 | 40.84 |
| 21 | 2 | 958.944 | 0.136 | 959.042 | 39.97 |
| 21 | 2 | 959.135 | 0.183 | 959.215 | 39.12 |
| 21 | 2 | 958.544 | 0.162 | 958.672 | 38.32 |
| 21 | 2 | 959.139 | 0.183 | 959.263 | 37.54 |
| 21 | 2 | 959.227 | 0.263 | 959.403 | 36.76 |
| 22 | 1 | 959.806 | 0.122 | 959.545 | 62.17 |
| 22 | 1 | 958.687 | 0.141 | 958.421 | 63.19 |
| 22 | 1 | 958.601 | 0.147 | 958.342 | 64.26 |
| 22 | 1 | 959.481 | 0.159 | 959.160 | 65.39 |
| 22 | 1 | 959.567 | 0.154 | 959.269 | 66.58 |
| 22 | 1 | 959.974 | 0.169 | 959.677 | 67.94 |
| 22 | 1 | 959.211 | 0.129 | 958.847 | 69.30 |
| 22 | 1 | 959.909 | 0.117 | 959.626 | 70.75 |
| 22 | 1 | 959.170 | 0.110 | 958.799 | 72.32 |
| 22 | 1 | 959.901 | 0.120 | 959.567 | 73.98 |
| 22 | 2 | 959.758 | 0.346 | 959.797 | 51.78 |
| 22 | 2 | 960.054 | 0.209 | 960.051 | 50.23 |
| 22 | 2 | 958.785 | 0.195 | 958.797 | 48.88 |
| 22 | 2 | 959.026 | 0.208 | 959.040 | 47.65 |
| 22 | 2 | 959.378 | 0.226 | 959.500 | 46.32 |
| 22 | 2 | 958.125 | 0.231 | 958.221 | 45.12 |
| 22 | 2 | 959.290 | 0.212 | 959.358 | 44.09 |
| 22 | 2 | 957.468 | 0.255 | 957.563 | 43.03 |
| 22 | 2 | 958.205 | 0.194 | 958.288 | 41.15 |
| 22 | 2 | 959.269 | 0.210 | 959.366 | 40.29 |
| 22 | 2 | 958.205 | 0.209 | 958.319 | 39.43 |
| 23 | 2 | 960.986 | 0.203 | 961.117 | 47.64 |
| 23 | 2 | 959.687 | 0.217 | 959.707 | 46.44 |
| 23 | 2 | 958.313 | 0.226 | 958.353 | 45.33 |
| 23 | 2 | 959.537 | 0.172 | 959.557 | 44.15 |
| 23 | 2 | 958.947 | 0.170 | 958.970 | 43.13 |
| 23 | 2 | 959.340 | 0.224 | 959.380 | 42.10 |
| 23 | 2 | 959.808 | 0.193 | 959.857 | 41.16 |

| 2003 - JULHO | | | | | |
|---|---|---|---|---|---|
| D | L | SDB | ER | SDC | HL |
| 23 | 2 | 959.059 | 0.212 | 959.123 | 40.27 |
| 23 | 2 | 958.872 | 0.174 | 958.966 | 39.41 |
| 23 | 2 | 958.761 | 0.182 | 958.860 | 38.58 |
| 23 | 2 | 959.899 | 0.161 | 959.972 | 37.78 |
| 24 | 1 | 958.169 | 0.157 | 958.058 | 61.51 |
| 24 | 1 | 959.407 | 0.157 | 959.369 | 62.58 |
| 24 | 1 | 960.003 | 0.146 | 960.011 | 63.64 |
| 24 | 1 | 959.053 | 0.134 | 958.705 | 64.79 |
| 24 | 1 | 958.814 | 0.120 | 958.447 | 65.95 |
| 24 | 1 | 959.133 | 0.179 | 958.728 | 67.13 |
| 24 | 1 | 959.300 | 0.153 | 958.897 | 68.48 |
| 24 | 1 | 959.594 | 0.151 | 959.201 | 69.78 |
| 24 | 1 | 959.191 | 0.154 | 958.776 | 71.21 |
| 24 | 1 | 959.413 | 0.173 | 959.031 | 72.72 |
| 24 | 1 | 959.459 | 0.151 | 959.075 | 74.33 |
| 24 | 2 | 960.556 | 0.138 | 960.559 | 48.45 |
| 24 | 2 | 959.123 | 0.123 | 959.122 | 47.11 |
| 24 | 2 | 959.161 | 0.161 | 959.162 | 45.79 |
| 24 | 2 | 958.658 | 0.140 | 958.666 | 44.60 |
| 24 | 2 | 959.052 | 0.158 | 959.059 | 43.50 |
| 24 | 2 | 959.382 | 0.137 | 959.396 | 42.38 |
| 24 | 2 | 959.908 | 0.162 | 959.962 | 41.21 |
| 24 | 2 | 959.210 | 0.165 | 959.278 | 40.25 |
| 24 | 2 | 958.258 | 0.155 | 958.328 | 39.32 |
| 24 | 2 | 959.589 | 0.155 | 959.668 | 38.42 |
| 24 | 2 | 958.885 | 0.189 | 958.983 | 37.56 |
| 25 | 1 | 957.774 | 0.132 | 957.551 | 60.22 |
| 25 | 1 | 959.323 | 0.146 | 959.030 | 61.13 |
| 25 | 1 | 959.185 | 0.134 | 958.833 | 62.15 |
| 25 | 1 | 959.700 | 0.142 | 959.340 | 63.13 |
| 25 | 1 | 959.632 | 0.169 | 959.254 | 64.20 |
| 25 | 1 | 959.247 | 0.143 | 958.879 | 65.26 |
| 25 | 1 | 959.228 | 0.146 | 958.879 | 66.39 |
| 25 | 1 | 959.068 | 0.142 | 958.754 | 67.67 |
| 25 | 1 | 959.107 | 0.170 | 958.822 | 68.92 |
| 25 | 1 | 960.296 | 0.160 | 960.071 | 70.25 |
| 25 | 1 | 959.881 | 0.141 | 959.730 | 71.75 |
| 25 | 2 | 958.633 | 0.182 | 958.705 | 46.85 |
| 25 | 2 | 959.302 | 0.202 | 959.375 | 45.58 |
| 25 | 2 | 959.402 | 0.170 | 959.461 | 44.40 |
| 25 | 2 | 958.802 | 0.160 | 958.848 | 43.28 |
| 25 | 2 | 959.179 | 0.168 | 959.222 | 42.13 |
| 25 | 2 | 958.735 | 0.174 | 958.782 | 41.12 |
| 25 | 2 | 959.169 | 0.187 | 959.239 | 40.12 |
| 25 | 2 | 959.915 | 0.239 | 960.028 | 39.16 |
| 25 | 2 | 959.635 | 0.182 | 959.756 | 38.25 |
| 25 | 2 | 958.512 | 0.204 | 958.641 | 37.22 |
| 25 | 2 | 959.537 | 0.215 | 959.690 | 36.40 |
| 30 | 1 | 959.681 | 0.127 | 959.603 | 63.46 |
| 30 | 1 | 959.181 | 0.141 | 959.101 | 64.49 |
| 30 | 1 | 959.366 | 0.178 | 959.225 | 65.51 |
| 30 | 1 | 958.576 | 0.145 | 958.423 | 66.56 |
| 30 | 1 | 959.205 | 0.142 | 959.031 | 67.64 |
| 30 | 1 | 959.342 | 0.114 | 959.193 | 68.84 |
| 30 | 1 | 959.151 | 0.135 | 958.970 | 70.23 |
| 30 | 1 | 958.831 | 0.135 | 958.700 | 71.78 |
| 30 | 1 | 959.295 | 0.135 | 959.164 | 73.26 |
| 30 | 1 | 959.349 | 0.156 | 959.312 | 74.93 |
| 30 | 1 | 960.219 | 0.165 | 960.139 | 76.55 |

| 2003 - AGOSTO | | | | | |
|---|---|---|---|---|---|
| D | L | SDB | ER | SDC | HL |
| 13 | 2 | 959.250 | 0.197 | 959.364 | 35.64 |
| 13 | 2 | 959.697 | 0.138 | 959.804 | 34.49 |
| 13 | 2 | 959.190 | 0.192 | 958.320 | 33.41 |
| 13 | 2 | 960.069 | 0.150 | 960.146 | 32.19 |
| 13 | 2 | 959.503 | 0.164 | 959.573 | 31.22 |
| 13 | 2 | 959.303 | 0.177 | 959.374 | 30.29 |
| 13 | 2 | 960.236 | 0.152 | 960.293 | 29.34 |
| 13 | 2 | 958.985 | 0.146 | 959.039 | 28.46 |

|       | 2003 - AGOSTO |       |         |       |
|-------|------|---------|-------|---------|-------|
| D | L | SDB | ER | SDC | HL |
| 13 | 2 | 958.621 | 0.160 | 958.683 | 27.65 |
| 13 | 2 | 959.128 | 0.155 | 959.189 | 26.87 |
| 13 | 2 | 959.316 | 0.194 | 959.406 | 26.12 |
| 14 | 2 | 960.382 | 0.167 | 960.485 | 31.55 |
| 14 | 2 | 960.235 | 0.179 | 960.335 | 30.59 |
| 14 | 2 | 958.429 | 0.175 | 958.530 | 29.70 |
| 14 | 2 | 959.157 | 0.170 | 959.247 | 28.82 |
| 14 | 2 | 959.318 | 0.169 | 959.405 | 27.99 |
| 14 | 2 | 958.046 | 0.208 | 958.143 | 27.19 |
| 14 | 2 | 960.287 | 0.167 | 960.376 | 26.37 |
| 14 | 2 | 958.678 | 0.151 | 958.796 | 25.63 |
| 14 | 2 | 957.585 | 0.191 | 957.720 | 24.87 |
| 14 | 2 | 959.216 | 0.182 | 959.367 | 24.17 |
| 14 | 2 | 959.123 | 0.157 | 959.283 | 23.49 |
| 18 | 1 | 958.762 | 0.129 | 958.652 | 72.56 |
| 18 | 1 | 959.789 | 0.195 | 959.777 | 73.70 |
| 18 | 1 | 958.425 | 0.167 | 958.433 | 76.56 |
| 18 | 1 | 959.476 | 0.143 | 959.351 | 78.04 |
| 19 | 1 | 959.488 | 0.130 | 959.363 | 65.93 |
| 19 | 1 | 958.735 | 0.149 | 958.583 | 66.69 |
| 19 | 1 | 960.112 | 0.120 | 959.957 | 67.47 |
| 19 | 1 | 958.817 | 0.130 | 958.652 | 68.30 |
| 19 | 1 | 959.641 | 0.122 | 959.503 | 69.17 |
| 19 | 1 | 959.631 | 0.146 | 959.498 | 70.11 |
| 19 | 1 | 959.836 | 0.176 | 959.727 | 71.04 |
| 19 | 1 | 960.023 | 0.163 | 959.937 | 72.04 |
| 19 | 1 | 959.889 | 0.180 | 959.825 | 73.10 |
| 19 | 1 | 958.664 | 0.231 | 958.606 | 74.15 |
| 19 | 1 | 959.583 | 0.193 | 959.594 | 75.28 |
| 19 | 1 | 959.549 | 0.166 | 959.526 | 76.54 |
| 19 | 1 | 959.410 | 0.182 | 959.226 | 78.05 |
| 19 | 2 | 960.171 | 0.274 | 960.285 | 28.42 |
| 19 | 2 | 958.417 | 0.180 | 958.548 | 27.55 |
| 19 | 2 | 959.393 | 0.164 | 959.486 | 26.56 |
| 19 | 2 | 958.220 | 0.178 | 958.320 | 25.78 |
| 19 | 2 | 959.030 | 0.150 | 959.127 | 25.02 |
| 19 | 2 | 959.428 | 0.152 | 959.512 | 24.23 |
| 19 | 2 | 959.716 | 0.178 | 959.812 | 23.48 |
| 19 | 2 | 959.144 | 0.146 | 959.232 | 22.80 |
| 19 | 2 | 959.713 | 0.187 | 959.818 | 22.12 |
| 19 | 2 | 958.552 | 0.177 | 958.664 | 21.37 |
| 19 | 2 | 958.918 | 0.164 | 959.037 | 20.69 |
| 19 | 2 | 959.332 | 0.161 | 959.446 | 20.05 |
| 19 | 2 | 958.814 | 0.140 | 958.931 | 19.42 |
| 19 | 2 | 959.237 | 0.169 | 959.383 | 18.81 |
| 20 | 1 | 959.275 | 0.129 | 959.103 | 69.32 |
| 20 | 1 | 958.703 | 0.144 | 958.583 | 70.36 |
| 20 | 1 | 959.194 | 0.152 | 959.102 | 71.30 |
| 20 | 1 | 960.001 | 0.164 | 959.939 | 72.30 |
| 20 | 1 | 960.350 | 0.175 | 960.159 | 73.48 |
| 20 | 1 | 959.304 | 0.202 | 959.120 | 74.58 |
| 20 | 1 | 959.396 | 0.205 | 959.181 | 75.78 |
| 20 | 1 | 959.523 | 0.141 | 959.375 | 81.59 |
| 20 | 1 | 959.937 | 0.150 | 959.838 | 83.37 |
| 20 | 2 | 958.246 | 0.191 | 958.162 | 32.03 |
| 20 | 2 | 958.635 | 0.191 | 958.560 | 30.99 |
| 20 | 2 | 958.567 | 0.168 | 958.474 | 29.99 |
| 20 | 2 | 959.692 | 0.160 | 959.600 | 29.03 |
| 20 | 2 | 958.484 | 0.259 | 958.427 | 28.08 |
| 20 | 2 | 958.358 | 0.186 | 958.271 | 27.20 |
| 20 | 2 | 959.137 | 0.201 | 959.084 | 26.30 |
| 20 | 2 | 959.874 | 0.183 | 959.841 | 25.47 |
| 20 | 2 | 958.624 | 0.175 | 958.556 | 24.64 |
| 20 | 2 | 959.219 | 0.196 | 959.194 | 23.88 |
| 20 | 2 | 959.587 | 0.230 | 959.578 | 23.15 |
| 20 | 2 | 959.562 | 0.185 | 959.468 | 22.34 |
| 20 | 2 | 958.070 | 0.180 | 958.003 | 21.64 |
| 20 | 2 | 958.633 | 0.199 | 958.559 | 20.91 |
| 20 | 2 | 958.973 | 0.158 | 958.903 | 20.24 |
| 21 | 1 | 959.831 | 0.195 | 959.276 | 64.62 |
| 21 | 1 | 958.998 | 0.135 | 958.437 | 65.41 |

|       | 2003 - AGOSTO |       |         |       |
|-------|------|---------|-------|---------|-------|
| D | L | SDB | ER | SDC | HL |
| 21 | 1 | 960.378 | 0.156 | 959.802 | 66.12 |
| 21 | 1 | 959.309 | 0.134 | 958.747 | 66.90 |
| 21 | 1 | 959.197 | 0.132 | 958.634 | 67.82 |
| 21 | 1 | 959.063 | 0.142 | 958.517 | 68.72 |
| 21 | 1 | 959.483 | 0.163 | 958.997 | 69.86 |
| 21 | 1 | 959.593 | 0.172 | 959.096 | 70.86 |
| 21 | 1 | 958.848 | 0.162 | 958.336 | 71.85 |
| 21 | 1 | 958.486 | 0.137 | 957.933 | 72.93 |
| 21 | 1 | 958.535 | 0.155 | 957.960 | 74.02 |
| 21 | 2 | 959.846 | 0.173 | 959.750 | 26.67 |
| 21 | 2 | 959.449 | 0.198 | 958.949 | 25.31 |
| 21 | 2 | 959.317 | 0.165 | 959.201 | 24.45 |
| 21 | 2 | 958.610 | 0.171 | 958.502 | 23.65 |
| 21 | 2 | 959.308 | 0.168 | 959.199 | 22.87 |
| 21 | 2 | 960.237 | 0.138 | 960.124 | 22.08 |
| 21 | 2 | 959.221 | 0.154 | 959.113 | 21.37 |
| 21 | 2 | 959.267 | 0.170 | 959.185 | 20.64 |
| 21 | 2 | 958.827 | 0.191 | 958.768 | 19.89 |
| 21 | 2 | 959.092 | 0.167 | 959.051 | 19.26 |
| 21 | 2 | 958.828 | 0.186 | 958.804 | 18.64 |
| 22 | 1 | 959.361 | 0.129 | 958.802 | 65.72 |
| 22 | 1 | 959.282 | 0.123 | 958.716 | 66.47 |
| 22 | 1 | 959.842 | 0.146 | 959.293 | 67.29 |
| 22 | 1 | 959.584 | 0.149 | 959.021 | 68.19 |
| 22 | 1 | 959.108 | 0.144 | 958.550 | 69.04 |
| 22 | 1 | 959.640 | 0.195 | 959.096 | 70.15 |
| 22 | 1 | 959.370 | 0.148 | 958.822 | 71.10 |
| 22 | 1 | 959.567 | 0.175 | 958.999 | 72.11 |
| 22 | 1 | 960.021 | 0.151 | 959.430 | 73.14 |
| 22 | 1 | 958.905 | 0.164 | 958.310 | 74.20 |
| 22 | 1 | 958.506 | 0.159 | 957.907 | 75.34 |
| 22 | 2 | 958.752 | 0.156 | 958.628 | 24.84 |
| 22 | 2 | 958.576 | 0.188 | 958.485 | 24.02 |
| 22 | 2 | 958.824 | 0.182 | 958.729 | 23.22 |
| 22 | 2 | 959.515 | 0.157 | 959.414 | 22.42 |
| 22 | 2 | 958.964 | 0.160 | 958.856 | 21.66 |
| 22 | 2 | 959.006 | 0.176 | 958.926 | 20.97 |
| 22 | 2 | 958.858 | 0.170 | 958.778 | 20.25 |
| 22 | 2 | 958.822 | 0.161 | 958.762 | 19.56 |
| 22 | 2 | 958.331 | 0.178 | 958.292 | 18.91 |
| 22 | 2 | 958.885 | 0.187 | 958.861 | 18.25 |
| 22 | 2 | 959.679 | 0.158 | 959.675 | 17.63 |
| 25 | 2 | 959.290 | 0.141 | 959.237 | 27.19 |
| 25 | 2 | 959.500 | 0.142 | 959.440 | 26.25 |
| 25 | 2 | 959.285 | 0.155 | 959.228 | 25.34 |
| 25 | 2 | 959.447 | 0.171 | 958.429 | 24.48 |
| 25 | 2 | 959.332 | 0.147 | 959.284 | 23.62 |
| 25 | 2 | 959.424 | 0.149 | 959.419 | 22.83 |
| 25 | 2 | 959.045 | 0.159 | 959.059 | 21.86 |
| 25 | 2 | 958.522 | 0.163 | 958.485 | 21.00 |
| 25 | 2 | 959.077 | 0.160 | 959.123 | 19.60 |
| 25 | 2 | 959.009 | 0.144 | 959.043 | 18.73 |
| 25 | 2 | 959.187 | 0.202 | 959.210 | 18.09 |
| 25 | 2 | 959.297 | 0.158 | 959.322 | 17.47 |

|       | 2003 - SETEMBRO |       |         |       |
|-------|------|---------|-------|---------|-------|
| D | L | SDB | ER | SDC | HL |
| 04 | 1 | 958.036 | 0.124 | 957.638 | 72.16 |
| 04 | 1 | 959.538 | 0.124 | 959.108 | 73.05 |
| 04 | 1 | 959.452 | 0.138 | 959.053 | 74.01 |
| 04 | 1 | 959.932 | 0.114 | 959.545 | 74.98 |
| 04 | 1 | 959.400 | 0.134 | 959.000 | 76.06 |
| 04 | 1 | 960.043 | 0.125 | 959.660 | 77.18 |
| 04 | 1 | 959.593 | 0.126 | 959.208 | 78.30 |
| 04 | 1 | 959.313 | 0.139 | 958.956 | 79.62 |
| 04 | 1 | 959.368 | 0.142 | 959.015 | 80.94 |
| 04 | 1 | 959.924 | 0.131 | 959.614 | 82.44 |
| 04 | 1 | 960.055 | 0.155 | 959.762 | 83.92 |
| 04 | 1 | 959.898 | 0.153 | 959.622 | 85.59 |
| 04 | 2 | 959.099 | 0.133 | 958.958 | 18.43 |

| 2003 - SETEMBRO | | | | | | 2003 - SETEMBRO | | | | |
|---|---|---|---|---|---|---|---|---|---|---|
| D | L | SDB | ER | SDC | HL | D | L | SDB | ER | SDC | HL |
| 04 | 2 | 958.809 | 0.148 | 958.671 | 17.66 | 09 | 1 | 960.399 | 0.118 | 959.923 | 70.79 |
| 04 | 2 | 959.306 | 0.138 | 959.172 | 16.89 | 09 | 1 | 960.531 | 0.128 | 960.096 | 71.61 |
| 04 | 2 | 958.994 | 0.166 | 958.879 | 16.18 | 09 | 1 | 960.041 | 0.121 | 959.620 | 72.51 |
| 04 | 2 | 958.845 | 0.154 | 958.734 | 15.46 | 09 | 2 | 960.269 | 0.118 | 960.247 | 16.96 |
| 04 | 2 | 959.366 | 0.201 | 959.277 | 14.79 | 09 | 2 | 958.563 | 0.129 | 958.537 | 16.22 |
| 04 | 2 | 959.691 | 0.179 | 959.558 | 13.99 | 09 | 2 | 958.796 | 0.150 | 958.762 | 15.37 |
| 04 | 2 | 959.223 | 0.168 | 959.082 | 13.41 | 09 | 2 | 959.428 | 0.152 | 959.371 | 14.63 |
| 04 | 2 | 959.182 | 0.153 | 959.027 | 12.77 | 09 | 2 | 958.845 | 0.137 | 958.796 | 13.97 |
| 04 | 2 | 959.113 | 0.160 | 958.961 | 12.23 | 09 | 2 | 958.849 | 0.144 | 958.800 | 13.32 |
| 04 | 2 | 959.093 | 0.138 | 958.926 | 11.69 | 09 | 2 | 959.320 | 0.132 | 959.276 | 12.72 |
| 05 | 1 | 959.545 | 0.127 | 959.142 | 63.48 | 09 | 2 | 958.126 | 0.151 | 958.107 | 12.16 |
| 05 | 1 | 960.107 | 0.124 | 959.709 | 64.10 | 09 | 2 | 958.244 | 0.149 | 958.245 | 11.60 |
| 05 | 1 | 959.778 | 0.120 | 959.366 | 64.68 | 09 | 2 | 958.960 | 0.167 | 958.992 | 11.02 |
| 05 | 1 | 960.018 | 0.117 | 959.631 | 65.35 | 09 | 2 | 958.450 | 0.163 | 958.495 | 10.39 |
| 05 | 1 | 959.424 | 0.103 | 959.051 | 65.98 | 09 | 2 | 959.292 | 0.156 | 959.332 | 9.87 |
| 05 | 1 | 959.956 | 0.135 | 959.578 | 66.68 | 09 | 2 | 959.800 | 0.146 | 959.821 | 9.41 |
| 05 | 1 | 959.301 | 0.160 | 958.957 | 67.39 | 09 | 2 | 958.944 | 0.156 | 958.953 | 8.96 |
| 05 | 1 | 959.557 | 0.117 | 959.190 | 68.35 | 10 | 1 | 959.995 | 0.124 | 959.514 | 65.57 |
| 05 | 1 | 959.752 | 0.128 | 959.385 | 69.12 | 10 | 1 | 959.700 | 0.123 | 959.230 | 66.19 |
| 05 | 1 | 959.097 | 0.139 | 958.690 | 69.98 | 10 | 1 | 960.456 | 0.128 | 959.993 | 66.84 |
| 05 | 1 | 960.372 | 0.137 | 959.937 | 70.85 | 10 | 1 | 960.029 | 0.133 | 959.567 | 67.47 |
| 05 | 2 | 958.886 | 0.175 | 958.814 | 16.89 | 10 | 1 | 959.956 | 0.122 | 959.525 | 68.19 |
| 05 | 2 | 958.824 | 0.218 | 958.777 | 16.12 | 10 | 1 | 959.413 | 0.156 | 959.015 | 68.89 |
| 05 | 2 | 960.492 | 0.225 | 960.443 | 15.41 | 10 | 1 | 959.182 | 0.129 | 958.823 | 69.81 |
| 05 | 2 | 959.099 | 0.200 | 959.038 | 14.74 | 10 | 1 | 959.873 | 0.110 | 959.551 | 70.57 |
| 05 | 2 | 959.740 | 0.204 | 959.705 | 14.11 | 10 | 1 | 959.353 | 0.106 | 959.171 | 71.36 |
| 05 | 2 | 959.205 | 0.164 | 959.083 | 13.38 | 10 | 1 | 959.277 | 0.136 | 959.055 | 72.25 |
| 05 | 2 | 960.316 | 0.158 | 960.189 | 12.75 | 10 | 1 | 959.304 | 0.135 | 959.106 | 73.17 |
| 05 | 2 | 958.643 | 0.167 | 958.522 | 12.18 | 10 | 1 | 958.795 | 0.118 | 958.603 | 74.13 |
| 05 | 2 | 959.312 | 0.159 | 959.188 | 11.64 | 10 | 1 | 959.522 | 0.158 | 959.396 | 75.18 |
| 05 | 2 | 959.301 | 0.161 | 959.177 | 11.12 | 10 | 1 | 958.587 | 0.146 | 958.467 | 76.28 |
| 05 | 2 | 960.530 | 0.201 | 960.432 | 10.62 | 10 | 1 | 958.826 | 0.151 | 958.673 | 77.47 |
| 08 | 1 | 957.509 | 0.139 | 957.065 | 64.06 | 10 | 2 | 959.186 | 0.184 | 959.262 | 19.36 |
| 08 | 1 | 959.992 | 0.133 | 959.546 | 64.62 | 10 | 2 | 958.840 | 0.188 | 958.909 | 18.39 |
| 08 | 1 | 959.369 | 0.130 | 958.979 | 65.18 | 10 | 2 | 959.098 | 0.145 | 959.156 | 17.52 |
| 08 | 1 | 960.451 | 0.129 | 960.057 | 65.75 | 10 | 2 | 958.607 | 0.181 | 958.668 | 16.70 |
| 08 | 1 | 960.286 | 0.132 | 959.868 | 66.36 | 10 | 2 | 957.614 | 0.200 | 957.679 | 15.95 |
| 08 | 1 | 959.971 | 0.124 | 959.605 | 66.98 | 10 | 2 | 958.468 | 0.197 | 958.515 | 15.24 |
| 08 | 1 | 959.746 | 0.141 | 959.381 | 67.74 | 10 | 2 | 958.443 | 0.154 | 958.462 | 14.54 |
| 08 | 1 | 959.143 | 0.170 | 958.806 | 68.41 | 10 | 2 | 958.602 | 0.155 | 958.599 | 13.88 |
| 08 | 1 | 959.364 | 0.153 | 959.104 | 69.10 | 10 | 2 | 958.593 | 0.157 | 958.606 | 13.25 |
| 08 | 1 | 960.549 | 0.146 | 960.266 | 69.88 | 10 | 2 | 958.551 | 0.150 | 958.583 | 12.64 |
| 08 | 1 | 960.284 | 0.130 | 960.021 | 70.64 | 10 | 2 | 958.986 | 0.139 | 959.036 | 12.03 |
| 08 | 1 | 959.777 | 0.154 | 959.649 | 71.64 | 10 | 2 | 959.017 | 0.180 | 959.107 | 11.43 |
| 08 | 1 | 959.752 | 0.158 | 959.271 | 72.60 | 10 | 2 | 958.714 | 0.169 | 958.819 | 10.89 |
| 08 | 1 | 959.725 | 0.155 | 959.272 | 73.51 | 10 | 2 | 959.004 | 0.183 | 959.119 | 10.19 |
| 08 | 1 | 960.165 | 0.130 | 959.743 | 74.51 | 16 | 1 | 958.911 | 0.123 | 958.499 | 66.15 |
| 08 | 1 | 959.805 | 0.108 | 959.420 | 75.50 | 16 | 1 | 959.328 | 0.121 | 958.928 | 66.78 |
| 08 | 2 | 958.669 | 0.221 | 958.595 | 18.72 | 16 | 1 | 959.836 | 0.127 | 959.416 | 67.46 |
| 08 | 2 | 959.488 | 0.167 | 959.422 | 17.05 | 16 | 1 | 959.636 | 0.113 | 959.248 | 68.13 |
| 08 | 2 | 959.265 | 0.158 | 959.195 | 16.24 | 16 | 1 | 959.623 | 0.144 | 959.218 | 68.97 |
| 08 | 2 | 958.900 | 0.188 | 958.869 | 15.56 | 16 | 1 | 959.161 | 0.138 | 958.824 | 69.85 |
| 08 | 2 | 958.882 | 0.155 | 958.820 | 14.90 | 19 | 1 | 959.094 | 0.179 | 958.710 | 67.99 |
| 08 | 2 | 959.314 | 0.168 | 959.253 | 14.24 | 19 | 1 | 958.680 | 0.119 | 958.385 | 68.68 |
| 08 | 2 | 958.465 | 0.139 | 958.410 | 13.56 | 19 | 1 | 960.077 | 0.184 | 959.713 | 69.62 |
| 08 | 2 | 958.823 | 0.164 | 958.777 | 12.98 | 19 | 1 | 958.846 | 0.116 | 958.522 | 70.47 |
| 08 | 2 | 958.681 | 0.151 | 958.630 | 12.39 | 19 | 1 | 959.146 | 0.124 | 958.786 | 71.36 |
| 08 | 2 | 958.556 | 0.159 | 958.525 | 11.85 | 19 | 1 | 959.239 | 0.128 | 958.886 | 72.18 |
| 08 | 2 | 958.920 | 0.193 | 958.905 | 11.27 | 19 | 1 | 958.920 | 0.170 | 957.879 | 73.06 |
| 08 | 2 | 959.059 | 0.178 | 959.040 | 10.75 | 19 | 1 | 958.559 | 0.130 | 958.307 | 74.00 |
| 08 | 2 | 958.902 | 0.205 | 958.893 | 10.27 | 19 | 1 | 959.992 | 0.162 | 959.785 | 75.07 |
| 09 | 1 | 958.263 | 0.123 | 957.817 | 63.46 | 19 | 1 | 958.405 | 0.141 | 958.158 | 76.14 |
| 09 | 1 | 959.776 | 0.116 | 959.358 | 63.99 | 19 | 2 | 960.283 | 0.146 | 960.204 | 6.78 |
| 09 | 1 | 959.189 | 0.128 | 958.791 | 64.63 | 19 | 2 | 958.376 | 0.201 | 958.314 | 6.27 |
| 09 | 1 | 959.840 | 0.117 | 959.463 | 65.21 | 19 | 2 | 958.971 | 0.157 | 958.888 | 5.79 |
| 09 | 1 | 960.020 | 0.112 | 959.623 | 65.84 | 19 | 2 | 958.349 | 0.159 | 958.288 | 5.35 |
| 09 | 1 | 959.553 | 0.139 | 959.182 | 66.47 | 19 | 2 | 959.814 | 0.169 | 959.772 | 4.89 |
| 09 | 1 | 959.881 | 0.149 | 959.437 | 67.12 | 19 | 2 | 959.271 | 0.184 | 959.184 | 4.44 |
| 09 | 1 | 958.953 | 0.149 | 958.510 | 67.78 | 19 | 2 | 959.104 | 0.161 | 958.998 | 4.04 |
| 09 | 1 | 960.646 | 0.123 | 960.182 | 69.18 | 19 | 2 | 959.719 | 0.150 | 959.600 | 3.65 |
| 09 | 1 | 960.121 | 0.128 | 959.651 | 69.92 | 19 | 2 | 958.382 | 0.167 | 958.275 | 3.25 |

| 2003 - SETEMBRO | | | | | | 2003 - SETEMBRO | | | | |
|---|---|---|---|---|---|---|---|---|---|---|
| D | L | SDB | ER | SDC | HL | D | L | SDB | ER | SDC | HL |
| 19 | 2 | 959.634 | 0.129 | 959.544 | 2.91 | 24 | 1 | 959.325 | 0.110 | 959.019 | 66.66 |
| 19 | 2 | 958.685 | 0.180 | 958.631 | 2.58 | 24 | 1 | 959.383 | 0.141 | 959.119 | 67.35 |
| 22 | 1 | 960.233 | 0.087 | 959.957 | 63.53 | 24 | 1 | 959.393 | 0.140 | 959.176 | 68.03 |
| 22 | 1 | 959.012 | 0.100 | 958.686 | 64.06 | 24 | 1 | 959.903 | 0.124 | 959.646 | 68.98 |
| 22 | 1 | 959.630 | 0.101 | 959.359 | 64.57 | 24 | 1 | 959.746 | 0.150 | 959.468 | 69.79 |
| 22 | 1 | 958.778 | 0.113 | 958.408 | 65.28 | 24 | 1 | 959.706 | 0.146 | 959.407 | 70.64 |
| 22 | 1 | 959.703 | 0.120 | 959.322 | 65.86 | 24 | 1 | 959.506 | 0.121 | 959.236 | 71.54 |
| 22 | 1 | 959.955 | 0.117 | 959.626 | 66.46 | 24 | 1 | 959.434 | 0.124 | 959.130 | 72.46 |
| 22 | 1 | 959.583 | 0.115 | 959.273 | 67.20 | 24 | 1 | 959.942 | 0.129 | 959.628 | 73.41 |
| 22 | 1 | 959.719 | 0.095 | 959.418 | 67.91 | 24 | 1 | 959.924 | 0.122 | 959.637 | 74.58 |
| 22 | 1 | 959.517 | 0.108 | 959.238 | 68.60 | 24 | 1 | 959.621 | 0.125 | 959.376 | 75.70 |
| 22 | 1 | 959.465 | 0.096 | 959.230 | 69.33 | 24 | 2 | 959.439 | 0.153 | 959.502 | 6.95 |
| 22 | 1 | 959.243 | 0.103 | 959.009 | 70.08 | 24 | 2 | 959.612 | 0.170 | 959.670 | 6.41 |
| 22 | 1 | 958.733 | 0.133 | 958.582 | 70.90 | 24 | 2 | 958.923 | 0.165 | 958.982 | 5.92 |
| 22 | 1 | 959.478 | 0.115 | 959.292 | 71.77 | 24 | 2 | 958.568 | 0.149 | 958.634 | 5.46 |
| 22 | 1 | 959.417 | 0.114 | 959.271 | 72.71 | 24 | 2 | 959.132 | 0.140 | 959.203 | 5.00 |
| 22 | 1 | 958.494 | 0.121 | 958.461 | 73.68 | 24 | 2 | 959.241 | 0.138 | 959.318 | 4.56 |
| 22 | 2 | 959.403 | 0.123 | 959.332 | 10.80 | 24 | 2 | 959.064 | 0.144 | 959.132 | 4.12 |
| 22 | 2 | 959.146 | 0.108 | 959.036 | 10.13 | 24 | 2 | 958.894 | 0.142 | 958.968 | 3.71 |
| 22 | 2 | 958.632 | 0.125 | 958.523 | 9.53 | 24 | 2 | 958.411 | 0.141 | 958.491 | 3.34 |
| 22 | 2 | 959.067 | 0.115 | 958.935 | 8.89 | 24 | 2 | 959.987 | 0.157 | 960.083 | 2.97 |
| 22 | 2 | 958.151 | 0.153 | 958.029 | 8.31 | 24 | 2 | 958.264 | 0.197 | 958.365 | 2.60 |
| 22 | 2 | 958.526 | 0.158 | 958.414 | 7.73 | 24 | 2 | 959.871 | 0.154 | 959.966 | 2.27 |
| 22 | 2 | 957.828 | 0.139 | 957.719 | 7.21 | 24 | 2 | 957.563 | 0.185 | 957.694 | 1.62 |
| 22 | 2 | 958.937 | 0.114 | 958.809 | 6.66 | 24 | 2 | 958.965 | 0.173 | 959.101 | 1.31 |
| 22 | 2 | 958.679 | 0.122 | 958.570 | 6.17 | | | | | | |
| 22 | 2 | 959.098 | 0.118 | 958.983 | 5.70 | | | 2003 - OUTUBRO | | | |
| 22 | 2 | 958.584 | 0.118 | 958.484 | 5.24 | D | L | SDB | ER | SDC | HL |
| 22 | 2 | 959.032 | 0.156 | 958.958 | 4.79 | 02 | 1 | 958.729 | 0.125 | 958.368 | 57.02 |
| 22 | 2 | 958.758 | 0.117 | 958.642 | 4.36 | 02 | 1 | 959.186 | 0.110 | 958.880 | 57.39 |
| 22 | 2 | 959.246 | 0.126 | 959.186 | 3.97 | 02 | 1 | 959.347 | 0.112 | 959.070 | 57.86 |
| 22 | 2 | 957.928 | 0.167 | 957.872 | 3.57 | 02 | 1 | 959.204 | 0.128 | 959.126 | 58.29 |
| 22 | 2 | 958.815 | 0.126 | 958.738 | 3.16 | 02 | 1 | 958.927 | 0.131 | 958.901 | 58.76 |
| 23 | 1 | 959.062 | 0.185 | 958.749 | 64.70 | 02 | 1 | 959.050 | 0.134 | 958.944 | 59.19 |
| 23 | 1 | 959.632 | 0.130 | 959.270 | 65.27 | 02 | 1 | 959.028 | 0.133 | 958.855 | 59.66 |
| 23 | 1 | 959.070 | 0.184 | 958.730 | 65.90 | 02 | 1 | 959.211 | 0.127 | 959.185 | 60.17 |
| 23 | 1 | 958.805 | 0.137 | 958.522 | 66.55 | 02 | 2 | 958.317 | 0.155 | 958.260 | 3.69 |
| 23 | 1 | 959.502 | 0.119 | 959.282 | 67.21 | 02 | 2 | 959.432 | 0.135 | 959.384 | 3.29 |
| 23 | 1 | 959.214 | 0.124 | 959.041 | 67.89 | 02 | 2 | 958.540 | 0.161 | 958.505 | 2.86 |
| 23 | 1 | 959.247 | 0.169 | 959.123 | 68.63 | 02 | 2 | 957.927 | 0.145 | 957.932 | 2.51 |
| 23 | 1 | 959.914 | 0.187 | 959.829 | 69.42 | 02 | 2 | 959.455 | 0.138 | 959.446 | 2.13 |
| 23 | 1 | 959.374 | 0.158 | 959.304 | 70.34 | 02 | 2 | 959.215 | 0.145 | 959.224 | 1.77 |
| 23 | 1 | 959.866 | 0.129 | 959.907 | 71.22 | 02 | 2 | 959.138 | 0.124 | 959.139 | 1.45 |
| 23 | 1 | 959.654 | 0.148 | 959.596 | 72.28 | 02 | 2 | 959.680 | 0.137 | 959.679 | 1.13 |
| 23 | 1 | 959.624 | 0.142 | 959.587 | 73.23 | 02 | 2 | 959.048 | 0.157 | 959.072 | 0.85 |
| 23 | 1 | 959.242 | 0.116 | 959.928 | 74.46 | 02 | 2 | 959.360 | 0.156 | 959.343 | 0.56 |
| 23 | 1 | 958.646 | 0.167 | 958.361 | 75.54 | 02 | 2 | 959.737 | 0.137 | 959.723 | 0.30 |
| 23 | 1 | 959.211 | 0.178 | 958.963 | 76.66 | 03 | 2 | 957.701 | 0.172 | 957.688 | 4.03 |
| 23 | 2 | 960.738 | 0.161 | 960.744 | 9.21 | 03 | 2 | 959.523 | 0.160 | 959.524 | 3.60 |
| 23 | 2 | 958.479 | 0.133 | 958.449 | 8.59 | 03 | 2 | 958.576 | 0.151 | 958.624 | 3.13 |
| 23 | 2 | 959.031 | 0.157 | 959.038 | 8.02 | 03 | 2 | 959.521 | 0.133 | 959.481 | 2.68 |
| 23 | 2 | 959.166 | 0.153 | 959.213 | 7.39 | 03 | 2 | 958.470 | 0.145 | 958.418 | 2.31 |
| 23 | 2 | 959.682 | 0.166 | 959.764 | 6.87 | 03 | 2 | 959.402 | 0.124 | 959.334 | 1.99 |
| 23 | 2 | 958.617 | 0.172 | 958.686 | 6.39 | 03 | 2 | 959.844 | 0.152 | 959.796 | 1.64 |
| 23 | 2 | 958.915 | 0.148 | 958.937 | 5.89 | 03 | 2 | 958.966 | 0.128 | 958.907 | 1.33 |
| 23 | 2 | 960.759 | 0.179 | 960.813 | 5.44 | 03 | 2 | 959.511 | 0.131 | 959.480 | 1.01 |
| 23 | 2 | 958.825 | 0.256 | 958.857 | 4.86 | 03 | 2 | 959.114 | 0.141 | 959.120 | 0.74 |
| 23 | 2 | 957.806 | 0.155 | 957.833 | 4.43 | 03 | 2 | 958.956 | 0.139 | 959.011 | 0.45 |
| 23 | 2 | 958.130 | 0.194 | 958.172 | 4.02 | 06 | 1 | 957.879 | 0.116 | 957.729 | 55.99 |
| 23 | 2 | 958.980 | 0.144 | 958.973 | 3.62 | 06 | 1 | 958.891 | 0.109 | 958.615 | 56.37 |
| 23 | 2 | 958.512 | 0.166 | 958.555 | 3.24 | 06 | 1 | 959.447 | 0.117 | 959.176 | 56.76 |
| 23 | 2 | 959.029 | 0.153 | 959.017 | 2.84 | 06 | 1 | 959.240 | 0.149 | 958.988 | 57.15 |
| 23 | 2 | 957.868 | 0.194 | 957.909 | 2.47 | 06 | 1 | 959.996 | 0.127 | 959.758 | 57.58 |
| 23 | 2 | 957.480 | 0.186 | 957.530 | 2.09 | 06 | 1 | 960.126 | 0.141 | 959.922 | 58.03 |
| 23 | 2 | 958.292 | 0.219 | 958.355 | 1.76 | 06 | 1 | 959.513 | 0.143 | 959.461 | 58.49 |
| 24 | 1 | 958.225 | 0.106 | 957.959 | 63.17 | 06 | 1 | 958.988 | 0.125 | 958.585 | 59.08 |
| 24 | 1 | 958.548 | 0.142 | 958.201 | 63.73 | 06 | 1 | 960.384 | 0.129 | 960.028 | 59.59 |
| 24 | 1 | 959.494 | 0.109 | 959.147 | 64.25 | 06 | 1 | 958.673 | 0.119 | 958.345 | 60.16 |
| 24 | 1 | 959.730 | 0.086 | 959.377 | 64.79 | 06 | 1 | 959.853 | 0.153 | 959.533 | 60.73 |
| 24 | 1 | 960.218 | 0.124 | 959.852 | 65.37 | 06 | 1 | 959.625 | 0.131 | 959.337 | 61.33 |
| 24 | 1 | 959.351 | 0.135 | 958.987 | 66.03 | | | | | | |

| 2003 - OUTUBRO | | | | | |
|---|---|---|---|---|---|
| D | L | SDB | ER | SDC | HL |
| 06 | 1 | 959.511 | 0.119 | 959.298 | 61.98 |
| 06 | 1 | 959.211 | 0.143 | 959.018 | 62.67 |
| 06 | 1 | 958.946 | 0.131 | 958.780 | 63.41 |
| 06 | 1 | 959.672 | 0.138 | 959.537 | 64.16 |
| 06 | 2 | 960.074 | 0.139 | 960.075 | 3.63 |
| 06 | 2 | 958.941 | 0.131 | 958.937 | 3.25 |
| 06 | 2 | 959.131 | 0.129 | 959.141 | 2.89 |
| 06 | 2 | 958.892 | 0.128 | 958.928 | 2.54 |
| 06 | 2 | 958.935 | 0.142 | 958.990 | 2.21 |
| 06 | 2 | 958.732 | 0.175 | 958.822 | 1.88 |
| 06 | 2 | 958.771 | 0.140 | 958.743 | 1.52 |
| 06 | 2 | 959.701 | 0.125 | 959.655 | 1.22 |
| 06 | 2 | 959.128 | 0.137 | 959.089 | 0.94 |
| 06 | 2 | 959.235 | 0.141 | 959.223 | 0.62 |
| 06 | 2 | 958.669 | 0.141 | 958.676 | 0.37 |
| 06 | 2 | 959.058 | 0.130 | 959.080 | 0.11 |
| 06 | 2 | 958.625 | 0.126 | 958.662 | 0.00 |
| 06 | 2 | 958.991 | 0.146 | 959.067 | 0.00 |
| 09 | 2 | 958.663 | 0.135 | 958.786 | 0.00 |
| 09 | 2 | 959.201 | 0.141 | 959.335 | 0.00 |
| 09 | 2 | 958.885 | 0.129 | 958.937 | 0.00 |
| 09 | 2 | 959.551 | 0.151 | 959.594 | 0.00 |
| 09 | 2 | 958.730 | 0.125 | 958.757 | 0.00 |
| 09 | 2 | 959.142 | 0.148 | 959.213 | 0.00 |
| 09 | 2 | 959.254 | 0.132 | 959.354 | 0.00 |
| 09 | 2 | 959.076 | 0.127 | 959.163 | 0.00 |
| 09 | 2 | 959.454 | 0.142 | 959.568 | 0.00 |
| 09 | 2 | 958.735 | 0.146 | 958.867 | 0.00 |
| 16 | 2 | 958.112 | 0.155 | 957.993 | 0.00 |
| 16 | 2 | 959.657 | 0.155 | 959.545 | 0.00 |
| 16 | 2 | 958.833 | 0.137 | 958.744 | 0.00 |
| 16 | 2 | 959.534 | 0.141 | 959.451 | 0.00 |
| 16 | 2 | 958.482 | 0.135 | 958.406 | 0.00 |
| 16 | 2 | 959.312 | 0.131 | 959.165 | 0.00 |
| 16 | 2 | 959.346 | 0.142 | 959.200 | 0.00 |
| 16 | 2 | 959.328 | 0.133 | 959.169 | 0.00 |
| 16 | 2 | 959.402 | 0.148 | 959.242 | 0.00 |
| 16 | 2 | 959.401 | 0.144 | 959.249 | 0.00 |
| 16 | 2 | 959.576 | 0.168 | 959.433 | 0.00 |
| 17 | 1 | 958.070 | 0.197 | 958.006 | 54.26 |
| 17 | 1 | 958.495 | 0.149 | 958.397 | 54.89 |
| 17 | 1 | 959.330 | 0.160 | 959.188 | 55.50 |
| 17 | 1 | 959.174 | 0.151 | 959.131 | 56.15 |
| 17 | 1 | 959.333 | 0.132 | 959.125 | 56.93 |
| 17 | 1 | 958.736 | 0.139 | 958.548 | 57.64 |
| 17 | 1 | 959.511 | 0.172 | 959.344 | 58.39 |
| 17 | 1 | 958.917 | 0.157 | 958.818 | 59.24 |
| 17 | 1 | 960.088 | 0.173 | 960.007 | 60.10 |
| 17 | 2 | 958.919 | 0.166 | 958.859 | 0.00 |
| 17 | 2 | 959.325 | 0.135 | 959.244 | 0.00 |
| 17 | 2 | 959.122 | 0.136 | 959.032 | 0.00 |
| 17 | 2 | 959.336 | 0.117 | 959.216 | 0.00 |
| 17 | 2 | 959.455 | 0.148 | 959.333 | 0.00 |
| 17 | 2 | 959.443 | 0.152 | 959.319 | 0.00 |
| 17 | 2 | 959.099 | 0.139 | 958.968 | 0.00 |
| 17 | 2 | 959.482 | 0.140 | 959.337 | 0.00 |
| 17 | 2 | 959.423 | 0.137 | 959.279 | 0.00 |
| 17 | 2 | 959.055 | 0.155 | 958.918 | 0.00 |
| 17 | 2 | 958.411 | 0.167 | 958.295 | 0.00 |
| 21 | 2 | 959.156 | 0.130 | 959.228 | 5.10 |
| 21 | 2 | 959.201 | 0.127 | 959.258 | 4.63 |
| 21 | 2 | 958.780 | 0.125 | 958.827 | 4.18 |
| 21 | 2 | 958.986 | 0.126 | 959.020 | 3.77 |
| 21 | 2 | 959.027 | 0.138 | 959.045 | 3.33 |
| 21 | 2 | 958.943 | 0.124 | 958.953 | 2.94 |
| 21 | 2 | 959.060 | 0.140 | 959.076 | 2.57 |
| 21 | 2 | 959.228 | 0.149 | 959.261 | 2.14 |
| 21 | 2 | 958.526 | 0.160 | 958.571 | 1.82 |
| 21 | 2 | 958.803 | 0.136 | 958.848 | 1.51 |
| 21 | 2 | 958.734 | 0.178 | 958.817 | 1.24 |
| 24 | 2 | 959.904 | 0.137 | 959.965 | 3.77 |

| 2003 - OUTUBRO | | | | | |
|---|---|---|---|---|---|
| D | L | SDB | ER | SDC | HL |
| 24 | 2 | 958.951 | 0.132 | 958.947 | 3.24 |
| 24 | 2 | 959.519 | 0.151 | 959.495 | 2.88 |
| 24 | 2 | 959.449 | 0.123 | 959.400 | 2.52 |
| 24 | 2 | 958.667 | 0.143 | 958.618 | 2.20 |
| 24 | 2 | 959.250 | 0.161 | 959.225 | 1.84 |
| 24 | 2 | 958.050 | 0.164 | 958.018 | 1.56 |
| 24 | 2 | 959.646 | 0.154 | 959.664 | 1.29 |
| 28 | 1 | 958.566 | 0.165 | 958.309 | 41.09 |
| 28 | 1 | 959.250 | 0.114 | 958.911 | 41.25 |
| 28 | 1 | 959.026 | 0.099 | 958.707 | 41.44 |
| 28 | 1 | 959.126 | 0.120 | 958.782 | 41.64 |
| 28 | 1 | 959.003 | 0.119 | 958.657 | 41.86 |
| 28 | 1 | 959.745 | 0.139 | 959.453 | 42.09 |
| 28 | 1 | 959.360 | 0.110 | 959.028 | 42.32 |
| 28 | 1 | 959.194 | 0.118 | 958.814 | 42.57 |
| 28 | 1 | 959.799 | 0.100 | 959.411 | 42.84 |
| 28 | 1 | 958.940 | 0.123 | 958.575 | 43.13 |
| 28 | 1 | 959.783 | 0.145 | 959.491 | 43.46 |
| 28 | 1 | 959.988 | 0.142 | 959.733 | 43.80 |
| 28 | 1 | 959.751 | 0.143 | 959.553 | 44.17 |
| 28 | 1 | 959.676 | 0.148 | 959.538 | 44.54 |
| 29 | 1 | 958.567 | 0.127 | 958.243 | 42.49 |
| 29 | 1 | 958.613 | 0.105 | 958.476 | 42.80 |
| 29 | 1 | 958.091 | 0.162 | 958.084 | 43.16 |
| 29 | 2 | 960.058 | 0.118 | 960.029 | 0.00 |
| 29 | 2 | 958.204 | 0.143 | 958.221 | 0.00 |
| 29 | 2 | 959.424 | 0.131 | 959.422 | 0.00 |
| 29 | 2 | 958.283 | 0.127 | 958.298 | 0.00 |
| 29 | 2 | 959.440 | 0.131 | 959.422 | 0.00 |
| 29 | 2 | 958.713 | 0.151 | 958.672 | 0.00 |
| 29 | 2 | 958.471 | 0.142 | 958.428 | 0.00 |
| 29 | 2 | 959.653 | 0.131 | 959.611 | 0.00 |
| 29 | 2 | 958.871 | 0.126 | 958.841 | 0.00 |
| 29 | 2 | 959.064 | 0.129 | 959.021 | 0.00 |
| 29 | 2 | 959.121 | 0.134 | 959.118 | 0.00 |
| 29 | 2 | 958.935 | 0.117 | 958.909 | 0.00 |
| 29 | 2 | 958.948 | 0.168 | 958.931 | 0.00 |
| 29 | 2 | 958.617 | 0.192 | 958.649 | 0.00 |
| 30 | 2 | 959.511 | 0.117 | 959.475 | 0.47 |
| 30 | 2 | 958.754 | 0.153 | 958.728 | 0.30 |
| 30 | 2 | 958.502 | 0.150 | 958.483 | 0.14 |
| 30 | 2 | 958.898 | 0.149 | 958.889 | 0.00 |
| 30 | 2 | 958.513 | 0.142 | 958.500 | 0.00 |
| 30 | 2 | 959.893 | 0.132 | 959.853 | 0.00 |
| 30 | 2 | 958.934 | 0.148 | 958.877 | 0.00 |
| 30 | 2 | 959.464 | 0.150 | 959.393 | 0.00 |
| 30 | 2 | 959.648 | 0.122 | 959.558 | 0.00 |
| 30 | 2 | 959.104 | 0.137 | 959.034 | 0.00 |
| 30 | 2 | 959.103 | 0.148 | 959.050 | 0.00 |
| 30 | 2 | 959.079 | 0.150 | 959.056 | 0.00 |
| 31 | 1 | 959.068 | 0.149 | 958.713 | 38.17 |
| 31 | 1 | 960.058 | 0.128 | 959.769 | 38.26 |
| 31 | 1 | 958.985 | 0.124 | 958.635 | 38.35 |
| 31 | 1 | 959.393 | 0.095 | 959.060 | 38.45 |
| 31 | 1 | 959.934 | 0.112 | 959.636 | 38.58 |
| 31 | 1 | 959.611 | 0.106 | 959.468 | 38.70 |
| 31 | 1 | 959.605 | 0.136 | 959.428 | 38.89 |
| 31 | 1 | 959.372 | 0.144 | 959.231 | 39.04 |
| 31 | 1 | 958.042 | 0.176 | 959.491 | 39.24 |
| 31 | 1 | 958.042 | 0.140 | 957.929 | 39.41 |
| 31 | 1 | 959.679 | 0.164 | 959.505 | 39.61 |
| 31 | 1 | 959.548 | 0.128 | 959.270 | 39.81 |
| 31 | 1 | 959.097 | 0.130 | 958.801 | 40.02 |
| 31 | 1 | 959.608 | 0.113 | 959.391 | 40.26 |
| 31 | 1 | 959.421 | 0.106 | 959.275 | 40.53 |
| 31 | 2 | 959.281 | 0.147 | 959.272 | 2.03 |
| 31 | 2 | 959.339 | 0.127 | 959.307 | 1.77 |
| 31 | 2 | 959.176 | 0.118 | 959.131 | 1.51 |
| 31 | 2 | 958.791 | 0.125 | 958.728 | 1.27 |
| 31 | 2 | 958.782 | 0.122 | 958.722 | 1.04 |
| 31 | 2 | 959.317 | 0.122 | 959.262 | 0.83 |

```
     2003 - OUTUBRO                                  2003 - NOVEMBRO
 D L   SDB     ER     SDC    HL              D  L   SDB     ER     SDC    HL
31 2 959.745 0.141 959.694  0.64             12 2 958.062 0.121 958.080  0.41
31 2 960.340 0.131 960.298  0.45             12 2 958.084 0.139 958.115  0.37
31 2 959.883 0.131 959.845  0.27             12 2 958.559 0.126 958.602  0.33
31 2 960.058 0.144 960.029  0.12             12 2 957.958 0.133 958.003  0.30
31 2 959.710 0.123 959.671  0.00             12 2 958.617 0.145 958.701  0.28
31 2 958.735 0.149 958.712  0.00             12 2 959.217 0.118 959.264  0.27
                                             12 2 959.087 0.140 959.154  0.26
     2003 - NOVEMBRO                         12 2 959.293 0.150 959.374  0.26
 D L   SDB     ER     SDC    HL              25 1 958.437 0.130 958.213 19.07
07 2 959.937 0.156 960.011  1.09             25 1 958.236 0.103 958.059 18.99
07 2 959.770 0.162 959.768  0.92             25 1 958.397 0.135 958.322 18.85
07 2 959.934 0.126 959.917  0.77             25 1 959.243 0.123 959.203 18.78
07 2 958.850 0.170 958.829  0.63             25 1 958.905 0.126 958.832 18.72
07 2 958.829 0.156 958.821  0.45             25 1 959.051 0.134 958.995 18.66
07 2 958.868 0.172 958.872  0.34             25 1 958.474 0.153 958.483 18.55
07 2 958.166 0.170 958.168  0.24             25 1 958.541 0.148 958.547 18.51
07 2 957.945 0.178 957.962  0.14             25 1 958.892 0.163 958.880 18.48
07 2 959.406 0.158 959.462  0.00             25 1 959.409 0.169 959.406 18.45
07 2 959.239 0.155 959.271  0.00             25 1 958.589 0.169 958.602 18.43
07 2 959.390 0.175 959.417  0.00             25 2 958.955 0.123 959.093  2.06
07 2 958.313 0.136 958.320  0.00             25 2 958.037 0.127 958.181  2.15
07 2 959.222 0.156 959.227  0.00             25 2 958.557 0.135 958.639  2.26
10 1 958.201 0.112 957.932 30.93             25 2 959.000 0.111 959.095  2.35
10 1 958.542 0.131 958.272 30.94             25 2 959.007 0.114 959.092  2.45
10 1 959.434 0.130 959.200 30.97             25 2 959.354 0.145 959.474  2.54
10 1 959.170 0.115 958.957 31.00             25 2 959.032 0.152 959.166  2.64
10 1 959.668 0.115 959.482 31.04             25 2 959.549 0.148 959.673  2.75
10 1 959.166 0.122 959.043 31.09             25 2 959.252 0.130 959.329  2.88
10 1 959.700 0.188 959.561 31.14             25 2 958.701 0.138 958.767  2.98
10 1 959.596 0.124 959.491 31.21             25 2 958.729 0.143 958.779  3.10
10 1 959.031 0.126 958.958 31.28             25 2 958.508 0.125 958.556  3.22
10 1 960.056 0.134 960.058 31.36             26 1 957.865 0.127 957.624 17.74
10 1 960.123 0.177 960.082 31.45             26 1 959.117 0.158 958.914 17.68
10 1 960.495 0.184 960.501 31.55             26 1 958.705 0.136 958.533 17.63
10 1 960.323 0.138 960.136 31.67             26 1 958.989 0.128 958.866 17.59
10 1 959.349 0.140 959.246 31.81             26 1 958.104 0.152 957.981 17.55
10 1 959.835 0.166 959.800 31.95             26 1 958.800 0.142 958.729 17.52
11 2 958.644 0.151 958.601  0.86             26 1 958.282 0.126 958.208 17.50
11 2 959.245 0.162 959.194  0.75             26 1 958.916 0.149 958.862 17.49
11 2 959.737 0.138 959.690  0.65             26 1 959.279 0.143 959.243 17.49
11 2 958.742 0.171 958.709  0.57
11 2 958.699 0.151 958.644  0.48                  2004 - JANEIRO
11 2 958.775 0.147 958.711  0.41              D L   SDB     ER     SDC    HL
11 2 958.185 0.154 958.131  0.36             06 1 959.570 0.119 959.570 17.49
11 2 958.662 0.146 958.606  0.30             06 1 959.822 0.113 959.822 17.49
11 2 959.030 0.125 958.973  0.25             06 1 959.970 0.092 959.970 17.49
11 2 958.517 0.134 958.467  0.22             06 1 959.472 0.160 959.472 17.49
11 2 958.931 0.151 958.950  0.16             06 1 959.295 0.124 959.295 17.49
11 2 958.097 0.160 958.128  0.15             06 1 959.337 0.137 959.337 17.49
11 2 958.562 0.151 958.596  0.15             06 1 959.707 0.136 959.707 17.49
11 2 958.753 0.151 958.797  0.15             06 1 959.228 0.152 959.228 17.49
11 2 958.947 0.136 958.981  0.16             06 1 959.199 0.152 959.199 17.49
12 1 957.771 0.091 957.419 29.42             06 1 959.139 0.174 959.139 17.49
12 1 958.842 0.103 958.532 29.44             06 1 960.186 0.139 960.186 17.49
12 1 959.509 0.083 959.276 29.46             06 1 959.022 0.144 959.022 17.49
12 1 959.486 0.102 959.286 29.49             06 1 959.030 0.127 959.030 17.49
12 1 960.038 0.101 959.785 29.54             06 1 959.382 0.146 959.382 17.49
12 1 960.862 0.110 960.644 29.58             06 1 958.809 0.148 958.809 17.49
12 1 959.514 0.131 959.297 29.65             06 1 958.992 0.140 958.992 17.49
12 1 959.018 0.102 958.857 29.71             06 1 959.404 0.124 959.404 17.49
12 1 960.341 0.117 960.211 29.79             06 1 958.848 0.134 958.848 17.49
12 1 959.573 0.119 959.474 29.88             06 1 958.870 0.112 958.870 17.49
12 1 959.128 0.115 958.905 29.99             06 1 958.826 0.119 958.826 17.49
12 1 959.705 0.124 959.528 30.10             06 1 959.053 0.096 959.053 17.49
12 1 959.176 0.113 959.055 30.22             06 1 958.968 0.113 958.968 17.49
12 2 960.340 0.145 960.411  0.77             07 1 958.208 0.139 958.208 17.49
12 2 958.925 0.142 958.975  0.67             07 1 958.525 0.106 958.525 17.49
12 2 958.830 0.142 958.875  0.59             07 1 959.609 0.114 959.609 17.49
12 2 958.154 0.110 958.173  0.53             07 1 959.741 0.109 959.741 17.49
12 2 958.082 0.106 958.109  0.47             07 1 959.405 0.136 959.405 17.49
```

| 2004 - JANEIRO | | | | | | 2004 - JANEIRO | | | | |
|---|---|---|---|---|---|---|---|---|---|---|
| D | L | SDB | ER | SDC | HL | D | L | SDB | ER | SDC | HL |
| 07 | 1 | 958.886 | 0.143 | 958.886 | 17.49 | 21 | 1 | 959.145 | 0.157 | 959.145 | 17.49 |
| 07 | 1 | 959.766 | 0.131 | 959.766 | 17.49 | 21 | 1 | 959.282 | 0.152 | 959.282 | 17.49 |
| 07 | 1 | 958.963 | 0.148 | 958.963 | 17.49 | 21 | 1 | 958.787 | 0.141 | 958.787 | 17.49 |
| 07 | 1 | 959.429 | 0.112 | 959.429 | 17.49 | 21 | 1 | 959.020 | 0.155 | 959.020 | 17.49 |
| 07 | 1 | 959.372 | 0.140 | 959.372 | 17.49 | 21 | 1 | 960.181 | 0.102 | 960.181 | 17.49 |
| 12 | 1 | 958.865 | 0.122 | 958.865 | 17.49 | 21 | 1 | 958.903 | 0.125 | 958.903 | 17.49 |
| 12 | 1 | 959.444 | 0.165 | 959.444 | 17.49 | 21 | 1 | 959.410 | 0.126 | 959.410 | 17.49 |
| 12 | 1 | 959.592 | 0.130 | 959.592 | 17.49 | 21 | 1 | 959.120 | 0.109 | 959.120 | 17.49 |
| 12 | 1 | 959.808 | 0.132 | 959.808 | 17.49 | 21 | 1 | 959.174 | 0.139 | 959.174 | 17.49 |
| 12 | 1 | 959.097 | 0.094 | 959.097 | 17.49 | 21 | 1 | 958.203 | 0.127 | 958.203 | 17.49 |
| 12 | 1 | 959.431 | 0.115 | 959.431 | 17.49 | 21 | 1 | 958.295 | 0.133 | 958.295 | 17.49 |
| 12 | 1 | 959.521 | 0.143 | 959.521 | 17.49 | 21 | 1 | 958.951 | 0.128 | 958.951 | 17.49 |
| 12 | 1 | 960.160 | 0.119 | 960.160 | 17.49 | 21 | 1 | 958.859 | 0.131 | 958.859 | 17.49 |
| 12 | 1 | 958.811 | 0.163 | 958.811 | 17.49 | 21 | 1 | 957.431 | 0.155 | 957.431 | 17.49 |
| 12 | 1 | 959.646 | 0.148 | 959.646 | 17.49 | 21 | 1 | 958.789 | 0.151 | 958.789 | 17.49 |
| 12 | 1 | 959.996 | 0.125 | 959.996 | 17.49 | 21 | 1 | 958.292 | 0.181 | 958.292 | 17.49 |
| 12 | 1 | 960.096 | 0.126 | 960.096 | 17.49 | 21 | 1 | 957.968 | 0.310 | 957.968 | 17.49 |
| 12 | 1 | 959.704 | 0.132 | 959.704 | 17.49 | 29 | 1 | 958.819 | 0.139 | 958.819 | 17.49 |
| 12 | 1 | 959.523 | 0.137 | 959.523 | 17.49 | 29 | 1 | 958.454 | 0.108 | 958.454 | 17.49 |
| 12 | 1 | 960.177 | 0.129 | 960.177 | 17.49 | 29 | 1 | 959.160 | 0.131 | 959.160 | 17.49 |
| 12 | 1 | 959.429 | 0.164 | 959.429 | 17.49 | 29 | 1 | 959.113 | 0.114 | 959.113 | 17.49 |
| 12 | 1 | 959.210 | 0.198 | 959.210 | 17.49 | 29 | 1 | 958.463 | 0.136 | 958.463 | 17.49 |
| 12 | 1 | 958.987 | 0.107 | 958.987 | 17.49 | 29 | 1 | 958.868 | 0.133 | 958.868 | 17.49 |
| 12 | 1 | 959.397 | 0.124 | 959.397 | 17.49 | 29 | 1 | 959.013 | 0.132 | 959.013 | 17.49 |
| 12 | 1 | 958.308 | 0.164 | 958.308 | 17.49 | 29 | 1 | 959.511 | 0.137 | 959.511 | 17.49 |
| 14 | 1 | 958.629 | 0.121 | 958.629 | 17.49 | 29 | 1 | 957.915 | 0.125 | 957.915 | 17.49 |
| 14 | 1 | 959.042 | 0.174 | 959.042 | 17.49 | 29 | 1 | 958.427 | 0.149 | 958.427 | 17.49 |
| 14 | 1 | 959.689 | 0.150 | 959.689 | 17.49 | 29 | 1 | 959.273 | 0.128 | 959.273 | 17.49 |
| 14 | 1 | 959.089 | 0.151 | 959.089 | 17.49 | 29 | 1 | 958.971 | 0.130 | 958.971 | 17.49 |
| 14 | 1 | 958.579 | 0.171 | 958.579 | 17.49 | 29 | 1 | 959.198 | 0.167 | 959.198 | 17.49 |
| 15 | 1 | 958.650 | 0.167 | 958.650 | 17.49 | 29 | 1 | 958.838 | 0.122 | 958.838 | 17.49 |
| 15 | 1 | 958.918 | 0.180 | 958.918 | 17.49 | 29 | 1 | 959.419 | 0.131 | 959.419 | 17.49 |
| 15 | 1 | 958.713 | 0.139 | 958.713 | 17.49 | 29 | 1 | 959.380 | 0.113 | 959.380 | 17.49 |
| 15 | 1 | 959.942 | 0.130 | 959.942 | 17.49 | 29 | 1 | 958.737 | 0.117 | 958.737 | 17.49 |
| 15 | 1 | 959.017 | 0.152 | 959.017 | 17.49 | 29 | 1 | 959.845 | 0.145 | 959.845 | 17.49 |
| 15 | 1 | 958.901 | 0.155 | 958.901 | 17.49 | 29 | 1 | 959.472 | 0.181 | 959.472 | 17.49 |
| 15 | 1 | 959.439 | 0.162 | 959.439 | 17.49 | 29 | 1 | 959.832 | 0.135 | 959.832 | 17.49 |
| 15 | 1 | 958.445 | 0.134 | 958.445 | 17.49 | 29 | 1 | 958.383 | 0.125 | 958.383 | 17.49 |
| 15 | 1 | 958.977 | 0.138 | 958.977 | 17.49 | 29 | 1 | 959.127 | 0.129 | 959.127 | 17.49 |
| 15 | 1 | 959.166 | 0.129 | 959.166 | 17.49 | 29 | 1 | 959.133 | 0.134 | 959.133 | 17.49 |
| 15 | 1 | 959.105 | 0.121 | 959.105 | 17.49 | 30 | 1 | 959.082 | 0.116 | 959.082 | 17.49 |
| 19 | 1 | 959.683 | 0.115 | 959.683 | 17.49 | 30 | 1 | 959.448 | 0.137 | 959.448 | 17.49 |
| 19 | 1 | 959.320 | 0.113 | 959.320 | 17.49 | 30 | 1 | 959.142 | 0.113 | 959.142 | 17.49 |
| 19 | 1 | 959.138 | 0.100 | 959.138 | 17.49 | 30 | 1 | 959.249 | 0.155 | 959.249 | 17.49 |
| 19 | 1 | 959.229 | 0.108 | 959.229 | 17.49 | 30 | 1 | 958.525 | 0.162 | 958.525 | 17.49 |
| 19 | 1 | 959.967 | 0.129 | 959.967 | 17.49 | 30 | 1 | 960.055 | 0.179 | 960.055 | 17.49 |
| 19 | 1 | 959.064 | 0.123 | 959.064 | 17.49 | 30 | 1 | 958.852 | 0.153 | 958.852 | 17.49 |
| 19 | 1 | 959.344 | 0.131 | 959.344 | 17.49 | 30 | 1 | 960.403 | 0.149 | 960.403 | 17.49 |
| 19 | 1 | 958.880 | 0.154 | 958.880 | 17.49 | 30 | 1 | 959.311 | 0.205 | 959.311 | 17.49 |
| 19 | 1 | 958.105 | 0.144 | 958.105 | 17.49 | 30 | 1 | 959.491 | 0.138 | 959.491 | 17.49 |
| 19 | 1 | 958.664 | 0.144 | 958.664 | 17.49 | 30 | 1 | 959.305 | 0.157 | 959.305 | 17.49 |
| 19 | 1 | 958.844 | 0.107 | 958.844 | 17.49 | 30 | 1 | 959.340 | 0.133 | 959.340 | 17.49 |
| 19 | 1 | 959.704 | 0.103 | 959.704 | 17.49 | 30 | 1 | 960.184 | 0.160 | 960.184 | 17.49 |
| 19 | 1 | 958.714 | 0.123 | 958.714 | 17.49 | 30 | 1 | 958.889 | 0.142 | 958.889 | 17.49 |
| 19 | 1 | 960.244 | 0.150 | 960.244 | 17.49 | 30 | 1 | 959.649 | 0.165 | 959.649 | 17.49 |
| 19 | 1 | 960.012 | 0.141 | 960.012 | 17.49 | 30 | 1 | 959.359 | 0.124 | 959.359 | 17.49 |
| 19 | 1 | 959.030 | 0.129 | 959.030 | 17.49 | 30 | 1 | 958.747 | 0.148 | 958.747 | 17.49 |
| 19 | 1 | 958.624 | 0.141 | 958.624 | 17.49 | 30 | 1 | 958.468 | 0.123 | 958.468 | 17.49 |
| 19 | 1 | 959.223 | 0.133 | 959.223 | 17.49 | 30 | 1 | 959.463 | 0.125 | 959.463 | 17.49 |
| 19 | 1 | 958.788 | 0.155 | 958.788 | 17.49 | 30 | 1 | 960.517 | 0.163 | 960.517 | 17.49 |
| 19 | 1 | 958.902 | 0.194 | 958.902 | 17.49 | 30 | 1 | 959.218 | 0.126 | 959.218 | 17.49 |
| 21 | 1 | 958.066 | 0.151 | 958.066 | 17.49 | 30 | 1 | 958.842 | 0.117 | 958.842 | 17.49 |
| 21 | 1 | 958.637 | 0.131 | 958.637 | 17.49 | 30 | 1 | 959.571 | 0.140 | 959.571 | 17.49 |
| 21 | 1 | 959.061 | 0.146 | 959.061 | 17.49 | | | | | | |
| 21 | 1 | 959.062 | 0.135 | 959.062 | 17.49 | | | | | | |
| 21 | 1 | 959.447 | 0.131 | 959.447 | 17.49 | 2004 - FEVEREIRO | | | | | |
| 21 | 1 | 959.036 | 0.147 | 959.036 | 17.49 | D | L | SDB | ER | SDC | HL |
| 21 | 1 | 960.013 | 0.179 | 960.013 | 17.49 | 02 | 1 | 960.085 | 0.123 | 960.085 | 17.49 |
| 21 | 1 | 958.967 | 0.170 | 958.967 | 17.49 | 02 | 1 | 958.680 | 0.121 | 958.680 | 17.49 |
| 21 | 1 | 959.338 | 0.166 | 959.338 | 17.49 | 02 | 1 | 959.231 | 0.118 | 959.231 | 17.49 |
| 21 | 1 | 959.394 | 0.184 | 959.394 | 17.49 | 02 | 1 | 959.236 | 0.131 | 959.236 | 17.49 |

| 2004 - FEVEREIRO | | | | |
|---|---|---|---|---|
| D L | SDB | ER | SDC | HL |
| 02 1 | 959.006 | 0.125 | 959.006 | 17.49 |
| 02 1 | 959.131 | 0.103 | 959.131 | 17.49 |
| 02 1 | 959.333 | 0.146 | 959.333 | 17.49 |
| 02 1 | 958.594 | 0.141 | 958.594 | 17.49 |
| 02 1 | 959.701 | 0.129 | 959.701 | 17.49 |
| 02 1 | 959.286 | 0.108 | 959.286 | 17.49 |
| 02 1 | 959.382 | 0.129 | 959.382 | 17.49 |
| 02 1 | 958.911 | 0.135 | 958.911 | 17.49 |
| 03 1 | 959.675 | 0.152 | 959.675 | 17.49 |
| 03 1 | 959.780 | 0.119 | 959.780 | 17.49 |
| 03 1 | 959.712 | 0.134 | 959.712 | 17.49 |
| 03 1 | 959.792 | 0.142 | 959.792 | 17.49 |
| 03 1 | 960.712 | 0.108 | 960.712 | 17.49 |
| 03 1 | 959.361 | 0.112 | 959.361 | 17.49 |
| 03 1 | 959.475 | 0.141 | 959.475 | 17.49 |
| 03 1 | 959.762 | 0.128 | 959.762 | 17.49 |
| 03 1 | 959.305 | 0.121 | 959.305 | 17.49 |
| 03 1 | 960.269 | 0.136 | 960.269 | 17.49 |
| 03 1 | 959.352 | 0.103 | 959.352 | 17.49 |
| 03 1 | 959.555 | 0.123 | 959.555 | 17.49 |
| 03 1 | 959.676 | 0.121 | 959.676 | 17.49 |
| 03 1 | 959.268 | 0.138 | 959.268 | 17.49 |
| 03 1 | 959.834 | 0.108 | 959.834 | 17.49 |
| 03 1 | 958.724 | 0.100 | 958.724 | 17.49 |
| 03 1 | 959.516 | 0.111 | 959.516 | 17.49 |
| 03 1 | 958.845 | 0.120 | 958.845 | 17.49 |
| 03 1 | 958.585 | 0.105 | 958.585 | 17.49 |
| 03 1 | 959.518 | 0.113 | 959.518 | 17.49 |
| 03 1 | 958.772 | 0.150 | 958.772 | 17.49 |
| 03 1 | 958.731 | 0.110 | 958.731 | 17.49 |
| 03 1 | 959.066 | 0.139 | 959.066 | 17.49 |
| 03 1 | 958.846 | 0.133 | 958.846 | 17.49 |
| 03 1 | 958.931 | 0.125 | 958.931 | 17.49 |
| 03 1 | 958.497 | 0.146 | 958.497 | 17.49 |
| 03 1 | 959.579 | 0.130 | 959.579 | 17.49 |
| 03 1 | 958.196 | 0.157 | 958.196 | 17.49 |
| 03 1 | 959.502 | 0.132 | 959.502 | 17.49 |
| 04 1 | 959.853 | 0.137 | 959.853 | 17.49 |
| 04 1 | 959.434 | 0.133 | 959.434 | 17.49 |
| 04 1 | 959.195 | 0.119 | 959.195 | 17.49 |
| 04 1 | 959.568 | 0.142 | 959.568 | 17.49 |
| 04 1 | 959.459 | 0.186 | 959.459 | 17.49 |
| 04 1 | 959.339 | 0.140 | 959.339 | 17.49 |
| 04 1 | 959.831 | 0.159 | 959.831 | 17.49 |
| 04 1 | 959.465 | 0.124 | 959.465 | 17.49 |
| 04 1 | 959.693 | 0.152 | 959.693 | 17.49 |
| 04 1 | 959.873 | 0.125 | 959.873 | 17.49 |
| 04 1 | 959.313 | 0.146 | 959.313 | 17.49 |
| 04 1 | 959.181 | 0.140 | 959.181 | 17.49 |
| 04 1 | 958.613 | 0.139 | 958.613 | 17.49 |
| 04 1 | 959.512 | 0.119 | 959.512 | 17.49 |
| 04 1 | 959.354 | 0.114 | 959.354 | 17.49 |
| 04 1 | 959.274 | 0.119 | 959.274 | 17.49 |
| 04 1 | 958.979 | 0.117 | 958.979 | 17.49 |
| 04 1 | 958.649 | 0.137 | 958.649 | 17.49 |
| 04 1 | 958.602 | 0.112 | 958.602 | 17.49 |
| 04 1 | 959.326 | 0.123 | 959.326 | 17.49 |
| 04 1 | 959.061 | 0.113 | 959.061 | 17.49 |
| 04 1 | 958.445 | 0.125 | 958.445 | 17.49 |
| 04 1 | 959.436 | 0.159 | 959.436 | 17.49 |
| 04 1 | 959.401 | 0.136 | 959.401 | 17.49 |
| 04 1 | 958.940 | 0.125 | 958.940 | 17.49 |
| 04 1 | 959.540 | 0.141 | 959.540 | 17.49 |
| 04 1 | 958.691 | 0.146 | 958.691 | 17.49 |
| 05 1 | 958.665 | 0.137 | 958.665 | 17.49 |
| 05 1 | 959.254 | 0.139 | 959.254 | 17.49 |
| 05 1 | 959.071 | 0.137 | 959.071 | 17.49 |
| 05 1 | 959.093 | 0.139 | 959.093 | 17.49 |
| 05 1 | 958.972 | 0.108 | 958.972 | 17.49 |
| 05 1 | 959.471 | 0.118 | 959.471 | 17.49 |
| 05 1 | 958.626 | 0.140 | 958.626 | 17.49 |

| 2004 - FEVEREIRO | | | | |
|---|---|---|---|---|
| D L | SDB | ER | SDC | HL |
| 05 1 | 958.226 | 0.122 | 958.226 | 17.49 |
| 05 1 | 959.097 | 0.105 | 959.097 | 17.49 |
| 05 1 | 959.203 | 0.153 | 959.203 | 17.49 |
| 05 1 | 958.965 | 0.148 | 958.965 | 17.49 |
| 05 1 | 959.081 | 0.127 | 959.081 | 17.49 |
| 05 1 | 959.698 | 0.122 | 959.698 | 17.49 |
| 05 1 | 959.126 | 0.169 | 959.126 | 17.49 |
| 05 1 | 960.134 | 0.137 | 960.134 | 17.49 |
| 05 1 | 959.462 | 0.115 | 959.462 | 17.49 |
| 05 1 | 959.419 | 0.126 | 959.419 | 17.49 |
| 05 1 | 959.236 | 0.111 | 959.236 | 17.49 |
| 05 1 | 958.339 | 0.112 | 958.339 | 17.49 |
| 05 1 | 959.203 | 0.111 | 959.203 | 17.49 |
| 05 1 | 959.671 | 0.124 | 959.671 | 17.49 |
| 05 1 | 959.132 | 0.147 | 959.132 | 17.49 |
| 13 1 | 958.908 | 0.161 | 958.908 | 17.49 |
| 13 1 | 958.868 | 0.132 | 958.868 | 17.49 |
| 13 1 | 958.780 | 0.126 | 958.780 | 17.49 |
| 13 1 | 959.337 | 0.107 | 959.337 | 17.49 |
| 13 1 | 959.021 | 0.116 | 959.021 | 17.49 |
| 13 1 | 959.592 | 0.138 | 959.592 | 17.49 |
| 13 1 | 958.315 | 0.128 | 958.315 | 17.49 |
| 13 1 | 959.285 | 0.115 | 959.285 | 17.49 |
| 13 1 | 959.174 | 0.166 | 959.174 | 17.49 |
| 13 1 | 959.177 | 0.124 | 959.177 | 17.49 |
| 13 1 | 959.088 | 0.109 | 959.088 | 17.49 |
| 13 1 | 959.845 | 0.136 | 959.845 | 17.49 |
| 13 1 | 960.096 | 0.128 | 960.096 | 17.49 |
| 13 1 | 959.441 | 0.128 | 959.441 | 17.49 |
| 13 1 | 959.146 | 0.138 | 959.146 | 17.49 |
| 13 1 | 959.302 | 0.152 | 959.302 | 17.49 |
| 13 1 | 958.900 | 0.133 | 958.900 | 17.49 |
| 13 1 | 959.138 | 0.157 | 959.138 | 17.49 |
| 13 1 | 959.830 | 0.129 | 959.830 | 17.49 |
| 13 1 | 959.336 | 0.120 | 959.336 | 17.49 |
| 13 1 | 958.903 | 0.135 | 958.903 | 17.49 |
| 13 1 | 960.389 | 0.133 | 960.389 | 17.49 |
| 13 1 | 959.457 | 0.111 | 959.457 | 17.49 |
| 13 1 | 959.460 | 0.110 | 959.460 | 17.49 |

| 2004 - MARCO | | | | |
|---|---|---|---|---|
| D L | SDB | ER | SDC | HL |
| 02 1 | 957.530 | 0.115 | 957.530 | 17.49 |
| 02 1 | 958.579 | 0.101 | 958.579 | 17.49 |
| 02 1 | 959.282 | 0.105 | 959.282 | 17.49 |
| 02 1 | 958.829 | 0.121 | 958.829 | 17.49 |
| 02 1 | 960.077 | 0.105 | 960.077 | 17.49 |
| 02 1 | 959.776 | 0.107 | 959.776 | 17.49 |
| 02 1 | 959.458 | 0.111 | 959.458 | 17.49 |
| 02 1 | 959.444 | 0.121 | 959.444 | 17.49 |
| 03 1 | 958.804 | 0.093 | 958.804 | 17.49 |
| 03 1 | 958.901 | 0.131 | 958.901 | 17.49 |
| 03 1 | 960.385 | 0.124 | 960.385 | 17.49 |
| 03 1 | 959.251 | 0.131 | 959.251 | 17.49 |
| 03 1 | 959.601 | 0.139 | 959.601 | 17.49 |
| 03 1 | 958.573 | 0.137 | 958.573 | 17.49 |
| 03 1 | 959.031 | 0.116 | 959.031 | 17.49 |
| 03 1 | 959.455 | 0.110 | 959.455 | 17.49 |
| 03 1 | 958.767 | 0.107 | 958.767 | 17.49 |
| 03 1 | 958.530 | 0.144 | 958.530 | 17.49 |
| 03 1 | 959.264 | 0.133 | 959.264 | 17.49 |
| 03 1 | 959.939 | 0.146 | 959.939 | 17.49 |
| 03 1 | 959.368 | 0.133 | 959.368 | 17.49 |
| 03 1 | 958.896 | 0.117 | 958.896 | 17.49 |
| 03 1 | 959.024 | 0.132 | 959.024 | 17.49 |
| 03 1 | 959.177 | 0.149 | 959.177 | 17.49 |
| 03 1 | 959.112 | 0.121 | 959.112 | 17.49 |
| 03 1 | 958.664 | 0.147 | 958.664 | 17.49 |
| 03 1 | 958.865 | 0.160 | 958.865 | 17.49 |
| 03 1 | 961.163 | 0.167 | 961.163 | 17.49 |

| 2004 - MARCO | | | | | | 2004 - MARCO | | | | |
|---|---|---|---|---|---|---|---|---|---|---|
| D | L | SDB | ER | SDC | HL | D | L | SDB | ER | SDC | HL |
| 03 | 1 | 959.816 | 0.137 | 959.816 | 17.49 | 12 | 1 | 958.684 | 0.169 | 958.684 | 17.49 |
| 03 | 1 | 959.111 | 0.119 | 959.111 | 17.49 | 12 | 1 | 959.406 | 0.168 | 959.406 | 17.49 |
| 08 | 1 | 957.739 | 0.096 | 957.739 | 17.49 | 12 | 1 | 959.104 | 0.147 | 959.104 | 17.49 |
| 08 | 1 | 958.641 | 0.107 | 958.641 | 17.49 | 16 | 1 | 959.301 | 0.184 | 959.301 | 17.49 |
| 08 | 1 | 959.763 | 0.113 | 959.763 | 17.49 | 16 | 1 | 959.986 | 0.250 | 959.986 | 17.49 |
| 08 | 1 | 959.171 | 0.123 | 959.171 | 17.49 | 16 | 1 | 960.670 | 0.110 | 960.670 | 17.49 |
| 08 | 1 | 958.985 | 0.101 | 958.985 | 17.49 | 16 | 1 | 958.107 | 0.219 | 958.107 | 17.49 |
| 08 | 1 | 959.074 | 0.111 | 959.074 | 17.49 | 16 | 1 | 958.497 | 0.144 | 958.497 | 17.49 |
| 08 | 1 | 958.431 | 0.166 | 958.431 | 17.49 | 16 | 1 | 959.088 | 0.115 | 959.088 | 17.49 |
| 08 | 1 | 959.335 | 0.176 | 959.335 | 17.49 | 16 | 1 | 959.199 | 0.122 | 959.199 | 17.49 |
| 08 | 1 | 959.591 | 0.136 | 959.591 | 17.49 | 16 | 1 | 959.442 | 0.141 | 959.442 | 17.49 |
| 08 | 1 | 958.372 | 0.143 | 958.372 | 17.49 | 16 | 1 | 957.950 | 0.347 | 957.950 | 17.49 |
| 08 | 1 | 959.603 | 0.206 | 959.603 | 17.49 | 17 | 1 | 959.623 | 0.132 | 959.623 | 17.49 |
| 08 | 1 | 959.748 | 0.110 | 959.748 | 17.49 | 17 | 1 | 959.398 | 0.136 | 959.398 | 17.49 |
| 08 | 1 | 959.506 | 0.129 | 959.506 | 17.49 | 17 | 1 | 959.811 | 0.123 | 959.811 | 17.49 |
| 08 | 1 | 958.062 | 0.149 | 958.062 | 17.49 | 17 | 1 | 960.088 | 0.138 | 960.088 | 17.49 |
| 08 | 1 | 959.156 | 0.187 | 959.156 | 17.49 | 17 | 1 | 959.432 | 0.115 | 959.432 | 17.49 |
| 08 | 1 | 959.181 | 0.237 | 959.181 | 17.49 | 17 | 1 | 959.267 | 0.145 | 959.267 | 17.49 |
| 08 | 1 | 960.932 | 0.186 | 960.932 | 17.49 | 17 | 1 | 959.186 | 0.128 | 959.186 | 17.49 |
| 09 | 1 | 958.811 | 0.124 | 958.811 | 17.49 | 17 | 1 | 959.794 | 0.163 | 959.794 | 17.49 |
| 09 | 1 | 959.352 | 0.119 | 959.352 | 17.49 | 17 | 1 | 959.916 | 0.217 | 959.916 | 17.49 |
| 09 | 1 | 959.153 | 0.105 | 959.153 | 17.49 | 17 | 1 | 959.157 | 0.159 | 959.157 | 17.49 |
| 09 | 1 | 959.574 | 0.101 | 959.574 | 17.49 | 17 | 1 | 958.996 | 0.159 | 958.996 | 17.49 |
| 09 | 1 | 959.308 | 0.127 | 959.308 | 17.49 | 17 | 1 | 958.639 | 0.169 | 958.639 | 17.49 |
| 09 | 1 | 959.338 | 0.154 | 959.338 | 17.49 | 17 | 1 | 959.520 | 0.154 | 959.520 | 17.49 |
| 09 | 1 | 959.186 | 0.135 | 959.186 | 17.49 | 17 | 1 | 959.336 | 0.162 | 959.336 | 17.49 |
| 09 | 1 | 958.856 | 0.122 | 958.856 | 17.49 | 17 | 1 | 959.759 | 0.170 | 959.759 | 17.49 |
| 09 | 1 | 958.945 | 0.148 | 958.945 | 17.49 | 17 | 1 | 959.145 | 0.143 | 959.145 | 17.49 |
| 09 | 1 | 959.702 | 0.112 | 959.702 | 17.49 | 17 | 1 | 959.246 | 0.155 | 959.246 | 17.49 |
| 09 | 1 | 958.951 | 0.156 | 958.951 | 17.49 | 17 | 1 | 959.734 | 0.146 | 959.734 | 17.49 |
| 09 | 1 | 960.121 | 0.159 | 960.121 | 17.49 | 17 | 1 | 959.123 | 0.164 | 959.123 | 17.49 |
| 09 | 1 | 958.365 | 0.128 | 958.365 | 17.49 | 17 | 1 | 959.558 | 0.143 | 959.558 | 17.49 |
| 09 | 1 | 958.721 | 0.124 | 958.721 | 17.49 | 18 | 1 | 957.642 | 0.134 | 957.642 | 17.49 |
| 09 | 1 | 959.279 | 0.127 | 959.279 | 17.49 | 18 | 1 | 959.171 | 0.157 | 959.171 | 17.49 |
| 09 | 1 | 958.940 | 0.123 | 958.940 | 17.49 | 18 | 1 | 958.346 | 0.161 | 958.346 | 17.49 |
| 09 | 1 | 959.077 | 0.165 | 959.077 | 17.49 | 18 | 1 | 958.837 | 0.136 | 958.837 | 17.49 |
| 09 | 1 | 960.288 | 0.148 | 960.288 | 17.49 | 18 | 1 | 958.094 | 0.156 | 958.094 | 17.49 |
| 09 | 1 | 958.693 | 0.145 | 958.693 | 17.49 | 18 | 1 | 958.125 | 0.148 | 958.125 | 17.49 |
| 09 | 1 | 959.284 | 0.130 | 959.284 | 17.49 | 18 | 1 | 958.645 | 0.136 | 958.645 | 17.49 |
| 09 | 1 | 958.673 | 0.131 | 958.673 | 17.49 | 18 | 1 | 959.009 | 0.132 | 959.009 | 17.49 |
| 09 | 1 | 959.021 | 0.141 | 959.021 | 17.49 | 18 | 1 | 958.553 | 0.147 | 958.553 | 17.49 |
| 09 | 1 | 958.555 | 0.136 | 958.555 | 17.49 | 22 | 1 | 960.033 | 0.173 | 960.033 | 17.49 |
| 09 | 1 | 958.899 | 0.154 | 958.899 | 17.49 | 22 | 1 | 959.106 | 0.157 | 959.106 | 17.49 |
| 10 | 1 | 957.685 | 0.132 | 957.685 | 17.49 | 22 | 1 | 959.307 | 0.171 | 959.307 | 17.49 |
| 10 | 1 | 958.785 | 0.100 | 958.785 | 17.49 | 22 | 1 | 958.562 | 0.173 | 958.562 | 17.49 |
| 10 | 1 | 958.990 | 0.093 | 958.990 | 17.49 | 22 | 1 | 958.861 | 0.153 | 958.861 | 17.49 |
| 10 | 1 | 959.092 | 0.103 | 959.092 | 17.49 | 22 | 1 | 959.169 | 0.153 | 959.169 | 17.49 |
| 10 | 1 | 959.197 | 0.110 | 959.197 | 17.49 | 22 | 1 | 958.360 | 0.159 | 958.360 | 17.49 |
| 10 | 1 | 959.103 | 0.094 | 959.103 | 17.49 | 22 | 1 | 958.971 | 0.166 | 958.971 | 17.49 |
| 10 | 1 | 958.712 | 0.131 | 958.712 | 17.49 | 22 | 1 | 959.266 | 0.172 | 959.266 | 17.49 |
| 10 | 1 | 959.260 | 0.121 | 959.260 | 17.49 | 22 | 1 | 958.246 | 0.296 | 958.246 | 17.49 |
| 10 | 1 | 959.868 | 0.120 | 959.868 | 17.49 | 22 | 1 | 959.621 | 0.187 | 959.621 | 17.49 |
| 10 | 1 | 958.110 | 0.102 | 958.110 | 17.49 | 22 | 1 | 960.437 | 0.215 | 960.437 | 17.49 |
| 10 | 1 | 959.604 | 0.132 | 959.604 | 17.49 | 22 | 1 | 959.121 | 0.205 | 959.121 | 17.49 |
| 10 | 1 | 959.167 | 0.147 | 959.167 | 17.49 | 26 | 1 | 958.368 | 0.158 | 958.368 | 17.49 |
| 10 | 1 | 960.057 | 0.120 | 960.057 | 17.49 | 26 | 1 | 959.231 | 0.161 | 959.231 | 17.49 |
| 10 | 1 | 958.826 | 0.106 | 958.826 | 17.49 | 26 | 1 | 958.953 | 0.153 | 958.953 | 17.49 |
| 10 | 1 | 959.118 | 0.111 | 959.118 | 17.49 | 26 | 1 | 959.414 | 0.122 | 959.414 | 17.49 |
| 10 | 1 | 959.032 | 0.117 | 959.032 | 17.49 | 26 | 1 | 959.551 | 0.138 | 959.551 | 17.49 |
| 10 | 1 | 958.803 | 0.124 | 958.803 | 17.49 | 26 | 1 | 959.603 | 0.133 | 959.603 | 17.49 |
| 10 | 1 | 958.530 | 0.122 | 958.530 | 17.49 | 26 | 1 | 959.219 | 0.136 | 959.219 | 17.49 |
| 10 | 1 | 958.715 | 0.148 | 958.715 | 17.49 | 26 | 1 | 959.565 | 0.124 | 959.565 | 17.49 |
| 10 | 1 | 958.689 | 0.138 | 958.689 | 17.49 | 26 | 1 | 959.456 | 0.118 | 959.456 | 17.49 |
| 10 | 1 | 959.170 | 0.158 | 959.170 | 17.49 | 26 | 1 | 959.707 | 0.141 | 959.707 | 17.49 |
| 10 | 1 | 958.614 | 0.136 | 958.614 | 17.49 | 26 | 1 | 960.062 | 0.126 | 960.062 | 17.49 |
| 10 | 1 | 958.687 | 0.172 | 958.687 | 17.49 | 26 | 1 | 959.852 | 0.143 | 959.852 | 17.49 |
| 10 | 1 | 958.578 | 0.142 | 958.578 | 17.49 | 26 | 1 | 958.581 | 0.135 | 958.581 | 17.49 |
| 10 | 1 | 959.032 | 0.149 | 959.032 | 17.49 | 26 | 1 | 959.351 | 0.133 | 959.351 | 17.49 |
| 10 | 1 | 958.776 | 0.136 | 958.776 | 17.49 | 26 | 1 | 959.424 | 0.137 | 959.424 | 17.49 |
| 12 | 1 | 958.289 | 0.144 | 958.289 | 17.49 | 26 | 1 | 958.922 | 0.157 | 958.922 | 17.49 |
| 12 | 1 | 958.797 | 0.137 | 958.797 | 17.49 | 26 | 1 | 960.260 | 0.183 | 960.260 | 17.49 |

| 2004 - MARCO | | | | | | 2004 - ABRIL | | | | |
|---|---|---|---|---|---|---|---|---|---|---|
| D | L | SDB | ER | SDC | HL | D | L | SDB | ER | SDC | HL |
| 26 | 1 | 959.866 | 0.157 | 959.866 | 17.49 | 02 | 1 | 958.372 | 0.145 | 958.372 | 17.49 |
| 26 | 1 | 959.764 | 0.167 | 959.764 | 17.49 | 02 | 1 | 958.375 | 0.114 | 958.375 | 17.49 |
| 26 | 1 | 959.009 | 0.168 | 959.009 | 17.49 | 02 | 1 | 958.341 | 0.141 | 958.341 | 17.49 |
| 26 | 1 | 959.284 | 0.163 | 959.284 | 17.49 | 02 | 1 | 959.089 | 0.114 | 959.089 | 17.49 |
| 26 | 1 | 959.202 | 0.135 | 959.202 | 17.49 | 02 | 1 | 958.854 | 0.119 | 958.854 | 17.49 |
| 26 | 1 | 959.171 | 0.125 | 959.171 | 17.49 | 02 | 1 | 958.195 | 0.221 | 958.195 | 17.49 |
| 26 | 1 | 958.544 | 0.144 | 958.544 | 17.49 | 02 | 1 | 959.124 | 0.128 | 959.124 | 17.49 |
| 26 | 1 | 958.963 | 0.130 | 958.963 | 17.49 | 02 | 1 | 959.605 | 0.137 | 959.605 | 17.49 |
| 26 | 1 | 959.171 | 0.130 | 959.171 | 17.49 | 02 | 1 | 959.364 | 0.127 | 959.364 | 17.49 |
| 29 | 1 | 959.377 | 0.138 | 959.377 | 17.49 | 02 | 1 | 959.456 | 0.132 | 959.456 | 17.49 |
| 29 | 1 | 959.915 | 0.128 | 959.915 | 17.49 | 02 | 1 | 959.121 | 0.155 | 959.121 | 17.49 |
| 29 | 1 | 959.261 | 0.128 | 959.261 | 17.49 | 02 | 1 | 959.799 | 0.150 | 959.799 | 17.49 |
| 29 | 1 | 958.950 | 0.101 | 958.950 | 17.49 | 02 | 1 | 959.888 | 0.139 | 959.888 | 17.49 |
| 29 | 1 | 959.510 | 0.107 | 959.510 | 17.49 | 02 | 1 | 960.080 | 0.101 | 960.080 | 17.49 |
| 29 | 1 | 959.870 | 0.111 | 959.870 | 17.49 | 02 | 1 | 959.134 | 0.099 | 959.134 | 17.49 |
| 29 | 1 | 959.782 | 0.111 | 959.782 | 17.49 | 02 | 1 | 959.437 | 0.126 | 959.437 | 17.49 |
| 29 | 1 | 959.815 | 0.098 | 959.815 | 17.49 | 02 | 1 | 958.649 | 0.152 | 958.649 | 17.49 |
| 29 | 1 | 959.401 | 0.159 | 959.401 | 17.49 | 02 | 1 | 959.207 | 0.113 | 959.207 | 17.49 |
| 29 | 1 | 960.080 | 0.110 | 960.080 | 17.49 | 02 | 1 | 961.091 | 0.139 | 961.091 | 17.49 |
| 29 | 1 | 960.405 | 0.143 | 960.405 | 17.49 | 02 | 1 | 959.350 | 0.133 | 959.350 | 17.49 |
| 29 | 1 | 959.291 | 0.167 | 959.291 | 17.49 | 02 | 1 | 959.107 | 0.105 | 959.107 | 17.49 |
| 29 | 1 | 959.693 | 0.165 | 959.693 | 17.49 | 02 | 1 | 959.442 | 0.140 | 959.442 | 17.49 |
| 29 | 1 | 959.978 | 0.133 | 959.978 | 17.49 | 02 | 1 | 959.384 | 0.136 | 959.384 | 17.49 |
| 29 | 1 | 959.622 | 0.122 | 959.622 | 17.49 | 06 | 1 | 957.819 | 0.112 | 957.819 | 17.49 |
| 29 | 1 | 959.734 | 0.134 | 959.734 | 17.49 | 06 | 1 | 958.823 | 0.123 | 958.823 | 17.49 |
| 29 | 1 | 958.855 | 0.128 | 958.855 | 17.49 | 06 | 1 | 959.013 | 0.108 | 959.013 | 17.49 |
| 29 | 1 | 958.820 | 0.120 | 958.820 | 17.49 | 06 | 1 | 958.842 | 0.121 | 958.842 | 17.49 |
| 29 | 1 | 959.509 | 0.133 | 959.509 | 17.49 | 06 | 1 | 958.841 | 0.108 | 958.841 | 17.49 |
| 29 | 1 | 959.356 | 0.130 | 959.356 | 17.49 | 06 | 1 | 959.212 | 0.120 | 959.212 | 17.49 |
| 29 | 1 | 958.999 | 0.164 | 958.999 | 17.49 | 06 | 1 | 959.330 | 0.113 | 959.330 | 17.49 |
| 29 | 1 | 958.664 | 0.137 | 958.664 | 17.49 | 06 | 1 | 959.407 | 0.119 | 959.407 | 17.49 |
| 29 | 1 | 960.592 | 0.131 | 960.592 | 17.49 | 06 | 1 | 959.030 | 0.122 | 959.030 | 17.49 |
| 29 | 1 | 959.278 | 0.120 | 959.278 | 17.49 | 06 | 1 | 959.432 | 0.139 | 959.432 | 17.49 |
| 29 | 1 | 959.320 | 0.117 | 959.320 | 17.49 | 06 | 1 | 958.960 | 0.154 | 958.960 | 17.49 |
| 29 | 1 | 959.355 | 0.112 | 959.355 | 17.49 | 06 | 1 | 959.273 | 0.180 | 959.273 | 17.49 |
| 29 | 1 | 959.079 | 0.149 | 959.079 | 17.49 | 06 | 1 | 960.333 | 0.144 | 960.333 | 17.49 |
| 29 | 1 | 959.322 | 0.172 | 959.322 | 17.49 | 06 | 1 | 959.109 | 0.142 | 959.109 | 17.49 |
| 30 | 1 | 958.180 | 0.155 | 958.180 | 17.49 | 06 | 1 | 958.303 | 0.192 | 958.303 | 17.49 |
| 30 | 1 | 958.869 | 0.142 | 958.869 | 17.49 | 06 | 1 | 958.463 | 0.138 | 958.463 | 17.49 |
| 30 | 1 | 959.427 | 0.117 | 959.427 | 17.49 | 06 | 1 | 959.175 | 0.161 | 959.175 | 17.49 |
| 30 | 1 | 959.310 | 0.121 | 959.310 | 17.49 | 06 | 1 | 959.207 | 0.128 | 959.207 | 17.49 |
| 30 | 1 | 958.821 | 0.110 | 958.821 | 17.49 | 06 | 1 | 959.320 | 0.115 | 959.320 | 17.49 |
| 30 | 1 | 960.088 | 0.131 | 960.088 | 17.49 | 06 | 1 | 958.651 | 0.126 | 958.651 | 17.49 |
| 30 | 1 | 959.212 | 0.146 | 959.212 | 17.49 | 06 | 1 | 959.447 | 0.126 | 959.447 | 17.49 |
| 30 | 1 | 959.779 | 0.101 | 959.779 | 17.49 | 06 | 1 | 959.463 | 0.146 | 959.463 | 17.49 |
| 30 | 1 | 959.500 | 0.113 | 959.500 | 17.49 | 06 | 1 | 959.116 | 0.141 | 959.116 | 17.49 |
| 30 | 1 | 959.569 | 0.137 | 959.569 | 17.49 | 07 | 1 | 958.043 | 0.121 | 958.043 | 17.49 |
| 30 | 1 | 959.578 | 0.119 | 959.578 | 17.49 | 07 | 1 | 959.439 | 0.123 | 959.439 | 17.49 |
| 30 | 1 | 958.583 | 0.131 | 958.583 | 17.49 | 07 | 1 | 958.497 | 0.182 | 958.497 | 17.49 |
| 30 | 1 | 959.527 | 0.144 | 959.527 | 17.49 | 07 | 1 | 959.818 | 0.118 | 959.818 | 17.49 |
| 30 | 1 | 959.285 | 0.141 | 959.285 | 17.49 | 07 | 1 | 959.994 | 0.120 | 959.994 | 17.49 |
| 30 | 1 | 960.999 | 0.140 | 960.999 | 17.49 | 07 | 1 | 960.008 | 0.120 | 960.008 | 17.49 |
| 30 | 1 | 959.319 | 0.116 | 959.319 | 17.49 | 07 | 1 | 960.123 | 0.116 | 960.123 | 17.49 |
| 30 | 1 | 958.275 | 0.154 | 958.275 | 17.49 | 07 | 1 | 959.133 | 0.180 | 959.133 | 17.49 |
| 30 | 1 | 958.378 | 0.141 | 958.378 | 17.49 | 07 | 1 | 959.642 | 0.141 | 959.642 | 17.49 |
| 30 | 1 | 958.871 | 0.139 | 958.871 | 17.49 | 07 | 1 | 960.341 | 0.175 | 960.341 | 17.49 |
| 31 | 1 | 958.029 | 0.142 | 958.029 | 17.49 | 07 | 1 | 959.556 | 0.146 | 959.556 | 17.49 |
| 31 | 1 | 958.840 | 0.117 | 958.840 | 17.49 | 07 | 1 | 958.934 | 0.205 | 958.934 | 17.49 |
| 31 | 1 | 959.510 | 0.114 | 959.510 | 17.49 | 07 | 1 | 959.070 | 0.139 | 959.070 | 17.49 |
| 31 | 1 | 959.115 | 0.098 | 959.115 | 17.49 | 07 | 1 | 959.221 | 0.155 | 959.221 | 17.49 |
| 31 | 1 | 959.468 | 0.104 | 959.468 | 17.49 | 07 | 1 | 958.773 | 0.166 | 958.773 | 17.49 |
| 31 | 1 | 959.755 | 0.117 | 959.755 | 17.49 | 07 | 1 | 958.267 | 0.206 | 958.267 | 17.49 |
| 31 | 1 | 959.822 | 0.110 | 959.822 | 17.49 | 07 | 1 | 958.193 | 0.157 | 958.193 | 17.49 |
| 31 | 1 | 959.567 | 0.144 | 959.567 | 17.49 | 07 | 1 | 957.617 | 0.146 | 957.617 | 17.49 |
| 31 | 1 | 959.210 | 0.139 | 959.210 | 17.49 | 07 | 1 | 958.898 | 0.160 | 958.898 | 17.49 |
| 31 | 1 | 959.140 | 0.132 | 959.140 | 17.49 | 07 | 1 | 958.789 | 0.159 | 958.789 | 17.49 |
| 31 | 1 | 959.889 | 0.114 | 959.889 | 17.49 | 07 | 1 | 958.430 | 0.129 | 958.430 | 17.49 |
| 31 | 1 | 959.717 | 0.124 | 959.717 | 17.49 | 07 | 1 | 958.631 | 0.162 | 958.631 | 17.49 |
|  |  |  |  |  |  | 07 | 1 | 958.833 | 0.149 | 958.833 | 17.49 |
|  |  |  |  |  |  | 08 | 1 | 958.875 | 0.143 | 958.875 | 17.49 |
|  |  |  |  |  |  | 08 | 1 | 959.093 | 0.123 | 959.093 | 17.49 |

| 2004 - ABRIL | | | | | | 2004 - MAIO | | | | |
|---|---|---|---|---|---|---|---|---|---|---|
| D | L | SDB | ER | SDC | HL | D | L | SDB | ER | SDC | HL |
| 08 | 1 | 958.920 | 0.141 | 958.920 | 17.49 | 03 | 1 | 959.478 | 0.143 | 959.478 | 17.49 |
| 08 | 1 | 959.086 | 0.178 | 959.086 | 17.49 | 03 | 1 | 958.926 | 0.124 | 958.926 | 17.49 |
| 08 | 1 | 958.689 | 0.129 | 958.689 | 17.49 | 03 | 1 | 959.320 | 0.131 | 959.320 | 17.49 |
| 08 | 1 | 958.091 | 0.137 | 958.091 | 17.49 | 03 | 1 | 959.050 | 0.144 | 959.050 | 17.49 |
| 08 | 1 | 958.677 | 0.132 | 958.677 | 17.49 | 03 | 1 | 960.006 | 0.122 | 960.006 | 17.49 |
| 08 | 1 | 958.995 | 0.115 | 958.995 | 17.49 | 03 | 1 | 959.032 | 0.148 | 959.032 | 17.49 |
| 19 | 1 | 958.086 | 0.123 | 958.086 | 17.49 | 03 | 1 | 959.748 | 0.145 | 959.748 | 17.49 |
| 19 | 1 | 959.393 | 0.112 | 959.393 | 17.49 | 03 | 1 | 959.553 | 0.187 | 959.553 | 17.49 |
| 19 | 1 | 958.914 | 0.112 | 958.914 | 17.49 | 03 | 1 | 959.677 | 0.130 | 959.677 | 17.49 |
| 19 | 1 | 960.113 | 0.123 | 960.113 | 17.49 | 03 | 1 | 960.002 | 0.166 | 960.002 | 17.49 |
| 19 | 1 | 960.098 | 0.117 | 960.098 | 17.49 | 03 | 1 | 958.908 | 0.184 | 958.908 | 17.49 |
| 19 | 1 | 959.849 | 0.121 | 959.849 | 17.49 | 03 | 1 | 959.366 | 0.151 | 959.366 | 17.49 |
| 19 | 1 | 959.915 | 0.132 | 959.915 | 17.49 | 03 | 1 | 959.708 | 0.173 | 959.708 | 17.49 |
| 19 | 1 | 959.799 | 0.125 | 959.799 | 17.49 | 03 | 1 | 959.638 | 0.194 | 959.638 | 17.49 |
| 19 | 1 | 960.534 | 0.142 | 960.534 | 17.49 | 03 | 1 | 959.806 | 0.143 | 959.806 | 17.49 |
| 19 | 1 | 960.252 | 0.157 | 960.252 | 17.49 | 03 | 1 | 958.563 | 0.108 | 958.563 | 17.49 |
| 19 | 1 | 959.583 | 0.170 | 959.583 | 17.49 | 03 | 1 | 958.778 | 0.126 | 958.778 | 17.49 |
| 22 | 1 | 958.347 | 0.108 | 958.347 | 17.49 | 03 | 1 | 959.220 | 0.106 | 959.220 | 17.49 |
| 22 | 1 | 960.300 | 0.138 | 960.300 | 17.49 | 03 | 1 | 958.720 | 0.135 | 958.720 | 17.49 |
| 22 | 1 | 959.524 | 0.152 | 959.524 | 17.49 | 03 | 1 | 958.843 | 0.127 | 958.843 | 17.49 |
| 22 | 1 | 958.946 | 0.132 | 958.946 | 17.49 | 03 | 1 | 959.049 | 0.125 | 959.049 | 17.49 |
| 22 | 1 | 959.120 | 0.126 | 959.120 | 17.49 | 03 | 1 | 959.107 | 0.127 | 959.107 | 17.49 |
| 22 | 1 | 959.402 | 0.123 | 959.402 | 17.49 | 03 | 1 | 958.301 | 0.135 | 958.301 | 17.49 |
| 22 | 1 | 959.404 | 0.111 | 959.404 | 17.49 | 03 | 1 | 958.560 | 0.145 | 958.560 | 17.49 |
| 22 | 1 | 959.473 | 0.129 | 959.473 | 17.49 | 03 | 1 | 959.403 | 0.164 | 959.403 | 17.49 |
| 22 | 1 | 959.866 | 0.123 | 959.866 | 17.49 | 03 | 1 | 959.245 | 0.256 | 959.245 | 17.49 |
| 22 | 1 | 958.990 | 0.125 | 958.990 | 17.49 | 04 | 1 | 957.773 | 0.131 | 957.773 | 17.49 |
| 22 | 1 | 959.568 | 0.176 | 959.568 | 17.49 | 04 | 1 | 959.002 | 0.143 | 959.002 | 17.49 |
| 22 | 1 | 958.805 | 0.144 | 958.805 | 17.49 | 04 | 1 | 958.852 | 0.138 | 958.852 | 17.49 |
| 22 | 1 | 958.893 | 0.134 | 958.893 | 17.49 | 04 | 1 | 959.127 | 0.152 | 959.127 | 17.49 |
| 22 | 1 | 959.269 | 0.156 | 959.269 | 17.49 | 04 | 1 | 959.670 | 0.186 | 959.670 | 17.49 |
| 22 | 1 | 959.599 | 0.148 | 959.599 | 17.49 | 04 | 1 | 959.673 | 0.157 | 959.673 | 17.49 |
| 22 | 1 | 959.181 | 0.130 | 959.181 | 17.49 | 04 | 1 | 957.733 | 0.197 | 957.733 | 17.49 |
| 22 | 1 | 959.309 | 0.142 | 959.309 | 17.49 | 04 | 1 | 958.779 | 0.169 | 958.779 | 17.49 |
| 22 | 1 | 959.336 | 0.149 | 959.336 | 17.49 | 04 | 1 | 959.147 | 0.138 | 959.147 | 17.49 |
| 28 | 1 | 959.639 | 0.134 | 959.639 | 17.49 | 04 | 1 | 959.733 | 0.128 | 959.733 | 17.49 |
| 28 | 1 | 959.138 | 0.130 | 959.138 | 17.49 | 04 | 1 | 960.289 | 0.141 | 960.289 | 17.49 |
| 28 | 1 | 958.720 | 0.141 | 958.720 | 17.49 | 04 | 1 | 958.826 | 0.166 | 958.826 | 17.49 |
| 28 | 1 | 959.027 | 0.151 | 959.027 | 17.49 | 04 | 1 | 959.542 | 0.131 | 959.542 | 17.49 |
| 28 | 1 | 958.746 | 0.135 | 958.746 | 17.49 | 04 | 1 | 958.590 | 0.190 | 958.590 | 17.49 |
| 28 | 1 | 959.061 | 0.151 | 959.061 | 17.49 | 04 | 1 | 959.237 | 0.147 | 959.237 | 17.49 |
| 28 | 1 | 958.402 | 0.138 | 958.402 | 17.49 | 04 | 1 | 959.105 | 0.137 | 959.105 | 17.49 |
| 28 | 1 | 959.368 | 0.123 | 959.368 | 17.49 | 04 | 1 | 958.967 | 0.151 | 958.967 | 17.49 |
| 28 | 1 | 958.916 | 0.164 | 958.916 | 17.49 | 04 | 1 | 958.757 | 0.156 | 958.757 | 17.49 |
| 28 | 1 | 959.099 | 0.143 | 959.099 | 17.49 | 04 | 1 | 958.997 | 0.180 | 958.997 | 17.49 |
| 28 | 1 | 958.832 | 0.182 | 958.832 | 17.49 | 04 | 1 | 958.599 | 0.169 | 958.599 | 17.49 |
| 28 | 1 | 959.377 | 0.193 | 959.377 | 17.49 | 05 | 1 | 958.824 | 0.134 | 958.824 | 17.49 |
| 29 | 1 | 960.457 | 0.131 | 960.457 | 17.49 | 05 | 1 | 959.002 | 0.137 | 959.002 | 17.49 |
| 29 | 1 | 959.271 | 0.136 | 959.271 | 17.49 | 05 | 1 | 958.656 | 0.148 | 958.656 | 17.49 |
| 29 | 1 | 959.758 | 0.156 | 959.758 | 17.49 | 05 | 1 | 959.392 | 0.103 | 959.392 | 17.49 |
| 29 | 1 | 960.578 | 0.163 | 960.578 | 17.49 | 05 | 1 | 959.581 | 0.103 | 959.581 | 17.49 |
| 29 | 1 | 959.314 | 0.126 | 959.314 | 17.49 | 05 | 1 | 959.868 | 0.104 | 959.868 | 17.49 |
| 29 | 1 | 959.397 | 0.140 | 959.397 | 17.49 | 05 | 1 | 959.421 | 0.119 | 959.421 | 17.49 |
| 29 | 1 | 959.600 | 0.142 | 959.600 | 17.49 | 05 | 1 | 959.866 | 0.124 | 959.866 | 17.49 |
| 29 | 1 | 958.861 | 0.146 | 958.861 | 17.49 | 05 | 1 | 959.410 | 0.144 | 959.410 | 17.49 |
| 29 | 1 | 958.944 | 0.136 | 958.944 | 17.49 | 05 | 1 | 958.877 | 0.134 | 958.877 | 17.49 |
| 29 | 1 | 958.942 | 0.121 | 958.942 | 17.49 | 05 | 1 | 959.303 | 0.188 | 959.303 | 17.49 |
| 29 | 1 | 959.590 | 0.150 | 959.590 | 17.49 | 05 | 1 | 958.488 | 0.162 | 958.488 | 17.49 |
| 30 | 1 | 958.711 | 0.121 | 958.711 | 17.49 | 05 | 1 | 959.219 | 0.160 | 959.219 | 17.49 |
| 30 | 1 | 959.591 | 0.127 | 959.591 | 17.49 | 05 | 1 | 959.442 | 0.137 | 959.442 | 17.49 |
| 30 | 1 | 958.943 | 0.140 | 958.943 | 17.49 | 05 | 1 | 960.388 | 0.403 | 960.388 | 17.49 |
| 30 | 1 | 958.856 | 0.143 | 958.856 | 17.49 | 05 | 1 | 958.481 | 0.251 | 958.481 | 17.49 |
| 30 | 1 | 959.299 | 0.171 | 959.299 | 17.49 | 05 | 1 | 957.480 | 0.292 | 957.480 | 17.49 |
| 30 | 1 | 959.415 | 0.159 | 959.415 | 17.49 | 05 | 1 | 959.815 | 0.168 | 959.815 | 17.49 |
| 30 | 1 | 959.683 | 0.141 | 959.683 | 17.49 | 05 | 1 | 958.593 | 0.189 | 958.593 | 17.49 |
| 30 | 1 | 960.148 | 0.131 | 960.148 | 17.49 | 05 | 1 | 959.141 | 0.189 | 959.141 | 17.49 |
| 30 | 1 | 960.145 | 0.140 | 960.145 | 17.49 | 11 | 1 | 958.855 | 0.114 | 958.855 | 17.49 |
| 30 | 1 | 959.860 | 0.158 | 959.860 | 17.49 | 11 | 1 | 958.954 | 0.118 | 958.954 | 17.49 |
|  |  |  |  |  |  | 11 | 1 | 959.195 | 0.119 | 959.195 | 17.49 |
|  |  |  |  |  |  | 11 | 1 | 959.805 | 0.156 | 959.805 | 17.49 |
|  |  |  |  |  |  | 11 | 1 | 958.951 | 0.120 | 958.951 | 17.49 |

| 2004 - MAIO | | | | | |
|---|---|---|---|---|---|
| D | L | SDB | ER | SDC | HL |
| 11 | 1 | 959.237 | 0.134 | 959.237 | 17.49 |
| 11 | 1 | 959.625 | 0.117 | 959.625 | 17.49 |
| 12 | 1 | 958.549 | 0.140 | 958.549 | 17.49 |
| 12 | 1 | 958.743 | 0.175 | 958.743 | 17.49 |
| 12 | 1 | 959.139 | 0.124 | 959.139 | 17.49 |
| 12 | 1 | 958.675 | 0.155 | 958.675 | 17.49 |
| 12 | 1 | 959.951 | 0.114 | 959.951 | 17.49 |
| 12 | 1 | 959.265 | 0.153 | 959.265 | 17.49 |
| 12 | 1 | 958.954 | 0.130 | 958.954 | 17.49 |
| 12 | 1 | 958.734 | 0.148 | 958.734 | 17.49 |
| 12 | 1 | 958.016 | 0.165 | 958.016 | 17.49 |
| 12 | 1 | 958.207 | 0.166 | 958.207 | 17.49 |
| 12 | 1 | 958.518 | 0.170 | 958.518 | 17.49 |
| 12 | 1 | 959.017 | 0.212 | 959.017 | 17.49 |
| 12 | 1 | 959.609 | 0.161 | 959.609 | 17.49 |
| 12 | 1 | 959.497 | 0.165 | 959.497 | 17.49 |
| 12 | 1 | 961.577 | 0.122 | 961.577 | 17.49 |
| 12 | 1 | 961.440 | 0.209 | 961.440 | 17.49 |
| 12 | 1 | 961.071 | 0.214 | 961.071 | 17.49 |
| 12 | 1 | 960.474 | 0.214 | 960.474 | 17.49 |
| 12 | 1 | 960.435 | 0.195 | 960.435 | 17.49 |
| 12 | 1 | 961.339 | 0.156 | 961.339 | 17.49 |
| 12 | 1 | 960.896 | 0.234 | 960.896 | 17.49 |
| 12 | 1 | 960.529 | 0.187 | 960.529 | 17.49 |
| 12 | 1 | 961.718 | 0.203 | 961.718 | 17.49 |
| 13 | 1 | 957.982 | 0.151 | 957.982 | 17.49 |
| 13 | 1 | 958.975 | 0.149 | 958.975 | 17.49 |
| 13 | 1 | 959.419 | 0.118 | 959.419 | 17.49 |
| 13 | 1 | 959.640 | 0.128 | 959.640 | 17.49 |
| 14 | 1 | 959.456 | 0.155 | 959.456 | 17.49 |
| 14 | 1 | 958.878 | 0.141 | 958.878 | 17.49 |
| 14 | 1 | 959.901 | 0.137 | 959.901 | 17.49 |
| 14 | 1 | 959.522 | 0.162 | 959.522 | 17.49 |
| 14 | 1 | 959.577 | 0.140 | 959.577 | 17.49 |
| 14 | 1 | 958.838 | 0.127 | 958.838 | 17.49 |
| 14 | 1 | 959.176 | 0.101 | 959.176 | 17.49 |
| 14 | 1 | 959.014 | 0.173 | 959.014 | 17.49 |
| 14 | 1 | 958.894 | 0.164 | 958.894 | 17.49 |
| 14 | 1 | 959.142 | 0.147 | 959.142 | 17.49 |
| 14 | 1 | 959.489 | 0.173 | 959.489 | 17.49 |
| 14 | 1 | 959.105 | 0.142 | 959.105 | 17.49 |
| 14 | 1 | 958.750 | 0.129 | 958.750 | 17.49 |
| 17 | 1 | 958.304 | 0.193 | 958.304 | 17.49 |
| 17 | 1 | 958.557 | 0.186 | 958.557 | 17.49 |
| 17 | 1 | 959.863 | 0.164 | 959.863 | 17.49 |
| 17 | 1 | 959.644 | 0.161 | 959.644 | 17.49 |
| 17 | 1 | 959.694 | 0.133 | 959.694 | 17.49 |
| 17 | 1 | 959.935 | 0.119 | 959.935 | 17.49 |
| 17 | 1 | 959.999 | 0.175 | 959.999 | 17.49 |
| 17 | 1 | 959.490 | 0.159 | 959.490 | 17.49 |
| 17 | 1 | 959.348 | 0.137 | 959.348 | 17.49 |
| 17 | 1 | 958.619 | 0.174 | 958.619 | 17.49 |
| 17 | 1 | 959.701 | 0.128 | 959.701 | 17.49 |
| 17 | 1 | 958.797 | 0.168 | 958.797 | 17.49 |
| 17 | 1 | 959.234 | 0.210 | 959.234 | 17.49 |
| 24 | 1 | 960.408 | 0.148 | 960.408 | 17.49 |
| 24 | 1 | 959.473 | 0.137 | 959.473 | 17.49 |
| 24 | 1 | 958.977 | 0.164 | 958.977 | 17.49 |
| 24 | 1 | 959.542 | 0.121 | 959.542 | 17.49 |
| 24 | 1 | 958.505 | 0.196 | 958.505 | 17.49 |
| 24 | 1 | 958.964 | 0.152 | 958.964 | 17.49 |
| 24 | 1 | 958.989 | 0.165 | 958.989 | 17.49 |
| 24 | 1 | 958.875 | 0.182 | 958.875 | 17.49 |
| 24 | 1 | 959.412 | 0.159 | 959.412 | 17.49 |
| 24 | 1 | 958.934 | 0.250 | 958.934 | 17.49 |
| 27 | 1 | 959.216 | 0.131 | 959.216 | 17.49 |
| 27 | 1 | 958.908 | 0.127 | 958.908 | 17.49 |
| 27 | 1 | 959.680 | 0.123 | 959.680 | 17.49 |
| 27 | 1 | 959.268 | 0.122 | 959.268 | 17.49 |
| 27 | 1 | 957.851 | 0.156 | 957.851 | 17.49 |
| 27 | 1 | 958.376 | 0.128 | 958.376 | 17.49 |

| 2004 - MAIO | | | | | |
|---|---|---|---|---|---|
| D | L | SDB | ER | SDC | HL |
| 27 | 1 | 959.246 | 0.161 | 959.246 | 17.49 |
| 27 | 1 | 958.412 | 0.140 | 958.412 | 17.49 |
| 27 | 1 | 959.399 | 0.140 | 959.399 | 17.49 |
| 27 | 1 | 958.637 | 0.131 | 958.637 | 17.49 |
| 27 | 1 | 959.477 | 0.174 | 959.477 | 17.49 |
| 27 | 1 | 959.538 | 0.145 | 959.538 | 17.49 |
| 27 | 1 | 959.168 | 0.140 | 959.168 | 17.49 |
| 28 | 1 | 960.494 | 0.144 | 960.494 | 17.49 |
| 28 | 1 | 959.445 | 0.116 | 959.445 | 17.49 |
| 28 | 1 | 959.339 | 0.120 | 959.339 | 17.49 |
| 28 | 1 | 960.028 | 0.140 | 960.028 | 17.49 |
| 28 | 1 | 959.074 | 0.162 | 959.074 | 17.49 |
| 28 | 1 | 959.440 | 0.176 | 959.440 | 17.49 |

| 2004 - JUNHO | | | | | |
|---|---|---|---|---|---|
| D | L | SDB | ER | SDC | HL |
| 02 | 1 | 959.740 | 0.150 | 959.740 | 17.49 |
| 02 | 1 | 959.288 | 0.171 | 959.288 | 17.49 |
| 02 | 1 | 958.446 | 0.171 | 958.446 | 17.49 |
| 02 | 1 | 958.852 | 0.147 | 958.852 | 17.49 |
| 02 | 1 | 958.864 | 0.156 | 958.864 | 17.49 |
| 07 | 1 | 959.266 | 0.140 | 959.266 | 17.49 |
| 07 | 1 | 959.679 | 0.136 | 959.679 | 17.49 |
| 07 | 1 | 959.456 | 0.126 | 959.456 | 17.49 |
| 07 | 1 | 960.032 | 0.130 | 960.032 | 17.49 |
| 07 | 1 | 960.175 | 0.126 | 960.175 | 17.49 |
| 07 | 1 | 959.594 | 0.145 | 959.594 | 17.49 |
| 07 | 1 | 959.690 | 0.122 | 959.690 | 17.49 |
| 07 | 1 | 960.151 | 0.149 | 960.151 | 17.49 |
| 07 | 1 | 960.392 | 0.135 | 960.392 | 17.49 |
| 07 | 1 | 959.496 | 0.161 | 959.496 | 17.49 |
| 07 | 1 | 960.132 | 0.119 | 960.132 | 17.49 |
| 07 | 1 | 959.717 | 0.143 | 959.717 | 17.49 |
| 08 | 1 | 960.035 | 0.219 | 960.035 | 17.49 |
| 08 | 1 | 959.699 | 0.174 | 959.699 | 17.49 |
| 08 | 1 | 958.810 | 0.159 | 958.810 | 17.49 |
| 08 | 1 | 958.881 | 0.217 | 958.881 | 17.49 |
| 08 | 1 | 959.286 | 0.154 | 959.286 | 17.49 |
| 08 | 1 | 957.853 | 0.254 | 957.853 | 17.49 |
| 08 | 1 | 957.622 | 0.264 | 957.622 | 17.49 |
| 08 | 1 | 957.936 | 0.290 | 957.936 | 17.49 |
| 09 | 1 | 959.988 | 0.108 | 959.988 | 17.49 |
| 09 | 1 | 959.780 | 0.168 | 959.780 | 17.49 |
| 11 | 1 | 958.196 | 0.211 | 958.196 | 17.49 |
| 11 | 1 | 959.814 | 0.188 | 959.814 | 17.49 |
| 11 | 1 | 959.221 | 0.174 | 959.221 | 17.49 |
| 11 | 1 | 959.301 | 0.177 | 959.301 | 17.49 |
| 11 | 1 | 959.845 | 0.137 | 959.845 | 17.49 |
| 11 | 1 | 960.980 | 0.141 | 960.980 | 17.49 |
| 11 | 1 | 959.533 | 0.125 | 959.533 | 17.49 |
| 11 | 1 | 960.229 | 0.099 | 960.229 | 17.49 |
| 11 | 1 | 959.620 | 0.088 | 959.620 | 17.49 |
| 11 | 1 | 959.998 | 0.077 | 959.998 | 17.49 |
| 11 | 1 | 958.427 | 0.128 | 958.427 | 17.49 |
| 11 | 1 | 959.001 | 0.152 | 959.001 | 17.49 |
| 11 | 1 | 958.827 | 0.162 | 958.827 | 17.49 |
| 11 | 1 | 959.086 | 0.231 | 959.086 | 17.49 |
| 11 | 1 | 959.850 | 0.159 | 959.850 | 17.49 |
| 11 | 1 | 959.273 | 0.170 | 959.273 | 17.49 |
| 11 | 1 | 958.783 | 0.161 | 958.783 | 17.49 |
| 11 | 1 | 959.314 | 0.159 | 959.314 | 17.49 |
| 11 | 1 | 959.461 | 0.180 | 959.461 | 17.49 |
| 14 | 1 | 958.611 | 0.158 | 958.611 | 17.49 |
| 14 | 1 | 960.282 | 0.195 | 960.282 | 17.49 |
| 14 | 1 | 959.332 | 0.186 | 959.332 | 17.49 |
| 14 | 1 | 957.729 | 0.198 | 957.729 | 17.49 |
| 14 | 1 | 959.849 | 0.144 | 959.849 | 17.49 |
| 14 | 1 | 959.642 | 0.107 | 959.642 | 17.49 |
| 14 | 1 | 959.820 | 0.155 | 959.820 | 17.49 |
| 16 | 1 | 958.503 | 0.173 | 958.503 | 17.49 |

| 2004 - JUNHO | | | | | | 2004 - JUNHO | | | | | |
|---|---|---|---|---|---|---|---|---|---|---|---|
| D | L | SDB | ER | SDC | HL | D | L | SDB | ER | SDC | HL |
| 16 | 1 | 959.629 | 0.126 | 959.629 | 17.49 | 22 | 1 | 958.626 | 0.175 | 958.626 | 17.49 |
| 16 | 1 | 959.241 | 0.185 | 959.241 | 17.49 | 22 | 1 | 959.333 | 0.163 | 959.333 | 17.49 |
| 16 | 1 | 959.577 | 0.151 | 959.577 | 17.49 | 22 | 1 | 959.396 | 0.179 | 959.396 | 17.49 |
| 16 | 1 | 959.179 | 0.158 | 959.179 | 17.49 | 22 | 1 | 960.200 | 0.184 | 960.200 | 17.49 |
| 16 | 1 | 959.012 | 0.133 | 959.012 | 17.49 | 22 | 1 | 958.836 | 0.136 | 958.836 | 17.49 |
| 16 | 1 | 959.353 | 0.139 | 959.353 | 17.49 | 22 | 1 | 958.735 | 0.168 | 958.735 | 17.49 |
| 16 | 1 | 959.647 | 0.170 | 959.647 | 17.49 | 22 | 1 | 959.595 | 0.153 | 959.595 | 17.49 |
| 16 | 1 | 959.163 | 0.180 | 959.163 | 17.49 | 22 | 1 | 958.939 | 0.135 | 958.939 | 17.49 |
| 16 | 1 | 959.590 | 0.133 | 959.590 | 17.49 | 22 | 1 | 959.153 | 0.182 | 959.153 | 17.49 |
| 16 | 1 | 959.082 | 0.166 | 959.082 | 17.49 | 22 | 1 | 959.029 | 0.136 | 959.029 | 17.49 |
| 16 | 1 | 959.804 | 0.319 | 959.804 | 17.49 | 22 | 1 | 958.137 | 0.180 | 958.137 | 17.49 |
| 16 | 1 | 958.480 | 0.248 | 958.480 | 17.49 | 22 | 1 | 959.199 | 0.220 | 959.199 | 17.49 |
| 16 | 1 | 958.923 | 0.301 | 958.923 | 17.49 | 22 | 1 | 958.704 | 0.197 | 958.704 | 17.49 |
| 16 | 1 | 959.175 | 0.193 | 959.175 | 17.49 | 25 | 1 | 959.191 | 0.142 | 959.191 | 17.49 |
| 16 | 1 | 959.844 | 0.198 | 959.844 | 17.49 | 25 | 1 | 958.420 | 0.157 | 958.420 | 17.49 |
| 16 | 1 | 959.269 | 0.173 | 959.269 | 17.49 | 25 | 1 | 958.957 | 0.139 | 958.957 | 17.49 |
| 16 | 1 | 957.896 | 0.291 | 957.896 | 17.49 | 25 | 1 | 959.251 | 0.195 | 959.251 | 17.49 |
| 16 | 1 | 958.865 | 0.151 | 958.865 | 17.49 | 25 | 1 | 958.960 | 0.210 | 958.960 | 17.49 |
| 16 | 1 | 958.686 | 0.217 | 958.686 | 17.49 | 25 | 1 | 959.203 | 0.123 | 959.203 | 17.49 |
| 16 | 1 | 958.865 | 0.208 | 958.865 | 17.49 | 25 | 1 | 959.123 | 0.153 | 959.123 | 17.49 |
| 16 | 1 | 958.456 | 0.260 | 958.456 | 17.49 | 25 | 1 | 959.389 | 0.157 | 959.389 | 17.49 |
| 17 | 1 | 957.017 | 0.185 | 957.017 | 17.49 | 25 | 1 | 959.369 | 0.204 | 959.369 | 17.49 |
| 17 | 1 | 959.623 | 0.186 | 959.623 | 17.49 | 25 | 1 | 958.366 | 0.168 | 958.366 | 17.49 |
| 17 | 1 | 959.195 | 0.182 | 959.195 | 17.49 | 25 | 1 | 959.637 | 0.197 | 959.637 | 17.49 |
| 17 | 1 | 958.926 | 0.174 | 958.926 | 17.49 | 28 | 1 | 958.366 | 0.170 | 958.366 | 17.49 |
| 17 | 1 | 961.656 | 0.205 | 961.656 | 17.49 | 28 | 1 | 958.970 | 0.137 | 958.970 | 17.49 |
| 17 | 1 | 959.766 | 0.182 | 959.766 | 17.49 | 28 | 1 | 960.038 | 0.147 | 960.038 | 17.49 |
| 17 | 1 | 958.682 | 0.176 | 958.682 | 17.49 | 28 | 1 | 959.356 | 0.090 | 959.356 | 17.49 |
| 17 | 1 | 959.443 | 0.226 | 959.443 | 17.49 | 28 | 1 | 960.250 | 0.098 | 960.250 | 17.49 |
| 18 | 1 | 958.710 | 0.144 | 958.710 | 17.49 | 28 | 1 | 959.613 | 0.110 | 959.613 | 17.49 |
| 18 | 1 | 959.178 | 0.123 | 959.178 | 17.49 | 28 | 1 | 959.549 | 0.115 | 959.549 | 17.49 |
| 18 | 1 | 959.053 | 0.133 | 959.053 | 17.49 | 29 | 1 | 958.803 | 0.161 | 958.803 | 17.49 |
| 18 | 1 | 959.842 | 0.127 | 959.842 | 17.49 | 29 | 1 | 959.208 | 0.121 | 959.208 | 17.49 |
| 18 | 1 | 959.495 | 0.187 | 959.495 | 17.49 | 29 | 1 | 959.298 | 0.135 | 959.298 | 17.49 |
| 18 | 1 | 959.527 | 0.134 | 959.527 | 17.49 | 29 | 1 | 958.804 | 0.146 | 958.804 | 17.49 |
| 18 | 1 | 959.660 | 0.110 | 959.660 | 17.49 | 29 | 1 | 959.190 | 0.164 | 959.190 | 17.49 |
| 18 | 1 | 960.081 | 0.132 | 960.081 | 17.49 | 29 | 1 | 959.351 | 0.126 | 959.351 | 17.49 |
| 18 | 1 | 959.112 | 0.112 | 959.112 | 17.49 | 29 | 1 | 959.609 | 0.120 | 959.609 | 17.49 |
| 18 | 1 | 959.485 | 0.101 | 959.485 | 17.49 | 29 | 1 | 959.397 | 0.139 | 959.397 | 17.49 |
| 18 | 1 | 959.211 | 0.144 | 959.211 | 17.49 | 29 | 1 | 959.172 | 0.162 | 959.172 | 17.49 |
| 18 | 1 | 959.097 | 0.135 | 959.097 | 17.49 | 29 | 1 | 960.266 | 0.149 | 960.266 | 17.49 |
| 18 | 1 | 959.094 | 0.130 | 959.094 | 17.49 | 29 | 1 | 959.010 | 0.175 | 959.010 | 17.49 |
| 21 | 1 | 958.187 | 0.181 | 958.187 | 17.49 | 29 | 1 | 959.000 | 0.164 | 959.000 | 17.49 |
| 21 | 1 | 959.635 | 0.133 | 959.635 | 17.49 | 29 | 1 | 958.963 | 0.124 | 958.963 | 17.49 |
| 21 | 1 | 958.788 | 0.143 | 958.788 | 17.49 | 29 | 1 | 958.508 | 0.128 | 958.508 | 17.49 |
| 21 | 1 | 959.177 | 0.147 | 959.177 | 17.49 | 29 | 1 | 959.078 | 0.201 | 959.078 | 17.49 |
| 21 | 1 | 959.687 | 0.144 | 959.687 | 17.49 | 29 | 1 | 959.050 | 0.170 | 959.050 | 17.49 |
| 21 | 1 | 959.265 | 0.145 | 959.265 | 17.49 | 29 | 1 | 959.787 | 0.176 | 959.787 | 17.49 |
| 21 | 1 | 959.746 | 0.121 | 959.746 | 17.49 | 29 | 1 | 958.428 | 0.207 | 958.428 | 17.49 |
| 21 | 1 | 959.462 | 0.175 | 959.462 | 17.49 | 30 | 1 | 957.847 | 0.148 | 957.847 | 17.49 |
| 21 | 1 | 958.854 | 0.184 | 958.854 | 17.49 | 30 | 1 | 958.178 | 0.158 | 958.178 | 17.49 |
| 21 | 1 | 958.921 | 0.166 | 958.921 | 17.49 | 30 | 1 | 958.967 | 0.138 | 958.967 | 17.49 |
| 21 | 1 | 959.512 | 0.161 | 959.512 | 17.49 | 30 | 1 | 959.633 | 0.192 | 959.633 | 17.49 |
| 21 | 1 | 958.925 | 0.154 | 958.925 | 17.49 | 30 | 1 | 959.966 | 0.137 | 959.966 | 17.49 |
| 21 | 1 | 958.680 | 0.269 | 958.680 | 17.49 | 30 | 1 | 959.121 | 0.176 | 959.121 | 17.49 |
| 21 | 1 | 959.052 | 0.171 | 959.052 | 17.49 | 30 | 1 | 959.398 | 0.171 | 959.398 | 17.49 |
| 21 | 1 | 958.052 | 0.232 | 958.052 | 17.49 | 30 | 1 | 959.826 | 0.120 | 959.826 | 17.49 |
| 21 | 1 | 958.747 | 0.203 | 958.747 | 17.49 | 30 | 1 | 959.401 | 0.181 | 959.401 | 17.49 |
| 21 | 1 | 957.890 | 0.222 | 957.890 | 17.49 | 30 | 1 | 960.110 | 0.228 | 960.110 | 17.49 |
| 21 | 1 | 959.773 | 0.216 | 959.773 | 17.49 | 30 | 1 | 959.220 | 0.161 | 959.220 | 17.49 |
| 21 | 1 | 959.772 | 0.232 | 959.772 | 17.49 | 30 | 1 | 958.303 | 0.166 | 958.303 | 17.49 |
| 21 | 1 | 958.973 | 0.188 | 958.973 | 17.49 | 30 | 1 | 959.267 | 0.119 | 959.267 | 17.49 |
| 21 | 1 | 959.657 | 0.189 | 959.657 | 17.49 | 30 | 1 | 959.542 | 0.167 | 959.542 | 17.49 |
| 22 | 1 | 959.727 | 0.179 | 959.727 | 17.49 | 30 | 1 | 959.368 | 0.152 | 959.368 | 17.49 |
| 22 | 1 | 959.598 | 0.134 | 959.598 | 17.49 | 30 | 1 | 959.340 | 0.184 | 959.340 | 17.49 |
| 22 | 1 | 959.384 | 0.144 | 959.384 | 17.49 | 30 | 1 | 958.476 | 0.170 | 958.476 | 17.49 |
| 22 | 1 | 959.010 | 0.150 | 959.010 | 17.49 | 30 | 1 | 959.136 | 0.205 | 959.136 | 17.49 |
| 22 | 1 | 959.144 | 0.170 | 959.144 | 17.49 | 30 | 1 | 958.002 | 0.352 | 958.002 | 17.49 |
| 22 | 1 | 959.406 | 0.167 | 959.406 | 17.49 | | | | | | |
| 22 | 1 | 959.226 | 0.184 | 959.226 | 17.49 | | | | | | |
| 22 | 1 | 959.180 | 0.153 | 959.180 | 17.49 | | | | | | |

```
         2004 - JULHO                              2004 - JULHO
D  L     SDB     ER     SDC     HL       D  L     SDB     ER     SDC     HL
02 1   959.104  0.128  959.104  17.49    14 1   959.530  0.119  959.530  17.49
02 1   959.443  0.107  959.443  17.49    14 1   959.107  0.097  959.107  17.49
02 1   959.152  0.106  959.152  17.49    14 1   959.298  0.107  959.298  17.49
02 1   959.436  0.110  959.436  17.49    14 1   957.787  0.219  957.787  17.49
02 1   959.495  0.134  959.495  17.49    14 1   959.146  0.129  959.146  17.49
02 1   959.137  0.184  959.137  17.49    14 1   959.833  0.146  959.833  17.49
02 1   959.724  0.093  959.724  17.49    14 1   959.495  0.114  959.495  17.49
02 1   959.707  0.104  959.707  17.49    14 1   959.121  0.162  959.121  17.49
02 1   959.740  0.101  959.740  17.49    14 1   959.137  0.163  959.137  17.49
02 1   959.709  0.113  959.709  17.49    14 1   958.737  0.174  958.737  17.49
02 1   959.642  0.097  959.642  17.49    15 1   959.584  0.143  959.584  17.49
02 1   959.545  0.175  959.545  17.49    15 1   960.149  0.120  960.149  17.49
02 1   958.279  0.245  958.279  17.49    15 1   959.715  0.113  959.715  17.49
02 1   958.909  0.320  958.909  17.49    15 1   959.997  0.111  959.997  17.49
02 1   959.323  0.160  959.323  17.49    15 1   959.585  0.105  959.585  17.49
02 1   959.058  0.156  959.058  17.49    15 1   960.187  0.102  960.187  17.49
02 1   959.455  0.156  959.455  17.49    15 1   959.591  0.111  959.591  17.49
02 1   959.013  0.158  959.013  17.49    15 1   959.691  0.115  959.691  17.49
02 1   958.824  0.156  958.824  17.49    15 1   959.621  0.110  959.621  17.49
02 1   957.504  0.218  957.504  17.49    15 1   960.158  0.091  960.158  17.49
07 1   959.414  0.151  959.414  17.49    15 1   959.520  0.119  959.520  17.49
07 1   959.445  0.155  959.445  17.49    15 1   959.294  0.115  959.294  17.49
07 1   960.042  0.149  960.042  17.49    15 1   959.224  0.133  959.224  17.49
07 1   959.136  0.142  959.136  17.49    15 1   958.563  0.158  958.563  17.49
07 1   959.488  0.137  959.488  17.49    15 1   959.415  0.124  959.415  17.49
07 1   960.696  0.121  960.696  17.49    15 1   959.110  0.166  959.110  17.49
07 1   957.824  0.257  957.824  17.49    15 1   959.564  0.141  959.564  17.49
07 1   959.696  0.119  959.696  17.49    15 1   959.166  0.153  959.166  17.49
07 1   959.048  0.153  959.048  17.49    15 1   959.115  0.166  959.115  17.49
07 1   959.431  0.250  959.431  17.49    15 1   959.674  0.152  959.674  17.49
07 1   959.031  0.170  959.031  17.49    16 1   959.903  0.139  959.903  17.49
07 1   958.896  0.141  958.896  17.49    16 1   958.894  0.167  958.894  17.49
07 1   959.554  0.160  959.554  17.49    16 1   959.305  0.120  959.305  17.49
07 1   958.848  0.128  958.848  17.49    16 1   959.983  0.125  959.983  17.49
07 1   959.312  0.147  959.312  17.49    16 1   958.845  0.147  958.845  17.49
07 1   958.857  0.149  958.857  17.49    16 1   959.832  0.113  959.832  17.49
07 1   958.415  0.162  958.415  17.49    16 1   959.066  0.117  959.066  17.49
07 1   959.582  0.147  959.582  17.49    16 1   959.737  0.117  959.737  17.49
07 1   959.080  0.199  959.080  17.49    16 1   959.492  0.128  959.492  17.49
13 1   958.764  0.185  958.764  17.49    16 1   959.990  0.126  959.990  17.49
13 1   959.652  0.139  959.652  17.49    16 1   959.715  0.109  959.715  17.49
13 1   959.913  0.151  959.913  17.49    16 1   958.698  0.134  958.698  17.49
13 1   959.496  0.126  959.496  17.49    16 1   958.714  0.135  958.714  17.49
13 1   959.829  0.139  959.829  17.49    16 1   959.206  0.117  959.206  17.49
13 1   959.935  0.123  959.935  17.49    16 1   958.849  0.130  958.849  17.49
13 1   959.534  0.142  959.534  17.49    16 1   958.513  0.119  958.513  17.49
13 1   958.661  0.134  958.661  17.49    16 1   959.080  0.128  959.080  17.49
13 1   959.809  0.153  959.809  17.49    16 1   959.763  0.130  959.763  17.49
13 1   958.318  0.131  958.318  17.49    16 1   959.104  0.146  959.104  17.49
13 1   961.141  0.111  961.141  17.49    16 1   959.500  0.138  959.500  17.49
13 1   959.764  0.136  959.764  17.49    16 1   958.993  0.140  958.993  17.49
13 1   959.003  0.136  959.003  17.49    23 1   960.692  0.162  960.692  17.49
13 1   959.346  0.155  959.346  17.49    23 1   958.971  0.144  958.971  17.49
13 1   959.315  0.186  959.315  17.49    23 1   959.642  0.119  959.642  17.49
13 1   958.816  0.180  958.816  17.49    23 1   959.060  0.146  959.060  17.49
13 1   960.642  0.179  960.642  17.49    23 1   959.405  0.138  959.405  17.49
13 1   959.003  0.140  959.003  17.49    23 1   959.432  0.135  959.432  17.49
13 1   958.869  0.164  958.869  17.49    23 1   960.063  0.143  960.063  17.49
13 1   958.590  0.197  958.590  17.49    23 1   959.647  0.147  959.647  17.49
13 1   958.270  0.188  958.270  17.49    23 1   959.754  0.145  959.754  17.49
14 1   959.838  0.123  959.838  17.49    26 1   959.381  0.162  959.381  17.49
14 1   959.059  0.159  959.059  17.49    26 1   960.449  0.113  960.449  17.49
14 1   959.487  0.135  959.487  17.49    26 1   959.445  0.146  959.445  17.49
14 1   959.420  0.160  959.420  17.49    26 1   959.711  0.141  959.711  17.49
14 1   960.210  0.138  960.210  17.49    26 1   959.939  0.131  959.939  17.49
14 1   959.797  0.129  959.797  17.49    26 1   959.661  0.168  959.661  17.49
14 1   960.017  0.123  960.017  17.49    26 1   959.879  0.105  959.879  17.49
14 1   960.333  0.136  960.333  17.49    26 1   960.283  0.133  960.283  17.49
14 1   959.188  0.161  959.188  17.49    26 1   958.929  0.185  958.929  17.49
14 1   960.254  0.194  960.254  17.49    26 1   959.557  0.160  959.557  17.49
14 1   959.753  0.102  959.753  17.49    26 1   959.000  0.193  959.000  17.49
```

| 2004 - JULHO | | | | | |
|---|---|---|---|---|---|
| D | L | SDB | ER | SDC | HL |
| 26 | 1 | 960.571 | 0.173 | 960.571 | 17.49 |
| 26 | 1 | 959.644 | 0.155 | 959.644 | 17.49 |
| 26 | 1 | 959.899 | 0.152 | 959.899 | 17.49 |
| 26 | 1 | 959.315 | 0.172 | 959.315 | 17.49 |
| 26 | 1 | 959.674 | 0.204 | 959.674 | 17.49 |
| 26 | 1 | 958.731 | 0.144 | 958.731 | 17.49 |
| 26 | 1 | 960.159 | 0.196 | 960.159 | 17.49 |
| 26 | 1 | 959.584 | 0.182 | 959.584 | 17.49 |
| 26 | 1 | 958.931 | 0.129 | 958.931 | 17.49 |
| 26 | 1 | 959.440 | 0.177 | 959.440 | 17.49 |
| 26 | 1 | 958.674 | 0.237 | 958.674 | 17.49 |
| 26 | 1 | 958.544 | 0.152 | 958.544 | 17.49 |
| 27 | 1 | 959.493 | 0.168 | 959.493 | 17.49 |
| 27 | 1 | 959.843 | 0.117 | 959.843 | 17.49 |
| 27 | 1 | 959.790 | 0.169 | 959.790 | 17.49 |
| 27 | 1 | 960.614 | 0.141 | 960.614 | 17.49 |
| 27 | 1 | 959.252 | 0.153 | 959.252 | 17.49 |
| 27 | 1 | 958.644 | 0.170 | 958.644 | 17.49 |
| 27 | 1 | 959.010 | 0.114 | 959.010 | 17.49 |
| 27 | 1 | 959.127 | 0.145 | 959.127 | 17.49 |
| 27 | 1 | 959.913 | 0.137 | 959.913 | 17.49 |
| 27 | 1 | 960.262 | 0.137 | 960.262 | 17.49 |
| 27 | 1 | 959.863 | 0.166 | 959.863 | 17.49 |
| 27 | 1 | 959.178 | 0.164 | 959.178 | 17.49 |
| 27 | 1 | 961.425 | 0.213 | 961.425 | 17.49 |
| 27 | 1 | 959.032 | 0.169 | 959.032 | 17.49 |
| 27 | 1 | 958.563 | 0.223 | 958.563 | 17.49 |
| 27 | 1 | 959.787 | 0.179 | 959.787 | 17.49 |
| 27 | 1 | 959.558 | 0.147 | 959.558 | 17.49 |
| 27 | 1 | 959.157 | 0.125 | 959.157 | 17.49 |
| 27 | 1 | 959.887 | 0.167 | 959.887 | 17.49 |
| 27 | 1 | 960.091 | 0.146 | 960.091 | 17.49 |
| 27 | 1 | 959.542 | 0.130 | 959.542 | 17.49 |
| 27 | 1 | 959.844 | 0.138 | 959.844 | 17.49 |
| 27 | 1 | 959.767 | 0.159 | 959.767 | 17.49 |
| 27 | 1 | 958.986 | 0.125 | 958.986 | 17.49 |
| 27 | 1 | 959.890 | 0.128 | 959.890 | 17.49 |
| 27 | 1 | 959.331 | 0.129 | 959.331 | 17.49 |
| 28 | 1 | 958.230 | 0.148 | 958.230 | 17.49 |
| 28 | 1 | 958.591 | 0.127 | 958.591 | 17.49 |
| 28 | 1 | 959.391 | 0.193 | 959.391 | 17.49 |
| 28 | 1 | 959.909 | 0.153 | 959.909 | 17.49 |
| 28 | 1 | 959.264 | 0.182 | 959.264 | 17.49 |
| 28 | 1 | 959.135 | 0.163 | 959.135 | 17.49 |
| 28 | 1 | 960.155 | 0.147 | 960.155 | 17.49 |
| 28 | 1 | 960.379 | 0.199 | 960.379 | 17.49 |
| 28 | 1 | 958.799 | 0.219 | 958.799 | 17.49 |
| 28 | 1 | 958.586 | 0.189 | 958.586 | 17.49 |
| 28 | 1 | 959.308 | 0.198 | 959.308 | 17.49 |
| 28 | 1 | 958.872 | 0.184 | 958.872 | 17.49 |
| 28 | 1 | 958.702 | 0.194 | 958.702 | 17.49 |
| 28 | 1 | 959.178 | 0.177 | 959.178 | 17.49 |
| 28 | 1 | 958.851 | 0.171 | 958.851 | 17.49 |
| 28 | 1 | 959.688 | 0.182 | 959.688 | 17.49 |
| 28 | 1 | 959.455 | 0.242 | 959.455 | 17.49 |
| 28 | 1 | 958.846 | 0.151 | 958.846 | 17.49 |
| 28 | 1 | 958.663 | 0.182 | 958.663 | 17.49 |
| 28 | 1 | 958.862 | 0.187 | 958.862 | 17.49 |
| 29 | 1 | 957.782 | 0.151 | 957.782 | 17.49 |
| 29 | 1 | 959.422 | 0.148 | 959.422 | 17.49 |
| 29 | 1 | 960.115 | 0.133 | 960.115 | 17.49 |
| 29 | 1 | 960.096 | 0.111 | 960.096 | 17.49 |
| 29 | 1 | 959.318 | 0.142 | 959.318 | 17.49 |
| 29 | 1 | 959.368 | 0.140 | 959.368 | 17.49 |
| 29 | 1 | 959.460 | 0.154 | 959.460 | 17.49 |
| 29 | 1 | 958.840 | 0.132 | 958.840 | 17.49 |
| 29 | 1 | 959.794 | 0.123 | 959.794 | 17.49 |
| 29 | 1 | 959.336 | 0.154 | 959.336 | 17.49 |
| 29 | 1 | 959.775 | 0.170 | 959.775 | 17.49 |
| 30 | 1 | 958.856 | 0.106 | 958.856 | 17.49 |
| 30 | 1 | 959.033 | 0.128 | 959.033 | 17.49 |

| 2004 - JULHO | | | | | |
|---|---|---|---|---|---|
| D | L | SDB | ER | SDC | HL |
| 30 | 1 | 959.867 | 0.131 | 959.867 | 17.49 |
| 30 | 1 | 959.307 | 0.139 | 959.307 | 17.49 |
| 30 | 1 | 959.256 | 0.135 | 959.256 | 17.49 |
| 30 | 1 | 959.012 | 0.141 | 959.012 | 17.49 |
| 30 | 1 | 959.880 | 0.154 | 959.880 | 17.49 |
| 30 | 1 | 960.380 | 0.188 | 960.380 | 17.49 |
| 30 | 1 | 960.104 | 0.164 | 960.104 | 17.49 |
| 30 | 1 | 959.799 | 0.150 | 959.799 | 17.49 |
| 30 | 1 | 959.360 | 0.123 | 959.360 | 17.49 |
| 30 | 1 | 958.906 | 0.155 | 958.906 | 17.49 |
| 30 | 1 | 959.927 | 0.149 | 959.927 | 17.49 |
| 30 | 1 | 960.019 | 0.134 | 960.019 | 17.49 |
| 30 | 1 | 959.556 | 0.136 | 959.556 | 17.49 |
| 30 | 1 | 959.140 | 0.134 | 959.140 | 17.49 |
| 30 | 1 | 959.346 | 0.153 | 959.346 | 17.49 |
| 30 | 1 | 959.869 | 0.166 | 959.869 | 17.49 |

| 2004 - AGOSTO | | | | | |
|---|---|---|---|---|---|
| D | L | SDB | ER | SDC | HL |
| 04 | 1 | 958.827 | 0.114 | 958.827 | 17.49 |
| 04 | 1 | 959.032 | 0.122 | 959.032 | 17.49 |
| 04 | 1 | 959.487 | 0.118 | 959.487 | 17.49 |
| 04 | 1 | 959.845 | 0.134 | 959.845 | 17.49 |
| 04 | 1 | 960.068 | 0.124 | 960.068 | 17.49 |
| 04 | 1 | 960.133 | 0.138 | 960.133 | 17.49 |
| 04 | 1 | 958.888 | 0.140 | 958.888 | 17.49 |
| 04 | 1 | 959.645 | 0.148 | 959.645 | 17.49 |
| 04 | 1 | 959.276 | 0.131 | 959.276 | 17.49 |
| 04 | 1 | 958.573 | 0.169 | 958.573 | 17.49 |
| 04 | 1 | 959.926 | 0.177 | 959.926 | 17.49 |
| 04 | 1 | 959.176 | 0.191 | 959.176 | 17.49 |
| 04 | 1 | 960.589 | 0.137 | 960.589 | 17.49 |
| 04 | 1 | 959.555 | 0.159 | 959.555 | 17.49 |
| 04 | 1 | 959.447 | 0.189 | 959.447 | 17.49 |
| 04 | 1 | 959.915 | 0.115 | 959.915 | 17.49 |
| 04 | 1 | 959.541 | 0.125 | 959.541 | 17.49 |
| 04 | 1 | 959.495 | 0.138 | 959.495 | 17.49 |
| 04 | 1 | 958.894 | 0.151 | 958.894 | 17.49 |
| 04 | 1 | 959.497 | 0.120 | 959.497 | 17.49 |
| 04 | 1 | 959.423 | 0.117 | 959.423 | 17.49 |
| 04 | 1 | 959.721 | 0.143 | 959.721 | 17.49 |
| 04 | 1 | 959.324 | 0.143 | 959.324 | 17.49 |
| 04 | 1 | 958.820 | 0.144 | 958.820 | 17.49 |
| 04 | 1 | 959.347 | 0.157 | 959.347 | 17.49 |
| 05 | 1 | 958.442 | 0.103 | 958.442 | 17.49 |
| 05 | 1 | 958.940 | 0.129 | 958.940 | 17.49 |
| 05 | 1 | 959.414 | 0.132 | 959.414 | 17.49 |
| 05 | 1 | 959.618 | 0.140 | 959.618 | 17.49 |
| 05 | 1 | 959.417 | 0.130 | 959.417 | 17.49 |
| 05 | 1 | 959.581 | 0.136 | 959.581 | 17.49 |
| 05 | 1 | 959.968 | 0.148 | 959.968 | 17.49 |
| 05 | 1 | 959.761 | 0.132 | 959.761 | 17.49 |
| 05 | 1 | 959.613 | 0.148 | 959.613 | 17.49 |
| 05 | 1 | 960.304 | 0.137 | 960.304 | 17.49 |
| 05 | 1 | 960.276 | 0.163 | 960.276 | 17.49 |
| 05 | 1 | 959.621 | 0.141 | 959.621 | 17.49 |
| 05 | 1 | 959.100 | 0.164 | 959.100 | 17.49 |
| 05 | 1 | 959.829 | 0.157 | 959.829 | 17.49 |
| 05 | 1 | 959.537 | 0.155 | 959.537 | 17.49 |
| 05 | 1 | 959.569 | 0.167 | 959.569 | 17.49 |
| 05 | 1 | 959.159 | 0.157 | 959.159 | 17.49 |
| 05 | 1 | 959.675 | 0.154 | 959.675 | 17.49 |
| 05 | 1 | 959.172 | 0.154 | 959.172 | 17.49 |
| 05 | 1 | 958.614 | 0.171 | 958.614 | 17.49 |
| 06 | 1 | 960.363 | 0.180 | 960.363 | 17.49 |
| 06 | 1 | 959.193 | 0.135 | 959.193 | 17.49 |
| 06 | 1 | 960.186 | 0.142 | 960.186 | 17.49 |
| 06 | 1 | 958.830 | 0.159 | 958.830 | 17.49 |
| 06 | 1 | 959.746 | 0.127 | 959.746 | 17.49 |
| 06 | 1 | 960.106 | 0.165 | 960.106 | 17.49 |

| 2004 - AGOSTO | | | | |
|---|---|---|---|---|
| D | L | SDB | ER | SDC | HL |
| 06 | 1 | 959.205 | 0.146 | 959.205 | 17.49 |
| 06 | 1 | 959.448 | 0.157 | 959.448 | 17.49 |
| 06 | 1 | 959.579 | 0.137 | 959.579 | 17.49 |
| 06 | 1 | 959.433 | 0.139 | 959.433 | 17.49 |
| 16 | 1 | 957.040 | 0.266 | 957.040 | 17.49 |
| 16 | 1 | 957.340 | 0.217 | 957.340 | 17.49 |
| 16 | 1 | 958.931 | 0.196 | 958.931 | 17.49 |
| 16 | 1 | 960.426 | 0.136 | 960.426 | 17.49 |
| 16 | 1 | 960.032 | 0.135 | 960.032 | 17.49 |
| 16 | 1 | 959.754 | 0.190 | 959.754 | 17.49 |
| 16 | 1 | 960.701 | 0.194 | 960.701 | 17.49 |
| 16 | 1 | 959.399 | 0.151 | 959.399 | 17.49 |
| 16 | 1 | 959.810 | 0.139 | 959.810 | 17.49 |
| 16 | 1 | 959.011 | 0.199 | 959.011 | 17.49 |
| 16 | 1 | 960.849 | 0.157 | 960.849 | 17.49 |
| 16 | 1 | 959.721 | 0.142 | 959.721 | 17.49 |
| 16 | 1 | 959.393 | 0.148 | 959.393 | 17.49 |
| 16 | 1 | 959.153 | 0.149 | 959.153 | 17.49 |
| 16 | 1 | 959.151 | 0.169 | 959.151 | 17.49 |
| 16 | 1 | 959.484 | 0.133 | 959.484 | 17.49 |
| 16 | 1 | 958.650 | 0.156 | 958.650 | 17.49 |
| 16 | 1 | 959.006 | 0.132 | 959.006 | 17.49 |
| 16 | 1 | 959.355 | 0.142 | 959.355 | 17.49 |
| 16 | 1 | 958.908 | 0.158 | 958.908 | 17.49 |
| 16 | 1 | 959.144 | 0.142 | 959.144 | 17.49 |
| 16 | 1 | 959.280 | 0.141 | 959.280 | 17.49 |
| 16 | 1 | 959.042 | 0.151 | 959.042 | 17.49 |
| 16 | 1 | 958.416 | 0.156 | 958.416 | 17.49 |
| 17 | 1 | 958.414 | 0.113 | 958.414 | 17.49 |
| 17 | 1 | 958.751 | 0.152 | 958.751 | 17.49 |
| 17 | 1 | 958.703 | 0.131 | 958.703 | 17.49 |
| 17 | 1 | 959.479 | 0.154 | 959.479 | 17.49 |
| 17 | 1 | 960.119 | 0.122 | 960.119 | 17.49 |
| 17 | 1 | 959.799 | 0.140 | 959.799 | 17.49 |
| 17 | 1 | 958.692 | 0.160 | 958.692 | 17.49 |
| 17 | 1 | 958.993 | 0.412 | 958.993 | 17.49 |
| 17 | 1 | 959.479 | 0.116 | 959.479 | 17.49 |
| 17 | 1 | 959.076 | 0.125 | 959.076 | 17.49 |
| 17 | 1 | 960.948 | 0.144 | 960.948 | 17.49 |
| 17 | 1 | 959.668 | 0.141 | 959.668 | 17.49 |
| 17 | 1 | 959.667 | 0.140 | 959.667 | 17.49 |
| 17 | 1 | 960.806 | 0.177 | 960.806 | 17.49 |
| 17 | 1 | 960.508 | 0.127 | 960.508 | 17.49 |
| 17 | 1 | 959.018 | 0.148 | 959.018 | 17.49 |
| 17 | 1 | 958.960 | 0.187 | 958.960 | 17.49 |
| 17 | 1 | 958.348 | 0.204 | 958.348 | 17.49 |
| 17 | 1 | 958.886 | 0.180 | 958.886 | 17.49 |
| 17 | 1 | 959.722 | 0.182 | 959.722 | 17.49 |
| 17 | 1 | 958.509 | 0.162 | 958.509 | 17.49 |
| 17 | 1 | 958.509 | 0.178 | 958.509 | 17.49 |
| 17 | 1 | 958.312 | 0.213 | 958.312 | 17.49 |
| 17 | 1 | 959.225 | 0.194 | 959.225 | 17.49 |
| 17 | 1 | 958.915 | 0.187 | 958.915 | 17.49 |
| 17 | 1 | 958.094 | 0.186 | 958.094 | 17.49 |
| 18 | 1 | 958.815 | 0.150 | 958.815 | 17.49 |
| 18 | 1 | 958.873 | 0.174 | 958.873 | 17.49 |
| 18 | 1 | 959.539 | 0.149 | 959.539 | 17.49 |
| 18 | 1 | 959.186 | 0.153 | 959.186 | 17.49 |
| 18 | 1 | 959.192 | 0.185 | 959.192 | 17.49 |
| 18 | 1 | 959.535 | 0.170 | 959.535 | 17.49 |
| 18 | 1 | 959.399 | 0.190 | 959.399 | 17.49 |
| 18 | 1 | 959.911 | 0.168 | 959.911 | 17.49 |
| 18 | 1 | 959.360 | 0.181 | 959.360 | 17.49 |
| 18 | 1 | 959.134 | 0.208 | 959.134 | 17.49 |
| 18 | 1 | 960.404 | 0.184 | 960.404 | 17.49 |
| 18 | 1 | 958.672 | 0.208 | 958.672 | 17.49 |
| 18 | 1 | 960.974 | 0.147 | 960.974 | 17.49 |
| 18 | 1 | 958.463 | 0.166 | 958.463 | 17.49 |
| 18 | 1 | 960.086 | 0.218 | 960.086 | 17.49 |
| 18 | 1 | 959.292 | 0.180 | 959.292 | 17.49 |
| 18 | 1 | 959.517 | 0.158 | 959.517 | 17.49 |
| 18 | 1 | 958.957 | 0.181 | 958.957 | 17.49 |
| 18 | 1 | 958.750 | 0.143 | 958.750 | 17.49 |
| 18 | 1 | 960.031 | 0.152 | 960.031 | 17.49 |
| 18 | 1 | 958.669 | 0.142 | 958.669 | 17.49 |
| 18 | 1 | 958.827 | 0.128 | 958.827 | 17.49 |
| 18 | 1 | 959.923 | 0.177 | 959.923 | 17.49 |
| 18 | 1 | 959.367 | 0.185 | 959.367 | 17.49 |
| 18 | 1 | 959.012 | 0.157 | 959.012 | 17.49 |
| 18 | 1 | 958.365 | 0.132 | 958.365 | 17.49 |
| 19 | 1 | 958.826 | 0.126 | 958.826 | 17.49 |
| 19 | 1 | 959.392 | 0.101 | 959.392 | 17.49 |
| 19 | 1 | 959.109 | 0.150 | 959.109 | 17.49 |
| 19 | 1 | 959.441 | 0.150 | 959.441 | 17.49 |
| 19 | 1 | 959.698 | 0.141 | 959.698 | 17.49 |
| 19 | 1 | 959.051 | 0.136 | 959.051 | 17.49 |
| 19 | 1 | 959.161 | 0.126 | 959.161 | 17.49 |
| 19 | 1 | 959.119 | 0.128 | 959.119 | 17.49 |
| 19 | 1 | 959.645 | 0.120 | 959.645 | 17.49 |
| 19 | 1 | 959.912 | 0.134 | 959.912 | 17.49 |
| 19 | 1 | 959.397 | 0.129 | 959.397 | 17.49 |
| 19 | 1 | 959.861 | 0.114 | 959.861 | 17.49 |
| 19 | 1 | 959.504 | 0.116 | 959.504 | 17.49 |
| 19 | 1 | 959.644 | 0.135 | 959.644 | 17.49 |
| 19 | 1 | 959.802 | 0.136 | 959.802 | 17.49 |
| 19 | 1 | 959.611 | 0.146 | 959.611 | 17.49 |
| 19 | 1 | 959.406 | 0.165 | 959.406 | 17.49 |
| 19 | 1 | 959.793 | 0.123 | 959.793 | 17.49 |
| 19 | 1 | 959.202 | 0.118 | 959.202 | 17.49 |
| 23 | 1 | 958.399 | 0.161 | 958.399 | 17.49 |
| 23 | 1 | 959.193 | 0.122 | 959.193 | 17.49 |
| 23 | 1 | 958.892 | 0.136 | 958.892 | 17.49 |
| 23 | 1 | 959.636 | 0.127 | 959.636 | 17.49 |
| 23 | 1 | 959.778 | 0.138 | 959.778 | 17.49 |
| 23 | 1 | 959.422 | 0.151 | 959.422 | 17.49 |
| 23 | 1 | 958.782 | 0.153 | 958.782 | 17.49 |
| 23 | 1 | 959.350 | 0.138 | 959.350 | 17.49 |
| 23 | 1 | 959.260 | 0.157 | 959.260 | 17.49 |
| 23 | 1 | 959.904 | 0.184 | 959.904 | 17.49 |
| 25 | 1 | 960.447 | 0.158 | 960.447 | 17.49 |
| 25 | 1 | 959.581 | 0.164 | 959.581 | 17.49 |
| 25 | 1 | 959.184 | 0.176 | 959.184 | 17.49 |
| 25 | 1 | 958.488 | 0.147 | 958.488 | 17.49 |
| 25 | 1 | 959.446 | 0.196 | 959.446 | 17.49 |
| 25 | 1 | 958.706 | 0.190 | 958.706 | 17.49 |
| 25 | 1 | 958.890 | 0.175 | 958.890 | 17.49 |
| 25 | 1 | 959.063 | 0.141 | 959.063 | 17.49 |
| 25 | 1 | 958.689 | 0.156 | 958.689 | 17.49 |
| 25 | 1 | 958.384 | 0.150 | 958.384 | 17.49 |
| 25 | 1 | 959.303 | 0.179 | 959.303 | 17.49 |
| 25 | 1 | 958.648 | 0.170 | 958.648 | 17.49 |
| 25 | 1 | 959.341 | 0.155 | 959.341 | 17.49 |
| 25 | 1 | 959.448 | 0.162 | 959.448 | 17.49 |
| 25 | 1 | 959.068 | 0.136 | 959.068 | 17.49 |
| 30 | 1 | 959.031 | 0.152 | 959.031 | 17.49 |
| 30 | 1 | 959.189 | 0.150 | 959.189 | 17.49 |
| 30 | 1 | 959.992 | 0.144 | 959.992 | 17.49 |
| 30 | 1 | 960.612 | 0.141 | 960.612 | 17.49 |
| 30 | 1 | 959.640 | 0.129 | 959.640 | 17.49 |
| 30 | 1 | 959.587 | 0.154 | 959.587 | 17.49 |
| 30 | 1 | 959.189 | 0.150 | 959.189 | 17.49 |
| 30 | 1 | 958.592 | 0.148 | 958.592 | 17.49 |
| 30 | 1 | 959.610 | 0.151 | 959.610 | 17.49 |
| 30 | 1 | 960.023 | 0.171 | 960.023 | 17.49 |
| 30 | 1 | 959.983 | 0.166 | 959.983 | 17.49 |
| 30 | 1 | 959.855 | 0.175 | 959.855 | 17.49 |
| 30 | 1 | 958.946 | 0.155 | 958.946 | 17.49 |
| 30 | 1 | 958.735 | 0.175 | 958.735 | 17.49 |
| 30 | 1 | 959.977 | 0.132 | 959.977 | 17.49 |
| 30 | 1 | 958.879 | 0.156 | 958.879 | 17.49 |
| 30 | 1 | 959.718 | 0.137 | 959.718 | 17.49 |
| 30 | 1 | 959.153 | 0.099 | 959.153 | 17.49 |

| 2004 - AGOSTO | | | | | | 2004 - SETEMBRO | | | | |
|---|---|---|---|---|---|---|---|---|---|---|
| D | L | SDB | ER | SDC | HL | D | L | SDB | ER | SDC | HL |
| 30 | 1 | 958.313 | 0.105 | 958.313 | 17.49 | 03 | 1 | 959.381 | 0.114 | 959.381 | 17.49 |
| 30 | 1 | 958.574 | 0.140 | 958.574 | 17.49 | 03 | 1 | 959.990 | 0.130 | 959.990 | 17.49 |
| 30 | 1 | 958.933 | 0.132 | 958.933 | 17.49 | 03 | 1 | 959.686 | 0.136 | 959.686 | 17.49 |
| 30 | 1 | 958.426 | 0.158 | 958.426 | 17.49 | 03 | 1 | 959.175 | 0.133 | 959.175 | 17.49 |
| 30 | 1 | 958.353 | 0.139 | 958.353 | 17.49 | 03 | 1 | 959.336 | 0.166 | 959.336 | 17.49 |
| 30 | 1 | 959.202 | 0.132 | 959.202 | 17.49 | 03 | 1 | 959.085 | 0.123 | 959.085 | 17.49 |
| 30 | 1 | 959.304 | 0.140 | 959.304 | 17.49 | 06 | 1 | 957.401 | 0.144 | 957.401 | 17.49 |
| 30 | 1 | 959.509 | 0.143 | 959.509 | 17.49 | 06 | 1 | 958.458 | 0.154 | 958.458 | 17.49 |
| 31 | 1 | 957.801 | 0.119 | 957.801 | 17.49 | 06 | 1 | 958.954 | 0.159 | 958.954 | 17.49 |
| 31 | 1 | 958.671 | 0.128 | 958.671 | 17.49 | 06 | 1 | 959.190 | 0.142 | 959.190 | 17.49 |
| 31 | 1 | 959.091 | 0.112 | 959.091 | 17.49 | 06 | 1 | 958.519 | 0.155 | 958.519 | 17.49 |
| 31 | 1 | 960.167 | 0.139 | 960.167 | 17.49 | 06 | 1 | 958.965 | 0.165 | 958.965 | 17.49 |
| 31 | 1 | 959.809 | 0.135 | 959.809 | 17.49 | 06 | 1 | 959.750 | 0.118 | 959.750 | 17.49 |
| 31 | 1 | 960.609 | 0.135 | 960.609 | 17.49 | 06 | 1 | 958.860 | 0.160 | 958.860 | 17.49 |
| 31 | 1 | 960.007 | 0.155 | 960.007 | 17.49 | 06 | 1 | 958.987 | 0.176 | 958.987 | 17.49 |
| 31 | 1 | 958.383 | 0.159 | 958.383 | 17.49 | 06 | 1 | 957.735 | 0.193 | 957.735 | 17.49 |
| 31 | 1 | 959.532 | 0.136 | 959.532 | 17.49 | 06 | 1 | 959.254 | 0.176 | 959.254 | 17.49 |
| 31 | 1 | 959.971 | 0.151 | 959.971 | 17.49 | 06 | 1 | 959.866 | 0.201 | 959.866 | 17.49 |
| 31 | 1 | 959.699 | 0.135 | 959.699 | 17.49 | 06 | 1 | 959.835 | 0.209 | 959.835 | 17.49 |
| 31 | 1 | 959.467 | 0.173 | 959.467 | 17.49 | 06 | 1 | 960.626 | 0.128 | 960.626 | 17.49 |
| 31 | 1 | 959.443 | 0.173 | 959.443 | 17.49 | 06 | 1 | 959.885 | 0.140 | 959.885 | 17.49 |
| 31 | 1 | 959.355 | 0.167 | 959.355 | 17.49 | 06 | 1 | 958.951 | 0.168 | 958.951 | 17.49 |
| 31 | 1 | 959.118 | 0.131 | 959.118 | 17.49 | 06 | 1 | 959.358 | 0.173 | 959.358 | 17.49 |
| 31 | 1 | 960.342 | 0.163 | 960.342 | 17.49 | 06 | 1 | 959.875 | 0.143 | 959.875 | 17.49 |
| 31 | 1 | 959.935 | 0.234 | 959.935 | 17.49 | 06 | 1 | 958.611 | 0.154 | 958.611 | 17.49 |
| 31 | 1 | 958.791 | 0.115 | 958.791 | 17.49 | 06 | 1 | 958.933 | 0.183 | 958.933 | 17.49 |
| 31 | 1 | 959.130 | 0.155 | 959.130 | 17.49 | 06 | 1 | 957.900 | 0.203 | 957.900 | 17.49 |
| 31 | 1 | 959.319 | 0.143 | 959.319 | 17.49 | 06 | 1 | 958.555 | 0.188 | 958.555 | 17.49 |
| 31 | 1 | 958.733 | 0.128 | 958.733 | 17.49 | 06 | 1 | 958.805 | 0.122 | 958.805 | 17.49 |
| 31 | 1 | 958.539 | 0.157 | 958.539 | 17.49 | 06 | 1 | 958.776 | 0.144 | 958.776 | 17.49 |
| 31 | 1 | 959.383 | 0.129 | 959.383 | 17.49 | 06 | 1 | 959.153 | 0.228 | 959.153 | 17.49 |
| 31 | 1 | 959.418 | 0.146 | 959.418 | 17.49 | 06 | 1 | 959.278 | 0.160 | 959.278 | 17.49 |
| 31 | 1 | 958.957 | 0.146 | 958.957 | 17.49 | 06 | 1 | 959.526 | 0.186 | 959.526 | 17.49 |
| 31 | 1 | 958.745 | 0.156 | 958.745 | 17.49 | 06 | 1 | 958.024 | 0.174 | 958.024 | 17.49 |
| 31 | 1 | 959.363 | 0.169 | 959.363 | 17.49 | 06 | 1 | 958.055 | 0.175 | 958.055 | 17.49 |
| | | | | | | 06 | 1 | 959.787 | 0.188 | 959.787 | 17.49 |
| | | | | | | 08 | 1 | 958.214 | 0.150 | 958.214 | 17.49 |
| | | 2004 - SETEMBRO | | | | 08 | 1 | 958.789 | 0.174 | 958.789 | 17.49 |
| D | L | SDB | ER | SDC | HL | 08 | 1 | 959.299 | 0.141 | 959.299 | 17.49 |
| 01 | 1 | 959.246 | 0.119 | 959.246 | 17.49 | 08 | 1 | 959.682 | 0.151 | 959.682 | 17.49 |
| 01 | 1 | 959.052 | 0.160 | 959.052 | 17.49 | 08 | 1 | 959.494 | 0.162 | 959.494 | 17.49 |
| 01 | 1 | 959.760 | 0.154 | 959.760 | 17.49 | 08 | 1 | 959.437 | 0.150 | 959.437 | 17.49 |
| 01 | 1 | 959.130 | 0.150 | 959.130 | 17.49 | 08 | 1 | 958.822 | 0.152 | 958.822 | 17.49 |
| 01 | 1 | 960.106 | 0.125 | 960.106 | 17.49 | 08 | 1 | 960.018 | 0.139 | 960.018 | 17.49 |
| 01 | 1 | 958.820 | 0.116 | 958.820 | 17.49 | 08 | 1 | 959.782 | 0.131 | 959.782 | 17.49 |
| 01 | 1 | 959.636 | 0.139 | 959.636 | 17.49 | 08 | 1 | 959.847 | 0.171 | 959.847 | 17.49 |
| 01 | 1 | 958.616 | 0.174 | 958.616 | 17.49 | 08 | 1 | 958.835 | 0.178 | 958.835 | 17.49 |
| 01 | 1 | 959.233 | 0.204 | 959.233 | 17.49 | 08 | 1 | 959.564 | 0.137 | 959.564 | 17.49 |
| 01 | 1 | 957.925 | 0.279 | 957.925 | 17.49 | 08 | 1 | 959.197 | 0.184 | 959.197 | 17.49 |
| 01 | 1 | 960.906 | 0.116 | 960.906 | 17.49 | 08 | 1 | 961.848 | 0.181 | 961.848 | 17.49 |
| 01 | 1 | 959.546 | 0.121 | 959.546 | 17.49 | 08 | 1 | 959.693 | 0.157 | 959.693 | 17.49 |
| 01 | 1 | 959.464 | 0.125 | 959.464 | 17.49 | 08 | 1 | 960.219 | 0.193 | 960.219 | 17.49 |
| 01 | 1 | 960.452 | 0.143 | 960.452 | 17.49 | 08 | 1 | 958.634 | 0.129 | 958.634 | 17.49 |
| 01 | 1 | 959.104 | 0.124 | 959.104 | 17.49 | 08 | 1 | 957.258 | 0.153 | 957.258 | 17.49 |
| 01 | 1 | 958.411 | 0.172 | 958.411 | 17.49 | 08 | 1 | 958.759 | 0.148 | 958.759 | 17.49 |
| 01 | 1 | 959.165 | 0.157 | 959.165 | 17.49 | 08 | 1 | 959.206 | 0.164 | 959.206 | 17.49 |
| 01 | 1 | 959.362 | 0.212 | 959.362 | 17.49 | 08 | 1 | 959.862 | 0.162 | 959.862 | 17.49 |
| 01 | 1 | 958.630 | 0.134 | 958.630 | 17.49 | 08 | 1 | 959.386 | 0.144 | 959.386 | 17.49 |
| 01 | 1 | 959.759 | 0.123 | 959.759 | 17.49 | 08 | 1 | 959.099 | 0.149 | 959.099 | 17.49 |
| 01 | 1 | 959.624 | 0.143 | 959.624 | 17.49 | 08 | 1 | 959.148 | 0.125 | 959.148 | 17.49 |
| 01 | 1 | 959.377 | 0.125 | 959.377 | 17.49 | 08 | 1 | 958.673 | 0.156 | 958.673 | 17.49 |
| 01 | 1 | 959.611 | 0.144 | 959.611 | 17.49 | 08 | 1 | 958.981 | 0.173 | 958.981 | 17.49 |
| 01 | 1 | 959.418 | 0.156 | 959.418 | 17.49 | 08 | 1 | 958.412 | 0.169 | 958.412 | 17.49 |
| 01 | 1 | 959.367 | 0.148 | 959.367 | 17.49 | 08 | 1 | 959.585 | 0.152 | 959.585 | 17.49 |
| 01 | 1 | 959.546 | 0.115 | 959.546 | 17.49 | 09 | 1 | 960.420 | 0.163 | 960.420 | 17.49 |
| 03 | 1 | 958.210 | 0.092 | 958.210 | 17.49 | 09 | 1 | 959.077 | 0.162 | 959.077 | 17.49 |
| 03 | 1 | 959.112 | 0.107 | 959.112 | 17.49 | 09 | 1 | 958.963 | 0.145 | 958.963 | 17.49 |
| 03 | 1 | 959.650 | 0.097 | 959.650 | 17.49 | 09 | 1 | 959.256 | 0.152 | 959.256 | 17.49 |
| 03 | 1 | 958.762 | 0.141 | 958.762 | 17.49 | 09 | 1 | 959.255 | 0.151 | 959.255 | 17.49 |
| 03 | 1 | 959.171 | 0.114 | 959.171 | 17.49 | 09 | 1 | 958.229 | 0.163 | 958.229 | 17.49 |
| 03 | 1 | 959.309 | 0.112 | 959.309 | 17.49 | 09 | 1 | 959.380 | 0.160 | 959.380 | 17.49 |

| 2004 - SETEMBRO | | | | |
|---|---|---|---|---|
| D  L | SDB | ER | SDC | HL |
| 09  1 | 958.955 | 0.163 | 958.955 | 17.49 |
| 09  1 | 959.038 | 0.155 | 959.038 | 17.49 |
| 09  1 | 959.528 | 0.149 | 959.528 | 17.49 |
| 09  1 | 959.420 | 0.101 | 959.420 | 17.49 |
| 09  1 | 959.124 | 0.118 | 959.124 | 17.49 |
| 09  1 | 958.904 | 0.108 | 958.904 | 17.49 |
| 09  1 | 959.616 | 0.146 | 959.616 | 17.49 |
| 09  1 | 959.517 | 0.154 | 959.517 | 17.49 |
| 10  1 | 958.265 | 0.150 | 958.265 | 17.49 |
| 10  1 | 958.881 | 0.123 | 958.881 | 17.49 |
| 10  1 | 959.078 | 0.122 | 959.078 | 17.49 |
| 10  1 | 959.612 | 0.133 | 959.612 | 17.49 |
| 10  1 | 958.980 | 0.139 | 958.980 | 17.49 |
| 10  1 | 959.240 | 0.148 | 959.240 | 17.49 |
| 10  1 | 959.680 | 0.153 | 959.680 | 17.49 |
| 10  1 | 959.446 | 0.187 | 959.446 | 17.49 |
| 10  1 | 959.714 | 0.162 | 959.714 | 17.49 |
| 10  1 | 957.671 | 0.166 | 957.671 | 17.49 |
| 10  1 | 958.764 | 0.154 | 958.764 | 17.49 |
| 10  1 | 959.228 | 0.135 | 959.228 | 17.49 |
| 10  1 | 958.761 | 0.147 | 958.761 | 17.49 |
| 10  1 | 959.261 | 0.153 | 959.261 | 17.49 |
| 10  1 | 958.783 | 0.131 | 958.783 | 17.49 |
| 10  1 | 958.575 | 0.139 | 958.575 | 17.49 |
| 10  1 | 958.458 | 0.175 | 958.458 | 17.49 |
| 10  1 | 958.770 | 0.154 | 958.770 | 17.49 |
| 10  1 | 959.687 | 0.145 | 959.687 | 17.49 |
| 10  1 | 959.434 | 0.164 | 959.434 | 17.49 |
| 10  1 | 958.753 | 0.154 | 958.753 | 17.49 |
| 10  1 | 959.137 | 0.161 | 959.137 | 17.49 |
| 10  1 | 958.601 | 0.199 | 958.601 | 17.49 |
| 10  1 | 958.694 | 0.181 | 958.694 | 17.49 |

| 2006 - JUNHO | | | | |
|---|---|---|---|---|
| D  L | SDB | ER | SDC | HL |
| 08  1 | 959.512 | 0.156 | 959.512 | 17.49 |
| 08  1 | 959.104 | 0.204 | 959.104 | 17.49 |
| 08  1 | 959.265 | 0.124 | 959.265 | 17.49 |
| 08  1 | 959.446 | 0.125 | 959.446 | 17.49 |
| 08  1 | 959.966 | 0.126 | 959.966 | 17.49 |
| 08  1 | 960.237 | 0.119 | 960.237 | 17.49 |
| 09  1 | 958.634 | 0.162 | 958.634 | 17.49 |
| 09  1 | 959.404 | 0.140 | 959.404 | 17.49 |
| 09  1 | 959.009 | 0.137 | 959.009 | 17.49 |
| 09  1 | 959.730 | 0.141 | 959.730 | 17.49 |
| 09  1 | 959.801 | 0.121 | 959.801 | 17.49 |
| 09  1 | 960.317 | 0.126 | 960.317 | 17.49 |
| 09  1 | 959.677 | 0.142 | 959.677 | 17.49 |
| 09  1 | 959.927 | 0.148 | 959.927 | 17.49 |
| 09  1 | 959.191 | 0.136 | 959.191 | 17.49 |
| 09  1 | 959.323 | 0.194 | 959.323 | 17.49 |
| 09  1 | 959.160 | 0.187 | 959.160 | 17.49 |
| 09  1 | 959.149 | 0.233 | 959.149 | 17.49 |
| 09  1 | 959.787 | 0.195 | 959.787 | 17.49 |
| 09  1 | 958.959 | 0.299 | 958.959 | 17.49 |
| 09  1 | 959.111 | 0.270 | 959.111 | 17.49 |
| 09  1 | 958.974 | 0.259 | 958.974 | 17.49 |
| 14  1 | 959.116 | 0.111 | 959.116 | 17.49 |
| 14  1 | 958.992 | 0.135 | 958.992 | 17.49 |
| 14  1 | 959.242 | 0.189 | 959.242 | 17.49 |
| 14  1 | 958.688 | 0.159 | 958.688 | 17.49 |
| 14  1 | 958.829 | 0.145 | 958.829 | 17.49 |
| 14  1 | 959.944 | 0.144 | 959.944 | 17.49 |
| 14  1 | 960.308 | 0.123 | 960.308 | 17.49 |
| 14  1 | 960.328 | 0.110 | 960.328 | 17.49 |
| 14  1 | 960.082 | 0.164 | 960.082 | 17.49 |
| 14  1 | 960.052 | 0.193 | 960.052 | 17.49 |
| 14  1 | 959.272 | 0.216 | 959.272 | 17.49 |
| 14  1 | 959.163 | 0.150 | 959.163 | 17.49 |
| 14  1 | 959.776 | 0.193 | 959.776 | 17.49 |

| 2006 - JUNHO | | | | |
|---|---|---|---|---|
| D  L | SDB | ER | SDC | HL |
| 14  1 | 958.653 | 0.207 | 958.653 | 17.49 |
| 20  1 | 959.789 | 0.146 | 959.789 | 17.49 |
| 20  1 | 958.123 | 0.173 | 958.123 | 17.49 |
| 20  1 | 959.638 | 0.198 | 959.638 | 17.49 |
| 20  1 | 958.302 | 0.208 | 958.302 | 17.49 |
| 20  1 | 959.524 | 0.174 | 959.524 | 17.49 |
| 20  1 | 958.553 | 0.265 | 958.553 | 17.49 |
| 20  1 | 959.382 | 0.436 | 959.382 | 17.49 |
| 21  1 | 958.358 | 0.140 | 958.358 | 17.49 |
| 21  1 | 959.820 | 0.207 | 959.820 | 17.49 |
| 21  1 | 959.826 | 0.132 | 959.826 | 17.49 |
| 21  1 | 959.424 | 0.118 | 959.424 | 17.49 |
| 21  1 | 960.122 | 0.110 | 960.122 | 17.49 |
| 21  1 | 959.443 | 0.132 | 959.443 | 17.49 |
| 21  1 | 959.933 | 0.121 | 959.933 | 17.49 |
| 21  1 | 959.500 | 0.121 | 959.500 | 17.49 |
| 21  1 | 959.848 | 0.136 | 959.848 | 17.49 |
| 21  1 | 960.063 | 0.106 | 960.063 | 17.49 |
| 21  1 | 959.514 | 0.101 | 959.514 | 17.49 |
| 21  1 | 959.862 | 0.140 | 959.862 | 17.49 |
| 21  1 | 959.719 | 0.135 | 959.719 | 17.49 |
| 21  1 | 958.877 | 0.175 | 958.877 | 17.49 |
| 21  1 | 959.054 | 0.144 | 959.054 | 17.49 |
| 23  1 | 959.408 | 0.128 | 959.408 | 17.49 |
| 23  1 | 959.843 | 0.145 | 959.843 | 17.49 |
| 23  1 | 958.749 | 0.131 | 958.749 | 17.49 |
| 23  1 | 959.114 | 0.138 | 959.114 | 17.49 |
| 23  1 | 959.675 | 0.166 | 959.675 | 17.49 |
| 23  1 | 958.768 | 0.148 | 958.768 | 17.49 |
| 23  1 | 959.171 | 0.139 | 959.171 | 17.49 |
| 23  1 | 959.323 | 0.165 | 959.323 | 17.49 |
| 23  1 | 959.773 | 0.168 | 959.773 | 17.49 |
| 23  1 | 959.010 | 0.279 | 959.010 | 17.49 |
| 23  1 | 959.120 | 0.208 | 959.120 | 17.49 |
| 23  1 | 959.035 | 0.174 | 959.035 | 17.49 |
| 23  1 | 959.280 | 0.325 | 959.280 | 17.49 |
| 30  1 | 959.826 | 0.144 | 959.826 | 17.49 |
| 30  1 | 958.946 | 0.125 | 958.946 | 17.49 |
| 30  1 | 958.865 | 0.163 | 958.865 | 17.49 |
| 30  1 | 959.550 | 0.156 | 959.550 | 17.49 |
| 30  1 | 959.985 | 0.132 | 959.985 | 17.49 |
| 30  1 | 960.475 | 0.136 | 960.475 | 17.49 |
| 30  1 | 960.113 | 0.148 | 960.113 | 17.49 |
| 30  1 | 960.195 | 0.119 | 960.195 | 17.49 |
| 30  1 | 960.285 | 0.160 | 960.285 | 17.49 |
| 30  1 | 960.848 | 0.158 | 960.848 | 17.49 |

| 2006 - JULHO | | | | |
|---|---|---|---|---|
| D  L | SDB | ER | SDC | HL |
| 04  1 | 959.391 | 0.190 | 959.391 | 17.49 |
| 04  1 | 958.490 | 0.184 | 958.490 | 17.49 |
| 04  1 | 960.013 | 0.129 | 960.013 | 17.49 |
| 04  1 | 959.124 | 0.219 | 959.124 | 17.49 |
| 04  1 | 959.148 | 0.116 | 959.148 | 17.49 |
| 04  1 | 959.159 | 0.144 | 959.159 | 17.49 |
| 04  1 | 959.833 | 0.152 | 959.833 | 17.49 |
| 04  1 | 959.956 | 0.142 | 959.956 | 17.49 |
| 04  1 | 960.079 | 0.157 | 960.079 | 17.49 |
| 04  1 | 959.470 | 0.158 | 959.470 | 17.49 |
| 04  1 | 959.286 | 0.149 | 959.286 | 17.49 |
| 04  1 | 958.127 | 0.114 | 958.127 | 17.49 |
| 04  1 | 959.385 | 0.134 | 959.385 | 17.49 |
| 04  1 | 959.228 | 0.143 | 959.228 | 17.49 |
| 04  1 | 959.453 | 0.141 | 959.453 | 17.49 |
| 04  1 | 958.920 | 0.138 | 958.920 | 17.49 |
| 04  1 | 958.174 | 0.183 | 958.174 | 17.49 |
| 04  1 | 958.466 | 0.210 | 958.466 | 17.49 |
| 04  1 | 958.525 | 0.150 | 958.525 | 17.49 |
| 04  1 | 960.026 | 0.219 | 960.026 | 17.49 |
| 04  1 | 957.268 | 0.181 | 957.268 | 17.49 |

| 2006 - JULHO | | | | | | 2006 - JULHO | | | | |
|---|---|---|---|---|---|---|---|---|---|---|
| D | L | SDB | ER | SDC | HL | D | L | SDB | ER | SDC | HL |
| 04 | 1 | 959.271 | 0.221 | 959.271 | 17.49 | 12 | 1 | 959.506 | 0.176 | 959.506 | 17.49 |
| 05 | 1 | 958.218 | 0.109 | 958.218 | 17.49 | 12 | 1 | 959.371 | 0.134 | 959.371 | 17.49 |
| 05 | 1 | 959.593 | 0.137 | 959.593 | 17.49 | 12 | 1 | 959.915 | 0.140 | 959.915 | 17.49 |
| 05 | 1 | 959.788 | 0.124 | 959.788 | 17.49 | 12 | 1 | 960.174 | 0.133 | 960.174 | 17.49 |
| 05 | 1 | 960.173 | 0.124 | 960.173 | 17.49 | 12 | 1 | 959.970 | 0.146 | 959.970 | 17.49 |
| 05 | 1 | 959.223 | 0.115 | 959.223 | 17.49 | 12 | 1 | 959.525 | 0.148 | 959.525 | 17.49 |
| 05 | 1 | 959.889 | 0.148 | 959.889 | 17.49 | 12 | 1 | 959.853 | 0.193 | 959.853 | 17.49 |
| 05 | 1 | 960.104 | 0.151 | 960.104 | 17.49 | 12 | 1 | 960.265 | 0.141 | 960.265 | 17.49 |
| 05 | 1 | 959.325 | 0.171 | 959.325 | 17.49 | 12 | 1 | 959.969 | 0.133 | 959.969 | 17.49 |
| 05 | 1 | 960.191 | 0.201 | 960.191 | 17.49 | 12 | 1 | 959.556 | 0.151 | 959.556 | 17.49 |
| 05 | 1 | 960.290 | 0.128 | 960.290 | 17.49 | 12 | 1 | 960.668 | 0.159 | 960.668 | 17.49 |
| 05 | 1 | 959.948 | 0.142 | 959.948 | 17.49 | 12 | 1 | 960.325 | 0.146 | 960.325 | 17.49 |
| 05 | 1 | 959.956 | 0.132 | 959.956 | 17.49 | 12 | 1 | 959.268 | 0.182 | 959.268 | 17.49 |
| 05 | 1 | 959.277 | 0.115 | 959.277 | 17.49 | 12 | 1 | 959.349 | 0.197 | 959.349 | 17.49 |
| 05 | 1 | 959.720 | 0.143 | 959.720 | 17.49 | 12 | 1 | 959.638 | 0.252 | 959.638 | 17.49 |
| 05 | 1 | 958.787 | 0.116 | 958.787 | 17.49 | 12 | 1 | 958.220 | 0.298 | 958.220 | 17.49 |
| 05 | 1 | 959.184 | 0.107 | 959.184 | 17.49 | 12 | 1 | 959.491 | 0.184 | 959.491 | 17.49 |
| 05 | 1 | 958.491 | 0.128 | 958.491 | 17.49 | 12 | 1 | 958.322 | 0.196 | 958.322 | 17.49 |
| 05 | 1 | 959.241 | 0.166 | 959.241 | 17.49 | 12 | 1 | 959.112 | 0.190 | 959.112 | 17.49 |
| 05 | 1 | 958.676 | 0.154 | 958.676 | 17.49 | 12 | 1 | 957.537 | 0.260 | 957.537 | 17.49 |
| 05 | 1 | 959.499 | 0.168 | 959.499 | 17.49 | 12 | 1 | 958.177 | 0.262 | 958.177 | 17.49 |
| 07 | 1 | 958.307 | 0.149 | 958.307 | 17.49 | 12 | 1 | 957.943 | 0.220 | 957.943 | 17.49 |
| 07 | 1 | 958.695 | 0.138 | 958.695 | 17.49 | 13 | 1 | 959.714 | 0.133 | 959.714 | 17.49 |
| 07 | 1 | 959.083 | 0.156 | 959.083 | 17.49 | 13 | 1 | 958.949 | 0.142 | 958.949 | 17.49 |
| 07 | 1 | 959.174 | 0.149 | 959.174 | 17.49 | 13 | 1 | 959.763 | 0.114 | 959.763 | 17.49 |
| 07 | 1 | 959.517 | 0.147 | 959.517 | 17.49 | 13 | 1 | 959.870 | 0.123 | 959.870 | 17.49 |
| 07 | 1 | 959.792 | 0.132 | 959.792 | 17.49 | 13 | 1 | 958.908 | 0.218 | 958.908 | 17.49 |
| 07 | 1 | 960.126 | 0.177 | 960.126 | 17.49 | 13 | 1 | 960.272 | 0.147 | 960.272 | 17.49 |
| 07 | 1 | 960.250 | 0.157 | 960.250 | 17.49 | 13 | 1 | 959.611 | 0.187 | 959.611 | 17.49 |
| 07 | 1 | 960.725 | 0.147 | 960.725 | 17.49 | 13 | 1 | 960.498 | 0.160 | 960.498 | 17.49 |
| 07 | 1 | 961.106 | 0.170 | 961.106 | 17.49 | 13 | 1 | 960.324 | 0.148 | 960.324 | 17.49 |
| 07 | 1 | 961.643 | 0.206 | 961.643 | 17.49 | 13 | 1 | 960.986 | 0.157 | 960.986 | 17.49 |
| 07 | 1 | 959.464 | 0.221 | 959.464 | 17.49 | 13 | 1 | 959.373 | 0.175 | 959.373 | 17.49 |
| 07 | 1 | 959.459 | 0.162 | 959.459 | 17.49 | 13 | 1 | 959.022 | 0.172 | 959.022 | 17.49 |
| 07 | 1 | 957.047 | 0.241 | 957.047 | 17.49 | 13 | 1 | 959.635 | 0.173 | 959.635 | 17.49 |
| 07 | 1 | 957.343 | 0.216 | 957.343 | 17.49 | 13 | 1 | 959.334 | 0.184 | 959.334 | 17.49 |
| 07 | 1 | 959.273 | 0.197 | 959.273 | 17.49 | 13 | 1 | 957.423 | 0.189 | 957.423 | 17.49 |
| 07 | 1 | 958.973 | 0.219 | 958.973 | 17.49 | 13 | 1 | 959.231 | 0.138 | 959.231 | 17.49 |
| 07 | 1 | 960.284 | 0.202 | 960.284 | 17.49 | 13 | 1 | 958.707 | 0.166 | 958.707 | 17.49 |
| 07 | 1 | 958.834 | 0.286 | 958.834 | 17.49 | 13 | 1 | 958.595 | 0.178 | 958.595 | 17.49 |
| 07 | 1 | 958.953 | 0.235 | 958.953 | 17.49 | 13 | 1 | 958.728 | 0.165 | 958.728 | 17.49 |
| 07 | 1 | 957.258 | 0.345 | 957.258 | 17.49 | 14 | 1 | 958.163 | 0.117 | 958.163 | 17.49 |
| 10 | 1 | 958.528 | 0.127 | 958.528 | 17.49 | 14 | 1 | 959.924 | 0.127 | 959.924 | 17.49 |
| 10 | 1 | 959.327 | 0.144 | 959.327 | 17.49 | 14 | 1 | 960.024 | 0.197 | 960.024 | 17.49 |
| 10 | 1 | 959.794 | 0.186 | 959.794 | 17.49 | 14 | 1 | 959.857 | 0.160 | 959.857 | 17.49 |
| 10 | 1 | 960.418 | 0.156 | 960.418 | 17.49 | 14 | 1 | 960.403 | 0.213 | 960.403 | 17.49 |
| 10 | 1 | 959.599 | 0.141 | 959.599 | 17.49 | 14 | 1 | 960.021 | 0.180 | 960.021 | 17.49 |
| 10 | 1 | 959.426 | 0.127 | 959.426 | 17.49 | 14 | 1 | 961.329 | 0.174 | 961.329 | 17.49 |
| 10 | 1 | 959.234 | 0.176 | 959.234 | 17.49 | 14 | 1 | 960.461 | 0.146 | 960.461 | 17.49 |
| 11 | 1 | 959.408 | 0.144 | 959.408 | 17.49 | 14 | 1 | 959.466 | 0.184 | 959.466 | 17.49 |
| 11 | 1 | 959.141 | 0.175 | 959.141 | 17.49 | 14 | 1 | 957.924 | 0.307 | 957.924 | 17.49 |
| 11 | 1 | 959.662 | 0.160 | 959.662 | 17.49 | 14 | 1 | 958.675 | 0.264 | 958.675 | 17.49 |
| 11 | 1 | 960.320 | 0.148 | 960.320 | 17.49 | 14 | 1 | 958.766 | 0.218 | 958.766 | 17.49 |
| 11 | 1 | 960.020 | 0.120 | 960.020 | 17.49 | 14 | 1 | 958.762 | 0.197 | 958.762 | 17.49 |
| 11 | 1 | 960.151 | 0.182 | 960.151 | 17.49 | 14 | 1 | 958.906 | 0.284 | 958.906 | 17.49 |
| 11 | 1 | 959.947 | 0.126 | 959.947 | 17.49 | 14 | 1 | 958.302 | 0.391 | 958.302 | 17.49 |
| 11 | 1 | 959.848 | 0.138 | 959.848 | 17.49 | 14 | 1 | 960.036 | 0.470 | 960.036 | 17.49 |
| 11 | 1 | 960.182 | 0.156 | 960.182 | 17.49 | 17 | 1 | 959.948 | 0.116 | 959.948 | 17.49 |
| 11 | 1 | 959.734 | 0.184 | 959.734 | 17.49 | 17 | 1 | 960.272 | 0.166 | 960.272 | 17.49 |
| 11 | 1 | 960.030 | 0.148 | 960.030 | 17.49 | 17 | 1 | 960.517 | 0.163 | 960.517 | 17.49 |
| 11 | 1 | 959.074 | 0.143 | 959.074 | 17.49 | 17 | 1 | 960.200 | 0.149 | 960.200 | 17.49 |
| 11 | 1 | 958.501 | 0.179 | 958.501 | 17.49 | 17 | 1 | 959.236 | 0.130 | 959.236 | 17.49 |
| 11 | 1 | 959.626 | 0.132 | 959.626 | 17.49 | 17 | 1 | 960.549 | 0.127 | 960.549 | 17.49 |
| 11 | 1 | 958.997 | 0.154 | 958.997 | 17.49 | 18 | 1 | 958.756 | 0.148 | 958.756 | 17.49 |
| 11 | 1 | 957.837 | 0.129 | 957.837 | 17.49 | 18 | 1 | 959.746 | 0.131 | 959.746 | 17.49 |
| 11 | 1 | 957.785 | 0.125 | 957.785 | 17.49 | 18 | 1 | 959.309 | 0.123 | 959.309 | 17.49 |
| 11 | 1 | 959.470 | 0.152 | 959.470 | 17.49 | 18 | 1 | 959.604 | 0.130 | 959.604 | 17.49 |
| 11 | 1 | 958.937 | 0.203 | 958.937 | 17.49 | 18 | 1 | 960.459 | 0.142 | 960.459 | 17.49 |
| 11 | 1 | 959.187 | 0.155 | 959.187 | 17.49 | 18 | 1 | 959.590 | 0.116 | 959.590 | 17.49 |
| 11 | 1 | 957.380 | 0.226 | 957.380 | 17.49 | 18 | 1 | 959.659 | 0.150 | 959.659 | 17.49 |
| 11 | 1 | 957.349 | 0.216 | 957.349 | 17.49 | 18 | 1 | 960.261 | 0.121 | 960.261 | 17.49 |

| 2006 - JULHO | | | | | | 2006 - JULHO | | | | |
|---|---|---|---|---|---|---|---|---|---|---|
| D | L | SDB | ER | SDC | HL | D | L | SDB | ER | SDC | HL |
| 18 | 1 | 959.307 | 0.186 | 959.307 | 17.49 | 21 | 1 | 959.470 | 0.175 | 959.470 | 17.49 |
| 18 | 1 | 959.540 | 0.143 | 959.540 | 17.49 | 21 | 1 | 959.056 | 0.161 | 959.056 | 17.49 |
| 18 | 1 | 960.071 | 0.145 | 960.071 | 17.49 | 21 | 1 | 959.363 | 0.216 | 959.363 | 17.49 |
| 18 | 1 | 959.040 | 0.147 | 959.040 | 17.49 | 21 | 1 | 959.564 | 0.172 | 959.564 | 17.49 |
| 18 | 1 | 958.105 | 0.189 | 958.105 | 17.49 | 21 | 1 | 958.043 | 0.215 | 958.043 | 17.49 |
| 18 | 1 | 958.758 | 0.208 | 958.758 | 17.49 | 21 | 1 | 959.313 | 0.211 | 959.313 | 17.49 |
| 18 | 1 | 959.544 | 0.178 | 959.544 | 17.49 | 24 | 1 | 959.172 | 0.157 | 959.172 | 17.49 |
| 18 | 1 | 959.153 | 0.170 | 959.153 | 17.49 | 24 | 1 | 959.969 | 0.179 | 959.969 | 17.49 |
| 18 | 1 | 958.628 | 0.217 | 958.628 | 17.49 | 24 | 1 | 959.593 | 0.148 | 959.593 | 17.49 |
| 18 | 1 | 958.752 | 0.204 | 958.752 | 17.49 | 24 | 1 | 958.393 | 0.171 | 958.393 | 17.49 |
| 18 | 1 | 957.362 | 0.217 | 957.362 | 17.49 | 24 | 1 | 959.521 | 0.150 | 959.521 | 17.49 |
| 18 | 1 | 957.800 | 0.232 | 957.800 | 17.49 | 24 | 1 | 959.652 | 0.144 | 959.652 | 17.49 |
| 18 | 1 | 959.228 | 0.214 | 959.228 | 17.49 | 24 | 1 | 959.811 | 0.147 | 959.811 | 17.49 |
| 18 | 1 | 958.317 | 0.278 | 958.317 | 17.49 | 24 | 1 | 960.238 | 0.155 | 960.238 | 17.49 |
| 18 | 1 | 958.077 | 0.368 | 958.077 | 17.49 | 24 | 1 | 960.251 | 0.155 | 960.251 | 17.49 |
| 19 | 1 | 959.314 | 0.153 | 959.314 | 17.49 | 24 | 1 | 960.761 | 0.144 | 960.761 | 17.49 |
| 19 | 1 | 958.951 | 0.126 | 958.951 | 17.49 | 24 | 1 | 959.115 | 0.214 | 959.115 | 17.49 |
| 19 | 1 | 960.268 | 0.143 | 960.268 | 17.49 | 24 | 1 | 958.749 | 0.202 | 958.749 | 17.49 |
| 19 | 1 | 959.498 | 0.189 | 959.498 | 17.49 | 24 | 1 | 958.136 | 0.181 | 958.136 | 17.49 |
| 19 | 1 | 959.622 | 0.152 | 959.622 | 17.49 | 24 | 1 | 959.858 | 0.149 | 959.858 | 17.49 |
| 19 | 1 | 959.086 | 0.148 | 959.086 | 17.49 | 24 | 1 | 958.997 | 0.165 | 958.997 | 17.49 |
| 19 | 1 | 960.414 | 0.166 | 960.414 | 17.49 | 24 | 1 | 958.536 | 0.178 | 958.536 | 17.49 |
| 19 | 1 | 960.567 | 0.132 | 960.567 | 17.49 | 24 | 1 | 959.092 | 0.159 | 959.092 | 17.49 |
| 19 | 1 | 960.030 | 0.186 | 960.030 | 17.49 | 24 | 1 | 959.705 | 0.204 | 959.705 | 17.49 |
| 19 | 1 | 961.355 | 0.163 | 961.355 | 17.49 | 24 | 1 | 960.032 | 0.180 | 960.032 | 17.49 |
| 19 | 1 | 958.844 | 0.178 | 958.844 | 17.49 | 24 | 1 | 959.097 | 0.159 | 959.097 | 17.49 |
| 19 | 1 | 958.562 | 0.224 | 958.562 | 17.49 | 24 | 1 | 959.916 | 0.207 | 959.916 | 17.49 |
| 19 | 1 | 959.888 | 0.177 | 959.888 | 17.49 | 24 | 1 | 959.229 | 0.250 | 959.229 | 17.49 |
| 19 | 1 | 958.543 | 0.179 | 958.543 | 17.49 | 24 | 1 | 959.503 | 0.208 | 959.503 | 17.49 |
| 19 | 1 | 959.502 | 0.170 | 959.502 | 17.49 | 24 | 1 | 959.566 | 0.230 | 959.566 | 17.49 |
| 19 | 1 | 959.412 | 0.183 | 959.412 | 17.49 | 25 | 1 | 958.349 | 0.150 | 958.349 | 17.49 |
| 19 | 1 | 958.341 | 0.202 | 958.341 | 17.49 | 25 | 1 | 959.222 | 0.172 | 959.222 | 17.49 |
| 19 | 1 | 958.684 | 0.183 | 958.684 | 17.49 | 25 | 1 | 959.407 | 0.177 | 959.407 | 17.49 |
| 19 | 1 | 958.991 | 0.273 | 958.991 | 17.49 | 25 | 1 | 959.635 | 0.138 | 959.635 | 17.49 |
| 19 | 1 | 959.368 | 0.261 | 959.368 | 17.49 | 25 | 1 | 960.175 | 0.178 | 960.175 | 17.49 |
| 20 | 1 | 959.832 | 0.130 | 959.832 | 17.49 | 25 | 1 | 960.816 | 0.165 | 960.816 | 17.49 |
| 20 | 1 | 958.888 | 0.114 | 958.888 | 17.49 | 25 | 1 | 959.566 | 0.172 | 959.566 | 17.49 |
| 20 | 1 | 959.843 | 0.159 | 959.843 | 17.49 | 25 | 1 | 960.268 | 0.152 | 960.268 | 17.49 |
| 20 | 1 | 959.565 | 0.138 | 959.565 | 17.49 | 25 | 1 | 958.738 | 0.169 | 958.738 | 17.49 |
| 20 | 1 | 959.675 | 0.136 | 959.675 | 17.49 | 25 | 1 | 960.396 | 0.229 | 960.396 | 17.49 |
| 20 | 1 | 959.254 | 0.121 | 959.254 | 17.49 | 25 | 1 | 959.945 | 0.197 | 959.945 | 17.49 |
| 20 | 1 | 960.442 | 0.135 | 960.442 | 17.49 | 25 | 1 | 959.441 | 0.132 | 959.441 | 17.49 |
| 20 | 1 | 959.833 | 0.174 | 959.833 | 17.49 | 25 | 1 | 959.354 | 0.150 | 959.354 | 17.49 |
| 20 | 1 | 960.521 | 0.209 | 960.521 | 17.49 | 25 | 1 | 958.051 | 0.168 | 958.051 | 17.49 |
| 20 | 1 | 960.017 | 0.153 | 960.017 | 17.49 | 25 | 1 | 959.845 | 0.144 | 959.845 | 17.49 |
| 20 | 1 | 960.072 | 0.129 | 960.072 | 17.49 | 25 | 1 | 958.659 | 0.164 | 958.659 | 17.49 |
| 20 | 1 | 958.446 | 0.205 | 958.446 | 17.49 | 25 | 1 | 959.684 | 0.133 | 959.684 | 17.49 |
| 20 | 1 | 959.556 | 0.159 | 959.556 | 17.49 | 25 | 1 | 959.506 | 0.149 | 959.506 | 17.49 |
| 20 | 1 | 958.714 | 0.243 | 958.714 | 17.49 | 25 | 1 | 959.086 | 0.185 | 959.086 | 17.49 |
| 20 | 1 | 959.332 | 0.167 | 959.332 | 17.49 | 25 | 1 | 959.601 | 0.247 | 959.601 | 17.49 |
| 20 | 1 | 959.818 | 0.159 | 959.818 | 17.49 | 25 | 1 | 959.160 | 0.229 | 959.160 | 17.49 |
| 20 | 1 | 958.646 | 0.177 | 958.646 | 17.49 | 25 | 1 | 959.104 | 0.264 | 959.104 | 17.49 |
| 20 | 1 | 959.271 | 0.185 | 959.271 | 17.49 | 25 | 1 | 959.619 | 0.242 | 959.619 | 17.49 |
| 20 | 1 | 959.552 | 0.194 | 959.552 | 17.49 | 26 | 1 | 958.461 | 0.169 | 958.461 | 17.49 |
| 20 | 1 | 959.828 | 0.233 | 959.828 | 17.49 | 26 | 1 | 959.563 | 0.119 | 959.563 | 17.49 |
| 20 | 1 | 958.871 | 0.215 | 958.871 | 17.49 | 26 | 1 | 958.817 | 0.126 | 958.817 | 17.49 |
| 20 | 1 | 959.142 | 0.255 | 959.142 | 17.49 | 26 | 1 | 961.013 | 0.189 | 961.013 | 17.49 |
| 21 | 1 | 958.721 | 0.184 | 958.721 | 17.49 | 26 | 1 | 958.992 | 0.163 | 958.992 | 17.49 |
| 21 | 1 | 957.324 | 0.211 | 957.324 | 17.49 | 26 | 1 | 957.926 | 0.140 | 957.926 | 17.49 |
| 21 | 1 | 959.556 | 0.207 | 959.556 | 17.49 | 26 | 1 | 959.618 | 0.140 | 959.618 | 17.49 |
| 21 | 1 | 959.901 | 0.201 | 959.901 | 17.49 | 26 | 1 | 958.945 | 0.165 | 958.945 | 17.49 |
| 21 | 1 | 958.914 | 0.170 | 958.914 | 17.49 | 26 | 1 | 959.082 | 0.169 | 959.082 | 17.49 |
| 21 | 1 | 958.763 | 0.192 | 958.763 | 17.49 | 26 | 1 | 958.881 | 0.152 | 958.881 | 17.49 |
| 21 | 1 | 960.008 | 0.181 | 960.008 | 17.49 | 26 | 1 | 959.997 | 0.152 | 959.997 | 17.49 |
| 21 | 1 | 958.835 | 0.149 | 958.835 | 17.49 | 26 | 1 | 959.181 | 0.197 | 959.181 | 17.49 |
| 21 | 1 | 960.345 | 0.168 | 960.345 | 17.49 | 26 | 1 | 958.822 | 0.159 | 958.822 | 17.49 |
| 21 | 1 | 960.209 | 0.171 | 960.209 | 17.49 | 27 | 1 | 959.606 | 0.154 | 959.606 | 17.49 |
| 21 | 1 | 959.551 | 0.164 | 959.551 | 17.49 | 27 | 1 | 958.934 | 0.143 | 958.934 | 17.49 |
| 21 | 1 | 961.979 | 0.193 | 961.979 | 17.49 | 27 | 1 | 959.354 | 0.162 | 959.354 | 17.49 |
| 21 | 1 | 959.679 | 0.172 | 959.679 | 17.49 | 27 | 1 | 958.616 | 0.263 | 958.616 | 17.49 |
| 21 | 1 | 958.938 | 0.161 | 958.938 | 17.49 | 27 | 1 | 959.474 | 0.131 | 959.474 | 17.49 |

| 2006 - JULHO | | | | | | | 2006 - AGOSTO | | | | |
|---|---|---|---|---|---|---|---|---|---|---|---|
| D | L | SDB | ER | SDC | HL | D | L | SDB | ER | SDC | HL |
| 27 | 1 | 959.170 | 0.164 | 959.170 | 17.49 | 11 | 1 | 958.840 | 0.121 | 958.840 | 17.49 |
| 27 | 1 | 960.035 | 0.153 | 960.035 | 17.49 | 11 | 1 | 959.460 | 0.127 | 959.460 | 17.49 |
| 27 | 1 | 959.096 | 0.174 | 959.096 | 17.49 | 11 | 1 | 959.480 | 0.162 | 959.480 | 17.49 |
| 27 | 1 | 960.136 | 0.174 | 960.136 | 17.49 | 11 | 1 | 959.922 | 0.134 | 959.922 | 17.49 |
| 27 | 1 | 959.254 | 0.159 | 959.254 | 17.49 | 11 | 1 | 960.123 | 0.190 | 960.123 | 17.49 |
| 27 | 1 | 959.497 | 0.162 | 959.497 | 17.49 | 11 | 1 | 960.032 | 0.148 | 960.032 | 17.49 |
| 27 | 1 | 958.913 | 0.149 | 958.913 | 17.49 | 11 | 1 | 960.531 | 0.261 | 960.531 | 17.49 |
| 27 | 1 | 958.092 | 0.180 | 958.092 | 17.49 | 11 | 1 | 959.303 | 0.180 | 959.303 | 17.49 |
| 27 | 1 | 959.570 | 0.208 | 959.570 | 17.49 | 11 | 1 | 960.004 | 0.152 | 960.004 | 17.49 |
| 27 | 1 | 958.970 | 0.199 | 958.970 | 17.49 | 11 | 1 | 959.581 | 0.248 | 959.581 | 17.49 |
| 27 | 1 | 959.235 | 0.191 | 959.235 | 17.49 | 11 | 1 | 959.493 | 0.207 | 959.493 | 17.49 |
| 27 | 1 | 958.614 | 0.213 | 958.614 | 17.49 | 11 | 1 | 960.184 | 0.202 | 960.184 | 17.49 |
| 28 | 1 | 959.715 | 0.159 | 959.715 | 17.49 | 11 | 1 | 958.187 | 0.240 | 958.187 | 17.49 |
| 28 | 1 | 958.552 | 0.184 | 958.552 | 17.49 | 11 | 1 | 957.193 | 0.254 | 957.193 | 17.49 |
| 28 | 1 | 959.076 | 0.098 | 959.076 | 17.49 | 14 | 1 | 959.569 | 0.111 | 959.569 | 17.49 |
| 28 | 1 | 959.693 | 0.107 | 959.693 | 17.49 | 14 | 1 | 959.660 | 0.118 | 959.660 | 17.49 |
| 28 | 1 | 959.444 | 0.113 | 959.444 | 17.49 | 14 | 1 | 959.684 | 0.113 | 959.684 | 17.49 |
| 28 | 1 | 959.577 | 0.138 | 959.577 | 17.49 | 14 | 1 | 958.972 | 0.124 | 958.972 | 17.49 |
| 28 | 1 | 958.794 | 0.106 | 958.794 | 17.49 | 14 | 1 | 959.750 | 0.094 | 959.750 | 17.49 |
| 28 | 1 | 959.146 | 0.123 | 959.146 | 17.49 | 14 | 1 | 959.391 | 0.142 | 959.391 | 17.49 |
| 28 | 1 | 959.727 | 0.112 | 959.727 | 17.49 | 14 | 1 | 959.019 | 0.133 | 959.019 | 17.49 |
| 28 | 1 | 959.448 | 0.108 | 959.448 | 17.49 | 14 | 1 | 959.370 | 0.152 | 959.370 | 17.49 |
| 28 | 1 | 959.508 | 0.127 | 959.508 | 17.49 | 14 | 1 | 959.188 | 0.159 | 959.188 | 17.49 |
| | | | | | | 14 | 1 | 959.711 | 0.158 | 959.711 | 17.49 |
| | | | | | | 14 | 1 | 960.067 | 0.170 | 960.067 | 17.49 |
| | | 2006 - AGOSTO | | | | 14 | 1 | 959.670 | 0.139 | 959.670 | 17.49 |
| D | L | SDB | ER | SDC | HL | 14 | 1 | 959.720 | 0.168 | 959.720 | 17.49 |
| 08 | 1 | 959.400 | 0.108 | 959.400 | 17.49 | 14 | 1 | 958.727 | 0.188 | 958.727 | 17.49 |
| 08 | 1 | 959.933 | 0.112 | 959.933 | 17.49 | 14 | 1 | 958.832 | 0.197 | 958.832 | 17.49 |
| 08 | 1 | 959.401 | 0.123 | 959.401 | 17.49 | 14 | 1 | 957.743 | 0.277 | 957.743 | 17.49 |
| 08 | 1 | 959.478 | 0.121 | 959.478 | 17.49 | 14 | 1 | 957.351 | 0.226 | 957.351 | 17.49 |
| 08 | 1 | 960.001 | 0.163 | 960.001 | 17.49 | 14 | 1 | 957.969 | 0.260 | 957.969 | 17.49 |
| 08 | 1 | 958.987 | 0.160 | 958.987 | 17.49 | 14 | 1 | 959.427 | 0.162 | 959.427 | 17.49 |
| 08 | 1 | 959.482 | 0.149 | 959.482 | 17.49 | 14 | 1 | 958.413 | 0.299 | 958.413 | 17.49 |
| 08 | 1 | 960.601 | 0.137 | 960.601 | 17.49 | 14 | 1 | 959.002 | 0.174 | 959.002 | 17.49 |
| 08 | 1 | 958.905 | 0.185 | 958.905 | 17.49 | 14 | 1 | 959.537 | 0.192 | 959.537 | 17.49 |
| 08 | 1 | 960.387 | 0.169 | 960.387 | 17.49 | 14 | 1 | 958.470 | 0.199 | 958.470 | 17.49 |
| 08 | 1 | 958.876 | 0.177 | 958.876 | 17.49 | 14 | 1 | 958.369 | 0.172 | 958.369 | 17.49 |
| 08 | 1 | 958.455 | 0.253 | 958.455 | 17.49 | 14 | 1 | 958.688 | 0.258 | 958.688 | 17.49 |
| 08 | 1 | 959.813 | 0.135 | 959.813 | 17.49 | 14 | 1 | 958.028 | 0.220 | 958.028 | 17.49 |
| 08 | 1 | 958.825 | 0.292 | 958.825 | 17.49 | 14 | 1 | 958.520 | 0.192 | 958.520 | 17.49 |
| 08 | 1 | 958.558 | 0.214 | 958.558 | 17.49 | 14 | 1 | 958.317 | 0.268 | 958.317 | 17.49 |
| 08 | 1 | 959.017 | 0.186 | 959.017 | 17.49 | 14 | 1 | 958.143 | 0.219 | 958.143 | 17.49 |
| 08 | 1 | 959.688 | 0.196 | 959.688 | 17.49 | 15 | 1 | 958.764 | 0.147 | 958.764 | 17.49 |
| 08 | 1 | 959.164 | 0.213 | 959.164 | 17.49 | 15 | 1 | 959.014 | 0.142 | 959.014 | 17.49 |
| 08 | 1 | 959.053 | 0.197 | 959.053 | 17.49 | 15 | 1 | 959.185 | 0.134 | 959.185 | 17.49 |
| 08 | 1 | 959.847 | 0.220 | 959.847 | 17.49 | 15 | 1 | 959.150 | 0.213 | 959.150 | 17.49 |
| 08 | 1 | 959.834 | 0.264 | 959.834 | 17.49 | 15 | 1 | 958.919 | 0.199 | 958.919 | 17.49 |
| 08 | 1 | 958.625 | 0.279 | 958.625 | 17.49 | 15 | 1 | 959.723 | 0.200 | 959.723 | 17.49 |
| 09 | 1 | 958.089 | 0.128 | 958.089 | 17.49 | 15 | 1 | 959.242 | 0.210 | 959.242 | 17.49 |
| 09 | 1 | 959.440 | 0.111 | 959.440 | 17.49 | 15 | 1 | 958.818 | 0.248 | 958.818 | 17.49 |
| 09 | 1 | 960.320 | 0.097 | 960.320 | 17.49 | 15 | 1 | 958.670 | 0.149 | 958.670 | 17.49 |
| 09 | 1 | 959.487 | 0.136 | 959.487 | 17.49 | 15 | 1 | 959.514 | 0.176 | 959.514 | 17.49 |
| 09 | 1 | 959.884 | 0.161 | 959.884 | 17.49 | 15 | 1 | 958.845 | 0.391 | 958.845 | 17.49 |
| 09 | 1 | 959.779 | 0.163 | 959.779 | 17.49 | 15 | 1 | 958.947 | 0.225 | 958.947 | 17.49 |
| 09 | 1 | 959.902 | 0.166 | 959.902 | 17.49 | 15 | 1 | 957.953 | 0.221 | 957.953 | 17.49 |
| 09 | 1 | 961.629 | 0.170 | 961.629 | 17.49 | 15 | 1 | 958.835 | 0.228 | 958.835 | 17.49 |
| 09 | 1 | 959.398 | 0.206 | 959.398 | 17.49 | 15 | 1 | 959.395 | 0.293 | 959.395 | 17.49 |
| 09 | 1 | 959.012 | 0.145 | 959.012 | 17.49 | 15 | 1 | 957.313 | 0.273 | 957.313 | 17.49 |
| 09 | 1 | 958.572 | 0.125 | 958.572 | 17.49 | 15 | 1 | 958.123 | 0.202 | 958.123 | 17.49 |
| 09 | 1 | 959.541 | 0.156 | 959.541 | 17.49 | 16 | 1 | 959.881 | 0.183 | 959.881 | 17.49 |
| 09 | 1 | 959.403 | 0.197 | 959.403 | 17.49 | 16 | 1 | 959.290 | 0.137 | 959.290 | 17.49 |
| 09 | 1 | 959.208 | 0.144 | 959.208 | 17.49 | 16 | 1 | 959.761 | 0.137 | 959.761 | 17.49 |
| 09 | 1 | 958.664 | 0.131 | 958.664 | 17.49 | 16 | 1 | 959.182 | 0.169 | 959.182 | 17.49 |
| 09 | 1 | 960.054 | 0.193 | 960.054 | 17.49 | 16 | 1 | 958.546 | 0.173 | 958.546 | 17.49 |
| 09 | 1 | 959.843 | 0.235 | 959.843 | 17.49 | 16 | 1 | 959.031 | 0.176 | 959.031 | 17.49 |
| 09 | 1 | 959.648 | 0.205 | 959.648 | 17.49 | 16 | 1 | 958.636 | 0.184 | 958.636 | 17.49 |
| 11 | 1 | 959.353 | 0.118 | 959.353 | 17.49 | 16 | 1 | 959.158 | 0.175 | 959.158 | 17.49 |
| 11 | 1 | 959.113 | 0.118 | 959.113 | 17.49 | 16 | 1 | 959.006 | 0.272 | 959.006 | 17.49 |
| 11 | 1 | 959.116 | 0.141 | 959.116 | 17.49 | 16 | 1 | 959.849 | 0.230 | 959.849 | 17.49 |
| 11 | 1 | 959.285 | 0.112 | 959.285 | 17.49 | 16 | 1 | 958.724 | 0.170 | 958.724 | 17.49 |

| 2006 - AGOSTO | | | | | |
|---|---|---|---|---|---|
| D | L | SDB | ER | SDC | HL |
| 16 | 1 | 958.445 | 0.199 | 958.445 | 17.49 |
| 16 | 1 | 958.345 | 0.202 | 958.345 | 17.49 |
| 16 | 1 | 959.086 | 0.153 | 959.086 | 17.49 |
| 16 | 1 | 959.015 | 0.183 | 959.015 | 17.49 |
| 16 | 1 | 958.076 | 0.394 | 958.076 | 17.49 |
| 17 | 1 | 959.167 | 0.163 | 959.167 | 17.49 |
| 17 | 1 | 958.313 | 0.158 | 958.313 | 17.49 |
| 17 | 1 | 959.128 | 0.128 | 959.128 | 17.49 |
| 17 | 1 | 959.053 | 0.156 | 959.053 | 17.49 |
| 17 | 1 | 958.827 | 0.142 | 958.827 | 17.49 |
| 17 | 1 | 959.064 | 0.142 | 959.064 | 17.49 |
| 17 | 1 | 959.408 | 0.137 | 959.408 | 17.49 |
| 17 | 1 | 959.368 | 0.155 | 959.368 | 17.49 |
| 17 | 1 | 958.759 | 0.255 | 958.759 | 17.49 |
| 23 | 1 | 958.639 | 0.183 | 958.639 | 17.49 |
| 23 | 1 | 958.679 | 0.140 | 958.679 | 17.49 |
| 23 | 1 | 959.492 | 0.142 | 959.492 | 17.49 |
| 23 | 1 | 958.720 | 0.158 | 958.720 | 17.49 |
| 23 | 1 | 959.829 | 0.155 | 959.829 | 17.49 |
| 23 | 1 | 958.980 | 0.158 | 958.980 | 17.49 |
| 23 | 1 | 960.018 | 0.147 | 960.018 | 17.49 |
| 23 | 1 | 960.011 | 0.165 | 960.011 | 17.49 |
| 23 | 1 | 959.897 | 0.150 | 959.897 | 17.49 |
| 23 | 1 | 959.282 | 0.160 | 959.282 | 17.49 |
| 23 | 1 | 958.741 | 0.174 | 958.741 | 17.49 |
| 23 | 1 | 959.764 | 0.201 | 959.764 | 17.49 |
| 23 | 1 | 958.613 | 0.190 | 958.613 | 17.49 |
| 23 | 1 | 959.253 | 0.143 | 959.253 | 17.49 |
| 23 | 1 | 958.659 | 0.207 | 958.659 | 17.49 |
| 23 | 1 | 958.897 | 0.206 | 958.897 | 17.49 |
| 23 | 1 | 958.998 | 0.142 | 958.998 | 17.49 |
| 23 | 1 | 959.352 | 0.203 | 959.352 | 17.49 |
| 23 | 1 | 957.876 | 0.199 | 957.876 | 17.49 |
| 23 | 1 | 957.450 | 0.256 | 957.450 | 17.49 |
| 23 | 1 | 958.940 | 0.243 | 958.940 | 17.49 |
| 25 | 1 | 959.072 | 0.203 | 959.072 | 17.49 |
| 25 | 1 | 958.759 | 0.171 | 958.759 | 17.49 |
| 25 | 1 | 958.941 | 0.174 | 958.941 | 17.49 |
| 25 | 1 | 959.128 | 0.165 | 959.128 | 17.49 |
| 25 | 1 | 958.847 | 0.156 | 958.847 | 17.49 |
| 25 | 1 | 959.160 | 0.153 | 959.160 | 17.49 |
| 25 | 1 | 960.039 | 0.189 | 960.039 | 17.49 |
| 25 | 1 | 959.789 | 0.166 | 959.789 | 17.49 |
| 25 | 1 | 958.542 | 0.168 | 958.542 | 17.49 |
| 25 | 1 | 959.710 | 0.322 | 959.710 | 17.49 |
| 25 | 1 | 959.809 | 0.285 | 959.809 | 17.49 |
| 25 | 1 | 959.172 | 0.267 | 959.172 | 17.49 |
| 25 | 1 | 961.359 | 0.143 | 961.359 | 17.49 |
| 25 | 1 | 959.175 | 0.147 | 959.175 | 17.49 |
| 25 | 1 | 957.746 | 0.211 | 957.746 | 17.49 |
| 25 | 1 | 957.759 | 0.285 | 957.759 | 17.49 |
| 25 | 1 | 957.233 | 0.326 | 957.233 | 17.49 |
| 25 | 1 | 957.173 | 0.459 | 957.173 | 17.49 |
| 25 | 1 | 960.529 | 0.347 | 960.529 | 17.49 |
| 25 | 1 | 960.039 | 0.245 | 960.039 | 17.49 |
| 25 | 1 | 959.179 | 0.206 | 959.179 | 17.49 |
| 31 | 1 | 959.370 | 0.194 | 959.370 | 17.49 |
| 31 | 1 | 958.214 | 0.159 | 958.214 | 17.49 |
| 31 | 1 | 958.692 | 0.157 | 958.692 | 17.49 |
| 31 | 1 | 958.787 | 0.172 | 958.787 | 17.49 |
| 31 | 1 | 959.360 | 0.190 | 959.360 | 17.49 |
| 31 | 1 | 959.063 | 0.167 | 959.063 | 17.49 |
| 31 | 1 | 960.734 | 0.176 | 960.734 | 17.49 |
| 31 | 1 | 959.496 | 0.176 | 959.496 | 17.49 |
| 31 | 1 | 959.469 | 0.176 | 959.469 | 17.49 |
| 31 | 1 | 959.505 | 0.203 | 959.505 | 17.49 |
| 31 | 1 | 957.880 | 0.178 | 957.880 | 17.49 |
| 31 | 1 | 959.674 | 0.142 | 959.674 | 17.49 |
| 31 | 1 | 959.191 | 0.197 | 959.191 | 17.49 |
| 31 | 1 | 959.833 | 0.183 | 959.833 | 17.49 |
| 31 | 1 | 958.833 | 0.175 | 958.833 | 17.49 |

| 2006 - AGOSTO | | | | | |
|---|---|---|---|---|---|
| D | L | SDB | ER | SDC | HL |
| 31 | 1 | 958.685 | 0.231 | 958.685 | 17.49 |
| 31 | 1 | 958.573 | 0.248 | 958.573 | 17.49 |
| 31 | 1 | 959.234 | 0.286 | 959.234 | 17.49 |
| 31 | 1 | 958.253 | 0.243 | 958.253 | 17.49 |
| 31 | 1 | 958.560 | 0.222 | 958.560 | 17.49 |
| 31 | 1 | 958.574 | 0.209 | 958.574 | 17.49 |
| 31 | 1 | 958.717 | 0.240 | 958.717 | 17.49 |
| 31 | 1 | 959.072 | 0.149 | 959.072 | 17.49 |

| 2006 - SETEMBRO | | | | | |
|---|---|---|---|---|---|
| D | L | SDB | ER | SDC | HL |
| 11 | 1 | 958.833 | 0.138 | 958.833 | 17.49 |
| 11 | 1 | 959.217 | 0.124 | 959.217 | 17.49 |
| 11 | 1 | 958.323 | 0.108 | 958.323 | 17.49 |
| 11 | 1 | 959.615 | 0.129 | 959.615 | 17.49 |
| 11 | 1 | 959.726 | 0.158 | 959.726 | 17.49 |
| 11 | 1 | 959.930 | 0.129 | 959.930 | 17.49 |
| 11 | 1 | 959.901 | 0.120 | 959.901 | 17.49 |
| 11 | 1 | 958.842 | 0.145 | 958.842 | 17.49 |
| 11 | 1 | 960.207 | 0.183 | 960.207 | 17.49 |
| 11 | 1 | 959.259 | 0.180 | 959.259 | 17.49 |
| 11 | 1 | 958.713 | 0.342 | 958.713 | 17.49 |
| 11 | 1 | 959.665 | 0.251 | 959.665 | 17.49 |
| 11 | 1 | 959.456 | 0.166 | 959.456 | 17.49 |
| 11 | 1 | 959.266 | 0.135 | 959.266 | 17.49 |
| 11 | 1 | 958.324 | 0.192 | 958.324 | 17.49 |
| 11 | 1 | 958.949 | 0.142 | 958.949 | 17.49 |
| 11 | 1 | 959.119 | 0.139 | 959.119 | 17.49 |
| 11 | 1 | 958.326 | 0.178 | 958.326 | 17.49 |
| 11 | 1 | 957.995 | 0.169 | 957.995 | 17.49 |
| 11 | 1 | 959.352 | 0.144 | 959.352 | 17.49 |
| 11 | 1 | 958.338 | 0.154 | 958.338 | 17.49 |
| 11 | 1 | 958.596 | 0.144 | 958.596 | 17.49 |
| 11 | 1 | 960.114 | 0.172 | 960.114 | 17.49 |
| 11 | 1 | 958.694 | 0.189 | 958.694 | 17.49 |
| 12 | 1 | 957.828 | 0.208 | 957.828 | 17.49 |
| 12 | 1 | 959.954 | 0.135 | 959.954 | 17.49 |
| 12 | 1 | 958.694 | 0.186 | 958.694 | 17.49 |
| 12 | 1 | 958.323 | 0.156 | 958.323 | 17.49 |
| 12 | 1 | 959.521 | 0.170 | 959.521 | 17.49 |
| 12 | 1 | 959.260 | 0.118 | 959.260 | 17.49 |
| 12 | 1 | 960.062 | 0.161 | 960.062 | 17.49 |
| 12 | 1 | 959.421 | 0.181 | 959.421 | 17.49 |
| 12 | 1 | 959.628 | 0.146 | 959.628 | 17.49 |
| 12 | 1 | 959.210 | 0.152 | 959.210 | 17.49 |
| 12 | 1 | 958.993 | 0.167 | 958.993 | 17.49 |
| 12 | 1 | 958.008 | 0.191 | 958.008 | 17.49 |
| 12 | 1 | 959.466 | 0.145 | 959.466 | 17.49 |
| 12 | 1 | 957.556 | 0.199 | 957.556 | 17.49 |
| 12 | 1 | 959.033 | 0.197 | 959.033 | 17.49 |
| 12 | 1 | 959.565 | 0.183 | 959.565 | 17.49 |
| 13 | 1 | 960.091 | 0.130 | 960.091 | 17.49 |
| 13 | 1 | 959.129 | 0.131 | 959.129 | 17.49 |
| 13 | 1 | 959.565 | 0.116 | 959.565 | 17.49 |
| 13 | 1 | 958.601 | 0.124 | 958.601 | 17.49 |
| 13 | 1 | 959.792 | 0.099 | 959.792 | 17.49 |
| 13 | 1 | 959.656 | 0.140 | 959.656 | 17.49 |
| 13 | 1 | 959.226 | 0.125 | 959.226 | 17.49 |
| 13 | 1 | 958.994 | 0.101 | 958.994 | 17.49 |
| 13 | 1 | 959.873 | 0.147 | 959.873 | 17.49 |
| 13 | 1 | 959.414 | 0.130 | 959.414 | 17.49 |
| 13 | 1 | 958.986 | 0.128 | 958.986 | 17.49 |
| 13 | 1 | 959.496 | 0.140 | 959.496 | 17.49 |
| 13 | 1 | 958.875 | 0.151 | 958.875 | 17.49 |
| 13 | 1 | 957.741 | 0.198 | 957.741 | 17.49 |
| 13 | 1 | 958.599 | 0.165 | 958.599 | 17.49 |
| 13 | 1 | 959.189 | 0.127 | 959.189 | 17.49 |
| 13 | 1 | 960.075 | 0.179 | 960.075 | 17.49 |
| 13 | 1 | 957.428 | 0.152 | 957.428 | 17.49 |
| 14 | 1 | 957.494 | 0.126 | 957.494 | 17.49 |

| 2006 - SETEMBRO | | | | | |
|---|---|---|---|---|---|
| D | L | SDB | ER | SDC | HL |
| 14 | 1 | 960.086 | 0.127 | 960.086 | 17.49 |
| 14 | 1 | 958.490 | 0.121 | 958.490 | 17.49 |
| 14 | 1 | 960.303 | 0.170 | 960.303 | 17.49 |
| 14 | 1 | 959.642 | 0.122 | 959.642 | 17.49 |
| 14 | 1 | 959.749 | 0.152 | 959.749 | 17.49 |
| 14 | 1 | 961.296 | 0.209 | 961.296 | 17.49 |
| 14 | 1 | 959.454 | 0.200 | 959.454 | 17.49 |
| 14 | 1 | 959.012 | 0.180 | 959.012 | 17.49 |
| 14 | 1 | 958.012 | 0.225 | 958.012 | 17.49 |
| 14 | 1 | 958.452 | 0.187 | 958.452 | 17.49 |
| 14 | 1 | 958.741 | 0.218 | 958.741 | 17.49 |
| 14 | 1 | 959.460 | 0.293 | 959.460 | 17.49 |
| 14 | 1 | 959.290 | 0.167 | 959.290 | 17.49 |
| 14 | 1 | 957.862 | 0.211 | 957.862 | 17.49 |
| 21 | 1 | 958.121 | 0.099 | 958.121 | 17.49 |
| 21 | 1 | 960.380 | 0.113 | 960.380 | 17.49 |
| 21 | 1 | 960.569 | 0.121 | 960.569 | 17.49 |
| 21 | 1 | 957.824 | 0.134 | 957.824 | 17.49 |
| 21 | 1 | 959.879 | 0.089 | 959.879 | 17.49 |
| 21 | 1 | 958.823 | 0.120 | 958.823 | 17.49 |
| 21 | 1 | 959.143 | 0.117 | 959.143 | 17.49 |
| 28 | 1 | 959.986 | 0.161 | 959.986 | 17.49 |
| 28 | 1 | 960.388 | 0.119 | 960.388 | 17.49 |
| 28 | 1 | 958.262 | 0.158 | 958.262 | 17.49 |
| 28 | 1 | 961.516 | 0.165 | 961.516 | 17.49 |
| 28 | 1 | 960.066 | 0.128 | 960.066 | 17.49 |
| 28 | 1 | 960.467 | 0.133 | 960.467 | 17.49 |
| 28 | 1 | 959.063 | 0.126 | 959.063 | 17.49 |
| 28 | 1 | 959.338 | 0.157 | 959.338 | 17.49 |
| 29 | 1 | 958.463 | 0.145 | 958.463 | 17.49 |
| 29 | 1 | 958.527 | 0.107 | 958.527 | 17.49 |
| 29 | 1 | 959.762 | 0.139 | 959.762 | 17.49 |
| 29 | 1 | 959.440 | 0.128 | 959.440 | 17.49 |
| 29 | 1 | 959.817 | 0.119 | 959.817 | 17.49 |
| 29 | 1 | 960.259 | 0.116 | 960.259 | 17.49 |

| 2006 - OUTUBRO | | | | | |
|---|---|---|---|---|---|
| D | L | SDB | ER | SDC | HL |
| 11 | 1 | 960.017 | 0.153 | 960.017 | 17.49 |
| 11 | 1 | 958.113 | 0.154 | 958.113 | 17.49 |
| 11 | 1 | 959.408 | 0.178 | 959.408 | 17.49 |
| 11 | 1 | 957.831 | 0.223 | 957.831 | 17.49 |
| 11 | 1 | 958.469 | 0.176 | 958.469 | 17.49 |
| 11 | 1 | 958.702 | 0.206 | 958.702 | 17.49 |
| 16 | 1 | 958.111 | 0.199 | 958.111 | 17.49 |
| 16 | 1 | 958.081 | 0.288 | 958.081 | 17.49 |
| 16 | 1 | 960.697 | 0.148 | 960.697 | 17.49 |
| 16 | 1 | 961.217 | 0.117 | 961.217 | 17.49 |
| 16 | 1 | 959.878 | 0.151 | 959.878 | 17.49 |
| 16 | 1 | 960.748 | 0.173 | 960.748 | 17.49 |
| 24 | 1 | 960.500 | 0.198 | 960.500 | 17.49 |
| 24 | 1 | 957.704 | 0.152 | 957.704 | 17.49 |
| 24 | 1 | 959.132 | 0.168 | 959.132 | 17.49 |
| 24 | 1 | 959.417 | 0.141 | 959.417 | 17.49 |
| 24 | 1 | 959.390 | 0.133 | 959.390 | 17.49 |
| 24 | 1 | 958.496 | 0.217 | 958.496 | 17.49 |
| 24 | 1 | 960.132 | 0.200 | 960.132 | 17.49 |
| 25 | 1 | 959.571 | 0.152 | 959.571 | 17.49 |
| 25 | 1 | 959.021 | 0.138 | 959.021 | 17.49 |
| 25 | 1 | 960.115 | 0.141 | 960.115 | 17.49 |
| 25 | 1 | 959.513 | 0.131 | 959.513 | 17.49 |
| 25 | 1 | 959.106 | 0.184 | 959.106 | 17.49 |
| 25 | 1 | 959.688 | 0.141 | 959.688 | 17.49 |
| 25 | 1 | 960.214 | 0.125 | 960.214 | 17.49 |
| 25 | 1 | 960.122 | 0.191 | 960.122 | 17.49 |
| 25 | 1 | 960.093 | 0.132 | 960.093 | 17.49 |
| 25 | 1 | 961.012 | 0.130 | 961.012 | 17.49 |
| 25 | 1 | 958.479 | 0.126 | 958.479 | 17.49 |
| 25 | 1 | 958.322 | 0.130 | 958.322 | 17.49 |
| 25 | 1 | 958.345 | 0.142 | 958.345 | 17.49 |

| 2006 - OUTUBRO | | | | | |
|---|---|---|---|---|---|
| D | L | SDB | ER | SDC | HL |
| 25 | 1 | 958.092 | 0.147 | 958.092 | 17.49 |
| 25 | 1 | 958.285 | 0.176 | 958.285 | 17.49 |
| 25 | 1 | 957.557 | 0.208 | 957.557 | 17.49 |
| 25 | 1 | 959.279 | 0.197 | 959.279 | 17.49 |
| 25 | 1 | 957.547 | 0.249 | 957.547 | 17.49 |
| 25 | 1 | 959.304 | 0.175 | 959.304 | 17.49 |
| 26 | 1 | 960.527 | 0.167 | 960.527 | 17.49 |
| 26 | 1 | 959.540 | 0.172 | 959.540 | 17.49 |
| 26 | 1 | 960.261 | 0.149 | 960.261 | 17.49 |
| 26 | 1 | 960.119 | 0.110 | 960.119 | 17.49 |
| 26 | 1 | 959.825 | 0.108 | 959.825 | 17.49 |
| 26 | 1 | 960.507 | 0.118 | 960.507 | 17.49 |
| 26 | 1 | 960.129 | 0.115 | 960.129 | 17.49 |
| 26 | 1 | 960.404 | 0.107 | 960.404 | 17.49 |
| 30 | 1 | 958.164 | 0.182 | 958.164 | 17.49 |
| 30 | 1 | 957.690 | 0.245 | 957.690 | 17.49 |
| 30 | 1 | 959.110 | 0.157 | 959.110 | 17.49 |
| 30 | 1 | 957.985 | 0.242 | 957.985 | 17.49 |
| 30 | 1 | 959.042 | 0.134 | 959.042 | 17.49 |
| 30 | 1 | 958.521 | 0.160 | 958.521 | 17.49 |
| 30 | 1 | 959.683 | 0.181 | 959.683 | 17.49 |
| 31 | 1 | 959.435 | 0.118 | 959.435 | 17.49 |
| 31 | 1 | 960.036 | 0.126 | 960.036 | 17.49 |
| 31 | 1 | 960.271 | 0.132 | 960.271 | 17.49 |
| 31 | 1 | 961.494 | 0.134 | 961.494 | 17.49 |
| 31 | 1 | 959.975 | 0.122 | 959.975 | 17.49 |
| 31 | 1 | 960.547 | 0.121 | 960.547 | 17.49 |
| 31 | 1 | 961.250 | 0.115 | 961.250 | 17.49 |
| 31 | 1 | 960.883 | 0.124 | 960.883 | 17.49 |
| 31 | 1 | 960.164 | 0.142 | 960.164 | 17.49 |
| 31 | 1 | 960.939 | 0.156 | 960.939 | 17.49 |
| 31 | 1 | 959.209 | 0.126 | 959.209 | 17.49 |
| 31 | 1 | 959.908 | 0.147 | 959.908 | 17.49 |
| 31 | 1 | 958.331 | 0.133 | 958.331 | 17.49 |
| 31 | 1 | 959.079 | 0.141 | 959.079 | 17.49 |
| 31 | 1 | 959.354 | 0.149 | 959.354 | 17.49 |
| 31 | 1 | 959.078 | 0.169 | 959.078 | 17.49 |
| 31 | 1 | 958.758 | 0.171 | 958.758 | 17.49 |
| 31 | 1 | 958.971 | 0.157 | 958.971 | 17.49 |
| 31 | 1 | 958.941 | 0.194 | 958.941 | 17.49 |
| 31 | 1 | 959.024 | 0.189 | 959.024 | 17.49 |

| 2006 - NOVEMBRO | | | | | |
|---|---|---|---|---|---|
| D | L | SDB | ER | SDC | HL |
| 01 | 1 | 960.367 | 0.102 | 960.367 | 17.49 |
| 01 | 1 | 960.009 | 0.111 | 960.009 | 17.49 |
| 01 | 1 | 960.060 | 0.095 | 960.060 | 17.49 |
| 01 | 1 | 961.223 | 0.099 | 961.223 | 17.49 |
| 01 | 1 | 960.671 | 0.109 | 960.671 | 17.49 |
| 01 | 1 | 960.419 | 0.116 | 960.419 | 17.49 |
| 01 | 1 | 958.693 | 0.134 | 958.693 | 17.49 |
| 01 | 1 | 960.255 | 0.127 | 960.255 | 17.49 |
| 01 | 1 | 959.660 | 0.152 | 959.660 | 17.49 |
| 01 | 1 | 959.762 | 0.173 | 959.762 | 17.49 |
| 01 | 1 | 958.886 | 0.159 | 958.886 | 17.49 |
| 01 | 1 | 957.383 | 0.146 | 957.383 | 17.49 |
| 01 | 1 | 958.639 | 0.154 | 958.639 | 17.49 |
| 01 | 1 | 959.700 | 0.122 | 959.700 | 17.49 |
| 01 | 1 | 959.287 | 0.127 | 959.287 | 17.49 |
| 01 | 1 | 959.134 | 0.113 | 959.134 | 17.49 |
| 01 | 1 | 959.682 | 0.157 | 959.682 | 17.49 |
| 01 | 1 | 959.549 | 0.185 | 959.549 | 17.49 |
| 01 | 1 | 958.576 | 0.161 | 958.576 | 17.49 |
| 06 | 1 | 959.724 | 0.143 | 959.724 | 17.49 |
| 06 | 1 | 959.880 | 0.132 | 959.880 | 17.49 |
| 06 | 1 | 958.904 | 0.202 | 958.904 | 17.49 |
| 06 | 1 | 958.156 | 0.173 | 958.156 | 17.49 |
| 06 | 1 | 959.778 | 0.136 | 959.778 | 17.49 |
| 06 | 1 | 959.183 | 0.123 | 959.183 | 17.49 |
| 06 | 1 | 960.228 | 0.156 | 960.228 | 17.49 |

| 2006 - NOVEMBRO | | | | |
|---|---|---|---|---|
| D | L | SDB | ER | SDC | HL |
| 06 | 1 | 960.864 | 0.156 | 960.864 | 17.49 |
| 06 | 1 | 958.372 | 0.180 | 958.372 | 17.49 |
| 06 | 1 | 960.007 | 0.144 | 960.007 | 17.49 |
| 06 | 1 | 958.901 | 0.162 | 958.901 | 17.49 |
| 06 | 1 | 959.765 | 0.140 | 959.765 | 17.49 |
| 06 | 1 | 959.017 | 0.150 | 959.017 | 17.49 |
| 06 | 1 | 958.855 | 0.178 | 958.855 | 17.49 |
| 06 | 1 | 958.250 | 0.221 | 958.250 | 17.49 |
| 13 | 1 | 959.359 | 0.195 | 959.359 | 17.49 |
| 13 | 1 | 959.426 | 0.184 | 959.426 | 17.49 |
| 13 | 1 | 959.730 | 0.188 | 959.730 | 17.49 |
| 13 | 1 | 960.169 | 0.154 | 960.169 | 17.49 |
| 13 | 1 | 960.768 | 0.192 | 960.768 | 17.49 |
| 16 | 1 | 959.777 | 0.112 | 959.777 | 17.49 |
| 16 | 1 | 958.941 | 0.123 | 958.941 | 17.49 |
| 16 | 1 | 960.691 | 0.122 | 960.691 | 17.49 |
| 16 | 1 | 960.879 | 0.131 | 960.879 | 17.49 |
| 16 | 1 | 958.679 | 0.111 | 958.679 | 17.49 |
| 16 | 1 | 959.810 | 0.095 | 959.810 | 17.49 |
| 16 | 1 | 960.125 | 0.111 | 960.125 | 17.49 |
| 16 | 1 | 959.300 | 0.123 | 959.300 | 17.49 |
| 16 | 1 | 958.146 | 0.169 | 958.146 | 17.49 |
| 16 | 1 | 958.823 | 0.118 | 958.823 | 17.49 |
| 16 | 1 | 959.201 | 0.107 | 959.201 | 17.49 |
| 16 | 1 | 959.105 | 0.145 | 959.105 | 17.49 |
| 16 | 1 | 959.083 | 0.158 | 959.083 | 17.49 |
| 16 | 1 | 958.060 | 0.119 | 958.060 | 17.49 |
| 17 | 1 | 960.459 | 0.110 | 960.459 | 17.49 |
| 17 | 1 | 959.258 | 0.103 | 959.258 | 17.49 |
| 17 | 1 | 959.568 | 0.094 | 959.568 | 17.49 |
| 17 | 1 | 959.226 | 0.118 | 959.226 | 17.49 |
| 17 | 1 | 959.515 | 0.129 | 959.515 | 17.49 |
| 17 | 1 | 959.517 | 0.151 | 959.517 | 17.49 |
| 17 | 1 | 957.532 | 0.182 | 957.532 | 17.49 |
| 22 | 1 | 959.248 | 0.128 | 959.248 | 17.49 |
| 22 | 1 | 958.785 | 0.115 | 958.785 | 17.49 |
| 22 | 1 | 958.631 | 0.115 | 958.631 | 17.49 |
| 22 | 1 | 958.657 | 0.129 | 958.657 | 17.49 |
| 22 | 1 | 959.232 | 0.172 | 959.232 | 17.49 |
| 22 | 1 | 959.424 | 0.150 | 959.424 | 17.49 |
| 22 | 1 | 959.481 | 0.134 | 959.481 | 17.49 |

| 2006 - DEZEMBRO | | | | |
|---|---|---|---|---|
| D | L | SDB | ER | SDC | HL |
| 04 | 1 | 958.591 | 0.132 | 958.591 | 17.49 |
| 04 | 1 | 958.833 | 0.127 | 958.833 | 17.49 |
| 04 | 1 | 959.042 | 0.113 | 959.042 | 17.49 |
| 04 | 1 | 959.069 | 0.126 | 959.069 | 17.49 |
| 04 | 1 | 959.797 | 0.117 | 959.797 | 17.49 |
| 04 | 1 | 959.981 | 0.163 | 959.981 | 17.49 |
| 04 | 1 | 959.523 | 0.148 | 959.523 | 17.49 |
| 04 | 1 | 959.730 | 0.123 | 959.730 | 17.49 |
| 04 | 1 | 959.909 | 0.142 | 959.909 | 17.49 |
| 04 | 1 | 961.176 | 0.120 | 961.176 | 17.49 |
| 04 | 1 | 958.915 | 0.118 | 958.915 | 17.49 |
| 04 | 1 | 958.160 | 0.637 | 958.160 | 17.49 |
| 04 | 1 | 959.311 | 0.125 | 959.311 | 17.49 |
| 04 | 1 | 960.119 | 0.132 | 960.119 | 17.49 |
| 04 | 1 | 959.745 | 0.132 | 959.745 | 17.49 |
| 04 | 1 | 958.879 | 0.156 | 958.879 | 17.49 |
| 04 | 1 | 959.076 | 0.156 | 959.076 | 17.49 |
| 08 | 1 | 958.218 | 0.147 | 958.218 | 17.49 |
| 08 | 1 | 959.329 | 0.126 | 959.329 | 17.49 |
| 08 | 1 | 959.425 | 0.199 | 959.425 | 17.49 |
| 08 | 1 | 959.426 | 0.177 | 959.426 | 17.49 |
| 18 | 1 | 958.947 | 0.118 | 958.947 | 17.49 |
| 18 | 1 | 959.349 | 0.143 | 959.349 | 17.49 |
| 18 | 1 | 958.953 | 0.159 | 958.953 | 17.49 |
| 18 | 1 | 958.910 | 0.171 | 958.910 | 17.49 |
| 18 | 1 | 959.013 | 0.151 | 959.013 | 17.49 |

| 2006 - DEZEMBRO | | | | |
|---|---|---|---|---|
| D | L | SDB | ER | SDC | HL |
| 18 | 1 | 958.605 | 0.176 | 958.605 | 17.49 |
| 18 | 1 | 960.296 | 0.160 | 960.296 | 17.49 |
| 18 | 1 | 958.872 | 0.174 | 958.872 | 17.49 |
| 18 | 1 | 959.745 | 0.223 | 959.745 | 17.49 |
| 18 | 1 | 960.104 | 0.200 | 960.104 | 17.49 |
| 18 | 1 | 958.933 | 0.271 | 958.933 | 17.49 |
| 18 | 1 | 959.234 | 0.210 | 959.234 | 17.49 |
| 18 | 1 | 957.746 | 0.173 | 957.746 | 17.49 |
| 18 | 1 | 959.859 | 0.152 | 959.859 | 17.49 |
| 18 | 1 | 959.717 | 0.155 | 959.717 | 17.49 |
| 18 | 1 | 959.250 | 0.152 | 959.250 | 17.49 |
| 18 | 1 | 961.693 | 0.137 | 961.693 | 17.49 |
| 18 | 1 | 959.939 | 0.127 | 959.939 | 17.49 |
| 18 | 1 | 960.382 | 0.138 | 960.382 | 17.49 |
| 18 | 1 | 959.386 | 0.134 | 959.386 | 17.49 |
| 18 | 1 | 958.052 | 0.318 | 958.052 | 17.49 |
| 18 | 1 | 959.923 | 0.127 | 959.923 | 17.49 |
| 18 | 1 | 959.353 | 0.126 | 959.353 | 17.49 |
| 18 | 1 | 959.068 | 0.130 | 959.068 | 17.49 |
| 19 | 1 | 958.634 | 0.113 | 958.634 | 17.49 |
| 19 | 1 | 958.264 | 0.104 | 958.264 | 17.49 |
| 19 | 1 | 958.939 | 0.107 | 958.939 | 17.49 |
| 19 | 1 | 959.970 | 0.108 | 959.970 | 17.49 |
| 19 | 1 | 959.860 | 0.121 | 959.860 | 17.49 |
| 19 | 1 | 958.717 | 0.139 | 958.717 | 17.49 |
| 19 | 1 | 959.640 | 0.116 | 959.640 | 17.49 |
| 19 | 1 | 959.896 | 0.170 | 959.896 | 17.49 |
| 19 | 1 | 958.650 | 0.170 | 958.650 | 17.49 |
| 19 | 1 | 958.731 | 0.174 | 958.731 | 17.49 |
| 19 | 1 | 957.399 | 0.155 | 957.399 | 17.49 |
| 19 | 1 | 959.436 | 0.165 | 959.436 | 17.49 |
| 19 | 1 | 959.503 | 0.152 | 959.503 | 17.49 |

| 2007 - JANEIRO | | | | |
|---|---|---|---|---|
| D | L | SDB | ER | SDC | HL |
| 11 | 1 | 959.003 | 0.158 | 959.003 | 17.49 |
| 11 | 1 | 959.260 | 0.133 | 959.260 | 17.49 |
| 11 | 1 | 959.232 | 0.456 | 959.232 | 17.49 |
| 11 | 1 | 958.144 | 0.148 | 958.144 | 17.49 |
| 11 | 1 | 959.411 | 0.149 | 959.411 | 17.49 |
| 11 | 1 | 958.334 | 0.144 | 958.334 | 17.49 |
| 11 | 1 | 960.024 | 0.126 | 960.024 | 17.49 |
| 11 | 1 | 959.625 | 0.140 | 959.625 | 17.49 |
| 11 | 1 | 958.521 | 0.140 | 958.521 | 17.49 |
| 11 | 1 | 959.468 | 0.159 | 959.468 | 17.49 |
| 11 | 1 | 958.527 | 0.150 | 958.527 | 17.49 |
| 19 | 1 | 959.912 | 0.105 | 959.912 | 17.49 |
| 19 | 1 | 959.340 | 0.090 | 959.340 | 17.49 |
| 19 | 1 | 959.206 | 0.123 | 959.206 | 17.49 |
| 19 | 1 | 960.464 | 0.104 | 960.464 | 17.49 |
| 19 | 1 | 959.645 | 0.120 | 959.645 | 17.49 |
| 19 | 1 | 960.017 | 0.158 | 960.017 | 17.49 |
| 19 | 1 | 958.853 | 0.124 | 958.853 | 17.49 |
| 19 | 1 | 959.692 | 0.104 | 959.692 | 17.49 |
| 19 | 1 | 958.735 | 0.108 | 958.735 | 17.49 |
| 19 | 1 | 958.895 | 0.141 | 958.895 | 17.49 |
| 19 | 1 | 959.093 | 0.117 | 959.093 | 17.49 |
| 19 | 1 | 960.331 | 0.140 | 960.331 | 17.49 |
| 19 | 1 | 958.747 | 0.125 | 958.747 | 17.49 |
| 19 | 1 | 958.595 | 0.132 | 958.595 | 17.49 |
| 19 | 1 | 959.239 | 0.166 | 959.239 | 17.49 |
| 25 | 1 | 958.652 | 0.111 | 958.652 | 17.49 |
| 25 | 1 | 960.616 | 0.149 | 960.616 | 17.49 |
| 25 | 1 | 960.070 | 0.134 | 960.070 | 17.49 |
| 25 | 1 | 960.469 | 0.127 | 960.469 | 17.49 |
| 25 | 1 | 959.742 | 0.120 | 959.742 | 17.49 |
| 25 | 1 | 959.302 | 0.107 | 959.302 | 17.49 |
| 25 | 1 | 959.308 | 0.109 | 959.308 | 17.49 |
| 25 | 1 | 958.974 | 0.140 | 958.974 | 17.49 |
| 25 | 1 | 959.088 | 0.119 | 959.088 | 17.49 |

| 2007 - JANEIRO | | | | | |
|---|---|---|---|---|---|
| D | L | SDB | ER | SDC | HL |
| 25 | 1 | 958.725 | 0.125 | 958.725 | 17.49 |
| 25 | 1 | 958.616 | 0.124 | 958.616 | 17.49 |
| 25 | 1 | 959.231 | 0.128 | 959.231 | 17.49 |
| 25 | 1 | 958.959 | 0.124 | 958.959 | 17.49 |
| 26 | 1 | 959.615 | 0.129 | 959.615 | 17.49 |
| 26 | 1 | 959.375 | 0.165 | 959.375 | 17.49 |
| 26 | 1 | 959.931 | 0.111 | 959.931 | 17.49 |
| 26 | 1 | 959.014 | 0.147 | 959.014 | 17.49 |
| 26 | 1 | 959.689 | 0.142 | 959.689 | 17.49 |
| 26 | 1 | 959.454 | 0.109 | 959.454 | 17.49 |
| 26 | 1 | 959.039 | 0.121 | 959.039 | 17.49 |
| 26 | 1 | 958.956 | 0.130 | 958.956 | 17.49 |
| 26 | 1 | 958.529 | 0.152 | 958.529 | 17.49 |
| 26 | 1 | 961.119 | 0.167 | 961.119 | 17.49 |
| 26 | 1 | 958.205 | 0.122 | 958.205 | 17.49 |
| 26 | 1 | 960.003 | 0.113 | 960.003 | 17.49 |
| 26 | 1 | 959.926 | 0.120 | 959.926 | 17.49 |
| 26 | 1 | 959.251 | 0.131 | 959.251 | 17.49 |
| 26 | 1 | 959.752 | 0.124 | 959.752 | 17.49 |
| 26 | 1 | 959.176 | 0.133 | 959.176 | 17.49 |
| 26 | 1 | 958.883 | 0.146 | 958.883 | 17.49 |
| 26 | 1 | 959.254 | 0.161 | 959.254 | 17.49 |
| 26 | 1 | 958.516 | 0.153 | 958.516 | 17.49 |
| 26 | 1 | 958.791 | 0.130 | 958.791 | 17.49 |
| 26 | 1 | 959.469 | 0.136 | 959.469 | 17.49 |

| 2007 - FEVEREIRO | | | | | |
|---|---|---|---|---|---|
| D | L | SDB | ER | SDC | HL |
| 01 | 1 | 960.093 | 0.143 | 960.093 | 17.49 |
| 01 | 1 | 959.465 | 0.159 | 959.465 | 17.49 |
| 01 | 1 | 960.024 | 0.132 | 960.024 | 17.49 |
| 01 | 1 | 960.520 | 0.118 | 960.520 | 17.49 |
| 01 | 1 | 959.142 | 0.144 | 959.142 | 17.49 |
| 01 | 1 | 959.296 | 0.127 | 959.296 | 17.49 |
| 01 | 1 | 959.350 | 0.117 | 959.350 | 17.49 |
| 01 | 1 | 957.854 | 0.122 | 957.854 | 17.49 |
| 01 | 1 | 958.756 | 0.159 | 958.756 | 17.49 |
| 01 | 1 | 958.865 | 0.158 | 958.865 | 17.49 |
| 01 | 1 | 959.673 | 0.127 | 959.673 | 17.49 |
| 01 | 1 | 959.383 | 0.175 | 959.383 | 17.49 |
| 01 | 1 | 959.039 | 0.117 | 959.039 | 17.49 |
| 01 | 1 | 959.242 | 0.115 | 959.242 | 17.49 |
| 05 | 1 | 959.523 | 0.126 | 959.523 | 17.49 |
| 05 | 1 | 959.747 | 0.140 | 959.747 | 17.49 |
| 05 | 1 | 959.278 | 0.141 | 959.278 | 17.49 |
| 05 | 1 | 959.818 | 0.141 | 959.818 | 17.49 |
| 05 | 1 | 959.291 | 0.125 | 959.291 | 17.49 |
| 05 | 1 | 959.097 | 0.150 | 959.097 | 17.49 |
| 05 | 1 | 959.266 | 0.162 | 959.266 | 17.49 |
| 05 | 1 | 958.436 | 0.138 | 958.436 | 17.49 |
| 05 | 1 | 958.979 | 0.203 | 958.979 | 17.49 |
| 05 | 1 | 961.334 | 0.123 | 961.334 | 17.49 |
| 05 | 1 | 959.663 | 0.127 | 959.663 | 17.49 |
| 05 | 1 | 958.805 | 0.141 | 958.805 | 17.49 |
| 05 | 1 | 960.005 | 0.143 | 960.005 | 17.49 |
| 05 | 1 | 960.081 | 0.147 | 960.081 | 17.49 |
| 05 | 1 | 959.686 | 0.138 | 959.686 | 17.49 |
| 05 | 1 | 958.953 | 0.135 | 958.953 | 17.49 |
| 05 | 1 | 959.469 | 0.121 | 959.469 | 17.49 |
| 05 | 1 | 959.391 | 0.122 | 959.391 | 17.49 |
| 05 | 1 | 959.604 | 0.138 | 959.604 | 17.49 |
| 05 | 1 | 958.947 | 0.150 | 958.947 | 17.49 |
| 09 | 1 | 959.006 | 0.160 | 959.006 | 17.49 |
| 09 | 1 | 959.241 | 0.140 | 959.241 | 17.49 |
| 09 | 1 | 959.458 | 0.128 | 959.458 | 17.49 |
| 09 | 1 | 960.430 | 0.129 | 960.430 | 17.49 |
| 09 | 1 | 960.237 | 0.151 | 960.237 | 17.49 |
| 09 | 1 | 960.152 | 0.149 | 960.152 | 17.49 |
| 09 | 1 | 960.195 | 0.178 | 960.195 | 17.49 |
| 09 | 1 | 960.131 | 0.160 | 960.131 | 17.49 |

| 2007 - FEVEREIRO | | | | | |
|---|---|---|---|---|---|
| D | L | SDB | ER | SDC | HL |
| 09 | 1 | 960.451 | 0.150 | 960.451 | 17.49 |
| 09 | 1 | 959.184 | 0.155 | 959.184 | 17.49 |
| 09 | 1 | 958.915 | 0.131 | 958.915 | 17.49 |
| 09 | 1 | 959.413 | 0.135 | 959.413 | 17.49 |
| 09 | 1 | 959.672 | 0.119 | 959.672 | 17.49 |
| 09 | 1 | 958.541 | 0.194 | 958.541 | 17.49 |
| 09 | 1 | 960.183 | 0.154 | 960.183 | 17.49 |
| 09 | 1 | 958.897 | 0.150 | 958.897 | 17.49 |
| 09 | 1 | 959.552 | 0.135 | 959.552 | 17.49 |
| 09 | 1 | 958.480 | 0.179 | 958.480 | 17.49 |
| 09 | 1 | 958.967 | 0.158 | 958.967 | 17.49 |
| 09 | 1 | 958.432 | 0.124 | 958.432 | 17.49 |
| 09 | 1 | 959.637 | 0.152 | 959.637 | 17.49 |
| 09 | 1 | 959.340 | 0.154 | 959.340 | 17.49 |
| 09 | 1 | 959.482 | 0.149 | 959.482 | 17.49 |
| 13 | 1 | 958.419 | 0.117 | 958.419 | 17.49 |
| 13 | 1 | 957.817 | 0.154 | 957.817 | 17.49 |
| 13 | 1 | 958.167 | 0.159 | 958.167 | 17.49 |
| 13 | 1 | 958.293 | 0.157 | 958.293 | 17.49 |
| 13 | 1 | 959.670 | 0.165 | 959.670 | 17.49 |
| 13 | 1 | 958.353 | 0.134 | 958.353 | 17.49 |
| 13 | 1 | 959.798 | 0.117 | 959.798 | 17.49 |
| 13 | 1 | 959.707 | 0.138 | 959.707 | 17.49 |
| 13 | 1 | 958.665 | 0.155 | 958.665 | 17.49 |
| 13 | 1 | 959.194 | 0.143 | 959.194 | 17.49 |
| 13 | 1 | 959.214 | 0.136 | 959.214 | 17.49 |
| 13 | 1 | 959.556 | 0.145 | 959.556 | 17.49 |
| 13 | 1 | 958.833 | 0.160 | 958.833 | 17.49 |
| 13 | 1 | 959.747 | 0.130 | 959.747 | 17.49 |
| 13 | 1 | 959.762 | 0.186 | 959.762 | 17.49 |
| 13 | 1 | 958.978 | 0.164 | 958.978 | 17.49 |
| 14 | 1 | 959.121 | 0.097 | 959.121 | 17.49 |
| 14 | 1 | 958.999 | 0.140 | 958.999 | 17.49 |
| 14 | 1 | 960.035 | 0.144 | 960.035 | 17.49 |
| 14 | 1 | 959.426 | 0.130 | 959.426 | 17.49 |
| 14 | 1 | 959.493 | 0.136 | 959.493 | 17.49 |
| 14 | 1 | 959.207 | 0.147 | 959.207 | 17.49 |
| 14 | 1 | 959.558 | 0.116 | 959.558 | 17.49 |
| 14 | 1 | 959.176 | 0.128 | 959.176 | 17.49 |
| 14 | 1 | 959.119 | 0.114 | 959.119 | 17.49 |
| 14 | 1 | 959.466 | 0.166 | 959.466 | 17.49 |
| 14 | 1 | 959.536 | 0.160 | 959.536 | 17.49 |
| 14 | 1 | 958.829 | 0.156 | 958.829 | 17.49 |
| 14 | 1 | 958.465 | 0.144 | 958.465 | 17.49 |
| 14 | 1 | 959.058 | 0.174 | 959.058 | 17.49 |
| 15 | 1 | 960.494 | 0.166 | 960.494 | 17.49 |
| 15 | 1 | 959.686 | 0.157 | 959.686 | 17.49 |
| 15 | 1 | 959.822 | 0.185 | 959.822 | 17.49 |
| 15 | 1 | 959.639 | 0.148 | 959.639 | 17.49 |
| 15 | 1 | 959.315 | 0.142 | 959.315 | 17.49 |
| 15 | 1 | 959.676 | 0.154 | 959.676 | 17.49 |
| 15 | 1 | 959.506 | 0.177 | 959.506 | 17.49 |
| 15 | 1 | 959.378 | 0.157 | 959.378 | 17.49 |
| 15 | 1 | 959.519 | 0.141 | 959.519 | 17.49 |
| 15 | 1 | 959.625 | 0.159 | 959.625 | 17.49 |
| 15 | 1 | 959.342 | 0.166 | 959.342 | 17.49 |
| 16 | 1 | 960.633 | 0.124 | 960.633 | 17.49 |
| 16 | 1 | 959.302 | 0.136 | 959.302 | 17.49 |
| 16 | 1 | 959.915 | 0.122 | 959.915 | 17.49 |
| 16 | 1 | 959.510 | 0.127 | 959.510 | 17.49 |
| 16 | 1 | 959.289 | 0.117 | 959.289 | 17.49 |
| 16 | 1 | 959.298 | 0.145 | 959.298 | 17.49 |
| 16 | 1 | 959.198 | 0.118 | 959.198 | 17.49 |
| 16 | 1 | 959.113 | 0.147 | 959.113 | 17.49 |
| 16 | 1 | 959.299 | 0.168 | 959.299 | 17.49 |
| 16 | 1 | 959.634 | 0.195 | 959.634 | 17.49 |
| 16 | 1 | 959.187 | 0.141 | 959.187 | 17.49 |
| 16 | 1 | 959.313 | 0.141 | 959.313 | 17.49 |
| 16 | 1 | 959.661 | 0.111 | 959.661 | 17.49 |
| 16 | 1 | 959.831 | 0.139 | 959.831 | 17.49 |
| 16 | 1 | 959.709 | 0.160 | 959.709 | 17.49 |

| 2007 - FEVEREIRO | | | | | | 2007 - MARCO | | | | |
|---|---|---|---|---|---|---|---|---|---|---|
| D | L | SDB | ER | SDC | HL | D | L | SDB | ER | SDC | HL |
| 16 | 1 | 960.216 | 0.150 | 960.216 | 17.49 | 01 | 1 | 957.863 | 0.133 | 957.863 | 17.49 |
| 16 | 1 | 959.568 | 0.248 | 959.568 | 17.49 | 01 | 1 | 958.525 | 0.134 | 958.525 | 17.49 |
| 16 | 1 | 959.832 | 0.183 | 959.832 | 17.49 | 01 | 1 | 959.455 | 0.105 | 959.455 | 17.49 |
| 16 | 1 | 959.731 | 0.154 | 959.731 | 17.49 | 01 | 1 | 958.998 | 0.146 | 958.998 | 17.49 |
| 16 | 1 | 960.083 | 0.294 | 960.083 | 17.49 | 01 | 1 | 958.719 | 0.134 | 958.719 | 17.49 |
| 16 | 1 | 959.915 | 0.180 | 959.915 | 17.49 | 01 | 1 | 959.292 | 0.146 | 959.292 | 17.49 |
| 22 | 1 | 959.651 | 0.099 | 959.651 | 17.49 | 01 | 1 | 959.636 | 0.137 | 959.636 | 17.49 |
| 22 | 1 | 958.583 | 0.087 | 958.583 | 17.49 | 01 | 1 | 960.404 | 0.173 | 960.404 | 17.49 |
| 22 | 1 | 958.929 | 0.095 | 958.929 | 17.49 | 01 | 1 | 959.139 | 0.152 | 959.139 | 17.49 |
| 22 | 1 | 959.349 | 0.113 | 959.349 | 17.49 | 01 | 1 | 959.094 | 0.174 | 959.094 | 17.49 |
| 22 | 1 | 958.471 | 0.150 | 958.471 | 17.49 | 02 | 1 | 959.420 | 0.112 | 959.420 | 17.49 |
| 22 | 1 | 959.700 | 0.137 | 959.700 | 17.49 | 02 | 1 | 960.408 | 0.129 | 960.408 | 17.49 |
| 22 | 1 | 958.801 | 0.141 | 958.801 | 17.49 | 02 | 1 | 960.728 | 0.116 | 960.728 | 17.49 |
| 22 | 1 | 959.971 | 0.117 | 959.971 | 17.49 | 02 | 1 | 959.591 | 0.120 | 959.591 | 17.49 |
| 22 | 1 | 959.092 | 0.121 | 959.092 | 17.49 | 02 | 1 | 958.827 | 0.154 | 958.827 | 17.49 |
| 22 | 1 | 959.217 | 0.123 | 959.217 | 17.49 | 05 | 1 | 959.861 | 0.131 | 959.861 | 17.49 |
| 22 | 1 | 960.433 | 0.143 | 960.433 | 17.49 | 05 | 1 | 959.303 | 0.150 | 959.303 | 17.49 |
| 22 | 1 | 959.751 | 0.134 | 959.751 | 17.49 | 05 | 1 | 959.728 | 0.118 | 959.728 | 17.49 |
| 22 | 1 | 959.935 | 0.149 | 959.935 | 17.49 | 05 | 1 | 959.193 | 0.099 | 959.193 | 17.49 |
| 22 | 1 | 958.694 | 0.123 | 958.694 | 17.49 | 05 | 1 | 959.587 | 0.131 | 959.587 | 17.49 |
| 22 | 1 | 959.018 | 0.147 | 959.018 | 17.49 | 05 | 1 | 960.214 | 0.133 | 960.214 | 17.49 |
| 22 | 1 | 959.818 | 0.178 | 959.818 | 17.49 | 05 | 1 | 959.371 | 0.154 | 959.371 | 17.49 |
| 22 | 1 | 960.102 | 0.148 | 960.102 | 17.49 | 05 | 1 | 960.271 | 0.139 | 960.271 | 17.49 |
| 22 | 1 | 959.626 | 0.149 | 959.626 | 17.49 | 05 | 1 | 958.923 | 0.112 | 958.923 | 17.49 |
| 22 | 1 | 959.158 | 0.148 | 959.158 | 17.49 | 05 | 1 | 959.515 | 0.131 | 959.515 | 17.49 |
| 22 | 1 | 959.000 | 0.129 | 959.000 | 17.49 | 05 | 1 | 960.256 | 0.176 | 960.256 | 17.49 |
| 22 | 1 | 959.713 | 0.138 | 959.713 | 17.49 | 05 | 1 | 959.642 | 0.166 | 959.642 | 17.49 |
| 23 | 1 | 958.894 | 0.150 | 958.894 | 17.49 | 05 | 1 | 959.747 | 0.141 | 959.747 | 17.49 |
| 23 | 1 | 959.709 | 0.117 | 959.709 | 17.49 | 05 | 1 | 959.446 | 0.175 | 959.446 | 17.49 |
| 23 | 1 | 959.252 | 0.100 | 959.252 | 17.49 | 05 | 1 | 959.881 | 0.137 | 959.881 | 17.49 |
| 23 | 1 | 959.965 | 0.126 | 959.965 | 17.49 | 05 | 1 | 960.872 | 0.140 | 960.872 | 17.49 |
| 23 | 1 | 959.712 | 0.111 | 959.712 | 17.49 | 05 | 1 | 959.085 | 0.170 | 959.085 | 17.49 |
| 23 | 1 | 959.382 | 0.104 | 959.382 | 17.49 | 05 | 1 | 959.436 | 0.147 | 959.436 | 17.49 |
| 23 | 1 | 959.655 | 0.141 | 959.655 | 17.49 | 05 | 1 | 959.268 | 0.177 | 959.268 | 17.49 |
| 23 | 1 | 960.122 | 0.136 | 960.122 | 17.49 | 05 | 1 | 958.375 | 0.194 | 958.375 | 17.49 |
| 23 | 1 | 959.958 | 0.124 | 959.958 | 17.49 | 05 | 1 | 958.445 | 0.153 | 958.445 | 17.49 |
| 23 | 1 | 959.775 | 0.138 | 959.775 | 17.49 | 05 | 1 | 959.349 | 0.168 | 959.349 | 17.49 |
| 23 | 1 | 959.894 | 0.141 | 959.894 | 17.49 | 05 | 1 | 958.694 | 0.159 | 958.694 | 17.49 |
| 23 | 1 | 959.444 | 0.153 | 959.444 | 17.49 | 05 | 1 | 959.210 | 0.152 | 959.210 | 17.49 |
| 23 | 1 | 958.889 | 0.145 | 958.889 | 17.49 | 05 | 1 | 959.955 | 0.189 | 959.955 | 17.49 |
| 23 | 1 | 959.554 | 0.162 | 959.554 | 17.49 | 05 | 1 | 958.963 | 0.193 | 958.963 | 17.49 |
| 23 | 1 | 961.314 | 0.144 | 961.314 | 17.49 | 06 | 1 | 959.029 | 0.117 | 959.029 | 17.49 |
| 23 | 1 | 960.104 | 0.120 | 960.104 | 17.49 | 06 | 1 | 959.214 | 0.123 | 959.214 | 17.49 |
| 23 | 1 | 959.414 | 0.109 | 959.414 | 17.49 | 06 | 1 | 959.560 | 0.149 | 959.560 | 17.49 |
| 23 | 1 | 959.506 | 0.129 | 959.506 | 17.49 | 06 | 1 | 959.236 | 0.124 | 959.236 | 17.49 |
| 23 | 1 | 959.583 | 0.594 | 959.583 | 17.49 | 06 | 1 | 959.431 | 0.114 | 959.431 | 17.49 |
| 23 | 1 | 959.959 | 0.146 | 959.959 | 17.49 | 06 | 1 | 959.345 | 0.119 | 959.345 | 17.49 |
| 23 | 1 | 959.535 | 0.186 | 959.535 | 17.49 | 06 | 1 | 959.011 | 0.117 | 959.011 | 17.49 |
| 23 | 1 | 959.456 | 0.121 | 959.456 | 17.49 | 06 | 1 | 959.580 | 0.146 | 959.580 | 17.49 |
| 26 | 1 | 958.985 | 0.105 | 958.985 | 17.49 | 06 | 1 | 958.742 | 0.143 | 958.742 | 17.49 |
| 26 | 1 | 959.221 | 0.119 | 959.221 | 17.49 | 06 | 1 | 959.020 | 0.139 | 959.020 | 17.49 |
| 26 | 1 | 958.975 | 0.164 | 958.975 | 17.49 | 07 | 1 | 959.399 | 0.117 | 959.399 | 17.49 |
| 26 | 1 | 959.160 | 0.160 | 959.160 | 17.49 | 07 | 1 | 959.238 | 0.126 | 959.238 | 17.49 |
| 26 | 1 | 958.939 | 0.168 | 958.939 | 17.49 | 07 | 1 | 959.604 | 0.144 | 959.604 | 17.49 |
| 26 | 1 | 958.776 | 0.143 | 958.776 | 17.49 | 07 | 1 | 960.239 | 0.142 | 960.239 | 17.49 |
| 26 | 1 | 959.541 | 0.169 | 959.541 | 17.49 | 07 | 1 | 959.260 | 0.142 | 959.260 | 17.49 |
| 26 | 1 | 958.843 | 0.151 | 958.843 | 17.49 | 07 | 1 | 959.876 | 0.163 | 959.876 | 17.49 |
| 26 | 1 | 959.468 | 0.157 | 959.468 | 17.49 | 07 | 1 | 959.870 | 0.126 | 959.870 | 17.49 |
| 27 | 1 | 957.911 | 0.117 | 957.911 | 17.49 | 07 | 1 | 959.370 | 0.124 | 959.370 | 17.49 |
| 27 | 1 | 958.684 | 0.128 | 958.684 | 17.49 | 07 | 1 | 959.575 | 0.121 | 959.575 | 17.49 |
| 27 | 1 | 959.281 | 0.123 | 959.281 | 17.49 | 07 | 1 | 959.269 | 0.123 | 959.269 | 17.49 |
| 27 | 1 | 959.247 | 0.095 | 959.247 | 17.49 | 07 | 1 | 960.381 | 0.141 | 960.381 | 17.49 |
| 27 | 1 | 958.921 | 0.099 | 958.921 | 17.49 | 07 | 1 | 958.576 | 0.190 | 958.576 | 17.49 |
| 27 | 1 | 958.838 | 0.112 | 958.838 | 17.49 | 09 | 1 | 958.736 | 0.167 | 958.736 | 17.49 |
| 27 | 1 | 958.962 | 0.112 | 958.962 | 17.49 | 09 | 1 | 958.670 | 0.157 | 958.670 | 17.49 |
| | | | | | | 09 | 1 | 959.421 | 0.131 | 959.421 | 17.49 |
| | | | | | | 09 | 1 | 959.164 | 0.135 | 959.164 | 17.49 |
| | | | | | | 09 | 1 | 957.772 | 0.136 | 957.772 | 17.49 |
| | | | | | | 09 | 1 | 959.474 | 0.149 | 959.474 | 17.49 |
| | | | | | | 09 | 1 | 959.696 | 0.129 | 959.696 | 17.49 |
| | | | | | | 09 | 1 | 959.424 | 0.150 | 959.424 | 17.49 |

| 2007 - MARCO | | | | | | 2007 - MARCO | | | | |
|---|---|---|---|---|---|---|---|---|---|---|
| D | L | SDB | ER | SDC | HL | D | L | SDB | ER | SDC | HL |
| 09 | 1 | 958.405 | 0.133 | 958.405 | 17.49 | 26 | 1 | 959.382 | 0.166 | 959.382 | 17.49 |
| 09 | 1 | 958.731 | 0.156 | 958.731 | 17.49 | 26 | 1 | 958.103 | 0.167 | 958.103 | 17.49 |
| 09 | 1 | 959.047 | 0.134 | 959.047 | 17.49 | 27 | 1 | 958.577 | 0.129 | 958.577 | 17.49 |
| 09 | 1 | 959.603 | 0.153 | 959.603 | 17.49 | 27 | 1 | 959.161 | 0.116 | 959.161 | 17.49 |
| 09 | 1 | 959.157 | 0.146 | 959.157 | 17.49 | 27 | 1 | 959.753 | 0.145 | 959.753 | 17.49 |
| 09 | 1 | 959.526 | 0.163 | 959.526 | 17.49 | 27 | 1 | 960.119 | 0.131 | 960.119 | 17.49 |
| 09 | 1 | 959.331 | 0.156 | 959.331 | 17.49 | 27 | 1 | 958.667 | 0.120 | 958.667 | 17.49 |
| 09 | 1 | 959.673 | 0.141 | 959.673 | 17.49 | 27 | 1 | 960.517 | 0.124 | 960.517 | 17.49 |
| 09 | 1 | 959.059 | 0.163 | 959.059 | 17.49 | 27 | 1 | 959.044 | 0.161 | 959.044 | 17.49 |
| 09 | 1 | 959.376 | 0.158 | 959.376 | 17.49 | 27 | 1 | 960.094 | 0.122 | 960.094 | 17.49 |
| 09 | 1 | 958.844 | 0.130 | 958.844 | 17.49 | 27 | 1 | 959.925 | 0.122 | 959.925 | 17.49 |
| 09 | 1 | 958.891 | 0.128 | 958.891 | 17.49 | 27 | 1 | 960.006 | 0.169 | 960.006 | 17.49 |
| 09 | 1 | 958.528 | 0.128 | 958.528 | 17.49 | 27 | 1 | 959.781 | 0.117 | 959.781 | 17.49 |
| 09 | 1 | 959.208 | 0.152 | 959.208 | 17.49 | 27 | 1 | 959.923 | 0.166 | 959.923 | 17.49 |
| 09 | 1 | 959.501 | 0.198 | 959.501 | 17.49 | 27 | 1 | 960.377 | 0.133 | 960.377 | 17.49 |
| 09 | 1 | 959.675 | 0.218 | 959.675 | 17.49 | 27 | 1 | 961.431 | 0.142 | 961.431 | 17.49 |
| 12 | 1 | 958.356 | 0.152 | 958.356 | 17.49 | 27 | 1 | 959.554 | 0.145 | 959.554 | 17.49 |
| 12 | 1 | 958.963 | 0.165 | 958.963 | 17.49 | 27 | 1 | 959.422 | 0.139 | 959.422 | 17.49 |
| 12 | 1 | 958.606 | 0.129 | 958.606 | 17.49 | 27 | 1 | 958.899 | 0.129 | 958.899 | 17.49 |
| 12 | 1 | 958.967 | 0.153 | 958.967 | 17.49 | 27 | 1 | 959.197 | 0.147 | 959.197 | 17.49 |
| 12 | 1 | 959.308 | 0.144 | 959.308 | 17.49 | 27 | 1 | 959.379 | 0.136 | 959.379 | 17.49 |
| 12 | 1 | 958.788 | 0.152 | 958.788 | 17.49 | 27 | 1 | 958.976 | 0.138 | 958.976 | 17.49 |
| 12 | 1 | 959.815 | 0.182 | 959.815 | 17.49 | 27 | 1 | 958.756 | 0.159 | 958.756 | 17.49 |
| 12 | 1 | 959.109 | 0.137 | 959.109 | 17.49 | 27 | 1 | 959.144 | 0.161 | 959.144 | 17.49 |
| 12 | 1 | 958.729 | 0.151 | 958.729 | 17.49 | 27 | 1 | 958.991 | 0.165 | 958.991 | 17.49 |
| 12 | 1 | 959.914 | 0.132 | 959.914 | 17.49 | 27 | 1 | 959.623 | 0.152 | 959.623 | 17.49 |
| 12 | 1 | 959.677 | 0.156 | 959.677 | 17.49 | 28 | 1 | 959.261 | 0.115 | 959.261 | 17.49 |
| 12 | 1 | 959.478 | 0.196 | 959.478 | 17.49 | 28 | 1 | 959.019 | 0.138 | 959.019 | 17.49 |
| 12 | 1 | 958.611 | 0.166 | 958.611 | 17.49 | 28 | 1 | 959.499 | 0.116 | 959.499 | 17.49 |
| 12 | 1 | 960.456 | 0.146 | 960.456 | 17.49 | 28 | 1 | 959.932 | 0.148 | 959.932 | 17.49 |
| 12 | 1 | 960.497 | 0.204 | 960.497 | 17.49 | 28 | 1 | 958.787 | 0.135 | 958.787 | 17.49 |
| 13 | 1 | 959.329 | 0.121 | 959.329 | 17.49 | 28 | 1 | 960.136 | 0.164 | 960.136 | 17.49 |
| 13 | 1 | 958.355 | 0.121 | 958.355 | 17.49 | 28 | 1 | 959.391 | 0.160 | 959.391 | 17.49 |
| 13 | 1 | 959.573 | 0.131 | 959.573 | 17.49 | 28 | 1 | 959.580 | 0.147 | 959.580 | 17.49 |
| 13 | 1 | 959.062 | 0.151 | 959.062 | 17.49 | 28 | 1 | 959.465 | 0.128 | 959.465 | 17.49 |
| 13 | 1 | 959.723 | 0.136 | 959.723 | 17.49 | 29 | 1 | 958.595 | 0.132 | 958.595 | 17.49 |
| 13 | 1 | 959.247 | 0.147 | 959.247 | 17.49 | 29 | 1 | 959.265 | 0.148 | 959.265 | 17.49 |
| 13 | 1 | 958.841 | 0.143 | 958.841 | 17.49 | 29 | 1 | 959.079 | 0.130 | 959.079 | 17.49 |
| 13 | 1 | 959.274 | 0.170 | 959.274 | 17.49 | 29 | 1 | 959.423 | 0.129 | 959.423 | 17.49 |
| 13 | 1 | 959.265 | 0.174 | 959.265 | 17.49 | 29 | 1 | 959.617 | 0.113 | 959.617 | 17.49 |
| 13 | 1 | 959.209 | 0.157 | 959.209 | 17.49 | 29 | 1 | 959.096 | 0.154 | 959.096 | 17.49 |
| 15 | 1 | 957.139 | 0.116 | 957.139 | 17.49 | 29 | 1 | 959.430 | 0.136 | 959.430 | 17.49 |
| 15 | 1 | 959.814 | 0.135 | 959.814 | 17.49 | 29 | 1 | 959.981 | 0.153 | 959.981 | 17.49 |
| 15 | 1 | 959.204 | 0.157 | 959.204 | 17.49 | 29 | 1 | 959.624 | 0.191 | 959.624 | 17.49 |
| 15 | 1 | 959.368 | 0.134 | 959.368 | 17.49 | 29 | 1 | 959.245 | 0.125 | 959.245 | 17.49 |
| 15 | 1 | 958.845 | 0.156 | 958.845 | 17.49 | | | | | | |
| 15 | 1 | 960.080 | 0.161 | 960.080 | 17.49 | | | | | | |
| 15 | 1 | 960.599 | 0.176 | 960.599 | 17.49 | 2007 - ABRIL | | | | | |
| 15 | 1 | 959.214 | 0.143 | 959.214 | 17.49 | D | L | SDB | ER | SDC | HL |
| 15 | 1 | 959.736 | 0.143 | 959.736 | 17.49 | 02 | 1 | 959.594 | 0.137 | 959.594 | 17.49 |
| 15 | 1 | 959.296 | 0.186 | 959.296 | 17.49 | 02 | 1 | 958.807 | 0.129 | 958.807 | 17.49 |
| 15 | 1 | 959.913 | 0.137 | 959.913 | 17.49 | 02 | 1 | 959.233 | 0.158 | 959.233 | 17.49 |
| 15 | 1 | 959.197 | 0.168 | 959.197 | 17.49 | 02 | 1 | 959.766 | 0.171 | 959.766 | 17.49 |
| 21 | 1 | 959.759 | 0.175 | 959.759 | 17.49 | 02 | 1 | 959.201 | 0.127 | 959.201 | 17.49 |
| 21 | 1 | 959.773 | 0.189 | 959.773 | 17.49 | 02 | 1 | 959.865 | 0.136 | 959.865 | 17.49 |
| 21 | 1 | 959.615 | 0.142 | 959.615 | 17.49 | 02 | 1 | 959.445 | 0.131 | 959.445 | 17.49 |
| 21 | 1 | 958.566 | 0.164 | 958.566 | 17.49 | 02 | 1 | 959.355 | 0.110 | 959.355 | 17.49 |
| 21 | 1 | 959.637 | 0.165 | 959.637 | 17.49 | 02 | 1 | 959.649 | 0.117 | 959.649 | 17.49 |
| 21 | 1 | 958.764 | 0.140 | 958.764 | 17.49 | 02 | 1 | 959.525 | 0.140 | 959.525 | 17.49 |
| 21 | 1 | 958.844 | 0.179 | 958.844 | 17.49 | 02 | 1 | 959.729 | 0.184 | 959.729 | 17.49 |
| 26 | 1 | 960.590 | 0.146 | 960.590 | 17.49 | 02 | 1 | 959.208 | 0.171 | 959.208 | 17.49 |
| 26 | 1 | 959.818 | 0.145 | 959.818 | 17.49 | 02 | 1 | 958.488 | 0.194 | 958.488 | 17.49 |
| 26 | 1 | 959.830 | 0.149 | 959.830 | 17.49 | 02 | 1 | 959.288 | 0.164 | 959.288 | 17.49 |
| 26 | 1 | 958.914 | 0.144 | 958.914 | 17.49 | 02 | 1 | 959.509 | 0.159 | 959.509 | 17.49 |
| 26 | 1 | 959.401 | 0.157 | 959.401 | 17.49 | 03 | 1 | 959.659 | 0.134 | 959.659 | 17.49 |
| 26 | 1 | 958.624 | 0.176 | 958.624 | 17.49 | 03 | 1 | 958.624 | 0.176 | 958.624 | 17.49 |
| 26 | 1 | 959.235 | 0.152 | 959.235 | 17.49 | 03 | 1 | 958.725 | 0.120 | 958.725 | 17.49 |
| 26 | 1 | 959.157 | 0.121 | 959.157 | 17.49 | 03 | 1 | 958.521 | 0.180 | 958.521 | 17.49 |
| 26 | 1 | 959.682 | 0.177 | 959.682 | 17.49 | 03 | 1 | 959.613 | 0.158 | 959.613 | 17.49 |
| 26 | 1 | 959.259 | 0.149 | 959.259 | 17.49 | 03 | 1 | 959.702 | 0.147 | 959.702 | 17.49 |
| 26 | 1 | 959.329 | 0.148 | 959.329 | 17.49 | 03 | 1 | 959.493 | 0.171 | 959.493 | 17.49 |

| 2007 - ABRIL | | | | | | 2007 - ABRIL | | | | |
|---|---|---|---|---|---|---|---|---|---|---|
| D | L | SDB | ER | SDC | HL | D | L | SDB | ER | SDC | HL |
| 03 | 1 | 959.468 | 0.145 | 959.468 | 17.49 | 17 | 1 | 961.232 | 0.152 | 961.232 | 17.49 |
| 03 | 1 | 959.307 | 0.168 | 959.307 | 17.49 | 17 | 1 | 959.785 | 0.163 | 959.785 | 17.49 |
| 04 | 1 | 958.756 | 0.141 | 958.756 | 17.49 | 17 | 1 | 959.608 | 0.156 | 959.608 | 17.49 |
| 04 | 1 | 959.130 | 0.149 | 959.130 | 17.49 | 17 | 1 | 959.189 | 0.148 | 959.189 | 17.49 |
| 04 | 1 | 959.936 | 0.134 | 959.936 | 17.49 | 17 | 1 | 959.544 | 0.143 | 959.544 | 17.49 |
| 04 | 1 | 959.465 | 0.171 | 959.465 | 17.49 | 17 | 1 | 959.890 | 0.158 | 959.890 | 17.49 |
| 09 | 1 | 960.293 | 0.146 | 960.293 | 17.49 | 17 | 1 | 960.009 | 0.215 | 960.009 | 17.49 |
| 09 | 1 | 959.900 | 0.117 | 959.900 | 17.49 | 18 | 1 | 960.004 | 0.150 | 960.004 | 17.49 |
| 09 | 1 | 958.993 | 0.138 | 958.993 | 17.49 | 18 | 1 | 959.609 | 0.112 | 959.609 | 17.49 |
| 09 | 1 | 959.434 | 0.103 | 959.434 | 17.49 | 18 | 1 | 959.407 | 0.159 | 959.407 | 17.49 |
| 09 | 1 | 959.582 | 0.145 | 959.582 | 17.49 | 18 | 1 | 959.145 | 0.154 | 959.145 | 17.49 |
| 09 | 1 | 959.195 | 0.131 | 959.195 | 17.49 | 18 | 1 | 958.734 | 0.181 | 958.734 | 17.49 |
| 09 | 1 | 959.016 | 0.123 | 959.016 | 17.49 | 18 | 1 | 958.511 | 0.176 | 958.511 | 17.49 |
| 09 | 1 | 959.624 | 0.145 | 959.624 | 17.49 | 18 | 1 | 958.521 | 0.204 | 958.521 | 17.49 |
| 09 | 1 | 959.087 | 0.127 | 959.087 | 17.49 | 18 | 1 | 959.758 | 0.150 | 959.758 | 17.49 |
| 09 | 1 | 960.044 | 0.156 | 960.044 | 17.49 | 18 | 1 | 961.309 | 0.172 | 961.309 | 17.49 |
| 09 | 1 | 959.396 | 0.140 | 959.396 | 17.49 | 18 | 1 | 959.615 | 0.219 | 959.615 | 17.49 |
| 09 | 1 | 959.806 | 0.172 | 959.806 | 17.49 | 18 | 1 | 959.065 | 0.142 | 959.065 | 17.49 |
| 11 | 1 | 958.810 | 0.127 | 958.810 | 17.49 | 18 | 1 | 957.151 | 0.531 | 957.151 | 17.49 |
| 11 | 1 | 959.745 | 0.130 | 959.745 | 17.49 | 18 | 1 | 960.041 | 0.165 | 960.041 | 17.49 |
| 11 | 1 | 959.719 | 0.121 | 959.719 | 17.49 | 18 | 1 | 959.822 | 0.139 | 959.822 | 17.49 |
| 11 | 1 | 959.308 | 0.106 | 959.308 | 17.49 | 18 | 1 | 959.424 | 0.190 | 959.424 | 17.49 |
| 11 | 1 | 959.064 | 0.125 | 959.064 | 17.49 | 18 | 1 | 959.791 | 0.181 | 959.791 | 17.49 |
| 11 | 1 | 960.176 | 0.135 | 960.176 | 17.49 | 18 | 1 | 959.629 | 0.181 | 959.629 | 17.49 |
| 12 | 1 | 958.248 | 0.123 | 958.246 | 17.49 | 19 | 1 | 958.246 | 0.154 | 958.246 | 17.49 |
| 12 | 1 | 957.381 | 0.115 | 957.381 | 17.49 | 19 | 1 | 957.901 | 0.139 | 957.901 | 17.49 |
| 12 | 1 | 959.243 | 0.131 | 959.243 | 17.49 | 19 | 1 | 958.167 | 0.157 | 958.167 | 17.49 |
| 12 | 1 | 958.800 | 0.093 | 958.800 | 17.49 | 19 | 1 | 959.324 | 0.152 | 959.324 | 17.49 |
| 12 | 1 | 959.893 | 0.124 | 959.893 | 17.49 | 19 | 1 | 958.411 | 0.169 | 958.411 | 17.49 |
| 12 | 1 | 960.352 | 0.124 | 960.352 | 17.49 | 19 | 1 | 959.250 | 0.131 | 959.250 | 17.49 |
| 12 | 1 | 959.875 | 0.116 | 959.875 | 17.49 | 19 | 1 | 959.098 | 0.164 | 959.098 | 17.49 |
| 12 | 1 | 960.017 | 0.155 | 960.017 | 17.49 | 19 | 1 | 960.323 | 0.143 | 960.323 | 17.49 |
| 12 | 1 | 960.858 | 0.150 | 960.858 | 17.49 | 19 | 1 | 959.066 | 0.161 | 959.066 | 17.49 |
| 12 | 1 | 959.355 | 0.153 | 959.355 | 17.49 | 19 | 1 | 958.831 | 0.237 | 958.831 | 17.49 |
| 12 | 1 | 959.638 | 0.125 | 959.638 | 17.49 | 19 | 1 | 960.377 | 0.161 | 960.377 | 17.49 |
| 12 | 1 | 961.600 | 0.159 | 961.600 | 17.49 | 19 | 1 | 960.380 | 0.147 | 960.380 | 17.49 |
| 12 | 1 | 959.026 | 0.164 | 959.026 | 17.49 | 19 | 1 | 959.929 | 0.175 | 959.929 | 17.49 |
| 12 | 1 | 959.728 | 0.196 | 959.728 | 17.49 | 19 | 1 | 959.691 | 0.139 | 959.691 | 17.49 |
| 12 | 1 | 959.312 | 0.203 | 959.312 | 17.49 | 19 | 1 | 959.492 | 0.241 | 959.492 | 17.49 |
| 12 | 1 | 959.027 | 0.168 | 959.027 | 17.49 | 19 | 1 | 959.274 | 0.144 | 959.274 | 17.49 |
| 12 | 1 | 957.535 | 0.212 | 957.535 | 17.49 | 19 | 1 | 960.303 | 0.128 | 960.303 | 17.49 |
| 12 | 1 | 958.398 | 0.309 | 958.398 | 17.49 | 19 | 1 | 959.773 | 0.178 | 959.773 | 17.49 |
| 12 | 1 | 958.174 | 0.192 | 958.174 | 17.49 | 19 | 1 | 959.907 | 0.143 | 959.907 | 17.49 |
| 12 | 1 | 959.476 | 0.241 | 959.476 | 17.49 | 19 | 1 | 960.738 | 0.160 | 960.738 | 17.49 |
| 13 | 1 | 960.218 | 0.099 | 960.218 | 17.49 | 19 | 1 | 958.498 | 0.518 | 958.498 | 17.49 |
| 13 | 1 | 959.255 | 0.129 | 959.255 | 17.49 | 19 | 1 | 959.300 | 0.156 | 959.300 | 17.49 |
| 13 | 1 | 959.575 | 0.139 | 959.575 | 17.49 | 24 | 1 | 958.415 | 0.197 | 958.415 | 17.49 |
| 13 | 1 | 958.609 | 0.108 | 958.609 | 17.49 | 24 | 1 | 959.009 | 0.158 | 959.009 | 17.49 |
| 13 | 1 | 959.192 | 0.114 | 959.192 | 17.49 | 24 | 1 | 958.771 | 0.180 | 958.771 | 17.49 |
| 13 | 1 | 959.747 | 0.099 | 959.747 | 17.49 | 25 | 1 | 958.760 | 0.117 | 958.760 | 17.49 |
| 13 | 1 | 959.524 | 0.135 | 959.524 | 17.49 | 25 | 1 | 958.967 | 0.117 | 958.967 | 17.49 |
| 13 | 1 | 959.140 | 0.157 | 959.140 | 17.49 | 25 | 1 | 959.127 | 0.131 | 959.127 | 17.49 |
| 13 | 1 | 959.957 | 0.142 | 959.957 | 17.49 | 25 | 1 | 959.152 | 0.131 | 959.152 | 17.49 |
| 16 | 1 | 958.684 | 0.093 | 958.684 | 17.49 | 25 | 1 | 958.750 | 0.116 | 958.750 | 17.49 |
| 16 | 1 | 959.591 | 0.130 | 959.591 | 17.49 | 25 | 1 | 959.770 | 0.118 | 959.770 | 17.49 |
| 16 | 1 | 959.355 | 0.119 | 959.355 | 17.49 | 25 | 1 | 959.145 | 0.108 | 959.145 | 17.49 |
| 16 | 1 | 959.954 | 0.139 | 959.954 | 17.49 | 25 | 1 | 959.944 | 0.142 | 959.944 | 17.49 |
| 16 | 1 | 959.641 | 0.127 | 959.641 | 17.49 | 25 | 1 | 959.919 | 0.146 | 959.919 | 17.49 |
| 16 | 1 | 959.375 | 0.158 | 959.375 | 17.49 | 25 | 1 | 959.838 | 0.115 | 959.838 | 17.49 |
| 16 | 1 | 958.886 | 0.154 | 958.886 | 17.49 | 25 | 1 | 960.461 | 0.127 | 960.461 | 17.49 |
| 16 | 1 | 958.855 | 0.164 | 958.855 | 17.49 | 26 | 1 | 959.577 | 0.097 | 959.577 | 17.49 |
| 16 | 1 | 959.251 | 0.195 | 959.251 | 17.49 | 26 | 1 | 960.163 | 0.112 | 960.163 | 17.49 |
| 17 | 1 | 958.355 | 0.177 | 958.355 | 17.49 | 26 | 1 | 959.070 | 0.095 | 959.070 | 17.49 |
| 17 | 1 | 959.350 | 0.123 | 959.350 | 17.49 | 26 | 1 | 959.765 | 0.096 | 959.765 | 17.49 |
| 17 | 1 | 958.730 | 0.152 | 958.730 | 17.49 | 26 | 1 | 959.258 | 0.110 | 959.258 | 17.49 |
| 17 | 1 | 958.712 | 0.112 | 958.712 | 17.49 | 26 | 1 | 958.791 | 0.122 | 958.791 | 17.49 |
| 17 | 1 | 958.455 | 0.159 | 958.455 | 17.49 | 26 | 1 | 960.031 | 0.111 | 960.031 | 17.49 |
| 17 | 1 | 959.269 | 0.147 | 959.269 | 17.49 | 26 | 1 | 959.837 | 0.104 | 959.837 | 17.49 |
| 17 | 1 | 959.168 | 0.154 | 959.168 | 17.49 | 26 | 1 | 960.110 | 0.103 | 960.110 | 17.49 |
| 17 | 1 | 958.456 | 0.154 | 958.456 | 17.49 | 26 | 1 | 960.050 | 0.123 | 960.050 | 17.49 |
| 17 | 1 | 958.763 | 0.150 | 958.763 | 17.49 | 26 | 1 | 960.018 | 0.106 | 960.018 | 17.49 |

| | | 2007 - ABRIL | | | |
|---|---|---|---|---|---|
| D | L | SDB | ER | SDC | HL |
| 26 | 1 | 960.325 | 0.125 | 960.325 | 17.49 |

| | | 2007 - MAIO | | | |
|---|---|---|---|---|---|
| D | L | SDB | ER | SDC | HL |
| 02 | 1 | 958.390 | 0.293 | 958.390 | 17.49 |
| 02 | 1 | 958.810 | 0.106 | 958.810 | 17.49 |
| 02 | 1 | 958.955 | 0.159 | 958.955 | 17.49 |
| 02 | 1 | 959.284 | 0.138 | 959.284 | 17.49 |
| 02 | 1 | 959.305 | 0.121 | 959.305 | 17.49 |
| 02 | 1 | 959.803 | 0.133 | 959.803 | 17.49 |
| 02 | 1 | 959.080 | 0.157 | 959.080 | 17.49 |
| 02 | 1 | 959.533 | 0.147 | 959.533 | 17.49 |
| 02 | 1 | 959.516 | 0.149 | 959.516 | 17.49 |
| 02 | 1 | 960.906 | 0.113 | 960.906 | 17.49 |
| 16 | 1 | 959.227 | 0.161 | 959.227 | 17.49 |
| 16 | 1 | 960.189 | 0.127 | 960.189 | 17.49 |
| 16 | 1 | 961.130 | 0.126 | 961.130 | 17.49 |
| 16 | 1 | 961.470 | 0.130 | 961.470 | 17.49 |
| 16 | 1 | 961.129 | 0.144 | 961.129 | 17.49 |
| 16 | 1 | 960.412 | 0.133 | 960.412 | 17.49 |
| 16 | 1 | 959.111 | 0.144 | 959.111 | 17.49 |
| 16 | 1 | 959.769 | 0.137 | 959.769 | 17.49 |
| 16 | 1 | 959.969 | 0.158 | 959.969 | 17.49 |
| 16 | 1 | 959.206 | 0.154 | 959.206 | 17.49 |
| 16 | 1 | 959.064 | 0.195 | 959.064 | 17.49 |
| 16 | 1 | 959.278 | 0.114 | 959.278 | 17.49 |
| 16 | 1 | 959.035 | 0.172 | 959.035 | 17.49 |
| 16 | 1 | 959.675 | 0.137 | 959.675 | 17.49 |
| 16 | 1 | 960.144 | 0.198 | 960.144 | 17.49 |
| 16 | 1 | 959.465 | 0.180 | 959.465 | 17.49 |
| 16 | 1 | 959.362 | 0.182 | 959.362 | 17.49 |
| 18 | 1 | 959.214 | 0.155 | 959.214 | 17.49 |
| 18 | 1 | 960.001 | 0.145 | 960.001 | 17.49 |
| 18 | 1 | 959.776 | 0.148 | 959.776 | 17.49 |
| 18 | 1 | 959.102 | 0.119 | 959.102 | 17.49 |
| 18 | 1 | 958.945 | 0.128 | 958.945 | 17.49 |
| 18 | 1 | 959.496 | 0.166 | 959.496 | 17.49 |

| | | 2007 - JUNHO | | | |
|---|---|---|---|---|---|
| D | L | SDB | ER | SDC | HL |
| 12 | 1 | 959.876 | 0.122 | 959.876 | 17.49 |
| 12 | 1 | 961.640 | 0.130 | 961.640 | 17.49 |
| 12 | 1 | 960.430 | 0.111 | 960.430 | 17.49 |
| 12 | 1 | 960.391 | 0.120 | 960.391 | 17.49 |
| 12 | 1 | 959.814 | 0.136 | 959.814 | 17.49 |
| 12 | 1 | 960.959 | 0.135 | 960.959 | 17.49 |
| 12 | 1 | 961.975 | 0.146 | 961.975 | 17.49 |
| 12 | 1 | 961.539 | 0.151 | 961.539 | 17.49 |
| 12 | 1 | 959.374 | 0.126 | 959.374 | 17.49 |
| 12 | 1 | 958.476 | 0.147 | 958.476 | 17.49 |
| 12 | 1 | 959.723 | 0.206 | 959.723 | 17.49 |
| 12 | 1 | 959.847 | 0.140 | 959.847 | 17.49 |
| 12 | 1 | 959.126 | 0.159 | 959.126 | 17.49 |
| 12 | 1 | 958.977 | 0.174 | 958.977 | 17.49 |
| 12 | 1 | 959.593 | 0.155 | 959.593 | 17.49 |
| 12 | 1 | 960.058 | 0.130 | 960.058 | 17.49 |
| 12 | 1 | 958.615 | 0.255 | 958.615 | 17.49 |
| 14 | 1 | 960.245 | 0.118 | 960.245 | 17.49 |
| 14 | 1 | 959.595 | 0.120 | 959.595 | 17.49 |
| 14 | 1 | 959.796 | 0.132 | 959.796 | 17.49 |
| 14 | 1 | 959.888 | 0.120 | 959.888 | 17.49 |
| 14 | 1 | 961.201 | 0.116 | 961.201 | 17.49 |
| 14 | 1 | 961.362 | 0.100 | 961.362 | 17.49 |
| 14 | 1 | 961.029 | 0.108 | 961.029 | 17.49 |
| 14 | 1 | 961.752 | 0.130 | 961.752 | 17.49 |
| 14 | 1 | 961.983 | 0.131 | 961.983 | 17.49 |
| 20 | 1 | 958.975 | 0.114 | 958.975 | 17.49 |
| 20 | 1 | 960.748 | 0.128 | 960.748 | 17.49 |
| 20 | 1 | 960.761 | 0.148 | 960.761 | 17.49 |

| | | 2007 - JUNHO | | | |
|---|---|---|---|---|---|
| D | L | SDB | ER | SDC | HL |
| 20 | 1 | 961.085 | 0.117 | 961.085 | 17.49 |
| 20 | 1 | 961.519 | 0.137 | 961.519 | 17.49 |
| 20 | 1 | 960.682 | 0.145 | 960.682 | 17.49 |
| 20 | 1 | 959.956 | 0.117 | 959.956 | 17.49 |
| 20 | 1 | 957.545 | 0.169 | 957.545 | 17.49 |
| 20 | 1 | 959.716 | 0.193 | 959.716 | 17.49 |
| 20 | 1 | 959.078 | 0.159 | 959.078 | 17.49 |
| 20 | 1 | 958.752 | 0.182 | 958.752 | 17.49 |
| 20 | 1 | 959.500 | 0.171 | 959.500 | 17.49 |
| 20 | 1 | 958.869 | 0.170 | 958.869 | 17.49 |
| 21 | 1 | 960.349 | 0.165 | 960.349 | 17.49 |
| 21 | 1 | 960.031 | 0.121 | 960.031 | 17.49 |
| 21 | 1 | 959.117 | 0.117 | 959.117 | 17.49 |
| 21 | 1 | 959.803 | 0.141 | 959.803 | 17.49 |
| 21 | 1 | 959.230 | 0.171 | 959.230 | 17.49 |
| 21 | 1 | 960.125 | 0.172 | 960.125 | 17.49 |
| 21 | 1 | 959.980 | 0.186 | 959.980 | 17.49 |
| 21 | 1 | 958.662 | 0.177 | 958.662 | 17.49 |
| 21 | 1 | 959.548 | 0.136 | 959.548 | 17.49 |
| 21 | 1 | 959.272 | 0.198 | 959.272 | 17.49 |
| 21 | 1 | 959.770 | 0.196 | 959.770 | 17.49 |
| 26 | 1 | 959.104 | 0.158 | 959.104 | 17.49 |
| 26 | 1 | 958.531 | 0.170 | 958.531 | 17.49 |
| 26 | 1 | 958.971 | 0.149 | 958.971 | 17.49 |
| 26 | 1 | 959.670 | 0.166 | 959.670 | 17.49 |
| 26 | 1 | 958.266 | 0.173 | 958.266 | 17.49 |
| 26 | 1 | 958.196 | 0.201 | 958.196 | 17.49 |
| 26 | 1 | 959.988 | 0.191 | 959.988 | 17.49 |
| 26 | 1 | 958.986 | 0.182 | 958.986 | 17.49 |
| 26 | 1 | 959.916 | 0.156 | 959.916 | 17.49 |
| 28 | 1 | 959.221 | 0.184 | 959.221 | 17.49 |
| 28 | 1 | 960.321 | 0.169 | 960.321 | 17.49 |
| 28 | 1 | 958.456 | 0.149 | 958.456 | 17.49 |
| 28 | 1 | 958.131 | 0.143 | 958.131 | 17.49 |
| 28 | 1 | 958.460 | 0.152 | 958.460 | 17.49 |

| | | 2007 - JULHO | | | |
|---|---|---|---|---|---|
| D | L | SDB | ER | SDC | HL |
| 03 | 1 | 961.178 | 0.134 | 961.178 | 17.49 |
| 03 | 1 | 959.835 | 0.167 | 959.835 | 17.49 |
| 03 | 1 | 958.958 | 0.164 | 958.958 | 17.49 |
| 03 | 1 | 959.172 | 0.155 | 959.172 | 17.49 |
| 03 | 1 | 959.622 | 0.170 | 959.622 | 17.49 |
| 03 | 1 | 958.978 | 0.162 | 958.978 | 17.49 |
| 03 | 1 | 959.578 | 0.190 | 959.578 | 17.49 |
| 03 | 1 | 959.434 | 0.160 | 959.434 | 17.49 |
| 03 | 1 | 958.735 | 0.173 | 958.735 | 17.49 |
| 03 | 1 | 959.516 | 0.194 | 959.516 | 17.49 |
| 04 | 1 | 958.294 | 0.140 | 958.294 | 17.49 |
| 04 | 1 | 960.334 | 0.135 | 960.334 | 17.49 |
| 04 | 1 | 960.072 | 0.121 | 960.072 | 17.49 |
| 04 | 1 | 960.173 | 0.109 | 960.173 | 17.49 |
| 04 | 1 | 960.321 | 0.114 | 960.321 | 17.49 |
| 04 | 1 | 960.603 | 0.141 | 960.603 | 17.49 |
| 04 | 1 | 958.324 | 0.155 | 958.324 | 17.49 |
| 04 | 1 | 958.318 | 0.161 | 958.318 | 17.49 |
| 04 | 1 | 959.071 | 0.140 | 959.071 | 17.49 |
| 04 | 1 | 958.536 | 0.122 | 958.536 | 17.49 |
| 04 | 1 | 958.866 | 0.141 | 958.866 | 17.49 |
| 04 | 1 | 958.927 | 0.173 | 958.927 | 17.49 |
| 04 | 1 | 959.265 | 0.142 | 959.265 | 17.49 |
| 04 | 1 | 960.015 | 0.173 | 960.015 | 17.49 |
| 05 | 1 | 960.144 | 0.169 | 960.144 | 17.49 |
| 05 | 1 | 959.232 | 0.143 | 959.232 | 17.49 |
| 05 | 1 | 957.966 | 0.184 | 957.966 | 17.49 |
| 05 | 1 | 959.213 | 0.166 | 959.213 | 17.49 |
| 05 | 1 | 959.493 | 0.225 | 959.493 | 17.49 |
| 05 | 1 | 958.758 | 0.186 | 958.758 | 17.49 |
| 05 | 1 | 958.392 | 0.172 | 958.392 | 17.49 |
| 05 | 1 | 959.823 | 0.193 | 959.823 | 17.49 |

| 2007 - JULHO | | | | |
|---|---|---|---|---|
| D | L | SDB | ER | SDC | HL |
| 06 | 1 | 957.601 | 0.171 | 957.601 | 17.49 |
| 06 | 1 | 957.424 | 0.169 | 957.424 | 17.49 |
| 06 | 1 | 958.844 | 0.168 | 958.844 | 17.49 |
| 06 | 1 | 960.678 | 0.170 | 960.678 | 17.49 |
| 06 | 1 | 959.696 | 0.149 | 959.696 | 17.49 |
| 06 | 1 | 958.865 | 0.166 | 958.865 | 17.49 |
| 06 | 1 | 959.463 | 0.157 | 959.463 | 17.49 |
| 06 | 1 | 959.186 | 0.183 | 959.186 | 17.49 |
| 09 | 1 | 958.783 | 0.104 | 958.783 | 17.49 |
| 09 | 1 | 959.081 | 0.138 | 959.081 | 17.49 |
| 09 | 1 | 958.701 | 0.136 | 958.701 | 17.49 |
| 09 | 1 | 959.411 | 0.131 | 959.411 | 17.49 |
| 09 | 1 | 958.938 | 0.139 | 958.938 | 17.49 |
| 09 | 1 | 959.162 | 0.177 | 959.162 | 17.49 |
| 09 | 1 | 959.646 | 0.162 | 959.646 | 17.49 |
| 09 | 1 | 958.719 | 0.154 | 958.719 | 17.49 |
| 09 | 1 | 959.642 | 0.160 | 959.642 | 17.49 |
| 09 | 1 | 959.010 | 0.120 | 959.010 | 17.49 |
| 10 | 1 | 958.816 | 0.175 | 958.816 | 17.49 |
| 10 | 1 | 959.417 | 0.142 | 959.417 | 17.49 |
| 10 | 1 | 958.408 | 0.171 | 958.408 | 17.49 |
| 10 | 1 | 959.459 | 0.120 | 959.459 | 17.49 |
| 10 | 1 | 958.898 | 0.147 | 958.898 | 17.49 |
| 10 | 1 | 958.727 | 0.155 | 958.727 | 17.49 |
| 10 | 1 | 958.590 | 0.152 | 958.590 | 17.49 |
| 10 | 1 | 958.650 | 0.129 | 958.650 | 17.49 |
| 10 | 1 | 958.659 | 0.169 | 958.659 | 17.49 |
| 10 | 1 | 959.570 | 0.160 | 959.570 | 17.49 |
| 10 | 1 | 959.079 | 0.186 | 959.079 | 17.49 |
| 11 | 1 | 959.303 | 0.132 | 959.303 | 17.49 |
| 11 | 1 | 960.620 | 0.159 | 960.620 | 17.49 |
| 11 | 1 | 960.342 | 0.112 | 960.342 | 17.49 |
| 11 | 1 | 961.746 | 0.147 | 961.746 | 17.49 |
| 11 | 1 | 961.177 | 0.118 | 961.177 | 17.49 |
| 20 | 1 | 960.576 | 0.108 | 960.576 | 17.49 |
| 20 | 1 | 960.328 | 0.112 | 960.328 | 17.49 |
| 20 | 1 | 960.647 | 0.134 | 960.647 | 17.49 |
| 20 | 1 | 959.413 | 0.145 | 959.413 | 17.49 |
| 20 | 1 | 960.980 | 0.137 | 960.980 | 17.49 |
| 20 | 1 | 959.771 | 0.167 | 959.771 | 17.49 |
| 20 | 1 | 958.274 | 0.166 | 958.274 | 17.49 |
| 20 | 1 | 957.316 | 0.197 | 957.316 | 17.49 |
| 20 | 1 | 958.224 | 0.168 | 958.224 | 17.49 |
| 20 | 1 | 959.919 | 0.124 | 959.919 | 17.49 |
| 20 | 1 | 958.877 | 0.145 | 958.877 | 17.49 |
| 20 | 1 | 959.436 | 0.163 | 959.436 | 17.49 |
| 20 | 1 | 960.272 | 0.188 | 960.272 | 17.49 |
| 20 | 1 | 959.502 | 0.189 | 959.502 | 17.49 |
| 20 | 1 | 959.310 | 0.229 | 959.310 | 17.49 |
| 31 | 1 | 959.669 | 0.160 | 959.669 | 17.49 |
| 31 | 1 | 959.551 | 0.148 | 959.551 | 17.49 |
| 31 | 1 | 958.515 | 0.174 | 958.515 | 17.49 |
| 31 | 1 | 959.639 | 0.130 | 959.639 | 17.49 |
| 31 | 1 | 959.278 | 0.154 | 959.278 | 17.49 |
| 31 | 1 | 957.762 | 0.135 | 957.762 | 17.49 |
| 31 | 1 | 959.203 | 0.154 | 959.203 | 17.49 |
| 31 | 1 | 959.778 | 0.159 | 959.778 | 17.49 |
| 31 | 1 | 959.506 | 0.129 | 959.506 | 17.49 |
| 31 | 1 | 959.833 | 0.152 | 959.833 | 17.49 |
| 31 | 1 | 958.136 | 0.164 | 958.136 | 17.49 |
| 31 | 1 | 958.522 | 0.224 | 958.522 | 17.49 |

| 2007 - AGOSTO | | | | |
|---|---|---|---|---|
| D | L | SDB | ER | SDC | HL |
| 02 | 1 | 959.907 | 0.131 | 959.907 | 17.49 |
| 02 | 1 | 961.184 | 0.105 | 961.184 | 17.49 |
| 02 | 1 | 960.579 | 0.125 | 960.579 | 17.49 |
| 02 | 1 | 960.427 | 0.103 | 960.427 | 17.49 |
| 02 | 1 | 960.947 | 0.140 | 960.947 | 17.49 |
| 02 | 1 | 961.612 | 0.145 | 961.612 | 17.49 |

| 2007 - AGOSTO | | | | |
|---|---|---|---|---|
| D | L | SDB | ER | SDC | HL |
| 03 | 1 | 960.181 | 0.096 | 960.181 | 17.49 |
| 03 | 1 | 959.807 | 0.111 | 959.807 | 17.49 |
| 03 | 1 | 960.028 | 0.126 | 960.028 | 17.49 |
| 03 | 1 | 960.237 | 0.115 | 960.237 | 17.49 |
| 03 | 1 | 959.696 | 0.106 | 959.696 | 17.49 |
| 03 | 1 | 959.876 | 0.091 | 959.876 | 17.49 |
| 03 | 1 | 960.226 | 0.122 | 960.226 | 17.49 |
| 03 | 1 | 961.159 | 0.130 | 961.159 | 17.49 |
| 03 | 1 | 958.360 | 0.141 | 958.360 | 17.49 |
| 03 | 1 | 958.456 | 0.184 | 958.456 | 17.49 |
| 03 | 1 | 958.585 | 0.184 | 958.585 | 17.49 |
| 03 | 1 | 959.015 | 0.162 | 959.015 | 17.49 |
| 03 | 1 | 958.442 | 0.145 | 958.442 | 17.49 |
| 03 | 1 | 958.311 | 0.145 | 958.311 | 17.49 |
| 03 | 1 | 958.840 | 0.164 | 958.840 | 17.49 |
| 03 | 1 | 959.728 | 0.166 | 959.728 | 17.49 |
| 03 | 1 | 959.116 | 0.186 | 959.116 | 17.49 |
| 07 | 1 | 957.703 | 0.178 | 957.703 | 17.49 |
| 07 | 1 | 959.197 | 0.138 | 959.197 | 17.49 |
| 07 | 1 | 959.658 | 0.148 | 959.658 | 17.49 |
| 07 | 1 | 958.726 | 0.120 | 958.726 | 17.49 |
| 07 | 1 | 959.024 | 0.169 | 959.024 | 17.49 |
| 07 | 1 | 960.100 | 0.149 | 960.100 | 17.49 |
| 07 | 1 | 958.819 | 0.134 | 958.819 | 17.49 |
| 07 | 1 | 958.950 | 0.183 | 958.950 | 17.49 |
| 07 | 1 | 959.260 | 0.160 | 959.260 | 17.49 |
| 07 | 1 | 959.819 | 0.155 | 959.819 | 17.49 |
| 07 | 1 | 959.148 | 0.186 | 959.148 | 17.49 |
| 07 | 1 | 958.903 | 0.188 | 958.903 | 17.49 |
| 08 | 1 | 958.555 | 0.116 | 958.555 | 17.49 |
| 08 | 1 | 959.302 | 0.093 | 959.302 | 17.49 |
| 08 | 1 | 960.076 | 0.106 | 960.076 | 17.49 |
| 08 | 1 | 960.306 | 0.133 | 960.306 | 17.49 |
| 08 | 1 | 960.053 | 0.123 | 960.053 | 17.49 |
| 08 | 1 | 959.561 | 0.139 | 959.561 | 17.49 |
| 08 | 1 | 960.329 | 0.118 | 960.329 | 17.49 |
| 08 | 1 | 960.729 | 0.140 | 960.729 | 17.49 |
| 08 | 1 | 960.675 | 0.148 | 960.675 | 17.49 |
| 08 | 1 | 960.598 | 0.142 | 960.598 | 17.49 |
| 08 | 1 | 961.075 | 0.123 | 961.075 | 17.49 |
| 08 | 1 | 961.752 | 0.165 | 961.752 | 17.49 |
| 08 | 1 | 958.821 | 0.209 | 958.821 | 17.49 |
| 08 | 1 | 958.309 | 0.179 | 958.309 | 17.49 |
| 08 | 1 | 959.213 | 0.129 | 959.213 | 17.49 |
| 08 | 1 | 960.247 | 0.140 | 960.247 | 17.49 |
| 08 | 1 | 959.383 | 0.171 | 959.383 | 17.49 |
| 08 | 1 | 959.018 | 0.134 | 959.018 | 17.49 |
| 08 | 1 | 959.909 | 0.142 | 959.909 | 17.49 |
| 08 | 1 | 958.462 | 0.121 | 958.462 | 17.49 |
| 08 | 1 | 959.906 | 0.167 | 959.906 | 17.49 |
| 08 | 1 | 958.808 | 0.161 | 958.808 | 17.49 |
| 10 | 1 | 959.459 | 0.089 | 959.459 | 17.49 |
| 10 | 1 | 959.808 | 0.123 | 959.808 | 17.49 |
| 10 | 1 | 959.462 | 0.126 | 959.462 | 17.49 |
| 10 | 1 | 959.627 | 0.116 | 959.627 | 17.49 |
| 10 | 1 | 959.719 | 0.153 | 959.719 | 17.49 |
| 10 | 1 | 959.846 | 0.119 | 959.846 | 17.49 |
| 10 | 1 | 959.605 | 0.151 | 959.605 | 17.49 |
| 10 | 1 | 959.017 | 0.153 | 959.017 | 17.49 |
| 10 | 1 | 960.259 | 0.200 | 960.259 | 17.49 |
| 10 | 1 | 958.571 | 0.108 | 958.571 | 17.49 |
| 10 | 1 | 958.891 | 0.130 | 958.891 | 17.49 |
| 10 | 1 | 958.897 | 0.133 | 958.897 | 17.49 |
| 10 | 1 | 959.645 | 0.154 | 959.645 | 17.49 |
| 10 | 1 | 959.894 | 0.148 | 959.894 | 17.49 |
| 10 | 1 | 958.837 | 0.146 | 958.837 | 17.49 |
| 10 | 1 | 958.569 | 0.183 | 958.569 | 17.49 |
| 10 | 1 | 959.001 | 0.158 | 959.001 | 17.49 |
| 10 | 1 | 958.351 | 0.168 | 958.351 | 17.49 |
| 10 | 1 | 959.589 | 0.173 | 959.589 | 17.49 |
| 10 | 1 | 959.201 | 0.206 | 959.201 | 17.49 |

| 2007 - AGOSTO | | | | | | 2007 - AGOSTO | | | | |
|---|---|---|---|---|---|---|---|---|---|---|
| D | L | SDB | ER | SDC | HL | D | L | SDB | ER | SDC | HL |
| 10 | 1 | 958.282 | 0.183 | 958.282 | 17.49 | 20 | 1 | 959.231 | 0.174 | 959.231 | 17.49 |
| 10 | 1 | 958.388 | 0.192 | 958.388 | 17.49 | 20 | 1 | 959.388 | 0.191 | 959.388 | 17.49 |
| 13 | 1 | 959.928 | 0.161 | 959.928 | 17.49 | 20 | 1 | 959.252 | 0.157 | 959.252 | 17.49 |
| 13 | 1 | 959.327 | 0.140 | 959.327 | 17.49 | 20 | 1 | 959.022 | 0.167 | 959.022 | 17.49 |
| 13 | 1 | 959.805 | 0.185 | 959.805 | 17.49 | 20 | 1 | 959.058 | 0.198 | 959.058 | 17.49 |
| 13 | 1 | 959.002 | 0.151 | 959.002 | 17.49 | 23 | 1 | 958.368 | 0.175 | 958.368 | 17.49 |
| 13 | 1 | 959.291 | 0.139 | 959.291 | 17.49 | 23 | 1 | 959.191 | 0.164 | 959.191 | 17.49 |
| 13 | 1 | 957.910 | 0.173 | 957.910 | 17.49 | 23 | 1 | 960.665 | 0.176 | 960.665 | 17.49 |
| 13 | 1 | 958.769 | 0.167 | 958.769 | 17.49 | 23 | 1 | 958.479 | 0.146 | 958.479 | 17.49 |
| 13 | 1 | 958.519 | 0.174 | 958.519 | 17.49 | 23 | 1 | 958.809 | 0.140 | 958.809 | 17.49 |
| 13 | 1 | 959.197 | 0.192 | 959.197 | 17.49 | 23 | 1 | 959.100 | 0.185 | 959.100 | 17.49 |
| 13 | 1 | 959.270 | 0.155 | 959.270 | 17.49 | 23 | 1 | 959.778 | 0.136 | 959.778 | 17.49 |
| 13 | 1 | 959.442 | 0.178 | 959.442 | 17.49 | 23 | 1 | 959.438 | 0.157 | 959.438 | 17.49 |
| 13 | 1 | 958.798 | 0.137 | 958.798 | 17.49 | 23 | 1 | 958.402 | 0.146 | 958.402 | 17.49 |
| 14 | 1 | 958.484 | 0.157 | 958.484 | 17.49 | 23 | 1 | 960.409 | 0.150 | 960.409 | 17.49 |
| 14 | 1 | 959.918 | 0.114 | 959.918 | 17.49 | 23 | 1 | 959.223 | 0.145 | 959.223 | 17.49 |
| 14 | 1 | 959.830 | 0.114 | 959.830 | 17.49 | 23 | 1 | 959.765 | 0.200 | 959.765 | 17.49 |
| 14 | 1 | 961.069 | 0.102 | 961.069 | 17.49 | 23 | 1 | 959.486 | 0.181 | 959.486 | 17.49 |
| 14 | 1 | 961.197 | 0.137 | 961.197 | 17.49 | 24 | 1 | 959.555 | 0.172 | 959.555 | 17.49 |
| 14 | 1 | 961.341 | 0.154 | 961.341 | 17.49 | 24 | 1 | 958.194 | 0.171 | 958.194 | 17.49 |
| 14 | 1 | 957.692 | 0.120 | 957.692 | 17.49 | 24 | 1 | 959.943 | 0.210 | 959.943 | 17.49 |
| 14 | 1 | 958.626 | 0.103 | 958.626 | 17.49 | 24 | 1 | 959.220 | 0.163 | 959.220 | 17.49 |
| 14 | 1 | 958.606 | 0.146 | 958.606 | 17.49 | 24 | 1 | 958.706 | 0.145 | 958.706 | 17.49 |
| 14 | 1 | 958.784 | 0.148 | 958.784 | 17.49 | 24 | 1 | 959.576 | 0.149 | 959.576 | 17.49 |
| 14 | 1 | 959.387 | 0.156 | 959.387 | 17.49 | 24 | 1 | 958.870 | 0.159 | 958.870 | 17.49 |
| 14 | 1 | 958.179 | 0.169 | 958.179 | 17.49 | 24 | 1 | 958.594 | 0.171 | 958.594 | 17.49 |
| 14 | 1 | 959.846 | 0.125 | 959.846 | 17.49 | 24 | 1 | 958.256 | 0.174 | 958.256 | 17.49 |
| 14 | 1 | 958.950 | 0.141 | 958.950 | 17.49 | 24 | 1 | 958.631 | 0.169 | 958.631 | 17.49 |
| 14 | 1 | 960.105 | 0.124 | 960.105 | 17.49 | 24 | 1 | 959.175 | 0.177 | 959.175 | 17.49 |
| 14 | 1 | 958.865 | 0.183 | 958.865 | 17.49 | 24 | 1 | 959.482 | 0.212 | 959.482 | 17.49 |
| 14 | 1 | 959.135 | 0.176 | 959.135 | 17.49 | 24 | 1 | 959.305 | 0.166 | 959.305 | 17.49 |
| 15 | 1 | 957.901 | 0.113 | 957.901 | 17.49 | 27 | 1 | 959.801 | 0.113 | 959.801 | 17.49 |
| 15 | 1 | 959.437 | 0.126 | 959.437 | 17.49 | 27 | 1 | 959.265 | 0.121 | 959.265 | 17.49 |
| 15 | 1 | 960.594 | 0.124 | 960.594 | 17.49 | 27 | 1 | 960.660 | 0.108 | 960.660 | 17.49 |
| 15 | 1 | 960.533 | 0.119 | 960.533 | 17.49 | 27 | 1 | 960.272 | 0.120 | 960.272 | 17.49 |
| 15 | 1 | 961.237 | 0.125 | 961.237 | 17.49 | 27 | 1 | 960.525 | 0.124 | 960.525 | 17.49 |
| 15 | 1 | 961.464 | 0.127 | 961.464 | 17.49 | 27 | 1 | 960.265 | 0.166 | 960.265 | 17.49 |
| 15 | 1 | 959.851 | 0.162 | 959.851 | 17.49 | 27 | 1 | 959.444 | 0.186 | 959.444 | 17.49 |
| 15 | 1 | 958.614 | 0.159 | 958.614 | 17.49 | 27 | 1 | 958.839 | 0.162 | 958.839 | 17.49 |
| 15 | 1 | 959.173 | 0.165 | 959.173 | 17.49 | 27 | 1 | 959.459 | 0.148 | 959.459 | 17.49 |
| 15 | 1 | 958.965 | 0.152 | 958.965 | 17.49 | 27 | 1 | 959.867 | 0.160 | 959.867 | 17.49 |
| 15 | 1 | 959.865 | 0.142 | 959.865 | 17.49 | 27 | 1 | 959.437 | 0.154 | 959.437 | 17.49 |
| 15 | 1 | 958.632 | 0.160 | 958.632 | 17.49 | 27 | 1 | 959.360 | 0.140 | 959.360 | 17.49 |
| 15 | 1 | 959.325 | 0.168 | 959.325 | 17.49 | 27 | 1 | 959.933 | 0.153 | 959.933 | 17.49 |
| 15 | 1 | 959.160 | 0.208 | 959.160 | 17.49 | 27 | 1 | 959.238 | 0.147 | 959.238 | 17.49 |
| 15 | 1 | 958.877 | 0.153 | 958.877 | 17.49 | 27 | 1 | 959.787 | 0.136 | 959.787 | 17.49 |
| 16 | 1 | 959.520 | 0.141 | 959.520 | 17.49 | 27 | 1 | 959.965 | 0.141 | 959.965 | 17.49 |
| 16 | 1 | 958.659 | 0.165 | 958.659 | 17.49 | 27 | 1 | 959.012 | 0.141 | 959.012 | 17.49 |
| 16 | 1 | 959.006 | 0.166 | 959.006 | 17.49 | | | | | | |
| 16 | 1 | 959.517 | 0.189 | 959.517 | 17.49 | | | | | | |
| 16 | 1 | 959.389 | 0.170 | 959.389 | 17.49 | 2007 - SETEMBRO | | | | | |
| 16 | 1 | 958.434 | 0.188 | 958.434 | 17.49 | D | L | SDB | ER | SDC | HL |
| 16 | 1 | 959.054 | 0.186 | 959.054 | 17.49 | 11 | 1 | 960.142 | 0.153 | 960.142 | 17.49 |
| 16 | 1 | 958.878 | 0.179 | 958.878 | 17.49 | 11 | 1 | 959.821 | 0.142 | 959.821 | 17.49 |
| 16 | 1 | 960.376 | 0.181 | 960.376 | 17.49 | 11 | 1 | 960.065 | 0.141 | 960.065 | 17.49 |
| 16 | 1 | 959.380 | 0.184 | 959.380 | 17.49 | 11 | 1 | 960.288 | 0.168 | 960.288 | 17.49 |
| 16 | 1 | 959.417 | 0.182 | 959.417 | 17.49 | 11 | 1 | 960.894 | 0.161 | 960.894 | 17.49 |
| 17 | 1 | 959.515 | 0.171 | 959.515 | 17.49 | 11 | 1 | 958.928 | 0.142 | 958.928 | 17.49 |
| 17 | 1 | 958.052 | 0.202 | 958.052 | 17.49 | 11 | 1 | 959.289 | 0.193 | 959.289 | 17.49 |
| 17 | 1 | 959.272 | 0.212 | 959.272 | 17.49 | 11 | 1 | 959.260 | 0.170 | 959.260 | 17.49 |
| 17 | 1 | 960.302 | 0.177 | 960.302 | 17.49 | 11 | 1 | 959.327 | 0.152 | 959.327 | 17.49 |
| 17 | 1 | 958.899 | 0.260 | 958.899 | 17.49 | 11 | 1 | 959.067 | 0.170 | 959.067 | 17.49 |
| 17 | 1 | 958.701 | 0.188 | 958.701 | 17.49 | 11 | 1 | 958.405 | 0.155 | 958.405 | 17.49 |
| 20 | 1 | 959.469 | 0.140 | 959.469 | 17.49 | 11 | 1 | 959.232 | 0.169 | 959.232 | 17.49 |
| 20 | 1 | 959.491 | 0.152 | 959.491 | 17.49 | 11 | 1 | 958.966 | 0.183 | 958.966 | 17.49 |
| 20 | 1 | 958.576 | 0.179 | 958.576 | 17.49 | 11 | 1 | 959.471 | 0.153 | 959.471 | 17.49 |
| 20 | 1 | 957.925 | 0.167 | 957.925 | 17.49 | 11 | 1 | 958.620 | 0.179 | 958.620 | 17.49 |
| 20 | 1 | 958.100 | 0.286 | 958.100 | 17.49 | 11 | 1 | 958.679 | 0.143 | 958.679 | 17.49 |
| 20 | 1 | 959.058 | 0.185 | 959.058 | 17.49 | 11 | 1 | 959.004 | 0.178 | 959.004 | 17.49 |
| 20 | 1 | 958.861 | 0.153 | 958.861 | 17.49 | 11 | 1 | 959.018 | 0.153 | 959.018 | 17.49 |
| 20 | 1 | 958.602 | 0.184 | 958.602 | 17.49 | 11 | 1 | 959.892 | 0.145 | 959.892 | 17.49 |

| 2007 - SETEMBRO ||||| | 2007 - OUTUBRO |||||
|---|---|---|---|---|---|---|---|---|---|---|
| D | L | SDB | ER | SDC | HL | D | L | SDB | ER | SDC | HL |
| 11 | 1 | 958.744 | 0.158 | 958.744 | 17.49 | 01 | 1 | 958.971 | 0.131 | 958.971 | 17.49 |
| 11 | 1 | 958.958 | 0.163 | 958.958 | 17.49 | 01 | 1 | 959.081 | 0.136 | 959.081 | 17.49 |
| 13 | 1 | 957.476 | 0.147 | 957.476 | 17.49 | 01 | 1 | 958.765 | 0.131 | 958.765 | 17.49 |
| 13 | 1 | 959.452 | 0.117 | 959.452 | 17.49 | 01 | 1 | 959.309 | 0.146 | 959.309 | 17.49 |
| 13 | 1 | 958.579 | 0.134 | 958.579 | 17.49 | 01 | 1 | 959.491 | 0.141 | 959.491 | 17.49 |
| 13 | 1 | 960.543 | 0.152 | 960.543 | 17.49 | 01 | 1 | 959.363 | 0.102 | 959.363 | 17.49 |
| 13 | 1 | 959.790 | 0.136 | 959.790 | 17.49 | 01 | 1 | 959.507 | 0.143 | 959.507 | 17.49 |
| 13 | 1 | 960.790 | 0.154 | 960.790 | 17.49 | 01 | 1 | 959.610 | 0.160 | 959.610 | 17.49 |
| 13 | 1 | 959.611 | 0.142 | 959.611 | 17.49 | 01 | 1 | 959.723 | 0.161 | 959.723 | 17.49 |
| 13 | 1 | 959.382 | 0.199 | 959.382 | 17.49 | 01 | 1 | 959.259 | 0.133 | 959.259 | 17.49 |
| 13 | 1 | 959.608 | 0.149 | 959.608 | 17.49 | 01 | 1 | 959.250 | 0.143 | 959.250 | 17.49 |
| 13 | 1 | 959.762 | 0.174 | 959.762 | 17.49 | 01 | 1 | 959.800 | 0.138 | 959.800 | 17.49 |
| 13 | 1 | 960.908 | 0.189 | 960.908 | 17.49 | 01 | 1 | 958.849 | 0.120 | 958.849 | 17.49 |
| 13 | 1 | 959.652 | 0.144 | 959.652 | 17.49 | 01 | 1 | 958.779 | 0.170 | 958.779 | 17.49 |
| 13 | 1 | 959.450 | 0.134 | 959.450 | 17.49 | 02 | 1 | 958.829 | 0.125 | 958.829 | 17.49 |
| 13 | 1 | 959.251 | 0.162 | 959.251 | 17.49 | 02 | 1 | 959.850 | 0.141 | 959.850 | 17.49 |
| 13 | 1 | 959.218 | 0.157 | 959.218 | 17.49 | 02 | 1 | 959.973 | 0.130 | 959.973 | 17.49 |
| 13 | 1 | 959.232 | 0.155 | 959.232 | 17.49 | 02 | 1 | 959.713 | 0.117 | 959.713 | 17.49 |
| 13 | 1 | 959.641 | 0.147 | 959.641 | 17.49 | 02 | 1 | 959.581 | 0.135 | 959.581 | 17.49 |
| 13 | 1 | 958.901 | 0.159 | 958.901 | 17.49 | 02 | 1 | 960.092 | 0.175 | 960.092 | 17.49 |
| 13 | 1 | 959.314 | 0.149 | 959.314 | 17.49 | 02 | 1 | 960.822 | 0.184 | 960.822 | 17.49 |
| 13 | 1 | 959.482 | 0.155 | 959.482 | 17.49 | 02 | 1 | 958.946 | 0.144 | 958.946 | 17.49 |
| 13 | 1 | 959.132 | 0.152 | 959.132 | 17.49 | 02 | 1 | 960.009 | 0.153 | 960.009 | 17.49 |
| 13 | 1 | 959.984 | 0.139 | 959.984 | 17.49 | 02 | 1 | 959.923 | 0.189 | 959.923 | 17.49 |
| 13 | 1 | 959.067 | 0.171 | 959.067 | 17.49 | 02 | 1 | 959.837 | 0.156 | 959.837 | 17.49 |
| 14 | 1 | 959.712 | 0.146 | 959.712 | 17.49 | 02 | 1 | 959.632 | 0.154 | 959.632 | 17.49 |
| 14 | 1 | 960.353 | 0.150 | 960.353 | 17.49 | 02 | 1 | 959.069 | 0.140 | 959.069 | 17.49 |
| 14 | 1 | 960.882 | 0.170 | 960.882 | 17.49 | 02 | 1 | 958.778 | 0.154 | 958.778 | 17.49 |
| 14 | 1 | 960.109 | 0.158 | 960.109 | 17.49 | 02 | 1 | 959.576 | 0.149 | 959.576 | 17.49 |
| 14 | 1 | 959.975 | 0.144 | 959.975 | 17.49 | 02 | 1 | 958.012 | 0.125 | 958.012 | 17.49 |
| 14 | 1 | 958.258 | 0.200 | 958.258 | 17.49 | 02 | 1 | 957.582 | 0.331 | 957.582 | 17.49 |
| 14 | 1 | 959.383 | 0.196 | 959.383 | 17.49 | 02 | 1 | 958.006 | 0.149 | 958.006 | 17.49 |
| 14 | 1 | 958.159 | 0.164 | 958.159 | 17.49 | 02 | 1 | 958.906 | 0.163 | 958.906 | 17.49 |
| 14 | 1 | 957.971 | 0.173 | 957.971 | 17.49 | 02 | 1 | 959.365 | 0.160 | 959.365 | 17.49 |
| 14 | 1 | 960.210 | 0.162 | 960.210 | 17.49 | 02 | 1 | 959.341 | 0.145 | 959.341 | 17.49 |
| 14 | 1 | 959.641 | 0.156 | 959.641 | 17.49 | 02 | 1 | 958.519 | 0.162 | 958.519 | 17.49 |
| 14 | 1 | 957.844 | 0.164 | 957.844 | 17.49 | 02 | 1 | 959.149 | 0.152 | 959.149 | 17.49 |
| 14 | 1 | 960.305 | 0.162 | 960.305 | 17.49 | 02 | 1 | 959.138 | 0.183 | 959.138 | 17.49 |
| 14 | 1 | 958.973 | 0.164 | 958.973 | 17.49 | 02 | 1 | 958.704 | 0.133 | 958.704 | 17.49 |
| 14 | 1 | 959.004 | 0.148 | 959.004 | 17.49 | 04 | 1 | 959.656 | 0.145 | 959.656 | 17.49 |
| 14 | 1 | 959.024 | 0.176 | 959.024 | 17.49 | 04 | 1 | 959.604 | 0.131 | 959.604 | 17.49 |
| 14 | 1 | 958.678 | 0.143 | 958.678 | 17.49 | 04 | 1 | 959.083 | 0.159 | 959.083 | 17.49 |
| 14 | 1 | 960.347 | 0.181 | 960.347 | 17.49 | 04 | 1 | 959.359 | 0.142 | 959.359 | 17.49 |
| 14 | 1 | 958.936 | 0.188 | 958.936 | 17.49 | 04 | 1 | 958.721 | 0.144 | 958.721 | 17.49 |
| 14 | 1 | 958.855 | 0.136 | 958.855 | 17.49 | 04 | 1 | 959.310 | 0.120 | 959.310 | 17.49 |
| 18 | 1 | 958.855 | 0.135 | 958.855 | 17.49 | 04 | 1 | 959.730 | 0.177 | 959.730 | 17.49 |
| 18 | 1 | 959.280 | 0.154 | 959.280 | 17.49 | 04 | 1 | 959.195 | 0.144 | 959.195 | 17.49 |
| 18 | 1 | 959.067 | 0.136 | 959.067 | 17.49 | 04 | 1 | 957.816 | 0.174 | 957.816 | 17.49 |
| 18 | 1 | 958.386 | 0.149 | 958.386 | 17.49 | 04 | 1 | 959.708 | 0.154 | 959.708 | 17.49 |
| 18 | 1 | 958.525 | 0.146 | 958.525 | 17.49 | 04 | 1 | 959.160 | 0.143 | 959.160 | 17.49 |
| 18 | 1 | 958.874 | 0.149 | 958.874 | 17.49 | 04 | 1 | 959.454 | 0.164 | 959.454 | 17.49 |
| 18 | 1 | 959.162 | 0.151 | 959.162 | 17.49 | 16 | 1 | 959.762 | 0.257 | 959.762 | 17.49 |
| 18 | 1 | 959.057 | 0.149 | 959.057 | 17.49 | 16 | 1 | 958.874 | 0.120 | 958.874 | 17.49 |
| 18 | 1 | 959.037 | 0.175 | 959.037 | 17.49 | 16 | 1 | 959.025 | 0.158 | 959.025 | 17.49 |
| 18 | 1 | 958.716 | 0.158 | 958.716 | 17.49 | 16 | 1 | 959.252 | 0.150 | 959.252 | 17.49 |
| 27 | 1 | 959.466 | 0.126 | 959.466 | 17.49 | 16 | 1 | 958.414 | 0.147 | 958.414 | 17.49 |
| 27 | 1 | 960.980 | 0.161 | 960.980 | 17.49 | 16 | 1 | 959.251 | 0.133 | 959.251 | 17.49 |
| 27 | 1 | 958.406 | 0.171 | 958.406 | 17.49 | 16 | 1 | 958.820 | 0.156 | 958.820 | 17.49 |
| 27 | 1 | 959.083 | 0.147 | 959.083 | 17.49 | 16 | 1 | 959.289 | 0.110 | 959.289 | 17.49 |
| 27 | 1 | 958.924 | 0.141 | 958.924 | 17.49 | 16 | 1 | 958.397 | 0.153 | 958.397 | 17.49 |
| 27 | 1 | 958.941 | 0.167 | 958.941 | 17.49 | 16 | 1 | 958.271 | 0.165 | 958.271 | 17.49 |
| 27 | 1 | 959.674 | 0.138 | 959.674 | 17.49 | 17 | 1 | 958.151 | 0.154 | 958.151 | 17.49 |
| 27 | 1 | 958.924 | 0.149 | 958.924 | 17.49 | 17 | 1 | 959.729 | 0.136 | 959.729 | 17.49 |
| 27 | 1 | 958.916 | 0.142 | 958.916 | 17.49 | 17 | 1 | 959.723 | 0.363 | 959.723 | 17.49 |
| 27 | 1 | 959.228 | 0.133 | 959.228 | 17.49 | 17 | 1 | 958.940 | 0.151 | 958.940 | 17.49 |
| 27 | 1 | 958.931 | 0.144 | 958.931 | 17.49 | 17 | 1 | 959.682 | 0.144 | 959.682 | 17.49 |
| 27 | 1 | 958.746 | 0.148 | 958.746 | 17.49 | 17 | 1 | 959.386 | 0.163 | 959.386 | 17.49 |
| 27 | 1 | 958.538 | 0.162 | 958.538 | 17.49 | 17 | 1 | 958.550 | 0.157 | 958.550 | 17.49 |
| 27 | 1 | 958.493 | 0.161 | 958.493 | 17.49 | 17 | 1 | 958.586 | 0.173 | 958.586 | 17.49 |
|   |   |         |       |         |       | 17 | 1 | 959.601 | 0.176 | 959.601 | 17.49 |
|   |   |         |       |         |       | 17 | 1 | 958.919 | 0.183 | 958.919 | 17.49 |

|  2007 - OUTUBRO  |||||  2007 - NOVEMBRO  |||||
|---|---|---|---|---|---|---|---|---|---|
| D | L | SDB | ER | SDC | HL | D | L | SDB | ER | SDC | HL |
| 17 | 1 | 959.173 | 0.134 | 959.173 | 17.49 | 19 | 1 | 957.306 | 0.153 | 957.306 | 17.49 |
| 17 | 1 | 958.131 | 0.166 | 958.131 | 17.49 | 19 | 1 | 957.698 | 0.128 | 957.698 | 17.49 |
| 17 | 1 | 958.711 | 0.151 | 958.711 | 17.49 | 19 | 1 | 959.544 | 0.136 | 959.544 | 17.49 |
| 17 | 1 | 958.436 | 0.146 | 958.436 | 17.49 | 19 | 1 | 960.106 | 0.171 | 960.106 | 17.49 |
| 17 | 1 | 959.411 | 0.182 | 959.411 | 17.49 | 19 | 1 | 959.423 | 0.175 | 959.423 | 17.49 |
| 17 | 1 | 959.275 | 0.161 | 959.275 | 17.49 | 19 | 1 | 959.763 | 0.182 | 959.763 | 17.49 |
| 17 | 1 | 958.914 | 0.156 | 958.914 | 17.49 | 23 | 1 | 959.220 | 0.409 | 959.220 | 17.49 |
| 17 | 1 | 959.328 | 0.166 | 959.328 | 17.49 | 23 | 1 | 961.108 | 0.174 | 961.108 | 17.49 |
| 17 | 1 | 958.611 | 0.166 | 958.611 | 17.49 | 23 | 1 | 959.720 | 0.178 | 959.720 | 17.49 |
| 17 | 1 | 958.952 | 0.165 | 958.952 | 17.49 | 23 | 1 | 960.939 | 0.167 | 960.939 | 17.49 |
| 17 | 1 | 958.577 | 0.268 | 958.577 | 17.49 | 23 | 1 | 958.664 | 0.168 | 958.664 | 17.49 |
| 22 | 1 | 959.245 | 0.141 | 959.245 | 17.49 | 23 | 1 | 958.696 | 0.199 | 958.696 | 17.49 |
| 22 | 1 | 959.372 | 0.162 | 959.372 | 17.49 | 23 | 1 | 960.774 | 0.192 | 960.774 | 17.49 |
| 22 | 1 | 958.965 | 0.135 | 958.965 | 17.49 | 23 | 1 | 960.547 | 0.159 | 960.547 | 17.49 |
| 22 | 1 | 958.736 | 0.160 | 958.736 | 17.49 | 23 | 1 | 959.294 | 0.172 | 959.294 | 17.49 |
| 22 | 1 | 958.825 | 0.143 | 958.825 | 17.49 | 23 | 1 | 960.333 | 0.160 | 960.333 | 17.49 |
| 22 | 1 | 959.180 | 0.132 | 959.180 | 17.49 | 23 | 1 | 960.134 | 0.181 | 960.134 | 17.49 |
| 22 | 1 | 958.972 | 0.153 | 958.972 | 17.49 | 23 | 1 | 960.227 | 0.145 | 960.227 | 17.49 |
| 22 | 1 | 958.643 | 0.124 | 958.643 | 17.49 | 23 | 1 | 960.723 | 0.161 | 960.723 | 17.49 |
| 22 | 1 | 959.887 | 0.156 | 959.887 | 17.49 | 23 | 1 | 959.455 | 0.152 | 959.455 | 17.49 |
| 22 | 1 | 958.556 | 0.161 | 958.556 | 17.49 | 23 | 1 | 959.986 | 0.156 | 959.986 | 17.49 |
| 22 | 1 | 958.181 | 0.151 | 958.181 | 17.49 | 23 | 1 | 960.244 | 0.148 | 960.244 | 17.49 |
| 30 | 1 | 958.490 | 0.155 | 958.490 | 17.49 | 23 | 1 | 959.776 | 0.186 | 959.776 | 17.49 |
| 30 | 1 | 958.777 | 0.235 | 958.777 | 17.49 | 23 | 1 | 959.881 | 0.178 | 959.881 | 17.49 |
| 30 | 1 | 958.211 | 0.120 | 958.211 | 17.49 | 23 | 1 | 959.972 | 0.157 | 959.972 | 17.49 |
| 30 | 1 | 959.554 | 0.098 | 959.554 | 17.49 | 23 | 1 | 959.040 | 0.188 | 959.040 | 17.49 |
| 30 | 1 | 958.857 | 0.124 | 958.857 | 17.49 | 26 | 1 | 959.674 | 0.141 | 959.674 | 17.49 |
| 30 | 1 | 958.815 | 0.125 | 958.815 | 17.49 | 26 | 1 | 958.143 | 0.137 | 958.143 | 17.49 |
| 30 | 1 | 958.197 | 0.137 | 958.197 | 17.49 | 26 | 1 | 959.290 | 0.111 | 959.290 | 17.49 |
| 30 | 1 | 959.318 | 0.146 | 959.318 | 17.49 | 26 | 1 | 958.465 | 0.133 | 958.465 | 17.49 |
| 30 | 1 | 959.709 | 0.144 | 959.709 | 17.49 | 26 | 1 | 958.442 | 0.121 | 958.442 | 17.49 |
| 30 | 1 | 959.199 | 0.159 | 959.199 | 17.49 | 26 | 1 | 958.240 | 0.103 | 958.240 | 17.49 |
| 30 | 1 | 959.866 | 0.140 | 959.866 | 17.49 | 26 | 1 | 959.793 | 0.190 | 959.793 | 17.49 |
| 30 | 1 | 959.663 | 0.176 | 959.663 | 17.49 | | | | | | |
| 30 | 1 | 958.337 | 0.171 | 958.337 | 17.49 | | | | | | |
| 30 | 1 | 958.611 | 0.177 | 958.611 | 17.49 |  2007 - DEZEMBRO  |||||
| 30 | 1 | 959.073 | 0.150 | 959.073 | 17.49 | D | L | SDB | ER | SDC | HL |
| 30 | 1 | 959.588 | 0.148 | 959.588 | 17.49 | 04 | 1 | 960.938 | 0.193 | 960.938 | 17.49 |
| 30 | 1 | 959.252 | 0.166 | 959.252 | 17.49 | 04 | 1 | 960.288 | 0.163 | 960.288 | 17.49 |
| 30 | 1 | 958.178 | 0.124 | 958.178 | 17.49 | 04 | 1 | 959.767 | 0.146 | 959.767 | 17.49 |
| 30 | 1 | 958.446 | 0.165 | 958.446 | 17.49 | 04 | 1 | 959.086 | 0.155 | 959.086 | 17.49 |
| 30 | 1 | 959.750 | 0.162 | 959.750 | 17.49 | 04 | 1 | 959.005 | 0.146 | 959.005 | 17.49 |
| 30 | 1 | 959.335 | 0.156 | 959.335 | 17.49 | 04 | 1 | 959.715 | 0.196 | 959.715 | 17.49 |
| 30 | 1 | 959.373 | 0.221 | 959.373 | 17.49 | 04 | 1 | 959.681 | 0.121 | 959.681 | 17.49 |
| 30 | 1 | 958.131 | 0.208 | 958.131 | 17.49 | 04 | 1 | 959.371 | 0.160 | 959.371 | 17.49 |
| 30 | 1 | 959.075 | 0.234 | 959.075 | 17.49 | 04 | 1 | 959.517 | 0.173 | 959.517 | 17.49 |
| 30 | 1 | 958.368 | 0.237 | 958.368 | 17.49 | 04 | 1 | 958.782 | 0.197 | 958.782 | 17.49 |
| 31 | 1 | 957.783 | 0.092 | 957.783 | 17.49 | 04 | 1 | 960.974 | 0.203 | 960.974 | 17.49 |
| 31 | 1 | 959.257 | 0.105 | 959.257 | 17.49 | 05 | 1 | 959.251 | 0.106 | 959.251 | 17.49 |
| 31 | 1 | 958.949 | 0.130 | 958.949 | 17.49 | 05 | 1 | 958.976 | 0.097 | 958.976 | 17.49 |
| 31 | 1 | 959.615 | 0.131 | 959.615 | 17.49 | 05 | 1 | 959.404 | 0.116 | 959.404 | 17.49 |
| 31 | 1 | 958.286 | 0.178 | 958.286 | 17.49 | 05 | 1 | 958.621 | 0.109 | 958.621 | 17.49 |
| 31 | 1 | 959.045 | 0.129 | 959.045 | 17.49 | 05 | 1 | 959.034 | 0.137 | 959.034 | 17.49 |
| 31 | 1 | 958.539 | 0.153 | 958.539 | 17.49 | 05 | 1 | 959.014 | 0.149 | 959.014 | 17.49 |
| 31 | 1 | 958.364 | 0.134 | 958.364 | 17.49 | 05 | 1 | 958.866 | 0.145 | 958.866 | 17.49 |
| 31 | 1 | 958.942 | 0.140 | 958.942 | 17.49 | 05 | 1 | 960.389 | 0.136 | 960.389 | 17.49 |
| 31 | 1 | 958.467 | 0.115 | 958.467 | 17.49 | 05 | 1 | 959.826 | 0.165 | 959.826 | 17.49 |
| 31 | 1 | 959.566 | 0.146 | 959.566 | 17.49 | 05 | 1 | 960.004 | 0.166 | 960.004 | 17.49 |
| 31 | 1 | 959.079 | 0.121 | 959.079 | 17.49 | 05 | 1 | 959.180 | 0.160 | 959.180 | 17.49 |
| 31 | 1 | 959.656 | 0.123 | 959.656 | 17.49 | 05 | 1 | 959.407 | 0.167 | 959.407 | 17.49 |
| 31 | 1 | 958.734 | 0.129 | 958.734 | 17.49 | 05 | 1 | 959.556 | 0.164 | 959.556 | 17.49 |
| 31 | 1 | 957.923 | 0.169 | 957.923 | 17.49 | 05 | 1 | 959.034 | 0.160 | 959.034 | 17.49 |
| 31 | 1 | 959.386 | 0.122 | 959.386 | 17.49 | 05 | 1 | 957.945 | 0.176 | 957.945 | 17.49 |
| 31 | 1 | 959.863 | 0.199 | 959.863 | 17.49 | 05 | 1 | 959.314 | 0.149 | 959.314 | 17.49 |
| 31 | 1 | 958.703 | 0.167 | 958.703 | 17.49 | 05 | 1 | 958.208 | 0.152 | 958.208 | 17.49 |
| 31 | 1 | 958.899 | 0.164 | 958.899 | 17.49 | 05 | 1 | 959.012 | 0.139 | 959.012 | 17.49 |
| 31 | 1 | 959.084 | 0.171 | 959.084 | 17.49 | 05 | 1 | 960.034 | 0.165 | 960.034 | 17.49 |
| 31 | 1 | 958.423 | 0.144 | 958.423 | 17.49 | 05 | 1 | 958.750 | 0.152 | 958.750 | 17.49 |
| 31 | 1 | 958.733 | 0.158 | 958.733 | 17.49 | 05 | 1 | 959.832 | 0.121 | 959.832 | 17.49 |
| 31 | 1 | 958.407 | 0.143 | 958.407 | 17.49 | 05 | 1 | 959.413 | 0.141 | 959.413 | 17.49 |
| 31 | 1 | 958.791 | 0.153 | 958.791 | 17.49 | 05 | 1 | 959.672 | 0.158 | 959.672 | 17.49 |

| 2007 - DEZEMBRO | | | | |
|---|---|---|---|---|
| D | L | SDB | ER | SDC | HL |
| 05 | 1 | 959.761 | 0.144 | 959.761 | 17.49 |
| 05 | 1 | 960.013 | 0.155 | 960.013 | 17.49 |
| 05 | 1 | 959.141 | 0.150 | 959.141 | 17.49 |
| 05 | 1 | 959.765 | 0.154 | 959.765 | 17.49 |
| 05 | 1 | 960.082 | 0.140 | 960.082 | 17.49 |
| 05 | 1 | 959.422 | 0.188 | 959.422 | 17.49 |
| 05 | 1 | 959.940 | 0.159 | 959.940 | 17.49 |
| 10 | 1 | 958.089 | 0.130 | 958.089 | 17.49 |
| 10 | 1 | 959.079 | 0.132 | 959.079 | 17.49 |
| 10 | 1 | 957.864 | 0.145 | 957.864 | 17.49 |
| 10 | 1 | 959.112 | 0.141 | 959.112 | 17.49 |
| 10 | 1 | 958.334 | 0.158 | 958.334 | 17.49 |
| 10 | 1 | 959.117 | 0.139 | 959.117 | 17.49 |
| 10 | 1 | 959.684 | 0.137 | 959.684 | 17.49 |
| 10 | 1 | 958.236 | 0.150 | 958.236 | 17.49 |
| 10 | 1 | 959.283 | 0.150 | 959.283 | 17.49 |
| 10 | 1 | 958.003 | 0.194 | 958.003 | 17.49 |
| 10 | 1 | 958.700 | 0.152 | 958.700 | 17.49 |
| 10 | 1 | 959.671 | 0.168 | 959.671 | 17.49 |
| 11 | 1 | 958.951 | 0.126 | 958.951 | 17.49 |
| 11 | 1 | 958.563 | 0.109 | 958.563 | 17.49 |
| 11 | 1 | 958.449 | 0.100 | 958.449 | 17.49 |
| 11 | 1 | 959.652 | 0.090 | 959.652 | 17.49 |
| 11 | 1 | 959.665 | 0.097 | 959.665 | 17.49 |
| 11 | 1 | 959.640 | 0.105 | 959.640 | 17.49 |
| 11 | 1 | 959.096 | 0.119 | 959.096 | 17.49 |
| 11 | 1 | 959.085 | 0.107 | 959.085 | 17.49 |
| 11 | 1 | 959.335 | 0.112 | 959.335 | 17.49 |
| 11 | 1 | 959.257 | 0.100 | 959.257 | 17.49 |
| 11 | 1 | 959.665 | 0.141 | 959.665 | 17.49 |
| 11 | 1 | 960.027 | 0.153 | 960.027 | 17.49 |
| 11 | 1 | 959.065 | 0.120 | 959.065 | 17.49 |
| 11 | 1 | 958.700 | 0.119 | 958.700 | 17.49 |
| 11 | 1 | 958.753 | 0.121 | 958.753 | 17.49 |
| 17 | 1 | 959.354 | 0.118 | 959.354 | 17.49 |
| 17 | 1 | 958.125 | 0.147 | 958.125 | 17.49 |
| 17 | 1 | 959.737 | 0.108 | 959.737 | 17.49 |
| 17 | 1 | 958.869 | 0.126 | 958.869 | 17.49 |
| 17 | 1 | 959.443 | 0.121 | 959.443 | 17.49 |
| 17 | 1 | 959.465 | 0.127 | 959.465 | 17.49 |
| 17 | 1 | 959.705 | 0.134 | 959.705 | 17.49 |
| 17 | 1 | 959.329 | 0.130 | 959.329 | 17.49 |
| 17 | 1 | 959.119 | 0.153 | 959.119 | 17.49 |
| 17 | 1 | 958.588 | 0.149 | 958.588 | 17.49 |
| 17 | 1 | 958.439 | 0.134 | 958.439 | 17.49 |
| 17 | 1 | 959.757 | 0.136 | 959.757 | 17.49 |
| 17 | 1 | 959.621 | 0.192 | 959.621 | 17.49 |
| 18 | 1 | 958.559 | 0.088 | 958.559 | 17.49 |
| 18 | 1 | 959.762 | 0.106 | 959.762 | 17.49 |
| 18 | 1 | 959.752 | 0.095 | 959.752 | 17.49 |
| 18 | 1 | 959.901 | 0.103 | 959.901 | 17.49 |
| 18 | 1 | 959.719 | 0.129 | 959.719 | 17.49 |
| 18 | 1 | 958.784 | 0.131 | 958.784 | 17.49 |
| 18 | 1 | 959.157 | 0.132 | 959.157 | 17.49 |
| 18 | 1 | 958.817 | 0.123 | 958.817 | 17.49 |
| 18 | 1 | 959.379 | 0.156 | 959.379 | 17.49 |
| 18 | 1 | 958.398 | 0.138 | 958.398 | 17.49 |
| 18 | 1 | 959.281 | 0.130 | 959.281 | 17.49 |
| 18 | 1 | 959.281 | 0.128 | 959.281 | 17.49 |
| 18 | 1 | 959.360 | 0.130 | 959.360 | 17.49 |
| 18 | 1 | 958.816 | 0.138 | 958.816 | 17.49 |

| 2008 - JANEIRO | | | | |
|---|---|---|---|---|
| D | L | SDB | ER | SDC | HL |
| 03 | 1 | 957.186 | 0.113 | 957.186 | 17.49 |
| 03 | 1 | 957.566 | 0.141 | 957.566 | 17.49 |
| 03 | 1 | 959.066 | 0.121 | 959.066 | 17.49 |
| 03 | 1 | 959.767 | 0.153 | 959.767 | 17.49 |
| 03 | 1 | 958.888 | 0.140 | 958.888 | 17.49 |
| 03 | 1 | 960.861 | 0.145 | 960.861 | 17.49 |

| 2008 - JANEIRO | | | | |
|---|---|---|---|---|
| D | L | SDB | ER | SDC | HL |
| 03 | 1 | 959.047 | 0.172 | 959.047 | 17.49 |
| 03 | 1 | 959.711 | 0.146 | 959.711 | 17.49 |
| 03 | 1 | 960.194 | 0.150 | 960.194 | 17.49 |
| 03 | 1 | 959.580 | 0.140 | 959.580 | 17.49 |
| 03 | 1 | 957.962 | 0.154 | 957.962 | 17.49 |
| 03 | 1 | 959.621 | 0.206 | 959.621 | 17.49 |
| 03 | 1 | 959.687 | 0.188 | 959.687 | 17.49 |
| 03 | 1 | 958.847 | 0.168 | 958.847 | 17.49 |
| 08 | 1 | 960.174 | 0.109 | 960.174 | 17.49 |
| 08 | 1 | 959.792 | 0.133 | 959.792 | 17.49 |
| 08 | 1 | 959.492 | 0.118 | 959.492 | 17.49 |
| 08 | 1 | 960.149 | 0.122 | 960.149 | 17.49 |
| 08 | 1 | 959.708 | 0.135 | 959.708 | 17.49 |
| 08 | 1 | 958.997 | 0.118 | 958.997 | 17.49 |
| 08 | 1 | 959.547 | 0.148 | 959.547 | 17.49 |
| 08 | 1 | 959.763 | 0.202 | 959.763 | 17.49 |
| 08 | 1 | 959.214 | 0.175 | 959.214 | 17.49 |
| 08 | 1 | 960.165 | 0.193 | 960.165 | 17.49 |
| 08 | 1 | 959.475 | 0.161 | 959.475 | 17.49 |
| 08 | 1 | 959.442 | 0.351 | 959.442 | 17.49 |
| 08 | 1 | 959.449 | 0.201 | 959.449 | 17.49 |
| 09 | 1 | 958.052 | 0.138 | 958.052 | 17.49 |
| 09 | 1 | 959.601 | 0.101 | 959.601 | 17.49 |
| 09 | 1 | 959.386 | 0.133 | 959.386 | 17.49 |
| 09 | 1 | 959.681 | 0.118 | 959.681 | 17.49 |
| 09 | 1 | 960.126 | 0.124 | 960.126 | 17.49 |
| 09 | 1 | 959.805 | 0.127 | 959.805 | 17.49 |
| 09 | 1 | 959.990 | 0.159 | 959.990 | 17.49 |
| 10 | 1 | 958.417 | 0.149 | 958.417 | 17.49 |
| 10 | 1 | 959.334 | 0.126 | 959.334 | 17.49 |
| 10 | 1 | 959.654 | 0.131 | 959.654 | 17.49 |
| 10 | 1 | 959.141 | 0.120 | 959.141 | 17.49 |
| 10 | 1 | 959.744 | 0.152 | 959.744 | 17.49 |
| 10 | 1 | 959.699 | 0.119 | 959.699 | 17.49 |
| 10 | 1 | 959.259 | 0.116 | 959.259 | 17.49 |
| 10 | 1 | 958.788 | 0.125 | 958.788 | 17.49 |
| 10 | 1 | 959.957 | 0.144 | 959.957 | 17.49 |
| 10 | 1 | 960.124 | 0.138 | 960.124 | 17.49 |
| 10 | 1 | 959.104 | 0.144 | 959.104 | 17.49 |
| 10 | 1 | 959.434 | 0.173 | 959.434 | 17.49 |
| 10 | 1 | 959.857 | 0.149 | 959.857 | 17.49 |
| 10 | 1 | 959.800 | 0.235 | 959.800 | 17.49 |
| 10 | 1 | 959.701 | 0.223 | 959.701 | 17.49 |
| 10 | 1 | 960.094 | 0.150 | 960.094 | 17.49 |
| 10 | 1 | 959.061 | 0.124 | 959.061 | 17.49 |
| 10 | 1 | 959.406 | 0.125 | 959.406 | 17.49 |
| 10 | 1 | 959.121 | 0.133 | 959.121 | 17.49 |
| 10 | 1 | 959.775 | 0.132 | 959.775 | 17.49 |
| 10 | 1 | 957.896 | 0.137 | 957.896 | 17.49 |
| 10 | 1 | 959.981 | 0.132 | 959.981 | 17.49 |
| 10 | 1 | 959.573 | 0.149 | 959.573 | 17.49 |
| 10 | 1 | 959.072 | 0.123 | 959.072 | 17.49 |
| 10 | 1 | 959.419 | 0.133 | 959.419 | 17.49 |
| 10 | 1 | 958.744 | 0.132 | 958.744 | 17.49 |
| 10 | 1 | 959.131 | 0.163 | 959.131 | 17.49 |
| 10 | 1 | 959.158 | 0.161 | 959.158 | 17.49 |
| 11 | 1 | 958.839 | 0.102 | 958.839 | 17.49 |
| 11 | 1 | 959.264 | 0.117 | 959.264 | 17.49 |
| 11 | 1 | 959.196 | 0.107 | 959.196 | 17.49 |
| 11 | 1 | 959.631 | 0.115 | 959.631 | 17.49 |
| 11 | 1 | 959.049 | 0.107 | 959.049 | 17.49 |
| 11 | 1 | 959.997 | 0.151 | 959.997 | 17.49 |
| 11 | 1 | 960.086 | 0.131 | 960.086 | 17.49 |
| 11 | 1 | 959.747 | 0.138 | 959.747 | 17.49 |
| 11 | 1 | 959.294 | 0.132 | 959.294 | 17.49 |
| 11 | 1 | 959.092 | 0.128 | 959.092 | 17.49 |
| 14 | 1 | 960.171 | 0.128 | 960.171 | 17.49 |
| 14 | 1 | 959.308 | 0.154 | 959.308 | 17.49 |
| 14 | 1 | 960.683 | 0.147 | 960.683 | 17.49 |
| 14 | 1 | 960.275 | 0.134 | 960.275 | 17.49 |
| 14 | 1 | 959.025 | 0.115 | 959.025 | 17.49 |

| 2008 - JANEIRO | | | | | |
|---|---|---|---|---|---|
| D | L | SDB | ER | SDC | HL |
| 14 | 1 | 959.183 | 0.130 | 959.183 | 17.49 |
| 14 | 1 | 959.205 | 0.164 | 959.205 | 17.49 |
| 14 | 1 | 959.043 | 0.140 | 959.043 | 17.49 |
| 14 | 1 | 959.278 | 0.163 | 959.278 | 17.49 |
| 14 | 1 | 959.368 | 0.143 | 959.368 | 17.49 |
| 14 | 1 | 959.583 | 0.172 | 959.583 | 17.49 |
| 14 | 1 | 959.554 | 0.147 | 959.554 | 17.49 |
| 16 | 1 | 959.220 | 0.125 | 959.220 | 17.49 |
| 16 | 1 | 959.148 | 0.130 | 959.148 | 17.49 |
| 16 | 1 | 959.915 | 0.245 | 959.915 | 17.49 |
| 16 | 1 | 959.803 | 0.116 | 959.803 | 17.49 |
| 16 | 1 | 960.368 | 0.143 | 960.368 | 17.49 |
| 16 | 1 | 959.277 | 0.141 | 959.277 | 17.49 |
| 16 | 1 | 960.103 | 0.125 | 960.103 | 17.49 |
| 16 | 1 | 959.321 | 0.152 | 959.321 | 17.49 |
| 16 | 1 | 959.225 | 0.151 | 959.225 | 17.49 |
| 16 | 1 | 959.763 | 0.134 | 959.763 | 17.49 |
| 16 | 1 | 959.414 | 0.150 | 959.414 | 17.49 |
| 16 | 1 | 959.518 | 0.127 | 959.518 | 17.49 |
| 16 | 1 | 959.973 | 0.163 | 959.973 | 17.49 |
| 16 | 1 | 959.260 | 0.150 | 959.260 | 17.49 |
| 16 | 1 | 959.982 | 0.113 | 959.982 | 17.49 |
| 16 | 1 | 959.686 | 0.165 | 959.686 | 17.49 |
| 16 | 1 | 960.189 | 0.130 | 960.189 | 17.49 |
| 16 | 1 | 960.058 | 0.134 | 960.058 | 17.49 |
| 16 | 1 | 959.664 | 0.168 | 959.664 | 17.49 |
| 16 | 1 | 959.749 | 0.150 | 959.749 | 17.49 |
| 16 | 1 | 959.705 | 0.128 | 959.705 | 17.49 |
| 16 | 1 | 959.311 | 0.174 | 959.311 | 17.49 |
| 16 | 1 | 958.759 | 0.144 | 958.759 | 17.49 |
| 16 | 1 | 960.075 | 0.157 | 960.075 | 17.49 |
| 16 | 1 | 958.534 | 0.162 | 958.534 | 17.49 |
| 16 | 1 | 958.970 | 0.209 | 958.970 | 17.49 |
| 16 | 1 | 959.476 | 0.165 | 959.476 | 17.49 |
| 16 | 1 | 959.515 | 0.170 | 959.515 | 17.49 |

| 2008 - FEVEREIRO | | | | | |
|---|---|---|---|---|---|
| D | L | SDB | ER | SDC | HL |
| 01 | 1 | 960.486 | 0.147 | 960.486 | 17.49 |
| 01 | 1 | 959.564 | 0.234 | 959.564 | 17.49 |
| 01 | 1 | 959.730 | 0.157 | 959.730 | 17.49 |
| 01 | 1 | 959.739 | 0.147 | 959.739 | 17.49 |
| 01 | 1 | 959.795 | 0.117 | 959.795 | 17.49 |
| 01 | 1 | 960.017 | 0.147 | 960.017 | 17.49 |
| 01 | 1 | 959.934 | 0.162 | 959.934 | 17.49 |
| 01 | 1 | 959.784 | 0.157 | 959.784 | 17.49 |
| 01 | 1 | 960.072 | 0.189 | 960.072 | 17.49 |
| 01 | 1 | 959.301 | 0.213 | 959.301 | 17.49 |
| 01 | 1 | 959.802 | 0.210 | 959.802 | 17.49 |
| 01 | 1 | 959.291 | 0.213 | 959.291 | 17.49 |
| 01 | 1 | 959.586 | 0.261 | 959.586 | 17.49 |
| 11 | 1 | 959.548 | 0.100 | 959.548 | 17.49 |
| 11 | 1 | 958.226 | 0.094 | 958.226 | 17.49 |
| 11 | 1 | 959.170 | 0.106 | 959.170 | 17.49 |
| 11 | 1 | 959.684 | 0.099 | 959.684 | 17.49 |
| 11 | 1 | 958.846 | 0.095 | 958.846 | 17.49 |
| 11 | 1 | 959.848 | 0.104 | 959.848 | 17.49 |
| 11 | 1 | 958.662 | 0.140 | 958.662 | 17.49 |
| 11 | 1 | 959.481 | 0.138 | 959.481 | 17.49 |
| 11 | 1 | 959.216 | 0.158 | 959.216 | 17.49 |
| 11 | 1 | 959.243 | 0.146 | 959.243 | 17.49 |
| 11 | 1 | 959.293 | 0.159 | 959.293 | 17.49 |
| 12 | 1 | 960.485 | 0.123 | 960.485 | 17.49 |
| 12 | 1 | 959.610 | 0.119 | 959.610 | 17.49 |
| 12 | 1 | 960.079 | 0.105 | 960.079 | 17.49 |
| 12 | 1 | 959.852 | 0.110 | 959.852 | 17.49 |
| 12 | 1 | 959.705 | 0.133 | 959.705 | 17.49 |
| 12 | 1 | 959.427 | 0.149 | 959.427 | 17.49 |
| 12 | 1 | 959.753 | 0.137 | 959.753 | 17.49 |
| 12 | 1 | 959.714 | 0.139 | 959.714 | 17.49 |

| 2008 - FEVEREIRO | | | | | |
|---|---|---|---|---|---|
| D | L | SDB | ER | SDC | HL |
| 12 | 1 | 959.505 | 0.147 | 959.505 | 17.49 |
| 12 | 1 | 960.006 | 0.159 | 960.006 | 17.49 |
| 12 | 1 | 959.126 | 0.190 | 959.126 | 17.49 |
| 14 | 1 | 958.340 | 0.095 | 958.340 | 17.49 |
| 14 | 1 | 958.901 | 0.106 | 958.901 | 17.49 |
| 14 | 1 | 960.104 | 0.116 | 960.104 | 17.49 |
| 14 | 1 | 959.054 | 0.113 | 959.054 | 17.49 |
| 14 | 1 | 958.675 | 0.159 | 958.675 | 17.49 |
| 14 | 1 | 959.754 | 0.133 | 959.754 | 17.49 |
| 14 | 1 | 958.167 | 0.244 | 958.167 | 17.49 |
| 14 | 1 | 959.787 | 0.123 | 959.787 | 17.49 |
| 14 | 1 | 959.158 | 0.123 | 959.158 | 17.49 |
| 14 | 1 | 958.671 | 0.135 | 958.671 | 17.49 |
| 14 | 1 | 958.971 | 0.140 | 958.971 | 17.49 |
| 14 | 1 | 958.929 | 0.166 | 958.929 | 17.49 |
| 15 | 1 | 957.803 | 0.119 | 957.803 | 17.49 |
| 15 | 1 | 958.575 | 0.114 | 958.575 | 17.49 |
| 15 | 1 | 958.449 | 0.125 | 958.449 | 17.49 |
| 15 | 1 | 959.856 | 0.098 | 959.856 | 17.49 |
| 15 | 1 | 958.697 | 0.186 | 958.697 | 17.49 |
| 15 | 1 | 959.313 | 0.079 | 959.313 | 17.49 |
| 15 | 1 | 959.289 | 0.101 | 959.289 | 17.49 |
| 15 | 1 | 959.178 | 0.121 | 959.178 | 17.49 |
| 15 | 1 | 958.386 | 0.136 | 958.386 | 17.49 |
| 15 | 1 | 959.299 | 0.108 | 959.299 | 17.49 |
| 15 | 1 | 959.955 | 0.101 | 959.955 | 17.49 |
| 15 | 1 | 958.947 | 0.131 | 958.947 | 17.49 |
| 18 | 1 | 957.579 | 0.113 | 957.579 | 17.49 |
| 18 | 1 | 959.670 | 0.114 | 959.670 | 17.49 |
| 18 | 1 | 959.168 | 0.103 | 959.168 | 17.49 |
| 18 | 1 | 958.375 | 0.102 | 958.375 | 17.49 |
| 18 | 1 | 958.668 | 0.134 | 958.668 | 17.49 |
| 18 | 1 | 959.554 | 0.111 | 959.554 | 17.49 |
| 18 | 1 | 960.019 | 0.114 | 960.019 | 17.49 |
| 18 | 1 | 959.876 | 0.134 | 959.876 | 17.49 |
| 18 | 1 | 958.878 | 0.119 | 958.878 | 17.49 |
| 18 | 1 | 959.281 | 0.131 | 959.281 | 17.49 |
| 18 | 1 | 959.032 | 0.126 | 959.032 | 17.49 |
| 18 | 1 | 960.092 | 0.166 | 960.092 | 17.49 |
| 18 | 1 | 959.311 | 0.156 | 959.311 | 17.49 |
| 19 | 1 | 958.025 | 0.113 | 958.025 | 17.49 |
| 19 | 1 | 958.853 | 0.103 | 958.853 | 17.49 |
| 19 | 1 | 959.911 | 0.118 | 959.911 | 17.49 |
| 19 | 1 | 959.385 | 0.129 | 959.385 | 17.49 |
| 19 | 1 | 959.834 | 0.095 | 959.834 | 17.49 |
| 19 | 1 | 960.074 | 0.107 | 960.074 | 17.49 |
| 19 | 1 | 958.784 | 0.136 | 958.784 | 17.49 |
| 19 | 1 | 959.592 | 0.106 | 959.592 | 17.49 |
| 19 | 1 | 959.196 | 0.131 | 959.196 | 17.49 |
| 19 | 1 | 960.266 | 0.150 | 960.266 | 17.49 |
| 19 | 1 | 959.099 | 0.146 | 959.099 | 17.49 |
| 19 | 1 | 959.130 | 0.129 | 959.130 | 17.49 |
| 19 | 1 | 959.598 | 0.147 | 959.598 | 17.49 |
| 19 | 1 | 959.499 | 0.154 | 959.499 | 17.49 |
| 19 | 1 | 959.903 | 0.165 | 959.903 | 17.49 |
| 20 | 1 | 958.131 | 0.113 | 958.131 | 17.49 |
| 20 | 1 | 958.862 | 0.129 | 958.862 | 17.49 |
| 20 | 1 | 959.380 | 0.135 | 959.380 | 17.49 |
| 20 | 1 | 958.702 | 0.119 | 958.702 | 17.49 |
| 20 | 1 | 958.988 | 0.138 | 958.988 | 17.49 |
| 20 | 1 | 958.836 | 0.142 | 958.836 | 17.49 |
| 20 | 1 | 959.583 | 0.150 | 959.583 | 17.49 |
| 20 | 1 | 959.366 | 0.142 | 959.366 | 17.49 |
| 20 | 1 | 958.037 | 0.209 | 958.037 | 17.49 |
| 27 | 1 | 959.842 | 0.107 | 959.842 | 17.49 |
| 27 | 1 | 959.035 | 0.128 | 959.035 | 17.49 |
| 27 | 1 | 960.051 | 0.155 | 960.051 | 17.49 |
| 27 | 1 | 960.688 | 0.144 | 960.688 | 17.49 |
| 27 | 1 | 959.265 | 0.145 | 959.265 | 17.49 |
| 27 | 1 | 958.708 | 0.135 | 958.708 | 17.49 |
| 27 | 1 | 959.439 | 0.168 | 959.439 | 17.49 |

| 2008 - FEVEREIRO | | | | | | 2008 - MARCO | | | | |
|---|---|---|---|---|---|---|---|---|---|---|
| D | L | SDB | ER | SDC | HL | D | L | SDB | ER | SDC | HL |
| 27 | 1 | 959.404 | 0.147 | 959.404 | 17.49 | 04 | 1 | 959.740 | 0.167 | 959.740 | 17.49 |
| 27 | 1 | 959.123 | 0.147 | 959.123 | 17.49 | 04 | 1 | 959.889 | 0.152 | 959.889 | 17.49 |
| 27 | 1 | 958.882 | 0.203 | 958.882 | 17.49 | 06 | 1 | 959.377 | 0.100 | 959.377 | 17.49 |
| 28 | 1 | 957.719 | 0.126 | 957.719 | 17.49 | 06 | 1 | 959.119 | 0.100 | 959.119 | 17.49 |
| 28 | 1 | 958.921 | 0.110 | 958.921 | 17.49 | 06 | 1 | 959.571 | 0.100 | 959.571 | 17.49 |
| 28 | 1 | 958.994 | 0.100 | 958.994 | 17.49 | 06 | 1 | 959.858 | 0.113 | 959.858 | 17.49 |
| 28 | 1 | 959.594 | 0.117 | 959.594 | 17.49 | 06 | 1 | 959.284 | 0.106 | 959.284 | 17.49 |
| 28 | 1 | 959.649 | 0.112 | 959.649 | 17.49 | 06 | 1 | 960.158 | 0.122 | 960.158 | 17.49 |
| 28 | 1 | 959.671 | 0.109 | 959.671 | 17.49 | 06 | 1 | 960.056 | 0.137 | 960.056 | 17.49 |
| 28 | 1 | 959.121 | 0.145 | 959.121 | 17.49 | 06 | 1 | 959.635 | 0.148 | 959.635 | 17.49 |
| 28 | 1 | 959.268 | 0.125 | 959.268 | 17.49 | 06 | 1 | 959.930 | 0.134 | 959.930 | 17.49 |
| 28 | 1 | 959.229 | 0.133 | 959.229 | 17.49 | 06 | 1 | 959.533 | 0.118 | 959.533 | 17.49 |
| 28 | 1 | 960.331 | 0.145 | 960.331 | 17.49 | 07 | 1 | 959.175 | 0.098 | 959.175 | 17.49 |
| 28 | 1 | 959.512 | 0.128 | 959.512 | 17.49 | 07 | 1 | 959.631 | 0.091 | 959.631 | 17.49 |
| 28 | 1 | 959.508 | 0.155 | 959.508 | 17.49 | 07 | 1 | 959.309 | 0.090 | 959.309 | 17.49 |
| 28 | 1 | 959.630 | 0.141 | 959.630 | 17.49 | 07 | 1 | 960.431 | 0.091 | 960.431 | 17.49 |
| 28 | 1 | 960.590 | 0.124 | 960.590 | 17.49 | 07 | 1 | 959.837 | 0.102 | 959.837 | 17.49 |
| 28 | 1 | 959.283 | 0.111 | 959.283 | 17.49 | 07 | 1 | 960.078 | 0.130 | 960.078 | 17.49 |
| 28 | 1 | 960.358 | 0.143 | 960.358 | 17.49 | 07 | 1 | 958.149 | 0.132 | 958.149 | 17.49 |
| 28 | 1 | 958.664 | 0.175 | 958.664 | 17.49 | 07 | 1 | 960.258 | 0.144 | 960.258 | 17.49 |
| 28 | 1 | 958.680 | 0.133 | 958.680 | 17.49 | 07 | 1 | 959.445 | 0.190 | 959.445 | 17.49 |
| 28 | 1 | 959.665 | 0.121 | 959.665 | 17.49 | 07 | 1 | 959.603 | 0.182 | 959.603 | 17.49 |
| 28 | 1 | 959.131 | 0.135 | 959.131 | 17.49 | 07 | 1 | 960.920 | 0.146 | 960.920 | 17.49 |
| 28 | 1 | 959.002 | 0.164 | 959.002 | 17.49 | 10 | 1 | 959.405 | 0.130 | 959.405 | 17.49 |
| 28 | 1 | 959.342 | 0.136 | 959.342 | 17.49 | 10 | 1 | 958.804 | 0.136 | 958.804 | 17.49 |
| 28 | 1 | 959.061 | 0.161 | 959.061 | 17.49 | 10 | 1 | 959.350 | 0.128 | 959.350 | 17.49 |
| 28 | 1 | 958.476 | 0.204 | 958.476 | 17.49 | 10 | 1 | 959.123 | 0.177 | 959.123 | 17.49 |
| 28 | 1 | 958.728 | 0.206 | 958.728 | 17.49 | 10 | 1 | 959.584 | 0.139 | 959.584 | 17.49 |
| | | | | | | 10 | 1 | 958.989 | 0.170 | 958.989 | 17.49 |
| | | | | | | 10 | 1 | 961.218 | 0.175 | 961.218 | 17.49 |
| | | 2008 - MARCO | | | | 10 | 1 | 958.877 | 0.157 | 958.877 | 17.49 |
| D | L | SDB | ER | SDC | HL | 10 | 1 | 959.640 | 0.142 | 959.640 | 17.49 |
| 03 | 1 | 957.702 | 0.081 | 957.702 | 17.49 | 10 | 1 | 959.901 | 0.133 | 959.901 | 17.49 |
| 03 | 1 | 959.182 | 0.088 | 959.182 | 17.49 | 10 | 1 | 959.153 | 0.131 | 959.153 | 17.49 |
| 03 | 1 | 959.773 | 0.098 | 959.773 | 17.49 | 10 | 1 | 959.102 | 0.165 | 959.102 | 17.49 |
| 03 | 1 | 959.730 | 0.133 | 959.730 | 17.49 | 10 | 1 | 958.437 | 0.149 | 958.437 | 17.49 |
| 03 | 1 | 959.941 | 0.110 | 959.941 | 17.49 | 10 | 1 | 959.136 | 0.146 | 959.136 | 17.49 |
| 03 | 1 | 960.272 | 0.089 | 960.272 | 17.49 | 10 | 1 | 959.262 | 0.163 | 959.262 | 17.49 |
| 03 | 1 | 959.827 | 0.079 | 959.827 | 17.49 | 10 | 1 | 958.999 | 0.179 | 958.999 | 17.49 |
| 03 | 1 | 960.134 | 0.120 | 960.134 | 17.49 | 10 | 1 | 959.690 | 0.168 | 959.690 | 17.49 |
| 03 | 1 | 959.012 | 0.115 | 959.012 | 17.49 | 10 | 1 | 958.986 | 0.167 | 958.986 | 17.49 |
| 03 | 1 | 959.424 | 0.116 | 959.424 | 17.49 | 11 | 1 | 959.329 | 0.095 | 959.329 | 17.49 |
| 03 | 1 | 960.358 | 0.122 | 960.358 | 17.49 | 11 | 1 | 959.173 | 0.097 | 959.173 | 17.49 |
| 03 | 1 | 959.428 | 0.115 | 959.428 | 17.49 | 11 | 1 | 960.057 | 0.095 | 960.057 | 17.49 |
| 03 | 1 | 959.097 | 0.121 | 959.097 | 17.49 | 11 | 1 | 959.907 | 0.122 | 959.907 | 17.49 |
| 03 | 1 | 961.959 | 0.133 | 961.959 | 17.49 | 11 | 1 | 960.020 | 0.160 | 960.020 | 17.49 |
| 03 | 1 | 960.577 | 0.099 | 960.577 | 17.49 | 11 | 1 | 960.210 | 0.124 | 960.210 | 17.49 |
| 03 | 1 | 959.020 | 0.097 | 959.020 | 17.49 | 11 | 1 | 959.562 | 0.128 | 959.562 | 17.49 |
| 03 | 1 | 959.333 | 0.101 | 959.333 | 17.49 | 11 | 1 | 959.257 | 0.123 | 959.257 | 17.49 |
| 03 | 1 | 958.820 | 0.106 | 958.820 | 17.49 | 11 | 1 | 959.669 | 0.155 | 959.669 | 17.49 |
| 03 | 1 | 959.275 | 0.115 | 959.275 | 17.49 | 20 | 1 | 960.091 | 0.139 | 960.091 | 17.49 |
| 03 | 1 | 959.573 | 0.114 | 959.573 | 17.49 | 20 | 1 | 958.590 | 0.162 | 958.590 | 17.49 |
| 03 | 1 | 958.980 | 0.114 | 958.980 | 17.49 | 20 | 1 | 958.643 | 0.136 | 958.643 | 17.49 |
| 03 | 1 | 959.654 | 0.116 | 959.654 | 17.49 | 20 | 1 | 958.963 | 0.155 | 958.963 | 17.49 |
| 03 | 1 | 959.633 | 0.133 | 959.633 | 17.49 | | | | | | |
| 03 | 1 | 958.728 | 0.146 | 958.728 | 17.49 | | | | | | |
| 03 | 1 | 959.792 | 0.149 | 959.792 | 17.49 | | | 2008 - ABRIL | | | |
| 03 | 1 | 959.044 | 0.149 | 959.044 | 17.49 | D | L | SDB | ER | SDC | HL |
| 04 | 1 | 958.154 | 0.122 | 958.154 | 17.49 | 11 | 1 | 959.339 | 0.116 | 959.339 | 17.49 |
| 04 | 1 | 958.357 | 0.110 | 958.357 | 17.49 | 11 | 1 | 959.490 | 0.146 | 959.490 | 17.49 |
| 04 | 1 | 959.381 | 0.133 | 959.381 | 17.49 | 11 | 1 | 959.135 | 0.148 | 959.135 | 17.49 |
| 04 | 1 | 958.900 | 0.112 | 958.900 | 17.49 | 11 | 1 | 959.170 | 0.099 | 959.170 | 17.49 |
| 04 | 1 | 959.716 | 0.113 | 959.716 | 17.49 | 11 | 1 | 959.292 | 0.129 | 959.292 | 17.49 |
| 04 | 1 | 959.734 | 0.110 | 959.734 | 17.49 | 11 | 1 | 959.334 | 0.145 | 959.334 | 17.49 |
| 04 | 1 | 959.567 | 0.145 | 959.567 | 17.49 | 11 | 1 | 959.032 | 0.134 | 959.032 | 17.49 |
| 04 | 1 | 959.403 | 0.125 | 959.403 | 17.49 | 11 | 1 | 959.126 | 0.156 | 959.126 | 17.49 |
| 04 | 1 | 960.101 | 0.141 | 960.101 | 17.49 | 11 | 1 | 959.002 | 0.207 | 959.002 | 17.49 |
| 04 | 1 | 959.898 | 0.146 | 959.898 | 17.49 | 24 | 1 | 958.859 | 0.103 | 958.859 | 17.49 |
| 04 | 1 | 959.430 | 0.152 | 959.430 | 17.49 | 24 | 1 | 958.785 | 0.113 | 958.785 | 17.49 |
| 04 | 1 | 959.485 | 0.169 | 959.485 | 17.49 | 24 | 1 | 958.984 | 0.118 | 958.984 | 17.49 |
| 04 | 1 | 959.050 | 0.192 | 959.050 | 17.49 | 24 | 1 | 959.837 | 0.115 | 959.837 | 17.49 |

| 2008 - ABRIL | | | | | | 2008 - MAIO | | | | |
|---|---|---|---|---|---|---|---|---|---|---|
| D | L | SDB | ER | SDC | HL | D | L | SDB | ER | SDC | HL |
| 24 | 1 | 959.960 | 0.117 | 959.960 | 17.49 | 07 | 1 | 959.508 | 0.135 | 959.508 | 17.49 |
| 24 | 1 | 960.489 | 0.146 | 960.489 | 17.49 | 08 | 1 | 958.671 | 0.120 | 958.671 | 17.49 |
| 24 | 1 | 959.862 | 0.171 | 959.862 | 17.49 | 08 | 1 | 959.341 | 0.101 | 959.341 | 17.49 |
| 24 | 1 | 959.239 | 0.242 | 959.239 | 17.49 | 08 | 1 | 959.337 | 0.108 | 959.337 | 17.49 |
| 24 | 1 | 960.016 | 0.216 | 960.016 | 17.49 | 08 | 1 | 960.943 | 0.133 | 960.943 | 17.49 |
| 24 | 1 | 959.652 | 0.097 | 959.652 | 17.49 | 08 | 1 | 961.076 | 0.105 | 961.076 | 17.49 |
| 24 | 1 | 959.265 | 0.105 | 959.265 | 17.49 | 08 | 1 | 959.958 | 0.148 | 959.958 | 17.49 |
| 24 | 1 | 959.562 | 0.140 | 959.562 | 17.49 | 08 | 1 | 960.000 | 0.102 | 960.000 | 17.49 |
| 24 | 1 | 959.318 | 0.134 | 959.318 | 17.49 | 08 | 1 | 958.367 | 0.143 | 958.367 | 17.49 |
| 24 | 1 | 959.415 | 0.150 | 959.415 | 17.49 | 08 | 1 | 959.100 | 0.142 | 959.100 | 17.49 |
| 25 | 1 | 958.428 | 0.134 | 958.428 | 17.49 | 08 | 1 | 959.865 | 0.135 | 959.865 | 17.49 |
| 25 | 1 | 958.893 | 0.109 | 958.893 | 17.49 | 08 | 1 | 959.849 | 0.160 | 959.849 | 17.49 |
| 25 | 1 | 959.601 | 0.092 | 959.601 | 17.49 | 08 | 1 | 958.855 | 0.161 | 958.855 | 17.49 |
| 25 | 1 | 960.633 | 0.112 | 960.633 | 17.49 | 08 | 1 | 959.446 | 0.157 | 959.446 | 17.49 |
| 25 | 1 | 960.611 | 0.130 | 960.611 | 17.49 | 08 | 1 | 958.339 | 0.146 | 958.339 | 17.49 |
| 25 | 1 | 960.132 | 0.165 | 960.132 | 17.49 | 08 | 1 | 959.759 | 0.160 | 959.759 | 17.49 |
| 25 | 1 | 959.382 | 0.172 | 959.382 | 17.49 | 08 | 1 | 958.970 | 0.153 | 958.970 | 17.49 |
| 25 | 1 | 958.973 | 0.138 | 958.973 | 17.49 | 08 | 1 | 958.312 | 0.153 | 958.312 | 17.49 |
| 25 | 1 | 959.288 | 0.145 | 959.288 | 17.49 | 08 | 1 | 959.685 | 0.142 | 959.685 | 17.49 |
| 25 | 1 | 958.406 | 0.117 | 958.406 | 17.49 | 08 | 1 | 959.042 | 0.169 | 959.042 | 17.49 |
| 25 | 1 | 959.535 | 0.115 | 959.535 | 17.49 | 08 | 1 | 959.191 | 0.198 | 959.191 | 17.49 |
| 25 | 1 | 959.476 | 0.107 | 959.476 | 17.49 | 08 | 1 | 959.199 | 0.188 | 959.199 | 17.49 |
| 25 | 1 | 958.658 | 0.125 | 958.658 | 17.49 | 15 | 1 | 959.272 | 0.110 | 959.272 | 17.49 |
| 25 | 1 | 959.478 | 0.161 | 959.478 | 17.49 | 15 | 1 | 959.223 | 0.097 | 959.223 | 17.49 |
| 25 | 1 | 959.268 | 0.147 | 959.268 | 17.49 | 15 | 1 | 959.678 | 0.107 | 959.678 | 17.49 |
| 25 | 1 | 959.071 | 0.132 | 959.071 | 17.49 | 15 | 1 | 960.657 | 0.110 | 960.657 | 17.49 |
| 28 | 1 | 958.566 | 0.095 | 958.566 | 17.49 | 15 | 1 | 959.732 | 0.142 | 959.732 | 17.49 |
| 28 | 1 | 958.911 | 0.116 | 958.911 | 17.49 | 15 | 1 | 959.920 | 0.161 | 959.920 | 17.49 |
| 28 | 1 | 959.680 | 0.106 | 959.680 | 17.49 | 15 | 1 | 959.083 | 0.125 | 959.083 | 17.49 |
| 28 | 1 | 960.034 | 0.148 | 960.034 | 17.49 | 15 | 1 | 958.695 | 0.126 | 958.695 | 17.49 |
| 28 | 1 | 960.368 | 0.122 | 960.368 | 17.49 | 15 | 1 | 959.309 | 0.129 | 959.309 | 17.49 |
| 28 | 1 | 959.616 | 0.131 | 959.616 | 17.49 | 15 | 1 | 959.566 | 0.131 | 959.566 | 17.49 |
| 28 | 1 | 959.433 | 0.116 | 959.433 | 17.49 | 15 | 1 | 959.925 | 0.169 | 959.925 | 17.49 |
| 28 | 1 | 960.068 | 0.134 | 960.068 | 17.49 | 15 | 1 | 959.337 | 0.147 | 959.337 | 17.49 |
| 28 | 1 | 958.390 | 0.125 | 958.390 | 17.49 | 15 | 1 | 958.697 | 0.127 | 958.697 | 17.49 |
| 28 | 1 | 959.150 | 0.129 | 959.150 | 17.49 | 15 | 1 | 959.231 | 0.176 | 959.231 | 17.49 |
| 28 | 1 | 959.761 | 0.154 | 959.761 | 17.49 | 15 | 1 | 958.376 | 0.209 | 958.376 | 17.49 |
| 28 | 1 | 960.179 | 0.158 | 960.179 | 17.49 | 15 | 1 | 958.862 | 0.230 | 958.862 | 17.49 |
| 28 | 1 | 959.159 | 0.169 | 959.159 | 17.49 | 16 | 1 | 961.210 | 0.129 | 961.210 | 17.49 |
| 28 | 1 | 959.549 | 0.146 | 959.549 | 17.49 | 16 | 1 | 959.914 | 0.114 | 959.914 | 17.49 |
| 28 | 1 | 958.740 | 0.132 | 958.740 | 17.49 | 16 | 1 | 959.370 | 0.114 | 959.370 | 17.49 |
| 28 | 1 | 958.718 | 0.105 | 958.718 | 17.49 | 16 | 1 | 959.934 | 0.125 | 959.934 | 17.49 |
| 28 | 1 | 959.327 | 0.133 | 959.327 | 17.49 | 16 | 1 | 959.768 | 0.144 | 959.768 | 17.49 |
| 28 | 1 | 959.254 | 0.159 | 959.254 | 17.49 | 16 | 1 | 959.727 | 0.177 | 959.727 | 17.49 |
| 28 | 1 | 959.334 | 0.142 | 959.334 | 17.49 | 20 | 1 | 959.623 | 0.125 | 959.623 | 17.49 |
| 28 | 1 | 960.220 | 0.146 | 960.220 | 17.49 | 20 | 1 | 959.594 | 0.113 | 959.594 | 17.49 |
| 28 | 1 | 959.268 | 0.149 | 959.268 | 17.49 | 20 | 1 | 959.914 | 0.115 | 959.914 | 17.49 |
| 28 | 1 | 958.888 | 0.159 | 958.888 | 17.49 | 20 | 1 | 959.668 | 0.117 | 959.668 | 17.49 |
| 28 | 1 | 959.362 | 0.154 | 959.362 | 17.49 | 20 | 1 | 959.699 | 0.106 | 959.699 | 17.49 |
| 28 | 1 | 959.319 | 0.211 | 959.319 | 17.49 | 20 | 1 | 959.863 | 0.126 | 959.863 | 17.49 |
| 29 | 1 | 959.134 | 0.104 | 959.134 | 17.49 | 20 | 1 | 959.822 | 0.120 | 959.822 | 17.49 |
| 29 | 1 | 959.346 | 0.142 | 959.346 | 17.49 | 20 | 1 | 960.285 | 0.112 | 960.285 | 17.49 |
| 29 | 1 | 959.902 | 0.093 | 959.902 | 17.49 | 20 | 1 | 960.972 | 0.139 | 960.972 | 17.49 |
| 29 | 1 | 959.595 | 0.118 | 959.595 | 17.49 | 20 | 1 | 960.641 | 0.162 | 960.641 | 17.49 |
| 29 | 1 | 959.867 | 0.111 | 959.867 | 17.49 | 20 | 1 | 959.677 | 0.146 | 959.677 | 17.49 |
| 29 | 1 | 959.396 | 0.099 | 959.396 | 17.49 | 20 | 1 | 959.869 | 0.155 | 959.869 | 17.49 |
| 29 | 1 | 960.211 | 0.095 | 960.211 | 17.49 | 20 | 1 | 958.805 | 0.134 | 958.805 | 17.49 |
| 29 | 1 | 959.929 | 0.103 | 959.929 | 17.49 | 20 | 1 | 959.356 | 0.127 | 959.356 | 17.49 |
| 29 | 1 | 960.176 | 0.117 | 960.176 | 17.49 | 20 | 1 | 959.521 | 0.156 | 959.521 | 17.49 |
| 29 | 1 | 960.512 | 0.102 | 960.512 | 17.49 | 20 | 1 | 959.330 | 0.152 | 959.330 | 17.49 |
| | | | | | | 20 | 1 | 959.271 | 0.170 | 959.271 | 17.49 |
| | | | | | | 20 | 1 | 959.678 | 0.196 | 959.678 | 17.49 |
| | | 2008 - MAIO | | | | 20 | 1 | 958.786 | 0.187 | 958.786 | 17.49 |
| D | L | SDB | ER | SDC | HL | 20 | 1 | 958.560 | 0.186 | 958.560 | 17.49 |
| 07 | 1 | 960.166 | 0.140 | 960.166 | 17.49 | 21 | 1 | 958.811 | 0.086 | 958.811 | 17.49 |
| 07 | 1 | 958.695 | 0.143 | 958.695 | 17.49 | 21 | 1 | 958.968 | 0.110 | 958.968 | 17.49 |
| 07 | 1 | 958.400 | 0.172 | 958.400 | 17.49 | 21 | 1 | 959.502 | 0.126 | 959.502 | 17.49 |
| 07 | 1 | 959.122 | 0.136 | 959.122 | 17.49 | 21 | 1 | 959.471 | 0.141 | 959.471 | 17.49 |
| 07 | 1 | 959.099 | 0.113 | 959.099 | 17.49 | 21 | 1 | 959.307 | 0.107 | 959.307 | 17.49 |
| 07 | 1 | 959.565 | 0.129 | 959.565 | 17.49 | 21 | 1 | 960.304 | 0.149 | 960.304 | 17.49 |
| 07 | 1 | 959.351 | 0.136 | 959.351 | 17.49 | 21 | 1 | 960.165 | 0.135 | 960.165 | 17.49 |

| 2008 - MAIO | | | | |
|---|---|---|---|---|
| D  L | SDB | ER | SDC | HL |
| 21  1 | 960.562 | 0.131 | 960.562 | 17.49 |
| 21  1 | 959.895 | 0.139 | 959.895 | 17.49 |
| 21  1 | 960.462 | 0.153 | 960.462 | 17.49 |
| 21  1 | 958.994 | 0.133 | 958.994 | 17.49 |
| 21  1 | 959.616 | 0.146 | 959.616 | 17.49 |
| 21  1 | 959.785 | 0.158 | 959.785 | 17.49 |
| 21  1 | 959.952 | 0.150 | 959.952 | 17.49 |
| 21  1 | 959.088 | 0.135 | 959.088 | 17.49 |
| 21  1 | 959.327 | 0.137 | 959.327 | 17.49 |
| 21  1 | 959.155 | 0.148 | 959.155 | 17.49 |

| 2008 - JUNHO | | | | |
|---|---|---|---|---|
| D  L | SDB | ER | SDC | HL |
| 03  1 | 959.040 | 0.123 | 959.040 | 17.49 |
| 03  1 | 959.746 | 0.114 | 959.746 | 17.49 |
| 03  1 | 960.043 | 0.135 | 960.043 | 17.49 |
| 03  1 | 960.709 | 0.154 | 960.709 | 17.49 |
| 03  1 | 960.088 | 0.126 | 960.088 | 17.49 |
| 03  1 | 959.962 | 0.148 | 959.962 | 17.49 |
| 03  1 | 960.161 | 0.124 | 960.161 | 17.49 |
| 03  1 | 960.411 | 0.159 | 960.411 | 17.49 |
| 03  1 | 959.502 | 0.155 | 959.502 | 17.49 |
| 03  1 | 959.720 | 0.159 | 959.720 | 17.49 |
| 03  1 | 958.885 | 0.158 | 958.885 | 17.49 |
| 03  1 | 959.409 | 0.153 | 959.409 | 17.49 |
| 03  1 | 959.907 | 0.162 | 959.907 | 17.49 |
| 03  1 | 958.863 | 0.148 | 958.863 | 17.49 |
| 03  1 | 958.971 | 0.195 | 958.971 | 17.49 |
| 06  1 | 959.589 | 0.129 | 959.589 | 17.49 |
| 06  1 | 959.206 | 0.112 | 959.206 | 17.49 |
| 06  1 | 959.040 | 0.117 | 959.040 | 17.49 |
| 06  1 | 959.432 | 0.113 | 959.432 | 17.49 |
| 06  1 | 959.644 | 0.130 | 959.644 | 17.49 |
| 06  1 | 959.486 | 0.138 | 959.486 | 17.49 |
| 06  1 | 959.200 | 0.175 | 959.200 | 17.49 |
| 06  1 | 959.884 | 0.196 | 959.884 | 17.49 |
| 09  1 | 957.987 | 0.161 | 957.987 | 17.49 |
| 09  1 | 959.745 | 0.146 | 959.745 | 17.49 |
| 09  1 | 958.880 | 0.169 | 958.880 | 17.49 |
| 09  1 | 959.748 | 0.113 | 959.748 | 17.49 |
| 09  1 | 958.966 | 0.154 | 958.966 | 17.49 |
| 09  1 | 958.637 | 0.198 | 958.637 | 17.49 |
| 12  1 | 959.578 | 0.112 | 959.578 | 17.49 |
| 12  1 | 959.659 | 0.106 | 959.659 | 17.49 |
| 12  1 | 960.685 | 0.110 | 960.685 | 17.49 |
| 12  1 | 960.867 | 0.098 | 960.867 | 17.49 |
| 12  1 | 961.029 | 0.133 | 961.029 | 17.49 |
| 12  1 | 961.267 | 0.137 | 961.267 | 17.49 |
| 12  1 | 960.565 | 0.115 | 960.565 | 17.49 |
| 12  1 | 961.477 | 0.121 | 961.477 | 17.49 |
| 12  1 | 959.542 | 0.127 | 959.542 | 17.49 |
| 12  1 | 959.229 | 0.103 | 959.229 | 17.49 |
| 12  1 | 959.003 | 0.155 | 959.003 | 17.49 |
| 12  1 | 958.369 | 0.143 | 958.369 | 17.49 |
| 12  1 | 959.298 | 0.131 | 959.298 | 17.49 |
| 12  1 | 958.446 | 0.149 | 958.446 | 17.49 |
| 12  1 | 958.887 | 0.185 | 958.887 | 17.49 |
| 12  1 | 958.806 | 0.188 | 958.806 | 17.49 |
| 18  1 | 960.143 | 0.133 | 960.143 | 17.49 |
| 18  1 | 960.062 | 0.123 | 960.062 | 17.49 |
| 18  1 | 959.607 | 0.140 | 959.607 | 17.49 |
| 18  1 | 960.946 | 0.117 | 960.946 | 17.49 |
| 18  1 | 960.041 | 0.115 | 960.041 | 17.49 |
| 18  1 | 961.036 | 0.127 | 961.036 | 17.49 |
| 18  1 | 961.715 | 0.141 | 961.715 | 17.49 |
| 18  1 | 961.501 | 0.137 | 961.501 | 17.49 |
| 18  1 | 959.533 | 0.205 | 959.533 | 17.49 |
| 18  1 | 958.605 | 0.134 | 958.605 | 17.49 |
| 18  1 | 959.087 | 0.177 | 959.087 | 17.49 |
| 18  1 | 959.002 | 0.209 | 959.002 | 17.49 |

| 2008 - JUNHO | | | | |
|---|---|---|---|---|
| D  L | SDB | ER | SDC | HL |
| 18  1 | 959.215 | 0.223 | 959.215 | 17.49 |
| 19  1 | 959.430 | 0.101 | 959.430 | 17.49 |
| 19  1 | 960.030 | 0.084 | 960.030 | 17.49 |
| 19  1 | 960.337 | 0.119 | 960.337 | 17.49 |
| 19  1 | 960.758 | 0.117 | 960.758 | 17.49 |
| 19  1 | 960.866 | 0.101 | 960.866 | 17.49 |
| 19  1 | 961.239 | 0.101 | 961.239 | 17.49 |
| 19  1 | 961.643 | 0.107 | 961.643 | 17.49 |
| 19  1 | 960.927 | 0.111 | 960.927 | 17.49 |
| 19  1 | 959.066 | 0.124 | 959.066 | 17.49 |
| 19  1 | 959.464 | 0.108 | 959.464 | 17.49 |
| 19  1 | 958.784 | 0.116 | 958.784 | 17.49 |
| 19  1 | 958.974 | 0.151 | 958.974 | 17.49 |
| 19  1 | 958.075 | 0.155 | 958.075 | 17.49 |
| 19  1 | 958.677 | 0.208 | 958.677 | 17.49 |
| 20  1 | 959.963 | 0.122 | 959.963 | 17.49 |
| 20  1 | 960.932 | 0.117 | 960.932 | 17.49 |
| 20  1 | 959.637 | 0.121 | 959.637 | 17.49 |
| 20  1 | 960.734 | 0.107 | 960.734 | 17.49 |
| 20  1 | 960.131 | 0.107 | 960.131 | 17.49 |
| 20  1 | 960.367 | 0.146 | 960.367 | 17.49 |
| 20  1 | 958.486 | 0.133 | 958.486 | 17.49 |
| 20  1 | 959.140 | 0.131 | 959.140 | 17.49 |
| 20  1 | 959.213 | 0.149 | 959.213 | 17.49 |
| 20  1 | 959.457 | 0.186 | 959.457 | 17.49 |
| 20  1 | 958.586 | 0.178 | 958.586 | 17.49 |
| 26  1 | 958.062 | 0.139 | 958.062 | 17.49 |
| 26  1 | 959.003 | 0.150 | 959.003 | 17.49 |
| 26  1 | 959.290 | 0.152 | 959.290 | 17.49 |
| 26  1 | 958.303 | 0.141 | 958.303 | 17.49 |
| 26  1 | 958.478 | 0.167 | 958.478 | 17.49 |
| 27  1 | 959.218 | 0.155 | 959.218 | 17.49 |
| 27  1 | 958.679 | 0.147 | 958.679 | 17.49 |
| 27  1 | 959.834 | 0.174 | 959.834 | 17.49 |
| 30  1 | 960.332 | 0.156 | 960.332 | 17.49 |
| 30  1 | 959.139 | 0.139 | 959.139 | 17.49 |
| 30  1 | 957.821 | 0.162 | 957.821 | 17.49 |
| 30  1 | 959.737 | 0.125 | 959.737 | 17.49 |
| 30  1 | 959.304 | 0.123 | 959.304 | 17.49 |
| 30  1 | 959.209 | 0.145 | 959.209 | 17.49 |
| 30  1 | 958.305 | 0.143 | 958.305 | 17.49 |
| 30  1 | 959.192 | 0.162 | 959.192 | 17.49 |
| 30  1 | 959.137 | 0.170 | 959.137 | 17.49 |

| 2008 - JULHO | | | | |
|---|---|---|---|---|
| D  L | SDB | ER | SDC | HL |
| 01  1 | 959.577 | 0.127 | 959.577 | 17.49 |
| 01  1 | 959.428 | 0.102 | 959.428 | 17.49 |
| 01  1 | 960.783 | 0.117 | 960.783 | 17.49 |
| 01  1 | 960.975 | 0.124 | 960.975 | 17.49 |
| 01  1 | 960.411 | 0.109 | 960.411 | 17.49 |
| 01  1 | 959.269 | 0.118 | 959.269 | 17.49 |
| 01  1 | 958.830 | 0.180 | 958.830 | 17.49 |
| 01  1 | 958.926 | 0.128 | 958.926 | 17.49 |
| 01  1 | 959.523 | 0.137 | 959.523 | 17.49 |
| 01  1 | 958.278 | 0.154 | 958.278 | 17.49 |
| 01  1 | 959.074 | 0.167 | 959.074 | 17.49 |
| 01  1 | 959.438 | 0.180 | 959.438 | 17.49 |
| 01  1 | 959.228 | 0.211 | 959.228 | 17.49 |
| 07  1 | 957.670 | 0.138 | 957.670 | 17.49 |
| 07  1 | 959.066 | 0.150 | 959.066 | 17.49 |
| 07  1 | 959.062 | 0.152 | 959.062 | 17.49 |
| 07  1 | 958.839 | 0.125 | 958.839 | 17.49 |
| 07  1 | 959.614 | 0.156 | 959.614 | 17.49 |
| 07  1 | 959.890 | 0.174 | 959.890 | 17.49 |
| 07  1 | 958.953 | 0.138 | 958.953 | 17.49 |
| 07  1 | 958.656 | 0.193 | 958.656 | 17.49 |
| 08  1 | 959.866 | 0.096 | 959.866 | 17.49 |
| 08  1 | 958.820 | 0.126 | 958.820 | 17.49 |
| 08  1 | 959.911 | 0.126 | 959.911 | 17.49 |

| 2008 - JULHO | | | | | | 2008 - JULHO | | | | |
|---|---|---|---|---|---|---|---|---|---|---|
| D | L | SDB | ER | SDC | HL | D | L | SDB | ER | SDC | HL |
| 08 | 1 | 960.312 | 0.126 | 960.312 | 17.49 | 18 | 1 | 958.942 | 0.120 | 958.942 | 17.49 |
| 08 | 1 | 960.744 | 0.106 | 960.744 | 17.49 | 18 | 1 | 959.300 | 0.114 | 959.300 | 17.49 |
| 08 | 1 | 960.518 | 0.096 | 960.518 | 17.49 | 18 | 1 | 958.839 | 0.129 | 958.839 | 17.49 |
| 08 | 1 | 959.309 | 0.146 | 959.309 | 17.49 | 18 | 1 | 958.979 | 0.153 | 958.979 | 17.49 |
| 08 | 1 | 958.373 | 0.172 | 958.373 | 17.49 | 18 | 1 | 958.820 | 0.130 | 958.820 | 17.49 |
| 08 | 1 | 958.797 | 0.155 | 958.797 | 17.49 | 18 | 1 | 958.523 | 0.158 | 958.523 | 17.49 |
| 08 | 1 | 959.543 | 0.162 | 959.543 | 17.49 | 18 | 1 | 958.737 | 0.136 | 958.737 | 17.49 |
| 08 | 1 | 959.444 | 0.151 | 959.444 | 17.49 | 18 | 1 | 959.505 | 0.149 | 959.505 | 17.49 |
| 08 | 1 | 959.875 | 0.163 | 959.875 | 17.49 | 18 | 1 | 959.338 | 0.154 | 959.338 | 17.49 |
| 08 | 1 | 958.194 | 0.217 | 958.194 | 17.49 | 18 | 1 | 959.007 | 0.150 | 959.007 | 17.49 |
| 08 | 1 | 958.726 | 0.164 | 958.726 | 17.49 | 18 | 1 | 959.246 | 0.149 | 959.246 | 17.49 |
| 09 | 1 | 960.281 | 0.091 | 960.281 | 17.49 | 21 | 1 | 959.952 | 0.125 | 959.952 | 17.49 |
| 09 | 1 | 958.912 | 0.113 | 958.912 | 17.49 | 21 | 1 | 958.480 | 0.124 | 958.480 | 17.49 |
| 09 | 1 | 959.252 | 0.106 | 959.252 | 17.49 | 21 | 1 | 958.616 | 0.148 | 958.616 | 17.49 |
| 09 | 1 | 959.216 | 0.102 | 959.216 | 17.49 | 21 | 1 | 960.989 | 0.172 | 960.989 | 17.49 |
| 09 | 1 | 958.839 | 0.126 | 958.839 | 17.49 | 21 | 1 | 959.472 | 0.155 | 959.472 | 17.49 |
| 09 | 1 | 958.648 | 0.104 | 958.648 | 17.49 | 21 | 1 | 958.625 | 0.157 | 958.625 | 17.49 |
| 09 | 1 | 959.220 | 0.135 | 959.220 | 17.49 | 21 | 1 | 958.435 | 0.149 | 958.435 | 17.49 |
| 09 | 1 | 958.601 | 0.151 | 958.601 | 17.49 | 21 | 1 | 959.797 | 0.135 | 959.797 | 17.49 |
| 14 | 1 | 959.149 | 0.163 | 959.149 | 17.49 | 21 | 1 | 959.906 | 0.181 | 959.906 | 17.49 |
| 14 | 1 | 959.637 | 0.146 | 959.637 | 17.49 | 21 | 1 | 958.295 | 0.161 | 958.295 | 17.49 |
| 14 | 1 | 958.844 | 0.112 | 958.844 | 17.49 | 21 | 1 | 958.587 | 0.160 | 958.587 | 17.49 |
| 14 | 1 | 959.139 | 0.119 | 959.139 | 17.49 | 21 | 1 | 959.305 | 0.201 | 959.305 | 17.49 |
| 14 | 1 | 959.251 | 0.118 | 959.251 | 17.49 | 22 | 1 | 960.290 | 0.104 | 960.290 | 17.49 |
| 14 | 1 | 958.795 | 0.143 | 958.795 | 17.49 | 22 | 1 | 960.334 | 0.094 | 960.334 | 17.49 |
| 14 | 1 | 959.098 | 0.160 | 959.098 | 17.49 | 22 | 1 | 960.485 | 0.093 | 960.485 | 17.49 |
| 14 | 1 | 959.175 | 0.151 | 959.175 | 17.49 | 22 | 1 | 959.915 | 0.098 | 959.915 | 17.49 |
| 14 | 1 | 959.127 | 0.151 | 959.127 | 17.49 | 22 | 1 | 960.962 | 0.077 | 960.962 | 17.49 |
| 14 | 1 | 959.877 | 0.153 | 959.877 | 17.49 | 22 | 1 | 960.883 | 0.098 | 960.883 | 17.49 |
| 14 | 1 | 959.039 | 0.137 | 959.039 | 17.49 | 22 | 1 | 961.561 | 0.089 | 961.561 | 17.49 |
| 15 | 1 | 958.111 | 0.147 | 958.111 | 17.49 | 22 | 1 | 961.344 | 0.092 | 961.344 | 17.49 |
| 15 | 1 | 958.976 | 0.159 | 958.976 | 17.49 | 22 | 1 | 960.583 | 0.083 | 960.583 | 17.49 |
| 15 | 1 | 960.005 | 0.173 | 960.005 | 17.49 | 22 | 1 | 960.644 | 0.091 | 960.644 | 17.49 |
| 15 | 1 | 959.311 | 0.145 | 959.311 | 17.49 | 22 | 1 | 958.209 | 0.149 | 958.209 | 17.49 |
| 15 | 1 | 959.060 | 0.132 | 959.060 | 17.49 | 22 | 1 | 958.297 | 0.114 | 958.297 | 17.49 |
| 15 | 1 | 959.083 | 0.142 | 959.083 | 17.49 | 22 | 1 | 959.407 | 0.151 | 959.407 | 17.49 |
| 15 | 1 | 959.028 | 0.172 | 959.028 | 17.49 | 22 | 1 | 959.704 | 0.141 | 959.704 | 17.49 |
| 15 | 1 | 959.811 | 0.148 | 959.811 | 17.49 | 22 | 1 | 959.587 | 0.138 | 959.587 | 17.49 |
| 15 | 1 | 959.406 | 0.144 | 959.406 | 17.49 | 22 | 1 | 960.044 | 0.147 | 960.044 | 17.49 |
| 15 | 1 | 958.528 | 0.197 | 958.528 | 17.49 | 22 | 1 | 959.160 | 0.128 | 959.160 | 17.49 |
| 15 | 1 | 958.166 | 0.187 | 958.166 | 17.49 | 22 | 1 | 958.607 | 0.143 | 958.607 | 17.49 |
| 16 | 1 | 958.757 | 0.152 | 958.757 | 17.49 | 22 | 1 | 959.030 | 0.145 | 959.030 | 17.49 |
| 16 | 1 | 959.573 | 0.156 | 959.573 | 17.49 | 22 | 1 | 959.484 | 0.160 | 959.484 | 17.49 |
| 16 | 1 | 958.879 | 0.111 | 958.879 | 17.49 | 23 | 1 | 959.480 | 0.106 | 959.480 | 17.49 |
| 16 | 1 | 958.925 | 0.160 | 958.925 | 17.49 | 23 | 1 | 958.997 | 0.116 | 958.997 | 17.49 |
| 16 | 1 | 959.304 | 0.151 | 959.304 | 17.49 | 23 | 1 | 959.857 | 0.094 | 959.857 | 17.49 |
| 16 | 1 | 958.690 | 0.174 | 958.690 | 17.49 | 23 | 1 | 958.488 | 0.109 | 958.488 | 17.49 |
| 16 | 1 | 958.178 | 0.158 | 958.178 | 17.49 | 23 | 1 | 960.637 | 0.122 | 960.637 | 17.49 |
| 16 | 1 | 957.893 | 0.134 | 957.893 | 17.49 | 23 | 1 | 958.383 | 0.096 | 958.383 | 17.49 |
| 16 | 1 | 959.871 | 0.133 | 959.871 | 17.49 | 23 | 1 | 959.363 | 0.126 | 959.363 | 17.49 |
| 16 | 1 | 958.984 | 0.147 | 958.984 | 17.49 | 23 | 1 | 958.595 | 0.174 | 958.595 | 17.49 |
| 16 | 1 | 959.245 | 0.134 | 959.245 | 17.49 | 23 | 1 | 958.776 | 0.268 | 958.776 | 17.49 |
| 17 | 1 | 959.750 | 0.136 | 959.750 | 17.49 | 23 | 1 | 959.229 | 0.159 | 959.229 | 17.49 |
| 17 | 1 | 959.524 | 0.134 | 959.524 | 17.49 | 23 | 1 | 959.759 | 0.137 | 959.759 | 17.49 |
| 17 | 1 | 960.228 | 0.132 | 960.228 | 17.49 | 23 | 1 | 959.151 | 0.173 | 959.151 | 17.49 |
| 17 | 1 | 960.436 | 0.125 | 960.436 | 17.49 | 25 | 1 | 959.975 | 0.105 | 959.975 | 17.49 |
| 17 | 1 | 959.707 | 0.130 | 959.707 | 17.49 | 25 | 1 | 958.667 | 0.128 | 958.667 | 17.49 |
| 17 | 1 | 960.080 | 0.164 | 960.080 | 17.49 | 25 | 1 | 959.257 | 0.161 | 959.257 | 17.49 |
| 17 | 1 | 960.939 | 0.134 | 960.939 | 17.49 | 25 | 1 | 959.507 | 0.133 | 959.507 | 17.49 |
| 17 | 1 | 960.527 | 0.107 | 960.527 | 17.49 | 25 | 1 | 958.322 | 0.159 | 958.322 | 17.49 |
| 17 | 1 | 959.037 | 0.140 | 959.037 | 17.49 | 25 | 1 | 959.090 | 0.136 | 959.090 | 17.49 |
| 17 | 1 | 958.875 | 0.123 | 958.875 | 17.49 | 25 | 1 | 958.745 | 0.124 | 958.745 | 17.49 |
| 17 | 1 | 959.542 | 0.113 | 959.542 | 17.49 | 25 | 1 | 959.628 | 0.166 | 959.628 | 17.49 |
| 17 | 1 | 958.918 | 0.143 | 958.918 | 17.49 | 25 | 1 | 958.077 | 0.142 | 958.077 | 17.49 |
| 17 | 1 | 958.460 | 0.155 | 958.460 | 17.49 | 25 | 1 | 959.135 | 0.148 | 959.135 | 17.49 |
| 17 | 1 | 958.965 | 0.159 | 958.965 | 17.49 | 25 | 1 | 958.392 | 0.113 | 958.392 | 17.49 |
| 17 | 1 | 958.386 | 0.130 | 958.386 | 17.49 | 29 | 1 | 959.881 | 0.110 | 959.881 | 17.49 |
| 17 | 1 | 958.648 | 0.188 | 958.648 | 17.49 | 29 | 1 | 959.877 | 0.100 | 959.877 | 17.49 |
| 17 | 1 | 959.220 | 0.146 | 959.220 | 17.49 | 29 | 1 | 960.907 | 0.114 | 960.907 | 17.49 |
| 17 | 1 | 959.161 | 0.137 | 959.161 | 17.49 | 29 | 1 | 961.296 | 0.112 | 961.296 | 17.49 |
| 17 | 1 | 959.261 | 0.178 | 959.261 | 17.49 | 29 | 1 | 960.717 | 0.114 | 960.717 | 17.49 |

| 2008 - JULHO | | | | | |
|---|---|---|---|---|---|
| D | L | SDB | ER | SDC | HL |
| 29 | 1 | 959.001 | 0.106 | 959.001 | 17.49 |
| 29 | 1 | 959.246 | 0.141 | 959.246 | 17.49 |
| 29 | 1 | 958.641 | 0.128 | 958.641 | 17.49 |
| 29 | 1 | 959.423 | 0.148 | 959.423 | 17.49 |
| 29 | 1 | 959.103 | 0.155 | 959.103 | 17.49 |
| 29 | 1 | 958.163 | 0.166 | 958.163 | 17.49 |
| 29 | 1 | 958.702 | 0.205 | 958.702 | 17.49 |
| 29 | 1 | 959.675 | 0.168 | 959.675 | 17.49 |
| 29 | 1 | 958.885 | 0.194 | 958.885 | 17.49 |
| 29 | 1 | 958.943 | 0.161 | 958.943 | 17.49 |
| 29 | 1 | 959.094 | 0.197 | 959.094 | 17.49 |
| 29 | 1 | 959.625 | 0.157 | 959.625 | 17.49 |
| 30 | 1 | 959.151 | 0.132 | 959.151 | 17.49 |
| 30 | 1 | 958.405 | 0.132 | 958.405 | 17.49 |
| 30 | 1 | 959.870 | 0.117 | 959.870 | 17.49 |
| 30 | 1 | 959.077 | 0.155 | 959.077 | 17.49 |
| 30 | 1 | 958.283 | 0.163 | 958.283 | 17.49 |
| 30 | 1 | 958.642 | 0.154 | 958.642 | 17.49 |
| 30 | 1 | 958.920 | 0.123 | 958.920 | 17.49 |
| 30 | 1 | 959.247 | 0.127 | 959.247 | 17.49 |
| 30 | 1 | 959.155 | 0.110 | 959.155 | 17.49 |
| 30 | 1 | 958.568 | 0.140 | 958.568 | 17.49 |
| 30 | 1 | 957.722 | 0.166 | 957.722 | 17.49 |

| 2008 - AGOSTO | | | | | |
|---|---|---|---|---|---|
| D | L | SDB | ER | SDC | HL |
| 01 | 1 | 959.390 | 0.097 | 959.390 | 17.49 |
| 01 | 1 | 959.402 | 0.128 | 959.402 | 17.49 |
| 01 | 1 | 958.782 | 0.131 | 958.782 | 17.49 |
| 01 | 1 | 959.627 | 0.107 | 959.627 | 17.49 |
| 01 | 1 | 959.457 | 0.107 | 959.457 | 17.49 |
| 01 | 1 | 958.828 | 0.174 | 958.828 | 17.49 |
| 01 | 1 | 959.645 | 0.139 | 959.645 | 17.49 |
| 01 | 1 | 959.581 | 0.129 | 959.581 | 17.49 |
| 01 | 1 | 959.400 | 0.103 | 959.400 | 17.49 |
| 01 | 1 | 958.954 | 0.118 | 958.954 | 17.49 |
| 01 | 1 | 958.026 | 0.132 | 958.026 | 17.49 |
| 05 | 1 | 961.055 | 0.100 | 961.055 | 17.49 |
| 05 | 1 | 961.165 | 0.111 | 961.165 | 17.49 |
| 05 | 1 | 960.681 | 0.148 | 960.681 | 17.49 |
| 05 | 1 | 961.529 | 0.137 | 961.529 | 17.49 |
| 05 | 1 | 960.244 | 0.139 | 960.244 | 17.49 |
| 05 | 1 | 960.747 | 0.121 | 960.747 | 17.49 |
| 05 | 1 | 961.806 | 0.126 | 961.806 | 17.49 |
| 05 | 1 | 959.414 | 0.160 | 959.414 | 17.49 |
| 05 | 1 | 958.981 | 0.116 | 958.981 | 17.49 |
| 05 | 1 | 959.040 | 0.137 | 959.040 | 17.49 |
| 05 | 1 | 959.131 | 0.148 | 959.131 | 17.49 |
| 05 | 1 | 959.416 | 0.126 | 959.416 | 17.49 |
| 05 | 1 | 958.315 | 0.151 | 958.315 | 17.49 |
| 05 | 1 | 958.873 | 0.182 | 958.873 | 17.49 |
| 05 | 1 | 958.602 | 0.206 | 958.602 | 17.49 |
| 07 | 1 | 959.548 | 0.153 | 959.548 | 17.49 |
| 07 | 1 | 960.324 | 0.132 | 960.324 | 17.49 |
| 07 | 1 | 961.256 | 0.139 | 961.256 | 17.49 |
| 07 | 1 | 960.134 | 0.131 | 960.134 | 17.49 |
| 07 | 1 | 960.883 | 0.120 | 960.883 | 17.49 |
| 07 | 1 | 961.401 | 0.170 | 961.401 | 17.49 |
| 07 | 1 | 959.609 | 0.182 | 959.609 | 17.49 |
| 07 | 1 | 959.668 | 0.166 | 959.668 | 17.49 |
| 07 | 1 | 958.813 | 0.111 | 958.813 | 17.49 |
| 07 | 1 | 959.316 | 0.133 | 959.316 | 17.49 |
| 07 | 1 | 959.460 | 0.163 | 959.460 | 17.49 |
| 07 | 1 | 959.542 | 0.117 | 959.542 | 17.49 |
| 07 | 1 | 959.292 | 0.123 | 959.292 | 17.49 |
| 07 | 1 | 959.725 | 0.123 | 959.725 | 17.49 |
| 07 | 1 | 959.210 | 0.142 | 959.210 | 17.49 |
| 07 | 1 | 958.730 | 0.138 | 958.730 | 17.49 |
| 12 | 1 | 959.257 | 0.157 | 959.257 | 17.49 |
| 12 | 1 | 960.016 | 0.131 | 960.016 | 17.49 |

| 2008 - AGOSTO | | | | | |
|---|---|---|---|---|---|
| D | L | SDB | ER | SDC | HL |
| 12 | 1 | 959.339 | 0.136 | 959.339 | 17.49 |
| 12 | 1 | 960.789 | 0.143 | 960.789 | 17.49 |
| 12 | 1 | 960.335 | 0.141 | 960.335 | 17.49 |
| 12 | 1 | 960.839 | 0.136 | 960.839 | 17.49 |
| 12 | 1 | 958.470 | 0.118 | 958.470 | 17.49 |
| 12 | 1 | 959.142 | 0.119 | 959.142 | 17.49 |
| 12 | 1 | 959.034 | 0.131 | 959.034 | 17.49 |
| 12 | 1 | 959.305 | 0.160 | 959.305 | 17.49 |
| 12 | 1 | 959.861 | 0.142 | 959.861 | 17.49 |
| 12 | 1 | 959.313 | 0.113 | 959.313 | 17.49 |
| 12 | 1 | 959.317 | 0.141 | 959.317 | 17.49 |
| 12 | 1 | 959.173 | 0.138 | 959.173 | 17.49 |
| 12 | 1 | 959.182 | 0.149 | 959.182 | 17.49 |
| 12 | 1 | 959.267 | 0.126 | 959.267 | 17.49 |
| 12 | 1 | 960.242 | 0.178 | 960.242 | 17.49 |
| 13 | 1 | 959.856 | 0.126 | 959.856 | 17.49 |
| 13 | 1 | 959.384 | 0.154 | 959.384 | 17.49 |
| 13 | 1 | 959.934 | 0.143 | 959.934 | 17.49 |
| 13 | 1 | 959.069 | 0.153 | 959.069 | 17.49 |
| 13 | 1 | 959.306 | 0.136 | 959.306 | 17.49 |
| 13 | 1 | 958.957 | 0.149 | 958.957 | 17.49 |
| 13 | 1 | 959.579 | 0.137 | 959.579 | 17.49 |
| 13 | 1 | 959.049 | 0.138 | 959.049 | 17.49 |
| 13 | 1 | 959.018 | 0.131 | 959.018 | 17.49 |
| 14 | 1 | 959.324 | 0.133 | 959.324 | 17.49 |
| 14 | 1 | 959.160 | 0.153 | 959.160 | 17.49 |
| 14 | 1 | 958.988 | 0.131 | 958.988 | 17.49 |
| 14 | 1 | 958.753 | 0.134 | 958.753 | 17.49 |
| 14 | 1 | 959.267 | 0.136 | 959.267 | 17.49 |
| 14 | 1 | 959.449 | 0.138 | 959.449 | 17.49 |
| 14 | 1 | 958.912 | 0.161 | 958.912 | 17.49 |
| 14 | 1 | 959.579 | 0.114 | 959.579 | 17.49 |
| 14 | 1 | 959.282 | 0.156 | 959.282 | 17.49 |
| 14 | 1 | 958.788 | 0.146 | 958.788 | 17.49 |
| 14 | 1 | 959.539 | 0.135 | 959.539 | 17.49 |
| 18 | 1 | 960.487 | 0.177 | 960.487 | 17.49 |
| 18 | 1 | 959.515 | 0.166 | 959.515 | 17.49 |
| 18 | 1 | 958.949 | 0.123 | 958.949 | 17.49 |
| 18 | 1 | 958.747 | 0.124 | 958.747 | 17.49 |
| 18 | 1 | 958.828 | 0.149 | 958.828 | 17.49 |
| 18 | 1 | 959.213 | 0.151 | 959.213 | 17.49 |
| 18 | 1 | 959.795 | 0.139 | 959.795 | 17.49 |
| 18 | 1 | 959.362 | 0.107 | 959.362 | 17.49 |
| 18 | 1 | 959.470 | 0.114 | 959.470 | 17.49 |
| 18 | 1 | 959.400 | 0.146 | 959.400 | 17.49 |
| 18 | 1 | 959.223 | 0.142 | 959.223 | 17.49 |
| 18 | 1 | 959.129 | 0.143 | 959.129 | 17.49 |
| 27 | 1 | 960.953 | 0.155 | 960.953 | 17.49 |
| 27 | 1 | 960.443 | 0.110 | 960.443 | 17.49 |
| 27 | 1 | 958.950 | 0.136 | 958.950 | 17.49 |
| 27 | 1 | 959.853 | 0.127 | 959.853 | 17.49 |
| 27 | 1 | 959.196 | 0.140 | 959.196 | 17.49 |
| 27 | 1 | 959.381 | 0.108 | 959.381 | 17.49 |
| 27 | 1 | 959.087 | 0.131 | 959.087 | 17.49 |
| 27 | 1 | 960.053 | 0.141 | 960.053 | 17.49 |
| 27 | 1 | 959.070 | 0.137 | 959.070 | 17.49 |
| 28 | 1 | 959.233 | 0.142 | 959.233 | 17.49 |
| 28 | 1 | 959.578 | 0.120 | 959.578 | 17.49 |
| 28 | 1 | 959.345 | 0.147 | 959.345 | 17.49 |
| 28 | 1 | 958.725 | 0.132 | 958.725 | 17.49 |
| 28 | 1 | 959.455 | 0.118 | 959.455 | 17.49 |
| 28 | 1 | 958.809 | 0.150 | 958.809 | 17.49 |
| 28 | 1 | 959.346 | 0.133 | 959.346 | 17.49 |
| 28 | 1 | 959.224 | 0.132 | 959.224 | 17.49 |
| 28 | 1 | 959.038 | 0.118 | 959.038 | 17.49 |
| 28 | 1 | 959.802 | 0.135 | 959.802 | 17.49 |
| 28 | 1 | 959.887 | 0.129 | 959.887 | 17.49 |
| 28 | 1 | 958.863 | 0.125 | 958.863 | 17.49 |

| 2008 - SETEMBRO | | | | | |
|---|---|---|---|---|---|
| D | L | SDB | ER | SDC | HL |
| 01 | 1 | 960.502 | 0.112 | 960.502 | 17.49 |
| 01 | 1 | 958.987 | 0.110 | 958.987 | 17.49 |
| 01 | 1 | 959.502 | 0.117 | 959.502 | 17.49 |
| 01 | 1 | 959.946 | 0.151 | 959.946 | 17.49 |
| 01 | 1 | 959.294 | 0.140 | 959.294 | 17.49 |
| 01 | 1 | 960.539 | 0.165 | 960.539 | 17.49 |
| 01 | 1 | 959.438 | 0.144 | 959.438 | 17.49 |
| 01 | 1 | 958.756 | 0.135 | 958.756 | 17.49 |
| 01 | 1 | 960.044 | 0.147 | 960.044 | 17.49 |
| 01 | 1 | 960.410 | 0.148 | 960.410 | 17.49 |
| 01 | 1 | 959.484 | 0.189 | 959.484 | 17.49 |
| 01 | 1 | 959.841 | 0.138 | 959.841 | 17.49 |
| 02 | 1 | 958.071 | 0.115 | 958.071 | 17.49 |
| 02 | 1 | 959.496 | 0.108 | 959.496 | 17.49 |
| 02 | 1 | 959.742 | 0.128 | 959.742 | 17.49 |
| 02 | 1 | 959.883 | 0.147 | 959.883 | 17.49 |
| 02 | 1 | 959.763 | 0.116 | 959.763 | 17.49 |
| 02 | 1 | 960.691 | 0.130 | 960.691 | 17.49 |
| 02 | 1 | 960.874 | 0.113 | 960.874 | 17.49 |
| 02 | 1 | 960.457 | 0.123 | 960.457 | 17.49 |
| 02 | 1 | 960.455 | 0.112 | 960.455 | 17.49 |
| 02 | 1 | 960.508 | 0.125 | 960.508 | 17.49 |
| 02 | 1 | 961.065 | 0.144 | 961.065 | 17.49 |
| 02 | 1 | 961.156 | 0.154 | 961.156 | 17.49 |
| 04 | 1 | 959.106 | 0.099 | 959.106 | 17.49 |
| 04 | 1 | 960.125 | 0.119 | 960.125 | 17.49 |
| 04 | 1 | 959.171 | 0.124 | 959.171 | 17.49 |
| 04 | 1 | 961.036 | 0.100 | 961.036 | 17.49 |
| 04 | 1 | 960.396 | 0.087 | 960.396 | 17.49 |
| 04 | 1 | 960.229 | 0.112 | 960.229 | 17.49 |
| 04 | 1 | 960.268 | 0.128 | 960.268 | 17.49 |
| 04 | 1 | 960.528 | 0.135 | 960.528 | 17.49 |
| 04 | 1 | 960.927 | 0.123 | 960.927 | 17.49 |
| 04 | 1 | 959.337 | 0.103 | 959.337 | 17.49 |
| 04 | 1 | 959.791 | 0.136 | 959.791 | 17.49 |
| 04 | 1 | 958.586 | 0.118 | 958.586 | 17.49 |
| 04 | 1 | 959.855 | 0.139 | 959.855 | 17.49 |
| 04 | 1 | 959.457 | 0.096 | 959.457 | 17.49 |
| 04 | 1 | 958.874 | 0.112 | 958.874 | 17.49 |
| 04 | 1 | 958.650 | 0.112 | 958.650 | 17.49 |
| 04 | 1 | 959.716 | 0.121 | 959.716 | 17.49 |
| 04 | 1 | 959.577 | 0.117 | 959.577 | 17.49 |
| 04 | 1 | 959.017 | 0.132 | 959.017 | 17.49 |
| 04 | 1 | 959.730 | 0.116 | 959.730 | 17.49 |
| 04 | 1 | 959.350 | 0.101 | 959.350 | 17.49 |
| 19 | 1 | 959.804 | 0.139 | 959.804 | 17.49 |
| 19 | 1 | 959.539 | 0.167 | 959.539 | 17.49 |
| 19 | 1 | 959.329 | 0.123 | 959.329 | 17.49 |
| 19 | 1 | 959.326 | 0.129 | 959.326 | 17.49 |
| 19 | 1 | 959.321 | 0.116 | 959.321 | 17.49 |
| 19 | 1 | 959.934 | 0.126 | 959.934 | 17.49 |
| 19 | 1 | 959.079 | 0.121 | 959.079 | 17.49 |
| 23 | 1 | 957.805 | 0.106 | 957.805 | 17.49 |
| 23 | 1 | 959.549 | 0.093 | 959.549 | 17.49 |
| 23 | 1 | 960.456 | 0.139 | 960.456 | 17.49 |
| 23 | 1 | 960.746 | 0.129 | 960.746 | 17.49 |
| 23 | 1 | 959.448 | 0.106 | 959.448 | 17.49 |
| 23 | 1 | 959.275 | 0.132 | 959.275 | 17.49 |
| 23 | 1 | 958.191 | 0.108 | 958.191 | 17.49 |
| 23 | 1 | 959.702 | 0.140 | 959.702 | 17.49 |
| 23 | 1 | 960.126 | 0.113 | 960.126 | 17.49 |
| 23 | 1 | 959.296 | 0.158 | 959.296 | 17.49 |
| 23 | 1 | 958.703 | 0.135 | 958.703 | 17.49 |
| 23 | 1 | 958.348 | 0.151 | 958.348 | 17.49 |
| 23 | 1 | 959.699 | 0.131 | 959.699 | 17.49 |
| 23 | 1 | 958.816 | 0.132 | 958.816 | 17.49 |
| 23 | 1 | 960.079 | 0.146 | 960.079 | 17.49 |
| 25 | 1 | 960.055 | 0.162 | 960.055 | 17.49 |
| 25 | 1 | 960.262 | 0.162 | 960.262 | 17.49 |
| 29 | 1 | 961.028 | 0.119 | 961.028 | 17.49 |
| 29 | 1 | 960.540 | 0.157 | 960.540 | 17.49 |

| 2008 - SETEMBRO | | | | | |
|---|---|---|---|---|---|
| D | L | SDB | ER | SDC | HL |
| 29 | 1 | 959.163 | 0.119 | 959.163 | 17.49 |
| 29 | 1 | 959.671 | 0.154 | 959.671 | 17.49 |
| 29 | 1 | 958.876 | 0.139 | 958.876 | 17.49 |
| 29 | 1 | 959.119 | 0.139 | 959.119 | 17.49 |
| 29 | 1 | 957.697 | 0.137 | 957.697 | 17.49 |
| 29 | 1 | 959.192 | 0.119 | 959.192 | 17.49 |
| 29 | 1 | 959.259 | 0.150 | 959.259 | 17.49 |

| 2008 - OUTUBRO | | | | | |
|---|---|---|---|---|---|
| D | L | SDB | ER | SDC | HL |
| 01 | 1 | 959.980 | 0.176 | 959.980 | 17.49 |
| 01 | 1 | 959.492 | 0.117 | 959.492 | 17.49 |
| 01 | 1 | 959.381 | 0.226 | 959.381 | 17.49 |
| 01 | 1 | 959.110 | 0.199 | 959.110 | 17.49 |
| 01 | 1 | 958.953 | 0.108 | 958.953 | 17.49 |
| 01 | 1 | 959.185 | 0.122 | 959.185 | 17.49 |
| 01 | 1 | 958.808 | 0.114 | 958.808 | 17.49 |
| 01 | 1 | 959.275 | 0.106 | 959.275 | 17.49 |
| 01 | 1 | 958.861 | 0.117 | 958.861 | 17.49 |
| 01 | 1 | 958.958 | 0.118 | 958.958 | 17.49 |
| 01 | 1 | 959.059 | 0.119 | 959.059 | 17.49 |
| 01 | 1 | 958.756 | 0.124 | 958.756 | 17.49 |
| 01 | 1 | 959.157 | 0.138 | 959.157 | 17.49 |
| 01 | 1 | 959.228 | 0.132 | 959.228 | 17.49 |
| 01 | 1 | 959.282 | 0.145 | 959.282 | 17.49 |
| 01 | 1 | 959.150 | 0.118 | 959.150 | 17.49 |
| 13 | 1 | 959.503 | 0.154 | 959.503 | 17.49 |
| 13 | 1 | 958.791 | 0.134 | 958.791 | 17.49 |
| 13 | 1 | 959.470 | 0.157 | 959.470 | 17.49 |
| 13 | 1 | 959.411 | 0.128 | 959.411 | 17.49 |
| 13 | 1 | 959.012 | 0.142 | 959.012 | 17.49 |
| 13 | 1 | 958.925 | 0.150 | 958.925 | 17.49 |
| 13 | 1 | 958.983 | 0.120 | 958.983 | 17.49 |
| 13 | 1 | 959.651 | 0.173 | 959.651 | 17.49 |
| 13 | 1 | 959.460 | 0.139 | 959.460 | 17.49 |
| 13 | 1 | 959.331 | 0.124 | 959.331 | 17.49 |
| 13 | 1 | 959.060 | 0.139 | 959.060 | 17.49 |
| 13 | 1 | 959.476 | 0.139 | 959.476 | 17.49 |
| 14 | 1 | 958.702 | 0.132 | 958.702 | 17.49 |
| 14 | 1 | 959.876 | 0.099 | 959.876 | 17.49 |
| 14 | 1 | 959.466 | 0.128 | 959.466 | 17.49 |
| 14 | 1 | 959.407 | 0.105 | 959.407 | 17.49 |
| 14 | 1 | 959.563 | 0.090 | 959.563 | 17.49 |
| 14 | 1 | 959.573 | 0.109 | 959.573 | 17.49 |
| 14 | 1 | 960.019 | 0.123 | 960.019 | 17.49 |
| 14 | 1 | 959.858 | 0.149 | 959.858 | 17.49 |
| 14 | 1 | 960.030 | 0.132 | 960.030 | 17.49 |
| 14 | 1 | 959.246 | 0.124 | 959.246 | 17.49 |
| 14 | 1 | 958.898 | 0.118 | 958.898 | 17.49 |
| 15 | 1 | 959.191 | 0.133 | 959.191 | 17.49 |
| 15 | 1 | 958.108 | 0.115 | 958.108 | 17.49 |
| 15 | 1 | 959.354 | 0.108 | 959.354 | 17.49 |
| 15 | 1 | 959.784 | 0.120 | 959.784 | 17.49 |
| 15 | 1 | 958.952 | 0.108 | 958.952 | 17.49 |
| 15 | 1 | 959.746 | 0.113 | 959.746 | 17.49 |
| 15 | 1 | 958.815 | 0.138 | 958.815 | 17.49 |
| 15 | 1 | 959.060 | 0.114 | 959.060 | 17.49 |
| 15 | 1 | 959.321 | 0.136 | 959.321 | 17.49 |
| 15 | 1 | 959.877 | 0.108 | 959.877 | 17.49 |
| 15 | 1 | 959.078 | 0.148 | 959.078 | 17.49 |
| 15 | 1 | 959.964 | 0.121 | 959.964 | 17.49 |
| 15 | 1 | 959.965 | 0.122 | 959.965 | 17.49 |
| 15 | 1 | 959.090 | 0.140 | 959.090 | 17.49 |
| 15 | 1 | 959.286 | 0.135 | 959.286 | 17.49 |
| 15 | 1 | 959.707 | 0.148 | 959.707 | 17.49 |
| 15 | 1 | 958.843 | 0.120 | 958.843 | 17.49 |
| 15 | 1 | 958.764 | 0.118 | 958.764 | 17.49 |
| 15 | 1 | 959.723 | 0.119 | 959.723 | 17.49 |
| 15 | 1 | 958.800 | 0.108 | 958.800 | 17.49 |
| 15 | 1 | 958.933 | 0.119 | 958.933 | 17.49 |

```
     2008 - OUTUBRO                                2008 - DEZEMBRO
D  L    SDB    ER     SDC    HL           D  L    SDB    ER     SDC    HL
15 1  958.853 0.144 958.853 17.49         02 1  958.559 0.141 958.559 17.49
15 1  958.982 0.124 958.982 17.49         02 1  959.283 0.105 959.283 17.49
15 1  959.709 0.146 959.709 17.49         02 1  959.858 0.112 959.858 17.49
15 1  959.396 0.136 959.396 17.49         04 1  960.200 0.138 960.200 17.49
16 1  960.316 0.104 960.316 17.49         04 1  960.006 0.107 960.006 17.49
16 1  959.341 0.133 959.341 17.49         04 1  959.823 0.122 959.823 17.49
16 1  960.265 0.139 960.265 17.49         04 1  959.212 0.116 959.212 17.49
16 1  959.521 0.116 959.521 17.49         04 1  959.910 0.123 959.910 17.49
16 1  959.769 0.135 959.769 17.49         04 1  959.909 0.114 959.909 17.49
16 1  959.208 0.130 959.208 17.49         04 1  958.999 0.136 958.999 17.49
                                          04 1  958.983 0.145 958.983 17.49
                                          04 1  959.112 0.138 959.112 17.49
     2008 - NOVEMBRO                      04 1  959.538 0.121 959.538 17.49
D  L    SDB    ER     SDC    HL           04 1  960.510 0.128 960.510 17.49
12 1  959.401 0.101 959.401 17.49         04 1  960.079 0.144 960.079 17.49
12 1  959.818 0.117 959.818 17.49         08 1  958.877 0.109 958.877 17.49
12 1  958.643 0.186 958.643 17.49         08 1  959.288 0.130 959.288 17.49
                                          08 1  959.389 0.111 959.389 17.49
                                          08 1  959.919 0.131 959.919 17.49
     2008 - DEZEMBRO                      08 1  959.848 0.162 959.848 17.49
D  L    SDB    ER     SDC    HL           08 1  960.231 0.121 960.231 17.49
01 1  959.994 0.137 959.994 17.49         08 1  959.338 0.122 959.338 17.49
01 1  959.251 0.164 959.251 17.49         08 1  959.495 0.113 959.495 17.49
01 1  959.682 0.162 959.682 17.49         08 1  959.841 0.115 959.841 17.49
01 1  959.797 0.122 959.797 17.49         08 1  959.568 0.151 959.568 17.49
01 1  961.263 0.109 961.263 17.49         08 1  959.012 0.130 959.012 17.49
01 1  960.513 0.131 960.513 17.49         08 1  959.580 0.156 959.580 17.49
01 1  959.745 0.096 959.745 17.49         08 1  957.809 0.131 957.809 17.49
01 1  959.691 0.097 959.691 17.49         08 1  958.623 0.144 958.623 17.49
01 1  959.527 0.111 959.527 17.49         08 1  960.783 0.111 960.783 17.49
01 1  959.365 0.112 959.365 17.49         08 1  959.917 0.128 959.917 17.49
01 1  959.485 0.121 959.485 17.49         08 1  959.835 0.122 959.835 17.49
01 1  959.495 0.136 959.495 17.49         08 1  960.347 0.109 960.347 17.49
01 1  959.847 0.169 959.847 17.49         08 1  959.536 0.116 959.536 17.49
01 1  960.094 0.134 960.094 17.49         08 1  960.697 0.175 960.697 17.49
01 1  959.965 0.148 959.965 17.49         08 1  959.580 0.123 959.580 17.49
01 1  960.079 0.173 960.079 17.49         08 1  959.751 0.126 959.751 17.49
01 1  959.682 0.101 959.682 17.49         08 1  960.081 0.121 960.081 17.49
01 1  958.965 0.126 958.965 17.49         08 1  959.598 0.118 959.598 17.49
01 1  959.288 0.127 959.288 17.49         08 1  958.933 0.134 958.933 17.49
01 1  959.509 0.132 959.509 17.49         09 1  958.996 0.138 958.996 17.49
01 1  960.094 0.141 960.094 17.49         09 1  958.986 0.110 958.986 17.49
01 1  959.217 0.131 959.217 17.49         09 1  959.416 0.094 959.416 17.49
01 1  960.603 0.161 960.603 17.49         09 1  959.127 0.137 959.127 17.49
01 1  959.557 0.156 959.557 17.49         09 1  959.720 0.100 959.720 17.49
02 1  959.081 0.105 959.081 17.49         09 1  959.088 0.116 959.088 17.49
02 1  959.553 0.092 959.553 17.49         09 1  959.417 0.150 959.417 17.49
02 1  960.016 0.095 960.016 17.49         09 1  958.536 0.170 958.536 17.49
02 1  959.451 0.126 959.451 17.49         09 1  958.905 0.121 958.905 17.49
02 1  960.107 0.108 960.107 17.49         09 1  960.045 0.146 960.045 17.49
02 1  959.766 0.122 959.766 17.49         09 1  960.295 0.131 960.295 17.49
02 1  959.829 0.098 959.829 17.49         09 1  960.373 0.097 960.373 17.49
02 1  959.837 0.099 959.837 17.49         09 1  960.020 0.119 960.020 17.49
02 1  959.754 0.135 959.754 17.49         09 1  959.706 0.113 959.706 17.49
02 1  959.366 0.120 959.366 17.49         09 1  959.386 0.110 959.386 17.49
02 1  959.699 0.130 959.699 17.49         09 1  959.114 0.166 959.114 17.49
02 1  959.214 0.132 959.214 17.49         09 1  959.917 0.133 959.917 17.49
02 1  958.335 0.130 958.335 17.49         09 1  959.461 0.114 959.461 17.49
02 1  960.176 0.180 960.176 17.49         09 1  960.256 0.124 960.256 17.49
02 1  960.921 0.100 960.921 17.49         09 1  959.783 0.097 959.783 17.49
02 1  959.524 0.111 959.524 17.49         09 1  959.674 0.123 959.674 17.49
02 1  959.068 0.127 959.068 17.49         09 1  959.259 0.134 959.259 17.49
02 1  958.857 0.103 958.857 17.49         09 1  959.686 0.129 959.686 17.49
02 1  959.316 0.110 959.316 17.49         09 1  958.783 0.116 958.783 17.49
02 1  959.478 0.133 959.478 17.49         09 1  959.673 0.119 959.673 17.49
02 1  959.253 0.114 959.253 17.49         10 1  958.817 0.080 958.817 17.49
02 1  960.083 0.129 960.083 17.49         10 1  959.803 0.116 959.803 17.49
02 1  959.885 0.117 959.885 17.49         10 1  959.644 0.130 959.644 17.49
02 1  958.804 0.146 958.804 17.49         10 1  960.225 0.116 960.225 17.49
02 1  959.620 0.113 959.620 17.49         10 1  959.412 0.123 959.412 17.49
02 1  959.334 0.128 959.334 17.49         10 1  959.938 0.118 959.938 17.49
```

| 2008 - DEZEMBRO | | | | |
|---|---|---|---|---|
| D | L | SDB | ER | SDC | HL |
| 10 | 1 | 960.244 | 0.132 | 960.244 | 17.49 |
| 10 | 1 | 960.088 | 0.147 | 960.088 | 17.49 |
| 10 | 1 | 961.286 | 0.147 | 961.286 | 17.49 |
| 10 | 1 | 959.716 | 0.155 | 959.716 | 17.49 |
| 10 | 1 | 959.726 | 0.156 | 959.726 | 17.49 |
| 10 | 1 | 959.119 | 0.121 | 959.119 | 17.49 |
| 10 | 1 | 960.062 | 0.141 | 960.062 | 17.49 |
| 10 | 1 | 959.084 | 0.138 | 959.084 | 17.49 |
| 10 | 1 | 959.266 | 0.152 | 959.266 | 17.49 |
| 29 | 1 | 960.924 | 0.122 | 960.924 | 17.49 |
| 29 | 1 | 960.238 | 0.166 | 960.238 | 17.49 |
| 29 | 1 | 959.578 | 0.140 | 959.578 | 17.49 |
| 29 | 1 | 959.809 | 0.127 | 959.809 | 17.49 |
| 29 | 1 | 959.467 | 0.118 | 959.467 | 17.49 |
| 29 | 1 | 960.046 | 0.158 | 960.046 | 17.49 |
| 29 | 1 | 960.238 | 0.128 | 960.238 | 17.49 |
| 29 | 1 | 959.556 | 0.175 | 959.556 | 17.49 |
| 29 | 1 | 959.359 | 0.126 | 959.359 | 17.49 |
| 29 | 1 | 959.066 | 0.107 | 959.066 | 17.49 |
| 29 | 1 | 959.225 | 0.125 | 959.225 | 17.49 |
| 29 | 1 | 958.033 | 0.146 | 958.033 | 17.49 |

| 2009 - JANEIRO | | | | |
|---|---|---|---|---|
| D | L | SDB | ER | SDC | HL |
| 12 | 1 | 959.358 | 0.104 | 959.358 | 17.49 |
| 12 | 1 | 959.219 | 0.109 | 959.219 | 17.49 |
| 12 | 1 | 959.204 | 0.104 | 959.204 | 17.49 |
| 12 | 1 | 959.878 | 0.127 | 959.878 | 17.49 |
| 12 | 1 | 959.328 | 0.114 | 959.328 | 17.49 |
| 12 | 1 | 959.719 | 0.125 | 959.719 | 17.49 |
| 12 | 1 | 959.362 | 0.127 | 959.362 | 17.49 |
| 12 | 1 | 959.005 | 0.157 | 959.005 | 17.49 |
| 12 | 1 | 959.660 | 0.145 | 959.660 | 17.49 |
| 12 | 1 | 959.881 | 0.130 | 959.881 | 17.49 |
| 12 | 1 | 960.237 | 0.149 | 960.237 | 17.49 |
| 12 | 1 | 959.600 | 0.139 | 959.600 | 17.49 |
| 12 | 1 | 960.509 | 0.154 | 960.509 | 17.49 |
| 12 | 1 | 960.099 | 0.140 | 960.099 | 17.49 |
| 12 | 1 | 959.447 | 0.158 | 959.447 | 17.49 |
| 12 | 1 | 959.543 | 0.143 | 959.543 | 17.49 |
| 12 | 1 | 958.510 | 0.131 | 958.510 | 17.49 |
| 12 | 1 | 958.602 | 0.120 | 958.602 | 17.49 |
| 12 | 1 | 958.083 | 0.136 | 958.083 | 17.49 |
| 12 | 1 | 960.064 | 0.123 | 960.064 | 17.49 |
| 12 | 1 | 958.333 | 0.143 | 958.333 | 17.49 |
| 12 | 1 | 958.409 | 0.118 | 958.409 | 17.49 |
| 12 | 1 | 958.668 | 0.150 | 958.668 | 17.49 |
| 13 | 1 | 959.712 | 0.075 | 959.712 | 17.49 |
| 13 | 1 | 959.422 | 0.111 | 959.422 | 17.49 |
| 13 | 1 | 958.865 | 0.094 | 958.865 | 17.49 |
| 13 | 1 | 959.717 | 0.092 | 959.717 | 17.49 |
| 13 | 1 | 959.747 | 0.080 | 959.747 | 17.49 |
| 13 | 1 | 959.364 | 0.117 | 959.364 | 17.49 |
| 13 | 1 | 959.004 | 0.116 | 959.004 | 17.49 |
| 13 | 1 | 959.105 | 0.124 | 959.105 | 17.49 |
| 13 | 1 | 959.187 | 0.096 | 959.187 | 17.49 |
| 13 | 1 | 960.093 | 0.131 | 960.093 | 17.49 |
| 13 | 1 | 959.644 | 0.099 | 959.644 | 17.49 |
| 13 | 1 | 959.660 | 0.113 | 959.660 | 17.49 |
| 13 | 1 | 959.503 | 0.121 | 959.503 | 17.49 |
| 13 | 1 | 959.484 | 0.122 | 959.484 | 17.49 |

| 2009 - AGOSTO | | | | |
|---|---|---|---|---|
| D | L | SDB | ER | SDC | HL |
| 17 | 1 | 959.952 | 0.122 | 959.952 | 17.49 |
| 17 | 1 | 959.481 | 0.097 | 959.481 | 17.49 |
| 17 | 1 | 959.109 | 0.103 | 959.109 | 17.49 |
| 17 | 1 | 959.968 | 0.142 | 959.968 | 17.49 |
| 17 | 1 | 960.288 | 0.122 | 960.288 | 17.49 |

| 2009 - AGOSTO | | | | |
|---|---|---|---|---|
| D | L | SDB | ER | SDC | HL |
| 17 | 1 | 959.788 | 0.131 | 959.788 | 17.49 |
| 17 | 1 | 959.563 | 0.153 | 959.563 | 17.49 |
| 17 | 1 | 959.963 | 0.125 | 959.963 | 17.49 |
| 17 | 1 | 959.497 | 0.157 | 959.497 | 17.49 |
| 17 | 1 | 959.487 | 0.130 | 959.487 | 17.49 |
| 17 | 1 | 960.275 | 0.175 | 960.275 | 17.49 |
| 17 | 1 | 958.858 | 0.242 | 958.858 | 17.49 |
| 18 | 1 | 957.828 | 0.107 | 957.828 | 17.49 |
| 18 | 1 | 959.700 | 0.097 | 959.700 | 17.49 |
| 18 | 1 | 960.000 | 0.094 | 960.000 | 17.49 |
| 18 | 1 | 959.894 | 0.120 | 959.894 | 17.49 |
| 18 | 1 | 959.073 | 0.101 | 959.073 | 17.49 |

| 2009 - OUTUBRO | | | | |
|---|---|---|---|---|
| D | L | SDB | ER | SDC | HL |
| 14 | 1 | 958.701 | 0.144 | 958.701 | 17.49 |
| 14 | 1 | 958.876 | 0.184 | 958.876 | 17.49 |
| 14 | 1 | 958.511 | 0.161 | 958.511 | 17.49 |
| 14 | 1 | 959.294 | 0.158 | 959.294 | 17.49 |
| 14 | 1 | 958.902 | 0.151 | 958.902 | 17.49 |
| 14 | 1 | 959.605 | 0.152 | 959.605 | 17.49 |
| 14 | 1 | 958.950 | 0.157 | 958.950 | 17.49 |
| 14 | 1 | 959.627 | 0.144 | 959.627 | 17.49 |
| 14 | 1 | 959.208 | 0.227 | 959.208 | 17.49 |
| 14 | 1 | 959.252 | 0.216 | 959.252 | 17.49 |
| 14 | 1 | 960.165 | 0.180 | 960.165 | 17.49 |
| 14 | 1 | 958.772 | 0.151 | 958.772 | 17.49 |
| 14 | 1 | 959.305 | 0.164 | 959.305 | 17.49 |

| 2009 - NOVEMBRO | | | | |
|---|---|---|---|---|
| D | L | SDB | ER | SDC | HL |
| 03 | 1 | 959.480 | 0.213 | 959.480 | 17.49 |
| 03 | 1 | 959.903 | 0.147 | 959.903 | 17.49 |
| 03 | 1 | 959.652 | 0.125 | 959.652 | 17.49 |
| 03 | 1 | 959.314 | 0.118 | 959.314 | 17.49 |
| 03 | 1 | 959.830 | 0.121 | 959.830 | 17.49 |
| 03 | 1 | 959.466 | 0.132 | 959.466 | 17.49 |
| 03 | 1 | 959.480 | 0.124 | 959.480 | 17.49 |
| 03 | 1 | 961.025 | 0.125 | 961.025 | 17.49 |
| 03 | 1 | 959.577 | 0.135 | 959.577 | 17.49 |
| 05 | 1 | 958.910 | 0.107 | 958.910 | 17.49 |
| 05 | 1 | 959.802 | 0.102 | 959.802 | 17.49 |
| 05 | 1 | 959.444 | 0.086 | 959.444 | 17.49 |
| 05 | 1 | 960.119 | 0.116 | 960.119 | 17.49 |
| 05 | 1 | 960.721 | 0.097 | 960.721 | 17.49 |
| 05 | 1 | 960.810 | 0.101 | 960.810 | 17.49 |
| 05 | 1 | 959.831 | 0.134 | 959.831 | 17.49 |
| 05 | 1 | 959.270 | 0.131 | 959.270 | 17.49 |
| 05 | 1 | 959.202 | 0.115 | 959.202 | 17.49 |
| 05 | 1 | 959.872 | 0.132 | 959.872 | 17.49 |
| 05 | 1 | 959.348 | 0.135 | 959.348 | 17.49 |
| 05 | 1 | 959.625 | 0.126 | 959.625 | 17.49 |
| 05 | 1 | 959.481 | 0.128 | 959.481 | 17.49 |
| 05 | 1 | 960.093 | 0.135 | 960.093 | 17.49 |
| 05 | 1 | 959.306 | 0.155 | 959.306 | 17.49 |
| 05 | 1 | 959.060 | 0.187 | 959.060 | 17.49 |
| 06 | 1 | 957.759 | 0.092 | 957.759 | 17.49 |
| 06 | 1 | 958.873 | 0.083 | 958.873 | 17.49 |
| 06 | 1 | 958.712 | 0.110 | 958.712 | 17.49 |
| 06 | 1 | 958.823 | 0.098 | 958.823 | 17.49 |
| 06 | 1 | 959.567 | 0.090 | 959.567 | 17.49 |
| 06 | 1 | 959.154 | 0.103 | 959.154 | 17.49 |
| 06 | 1 | 959.016 | 0.112 | 959.016 | 17.49 |
| 06 | 1 | 960.242 | 0.109 | 960.242 | 17.49 |
| 06 | 1 | 959.136 | 0.115 | 959.136 | 17.49 |
| 06 | 1 | 959.054 | 0.135 | 959.054 | 17.49 |
| 06 | 1 | 958.574 | 0.122 | 958.574 | 17.49 |
| 06 | 1 | 958.986 | 0.151 | 958.986 | 17.49 |
| 06 | 1 | 959.154 | 0.139 | 959.154 | 17.49 |